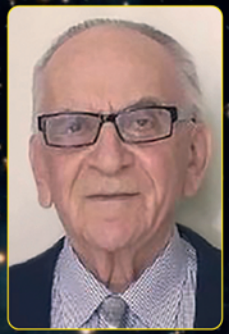

Professor
R.E. Gershberg
(Russia)
Crimean Astrophysical Observatory, Russian Academy of Sciences

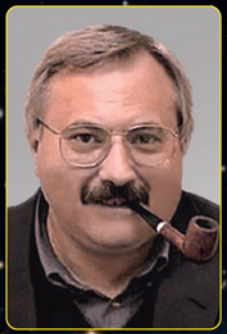

Professor
N.I. Kleeorin
(Israel)
Ben-Gurion University in the Negev, Beer-Sheva

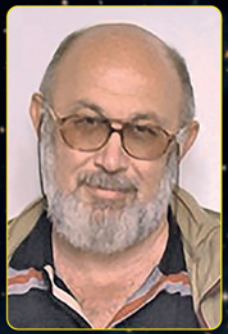

Professor
L.A. Pustilnik
(Israel)
Tel Aviv University, Ariel University, Israeli Space Agency

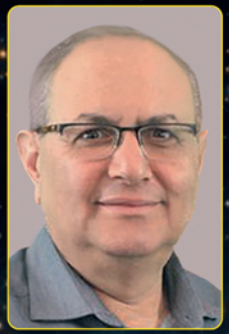

Professor
V.S. Airapetian
(USA)
PhD Senior Astrophysicist
American University, DC

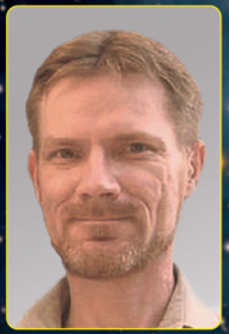

Astrophysicist
A.A. Shlyapnikov
(Ukraine)

In the monograph, the authors systematize and generalize the results of studying solar-type activity that is characteristic of a significant part of mid- and low-mass stars of the Galaxy, outline the characteristics of such stars in the quiescent state, during the sporadic flares and variations of magnetic activity over the course of stellar evolution. The observational data taken over the whole range of the electromagnetic spectrum from decametric radio waves to X-rays are described in detail. The current models of stellar flares and stellar dynamo models are considered in two theoretical chapters. The last part of the book describes the impact of stellar activity of mid- and low-mass stars on exoplanetary environments. The Catalog of Stars with Solar-Type Activity is provided in the supplementary section.

The book is intended for researchers involved in studying the physics of stars and the Sun, graduate students, and students specialized in heliophysics, astrophysics, and in the field of space physics.

PHYSICS OF MID- AND LOW-MASS STARS WITH SOLAR-TYPE ACTIVITY AND THEIR IMPACT ON EXOPLANETARY ENVIRONMENTS

R.E. Gershberg, N.I. Kleeorin,
L.A. Pustilnik, V.S. Airapetian,
A.A. Shlyapnikov

# PHYSICS OF MID- AND LOW-MASS STARS WITH SOLAR-TYPE ACTIVITY AND THEIR IMPACT ON EXOPLANETARY ENVIRONMENTS

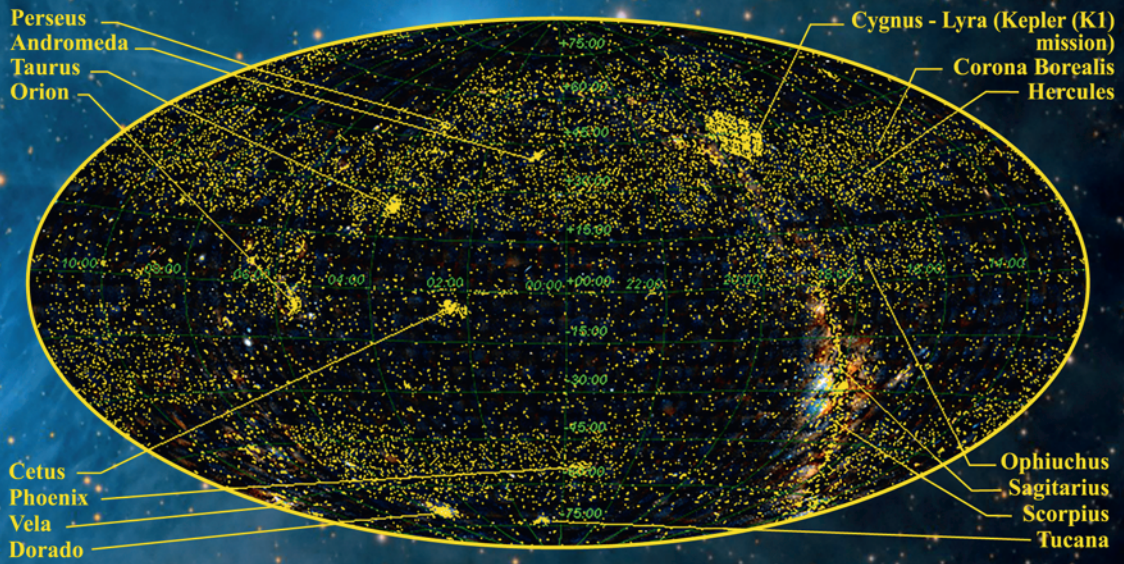

Distribution of 314618 stars with solar-type activity over the celestial sphere

**R.E. GERSHBERG**
**N.I. KLEEORIN**
**L.A. PUSTILNIK**
**V.S. AIRAPETIAN**
**A.A. SHLYAPNIKOV**

# PHYSICS OF MID- AND LOW-MASS STARS WITH SOLAR-TYPE ACTIVITY AND THEIR IMPACT ON EXOPLANETARY ENVIRONMENTS

In the monograph, the authors systematize and generalize the results of studying solar-type activity that is characteristic of a significant part of mid- and low-mass stars of the Galaxy, outline the characteristics of such stars in the quiescent state, during the sporadic flares and variations of magnetic activity over the course of stellar evolution. The observational data taken over the whole range of the electromagnetic spectrum from decametric radio waves to X-rays are described in detail. The current models of stellar flares and stellar dynamo models are considered in two theoretical chapters. The last part of the book describes the impact of stellar activity of mid- and low-mass stars on exoplanetary environments. The Catalog of Stars with Solar-Type Activity is provided in the supplementary section.

The book is intended for researchers involved in studying the physics of stars and the Sun, graduate students, and students specialized in heliophysics, astrophysics, and in the field of space physics.

The cover shows the distribution of 314618 stars with solar-type activity over the celestial sphere.

*In memory*
*of Solomon Borisovich Pikelner,*
*an eminent astrophysicist and marvelous person,*
*dedicated to the 100th anniversary of his birth*

# Contents

# Preface to the English Edition

This book is a translated version of the Russian monograph *Physics of Mid- and Low-Mass Stars with Solar-Type Activity* (Moscow, FIZMATLIT: 2021. 768 p.). It was complemented with Part 5 “Impact of stellar activity on exoplanetary environments”, including signatures on exoplanetary habitability, and with about seventy insertions in the text and corresponding references with some new results published after October 2020.

N.I. Kleeorin contributed to Chapter 4.3, L.A. Pistilnik wrote Chapter 2.6, V.S. Airapetian contributed to Part 5, A.A. Shlyapnikov compiled the extensive Catalog of Stars with Solar-Type Activity and provided its description in the Supplement section. The Catalog is hosted on the server of the Crimean Astrophysical observatory; it has regularly being complemented and by July 1, 2022 contained data on 314618 objects. The rest text of the monograph was written by R.E. Gershberg.

We thank Maria Katsova for providing information about recent publications on the monograph subject, to Marina Smirnova, Marina Butuzova, and Tatyana Sokolova for the help in preparing this edition.

December 2023 Authors

# Preface to the Original Russian Book

In 1958, the Commission on Variable Stars of the International Astronomical Union distinguished cool dwarf stars with sporadic flares of high-amplitude brightness into a separate group that was referred to as UV Cet-type variables. The discovery of high activity of these coolest types among known then stars was intriguing, and the hypotheses regarding their nature involved “new physics”. But over the past 60 years, the situation has dramatically changed. Instead of visual and photographic observations of the XXth century, all power of the modern all-wavelength astronomy has come: flare stars were studied within the whole optical range, in the near and far ultraviolet, in X-ray, infrared and radio range, whereas the largest specialized telescopes were involved, and a number of campaigns for simultaneous observations of such stars in different spectral regions were carried out. This resulted in identifying physical signatures and properties of activity of flare red dwarfs and the Sun, but on the UV Cet stars for a variety of reasons it is more pronounced. This activity is based on stellar magnetism, and its place in stellar evolution is ascertained. The basic conclusion that can be drawn from these studies is that the considered activity is the most widespread in the realm of stars and is typical of most stars at a certain evolutionary stage.

In this book, we describe in detail the history of observational, empirical, and theoretical studies of the considered stellar activity, hereinafter referred to as solar-type activity. Particular attention is paid to the extensive optical studies, ultraviolet observations with space telescopes, including the International Ultraviolet Explorer (IUE) and the Hubble Space Telescope (HST), observations with the Einstein, Chandra, and XMM-Newton X-ray observatories, as well as extensive photometric studies. The latter mainly involves long-term panorama photometry with the Kepler Space Telescope of simultaneously over 100 thousand stars that revealed superflares with energy hundreds and thousands of times more powerful than the largest flares observed on the Sun; observations with the space astrometric interferometer for astrophysics GAIA yielding the two-color photometry of about 1.7 billion objects. The ground-based spectral Sloan project has strikingly increased the volume of the Galaxy space where red dwarfs became available for studying, which led to important stellar astronomical and astrophysical discoveries. The book includes the results obtained with ROSAT, ASCA, WISE, ULTRACAM, SWIFT, SPITZER, EVRYSCOPE, TESS, and some other missions. Of particular importance is the discovery made with the 6-meter telescope of the Special Astrophysical Observatory RAS: throughout a flare of the red dwarf star, UV Cet, the sub-second flares with polarization of emission up to 35–40% were recorded, i.e., the conspicuous evidence for the synchrotron emission was acquired, whose searches in flares were vainly performed over the course of several decades.

In Part 1 of the book, we outline the properties of stars with solar-type activity in the quiescent state, in Part 2 — flares, in Part 3 — long-term variations of the considered stellar activity, in Part 4 — the observed and theoretical problems of stellar magnetism. In Supplement, we provide a description of the Catalog of Stars with Solar-Type Activity; the Catalog is regularly being complemented.

We are deeply thankful to Dr. M.M. Katsova for useful discussions and operative information on new publications concerning the monograph subject, to Drs. K.M. Kuzanyan and I.V. Rogachevsky for the discussion of issues on stellar dynamo, and to Olga Saletskaya and Marina Smirnova for the all-round help in preparing the manuscript.

October 2020 Authors

# Introduction

## A Brief History of Studying Flare Red Dwarfs

In 1924, Ejnar Hertzsprung noticed that one of the faint stars in Carina was brighter by $2^m$ on one of the photographs taken on the night of January 29. The rate of brightness increase suggested that this star did not belong to novae or oscillating RR Lyrae-type stars. Thus, Hertzsprung concluded that the effect could be produced by the fall of an asteroid on star. Apparently, it was the first recorded flare on stars of this type.

Studying the spectra of faint stars in Orion in December 1938, Wachmann (1939) found an unusual variable star with an abnormal spectrum. The spectrum was obtained using an objective prism, the spectra being discretely broadened in one direction at 0.02 mm every 6 min during an hour. During the first third of the exposure the stellar spectrum resembled that of a nucleus of a planetary nebula of the WR type: on a background of continuous radiation of the type of continuum of B or A stars one could see strong hydrogen emission lines $H_\gamma$, $H_\delta$, $H_\varepsilon$, and $H_\zeta$. Then, the brightness of the object decreased by at least one and a half stellar magnitudes, and in the band corresponding to the other part of the exposure mainly emission lines became visible. At the same time, the spectrum of the adjacent star displayed uniform darkening throughout its width. In the images obtained a month later, the star had a regular K spectrum without emission. Apparently, Wachmann was the first to record the spectrum of a flare on a UV Cet-type star.

In 1940, while examining the parallaxes of faint stars van Maanen (1940) noticed that the brightness of the M6e star Lalande 21258 (= WX UMa) was about $16^m$ on more than 20 plates, but in two images obtained on 11 May 1939 with an interval of about half an hour it was $14.2^m$ and $14.5^m$ (Fig. 1). Several years later, van Maanen (1945) detected a similar phenomenon in measuring the parallax of Ross 882 (= YZ CMi). Reporting this fact, he noted that both variables had low luminosity, belonged to a late spectral class, and should be objects of the same type.

In 1947, Luyten noted a very large proper motion, 3.37 arcsec per year, of the $14^m$ star L 726-8 and invited astronomers from several observatories to study this object. Carpenter found that the star was very red and determined its parallax. Page and Struve obtained spectrograms of the star, which appeared to be an M6 star with the signatures of hydrogen and calcium emission lines. Joy and Humason observed L 726-8 at the 100″ reflector and established its binarity. They estimated the difference of brightness ($\sim 0.5^m$) and angular separation ($\sim$ 1.5″). Van Biesbroeck and van den Bos carried out micrometric observations of the system. Determining the parallax within this cooperative study, on 7 December 1948 Carpenter obtained a plate on which one of the five images of the star was much brighter than the other four. Upon studying this image, Hughes and Luyten suggested that there was a strong flare on the weak component of the system: over 3 min its brightness increased by more than 12 times. Having collected preliminary communications of observers, Luyten (1949a) found that L 726-8 was a binary system with the smallest mass components. He estimated the energy release in the flare as $4 \cdot 10^{31}$ erg and noted its explosive character. He remembered similar observations by van Maanen. Then, Joy and Humason (1949) reported additional important data obtained at the 100″ reflector. In August–September 1948, they obtained spectrograms for each component of the system. In all spectrograms, except for the one of 25 September 1948, the spectrum of the M dwarf had very strong emission lines of hydrogen and calcium. The spectrum of 25 September 1948 strongly differed from the others (Fig. 2): absorption lines and bands were almost filled in by strong continuous radiation, which was the most explicit in the range of wavelengths shorter than $H_\delta$, bright hydrogen lines were strengthened as compared to calcium lines, also the lines $\lambda$ 4026 Å and $\lambda$ 4471 Å of neutral helium and $\lambda$ 4686 Å of ionized helium were seen in the emission, but there were no forbidden lines. Joy and Humason were sure that on 25 September 1948 they observed a similar phenomenon to that recorded by Carpenter in the direct image of 7 December 1948. But unlike Luyten, who assumed that the flare was due to the occurrence of emission lines in the weak component of the system, Joy and Humason concluded that the increased brightness was primarily caused by continuous radiation. Later, the weak component of the system L 726-8 was named as UV Cet.

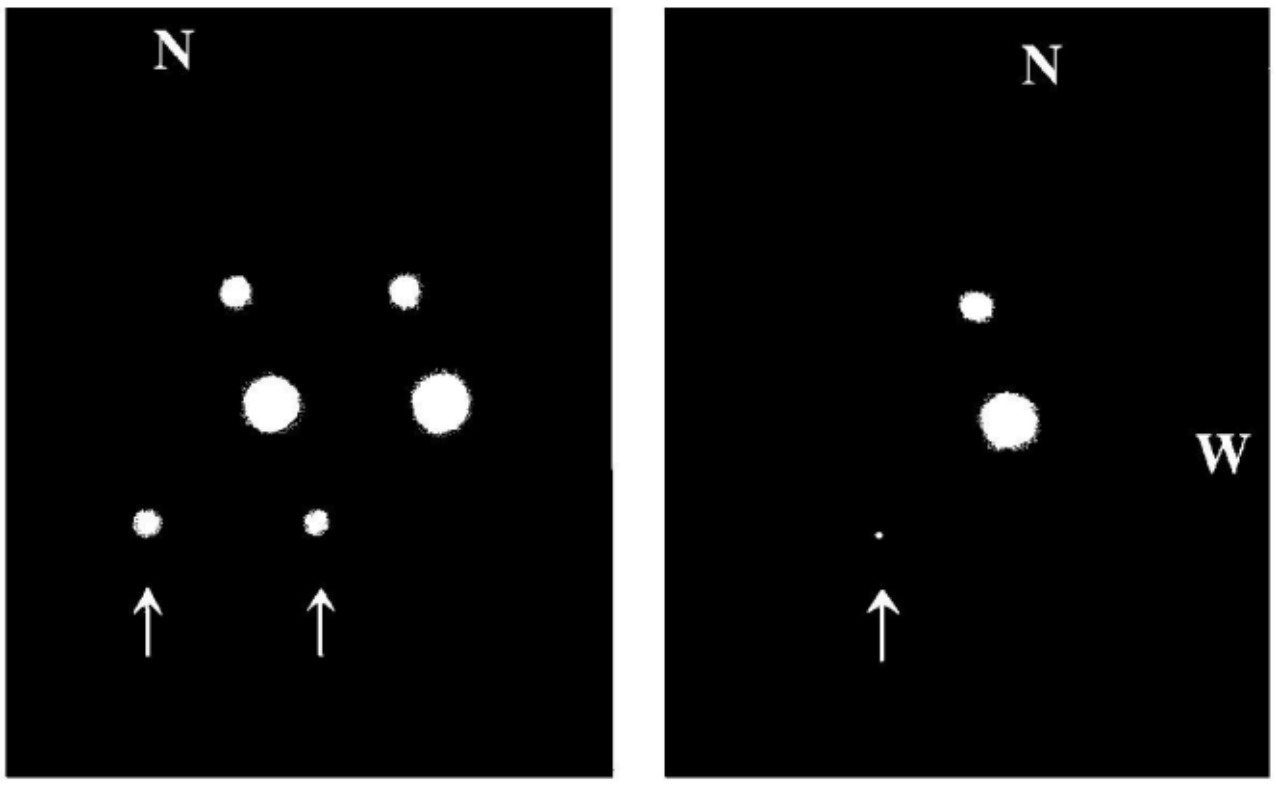

Fig. 1. Photos of WX UMa obtained by van Maanen (1940): stellar brightness of $14.2^m$ and $14.5^m$ (*left*) and normal brightness of about $16^m$ (*right*)

Cooperative research of UV Cet by American observers stimulated a thorough examination of dMe variable stars. The next decade was a period of fast accumulation of data. Over these years, the first photoelectric records of the light curves of flares were obtained (Gordon and Kron, 1949; Liller, 1952; Roques, 1953, 1954), numerous collections of negatives obtained in the course of long-term monitoring at different observatories were investigated (Luyten, 1949b; Shapley, 1951, 1954; van de Kamp and Lippincott, 1951; Lippincott, 1952, 1953; Hoffleit, 1952; Luyten and Hoffleit, 1954; Gaposhkin, 1955; Petit and Weber, 1956), series of visual observations were performed, and the first attempts to determine statistical characteristics of flare activity were undertaken (Petit, 1955, 1957; Oskanian, 1953, 1964), slit spectrograms of two stellar flares of the considered type were obtained (Herbig, 1956; Joy, 1958), the data on sporadic changes of spectra were collected for some red dwarf stars that suggested affiliation of these objects to flare stars (Joy, 1960; Thackeray, 1950; Popper, 1953; Münch and Münch, 1955; Bidelman, 1954). It should be emphasized that these studies were not focused on the identification of the nature of flares on the UV Cet-type stars. They mainly dealt with the description of the objects that were definitely or presumably attributed to this type or with the examination of general characteristics of this class of stellar variability and the place of UV Cet-type stars among other eruptive variables of late spectral classes. Numerous publications by Petit (1954, 1955, 1957, 1958, 1959, 1961), who initiated an international cooperation of observers of variable stars aimed at studying flare objects, made an important contribution to this field. Results of this decade of "initial accumulation" of data on the activity of red dwarfs were thoroughly described in the author's monograph (Gershberg, 1970a).

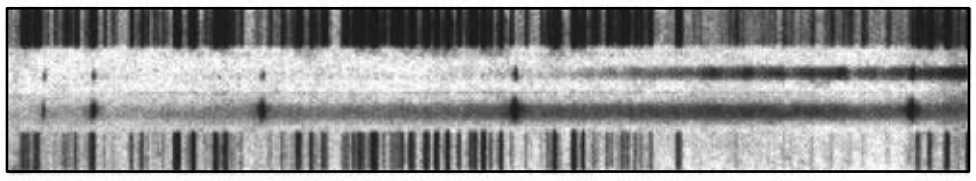

Fig. 2. Spectra of UV Cet obtained by Joy (1960): stellar spectrum in the normal state (*top*) and during the flare of 25 September 1948 (*bottom*)

In 1958, at the X General Assembly of the International Astronomical Union UV Cet-type stars were ranked as a special type of eruptive variables in the general classification of variable stars. They were defined as "dMe stars subjected to flare with the amplitude of $1^m$ to $6^m$. Maximum brightness is attained in seconds or several dozen seconds after the onset of the flare, the star returns to its normal brightness after several minutes or dozens of minutes. A typical representative — UV Cet." This purely photometric definition is reproduced in all editions of the General Catalog of Variable Stars.

Further advances in studying flare red dwarfs were to a great extent due to the development of new astrophysical methods and updating of traditional methods, which treated transience and irregularity of stellar flares as essential properties of the objects. Thus, since the 1960s photoelectric monitoring has become the basic source of data on photometric properties of flare stars: light curves of flares, their energy and time distribution, the properties of the quiescent state beyond flares. The photographic method of multiple exposures has become the basic way of detection of flare stars in the nearest stellar clusters and made it possible to estimate the prevalence of such phenomena in the stellar realm. At the same time, the first successful observations of stellar flares in the radio range were carried out at the Jodrell Bank Observatory. In the mid 1960s, the first high time resolution slit spectra of flares on UV Cet-type stars were recorded with the Shajn reflector in the Crimea and the Struve telescope in Texas. The analysis of these spectra together with the available data on radio emission of flares and features of chromospheres and photospheres of such stars led to a conclusion on the physical identity between the activity of flare stars and the Sun. Started in the 1970s, extraterrestrial studies of flare stars were advanced in the 1980s by the IUE satellite that recorded ultraviolet spectra of the chromosphere and the transition zone to the corona in quiescent and active states. Using the Einstein Observatory, X-ray emission of coronae of many flare stars was discovered. Finally, by the mid 1980s, the Very Large Antenna in New Mexico (USA) recorded microwave radio emission of flares and quiet coronae of flare stars. X-ray emission of many UV Cet-type stars was recorded by EXOSAT. In the early 1990s, the EUVE satellite provided the data on the far ultraviolet of stellar flares. However, the results of these and later studies from the X-ray telescopes ROSAT, BeppoSAX, ASCA, Chandra, XMM-Newton and from the Hubble Space Telescope (HST) and Far Ultraviolet Spectroscopy Explorer (FUSE) should not be considered as a history. They are included into the database of modern multiwavelength astrophysical observations of flare stars; they will be thoroughly considered in the following chapters and complemented with results acquired with the technique of the XXIst century.

The data on flare stars will be theoretically interpreted in the context of general notions on the activity of lower main-sequence stars eventually related to stellar magnetism. The current level of these ideas is still rather far from an ideal strict deductive theory, and observational and empirical studies are still governing the research of these variable stars. This fact determines the structure of the book: yet the list of experimental confirmations of a strict and complete physical theory cannot be compiled, but we can already deviate from the structure of two previous brief monographs devoted to this subject (Gershberg, 1970a, 1978), in which we consistently considered the results of various observational methods and on this basis developed and substantiated a general concept of activity of UV Cet-type stars. Here, we follow the accepted general concept of physical identity of the activity of the Sun and the lower main-sequence stars of middle and late spectral types to analyze and discuss various phenomena and structures on such stars using the whole body of relevant observational data.

The term “flare UV Cet-type stars” appeared after the detection of transient flares on red M dwarf stars. Today, it is clear that the low luminosity of the stars creates favorable conditions for the detection of the UV Cet-type activity but does not favor the occurrence of this activity, which, being physically identical to the solar activity, takes place on dwarf stars in a wide spectral range from G to M stars. Transformation of the phenomenological term designating the observed phenomenon into an astrophysical term reflecting the physical essence of the process evidently illustrates the progress of astronomical knowledge. In this book, the term “flare UV Cet-type stars” (as well as reduced variants “flare stars” and “UV Cet-type stars”) will be used for physical variables of the lower main sequence displaying solar-type activity. Finally, the same meaning will be put into the traditional term “flare red dwarfs”, though, strictly speaking, “flare UV Cet-type stars” also include less numerous orange dwarfs and even less frequent yellow G dwarfs, similar to our Sun.

Manifestations of the activity of lower main-sequence stars, as on the Sun, are rather numerous and diverse. Certainly, first of all these are the “principal” local nonstationary phenomena, sporadic flares involving all layers of the stellar atmosphere. Other manifestations are cool spots observed on the stellar surface and the variability of large-scale structures of stellar atmospheres, chromospheres, and coronae. Since there is experimental evidence of structural arrangement of chromospheres, while the observations of coronae are represented within the loop model, the upper layers of stellar atmospheres should also be considered in the context of stellar activity. Finally, the data on long-term cycles of stellar activity, similar to the 11-year solar cycle, are accumulated.

The physical identity of the ultimate cause of all observed phenomena of stellar activity of the considered type is doubtless. But observations reveal regular differences of average spectral types of dwarf stars that are characterized by different manifestations of this activity: flares and quiescent chromospheres of the coolest M stars are the most thoroughly investigated, early M and late K stars prevail among spotted dwarfs, long-term activity cycles are found mainly for early K and late G dwarfs. This regularity reflects observational selection: flares and chromospheric emission are easier identified on the background of weak photospheric radiation of the coolest dwarf stars. To detect spots, i.e., to record slight distortions of the light curve of a star, the level of its regular brightness should be determined with high accuracy, which requires a brighter photosphere. Radiation of the photosphere will hide chaotic brightness bursts caused by flares. Finally, long-term activity cycles are found mainly in high-resolution spectrometric observations, which are applicable only to even brighter stars. The X-ray emission of coronae, for which photospheric radiation is not a hindrance, has been detected from early F to late M dwarfs. The question on the degree to which observational selection masks the real dependence of the total level of various manifestations of stellar activity on their luminosity, mass, and age can be solved upon special analysis of the observational data. The most illustrative in this relation is the evolution of notions on the spectral content of carriers of the considered stellar variability. As mentioned, it was defined as being typical of the coolest M dwarfs, and the recent space observations have traced it toward A stars.

Thus, we propose a wider astrophysical definition of UV Cet-type stars as compared to the above purely photometric definition: UV Cet-type stars are lower main-sequence objects that display the phenomena typical of solar activity.

# Part 1

# Flare Stars in the Quiescent State

## 1.1. General Characteristics of Flare Stars

In the first half of the past century, flare stars of the UV Cet type were known only in the solar vicinity and in several closest open clusters. Being absolutely weak or very weak objects, as a rule, these stars could be studied in detail at distances of less than 20–30 parsecs. However, they represent a significant part of the stellar population: the spatial density of flare stars in the solar vicinity is 0.056 stars/$pc^3$, whereas the total density of the stars is only approximately twice as high (Shakhovskaya, 1995). The rapid decrease in the density of flare stars with increasing distance from the Sun is due to the observational selection of low-luminosity objects. This statement was confirmed by the special statistical study of Mirzoyan et al. (1988), who disproved the alternative statement on the existence of a real clustering of flare stars in the solar vicinity. Thus, the known UV Cet-type stars are a random sample of the stellar population of the Galaxy determined by the position of the Sun within our stellar system. Only recently, within the Sloan Digital Sky Survey, this constraint has been overcome, and now such variable stars are studied up to distances of hundreds of parsec, see below.

The first lists of such objects appeared in the middle of the past century contained 2-3 dozen stars. At the end of the century, Hawley et al. (1996) performed a spectral classification of about 2000 M dwarfs close to the Sun and found that 105 of them were M0–M3 emission dwarfs and 208 were M4–M8 emission stars. Taking into account this and other results of those years in the Crimea, the GKL99 catalog was compiled (Gershberg et al., 1999), containing 462 flare UV Cet-type stars and related objects in the vicinity of the Sun. The first row of Table 1 lists the distribution of objects of this Catalog according to spectral types. The recently published GTSh10 catalog (Gershberg, Terebizh, Shlyapnikov, 2011) contains data on 5535 stars with different manifestations of solar-type activity. The second row of Table 1 lists data on spectral types of flare stars in this catalog. The last row is filled according to the data from Supplement.

Table 1. The number of active stars in different catalogs

| Spectra | F | G0 – G9 | K0 – K3 | K4 – K8 | M0 – M3 | M4 – M8 |
|---|---|---|---|---|---|---|
| UV Cet-type stars in the GKL99 catalog | | 10 | 19 | 25 | 146 | 212 |
| Stars with solar-type activity in the GTSh10 catalog | 10 | 13 | 23 | 66 | 224 | 230 |
| Stars with solar-type activity in the CSSTA-3 catalog (see Supplement) | 1354 | 99785 | 109810 | 44018 | 39658 | 19993 |

Petit (1961), Joy and Abt (1974) had already discovered a fast increase of the fraction of emission objects while proceeding from early M dwarfs to late dwarfs. Thus, Joy and Abt concluded that all dwarfs later than M5.5 belonged to the emission type. But Giampapa (1983) discovered that nonemission stars prevailed among M6 and later dwarfs. A more detailed consideration of the problem revealed that late Me dwarfs were young objects with low spatial velocities, while late M stars were mainly objects with kinematic characteristics inherent in the stars of the old disk and halo (Giampapa and Liebert, 1986; Reid et al., 1995). According to Shakhovskaya (1995), the fraction of flare stars among dwarfs of the appropriate spectral types increased from 3% for early G to 30% for late M stars. According to Hawley et al. (1996), the fraction of emission objects among K6 was 1% and for M0–M3 it was close to 10%. The percentage monotonically increased up to 60% for M6 and then decreased; emission was observed only in 13 of 32 dwarfs of later subclasses. Such a complicated dependence is due to the combination of observational selection — flares and the strong chromospheres are more easily detected on spectrally later stars — with probable longer duration of the emission phase in lower-mass stars (Herbst and Miller, 1989; Hawley et al., 1996) and real decreasing activity of the coolest M stars.

Herbig (1956) already found a flare on the low-mass star VB 10 that was orders of magnitude stronger than the strongest solar flares. Later, Linsky et al. (1995) recorded an ultraviolet flare on this star. Then, Fleming et al. (2000) recorded an X-ray flare on it. The variable $H_\alpha$ emission is suspected on the very cool and rapidly rotating dwarf BRI 0021–0214 classified as > M9.5 (Tinney et al., 1997), while on the dM9.5e star 2MASSW J0149090+295613 a flare was recorded with an $H_\alpha$ amplitude of equivalent width near 30 (Liebert et al., 1999).

According to Tinney (1995), as to the latest M dwarfs, the Catalog of the Nearest Stars CNS3 (Gliese and Jahreiss, 1991) is essentially incomplete. The discovered spottedness of very low-mass stars (Terndrup et al., 1999; Krishnamurthi et al., 2001b) and X-ray flares on LHS 2065 (Schmitt and Liefke, 2002, 2004) and on LP 944-20 (Rutledge et al., 2000) evidence that this activity takes place on stars till the end of the main sequence and on brown dwarfs. Gizis (1998) discovered two M subdwarfs, Gl 781 A and Gl 455, with $H_\alpha$ emission: both were the components of binary systems and the properties of their chromospheres and coronae were similar to those of dMe dwarfs with solar metallicity. This is confirmed by the existence of chromospheres on components of the multiple system LHS 1070 (Leinert et al., 2000) and a powerful flare of the component B of this system with an amplitude of $\Delta B > \sim 8.2^m$ throughout which, following the estimate of Almeida et al. (2011), the total energy was about $2{\cdot}10^{33}$ erg, and the magnetic field strength reached 5.5 kG. Two other M6.5 active dwarfs were found by Gizis et al. (2000a) in the near multiple systems. Gizis et al. (2000b) systematically studied the activity of the weakest main-sequence stars. Having examined 60 M7–L dwarfs, they found that near M7 the occurrence rate of $H_\alpha$ in the emission reached 100%, then it decreased to 60% for L0 and to 8% for L4 dwarfs, but the luminosity ratio $L_{H\alpha}/L_{bol} = R_{H\alpha}$ for these ultracool stars did not reach the values characteristic of earlier M dwarfs, and the ratio started decreasing for M6 and continued for late L dwarfs. From the $H_\alpha$ variations in the considered sample Gizis et al. (2000b) concluded that flare activity was common for M7–M9.5 dwarfs and up to half of their fluxes in $H_\alpha$ could be due to flares.

In the context of the mentioned L dwarfs, it is worth noting that since the beginning of the new century, due to the advances of IR and space technique, there has been a significant step forward in studying the coolest stars — late M, L, and then T dwarfs, which are closely adjacent to planets.

In the review of Basri and Mohanty (2003), the authors listed numerous publications containing measurements of $H_\alpha$ emission carried out on the border of centuries and found that the ratio $L_{H\alpha}/L_{bol}$ decreased after the effective temperature 3200 K, abruptly dropped after 2400 K, and temperature has become a more important parameter for the presence of emission than rotation and Rossby number.

Barnard's star (GJ 699) is a M4 object with the largest known proper motion. For many years, it had been considered as a very old subdwarf star with no magnetic activity. But on the border of centuries, there was detected a weak X-ray radiation of the corona, fairly strong UV chromospheric emission, and the absence of Balmer lines in absorption and emission; this shows the existence of some chromosphere and conservation of magnetic activity of dwarf stars up to 7–12 billion years (Riedel et al., 2005). Paulson et al. (2006) derived two spectra of this star throughout the flare with the echelle spectrograph at the 2.7-meter Harlan J. Smith Telescope of the McDonald Observatory. The continuum peak in the blue region corresponded to the blackbody temperature somewhat higher than 8000 K, in the range from 3734 to 4271 Å they identified throughout the flare about 90 enhanced lines — with filled in absorption or emission — and emission Balmer lines from $H_\beta$ to $H_{11}$. They reproduced the Balmer line profiles at a density of $1.3 \cdot 10^{13}$ cm$^{-3}$ and at the Stark broadening, but the 30-minute exposure did not allow them to acquire estimates of mass motion, evaporation, and velocity field. The flare produced a deep chromospheric heating, which was evident from a strong blue continuum, filled in HeI $\lambda$ 5876 Å and HeII $\lambda$ 4686 Å lines, and the heated upper photosphere; this was shown by neutral metal lines. Using the 15-year spectral and photometric lines, Toledo-Padron et al. (2019) recently estimated a rotation period of Barnard's star of 145 ± 15 days, found its differential rotation, a very low level of chromospheric emission $\log R'_{HK}$ = –5.82, and an activity cycle of 10 ± 2 years.

Hambaryan et al. (2004) identified a strong X-ray source of $L_X/L_{bol} \sim 0.1$ with the M9 dwarf 1RXS J115928.5-524717 with a proper motion of 1.08″ per year.

In the broad pair of white and red dwarfs LHS 4039/4040, Scholz et al. (2004) revealed the third body of the M8.5 type with strong $H_\alpha$ emission and blue continuum; based on the ratio $F_{H\alpha}/F_{bol}$, this object is similar to the M9.5 star 2MASSW J0149090+295613 both during the flare and in the quiescent state. Later, Scholz R.-D. et al. (2005) classified three bright close point sources from the list 2MASS as M4.0e (L 449-1), M4.5e (L 43-72), and M5.0e (LP 949-15) and found $H_\alpha$ emission and X-rays; moreover, X-ray flares were recorded on two of them. According to Reid and Walkowicz (2006), the M4 dwarf LP 261-75 in a pair with the dwarf L6 had strong $H_\alpha$ emission and could be identified with the X-ray source 1RXS J095102.7+355824.

Schmidt et al. (2007) analyzed 152 M7-L8 dwarfs selected from the Catalog 2MASS within 20 pc from the Sun and in 11 of these objects they detected photometric variability from small brightness oscillations to large flares; the most noticeable flares were on the star M7 2MASS J1028404-143843 and on the L1 dwarf 2MASS J10224821+5825453. From a sample of stars, the authors found an average tangential velocity of 31.5 km/s with a dispersion of 20.7 km/s. $H_\alpha$ emission was detected in 63 out of 81 late M dwarfs and in 16 out of 69 L dwarfs. No correlation between the logarithm of ratio of radiation fluxes $\log(F_{H\alpha}/F_{bol})$ and the tangential velocity was found, but the peak of emission occurrence on M7 stars accompanied by decreasing to average L was confirmed.

Reiners and Basri (2006) detected the variability of $H_\alpha$ emission in the binary system of brown dwarfs 2MASSW L0 + L1/1.5. Rockenfeller et al. (2006a) carried out a multicolor photometric monitoring of 19 M2–M9 dwarfs of the field and, having compared the derived results with the published data on L dwarfs, concluded that variability is more typical of L than M dwarfs. This may be caused by a change in the physical nature of variability or by extension of relevant sources on the stellar surface.

The most complete data on $H_\alpha$ emission in M dwarfs have been acquired from observations within the huge Sloan Digital Sky Survey (SDSS) that includes not dozens but many dozens of thousands of M dwarfs (Fig. 3). Based on these observations, a team of astrophysicists in Seattle supervised by Susan Hawley carried out an extensive series of investigations of activity for M dwarfs. SDSS allowed one by 2-3 orders of magnitude to increase the distance up to which an astrophysical study of M dwarfs is possible. As a result, it contains data on 15 million dwarfs with masses from $0.1M_\odot$ to $0.8M_\odot$, among them one may expect up to 2 million flare stars, and it significantly supplements the available data regarding such stars. Due to a drastic change of the distance, up to which one may identify cool dwarfs, Silvestri et al. (2004) detected a decrease of the fraction of active stars among them with increasing distance above the Galaxy plane. West et al. (2006) analyzed more than 2600 dM7 stars, confirmed that the fraction of magnetoactive ones among them and the average $H_\alpha$ luminosity decreased with height above the galactic plane (Fig. 4), and associated these effects with the dynamic heating of thin-disk stars and fast attenuation of magnetic activity at an age of 6–7 billion years. Based on a huge array of SDSS, functions of luminosity and masses for M dwarfs have already been plotted (Bochanski et al., 2010). Making use of about 8000 stars from the SDSS survey, West et al. (2004) confirmed the occurrence maximum of $H_\alpha$ emission for M7 dwarfs, although it was less than 100%, and the ratio $F_{H\alpha}/F_{bol}$ was roughly constant in the interval of M0–M7 and slightly decreased in the interval of M8–L0 types.

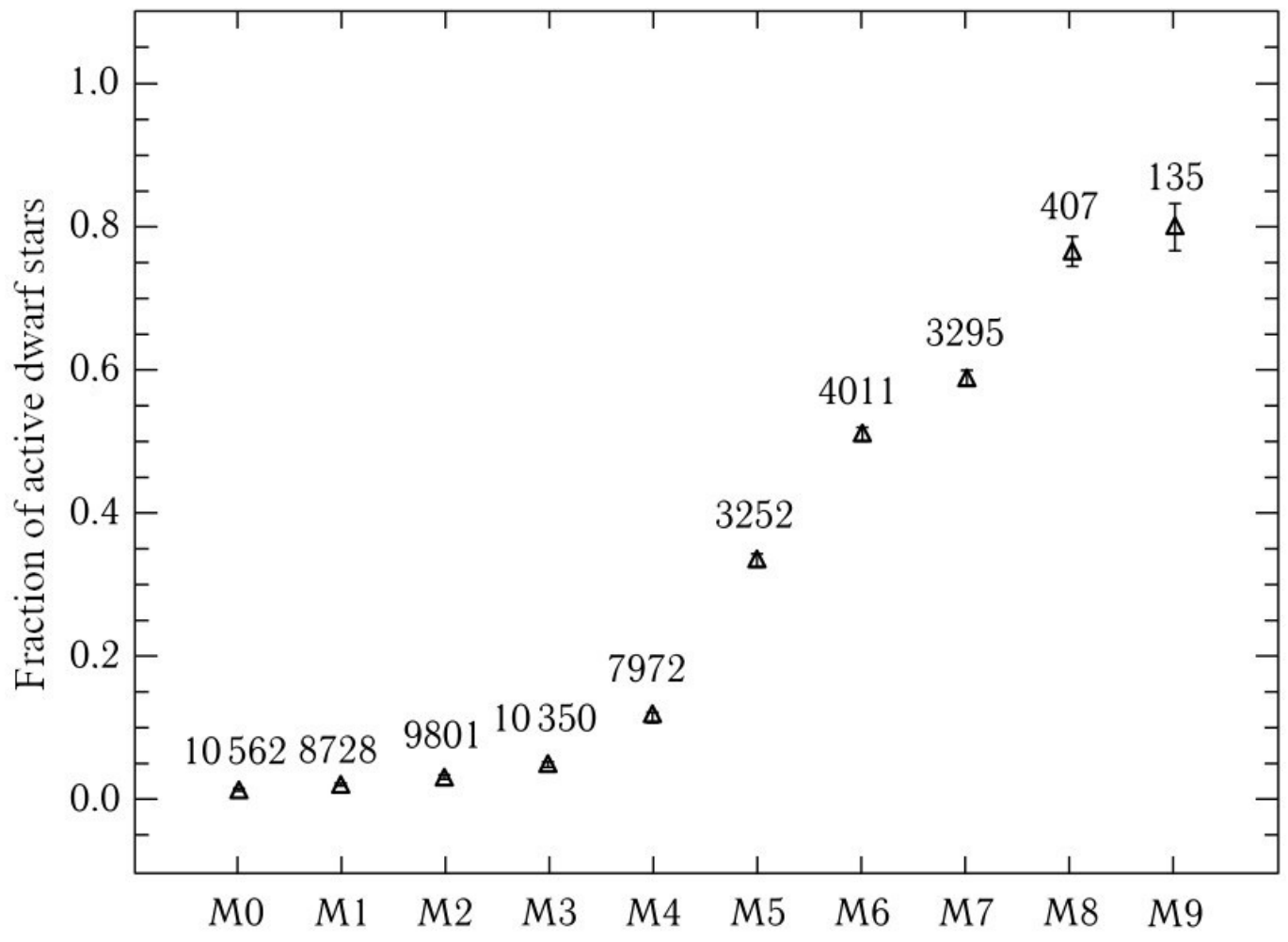

Fig. 3. Fraction of active dwarf stars of M0–M9 spectral types with emission, numbers indicate the quantity of appropriate objects (West et al., 2011)

Kowalski et al. (2009) applied the multiple photometry of SDSS to extract characteristics of flare activity for M dwarfs as a function of the spectral type, magnetic activity level, distance from the galactic plane, and line-of-sight direction. They studied more than 50000 light curves of M dwarfs in the ultraviolet and green rays and detected 271 flares strongly correlated with $H_\alpha$ emission in the quiescent state of stars. But flares without such emission were recorded on 8 stars. A fraction of flare stars significantly increases toward cooler dwarfs, and the majority of such stars are within the 300-pc layer above and under the Galaxy plane. Flares with the most powerful luminosities in the U band fall upon the early M dwarfs, and the best estimate of the lower frequency limit of flares with $\Delta U > 0.7^m$ on stars with $U < 22^m$ is 1.3 flares/(h· deg$^2$), but notably depends on the line-of sight direction.

Using the statistics of 41000 objects from the spectroscopic catalog of M dwarfs SDSS, Bochanski et al. (2011) found that based on the absolute magnitudes in the SDSS *r* band, the magnetoactive dwarfs were brighter than inactive ones of the same spectral types, whereas metal-poor M dwarfs were weaker than the appropriate stars with enhanced metallicity.

Following West et al. (2011), based on a sample of many dozens of thousands of M dwarfs, $H_\alpha$ emission is detected for 13%, $H_\beta$, $H_\gamma$, and CaII K emission — for 6%. From these observations, a noticeable decrease of the fraction of M dwarfs with $H_\alpha$ emission is seen with the distance from the Galaxy plane (see Fig. 4). Later, Pineda et al. (2013) considered the dependence of activity on the position of a star in the Galaxy.

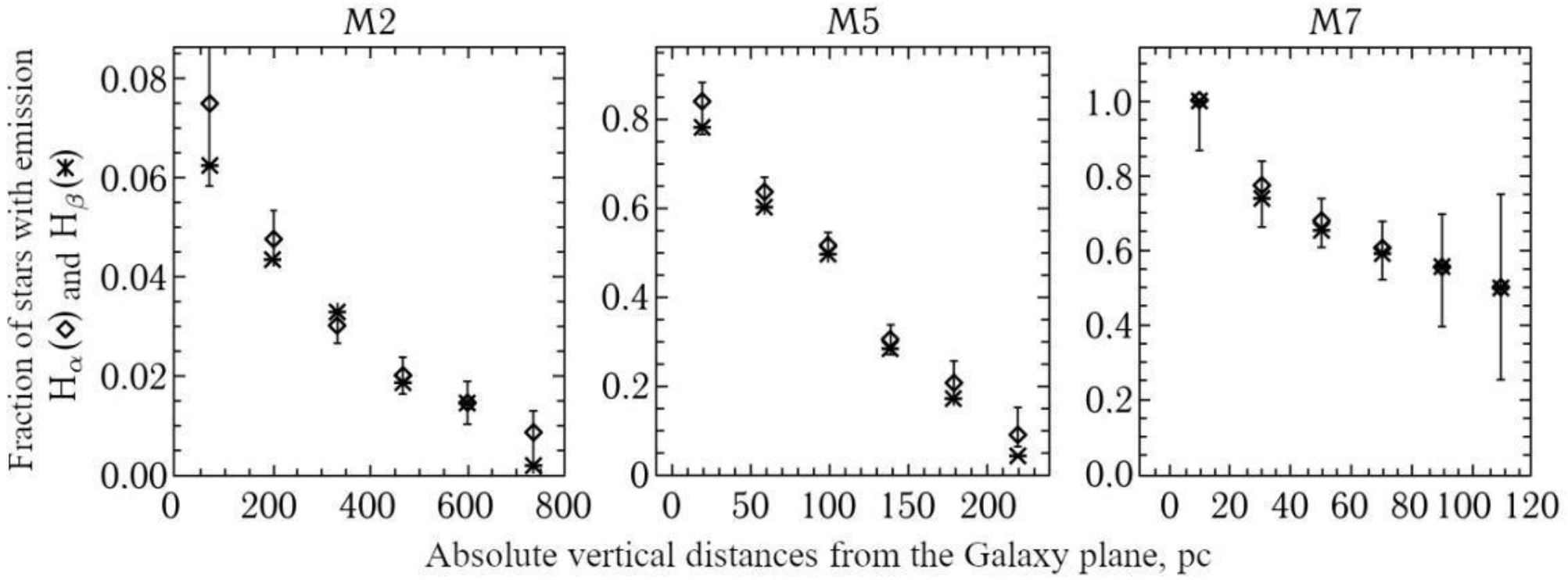

Fig. 4. Fraction of M dwarfs with emission at different absolute distances from the Galaxy plane (West et al., 2011)

Using four consecutive spectra, Liebert et al. (2003) recorded decreasing strong $H_\alpha$ emission for the L5 dwarf 2MASS J01443536-0716142, but no such emission was found in the 11-month-old spectrum. Whereas for another L5 dwarf 2MASSI J1315309-264951, stronger $H_\alpha$ emission was seen in four separate spectra. The temporal filling factor of flares of L5 dwarfs was about 1%, and several known constant $H_\alpha$ emitters should be either not older than 10–100 million years or objects with the unknown mechanism for pumping this emission.

Castro et al. (2011) detected the X-ray radiation for a member of the association TW Hydrae 2MASSW J1139511-315921 at the level $\log(L_X/L_{bol}) = -4.8$; this is by an order of magnitude lower than for another member of this association — the brown dwarf TWA 5B, but such divergence is typical of young brown dwarfs.

Bell et al. (2012) analyzed about 12000 spectra for 3500 magnetoactive M0–M9 stars from SDSS with the aim of studying variability of $H_\alpha$ emission on a time scale from minutes to weeks and found that stars with high ratios $L_{H\alpha}/L_{bol}$ occurred for all spectral types. More active stars with low variability were characterized by the duration of variations of more than an hour, whereas less active objects with high variability were characterized by the duration of variations of less than 15 minutes.

Schmidt et al. (2012a) analyzed the $H_\alpha$ emission for M and L dwarfs from SDSS and found that on M dwarfs active regions of the chromosphere covered a large area and had high density at high temperatures, whereas on L dwarfs the filling factor was not high and densities at high temperatures were not high. Later, Schmidt et al. (2015) compiled a sample of 11820 M7–L8 dwarfs from SDSS and Baryon Oscillation Sky Survey (BOSS), applying the photometry of 2MASS and Wide-field Infrared Sky Explorer (WISE)[1], prolonged the

[1] WISE — a broadband infrared space telescope of NASA launched in December 2009 with the aim of deriving a survey of the whole sky in four IR ranges: 3.4 microns — for galaxies and stars, 4.6 microns — for the detection of brown dwarfs, 12 microns — for asteroids, and 22 microns — for star-formation regions with the temperature from –230 to 270 °C. A five-mirror afocal telescope with a diameter of 40 cm and a focus of 1.36 m with a field of view of 47′, the cameras are equipped with the 1024 × 1024 pixel matrices with a field of 6 and 12″/pc, exposures 8.8 s. The telescope is equipped with

dependence of colors on spectral types up to ultracool dwarfs and estimated the course of activity along this dependence, considering $H_\alpha$ emission as its characteristic. They found that the fraction of active dwarfs grew along the sequence of M stars, reaching ~ 90% for L0 stars and then dropped to 50% around L5, although the ratio $\log(L_{H\alpha}/L_{bol})$ started decreasing for M4; on late L dwarfs no $H_\alpha$ emission was found. On the interval of M7–L3 the activity intensity decreased. Based on $H_\alpha$ observations, applying a one-dimensional model of the chromosphere, they found that in the range of M0–L7 the chromospheres of L dwarfs were cooler than that of M dwarfs and had a lower filling factor.

Jenkins et al. (2009) analyzed more than 300 M dwarfs and for 43% of them found $H_\alpha$ emission. Based on observations during 6 epochs, Riaz and Gizis (2007) found for the L5 dwarf 2MASSI J1315309-264951 $W_{H\alpha} \sim 25$ Å in the quiescent state and about 100 Å during a prolonged flare. The chromospheric character of hydrogen emission was confirmed by the NaI D line emission.

Stelzer et al. (2013) studied the X-ray and ultraviolet radiation of M dwarfs inside the circumsolar sphere with a radius of 10 pc and in a sample of young, up to 10 million years, stars in the association TW Hya. They found the following general pattern of activity of such stars. Surface fluxes in $H_\alpha$, in near and far ultraviolet, and in X-rays were interrelated by the power dependence. The activity indices defined as ratios of luminosity in the considered band to the bolometric luminosity revealed a divergence for a fixed spectral type up to 2-3 orders of magnitude, and it was maximal for M4; a spread of rotation velocities was maximal as well. The average activity index, exposing the saturation level, decreased from the X-rays through the far and near ultraviolet toward $H_\alpha$. A comparison between the full sample and properties of the members of the TW Hya association showed a drop by almost 3 orders of magnitude of luminosity in these bands on a time scale from 10 million to several billions of years, and in X-rays this drop was steeper than that in ultraviolet.

On the other hand, in the region of stars of earlier spectral types, Landini et al. (1986) detected a strong X-ray flare on the single G0 V star $\pi^1$ UMa, whose age is by an order of magnitude less than that of the Sun, and the quiet luminosity in the range of 0.2–4 keV exceeds by two orders of magnitude the corresponding solar luminosity. Mullan and Mathioudakis (2000) detected two EUV flares with amplitudes $A_{EUV} = 5$ and 7 on the F2 dwarf HD 2726, which $L_X = 4.9 \cdot 10^{30}$ erg/s, $L_X/L_{bol} = 10^{-3.98}$, and FeXVIII–XXIII lines dominated in the spectrum. Freire Ferrero et al. (2004) paid attention to the F5-6 dwarf HD 111456 which, being near the boundary of stars with solar-type activity, exhibited emission of cores of CaII lines, filled in $H_\alpha$ contour, strong UV lines of CII, CIV, SiII, SiIV, and hard X-rays. The absence of variability in optical and UV lines throughout several days was interpreted by a uniform distribution of active regions over the star or by a small inclination angle of the stellar rotational axis to the line-of-sight.

The significant inhomogeneity of the above sample of flare stars in the solar vicinity is primarily due to the accidental proximity of the stars of different origin and age to the Sun, which is evidenced by their heterogeneous kinematic characteristics (Kunkel, 1975a;

---

a tank of liquid hydrogen, which provides observations of objects with the temperature up to 7 K, the sensitivity of WISE is 500 times higher than that of IRAS. By October 2010, the liquid hydrogen supply had finished, and the instrument proceeded to operation in two ranges to study near-earth asteroids, comets, and other minor bodies of the Solar system. It has detected up to the current sleeping state within the NEOWISE program 33 thousand space objects. Over the period of half a year there were derived 1.5 million images of regions, 16 frames each.

Shakhovskaya, 1975; Hawley et al., 1996). Using these characteristics, Veeder (1974) ranked 15 of 33 flare stars as the old disk population, while Stauffer and Hartmann (1986) 9 of 25, respectively. Kinematic estimates of the age of different groups of UV Cet-type stars vary within $6 \cdot 10^6$–$10^{10}$ years. Apparently there are members of different young kinematic groups among flare stars in the solar vicinity: the Hyades, the Pleiades, and Sirius (Poveda et al., 1995). The heterogeneity of the sample of flare stars is also due to the fact that the stars are attributed to the UV Cet-type variables using several independent criteria: optical, ultraviolet, and X-ray flares, powerful quiet chromospheres and coronae, and noticeable effects of spottedness. The efficiency of different selection criteria varies for stars of different spectral type.

Over 30% of the known UV Cet-type stars are members of binary and multiple systems, with the rate of occurrence increasing toward lower-luminosity objects (Pettersen, 1991). Recently, Janson et al. (2012) confidently confirmed the estimate of Pettersen: 27 ± 3% M dwarfs are included into binary systems. One of the most active and well-studied pairs YY Gem consists of two practically similar M dwarfs, whereas the eclipsing system CM Dra, in which Kozhevnikova et al. (2004) during 155 hours recorded four flares with energy that exceeds the solar one by hundreds of times, includes components M1 and M4.5e. As Melikian et al. (2011) claimed, the activity level of UV Cet comprising the binary system depends on the nonconstant distance between components. But among all spectral types there are definitely single flare stars, thus the binarity is not a necessary condition for the considered activity type. Observational selection leads to the conclusion that the overwhelming majority of the known binary systems comprising the UV Cet-type stars consist of pairs of flare red dwarfs: it is more difficult to find a flaring component in the systems comprising red dwarfs and higher luminosity stars, if the pair is not very wide. Sometimes, flare components in such pairs are found only during flare events (Arsenijevic, 1985; Shakhovskoy, 1993; Huensch and Reimers, 1995). However, making use of observations with Kepler[1], contrary to a widespread opinion that invisible satellites are responsible for the recorded flares on hot stars, Balona recently came to the conclusion on the reality of flares on stars in the entire spectral range from M to A and on the approximate constancy of flare occurrence in this range.

Using the HATNet system, comprising seven short-focus telescopes with a diameter of 11 cm and F/1.8 mounted in Arizona, Hawaii, and Israel and equipped with the wide-field CCD photometers with a field from 8.2° to 10.6°, Hartman et al (2011) performed a photometric IR survey of 27560 K and M dwarfs of the field with proper motions of more than 30 ms per year, searching for the periodic brightness variability and prolonged flares with high amplitudes. As a result, they discovered 2120 potentially variable stars, including 60 flare stars with 64 recorded flares. Most of the detected variables were strongly spotted BY Dra stars. A fraction of stars with variability amplitudes of more than $0.01^m$ exponentially grew with the color index $V–K_S$, hence roughly a half of dwarfs of the field in the solar vicinity with $M < 0.2M_\odot$ were variable at this level.

[1] Kepler is a specialized wide-field telescope of the Schmidt system placed into orbit in March 2009 for the discovery and photometric study of exoplanets. The aperture diameter of the telescope is 95 cm, the primary mirror size is 1.4 m, in its focus there was mounted a mosaic comprising 42 CCD matrices with a size of $5 \times 2.5$ cm$^2$ with a total area of 95 megapixels; this mosaic provides observation of the field of 115 deg$^2$. The telescope was constantly directed to the region in constellations Cygnus and Lyra, recorded images of more than 150000 stars from 9 to $16^m$ in the range from 430 до 890 Å with the accuracy up to $0.001^m$, a query of matrices occurred every 6 s and information dropped out in 30 s. It is evident that the by-product of accumulation of such a large volume of continuous high-accuracy data on exoplanets is a possibility of studying photometric variability of stars.

A discussion of this issue will be continued in Part 3 of the book.

The distinctions between G0 dwarfs, whose structure is the most similar to the solar one, and the latest M dwarfs are very significant. Within this range of spectral types the effective temperatures vary from 6000 to 2500 K, stellar masses vary from $1M_\odot$ to $0.06M_\odot$, the radii from $1R_\odot$ to 0.1 $R_\odot$, and the luminosity from 1 to 0.0008 of the solar luminosity. In addition to these significant quantitative external distinctions, there are major qualitative distinctions of the inner structure. G dwarfs are characterized by thermonuclear burning of hydrogen and by radiative transfer of the energy in the central part, while the convective zone occupies only the outer layers (about 30% of the stellar radius). On the other hand, in the stars close to the spectral type M4 with a mass of $0.35M_\odot$ and an effective temperature of about 3200 K, there is a transition to completely convective structures accompanied by the disappearance of the boundary between the radiative core and the convective envelope on which, according to the modern concepts, the effective generation of the toroidal magnetic field occurs following the $\alpha-\omega$ dynamo mechanism. For even later dwarfs close to M9 the main sequence terminates: red dwarfs are replaced by brown dwarfs with a mass of about $0.07M_\odot$, the central stellar temperatures are not sufficient to sustain thermonuclear fusion of hydrogen, therefore they have no thermonuclear energy sources and “live” at the expense of gravitational compression. However, the transition from main-sequence stars to brown dwarfs can occur without significant change in the activity level, since all the diverse internal structures have convective transfer, at least in the outer regions of stars, and noticeable axial rotation, which result in the generation of magnetic fields of low- and mid-mass stars and all relevant activity phenomena. The recent discovery of the X-ray, radio, and $H_\alpha$ radiation from young brown dwarfs and the photometric variability of cool dwarfs apparently supports these considerations (Neuhauser and Comeron, 2001; Bouy et al., 2004; Berger et al., 2009, 2010; Schmidt and Cruz, 2006). The coolest star with the recorded flare of the considered type was the L2.5 dwarf ULAS J224940.13-011236.9 with an effective temperature of 1930 ± 100 К (Jackman et al., 2019). Stassun et al. (2011) specified a smooth course of global stellar parameters in the interval of M3–M6 despite noticeable variations of the inner structure while transiting to convective objects. Based on high-precision infrared interferometric observations with VLT and measurements of parallaxes with GAIA, Rabus et al. (2019) recently detected a discontinuity in the monotonous course of the temperature–radius dependence of M dwarfs in the region of 3200–3340 К, i.e., in the transition zone from partially to fully convective stars with a mass of $0.23M_\odot$.

## 1.2. Photospheres

Because of the immense interstellar distances the only stellar surface we can observe directly is that of the Sun. To reconstruct the properties of "visible stellar surface" from the total observed radiation one should take into account two main factors: systematic limb darkening and irregular surface inhomogeneities due to starspots. The regions of the stellar surface free of spots are called quiescent photospheres.

### 1.2.1. Quiescent Photospheres

The structure, and consequently the radiation, of a stationary quiescent stellar photosphere are determined by three main independent factors: the effective temperature, gravity, and chemical composition of the radiating substance. These parameters are determined by means of spectral analysis of stellar spectra and are involved in theoretical calculations of the models of stellar atmospheres. Thus far, both analytical and synthetic operations have disregarded stellar activity. Otherwise, if the three parameters of flare and nonflare stars are equal, it is suggested that their structures are identical. Of course, this suggestion is valid only in a certain approximation, and recent studies reached the limits of its applicability.

A wide range of effective temperatures of flare stars produces considerable differences both in the spectral composition and the intensity of continuous radiation within the deep subphotospheric layers, and in the continuous and selective absorption coefficients governing the form of the observed absorption spectrum of photospheres. Thus, the intensities of the absorption lines of hydrogen and metals are comparable in the spectra of G stars. In the spectra of K stars, the lines and line blends of metals prevail. In the spectra of the latest K or M0 stars, the bands of TiO molecules arise, which soon strengthen in the later spectral subtypes. In the late M stars, the optical spectrum of the photosphere is determined by the molecular bands of TiO and VO, which bind the excess oxygen remaining after the binding of almost all the carbon in the CO molecule, while the infrared spectrum depends on the bands of water vapor, which binds hydrogen not involved in the $H_2$ molecules. All these molecules practically do not leave the spectral intervals with an intrinsic continuum of the outgoing radiation (Allard et al., 1997).

Spectral classification of active and inactive dwarfs, i.e., the estimates of effective temperatures, is performed using the same spectral criteria, which evidences at least the essential similarity of the temperature structure of the two stellar types. It is known that initially the set of criteria of spectral classification was developed for the photographic range of the spectrum. Then, for cool stars the system was expanded to the near infrared region. Recently, Reid et al. (1995) proposed an algorithm for the spectral classification of cool dwarfs based on the strongest band of TiO $\lambda$ 7050 Å, which makes it possible to determine the spectral type of K7–M6 dwarf stars from the depth of the band to the accuracy of a half-subtype. This algorithm cannot be applied to hotter stars, since they do not have this band, while for cooler stars this band is saturated and weakens due to the increasing absorption of VO. However, the applicability range of this algorithm covers the vast majority of known UV Cet-type stars.

Recently, Kirkpatrick et al. (2011) summarized the consideration of the first hundred coolest stars in the range of spectral types M–Y discovered with the WISE instrument; among them 80 stars were T6 and later stars. These data were used for the preliminary estimation of the space density lower limit of such objects and for the discussion of constraints on the function of masses and the lower mass limit during the star formation.

* * *

For G0–M5 main-sequence stars, the logarithm of gravity smoothly changes from 4.4 to 4.8 and there is no reason to expect systematic distinctions in this parameter for UV Cet stars and quiescent dwarfs of the same spectral types.

The strong differences in the chemical composition as on hot chemically peculiar stars are not known in flare stars (Hartmann and Anderson, 1977; Mould, 1978). But X-ray observations (see Sect. 1.4.2) revealed a pronounced — up to an order of magnitude — poor abundance of hard elements in coronae of a series of active dwarfs with respect to the solar abundance and, apparently, with respect to the abundance of photospheres of these stars — the so-called FIP effect, which made up of enhanced abundance of elements with the first ionization potentials lower than 10 eV as compared to the abundance of elements with the first ionization potentials higher than this value; there are cases of the inverse FIP effect.

As to the general level of metallicity of the atmospheres of the stars under consideration, there are distinct spectral criteria for estimating the factor: as the metallicity decreases, the intensities of hydride bands intensify as compared to oxide bands. This effect is easily explained: as the metallicity of a substance decreases, so does the relative abundance of both Ca and O that determines the selective absorption in the bands of CaO and $H^-$, which to a considerable degree governs the continuous absorption. Thus, the depth of the CaO oxide line, which in the first approximation is proportional to the expression $n(\mathrm{Ca}) \times n(\mathrm{O})/n(\mathrm{H}^-)$, decreases. But the relative abundance of H at least does not decrease with the decrease in metallicity, thus the absorption in CaH bands that is proportional in the first approximation to the expression $n(\mathrm{Ca}) \times n(\mathrm{H})/n(\mathrm{H}^-)$ at least does not decrease. These considerations are equally applicable to active and inactive dwarfs. They formed a basis for the successful search for photometric effects of metallicity in the Hertzsprung–Russell diagram and in two-color infrared diagrams.

In the region of M dwarfs, the main sequence has a dispersion along the luminosity axis of about one and a half stellar magnitudes, and the exact photometry shows that the stars with metallicity of the young disk are the brightest in the infrared bands, the stars with metallicity of the old disk ($[M/H] < -0.5$) are weaker by $0.3^m$ –$0.6^m$ and the dwarfs with $[M/H] < -1$ are weaker by $0.8^m - 1.2^m$ (Leggett, 1992). (From CO bands in the near infrared region Viti et al. (2002) estimated $[M/H] = -1$ in the binary system CM Dra.) Although active and inactive dwarfs are mixed, dMe stars are, on average, brighter by $0.34^m$ than dM stars in the diagram ($M_V$, $R-I$) (Stauffer and Hartmann, 1986) and by $0.66^m$ in the diagram ($M_K$, $V-K$) (Hawley et al., 1996). This small shift against the background of considerable dispersion was noted by Kuiper in 1942 and by Veeder in 1974. It can only partially be explained by the fact that some dMe stars have not reached the main sequence or by the unaccounted binarity, but it is most likely due to the enhanced metallicity of some dMe stars. Specially selected two-color diagrams support this conclusion (Stauffer and Hartmann, 1986; Leggett, 1992).

Mann et al. (2013) suggested a new procedure of estimating metallicity for K7–M7 dwarfs in the range of $-1.1 < [\mathrm{Fe/H}] < +0.5$, using optical and near IR spectra and worked it out on a sample of 120 binary FGKM stars. Metallicity of M dwarfs was discussed in a number of reports at the $17^{th}$ Symposium on Cool Stars in Barselona in June 2012. Terrien et al. (2012) acquired an important in this respect result concerning the metallicity of components of the binary eclipsing system CM Dra: $[\mathrm{Fe/H}] = -0.30 \pm 0.12$, which consisted of two M dwarfs — M1 and M4.5. It is pivotal for the discussion of the lower main sequence. As for the statistical metallicity properties of M dwarfs, then following West et al. (2011), it decreased noticeably with the distance from the Galaxy plane where the fraction of older stars increased. Woolf and West (2012) noted that M dwarfs have the same problem that is detected for G and K dwarfs: the number of metal-poor M dwarfs is insufficient for constructing a simple closed model of the local galactic evolution, as the number of stars with [Fe/H] ~ −0.5 comprises only 1% of the number of stars with [Fe/H] = 0.

Applying the original technique, Malygin et al. (2012) conducted a comparison of the chemical composition of atmospheres of the two solar-type stars: the metal-poor star HD 10700 and the metal-rich star HD 1835. The metal abundance of the latter one was found to exceed the appropriate values of the former star by 0.42–0.50 dex for VI and SiI and up to 0.88–0.89 dex for CrI and FeI.

As stated above, the data on photospheric criteria of the activity of red dwarfs were obtained relatively recently. Hawley et al. (1996) discovered that the fine structure of the TiO bands in the infrared region of the spectrum correlates with the degree of activity of a star: TiO bands arising near the temperature minimum or deeper in the photosphere appear to depend on $L_{H\alpha}/L_{bol}$. This spectral peculiarity cannot be related to the metallicity fluctuations. These fine effects can provide useful constraints on the models of dwarf stars.

It should be noted here that kinematic characteristics largely account for the collective properties of stellar groups, while metallicity is a characteristic of an individual star, and there is no unique dependence between them. Nevertheless, while proceeding from the young disk to the old one and further to the halo, we observe an increase in the dispersion of velocities accompanied by decreasing metallicity as the stellar age increases.

* * *

An estimate of the enhanced lithium is of particular interest in the spectroscopic studies of flare stars since the abundance of this element is determined both by the general evolution of a star within which it burns out and its activity when the enhancement increases due to spallation reactions. It has long been known about an abnormal lithium abundance in the atmosphere of the flare star Gl 182 (= V 1005 Ori) (Bopp, 1974a; de la Rezza et al., 1981). Strictly speaking, it may presumably be not an exception, but an extreme case, as a correlation between the lithium abundance in the atmospheres of K dwarfs and their activity level was suspected recently (Zboril et al., 1997; Favata et al., 1997). An analogous correlation in 110 F–G stars was found by Cutispoto et al. (2003). But such anomalies of individual elements with low absolute abundance do not have effect on the structure of a star and its atmosphere in general.

The results of lithium studies in individual active stars are as follows.

Based on observations with VLT/UVES in Chile and with AAT in Australia, Pavlenko et al. (2007) estimated a high lithium abundance in the atmosphere of the brown dwarf LP 944-20: $\log N(\mathrm{Li}) = 3.25 \pm 0.25$. Koenig et al. (2005) studied 8 flare F–K stars and for the most active two stars, HN Peg and EK Dra, found excess lithium. Zapatero Osorio et al. (2005a) explored the brown dwarf GJ 569 Bab with the Keck telescope and found a significant decrease in lithium abundance on both components. (The infrared photometry of this system did not make it possible to unambiguously determine orientations of axes for components (Kenworthy and Scuderi, 2012).) Having studied the lithium abundance among members of the Pleiades, Xiong and Deng (2006) came to the conclusion that the dispersion of these values in stars of similar spectral types was substantially governed by the atmospheric effects of active stars, and the correlation of high lithium abundance and rotation reflected the correlation of abundance of this element and stellar activity. Christian et al. (2008) studied relative enhanced abundances of lithium isotopes in the atmosphere of the active dwarf GJ 117 and found $^6Li/^7Li = 0.05 \pm 0.02$. This ratio was consistent with the expected value caused by spallation reactions on the surface of the star. Baumann et al. (2010) considered the lithium abundance for 117 solar-type stars and found its substantial decrease with age, but did not confirm the hypothesis that stars with planets are to a greater extent deprived of this element. Takeda et al. (2010) examined the association of lithium abundance with rotation on 118 solar-type stars and confirmed that its surface abundance progressively dropped with decreasing rotation velocity. Lubin et al. (2010) studied the lithium abundance for the solar-type stars with low chromospheric activity suspected in the Maunder minimum state and found that on stars with $\log R'_{HK} < -5.0$ the enhanced lithium abundance was twice less as compared to the mean enhanced abundance for the stars from the sample. But besides the Maunder minimum effect, the reduced lithium abundance may be due to the age of the star, when it brings to an end its presence on the main sequence.

MacDonald and Mullan (2010) associated the cases of statistically significant differences in estimating stellar age based on the lithium abundance and on their position in the Hertzsprung–Russell diagram with effects of suppressing convection by magnetic fields. Following their estimates, the magnetic suppression of convection increases the age on HR isochrones for all stars, but the lithium age decreases for fully convective M stars and increases for stars with radiative cores. They found that the magnetoactive model is an alternative of the inner structure model in which variations of the mixing length are considered.

Mishenina et al. (2012) compared the enhanced lithium abundance and activity level in a sample of 150 slow rotators of FGK dwarfs and revealed a decrease in the lithium abundance with decreasing temperature and a significant divergence in this correlation due to the difference in age. The correlation between the enhanced lithium abundance, rotation rate, and chromospheric activity level was detected for stars with effective temperatures from 6000 to 5200 K, whereas at lower temperatures the correlation between the enhanced lithium abundance and $v\sin i$ disappeared. Xing and Xing (2012) found a correlation between the lithium abundance and X-ray luminosity for young solar-type stars: as the enhanced abundance of this element decreases, the X-ray radiation strengthens.

Katsova et al. (2016) explored the enhanced lithium abundance for G and K stars on which very powerful flares were detected within the space experiment Kepler and found that the significant divergence in the abundance of this element was primarily due to the differences in its abundance in the original interstellar medium.

Based on numerous high-dispersion spectra of Prox Cen, Pavlenko et al. (2017) found that the enhanced lithium abundance in this star was consistent with its expected burning in an old and fully convective low-mass star.

* * *

For approximate estimates for the atmospheres of dwarf stars of the lower main sequence, one can use values of effective temperatures, absolute luminosities of stars and stellar radii calculated from these values using the blackbody ratio

$$R_*/R_\odot = \left[L_{\rm bol}/\left(4\pi\sigma T_{\rm eff}^4 R_\odot^2\right)\right]^{1/2} = (L_{\rm bol}/L_\odot)^{1/2}(T_\odot/T_{\rm eff})^2 = 3{,}33\cdot 10^7\left(L_{\rm bol}/L_\odot\right)^{1/2} T_{\rm eff}^{-2}. \quad (1)$$

Table 2 lists the basic parameters of A9–L8 dwarfs acquired by independent methods: data extracted from de Jager and Nieuwenhuijzen (1987) are given in Columns 2 and 7; effective temperatures and sizes of K0–M5.5 stars from interferometric measurements performed by

Table 2. Effective temperatures, radii, and masses of dwarf stars

| Spectral type | $T_{eff}$, K | | | | | $R_*/R_\odot$ | | | $T_{eff}$, K (model) | $R_*/R_\odot$ (model) | $\log(L_*/L_\odot)$ (model) | Mass, $M$ (model) |
|---|---|---|---|---|---|---|---|---|---|---|---|---|
| 1 | 2 | 3 | 4 | 5 | 6 | 7 | 8 | 9 | 10 | 11 | 12 | 13 |
| A9 | | | | | | | | | 7440 | 1.84 | 0.97 | 1.67 |
| F5 | | | | | | | | | 6510 | 1.46 | 0.54 | 1.33 |
| F9 | | | | | | | | | 6060 | 1.23 | 0.26 | 1.14 |
| G0 | 5940 | | | | | 1.06 | | | 5920 | 1.12 | 0.14 | 1.08 |
| G2 | 5790 | | | | | 1.03 | | | 5770 | 1.01 | 0.01 | 1.02 |
| G4 | 5640 | | | | | 1.00 | | | 5680 | 0.986 | –0.04 | 0.99 |
| G8 | 5310 | | | | | 0.96 | | | 5490 | 0.909 | –0.17 | 0.94 |
| K0 | 5150 | 5347 | | | | 0.93 | 0.80 | | 5280 | 0.817 | –0.33 | 0.87 |
| K1 | 4990 | 5147 | | | | 0.91 | 0.81 | | 5170 | 0.814 | –0.36 | 0.85 |
| K2 | | 5013 | | | | | 0.72 | | 5040 | 0.763 | –0.47 | 0.78 |
| K3 | 4690 | 4680 | | | | 0.86 | 0.79 | | 4830 | 0.729 | –0.58 | 0.75 |
| K4 | 4540 | 4507 | | | | 0.82 | 0.74 | | 4600 | 0.726 | –0.67 | 0.72 |
| K5 | 4410 | 4436 | | | | 0.80 | 0.69 | | 4410 | 0.698 | –0.78 | 0.68 |
| K7 | 4150 | 3961 | | | | 0.72 | 0.60 | | 4070 | 0.654 | –0.98 | 0.63 |
| K9 | 3940 | | | | | 0.66 | | | 3940 | 0.552 | –1.18 | 0.56 |
| M0 | 3840 | 3907 | | | | 0.63 | 0.58 | | 3870 | 0.559 | –1.20 | 0.55 |
| M0.5 | | 3684 | | | | | 0.51 | | 3800 | 0.535 | –1.27 | 0.54 |

| Spectral type | $T_{\text{eff}}$, K | | | | | $R_*/R_\odot$ | | | $T_{\text{eff}}$, K (model) | $R_*/R_\odot$ (model) | $\log(L_*/L_\odot)$ (model) | Mass, $M$ (model) |
|---|---|---|---|---|---|---|---|---|---|---|---|---|
| 1 | 2 | 3 | 4 | 5 | 6 | 7 | 8 | 9 | 10 | 11 | 12 | 13 |
| M1 | 3660 | 3497 | | | | 0.56 | 0.40 | | 3700 | 0.496 | –1.40 | 0.49 |
| M1.5 | | 3674 | | | | | 0.50 | | 3650 | 0.460 | –1.47 | 0.47 |
| M2 | 3520 | 3464 | | | | 0.48 | 0.39 | | 3550 | 0.434 | –1.57 | 0.44 |
| M2.5 | | 3442 | | | | | 0.30 | | 3500 | 0.393 | –1.68 | 0.40 |
| M3 | 3400 | 3412 | | | | 0.41 | 0.41 | | 3410 | 0.369 | –1.78 | 0.36 |
| M3.5 | | 3104 | | | | | 0.32 | | 3250 | 0.291 | –2.07 | 0.26 |
| M4 | 3290 | 3222 | | | | 0.35 | 0.19 | | 3200 | 0.258 | –2.20 | 0.22 |
| M5 | 3170 | | 2800 | | | 0.29 | | 0.64 | 3030 | 0.199 | –2.52 | 0.16 |
| M5.5 | | 3054 | 2800 | | | | 0.14 | 0.68 | 3000 | 0.149 | –2.79 | 0.12 |
| M6 | 3030 | | 2819 | 2900 | | 0.24 | | 0.94 | 2850 | 0.127 | –3.02 | 0.10 |
| M7 | 2860 | | 2675 | 2615 | | 0.20 | | 0.50 | 2650 | 0.118 | –3.21 | 0.090 |
| M7.5 | | | 2587 | | | | | 0.56 | 2600 | 0.112 | –3.29 | 0.089 |
| M8 | 2670 | | | 2500 | | | | | 2500 | 0.111 | –3.36 | 0.082 |
| M9 | 2440 | | | 2425 | | | | | 2400 | 0.095 | –3.57 | 0.079 |
| M9.5 | | | | | 2050 | | | | 2320 | 0.104 | –3.55 | 0.078 |
| L0 | | | | 2325 | 1900 | | | | 2250 | 0.108 | –3.57 | 0.077 |
| L0.5 | | | | | 1950 | | | | | | | |
| L1 | | | | 2225 | 1900 | | | | 2100 | 0.107 | –3.70 | 0.076 |

| Spectral type | $T_{eff}$, K | | | | | $R_*/R_\odot$ | | | $T_{eff}$, K (model) | $R_*/R_\odot$ (model) | $\log(L_*/L_\odot)$ (model) | Mass, $M$ (model) |
|---|---|---|---|---|---|---|---|---|---|---|---|---|
| 1 | 2 | 3 | 4 | 5 | 6 | 7 | 8 | 9 | 10 | 11 | 12 | 13 |
| L2 | | | | 2100 | 1900 | | | | 1960 | 0.104 | –3.84 | 0.075 |
| L3 | | | | 1960 | 1800 | | | | 1830 | 0.102 | –3.98 | ... |
| L3.5 | | | | | 1800 | | | | | | | |
| L6 | | | | 1625 | 1700 | | | | 1490 | 0.099 | –4.36 | ... |
| L6.5 | | | | | 1700 | | | | | | | |
| L7 | | | | 1525 | 1700 | | | | 1410 | 0.099 | –4.46 | ... |
| L7.5 | | | | | 1400 | | | | | | | |
| L8 | | | | 1475 | 1400 | | | | 1350 | 0.097 | –4.55 | ... |

Boyajian et al. (2012) — in Columns 3 and 8; average temperatures and sizes for 13 M5–M7.5 dwarfs found from IR fluxes in the J band by Mohanty et al. (2004) — in Columns 4 and 9; effective temperatures from the multicolor photometry of the coolest dwarfs by Golimowski et al. (2004) — in Column 5; and effective temperatures in the range of M9.5–L8 extracted from the spectral analysis by Schweitzer et al. (2003) — in Column 6. The data of the later temperature scale by Rajpurohit et al. (2013) differ from those listed in Table 2 commonly by not more than 100 K. Houdebine et al. (2019) considered the literature data on 1910 K and M dwarfs and concluded that the divergence in the color correlation $(R - I)_c$ and effective temperature was about 40 K. But the most precise interferometric measurements of radii of components of the closest in the northern sky system 61 Cyg K5V + K7V yielded 0.665 ± 0.005 and 0.95 ± 0.008 in solar radius units (Kervella et al., 2008); this allows one to suspect that radii by de Jager and Nieuwenhuijzen (1987) are somewhat overestimated. The average stellar radii from a sample of 612 late K and M dwarfs found by Houdebine et al. (2016) are in good agreement with the results of Boyajian et al. (2012). Recently, Rabus et al. (2019) published the radii and temperatures of more than two dozens of M dwarfs from the infrared interferometric observations with VLT and from space measurements of parallaxes with GAIA: in the interval of 3200–3340 K they found a bump in stellar sizes from 0.42 to 0.18 solar radii.

In the mentioned studies, a high inner accuracy of the results is claimed, but as a comparison of appropriate columns shows, the outer accuracy, especially for stellar sizes, is far from being desirable. Moreover, there were noticed the considerable discrepancies of the acquired from observations and calculated based on the evolutionary theory parameters of M dwarfs: these are higher, up to 5–10%, and cooler than expected. Morales et al. (2008) showed that this effect was equally manifested in single stars and in components of the binary systems. There were attempts to associate these effects with stellar activity – with the suppression of convection by magnetic fields (Torres, 2013; Stassun et al., 2012; MacDonald and Mullan, 2013), as these discrepancies were actually absent in inactive stars. Feiden and Chaboyer (2012) theoretically considered an impact of the large-scale magnetic field on the stellar evolution and within the program, involving the only magnetic field parameter — its strength in the photosphere, found that even on a solar-type star the field increases the model radius by approximately 10%. Lopez-Morales (2007) found that, besides stellar magnetism, an increase of the stellar radius in dwarfs with a mass of more than $0.35M_{\odot}$ might be due to enhanced metallicity, which as well as magnetism could decrease the transparency of stellar substance. Helminiak et al. (2011) attributed small variations in radii of components in two eclipsing systems to activity variations. Morales et al. (2010) found that spots could yield the effect under discussion only having a high filling factor.

Columns 10–13 of Table 2 are derived using the data of GAIA DR2 (Mamajek, 2019).

In the atmosphere models of cool dwarfs, the dissociative equilibrium of more than one hundred of two- and three-atom molecules and the ionization equilibrium of dozens of atoms and ions are usually calculated. In addition to spontaneous and stimulated recombination and free–free transitions of tens of atoms, millions of atomic and molecular lines are taken into account (Allard and Hauschildt, 1995). These calculations are made in hydrostatic and LTE approximations and involve two principal difficulties: insufficiency of data on absorption coefficients of numerous molecules and dust particles in the photospheres of cool stars and the approximateness of the theory of convective energy transfer in subphotospheric layers. Over recent years, an amount of certain progress was achieved in both directions (Allard et al., 1997). Based on two- and three-dimensional radiative and hydrodynamic calculations, Ludwig et al. (2002) came to a fairly unexpected conclusion concerning the expected absence of qualitative differences between the granulation patterns on the Sun and on a dwarf with effective temperature of 2800 K and quantitative distinctions of the patterns in the intensity contrast, the horizontal scale of inhomogeneities, and the characteristic velocity of motion, which are consistent with the predictions of the mixing length theory. But the relatively recent detection of dust disks around active red dwarfs makes one relate with more attention to the role of dust in such objects (Schuetz et al., 2004; Liu et al., 2004).

During the current decade, of particular importance are experimental advances in interferometry and infrared photometry and spectroscopy, which substantially contributed to the data on basic characteristics of cool and very cool stars. Parameters of red M dwarfs were specified, the characteristics of brown L dwarfs were determined, and the following temperature class of space objects, "methane" or T dwarfs, was intensively studied. The temperature of these bodies is several hundred Kelvin, their masses are tens of Jupiter's masses, i.e., these are intermediate objects between the coolest stars and planets (Burningham et al., 2009, 2011). But the most surprising thing is that $H_{\alpha}$ emission was detected in T dwarfs, although in their neutral, cool and dust atmospheres one may hardly expect processes which are related to electrodynamic processes of solar activity (Liebert, 2003b; Liebert and Burgasser, 2007).

According to Jones and Tsuji (1997), it is required to take into account the formation of dust in stellar atmospheres when constructing synthetic spectra within the range 6500–7500 Å for stars with an effective temperature below 3000 K. Later, Tsuji and Nakajima (2016) found that dust was formed in the M8.5 dwarf 2MASS J1835379+325954 with an effective temperature of 2275 K. Following the calculations by Chabrier et al. (2000), at the effective temperatures lower than 1300–1400 K in the near IR region of the spectra, the significant role is played by methane absorption – this is a transition from L dwarfs toward “methane” T dwarfs.

The role of dust in atmospheres of M and L stars was actively discussed at the 15th Cambridge Workshop on Cool Stars (Seifahrt et al., 2009).

Recently, Rajpurohit et al. (2012) performed a detailed analysis of the triple system LHS 1070 comprising M5.5, M9 and L0 dwarfs. (In the above, a powerful flare of the component B of this system has already been mentioned (Almeida et al., 2011)). Applying four models of cool stellar atmospheres, the dust formation was included into three of them, Rajpurohit et al. presented the basic spectral peculiarities of all three components of the system in the optical and IR wavelength ranges at the solar abundance of elements and at effective temperatures of 2950 K, 2400 K, and 2300 K: for A, B, and C components, respectively, log $g$ = 5.3, 5.5, and 5.5, $R_*/R_\odot$ = 0.14, 0.10, and 0.10, $\log(L_*/L_\odot)$ = –2.9, –3.5, and –3.7. Taking into account the used wealth of data and various theoretical models, these results may be considered as supporting ones.

According to theoretical calculations, while proceeding from L to T dwarfs the dust clouds close to the photosphere should disappear, and on the border of L/T one may expect the brightness variability due to the gaps between clouds. Girardin et al. (2013) carried out observations of nine brown dwarfs and in one of them, SDSS J1052+4422 (T0.5), identified periodic brightness fluctuations with a period of 3.0 h and amplitude of about $0.06^m$.

A general notion on calculations of synthetic spectra for cool stars is provided by Pavlenko et al. (2006) in which, to find the energy distribution in a spectrum of the M6 dwarf GJ 406, there were taken into account the lines of molecules of $H_2O$, TiO, CrH, FeH, CO and MgH, and the VALD atomic line list.

In the far IR range, Liseau et al. (2013) first measured the minimum temperature in the atmosphere of the star different from the Sun — $\alpha$ Cen A.

* * *

A general idea on the optical spectra of the photosphere of flare red dwarf stars is given by publications of Pettersen and Hawley (1987, 1989): these provide low-dispersion spectra of about three tens of G9–M5 stars obtained during observations with the 2.1-meter Otto Struve Telescope at McDonald Observatory, and the measured values of intensity bumps at the heads of molecular bands of TiO, MgH, CaOH and CaH. Later, Cincunegui and Mauas (2004) published the calibrated echelle spectra of 91 F–M dwarfs with different levels of chromospheric activity in the range of $\lambda$ 3860–6690 Å with a resolution of 26 400 derived with the 2.15-meter telescope CASLEO in the Argentina Andes. Spectral peculiarities of M, L, and T dwarfs in the range of 5.5–38 μm were described by Cushing et al. (2006) from observations with a low-dispersion spectrograph during the Spitzer space mission. They found the strongest molecular bands of $H_2O$ at 6.27 μm, $CH_4$ at 7.65 μm, and $NH_3$ at 10.5 μm. Recently, Yi et al. (2014) compiled a catalog of M dwarfs called the LAMOST[1] pilot survey comprising data on 67 082 such stars with their spectral classification, radial velocities, equivalent widths of the $H_\alpha$ line, numerous indices of molecular bands and the metallicity parameter. Physical parameters of the nearest known pair of T dwarfs $\varepsilon$ Ind Bab are estimated as follows: $L/L_\odot$ = –4.70±0.02 and –5.23±0.02, $T_{\rm eff}$ = 1320±20 K and 910±30 K, log $g$ = 5.25 and 5.50 at an age of the system of 4.0±0.3 billion years (King et al., 2010). Parameters of the cooler T8.5 dwarf

[1] LAMOST — the Large Sky Area Multi Object fibre Spectroscopic Telescope — the meridional wide-field spectral Schmidt telescope is located at the Xinglong station, province Hebei, China.

Wolf 940 B, comprising a pair with the dwarf M4e, were estimated by Leggett et al. (2010) as $L/L_{\odot} = -6.01 \pm 0.05$, $T_{\mathrm{eff}}$ = 585–625 K, log $g$ = 4.83–5.22 cm/s$^2$ at an age between 3 and 10 billion years, moreover lower values of $T_{\mathrm{eff}}$ and $g$ corresponded to a smaller age and mass of 24 Jupiter's masses, whereas higher values of $T_{\mathrm{eff}}$, $g$, and age — to a mass of 45 Jupiter's masses.

* * *

Although stellar surfaces cannot be observed directly, using astroseismology methods Kjeldsen et al. (1999) presumed a granulation effect on $\alpha$ Cen A stars, whose physical parameters are close to solar ones. Indirect information on the inhomogeneity of the photospheres of active stars is provided by polarimetric observations (Alekseev and Kozlova, 2002, 2003a).

Using the multicomponent stellar photosphere model, which accounts for center-limb variations and rotational broadening of lines, by the spectrum inversion method Frutigeret al. (2005) studied granulation on both components of the system $\alpha$ Cen AB. While inversion, there were taken into account the profiles of ascending and descending flows, line profile bisectors, mass conservation condition, stratification of thermodynamical parameters, $g$, vsin$i$, and abundance of elements. As a result, the temperature range and rate of convective motions in $\alpha$ Cen A (G2 V) were very close to solar ones, and in $\alpha$ Cen B (K1V), at the same rate of upward and downward substance motion, the horizontal rates appeared to be lower, and this should reduce the granule sizes as compared to solar ones.

Cranmer et al. (2014) considered stellar gravitation as a source of high-frequency fluctuations on light curves derived with Kepler, distinguishing this component from the whole variability affected by the combined action of granulation, acoustic oscillations, magnetic activity, and rotation. Having applied this technique to 508 "Kepler" stars, they obtained some constraints on other components of this "noise".

Bedding and Kjeldsen (2003) reported on the detection of analogs of solar 5-minute oscillations in $\alpha$ Cen A, then 42 frequencies of small rate fluctuations, and estimated the lifetime of such individual modes as 1–2 days; this is substantially shorter than on the Sun (Bedding et al., 2004). Later, Kjeldsen et al. (2005) reported on the detection of 37 pulsation modes in $\alpha$ Cen B, on the pulsations in $\delta$ Pav centered at 2.3 mHz, and by a new way they estimated oscillation amplitudes in $\alpha$ Cen A, $\alpha$ Cen B, $\delta$ Pav, $\beta$ Hyi, and the Sun. However, using the MOST (Microvariability and Oscillations of Stars)[1] facilities, Rucinski et al. (2004) during 30.5 days carried out almost continuous broadband photometry of the other solar-type star $\kappa^1$ Cet, photometric readouts were recorded approximately per minute, but their analysis did not yield acoustic oscillations in a range of 0.5–4 mHz up to $7\text{–}9 \cdot 10^{-6}$ stellar magnitude. Teixeira et al. (2009) derived other results with HARPS[2]: they recorded pulsations of the G8 dwarf $\tau$ Cet, similar to solar 5-minute oscillations, but with twice lower amplitude and a duration of about 4 minutes; using parameters of these pulsations, they made a high-accuracy estimate of the average density of the star and, taking the spectrophotometric radius into account, found its mass to be $0.783M_{\odot}$.

[1] The MOST facility is able to conduct a continuous, for over weeks, monitoring with a frequency of several readouts per minute, with an accuracy of up to several micromagnitudes.

[2] HARPS — High Accuracy Radial velocity Planetary Searcher — mounted in 2003 on the 3.6 m telescope ESO in La Silla, a measuring instrument for radial velocities with an accuracy of up to 1 m/s, a fiber-fed echelle spectrograph with a resolution on the sky of 1″, operating in a range of 3800–6900 Å and a spectral resolution of 115000, detector is a mosaic from two pixels of 15 μm, thermostabilization is up to 0.01 °C.

Using the MOST facility, Gai et al. (2008) studied the diffusion of helium and metals in the $\varepsilon$ Eri star: from the models with different input parameters they searched for the agreement with global stellar parameters — effective temperature, luminosity, size, and metallicity, found a great frequency sharing of 194 ± 1 μHz, an age of about 1 billion years, and concluded that the helium diffusion had a little effect on the inner stellar structure, whereas metal diffusion led to a higher temperature of the radiative core, higher speed of sound, higher frequencies, and their differences.

Menard et al. (2002) detected a linear polarization of M9, L3.5, and L8 dwarfs of up to 0.2%. Their closeness excluded an impact of interstellar polarization and led to higher probability of the presence of dust in the atmospheres of very cool stars, and the revealed dust disks near active red dwarfs made one relate with more attention to the role of dust in such objects (Schuetz et al., 2004; Liu et al., 2004).

Zapatero Osorio et al. (2005b) carried our polarimetric observations of 45 M4.5–L7.5 dwarfs and in 10 L and one M dwarfs detected a significant polarization attributed to uneven coverage of stellar photospheres by dust clouds.

Pandey et al. (2009) detected strong and variable radiation polarization of the active dwarf LO Peg in BVR bands and attributed it to scattering in the optically thin circumstellar envelope or in prominence-type structures.

Patel et al. (2016) carried out multicolor observations of linear polarization for 43 late active dwarfs and found a systematic decrease in its average values toward long waves: 0.16 ± 0.01, 0.080 ± 0.006, 056 ± 0.004, and 0.042 ± 0.003% in the B, V, R, and I bands. For three objects, the polarization was caused by the interstellar medium, for other ones — by magnetic amplification and scattering. A polarization mechanism was defined on the spectral index in the power representation of dependence of polarization degree on the wavelength, as it was done earlier by Huovelin et al. (1988). Apparently, only scattering could not interpret the observed polarization, whereas for V538 Aur only magnetic amplification was enough; in most objects both mechanisms worked. However, in some objects an additional polarization source may exist, and the magnetic amplification requires fields with a strength of 0.5–3.0 kG. Confirming the results by Huovelin et al. (1988), Patel et al. (2016) found an increase in the linear polarization degree with growing stellar magnetic activity, and this effect was more pronounced in the B band.

## 1.2.2. Stellar Rotation

A general pattern of rotation of active stars is fairly diverse and is to be further discussed on repeated occasions due to various manifestations of the considered activity and within gyrochronological estimates of stellar age. In this section, within this subject, we provide only general notions on stellar rotation and outline a number of illustrative examples.

Quiet M dwarfs are the most slowly rotating main-sequence stars. Their axial rotation rates do not usually exceed 2 km/s (Marcy and Chen, 1992). But the axial rotation rates of flare stars, as a rule, noticeably exceed this value: for 40% single UV Cet-type stars these are close or somewhat higher than 10 km/s. Out of 29 dMe stars considered by Stauffer and Hartmann (1986), 11 stars have rotation rates of more than 10 km/s, whereas out of 170 dM stars only one or two have such a fast rotation. However, the rotation rates can be considerably higher in nonsynchronized pairs. For the dM1.5e star Gl 890, whose binarity had not been established, the rotation rate is 70 km/s (Pettersen et al., 1987), for the K2 dwarf BD+08°102, which seems to form a pair with the white dwarf, vsin*i* ~ 90 km/s (Kellett et al., 1995). Following Bopp and Fekel (1977a) and Bopp and Espenak (1977), the strong chromospheric emission and spottedness arise at rotation rates of more than 5 km/s. (This criterion is consistent with the statement of Yang et al. (1987ab) on the fact that all M dwarfs are the components of binary systems with a period of less than 5 days and are emission stars.) Later, Jenkins et al. (2009) increased this critical value up to 7 km/s. But this is not a strict criterion: there are known emission dwarfs with a rotation rate of about 4 km/s (Torres et al., 1985). The stars with rotation rates of less than 1–2 km/s do not burst or have a very low activity level (Pettersen, 1991). However, if there is a fair correlation between the activity level and rotation rate for G and K dwarfs (Hartmann and Noyes, 1987), then for M dwarfs this correlation disappears or substantially weakens. Thus, for the flare star Proxima Cen vsin*i* = 0.5 km/s, for Gl 890 vsin*i* = 70 km/s, and values $\log(L_{H\alpha}/L_{bol})$ characterizing emission of the quiet chromosphere differ only by 0.4 for this stars. Note that a small value of vsin*i* for Proxima Cen is not associated with a small value of the angle *i* but indeed is due to its slow rotation because, according to Benedict et al. (1993), the axial rotation period is 42 days and, following Wargelin et al. (2010), a period of 83 days seems to be preferable one. Using the high-resolution spectra, out of 99 studied red dwarfs for 24 of them Delfosse et al. (1998) detected vsin*i* > 2 km/s. All these stars appeared to be later than M3.5 and, based on kinematic characteristics, younger than old disk populations. Following Basri (2001), after M5 there grows a fraction of fast rotating stars that approaches 100% for M9. Later, this statement was clarified.

Fast rotation is one of the basic global characteristics of stars with solar-type activity, since it is a finite energy source for the generation of stellar magnetic fields and the whole further sequence of phenomena of the considered activity. The evolution of stellar rotation generally proceeds from the fast rotation of young objects triggered by the compression of original gas and dust protostellar structure to the very slow rotation of old stars. This mainline direction of evolution is superimposed by the stellar binarity effect, which slows down the secular deceleration of rotation for components in a pair. Simple correlations between rotation and activity, being observed at relatively low rotation rates, at higher rates are superimposed by the saturation effect, when starting from some rotation rate and with its further growing the activity no longer increases and remains at the attained maximum level. The saturation rate for M dwarfs is 5 km/s, for K dwarfs — 10 km/s, and for G dwarfs —15 km/s, and the saturation level for M dwarfs is $L_X/L_{bol} \sim 10^{-3}$. The fastest M9.5 rotator BRI 0021-0214 has a rotation period of less than 3 hours and the 20-fold exceeding of the critical rotation rate, but has no emission in the $H_\alpha$ line; this phenomenon of decreasing activity at the highest rotation rates was named supersaturation. All the listed effects significantly depend on the global stellar parameters and on the observational range.

A pivotal role of rotation in the considered activity of lower main-sequence stars appears to be evident when performing a systematic comparison of $v_{rot}$ with a level of the permanent chromospheric and coronal emissions; this will be performed in the following chapters of the book. The data exhibiting the wealth and diversity of this characteristic of active stars are to be provided.

* * *

From the analysis of M dwarfs in the Pleiades and Hyades, Terndrup et al. (2000) conclude that for older Hyades the average rotation rates of stars with a mass of about $0.4M_\odot$ are 2.5 times lower than those in Pleiades, and this difference is responsible for a lower luminosity of chromospheres and coronae in the Hyades.

With the aim of studying an association of emission lines of the chromosphere and transition zone, Ambruster et al. (2000) examined spectra, taken with the Hubble telescope, of 6 single K0–K6 dwarfs with a primordial enhanced Li abundance, with proper motions of a group of the Pleiades and with rotation periods from 8 hours to 7 days. The rotation–activity ratios were constructed for all lines with a fairly high *S/N* ratio. Lines of the transition zone in spectra of Speedy Mic and AB Dor exhibited saturation.

Pizzolato et al. (2003) considered the impact of X-ray emission saturation on the correlation between stellar activity in this range and rotation. They explored 110 stars of the field and 149 stars from five open clusters with the detected by the ROSAT orbital telescope X-rays and known rotation periods and distinguished two radiation modes: in the first one the rotation period predicted well the general X-ray luminosity, in another one — a constant ratio of the saturated X-ray radiation to the bolometric luminosity was achieved. For different masses Pizzolato et al. estimated the critical rotation periods for proceeding from one mode to another one and found the characteristic time for X-ray Rossby numbers.

Mohanty and Basri (2003) derived the high-resolution spectra of a large sample of middle M and L dwarfs and found an increase of the rotation rate in this range toward cooler stars, and no slow rotators were found among L dwarfs (Basri, 2000); this might be due to an increase in the braking time of lower-mass stars or due to a shift of L dwarfs toward younger objects. Based on the measured values of $F_{\mathrm{H}\alpha}$ or $L_{\mathrm{H}\alpha}/L_{\mathrm{bol}}$ in the range of M4–M8.5, the saturation was observed as well as in earlier types, but in M5.5–M8.5 it occurred at $\sim$ 10 km/s, whereas in M4-M5 — at < 4 km/s. For M9 and later dwarfs, a dramatic decay of activity was observed, the rotation-activity ratio was disturbed, the $\mathrm{H}_\alpha$ emission level was far lower than that in earlier types, and this emission was not observed in the predominantly neutral and dust atmosphere.

At a given equivalent width, the $\mathrm{H}_\alpha$ flux of $F_{\mathrm{H}\alpha}$ rapidly drops from middle to late M dwarfs. The question as to whether the rotation has effect on the activity of M dwarfs may be answered as follows: the rotation so substantially affects the middle M dwarfs that if there is no activity, then it is a slow rotator.

Bailer-Jones and Mundt (1999) detected a periodic photometric variability of the L dwarf 2MASSW J1145572+231730 with $W_{\mathrm{H}\alpha}$ = 4.2 Å, a period of 7.2 hours and an equatorial velocity of 17 km/s and associated this modulation with magnetic formations on the stellar surface or with nontransparent dust clouds in the atmosphere, since the dust is actively formed at $T$ < 2000 K. Later, Bailer-Jones and Mundt (2001) carried out observations of 21 late M and L dwarfs of the field and in the Pleiades and $\sigma$ Ori clusters, for 11 of them detected variability from 0.09 to $0.055^m$ in the I band. For a fraction of these stars, a periodic variability was detected with periods from 3 to 13 hours; those with no periodicity were referred to stars with fast evolution of surface structures. Bailer-Jones and Mundt associated the circumstance that variability is more characteristic of L dwarfs than of late M dwarfs with active dust-formation in cooler stars. On the other hand, following Mohanty et al. (2002), at the reasoned speeds of sound and convective vortices, the length scale for the generation of the exciting field (which will later dissipate and heat the atmosphere) is fairly large. Thus, the neutral atmosphere is a very likely reason for decreasing activity for M9 and later dwarfs, independent of rotation.

Bailer-Jones (2004) measured the rotation rates for 16 M9–L7.5 dwarfs with VLT/UVES and found them in the range of 10–40 km/s, i.e., all of them were fast rotators. For the L1.5 dwarf 2M 1145+2317, there was estimated a rotation period of 11.2 ± 0.8 hours, a lower radius limit of $0.1R$, and age of more than half a billion years. A period of 2. 7 ± 0.1 h was confirmed for the dwarf 2M 1334+1940. In cases when the variability and periodicity had not been found, Bailer-Jones attributed it due to a small contrast of surface structures and in three cases of nonperiodic variability — due to nonuniformity of dust clouds.

Lane et al. (2007) detected a periodic variability in the I band of two very cool dwarfs located on either side of the boundary M/L: M9 stars TVLM 513-46546 with a period of 1.96 h and L3.5 stars 2MASS J00361617+1821104 with a period of about 3 h. Their radio emissions had the same periods. Therefore, an assumption was put forward that this variability may be produced by strong local magnetic fields, which may interact with evolving dust clouds. An existence of such clouds, partly overlapping the atmosphere, was confirmed by Littlefair et al. (2008) from observations of the fast rotator TVLM 513–46546 with VLT using the high-speed three-channel photometer ULTRACAM[1] (ULTRA fast CAMera): in the course of these observations, an anticorrelation of green and IR light curves was detected; this contradicts to the standard starspot model but may be agreed with dust structures. Earlier, they derived the analogous results during observations of the ultracool dwarf 2MASS 1300+1912 with the photometer ULTRACAM in the green band and in the narrow filter centered at the Na I line (Littlefair et al., 2006).

Reiners et al. (2007b) described anomalous activity in the eclipsing pair of brown dwarfs 2MASS J05352184–0546085 in which the primary component was cooler than the secondary one, but it rotated twice faster than the secondary component and had 7 times stronger $H_\alpha$ emission; an assembly of these properties may be figured out assuming a strong magnetic field generation on the primary component, which suppresses the convective transport, reduces the stellar temperature, and excites strong $H_\alpha$ emission.

Applying high-dispersion spectra, Reiners and Basri (2008b) studied the rotation of 45 L dwarfs and found rapidly and slower rotating objects. From $H_\alpha$ emission they estimate the chromospheric activity and conclude that the magnetic braking is a basic thing in evolution of the angular momentum for brown dwarfs as well, although the breaking time grows as the stellar mass decreases. Activity weakens as the temperature decreases, and all active objects are variable. For L dwarfs, there is no rotation–activity dependence, and the minimum rotation rate increases toward later spectral subtypes, rising from about zero for old M dwarfs of middle subtypes up to 20 km/s for middle L dwarfs. The breaking law depends on the temperature or mass, and this may interpret all results concerning rotation and yields a dependence of the angular momentum on age. A mechanism of the angular momentum loss becomes ineffective for less and older bodies.

Considering the coronae of young dwarfs AB Dor, Speedy Mic, and Rst 137B, being at the evolutionary phase that is characteristic of early post T Tau with maximum rotation, and the literature data on 22 other sources, Garcia-Alvarez et al. (2008) discussed the boundary between saturation and supersaturation. Based on the high-resolution X-ray spectra taken with Chandra, they considered characteristics of coronae as a function of $L_X/L_{bol}$ and temperature and found that this boundary occurred at the coronal temperature $T >\sim 10^7$ K and the coronal iron abundance revealed an inverse correlation with $L_X/L_{bol}$, slowly decreasing as this ratio grew and rapidly dropping when $L_X/L_{bol} >\sim 3 \cdot 10^{-4}$. Besides, they identified a growth of the [O/Fe] ratio as the Rossby number decreased and the achieved by this ratio value 0.5 on the boundary with supersaturation. Later, Garcia-Alvarez et al. (2011) considered the spottedness of two fast rotators in the moving group of $\beta$ Pic, HD 199143 (F7V), and CD-64°1208 (K7V), and estimated their rotation periods of 0.356 and 0.355 days, respectively.

[1] ULTRACAM — a portative high-speed photometer designed to study faint astronomical objects at high temporal resolutions (Dhillon et al., 2007). Owing to two dichroic filters and three CCD cameras, it simultaneously records three channels at an operating frequency of 500 Hz.

Jenkins et al. (2006) compiled the catalog of active southern stars and then, as already mentioned above, clarified the rotation rate for M dwarfs at which, as a rule, $H_\alpha$ emission appears (Jenkins et al., 2009). They carried out a comprehensive study of rotation rates for M dwarfs using the high-resolution spectra of 56 such stars. Taking the used literature data into account, they involved 80% of middle M dwarfs, confirmed a slow growth of rotation rates with decreasing effective temperature, found maximum in the distribution of v sin*i* of about 3 km/s and detected a variation of this distribution between early and late M dwarfs, which was associated with either variation in the field topology between partially and fully convective stars or the occurrence of dust at the temperatures lower than 2800 K in stellar atmospheres. The considered sample comprised 198 objects with v sin*i* < 10 km/s and 43% of it comprised $H_\alpha$ emission. Throughout a month, a spectrum of the GJ 1253 star varied from purely absorption to the one with a clear emission peak in this line.

Barnes (2010) suggested a nonlinear model of evolution of rotation of main-sequence stars, which comprises the Rossby number and two dimensionless constants extracted from solar data and data on open clusters. The model involves two limiting cases of rotation that are compatible with sequences of rapidly and slowly rotating stars in young open clusters; in each case there are different dependencies on mass and age, and the model describes evolution from one family to another through a separating gap; here the rate of rotation period variation is maximum, about 0.02 days per million years; this exceeds the present-day solar braking by a factor of 7. The time required for achieving the indicated gap is figured out within the model, with an initial rotation period of 0.12–3.4 days it comprises up to 180 million years for solar-mass stars and 2–4 billion years for stars of $0.3M_\odot$. The dispersion of initial rotation periods may add an error of up to 128 million years to these gyrochronological estimates.

Houdebine (2010a) carried out the high-precision, up to 0.3 km/s, measurements of rotation rates for 10 M1e and 90 M1 dwarfs, confirmed the previously detected by him bimodal distribution of these rates with average values of *P*/sin*i* 4.45 and 14.5 days and found that fast rotators were mainly M1e and several anomalous M1 stars. In the analogous investigation of 12 dMe4 and 9 dM4 stars, Houdebine (2012) detected a large fraction of fast rotators among them than among M1 dwarfs, for dM4 the rotation periods decreased with decreasing stellar radii. Having measured v sin*i* for 88 dM0 stars, one dM0e star, and one dM0(e), Houdebine and Mullan (2015) found that among such stars there were few fast rotators and far less than among dM2 stars. Due to a small number of fast rotators, the distribution of *P*/sin*i* was not bimodal like for dM2.

Wright et al. (2013) found a strong dependence of the X-ray luminosity on the Rossby number: $L_X/L_{bol} \propto \mathrm{Ro}^{-2.70}$.

Jackson and Jeffries (2010) studied the chromospheric activity for M dwarfs of the young open cluster NGC 2516. They found a noticeable growth of the fraction of fast rotators toward later stars: for 20% of M0–M1 stars v sin*i* > 15 km/s, whereas 90% of M4 dwarfs satisfied this condition. Activity indices extracted from observations of the CaII infrared triplet show a different dependence on the rotation period for partially and fully convective stars: for K3–M2.5 the chromospheric activity grows with decreasing Rossby number and achieves saturation at Ro < 0.1, whereas for cooler stars, which seem to be fully convective, and almost for all stars with Ro < 0.1 one can clearly see a decrease of the chromospheric activity toward cooler stars with its weakening by a factor of 2–3 in the interval from M2.5 to M4. But this activity weakening at small Rossby numbers is not seen in X-rays. No evidence for supersaturation in the chromosphere of any spectral types at Ro < 0.01 is detected. Jackson and Jeffries came to the conclusion that variations on the boundary of partially and fully convective stars were due to variations of the general magnetic field topology rather than a decrease of the dynamo-generated magnetic flux.

Jeffries et al. (2011) analyzed XMM–Newton observations of 97 rapidly rotating M dwarfs of the young open cluster NGC 2547 and found that for these stars, as well as for G and K dwarfs, the coronal activity grew with decreasing Rossby number, and activity was saturated at the level of $L_X/L_{bol} \sim 10^{-3}$ at logRo < −0.8. But supersaturation was not detected for M dwarfs, although many of the considered objects had logRo < −1.8. At such a Rossby number, there was a supersaturation for more massive stars: a critical value of 0.3 days for K dwarfs and 0.2 for M dwarfs.

Irwin et al. (2011) detected rotation periods for 41 fully convective M dwarfs of the field in the range of values from 0.28 to 154 days; most objects from this sample were referred to the thick disk population or halo, and their mean axial rotation period was 92 days.

Cook et al. (2014) identified 38 fast rotators — ultracool dwarfs with X-ray emission by two orders of magnitude lower than the saturation level; this is a manifestation of supersaturation at the anticorrelation between rotation and activity. A divergence of X-ray activity at the fixed rotation rates for these objects was by a factor of 3 greater than that for stars of earlier types. They suggested that the centrifugal throw-off of the corona did not occur and put forward an assumption that an additional parameter, governing supersaturation, was the presence of two dynamo modes yielding different magnetic topologies; and this provided a dispersion of X-ray emission.

Reiners et al. (2012) compiled a catalog of rotation and activity for a sample of 334 M0–M4.5 stars, i.e. on the boundary between partially and fully convective objects, extracted high-resolution spectra of 206 objects and determined their rotation rates, vsin*i*, and a value of $H_\alpha$ emission. They confirmed a steep growth of the fraction of active stars and an increase in the rotation rate on the boundary with fully convective stars; in the obtained sample all significantly rotating stars were active and all active stars considerably rotated. They did not observe any evidence for the transition from a rotationally dominated dynamo in partially convective stars to a rotation-independent turbulent dynamo in fully convective objects, and there was a unified rotation-activity ratio on both sides of the boundary of partially and fully convective stars.

Olmedo et al. (2013) studied rotation periods for 36 G stars based on Mg II h + k lines and found that variations of magnesium lines were 2.5 times greater than those of calcium lines.

Particular attention in studies of rotation is deserved by M dwarfs due to the transition to fully convective structures around M3.5.

Having measured vsin*i* for 72 dM3 and 8 dM3e stars, Houdebine and Mullan (2015) concluded on the peculiarities of rotation for M3 dwarfs. They found that the distribution of *P*/sin*i* for dM3 stars was substantially different from that for dM2 and dM4 stars, and the average rotation period of slow rotators of dM3 was 25.8 days, whereas for dM2 and dM4 it was 14.4 and 11.4 days, respectively. About dM3 the dynamo efficiency seemed to grow; this might be associated with a transition to fully convective structures. Having enlarged the sample to K4 dwarfs and added X-ray observational data into consideration, Houdebine et al. (2017) found that for M3 and M4 dwarfs there was no simple rotation–activity dependence, and different dynamo modes were required for high and low activity. They investigated the dependence of *S*, characterizing calcium emission of the chromosphere (see Subsect. 1.3.1.1), on the Rossby number and found that for K6, M2, M3, and M4 dwarfs, at a given Rossby number, the value of *S* was 3, 10, 20, and 90 times less than that for F–G–K dwarfs and concluded that the heating of the chromosphere by the corona increased by a factor of 100 between K4 and M4.

Using the 8.2-meter Subaru Telescope, Nogami et al. (2014) carried out a spectral investigation with a resolution of about 80 000 and $S/N \sim 70$ for two G dwarfs KIC 9766237 and KIC 9944137 on which in the course of Kepler observations flares with energy of about $10^{34}$ erg were recorded. As the extracted spectra lead to the parameters of these stars — their temperatures, rotation rates and rotation periods, strengths of photospheric magnetic fields, enhanced lithium abundance, spottedness degree, very close to the solar ones, one should admit that the slowly rotating solar-type stars undergo such powerful flares, or there are invisible satellites of the late spectral type, although no evidence for binarity was detected from spectral data.

West et al. (2019) considered spectroscopic observations and light curves of 238 close M dwarfs with available TESS data with the aim of searching for ratios between magnetic activity (on $H_\alpha$ emission), rotation period $P_{rot}$, and age of stars; the application of TESS data provided noticeable advantages as compared to earlier analogous studies, as TESS yields hundreds of M dwarfs with $P_{rot}$ in the range from less than a day to more than 100 days, whereas only fast rotators were previously detected from spectral estimates of $P_{rot}$. As a result, West et al. found that there was an association of magnetic activity with rotation for all spectral subtypes of M dwarfs. For early M dwarfs, the fraction of active ones decreased with increasing $P_{rot}$, and for late ones this fraction extended toward slower rotators and then abruptly dropped. All fast rotators were magnetoactive, but the signatures of activity were detected for several late and slow rotators. M1–M4 dwarfs with $P_{rot} < 26$ days and M5–M8 with $P_{rot} < 86$ days were magnetoactive: the magnetic activity level dropped with increasing $P_{rot}$ for early M dwarfs but persisted for fully convective stars up to $P_{rot} > 90$ days. This difference leads to an assumption that mechanisms of the interaction between the stellar rotation and the heating of the chromosphere are different in partially and fully convective dwarfs. According to the kinematic analysis, M dwarfs with prolonged $P_{rot}$ belong to older population, therefore it is reasonable to assume that they had higher rotation at earlier evolutionary stages and, consequently, a significant magnetic field due to which this rotation decreased with time.

* * *

The differential rotation value is important for the considered solar-type activity, aside from axial rotation velocity. This phenomenon was detected on the Sun by Scheiner in the XVII$^{th}$ century. Following up-to-date data, the rotation period of equatorial solar regions is 25.38 days, and that of near-polar latitudes is 34.3 days. For the differential solar rotation, there is a ratio

$$\Omega(\phi) = \Omega(0) - \Delta\Omega\sin^2\phi \text{ and } D_r = \Delta\Omega/\Omega(0),$$

where $\Omega(\phi)$ is the angular rotation rate at the latitude $\phi$, $\Omega(0)$ is the velocity at the equator, $\Delta\Omega$ is the difference of velocities at the equator and at the pole, $D_r$ is the differential rotation coefficient; for the Sun $D_r = 0.19$ and the absolute drift velocity of the spot-forming band is 3–4° per year.

An analogous phenomenon was detected on stars, and its investigation was started in the late 1980s (Hall, 1991). The solar-type rotation was shown for 84 binary and single stars of different spectral types, i.e., the rotation of equators was faster than that of near-polar regions; there was shown a decrease of this effect for rapidly rotating objects. Hall estimated coefficients $D_r$ and found their statistical correlation with axial rotation periods. These studies were stepped up since the beginning of space photometry. Moreover, there were also detected stars with antisolar differential rotation, when the rotation of circumpolar regions occurs faster than that of equatorial ones. A change of the drift directions seems to occur at some critical spectral type. According to the review by Alekseev (2019), the value of $D_r$ in modulus lies within the range of 0.01–0.09, and the drift velocities of the spot-forming bands are 0.2–2.9° per year.

Collier Cameron et al. (2002) and Petit et al. (2003) developed an algorithm for determining the differential rotation of the stellar surface based on a series of Doppler maps on which one can trace the migration of individual large spots. The application of this algorithm to the four sequences of echelle spectra of AB Doradus showed the possibility of determining velocity amplitudes and rotation periods for a great number of individual spots at low and middle latitudes. Particularly, the equator was found to be ahead of polar regions by one revolution approximately for 110 days. Small spots were detected to have a great dispersion of about an average ratio for large spots due to the effect of supergranulation turbulent fluctuations. Then Petit et al. (2004) included into analogous consideration the data derived by the Zeeman-Doppler imaging technique (see Subsect. 1.2.4.3), analyzed observations of young stars AB Dor, LQ Hya, PZ Tel, and RX J1508.6-4423 and in all cases found a sign of differential rotation that was coincident with the solar one and equators that surpassed polar regions by one revolution for 40–110 days, whereas for the Sun this value was 120 days. Having compared periodic brightness variations for LQ Hya over three epochs, Yu (2007) estimated the differential rotation of this star as $\Delta\Omega/\Omega = 0.025$. Lanza (2006, 2007) constructed a model for transiting the angular momentum in the convective region of the rapidly rotating star, based on series of AB Dor and LQ Hya revealed the variability of their equatorial velocities and surface differential rotation and within the mean-field theory (see Chapter 4.3) acquired notions on the magnetic field strength in their convective regions.

Based on high-resolution spectra, Reiners and Schmitt (2003a) determined the differential rotation for 32 F–K stars and found that it was more characteristic of slowly rotating stars, although there were deviations from the solid-body rotation for some fast rotators. They acquired indications against a significant differential rotation for the most active stars. Based on the high-quality spectra for 135 F and later stars, they found that the broadened profiles for 70 of them with v$\sin i$ > 45 km/s showed no effects of multiplicity and spottedness. The overwhelming majority of profiles corresponded to the solid-body rotation. This means that the differential rotation decreases for stars with v$\sin i$ >~ 50 km/s (Reiners and Schmitt, 2003b).

Kitchatinov and Rüdiger (2004) associated the antisolar differential rotation with the fast meridional flow, and Browning (2011) found that if the rotation rate of the fully convective star was high, then a solar pattern of the differential rotation emerged, but at the very low rates there appeared an antisolar pattern.

Barnes et al. (2005b) studied a dependence of the rotation rate of spots on the latitude in a homogeneous sample of young fast solar-type rotators. They considered 10 stars of spectral types G2–M2 and found a tendency for decreasing surface differential rotation as the effective temperature decreased. The approaching to solid-body rotation with increasing relative size of the convective region means that the dynamo mechanism acting in low-mass stars may substantially differ from the solar one. On the other hand, Reiners (2006) measured rotation rates for 147 F and later stars and for 28 of them found differential rotation but did not detect it for A stars, and associated the emergence of differential rotation with that of the convective region.

By means of the photometric satellite MOST, Croll et al. (2006) carried out continuous observations of $\varepsilon$ Eri during three revolutions and detected two spots with $\Delta m \sim 0.01^m$ at latitudes 20.0° and 31.5° with rotation periods of 11.35 and 11.55 days. Based on these data they estimated a differential rotation coefficient of 0.11, which corresponds to theoretical predictions for a solar-type star having a twice greater angular velocity. The inclination angle of the rotation axis was estimated to be 30°; this corresponds to estimates for the disk and planet orbit and leads to an equatorial velocity of 3.42 km/s. Using the same facility, throughout 5.8 days of observations for AD Leo Hunt-Walker et al. (2012) detected a 2.23-day rotational modulation of stellar brightness.

Kueker and Ruediger (2007) suggested a model of stellar differential rotation based on hydrodynamics of the mean field, which is strongly dependent on the spectral type and weakly dependent on the rotation rate. The calculations showed a strong differential rotation for F stars near the upper boundary of stellar convective regions with a short turnover time of convective vortices.

Following the review of Collier Cameron (2007), for the rapidly rotating young stars of spectral types from F to M of the main sequence, the differential rotation significantly increases as the effective temperature grows, and the tidal interaction suppresses differential rotation.

In the course of three-year observations with the MOST facility, Walker et al. (2007) studied differential rotation of the solar-type star $\kappa^1$ Cet and found its character to be fairly close to the solar one: spots on the star were localized in the range of latitudes 10–75°, the rotation period at the stellar equator was 8.77 days, the differential rotation coefficient was 0.09; this is less than the solar values but corresponds to a stellar age of 750 million years.

Jeffers and Donati (2008) studied the young fast rotator HD 171488, recording the Stokes I and Stokes V data with the spectropolarimeter at the Bernard Lyot Telescope. They found a strong and slightly decentralized polar structure that was greatly superior to high- and low-latitude spots, and the large-scale magnetic field topology showed a strong ring of the azimuthal field with dependence of polarization on the latitude and large regions of the radial field with negative polarity at all latitudes. An estimate of differential rotation parameters showed that the equator exceeded the pole for 12 days; this was then the maximum recorded differential rotation. Taking subsequent observations of this star into account and based on three observational epochs, Jeffers et al. (2011) underscored the predominance of polar and high-latitude spots and the large-scale topology with radial and azimuthal magnetic field components. In the course of these observations, no polarity reversal was observed as it was for the solar-type stars, and there were no variations in differential rotation.

Using the extensive HATNet data on studying exoplanets, Hartmann el at. (2010) explored photometric rotation periods for 368 stars in the Pleiades with masses between 0.4 and $1.3M_{\odot}$. They detected periodic variations for 74% of cluster members of these masses in the field of view and for 93% of cluster members with masses between 0.7 and $1.0M_{\odot}$. For stars with masses of more than $0.85M_{\odot}$ the extracted estimates of periods, vsin$i$, and radii were consistent with isotropic distribution of rotation axes and with the established scheme of differential rotation, but for lower masses sin $i$ systematically proved to be more than unity. Possibly, the scheme for inner transfer of the angular momentum for slowly rotating stars needs to be specified.

Do Nascimento et al. (2010) considered masses and age of convective envelopes for 117 solar analogs, their rotation rates, and lithium abundance and found no straightforward correlations between these values. Following their conclusion, the standard model of matter mixing in the convective region and lithium burn-up at the bottom of this region is insufficient for understanding observations and should be complemented taking differential rotation into account.

Kitchatinov and Olemskoy (2011) elaborated a computational model for stellar differential rotation based on hydrodynamics of the mean field, successfully applied it to the two mean-velocity rotators and two fast rotators, and estimated a series of models for stars of different masses and different metallicity with a rotation period of 1 day. Moreover, Kitchatinov and Olemskoy (2012) noticeably decreased the number of input parameters in the theory of differential rotation, leaving only two most important ones: surface temperature and rotation period. They confirmed a tendency for growing differential rotation as the temperature increases, whereas for F stars this growth was far steeper than for G and K stars. With increasing temperature and rotation rate, the meridional flow amplitude increased, for fast rotators it concentrated at the boundary layers of the convective region.

In the report "Differential rotation in theory and observations" Czesla et al. (2013) summarized the development of this direction and formulated concise issues for forthcoming investigations:

– to what extent the differential rotation is everpresent;

– whether it is variable over time, if it is so, then what are its time scales and amplitudes;

– whether the antisolar differential rotation is real;

– how strongly magnetic fields surpass differential rotation.

At the same time Kitchatinov (2013), having established the association between the differential rotation and the meridional flow in stars, came to the conclusion that the rate of such a flow was maximum in upper and lower boundary layers of the convective region, and the thickness of these layers decreased with increasing rotation rate. The differential rotation was produced by convection and meridional flows, transporting the angular momentum, and for its emergence the temperature difference at the pole and equator was important, which was due to the rotational anisotropy of convective transfer; the differential rotation was the most sensitive to this anisotropy. Convective flows of the angular momentum were directed radially inside at the slow rotation but changed their direction to the equator and parallel to the rotation axis with increasing rotation rate. The constructed by Kitchatinov model predicts some variations of differential rotation of the star of a given mass with the evolutionary rotation braking and its significant increase for high-mass stars.

At the Cool Stars 20 conference Metcalfe (2018) noted that the Kepler data on open stellar clusters confirmed the stellar-rotation braking and the weakening of stellar activity up to 2.5 billion years, but later, the angular momentum seemed not to decrease due to the termination of magnetic braking, although the chromospheric activity continued to weaken even at constant rotation; whereas the activity cycle duration smoothly increased till its extinction.

Using the observations with Kepler, Nielsen and Gizon (2018) explored 3901 stars and found that the rotation rate commonly anticorrelated with the activity cycle, like in the last phase of the solar cycle. The anticorrelation degree was the highest at the rotation period close to the solar one but practically disappeared for fast and very slow rotators. For the solar-like rotators, the rotation rate variations identified the asymmetry between the start and end of the cycle, like in solar butterfly diagrams.

From the Kepler light curves Popinchalk et al. (2021) studied the rotation periods of 713 M0-M8 dwarfs of different age and found the transition time from fast to slow rotation to be dependent on the spectral type: the redder and less massive objects remains active longer.

From the Gaia DR3 data, Distefano et al. (2022) compiled a catalog of 474,026 stars from the light curves of which they detected periodic oscillations attributed to the effects of dark spots and bright faculae, i.e., to the manifestations of stellar magnetism. About 430,000 cataloged objects are new discovered variables. The catalog provides 66 parameters for each star among which the most important are the rotation period, the photometric amplitude, and the Pearson correlation coefficient between brightness and color magnitude variations. Using this coefficient made it possible to separate the objects considered into those in which brightness variations are due to dark spots and those in which bright faculae are responsible for these variations.

### 1.2.3. Starspots

Dark spots on the Sun, the first manifestation of solar activity, were probably detected by the priests of ancient Babylon with the naked eye. Sunspots were mentioned in Chinese, Japanese, and Korean medieval chronicles. Scientific investigation of sunspots was started with the invention of telescopes in the early 17$^{th}$ century. The basic currently known characteristics of sunspots are as follows.

Sunspots are the regions where strong local magnetic fields up to several kilogauss emerge on the solar surface. These fields substantially suppress convective transfer — the basic mechanism of upward heat transfer within the convective zone of the Sun. As a result, the effective temperature in the spot center is 3800 K, which is 2000 K lower than that of the quiescent photosphere, therefore sunspots look dark on its background.

The spots are surrounded by facular areas — brighter regions of the photosphere with weaker magnetic fields. The disturbed chromosphere, the regions of enhanced brightness in the CaII H and K emission lines, is located above the spots and the faculae. These features, already found during the initial optical observations of the Sun, are called active regions. They demonstrate a noticeable variation with time. The subsequent radio, UV, and X-ray observations of the Sun have shown that active regions are related to the sources of increased emission in these wavelength ranges as well. In other words, active regions include the solar atmosphere throughout its height.

Due to the occurrence of dark spots, the summary flux deficit of the solar photosphere radiation reaches $0.002^{m}$, but there is some excess in overlapping with the additional emission of facular areas. Thus, the maximum bolometric luminosity of the Sun occurs at the epoch of the activity maximum, i.e., under the heaviest spottedness. However, a decrease in the luminosity of the visible solar hemisphere was confidently recorded within the ACRIM experiment (Willson et al., 1981), when a group of spots passed over the solar disk.

At the beginning of the 11-year solar cycle the spots appear at heliographic latitudes of about 35°, by the end of the cycle the regions of their appearance drift toward approximately the 5° latitude. The dynamo theory relates this visual effect to the differential rotation of the Sun, which generates a toroidal magnetic field and provides buoying of newly formed magnetic tubes upward to the solar surface.

The size of the smallest spots, pores, is of the order of 1000 km, which corresponds to the characteristic size of photospheric granules. Since the granules are convection cells near the solar surface, their properties are closely related to the convection characteristics. Large spots can be an order of magnitude larger, while the greatest spot groups cover up to 0.006 of the area of the visible solar hemisphere and their diameter achieves 10° in the heliocentric coordinate system. The number of spots quickly decreases as their sizes increase. The lifetime of pores is several hours, while large spots can exist for up to several months.

Large spots have distinct photometric structure: they consist of a darker central part, the umbra, and a somewhat lighter periphery, the penumbra. The degree of reduction of the flux intensity $I_{shad}/I_{photosphere}$ varies considerably with the wavelength: from 0.02 at about 4000 Å to 0.6–0.7 at about 2 μm. As a spot approaches the disk edge, this ratio decreases within the whole wavelength range due to a steeper temperature gradient in the spot as compared to the quiescent photosphere.

In the spot center, the magnetic field is vertical, but as observation moves away from the spot axis, a transverse field component arises. It is still not clear which factor determines the main characteristic of a spot — its magnetic flux. The magnetic field strength and the spot size are not correlated; the field strength satisfies the condition of hydrostatic equilibrium of the magnetic flux tube and the quiescent photosphere surrounding the spot.

In 1949, Kron (1952) discovered spots on red dwarfs from the slight distortions of the light curve of the YY Gem binary system, which differed from the curve typical of eclipse systems. Both components of the YY Gem system are dMe flare stars.

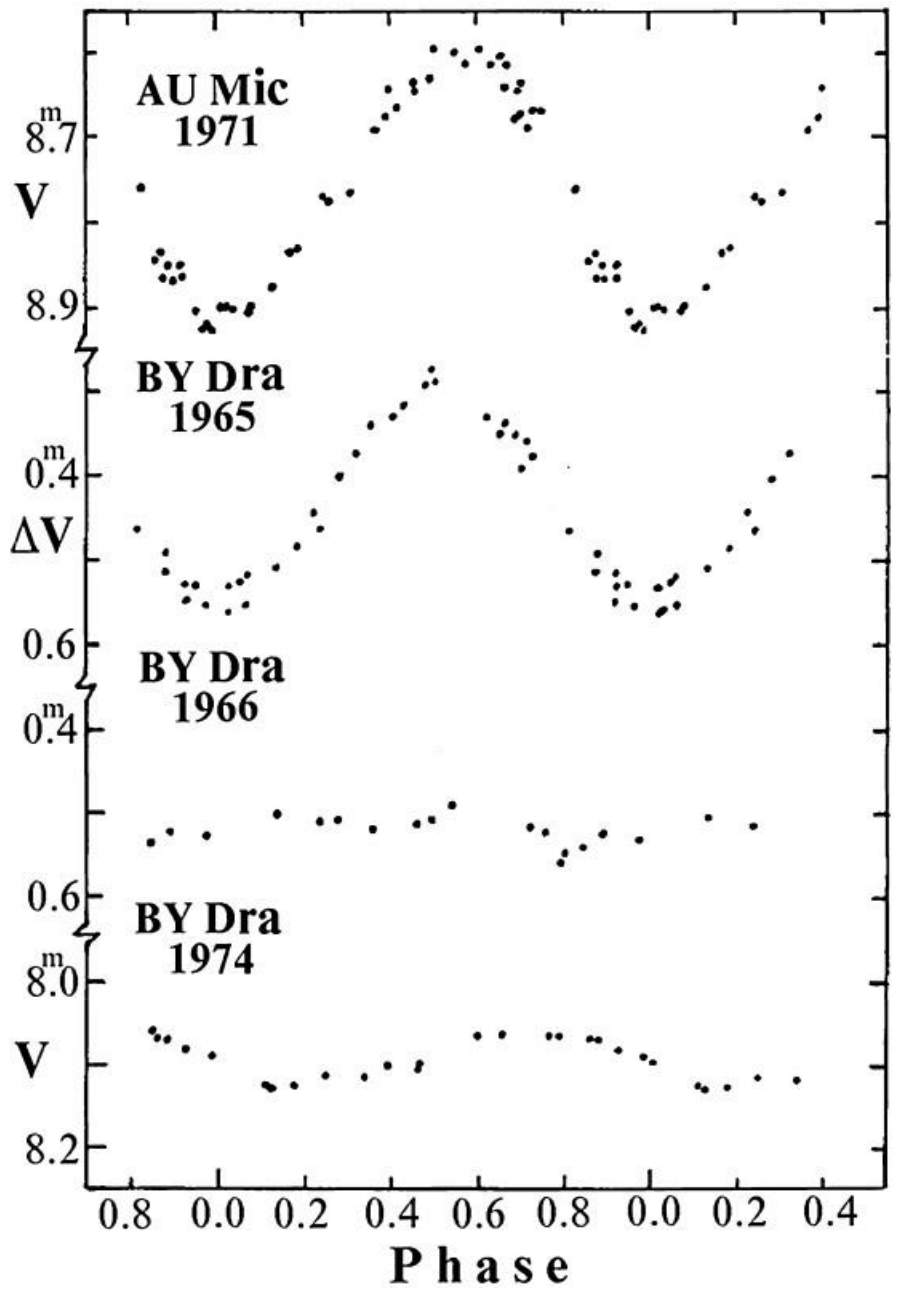

Fig. 5. Brightness variations of two red dwarfs due to their spottedness (Rodonò, 1980)

Hall (1994) stated that the idea of stellar spottedness was as old as searches for the cause of photometric variability of stars. The point is that spottedness is the first explanation of brightness variation offered by the discoverers of various types of variable stars starting from long-period variables. Pickering (1880) even proposed to explain all types of stellar variability by spottedness. Further studies soon disproved this global assumption and discredited the idea of stellar spottedness to the extent that when Kron did find the spottedness of the YY Gem system, this discovery did not impress “variable-star investigators”. Only 15 years later, when the interest in flare-emission dwarfs had become high enough and Chugainov (1966) found spottedness in the BY Dra binary system with similar components, the studies of this phenomenon became widespread. The concept of starspots on red dwarfs had indisputable advantages over the competing models of eclipses and stellar oscillations and soon became conventional. Typical light curves of spotted stars are shown in Fig. 5. The term “BY Dra-type stars” has been introduced for spotted stars, though now it is clear that it does not mean a new variability type, but a new manifestation of the activity of red dwarfs. Hall (1972) was one of the first to associate the spottedness of cool stars with the phenomena similar to solar activity.

Since stellar surfaces cannot be studied directly, the analogs of the solar activity regions and activity complexes are primarily used to analyze time variations of observed parameters and their correlations. It is obvious that in the case of surface inhomogeneities such time variations can occur only when the rotation axis of a star deviates noticeably from the line of sight. The surface inhomogeneities are investigated using spectral and photometric methods. The former requires dense series of high-resolution spectrograms for axial rotation period and a high signal-to-noise ratio (*S/N*). Thus, the method is applicable only to relatively bright objects, variable stars of the RS CVn type. Flare red dwarfs are too weak for the fine spectral analysis and, as a rule, are studied using less informative photometric methods.

Thus, the main photometric effect of red dwarf spottedness is periodical brightness changes in the optical range of wavelengths with amplitudes varying from one hundredth to two–three tenths of the stellar magnitude, which during 10–100 and more rotations of a star occur strictly synchronously with the axial rotation, while at larger time intervals the average brightness and its oscillation amplitude change sporadically, and phase shifts and slight variations of the period of these small-amplitude brightness variations occur. (Periodical brightness variations with an amplitude of $0.01^m$ were recorded during the high-precision observations of Proxima Cen with the Hubble Space Telescope (Benedict et al., 1993). Two-day ground-based observations of BD+22°4409 with an axial rotation period of about ten hours detected a change in the brightness minimum depth by $0.03^m$ (Robb and Cardinal, 1995).) Obviously all these effects can be related to the appearance, development, and decay of starspots, as well as to their drift over the surface of a differentially rotating star. At the same time, it is obvious that determining the characteristics of the surface inhomogeneity from variations of the observed integral flux is a rather ambiguous task.

As a first step in solving this problem, Krzeminski (1969) “gathered” all inhomogeneities in one spot, “decreased” its temperature by 350 K, which corresponds to two spectral subtypes, and estimated the spot size required for the observed photometric effect: the obtained value was about 10% of the stellar surface. His result gave an impetus to constructing a number of spottedness models for red dwarfs.

It should be noted that shortly before the discovery of spots on YY Gem Kron (1947) found spottedness in the AR Lac binary system. The systems have many differences: YY Gem consists of two red dwarfs of the main sequence, while AR Lac, one of the most studied of the RS CVn-type systems, consists of enhanced-brightness stars that had undergone a noticeable evolution and left the main sequence. RS CVn-type objects will not be thoroughly considered in this monograph, but since the methods for analyzing the spottedness of both types of variables were developed in parallel, they have much in common. Having studied many-year observations of close binary systems, Messina (2008) identified objects of both types in which color variations were defined by either cool spots or light faculae. In a recent comprehensive review on starspots of Strassmeier et al. (2009), the volume of data on spots for the RS CVn-type star significantly exceeded that on spots for dwarf stars.

In this Section, the results of an experimental study on starspots are outlined; the theoretical investigation of their magnetism is given in Sect. 4.3.4.

**1.2.3.1. Estimates of Parameters of Individual Starspots.** Quantitative characteristics of starspots from the observed light curve of cool stars have been the object of intense studies for about 50 years: more than 350 publications are devoted to this subject. To avoid the situation when “an extensive literature substitutes research by cross-referencing to contradicting opinions of numerous authors rather than provides answers to thrilling questions" (L.N. Gumilev), we shall mention only those that were important for conceptual progress of the studies.

Evans (1971) constructed the first theoretical light curve for a star with a spot at the equator for the case when the rotation axis of the star lies in the picture plane, while the spot boundaries are set by the sections of meridians and parallels. Reducing the problem to the two-parametric statement, he successfully presented observations of the CC Eri spotted star for each of the three observational seasons using the spots stretched along the equator for many tens of degrees.

Bopp and Evans (1973) proposed a computational layout for the light curves of spotted stars with arbitrarily oriented rotation axes and taking the limb darkening into account. Within this model, they successfully presented all the above observations by Chugainov, Krzeminski, Evans, and the data of Torres et al. (1972). To present the light curves with a continuous brightness change throughout the entire axial rotation period, Bopp and Evans used a hypothesis on large nonsetting high-latitude spots, while to explain slight variations of the color index (B−V) they used a hypothesis on the low temperature of spots. As a result, the light curves of CC Eri for three seasons and those of BY Dra for two seasons were successfully presented by 2000 K spots stretching up to the latitude of 65°, the total area varying from 4 to 20% of the stellar hemisphere. Later, using the formalism developed in this publication, Bopp and Fekel (1977b) analyzed the observational data for another flare-spotted star, FF And, and concluded that these observations could be presented by a dark-spot model ($\Delta T \sim 500$ K) with a longitudinal length of about 125° or by a "black spot" model ($\Delta T \sim 1800$ K) with a length of up to 70°.

Torres and Ferraz Mello (1973) developed the theory of light curves and color indices of stars with blackbody spots, reproduced the above observations of BY Dra and CC Eri, and employed the second dark spot to present the asymmetrical light curve of AU Mic, noting that the observations in the B and V bands were insufficient for an unambiguous selection of parameters of the two-spot model. Later, the two-spot approach was used to present a number of asymmetrical light curves of the RS CVn-type stars (Bopp and Noah, 1980).

Vogt (1975) made a correction to the Torres–Ferraz Mello model for the second component of a binary system and used the energy distribution in cooler star spectra as the energy distribution in the spot spectrum. (It is noteworthy that using comparative spectrophotometry of sunspots and late-type stars Badalyan and Obridko (1984) found that the sunspot should be attributed to M0, K5 or G8–K0 spectral types, respectively, depending on spectral details being compared, such as continuum and weak or strong lines.) Vogt developed Mullan's idea (1974) that the large longitudinal Evans spots were, in fact, the sequences of spot groups resulting from the dynamo mechanism. He noted that within this model the stellar brightness variability could be governed by differential changes in the spot belt.

Friedemann and Gürtler (1975) performed extensive calculations for theoretical light curves of a blackbody star with one blackbody round spot for various combinations of the spot radius, its position on the star, and orientation of the stellar rotation axis with respect to an observer. Using this model, they found that at equiprobable spatial orientation of rotation axes and equiprobable location of spots on the stars only for 72% of stars with the spot of radius 45° and only for 44% of stars with a spot of radius 25° the photometric effect should exceed $\Delta V = 0.1^m$.

In 1970, the light and dark hemispheres of BY Dra had the maximum brightness over 10 years of observations. Assuming that during this season the light stellar hemisphere had no spots and, following the Torres–Ferraz Mello formalism (1973), Chugainov (1976) analyzed its brightness and color index (B−V) variations over the decade and concluded that usually both hemispheres were spotted and at minimum stellar brightness the spots covered up to 60% of the area of the dark hemisphere. He explained the relatively small temperature differences between the quiescent photosphere and spots of about 400 K by the predominance of penumbrae in the spot areas.

Oskanyan et al. (1977) analyzed the two-decade photometry of BY Dra and concluded that for a period of about two years with dense observational series the stellar brightness could be presented by a train of six sinusoids with abrupt changes: average brightness of up to $0.3^m$, the amplitudes of periodical variations within $0.01^m$–$0.02^m$, the photometric period of up to 0.2 day, and the phase up to 130°. These features could be explained by disappearing old spots and new spots appearing at different latitudes, including the near-polar ones. Oskanyan et al. noticed that the estimate of the brightness of a spot-free star was extremely important for constructing the spottedness pattern; however, the values could be determined only under long-term photometric observations. It is not inconceivable that the underestimated level of the maximum brightness could sometimes lead to erroneous conclusions on the existence of hot spots.

Davidson and Neff (1977) carried out the first multicolor BVRI observations of BY Dra and concluded that their data agreed with the model of the dark spot 200–300 K cooler than the photosphere and stretched longitudinally to 200–220°.

Following Kopal's formalism proposed for the light curves of eclipse systems, Budding (1977) developed an analytical theory for constructing the light curves for stars with one round spot. Adding an automated iteration procedure, he reprocessed Evans' observations for CC Eri and together with his own observations for YY Gem presented the brightness curves of these spotted stars by estimating the optimal spot parameters and probable errors. Using the Budding formalism, Olah (1986) calculated the expected B−V color indices for the M0 dwarf considering the temperature differences $T_{\text{phot}} - T_{\text{spot}}$, the spot size, the inclination of the rotation axis with respect to the line of sight, spot latitudes, and limb-darkening coefficients. Later, the Budding ideology was further developed by Banks et al. (1991). Recently, Kipping (2012) generalized Budding's model (1977) for the analysis of high-accuracy Kepler observations of many thousands of stars, including nonlinearity of the photosphere and spot limb darkening, partial derivatives for all input parameters, evolution of depths of nonsetting spots. This model allowed one to estimate stellar rotation periods, areas of spots, the inclination of the rotation axis, and differential rotation.

Vogt (1981) noted that the above models could present the observed photometric effects only, but did not state and solve the strict inverse problem of imaging of inhomogeneous stellar surfaces. He proposed an algorithm that uncoupled the determination of temperature and spot geometry using the observations in the V and R bands and applied the Barnes–Evans relation associating the surface brightness and the color index. The Vogt algorithm yields the difference of color indices Δ(V−R) of the quiescent photosphere and a spot, as well as a certain function of the other parameters: the inclination of the rotation axis with respect to the line of sight, the limb-darkening coefficient, the size and shape of the spot. Given a number of additional assumptions, the function can be used to estimate the spot geometry. One of the assumptions implies a spot-free bright hemisphere, i.e., precise determination of the absolute maximum of stellar brightness. Assuming the simplest geometry of a single round spot, Vogt used the Budding technique (1977) to determine the optimal dimensions and positions of such spots on BY Dra and constructed the first pattern of its spottedness evolution. According to his calculations, in 1965 a dark spot of radius 30–50° was at a latitude of about 40°, then it drifted by 20–30° toward the pole, in 1970–75 it disappeared leaving small hot spots in its place, then in 1977 a somewhat smaller dark spot appeared again at about 40° latitude. From the variations of the axial rotation period Vogt estimated the velocity of differential rotation of the stellar photosphere, which was almost coincident with the appropriate value for the Sun. The estimated temperature of dark spots on BY Dra was 600 K lower than the photospheric temperature. According to Stauffer's statistical colorimetric study (1984), for the spotted stars in the Pleiades $\Delta T > 700$ K.

It should be noted that Oskanyan et al. (1977), Vogt (1981), and Poe and Eaton (1985) considered photometric effects for both cool dark and hot light spots. However, the subsequent observations steadily pointed to reddening of stars during the phase of maximum spottedness, which unequivocally evidenced the occurrence of dark and cool spots. A long-living "blue" spot found in the Walraven photometric system on one of the most active red dwarfs AU Mic is apparently an active region with strong Balmer emission or a result of numerous weak flares (Byrne, 1993a). The anticorrelation of brightness in the V band and the U−B color index of YZ CMi found by Amado (1997) could be due to the same cause.

A proper program for estimating the parameters of two spots with account of the energy distribution in the spectra of stars of different spectral types was developed at Villanova University (USA). The program was used to analyze a number of RS CVn-type stars with asymmetrical light curves (Dorren et al., 1981; Guinan et al., 1982; Dorren and Guinan, 1982b).

La Fauci and Rodonò (1983) developed a program to calculate theoretical light curves for stars with two different arbitrarily located blackbody round spots. They successfully presented the recorded light curves of the RS CVn-type star II Peg; the curves were often asymmetrical and appreciably changed from one season to another. First, the program calculates a dense grid of theoretical curves in the total space of the sought parameters, then using an iteration procedure it restricts the parameter range, and thus selects the best model. A pattern of spottedness symmetric with respect to the equator, i.e., including four spots instead of two, is considered.

Based on the Budding formalism and the Barnes–Evans relationships for a star with one or several round spots, Poe and Eaton (1985) developed a computer program for calculating theoretical light curves and color indices. The program accounted for the contribution of a secondary component to the total brightness of the system. It revealed a marked dependence of the calculated color indices on the limb-darkening coefficient. The analysis of BY Dra observations in 1965–1980 with this program led to a conclusion that the observed photometric history of the system could be presented by the evolution of one cool spot, whose radius and longitude varied from 17° to 58° and from 50° to 83°, respectively, and the area varied from 2 to 23% of the stellar surface.

To analyze the photometry of some red dwarfs and RS CVn-type stars within the joint programs involving IUE, Rodonò et al. (1986) elaborated an interactive calculation technique for two-spot models based on the Friedemann and Gürtler formalism (1975). The calculation involved two stages: first, using the trial-and-error method the parameters of one spot were selected with respect to the $\chi^2$ criterion, then the parameters of the second spot were estimated from remaining discrepancies. The large number of varying parameters provided a high-accuracy presentation of observations. Later, the program had become widely used for analyzing both symmetrical and asymmetrical light curves for both types of spotted stars (Fig. 6). On the assumption of equal temperature of spots, the observations shown in Fig. 6 were presented by the models with the following parameters: for BY Dra $\Delta T = 600$ K, the spot radii of 39° and 22°, the latitudes of 77° and 0°, the 100° difference of longitudes, the spot areas of 11 and 4% of the stellar surface, respectively; for AU Mic $\Delta T = 850$ K, spot radii of 37° and 22°, latitudes of 81° and 5°, a 90° longitude difference, and areas of 10 and 4% of the stellar surface.

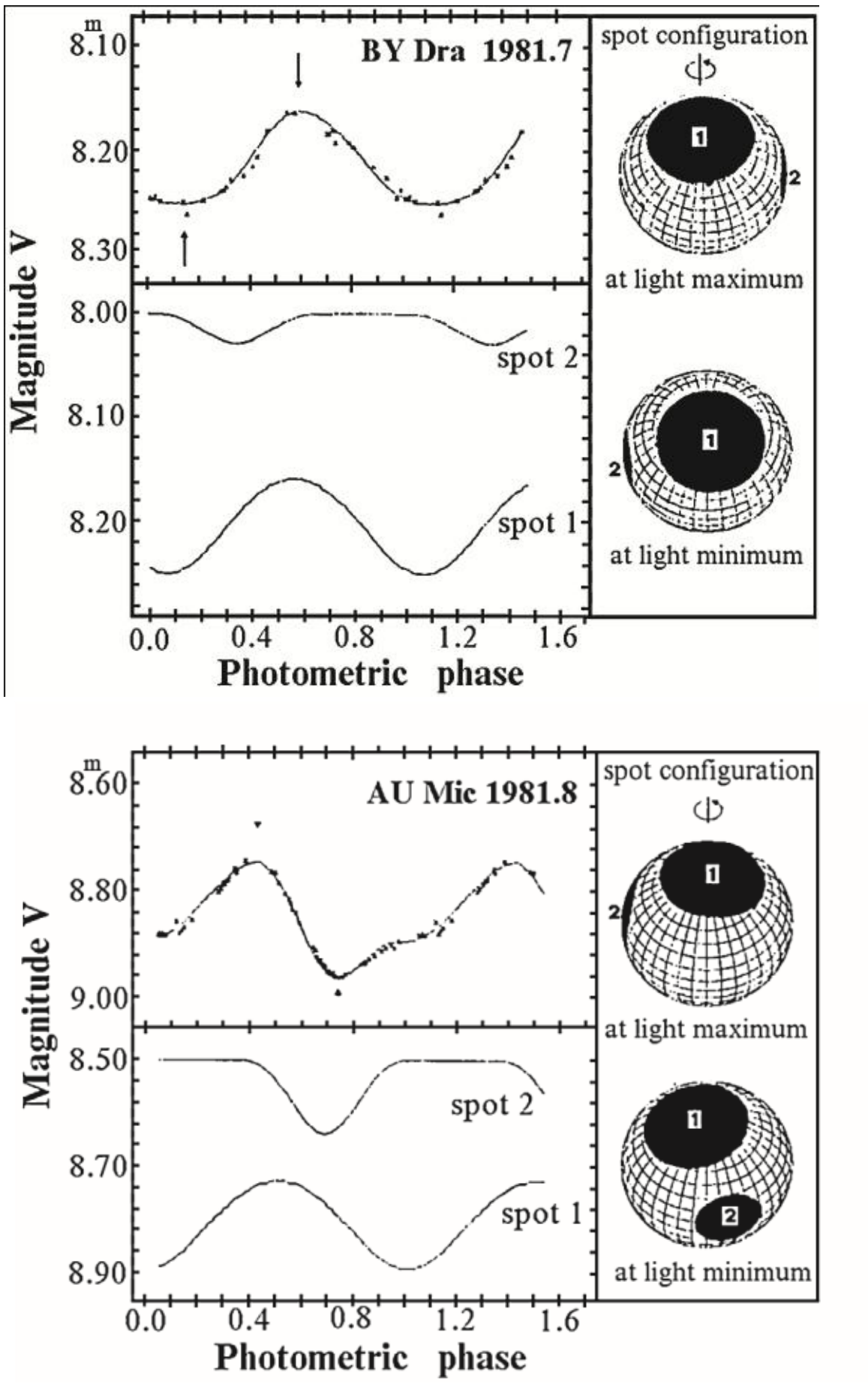

Fig. 6. The light curves of two red dwarfs presented within the two-spot model (Rodonò et al., 1986)

From the TESS photometry and spectral observations with the telescope at the Haute-Provence Observatory and the Keck telescope in 1995–2013, Xu et al. (2022) constructed a two-spot model for the G dwarf HD 134319. They estimated varying sizes of spots and their migration on the star's surface and found that the primary spot lives over 300 days, and with more evolving secondary spot it represents the lights curves of the star. Based on the spectral data, they noted the variability of chromospheric activity. Providing a quite detailed pattern of the evolution of this spot configuration, the authors note that this two-spot hypothesis is not reasoned enough.

Summing up the calculation results for starspot parameters obtained in some earlier studies and using the described technique, Rodonò (1986b) noted that within the brightness amplitude of $0.05^m$ –$0.35^m$ the radii of typical spots varied from 10° to 32° and the ratio $T_{spot}/T_{phot}$ varied from 0.70 to 0.86. These ratios lie between the values characteristic of sunspot penumbrae and umbrae and closely approach the latter. Recording the effects of smaller spots requires much more accurate photometry.

Budding and Zeilik (1987) developed an iterative algorithm for determining the parameters of the two-spot model that enabled a reliability assessment of the resulting solution and a possibility of searching for minimum $\chi^2$. This algorithm was applied to a number of systems of the RS CVn type. As a result, active longitudes and rather high latitudes were suspected for relatively small spots (Eaton, 1992).

Strassmeier (1988) developed an algorithm for simultaneous calculation of theoretical light and color curves and spectral-line profiles in the spectra of single and binary stars with any number of spots of arbitrary geometry from the digitized stellar surface map. The algorithm enables interactive adjustment of the spot map to photometric and spectral observational data.

Kjurkchieva and Shkodrov (1986) and Kjurkchieva (1987) elaborated an analytical theory of the light curves of spotted stars for the case when a round spot was completely hidden behind the stellar limb at a certain phase interval of axial rotation and the light curve had a horizontal section. If there is more than one clear minimum on the brightness curve, the algorithm makes it possible to attribute a proper spot to each minimum and to estimate the acceptability of the obtained set of parameters from the proximity of independent estimates of the inclination of the rotation axis. Later, Kjurkchieva tailored this procedure for the nonlinear limb-darkening coefficient and for a certain family of temperature profiles of spots (Kjurkchieva 1989, 1990).

Using the Kopal–Budding analytical approach, Vetesnik and Esghafa (1989) proposed an algorithm for calculating the light curves and the curves of radial velocities for different spottedness patterns: for a single round spot, for a finite number of round spots, and for spots with brightness varying along the radius. Somewhat earlier, Lodenquai and McTavish (1988) developed an algorithm for determining the parameters of the two-spot model with regard to the photometric umbra-penumbra structure of each spot and applied it to the five-year observational series of one of the RS CVn-type stars. Their model reflected the variations of light curves based on the migration of spots of invariant size over the star.

Reasoning from the suggestion that for a single spot passing over the stellar disk the brightness should be described by a negative sinusoid branch, Hall et al. (1989) proposed splitting of the observed light curves into sections and separated 7 to 9 individual spots with lifetimes from 3.5 months to 2 years in the light curve of the G8 V star V 478 Lyr. For RS CVn the radii of such spots appeared to be close to 14°, i.e., much less than in the relevant two-spot models.

The international symposium held in 1990 in Armagh (Byrne and Mullan, 1992) was focused on up-to-date photometric and spectral methods for studying surface inhomogeneities on cool stars and astrophysical results obtained using the methods. In the introductory paper, Eaton (1992), in particular, reported the results of analyzing the spottedness of several RS CVn-type eclipse systems for which the photometry of eclipses provided additional information about the spots.

Of particular interest were the results obtained under uniform processing of long-term series of uniform observations. Thus, Cutispoto analyzed the observations of the BY Dra star between 1981 and 1987 and concluded that within the two-spot models the light curves of the star could be presented by means of a large near-polar spot, which over the years drifted from 75° via 73° to 82° latitude and made a complete revolution over the longitude, and a 2–3-times smaller spot that over the same period drifted from 40° via 19° to 36° latitude and made three complete revolutions over the longitude (Rodonò, 1992).

Based on the long-term observational series, the latitudes of spots found from the stellar spottedness models were compared with the axial rotation periods during the appropriate seasons. This enabled the estimations of differential rotation for a number of stars. As opposed to Vogt's estimate (1981) for the differential rotation of BY Dra that practically coincided with the solar value, the new estimates for this star appeared to be much lower (Panov and Ivanova, 1993). In addition, the values decreased toward the later spectral types, which contradicted the expectations based on the dynamo theory. In one case, the angular velocity of the near-polar zone appeared to be even higher than in the equatorial zone. Different ad hoc schemes were used to explain these features (Rodonò, 1986c; Lanza et al., 1992; Mullan, 1992).

The progress of the two-spot presentation of the photometry of spotted stars had a decisive effect on the determination of the character of spottedness from the spectral line bisectors: Toner and LaBonte (1992) sought the characteristics of two spots that would provide the observed variations of the bisectors.

Kang and Wilson (1988) developed an iterative algorithm for estimating the parameters of two spots as applied to the RS CVn-type binary systems. The algorithm based on the least-squares method accounts for the effects of tidal and radiation interaction of the components of the systems.

Strassmeier and Bopp (1992) analyzed photometric observational series of the chromospherically active star VY Ari and found that over two seasons its light curve was a subject to changes at weekly intervals and could be presented by the following model of spottedness of the main component: there are always one large polar spot and from 2 to 4 large spots between 0° and 50° latitudes approximately at the same longitudes, some of them appeared over several days and their total area was equal to 10–15% of the stellar surface.

Having considered the long-term observational series of the BY Dra and EV Lac stars, Pettersen et al. (1992b) concluded that large near-polar spots existed on both stars for 14 and 10 years, respectively, but all spots on EV Lac were mostly on one hemisphere.

Scheible and Guinan (1994) studied the dG0 star EK Dra, which is in many respects similar to the very early Sun, and found that the observed light curve of the star could be presented by the two-spot model, the spots with a radius of about 20° being placed at middle latitudes and spaced in longitude for 140° with $\Delta T \sim 500$ K. Later, Jaervinen et al. (2005) analyzed the photometric observations of this star over 21 years and revealed a change-over of active longitudes, the flip-flop effect, with duration of the cycle of 4–4.5 years; the other cycle of 10.5 years is triggered by the migration of active longitudes. Moss (2005) constructed a theory of flip-flop effect within the dynamo of the nonlinear mean field of the fast late rotator; following his calculations, this effect strengthens in objects with intermediate thickness of dynamo-active envelopes.

Eker (1994) developed a new analytical algorithm for calculating the light curves and the color indices of spotted stars with an arbitrary number of round spots and in numerous model calculations proved the dependence of theoretical curves on the accepted law of limb darkening and the size of spots, as well as on the inclination of the rotation axis to the line of sight and the spot latitude. He proved that the limb darkening should be taken into account, in particular, for the color-index curves, and that the linear approximation was sufficient for the darkening coefficient. Furthermore, Eker thoroughly considered the models of uniform filling of the equatorial band by different numbers of spots of identical size. On the basis of this experience using the trial-and error method, within the two-spot model he presented the light curves of BY Dra and HK Lac, while for the latter he considered both the pure two-spot model and that supplemented by the equatorial spottedness band. In all cases, two spots under consideration were approximately symmetrical about the equator, approximating the solar situation, and the deviation of the theoretical light curve from the initial observational data did not exceed $0.0025^m$. Later, Eker (1995, 1996) improved the algorithm introducing the search for the best solution of spot parameters using automatic iterations and studied the unambiguity of the determination of the parameters of one spot from the light curves in the V and R bands.

Butler et al. (1996) considered the observations of the YY Gem system in four photometric bands and using the Budding–Zeilik algorithm (1987) for eclipse systems found that, on average, $\Delta T \sim 600$ K.

Using the analytical Budding approach (1977), Kővári and Bartus (1997) studied the stability and reliability of the spottedness models obtained using this algorithm. They found that the required error of the initial photometric data should not exceed $0.002^m$–$0.005^m$. The error in the estimate of the brightness of an unspotted star by $0.03^m$–$0.05^m$ resulted in an essential distortion of the distribution of spots, whereas the error in the inclination of the rotation axis by 10° resulted in considerable errors in the latitude and the size of spots and for an inclination of less than 20° it was impossible to get reliable values of the parameters.

Kővári (1999) showed that the light curves and B−V color indices of BY Dra over 30 years could be satisfactorily presented if the spottedness of both components of the systems was considered.

Studying the distribution of spots on young single solar-type stars LQ Hya, AB Dor, and EK Dra, Berdyugina and Jaervinen (2005) found that they had much in common with solar activity phenomena, particularly, the presence of cyclic activity on very young dwarfs made it possible to explore the evolution of stellar magnetic activity during the entire main-sequence phase. In 2004, Zboril (2005) presented his observations of LQ Hya by a model with one spot at high latitude, which was preserved for about three months.

Scholz A. et al. (2005) analyzed the IJH monitoring of very-low-mass stars in the Pleiades and for 9 of them detected rotation periods. Small amplitudes of brightness variations led to the conclusion on the contrast of spots and the photosphere at a level of 18–31% and on the filling factor at a level of 4–5%. From the latter, it was concluded that the low-mass stars might have a small number of spots or their symmetrical distribution, and these differences might be due to the transition from the envelope to the distributed dynamo.

Frasca et al. (2005) suggested a way of estimating the spot temperature and area from the light curves and line-depth ratios and constructed a grid of solutions for estimating $\chi^2$ in a wide range of temperatures.

Messina et al. (2006a) considered the constraints on temperatures, areas, and latitudes of spots put by amplitudes of brightness and color indices, while in Messina et al. (2006b) they treated the modeling of amplitudes of brightness and color by two-temperature photometric inhomogeneities on three magnetoactive K dwarfs, AB Dor, DX Leo, and LQ Hya, whereas the two-temperatureness might be implemented by both a spot and facula and a spot umbra and penumbra. Assigning $T_{\rm eff}$, log $g$ and $i$, they sought the regions with $\chi^2$ from 0 to 2 on the planes ($T_{\rm spot}$, spot latitude) and ($T_{\rm spot}$, spot area). This investigation may be treated as an attempt to transit from the determination of parameters for individual spots to their statistical characteristics on the star; this will be described in Subsect. 1.2.3.2.

Fang et al. (2010) performed photometric and spectral observations of the G dwarf AP 149 in the open cluster $\alpha$ Per and analyzed light curves within the two-spot model. They found longitudinal shifts of active regions in opposite directions on a time scale of a day and on a more prolonged scale. Variations of the excess H and K CaII emission correlated with spot regions, whereas there was no such correlation in $H_\beta$ emission.

Analyzing the photometric part of the all-wavelength — from ultraviolet to radio range — observational campaign for YY Gem in 1988, Butler et al. (2015) concluded that spots were located in the quadratures of the binary system at the intermediate latitudes of components.

Patel et al. (2013) analyzed 20-year observations of the binary K5 system V 1147 Tau and at 16 epochs found minimum level variations; this was indicative of spottedness variations in the range from 9 to 22%, whereas spots were primarily at two active longitudes, and changes of active longitudes were recorded. The $H_\alpha$ emission was detected, as well as a noticeable polarization in the BVRI bands.

The two-spot model of the stellar surface was applied for the interpretation of high-accuracy photometric observations of stars from spacecrafts (see, e.g., Lanza et al. (2014) and Santos et al. (2017)).

With the aim of estimating the latitudinal distribution of spots and differential rotation of stars, Santos et al. (2017) plotted the synthetic light curves with different inclination angles of the stellar rotation axis, limb darkening, rotation rates, one or two spots, their latitude, intensity contrast, and sizes. The periodogram analysis of the derived curves showed that one should consider a ratio between the values of the first and second harmonics of the rotation period, which is an appreciable function of the fraction of time when spots are visible governed by their latitude and/or the inclination angle of the stellar rotation axis. This ratio is small for more prolonged visibility of spots.

* * *

The current analytical and numerical calculation methods for theoretical light curves and color indices of spotted cool stars integrate the intensity over the surface of stars of variable brightness due to accidental spots and systematic limb darkening. They solve the direct problem of photometric imaging. The fact that analytical algorithms deal with round spots, while numerical algorithms usually consider rectangular spots is not a principal distinction. When solving the direct problem, one can take into account any type of limb-darkening coefficient and the spot temperature profile. Both algorithms present observations with high accuracy.

As to the inverse problem on the restoration of a stellar surface map from the observed integral flux, despite the availability of the algorithms varying from the simple trial-and-error method to the image restoration developed by Banks et al. (1991) and the matrix operations proposed by Wild et al. (1994), the current situation leaves much to be desired. Certainly, there is no doubt that starspots are cooler by hundreds of degrees than the photosphere and occupy tens of percent of the stellar surface. But the results of estimating parameters of individual spots give rise to some doubts.

The point is that despite the diversity of the applied computational algorithms, all the above studies of the spottedness of red dwarfs were focused on solving the problem stated by Krzeminski: to present the observed light curves by a minimum number of separate surface structures with determined individual parameters. But it is not obvious that the way to imaging a spotted stellar surface lies through solving the Krzeminski problem. Moreover, one can mention some principal shortcomings of the above models of spottedness of red dwarfs.

First, all considered solutions apply the idea of high-latitude or polar spots to interpret the light curves without noticeable horizontal section with maximum stellar brightness, which is valid for practically all observed light curves. However, there are no such spots on the Sun. Schuessler and Solanki (1992) and Granzer et al. (2000, 2004) showed that at fast rotation of a star, when the Coriolis force exceeded the buoyancy force of the magnetic flux tube, the spots should shift to the pole. But only a few red dwarfs are fast rotators, though on one of them,

Gl 890 = HK Aqr with vsin*i* = 70km/s, spots probably exist at high latitudes (Young et al., 1990). The results of the Doppler imaging of a number of the RS CVn-type stars and about ten cool dwarfs are usually cited in support of the existence of high-latitude and polar spots (Strassmeier, 2009). In fact, many such variables are fast rotators and the close binarity of these systems can result in phenomena that differ qualitatively from those on the Sun.

Secondly, the above photometric imaging algorithms often involve the search for hierarchical models, i.e., the surface structures in which one can pick out a main spot and estimate its parameters, subtract its photometric effect from the observed light curve, and estimate the parameters of a second-rank spot (see, e.g., Rodonò et al., 1986; Berrios-Salas et al., 1989; Banks et al., 1991). In some cases this procedure was continued till the third spot. Each step of the procedure requires 3–4 additional free parameters, thus it is obvious that these approximations can lead to a rather high reproduction of observations. However, as in the case of high-latitude spots, the resulting pattern does not resemble the solar spottedness even remotely. This is not only an oversimplification of the pattern of solar spottedness, it is not even reduced to this pattern in principle. Otherwise, the ideology of the above algorithms does not account for the a priori information that is of major importance for image restoration: it does not involve the requirement of the class of solutions, which includes the observed pattern of solar spottedness.

Thirdly, all one- and two-spot models have a pure photometric shortcoming. If a spot is placed at not very high latitude, it is obvious that it will determine two measured photometric values: total attenuation of the stellar brightness, i.e., its maximum brightness in the epoch under consideration, and the amplitude of the rotational brightness modulation. Thus, there should be an unequivocal relation between the two values. An increase in the number of spots and limb darkening should somehow wash away the expected dependence. But the observations do not justify the expectations. Thus, Alekseev and Gershberg (1996c) collected the data on 140 observation seasons of 13 red dwarfs and established that in 127 successive seasons the signs of variations of maximum brightness and of the amplitude of periodical oscillations coincided only in 32 cases.

The listed shortcomings of the available algorithms of the photometric study of spotted red dwarf stars have stimulated principally new approaches to the problem.

* * *

Saar and Neff (1990) proposed a purely spectral method for estimating the temperature and size of starspots by the comparison of values of bumps in the heads of two molecular bands $\lambda$ 7100 Å and $\lambda$ 8860 Å, having different temperature dependence, in the spectrum of an active star with analogous values in stellar spectra where the absence of spots is assumed. For the BD +26°730 star this photometry-independent method yielded $\Delta T$ = 750 K and the spotted region of 20% of the stellar disk. According to similar later studies, for this star $\Delta T$ = 1330 K and the surface filling factor by the spots $f$ = 0.51, for EQ Vir $\Delta T$ = 830 K and $f$ = 0.43, and for LQ Hya $\Delta T$ = 1540 K and $f$ = 0.50 (Saar et al., 2001). O'Neal et al. (2004) continued these observations and from TiO bands they determined the temperatures and areas of spots on EQ Vir (3350 K) and EK Dra (3800 K).

This method essentially solves spectrally the Krzeminski problem and does not allow imaging the spots.

**1.2.3.2. Zonal Spottedness Model.** Contrary to the considered above methods for the analysis of the spotted photosphere in which a task for determining the parameters of individual spots was posed, Alekseev and Gershberg (1996a, b, c, 1997b) suggested searching for the general properties of the spotted stellar regions, the so-called zonal spottedness model.

It is obvious that in the course of photometric observations of a rotating star with a set of discrete dark spots along some latitude the same effect will be provided by two symmetric about the equator dark bands with a longitudinally variable width. This so-called zonal spottedness model is described by four free parameters: the distance of dark bands from the equator, extreme values of its width, a ratio between the surface brightness of spots and the quiescent photosphere of a star; furthermore, it is assumed that between extrema the widths of bands vary linearly along the longitude. In addition to the analogy with a washed-away pattern of the solar spottedness, the pure stellar arguments in favor of this model are the conclusions by Evans (1971), Bopp and Evans (1973), Vogt (1975), Bopp and Fekel (1977b) and Davidson and Neff (1977) on the extension of starspots along the equator for tens of degrees, as well as the reasoning of Mullan (1974) on the formation of such a structure through the dynamo mechanism. A geometrically similar scheme was considered by Eaton and Hall (1979) in analyzing the spottedness of the components of the RS CVn-type systems: in two bands parallel to the equator, they specified such a distribution of small numerous dark spots that a summary effect of darkening along the bands was described by one cosinusoidal period shifted along the ordinate axis for a unity. The photometric equivalence of a nonsetting high-latitude or near-polar spot and a nonuniform equatorial band of small spots was noted by Vogt (1981), Rodonò et al. (1986), Pettersen et al. (1992b), and Panov and Ivanova (1993). As they failed to find the rotational modulation of brightness of the dMe star FK Aqr in 1983, when its brightness was somewhat lower than in 1979, Byrne et al. (1990) formulated a hypothesis on uniform distribution of spots along the stellar longitude. Such a uniform distribution of spots along the equator and two active longitudes with large spotted regions were found by Messina et al. (1999b) on the young G dwarf HD 134319, using the algorithm of Lanza et al. (1998) for reconstructing an image of the stellar surface by the method of maximum entropy and by Tikhonov's regularization.

The computational algorithm for the direct problem within the zonal spottedness model based on general correlations of Dorren (1987) was developed to analyze the observations performed in the BVRI photometric bands. The first version of the inverse problem within the zonal spottedness model was presented in detail in the monograph by Alekseev (2001). Assuming that the spots radiate as cooler stars, the expected relationships between the intensity ratios $\beta_\lambda = (I_{spot}/I_{phot})_\lambda$ were determined from the stars with well-known photometric data and radii:

$$\beta_B = \beta_V^{1.7}, \quad \beta_R = \beta_V^{0.70} \quad \text{and} \quad \beta_I = \beta_V^{0.40} \; . \tag{2}$$

The listed exponents are valid to the accuracy of one–two hundredths for G8–M6.5 stars. These relationships replace the Barnes–Evans relationships or the assumption on the Planck energy distribution in the radiation of spots that were used in the above algorithms. The limb-darkening coefficients obtained by van Hamme (1993) were used in the calculations. The measurements for brightness maxima and minima were taken in four photometric BVRI bands, the four parameters of the zonal spottedness model were determined from the excessive system of 8 equations using the least squares method.

Table 3 presents the calculation results for the parameters of the zonal spottedness models of EV Lac for 23 observational seasons (Alekseev and Gershberg, 1996c; Alekseev, 2000): observational seasons, the difference of the maximum stellar brightness and the absolute brightness maximum, for which its maximum brightness over all considered observation seasons is taken, the amplitude of periodical brightness variations $\Delta V$, the distance along the latitude $\varphi_0$ from the equator to the spottedness band, the maximum width of the spottedness band $\Delta\varphi$ and its minimum value $\Delta\varphi_{min}$ in fractions of $\Delta\varphi$, the ratio of intensities $\beta_V = (I_{spot}/I_{photosphere})_V$ in the V photometric band and the spottedness degree of darker and lighter hemispheres in percent of the area of the stellar hemisphere. The formal calculation accuracy of $\varphi_0$ and $\Delta\varphi$ is tenths of a degree. The same result $\varphi_0 = 0$ for all observations evidences that for EV Lac the sought four- parameter spottedness model degenerates to a three-parameter model: two sought bands symmetric about the equator merge into one equatorial band. The values of $\beta_V$ found vary from 0.37 to 0.62 with an average of 0.53. Following Pettersen (1976), we take the efficient temperature of EV Lac equal to 3300 K and obtain for $\beta_V$ the following differences between the photospheric temperature and the estimates of blackbody spot temperatures: from 370 K to 190 K with an average of 240 K.

Table 3. Zonal spottedness model for EV Lac (Alekseev, 2000)

| Epoch | $\Delta V_{max}$ | $\Delta V$ | $\varphi_0$ | $\Delta\varphi$ | $\Delta\varphi_{min}$ | $\beta_V$ | $S_{max}$ | $S_{min}$ |
|---|---|---|---|---|---|---|---|---|
| 1971.6 | $0.13^m$ | $0.11^m$ | 0 | 20.0° | 0.42 | 0.54 | 29.2 | 19.4 |
| 1972.6 | 0.05 | 0.14 | 0 | 16.5 | 0.07 | 0.53 | 21.8 | 8.6 |
| 1973.7 | 0.00 | 0.12 | 0 | 13.6 | 0.00 | 0.57 | 17.6 | 5.8 |
| 1974.6 | 0.11 | 0.07 | 0 | 13.9 | 0.46 | 0.51 | 20.8 | 14.2 |
| 1975.6 | 0.30 | 0.01 | 0 | 16.4 | 0.90 | 0.37 | 27.6 | 26.0 |
| 1976.6 | 0.04 | 0.02 | 0 | 6.1 | 0.54 | 0.62 | 9.4 | 7.0 |
| 1979.6 | 0.08 | 0.08 | 0 | 13.4 | 0.33 | 0.53 | 19.2 | 11.6 |
| 1980.7 | 0.06 | 0.08 | 0 | 12.5 | 0.25 | 0.55 | 17.6 | 9.4 |
| 1981.7 | 0.10 | 0.06 | 0 | 12.8 | 0.47 | 0.52 | 19.2 | 13.2 |
| 1983.7 | 0.08 | 0.06 | 0 | 11.9 | 0.41 | 0.54 | 17.6 | 11.4 |
| 1984.7 | 0.06 | 0.11 | 0 | 14.7 | 0.16 | 0.53 | 20.0 | 9.4 |
| 1985.5 | 0.11 | 0.05 | 0 | 12.6 | 0.55 | 0.52 | 19.4 | 14.4 |
| 1986.7 | 0.06 | 0.12 | 0 | 15.5 | 0.14 | 0.53 | 21.0 | 9.4 |
| 1987.7 | 0.09 | 0.06 | 0 | 12.4 | 0.44 | 0.53 | 18.4 | 13.0 |

| Epoch | $\Delta V_{max}$ | $\Delta V$ | $\varphi_0$ | $\Delta\varphi$ | $\Delta\varphi_{min}$ | $\beta_V$ | $S_{max}$ | $S_{min}$ |
|---|---|---|---|---|---|---|---|---|
| 1988.8 | 0.13 | 0.00 | 0 | 9.9 | 0.94 | 0.53 | 17.0 | 16.4 |
| 1990.0 | 0.10 | 0.08 | 0 | 14.2 | 0.39 | 0.51 | 20.8 | 13.2 |
| 1991.7 | 0.11 | 0.02 | 0 | 10.3 | 0.76 | 0.54 | 16.8 | 14.6 |
| 1992.7 | 0.11 | 0.06 | 0 | 13.2 | 0.50 | 0.51 | 20.0 | 14.2 |
| 1993.7 | 0.12 | 0.02 | 0 | 10.9 | 0.76 | 0.53 | 17.8 | 15.6 |
| 1994.7 | 0.09 | 0.02 | 0 | 9.3 | 0.71 | 0.56 | 15.0 | 12.6 |
| 1995.7 | 0.12 | 0.04 | 0 | 13.9 | 0.66 | 0.58 | 22.0 | 17.8 |
| 1996.7 | 0.10 | 0.06 | 0 | 14.4 | 0.51 | 0.58 | 21.8 | 15.6 |
| 1998.8 | 0.13 | 0.04 | 0 | 14.5 | 0.68 | 0.57 | 23.2 | 19.2 |

Using the above calculation technique for the parameters of the zonal spottedness model, Alekseev (2001) uniformly analyzed the observations for 25 dwarfs over more than 340 observational epochs. Since in the early 2000s, the total number of known objects with the evidences of spottedness was over one hundred, thus, the sample analyzed was sufficiently representative. It covered dG1e-dM4.5e spectral types with rotation rates from a few to 25 km/s.

The results prove that the observed variety of the light-curve parameters of spotted red dwarfs at maximum and minimum, i.e., the amplitudes of the rotational brightness modulation and the ratio of amplitudes $\Delta B/\Delta V$, $\Delta R/\Delta V$, and $\Delta I/\Delta V$, can be presented within the simplest four-parameter model of zonal spottedness with the differences O–C that do not exceed the observation errors at $\varphi_0$ = 0–55°, $\Delta\varphi$ = 0.5°–34°, $\Delta\varphi_{min}$ = 0.00–0.95, and $\beta_V$ = 0.03–0.58. The models describe the general stellar spottedness that covers from 2 to 50% of the total stellar surface. (It should be noted that the spottedness parameters of LQ Hya obtained within the framework of the zonal model do not contradict the results of the above Doppler imaging.)

The zonal model presents the observations without applying the hypothesis on large cool near-polar spots, although when calculating the model parameters only the natural restriction $\varphi_0 + \Delta\varphi \leq 90°$ was used and the near-polar spots were not excluded in advance.

Calculation of the parameters of the zonal model does not involve any hierarchical considerations.

For 10 of the 25 considered stars the photometric imaging was performed before using traditional models. Comparison of the results of imaging for spot area and temperature calculated with the values obtained using the zonal models shows satisfactory agreement.

Since the average stellar brightness and the amplitude of periodical oscillations within the zonal spottedness model are determined by two varying parameters — the width of the spottedness band $\Delta\varphi$ and the degree of its uniformity $\Delta\varphi_{\min}$, then the increase in the average brightness of a star from one season to another can be accompanied by an increase or a decrease of the amplitude of the rotational variation of its brightness observed in reality.

The obtained individual and average parameters of the zonal models were compared with the global characteristics of the stars: absolute magnitudes $M_V$ , rotational velocities $v_{rot}$ and the Rossby numbers Ro that are equal to the ratio of the axial rotation period of the star to the characteristic circulation time of convective vortices. Although the mass of a main sequence star is the main factor governing its absolute luminosity and the circulation velocity of convective vortices, and therefore the Rossby number functionally depends on $M_V$ and $v_{rot}$, it is appropriate to compare the found parameters of the zonal models with these global characteristics independently.

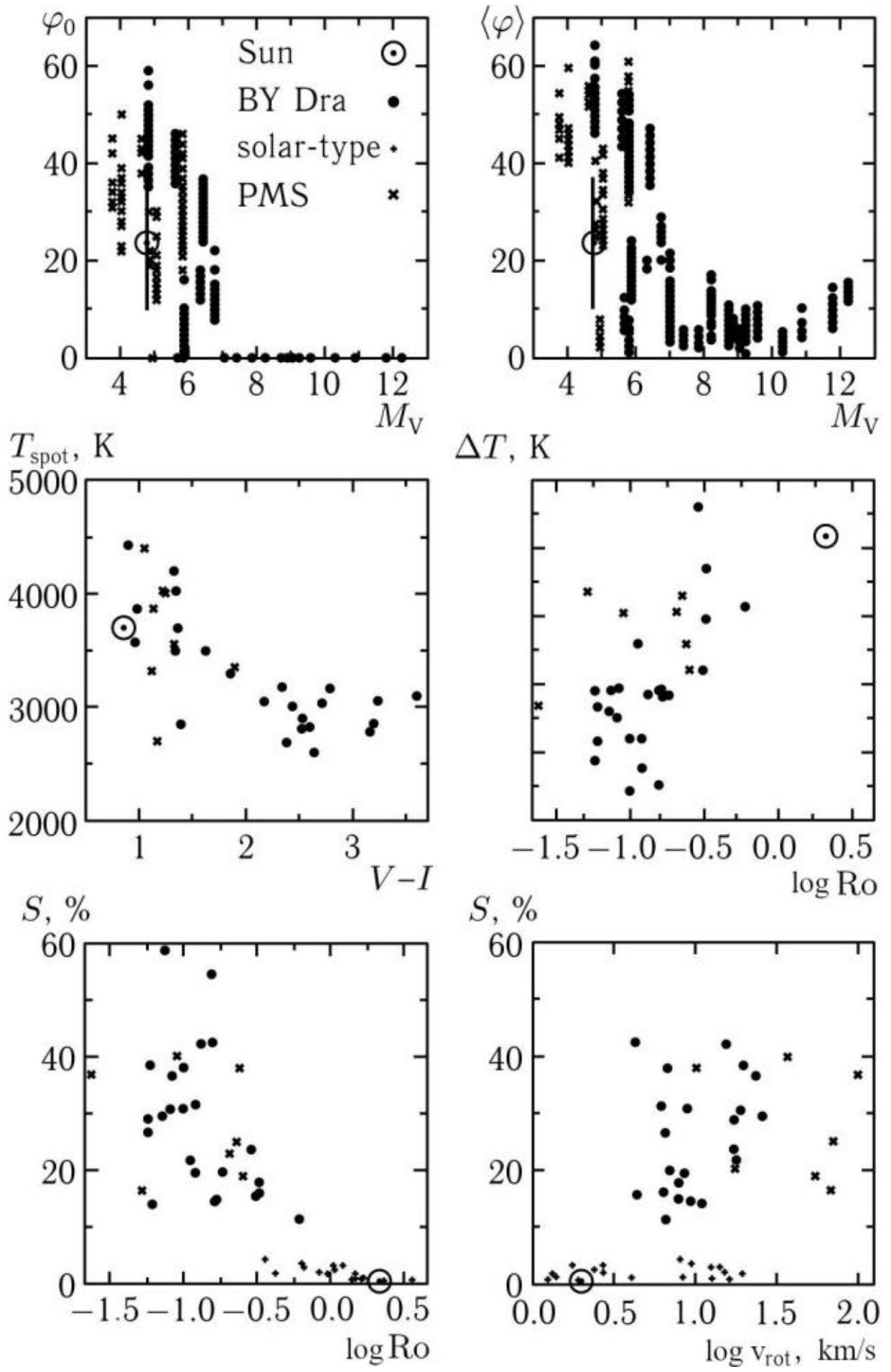

Fig. 7. Comparison of the calculated parameters of the zonal models of red dwarfs (the latitude $\varphi_0$, the average spot width $\langle\varphi\rangle$, the spot temperature $T_{spot}$, the difference between the temperature of photospheres and spots $\Delta T$, the degree of spottedness of stellar surfaces $S$ with the global stellar parameters (the absolute stellar magnitude $M_V$, the rotational velocity $v_{rot}$, and the Rossby numbers) (Alekseev et al., 2019a, b)

Figure 7 shows six plots outlining some physically meaningful correlations with the global characteristics of the zonal model parameters: the latitude $\varphi_0$, the average spot latitude $\langle\varphi\rangle = \varphi_0 + (\Delta\varphi)/2$, the spot temperature, the temperature difference of the photospheres and spots, and the average degree of spottedness $S$ (%); the encircled dot marks the middle position of the Sun in these diagrams, the vertical sections in the two upper charts represent the ranges of the solar values $\varphi_0$ and $\langle\varphi\rangle$.

The following conclusions were drawn from the data presented in Fig. 7.

For the stars with $M_V > 7^m$, the calculations revealed merging of two spottedness bands assumed symmetric about the equator. For half of the hotter stars the calculations yielded two bands, as on the Sun. The stars with split spottedness bands demonstrate a tendency to the growth of $\varphi_0$ for brighter stars, including the Sun. The star BE Cet, the most similar to the Sun, displays the least value of $\Delta\varphi$, which corresponds to the situation that is the most similar to the solar one.

Comparison of $\langle\varphi_0\rangle$ and $M_V$ shows that on all the stars the spotted regions are localized at low and middle latitudes. One can clearly see that on cooler stars spots tend to approach the equator: for late K and M dwarfs the average latitude of spots does not exceed 20°. On hotter stars, spots are shifted to middle latitudes and occupy a greater latitude range.

The calculation results for the zonal models show that the temperatures of spotted regions vary from 4000 K for solar-type stars to 2500–3000 K for the coolest M dwarfs. The correlation coefficient of the values on the relevant diagram is $r(T_{spot}, M_V) = 0.69 \pm 0.08$. Note that the temperatures of spots detected by our purely photometric way are in agreement with estimates of these values found by Berdyugina (2005), do not contradict the temperatures obtained later by Herbst et al. (2021) and by several hundred of K less than those found by Johnson et al. (2021). However all non-Crimean methods demand to use the additional important astrophysical data. Within the magnetic hydrodynamics, this issue is discussed in what follows in Item 7 in Sect. 4.3.4.

Comparison of the temperature differences of the photosphere and the spot $\Delta T$ with global stellar parameters showed that, on average, this difference achieved 2000 K for hot and 300 K for cool stars. In addition, one can suggest the statistical growth of $\Delta T$ with the growth of the Rossby number: the correlation factor is $r(\Delta T, \log\mathrm{Ro}) = 0.67 \pm 0.05$.

The maximum areas of the spotted regions tend to grow with increasing stellar rotation rate and with decrease of the Rossby number. However, taking into account the above dependence $\mathrm{Ro}(v_{rot})$, the two latter plots in Fig. 7 cannot be considered independent.

Thus, in all the plots of Fig. 7 the solar spottedness parameters fall into the regions occupied by the parameters of the stellar zonal spottedness models or into the natural extension of these regions. Otherwise, there is a tendency to join the parameters of the calculated stellar zonal models with the solar spottedness characteristics. This fact suggests that the zonal spottedness models actually reflect the essential properties of the surface inhomogeneity of red dwarfs.

The subsequent multicolor photometric observations in the Crimea were also successfully represented within the zonal spottedness model (Alekseev and Kozlova, 2001, 2002, 2003a, b). We would like to emphasize the qualitatively new result obtained by Livshits et al. (2003) and Katsova et al. (2003): for several most heavily spotted stars, for which the zonal models yielded separate spottedness belts, they found a systematic shift of the lower borders of these belts to the equator as the activity cycles developed simultaneously with an increase of the belt areas (see Fig. 8). Otherwise, within the concept of Crimean zonal spottedness models, for EK Dra, VY Ari, V 775 Her, V 833 Tau, and LQ Hya they plotted diagrams of the spot drifts that were analogous to the Maunder solar butterflies and determined differential rotation of these stars. The rate of decrease of $\phi_0$ is 2–3 times lower than that on the Sun.

Bruevich and Alekseev (2007) estimated a degree of spottedness for a number of stars with activity level that is close to the solar one and found that its value increased from the solar level by 0.3% to 1–5% for stars from the HK project and then abruptly increased up to 20–35%. They revealed a close relation between the area of spots and X-ray radiation power of stars with different activity levels.

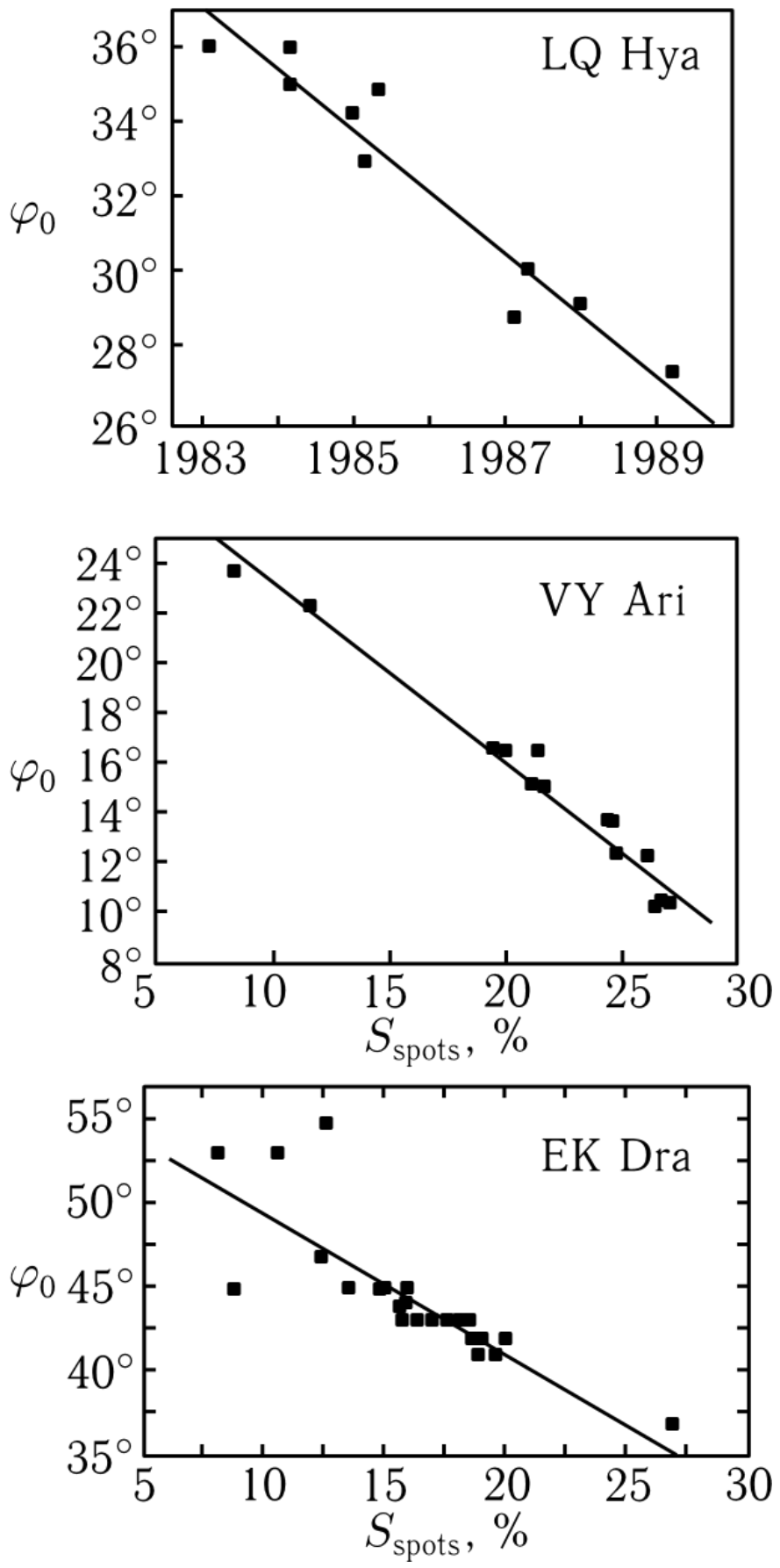

Fig. 8. Stellar cycle evidences: the analogs of the Maunder solar butterflies obtained using the zonal spottedness models (Livshits et al., 2003) and Katsova et al. (2003)

Alekseev (2006) successfully applied the concept of zonal spottedness to variables of other types: RS CVn, FK Com, and T Tau. A good reason for demonstrating the applicability of the zonal model to stars which are different from red dwarfs was given by observations by Grankin and Artemenko (2009) (Fig. 9). These plots exhibit light curves throughout 8 consecutive observational seasons of one of the most active and bright T Tau stars — V 410 Tau. According to the concept of zonal spottedness, throughout the first four seasons light curves fitted the spotted star with substantial nonuniformity of the longitudinal distribution of spots, whereas in two last seasons this nonuniformity basically disappeared. It is of importance that at decreasing amplitude of periodic brightness fluctuations of the star from $\Delta V > 0.6^m$ throughout first seasons up to $\Delta V < 0.1^m$ during the last ones, the average brightness level did not basically change. Within the zonal model this corresponds to the migration of spots up to uniform distribution without significant variations in their sizes. The same reason — the axisymmetric distribution of spots — was attracted by Jackson and Jeffries (2012) while considering several hundreds of low-mass stars of the open cluster NGC 2516 and answering the question why some young cool stars show spot modulation and other ones do not.

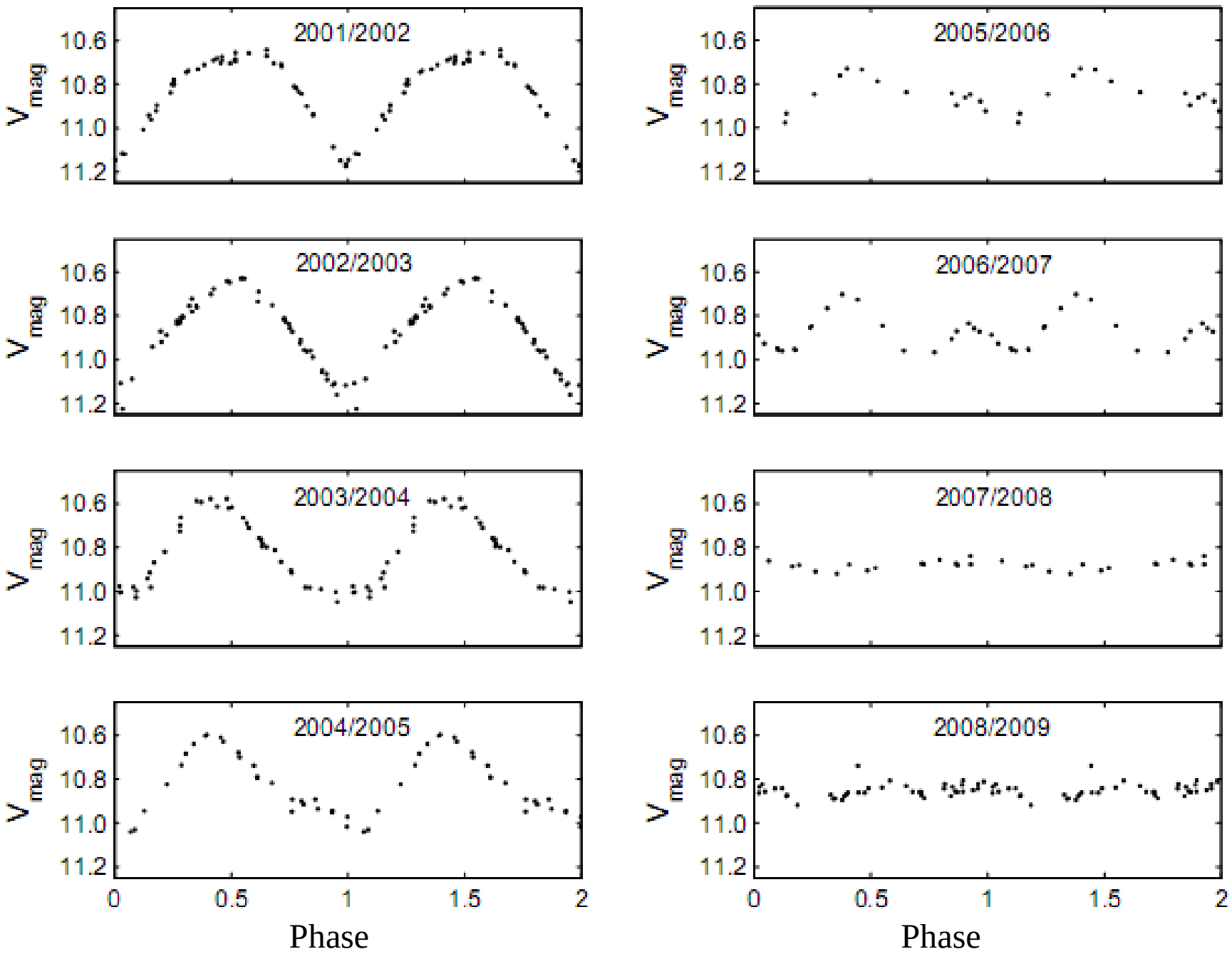

Fig. 9. Light curves of one of the brightest and active T Tau stars V 410 Tau throughout eight consecutive observational seasons. Within the zonal spottedness model, the transition from light curves during the first seasons to the light curves during the last seasons corresponds to the transition from the nonuniform distribution of spots along the longitude to the uniform one (Grankin and Artemenko, 2009)

Being a model with four free parameters, the zonal spottedness model described only symmetric light curves. Trying to get rid of this shortcoming, Alekseev (2008) considered the zonal spottedness model with a more complicated, two-peaked, change of the filling factor along longitudes: he introduced two additional parameters for describing the dependence of the spottedness band on the longitude — this physically means the consideration of two simultaneously active longitudes on the star — and presented the extreme points of light curves for 25 stars at 679 epochs with an accuracy of not worse than $0.01^m$. However, the newly calculated four input parameters of models did not basically change. This follows from the ratios between similar parameters averaged throughout all epochs of all stars: regression coefficients between these values differ from unity by 0.003 to 0.035. Thus, taking the second active longitude into account, causing some improvement of knowledge on the observed light curves, does not essentially change values of input parameters of models.

Within the described above revised zonal spottedness model, Alekseev and Kozhevnikova (2017, 2018) considered variations in this spottedness for 12 M and 13 G-K dwarfs for decades and for 11 stars found a drift of spots to the pole, and for 6 stars suspected cycles of activity lasting from 25 to 40 years. The detected velocities of spot drifts are 2–3 times less in latitude than those on the Sun.

Recently, Alekseev and Gershberg (2021) showed that numerous calculations of stellar zonal spottedness models make it possible to confidently construct a dependence of the spot temperature on the effective temperature of the stellar photosphere, which should significantly simplify the estimates of other spottedness parameters.

**1.2.3.3. Doppler Imaging.** The spectral method of the Doppler imaging has gained a wide distribution in studying stellar spottedness so far. Within this method the stellar surface splits into tens of regions, each of them is identified with a varying, due to stellar rotation, position inside the spectral line. A spot with decreased intensity causes a decrease in intensity of this detail of the spectral line; the longitude of the considered region is determined by the motion of the detail differentiating this region along the line profile; while the latitude is determined by the ratio between the maximum amplitude of the varying region of the line in longitude, which is evident at the stellar equator, to the amplitude of the varying region of the line, which is to be zero at the pole. The joint consideration of a number of spectrograms acquired at various rotation phases allows one to distinguish regions with decreased intensity on the stellar surface, i.e., to determine coordinates, sizes, and spot temperatures. There exist several algorithms of the Doppler imaging method, but it may be applied to fairly bright stars with pronounced rotation in order to reliably estimate the intensity of a comparatively narrow region of the spectral line well broadened by rotation. This constraint is rather substantial for weak flare stars.

According to Hatzes (1995), the Doppler imaging of the YY Gem system, comprising two dMe stars, led to the conclusion on the noticeable spottedness of both components at latitudes of about 45°, weak equatorial spots, and the absence of spots at latitudes of more than 60°.

Saar et al. (1992) constructed a temperature map of the surface of the single K2 dwarf LQ Hya, detected dark spots with $\Delta T \sim 300$ K at middle latitudes but found no high-latitude spots. Strassmeier et al. (1993) and Rice and Strassmeier (1996) addressed again to the LQ Hya star in which Jetsu (1993) detected differential rotation. They found that each of nine spectral lines used in the course of imaging yielded a dark spottedness band in the latitudinal range from −10° to +35° with the average temperature difference $\Delta T = 400$–$500$ K, whereas, concerning the existence of spots at high latitudes, different spectral lines gave controversial results. A shift of the middle of the spottedness band from the stellar equator seems to be an artifact caused by the inclination of the stellar rotation axis to the picture plane, and indeed the spottedness regions form the structure that is symmetrical to the equator. From observations two years later they found again a broad equatorial spottedness band for this star at a latitude from −10° to +35° (or two symmetrical about the equator bands with a small gap) with $\Delta T < 700$ K and weak, less confident, polar spots (Rice and Strassmeier, 1998). Using a new technique for reconstructing images, Berdyugina et al. (2001) reconsidered observations of LQ Hya in 1993–1999 and based on nine spectral lines found that there were always large active regions on the star in the latitudinal range from 50° to 70° and small spots of about 40° at $\Delta T \sim 900$ K. Kővári et al. (2004) studied LQ Hya, using 28 temperature maps throughout 1996–2000, and revealed predominant spottedness at low and middle latitudes from −20° to +50° without a single polar spot and confirmed conclusions drawn by Berdyugina concerning active longitudes on the star.

Somewhat earlier, Strassmeier and Rice (1998) analyzed 12 lines in the spectrum of the EK Dra star, which is very close in parameters to the Sun but has an age of about 70 million years and vsin$i$ ~ 17 km/s. Applying all lines, they confidently found spots near the equator and at middle latitudes of both hemispheres, whereas polar spots were detected in 9 of 12 lines, and their MHD calculations yielded the expected latitudes of flux tubes of the toroidal field only in the range of 25–65°. Jaervinen et al. (2006) performed the Doppler imaging of EK Dra from observations throughout a year and found spots at high and low latitudes with $\Delta T$ = 800 K, furthermore, carried out the inversion of Doppler maps into a light curve.

Stout-Batalha and Vogt (1999) plotted surface images for the two rapidly rotating K dwarfs in the Pleiades and in each of them found spots at high latitudes of 70–77° at $\Delta T$ ~ 800–900 K and no low-latitude spots. Similar results were previously obtained by Ramseyer et al. (1995) for the K2 dwarf V 471 Tau comprised into a binary system with the white dwarf and having an axial rotation rate of 91 km/s. On the K star, they detected a wide spottedness band in the latitudinal range of 60–80° and small spotted regions at about 40°.

Barnes et al. (1998) and Jeffers et al. (2002) analyzed two rapidly rotating G dwarfs in the $\alpha$ Per cluster; to study the surface of these rather weak stars of ~ $11.5^m$, they elaborated the Doppler imaging method using a number of spectral lines from echellegrams and found that there existed spottedness areas at low latitudes and near-polar spots on each of these dwarfs. By the analogous method, using about 2000 photospheric spectral lines, Lister et al. (1999, 2001) carried out an imaging of the rapidly rotating single K dwarf LO Peg. They also came to the conclusion on the existence of the intensively spotted band at a latitude of 25° ± 10°, on the absence of spots at middle latitudes and on the presence of high-latitude spots, not covering the pole, since in the center of the line profile no appropriate flat area was detected.

For RE 1816+541, Barnes and Collier Cameron (2001) detected spots at all latitudes but without polar spots. For the fast G rotator He 699 Jeffries et al. (2002) found a decentralized polar spot and small structures at low latitudes, which were responsible for variations of the photometric light curve. Zboril (2003) proposed two variants of spottedness for YZ CMi to interpret light curves and color throughout three seasons in the 1970–1990s: one high-latitude spot and three spots at the equator; $\Delta T$ ~ 500 K, a filling factor of 10–15 and 5% at an axis inclination of 60° and 25% at an inclination of 75°.

McIvor et al. (2003) provided arguments in favor of the openness of the field structure of polar spots. Using maps for AB Dor throughout 1988–1994, Jeffers et al. (2007) found a long-term polar cap, short-lived structures at low and high latitudes, and variations of differential rotation from year to year.

Barnes et al. (2004) studied the M1.5e dwarf HK Aqr in 2001–2002 and found latitudinal distributions of its spots during each season, without polar caps and with very low differential rotation. Later, Barnes (2005) explored the young fast rotator Speedy Mic (= BO Mic) and found strong spottedness from the equator toward high-latitude regions, while on latter ones it was not continuous but only at some definite longitudes. Due to differential rotation, the equator surpassed polar regions every 191 ± 17 days. Piluso et al. (2008) carried out the Doppler imaging for LO Peg from observations throughout four nights in September 2003 and found that high-latitude spots with a nonsymmetric polar cap, weakly varying since 1993, prevailed in its photospheric activity, whereas small spots at latitudes between 10° and 60° varied as compared to maps over 1993 and 1998. Tas (2011) carried out the UBVR photometry of the star in 2003–2009 and detected variations of surface activity during days and months and variations of the average and extreme brightness levels inside the observational season; the lifetime of the active longitude accounted for about 1.3 years, an activity cycle of about 4.8 years.

As already mentioned, Garcia-Alvarez et al. (2011) constructed spottedness maps for two fastest rotators in the young moving group $\beta$ Pic: HD 199143 (F7V) and CD–64 1208 (K7V); they found an agreement of the imaging for the former star by the maximum entropy method with Tikhonov's regularization and a growth of spottedness by 7% during 20 days. The maps of spottedness for the latter star agreed well longitudinally by two methods, but Tikhonov's regularization yielded the whole hemisphere to be spotted. In both cases, two active longitudes were detected. This phenomenon — the abruptly nonuniform longitudinal distribution of active phenomena, their concentration at two opposite longitudes — is well known from solar observations. Recently, Weber et al. (2013) suggested a theory of the emergence of such structures based on the properties of fine current tubes in the convective rotating envelope.

Using the Doppler tomography method, Ribeiro et al. (2010) detected two active longitudes in the hydrogen and ionized calcium lines for the emission M dwarf QS Vir in a pair with a white dwarf.

Applying the high-dispersion spectra of the young fast rotator BO Mic (Speedy Mic) derived with VLT, Wolter et al. (2005) performed the Doppler imaging at two epochs separated by 13 stellar revolutions, and in the course of each imaging they used spectra derived over one stellar revolution. The imaging showed five groups of spots and their fast evolution over two stellar revolutions. The Doppler imaging in two independent ranges yielded a fit of spot positions with an accuracy of up to 10°.

According to Savanov and Strassmeier (2005), it was required to take molecular lines of TiO, CO, OH, and CN into account for the Doppler imaging of photospheres with $T$ < 4250 K (K7–M0). Later, Strassmeier et al. (2007) elaborated a program of the Doppler imaging which yielded not traditional maps of temperatures and filling factors of spots but magnetic field maps based on Stokes parameters.

In the 2000s, the Doppler imaging technique was significantly developed by Berdyugina and colleagues (Jaervinen et al., 2008; Berdyugina et al., 2008; Jaervinen and Berdyugina, 2010) by including the Occam approximation into the construction of stellar temperature maps. Based on high-resolution spectra, the EK Dra star, a solar counterpart at an age of 30–50 million years, was the first studied by this technique (Jaervinen, 2007). On this star, spots with $\Delta T$ = 500 K were detected, their latitudes varied with time, but they showed a very low differential rotation of the same sign as that on the Sun. Then the review of G–K–M stars performed by this technique has detected for some of them a circular polarization in molecular lines from 0.1 to 1% and in chromospheric emission lines — up to 2%. The observed molecular circular polarization of M dwarfs pointed to the single-pole magnetic fields that covered up to 10% of the stellar disk; smaller signals from K dwarfs corresponded to the weaker and more complicated fields than those on M dwarfs.

Based on 12-year observations of V 889 Her, Jaervinen et al. (2008) suspected the 9-year activity period with a change between two active longitudes and one of these changes was recorded in the maximum phase of the global activity. As for the component of the IR CaII triplet, this star was far active than the Sun, the temperature of its spots was 1500 K lower than that of the quiescent photosphere. The differential rotation found was far less than the solar one but of the same sign.

Golovin et al. (2012) performed a photometric, polarimetric and spectroscopic analysis of the young active spotted K7 star FR Cnc throughout 2002–2008. In 2005, using the ASAS data, they recorded a dramatic decrease in the amplitude of photometric variations at the constant general average brightness level. This was interpreted as a result of fast redistribution of spots. From the photometry acquired at Terskol, two episodes of stellar brightenings were detected; one of them coincided in phase with a strong flare on 23.11.2006. Polarimetry in the BVR bands allows one to suspect the existence of an additional source of polarization. According to the kinematic parameters, the star is attributed to the young disk population and, possibly, to the moving group IC 2391. Based on the lithium line, its age is 10–120 million years. The optimization of the Doppler imaging leads to vsin*i* = 46.2 km/s and the absence of polar spots. The light curve constructed from Doppler maps turns out to be close to the observed one through the morphology and amplitude.

Despite numerous results of the Doppler imaging, the photometric imaging continues to be relevant. First, photometric investigations cover more prolonged time periods and thus are better suited for searching for activity cycles, differential rotation, and other long-term effects. Second, part of stars having low rotation rates are generally inaccessible for the Doppler imaging, and only photometric studies suit for searching for the statistical dependences of parameters of starspots in a wide range of global stellar characteristics. Third, the result effect of the Doppler imaging is greatly influenced by the selection of lines under investigation, the precise determination of the stellar rotation rate and inclination angle of the rotation axis, the selection of parameters of the atmosphere model, the contribution of the chromospheric activity, successful selection of the integration grid, etc. And finally, taking into account results of the Doppler imaging of the same input data but acquired by different methods and presented in the paper by Strassmeier et al. (1991), one may only be surprised at the repeated statements on the fact that results of the Doppler imaging weakly depend on the used method for solving the inverse problem. When considering it closely, the Doppler imaging is still a sort of art, whereas a more rough estimate of the spottedness parameters from photometric observations is free of the required numerous additional input data and ambiguousness in the selection of the calculation method.

**1.2.3.4. Matrix Light-Curve Inversion.** In 1989, Wild (1991) suggested and then Harmon and Crews (2000) developed a purely photometric method of stellar imaging by the method of matrix light-curve inversion (MLI). This method does not previously imply neither the number of spots nor their shape, and the stellar surface is split, as in the Doppler imaging, into numerous small areas. The uniqueness of solving the distribution of their brightness — this incorrectly posed problem — is sought with the help of a particular regularization procedure. At the appropriate inclination of the stellar rotation axis to the line of sight and light curve recorded at different wavelengths, the method allows one to detect high-latitude spots which are often identified by the Doppler imaging. Being inferior to the Doppler imaging through resolution, MLI has advantage in the possibility of using numerous archival data, in the relative simplicity in deriving new observations, and in the absence of rigid requirements for stellar rotation rates and spectral resolution.

Using the matrix light-curve transform, Harmon and Saranathan (2004) plotted a map of spots for the young fast rotator LO Peg. Barnes et al. (2005a) carried out continuous observations of this star during seven nights and detected a strong polar cap with branches down to middle latitudes but not lower than 15°. The obtained differential rotation is consistent with the additional equator rotation over 181 ± 35 days. Such differential rotation is lower than that for G and early K dwarfs; this allows one to suspect a decrease in this rotation toward stars at the PMS stage.

The matrix light-curve inversion method proved to be widely applied when the panorama data became available from the Kepler satellite. Savanov and Strassmeier (2008) elaborated an original matrix light-curve inversion program. To solve the inverse problem, a statistical approach suggested by Terebizh (2004) was included into this program. The program by Savanov and Strassmeier solved the inverse problem of reconstructing temperature inhomogeneities of the stellar surface through the light curve in the two-temperature approximation.

**1.2.3.5. Panorama Photometry - Starspots.** Progress in the creation of wide-angle optical telescopes and two-dimension light receivers allowed one to proceed from observations of individual space sources to the registration of objects in the noticeable areas of the sky, i.e., to the panorama optical photometry, though in the X-rays such observations had already been carried out at the Einstein Observatory. Qualitative differences of the panorama photometry technique and data derived with it from the appropriate characteristics of the traditional photometry made it advisable to distinguish these studies into a separate section.

Moderate CoRoT[1] was a precursor of the mentioned large space experiment Kepler. It was initially intended to search for exoplanets but provided a possibility to investigate the stellar surface inhomogeneity.

Obviously, the transit of a spot across the stellar disk due to its axial rotation and the transit of a planet against the background of a star due to the Kepler motion of the planet may cause for an outer observer a similar low-amplitude attenuation of the stellar brightness that can be detected by high-precision photometry. But the transit of a spot lasts for a half-period of stellar rotation, i.e., from fractions to several days. This picture may change from season to season due to evolution of a spot, whereas the transit of a planet lasts for hours and should be reproduced with accuracy that is characteristic of the celestial mechanics.

From observations with CoRoT, Mosser et al. (2009) studied light curves for the four bright F5–G0 stars with short-lived small spots, made a high-accuracy estimate of stellar rotation periods and the lifetime of spots. From a ratio of these values and estimates of the rotation-axis inclination they concluded on the presence of differential rotation.

Lanza et al. (2009) studied the photometric peculiarities of the G7V star CoRoT-Exo-2a, its age was estimated to be 500 million years, and throughout 142 days obtained a continuous light curve with an unprecedented then accuracy. Applying the regularization by the maximum entropy method, they searched for the effects of cool spots and hot faculae and resulted in a model with two active longitudes, the distance between them varied by 80° in the course of observations, and differential rotation was estimated to be about 1%. The total spotted area showed fluctuations with a period of 28.9 days, and this period might be due to the distribution of the subphotospheric Rossby wave. The contribution of faculae proved to be far less than that on the modern Sun.

Huber et al. (2010) analyzed about 30 revolutions of the CoRoT-2 star with 79 transits of the planet across its disk, presented the spot distribution on the eclipsed and noneclipsed parts of the disk and its evolution. On average, the eclipsed part turned out to be less bright by 5 % than the noneclipsed one. If one takes the contrast of sunspots into account, then the total surface of spots is 19% of the stellar surface and up to 40% of the eclipsed surface. The authors detected a significant differential rotation with $\Delta\Omega > 0.1$ rad/day and two active longitudes separated approximately by 180°. Silva-Valio et al. (2010) and Silva-Valio and Lanza (2011) analyzed the same observations of the CoRoT-2 star throughout 135 days and found a region, free of spots, near the equator, estimated a stellar rotation period of 4.48 days, a differential rotation in latitude of 0.042 rad/day, from the low-amplitude variations estimated characteristics of starspots: their number — 9, radius — up to 0.172*R*, spots occupied 10–20% of the stellar surface, $T_{spot}$ = 4700 K, and $\Delta T$ = 925 K.

[1] CoRoT (COnvection ROtation and planetary Transits) — a European satellite, launched at the polar orbit in December 2006 from the Baikonur cosmodrome with the aim of searching for exoplanets by the transit method and for astroseismology studies. It has an afocal two-mirror telescope with a 27-centimeter entrance pupil, the objective plots an image at four CCDs of 2 K × 4 K in a focus of 1.2 m.

Using the mentioned program, Savanov (2010) analyzed the 142-day monitoring from the CoRoT satellite of the young active solar-type star CoRoT-Exo-2a. Within the two-temperature approximation of the stellar surface model, he singled out five time intervals with a duration from 55 to 15 days and relevant changes of active longitudes, the longitudinal migrations of active regions could be evidence for stellar differential rotation.

Using the light curves in the V band derived within the ASAS-3 project, Savanov and Dmitrienko (2011a) explored the spottedness of chromospherically active stars GSC 08923-01147 and GSC 08933-01802 and found that their brightness varied from $0.45^m$ to $0.55^m$ and spots covered up to one third of the surface of the former and 0.2–0.3 of the latter. Active longitudes were separated approximately by half a period, and at their maximum separation the amplitude of brightness variations was minimum; however, sometimes there existed only one active region on the latter star.

Savanov and Dmitrienko performed a large cycle of investigations of stellar spottedness from Kepler observations.

Savanov (2011a) analyzed almost 5-week continuous observations of stars KOI 877 and KOI 896. Two planets were detected on each of them, and Savanov found evidence for two active longitudes separated by 165° and 135° on each star. Two spots were detected on each one. On the former star these were comparable in size, on the latter star — the area of one spot was greatly superior to the area of another one. The total spottedness of stellar disks was close to solar one: from 0.6 to 1.1% of the full visible surface. During the observations on KOI 877, there seemed to be a change of active longitudes. Later, based on observations of these stars throughout four years, Savanov and Dmitrienko (2015) concluded that there were three states of activity on the former star, and there were always two active regions on the latter one. The differential rotation estimates showed that its value for the former star was comparable with the solar value and for the latter star it was twice higher.

Savanov (2011) considered the data from Kepler regarding the K2 dwarf KIC 8429280 throughout 105 revolutions and detected two active longitudes separated by 180°. Motions of these longitudes over the star turned out to be complicated and unstable: at one time they shifted with the stellar rotation, at the other time — against its rotation. The total spotted area reached 4% of the visible disk, and the periodicity in variations of this value was not less than 90 days. Based on 138-day Kepler observations of the same young and rapidly rotating K2 dwarf, Frasca et al. (2011) detected seven spots on it and a significant solar-type differential rotation. Spots were located on three bands: at the equator and at latitudes of ± (50–60°). Optical observations of this star revealed strong hydrogen and calcium emission in the cores of optically thick emission lines formed in the structures of the solar faculae type.

Using 44.5-day Kepler observations, Savanov and Dmitrienko (2011b) studied activity of the fully convective M dwarf GJ 1243 and on this fast rotator with a rotation period of 0.593 days found two active longitudes separated by 203° and very stable positions of spots. At an inclination angle of the stellar rotation axis to the line-of-sight of $i$ = 60° and 30°, the spotted area accounted for 3.2 and 5.6%. An analogous analysis of 124-day Kepler observations of the slower rotating M dwarf LHS 6351 resulted in detecting two active longitudes shifted with velocities from 0.006 to 0.014 rad/day, and the total spottedness from 1.2 to 0.92% at the angle $i$ = 60° and from 1.8 to 1.0 at $i$ = 30° (Savanov and Dmitrienko, 2012).

Fröhlich et al. (2012) considered Kepler observations throughout 229 days for the two young solar-type stars G1.5V KIC 7985370 and KIC 7765135. The light-curve inversion of the former required, at least, seven spots, and that of the latter — nine spots. Optical observations revealed for both targets a high level of chromospheric activity, as for G stars in the Pleiades. The chromospheric activity arose in structures of the solar faculae type and detected a solar-type differential rotation.

Harrison et al. (2012) considered light curves of 849 stars with an effective temperature of less than 5200 K recorded with Kepler and for 265 of them they detected periodic brightness variations due to the rotation with periods from 0.31 to 126.5 days. A significant number of stars with rotation periods close to solar ones revealed an activity level by two orders of magnitude higher than the solar one. The modeling of light curves showed that active regions on these cool stars were preferentially located near the rotation poles or in two low-latitude groups on different hemispheres.

Analyzing 1141-day data on the M dwarf of the central star in the compact Kepler 32 planetary system with five planets, Savanov and Dmitrienko (2013) detected variations of the photospheric stellar rotation period, which correspond to the latitude drift of active regions, a change of active longitudes (flip-flop effect) in 200–300 days and estimated an area of spotted regions of 0.3–1.7% of the visible stellar hemisphere.

Based on Kepler observations throughout two years, Roettenbacher et al. (2013) studied the target KIC 5110407. They considered 172 stellar rotations, assigning four values of the inclination angle of the stellar rotation axis to the line-of-sight, and in all cases presented observations with an accuracy of up to $0.002^m$ by the light-curve inversion method. In the course of all observational epochs, at least one large group of spots was detected, which yielded a modulation of the flux from the star with an amplitude in the Kepler band from 0.01 to 0.1. The stellar differential rotation extracted from these data proved to be significantly lower than the solar one. A half-hour time resolution was too low for the systematic study of stellar flares, and nonetheless Roettenbacher et al. recorded 17 flares.

Using the Kepler data, Notsu et al. (2013) studied stellar rotation and found that the majority revealed the quasiperiodic brightness modulation with frequencies of one or several tens of days, which could be interpreted by the rotation of stars with large spots. Stars with relatively slow rotation prove to flare with energies similar to fast rotators, though, on average, the occurrence of flares on them is lower than that on fast rotators; the energy of superflares is associated with the total area of spots, and this association resembles solar one.

Gizis et al. (2013) studied 15-month Kepler observations of the L1 dwarf WISER J190648.47+401106.8 supported by VLA in the radio and Gemini North in the optical range and found that this magnetoactive old-disk star had constant radio emission and variable $H_\alpha$ emission. In the optical range there was detected a period of 8.9 h with an amplitude of 1.5%, which persisted throughout a year. The light curve was modeled with one high-latitude spot. A powerful flare with energy in the white light $E = 10^{32}$ erg was recorded spectroscopically; following the Kepler data, such flares on this star occur once-twice a month.

Using three methods for estimating spot areas, Savanov (2015) considered properties of spots on 737 stars having exoplanets. He found no indications that magnetic activity of stars with planets had pronounced peculiarities, which would distinguish them from activity of stars of a more extensive sample. The spottedness of stars with exoplanets in the overwhelming number of cases does not exceed 5%. Stars with effective temperatures lower than 5750 K reveal a monotonous decrease of spottedness with increasing rotation period.

Savanov and Dmitrienko (2015b) considered activity and spots on the surface of 279 G stars on which more than 1500 superflares with energies in the range $10^{33}$–$10^{36}$ erg were detected with Kepler. They showed that the range of flare energy was implemented on the entire interval of rotation periods, though in the diagram flare energy–rotation period there might be a bimodal distribution with practically similar maxima for objects with rotation periods of less or more than 10 days. According to the area of cool spots, the considered stars were distributed into three groups, but the energy range was about similar. Savanov and Dmitrienko confirm that the flare activity is not directly associated with near-polar active regions. An analysis of stars with more than 20 superflares shows that events that differed in energy up to two orders of magnitude occurred at small variations of spottedness on the same star. The spottedness variability by a factor of 5–6 was recorded on two objects KIC 10422252 and KIC 11764567.

Based on the Kepler data, Savanov et al. (2016) studied spots and activity of the M dwarf KIC 1572802: almost 60,000 measurements over 1460 days. Two peaks of rotation periods, 0.37088 and 0.37100 days, provided evidence for a differentially rotating star: $\Delta\Omega/\Omega = 0.0056 \pm 0.0010$, which was significantly lower than that on the Sun. From the maps of temperature inhomogeneities, the positions of active regions were determined, the flip-flop process lasting for 7 days was recorded, and the total area of spots was estimated to be 7% of the stellar surface, which significantly exceeds the average spottedness of stars in the temperature range of 3500–4500 K. The detected value of the Rossby number Ro = 0.011 corresponds to the saturation mode of X-ray radiation.

Based on data taken with Kepler and the space telescope GALEX[1], Dmitrienko and Savanov (2017) analyzed spottedness and activity for 1570 M dwarfs. The considered objects were distributed into four age groups. Stars with spottedness from 0.01 to 0.1 are included into a group of the youngest and most active objects with an age of less than 100 million years. Depending on the rotation period, the considered objects were distributed into three groups and a group with periods of up to 12 days also comprised the youngest and most active objects. The X-ray radiation saturation arose at the Rossby number Ro = 0.13. The constancy of spottedness was detected at an age of more than 900 million years. The youngest and most spotted dwarfs proved to be brighter in the near ultraviolet, and with increasing spottedness the difference of fluxes in the near and far ultraviolet decreased.

Savanov and Dmitrienko (2017a) estimated spottedness for 2846 solar-type stars with $T_{\rm eff}$ = 5700–5800 K and log $g$ = 4.4–4.5. For the stars of the whole sample, the average spottedness accounts for 0.004, which is comparable with that of the Sun; for 279 most active of them it exceeds 0.007. The activity of the overwhelming majority of objects from the sample does not depend on age; it decreases with age in the group with maximum spottedness. The spottedness and chromospheric activity level seem to undergo a dramatic attenuation for stars older than 4 billion years.

Using the Kepler data, Savanov and Dmitrienko (2017b, 2018) considered spottedness and activity for 759 members of the Pleiades cluster and 47 members of the Hyades. For the Pleiades stars, a break of spottedness was detected at a temperature of 6100 K, as well as no dependence on the rotation rate. For 27 stars with masses close to solar ones, the average spottedness accounts for 0.031 ± 0.003, their average rotation periods are 4.3 days, and their activity is significantly higher than the solar one. The spottedness of the Hyades stars turns out to be noticeably lower than that of the Pleiades, the activity of 7 of them is quite close to solar activity, and their rotation periods are 8.6 days.

---

[1] Galaxy Evolution Explorer (GALEX) is a spacecraft launched into orbit in April 2003 by NASA and intended for studies in two ultraviolet bands — in the near ultraviolet at 1750–2750 Å (NUV) and in the far ultraviolet at 1350–1750 Å (FUV). The 50 cm Ritchey–Chrétien telescope GALEX was equipped with a dichroic splitter-corrector that passed NUV, reflected FUV, and increased the field with a resolution of about 5″ to 1.2°. For the spectral observations, a transparent prism was injected into the beam and yielded a resolution of 100–250. The GALEX telescope carried out observations with a resolution of 0.01 s.

The problem of starspots was one the prominent at the Cool Stars 20 conference which was held in Boston in July–August 2018.

From observations of eclipses of active K4 dwarf HAT–P–11 spots by a planet carried out with short exposures with Kepler, Morris et al. (2018) found that in the activity maximum the distribution of spots along the stellar latitude was the same as that for the Sun, their average latitude was 16 ± 1°, the resulting area was about 3%, the rotation period was 29 days, and chromospheric emission was consistent with a cycle of about 10 years.

Gamilton et al. (2018) compiled three families of starspots to consider the influence of their sizes and temperature contrast on the energy interaction with the chromosphere: short-lived spots not affecting the temperature structure of the photosphere, spots with deep roots, causing thermal restructuring of the photosphere, and small spots, leading to the surface restructuring of the surrounding photosphere.

From the Kepler data, Maehara et al. (2018) considered an association of spottedness with flare activity for G, K, and M stars with superflares and found that a fraction of stars with high amplitudes of rotating variability was due to large spots, it decreased with growing rotation period, and there was a good correlation between bolometric energy of the largest superflare on the star and magnetic energy near spots. Mean frequencies of a given bolometric power were approximately proportional to the area of spots. These results suggest that the flare activity level is significantly governed by the area of spots; this was previously noted by Katsova and Livshits while analyzing calcium emission of active stars.

Using the Kepler data, Namekata et al. (2018) constructed light curves of the solar-type stars with superflares and from minima of these curves they estimated the lifetime of groups of spots to be 50–300 days at a filling factor of 0.1–2% of the stellar hemisphere surface. The time of emergence and disruption of these spots is consistent with solar values or somewhat lower.

Moreover, Namekata et al. (2020a) analyzed brightness of the solar-type star Kepler 17, using the Kepler data and its eclipses by the hot Jupiter exoplanets, which occurred every 1.5 days. They came to the conclusion on the existence of numerous large spots on the star, but the results on two independent methods are noticeably different.

* * *

We provide some more results on studying spottedness of active stars by different methods.

For the photometric modeling of spots, Croll (2006) suggested the Markov chain Monte Carlo method. In the late 2005, with the photometric satellite MOST, Croll et al. (2006) carried out continuous observations of $\varepsilon$ Eri throughout three stellar revolutions and detected two spots with $\Delta m \sim 0.01^m$ at latitudes 20.0° and 31.5° with rotation periods of 11.35 and 11.55 days; from these data they estimated a differential rotation coefficient of 0.11, which corresponds to theoretical predictions for the solar-type star having a twice higher angular rotation. The inclination angle of the rotation axis was estimated to be 30°, which is consistent with estimates for the disk and planetary orbit and leads to an equatorial velocity of 3.42 km/s.

Analyzing the high-accuracy light curves of $\varepsilon$ Eri and $\kappa^1$ Cet derived with MOST, Gondoin (2008) expressed a suspicion that their variations could govern not only dark starspots but bright faculae that were especially noticeable near the disk limb. In continuation of these studies, Savanov (2009) estimated the relative area of faculae and cool spots as a small value and concluded that long-term brightness variations of stars younger than the Sun were generally due to variable spots, whereas the contribution of facular fields got more noticeable for older stars. Reporting at the Cool Stars 20 conference, Reinhold et al. (2018) somewhat extended this idea: in the variability of active stars the spots dominate, whereas for less active stars these are bright faculae. The transition from one type of variability to another one is implemented at the Vaughan–Preston gap. Series of chromospheric activity become purely sinusoidal with small amplitudes for slowly rotating inactive stars, which is associated with such dynamo that cannot generate quite strong magnetic fields to support long-lived large spots but enough to generate distinct chromospheric cycles.

In the course of searching for exoplanets at the Apache Point Observatory, Kundurthy et al. (2011) recorded one stellar flare and one spot structure on GJ 1214.

During the high-precision photometry with MOST, Lanza et al. (2011) ascertained the series of radial velocities of the active dwarf HD 189733, estimated its differential rotation $\Delta\Omega/\Omega = 0.23 \pm 0.10$, traced the evolution of several active regions with lifetimes of 2–5 days and with an area of 0.1–0.3% of the disk area.

Frasca et al. (2010) analyzed a photometric monitoring of the young solar-type rotator HD 171488 (V 889 Her) with $P_{\rm rot} = 1.337$ days and presented B and V light curves by two large high-latitude spots and bright faculae with a difference of photosphere temperatures and spots of about 1500 K.

The first direct spectral evidence of cool spots on the solar-type star was presumably inferred by Campbell and Cayrel (1984). By cross-correlation they detected molecular lines of TiO and CaH that are characteristic of solar spots in the spectrum of the G2 dwarf HD1835 (= BE Cet) recorded with a signal-to-noise ratio of $\sim$ 1000.

Throughout four seasons, Toner and Gray (1988) spectrally patrolled the G8 dwarf $\xi$ Boo A and detected variations of asymmetry and equivalent widths of absorption lines with the axial rotation period of the star. The obtained data were initially interpreted by them as a result of passing of the surface inhomogeneity, the so-called "starpatch" covering about 10% of the disk, over the stellar disk, with the temperature 200 K lower and the dispersion of granulation velocities by factors of 1.5–2 higher than that in the quiescent photosphere. Later, Toner and LaBonte (1991) considered another interpretation of the detected spectral peculiarities of $\xi$ Boo A, assuming horizontal flows of matter analogous to the Evershed flow around sunspots. According to calculations, the starpatch covering 10% of the disk at a latitude of about 30° is darker by 10–20% than the surrounding photosphere with a characteristic flow velocity of about 8 km/s and the same dispersion of the velocity explains the observed asymmetry and broadening of spectral lines. Different ratios of umbra and penumbra in the starpatch structure were considered. However, there is no certainty that the starpatches were really a new type of surface inhomogeneity rather than a new manifestation of ordinary starspots or activity centers revealed by the new research technique.

Finally, it is appropriate to note that the long time span of the constant spotted region phase can be due to both the considerably long starspot lifetime and the existence of active longitudes with regenerating starspots (Rojzman and Lorents, 1991).

Here we should note the G0 star V 889 Her, which has the lithium abundance by a factor of 140 higher than that of the Sun and rotates 20 times faster, i.e., it is a very young Sun. It reveals noticeable starspots in the Doppler imaging and large variations of brightness and radial velocities. The first Doppler imaging of this dwarf performed on four lines by Strassmeier et al. (2003) exhibits a cool polar cap, high-latitude spots with $\Delta T$ = 500–1600 K and unconfident details at middle and low latitudes. A study of Huber et al. (2009) showed that these large variations were indeed due to strong spottedness, which maintained the structure and localization at least throughout a year. Furthermore, in the course of the 5-year multicolor monitoring and 16-year photometric series in the $R_C$ band of the fully convective ultrafast rotator M4 V 374 Peg with a rotation period of 0.44 days, Vida et al. (2010, 2016) confirmed its almost solid-body rotation, spottedness stability, detected two active longitudes and constructed a spottedness model with $\Delta T$ = 200 K. Moreover, they detected small brightness fluctuations from night to night and numerous flares near one of active longitudes with significantly decreasing frequency throughout a month; spectral observations made them suspect vertical ejections with velocities of up to 675 km/s of both signs. Analyzing the multicolor patrol of the dM1.5e star EY Dra, they detected a change of active longitudes, a rotation period of about half a day and an activity period of about 350 days. Whereas Kővári et al. (2010) found no significant differential rotation on V889 Her.

As an analog of the quite rare phenomenon on the Sun, it is worth to mention the system V 405 And. Activity was observed on one of its components throughout three revolutions and it was interpreted as localized on the transequatorial magnetic loop, which connects active regions on two hemispheres of the star (Vida et al., 2009).

Jackson and Jeffries (2013) suggested a model for predicting the distribution of light curve amplitudes of the assembly of low-mass magnetoactive stars with the arbitrary orientation of their rotation axes, with a given characteristic size of starspots and their random distribution over the star. To make a model complete, it was required to add a filling factor of spots and their temperature. They estimated models for a filling factor of 0.4 ± 0.1 and with ratios between temperatures of spots and the quiescent photosphere of 0.7 ± 0.05, and, comparing the calculated models with data on M dwarfs of the open cluster NGC 2516, found the characteristic size of spots to be 25,000 km, which is close to the group of solar spots but 2–5 times less than the sizes on G stars detected from passages of exoplanets.

With the Kepler spacecraft, Davenport et al. (2015) performed a high-precision monitoring of the dM4e star GJ1243 and in its light curve detected modulation with a depth of 2.2% with a period of 0.59259 ± 0.00021 days. They interpreted this fact by the existence of two spots on the star, whereas the first spot (or a group of spots) proved to be very stable throughout six years, while the second spot appeared in three regions of the light curve and disappeared over 100–500 days; there was detected a drift in longitude or the differential rotation with $\Delta\Omega$ = 0.012 ± 0.002 rad/day, which was one of the smallest measured values of this parameter.

Throughout a year by means of TESS, Ioannidis and Schmitt (2020) carried out a photometric study of the young fast rotator AB Dor, involving almost 600 its revolutions. The enhanced activity of the star was valid for 11% of time; spots were located at definite longitudes and from low to high latitudes. Positions of spots were governed by both the stellar differential rotation and their lifetime of 10–20 days. On the low-spotted hemisphere of the star, the flare occurrence was 60% less than that on the more spotted hemisphere, but the fact that their number did not fall to zero is interpreted as a presence of high-latitude spots, which are absent on the Sun.

**1.2.3.6. Some Problems of Starspot Physics.** Starspots are concerned with two fundamental problems of stellar physics: stellar magnetism and the radiation deficit of the spotted photosphere. Experimental data for the first problem will be considered further in this chapter. For many decades, the Parker hypothesis (1955a) on the emergence of tubes of the toroidal magnetic field was widely used in solving the problem of sunspot formation, though there is another point of view on the decisive role of convective motions (Getling, 2001). As to the second fundamental problem, its acuteness grows sharply on proceeding from the Sun, on which the area of spots hardly reaches 0.5% of the surface, to red dwarfs, on which spotted regions cover tens of percent of their surfaces and the radiation deficit in the optical range reaches $0.5^m$ and more (Phillips and Hartmann, 1978; Hartmann et al., 1981; Bondar, 1996). Expanding the considerations of de Jager (1968) who stated that the radiation deficit of a sunspot could cover the energy requirements of a strong solar flare, Mullan (1975b) advanced an idea on the removal of the radiation deficit of stellar photospheres by Alfven waves. Hartmann and Rosner (1979) considered several opportunities for the energy channel of the photospheric radiation deficit: the transmission into another spectrum region or into an unobserved energy form carried upward (Alfvén waves?); pumping-over to the adjacent photospheric regions; conservation inside and redistribution with time. However, they were unable to choose one of them. Spruit (1992) concluded that this energy remained in the subphotospheric layers and was spent for a slight heating of the convective zone. According to his estimates, such a heating does not lead to detectable changes in the stellar structure. Besides, it is not clear for which values of the deficit this conclusion is valid.

The described above uniform construction of zonal spottedness models carried out in the Crimea for 24 red dwarfs during more than 340 epochs made it possible to estimate the bolometric deficit of the radiation of spotted photospheres of dwarf stars through the ratio

$$\Delta L_{\rm bol}/L_{\rm bol} = S(T^4_{\rm phot} - T^4_{\rm spot})/T^4_{\rm phot} \quad (3)$$

where $S$ is the portion of the stellar surface covered by spots, and $T_{\rm phot}$ and $T_{\rm spot}$ are blackbody temperatures of the photosphere and spots, respectively. As shown by calculations of Alekseev et al. (2001), for the most active dKe stars this deficit achieved 30% of bolometric luminosity and in absolute units varied from $3 \cdot 10^{29}$ to $5 \cdot 10^{32}$ erg/s.

The problem of radiation deficit is related to another problem of sunspots and starspots: the uncertainty in the equation of the energy of matter in spots does not allow one to precalculate their effective temperature and construct the model of internal structure (Mullan, 1992). On the other hand, the convection theory is related to two characteristics of spots — their lifetimes and minimal sizes. If it is assumed that the lifetimes of spots are determined by the diffusion decay of magnetic flux tubes and the diffusion is determined by the convective turbulence, then the expected and observed lifetimes of sunspots and starspots are rather close. This circumstance supports the validity of the existing phenomenological models of stellar convection.

In addition to the above fundamental problems of starspots, a number of special questions are still to be answered. Are large spotted regions indeed giant spots (or close pairs of spots of different polarity) that cover up to tens of percent of a stellar surface or groups of smaller spots as the largest groups of sunspots? Having considered the stabilization conditions of sunspots, Mullan (1983) concluded that large spotted regions on red dwarfs should be formed by a great number of small spots that practically had no penumbra. Which factor determines the maximum size of starspots: total depth of the convective zone or giant convective cells that for whatever reason did not degenerate into granulation cells or the size of supergranules? Can large spots coexist with smaller spots? How small are the latter? How large are the smallest spots on red dwarfs?

It is believed that the magnetic flux tubes in the solar-type shell dynamo are formed on the interface of the convective zone and the radiative core. If a distributed dynamo starts acting on low-mass stars, do the spots of different size emerge from the same depth? How great is the depth?

Does differential rotation of a star destroy large starspots? If so, what pieces do they fall into? Can small spots merge into large spots?

What do the determined spot temperatures correspond to? Do the giant spots have an umbra-penumbra-type structure?

Although the list of unsolved questions is long, the fact of considerable spottedness of flare stars is sufficient to make an important physical conclusion: the magnetohydrodynamic situation on such stars should, as a rule, differ significantly from that on the Sun, where, as mentioned, the total area of the spotted surface does not exceed 0.5% of the surface. Due to the close location of starspots they form more often the so-called $\delta$-configurations (Zirin and Liggett, 1987) and in their vicinity strong flares should occur more often than on the Sun. This is confirmed by observations. It is not clear whether isolated active regions are preserved at such a high density of spots and whether under these conditions the sympathetic flares become typical rather than a rare phenomenon, as occurs on the Sun.

A significant difference of starspots from sunspots is their existence at high latitudes. This circumstance resulted in a number of theoretical models up to an idea on the distributed dynamo in rapidly rotating stars, which may substantially differ from the solar dynamo (Yadav et al., 2015).

* * *

The most comprehensive statistics of main-sequence FGKM-type stars with spots is provided in Chahal et al. (2022). They compiled a catalog of BY Dra-type variables that consists of 78954 objects and includes their effective temperatures, radii, luminosities, masses, rotation periods, and photometric magnetic indices $S_{ph}$ in the g and r bands. Over a half of cataloged objects are rapid rotators K dwarfs and 94% of objects have rotation periods of less than 10 days. The paper discusses the correlations of the enumerated parameters within the notions about distinctions in stellar structures of different temperatures, ages, and magnetism of different saturation degree.

Bicz et al. (2022) elaborated a program to model the light curves of spotted stars to estimate the number of spots along with their parameters. Using the TESS observations of M dwarfs, they estimated the presence of two spots on GJ 1243 with mean temperature about 2800 K and spottedness 3% of the stellar surface and two spots on V374 Peg with a mean temperature of about 3000 K and spottedness about 6% of the stellar surface. For two observations of YZ CMi separated in time by one and a half year, they found a three-spot model with a mean temperature of about 3000 K and spottedness about 9% of the stellar surface and a four-spot model with mean temperature about 2800 K and spottedness about 7% of the stellar surface. Another program elaborated by them is capable of automatically finding flares throughout the light curves. They detected dozens of flares with energy of $10^{31}$–$10^{34}$ erg with ignition time from 4 to 77 minutes and extinction time from 12 to 273 minutes.

* * *

Thus, we should note that starspots are one of the most impressive phenomena of the solar-type stellar activity; they are studied by a variety of different astrophysical methods. In the 2000s, the reviews on various aspects of these phenomena were published, among them Collier Cameron (2001), Berdyugina (2005), Strassmeier (2009). But after these publications the results derived in the course of the space experiment Kepler have brought qualitatively new important data. The high-precision panorama photometry from this facility allowed Balona et al. (2016) to detect spots and flares on A stars, extending the range of objects with the considered nonstationarity toward the region of hotter stars. The existence of spots on A stars points to the presence of magnetic fields on them.

## 1.2.4. Magnetic Fields

The structure of the solar magnetic field at the photospheric level is very complicated. Apparently, it concentrates in small discrete flux tubes with diameters of 100–200 km that are irresolvable in ground-based observations. The tubes come to the solar surface and form different visible structures: thick ropes of tubes form dark sunspots, and their small groups are responsible for such regions of increased brightness as light faculae and the knots of the chromospheric network; the field strength in the tubes is 1–2 kG. The total magnetic field of the Sun is rather weak, its strength is of the order of 1 G. Main magnetic structures are local fields of sunspots, with a strength of 1.5 to 3–4 kG. Since sunspots are located at middle and low heliographic latitudes, the contribution of the regions with latitudes of above 50° to the total solar magnetic flux does not exceed 10%. Sunspots are surrounded by facular areas, the regions with total radiation higher by several percent than the radiation of the quiescent photosphere. The facular areas are tens of times larger than the spots and live 2–3 times longer; therefore facular areas are observed in the absence of spots as well. The main contribution to the total magnetic flux on the Sun is produced by fields of 1–2 kG and the total area of the magnetized surface is 1–2%, in the faculae the filling factor increases to 5–25%. The structure of the magnetic field in the solar photosphere changes over a day or even faster. Apparently, the described complicated pattern is only a section on the photospheric level of the three-dimensional structure of the solar magnetic field. All these circumstances should be taken into account in discussing the magnetism of red dwarfs.

Three dwarfs were mentioned in Babcock's Catalog of Magnetic Stars (1958): $\varepsilon$ Eri and 61 Cyg A among the stars with narrow lines showing a low Zeeman effect or its absence, and HD 88230 among the stars in which a magnetic field is probable, but have not been confidently established. After the publication of the Catalog the magnetism of the main sequence cool dwarf stars was not studied; the interest in this problem resumed only 15 years later, after the discovery of spottedness of the stars.

As already mentioned, the effect of sunspots on the solar luminosity is negligible. They are of particular interest to heliophysics as places where strong magnetic fields are localized. Therefore, when huge (with respect to solar scales) spottedness areas were found on red dwarfs at which flare activity was stronger by orders of magnitude than that on the Sun, a number of estimates of the expected properties of starspots was published long before the magnetometric techniques reached a level sufficient for experimental study of the structures.

Assuming that the deficit of radiative energy due to spots converts into the magnetic field energy that is then radiated in the flares, Evans and Bopp (1974) estimated the characteristic field strength as tens of kilogauss.

It is known that, the sizes of sunspots correspond to the characteristic sizes of supergranules (Svestke, 1967). Assuming that a similar situation occurs on˘ red dwarfs, Mullan (1973, 1974) calculated the size of such structures on stars and concluded that spots on them should achieve 50–60°. Rucinski (1979) estimated the size of granules and supergranules in a wide range of spectra and found that the size of supergranules decreased monotonically from G2 to M stars, the size of granules was maximum for K4 dwarfs, and by M8 both characteristic sizes decreased to 100–140 km. According to Mullan's calculations within his theory of cellular convection and the hypothesis on the removal of the radiation deficit from spots by Alfvén waves, one should expect magnetic fields with a strength of up to 20 kG and a temperature below 2000 K on the surface of such spots. Given such strong magnetic fields in starspots, one should expect magnetic fields of 5 kG and higher in the stellar active regions.

Such fields could ensure the heating of rather dense chromospheres by Alfven waves and finally the existence of dMe as opposed to dM stars (Mullan, 1975a).

Assuming that there was Biermann's battery effect in dMe stars, Worden (1974) estimated the expected strength of the magnetic field on such stars as $10^4$–$10^5$ G. Accepting these estimates, Mullan (1975c), however, showed significant advantages of the dynamo mechanism over the battery effect.

The first magnetometric observations of F–G dwarfs with calcium emission after Babcock were undertaken by Boesgaard (1974). Among 8 objects for which Zeeman spectrograms were obtained, only for the two coolest (G8 star $\xi$ Boo A and K0 star 70 Oph A) was a longitudinal magnetic field of 140 and 115 G found at the level of $4\sigma$ and $9\sigma$, respectively. However, the observation results for $\xi$ Boo A were not confirmed later (Boesgaard et al., 1975).

Zeeman observations of the $H_\alpha$ region in the spectrum of the BY Dra star suggest that the found profile splitting can be associated with a magnetic field of up to 40 kG, although the absorption spectrum did not evidence a field of over several kilogauss (Anderson et al., 1976). These results were not justified later (Anderson, 1979; Vogt, 1980).

Mullan and Bell (1976) developed the theory of polarization of stellar radiation due to a large number of magnetosensitive absorption lines falling into the UBVR photometry bands. Applying these calculations to the observations of BY Dra, they estimated the photospheric field as 10 kG; however, later the initial polarimetric results were not confirmed.

All methods of direct measurements of magnetic fields on stars are based on different manifestations of the Zeeman effect, which consists in splitting of spectral lines into multiplets with different polarization of components.

**1.2.4.1. Zeeman Spectropolarimetry.** Vogt (1980) observed about twenty red dwarf stars with the sensitivity of Zeeman spectropolarimetry increased by two orders of magnitude using the Reticon detector, but did not reveal an effective magnetic field with a strength confidently exceeding the measurement error of 100–160 G either in the absorption lines or in $H_\alpha$ emission. Having acquired this negative result, he thoroughly studied all theoretical predictions for strong fields on the stars under consideration and noted that they were insufficiently rigorous and rather ambiguous. Thus, he concluded that his result admitted the existence of general incoherent magnetic fields on the stars with a strength of up to one kilogauss and magnetic fields of spots of up to 10–15 kG. However, the concept proposed by Mullan for dMe stars as magnetic dM stars remains valid for this essential revision of the expected magnetic properties of red dwarfs.

The significant achievements in magnetometric polarized-light studies of hot magnetic stars cooled off the interest of observers of cool stars, and over almost two decades only a few studies were undertaken in this field.

Brown and Landstreet (1981) measured the longitudinal magnetic field using Zeeman polarimetry in many spectral lines. They installed a diaphragm in the focus of the Palomar coude spectrograph, which cut out the absorption lines of the spectrum of a K star, and the Fabry lens accumulated the transmitted radiation on a photomultiplier. The spectrum was scanned by swinging a flat-parallel quartz plate. They observed $\xi$ Boo A, 70 Oph A, two K5 dwarfs HD 131977 A and 61 Cyg A, and two dMe stars, but did not reveal a field over twenty gauss; according to their estimates for $\xi$ Boo A, B = +1 ± 12 G. Thus, they reduced the upper limit of the longitudinal field found by Vogt (1980) by an order of magnitude.

Borra et al. (1984) used the CORAMAG spectropolarimeter updated for a Cassegrain focus. The polarization optics was inserted in the parallel beam, and the right- and left-polarized radiation was recorded alternately with a switching frequency of 100 Hz. The device was able to record many lines simultaneously. Over 11 nights they made 112 observations of the longitudinal field of $\xi$ Boo A. Once they recorded a field of 25.0 ± 6.4 G, then four nights later, of 72 ± 30 G; on the other nights no significant field was found. Comparing these results with the measurements of the field of this star from the Zeeman broadening, Borra et al. concluded that there should exist several large magnetic regions composed of several hundreds of smaller two-pole structures.

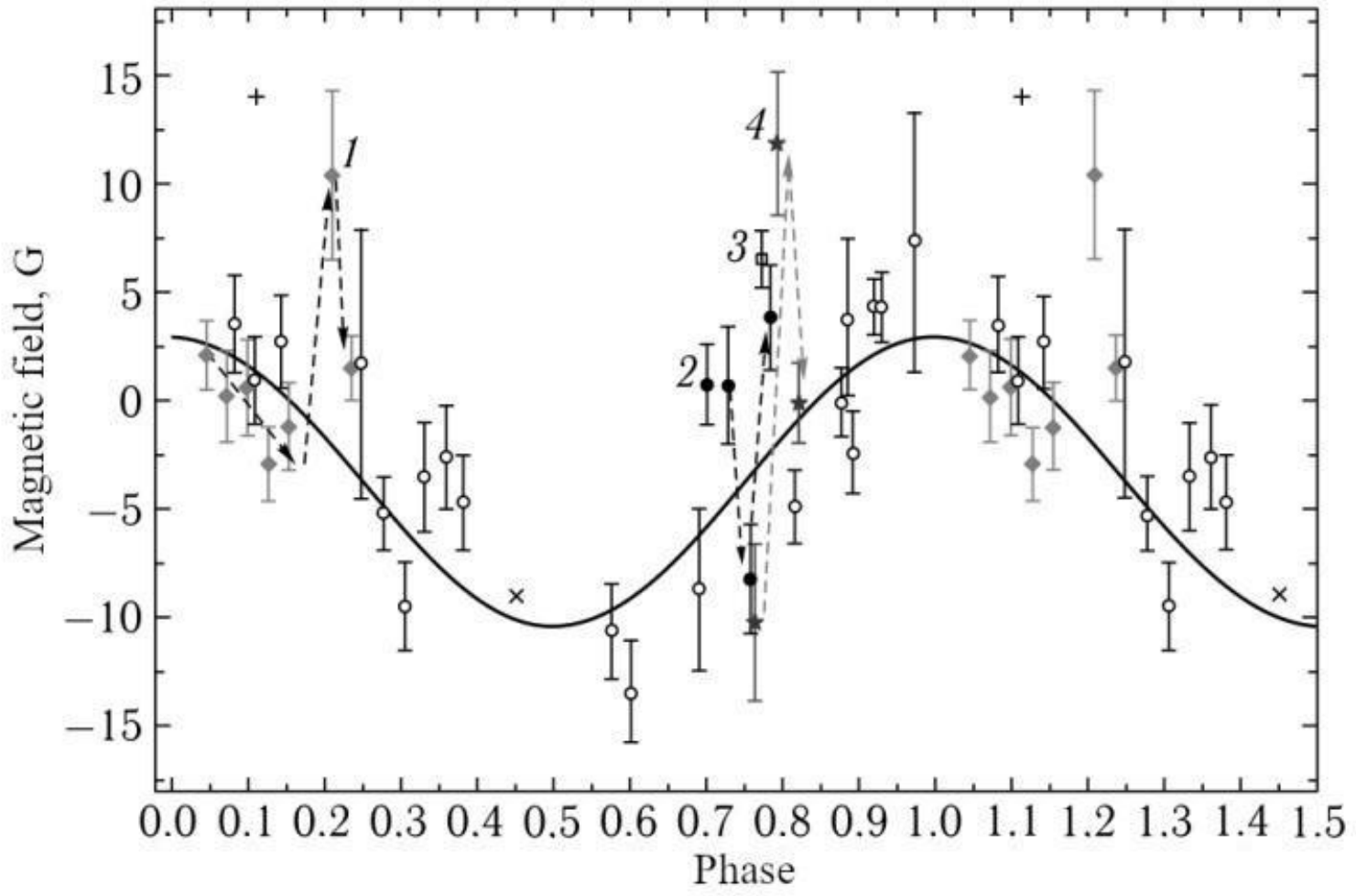

Fig. 10. Magnetic field variations of 61 Cyg A phased with a rotation period of 36.618 ± 0.061 days. Filled symbols indicate part of observations at CrAO carried out during consecutive nights. Dashed arrows connect a sequence of observations when a process of outflow of unipolar spots took place. Bias and straight crosses denote sporadic observations of Borra et al. (1984) and Brown and Landstreet (A201981), respectively

In the Crimea, a Stokesmeter with a CCD matrix as a detector was produced and mounted on the 2.6-meter Shajn telescope for spectropolarimetric studies. To avoid nonuniform pixel sensitivity, the spectra were recorded in two positions of a quarter-wave plate differing by a 90-degree turn. Using this device, $\xi$ Boo A was observed for 10 nights, but only during two of them was the total magnetic field with a strength of 46 ± 14 and 55 ± 15 G found (Hubrig et al., 1994). Then Plachinda and Tarasova (1999) used this device to observe several F–G stars. The observations with a spectral resolution of 30000 and a signal-to-noise ratio of 300–450 provided an observation accuracy of up to several gauss. But on $\beta$ Com, the only G0 dwarf in the observational program, no significant magnetic field was found. The first real result was obtained by Plachinda and Tarasova (2000) when they found distinct periodical variations of the total field of $\xi$ Boo A from +30 to −10 G with a period of about 6 days that coincided with the axial rotation period of the star. The results obtained by Brown and Landstreet (1981) and Borra et al. (1984) fit rather well the phase curve constructed by Plachinda and Tarasova. Then, Tarasova et al. (2001) measured the total magnetic field for three solar-type stars – $\varepsilon$ Eri, $\chi^1$ Ori, and 61 Cyg A – and obtained the following results: during two of nine observational nights for $\varepsilon$ Eri they recorded a field of −10 and +21 G at the level of 4–5$\sigma$, during two of five observational nights for $\chi^1$ Ori they recorded a field of −8 and +11 G at the level of 2$\sigma$, and during two successive nights of 15 observational nights for 61 Cyg A they recorded a field of about 13 G at the level of 4–5$\sigma$. It should be noted that close values of magnetic fields for $\varepsilon$ Eri, $\xi$ Boo A, and 61 Cyg A exceed the relevant solar value by an order of magnitude, although the first two of the three stars are younger and more active than the Sun, the last one is older and has a similar activity level to the Sun.

In the course of further analysis of data from the magnetometric observations of 61 Cyg A, Plachinda (2004) and later Baklanova et al. (2011) found a rather confident change of the total magnetic field of the star within +4–−13 G that was synchronous with the stellar rotation (Fig. 10). But against the background of a smooth sinusoidal curve they found 3 or 4 sharp spikes that decayed over 2–4 days. The authors interpreted these facts as a result of the appearance of a large magnetic structure on the visible stellar hemisphere. First, a structure similar to the leading spot of bipolar group with a noticeable magnetic flux emerged. Then, an analogous structure of the type of a tail spot of the group with magnetic flux of the opposite sign appeared tens of hours later. Numerical modeling has shown that to obtain the observed effect it suffices to get such structures at low latitude with a field strength of up to 3–4 kG and a characteristic size of about $10^{10}$ cm.

**1.2.4.2. Robinson's Spectrophotometry.** In the article that appeared simultaneously with the above Vogt publication (1980), Robinson (1980) proposed a new method based on the magnetometric principle for studying cool stars from the Zeeman broadening of lines in nonpolarized spectra, which significantly expanded the possibilities of the studies. This method presents the observed profile of the absorption line as a sum: $F_{\mathrm{observ}} = f F_m(B) + (1 - f) F_q(B = 0)$, where $f$ is the fraction of the stellar surface occupied by the magnetic region, $F_m(B)$ is the profile of the line occurring in the magnetic area with the field strength $B$, and $F_q$ is the profile of the line appearing in the undisturbed photosphere. Such an expansion of the observed line profiles is possible when considering the lines with different magnetic field sensitivity, i.e., different Lande factors. The paper in which Robinson used the Fourier transforms of the profiles and the assumption on normal Zeeman triplet of optically thin lines formed a basis for real measurements of magnetic fields on late dwarfs. But these observational techniques and the ideology of data processing and calculations of $F_m(B)$ have evolved greatly over the past years.

The first observations applying the Robinson method were carried out in the $\lambda$ 6843 Å and $\lambda$ 6810 Å lines of neutral iron that belong to the same triplet (Robinson et al., 1980). Observations of the chromospherically active G8V star $\xi$ Boo A and the sunspot revealed the expected differences in the profiles of these lines, whereas on the quiet solar surface these profiles, as one would expect, were identical. According to these observations, the magnetic field of up to 2900 ± 550 G occupied up to 45% of the surface of $\xi$ Boo A, whereas on the K0V star 70 Oph A the field of strength of up to 1800 ± 350 G occupied up to 10% of the surface. The estimates of the field strength in sunspots and faculae yielded results close to those of traditional magnetometric observations of the Sun. The later data by Brown and Landstreet (1981) and Borra et al. (1984) on small longitudinal fields did not contradict the revealed values of kilogauss and provided good mutual compensation of circular polarization from numerous local stellar magnetic fields with different polarity. Since a high strength of the magnetic field is observed not only in relatively small sunspots, but also in more extensive facular areas of active regions and in the bright chromospheric network, at first it was not clear whether the magnetized regions of the stars should be identified with large starspots or extended structures of higher brightness.

Marcy's study (1981) of another couple of FeI lines – $\lambda$ 6173 Å and $\lambda$ 6241 Å – did not confirm the presence of a noticeable magnetic field on $\xi$ Boo A, although using the observations of a sunspot he tested the validity of the measurement technique. Marcy concluded that the point was the variability of the magnetic field of $\xi$ Boo A. This conclusion was supported by the known evidence for the variability of emission of the CaII K line, and the absorption of the HeI $\lambda$ 10830 Å line, and the X-ray emission from the corona. Qualitative estimates showed that the detectable field had to disappear if its strength reduced to $B$ = 1000 G or the filling factor $f$ reduced to 0.06 of the stellar hemisphere surface.

Marcy (1984) compared different computational versions of the analysis of the Zeeman broadening, considered probable sources of errors in the estimates of the field parameters, and summed up the results of his magnetometric observations of the FeI lines $\lambda$ 6173 Å and $\lambda$ 6241 Å at the Lick Observatory. For 19 out of 29 program G0–K5 dwarfs he revealed magnetic fields within 600–3000 G and the filling factor of the visible hemisphere varying from 50% to 89%. Such a high factor is typical of sunspots, i.e., the stars should be considered as completely covered by spots. Marcy found that the regions with magnetic fields on K dwarfs were definitely larger than those on G stars, but he did not reveal a dependence between the field strength and the spectral type. The magnetic fields measured by Marcy on $\xi$ Boo A and 70 Oph A were noticeably weaker but had a higher filling factor than those found by Robinson et al. (1980). For two chromospherically active dwarfs Marcy suggested field variations within a day. G dwarfs prevailed among the stars that did not show any evidence of a magnetic field. The magnetic flux proved a statistical dependence on the rotation rate of the star as $v_{rot}^{0.5-1.0}$.

Gray (1984) generalized the Robinson algorithm for simultaneous analysis of several spectral lines taking into account some effects of radiative transfer in the lines. Using the 2.1-meter Struve telescope of the McDonald Observatory for magnetometric observations of 18 F–G–K dwarfs, he found on seven G6 and even later stars magnetic fields of 1.9–2.4 kG and a filling factor of 25–40%.

The brightest flare star AD Leo was the first dMe star on which the magnetic field was found and measured from Zeeman broadening. The magnetometric observations were performed by Saar and Linsky (1985) on a Fourier spectrometer mounted on the 4-m Mayall telescope at the Kitt Peak Observatory. The observations were performed in the infrared region of the spectrum near 2.22 μm, where Zeeman splitting proportional to the squared wavelength was more noticeable. Saar and Linsky used five lines of neutral titanium and found magnetic fields with $B = 3800 \pm 260$ G and $f = 0.73 \pm 0.06$ (Fig. 11). The observational results for a sunspot and the inactive star 61 Cyg A illustrate the noticeable magnetism of AD Leo. Saar and Linsky showed that the revealed field strengths corresponded to the gas pressure of the quiescent photosphere, while the greater area of the magnetized region should be associated with high-efficiency generation of the magnetic flux from a rapidly rotating star through the dynamo mechanism.

Then, Saar et al. (1986a) published their magnetometric observations for the known flare and spotted dK5e star EQ Vir. Zeeman broadening was studied using the 4.5-m multimirror telescope from the magnetosensitive line FeI $\lambda$ 6173 Å and nearly NaI, CaI, and FeI lines with lower Lande factors. For the first time, radiative transfer in the lines was taken into account, while the blends were thoroughly considered by comparing the lines with the spectra of the inactive star 61 Cyg A. The calculations were performed for the LTE Milne–Eddington atmospheric model on the assumption that the atmosphere parameters were identical in the magnetic and nonmagnetic regions. As a result, the strength of the field covering up to 80±15% of the stellar surface was estimated as 2500 ± 300 G. These results combined with the earlier data on the field absence obtained from the polarized-light measurements evidenced the complicated topology of the magnetic field and numerous small bipolar structures. Similarly to AD Leo, the field strength corresponded to the gas pressure in the photosphere, while the greater average field $\langle B \rangle = fB$ corresponded to the greater rotational velocity of the star.

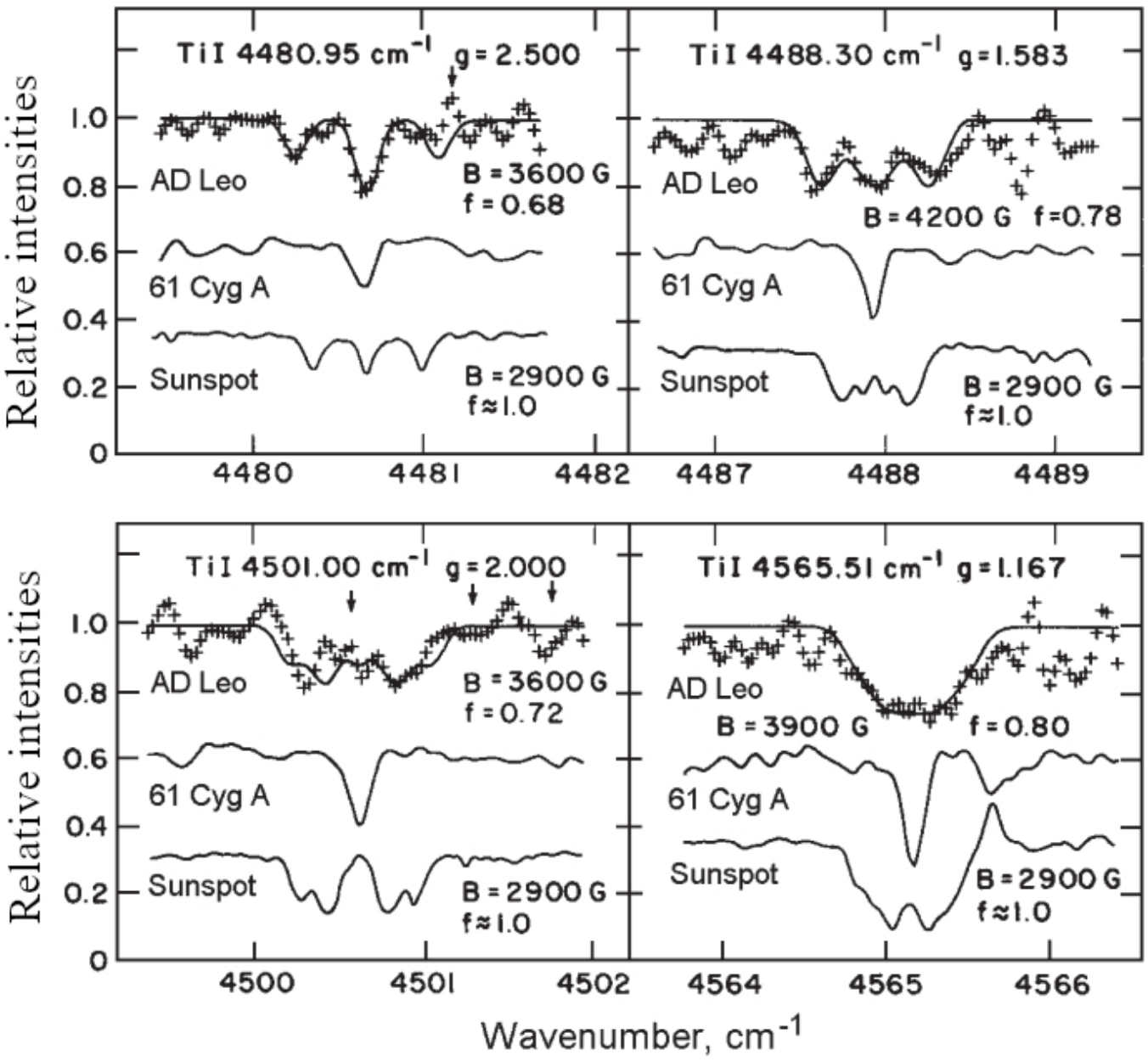

Fig. 11. Spectra of the flare star AD Leo, low-active star 61 Cyg A, and the solar spot in the region of infrared lines of the neutral titanium with different magnetic sensitivity (Saar and Linsky, 1985)

Using the above technique, Saar et al. (1986b) analyzed 11 spectra of the active K2 dwarf $\varepsilon$ Eri obtained within two months in 1984. For the $\lambda$ 6173 Å line, which is the most sensitive to the magnetic field, the comparison profile was obtained from the inactive star spectrum. As opposed to considerable changes in the field strength and in the filling factor at approximate steadiness of the magnetic flux found by Marcy (1984) from 10 measurements made over two years, Saar et al. established variations of the field within 1.7–2.3 kG and the filling factor within 7–15%, but did not confirm the magnetic-flux constancy suggested by Gray (1984). They found that the changes of the measured parameters occurred at times that were shorter than the rotational period. When these measurements were supplemented by the results of magnetometric analysis for five other G–K dwarfs, the assumption on the proximity of measured field strength and their expected values, given equal field and gas pressure in the quiescent photosphere, was confirmed again (Saar and Linsky, 1986).

Gondoin et al. (1985) thoroughly studied the spectra of G8 stars in photometric H and K bands and found that only 4 nonblended lines of neutral iron near 1.56 μm matched the observations of the Zeeman broadening. But no manifestations of magnetic field were found on $\xi$ Boo A. Explaining this result by the field variability, they noted that the observations in the optical and infrared ranges could result in incongruous conclusions because the optical range was characterized by a low contribution of spots in the total radiation and fields dominating in the faculae, whereas in the infrared range the relative contribution of spots increased and for substantial spot fields they could become dominant in the total signal of the magnetic field.

Applying the Fourier spectrometer technique tested in the observations of AD Leo (Saar and Linsky, 1985), Saar et al. (1987) carried out magnetometric studies of a number of M dwarfs. The magnetic-field-insensitive FeI line was used to determine the parameters of nonmagnetic broadening of the lines and these parameters were used to calculate theoretical magnetosensitive profiles. Based on the observational results, Saar et al. established that five of the six dM stars showed no sign of the field, i.e., $B < 700$ G or $f < 0.20$, but they obtained $B = 2.5$ kG and $f = 0.20$ for the dM1 star Gl 229 and $B = 2.5$– 5.2 kG and $f = 0.6$–0.9 for the flare dMe stars AD Leo, EQ Vir, BY Dra, AU Mic, and EV Lac. Combining these results with the measurements for magnetic fields on K dwarfs, Saar et al. confirmed the compliance of $B$ with the gas pressure in the photosphere within 1–5 kG and the distinct growth of $B$ for the stars of later spectral classes. The filling factor $f$ correlates with the angular rotation rate of the star $\Omega$ rather than with the spectral type. The correlation of $\Omega$ with the averaged field $fB$ is even closer. Probably, for axial rotation periods longer than 4 days $f$ grows linearly together with $\Omega$, then $f$ is saturated at a level of 0.8. This relation and the observed strong fields on completely convective stars AD Leo and EV Lac did not fit the dynamo theory proposed by Durney and Robinson (1982). The saturation of $f$ was suggested by the Skumanich and McGregor dynamo theory (1986), $f$ being in a better correlation with the Rossby number than with $\Omega$, whereas strong fields in completely convective stars required use of the distributed dynamo idea. Nevertheless, some important conclusions were based on the established correlations. First, if the strength of magnetic fields is governed by the condition of hydrostatic equilibrium of the field flux tubes and the ambient photosphere, the values of $B$ are independent of the specific dynamo mechanism. Secondly, the efficiency of the acting dynamo mechanism and the variations of stellar magnetism are manifested in the changes of $f$ (Saar, 1987). Later, Saar (1994), using new observational data and improved analysis technique, determined more precisely the parameters of the magnetic field of active M dwarfs AU Mic, AD Leo, and EV Lac: $B = 4.0$–4.3 kG and $f = 0.55$–0.85.

Hartmann (1987) analyzed different estimates of magnetic-field parameters from the Zeeman broadening and concluded that since the lines with different relative depth were compared, both $B$ and $f$ could be systematically overestimated, which did not occur if the lines of one multiplet were compared. On the other hand, comparison of lines in the spectra of different stars can yield erroneous conclusions, since the effect of Zeeman broadening is influenced by the micro- and macroturbulence velocities and stellar rotation rates, which can hardly be identical for different stars. Finally, downward of the main sequence the lines of neutral metals become stronger and wider, therefore on M dwarfs only the fields that are stronger than those on the Sun can be found. Saar (1987) noticed some shortcomings of the first estimates of $B$ and $f$ of the field made using the Zeeman effect, which particularly disregarded weak blends distorting the profiles of most lines in the spectra of late stars. He believed that Marcy (1984) could discover the magnetic field in the inactive star 61 Cyg A from the line $\lambda$ 6173 Å because he disregarded CN and TiO blends. According to Saar (1987), the obtained high values of $f$ evidenced that the magnetized regions were the analogs of the solar faculae and the bright network rather than cool spots. This explained the significant variations of $f$ at low photometric variability of many active dwarfs. By studying 29 G0– M4.5 dwarfs Linsky and Saar (1987) found an anticorrelation of $B$ and $T_{eff}$, an independence of $B$ from $\Omega$ and the Rossby number, as well as an independence of $f$ from $T_{eff}$.

Using the 2.5-m telescope of the Mount Wilson Observatory, Bruning et al. (1987) studied seven late K and early M dwarfs. On six of them the magnetic fields varied from 750 to 2000 G and $f$ varied from 0.20 to 0.65. Using the nonmagnetic line FeI $\lambda$ 6180 Å, they found the parameters of nonmagnetic broadening of the lines, then using the line $\lambda$ 6173 Å and the criterion $\chi^2$ they determined the optimal values of *B* and *f*.

Saar (1988) used and thoroughly analyzed the analytical solution of the transfer equation with the source function linearly depending on the optical depth and on the assumption of constant ratio of opacities in the line and in continuum throughout the atmosphere depth. Unlike Saar, Marcy and Basri (1989) developed the theory of magnetometric observations by introducing the numerical LTE stellar atmosphere model in which all parameters could vary with depth, and the transfer equations for individual Stokes parameters. From the magnetic-field-insensitive line of neutral iron $\lambda$ 7748 Å, they determined the abundance of iron and the turbulent velocity in the stellar atmosphere and then analyzed the line $\lambda$ 8468 Å sensitive to the field, adjusting the observed profile to the theoretical one by varying *B* and *f*. The observations were executed at the 3-m Lick telescope. Six of the 11 examined late G and K dwarfs distinctly displayed the Zeeman broadening, the strongest effect was found for $\xi$ Boo A ($B$ = 1600 G, $f$ = 0.22) and $\varepsilon$ Eri ($B$ = 1000 G, $f$ = 0.30). By the example of two other stars it was shown that among the late K dwarfs there were definitely stars with very low field strength, although the old and slowly rotating star 61 Cyg A, on which Saar and Linsky (1985) did not reveal any sign of magnetism, demonstrated again a clear presence of the field. A field of above 1600 G and a filling factor of over 30% were not found, though higher values of both parameters were measured before.

Saar (1991) performed magnetometric observations of two interesting K dwarfs. The rapidly rotating BD+26°730 star is of interest because its 60-year activity cycle was photometrically established and it is observed almost from the pole, thus the observed variations are primarily due to the proper evolution of the surface structures rather than the rotation of the star itself. The analysis of the profiles of the magnetosensitive line $\lambda$ 6173 Å and nearly low-sensitivity FeI, CaI, and NiI lines in the spectrum of the star and of the inactive dwarf HD 32147 revealed a good agreement between the lines profiles with small Lande factor and noticeable broadening of the $\lambda$ 6173 Å line, which can be attributed to a field with $B$ = 2600 G and $f$ = 0.5. Over 8 years, the brightness of this star decreased by $\Delta B = 0.15^m$, but the noticeable increase in its spottedness was not accompanied by any noticeable change of the intensity of the $H_\alpha$ line and ultraviolet lines of its chromosphere and transition zone. For another interesting star HD 17925, the youngest star for which magnetometric observations were successful, Saar (1991) found $B$ = 1500 G and $f$ = 0.35.

Combining the principle of Doppler imaging with the magnetometric analysis based on nonpolarized spectra, Saar et al. (1992) studied the spectra of the rapidly rotating K2 dwarf LQ Hya during seven phases of the rotational period. They considered the profile variations of five FeI, CaI, and NiI lines, including the magnetosensitive FeI line $\lambda$ 6173 Å and constructed temperature and magnetic maps. The maps demonstrated a rather good correspondence of dark and magnetic regions with $fB$ = 1.0 kG. To match this result with the above pattern, Saar et al. suggested that light and nonmagnetic regions were the regular photosphere, while magnetometric observations showed photometrical suppression of light faculae by dark spots in the center of which $fB$ reached 2.5 kG.

The first evidences of the complicated distribution of the magnetic field strength were obtained by Saar (1992) in analyzing the infrared TiI lines in the spectrum of AD Leo using the technique described above (Saar and Linsky, 1985). The obtained line profiles could not be represented by one pair of $f$ and $B$ describing the two-component photosphere with uniform magnetic and nonmagnetic regions throughout the star: $B$ = 3.5 kG and $f$ = 0.30 in the 0.28 phase, $B$ = 2.7 kG and $f$ = 0.45 in the 0.60 phase, and $B$ = 3.0 and $f$ = 0.40 in the 0.91 phase (Saar et al., 1994c). The observations could be satisfactorily presented either by two types of magnetic regions with $B_1 = 2.4$ kG, $f_1 = 0.45$ for faculae and $B_2 = 5.0$ kG, $f_2 = 0.30$ for spots, or by a magnetic field with a vertical strength gradient, which was first considered by Grossmann-Doerth and Solanki (1990) in connection with the measurements of stellar fields. One should remember that large sunspots are characterized by high field strength and low vertical gradient, whereas individual thin flux tubes are distinguished by low dispersion of the strength at a fixed level and a considerable vertical gradient.

For the magnetic fields of about 4 kG found for active M dwarfs by Saar (1994), the splitting of the infrared lines into the $\pi$ and $\sigma$ components should be distinctly seen in the high-resolution spectra. Reasoning from these considerations, Johns-Krull and Valenti (1996) observed two active M4.5e dwarfs EV Lac and Gl 729 with the 2.7-m Harlan J. Smith telescope at the McDonald Observatory. They observed the FeI $\lambda$ 8468 Å line with a resolution of 120000, the signal-to-noise ratio varied from 170 to 280, and an exposure of several hours. In parallel, they observed inactive M dwarfs. The resulting spectra made it possible to "see" immediately the Zeeman splitting: beyond the magnetosensitive line $\lambda$ 8468 Å the spectra of inactive stars Gl 725 B and Gl 876 "lead" the EV Lac spectrum but the profile of the line itself differed sharply, its core was much shallower and $\sigma$ components were clearly seen on both sides (see Fig. 12). For qualitative analysis of the spectra following the Allard and Hauschildt model (1995), theoretical profiles of this line for the atmosphere with and without field were calculated for $T_{\rm eff} = 3100$ K and log $g = 5.0$ using the transfer equations for individual Stokes parameters. As a result, it was established that the parameters of magnetic fields of active stars were determined ambiguously: for EV Lac $B = 4.2$ kG and $f = 0.4$ from the profile wings and $B = 3.4$ kG and $f = 0.6$ for the entire profile; analogously for Gl 729 $B = 2.8$ kG and $f = 0.4$ from the line wings and $B = 2.4$ kG and $f = 0.6$ from the entire profile. As Saar (1992) did in studying AD Leo, Johns-Krull and Valenti concluded that the absence of common estimates suggested nonuniform magnetic fields on the stellar surface or noticeable vertical gradients. But the difference $\Delta B \sim 1$ kG for two flare stars of identical spectral type suggests that the idea of gas pressure in the photosphere completely governing the field strength should be defined more precisely. Probably, this is valid for G and K dwarfs, but for M stars where, with increasing stellar activity due to increasing depth of the convective zone and high rotation rate, $f$ approaches unity, both $f$ and $B$ increase. Following the estimate of Kochukhov et al. (2001) obtained by the same method at the 3.6-meter ESO telescope with a resolution of 140000 and $S/N \sim 200$–250, for the M0 star GJ 1049 $B \sim 5$ kG and $f$ is at least 0.5.

Making use of 16 infrared lines of neutral iron near 1.56 μm, Valenti et al. (1995) analyzed the photospheric magnetic field of the K2V star $\varepsilon$ Eri from observations with a Fourier spectrometer at the 4-m Mayall telescope. To present the recorded profiles, the Stokes parameters were calculated by varying five free parameters: effective temperature, relative abundance of iron, macroturbulent velocity, $B$, and $f$. The temperature variations in the stellar atmosphere were assumed identical for magnetic and nonmagnetic regions and were taken from the linearly scaled model of the solar atmosphere. As a result, Valenti et al. found that the parameters $B = 1.44$ kG and $f = 0.088$ ensure the best representation within the two-component model. The corresponding value of $fB = 0.13$ kG is much lower than the previous estimates, but the thorough analysis of probable errors showed that the last estimate was the most reliable. The high reliability of the result is primarily due to the fact that in the obtained spectra the $\sigma$ component of the line $\lambda$ 1.56 μm with the Lande factor equal to 3.0 is distinctly separated from the $\pi$ component. Apparently, probable errors of the obtained parameters do not exceed 15% for $B$ and 35% for $f$ owing to the identical atmospheric model accepted for magnetic and nonmagnetic regions. Only somewhat lower estimates for the upper limits of *B* were obtained for 40 Eri and *σ* Dra.

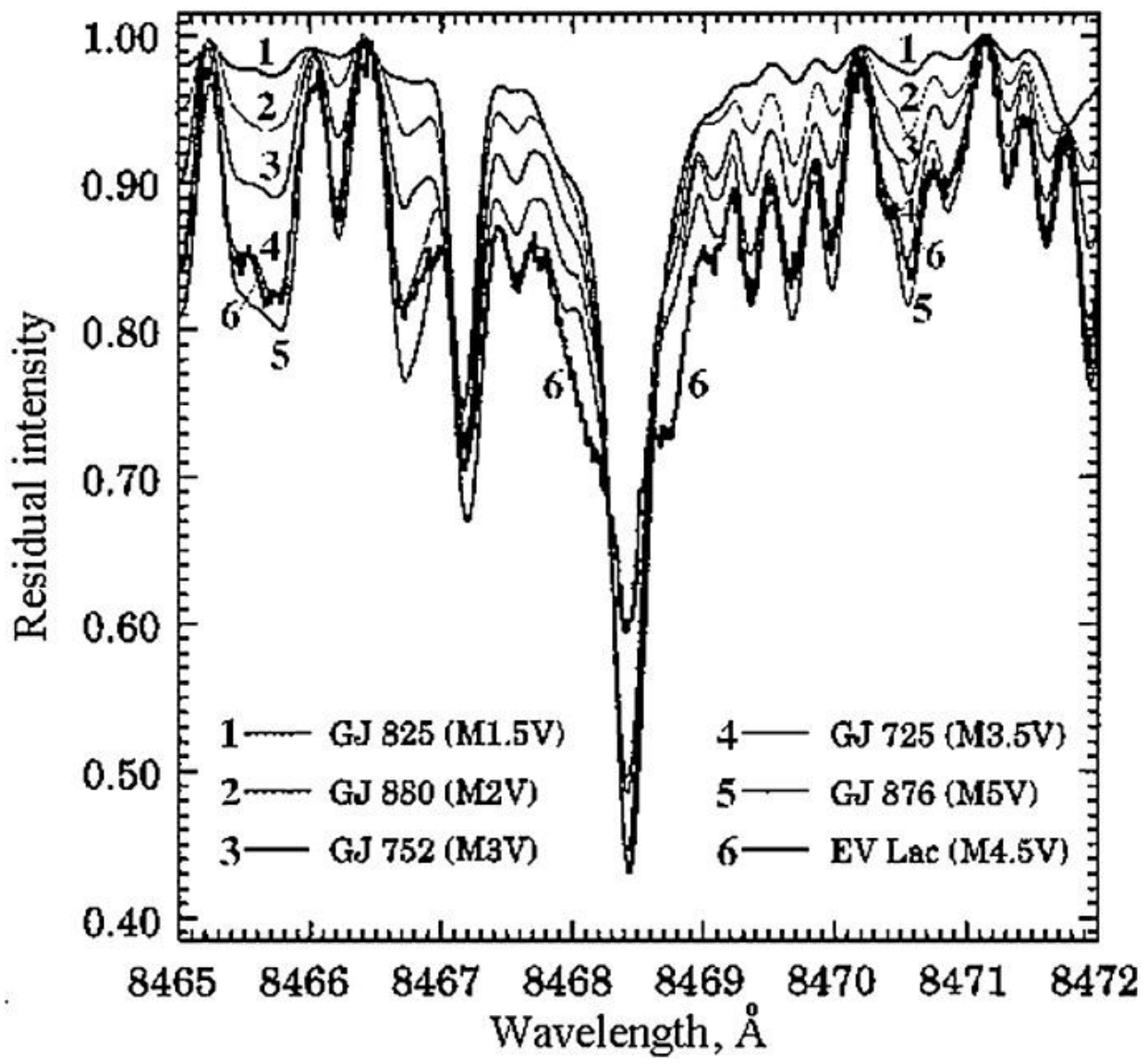

Fig. 12. Spectra of the EV Lac flare star and inactive M dwarfs in the region of the magnetosensitive line $\lambda$ 8468 Å (Johns-Krull and Valenti, 1996)

Using space technologies, an echelle spectrograph was developed for magnetometric studies in the infrared range. The spectrograph had an InSb detector whose sensitivity was a hundred times higher than that of the Fourier spectrograph used before. The first observational results obtained with this device were reported by Saar (1996a). Analysis of the observations of the K5Ve star Gl 171.2 A (=BD + 26°730 = V 833 Tau) within the framework of the two-component model of its surface with regard to the entire pattern of the Zeeman splitting and the calculation of the radiative transfer considering the magneto-optical effect in the Milne–Eddington atmosphere yielded the estimates *B* = 2.8 kG and *f* = 0.60, which complied with the previous estimates based on optical lines (*B* = 2.6 kG and *f* = 0.50). This correspondence evidences that some earlier optical observations were sufficiently precise, but the significant spottedness of this star provides an ambiguous answer to the question on the regions to which the obtained measurements belong. Thus, if the contrast between the spot and photosphere is large in the optical range, the radiation of spots can be neglected and the magnetic signal is produced by faculae and the bright network, while in the infrared region this contrast is weak and the contribution of spots to the magnetic signal becomes substantial or even dominant. Thus, from the observations of LQ Hya Saar (1996a) obtained *B* = 3.5 kG and *f* = 0.7. However, the equivalent width of one of the infrared titanium lines appeared to be four times greater than that in the less active *ε* Eri that belongs to the same spectral type. Such a significant amplification of the line can be explained only by the fact that it was formed at a lower temperature, i.e., the magnetic signal in this case is emitted by the spots and for them $f \sim 0.5$. These considerations require a three-component scheme: quiescent photosphere, faculae, and spots. For three dMe stars DT Vir, AD Leo, and YZ CMi, preliminary estimates yielded $B > 3$ kG and high values of *f*. The data on AD Leo and YZ CMi do not fit the two-component scheme either: there is a narrow $\pi$ component and very broad $\sigma$ components, which can be explained by the nonuniformity or appreciable vertical gradient of the magnetic field.

Table 4. Magnetic parameters of dwarf stars (Saar, 1996b)

| Star | Spectral type | Axial rotation period, days | $F$, % | $B$, kG | $\log F_X$, erg/(cm$^2 \cdot$ s) | $\log F_{CIV}$, erg/(cm$^2 \cdot$ s) |
|---|---|---|---|---|---|---|
| Sun | G2 V | 25.4 | 1.5 | 1.5 | 4.8 | 4.0 |
| HD 115383 | G0 V | 4.9 | 19 | 1.0 | 6.2 | 5.1 |
| HD 20630 | G5 V | 9.4 | 20 | 1.8 | 5.9 | 4.8 |
| Ksi Boo A | G8 V | 6.2 | 18 | 1.9 | 6.4 | 4.9 |
| HD 131511 | K1 V | 9 | 6 | 1.7 | 5.8 | 4.7 |
| HD 26965 | K1 V | 37 | <2.7 | 1.7 | 5.3 | 3.7 |
| HD 185114 | K1 V | 27.2 | <1.9 | 1.36 | 5.0 | 4.0 |
| HD 22049 | K2 V | 11.3 | 8.8 | 1.44 | 5.7 | 4.6 |
| HD 17925 | 2 V | 6.76 | 35 | 1.5 | 6.1 | 5.0 |
| LQ Hya | K2 Ve | 1.60 | 70 | 3.5 | 7.3 | 5.6 |
| Gl 171.2 A | K5 Ve | 1.85 | 50 | 2.8 | 7.1 | 5.7 |
| EQ Vir | K5 Ve | 3.9 | 55 | 2.5 | 7.0 | 5.3 |
| DT Vir | M2 Ve | 1.54 | 50 | 3.0 | 6.9 | |
| AD Leo | M3.5 Ve | 2.6 | 60 | 4.0 | 6.6 | 5.5 |
| YZ CMi | M4.5 Ve | 2.78 | 67 | 4.2 | 6.8 | 5.3 |
| EV Lac | M4.5 Ve | 4.38 | 50 | 3.8 | 6.7 | |
| Gl 729 | M4.5 Ve | | 50 | 2.6 | 6.2 | |

In 1995, Saar (1996b) summed up the results of 15 years of observations of magnetic fields on dwarf stars of the lower main sequence. Over these years new observational techniques were developed and data analysis became more complete, thus Saar ignored all observations that were performed with relatively low signal-to-noise ratios, without regard to the radiative transfer in the lines and the integration on the stellar disk, the results for K stars obtained for the line $\lambda$ 8648 Å and inconsistent with infrared data, and the results on Zeeman intensification of lines that yield a low-accuracy product of $fB^2$. The results of such a strict selection are presented in Table 4. These data confirmed and defined more precisely the statistical regularities derived earlier. Thus, for G–K stars and at least for the low-activity M dwarf Gl 729 there is a relationship $B \leq B_{eq}$, where $B_{eq} = (8\pi P_{ph})^{1/2}$, where $P_{ph}$ is the gas pressure in the quiescent photosphere, for the very spotted K2 star LQ Hya and active dKe and dMe dwarfs $B > B_{eq}$. As mentioned above, for LQ Hya this is associated with the contribution of spots to the magnetic signal, when it is natural to expect that $B_{spots} > B_{faculae} \sim B_{eq}$. Apparently, a similar situation is characteristic of emission K and M dwarfs.

Furthermore, the earlier proposed concept of the critical rotation rate remains valid. For slowly rotating stars, the relationship $B \leq B_{eq}$ is valid and $f$ depends strongly on the rotation rate: it is proportional to $P_{rot}^{-1.8}$ for $P_{rot} > 3$ days and for faster rotation a saturation at the level of $f \sim 0.6$ occurs. Instead of the axial rotation period as a global stellar parameter one can consider the Rossby number. In this case, the following statistical relations are valid: either $f$ is proportional to $Ro^{-1.3}$ at $Ro^{-1} < 8$ and $f \sim 0.6$ for $Ro^{-1} > 8$, or $\log f = -0.26–0.85$ Ro. In the latter relationship, the dependence of $f$ on Ro is identical to that obtained by Montesinos and Jordan (1993) from an earlier sampling of magnetometric measurements, but the absolute value of $f$ is approximately half as large. Saar explained this general tendency of a systematic decrease of $B$ and in particular of $f$ by the fact that an improved analysis technique could reveal more subtle effects responsible for some parts of the profile broadening that were first completely attributed to the magnetic field.

A principal disadvantage of the Saar analysis (1996a, b) is the hypothesis on identical thermodynamic structures of the magnetic regions and normal photospheres.

To determine the magnetic field parameters, Rueedi et al. (1997) analyzed the optical spectra of 13 G1-K5 dwarfs of low and moderate activity and found $fB = 165 \pm 30$ G only for $\varepsilon$ Eri. Their analysis was based on the numerical solution of the Unno–Rachkovsky equations for many lines with regard to magneto-optical effects. The results are in good agreement with the conclusions of Valenti et al. (1995), but Rueedi et al. doubted the significantly higher values of $fB$ obtained before and the validity of separate estimates of $f$ and $B$ from optical spectra with a resolution of less than 100000 and a signal-to-noise ratio less than 250. For $\varepsilon$ Eri, they also obtained constraints on the temperature of magnetic regions: it should be the same as or higher than that in nonmagnetic regions, i.e., the obtained magnetic-field parameters were attributed to faculae.

**1.2.4.3. Zeeman-Doppler Imaging.** Donati et al. (1997) applied a new Zeeman-Doppler spectropolarimetric method (ZDI) for magnetometric studies. Within this method, circular polarization is measured in separate regions of the line profile that correspond to different sectors of the stellar surface for stars with significant rotation. This method principally provides information on the three-dimension field structure because it allows one to determine radial, meridional, and azimuthal components, using profiles of a large number of spectral lines in the circularly polarized light. This technique makes it possible to directly explore the evolution of magnetic fields of stars, their cycles, but it does not perceive the strongest fields because "does not discern" dark spots. Within this method, a sought-for effect accounts for thousandths of fractions of the measured values, therefore it may be applied to only fairly bright stars in which a high signal-to-noise ratio may be achieved during a small phase interval. Among two dozen objects explored by Donati et al. (1997) there were two K dwarfs LQ Hya and CC Eri. Up to two thousand lines were recorded in their spectra. Analysis of spectra showed that magnetic fields on these stars comprised separate magnetic regions of different polarity and different orientation, and their surfaces were definitely not hotter than the quiescent photosphere, i.e., these were spots rather than faculae. Later, from the more complete data, Donati (1999) came to the conclusion that on the surface of LQ Hya there existed large-scale structures of the azimuthal field with a strength of hundreds of gauss and with different signs at different latitudes, and such structures persisted for many years.

Using the Zeeman-Doppler images of the surface magnetic fields of young fast rotators AB Dor and LQ Hya for several years, Jardine et al. (2000, 2002) constructed coronal potential fields and detected that such large-scale fields had not essentially changed for three years. Open regions of these fields come from two regions at middle latitudes at opposite longitudes. A stellar wind may form in them. Closed regions form tori near each pole, they may support prominences. Earlier, Jardine et al. (2001) found conditions for supporting prominences near the rotating stars with an arbitrary potential magnetic field. Such conditions existed both inside and outside the corotation radius of a single star for even the simplest field and at significant distance above the equatorial plane, as observed for AB Dor and PZ Tel, but the field should exhibit noticeable differences from a simple dipole at a distance of more than $4R_*$.

Other results of the Zeeman-Doppler imaging for LQ Hya were acquired by McIvor et al. (2004) from observations in December 2000 and December 2001. Extrapolating surface fields of the star potentially derived by ZDI, they found that in 2000 the large-scale field of the corona resembled an inclined dipole with two emersions, separated by 180° in longitude, at middle latitudes. A year later, these emersions moved toward poles, and the large-scale field resembled a smoothed dipole with a significant contribution of a number of small-scale west-east arcades. This situation differs substantially from that on the Sun where numerous latitude dipoles are valid near the cycle maximum, when the large-scale field resembles an inclined dipole. Despite a dramatic variation of the field structure, the emission measure EM changed a little and accounted for $10^{51.15}$ cm$^{-3}$.

McIvor et al. (2003) considered three possible models of polar spots on AB Dor: a unipolar spot, a bipolar region, and a nest of spots of different polarity. Having added these models to a map of the surface field for AB Dor determined by the ZDI technique, they observed the way these additions could change the resulting structure — the value and rotational modulation of the X-ray EM and the presence of prominences captured near the corotation radius, and found that only a single spot might yield a conspicuous effect, resulting in the open polar field.

Donati et al. (2003) reviewed the results of 5-year spectropolarimetric ZDI observations of AB Dor and LQ Hya with the Anglo-Australian Telescope. All the derived data were indicative of the continuous presence of large-scale, preferentially azimuthal fields on the surfaces of these stars, and they interpreted this fact as evidence of a dynamo mechanism that is valid over the whole thickness of the convective zone rather than constrained near its lower boundary similar to the Sun. The detected radial and azimuthal fields were identified with poloidal and toroidal components of the large-scale field. The authors suggested that the relative fractions of magnetic energy in these field components and polarity of the axisymmetric field component were variable and could be potentially useful for the diagnostics of magnetic cycles for AB Dor and LQ Hya.

To understand the existence of high-latitude magnetic fluxes, Mackay et al. (2004) estimated the emergence of new magnetic dipoles and the transfer of such fluxes of mixed polarity and found that, as to AB Dor, this might be possible during their emersion on the stellar surface at latitudes 50–70° and at a tenfold increase of the meridional flow rate up to 100 m/s.

From spectroscopic observations with the 3.9-meter Anglo-Australian Telescope, Slee et al. (2004) detected a strong magnetic field signal based on 4000 spectral lines of the CC Eri star.

Within spectropolarimetric observations, Petit et al. (2005) studied the magnetic field of $\xi$ Boo A. By the direct modeling of the Stokes V profile they found that the global magnetic field comprised two components: the inclined dipole with a strength of 40 G and an inclination of 35° to the rotation axis and the large-scale toroidal field with a strength of 120 G. Throughout 40 days there occurred an increase of the field strength and dipole inclination.

Marsden et al. (2006) carried out spectropolarimetry of the young G dwarf V899 Her with the 3.9-meter Anglo-Australian Telescope. Using the ZDI method, they plotted maps of brightness and topology of the magnetic field. Furthermore, they derived a large polar spot and weaker magnetic structures at low and middle latitudes. A reconstruction of the surface magnetic field revealed regions of the radial field at all latitudes except for polar ones and the azimuthal field that formed an almost full ring at latitudes 60–70°. Following DI, the surface differential rotation of the star was 7 times higher than that of the Sun.

Jardine et al. (2010) presented a general layout of modeling magnetic fields of stellar coronae from maps of spots (DI) and maps of the surface magnetic field (ZDI). First, using the ZDI technique, the surface magnetic field of a star is determined. It usually reveals a complicated distribution of spots, often hardly similar to solar one: with spots and elements of mixed polarity at high latitudes up to the pole. From these data, the potential or nonpotential 3D field is estimated, which is supplemented with fields of separate spots detected by the DI technique. In the corona, all the derived force lines are "hanged" by isothermal coronal plasma in the hydrostatic equilibrium condition. Supposing proportionality of the gas pressure to the magnetic one, one can find a coefficient of proportionality from the X-ray radiation of the corona. The obtained solution may be used as an input one in the MHD problem, which yields a common solution of the stellar corona and stellar wind.

**1.2.4.4. Infrared Molecular Magnetometry.** In the 2000s, there was formed a new direction in measuring magnetic fields of cool stars — infrared molecular magnetometry. Fundamentals of the general theory of the Zeeman effect in two-atom molecules of TiO and FeH were developed by Valenti et al. (2001) and Berdyugina and Solanki (2002). Lines in the FeH bands of about 1 μm in high-dispersion spectra of M and L dwarfs were far less blended than in other molecular bands in the optical and near IR ranges, and revealed remarkable sensitivity to the magnetic field.

The first astrophysical findings in this direction of measuring stellar magnetic fields were reported by Reiners and Basri (2006). They considered the inactive M dwarf GJ 1227, which basically has no magnetic field, and the active star EV Lac with $fB \sim 3.9$ kG and suggested an algorithm of linear interpolation of magnetosensitive characteristics for estimating magnetic field fluxes on other stars. Using this technique, Reiners and Basri (2007) estimated values of $fB$ from observations of 24 M2–M9 stars with the Keck I telescope. Values of v$\sin i$ and intensity of $H_\alpha$ emission were estimated simultaneously. The derived results are listed in Table 5. A comparison with data from Table 4 shows that molecular magnetometry allows one to extend toward cooler stars, and for four common stars the estimates of $fB$ in Table 5 exceed the appropriate values in Table 4 by 20–110%. It follows from Table 5 that the magnetic field forms on M dwarfs of all subtypes, but M dwarfs in these samples are very few for any conclusions; for M dwarfs of middle subtypes, a slow rotation corresponds to weak fields; and for late M stars, rotation is always measurable and the strongest fields up to 3–4 kG are characteristic of the fastest rotators. The $H_\alpha$ emission "substitutes" the estimates of the field, although the ratio $L_{H\alpha}/L_{bol}$ and *fB* vary with effective temperature. But a drop of $L_{H\alpha}/L_{bol}$ at the end of the main sequence is not accompanied by the extinction of the field.

Table 5. Magnetic parameters of dwarf stars (Reiners and Basri, 2007)

| Stars | Sp | $T_{eff}$, (K) | vsin *i*, km·s$^{-1}$ | $W_{H\alpha}$, Å | log($L_{H\alpha}/L_{bol}$) | log($L_X/L_{bol}$) | *fB* (kG) |
|---|---|---|---|---|---|---|---|
| Gl 70 | M2.0 | 3580 | ≤3 | < 0.20 | < −4.91 | < −4.44 | < 0.1 |
| Gl 729 | M3.5 | 3410 | 4 | 4.47 | −3.66 | −3.50 | 2.2±0.1 |
| Gl 873 | M3.5 | 3410 | ≤3 | 5.98 | −3.53 | −3.07 | 3.9 |
| AD Leo | M3.5 | 3410 | ≈3 | 3.40 | −3.78 | −3.02 | 2.9±0.2 |
| Gl 876 | M4.0 | 3360 | ≤3 | <0.21 | <−4.97 | −5.23 | <0.2 |
| GJ 1005A | M4.0 | 3360 | ≤3 | <0.21 | <−4.95 | −5.05 | <0.1 |
| GJ 299 | M4.5 | 3300 | ≤3 | <0.33 | <−4.84 | <−5.55 | 0.5±0.2 |
| GJ 1227 | M4.5 | 3300 | ≤3 | <0.32 | <−4.85 | <−3.86 | <0.1 |
| GJ 1224 | M4.5 | 3300 | ≤3 | 4.13 | −3.74 | −3.06 | 2.7±0.1 |
| YZ CMi | M4.5 | 3300 | 5 | 8.42 | −3.43 | −3.02 | >3.9 |
| Gl 905 | M5.0 | 3240 | ≤3 | <0.30 | <−4.95 | −3.75 | <0.1 |
| GJ 1057 | M5.0 | 3240 | ≤3 | 1.06 | −4.41 | <−3.87 | <0.2 |
| GJ 1245B | M5.5 | 3150 | 7 | 4.26 | −3.76 | −3.58 | 1.7±0.2 |
| GJ 1286 | M5.5 | 3150 | ≤3 | 1.50 | −4.21 | <−3.77 | 0.4±0.2 |
| GJ 1002 | M5.5 | 3150 | ≤3 | <0.36 | <−4.83 | <−5.24 | 0.0 |
| Gl 406 | M5.5 | 3150 | 3 | 9.93 | −3.39 | −2.77 | 2.4±0.1 |
| GJ 1111 | M6.0 | 2840 | 13 | 8.12 | −3.92 | −3.88 | 1.7±0.2 |
| VB 8 | M7.0 | 2620 | 5 | 7.09 | −4.27 | −3.47 | 2.3±0.2 |
| LHS 3003 | M7.0 | 2620 | 6 | 7.60 | −4.24 | −4.01 | 1.5±0.2 |
| LHS 2645 | M7.5 | 2540 | 8 | 4.26 | −4.67 | | 2.1±0.2 |
| LP 412−31 | M8.0 | 2480 | 9 | 23.05 | −3.89 | | >3.9 |
| VB 10 | M8.0 | 2480 | 6 | 6.53 | −4.44 | −4.90 | 1.3±0.2 |
| LHS 2924 | M9.0 | 2390 | 10 | 5.76 | −4.70 | −4.35 | 1.6±0.2 |
| LHS 2065 | M9.0 | 2390 | 12 | 29.05 | −4.00 | −3.50 | >3.9 |

With UVES/VLT, Reiners et al. (2007a) determined the average magnetic field of CN Leo, using the molecular FeH band as $fB$ = 2.2 kG and found it to be variable on time scales up to 6 hours.

Later, within the described technique, Reiners and Basri (2008a), using the FeH lines, measured the magnetic flux of Prox Cen: $450 < fB < 750$. This value proved to be lower than that for the most active flare stars but in accordance with its slow rotation, flare activity level, and emission level of this star. By means of the Keck telescope, Reiners et al. (2009a) measured magnetic fluxes of 7 fast rotators in which the X-ray and $H_\alpha$ emissions are indicative of saturation, i.e., such a high level that does not depend on rotation. For these stars, the Rossby number was about 0.01 and *fB* was from 2 to > 3.9 kG, filling factors of the field and emission were close to 1, i.e., the magnetic flux was saturated.

Shulyak et al. (2011a) performed magnetometric studies in molecular lines of FeH with the cooled high-dispersion infrared echelle spectrograph (CRIRES) mounted on VLT at ESO. They carried out observations of 10 close binary and multiple systems comprising M dwarfs of early and late subtypes, i.e., stars on both sides of the transition from partially to fully convective stars. From high-resolution spectra, the rotation rates, magnetic field strengths, and metallicities were determined. The earlier conclusion was confirmed that low-mass M dwarfs decelerate not so quickly as earlier M stars. Using four different strategies for accounting for poorly known parameters of stellar atmospheres, they estimated magnetic fields between 1.5 and 3 kG on primary components of GJ 852, GJ 234, LP 717-36, and GJ 3322, and on the secondary component of GJ 852; they confirmed a field of 2 kG on the primary component of the GJ 2005 triple system. Given v*sini* > 10 km/s, one fails to explain the field strength, but it grows with increasing v*sini* for stars with low and middle rotation rates. Apart from a detailed consideration of a molecule of FeH, Berdyugina (2002) suggested and elaborated a technique for the analysis of stellar (and solar) spots, using molecular bands of TiO $\lambda$ 7055 Å.

* * *

Kochukhov et al. (2009) carried out magnetometric observations of 35 K7–M6 dwarfs with VLT/CRIRES at ESO. At the spectral resolution $10^5$ they worked with lines of Ti I $\lambda$ 2227 nm and Na I $\lambda$ 2208 nm. In the spectrum of YZ CMi, one can see a splitting of the titanium line, which corresponds to a field of 3–4 kG, and in the spectrum of GJ 1049 — a broadening of this line corresponding to a field of 1–2 kG. In spectra of both stars and AD Leo, a splitting of the strong sodium IR line is seen well. Profiles of this line are presented by combinations of components with different strengths and filling factors of the field; such a modeling yields $\Sigma fB$ = 4.5 kG for YZ CMi, 3.2 kG for AD Leo, 4.3 kG for GJ 398, and 2.9 kG for GJ 1049, whereas for the first two the accuracy of estimates is by an order of magnitude better than in Reiners and Basri (2007). These estimates are noticeably higher than those in the Zeeman-Doppler imaging of Morin et al. (2008b), which may be associated with worse spatial resolution of ZDI observations.

Shulyak et al. (2010) performed magnetometric studies of M dwarfs simultaneously, using atomic and molecular FeH lines, and confirmed the existence of strong fields on YZ CMi, EV Lac, AD Leo, and GJ 1224 but 15–30% weaker than the appropriate estimates of Reiners and Basri (2007). On the other hand, field estimates for the FeI $\lambda$ 8468 Å line were systematically higher than those for FeH lines (Shulyak et al., 2011a). Later, Shulyak et al. (2014) carried out magnetometric observations of four M dwarfs on the boundary of partially and fully convective stars GJ 388, GJ 729, GJ 285, and GJ 406, found a maximum average field for GJ 285 of 3.5 kG at its maximum value of 7–7.5 kG, found no difference in the distribution of fields on partially and fully convective stars, but for the most active of these stars with maximum X-rays and average surface magnetic field, the rotation was fast.

* * *

Of particular attention is a series of magnetometric studies carried out with two identical specialized spectrographs ESPADOnS and NARVAL (Donati et al., 2008b)[1].

Using ESPADOnS, Phan-Bao et al. (2006) conducted spectropolarimetric observations of two M dwarfs with noticeable differences in rotation rates: EV Lac (M3.5) and HH And (M5.5). They revealed a longitudinal magnetic field from 18 to −40 G for EV Lac, whereas this variation occurred less than during 50 minutes and no flares were observed during this time. Only an upper limit of the field of 5 G was obtained for HH And.

From observations with NARVALe, Morin et al. (2008a, 2008b, 2009) reported the first results of a spectropolarimetric analysis of five M3–M4.5 dwarfs with fast rotation: AD Leo, EV Lac, EQ Peg, YZ CMi, and V 374 Peg with the aim of determining the dynamo effect on both sides from the boundary between fully and partially convective stars. The Zeeman-Doppler imaging (ZDI) technique led to the conclusion that all considered stars had axisymmetric large-scale poloidal fields without a toroidal component, three of them observed during different years showed stability throughout a year; the magnetic topology of the fully convective stars differed significantly from hotter G and K stars, usually having a strong toroidal component in the form of the azimuthal ring of the field, which was coaxial with stellar rotation. From the rotation modulation of nonpolarized line profiles they concluded that the contrast and filling factor of the available spots on the fast rotator V374 Peg were far lower than those on not fully convective active stars with the same rotation period. Based on separate circularly polarized profiles, the large-scale topology of the magnetic field was found to be stable throughout a year and revealed a very weak differential rotation, which was 10 times lower than the solar one. The quantitative results of this study are presented in Table 6. Here $R_X = L_X/L_{bol}$, the last four columns list fractions of magnetic energy in appropriate components of the detected fields.

Donati et al. (2008a, 2009) determined rotation periods and constructed the large-scale magnetic topology of six early M dwarfs. They found that such dwarfs comprised preferentially large-scale fields with domination of the toroidal and nonaxisymmetric poloidal configuration with significant differential rotation and long-term variability; only the most low-mass star of this sample has almost fully poloidal, generally axisymmetric large-scale field, which resembles fields of middle M dwarfs. This abrupt change in the large-scale topology around M3 is not seen in the X-ray luminosity governed by the general magnetic flux, and it means that the dynamo process becomes more effective in producing the large-scale field after M3 at the same magnetic flux. They concluded that the abrupt variation of magnetic field topology around M3 was associated with a fast decrease of radiative cores for low-mass stars and simultaneously a fast increase of the convective turnover time. This circumstance can be partially responsible for the reduced magnetic braking of the fully convective stars.

[1] For spectropolarimetric observations of cool low-mass stars, the spectrographs-twins were designed: ESPADOnS for 3.6-m CHFT in Hawaii and NARVAL for the 2-m Bernard Lyot Telescope in Pic du Midi. These are echelle achromatic instruments for the range 3700–10000 Å with a spectral resolution of 65000, with fibre supply are mounted on the bench of the Cassegrain focus and may simultaneously record two orthogonal components of polarization with a CCD detector with total effectiveness, including the telescope and detector, 10–15%.

Thus, there were found noticeable differences between partially and fully convective low-mass stars as regards the topology and characteristic sizes of surface magnetic fields and differential rotation; this allows one to suspect different dynamo mechanisms in these two groups of stars.

Table 6. Characteristics of magnetic fields of rapidly rotating M dwarfs (Morin et al., 2008b)

| Stars (a year of observations) | $M/M_{\odot}$ | $P_{rot}$, days | Ro, $10^{-2}$ | $\log R_X$ | *f* | <*B*>, kG | Fractions of magnetic energy in components, % | | | |
|---|---|---|---|---|---|---|---|---|---|---|
| | | | | | | | Poloidal | Dipole | Quadrupole | Octupole |
| EV Lac(06) | 0.32 | 4.38 | 6.8 | -3.3 | 0.11 | 0.57 | 87 | 60 | 13 | 3 |
| (07) | – | – | – | – | 0.10 | 0.49 | 98 | 75 | 10 | 3 |
| YZ CMi(07) | 0.31 | 2.77 | 4.2 | -3.1 | 0.11 | 0.56 | 92 | 69 | 10 | 5 |
| (08) | – | – | – | – | 0.11 | 0.55 | 97 | 72 | 11 | 8 |
| AD Leo(07) | 0.42 | 2.24 | 4.7 | -3.2 | 0.14 | 0.19 | 99 | 56 | 12 | 5 |
| (08) | – | – | – | – | 0.14 | 0.18 | 95 | 63 | 9 | 3 |
| EQ Peg A (06) | 0.39 | 1.06 | 2.0 | -3.0 | 0.11 | 0.48 | 85 | 70 | 6 | 6 |
| EQ Peg B(06) | 0.25 | 0.4 | 0.5 | -3.3 | | 0.45 | 97 | 79 | 8 | 5 |
| V374 Peg (05) | 0.28 | 0.45 | 0.6 | -3.2 | | 0.78 | 96 | 72 | 12 | 7 |
| (06) | – | – | – | – | | 0.64 | 96 | 70 | 17 | 4 |

While summing up the results of the campaign (Fig. 13), Morin et al. (2010) reported that in the course of the analysis of 11 fully convective M5–M8 dwarfs the qualitatively different magnetic topologies were detected: strong axisymmetric dipolar (strong dipolar — SD) and weak fields with significant nonaxisymmetric components and sometimes with a significant toroidal component (weak multipolar — WM). A comparison of absolute fluxes showed that in the second case the large-scale components comprised less energy, similar to early M dwarfs. But stars in both groups had similar stellar parameters and there was no evidence for their separation in the mass–rotation ratio, and the transition within one star from one structure to another was suspected. They found a significant toroidal axisymmetric component of the field for the M8 dwarf VB10 but detected no field on the M7 star VB 8. A concept of existence of two types of magnetic structures and, consequently, two types of dynamo for the most low-mass M dwarfs was developed by Morin et al. (2011b).

Phan-Bao et al. (2009) reported the results of observations of the M4 dwarf Gl 490 B throughout three nights with ESPaDOnS. They detected the large-scale axisymmetric poloidal field with a flux of $fB$ = 3.2 kG; this was significantly higher than the average value 0.68 kG. This divergence meant that a great part of magnetic energy was contained in small-scale structures. The light curve of $H_{\alpha}$ emission provided evidence for the rotational modulation, i.e., the existence of local structures in the chromosphere.

Based on observations with NARVALe throughout three seasons, Petit et al. (2009) traced variations of the large-scale magnetic field of the solar-type star — the fast rotator HD 190771 with a rotation period of 8.8 days. During observations between 2007 and 2008, there occurred a polarity reversal of the axisymmetric component of the field, whereas between seasons of 2008 and 2009, a fraction of magnetic energy within the toroidal component abruptly reduced.

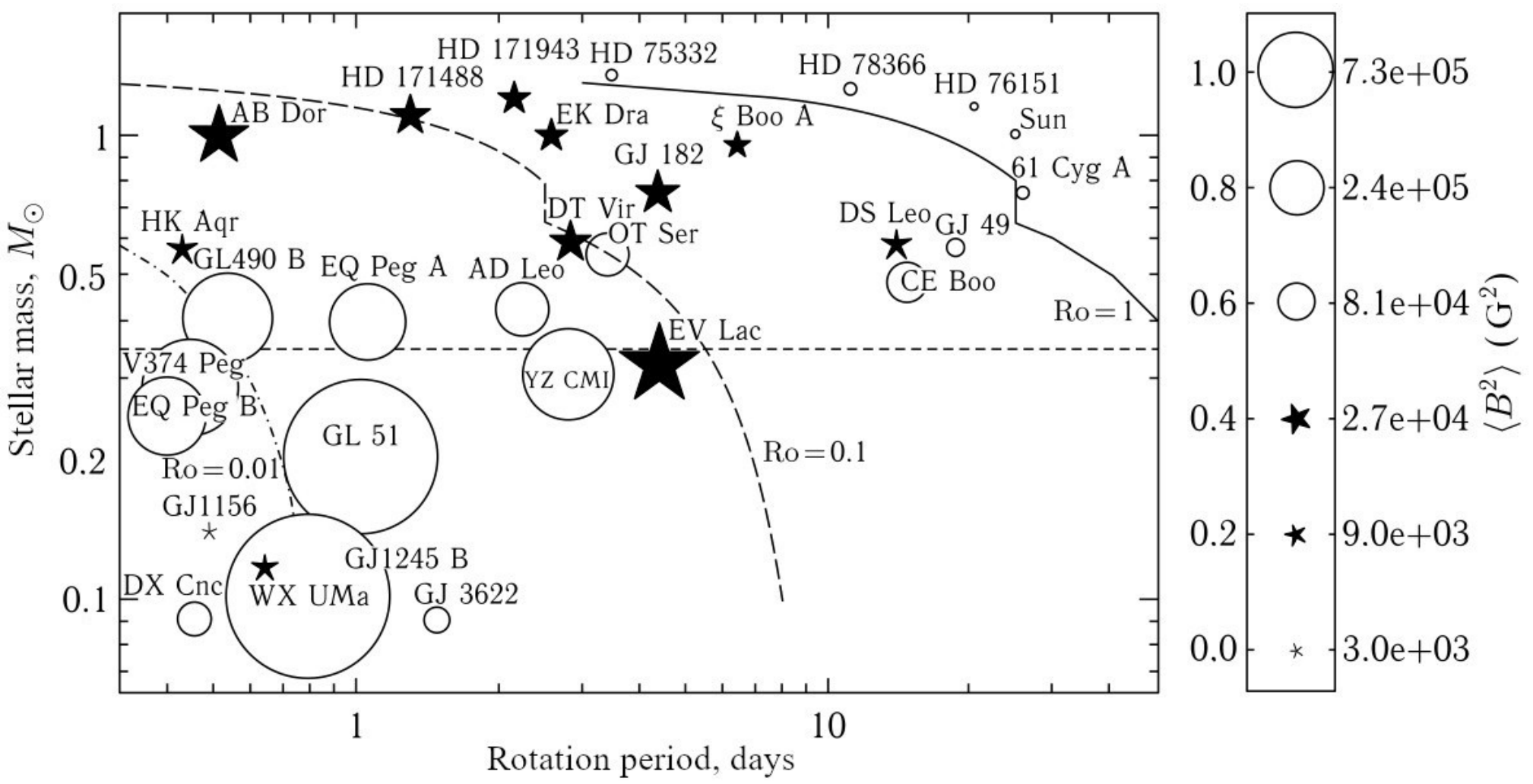

Fig. 13. Basic properties of the large-scale magnetic topology of cool stars. The sizes of symbols correspond to the density of magnetic energy, circles — an axisymmetric component, asterisks — a nonaxisymmetric component. Light curves indicate the position of Rossby numbers equal to 1, 0.1, and 0.01. The smallest and the largest symbols correspond to average strengths of the field of 3 G and 1.5 kG. The horizontal dashed line marks a boundary between partially and fully convective stars at a level of $0.35M_\odot$ (after Donati (2011) and Morin (2012))

Between 2007 and 2011, using the NARVAL spectropolarimeter at the Bernard Lyot Telescope, Morgenthaler et al. (2011) studied 19 stars in the range from 0.6 to $1.4M_\odot$ and with rotation periods from 3.4 to 43 days. Throughout four years, all the considered stars revealed variability of different types. Using the ZDI technique, the large-scale maps of stellar magnetic fields were plotted, and their long-term evolution was explored. Stars that have shown at least one reversal in the course of the monitoring generally rotated faster and their masses were equal to or somewhat higher than that of the Sun. During the monitoring, a polarity reversal was recorded for several stars, i.e., the magnetic cycles, significantly shorter than the cycles determined by the monitoring of chromospheric calcium lines; this provided evidence for the coexistence of several magnetic characteristic times on the same star.

Based on 76 spectra derived with the NARVAL spectrograph throughout seven epochs in the same 2007–2011, Morgenthaler et al. (2013) carried out a long-term monitoring of the magnetic field of the solar-type star $\xi$ Boo A. From approximately 6100 photospheric lines[1]

[1] Plachinda (2013) combined two different methods for implementing ZDI: the LSD method (Least-Squares Deconvolution), when based on a number of spectral lines one can determine a single weighted average contour, and SL method (Single Line) in the frame of which each spectral line is used separately. The LSD method has a high penetrating power but uses an assumption on the similar magnetic situation

they acquired average line profiles in the polarized light and by the ZDI technique modeled the average magnetic field of each spectrum. Moreover, they measured the widths of several magnetosensitive spectral lines, radial velocities, the asymmetry of lines and chromospheric emission in the CaII H and K cores. The modeling showed that in most active epochs of 2007 and 2011, the large-scale field of $\xi$ Boo A was to a great extent asymmetric with a predominance of the polarized component, which maintained continuous polarization and a significant fraction of magnetic energy of the large-scale field in all epochs. Magnetic topologies during these activity maxima were very similar, suggesting a short cyclicity of the large-scale field. The average magnetic flux varied within a factor of 2, whereas the chromospheric flux — only within a factor of 1.2. There was a correlation between CaII, $H_\alpha$, and widths of magnetosensitive lines caused by rotational modulation and variability. A large latitudinal shift, 0.3 rad/day, resulted in significant differential rotation.

In *Publications of the European Astronomical Society*, Morin (2012) presented a comprehensive review on magnetism of low-mass stars in which provided theoretical fundamentals of the modern magnetometry of such objects and summarized the listed above experimental results on fully convective stars and down to brown dwarfs as to the structure, magnetic field strength, and rotation pattern.

The spectra derived with ESPADOnS and NARVAL by the mid 2014 are published by Petit et al. (2014).

Bychkov et al. (2015) conducted a detailed analysis of the performed by Donati and Morin with colleagues magnetometric observations of the flare stars OT Ser and YZ CMi and found that the values of magnetic field strengths could be represented by sums of two sinusoids with average values of 65 and more than 400 G at the rotation periods of 3.424 and 2.7773 days, respectively. Suggesting that the accumulated magnetic energy of a star is revealed in stellar flares, following the statistics of flares, they estimated a ratio of the strength of the magnetic field generator to the stellar rotation energy as $2 \cdot 10^{-4}$ for YZ CMi and $10^{-6}$ for OT Ser.

Berdyugina et al. (2017) first measured a magnetic field on the surface of the active brown M8.5 dwarf LSR J1835 + 3259. This rapidly rotating star with a period of 2.8 hours and effective temperature of 2800 K, log $g$ = 4.5, a mass of 55 ± 4 Jupiter masses and age of 22 ± 4 million years revealed variable radio and optical emission. Observations were carried out with a spectropolarimeter at the Keck telescope on August 22 and 23, 2012. On the first night, when the active region of the star was turned to the Earth, through the circular polarization of the sodium lines $\lambda$ 8190 Å a magnetic field strength was estimated to be 5.1 kG and the filling factor was at least 11% of the visible hemisphere. Line profiles allowed one to represent the stellar active region by hot plasma loops with temperatures higher than 7000 K, with the stratification of radio and optical sources and the measured field at their base. The magnetism of the star was suggested to be associated with its youth and may be due to the magnetic accretion, similar to more massive T Tau stars.

* * *

Within MHD, Zaqarashvili et al. (2011) considered the large-scale waves on rapidly rotating tachoclines in the presence of the toroidal magnetic field and found that the low-frequency magnetic Rossby waves were localized at the poles, whereas the high-frequency Poincare waves were concentrated near the equator. The nonstable magnetic Rossby waves may result in increasing magnetic fluxes at high latitudes on rapidly rotating stars, which leads to near-pole spots and short-period magnetic activity.

in the regions of formation of all lines, whereas within the SL method there is no such an assumption but its penetrating power is less.

At the 17th Cool Stars symposium (Barselona, Spain, 2012), 12 leading specialists made a four-page report on various aspects of the modern study of stellar magnetism (Morin et al., 2013). In paragraphs considering starspots, they noted two effective then methods for measuring magnetic fields — the Zeeman broadening of spectral lines and the Zeeman–Doppler imaging of rapidly rotating stars — and an effect of the magnetic field on the fundamental parameters of low-mass stars, imaging of starspots and their magnetic fields through molecular IR lines, and the MHD consideration of the surface convection in cool stars.

**1.2.4.5. Statistical Magnetometric Methods.** In the late 1970s and early 1980s, when the described above methods of direct magnetometric observations were being elaborated, the statistical methods for estimating parameters of stellar magnetic fields were suggested. They have gone out of use so far but are of particular historical interest.

The Stenflo–Lindegren statistical method. To study the magnetic properties of spatially unresolved structures on the solar surface, Stenflo and Lindegren (1977) proposed a method of the statistical analysis of many spectral lines. The essence of the method is that a measured parameter of the line profile, e.g., its depth, width at a certain fixed level or the equivalent width, is written as a sum of functions of known atomic constants and unknown thermodynamic and magnetic parameters in the region of line formation. From these relations one can find unknown parameters by studying a set of nonblended lines using the regression method. Mathys and Solanki (1987) and Solanki and Mathys (1987) used this approach to analyze magnetic fields of stars. As in the Robinson method, this statistical method provides an estimate of the effective strength of the magnetic field and the area it occupies on a star. In addition, it allows thermodynamic parameters of magnetized and nonmagnetized regions of the photosphere to be separated. Thus, the obtained results depend weakly on the availability of weak blends.

The first results of studying the magnetic fields on lower main-sequence stars by the statistical method were reported by Mathys and Solanki (1987) and Solanki and Mathys (1987). They observed three stars with a resolution of 100000 and a signal-to-noise ratio of several hundred and analyzed many tens of optical FeI lines. For $\varepsilon$ Eri, the field strength determined with the greatest reliability was 1800–2500 G and the filling factor was 0.15–0.20. Strictly speaking, here, as in the Robinson method, it is not the value of $f$ itself that is determined but rather the product $f\delta_c$, where $\delta_c$ is the continuum brightness contrast between the magnetic and nonmagnetic parts of the photosphere. Analysis of these observations resulted in the conclusion: magnetic regions are hotter than nonmagnetic ones, i.e., they are similar to solar faculae.

Mathys and Solanki (1989) found $fB$ and $B$ for four stars and in two of them, also studied by Saar (1988), they found much higher values. They did not manage to find an unambiguous cause of the discrepancy but suggested that, if magnetic regions were hotter than nonmagnetic ones, the estimates of $f$ obtained on the assumption of identical magnetic and nonmagnetic atmospheres were overestimated. Grossmann-Doerth and Solanki (1990) paid attention to the fact that in magnetometric studies different observers often used various lines of neutral iron $\lambda$ 5250, 6173, 8468, and 15648 Å formed at noticeably different depths. If one considers that the vertical gradient of a stellar magnetic field is analogous to that occurring in the flux tubes of solar faculae, i.e., with the field pressure decreasing exponentially, in the estimates for $B$ one can expect a discrepancy of up to 1 kG, but this circumstance should not distort an estimate of the magnetic flux $fB$.

Savanov and Savel'eva (1997) refined the Mathys and Solanki algorithm (1989) and estimated magnetic-field parameters for 10 G–K dwarfs. They estimated *fB* for four stars: for $\xi$ Boo A *fB* = 990 ± 250 G, for 70 Oph A *fB* = 1700 ± 500 G, for $\varepsilon$ Eri *fB* = 1200 ± 500 G, and for HD 2047 *fB* = 3200 ± 1100 G. For $\xi$ Boo A, two spectra were obtained with an interval of one hour, but only one of them provided a magnetic-field signal, which is probably explained by the motion of a magnetoactive region over the stellar disk due to rotation. However, it is not excluded that simple expansion of the absorption line parameters in the magnetic field into the sum of simple functions is not sufficient to replace the allowance for the radiative transfer effects and averaging over the stellar disk required strict solution of the problem. It does not provide the required precision of the estimate of stellar field parameters.

Zeeman enhancement of equivalent widths of lines. Another statistical method for estimating stellar magnetic fields was developed by Basri et al. (1992). The essence of the method is as follows. First, inactive stars with temperature and iron abundance close to those of active stars under consideration are selected. Two to three dozen nonblended lines of neutral iron with different magnetic sensitivity are chosen in the spectra of inactive stars. The expected equivalent widths of these lines for nonmagnetic and magnetic atmospheres with a field of 2 kG are calculated from the known atmosphere models. The ratio of an equivalent width of a line in the magnetized atmosphere to the appropriate value in the nonmagnetic atmosphere is taken as a factor of Zeeman enhancement of the line. Then the factors are compared in the spectra of real active stars with those theoretically calculated, which enables an estimation of the stellar field parameters. This method requires high-resolution spectra with high signal-to-noise ratio, the spectra of inactive stars should be "rotated up" to the value of vsin*i* typical of an active star. Basri and Marcy (1994) applied this method to study five chromospherically active stars. The axial rotation period of four of them was less than seven days, 11.3 days for $\varepsilon$ Eri, and for the compared stars the period was about 30 days. In observations at the 3-meter Lick telescope with an echelle spectrograph and during 1–2 h exposures, the spectra within 3800–9000 Å were recorded; 23 iron lines within 5320–8760 Å were used. As a result, with the highest confidence the field for EQ Vir was found, *fB* = 1.73 kG, which is in rather good correspondence with the estimates based on the Zeeman broadening. For $\varepsilon$ Eri and HD 17925 *fB* < 0.5 kG, which is the detection limit of the method. For LQ Hya and Gl 171.2 A the data scattered considerably, which was probably caused by the significant spottedness of these red dwarfs. In principle, the method of Zeeman enhancement of lines should work for weaker stars better than the method of Zeeman broadening.

* * *

Great efforts were undertaken to search for stellar magnetic fields with broadband polarimetry (Huovelin and Saar, 1991; Saar and Huovelin, 1993). The authors faced some principal difficulties concerned with the necessity of obtaining extremely high-precision observations and separating the magnetic-field effects from the Rayleigh and Thompson scattering. But Alekseev (2003) carried out polarimetric observations of LQ Hya, VY Ari, and EK Dra with the UBVRI photometer-polarimeter of AZT-11 in the Crimea and detected a proper linear polarization in the U band. The modeling of this effect based on the Zeeman effect resulted in the conclusion that all these stars had strong local magnetic fields up to 2 kG with filling factors up to 40%, and magnetized regions coincided with cool photospheric spots detected from photoelectric observations.

* * *

In the course of observations with the Kepler spacecraft, Balona (2015) found that the frequency of flares was approximately similar on stars of spectral types from M to A. The detection of spots and flares on A stars provided evidence for the presence of strong magnetic fields on them.

* * *

Let us sum up the results.

The many-year study of the Sun has shown that photospheric magnetic fields are a crucial factor that governs different nonstationary processes on the solar surface. This circumstance stimulated prolonged efforts directed at acquiring information on magnetic fields of lower main-sequence stars. But magnetometric observations of cool stars proved to be one of the finest experimental studies in both deriving reliable data, which are evidenced by numerous listed above instrument designs, and their correct analysis. Nevertheless, a series of results of such studies may be considered as reliably established. These results are as follows.

Similar to the Sun, lower main-sequence stars have no global strong magnetic fields that are characteristic of some stars of early spectral types. The global fields of the stars under consideration are not stronger than one–two tens of gauss.

But many middle- and low-mass dwarfs, the Sun as well, have local magnetic fields of numerous small different-polarity structures with the strength of hundreds of gauss to many kilogauss, though on some of the explored G–K–M dwarfs the fields are weaker than the detection threshold.

The interpretation of measured parameters of local magnetic fields is not evident and requires additional simulation. Apparently, the measurements in the optical region, in particular on G and early K dwarfs, yield the magnetic parameters of photospheric faculae, whereas in the infrared region, in particular for M stars, the fields of starspots start being revealed. Thus far, there is no generally accepted algorithm for calculating the three-component model including a normal photosphere, faculae, and spots. However, recently Johnson et al. (2021) published the first analysis of the expected photometric peculiarities of magnetoactive G–M stars with hot faculae and cool spots. The developed theory is based on the light curves in different photometric bands and is built on distinctions in visibility of spots and faculae at different distances from the disk center, and on the effect of limb darkening.

In comparing the results of magnetometric observations in different spectral regions one should bear in mind not only the difference in the contrast between the spot and photosphere, but also the fact that the coefficients of continuous absorption of stellar photospheres noticeably vary in different regions: they are almost equal at 0.62 and 2.2 μm, higher at 0.85μm and much lower at 1.56 μm. Thus, at 1.56 μm one can expect greater $B$ and lower $f$, whereas at 0.85μm $B$ is lower and $f$ is higher.

The strength of local magnetic fields of yellow and not too late red dwarfs is defined by the condition of equal pressures of the field and the ambient photosphere. Therefore, they systematically increase from many hundreds of gauss for G stars to many kilogauss for M dwarfs, "forget" the specific generation mechanism, and do not correlate with the surface filling factor and the rotation rate of the star.

The total efficiency of the field-generation mechanism is determined by the depth of the convective zone and the rotational velocity of a star. The total efficiency and the temporal variations of the magnetic flux are manifested in the filling factor that systematically increases from several hundredths for G stars and approaches unity for M dwarfs. As the rotation rate increases, so does the total magnetic flux or the average field strength $fB$ that is proportional to the flux.

The observational data thus far have not provided direct confirmation of the existence of magnetic fields of tens of kilogauss in starspots, though the facular areas of several kilogauss can be considered as an indirect argument in favor of such strong spot fields.

Messina et al. (2001, 2003) considered the reasons why *fB* should correlate with the maximum amplitudes of rotational modulation of stellar brightness in the V band. They found a correlation of *fB* with the rotational periods and Rossby numbers of active stars of the field and of six clusters of different age. At the photospheric level, this result closes the earlier revealed rotation–activity correlations at chromospheric and coronal levels and in the transition zone; this will be discussed in the following chapters. The maximum amplitudes are the highest for K dwarfs of the age of the Pleiades and the $\alpha$ Per cluster and decrease as the age of the cluster increases. As for the activity in the upper layers of stellar atmospheres, the proposed photospheric activity index suggests the effect of activity saturation for the fastest rotators with the rotational periods of less than 0.35 days and Rossby numbers less than 0.02.

## 1.3. Chromospheres and Transition Zones

In the very first physical models, stars were considered as opaque gravity-bound formations of ideal gas. Under the conditions of radiative equilibrium the bodies should have maximum temperatures and densities in their centers and minimum temperatures and densities on the surfaces. However, solar studies showed that this model was incorrect: as matter comes up from the subphotosphere its temperature decreases, but upon reaching about 4200 K at a certain level it starts increasing again. Thus, at a height of about 300 km above the temperature minimum an effective ionization of hydrogen starts. These layers of the solar atmosphere were discovered first from the emission of the bright red line of hydrogen $H_\alpha$ during solar eclipses and got the name “chromosphere”.

The contradiction between the initial stellar model and the observations of the Sun was removed when the convective energy transfer was found to play an important role in subphotospheric layers of medium- and low-mass stars and to generate the nonradiative energy flux released in the atmosphere. This energy flux, as acoustic and/or hydromagnetic waves, interacts with atmospheric matter. The density of matter decreases with height following the barometric law approximately, while the density of the nonradiative energy flux decreases slowly enough because of the weak absorption and weak divergence of the flux. Thus, if the absorption coefficient changes slightly, the condition of the steadiness of the atmosphere, at which the amount of the absorbed mechanical energy should be equal to the amount of emitted radiative energy, can be realized only in the case of a noticeable increase of the temperature of matter with height. As a result, there is a temperature inversion: above the photosphere with an effective temperature of about 5700 K and a cooler temperature minimum there is a chromosphere with a temperature of 5000–20000 K, then there are a very thin transition zone with the temperature varying from 20000 to $10^6$ K and the corona with a temperature of above $10^6$ K. The temperature jumps between the adjacent atmosphere layers are eventually due to the discreteness of states of hydrogen and helium atoms, which mainly govern the thermodynamic properties of the space plasma. Certainly, these simple considerations can provide only a general physical explanation of the temperature inversion in stellar atmospheres, but the real pattern is much more complicated: it should include different types of hydromagnetic waves, which interact with matter and with each other in different ways, the heat conduction of the hot corona toward the cooler underlying layers, variations at different spatial and temporal scales of the initial flux of nonradiative energy, and its significant deviations from the uniform distribution over the stellar surface owing to the complicated structure of the magnetic field.

Initially, the solar chromosphere was observed only in the limb during short-term solar eclipses. With the invention of Lyot-type coronagraphs and spectroheliographs a regular monitoring of optical characteristics of the chromosphere was started. Today, the solar chromosphere is observed in optical and radio ranges by a broad network of ground observatories and in ultraviolet from spacecrafts, which provides simultaneous observations of emission lines of the chromosphere and the transition zone.

In brief, the results of studying the solar chromosphere are as follows.

The basic feature of the solar chromosphere is the extreme nonuniformity of all characteristics. In the lower layers of the chromosphere, to a height of about 1500 km, numerous emission lines are observed that correspond to absorption lines of the photosphere. On increasing height these lines weaken, but the emission intensity of ionized calcium and hydrogen increases and helium emission arises. In the upper chromosphere expanding from 4000 to 10000 km, calcium emission disappears, hydrogen emission weakens, and helium emission strengthens. The density of matter in the chromosphere varies from $10^{15}$ cm$^{-3}$ at the base to $10^9$ cm$^{-3}$ near the upper limit, but at the chromosphere base the ionization degree of matter is only 0.1%, near the upper limit it is complete, thus the range of electron densities is much narrower, $10^{12}$–$10^9$ cm$^{-3}$. The systematic changes with height are superimposed by the considerable variations on the solar surface, since the chromosphere is not a continuous plane-parallel structure but an ensemble of numerous plasma loops. At the limb in the lower chromosphere these details are indistinguishable; these loops are visible as separate blobs in the mid-chromosphere, while the upper chromosphere contains only short-lived vertical 3000–km thick structures, spicules. Simultaneously, there are tens of thousands of spicules on the Sun, but they occupy less than 1% of the solar surface, and there is hotter coronal gas between them. Bright structures, flocculi in the active regions and a bright chromospheric network, whose cells outline the supergranules, are seen in the lines of ionized calcium in spectroheliograms. During the epoch of the solar maximum activity in 1980, when the mean total area of sunspots was close to 0.2% of the solar disk area, the mean total area of calcium flocculi was about 3% of the disk area. The flocculi of the mid-chromosphere were clearly seen in calcium spectroheliograms, when passing to the upper chromosphere at the level of formation of the emission of $Ly_\alpha$, grow by approximately 40% (Lean, 1992). In the center of the $H_\alpha$ line near the active regions one can clearly see dark filaments. Characteristic motions in the quiescent chromosphere are the rise of gas in spicules with a velocity of up to 20 km/s and a spread of matter from the centers of supergranules to their periphery with a velocity of about 0.5 km/s. Close neighboring active regions on the Sun form activity complexes. Usually these structures are placed on the solar disk in a nonrandom way; there are regions of preferable occurrence that persist for long periods, the so-called active longitudes. During the epochs of activity maxima the neighboring activity complexes are overlapped and are difficult to distinguish.

As stated above, the solar atmosphere comprises a thin transition zone, wherein the temperature increases from chromospheric to coronal values. Its thickness is much less than the pressure height scale, only of the order of hundreds of kilometers, but this region provides strong emission in ultraviolet lines of the high ionization stages of carbon, nitrogen, silicon, and some other elements. Both photospheric spots and chromospheric flocculi and the emission lines above the spots in the transition zone depend on the solar-cycle phase. Thus, during the epoch of solar maximum the spots were not seen in the light of CIV lines, whereas a ten-fold increase in the intensity of lines of the transition zone was found in the vicinity of the solar minimum straight above the spot umbra (Mullan, 1992). From the Earth, the transition zone of the Sun is studied in the range of millimeter and centimeter wavelengths using radio-astronomy methods.

It is obvious that stellar chromospheres and transition zones cannot be studied as thoroughly as those of the Sun, since only integral stellar characteristics can be observed, but the multiplicity of observed stellar chromospheres to a certain extent compensates for this limitation.

## 1.3.1. Optical and Ultraviolet Spectra of Chromospheres and Transition Zones

Stellar chromospheres were discovered with the beginning of spectral studies of stars of middle and late spectral types using the strong emission of the resonance CaII H and K doublet in the violet part of spectra. In addition to these bright lines, strong hydrogen emission was already found in the first spectra of flare stars. The notation dKe and dMe is used to notify the presence of hydrogen emission in the spectra of K–M dwarfs. The overwhelming majority of studies carried out so far are based on the analysis of hydrogen and calcium emissions recorded from the Earth and ultraviolet spectra obtained from space. Few, but fundamentally new, ultraviolet data were obtained from the space telescope Hubble and within the FUSE experiment. The publications by Pettersen and Hawley (1987, 1989) provided a general view of optical spectra of the chromospheres of flare stars. Based on the observations of about thirty objects with the 2.1-meter Struve telescope, they dealt with the energy distribution in the spectra of G9–M5 stars in the broad wavelength range from 3900 to 9000 Å, the Balmer decrement from $H_\alpha$ to $H_9$, and absolute surface fluxes in the $H_\beta$ and CaII lines.

**1.3.1.1. Calcium Emission in the H and K Lines.** Chromospheric CaII H and K emission lines seen against the background of broad and deep absorption profiles are one of the most well-studied details of the solar spectrum. This is explained by their accessibility for ground-based observers and high sensitivity of photoemulsions in this wavelength region. Being excited by electron collisions, the lines provide important information on the temperature structure of atmospheres. Depending on the photospheric base — a quiescent photosphere, a spot or a facula — the profiles of these lines differ significantly, even being averaged over the solar disk they differ for different phases of the solar cycle.

Long lists of dwarf stars with the calcium emission were published in the 1940s–1950s (Joy and Wilson, 1949; Bidelman, 1954). Later, Glebocki et al. (1980) made a compilation of the lists. The atlases by Pasquini et al. (1988) and Rebolo et al. (1989) contain the records of several tens of spectra of F4–K5 dwarfs with different activity level in the vicinity of the CaII H and K lines with a resolution of 30000–100000 and a high *S/N* ratio. The spectra were recorded by the echelle spectrograph located at the coude focus of the 1.4-meter telescope of the South European Observatory and their quality is rather close to that of solar spectrum records.

In 1966, O. Wilson started a long-term program on systematic measurements of the intensities of emission cores of the CaII H and K lines in the spectra of about one hundred medium- and low-mass stars. Instead of the eye estimates of emission intensity used in his previous study (Wilson, 1963), in 1966 photoelectric measurements were started. A scanner with four sequentially opening entrance slits installed at the coude focus of the 2.5-meter telescope of the Mount Wilson Observatory recorded the fluxes in line cores and in the broad bands of the adjacent continuum. Based on the results of 11-year spectral observations, Wilson concluded that all stellar chromospheres, from the weakest to the strongest, demonstrated significant variations of calcium emission fluxes for the times from one day to several months and these variations increased together with the fluxes (Wilson, 1978). The studies initiated by Wilson were actively continued. His results gave an impetus to different astrophysical directions, from registration of individual flares, the estimates of axial rotation periods and characteristic lifetimes of active regions to the discovery of the analogs of the 11-year solar cycle and the conclusions on the decay rate of the stellar magnetism at evolutionary times. Below, we consider the results of spectral monitoring of calcium emission that are mainly related to the physics of stellar chromospheres, conclusions on cyclic and evolutionary changes will be presented in Part 3.

In 1980, a new four-channel HK-photometer was installed on the 1.5-m telescope at the Mount Wilson Observatory. Between July and October 1980, Wilson's team carried out intensive observations of 46 dwarfs (Fig. 14) and made the following conclusions. Calcium-emission intensities modulated with periods of days for 19 stars; the modulation was due to nonuniform distribution of the chromospheric emission over the surfaces of rotating stars. It is noteworthy that thus-obtained direct estimates of axial rotation periods are independent of the orientation of rotation axes and include slowly rotating stars for which the traditional spectroscopic method is inappropriate. In many cases, the phases of rotational modulation lasted over the whole observational period, which evidences the long-term asymmetric distribution of active regions along the longitude. The comparison of the found periods with the emission intensity made it possible to suggest that the level of chromospheric activity of stars depended mainly or even solely on their rotation (Vaughan et al., 1981; Baliunas et al., 1983). Stimets and Giles (1980) concluded that the analysis of nonuniformly distributed HK-photometer measurements through the autocorrelation method had certain advantages over the power-spectrum method, thus Vaughan et al. (1981) used the former.

Baliunas et al. (1981) studied the intensities of the CaII H and K lines in the spectrum of the active star $\varepsilon$ Eri and found that the variations of both lines correlated at the level of 5% from one night to another and at the level of 7% at a time interval of 15 min. The total energy of these fast variations was about $2 \cdot 10^{30}$ erg and, since the localization of the source of the radiation variability on a small part of the disk was quite probable, one could not exclude that stellar flares were responsible for these variations.

On the basis of dense observational series of about 100 program stars obtained with the HK-photometer during three seasons by the team of O. Wilson, Baliunas et al. (1985) made a spectral analysis of the time series of intensity of the calcium emission. They found that the periodicity of this intensity varied from one season to another and even within the same season for 12 stars, and that two different periods coexisted during at least one season for 10 stars. The result can be explained either by the existence of active regions at different latitudes of a star with noticeable differential rotation or by the phase shift due to the emergence of a new active region at the longitude where no active regions existed before. For four stars, for which the data were collected over 18 years, the effect of differential rotation was the greatest. On the G8 star HD 101501, cyclic changes of the calcium emission with the characteristic time of about 7.5 years and a differential rotation of 10% were found. On the G0 dwarf HD 114710, no cyclic changes were detected but the value of the differential rotation was determined as 21%. On the G1 star HD 190406 and the G0 star HD 206860, cyclic variations with the characteristic times of 2.6 and 5.3 years and differential rotations of 11 and 5% were established. During the third observational season two simultaneous periods of variation in the calcium emission were found on HD 190406, the second period coincided with the period found during the first observational season. On the whole, the pattern is consistent with the solar one when spots of a cycle coming to an end are at low latitudes and have one rotational period, while coexistent spots of a new cycle appear at higher latitudes and owing to the differential rotation for the Sun have another rotational period. On HD 206860, with an axial rotation period of about 5 days the active region was preserved during three observational seasons. It should be noted that the differential rotation for the Sun does not exceed 3%. On the basis of 10-year observations of HD 114710, Donahue and Baliunas (1992) concluded that the star had two longitudinal activity zones, but, as opposed to the Sun, the axial rotation period increased as the chromospheric emission in the cycle weakened. Considering the rotation character of 22 stars, for which rotational periods were determined during several seasons, Donahue and Baliunas (1994) found on 12 of them the characteristics of solar "butterflies", on six stars there was an inverse pattern, when one could expect a drift of active regions during one cycle to the pole, and for four stars they suspected a change in the direction of the drift of active zones during the cycle.

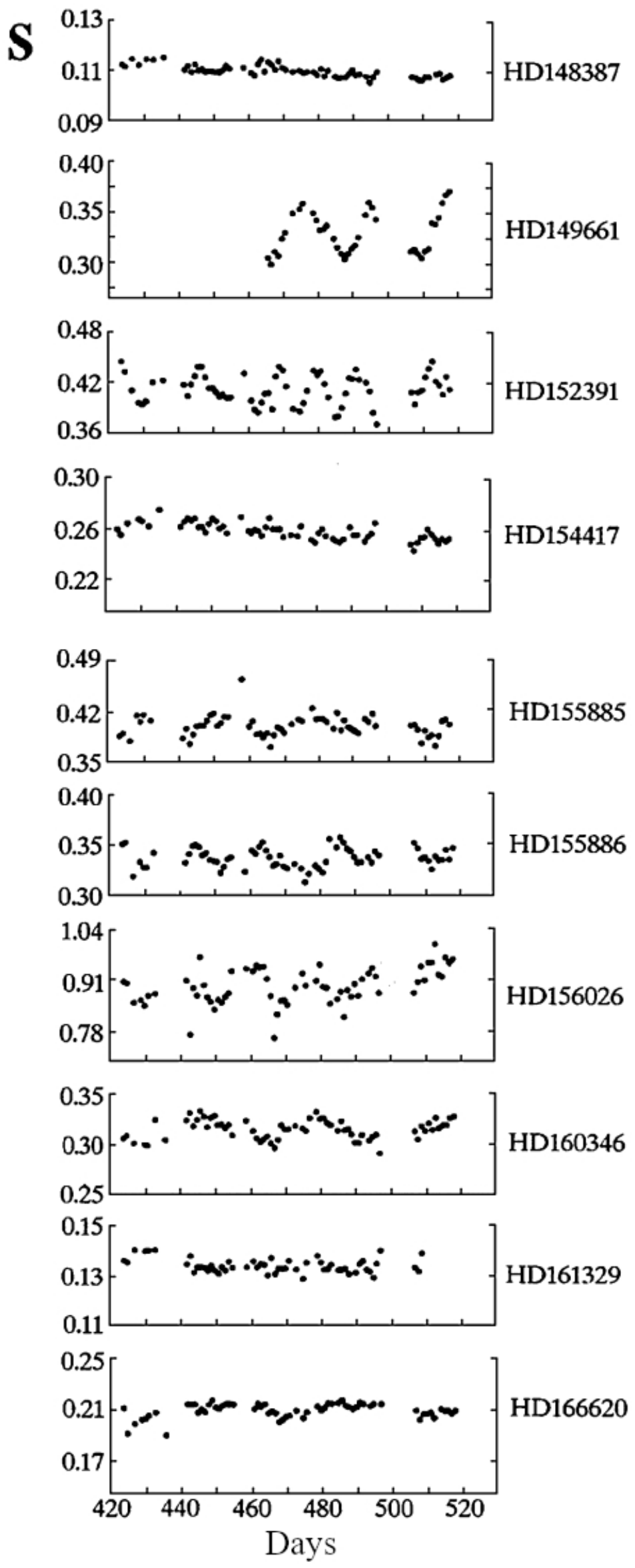

Fig. 14. Time variation of *S*, the ratios of fluxes in the cores of the CaII H and K emission lines to fluxes in the adjacent regions of the photospheric spectrum, measured in the course of nightly monitoring in 1980 using the HK-photometer for F–K dwarfs (Baliunas et al., 1983)

In 1991, 18 participants of the O. Wilson program published over 65000 measurements for about 1300 stars obtained in 1966–83 (Duncan and 17 coauthors, 1991). In 1995, 27 participants of the program published the results of 25-year studies (Baliunas et al., 1995). Later, Write et al. (2004) published about 18000 measurements of S for 1228 F–M stars obtained from the archives on searching for exoplanets of the Keck and Lick observatories.

* * *

According to Blanco et al. (1974), in the sample of stars of the same age the maximum ratio of the CaII K line luminosity to the stellar bolometric luminosity falls at the spectral type K0.

Linsky et al. (1979a) obtained at the 4-m telescope of the Kitt Peak Observatory the high-dispersion spectra of 17 G0–M2 dwarfs in the region of CaII H and K lines, calibrated them in absolute units, and found the ratios of fluxes in these lines to bolometric stellar luminosities — $R_{HK}$. These ratios were lower for the stars older than the Sun, $\alpha$ Cen A and $\alpha$ Cen B, were the same for 61 Cyg A and 61 Cyg B, and were greater for younger G dwarfs. Later, using the same equipment, Giampapa et al. (1981a) observed seven K–M dwarfs. The ratios of the fluxes in the calcium lines to the bolometric luminosities varied from $3 \cdot 10^{-5}$ to $9 \cdot 10^{-5}$ for dwarfs with hydrogen emission and are lower by an order of magnitude for dwarfs without the emission, the latter values being close to the solar value of $8 \cdot 10^{-6}$. Since the emissions in the CaII H and K lines are the first indicator of the existence of the stellar chromosphere, the search for stars without chromospheres is reduced to the search for stars without calcium emission. Out of four late M dwarfs without traces of the $H_\alpha$ line considered by Mathioudakis and Doyle (1991a), one had very weak calcium emission and three stars did not display any.

Middelkoop (1982) calibrated the values of $S$ — ratios of fluxes in the CaII H and K lines and in the adjacent continuum — measured by the team of the Mount Wilson Observatory into the absolute fluxes from a unit of the stellar surface $F_{HK}$. In so doing, he found the dependence of $F_{HK}$ on rotation, which is common for stars of different spectral types, for single stars, and components of binary systems. The latter fact means that the high chromospheric activity of components is due to the fast rotation determined by the tidal interaction in the systems rather than directly by the tidal heating of the atmospheres.

Based on the analysis of several hundreds of F–G–K stars, Mewe et al. (1981) and Schrijver (1983) showed that the radiation fluxes in the CaII H and K lines within each spectral interval of F–G–K stars overlapped by more than an order of magnitude, but at the same time they reliably recorded the lower level of such fluxes $F_{HK}^{min}$(Sp), the so-called basal chromospheric level, which was definitely higher than the detection threshold of this emission. The differences $\Delta F_{HK}$ between $F_{HK}$ of specific stars and the appropriate $F_{HK}^{min}$(Sp) were more closely related to other indicators of the stellar activity than the fluxes $F_{HK}$ themselves. Hence, Mewe et al. (1981) suggested that $F_{HK}^{min}$(Sp) was caused by nonmagnetic heating of atmospheres. The intensity of calcium emission on the solar surface with minimum magnetic activity in the center of supergranules far from the facular areas falls exactly on the basal level of the chromospheric emission. Rutten (1984) carried out additional observations with the HK-photometer. The observational results made it possible to revise the Middelkoop calibration and conclude that the basal level was independent of metallicity. Later, Rutten (1986, 1987) found the dependence on the color index in the linear relation coefficient between $F_{HK}$ and the logarithms of axial rotation periods. He also established an age-independent correlation between $\Delta F_{HK}$ and the axial rotation period or the angular velocity common for single dwarfs, members of binary systems, and most giants. The dependence of the activity level on rotation and its independence of age showed that the magnetic field responsible for the activity was generated by the dynamo mechanism and was not a relict field. Schrijver et al. (1989a) considered the Mount-Wilson measurements for the emission in the CaII H and K lines, the data on ultraviolet emission of MgII h and k, and the data on stellar variability and rotation. They concluded that the basal level of chromospheric activity was not linked to the magnetic activity and was characteristic of the entire stellar surface, whereas the magnetic component was present in active regions and in the chromospheric network. The extrapolation of the relations between the activity level and stellar variability showed that such variability disappeared near the basal level of chromospheric activity, i.e., the active regions and the chromospheric network did not exist on the lowest-activity stars. Stellar rotation was negligible near the basal level of chromospheric activity, which meant that it was not likely that the dynamo mechanism was active on the stars with basal chromospheric activity.

Based on the observations of 26 F5–K3 dwarfs in the CaII H and K lines, Schrijver et al. (1992) determined more precisely the basal level of the chromospheric activity: it was much above the level provided by the purely radiative atmosphere, which suggests the existence of nonradiative heating even for the lowest-activity stars The basal level of chromospheric activity can cover the entire stellar surface, but one should not exclude the existence of time-variable cool and hot zones, if the generation of acoustic waves below the photosphere is nonuniform.

The close relation between the chromospheric activity and local magnetic fields, which is known from solar studies, formed a basis for the concept of the secular weakening of stellar magnetic fields caused eventually by the deceleration of rotation. However, in addition to rotation, the magnetic field depends on another independent parameter: stellar mass, the mass of the convective zone where the field is generated, the mass-dependent spectral type of a star, or the time of overturn of convective vortices in the convective zone that depends also on the spectral type. All this led to the idea on the dependence of the chromospheric activity level on the Rossby number, the ratio of the axial rotation period to the theoretically calculated overturn time of convective vortices. Using the axial rotation periods of more than 40 stars found from the variation of their CaII H and K emission, Noyes et al. (1984a) found a correlation of such periods with the emission level averaged over 15 years. A particularly close correlation was established between ratios of fluxes in the CaII H and K lines, revised to take into account the contribution of the radiative atmosphere, to the bolometric luminosities of the stars, the values of $R'_{\mathrm{HR}}$ and the Rossby numbers (Fig. 15). The dependence turned out to be so close that it was used to estimate the axial rotation period from the observed luminosity of calcium emission. Quast and Torres (1986) made use of this fact: from the intensity of calcium emission they determined the axial rotation periods of several stars and then compared the periods and the color index B−V. The stars attributed to young and old objects according to their kinematic parameters are clearly distinguished in the constructed plots. Kim and Demarque (1996) refined the results of Noyes et al. (1984a) by estimating the time of passage of convective elements throughout the whole convective zone, while Noyes et al. used only the time of passage of the local height scale at the bottom of the zone. Within this approach, the saturation effect remained: at low Ro, i.e., for fast rotators, the dependence of the activity level on the rotation rate disappeared (Montesinos, 2001).

Pasquini (1992) compared the high-accuracy profiles of the CaII K line in the spectra of G0-G5 stars with the profiles of this line obtained from the whole Sun and separate structures on its surface. General tendencies and relationships observed on the Sun throughout the cycle — line intensity, width, and asymmetry — take place on solar-type stars with differing chromospheric activity levels. Apparently, the stars cover a broader activity range than the Sun during the cycle. The spectrum of the most active stars suggests the presence of structures of the solar faculae type that should cover most of their surfaces. Despite the similarity of the spectra of solar faculae and active stars, in interpreting the latter it is impossible to choose the domination of the facula brightness or the filling factor of faculae of the stellar surface.

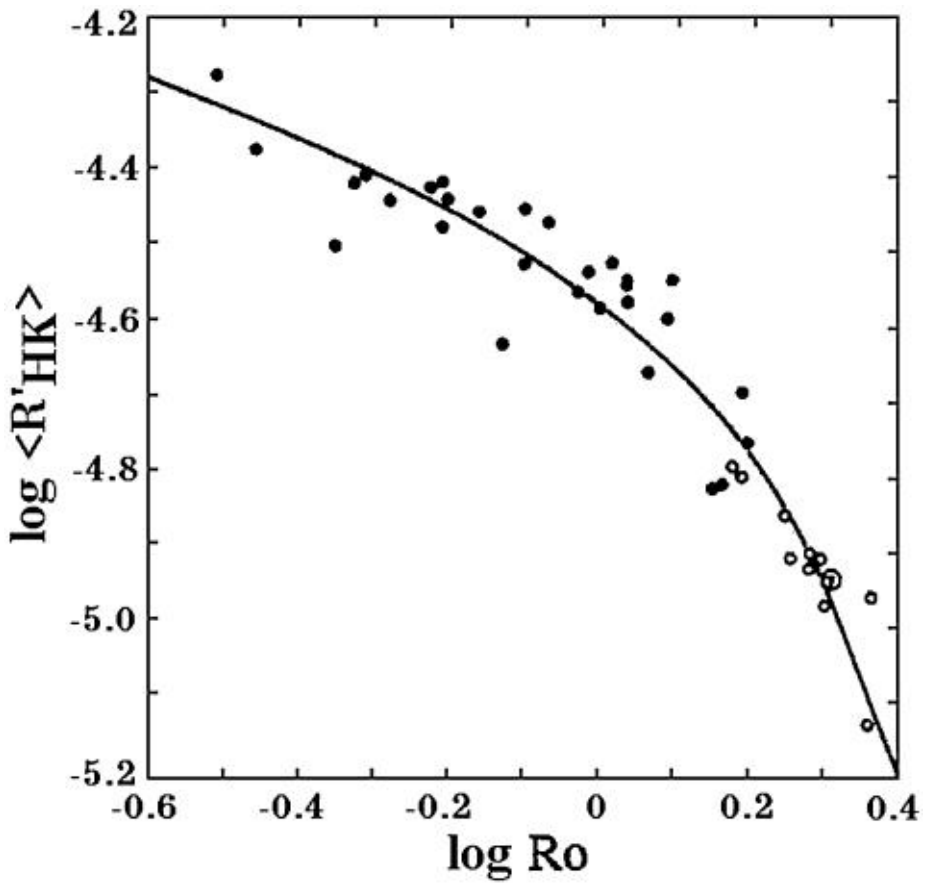

Fig. 15. Dependence of the chromospheric luminosity in the CaII H and K lines related to bolometric luminosity on the Rossby number (Noyes et al., 1984a)

For four years, Garcia Lopez et al. (1992) monitored the CaII H line profiles in the spectra of six G–K stars — $\xi$ Boo A, $\alpha$ Cen B, $\varepsilon$ Ind, $\varepsilon$ Eri, $\kappa$ Cet, and 70 Oph A — with a resolution of 80000 and $S/N$ = 200. They found that both the emission flux of the central component of the line and its width were variable. They revealed the shifts of the emission core corresponding to the velocities of hundreds of meters per second and of the absorption detail with lower velocities. The comparison of wavelengths of $H_3$ and the emission core provided the first evidence of the vertical ascending and descending motions in the chromosphere, which were stable for several years on the star.

Using the Calar Alto 2.2-meter telescope and the 2.5-meter Isaac Newton telescope (Canary Islands), Montes et al. (1994) observed about 50 chronometrically active binary systems, including more than a dozen systems composed of active dwarfs with orbital periods of 0.7–6 days. They found that the K line at the level of half-intensity ($W_0$) and at the emission base ($W_1$) was broader than those corresponding to the Wilson–Bappu relationship: $W_0$ was more affected by the stellar rotation rate, while $W_1$ depended more on the emission intensity.

The qualitative relations between the magnetic field strength of individual structures on the solar surface and the brightness of calcium flocculi were established in the late 1950s from the comparison of spectroheliograms and magnetograms of the Sun. Skumanich et al. (1975) found a linear relation between the CaII emission and the flux module of the local magnetic field for the quiet Sun. Schrijver et al. (1989b) found quantitative relations between the intensity of CaII K and the magnetic flux density for the quiet Sun and its active regions that are valid for a broad range of the structures on the solar surface:

$$\Delta F_{\mathrm{CaII}} = 0.55\langle fB\rangle^{0.62\pm0.14} \qquad (4)$$

where $\Delta F_{\mathrm{CaII}}$ is the excess over the minimum nonmagnetic luminosity of the chromosphere. Later, Saar (1991) determined more precisely the exponent as 0.56±0.03. The minimum luminosity level of the solar chromosphere measured in the centers of supergranules in very quiescent solar regions was equal to the minimum flux density of the solar-type stars and, on average, only 30% lower than the minimum flux level in the sample of red dwarfs. For the chromospheric emission of CaII, this dependence is valid up to $\langle fB\rangle$ ~ 300 G and saturation occurs at greater fluxes when the maximum efficiency of the magnetic heating is achieved or the filling factor approaches unity at the level where the optical thickness in the cores of the CaII lines achieves unity. The found saturation level coincides with the averaged magnetic flux from a solar active region. This coincidence can be explained by the physics of the atmosphere: the point is the mechanism of the radiative transfer or magnetic heating rather than stellar age or rotation rate.

An example of synchronous change of the magnetic flux and the calcium emission on the scale of a long-term activity cycle was found for the K5V star $\kappa^1$ Cet: according to Saar and Baliunas (1992b), from four points obtained in 1984–88 the value of $\Delta F_{\rm CaII}$ is proportional to $(fB)^{0.4\pm0.25}$, which does not contradict the correlation found from individual measurements of different stars with an exponent of 0.56 ± 0.03. Earlier, Saar et al. (1986b) found the weak correlation of $f$ and $F_{\rm HK}$ for the K2 dwarf $\varepsilon$ Eri.

Frasca et al. (2000) detected an anticorrelation of calcium flocculi and spots for the young solar-type star HD 206860 and based on these data constructed a rough 3D model of the outer atmosphere of this star.

Apart from the team from Mount Wilson, the long-term observations of stars were carried out at the Lowell Observatory in Strömgren b and y bands in 1984–2000 and at the Fairborn Observatory in 1993. Based on the latter, Radick et al. (2004) believed that the accumulated data allow one to make a conclusion on the general properties of long-term activity and variability of many stars considered in reviews. In particular, they found that the present-day Sun is a typical star in this relation for its age; its behavior may be somewhat more regular and quiet, since there are objects with stronger and more chaotic variations. The chromospheric activity of the Sun is somewhat higher than the statistical power relation between stellar brightness and chromospheric variability, but its brightness variations over the 11-year cycle proves to be somewhat lower as compared to stars of the same activity level.

Hall and Lockwood (2004) carried out 3700 measurements of the CaII H and K lines for 57 stars and 10 of them turned out in the flat state, i.e., they could be suspected to be objects in the Maunder minimum. But not always the emission of these stars is weaker than that in the solar minimum, whereas the solar minimum lies near the lowest level observed for stars with cyclic activity. However, many flat-stars have the HK-activity level, which is comparable or even higher than the solar minimum, and there is no data on a significant decrease in magnetic activity for flat stars.

Based on a substantial series of high-dispersion spectra of the rapidly rotating K dwarf BO Mic, Wolter and Schmitt (2005) detected significant variations of emission cores of the CaII H and K lines, which were attributed to a growth and disintegration of active regions and to their rotational modulation. Equivalent widths of the CaII K line were in antiphase with stellar photospheric brightness, which means the coincidence of positions of dark spots and bright chromospheric structures.

While exploring the infrared range, in addition to the CaII H and K lines, the IR triplet of CaII $\lambda$ 8498, 8542, and 8662 Å was investigated. Andretta et al. (2005) performed the nonLTE calculations of these lines for the high-dispersion photospheric spectra and showed a necessity of such calculations for using $R_{\rm IRT}$ as an indicator of chromospheric activity.

Rauscher and Marcy (2006) studied profiles of the CaII H and K emission lines in high-resolution spectra of 147 K7–M5 stars derived with the Keck telescope during six years. Within this comparatively narrow mass range from 0.30 to $0.55 M_\odot$ the average values of FWHM, equivalent widths, and luminosities of calcium lines increased by a factor of 3 toward high masses. The obtained profiles are represented by two Gaussian curves: chromospheric emission and central nonLTE absorption. Furthermore, for the majority of sample stars the central absorption is shifted by approximately 0.1 km/s toward the red side with respect to emission. FWHM of emission calcium lines grows with luminosity, resembling the Wilson–Bappu effect in the F–G–K stars. The K line is somewhat wider than the H line, and its absorption component is shifted somewhat more toward the red region, pointing to some differences in formation levels of these lines. The equivalent width of $H_\alpha$ correlates with equivalent widths of calcium lines only in the case when they exceed 2 Å, which means the existence of the magnetic threshold. Higher than this threshold, the upper and lower chromospheres become thermally linked.

Using the 3.6-meter Italian Galileo National Telescope, Busa et al. (2007) derived the high-resolution spectra in the range from 4960 to 10 110 Å of 42 active stars with values of $\log R'_{HK}$ from −3 to −5 and measured two indicators of chromospheric activity from the CaII IRT: residual equivalent widths and $R_{IRT}$. Comparing these values with $R'_{HK}$, they detected their good correlation.

Schröder et al. (2009) published values of *S* for 481 solar-type stars obtained by a specially developed method for fast rotators.

Zerjal et al. (2013) reported on preparing a catalog of spectra for stars of the southern sky in the CaII IRT region, including about half a million of middle-resolution spectra. Through the measured equivalent widths with $S/N > 20$ they distinguished more than 14 000 active stars.

Based on the extensive Gaia DR3 data, Lanzafame et al. (2022) performed a systematic analysis of luminosity in the stars of the Ca II infrared triplet. The position of these stars in the color-magnitude diagram and correlation with the amplitude of the photometric rotational modulation is considered. They found that the highest level of IRT activity is associated with PMS stars and RS CVn binary systems and some evidence of a bimodal distribution in MS stars with $T_{eff} > 5000$ K. Stars with 3500 K $< T_{eff} <$ 5000 K are found to be either very active PMS stars or active MS stars with a unimodal distribution. A dramatic change in activity distribution is found for $T_{eff} < 3500$ K with dominance of low-activity stars close to the transition between partially- and fully-convective stars and a rise in activity down into fully-convective regime.

Mittag et al. (2013) proposed an algorithm of recalculating values of *S* into absolute values of calcium fluxes, for which very large samples of stars are accessible.

**1.3.1.2. Hydrogen Emission.** As already noted, the hydrogen emission lines in the spectrum of a red dwarf are the first evidence of its affiliation to the UV Cet-type flare stars. This emission is seen even in low-resolution spectra and is the sufficient criterion for attributing a red dwarf to flare stars: the experience of many observers showed that, as a rule, 10–15-hour photoelectric monitoring in the U band is sufficient to record a flare on an emission dwarf. Furthermore, Torres et al. (1983) for several years ran photometric observations of 90 red dwarfs to find periodical brightness variations: 11 of 20 stars with $H_\alpha$ emission and 2 of 30 dwarfs with only calcium emission displayed the variations, whereas for none of the 40 nonemission stars were brightness variations observed. Thus, the existence of hydrogen emission to a high confidence ensures flare activity and spottedness of red dwarfs.

The $H_\alpha$ emission is invisible on the disk in the spectrum of the quiet Sun. In the spectra of bright flocculi, this strong absorption line is only slightly flooded in the center and the equivalent width of the flooding emission can be estimated as 0.02–0.03 Å. A similar effect is observed in the spectra of relatively similar to the Sun chromospherically active G–K stars (Zarro, 1983). The hydrogen emission in the solar-type stars was first studied in detail by Herbig (1985). Using the coude spectrograph of the Lick telescope, he obtained spectrograms of about 40 F8–G3 stars. Based on the high ratio *S/N* ~ 200–350, he established weak emission components with $W_{H\alpha}$ ~ 0.1 Å in the center of the $H_\alpha$ absorption lines, which were in good correlation with calcium emission fluxes. According to his estimate, the ratios of fluxes in the main chromospheric channels of radiative losses of these stars are $F_{HK} : F_{IRCaII} : F_{H\alpha} = 1 : 0.81 : 0.36$.

If, for the Sun and solar-type stars, $W_{H\alpha}$ is usually a few hundredths of an angstrom, in the spectra of dKe–dMe stars it is of an order of several angstroms. This difference of two orders of magnitude is to a considerable extent due to a lower intensity of the photospheric continuum in the spectra of later stars, since for the effective temperatures presented in Table 2, for the wavelength $H_\alpha$ the ratios of the Planck functions *B*(G2)/*B*(Sp) is 12 for the M2 spectra and 50 for M7 spectra. These qualitative reasonings can be refined using the results of Hawley et al. (1996) for spectral types and equivalent widths of 321 red dwarfs. Since during their spectral observations that underlie the measurements of $W_{H\alpha}$, no strict photometric monitoring was run, it is probable that some of the published values of $W_{H\alpha}$ were obtained during flares; in this case $W_{H\alpha}$ usually increased by a factor of 2–3. Reasoning from the average frequency and the average duration of stellar flares (see Chapter 2.2), one can estimate that among 321 red dwarf stars the values from 1 to 5 could be obtained during flares. Thus, the exclusion of the 6 highest values from consideration should nullify the distorting effect of flares on the estimates of parameters of quiescent stellar chromospheres.

Table 6. Characteristics of the $H_\alpha$ emission line in the spectra of late K and M dwarfs (measured by Hawley et al. (1996))

A. $W_{H\alpha}$ distribution

| $W_{H\alpha}$, Å | < 2 | 2–3 | 3–4 | 4–5 | 5–6 | 6–7 | 7–8 | 8–9 | 9–10 | > 10 |
|---|---|---|---|---|---|---|---|---|---|---|
| *N* | 54 | 55 | 59 | 55 | 37 | 20 | 15 | 6 | 7 | 13 |

B. $W'_{H\alpha}$ distribution

| $W'_{H\alpha}$, 0.08 Å | < 1 | 1–2 | 2–3 | 3–4 | 4–5 | 5–6 | 6–7 | 7–8 | > 8 |
|---|---|---|---|---|---|---|---|---|---|
| *N* | 25 | 59 | 87 | 74 | 39 | 19 | 8 | 3 | 1 |

C. $\langle W'_{H\alpha}\rangle$ versus spectral type

| Spectral type | K5–M0.5 | M1–M2.5 | M3 | M3.5 | M4 | M4.5 | M5–M9 |
|---|---|---|---|---|---|---|---|
| $\langle W'_{H\alpha}\rangle$, 0.01 Å | 32 ± 10 | 23 ± 12 | 25 ± 10 | 26 ± 13 | 24 ± 11 | 22 ± 11 | 19 ± 11 |
| *N* | 25 | 29 | 30 | 58 | 70 | 53 | 50 |

Part A of Table 6 presents the distribution of 315 values of $W_{H\alpha}$ measured by Hawley et al. (1996): the distribution density of $W_{H\alpha}$ from a detection threshold of 1 Å to 5 Å is approximately constant, while the number of higher values smoothly decreases to a minimum of about 12 Å. Part B presents the distribution of the equivalent widths $W'_{H\alpha} = W_{H\alpha} \cdot B(\mathrm{Sp})/B(\mathrm{G2})$, where *B* is the Planck function in the $H_\alpha$ wavelength, reduced to the solar atmosphere. All $W'_{H\alpha}$ values of the sample under consideration are within 0.02–0.74 Å, and 70% of them are within 0.08–0.32 Å. The average for the sample $\langle W'_{H\alpha}\rangle = 0.24$ Å corresponds to an absolute flux of $1.6 \cdot 10^6$ erg/(cm$^2 \cdot$ s). Part C presents the distribution of average values of $\langle W'_{H\alpha}\rangle$ along spectral types. The table shows that the dependence of $\langle W'_{H\alpha}\rangle$ on the effective stellar temperature is rather weak and there is only a small systematic decrease of $\langle W'_{H\alpha}\rangle$ with transition to later spectra, but within each spectral type the dispersion of the values is fairly large. Thus, the characteristic absolute surface brightness of stellar chromospheres in the $H_\alpha$ line averaged over the disk is 10 times higher than the appropriate value for solar flocculi.

Using the echelle spectrograph of the 10-meter Keck telescope, Basri (2001) studied the region $H_\alpha$ of more than 60 late M and L dwarfs. He noticed clear emission from all M5–M9.5 stars of the sample and fast weakening of emission after M9.5. He found that measured $W_{H\alpha}$ did not demonstrate the dependence on the rotation rate, while the ratios $L_{H\alpha}/L_{bol}$ systematically and rapidly decreased toward the late spectral types.

* * *

The investigation of physical conditions in the chromospheres of active dwarfs from hydrogen emission was started by Wilson (1961), who obtained the spectrum of EV Lac with a dispersion of 9 Å/mm using the Palomar telescope in 1960. During the 6-h exposure no appreciable brightening of the star was noticed, thus it was concluded that the resulting spectrum belonged to the quiescent state. The hydrogen emission lines in the spectrum were evidently broader than those of helium and metals. Thus, assuming a purely thermal mechanism of their broadening, Wilson estimated the upper temperature limit in the chromosphere as 14000 K.

The first systematic consideration of hydrogen emission from the chromospheres of flare stars was performed in the Crimea (Gershberg, 1970b). The results of spectral observations of five emission red dwarfs with the 2.6-m Shajn telescope allowed the relative intensities and absolute luminosities of the Balmer lines and the width of the $H_\alpha$ profiles to be estimated. By comparing these data with the emission spectra of the solar chromosphere obtained at different phases of the solar eclipse it was shown that the density of matter in the layers of stellar chromospheres, in which the hydrogen emission was formed, was ten times higher than the appropriate values on the Sun. Since the structure of the isothermal atmosphere is determined by the ratio $M/R$, which for flare stars varies from 0.42 to 1.4 of the appropriate solar ratio, the general geometric structure of the chromospheres in both cases should be similar as well. Therefore, to heat a denser atmosphere to the chromospheric temperature, the power of the heating mechanism, the density of upward flux of nonradiative energy, and/or thermal conductivity from the corona (see Chapter 1.5) of flare stars should be higher at least by an order of magnitude than for the Sun.

Within the framework of the Sobolev concept of moving stellar envelopes, the theory of the Balmer decrement was developed for the case of purely collisional excitation of hydrogen lines in the isothermal gas free of external radiative excitation. Despite the considerable distinctions of this model from the stellar chromosphere, these calculations were applied to the Balmer decrements recorded in the quiescent states of nine flare stars. This resulted in the first estimates of the characteristic electron densities of stellar chromospheres: $(1–4) \cdot 10^{12}$ cm$^{-3}$ at 10000 K (Gershberg, 1974a). Then Grinin (1979) added photospheric radiation, but this almost did not influence the estimates of electron density. Though the concept of moving envelopes satisfactorily presented the observed decrement, it required noticeable internal movements in the radiating medium with velocities of up to 20–30 km/s, which had not been supported by observations. This fact raised doubts about the validity of the density estimates. The theory of the Balmer decrement elaborated later was free of the assumption on the substantial motion in the radiating medium: the quantum exit was achieved not at the expense of the differential motion of matter, but due to the quantum frequency drift at multiple scattering and their exit in the wings of profile lines. Application of the theory to the observed decrements yielded density estimates that exceeded the above estimates approximately by a factor of 3 (Katsova, 1990).

The observations of 9 flare stars performed in the Crimea resulted in the measurement of the absolute radiation fluxes of the M2–M6 chromospheres in the $H_\gamma$ line. All the fluxes turned out to be close to $5 \cdot 10^5$ erg/(cm$^2 \cdot$ s) to an accuracy of 2. The effective thicknesses of radiating layers estimated from the fluxes vary from several hundreds to one thousand kilometers.

Shakhovskaya (1974b) increased the number of studied spectra of emission red dwarfs at the 2.6-meter Shajn reflector to 43. She found that for M0–M5 spectral types the ratios of intensities $I_{H\gamma}/I_{H\beta}$ and $I_{H\delta}/I_{H\beta}$ were practically independent of the absolute luminosity (Fig. 16).

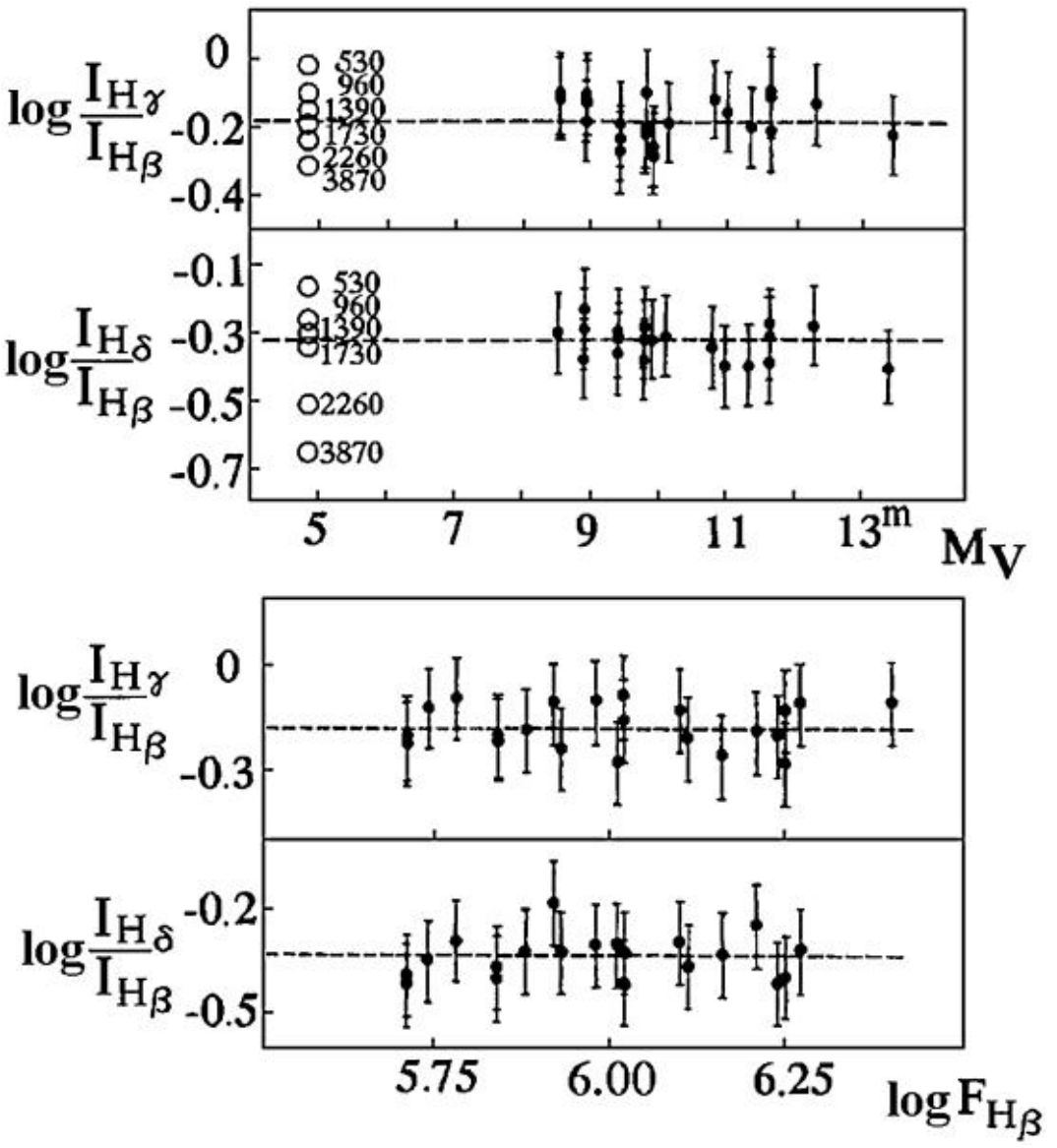

Fig. 16. Intensity ratios of the Balmer lines in the spectra of emission red dwarfs, depending on the absolute magnitude MV and absolute intensity of the $H_\beta$ line. Open circles show the appropriate values in the spectrum of the solar chromosphere obtained at different phases of the solar eclipse; numbers denote effective heights above the solar photosphere level (Shakhovskaya, 1974)

The absolute flux in the $H_\beta$ line for this sample of stars also lies within a relatively narrow range of values and there is no systematic change of the relative intensities of the Balmer lines depending on the flux (see Fig. 16). However, later, Katsova (1990) suspected that the ratio $I_{H8}/I_{H\gamma}$ grows systematically with an increase in the X-ray luminosity of the star. Shakhovskaya (1974b) found a systematic decrease of the ratio of the intensity of CaII emission lines to the intensity of the hydrogen emission for later M stars.

Worden and Peterson (1976) obtained spectra of 8 dKe-dMe stars in the red part of the spectrum with the 4-meter telescope of the Kitt Peak Observatory and found that the double-peaked profile of the $H_\alpha$ line was common for emission red dwarfs. Having considered various explanations of the profile formation, they concluded that the most probable model was the optically thick line in the center with the electron density of the radiating medium less than $10^{13}$ cm$^{-3}$. Pettersen and Coleman (1981) estimated the distance between the peaks of the double-peaked $H_\alpha$ profile in the spectrum of AD Leo as 0.6 Å for the profile half-width of 1.35 Å and its equivalent width of about 3.1 Å. The profiles of the $H_\alpha$ line and several other high-resolution Balmer lines, and the estimates of their equivalent widths and FWHM values in the spectra of 17 red dwarfs were published by Worden et al. (1981). Pettersen et al. (1984b) studied the $H_\alpha$ profiles with a resolution of 0.45 Å in the spectra of EV Lac, EQ Peg A, and V 1054 Oph. They found that the self-absorption outlined in the EV Lac profile, which was not observed for EQ Peg A, while the profile of this line in the spectrum of V 1054 Oph could be presented as a superposition of two emission profiles with self-absorption of up to 10–15% of the maximum level. According to their estimates, the electron density in the chromospheres varies from $10^{12}$ to $10^{13}$ cm$^{-3}$. Later, Pettersen (1989b) studied the $H_\alpha$ profiles of eight K5–M1 stars in the quiescent state with equal or even higher resolution (up to 0.20 Å): V 833 Tau, EQ Vir, BY Dra, 61 Cyg, DK Leo, V 1005 Ori, AU Mic, and YY Gem. For all single stars (AU Mic, V 1005 Ori, DK Leo, and EQ Vir) they revealed double-peaked profiles with FWHM = 1.4–1.5 Å, but for the components of YY Gem FWHM was 1.85 Å and rapid rotation ($v\sin i$ = 40 km/s) filled in the central self-absorption. The resulting splits in the double-peaked profiles were equal to 0.55–1.0 Å. Using these values, the measured intensities, and the Cram–Mullan (1979) theoretical relations (see below), Pettersen estimated the electron density of the chromospheres as $5 \cdot 10^{11}$–$5 \cdot 10^{12}$ cm$^{-3}$.

As stated above, the Balmer emission is a sufficient criterion for attributing a red dwarf to flare stars, but this is not a necessary condition. There are red dwarfs without hydrogen emission, on which flares characteristic of the UV Cet-type stars were recorded. One of them, GX And, is spectrophotometrically very close to AD Leo, but the level of its flare activity is lower approximately by a factor of 25. Pettersen and Coleman (1981) studied the star in the red region with high spectral resolution and found rather strong absorption of $H_\alpha$. Cram and Mullan (1979) interpreted the existence of such objects as follows. In a cool dwarf with a negligibly low nonradiative energy flux, the column density of neutral hydrogen atoms at the second level is very low, which results in weak hydrogen absorption lines in the photosphere spectrum with equivalent widths of $H_\alpha$ of less than 0.1 Å. For the dwarfs with growing temperature gradient, which leads to the formation of the chromosphere, the column density $n_2 l$ increases, and so does the absorption in the Balmer lines to $W_{\mathrm{H}\alpha} \sim 0.7$ Å. GX And is characterized by a similar situation. As the temperature gradient and the mass of the chromospheric layer further increase, the column density of collision-excited atoms grows at higher levels, and the Balmer emission arises first in the outer wings of the absorption profiles, then as a pure emission double-peaked profile. In this case, the electron density in the upper chromosphere exceeds $10^{11}$ cm$^{-3}$. The prevalence of collision processes over photoionization in the formation of the $H_\alpha$ line for this electron density was noted by Fosbury (1974), this is a qualitative distinction of dMe stars from the Sun.

The study of Cram and Mullan (1979) was an important step in understanding the chromospheres of cool stars. They proved that the difference between dMe and dM stars consisted not in the presence/absence of chromospheres, but in the fact that dMe stars had rather strong chromospheres, which could be characterized by a certain average increased density in the uniform models or by the filling factor of active regions of stellar surfaces.

Using the Lick telescope, Young et al. (1984) obtained the spectra of 77 M dwarfs in the region of the $H_\alpha$ line, among which they distinguished two single stars, Gl 410 and Gl 179, whose absorption profiles had an intermediate form between those characteristic of dM and dMe stars. Similar features were noticed for Gl 15 B and Gl 425 A, therefore Young et al. called them "marginal dMe stars". Using the term "marginal BY Dra-stars" for the same group of K–M dwarfs, Saar and Bopp (1992) added a dozen potential members of this group. These intermediate profiles can be represented as absorption profiles of dM stars filled in by the emission of 0.30–0.45 Å, which, according to Young et al., is formed in the active regions of the stars, and the stars correspond to the transition from active dMe stars to inactive dM stars. Giampapa (1985) estimated the surface filling factor of dM stars by faculae as 0.25. But the presence of strong absorption of $H_\alpha$ evidenced noticeable nonradiative heating of the stellar atmosphere.

Stauffer and Hartmann (1986) thoroughly studied over 200 M dwarfs. Using the echelle spectra with a resolution of 0.15 Å, they measured the equivalent widths of the $H_\alpha$ line to an accuracy of up to 0.03 Å and the Gaussian half-widths to an accuracy of up to 10%. In their sample, the oldest and the lowest-activity stars had weak absorption $H_\alpha$ lines, then, according to Cram and Mullan (1979), as the chromospheric activity strengthened, this absorption line increased and then transferred into emission. The majority of stars studied by Stauffer and Hartmann have $H_\alpha$ absorption lines with an equivalent width of not more than 0.7 Å systematically decreasing for cooler stars, while the average equivalent widths of the $H_\alpha$ emission lines grow toward cooler stars. Nine stars of the sample do not have any $H_\alpha$ lines. The dependence of $W_{H\alpha}$ on the spectral type demonstrates a scatter that noticeably exceeds the measurement accuracy of the compared values. Stauffer and Hartmann associated this scatter with the difference in metallicity of the stars of different age: metal-poor stars had weaker chromospheres.

Cutispoto and Giampapa (1988) performed spectral monitoring of the $H_\alpha$ line of three dM stars with a resolution of about 40000. They confirmed the Cram–Mullan concept and first found the low-amplitude variations of $W_{H\alpha}$ in the spectra of this type of stars (Gl 15 A, Gl 411, and Gl 526), which did not fit the idea on stationary heating of the atmospheres of such stars and required a certain variable component. Later, the low-amplitude variations of the $H_\alpha$ absorption line in the spectra of several K–M dwarfs were suspected by Panagi et al. (1991) based on the observations with the 4.2-meter William Herschel telescope with a resolution of about $10^5$. Recently, France et al. (2013) studied six M dwarfs with $H_\alpha$ absorption from spectra derived with HST. For all of them they found fluxes in emission lines in the ultraviolet with a variability of 50–500% on time scales of $10^2$–$10^3$ s.

According to the calculations by Cram and Giampapa (1987), the observed profiles can appear both in summing up the radiation of heterogeneous sections of the stellar surface and when the general density of the chromosphere increases. A comparison of the $H_\alpha$ lines and the calcium emission made them conclude that the chromosphere with a nonuniform surface was more probable. However, later Giampapa (1992) used the observations of Stauffer and Hartmann (1986) to revise the factors governing the observed broad activity range of red dwarfs, in particular, the considerable diversity of the $H_\alpha$ line: the variations of the filling factor of their surfaces by the active regions or the changing structure of chromospheres with height. He demonstrated that if any surface inhomogeneities of the chromospheres played a crucial role, the widths of the $H_\alpha$ absorption line in the integral stellar spectra should either be independent of their equivalent widths or increase as the equivalent widths decrease. But the data of Stauffer and Hartmann provide a definitely different pattern: systematic narrowing of the $H_\alpha$ absorption in the course of the decrease of their equivalent width. This result disproves the nonuniform model with the stellar surface including emission and absorption sections. This argument supports the pivotal role of changes of the chromosphere structure with height in providing the observed diversity of spectra of M stars. However, the obvious scheme of stars with different degrees of coverage by active regions is still widely used (see Subsect. 1.3.2.2).

Young et al. (1989) analyzed observations of the $H_\alpha$ line in the spectra of M stars obtained by Stauffer and Hartmann (1986), Bopp (1987), and Young et al. (1987a) with a resolution of 0.15–0.5 Å. In all samples, the $H_\alpha$ emission widths were almost equal: FWHM = 1.3 Å.

Selecting for dMe stars the dM stars of appropriate spectral types with maximum absorption in the $H_\alpha$ line, Young et al. (1989) determined the excess emission for each star and considered it as a measure of nonradiative heating of the chromosphere. This approach differs from the Cram and Mullan concept (1979), in which the chromosphere is considered responsible for the appreciable part of absorption in the $H_\alpha$ line. But practically, the estimates of the excessive emission systematically differ in the two approaches by only a small value that depends on the spectral type. The excesses of $W_{H\alpha}$ found by Young et al. (1989) varied from 1 to 7 Å and the excessive absolute luminosities $\Delta L_{H\alpha}$ calculated from them systematically reduce for later stars. The most reliable correlation was found between this luminosity and the bolometric luminosity of a star, which suggests a very weak dependence of $\Delta L_{H\alpha}$, which is equal to the product of the brightness of active regions and the filling factor, on stellar mass: $\Delta L_{H\alpha} \sim M_*^{0.3}$.

Herbst and Miller (1989) measured the equivalent widths of the $H_\alpha$ lines of 118 K–M dwarfs using the filter technique at the Van Vleck Observatory. Considering the earlier spectral and photometric measurements of these values, they analyzed the chromospheric activity of 380 red dwarfs. In the diagram ($W_{H\alpha}$, R−I) one can clearly see the “main sequence” with the $H_\alpha$ absorption line and about 70 emission stars with a great dispersion of $W_{H\alpha}$ values that, on average, grow toward later stars. The calculated luminosities $L_{H\alpha}$ and the surface fluxes $F_{H\alpha}$ of the emission stars systematically decrease toward cooler stars, whereas the upper limit of the ratio $R_{H\alpha} = L_{H\alpha}/L_{bol}$ is practically constant and equal to $10^{-3.9}$ for stars of all the considered range of spectra. According to Delfosse et al. (1998), for M dwarfs this saturation level was $10^{-3.5}$.

Using a less precise method, Giampapa and Liebert (1986) and Liebert et al. (1992) continued measurements of Stauffer and Hartmann (1986) for the weakest M dwarfs and found that the dispersion of values of the $H_\alpha$ equivalent widths increased with growing absolute magnitude. The coolest star with $H_\alpha$ absorption, Gl 283 B, has $M_V = 16.8^m$, and stars with $M_V > 17^m$ have the greatest $W_{H\alpha}$. As opposed to the prevailing opinion, they found that among M dwarfs later than M5.5 there were stars without $H_\alpha$ emission and among dwarfs with $M_V > 15^m$ the number of dMe and dM stars was comparable. The objects with kinematics of the young disk prevailed among dMe stars, while among dM those with kinematics of the old disk and halo dominated.

Robinson et al. (1990) used the 3.9-meter Anglo-Australian telescope for the high-resolution observations of $H_\alpha$ and CaII H and K in the spectra of 50 K5–M5 dwarfs. They published records of the obtained spectra and determined more precisely the contribution of photospheres into the measured emission fluxes and the basal levels of chromospheric activity. Based on observations of the $H_\alpha$ line, they proved the existence of the lower and upper limits of $W_{H\alpha}$ for different spectral types: the maximum absorption of this line decreased, while the maximum value of the emission increased toward later spectral types. They suggested qualitative estimates of the chromospheric fluxes for four typical levels of chromospheric activity:

$$F_K = 5 \cdot 10^3 \text{ erg/(cm}^2 \cdot \text{s) and } W_{H\alpha} = -0.3 \text{ Å},$$

$$F_K = 5 \cdot 10^4 \text{ erg/(cm}^2 \cdot \text{s) and } W_{H\alpha} = -0.5 \text{ Å},$$

$$F_K = 10^5 \text{ erg/(cm}^2 \cdot \text{s) and } W_{H\alpha} = 0 \text{ Å},$$

$$F_K = 10^6 \text{ erg/(cm}^2 \cdot \text{s) and } W_{H\alpha} = 5 \text{ Å}.$$

It is worth noting an approximate character of the mentioned conformities. The point is that CaII and $H_\alpha$ emissions provide evidence for the chromospheric activity of stars but from several different sides: the calcium lines are more sensitive to variations of flocculi, whereas the hydrogen emission — to filaments. The $H_\alpha$ emission is less sensitive to activity cycles and rotational modulation, i.e., the average chromospheric flux characterizes better than the calcium emission. Therefore, somewhat other ratios of these emissions were provided in a series of subsequent studies. Moreover, as it is noted by Meunier et al. (2022), these emissions can occur from various sources: active regions and filaments, which can also cause different ratios of their intensities.

Pasquini and Pallavicini (1991) carried out observations with high *S/N* ratio using the echelle coude spectrograph at the 1.5-m telescope of the South European Observatory. They studied about 70 F6–K5 dwarfs with different activity levels in the regions of $H_\alpha$ and CaII H and K lines. They compared the calculated surface fluxes, confirmed the results of Shakhovskaya (1974b) on the nonconstancy of a relation of the calcium and hydrogen emissions, detected a systematic growth of the ratio $F_{H\alpha}/F_{HK}$ toward cooler stars, and established the equality of these fluxes for K3–K5.

Strassmeier et al. (1993) studied the chromosphere of one of the most active single K2 stars LQ Hya: its surface fluxes in H and K lines and the IR triplet of CaII and $H_\varepsilon$ are close to those recorded in binary RS CVn-type systems, and even averaged over the disk intensities of the calcium emission lines and emission equivalent widths significantly exceed the appropriate values in solar faculae. The $H_\alpha$ line profiles are double-peaked and the “blue” peak is always higher than the “red” one as in flare dMe stars DK Leo, V 1005 Ori, and AU Mic. This asymmetry seems to be due to the proper velocity field in the chromosphere. Applying the Cram and Mullan theory (1979) for the isothermal chromosphere, Strassmeier et al. estimated the matter density at the formation level of the $H_\alpha$ line on splitting profile peaks as $(1–5) \cdot 10^{11}$ $cm^{-3}$; this value is coincident with the boundary density value for the transition from radiative to shock controlled $H_\alpha$ emission. At the maximum stellar brightness phase there is the minimum value of FWHM, the minimum splitting of its peaks, and, consequently, the chromosphere density above spots is lower than that above the undisturbed photosphere.

With the 2.2-meter telescope in Calar Alto and the 2.5-meter Isaac Newton telescope in the Canary Islands, Fernandez-Figueroa et al. (1994) and Montes et al. (1995a, b, 1996) observed about 50 chromospherically active binary systems, including more than a dozen binary systems composed of active dwarfs with orbital periods varying from 0.7 to 6 days. The observations were performed in the regions of $H_\alpha$ and CaII H and K with a resolution of 0.1–1.0 Å. To estimate the level of chromospheric activity, the researchers used the method developed by Young et al. (1989): they subtracted the spectra of inactive stars of close spectral types and luminosities from the observed spectra of active systems. They concluded that the thus-obtained excess of the $H_\alpha$ emission was a preferable index of chromospheric activity. They found a systematic decrease of the index with increasing axial rotational period and the Rossby number close to linear correlation of this $H_\alpha$ index with analogous indices with respect to CaII H and K lines, as well as nonlinear correlation of the index with the radiation in the CIV lines and in X-rays. For equal axial rotation periods of the components of binary systems, they obtained a stronger calcium emission than in single stars. In five systems, they revealed a strong emission of $H_\varepsilon$, which was probably pumped by the strong emission in the CaII H line.

Using the echelle spectrograph of the Keck telescope, Basri and Marcy (1995) obtained the spectra of 8 stars near the lower limit of the main sequence. They found that in the spectrum of the very late and rapidly rotating ($v\sin i \sim 40$ km/s) star BRI 0021-0214, a candidate for brown dwarfs, classified as M9.5+, the $H_\alpha$ emission and absorption lines were not seen. This can provide evidence for the fast weakening of magnetic heating of the stars with masses lower than $0.09M_\odot$ and a qualitative change of the mechanism of the loss of angular momentum for such stars, rather than a very cool atmosphere with negligible column density of hydrogen atoms at the second level. However, Tinney et al. (1997) suspected variable $H_\alpha$ emission on this star. With the same telescope Basri et al. (1996) carried out high-dispersion observations of 17 dwarfs of M6.5 and later spectral type with vsin*i* varying from < 5 to 40 km/s. In all cases, they detected the $H_\alpha$ emission with an equivalent width of 0.4–20 Å. The surface fluxes calculated using these values were noticeably weaker than for earlier M dwarfs, which conflicts with the already known systematic growth of $F_{H\alpha}$ from F up to middle M dwarfs. Thus, rapid rotation of late M dwarfs does not guarantee stronger $H_\alpha$ emission, which is probably due to the conversion of the envelope dynamo into the distributed dynamo that is less sensitive to rotation.

Of particular interest are observations of the Gl 890 (= HK Aqr) star, one of the fastest rotators among red dwarfs with vsin*i* ~ 70 km/s. Young et al. (1990) obtained a relatively dense observational series for this star and found an obvious asymmetry of the $H_\alpha$ line profile. Over a half-period of its axial rotation they revealed that the position of the bisector of the line changed as its equivalent width changed from 2.0 to 1.2 Å. They found a close correlation of the $H_\alpha$ emission with MgII h and k, as well as the localization of the minima of these emissions in the region of the broad minimum of stellar brightness in the V band, which meant a noticeable mismatch of active and spotted regions on the star. The mismatch of the maximum of the equivalent width of the $H_\alpha$ emission and the minimum of stellar brightness was confirmed by Byrne and Mathioudakis (1993) in the course of special photometric and spectral monitoring of the star in August 1991. Byrne and McKay (1990) analyzed its ultraviolet spectra and did not detect a noticeable increase in the emission, which should have been expected for such a fast rotation. The fluxes in the MgII h and k lines and the FeII $\lambda$ 2600 Å blends showed a distinct anticorrelation with the spottedness of the photosphere, and this phase relation preserved for at least two years. They interpreted the detected decrease in the brightness of the MgII h and k lines at a certain phase as an eclipse of the star by the low-mass component. However, Doyle and Collier Cameron (1990) noted simultaneous weakening of $H_\alpha$, which was not observed in the broadband photometry, and concluded that the eclipse was caused by a cloud of neutral gas rather than by a small star. Later, this conclusion on the eclipse by the gaseous cloud was supported by Byrne et al. (1992b), who found that the emission of the CaII K line was also in antiphase with the stellar brightness. Further observations of the star required the combination of the bright active region in $H_\alpha$ preserved at least for several days and the flare activity of the star (Byrne et al., 1994). During three nights, Byrne et al. (1996) carried out spectral monitoring of the star with a resolution of 39000 using the 3.9-meter Anglo-Australian telescope. They obtained about 50 echelle spectra in the wavelength range from 4813 to 6810 Å with 5-minute exposures; $H_\alpha$, $H_\beta$, and HeI $D_3$ were recorded simultaneously on the CCD matrix. Upon analysis of $H_\alpha$ profiles they concluded that from time to time absorption details passed along the chromospheric emission profile from the blue wing to the red one with velocities not higher than vsin*i*, and this transition lasted for less than a half-revolution of the star. The simplest interpretation of these observations is that up to 7 structures of a type of low-latitude prominences pass in the stellar corona, they are preserved for at least 5 revolutions of the star and are projected on its disk. The fact that the star Gl 890 has one of the greatest axial rotation rates but does not demonstrate a noticeably increased activity level as compared to other dMe stars, according to Skumanich and McGregor (1986), can be caused by the saturation of magnetic activity.

By now, at least two late fast rotating dwarfs have been found on which prominence-type phenomena are suspected. These are the K5 stars BD+22°4409 with vsin*i* = 69 km/s and RE1816541 with vsin*i* = 61km/s, its axial rotation period is practically equal to that of HK Aqr, about 11h (Byrne, 1997; Eibe et al., 1999). Probably, the electromagnetic forces retain such structures above the stellar equator at a distance of about two radii. According to Ferreira (2001), such structures can be maintained above the middle latitudes as well.

Comparing the spectral and photometric observations of DH Leo in 1984 and 1987, Newmark et al. (1990) noted that as the spot area decreased by 50%, the intensity of Balmer lines decreased only by 20%, which suggested that 60% of the chromospheric emission on the star occurred in the quiescent chromosphere, not related to the spotted areas. The measured ratios of the fluxes in the Balmer lines are $F_{H\alpha} : F_{H\beta} : F_{H\gamma} : F_{H\delta} = 2.07 : 1 : 0.67 : 0.43$, which is close to the ratios of the solar chromosphere. The most complete Balmer decrement was measured for the flare star Gl 431 by Doyle et al. (1990b):

$$F_{H\alpha} : F_{H\beta} : F_{H\gamma} : F_{H\delta} : F_{H8} : F_{H9} : F_{10} : F_{11} : F_{12} : F_{13} : F_{14} =$$

$$4.7 : 1.7 : 1 : 0.57 : 0.34 : 0.21 : 0.14 : 0.13 : 0.10 : 0.06 : 0.03.$$

Having considered the SDSS spectra of about 8000 late M dwarfs, West et al. (2004) confirmed the $H_\alpha$ emission maximum of about M8 but found that not all M7–M8 dwarfs had this emission, and the ratio $L_{H\alpha}/L_{bol}$ was practically constant within M0–M5 and then decreased for cooler stars.

Lyra and Porto de Mello (2005) performed calibration of $H_\alpha$ emission both as a chromospheric diagnostics and indicator of age, using the 9-year observations of 175 stars close to the Sun and stars with well determined ages up to 2 billion years, which belong to clusters and kinematic groups.

Based on the echelle middle-resolution spectra, which simultaneously overlap the calcium and $H_\alpha$ emission, Cincunegui et al. (2007) compared the calculated from the values of $S$ fluxes in calcium lines and $H_\alpha$ emission fluxes. From 109 F6–M5 stars of various activity levels it was revealed that between hydrogen and calcium indicators of activity there were quite diverse ratios, and one could not consider them as substitutable. Later, Gomes da Silva et al. (2011, 2014) studied correlations between $\log R_{HK}$ and $\log R_{H\alpha}$ from the 9-year sample of 271 F–G–K stars and detected their great diversity: about 20% of the sample showed a strong positive correlation and 3% — a strong negative one. The former group comprised predominantly more active and cooler stars, the latter — high-metallicity stars. These results are close to those obtained by Cincunegui et al. (2007). From calcium lines there were detected 69 activity cycles and from the hydrogen line — 9.

Walkowicz and Hawley (2009) compared emissions in calcium and Balmer lines in the spectra of close M3 dwarfs and found a strong positive correlation between simultaneously measured lines in active stars, although during different-time measurements this correlation weakened or disappeared. But at the low and intermediate activity levels the correlation became doubtful, when there was $H_\alpha$ absorption at the calcium emission. Therefore, observations of only $H_\alpha$ line do not allow one to distinguish the middle-activity and inactive stars, and the high-emission stars are the most active.

Berger et al. (2010) performed simultaneous observations in the X-rays, radio, and $H_\alpha$ for three M9.5–L2.5 stars: BRI 0021–0214, LSR 060230.4+391059, and 2MASS J052338.2–140302. The hydrogen emission was recorded for the first and third stars, X-rays and radio were recorded for none of these objects.

Houdebine et al. (2009a) derived high-resolution spectra of dMe-, dM(e)-, dM, and dK stars, recorded hydrogen lines from $H_\alpha$ to $Pa_8$, CaII H, CaII IRT, NaI D, and HeI $D_3$ and traced their systematic variations for stars with different activity levels. The fairly constant value of FHMW ~ 1.5 Å was consistent with the constant temperature jump in the chromosphere. The $Pa_{\varepsilon}$ line was first recorded for six dwarfs, and for AU Mic it had a form of weak absorption with weak emission wings. Later, from the high-dispersion spectra Houdebine et al. (2012) established that there was no emission in CaII cores for dM1 stars Gl 745 A and Gl 745 B, and $W_{H\alpha}$ accounted for 0.171 and 0.188 Å, respectively. For the dM1 star Gl 63 $W_{H\alpha}$ = 0.199 Å. These are objects with the lowest activity among M dwarfs.

Then, considering equivalent widths of $H_{\alpha}$ and CaII H and K as independent values, Houdebine (2010b) suggested a method for constructing two-component model chromospheres with active and quiescent regions based on the ratio of these widths and applied it to nine dM1 stars with metallicity close to the solar one. For all objects the plotted models yielded a good reconstruction of profiles. Furthermore, for seven stars the profiles lay between the maximum of absorption and emission, and for two stars — purely emission $H_{\alpha}$ profiles and the CaII profile, which was coincident with that of the dM1e star. In a series of cases there is an ambiguous solution. The typical size of the filling factor is 20–40%. With increasing activity the pressure in the chromosphere grows rather than the filling factor.

Later, Houdebine (2012) compared several hundred measurements of the CaII and $H_{\alpha}$ resonance lines for many tens of dK5 and dM4 stars and their values of $L_X$ and found that equivalent widths of calcium lines did not depend on $M_V$, whereas for dM1 stars the magnetic activity level noticeably depended on $M_V$. Furthermore, according to Houdebine, the power correlation indices $W_{CaII}$ $(P/\sin i)$ are equal to −0.80, −1.53, and −3.72 for dK5, dM1, and dM4, whereas the luminosity $L_X$ increases faster than the calcium emission with growing rotation rate.

Lee et al. (2010) studied the variability of $H_{\alpha}$ emission for 343 M3.5–M8.5 dwarfs with exposures of about 5 minutes. The eighty percent of the sample revealed a statistically significant variability within the whole time scale covered by observations with amplitudes of 1.25–4 and the frequency of flares with an amplitude of the stellar magnitude of about 0.05 per hour. The increase of variability toward later spectral types was found despite the general decrease of the ratio $L_{H\alpha}/L_{bol}$: this ratio reached 0.5 dex for M7–M8 objects and only 0.15 dex for M4–M5. Within the whole assembly of variability events, a growth by almost an order of magnitude of the number of such events was detected in the interval of 10–30 minutes, as well as the retention of this value during more prolonged events.

From 15 high-dispersion spectra of LQ Hya derived in the $H_{\alpha}$ region in spring 2000, Frasca et al. (2008) detected a strong variability of this line, found a closeness in the longitude of active chromospheric regions and photospheric spots acquired by the Doppler imaging. During the observations in 2000, the star was far more active than in 1991 (Strassmeier et al., 1993): strong emission is consistent with higher electron density, the rotational modulation of chromospheric characteristics is more pronounced in more extended active regions.

Osten et al. (2011) performed the Doppler imaging of the close binary system FS Aur-79, which consists of active dwarfs K7 and M3e, and detected two emission regions associated with prominences on the primordial component, and extended chromospheric activity on the secondary component. Profiles of the $H_{\alpha}$ and $H_{\beta}$ emission lines, their equivalent widths, and ratios of these widths as functions of orbital phases showed that emission mainly arose above the secondary component and between components; photometry revealed starspots near active chromospheric regions, and the absence of the lithium line allowed one to yield a lower age threshold of at least 500 million years.

Da Silva et al. (2011) compared indicators of the long-term activity of 23 M dwarfs and found that between $S_{CaII}$ and indicator of NaI the correlation was very close and did not depend on the stellar activity level; between $S_{CaII}$ and $H_{\alpha}$ the correlation depended on the stellar activity level, and between $S_{CaII}$ and He I there was no close correlation. Schlieder et al. (2012) found that the NaI 8200 Å line could serve as an indicator for separating M dwarfs into objects younger and older than 100 million years.

Miles-Paez et al. (2017) studied about a hundred L0–T8 dwarfs from their own observations with the Spitzer telescope and literature data and concluded that 94% of L0–L3.5 dwarfs show $H_\alpha$ emission, whereas for L4 and later stars this fraction drops to 20%. Since the photometric variability on the whole interval of L0–T8 is approximately constant, 30–55%, then they conclude that this variability does not correlate with hydrogen emission and is due to dust clouds rather than magnetic activity.

Recently, from high-resolution spectra of 331 M dwarfs derived with the Calar Alto telescope, Schöfer et al. (2022) studied activity indicators at visible and near infrared wavelengths: they explored equivalent widths of $H_\alpha$, $D_3$ lines and the neutral helium $\lambda$10833 Å, $Pa_\beta$, sodium D lines, calcium infrared triplet, and photospheric molecular bands of TiO and VO. They found that in each spectral subtype of slowly rotating stars, the $H_\alpha$ absorption is strong, the HeI $D_3$ line, sodium D lines, and the calcium infrared triplet (IRT) correlate well with this line; the infrared line of HeI and $Pa_\beta$ does not correlate with other indicators. The TiO bands reveal the activity effect, whereas the VO bands do not. The activity is shown for 29% of the sample stars, but for earlier than M-3.5 this fraction is lower than 20% and grows toward later stars. The rotational modulation effect is the most pronounced in the TiO bands, in $H_\alpha$ lines, and calcium IRT.

The $H_\alpha$ emission was revealed on L dwarfs as well, but the value of $\log(L_{H\alpha}/L_{bol})$ decreased from −3.8 for M4 to −5.7 for L3 (Jackman et al., 2019). The contribution into the luminosity of L dwarfs is presumably made by both the stellar chromosphere and auroral radiation.

**1.3.1.3. Other Emission Lines in the Optical Range.** In the mentioned high-dispersion spectrogram of EV Lac obtained by Wilson (1961) in 1960 at the Palomar telescope, in addition to 22 absorption lines and the lines of hydrogen and calcium emission, there are the emission lines of FeI $\lambda$ 3719.9 Å and $\lambda$ 3859.9 Å, HeI $\lambda$ 3888.6 Å and $\lambda$ 4471.5 Å, SiI $\lambda$ 3905.5 Å, and CaI $\lambda$ 4226.7 Å. The lines of metals are narrow, those of helium are slightly broadened, and the lines of hydrogen are noticeably broadened. Worden and Peterson (1976) recorded the emission of the neutral helium $D_3$ line in the spectra of several flare stars, whereas in the course of one-week observations they recorded a five-fold slow increase of the equivalent width of the helium $D_3$ line under simultaneous weakening of $H_\alpha$ emission and emission cores of the sodium D lines.

During the observations at the Kitt Peak Observatory, Giampapa et al. (1978) obtained a series of spectrograms of the quiescent state of the star AD Leo in the red region with a resolution of 0.22 Å. In the summed spectrum, which corresponds to the 5-hour exposure, they found emissions of $H_\alpha$, neutral helium lines $\lambda$ 5876 Å and $\lambda$ 6678 Å, and sodium D lines with equivalent widths of 1.4, 0.31, 0.058, and 0.7 Å, respectively. The ratio of intensities of the triplet and singlet helium lines evidenced the collisional character of their excitation in the upper chromosphere at 20000–50000 K.

The sodium D line profiles and parameters of their emission cores in the spectra of seven red dwarfs are presented in Worden et al. (1981).

The helium $D_3$ absorption line is absent in the spectrum of the quiescent photosphere of the Sun but can be seen in active regions, and therefore is an indicator of local magnetic activity. Danks and Lambert (1985) obtained the high-resolution spectra of 20 F–G–K stars in the region of the helium $D_3$ line and found a correlation of equivalent widths of this absorption line with the chromospheric emission of CaII H and K and the coronal emission in soft X-rays.

The correlations are close to linear, which can occur if they are due to the filling factor of stellar surfaces by faculae. The rotational modulation of the $D_3$ line is found in the spectrum of $\chi^1$ Ori and suspected for $\kappa$ Cet and $\varepsilon$ Eri. Cutispoto and Giampapa (1988) performed spectral monitoring of the helium $D_3$ line in the spectra of $\chi^1$ Ori, $\xi$ Boo A, and 70 Oph A with a resolution of 80000 at Kitt Peak. For $\chi^1$ Ori and 70 Oph A they found variations of equivalent widths of the line, exceeding by 2–4 times the measurement errors, which supported the results of Danks and Lambert (1985).

Shcherbakov (1979) obtained the first spectra of the flare stars AD Leo, BY Dra, EQ Peg, and EV Lac in the infrared region. He found a considerable weakening of components of the CaII absorption triplet and associated it with their filling-in by chromospheric emission. During higher spectral resolution observations, Pettersen and Coleman (1981) directly revealed the emission cores in all components of the CaII infrared triplet in the spectrum of AD Leo. For earlier chromospherically active stars $\xi$ Boo A, 70 Oph A, and $\varepsilon$ Eri the filling-in of the core of the $\lambda$ 8542 Å line was found by Linsky et al. (1979b). The nonLTE calculations of these lines by Andretta et al. (2005) were mentioned above.

Table 7. Relative contributions of different emissions into the resulting radiative losses of chromospheres (according to Pettersen, 1987)

| Spectral type | Stars | Radiative losses for | | | |
|---|---|---|---|---|---|
| | | Balmer lines | CaII HK + IR triplet | MgII hk | other lines |
| dM6e | UV Cet | 0.77 | 0.05 | 0.07 | 0.11 |
| dM5e | YZ CMi, AT Mic | 0.60 | 0.10 | 0.15 | 0.15 |
| dM4e | EV Lac, EQ Peg | 0.53 | 0.21 | 0.11 | 0.15 |
| dM1e | AU Mic, YY Gem | 0.43 | 0.29 | 0.14 | 0.14 |
| dM0e | V 1005 Ori | 0.30 | 0.36 | 0.25 | 0.09 |
| dK5e | EQ Vir, BY Dra | 0.20 | 0.40 | 0.20 | 0.20 |
| dG2 | Sun | 0.13 | 0.64 | 0.23 | |

Then, analogous data on chromospheric lines of helium, calcium, and sodium in the red region of the spectrum were obtained and described by Pettersen et al. (1984b). According to Pettersen (1988), for dKe and early dMe stars the surface fluxes in the CaII H and K lines exceed the fluxes in the infrared triplet, whereas for late dMe stars the ratio is inverse (Table 7). Later, Pettersen (1989b) observed eight dK5–dM1 stars with the 2.7-meter and 2.1-meter telescopes of the McDonald Observatory and the main McMath Solar Telescope with a resolution of 0.46–0.20 Å. He studied regions of the sodium and helium D lines, $H_\alpha$, HeI $\lambda$ 6678 Å, LiI $\lambda$ 6707 Å, and the CaII infrared triplet. The accompanying photometric monitoring identified spectra related to the quiescent state of the stars. All considered stars, except for 61 Cyg, were known as spotted dwarfs, but in the above lines the rotational modulation effect was not noticed, though the monthly and annual changes of the line intensities were found. The weak emission of the sodium $D_2$ line was found in the spectra of V 1005 Ori and AU Mic, in the spectrum of BY Dra the absorption of sodium D lines was to a great extent filled in by emission, and the HeI $D_3$ emission was seen in the spectra of V 1005 Ori and DK Leo. The HeI $\lambda$ 6678 Å line was not found in either of the stars. The lithium absorption line was very strong in the spectrum of V 1005 Ori: $W_{\rm Li} \sim 0.3$ Å, which is by an order of magnitude weaker in the spectrum of V 833 Tau, but it was absent in the spectra of EQ Vir, DK Leo, YY Gem, and AU Mic. Distinct emission of all the three components of the CaII infrared triplet in the center of broad absorption profiles was recorded in the spectra of all considered stars, except for YY Gem, where it was filled in by fast rotation, and the inactive 61 Cyg. The central emission in the components of the calcium infrared triplet is the most obvious from the comparison of spectra of pairs of stars of the same spectral type and different level of chromospheric activity. Foing et al. (1989) made this comparison from the observations of F9-K5 dwarfs with a resolution of 80000 using the echelle spectrograph at the coude focus of the 1.4-meter telescope of the South European Observatory. Using a similar procedure, Latorre et al. (2001) found the emission of the CaII infrared triplet, as well as $H_\alpha$ and $H_\beta$, in the spectra of each component of the OU Gem system.

Garcia Lopez et al. (1993) studied the helium $D_3$ line in the spectra of 145 F–G stars. This triplet is weaker than the infrared triplet of $\lambda$ 10830 Å, but is more accessible for observations. They confirmed that the HeI $D_3$ line was a suitable indicator of chromospheric activity for early F stars. Saar et al. (1997) studied the high-resolution spectra of 76 G and K dwarfs and the dependence of the helium $D_3$ line on rotation. They found that for $P_{\rm rot} > 4$ days the flux adsorbed in the line was $F_{\rm D3} \sim P_{\rm rot}{}^{-1.2}$, and at faster rotation its behavior depended on the spectral type: the flux was almost constant for G stars, decreased for K stars, and transformed into emission for late K dwarfs. For $P_{\rm rot} > 4$ days, $F_{\rm D3} \sim \Delta F_{\rm HK}{}^{1.5} \sim F_{\rm CIV}{}^{0.7} \sim F_{\rm X}{}^{0.6}$, which makes it possible to consider this line as formed in the upper chromosphere.

In the course of 12-year observations of 61 Cyg A, Larson et al. (1993) noticed a clear variability of the core of the component of the CaII infrared triplet $\lambda$ 8662 Å. These variations allowed the axial rotation period of the star to be determined as 36.21 days, which was consistent with the estimate based on CaII H and K lines. The stability of the 36-day period for many years suggests the long-term existence of active regions on the star or their regular recovery at active longitudes.

Emission was found in the center of the broad absorption line of CaI $\lambda$ 4227 Å of the flare stars Gl 234 AB, Gl 375, Gl 431, and AD Leo (Fosbury, 1974; Doyle et al., 1990b; Mathioudakis and Doyle, 1991a).

Using the echelle spectrograph with a resolution of about $10^5$ at the 4.2-meter William Herschel telescope in the Canary Islands, Panagi et al. (1991) observed the spectra of 11 K–M dwarfs. They obtained and published the profiles of the $H_\alpha$ line, sodium D lines, and components of the CaII infrared triplet. A comparison of the suspected variations of depths of the sodium D lines with the previous observations led to the assumption on activity at the temperature minimum level. Andretta et al. (1997) noticed the perspectives of studying the sodium D lines for the diagnostics of the lower and middle chromospheres of M dwarfs. Then, Short and Doyle (1998) obtained and analyzed the high-dispersion spectra in the regions of $H_\alpha$ and sodium D lines for five M dwarfs with different activity levels. They found a certain qualitative similarity in the behavior of the sodium and $H_\alpha$ lines: as the activity level increases, first the absorption cores of D lines strengthen, and then the emission of these lines develops. For dMe stars, the chromospheric models constructed on the basis of the sodium and $H_\alpha$ lines are very close, while for low-activity dwarfs, the column density at the level of the transition zone for sodium lines is higher by an order of magnitude than that obtained for the $H_\alpha$ line.

Thatcher and Robinson (1993) observed the spectra of early K stars with a resolution of 55000 to determine a number of indicators of the chromospheric activity at different levels of the stellar atmosphere. Each of the stars was spectrographed during one or two successive nights in the regions of sodium D lines, CaI $\lambda$ 4227 Å, the MgI green triplet and $\lambda$ 4571 Å, the CaII H, K, and IR triplet lines, $H_\alpha$ and $H_\beta$. As the indicators, they considered the integral fluxes in the calcium and $H_\alpha$ lines calculated as the differences between the profiles of these lines for active and inactive stars, the equivalent widths of $H_\alpha$ and $H_\beta$ lines, and the depths of their cores. Except for the equivalent widths of the Balmer lines, all the above indicators correlated well.

Byrne et al. (1998) obtained the high-dispersion spectra with a high *S/N* ratio for 14 K4–M5 dwarfs with different activity levels and found that the behavior of the HeI $\lambda$ 10830 Å line was basically similar to that of the $H_\alpha$ line: for the low-activity stars this line was in absorption, with increasing activity, particularly with growing $L_X$, the absorption weakened, while on AT Mic, one of the most active stars, it transformed into emission, though there was no certainty about the absence of flares during the exposure.

Schmitt and Wichmann (2001) found the coronal forbidden line of FeXIII $\lambda$ 3388 Å in the optical spectrum of the M6 dwarf CN Leo. Later, Fuhrmeister et al. (2004) undertook a search for this line in spectra of 15 M dwarfs of different activity levels and detected it in spectra of the strong flare LHS 2076. The behavior over time of this line for CN Leo showed a high variability level on the scale of hours, which could be due to the heating of the corona by microflares.

Pavlenko et al. (2017) performed a comprehensive analysis of emission lines from 147 optical high-dispersion spectra of Prox Cen derived with HARPS and came to the conclusion on the complex structure of high-temperature regions of their formation: numerous strong lines arise in the chromosphere, emission cores of strong absorption metal lines are formed there, the Balmer emission lines of flares lighten above the chromosphere, and higher the wind emission lines are formed with a shift of −30 km/s; here the HeI λ 4026 Å line appears.

**1.3.1.4. Ultraviolet Spectra.** The ultraviolet radiation of the quiet Sun comes from its chromosphere and the region transitional to the corona. But its source does not form a uniformly luminous layer at a certain level of the solar atmosphere. The ultraviolet radiation emanates mainly from the chromospheric network outlining the boundaries of supergranules: flows in the photosphere rake magnetic flux tubes toward the boundaries. The changes in the chromospheric emission averaged over the solar disk during the 11-year cycle are in close correlation with a fraction of the surface covered by faculae.

In the early 1978, the American–European satellite IUE was orbited. It was the first efficient facility for studying ultraviolet spectra of weak stars and other celestial bodies. The primary mirror of the telescope was 45 cm in diameter; the Ritchey–Chretien optical design had a field of view of 16 arcmin. The telescope was equipped with two echelle spectrographs designed for observations in the alternative mode in the ranges of 1150–2000 Å and 1825–3200 Å. Being equipped with the echelle and cross-dispersion diffraction grating, spectrographs provided a resolution of $10^4$. When the echelle was shut down by flat mirrors and only the grating was used, the resolution was 6 Å. Each spectrograph had two entrance diaphragms: a round diaphragm with a diameter of 3″ and an oval diaphragm with the axes of 10″ and 20″; the latter enabled several spectra to be obtained by a consecutive shifting of the image. The entrance diaphragms were drilled in the mirror plate that reflects the image of the field of view onto the system of fine guiding. The spectrograph detectors were Uvicon television systems with the accumulation of images. In front of the detectors, the multichannel converters of the ultraviolet into visible radiation were installed.

IUE was used to study red dwarfs for more than 15 years and provided data for an extensive databank on the ultraviolet radiation of these objects.

Carpenter and Wing (1979) were among the first to obtain the ultraviolet spectra of flare stars with IUE. In the spectra of UV Cet, Proxima Cen, and YZ CMi, they noted the strong CIV resonance doublet and weak OI and SiII lines in the short-wavelength range, and the strong MgII and FeII emission in the long-wavelength range. Hartmann et al. (1979) obtained the spectra in the range of 1215–1820 Å of the EQ Peg system composed of dM3.5e and dM4.5e stars, and the G8V star $\xi$ Boo A. In the recorded spectrograms, one can see the high-temperature lines of NV, SiIV, and CIV arising in the transition zone, the chromospheric CI, OI, CII, and SiII lines and the HeII $\lambda$ 1640 Å line, whose origin required a particular consideration. The high-temperature lines in the spectra of both objects had similar absolute fluxes per unit of the stellar surface area, which implied a weak dependence of the properties of the transition zone on the effective stellar temperature. These fluxes of emission lines averaged over stellar surfaces are comparable with the appropriate values in the active solar regions or should noticeably exceed them in the case of unevenness of the structure of chromospheres and transition zones.

In 1979, Butler et al. (1981) obtained the spectra of three M dwarfs with different levels of flare activity in the IUE short-wavelength range: one of the most active UV Cet-type stars AU Mic, the active star Gl 867 A, and the low-activity star Gl 825. During one exposure of Gl 867 A a flare occurred. In the spectra of AU Mic and Gl 867 A, the N V, O I, C II, Si IV, C IV, He II+Fe II, C I, and Si II emission lines were identified and their absolute fluxes were measured, but in the spectrum of Gl 825 no emission lines were found, though the flare activity had been recorded on this dwarf before.

In the long-wavelength part of the ultraviolet spectrum of Proxima Cen, Haisch and Linsky (1980) identified the strongest MgII h and k emission doublet and less confidently a number of other metal lines– FeII, CrII, MnII, FeI, CuI, and TiII. Taking into account the earlier published observations of Carpenter and Wing (1979) and Hartmann et al. (1979), Haisch and Linsky suggested that the increased NV emission as compared to the spectrum of the quiet Sun and weakened OI and SiII emissions were generally characteristic of active red dwarfs.

In considering the high-dispersion spectra of a number of G–K stars of different luminosity in the region of the resonance magnesium doublet obtained at IUE with a resolution of 0.20 Å, Basri and Linsky (1979) did not reveal self-absorption in the emission cores in the spectra of $\xi$ Boo A and $\varepsilon$ Eri.

Soon, Linsky et al. (1982) published an extensive study on ultraviolet spectra of cool dwarfs based on their observations of seven emission objects (AU Mic, EQ Peg, AT Mic, YZ CMi, Proxima Cen, UV Cet, and EQ Vir) and several nonemission K–M dwarfs within the whole IUE wavelength range. As in the previous observations at IUE, to eliminate the flare effect, the stellar images were shifted along the slit. Figure 17 illustrates a general idea on the character of the low-dispersion spectra obtained: in the wavelength region shorter than 1600 Å these are purely emissive spectra and in the longer-wavelength region the emissions are seen above the weak photospheric background. The main emissions are identified in Fig. 17. Almost simultaneously with the ultraviolet observations, the spectra of the considered stars in the optical region from $H_\alpha$ to CaII H and K were obtained. The following conclusions were made on the basis of the analysis of all these data.

Among the emission lines arising in the transition zone, the strongest are the lines of the ultraviolet resonance doublet of CIV $\lambda$ 1548/51 Å, and the ratio of fluxes in these lines to the bolometric luminosity $R_{CIV}$ characterizes energy losses for the heating of the region. $R_{CIV}$ is much higher for active stars than for inactive dwarfs and noticeably increases for cooler stars. The different dependence of $R_{hk}$ and $R_{CIV}$ on the effective stellar temperature suggests different heating mechanisms of the chromospheres and transition zones.

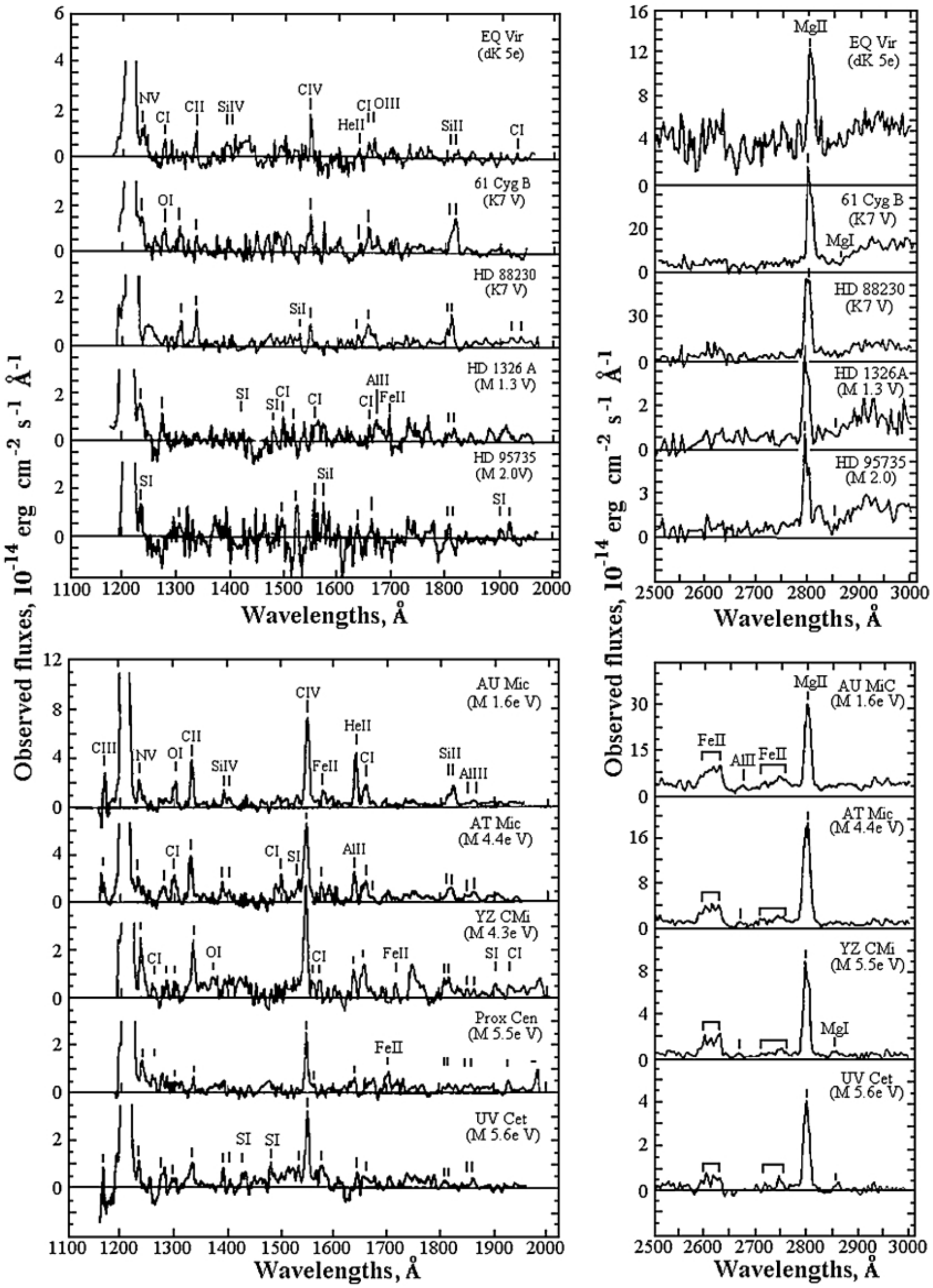

Fig. 17. The spectra of 10 K–M dwarfs obtained at IUE and calibrated in absolute units (Linsky et al., 1982)

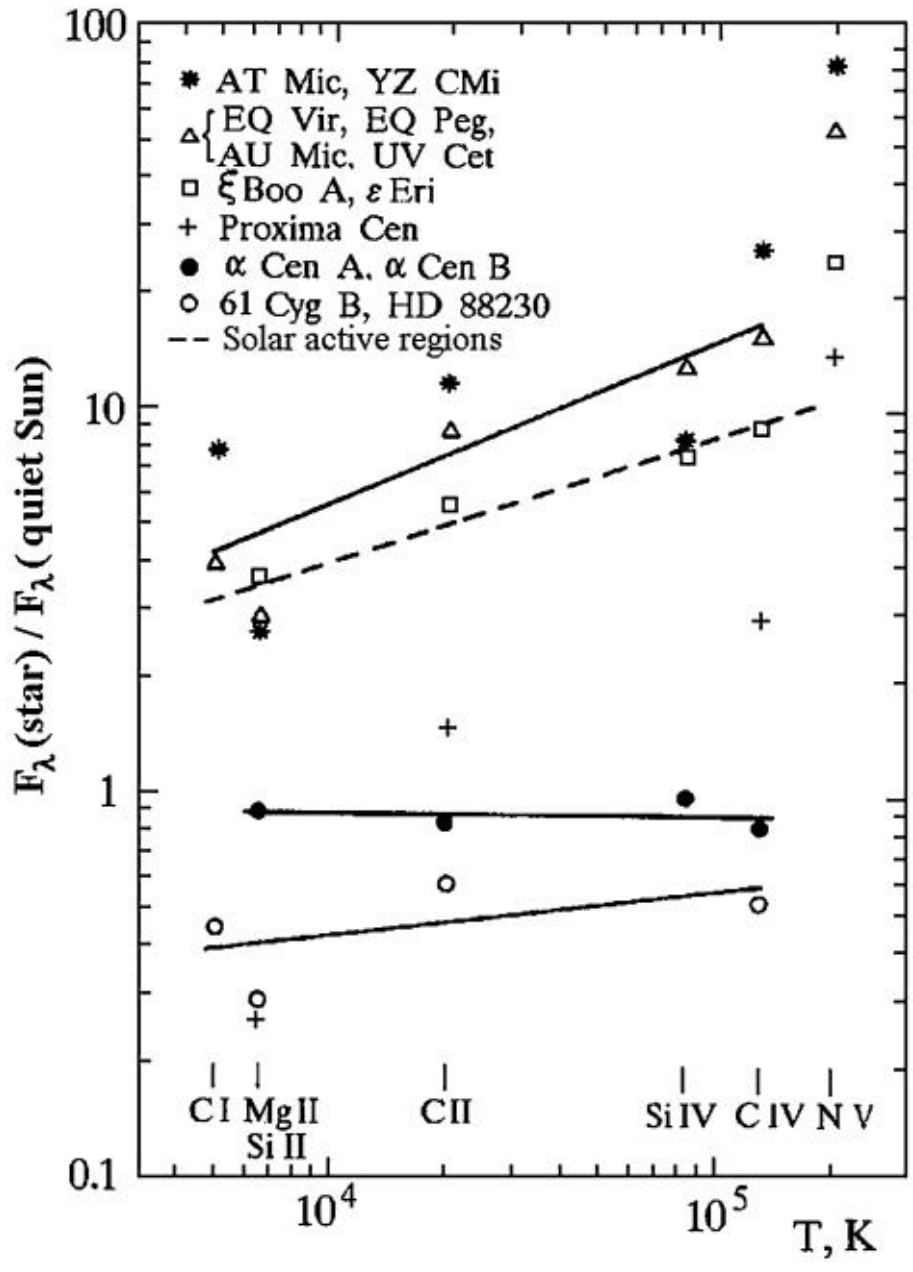

Fig. 18. The ratios of fluxes in ultraviolet lines per unit of the stellar surface to the appropriate solar fluxes (Linsky et al., 1982)

Figure 18 shows the ratios of fluxes on the surfaces of cool stars in ultraviolet lines arising at different temperatures to the appropriate fluxes from the solar surface. Depending on the value of $R_{\mathrm{CIV}}$ the cool dwarfs are divided into 6 groups: 3 groups of dMe stars, active G–K dwarfs $\xi$ Boo A and $\varepsilon$ Eri, dM stars 61 Cyg B and HD 88230, and the quiescent G–K dwarfs $\alpha$ Cen A and $\alpha$ Cen B. One can see that all fluxes on dM stars are approximately three times lower than those from the quiet Sun, whereas for quiescent G–K dwarfs they are the same as for the Sun. The fluxes on active G–K stars are the same as those in the active regions of the Sun. Finally, as a rule, active dMe stars have even larger fluxes with qualitatively identical or even stronger dependence on the temperature of emission formation. Otherwise, the assumption that the stars are completely covered by structures similar to active solar regions is insufficient to explain their luminosity in these lines. Considering that some dMe stars have spotted chromospheres, the radiative intensities of their active regions should be even higher. Probably, the cause is a denser packing of the thin magnetic flux tubes, which occupy not more than 25% of the solar facular areas. The change of the emission measure with the temperature, at which different lines are formed, is similar in all cases, but the values of these measures for dMe stars are approximately 30 times larger and for dM dwarfs three times smaller than those for the quiet Sun.

Among ultraviolet emissions, a particular place in the spectra of cool stars is occupied by the line HeII $\lambda$ 1640 Å, because, unlike the other lines of the transition zone, it can be excited by recombinations after the second ionization of helium by soft X-rays and therefore reflects the conditions existing in the solar corona rather than those in the transition zone. Later, for solar-type stars Rego et al. (1983) found a rather close correlation of the fluxes in the line $\lambda$ 1640 Å and in soft X-rays suspected earlier by Hartmann et al. (1979), which proved coronal excitation of the line. For stars differing from the Sun, this correlation is not valid due to the different contribution of the FeII emission to the blend $\lambda$ 1640 Å.

The ratio of HeII/CIV fluxes varies from 0.15 to 0.6 and is independent of the effective temperature of a star, the flux in the CIV lines, and stellar activity. As a convenient activity diagnostics parameter one should mention the SiII $\lambda$ 1808 and $\lambda$ 1817 Å lines: they are in good correlation with the magnesium doublet but are recorded simultaneously with the main lines of the transition zone.

Along with the considerable similarity of the ultraviolet spectra of cool dwarfs, whose photospheric spectra differ noticeably, one should note the weakening of the lines of SiII $\lambda$ 1808/17 Å and HeII $\lambda$ 1640 Å for active M stars, which can be caused by a sharper temperature gradient and/or higher atmospheric pressure. The appropriate line for estimating the density of matter is the intercombination line of SiII $\lambda$ 1892 Å.

Assuming the total radiation of the strong ultraviolet lines of OI and SiII proportional to all radiative losses of the chromosphere and the total radiation of the SiIV, CIV, and NV lines proportional to all radiative losses of the transition zone, Oranje et al. (1982) compared these resulting fluxes and found a rather close relation

$$F_{\rm tr} = 4.3 \times 10^{-3} F_{\rm chrom}^{1.44\pm0.02} \, . \qquad (5)$$

Using a more complete sample of A–M stars, Oranje (1986) revealed close relations that could hardly be distinguished from the linear correlation between the intensities of different chromospheric lines and between the intensities of the transition zone lines, but in comparing the emissions of these two types he found the relationships similar to (5). In particular, the fluxes in the ultraviolet chromospheric lines of CI λ 1657 Å, OI λ 1305 Å, and SiII λ 1808/17 Å were proportional to $F_{\rm MgII}$ to the power of 1.14–1.22. The fluxes in the transition zone lines of CII λ 1335 Å, SiIV λ 1400 Å, CIV λ 1548/51 Å, and NV λ 1240 Å were proportional to the flux in the MgII lines to the power of 1.6–1.7. All the found relations had a correlation coefficient above 0.9. But the emission K–M dwarfs noticeably deviated from these general relationships, probably because the Balmer lines played the role of the main source of radiative losses in their atmospheres rather than magnesium and calcium lines. There was a nonlinear dependence between coronal and chromospheric radiation similar to (5), see below.

On studying ultraviolet spectra of close binary systems with solar-type components, Vilhu and Rucinski (1983) discovered the saturation effect in the "rotation–activity" dependence: for $P_{\rm rot} < 3$ days, the intensities of the transition-zone lines did not depend on $P_{\rm rot}$ and the spectral type. This effect was later studied in different wavelength ranges. Thus, Vilhu et al. (1986) found the upper limits of the ratios $F_{\rm MgII}/F_{\rm bol} < 2 \cdot 10^{-4}$ and $F_{\rm CIV}/F_{\rm bol} < 2 \cdot 10^{-5}$ for the main-sequence stars of G0 and later spectral types. First, this saturation effect was discussed within the purely geometric considerations on the filling of the entire stellar surface by active regions. But in this case one cannot expect the surface fluxes that would exceed the solar values by more than one and a half or two orders of magnitude. Apparently, the found limits are due to the feedback between the rotation of stars with convective envelopes and magnetic fields generated in them: strong fields suppress differential rotation, which restricts the magnetic field strength.

Combining the IUE, HST, and EUVE observational results of six stars of late spectral types from the Sun to AU Mic and using the technique of emission measure as a temperature function, Doyle (1996a) found that within the temperature range of $10^4$–$10^7$ K all radiative losses of the chromosphere of a cool dwarf, except for those of the lines and continua of hydrogen, can be estimated practically as a linear function of the measured luminosity of the CIV $\lambda$ 1548 Å line. Considering this fact, Doyle (1996b) doubted the validity of these saturation levels obtained before on the basis of individual lines. He preferred the estimates of total radiative losses from the measured fluxes of $F_{CIV}$ and the ratio $F_{tot}/F_{CIV}$ found for the Sun and a number of stars. For F–M dwarfs with axial rotation periods from 0.3 to 50 days, Doyle found quite a good correlation between the periods $P_{rot}$ and thus found the fluxes $F_{tot}$ as

$$\log F_{tot} = 8.32 - 1.42 \log P_{rot} \quad (6)$$

with a correlation coefficient of 0.87. The fastest rotators did not demonstrate saturation; they had only a noticeable dispersion of points. An even better correlation was found between the ratio $R_{tot} = F_{tot}/F_{bol}$ and the Rossby numbers

$$\log R_{tot} = -3.82 - 1.45 \log \mathrm{Ro} \quad (7)$$

with a correlation coefficient of 0.92. However, one can clearly see here the saturation at $\log \mathrm{Ro} < -1$, which corresponds to the axial rotation period of about two days. In this connection, Doyle suggested that the saturation should not be related to the surface filling by active regions or to the rearrangement of the dynamo mechanism mode, but only to the disregard for radiative losses for hydrogen, which should be large for fast rotators because of the increased temperature minimum and increased density of the chromosphere.

Ayres et al. (1983a) compared high-dispersion ultraviolet spectra of three chromospherically active stars $\chi^1$ Ori (G0V), $\xi$ Boo A (G8V), and $\varepsilon$ Eri (K2V) and two inactive stars $\alpha$ Cen A (G2V) and $\alpha$ Cen B (K1V). Active stars displayed surface fluxes in the lines that exceeded the appropriate fluxes of quiescent stars by an order of magnitude, but the main line profiles were practically identical. There was a similar difference in the spectra of regions of the solar photosphere with strong and weak magnetic fields. This called into question the idea that the difference in the spectra of two star groups was due to the differences in filling factors, though the differing ratios of the intensities of the SiII $\lambda$ 1892 Å line to the intensities of the CIV $\lambda$ 1548/51 Å and SiIV $\lambda$ 1394 Å lines evidenced certain qualitative changes in the chromospheres. As in the spectra of the transition zone of the Sun, the described stellar spectra illustrate a systematic and stable red shift of the SiIV and CIV lines by 4–8 km/s with respect to the chromospheric lines. This seems to be a result of some vibrational mode in small loops of magnetic flux tubes, which occurs under different conditions at the bases of such loops (Mariska, 1987).

Boesgaard and Simon (1984) observed the young G0 star $\chi^1$ Ori with IUE and found the increased surface fluxes as compared to the spectrum of the quiet Sun: in the MgII lines by a factor of 2.4, up to 6 in the other chromospheric lines, and up to 11–22 in the transition-zone lines.

To study the profiles of ultraviolet emission lines, Ayres et al. (1983b) observed AU Mic with IUE and obtained a high-resolution 18-h exposure of the short-wavelength part and 4-h exposure of the long-wavelength part of its spectrum. Regular photometric monitoring did not reveal noticeable flares. Symmetric narrow emission lines without noticeable wavelength shifts were recorded. The values of FWHM were within 74–30 km/s, the latter value was close to the width of the instrumental profile. The widths of the transition-zone lines were almost the same as for earlier dwarfs, which evidenced that the kinematics and heating mechanism of the stars with different luminosities, from solar-active regions to M dwarfs, were common. But the width of the MgII k line corrected for the instrumental profile was half that of G–K dwarfs.

Based on IUE observations of the flare star BD+26°730 with a small axial rotation period of ($1.8^d$), Bopp et al. (1983) found that the absolute fluxes of ultraviolet lines in its spectrum were greater than in all earlier studied dKe-dMe stars.

The high-resolution observations with IUE made it possible to derive the Mg II h and k line profiles and compare them with characteristics of the CaII H and K lines. Thus, Blanco et al. (1982) detected this emission, starting from the spectral type A7, i.e., much earlier than the calcium emission occurred. They found an independence of the ratio of $R_h$ to the effective stellar temperature, an exponential decrease of the flux in these lines with growing axial rotation period, and a decrease of the ratio of the Mg II/CaII fluxes with increasing radiative losses of stellar chromospheres. For the F2 dwarf 78 UMa, Simon (1986) recorded the high-temperature ultraviolet emission lines with surface fluxes exceeding those from the Sun by a factor of 4.

Oranje and Zwaan (1985) found a clear linear dependence between the $F_{hk}$ and $F_{HK}$ fluxes. Later, Schrijver et al. (1992) compared the observations of 26 F5–K3 dwarfs in the CaII H and K and MgII h and k lines made at intervals of not more than 36 h and refined the linear relations between the appropriate surface fluxes. They determined the basal level of chromospheric activity in calcium emission and showed that in previous studies the points were dispersed due to the nonsimultaneity of observations and the uncertainties in the absolute calibration.

Haisch and Basri (1985) compared the ultraviolet spectra of the Sun and 14 G dwarfs and found that the continuum in the region $\lambda > 1600$ Å provided information on the temperature minimum $T_{min}$. These values displayed noticeable dispersion for the stars of the same spectral types: more active stars had the increased $T_{min}$. The temperature minimum on the Sun had spatial structures up to 1, the differences in the emission from this level could be explained by different filling factors of the solar photosphere by faculae.

Bromage et al. (1986) compared the ultraviolet spectra of the quiet Sun and the flare star AT Mic and found that in the stellar spectrum averaged over the surface fluxes in the transition-zone lines and the HeII $\lambda$ 1640 Å line were strengthened by a factor of 10–40, whereas in the chromospheric lines, only by several times, while the MgII h and k doublet was weakened by a factor of 1.5.

From the set of lines and their widths determined during the 48-h IUE monitoring of the flare star AT Mic with photometric support from South Africa, New Zealand, and Chili, Elgaroy et al. (1988) established that its spectra were very similar to those of AU Mic (Ayres et al., 1983b). In the spectra of AT Mic obtained definitely or hypothetically without flares, more than a dozen lines of the ionized iron within the range of 2580–2632 Å were identified. For the range 1300–1820 Å they estimated the average surface fluxes in the lines that were stronger, as a rule, by a factor of 2–5, than those in the very active regions on the Sun. The FWHM of the MgII, FeII, CII, SiIV, and CIV lines (in order of increasing formation temperatures) systematically increased from 36 to 88 km/s and their shifts with respect to the chromospheric lines systematically decreased from +5 to −12 km/s. Measured widths of the MgII lines in the spectra of AU Mic (Ayres et al., 1983b), AT Mic (Elgaroy et al., 1988), and AD Leo (Ambruster et al., 1989b) suggest that the Wilson–Bappu relation for the widths of the emission components of these lines and the stellar absolute luminosities is valid up to red dwarfs. However, later observations by Elgaroy et al. (1990) revealed that for a number of red dwarfs this relation was not valid: observed widths are too narrow, probably due to the small optical thickness.

Table 8. Absolute radiative losses for different emissions from a unit of the surface of AD Leo of $10^4$ erg/(cm$^2$ · s) (Sundland et al., 1988)

| | |
|---|---|
| CaII IR triplet | 29 |
| H$_\alpha$ | 118 |
| H$_\beta$ | 59 |
| H$_\gamma$ | 24 |
| H$_\delta$ | 20 |
| CaII H + H$_\varepsilon$ | 27 |
| CaII K | 18 |
| MgII h+k | 54 |
| FeII $\lambda$ 2751 Å blend | 9 |
| FeII $\lambda$ 2610 Å blend | 30 |
| CIV $\lambda$ 1548/51 Å | 13 |
| SiIV $\lambda$ 1394 Å | 2 |
| CII $\lambda$ 1335 Å | 4 |
| NV $\lambda$ 1240 Å | 3 |

On the basis of optical and ultraviolet observations of AD Leo Sundland et al. (1988) constructed a general spectrum of the quiescent star from the infrared triplet of calcium up to the NV line in the far ultraviolet. As follows from Table 8, the chromospheric lines yield $4 \cdot 10^6$ erg/(cm$^2$ · s), while the lines of the transition zone are $3 \cdot 10^5$ erg/(cm$^2$· s). But this estimate disregards numerous lines of the transition zone in the far ultraviolet. Using the solar spectrum in this region, from four measured lines of the transition zone Sundland et al. estimated the total luminosity of lines of the transition zone of AD Leo as $10^6$–$10^7$ erg/(cm$^2$· s).

Vilhu et al. (1989) measured the flux in the MgII h and k lines in the spectrum of the binary eclipse star CM Dra, whose components were dM4 stars, and they seemed to be completely convective. Considering these measurements and published data, they determined more precisely the saturation levels of the CIV and MgII emissions in the range of spectral types F–M for normal stars with convective envelopes: in the CIV lines it was $\sim 10^6$ erg/(cm$^2$· s) up to G0 and then decreased monotonically to $2 \cdot 10^5$ erg/(cm$^2$· s) toward late M stars; in the MgII lines $\sim 10^7$ erg/(cm$^2$ · s) up to G7 and then decreased monotonically to $8 \cdot 10^5$ erg/(cm$^2$ · s) toward late M stars. For the latter, the saturation was achieved near the axial rotation period of about 5 days, for G stars of about 3 days. It was suggested that under saturation the whole stellar surface was filled with magnetic fields, whose strength was determined by gas pressure in the photosphere.

To estimate the total radiative losses of the upper stellar atmospheres at $2 \cdot 10^5$–$2 \cdot 10^6$ K, Doyle (1989a) used the observational ratio of total losses to the luminosity of the CIV $\lambda$ 1548 Å line on the Sun and measured fluxes in this line on stars. He obtained total losses of dKe-dMe stars in the region with the temperature above $10^4$ K at the level of (3–4) $\cdot 10^7$ erg/(cm$^2 \cdot$ s), which was slightly higher than the appropriate value in the solar-active regions.

Stellar observations enable comparisons of the objects of different age and with different rotation in the broad range of effective temperatures, while the investigation of bright and extensive structures on the Sun enables observations with a high spectral resolution, high *S/N* ratio, and in the regions with different levels of magnetic activity. Thus, Cappelli et al. (1989) compared eight main emissions in the short-wavelength part of the ultraviolet spectra of 45 F5–K5 stars and in 10 different structures of the solar surface — quiet regions, faculae of different strength, and flares — and found a considerable similarity. In particular, the solar data satisfied well the flux–flux ratios derived from the stellar data and covering three orders of magnitude: surface fluxes of slowly rotating stars were analogous to solar regions with low magnetic activity, while the fluxes of active rapidly rotating stars were similar to the strongest solar faculae and even flares. The weakest fluxes from solar structures were weaker than those from the least active stars, which suggested that even on the most inactive stars several percent of the surface could be covered by active regions. The filling factor of the most active objects was close to unity, or the active regions on them were brighter than the brightest solar analogs. But the continuous distribution of the solar structures over the activity level prevented unambiguous estimates of the surface filling factor by active regions.

Byrne and Doyle (1989) performed ultraviolet observations of two dM stars Gl 784 and Gl 825 and found that the surface fluxes in the chromospheric lines, in Ly$_\alpha$, and in the transition-zone lines were somewhat lower than those on the quiet Sun. This means that the ratios $F_*/F_\odot$ are independent of the temperature of emission formation, whereas for dMe stars these ratios grow rapidly toward the increased temperature of line formation. The fluxes in the MgII lines for these dM stars are much less than for dMe stars and the quiet Sun, but they continue to be the main coolers of the chromosphere surpassing the efficiency of Ly$_\alpha$. The fluxes in nonhydrogen lines in the spectra of these dM stars are close to the appropriate fluxes in the spectra of inactive K dwarfs 61 Cyg A and B. Thus, Byrne and Doyle concluded that the transition from dM to dMe stars was due to a growth of the surface filling factor by active regions. Then, Doyle et al. (1990a) compared the fluxes in the MgII h and k and Ly$_\alpha$ lines in the spectra of 21 M dwarfs and found that the losses for Ly$_\alpha$ in dMe stars were almost twice as high as those in the MgII lines, they amounted to about 25% of the total radiative losses of the chromosphere and were comparable with the losses for H$_\alpha$. However, such a situation can occur only if there is a temperature plateau in the lower part of the transition zone or with different filling factors for Ly$_\alpha$ and H$_\alpha$. For dM stars Ly$_\alpha$ and H$_\alpha$ provide a comparable contribution to the radiative losses. The fluxes in Ly$_\alpha$ of dMe stars are higher by almost an order of magnitude than those in dM dwarfs.

From ultraviolet spectra of the dM(e) stars Gl 105 B, Gl 447, and Gl 793 Byrne (1993c) measured surface fluxes in the Ly$_\alpha$, CIV, and MgII lines and analyzed them together with analogous data for dM (Gl 784 and Gl 825), dM(e) (Gl 900), and dMe stars (YZ CMi and AU Mic). He found that $F_{\rm MgII}$ of the dM(e) star Gl 900 was actually intermediate between the appropriate values of the dM and dMe stars, but $F_{\rm MgII}$ of newly measured dM(e) dwarfs was lower than that in dM stars, for example for Gl 105 B, by an order of magnitude. A similar situation occurs with the fluxes in two other lines. This means that dM(e) stars, Byrne's "zero $H_\alpha$ stars", cannot be considered as stars with nonuniform surfaces, partly analogous to dM and dMe stars, or as stars with uniform surface and nonradiative heating intermediate between the dM and dMe stars.

Landsman and Simon (1991) analyzed normalized fluxes in the $Ly_\alpha$ line, i.e., the ratios $R_{Ly\alpha}$, for more than 260 stars and found that the emission in this line appeared on late A and early F stars, though they had rather thin convective zones. For flare stars UV Cet, EV Lac, and AT Mic the values of $R_{Ly\alpha}$ are among the highest, about $3 \cdot 10^{-4}$.

Haisch et al. (1990a) observed the YY Gem system in the ultraviolet and X-rays. The level of MgII emission appeared to be very stable over three years, which is apparently due to the uniform distribution of the regions radiating the lines along the longitude or even saturation, since, according to Doyle (1987), the stars with the short axial rotation period, $0.8^d$, should have a 100% surface filling factor for magnetic regions. However, the transition-zone lines have rotational modulation and change over a greater time scale.

Applying the method of phase-dispersion minimization to analyze the time series of irregular sampling, Hallam et al. (1991) considered the IUE spectra of F–K dwarfs to find the rotational modulation of the major chromospheric and transition zone lines. For the young active K2 dwarf $\varepsilon$ Eri, they found modulation of the ultraviolet line fluxes with a period of about 2.8 days with preserved active longitudes during several hundred revolutions. This period is four times shorter than that found from the calcium monitoring at the Mount Wilson Observatory and could be its harmonics. Less certain evidence of rotational modulation of ultraviolet fluxes with a period of about 29 days was found for the old G2 dwarf $\alpha$ Cen A.

As in studying the fluxes in the CaII H and K lines, the basal level was found for the MgII, SiII, CII, SiIV, CIV lines, although in the transition-zone lines this level was not as far from the detection threshold, as in the chromospheric lines (Rutten et al., 1991). Mathioudakis and Doyle (1992) extended the basal level of the chromospheric radiation of MgII to M dwarfs, and Byrne (1993c) measured the surface flux in the MgII h and k lines in the spectrum of Gl 105 B: the $F_{hk}$ flux is the lowest for M dwarfs — $8 \cdot 10^3$ erg/(cm$^2$ · s), which is 70 times lower than that for the Sun.

Using the GHRS spectrograph at the Hubble telescope, Peterson and Schrijver (2001) studied the MgII h and k lines for G dwarfs with metallicity that is 300 times lower than the solar metallicity. In all objects they noticed magnesium emission with split profiles, but the fluxes in the lines were within a narrower range than the metallicity range.

In 1986, Quin et al. obtained 40 ultraviolet spectra of one of the most active flare stars, AU Mic, for two full revolutions (Quin et al., 1993). The average surface fluxes calculated using the data are presented in Table 9. The accuracy of the estimate of absolute fluxes in the strongest lines was assessed as 20–30%. The fluxes in the chromospheric and transition zone lines display a considerable dispersion, which is probably due to the frequent low-amplitude flares in the active regions with a filling factor close to unity. Against the background of the dispersion, no rotational modulation of the line intensity was found. From three density-sensitive intensity ratios of the lines (SiIV, CIII $\lambda$ 1176 Å and AlIII to CIII $\lambda$ 1908 Å), they obtained well-fitting lower-pressure limits in the line formation region. Using the method of differential emission measures, they estimated the total radiative losses within $10^4$–$10^7$ K for the pressure $P_e = 3 \cdot 10^{15}$ cm$^{-3}$· K and for solar content of the elements. According to Quin et al., the total radiative losses in the atmospheres of AU Mic and the Sun are presented in Table 10.

The total radiative losses of the atmosphere are about 1% of the AU Mic bolometric luminosity and 0.003% of the solar bolometric luminosity. The distribution of losses among different atmospheric levels is fairly different: $F_{\text{chromosphere}} : F_{\text{transition zone}} : F_{\text{corona}}$ is 1:0.6:0.7 on AU Mic and 1:0.4:0.03 on the Sun.

Table 9. Average surface fluxes of AU Mic (after Quin et al., 1993)

| Line | $\lambda$, Å | $F$, $10^4$ erg/(cm$^2 \cdot$ s) |
|---|---|---|
| CIII | 1176 | 8.8 |
| Ly$_\alpha$ | 1215 | 140 |
| NV | 1240 | 4.2 |
| SiII | 1264 | 1.1 |
| OI | 1307 | 6.0 |
| CII | 1335 | 8.8 |
| SiIV | 1400 | 7.0 |
| CIV | 1550 | 13.7 |
| HeII | 1640 | 10.5 |
| CI | 1660 | 5.3 |
| AlII | 1670 | 2.8 |
| SiII | 1820 | 9.7 |
| AlIII | 1860 | 1.4 |
| CIII | 1908 | < 0.7 |
| MgII | 2800 | 94.8 |
| FeII | 2800 | 50.8 |

Landsman and Simon (1993) compiled the catalog of measured fluxes in the Ly$_\alpha$ line for 275 stars; some of them were revised with respect to the interstellar absorption, which increased the values of the fluxes by a factor of 2–5. In almost all cases the radiative losses in Ly$_\alpha$ were comparable with those in the magnesium doublet, while the corrected relative fluxes $R_{\text{Ly}\alpha}$ were in close correlation with the chromospheric emission of CII $\lambda$ 1335 Å formed in the upper chromosphere at a temperature of about 15000 K.

In the course of the cooperative campaign in 1988, Butler et al. (1994) found for the YY Gem system a close similarity of the light curve in the V band and in the MgII h and k lines near the secondary minimum, which meant fairly complete and uniform filling of the stellar surface by the regions radiating in these lines. In the $Ly_{\alpha}$ line the eclipse was noticeably longer than the optical one, which can be caused either by greater extension of the radiating region of the eclipsed component or by greater effective part of the eclipsing component at the expense of circumstellar matter.

Table 10. Total radiative losses in the atmospheres of AU Mic and the Sun (Quin et al., 1993)

| Region of the stellar atmosphere | Temperature range | Radiative losses, $10^6$ erg/(cm$^2$ · s) | |
|---|---|---|---|
| | | on AU Mic | on the Sun |
| Chromosphere | $4.0 < \log T < 4.3$ | 34.7 | 2.0 |
| Transition zone | $4.4 < \log T < 6.1$ | 20.4 | 0.8 |
| Corona | $6.2 < \log T < 7.6$ | 24.5 | 0.06 |
| Total losses | $4.0 < \log T < 7.6$ | 80 | 3 |

Doyle et al. (1994) obtained the spectra in the ultraviolet and visible wavelength ranges of four M dwarfs with very low chromospheric activity. Only Gl 821 had weak calcium emission with an absolute flux on the surface of $8 \cdot 10^3$ erg/(cm$^2$·s) and only Gl 813 had weak magnesium emission with a flux of $4 \cdot 10^3$ erg/(cm$^2$· s), half that from Gl 105 B that was considered before as the most inactive dM(e) star. None of the stars had $H_{\alpha}$ emission and all stars had low rotation rate.

In 1996, using GHRS/HTS with the low-resolution G140L grating, Vilhu et al. (2001) performed observations of the FeXXI $\lambda$ 1354 Å line in the spectrum of the active dwarf AB Dor. The observations covered a half of the stellar revolution with 11.5 min time resolution and detected EM in this line for $T_e = 10^7$ K that was higher than that derived during earlier ROSAT, ASCA/EUVE and later XMM-Newton observations.

For a long time there has existed an opinion that for M7–M9 stars the scintillation in the transition zone occurs only during the flares. But Hawley and Johns-Krull (2004) carried out special observations with the Hubble telescope for stars VB 8, VB 10, and LHS 2065 and detected that their CIV, SiIV, and HeII lines were variable but always present.

* * *

Simultaneous magnetometric and ultraviolet observations of $\xi$ Boo A revealed strong modulation of the transition-zone lines in the phase with the magnetic flux (Saar et al., 1988). Linsky et al. (1994b) compared the fluxes (IUE) in ultraviolet CIV and MgII lines with the magnetic fluxes $fB$ (McMath telescope) from the solar-type stars 59 Vir, $\xi$ Boo A, and HD 131511. All three stars demonstrated appreciable variations of the line intensities, but the variations of 59 Vir had no rotational modulation. The comparison of the measured fluxes in the CIV lines and the $fB$ value revealed a clear correlation $F_{\rm CIV} \sim (fB)^{0.66\pm0.05}$, which is valid for the solar data presented by Schrijver (1990) and all considered stars, except for 59 Vir.

* * *

The next spacecraft that was successfully used to study stellar atmospheres after IUE was the Wide Field Camera (WFC)[1]. WFC discovered the active system RE 0618+75, the binary system with a period of 0.54 days of two very rapidly rotating dM3e stars. Its orbit, spatial orientation, and global characteristics of components were determined from optical observations (Jeffries et al., 1993). Though it is a fast rotator, its activity level does not exceed that of slower rotating dMe dwarfs, which supports the idea of the existence of the activity maximum stipulated either by the fact that the surface filling factor for active regions approached unity, or by the reverse negative feedback in the dynamo mechanism.

In the course of WFC observations, Jeffries and Bromage (1993) found strong EUV radiation of the late dwarf Gl 841 A. The subsequent optical observations revealed the binarity of the source with a period of about 1.12 days, the emission in $H_\alpha$ and CaII in each component and their flare activity.

Wood et al. (1994) made a statistical analysis of all nondegenerated stars within 10 pc recorded during the WFC all-sky survey. Among 220 objects known in these vicinities of the Sun, 41 objects were found with confidence and 14 were suspected at least in one photometric band. Luminosities of the objects were calculated in the $S_1$ and $S_2$ bands and compared with X-ray luminosities from the IPC Einstein Observatory data (see Sect. 1.4.1). The luminosity functions calculated for different spectral types demonstrated similarity for the stars of middle spectral types and a noticeable dispersion for M dwarfs: in EUV their luminosities were noticeably lower, which is probably due to the saturation effect of the ratio $R_{\rm EUV}$. Most members of binary systems do not differ from single stars of the appropriate spectral types in $L_{\rm EUV}$, but in the systems AT Mic, $\xi$ UMa, FK/ FL Aqr, Wolf 630, $\xi$ Boo, and EQ Peg there is an increased EUV radiation. Brighter sources have a higher ratio $L_{\rm X}/L_{\rm EUV}$, which can be explained by a higher coronal temperature. For AU Mic, the brightest coronal source in the radius of 10 pc from the Sun, this temperature achieves $4 \cdot 10^7$ K. The dependence $L_{S1}$ and $L_{S2}$ on rotation is steeper than $L_{\rm X}(v_{\rm rot})$.

* * *

The Hubble Space Telescope (HST) that started its operation in April 1990 opened new opportunities for obtaining ultraviolet spectra of flare stars with much higher resolution and *S/N* ratio than before.

In the course of cooperative observations of AD Leo in May 1991, Saar et al. (1994c) obtained spectral data from HST. The constancy of fluxes in the lines for the times 5, 10, and 30 s was analyzed. All lines showed variability at all these times. In a certain conventional definition of a flare, 35% of the flux in the SiIV lines can be attributed to the constant level, 40% to flares, and 25% to microflares or active regions. If another definition of a flare is used, the latter part includes 10–20% of energy. But in the chromospheric CI line, 80% of the energy belongs to stationary radiation. Later, Saar and Bookbinder (1998a) performed analogous observations of two young dwarfs in the range 1380–1670 Å: G0 star HD 129333 with an axial rotation period of 2.8 days and K2 star LQ Hya with a period of 1.8 days. For a time resolution of 1s they found low-amplitude flares, which amounted to 8% of the radiation from the transition zone of HD 129333 and 11% in LQ Hya, in the summary flux of the CIV+SiIV lines.

---

[1] Built in the United Kingdom and mounted together with the X-ray telescope ROSAT on the German–UK–USA satellite, it was launched into orbit in June 1990. The wide-field camera is a "nested" telescope with grazing-incidence Wolter–Schwartzschild Type I mirrors and an aperture of 58 cm, a field of view of 5° in diameter, a spatial resolution of 1 arcmin in the center of the field of view and 3 arcmin at the edge, and a position accuracy from 20 arcsec to 1 arcmin. During sky surveys, the apparatus rotated about the axis perpendicular to the Sun, shifting the swath by one degree every day, thus each object was observed for 5 days every 96 min with an exposure of up to 80 s. In addition to survey observations, the apparatus was used to make more than 6000 long-duration exposures of individual pointing. The detectors were multichannel plates with a curved focal surface sensitized by cesium iodine. The filters layered on a thin film and separating the bands 90–210 Å ($S_1$), 62–111 Å ($S_2$), 56–83 Å ($P_1$), and 17–24 Å ($P_2$) were mounted in front of the detector; in survey observations the filters were changed once a day. Thus, the apparatus was suitable for studying the upper transition zones and the lower coronae of cool stars.

Linsky et al. (1994a) studied the red shift of the transition-zone lines with respect to photospheric lines for the stars of late spectral types. It is known that on the Sun such shifts are observed above the regions with strong magnetic fields and are caused by the plasma flows descending along the magnetic flux tubes. Linsky et al. considered one of the most active red dwarfs, AU Mic. In its spectra obtained with a resolution of 20000 in the region of the CIV and SiIV lines, they recorded such a shift of 4.1±2.2 km/s. In studying the components of the $\alpha$ Cen AB system, Wood et al. (1997) found that as on the Sun, the red shift of the transition zone lines grew with increasing temperature of line formation to $\log T$ = 5.1, and then decreased. From the ratio of the OIII line intensities, sensitive to the electronic density, at the level of $\log T$ = 5.14 they obtained the estimates $\log n_e$ = 9.65 and 9.50 for $\alpha$ Cen A and B, respectively. These estimates agree with the values of electron pressure in the coronae obtained from EUVE data (see below).

Linsky and Wood (1994) observed the flare star AU Mic with 162-s exposures: 17 exposures in the region of the CIV doublet and 22 exposures in the region of the SiIV doublet. The CIV line profiles showed no time variations, but their wings up to 200 km/s exceeded by a factor of three the wings of the instrumental profile. The observed profiles of both lines were presented by the sums of two Gaussians with widths of about 30 and 170 km/s with negligible relative shifts of their centers. The widths found noticeably exceed the expected Doppler widths for the temperature of formation of the lines. Narrow components resemble these line profiles observed in solar active and quiet regions, while the wide components resemble those found in the so-called explosive phenomena on the Sun. They arise on the solar disk with a frequency of several hundred per second in small regions of several arcseconds and have a lifetime of about a minute. They occupy about 1% of the solar surface, generate about 5% of the radiation in the CIV and SiIV lines, while the broad components in the spectrum of AU Mic yield up to 40% of fluxes in these lines and should occupy about 12% of the stellar surface. The physical essence of the explosive phenomena is not completely clear, probably new magnetic fluxes emerge in old magnetic structures, thus the magnetic fields are annihilated there. The surface fluxes in CIV lines are up to $6 \cdot 10^5$ erg/(cm$^2 \cdot$ s), which is three times higher than the saturation level estimated by Vilhu (1987). Total radiative losses in the most active regions are estimated as several $10^9$ erg/(cm$^2 \cdot$ s) and the appropriate ratio $R_{\rm tot}$ = 0.23 is one of the record ratios for the magnetic heating of stellar atmospheres. (The event corresponding to the explosive events on EV Lac in the CIV $\lambda$ 1548/51 Å line recorded with the space astrophysical station Astron on 6 February 1986 will be described in Sect. 2.4.3.)

Maran et al. (1994) obtained the spectra of AU Mic with a resolution of about 10000; the observations were performed in the mode of high-speed spectrography with exposures of 0.4 s over 30 min at each of seven subsequent turns around the Earth. They studied the wavelength range of 1345–1375 Å, containing the CI, OI, OV, ClI, and FeXXI lines, which are formed at $10^4$– $10^7$ K. Parallel IUE observations proved that during the HST observations the star was quiet. Of particular interest is the FeXXI $\lambda$ 1354 Å line formed at $T \sim 10^7$ K: it does not display the traces of shifts and asymmetry and its width allows for a nonthermal velocity of not more than 38 km/s. Based on this line and the simultaneous IUE data, the complete distribution of *EM* in the entire temperature range of the upper atmosphere of the star was first constructed from the observations. A comparison of the intensities of the OV $\lambda$ 1371 Å and $\lambda$ 1218 Å lines measured by Woodgate et al. (1992) provided an estimate of the electron density of $5 \cdot 10^{10}\, cm^{-3}$ at $T \sim 250000$ K. Comparison of the obtained spectrum of AU Mic with the spectra of the solar flare and the active region proved its similarity to the flare spectrum.

To analyze the $Ly_\alpha$ line profile observed with HST in the spectra of cool dwarfs, Gayley (1994) developed the approximate analytical theory of line wings within the concept of radiative transfer with the partial frequency redistribution for a simplified chromospheric model. The theory makes it possible to determine the electron density and its gradient with depth in a certain typical chromospheric layer from the measured surface flux at a distance of 1 Å from the line center and from the spectrophotometric gradient in this point. Applying this theory to the observations of the flare star AU Mic (Woodgate et al., 1992), where a resolution of 0.1 Å at high *S/N* was achieved, Gayley estimated the range of electron density as $9 \cdot 10^{10}$–$3 \cdot 10^{10}$ $cm^{-3}$ in the layer forming the wings of the line and expanding with respect to the ionization degree of hydrogen from 0.1 to 0.001. This implies that the profile of $Ly_\alpha$ is the most sensitive to the electron density in a hydrogen ionization region of 0.05–0.10, which corresponds to the temperature plateau in the chromosphere.

Linsky et al. (1995) observed ultraviolet spectra of the components of the binary system Gl 752: the bright component dM3.5 Gl 752 A and the faint component dM8e Gl 752 B (= VB 10). This pair is of particular interest, since one component is earlier and the other is later than the spectral type M5, the boundary separating the stars with convective envelopes and completely convective stars; the star Gl 752 B is close to the lower limit of the main sequence, where thermonuclear burn of hydrogen ceases and red dwarfs are replaced by brown dwarfs. The observations were performed with a resolution of 0.17 Å in the wavelength range of 1160–1718 Å for Gl 752 Å and 1287– 1575 Å for Gl 752 B. In the spectrum of Gl 752 A, one can clearly see the lines CIII $\lambda$ 1176 Å, SiIII $\lambda$ 1207 Å, NV $\lambda$ 1240 Å, OI $\lambda$ 1304 Å, CII $\lambda$ 1335 Å, SiIV $\lambda$ 1400 Å, CIV $\lambda$ 1550 Å, and HeII $\lambda$ 1640 Å. But no lines are seen in the spectrum of Gl 752 B obtained by summing 10 exposures with a total duration of 54 min (see Fig. 58).

Saar and Bookbinder (1998b) analyzed the spectra of the eclipse system CM Dra in the MgII region obtained with one-second resolution. They noticed several flares. After subtracting them, on the light curve they found the differences in the wings of entrance and exit point from the eclipse that evidenced the inhomogeneity of the chromosphere, while the differences between eclipses suggested fast evolution of these inhomogeneities. During HST observations of three G dwarfs in the $\alpha$ Per cluster and in the Pleiades, Ayres (1999) found chaotic variability in the SiIV $\lambda$ 1393 Å line for times of tens of minutes on all the stars.

Wood and Linsky (1998) thoroughly analyzed the $Ly_\alpha$ line profile to study the interaction of the stellar wind with interstellar medium, discovered the hot "hydrogen wall" of neutral gas, and estimated the pressure in the stellar wind of several stars, whose parameters were close to the solar ones (see Sect. 1.4.4).

When the high-resolution diffraction spectrograph GHRS/HST was replaced for the echelle spectrograph STIS HST, Pagano et al. (2000) obtained the spectra of AU Mic in the range of 1170–1730 Å with a resolution of 46000 in four turns of the telescope around the Earth. During 9200 s of the summary exposure of 10100 s the star was in the quiescent state, 142 emission lines of 28 atoms and ions arising at different levels of the stellar atmosphere were identified in its spectra. The spectra were used to study the atmosphere of one of the most active dMe stars. A comparison of the relative intensities of six lines of the CIII (4) multiplet of about 1176 Å in the AU Mic spectrum and in the spectra of solar faculae, in the limb, and the spot umbra showed that the best agreement occurred for facula spectra, which meant that the lines in the AU Mic spectrum were closer to the solar-active regions than to the quiet Sun. Continuing the studies of Linsky and Wood (1994), who found broad and narrow components in the profiles of the CIV and SiIV lines, Pagano et al. found an analogous effect in other transition zone lines: narrow and broad components were shifted with respect to the photospheric spectrum for +2.2±1.1 km/s and +5.4±2.3 km/s, respectively. But the earlier suggested change of the shift with the temperature of line formation was not confirmed. The fluxes in narrow and broad components are comparable, FWHM = 31 and 103 km/s for narrow and broad components, respectively, and the latter is close to the appropriate value in the mentioned explosive phenomena on the Sun. Analysis of the FeXXI $\lambda$ 1354 Å coronal line profile showed that the nonthermal velocity component of the ions radiating in plasma at $T \sim 10^7$ K was about 70 km/s, almost half their thermal velocity. Thus, these motions are subsonic and the dissipation of shock waves is not significant for the corona heating. The radial velocity of the line does not differ from the velocity of photospheric lines, which suggests that hot plasma is retained by strong magnetic fields. Analysis of the HeII $\lambda$ 1640 Å emission line suggests that it was excited by collisions (about 40%) and by cascade transitions after the recombination (60%), while the nonthermal velocity component was 48 km/s. The differential emission measure (DEM) curve constructed for AU Mic systematically differed from the solar curve: it reached a minimum at $\log T = 4.7$, this value was maintained till $\log T = 5.4$ or 6.4, and then increased by an order of magnitude near $\log T = 7.0$. The data does not allow one to estimate the temperature at which DEM starts decreasing, while on the Sun this occurs already at $\log T = 6.4$. From the DEM curve found for AU Mic, Pagano et al. (2000) estimated the total radiation for the range of $\log T = 4.1$–7.0 as $2 \cdot 10^8$ erg/(cm$^2 \cdot$ s) or $2 \cdot 10^{-2} L_{\rm bol}$. From the ratio of components of the intercombinatory multiplet OIV $\lambda$ 1401/1407 Å they estimated the density at the level of $\log T = 5.25$ as $\log n_{\rm e} = 10.8$, from other lines at the same level the density was 30 times higher. Obviously this suggests nonuniformity of the medium: low-density structures cover 10–20% of the surface, high-density structures occupy about 0.1%, but yield a comparable contribution to the total emission measure and the summarized profiles of the emission lines.

Pagano et al. (2004) analyzed the echelle spectrum of $\alpha$ Cen A obtained with a resolution of 2.6 km/s at STIS HST. Within the range of 1140–1670 Å they identified and measured 671 emission lines of 37 different ions and molecules of CO and $H_2$. They presented the profiles of the strongest emissions of the transition zone SiIII $\lambda$ 1206 Å, NV $\lambda$ 1238 Å, SiIV $\lambda$ 1393 and 1402 Å, and CIV $\lambda$ 1548 and 1502 Å by two Gaussians with average widths of about 43 and 72 km/s and with an average contribution of the wide component of about 45% to the general profile width. The Gaussians were shifted by several km/s with respect to the photospheric lines to the long-wavelength part, with a shift of narrow components being slightly greater. The widths of the nonthermal components increased from 7.5 to 39 km/s when the emission formation temperature increased from 6000 to 200000 K. Pagano et al. concluded that the obtained spectrum of $\alpha$ Cen A fit the best the spectrum of the Sun as a star.

Using STIS data, Ayres et al. (2003) studied the forbidden coronal lines in the spectra of late stars, including 12 F7–M5.5 dwarfs. They considered all possible lines of different elements and concluded that only the emissions of FeXII $\lambda$ 1242 and 1349 Å and of FeXXI $\lambda$ 1354 Å with formation temperatures of $2 \cdot 10^6$ and $10^7$ K, respectively, were promising for the diagnostics of coronal plasma. They found with confidence the emission in Proxima Cen, AD Leo, EV Lac, AU Mic, $\xi$ Boo, $\kappa$ Cet, $\chi^1$ Ori, and $\zeta$ Dor with the average FWHM of about 110 km/s but did not reveal it in $\varepsilon$ Eri, 70 Oph, $\tau$ Cet, and $\alpha$ Cen. The emission of FeXII found in seven stars of the sample had FWHM equal to 40–56 km/s. This means that in both cases the widths of the lines were close to the thermal widths. Comparison of the fluxes in these coronal lines with $L_X$ (0.2–2 keV) revealed that $L_{FeXII} \sim L_X^{1/2}$ whereas there was a linear correlation between $L_{FeXXI}$ and $L_X$ up to the minimum activity level $L_X/L_{bol} \sim 10^{-5}$. The absence of a noticeable Doppler shift of the lines suggests that radiating plasma is localized in magnetic loops rather than in the hot wind.

From the HST/STIS data Jordan et al. (2001) obtained slightly different results. In the spectra of $\varepsilon$ Eri, 70 Oph A, and $\kappa^1$ Cet they found the forbidden lines of Fe XII $\lambda$ 1242 and 1349 Å. For $\varepsilon$ Eri from the measured line parameters they estimated the pressure as $< 7 \cdot 10^{15}$ cm$^{-3}$ · K, the column emission measure as $9 \cdot 10^{27}$ cm$^{-5}$, and the magnetic field as $B \sim 20$ G.

Brandt et al. (2001) derived HST/GHRS spectra of the fast rotator AB Dor and compiled an atlas including 78 emission lines. Lines of the quiescent chromosphere and transition zone showed narrow cores overlapping the very wide wings, which could be due to the gas emission that was in corotation with the star and expanding near the Kepler corotation radius. Another solution yielded a large prominence in the $H_\alpha$ rays that was at the corotation distance. Parameters of the plasma were estimated as electron density of (2–3) $\cdot 10^{12}$ cm$^{-3}$, the electron temperature 30 000 K, and the minimum between $\log T = 5$ and 5.5 was valid in the distribution of EM.

Christian and Mathioudakis (2002) studied the echelle optical spectra of a number of stars of late spectral classes with strong emission in the EUV region and found that the equivalent widths of $H_\alpha$ emission achieved 8 Å, some of them had strong emission in the helium D line, the rotation rate from 5 to 80 km/s, and $L_{EUV}/L_{bol}$ up to $10^{-3}$.

* * *

Judge et al. (2004) performed a quantitative comparison of the UV spectra of the Sun near the minimum phase of the 11-year cycle, the solar-type star $\alpha$ Cen A, and the extremely low-active dwarf G8 $\tau$ Cet, which sometimes reveals rotational modulation rather than the long-term cyclic variations, and may be in the phase of the Maunder minimum. They detected similarity of the lower and middle chromosphere, but the lines of the upper chromosphere and middle transition zone for $\tau$ Cet had half the density of fluxes, significantly narrower lines, and the transition zone lines did not reveal a red shift.

Using the high-resolution HST spectra of 13 dwarf stars of the F9–G5 type of different age, different magnetic field structure, and different rotation periods, Linsky et al. (2012a) studied a dependence of the centroid of velocities of the ultraviolet emission lines on the rotation period in the range from 1.47 days for the Pleiades HII 314 to 28 days for $\alpha$ Cen A. They found a systematic increase in the red shift for faster rotators, and the fastest rotator of the sample, HII 314, displayed an increasing red shift at all temperatures above $\log T = 4.6$. Starting from a rotation period of 2 days, the differences in the corona structure were observed and it was suggested on the difference in heating mechanisms of the corona for HII 314 and coronae of other sample stars. From the HST observations in the far ultraviolet of 1150–1500 Å, Linsky et al. (2012b) found a clear tendency for increasing brightness temperature in all this range with decreasing rotation period, and for the fastest rotators this temperature was close to that of the solar faculae.

* * *

Orbited in June 1992, the American spacecraft Extreme Ultraviolet Explorer (EUVE) was designed for the investigation of the electromagnetic radiation in the range of 50–760 Å. After the long-term ultraviolet studies with IUE and operation of HST and the X-ray studies at the Einstein Observatory, EXOSAT, and ROSAT (see Sect. 1.4.1), this intermediate range was addressed only by the survey observations with the Wide Field Camera (WFC) and by a few spectral observations of the brightest stars with coronae using the diffraction spectrograph TGS EXOSAT. The working range of EUVE in the temperature range of $2\cdot10^5$–$3\cdot10^6$ K covered plasma radiation from the transition zones to the lower coronae of cool stars. EUVE included 4 grazing-incidence telescopes: 3 coaligned scanning telescopes with 4 photometric bands centered at 100, 180, 400, and 550 Å, and a three-channel telescope-spectrometer mounted perpendicular to them and centered at the ranges of 70–190, 140–380, and 280–760 Å with resolutions of 0.5, 1, and 2 Å, respectively, designed for deep spectral surveys and for simultaneous photometric measurements in the bands centered at 100 and 180 Å. Each channel of the telescope-spectrometer had a field of view of 5.25° along dispersion and 2.1° in the perpendicular direction. The diffraction gratings with variable pitch were mounted in the convergent beam. Multichannel plates with diameter of 50 mm and a pixel dimension of 29 μm were used as detectors.

By August 1993, the all-sky survey by scanning telescopes was completed. In the course of the project, each object was observed for 10–20 s in each of several successive turns of the apparatus around the Earth. As a result, the total exposure for the objects placed on the ecliptic was about 400 s and that of the near-polar objects was about 20000 s. Thus, by May 1995 about 740 sources of EUV radiation were recorded, among them 270 were F–M stars. Some of them were previously unknown active dwarfs, candidates for the objects of the UV Cet type.

Ball and Bromage (1995) performed photometric monitoring of four such candidates from the results of the EUVE survey and revealed flare activity in all of them. In total, they recorded 14 flares with amplitudes of $0.3$–$4^m$ in the U band.

With the help of EUVE it was established that 29 of the 47 flare stars listed by Pettersen (1991) were EUV sources, 25 were dMe stars, and 13 were dM stars, and only one of 13 dM stars was a EUV source. Apparently, the level of saturation of K–M dwarfs in the EUV range achieved $10^7$ erg/(cm$^2\cdot$ s) (Vedder et al., 1994).

Mathioudakis et al. (1994) studied 19 stars of late spectral type with ultimately low chromospheric activity in the EUV region and found the flux from the K dwarf Gl 33 in the band 60–180 Å. This result is consistent with the calculations of Mullan and Cheng (1994), who proved that under acoustic heating coronae with temperatures of about $10^6$ K could exist and the surface flux from them could reach $10^5$ erg/(cm$^2\cdot$ s).

Mathioudakis et al. (1995a) compared the level of extreme ultraviolet (EUV) fluxes from 74 main-sequence stars observed at EUVE with their rotation and found that, similarly to the chromospheric activity in MgII h and k lines, the Rossby number better characterized the level of EUV activity than the stellar rotation period in the sample of stars with different effective temperatures. But for EUV the saturation occurred at lower Rossby numbers than for the chromospheric activity of magnesium.

Observing the flare star DH Leo at EUVE, Stern and Drake (1996) found the modulation of its EUV radiation with a period of 1.05 days, which corresponded to the photometric period of the star.

EUVE recorded quiet EUV radiation from the very-low-mass star VB 8 (Drake et al., 1996). In combination with the X-ray data from the Einstein Observatory and ROSAT, these observations enabled the estimates of the coronal temperature as several million kelvins and only a slight amplitude of EM variations on the interval of about 10 years. This evidences that the activity energy source of the star is a turbulent dynamo rather than a solar-type dynamo.

Tsikoudi and Kellett (1997) studied the EUV radiation of 127 active late stars and recorded 49 of them in one or two (S1 and S2) WFC bands. In addition to 35 flares on 23 stars, on almost half of the objects they found low-amplitude variations at times of 1–2 h to a day that were independent of rotational or orbital modulation; most of them were dKe and dMe stars.

Analyzing EUVE observations of AU Mic derived in July 1993, Del Zanna et al. (1996) found that if one assumes the cosmic iron abundance, then a high-temperature component is required for the observations in the range of 80–120 Å; if the iron abundance is low, then such a component is not required. But the plasma with a temperature of $6 \cdot 10^7$ K is required in a flare.

* * *

The scientific program of the 5-day US–German space experiment ORFEUS carried out in September 1993 from a shuttle included far-ultraviolet studies of active dwarfs. ORFEUS was equipped with a 1-meter telescope, its mirror due to special coverage could operate at up to 500 Å, and with a spectrometer with a resolution of 5000 on the variable-pitch diffraction grating. The K–M dwarfs $\varepsilon$ Eri, AU Mic, $\kappa$ Cet, BY Dra, and $\alpha$ Cen A and B were studied within the program. The emission lines of the transition zones were recorded in their spectra. The electron density was estimated with the spectrum of $\varepsilon$ Eri in the region of formation of the lines of $10^9$–$10^{10}$ cm$^{-3}$ from the ratio of the intensities of the CIII $\lambda$ 1175 Å and $\lambda$ 977 Å lines. The estimates are close to the appropriate solar values (Schmitt et al., 1996a).

* * *

In 1999, NASA orbited the Far Ultraviolet Spectroscopy Explorer (FUSE) to study the radiation of space objects in the wavelength range of 900–1200 Å that was inaccessible to IUE and HST spectrographs. FUSE consists of 4 coaligned prime-focus telescopes with off-axis parabolic mirrors of $352 \times 387$ mm$^2$ feeding four Rowland spectrographs with spherical holographic gratings.

From the EUVE, STIS HST, and FUSE observations Del Zanna et al. (2002) performed an analysis of the upper atmosphere — the transition zone and corona — of AU Mic in the quiescent state, paying attention to the limitations due to the uncertainties in available atom constants and insufficiently strict selection of lines under analysis. According to their calculations, the best fit for DEM yielded the model with $P_e = 10^{16}$ cm$^{-3} \cdot$ K.

Using the FUSE apparatus, in the spectra of seven A stars Simon et al. (2002) studied the regions of subcoronal lines of CIII $\lambda$ 977 and 1175 Å and OVI $\lambda$ 1032 and 1037 Å forming in a temperature range of 50 000–300 000 K. These lines were detected only in the coolest from the considered stars with $T < 8200$ K. Thus, Simon et al. (2002) came to the conclusion on the correctness of predicting, through the modern models, the transition zone from the convective to radiative envelopes of stars and on the extension of this transition in the range of 100 K. Later, using FUSE, Neff and Simon (2008) carried out the observations of 14 A stars in the temperature range from 7720 to 10 000 K with the aim of searching for the OVI emission lines as an evidence for occurrence of the convective region and for 11 of them detected this emission. However, the widths of these lines and the ratios $L_X/L_{OVI}$ provided evidence for their belonging to invisible active K or M companions of these bright A stars.

Using FUSE, Redfield et al. (2002) carried out the observations of five dwarfs, $\alpha$ Cen A, $\alpha$ Cen B, $\varepsilon$ Eri, AU Mic, and AB Dor, with a resolution of 20 000 and detected more than 40 emission lines and blends of HeII, NII, NIII, CII, CIII, SiIII, SiIV, SIII, SIV, SVI, FeIII, OVI, FeXVIII, and FeXIX within the range of 912–1180 Å. Most of these lines emerge in the transition zone at the temperatures of formation between 50 000 and 500 000 K, chromospheric lines originate at lower temperatures, and the forbidden lines of highly ionized iron appear at a coronal temperature of $10^7$ K. The strongest lines of the range, CIII $\lambda$ 977 Å and $\lambda$ 1176 Å and OVI $\lambda$ 1032 Å were stronger by one–two orders of magnitude than other emissions; they were recorded within the ORFEUS experiment as well. The electron density was estimated as (3−4) $\cdot 10^9$ $cm^{-3}$ from the ratios of the intensities of the CIII lines in the spectra of $\alpha$ Cen A and B. The values are in agreement with the data obtained by Wood et al. (1997) from HST data. Similarly to the HST observations of the CIV and SiIV lines in the spectrum of AU Mic, Redfield et al. (2002) presented the profiles of strong lines by two Gaussians, narrow components with characteristic velocities of 20–30 km/s and broad components with velocities of up to 100 km/s, and confirmed the tendency for increasing contribution of the broad component with increasing activity of the star and decreasing contribution with growing temperature of formation of the appropriate emission.

Continuing the Ayres et al. (2003) study of forbidden coronal emission from the STIS HST observations, Redfield et al. (2003) carried out an analogous study using the FUSE data. They considered seven F8–M5.5 dwarfs: EK Dra, $\alpha$ Cen A and B, $\varepsilon$ Eri, TW Hya, AU Mic, and Proxima Cen. Having considered more than 50 candidates, they concluded that stellar spectra contained forbidden coronal lines of FeXVIII $\lambda$ 974 Å and FeXIX $\lambda$ 1118 Å with formation temperatures of 6 and 8 MK, respectively. The line $\lambda$ 974 Å free of blending was found in the spectra of AU Mic, EK Dra and $\varepsilon$ Eri, in all cases the line width was thermal and without a noticeable Doppler shift. In the spectra of these stars they found the emission of $\lambda$ 1118 Å but noticeably blended by the CI line. A distinct correlation close to linear was found between the luminosities of $L_{\rm FeXV\ III}$ and $L_{\rm X}$.

Karmakar et al. (2016) noted an increase in the X-ray and ultraviolet radiation in the spot regions of the fast rotator LO Peg.

## 1.3.2. Models of Stellar Chromospheres

Theoretical consideration of solar and stellar chromospheres basically reduces to the calculation of the self-consistent problem on the energy balance of the medium heated by a flux of nonradiative energy and completely cooled by radiation from the whole volume. The independent variable in this problem is the power of the input energy flux, the parameters are the conditions at the interface with the photosphere: gravity, density, temperature, and chemical composition of matter. The solution of this problem involves cumbersome calculations of radiative losses of the medium that has a complex chemical composition and considerable gradients of temperature, density, and ionization state, as well as an appreciable optical thickness in frequencies of many lines. But depending on the specific form of the nonradiative energy flux arriving from below, different hydrodynamic and hydromagnetic disturbances can dissipate and transform one into another in the chromosphere, and the calculation of such a stationary system of disturbances involves considerable difficulties. Therefore the observations can be used to solve the inverse problem: the power and the form of the arriving flux of nonradiative energy are estimated from physical parameters of observed chromospheres. Parameters of stellar chromospheres were first determined in the 1960s, and above we enlisted the estimates obtained for the electron density, electron temperature, and the radiation intensities in different lines.

In the first chromospheric models of K5–M3 dwarfs constructed before the quantitative values of the above parameters were obtained, Kandel (1967) considered the stability of such structures on the “mass–temperature” plane, where the temperature parameter was the value at the hydrogen temperature plateau and the mass parameter was the column density above the plateau. (Under thermodynamic analysis of the solar chromosphere Athay and Thomas (1956) found that there are at least two plateaux in the temperature distribution with height as a consequence of the presence of such “cooling” temperature stabilizers as neutral atoms of hydrogen and helium ions in the medium.) On the plane, Kandel calculated the net of hydrostatically equilibrium plane-parallel structures with microturbulent velocity and calculated for each model the expected profiles of the CaII H and K emission lines with partial allowance for the non-LTE effects. Having compared the resulting values with the observations, he restricted the region where stellar chromospheres could exist, but excluded too hot and too dense models, in which hydrogen emission appeared: within the accepted scheme the correct calculation of the model was impossible. This fact reduced the significance of his study as an insight into UV Cet-type stars.

**1.3.2.1. Semiempirical Homogeneous Models.** The lack of a strict chromospheric theory favored the wide spread of semiempirical models based almost exclusively on the observed characteristics of continuous and linear stellar spectra. Semiempirical models are calculated by selecting such a temperature distribution with height at which the calculated continuum and line profiles best fit those observed. The distribution of the turbulent velocity with height is selected from the widths of the lines formed at different heights, while the change of density is taken in agreement with the hydrostatic equilibrium condition. Since it is assumed that the chromosphere parameters are independent of the considered region of the stellar surface, in this sense the models can be called homogeneous or one-component models. Since the mechanism for heating of chromospheres is not completely clear, the calculations of semiempirical chromospheric models disregard the energy balance.

For the first time, such calculations for the solar atmosphere were performed using the results of extensive spectral observations from the far ultraviolet to the microwave range (Vernazza et al., 1973). For the extensive set of spectral data they managed to find a section of continuous spectrum or a spectral line that are formed at each height in the photosphere and the chromosphere. As a result, they determined the temperature change throughout the atmosphere height with sufficient reliability and found a plateau with a slow increase of the temperature from 6000 to 7000 K predicted by Athay and Thomas (1956) in the range of heights from 1000 to 2000 km.

Somewhat later, a similar method for the calculation of semiempirical models was applied to the chromospheres of active stars, based on the progress achieved in the understanding of physical distinctions between the quiescent solar chromosphere and flocculi. First, it was suggested that the increased luminosity of the chromosphere in active regions is directly due to a steeper temperature gradient, then the idea of such luminosity due to the lower level of the chromosphere in active regions was discussed. In both schemes, chromospheric temperature should be accompanied by higher electron density. But the primary physical distinction of active regions from the quiet Sun is increased magnetic fields, whose closed flux tubes result both in an increase of the temperature gradient and a lowering of the middle level of the chromosphere. Since flare UV Cet stars are magnetic red dwarfs, it is natural that chromospheric models for red dwarfs are constructed reasoning from the models of active regions of the Sun.

Many researchers have calculated so far the chromospheric models of cool stars. Of course, due to the absence of a sufficient number of high-quality observational data the reliability of stellar models is significantly lower than for solar models. But the calculation scheme used within the concept of radiative transfer in frequencies of optically thick lines is the same. First, one selects or calculates a proper photospheric model corresponding to the spectral type and luminosity of a star under consideration, postulates a certain temperature distribution with height in the atmosphere conjugated to the upper limit of the photosphere, and solves the non-LTE equations of state of hydrogen atoms governing the ionization state of the medium. Then the profiles of emitted fluxes in lines are calculated and compared with the observational results, which allows one to correct the temperature distribution in the course of further iterations. In this case, the absolute density can vary because pressure at the upper boundary of the chromosphere or the mass of matter above the boundary is a free parameter of the model.

Ayres et al. (1976) were among those who started this kind of a study. They calculated the plane-parallel one-component hydrostatic models of upper photospheres and lower chromospheres of two dwarfs similar to the Sun – $\alpha$ Cen A (G2) and $\alpha$ Cen B (K1) – from the profiles of the CaII K lines. Both stars in the upper photosphere demonstrate deviations from the radiative equilibrium compared to analogous phenomena on the Sun. For both stars, the column density at the level of 8000 K, at the bottom of the transition zone, is the same as on the Sun. Then, Kelch (1978) calculated the photospheric and chromospheric models of two active stars – 70 Oph A and $\varepsilon$ Eri – from the measured profiles of the K CaII lines and the fluxes in the MgII h and k lines. His calculations proved that the agreement between the calculations and the observations can be improved, if instead of the earlier assumed constancy of the temperature gradient $dT/d\log m$, where $m$ is the column density, throughout the height of the chromosphere one assumes the same temperature change only in the lower chromosphere and isothermicity in its upper layers, thus approximating the temperature plateau known from the solar chromosphere. In Kelch's models, the temperature gradient is higher by a factor of 1.5–2 than in the chromospheric models of the Sun and inactive stars. Later, using the data on CII, MgII, SiII, and SiIII ultraviolet lines recorded from the IUE satellite in the spectrum of $\varepsilon$ Eri, Simon et al. (1980) performed more detailed calculations of the upper chromosphere of the dwarf. They obtained an identical temperature change below the temperature plateau, while the plateau was somewhat cooler, and their model was generally closer to the model of bright chromospheric points on the Sun.

From the high-dispersion profiles of the CaII K lines Kelch et al. (1979) constructed the models of photospheres and lower chromospheres of eight F0–M0 main sequence stars, including the pairs of stars of identical spectral types, but with a different level of chromospheric activity. The models proved that active stars displayed nonradiative heating already in the upper layers of the photosphere, which resembles the situation in solar faculae. Such stars have a greater temperature gradient in the lower chromosphere and a hotter and deeper temperature minimum as compared to inactive stars of the same spectral types. Thus, the best temperature gradient $dT/d\log m$ for the lower chromosphere model of the flare star EQ Vir is 1800 K, while on the quiet Sun this value is 900 K; the model for the inactive star 61 Cyg B yielded a close value, whereas in the active solar regions it is 1600 K. As nonradiative heating increases, the distance between the CaII K emission peaks – $\Delta\lambda_{K2}$ – decreases and the emission width at the base – $\Delta\lambda_{K1}$ – increases. According to the observations, the models yield the $H_\alpha$ absorption line in the spectrum of the dM0 star 61 Cyg B and the $H_\alpha$ emission line in the spectrum of the flare star EQ Vir.

From the CaII K line profiles Giampapa et al. (1982a) constructed the chromospheric models of emission dwarfs EQ Vir, Gl 616.2, and YZ CMi and nonemission M stars Gl 393 and Gl 411. The models were calculated for a plane-parallel configuration with hydrostatic equilibrium and constant temperature gradients $dT/d\log m$ from the temperature minimum to 6000 K and from 6000 K to 9000 K at turbulent velocities from 1 to 2 km/s. Within the calculated "calcium" chromospheric models that represented well the CaII K line they managed to obtain the $H_\alpha$ emission line for the dMe dwarf YZ CMi and $H_\alpha$ absorption line for the dM star Gl 411 that belonged to the halo population, but failed to match the measured fluxes in the MgII h and k lines. To eliminate the contradiction, they assumed that there were different filling factors of chromospheres for calcium and magnesium flocculi. Since they are formed at somewhat different heights, their expected relation to the divergent magnetic field flux tubes can produce this effect. From the comparison of absolute fluxes in the CaII and MgII lines of the stars with a different level of chromospheric activity they concluded that the average filling factor in calcium lines was 0.13, while in magnesium lines it was 0.31.

According to Cram and Mullan (1979), in the calculations of stellar chromospheres the temperature change with height was set by the "joint" scheme: two sections of constant values of the gradients $dT/d\log m$. This method of specification of the chromospheric model artificially reduced the number of free parameters. But the ideology of the modeling method was the same: the model was chosen based on the criterion of the agreement between calculated and observed spectral characteristics.

To avoid laborious calculations within the models of stellar chromospheres through the approach of Ayres et al. (1976), Cram and Giampapa (1987) elaborated a simplified theory of the formation of hydrogen and CaII lines in the atmospheres of cool dwarfs. The simplifications consist in the replacement of the temperature-inhomogeneous chromosphere by an isothermal one and in the appropriate reduction of the calculations of the emitted fluxes in optically thick lines. Using this model, they calculated the expected values of the $F_{HK}$ fluxes and $H_\alpha$ equivalent widths for different values of column density. The nonlinear dependence of $W_{H\alpha}$ and the monotonic dependence of $F_{HK}$ on column density led to a U-like curve on the plane ($F_{HK}$, $W_{H\alpha}$) for the broad range of column-density values, the curve being dependent on the effective temperature of a star and the temperature of its chromosphere. The $H_\alpha$ equivalent width measured by Stauffer and Hartmann (1986) did not contradict these theoretical expectations, but the comparison of $W_{H\alpha}$ and $F_K$ in a larger sample of M stars revealed considerable dispersion of the points, which can be associated with different levels of formation of the lines in the chromosphere or with different surface filling factors of the emission regions in these lines (Giampapa et al., 1989). Thus, the strong $H_\alpha$ absorption in the spectra of dM stars is a chromospheric indicator that is probably even more sensitive than CaII emission, and M dwarfs without a chromosphere should have weak $H_\alpha$ absorption and no CaII emission. To find such objects, Fleming and Giampapa (1989) performed spectral observations with a resolution of 0.11 Å of the group of late M stars to $M_V = 17.8^m$ at the multimirror 4.5-meter telescope. For VB 8, the weakest emission star in the sample, the ratio $R_K$ is equal to $4 \cdot 10^{-6}$, which is lower than for typical dMe stars and even for the quiet Sun and is equal to the values for M dwarfs with $H_\alpha$ absorption. For the star LHS 2, which is absolutely brighter by $2.4^m$, the ratio $R_K$ is three times lower and the $H_\alpha$ line is not found. Discovery of stars with weak $H_\alpha$ absorption and without calcium emission would be important for the whole concept of magnetism and activity of M dwarfs.

Jordan et al. (1987) proposed a principally new calculation method for the models of stellar atmospheres, which was tested before on the Sun, and applied it to the ultraviolet observations of five G0–K2 dwarfs described by Ayres et al. (1983a). As opposed to the above models based on the calculations within the theory of transfer of the profiles of optically thick lines, this method is based on the analysis of optically thin lines excited by electron collisions. The column emission measures are calculated from the surface fluxes in the lines with different formation temperatures — MgII, SiII, CII, SiIII], SiIV, CIV, and NV. The EM change is constructed within $10^4$–$3 \cdot 10^5$ K. The application of X-ray data allows this change to be expanded to coronal temperatures. Then, from the density-sensitive ratios of the intensities of the emission lines one estimates the absolute electron density at the level of formation of the lines and constructs a complete hydrostatically equilibrium model of stellar atmosphere. Jordan et al. constructed such models covering the upper chromospheres and coronae for the five dwarfs: they encompassed the range of models from the quiet Sun to well-developed active solar regions. The estimates of the column density and pressure in the chromospheres were much higher than those in the earlier chromospheric models of these stars based on the CaII K line profile.

Mathioudakis et al. (1991) performed ultraviolet observations of the star Gl 182 (= V 1005 Ori) with an abnormally high abundance of lithium. From the ratio of the intensities of the CIII $\lambda$ 1176 Å and $\lambda$ 1908 Å and SiIV $\lambda$ 1396 Å lines that are sensitive to the electron density of radiating matter they estimated density in the stellar atmosphere at the level of $\log T = 4.8$ as $7 \cdot 10^9$ cm$^{-3}$. By analyzing differential emission measures they estimated the total radiative losses of the quiescent atmosphere of the star within the range of $4.3 < \log T < 5.4$ as $3 \cdot 10^6$ erg/(cm$^2 \cdot$ s). Then, Byrne and Lanzafame (1994) studied the HeI line $\lambda$ 10830 Å in the spectrum of this star with a resolution of 7000. To interpret the measured equivalent width of 0.17 Å, they constructed a grid of models with nonLTE radiative transfer for plasma from hydrogen, helium, and silicon. The dependence of $W_{\lambda 10830}$ on pressure resembles the situation with the $H_\lambda$ line: first, as pressure increases, so does the equivalent width of the helium absorption line, which then turns into an emission line. Later, these calculations were supported by the observations of Byrne et al. (1998). In the range of pressures acceptable for the observed fluxes in the Gl 182 $H_\alpha$ and SiII lines they determined an optimal pressure of 0.22 dyn/cm$^2$ and the temperature plateau as 8450 K. The calculated model agrees with the results of Mathioudakis et al. (1991) for CIII] lines.

From IUE observations of ultraviolet lines in the spectra of three inactive G–K dwarfs $\tau$ Cet, $\delta$ Pav, and 61 Cyg A Fernandez-Figueroa et al. (1983) calculated the emission-measure distribution with temperature and constructed models of the transition zones assuming their uniform distribution over the stellar surface. The total radiative losses in the atmosphere of Gl 380 estimated by Byrne and Doyle (1990)) within the temperature range of $4.3 < \log T < 5.4$ using the emission-measure distribution vs. temperature practically coincide with the appropriate value for the quiet Sun. From the ratio of the intensity of the intercombination lines CIII $\lambda$ 1908 Å and SiIII $\lambda$ 1892 Å in the spectrum of the star they estimated its electron density as $\log n_e \sim 9.9$ at a level of $\log T = 4.7$, which is close to the solar value and is much lower than that on dMe stars.

Thatcher et al. (1991) formulated the problem of constructing the models of stellar atmospheres in the context of the solar atmosphere model by Vernazza et al. (1981), that is, using a set of different lines formed at different levels of the atmosphere, as opposed to the pure "calcium" models of Kelch (1978), Kelch et al. (1979), Linsky et al. (1979b), and Giampapa et al. (1982a) and pure "hydrogen" models by Cram and Mullan (1979, 1985). Thatcher et al. constructed the model of the lower chromosphere of the K2 dwarf $\varepsilon$ Eri from the K line profiles and two components of CaII IR triplet, Na D lines, and $H_\alpha$ and $H_\beta$. The model was constructed using the solution of nonLTE equations of statistical equilibrium and equations of radiative transfer under hydrostatic equilibrium. Within the calculated model Na D lines were an important diagnostics for the upper photosphere, while the CaII IR triplet provided the localization of the temperature minimum. The depth of $H_\alpha$ and $H_\beta$ appeared to be sensitive to the gradients of the transition zone and to the pressure in its base. On the whole, the chromospheric model of $\varepsilon$ Eri developed by Thatcher et al. yields a lower temperature gradient $dT/d\log m$ and a smaller extension in depth of the temperature plateau than the models constructed by Kelch (1978) and Simon et al. (1980). However, the temperature at the plateau in the former is intermediate between the values provided by the latter models. The model of Thatcher et al. yields a slightly increased level of the temperature minimum and the base of the transition zone localized at the level of log$m$ = −4.54. To obtain a more complete solution of the problem, Thatcher and Robinson (1993) observed the spectra of early K stars with a resolution of 55000. Each star was spectrographed in one or two successive nights in the regions of the Na D lines, CaI $\lambda$ 4227 Å, green triplet and MgI $\lambda$ 4571 Å, CaII H, K, and IR triplet lines, $H_\alpha$ and $H_\beta$. As indicators of the chromospheric activity at different levels of stellar atmospheres they considered integral fluxes in the calcium and in $H_\alpha$ lines calculated as differences between the profiles of these lines for active and inactive stars, equivalent widths of $H_\alpha$ and $H_\beta$ lines, and the depths of their cores. The analysis of the obtained data proved that the CaI and NaI lines were suitable for modeling the atmospheres of inactive and moderately active stars. The MgI $\lambda$ 4571 Å line is important for the region of temperature minimum. The model should present UV and IR lines of ionized calcium in parallel. They should comply with the depths and fluxes in the Balmer lines, whose central cores are filled in due to the high pressure in the transition zone, while the emission wings are due to the high-temperature plateau. The Balmer decrement can serve as an activity indicator. But the equivalent widths of the Balmer lines are poor activity indicators.

For the extremely weak chromospheric calcium emission Doyle et al. (1994) constructed the models of late dwarf atmospheres without $H_\alpha$ emission. Their calculations proved that a very weak flux, of about $10^3$ erg/(cm$^2\cdot$ s), could leave the atmosphere with a temperature minimum of about 2600 K. Later, Doyle et al. (1998), for the M dwarf Gl 105 B with very low surface flux $F_{HK} \sim 6 \cdot 10^3$ erg/(cm$^2\cdot$ s) and low luminosity log$L_X$ < 26.1, calculated the chromospheric model with a steep temperature rise after the temperature minimum at 2650 K to the thin chromosphere and with total chromospheric radiative losses at the level of $10^5$ erg/(cm$^2\cdot$ s).

Houdebine and Doyle (1994a) thoroughly analyzed the effect of different parameters of stellar chromospheres of active M dwarfs on the hydrogen emission spectrum. They found that to reproduce the observed ratios of $Ly_\alpha/H_\alpha$ fluxes, $H_\alpha$ and $H_\beta$ line profiles with the self-absorption, and the widths of these lines in hydrostatic models with the constant temperature gradient $dT/d$log$m$, this chromospheric gradient should be rather high and expand up to 8200 K, whereas the transition zone should be very thin but of considerable column density, have high pressure and be placed at a level of log$m \sim -3$. The break in the temperature change separates the formation regions of the Lyman and other hydrogen series. The turbulence and rotation have a slight influence on the line profiles, though the rotation can fill in the double-peaked structure of $H_\alpha$. The profiles depend weakly on the effective temperature. These general results were applied to the detailed analysis of the hydrogen emission of one of the most active dMe stars, AU Mic. Houdebine and Doyle proved that, in addition to the observed $H_\alpha$ and $H_\beta$ profiles and the flux ratio $F_{Ly\alpha}/F_{H\alpha}$, the model reproduced the Balmer decrement, the Balmer jump, and the width of the $Ly_\alpha$ line as well, but overestimated the equivalent width of $H_\alpha$ by a factor of 3. This contradiction can be removed by the assumption on the inhomogeneity of the chromosphere with a filling factor of 0.3. Analogous considerations of helium emission lead to the idea of inhomogeneity of the transition zone. The resulting structure of the stellar chromosphere approaches the existing models of solar flares but requires a strong and continuous flux of nonradiative energy. In this case, the Lyman radiation ensures radiative losses of the transition zone, and the Balmer radiation — those of the chromosphere, with the Lyman series heating the upper chromosphere and the Balmer one — the lower chromosphere and the region of the temperature minimum.

Continuing the studies of the atmospheres of M dwarfs with minimum and maximum chromospheres, Houdebine et al. (1995) performed detailed calculations of a number of intermediate models to determine the expected spectral features between the extremities over the whole activity range. In other words, they analyzed the general scheme of variation of the $H_\alpha$ line profile in the spectra of M dwarfs with the chromospheres of varying activity outlined by Cram and Mullan (1979) for the following typical situations:

– high activity with strong emission and weak self-absorption;
– intermediate activity with emission wings and an absorption core;
– intermediate activity with strong and wide absorption;
– low activity with weak and narrow absorption;
– zero activity with indistinguishable $H_\alpha$ line – zero $H_\alpha$ stars.

The models were calculated on the following assumptions: identical photosphere up to the temperature minimum corresponding to the M4 dwarf; temperature minima of 3000 and 2660 K, hydrostatically equilibrium atmosphere, constant temperature gradient $dT/d\log m$ from the temperature minimum to 8200 K and a transition zone from this temperature to $3 \cdot 10^5$ K. The column density above the chromosphere varied from $\log m_o = -3$ to the level of a zero $H_\alpha$ star. The calculations proved that the lines of all hydrogen series in the most active stars should demonstrate emission. As the activity level decreases, the Brackett lines are the first to convert into absorption, then this occurs with the Paschen and Balmer lines. After these lines disappear, Lyman absorption lines should remain in the spectrum. The transition from the Balmer emission to absorption should occur when $m_0$ decreases by only one order of magnitude. The values of $m_0$, at which qualitative changes occur in the hydrogen spectrum, strongly depend on the value of the temperature minimum. Having compared the calculated line profiles and the equivalent widths obtained within the above model atmospheres with the observations, Houdebine et al. (1995) plotted the objects with different chromospheres on the "$\log m_0$–electron density" plane (see Fig. 19).

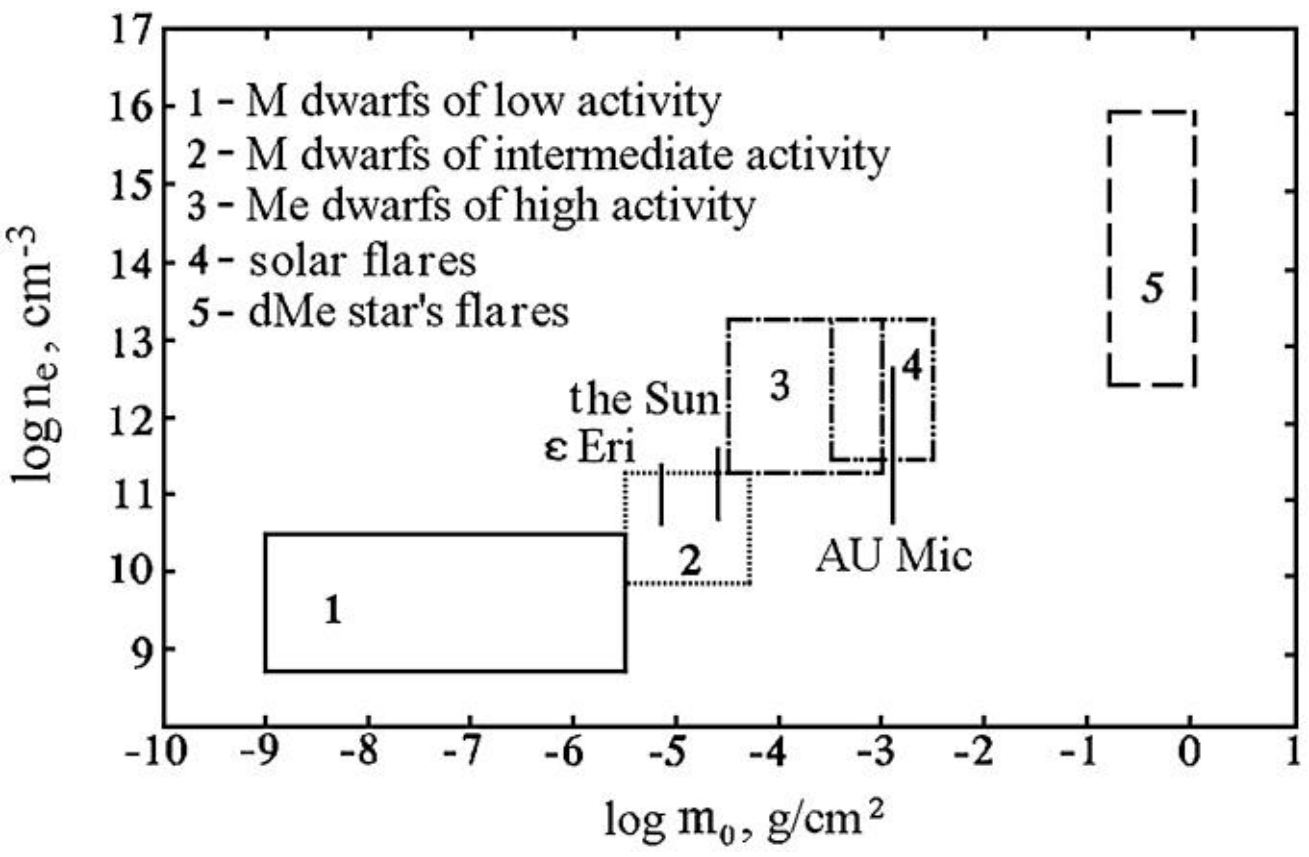

Fig. 19. Localization of chromospheric radiation sources and flares on the electron density–$\log m_0$ plane (Houdebine et al., 1995)

Then, Houdebine et al. (1996) considered the continuous radiation of stellar chromospheres in the broad range of chromospheric activity from the basal to a very high level in M dwarfs with the same temperature minima. The calculations showed that this continuous radiation depended first on pressure in the transition zone and the value of the temperature minimum, and therefore could be used for the diagnostics of the stellar chromosphere throughout its depth. Houdebine et al. found that the ultraviolet continuum was the main channel of chromospheric radiative losses, in particular, the losses for $H_\alpha$ were, as a rule, only a small portion of the total losses for the hydrogen lines and continua. Thus, the earlier observed saturation in MgII and some other lines does not mean saturation of magnetic heating of the stellar atmosphere. In the region $\lambda$ < 915 Å the Lyman continuum dominates, while in the range of 915 Å < $\lambda$ < 3650 Å so does the radiation of the transition zone and the temperature minimum, and in the range 3650 Å < $\lambda$ < 40 μm the temperature minimum dominates. The fluxes calculated by Houdebine et al. in the UBVRIJKL bands show the expected excess in the U band and probably in the B band. The stars with strong $H_\alpha$ emission should have an excess of U−B color rather than in B−V, which corresponds to the observations. Finally, the stars with $H_\alpha$ emission should be systematically brighter than the stars of the same spectral types with $H_\alpha$ absorption, which Houdebine et al. considered as evidence of the youth of dMe stars and their relation to the T Tau-type stars.

Mauas and Falchi (1994) developed a semiempirical atmospheric model of the quiescent flare star AD Leo, selecting the model parameters in such a way as to present simultaneously a large number of observed values: stellar continuum from UBVR to 25 μm, the Balmer line profiles, Na D doublet, Mg b triplet, CaI $\lambda$ 4227 Å, CaII K line, and the component of the IR triplet of this ion $\lambda$ 8499 Å, fluxes in $Ly_\alpha$ and MgII h and k. Eight elements were considered in the calculations under the assumption of solar abundance and turbulent velocities of 1–2 km/s. In the resulting model, the flux in $Ly_\alpha$ was three times higher than the observed one, the observed and calculated fluxes in the magnesium doublet coincided, the flux in the CaII K line was 5 times lower than the expected value, fluxes in the Balmer lines differed from the measured fluxes by −34, +13, +17, and +27% for $H_\alpha$, $H_\beta$, $H_\gamma$, and $H_\delta$, respectively. Some of these discrepancies could be attributed to different levels of stellar activity at the moments of nonsimultaneous observations in different parts of the spectrum.

Then, Mauas et al. (1997) carried out calculations of semiempirical chromospheric models of two extremely low activity stars – Gl 588 and Gl 628 – using broadband photometry, the hydrogen, calcium, and Na line profiles, and fluxes in the MgII lines. Their models satisfactorily presented the continuum, profiles, and equivalent widths of the CaII K lines, $H_\alpha$, $H_\beta$, and Na D, as well as fluxes in the magnesium doublet. The models of these M stars with a $H_\alpha$ absorption line were compared with the earlier constructed chromospheric model of AD Leo with ten-times stronger fluxes in MgII and CaII and a hundred-times stronger X-ray flux. The comparison proved that the chromospheric temperature was higher by 2000 K at a height of $\log m_0 = -4$ in the upper chromosphere of AD Leo, the electron density was higher by an order of magnitude, and the transition zone started at a lower height.

Jevremovic et al. (2000) considered the dependence of the chromospheric models of M dwarfs on microturbulent velocity at a fixed temperature profile. They developed 12 different models with fixed velocities throughout the chromosphere, with velocities equal to a certain portion of the local speed of sound, and with the velocity profiles composed of three linear sections. The main conclusion is that with a growth of the microturbulent velocity the electron density of the medium decreases, the level of the central minimum decreases in the profile of the $H_\alpha$ emission line, while $H_\beta$ turns into absorption, and the NaI D line profiles narrow. They estimated the column density for the transition zone and for the temperature minimum of CR Dra star, $\log m_0 = -4.3$ and $-1$, respectively. Then, Jevremovic et al. (2001) constructed a grid of hydrostatic chromospheric models for AD Leo varying the depth of the temperature minimum and the bases of the transition zone given $dT/d\log m$=const. The observations of the $H_\alpha$ line profile in the spectrum of the quiescent star were best presented at the temperature minimum of $\log m_0 = -1$ and at the base of the transition zone of $\log m = -4$.

**1.3.2.2. Surface Inhomogeneities at the Chromospheric Level.** The chromospheres of active stars, as well as the solar chromosphere, should have considerable surface inhomogeneities, which are confirmed by observations.

In the late 1960s, Krzeminski and Kraft (1967) and Krzeminski (1969) published the results of photometric observations of nine red dwarfs. Spottedness was found only on dMe but not on dM stars. Based on this result and subsequent observations, Martins (1975) advanced the concept of "activity centers" on stars by analogy with active regions or larger activity complexes on the Sun. The conclusion on the spottedness of dMe stars was repeatedly confirmed in later observations (Bopp and Espenak, 1977; Bopp et al., 1981). All the observations suggested the inhomogeneity of stellar chromospheres, but parallel monitoring provided direct evidence.

In 1965, when the spottedness of BY Dra was the greatest, the CaII emission equivalent widths were several times greater than those during the following years (Gershberg and Shakhovskaya, 1974).

During spectral monitoring of seven flare stars by the Struve telescope (McDonald Observatory) Bopp (1974b,c) revealed considerable weakening of the Balmer and calcium emissions over an interval of about a day. Since the parallel photometric monitoring revealed no flares and the axial rotation period of the stars was several days, Bopp associated these emission fluctuations with the inhomogeneities of the stellar chromospheres. From the observations of the eclipse YY Gem system Bopp (1974d) and Ferland and Bopp (1976) estimated the longitudinal extension of the emission regions emitting Balmer lines as 20–60°. Later, based on the spectra of YY Gem obtained out of flares, Kodaira and Ichimura (1980) assumed that there was a 2-sector structure of the $H_\beta$ emission regions on the main component of the system and a 4-sector structure on the secondary component, for the latter the contrast of active regions at active longitudes being much greater than for the main component. During the following years the sectorial structure of the main component demonstrated changes on the time scale of several months (Kodaira, 1986).

The anticorrelation of the equivalent width of the $H_\alpha$ emission and the brightness of CC Eri were found by Busko et al. (1977), who naturally linked it with the chromospheric active region near the dark spot.

The filling factor of the chromosphere by the hydrogen emission was first directly estimated by Grinin (1979) in analyzing physical conditions in active regions of flare stars: 6% on AD Leo and 14% on EQ Peg A.

Specific examples of the activity centers were found by Dorren and Guinan (1982b) during photometric studies of five single dwarfs, for which the variations in the CaII emission with time scales of about a week and several years were found in the course of long-term spectral monitoring. On three stars they found low-amplitude brightness variations in the band close to B, and on two of them (12 Oph and 61 Cyg A) brightness and $H_\alpha$ and CaII emission were in anticorrelation, dark photosphere spots and active chromospheric regions typical of the Sun. Similar results were obtained by Chugainov (1983) for $\xi$ Boo AB and HD 1835, Reglero et al. (1986) for the G0V star HD 206860, Barden et al. (1986) for DH Leo, and Strassmeier et al. (1989) for HD 80715.

Radick et al. (1983) carried out photometric observations of 11 F3–K4 dwarfs of those spectrally monitored by Wilson's team and found the anticorrelation of brightness and calcium emission for two or three of them. Later, they studied five fast rotators F8–G2 in the Hyades and found the same effect but with very low variability amplitudes: $0.01^m$ in the total optical brightness and $0.03^m$ in the CaII line flux (Radick et al., 1987).

Kodaira and Ichimura (1982) revealed periodic variations of the intensity of $H_\beta$ emission in the spectrum of the YY Gem system, whose maximum coincided with the brightness minimum due to spottedness. Andersen et al. (1986) found a distinct anticorrelation of chromospheric ultraviolet lines and HeII line $\lambda$ 1640 Å with the brightness of EV Lac with simultaneous absence of rotational modulation of the intensities of transition-zone lines. One possible explanation is the difference in the filling factors at different heights of the stellar atmosphere. Gershberg et al. (1991b) discovered the anticorrelation of the $H_\alpha$ intensity and brightness of this star in one season and the absence of such dependence during the next year, one year later a similar correlation was found by McMillan and Herbst (1991). The same situation for V 775 Her, VY Ari, and LQ Hya was detected by Alekseev and Kozlova (2000, 2001, 2002, 2003a, b). Byrne (1986) traced a distinct rotational modulation of the $H_\alpha$ intensity in the dK5e star HDE 319139, on which one of the strongest hydrogen emissions is observed.

Analysis of fast variations of polarization and calcium emission found during almost simultaneous observations of $\xi$ Boo A and HD 206860 made it possible to suggest at least three active regions on each star (Huovelin et al., 1988).

Hallam and Wolff (1981) carried out independent periodogram analysis of the intensities of one of the three emission lines $Ly_\alpha$, SiII, and MgII in the ultraviolet spectra of six F8-K7 dwarfs: 111 Tau, $\alpha$ Cen A, $\alpha$ Cen B, $\varepsilon$ Eri, 61 Cyg A, and 61 Cyg B. For each star up to a dozen spectra were obtained. The periods of close duration and phases found from different lines were identified with axial rotation periods. Thus-found periods from 12 to 47 days are in good agreement with the appropriate values estimated from the calcium emission. On some stars during the period one intensity maximum of one emission and two maxima of the other emission (at the opposite longitudes) were recorded. The deepest modulation was found in the $Ly_\alpha$ line, while the weakest in MgII. Later, Hallam et al. (1991) applied an improved method of search for the periodicity in nonuniformly distributed data and found the period: 2.8 days for $\varepsilon$ Eri and 29 days for $\alpha$ Cen A.

After three four-day monitoring series of the young G0 star $\chi^1$ Ori Boesgaard and Simon (1984) found a distinct modulation of the intensity of CIV lines that preserved the phase for about a year, with an axial rotation period of 5.1 days, which was determined from the variations of calcium emission. The behavior of the HeII $\lambda$ 1640 Å emission was analogous, and both emissions retained the phase for about a year. The rotational modulation was found in the lines of all ions, except for MgII. An asymmetric CIV light curve can be interpreted as a result of luminosity of a large active region, which after a 1/4 revolution is followed by the second region, and the total high level of this emission flux may be due to the bright chromospheric network in addition to the active regions. Independent estimates of spottedness of the chromosphere in different lines yielded filling factors from 20 to 50% of the stellar surface.

The intensity of the high-temperature resonance CIV doublet in the spectrum of the G0 dwarf $\chi^1$ Ori over one day, which is equal to 0.22 of an axial rotation period, reduced by a factor of three, while the intensity of the CII $\lambda$ 1355 Å line by 50%, then the lines returned to the initial level (Simon, 1986).

In the spectrum of the single very fast rotating star Gl 890, Byrne and McKay (1990) found the rotational modulation of the intensity of MgII ultraviolet lines, while Young et al. (1990) found successive shifts of the centroid of the $H_\alpha$ line to the blue and red sides in the phase with weakening of the MgII line and positive correlation of the intensities of $H_\alpha$ line with the brightness of the star. But in later observations, Byrne and Mathioudakis (1993) recorded an asymmetric light curve, which was presented by the model with two spots of close size spaced in longitude for 90°. Within this interpretation of the light curve no positive correlation should be observed between the $H_\alpha$ emission and the spots, and Young et al. (1990) actually found the $H_\alpha$ intensity minimum in the vicinity of the broad brightness minimum in the V band.

Newmark et al. (1990) revealed a clear anticorrelation of equivalent widths of $H_\alpha$,$H_\beta$, and CaII H emissions and the brightness of DH Leo. As brightness weakened in the R band by 10% the equivalent widths increased by 30%, which meant that there was a real enhancement of emission near the starspots. But between the emission extremes of CaII H and K and CaII IR triplet there was a phase shift for a half-period.

Byrne et al. (1992a) found a distinct anticorrelation between the brightness of the CC Eri star and the intensity of MgII lines: the amplitude of variations of the brightness of these lines achieved 40% for the amplitude of the stellar brightness in the V band of 5%. The variations of the intensities of other high-temperature lines SiII, CII, and CIV were much smaller, they hardly exceeded random scatter.

Bopp and Ferland (1977) did not find rotational modulation of calcium emission in the BY Dra spectrum. Butler (1996) compared the spottedness and the rotational modulation of the emission lines of flocculi on BY Dra and AU Mic from the observations carried out in the 1980s (Butler et al., 1983, 1987) and established that there was a correlation between the spottedness and the high-excitation emissions on BY Dra, while the frequent flares on AU Mic prevented detection of the correlation. As to the MgII emission, there was no correlation probably because the emission covered almost the whole star, although during the one-week observations in October 1981 the correlation was about to appear (Butler et al., 1984). From the UBV observations in 1974–1980 Contadakis (1997) discovered alternate periods of strong and weak rotational modulation of stellar brightness with the duration of one–two months on BY Dra, in the periods of increased modulations there occurred a higher frequency of flares and enhanced $H_\alpha$ emission.

Stern and Drake (1996) studied three bright BY Dra-type systems in the extreme ultraviolet. On the light curve of FK Aqr over a quarter of a period, which is about a day, they recorded twice as high radiation that could be due to the active region on one of the system components. From the light curve of DH Leo encompassing approximately eight periods they suspected rotational modulation of brightness. For BF Lyn they noticed brightness variations of up to 50%, but observations of the system lasted for less than a period, so it was not clear whether the variations were associated with a flare, rotational modulation, or the evolution of the active region.

In 1986, an extensive campaign of comprehensive study of the active G8 star $\xi$ Boo A was undertaken. Magnetometric observations of the $\lambda$ 6173 Å line and spectral observations in the region of the helium $D_3$ line were carried out at the American National Solar Observatory McMath. Multicolor broadband polarimetric observations were performed with the five-channel photopolarimeter in the Crimea. Spectral observations in the ultraviolet and in the CaII H and K lines were carried by the IUE satellite and at the Mount Wilson Observatory, respectively (Saar et al., 1988). Analysis of the obtained data revealed clear synchronous changes in the magnetic flux, the intensity of the UV emission of CIV and CII lines, and CaII emission, which was the first evidence of the relation between the magnetic flux of the photospheric field and the emission of the outer stellar atmosphere. The absorption maximum in the helium $D_3$ line and the maximum of filling-in of the cores of Na D lines were also near the magnetic flux maximum. On the basis of comparison of magnetometric and polarimetric data, a hypothesis on the existence of four longitudinal sectors with increased magnetic fluxes on the star was advanced.

By analogy with the Sun one can expect that monitoring of stars with an asymmetric distribution of active regions on the disk will show a growth and decline of the ultraviolet emission, X-ray radiation, and magnetic flux at certain phases of the rotation period. This prediction was confirmed in the observations of $\varepsilon$ Eri (Saar et al., 1986b) and $\xi$ Boo A (Saar et al., 1986b, 1988).

Frasca et al. (2001) compared photometric and spectral observations of the G0 dwarf HD 206860 and found a distinct anticorrelation of the brightness with the fluxes in the CaII and $H_\alpha$ lines at a coincident axial rotation period of 4.74 days. Lopez-Santiago et al. (2003) obtained similar results in studying many lines of the echelle spectra of PW And.

Korhonen et al. (2010) found a correlation of the $H_\alpha$ emission and photospheric active regions on the fast M rotator EY Dra and, apparently, on V 374 Peg. Biazzo et al. (2007) found an analogous correlation for the G–K dwarfs HD 166, $\varepsilon$ Eri, and $\kappa^1$ Cet but detected no variations in the photosphere and chromosphere of $\chi^1$ Ori.

The list of cases when observations suggested surface inhomogeneity of the stellar chromosphere can be expanded. But long-term and homogenous observations of the fluxes in the CaII H and K lines in the spectra of a large number of stars started by O. Wilson (see Fig. 14) are more representative. These studies have already been referred to in Subsect. 1.3.1.1. and will be discussed in Part 3. Analysis of such observations yielded the estimates of the characteristic lifetime of active regions on the stars as 50 days and the characteristic lifetime of the active complex as about 200 days (Donahue et al., 1997). Application of the gapped wavelet analysis to the long-term observation series for the active stars HD 1835, HD 82885, HD 149661, and HD 190007 made it possible to discover local long-lived active regions that had a tendency, at least three of four stars, to resume in the narrow latitudinal region (Soon et al., 1999).

In the above one-component models, the luminosities of stellar chromospheres were eventually determined from the total mass of matter at the chromospheric temperature. An alternative approach implies the variation of the surface filling factor for active regions (Giampapa, 1980). Above, we mentioned the estimates by Giampapa et al. (1982a) for the average filling factors of chromospheres in the CaII and MgII lines as 0.13 and 0.31, respectively. But analyzing the $H_\alpha$ absorption line in the spectra of dM stars, Giampapa (1985) concluded that it was not sufficient to vary only this value, since the lower limit of the filling factor of dM star surfaces was rather high, at least 0.3. Thus, he advanced the hypothesis: the differences in dM and dMe stars were determined by the structure of magnetic fields rather than by the filling factor. On dMe stars the local fields form closed loops, whereas on dM stars these are open structures of the type of solar coronal holes. Almost at the same time, independently of Giampapa's conclusions, Byrne et al. (1985) formulated a similar idea: magnetic structures uniformly distributed over the surface were formed on the stars with axial rotation periods of more than 10 days, on faster rotating stars there was a considerable concentration of local magnetic fields, which resulted in more pronounced manifestations of activity.

Byrne and Doyle (1990) compared ultraviolet spectra of two dM stars, Gl 380 and Gl 411, and the Gl 900 star with very low emission. The latter is a "marginal dMe star", according to Young et al. (1984), but Doyle and Byrne proposed to replace this term by the notation dM(e). They found that the surface flux in the CIV $\lambda$ 1548/51 Å line of Gl 900 was higher by an order of magnitude than that of dM stars but was lower by a factor of 3–5 than that of dMe dwarfs. Identical intermediate fluxes were recorded in the $Ly_\alpha$, CII $\lambda$ 1335 Å, and SiII $\lambda$ 1817 Å lines. This agrees with the considerations of the dM→ dMe transition as an increase in the surface filling for solar-type active regions, and the weak $H_\alpha$ line of the Gl 900 star is a result of filling-in of the absorption line by emission. However, Turner et al. (1991) made another conclusion on the basis of comparing the intensities of the $H_\alpha$, CaII H, K, and the IR triplet lines of the inactive star Gl 1, Gl 735 active dwarf, and the intermediate-activity star Gl 887. Assuming that on the intermediate-activity star active regions have the same properties as on an active star, the other regions are as on an inactive star, they estimated the relative area of different regions on the intermediate-activity star from one of the above chromospheric lines. But the solution does not suit the other two chromospheric lines. Therefore, they concluded that the observed activity range of M dwarfs was due to the changes in the function of chromospheric heating over the whole star rather than variations of the filling factor for active regions. Later, Giampapa (1992) made a similar conclusion. However, the substantial contribution of microflares to the emission of the Balmer lines of the "quiescent" chromosphere found by Alekseev et al. (2003) makes us turn back to this discussion.

Cerruti-Sola et al. (1992) compared absolute profiles of the MgII h and k lines in the spectra of about 40 F6–K5 dwarfs and in the spectra of 22 different regions on the Sun – quiet regions, the bases of coronal holes, faculae, and flares – obtained with comparable spectral resolution. Local fluxes in these lines on the Sun overlapped the range of fluxes averaged over disks on the stars under consideration and the spectra of stars of different activity levels were in good correspondence with different solar regions. This suggests that different MgII emission levels in the spectra of stars of close spectral types are caused by different surface filling factors for magnetic fields and related MgII emission regions. Cerruti-Sola et al. also found that the data for different solar regions fitted the same relations of fluxes in different lines constructed for stars. The obtained results principally enable calculations of the surface filling factor for active regions, but the fact of immense luminosity dispersion in MgII lines in different solar structures makes this procedure rather formal.

Panagi and Mathioudakis (1993) collected and analyzed spectral observations in the $H_\alpha$, CaII H and K, and MgII h and k regions of about 600 K–M dwarfs. They noted a systematic growth of $W_{H\alpha}^{emission}$ toward late spectral types and a growth of $W_{H\alpha}^{absorption}$ toward earlier types, a systematic decrease of the luminosities $L_{CaII}$ and $L_{MgII}$ toward later dwarfs and similar by a less pronounced change of the surface fluxes of $F_{CaII}$ and $F_{MgII}$. The close correlation of the values of $L_{CaII}$ and $L_{MgII}$ evidences the formation of these emissions in the overlapping atmospheric regions with common surface inhomogeneities. Such a close correlation of luminosities in the $Ly_\alpha$ line and in the X-ray region suggests the closeness of the structures of surface inhomogeneities from the upper chromosphere to corona, whereas the comparison of $W_{H\alpha}^{emission}$ with $L_{CaII}$ reveals great dispersion, which Panagi and Mathioudakis associated with the essentially different structures of surface inhomogeneities at the appropriate levels of stellar atmospheres. In comparing the values of $W_{H\alpha}$ and fluxes in CaII, they found the U-like dependence $W_{H\alpha}(\log F_{CaII})$ theoretically precalculated by Cram and Giampapa (1987).

Andretta and Giampapa (1995) developed the method of estimating the filling factor for chromospheric active regions of F–G stars based on the observations of the helium $D_3$ lines and $\lambda$ 10830 Å and calculations of their expected intensities in stellar active regions. These lines appear in absorption in active solar regions, but are invisible or very weak in the quiescent photosphere and in sunspots. Therefore, there should be magnetic regions beyond starspots. To calculate possible maximum equivalent widths of the helium absorption lines they used the method applied earlier by Cram and Mullan (1979) to calculate the $H_\alpha$ equivalent widths on active dwarfs. They took the temperature structure of the solar quiescent atmosphere, following Vernazza et al. (1981), that was then shifted to deeper and denser atmosphere layers, for each shift they calculated the plane-parallel hydrostatic chromospheric model with regard to the nonLTE ionization of hydrogen, the contribution to the electron density of metals and molecules and helium ionization by the isothermal corona with exponentially decreasing pressure, the corona being in hydrostatic equilibrium with the chromosphere. The temperature structure was combined with the photospheric models of F and K stars. The calculations showed that the helium triplet lines arose under the scattering of the photospheric radiation at two levels: in the upper chromosphere at $T_e \sim 7000-8000$ K and in the region of the temperature plateau at $T_e \sim 20000$ K. In the upper chromosphere, the population of the lower levels of these lines occurs due to recombinations and a cascade transition after the ionization by coronal ionization. At the temperature plateau, this occurs due to collisional excitations, so that the role of this plateau starts dominating as the chromosphere moves deep inside to higher densities. The prevailing collisional excitation, as in the case with the $H_\alpha$ line, results in the appearance of the emission line $\lambda$ 10830 Å. Calculations showed that for F–G stars the maximum equivalent width of this absorption line was close to 0.45 Å. For the $D_3$ line such a situation can occur at the much higher density typical of flares. The comparison of the $H_\alpha$ and $D_3$ lines in solar structures of different activity level confirmed the validity of the method. Thus-found filling factors from the $D_3$ lines are within 0.03–0.32 for the F–G stars under consideration, from the $\lambda$ 10830 Å line within 0.02–0.62. For the five G dwarfs, for which the estimates of the photosphere filling factor for strong magnetic fields are available, the chromosphere filling factors from the $D_3$ line are systematically much lower than the "magnetic" filling factors, but the latter are rather close to the chromosphere filling factor from the $\lambda$ 10830 Å line. Houdebine et al. (2009b) extensively considered relations of the sodium and helium D lines, and CaII lines in spectra of M1 stars.

* * *

Let us briefly sum up the results of studying stellar chromospheres.

The concept of heating of the atmospheres of medium- and low-mass stars by nonradiative energy fluxes originally developed for the Sun did not contain a concrete definition of the physical nature of these fluxes, but it explained the main features of stellar chromospheres: strong emission of Balmer, calcium, magnesium, and other numerous UV lines, a broad range of absolute luminosities of the lines, extremely weak and the strongest stellar chromospheres, the correlation of their luminosities with masses and stellar rotation, systematic changes of the content of chromospheric emission along the spectral sequence. Accumulated observations evidence the essential surface inhomogeneity of these components of stellar atmospheres. The known structure of the solar chromosphere provides heuristic considerations on the possible character of such inhomogeneities of stellar chromospheres, but no satisfactory solution of this inverse problem, an estimate of the parameters of surface inhomogeneities from the observed integral fluxes, has been obtained yet. Consideration of different starting assumptions – the model of a plane mosaic of active and inactive elements at different chromospheric levels due to the divergence of magnetic flux tubes, the considerations on the broad brightness ranges of active regions and the surface filling factors on them – yielded no result. Probably, the reason is the lack of insight into the structure of the magnetic field that eventually governs the structure of the stellar chromosphere. Available homogeneous models of stellar chromospheres provide only the first approximation of these structures, and it is obvious that the parameters of the radiating medium will differ in the homogeneous and inhomogeneous models. This statement is illustrated by Fig. 20, which presents some results of the construction of semiempirical models of quiescent chromospheres. The solid curves show the models developed by Baranovskii et al. (2001a) based on the profiles and equivalent widths of the $H_\alpha$, $H_\beta$, and $H_\gamma$ lines in the spectrum of EV Lac observed with the 2.6-meter Shajn telescope in the Crimea. The calculations were carried out assuming that the chromosphere was either homogeneous or the active regions radiating the emission lines occupied 1/2 and 1/3 of the stellar surface, the three variants are marked by 1, 2, and 3, respectively. The second panel shows the temperature structures of the models, while the bottom panel illustrates the change of the electron density. In the calculations, the "joint" scheme of the temperature structure was not used, and the model of a homogeneous chromosphere has a distinct temperature plateau, where the main part of radiation in the Balmer lines is formed. Comparison of curves 1–3 in Fig. 20 shows that the structures of the calculated models of active regions differ from the homogeneous chromosphere model by the higher temperature of matter on the plateau, less extended isothermal region, and an earlier smooth rise to the high-temperature region. The $H_{\alpha}$ and $H_{\beta}$ emissions start forming practically at the same depths as in the model of a homogeneous chromosphere, but in the active region models they go higher and their optical thicknesses are 3–4 times greater, electron densities are increased in them by a factor of 2 to 4.

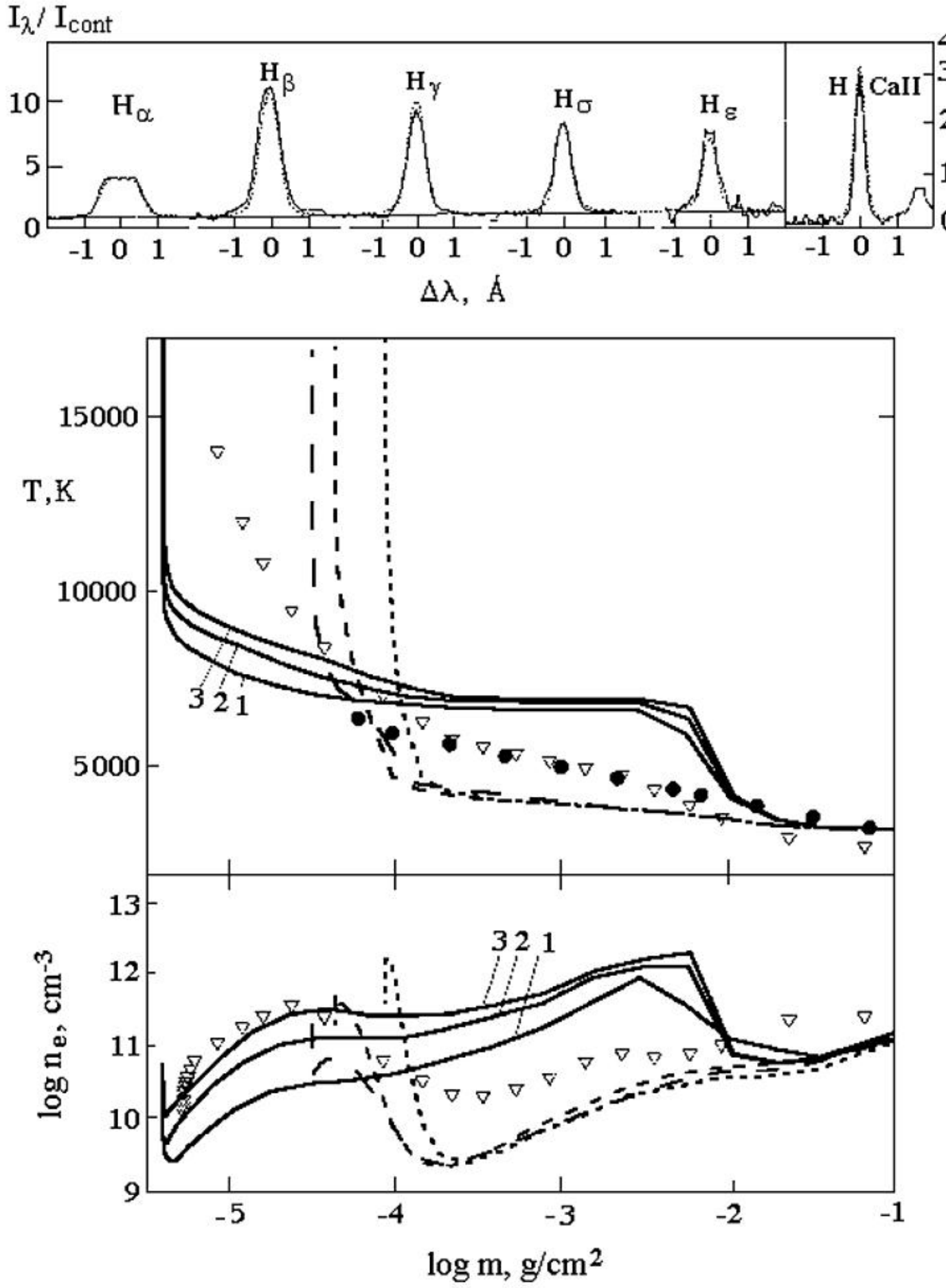

Fig. 20. *Top panel*: the observed emission line profiles in the spectrum of the summed radiation of active regions and three EV Lac microflares (Alekseev et al., 2003): dotted lines denote observations, solid lines — calculations. *Middle and bottom panels*: semiempirical models of active regions of the chromosphere of EV Lac: solid, dashed, and dotted curves — see the text; triangles illustrate the model of the quiescent chromosphere of AD Leo by Mauas and Falchi (1994), circles correspond to the model of the lower chromosphere of YZ CMi based on the calculations of Giampapa et al. (1982a)

Soon, these results were essentially supplemented by Alekseev et al. (2003) who analyzed the radiation from the chromosphere of EV Lac in 1994 using much better observational data — high-resolution echelle spectrograms from the Nordic Optical Telescope. The analysis showed that to present simultaneously recorded line profiles from $H_\alpha$ to $H_\varepsilon$ and H CaII, one should take into account not only the spatial filling factor of the chromosphere but, which is more important, the physical inhomogeneity of the radiation source. The discovered broad wings of emission lines show that in a quiescent star a considerable fraction of the line chromospheric emission is due to microflares whose amplitudes in broad photometric bands are lower than the registration threshold of individual bursts. Consideration of two 15-min intervals before and after the fast flare of 30 August 1994 (23:19 UT) with $\Delta U = 0.85^m$ showed that at each interval the recorded radiation could be presented as a sum of radiations of active regions and three different microflares, while the latter provided the main contribution to the radiation in the hydrogen lines. A presentation of the EV Lac spectrum out of a flare is shown in the top panel of Fig. 20. In the diagrams of temperature and density structures dashed lines with long and short dashes show the models of active regions constructed after subtracting the two above observed spectra of the appropriate contributions of microflares. The closeness of these curves to each other evidences sufficient correctness of the procedure of accounting of microflares. Dotted curves in Fig. 20 present the model of the active region on EV Lac during the flare of 30 August 1994. On the whole, the models of "pure" active regions, as compared to the models 1–3, display noticeably lower electron temperatures of active regions and the regions themselves go down to deeper atmospheric layers. In Fig. 20, triangles illustrate the model of the quiescent chromosphere of AD Leo by Mauas and Falchi (1994), circles correspond to the model of the lower chromosphere of YZ CMi based on the calculations by Giampapa et al. (1982a). As it follows from the plots, on the planes ($T_e$, log*m*) and (log$n_e$, log*m*) these models in the temperature range lower 10 000 K are localized inside the band occupied by the described models. One may associate a conspicuous closeness of the homogeneous model of the chromosphere of AD Leo to the models of active regions of EV Lac with the expected smaller contribution of microflares into the spectrum of AD Leo because the frequency of flares on this star is 2–3 times lower than that on EV Lac.

Vieytes et al. (2005, 2009) constructed the chromospheric models of the Sun as a star and nine solar-type stars with different activity levels with the fitting of CaII K and $H_\beta$ lines, including the asymmetry of profiles due to the microscopic motions, while the total radiative fluxes were estimated for the maintenance of the presented structures. They found that variations in activity had impact on the region of the temperature minimum for low-mass stars and on the chromospheric plateau — for others. Purely radiative losses were linearly associated with $S_{\rm CaII}$.

From the UV spectra of $\varepsilon$ Eri derived with HST/STIS and FUSE, Sim and Jordan (2005) redetermined the emission measure distribution and constructed a semiempirical model of the chromosphere and transition zone of the star. Calculating the ionization equilibrium, they took into account the dielectron recombination of C, N, O, and Si, and constraints on the FIP effect were obtained.

Using the echelle spectrograph UVES/VLT, Fuhrmeister et al. (2005) derived spectra of 323 M3.5–L dwarfs with $H_\alpha$ emission within the ranges 3030–3880 Å and 4580–6680 Å. They selected out of them five stars with effective temperatures from 3200 to 2500 K — AD Leo, CN Leo, YZ CMi, LHS 3003 and DX Cnc — and carried out calculations of semiempirical models of quiescent chromospheres within the PHOENIX program. The calculations suggested a coverage of the entire star by the chromosphere, the hydrostatic and ionization equilibrium, and the full frequency redistribution at the reradiation of quanta of $Ly_\alpha$ and CaII

H and K; the nonLTE effects of the first stages of ionization of H, He, C, N, O, Fe, Ti, Na, and Mg were taken into account. Fuhrmeister et al. found that the account of nonLTE effects of CNO had effect on the formation of hydrogen and sodium lines, whereas such effects of less abundant elements, Co, Ni, and Ti, affected only lines of these elements. For AD Leo, YZ CMi, and CN Leo, the model chromospheres were constructed by fitting the Balmer lines above $H_9$, NaI D line and many lines of FeI and MgI in the range from 3650 to 3870 Å, for LHS 3003 and DX Cnc — by fitting the $H_\alpha$, $H_\beta$, NaI D, and He I $\lambda$ 5875 Å lines. Due to using the FeI lines, the calculated models first allowed one to describe the transition from absorption lines with emission peaks in the center to the purely emission lines. But these homogeneous models, contrary to inhomogeneous models by Alekseev et al. (2003), could not represent the broad wings of hydrogen lines. For LHS 3003 and DX Cnc, Fuhrmeister et al. considered the influence of dust on the outward fluxes of emission lines. In Appendix to the work, a list of emission lines is given in two specified above wavelength ranges in spectra of 21 M and L dwarfs, the NaI D emission is traced up to L3.

Developing the ideology of Crimean astrophysicists who in the 1980-90s performed a dozen international cooperative observations of EV Lac, Hawley and her colleagues throughout two days in September 2001 carried out an observational campaign of this star from the X-rays to the radio range with the aim of ascertaining the structure of the stellar upper chromosphere and ongoing processes in the quiescent state and during flares (Osten et al., 2005, 2006b). These observations were conducted at the incomparably higher instrumental level: the Chandra X-ray apparatus was applied, as well as the ultraviolet telescope Hubble, the radio telescope VLA, and the 2.1 and 2.7-meter optical telescopes of the McDonald Observatory. As to the results concerning the quiescent transition zone chromosphere–corona, Osten et al. (2006b) detected their nonthermal broadening from the profiles of the UV and FUV lines, and the widths of these lines approached the maximum at the temperatures $< 10^5$ K, that is lower than on the Sun; the increased turbulent pressure in the transition zone should affect the structure and dynamics of the stellar atmosphere. The results of these observations of the quiet corona and flares over the whole depth of the atmosphere will be outlined in the following chapters.

**1.3.2.3. Magnetohydrodynamic Model Chromospheres.** The most comprehensive calculation of the chromosphere of the M dwarf was carried out by Wedemeyer et al. (2013), taking into account the MHD structures for the star with $T_{\rm eff} = 3240$ K and log $g = 4.5$, which corresponds to the parameters of AD Leo. The numerical three-dimension calculations were performed within the CO5BOLD program for the volume of $1950 \times 1950$ km$^2$ and from $-700$ to $+1000$ km, covering the upper convective region, the photosphere, and the chromosphere, including initial magnetic fields with a strength varying from 10 to 500 G and different topology. In the lower photosphere the magnetic fields are effectively frozen-in, rapidly entangled by the plasma motions to a strength of more than 1000 G and cannot be described by ideal flux tubes. Outgoing from the photosphere, the fields occupy the whole volume of the chromosphere and, hence, the chromosphere turns out to a great extent highly dynamic and inhomogeneous: magnetic fields and shock waves produce a complicated temperature and spatial structure, which is well seen in the plotted maps of intensities in the core of CaII K line and at mm wavelengths. The detected small-scale structures cannot be described within the one-dimension static model, which is widely applied when constructing semiempirical model atmospheres and interpreting observations. The constructed models ease an analysis of the small-scale processes which cannot be observed on stars but may be significant for understanding the atmospheres of M dwarfs and their activity. An example of the so-called “magnetic tornados”, recently discovered on the Sun, was first presented by Wedemeyer et al. (2013) in the model atmospheres of M dwarfs.

## 1.4. Stellar Coronae

The solar corona has been known since antiquity: during total solar eclipses a silver glow of the corona flares up for several minutes around the solar disk covered by the Moon, then it disappears without a trace when the smallest bright crescent of the Sun appears. Solar eclipses showing corona have always been the most impressive performances in the celestial theater.

Emission lines in the solar corona spectrum were long a mystery. Identification of the lines with multiply ionized metals by B. Edlen in 1942 became a scientific sensation. This discovery unambiguously led to the estimate of the temperature of coronal plasma as a million degrees. (G.A. Shajn was one of the first who formulated the hypothesis on very hot solar corona: during the eclipse of 1936 he did not find the CaII H and K absorption lines in the photospheric spectrum of the Sun scattered by the corona and explained this fact by the scattering on fast electrons of very hot plasma.) Today, the concept of a high-temperature corona is a fundamental of the solar physics. The coronal temperature is determined by several independent methods: from the vertical gradient of coronal density, from its blackbody radiation at meter radio waves, from high-temperature lines in the region of extreme ultraviolet, from their existence and widths. Such direct and independent observations are used in constructing the models of the solar corona, because here, as for the chromosphere, physical bases of high temperature are known, but since there is no complete certainty about the heating mechanism, in constructing coronal models the energy considerations are usually disregarded.

The structure of the solar corona is determined by frozen-in magnetic fields and, as a rule, is far from spherical symmetry. The global structure of the corona is subject to slow changes correlated with the 11-year activity cycle. The local magnetic fields of active regions are associated with small-scale formations, coronal condensations, which are well seen in the X-ray photographs of the Sun and in which matter density and temperature are somewhat higher than beyond the structures. In the regions adjacent to spots, the temperature of the coronal plasma achieves $10^7$ K. In observations with high spatial resolution, the coronal condensations are seen as a system of loops and arches (see Fig. 21). The characteristic length of the condensations is $10^9$– $10^{10}$ cm or 0.01–0.1$R_\odot$, magnetic fields in them achieve several hundred gauss, and at matter density of about $10^9$ cm$^{-3}$ the Alfven velocity is close to $3 \cdot 10^9$ cm/s. The regions of corona free of loops form the so-called coronal holes, in which magnetic fields are practically radial and the plasma temperature is slightly reduced. In other words, the solar corona is not a homogeneous hot atmosphere, but, as the chromosphere, an ensemble of plasma structures of different scales and configurations governed by the local magnetic fields. According to Schmitt (1996), 90% of the X-ray flux from the solar corona comes from several active regions that in total occupy only 1% of its disk.

Nowadays, the solar corona is observed from the Earth and from spacecrafts over the whole range of electromagnetic radiation. The X-ray radiation of the Sun as a star in the bands of 0.5–3 Å, 1–8 Å, 8–20 Å, and 44–60 Å was regularly recorded already for more than three solar cycles. The solar wind, the outmost layers of the expanding corona that, according to Parker (1963), cannot be in static equilibrium with the interstellar medium, is directly sounded from the spacecrafts.

Kahn (1969) used Parker's hydrodynamic theory of the solar wind (1963) as an initial point in constructing the first semiempirical model of the corona and stellar wind of a flare star. He used the results of Lovell (1969), who recorded the flare on YZ CMi on 19 January 1969 in the radio-frequency region and found a 4–5-min delay of the radio burst at 240 MHz

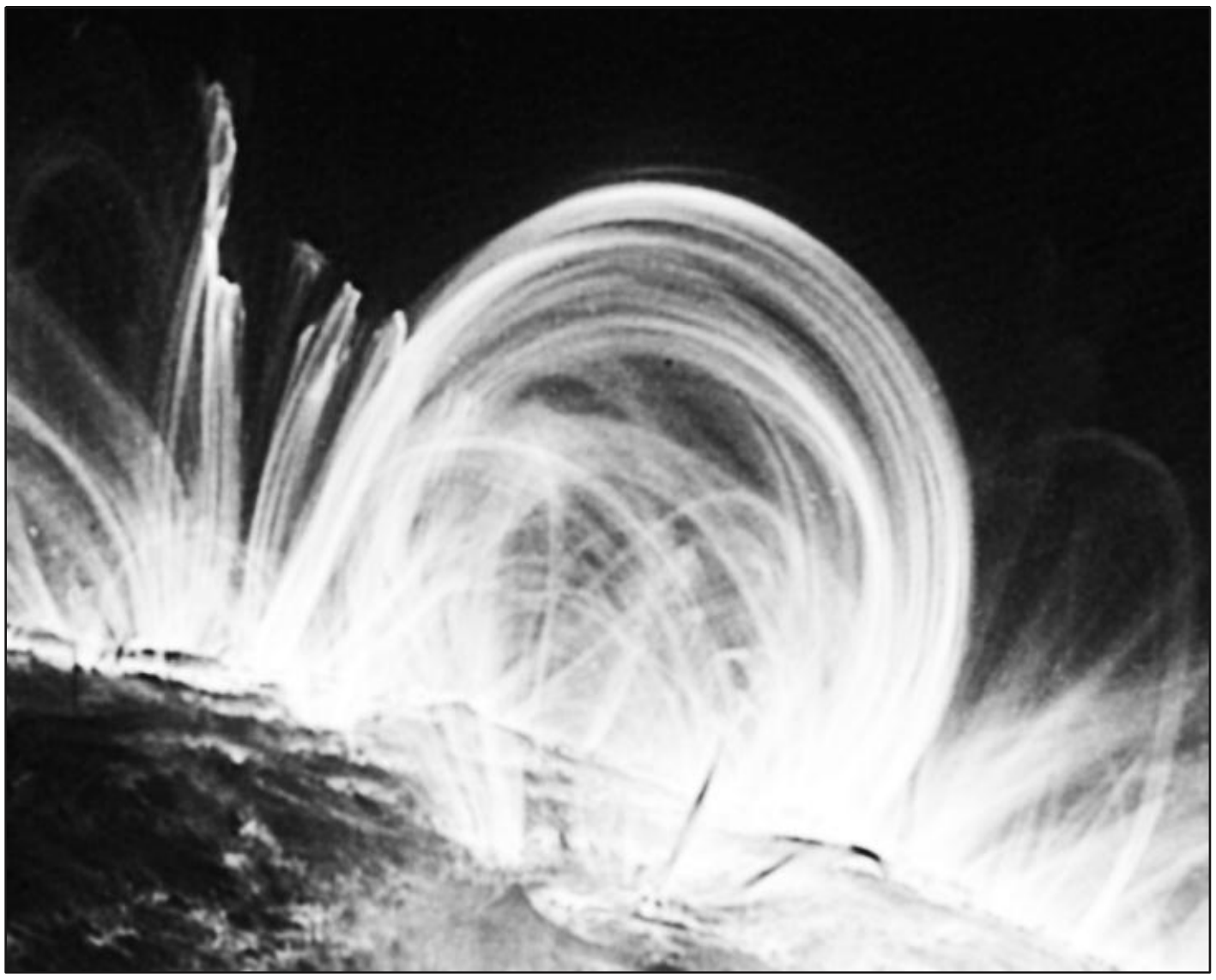

Fig. 21. The loop structure of the solar corona photographed on 6 November 1999 in the 171 Å line from TRACE (NASA Lockheed Martin Solar and Astrophysics Laboratory)

with respect to the burst at 408 MHz. Lovell interpreted the radio-range radiation as plasma oscillations in the stellar corona and associated the delay with the propagation of a shock wave in the medium with decreasing density. Considering these data and setting reliable values for the temperature of the stellar corona and the shock-wave propagation velocity, Kahn constructed the complete density and velocity profile of the expanding stellar corona and estimated its secular mass loss as $3 \cdot 10^{-12} M_\odot$/year.

Zirin (1976) proposed an optical method of studying stellar coronae: owing to hard radiation a noticeable population of the neutral helium metastable level $2s^3S$ appears in stellar atmospheres, which results in the formation of the absorption IR line $\lambda$ 10830 Å. Zirin discovered this variable intensity line in the spectra of $\varepsilon$ Eri, 70 Oph A, and 61 Cyg A. Later, Zarro and Zirin (1986) compared the measured equivalent widths of this He line in the spectra of about 70 F0-K7 stars with the appropriate ratios $L_X/L_{bol}$ and found a linear correlation between $W_{\lambda 10830\ \text{Å}}$ and $\log(L_X/L_{bol})$ for F7 and later stars. To interpret the relation, they used the considerations on the spottedness of stellar coronae. However, Lanzafame and Byrne (1995) analyzed the absorption profile of this He line in the spectrum of the M dwarf V 1005 Ori and showed that it could be reproduced within the semiempirical model of the chromosphere and the transition zone disregarding additional ionization by particle beams or X-ray radiation from the corona.

* * *

Elegant prominences of cold gas floating in the hot corona are one of the most beautiful scenes on the Sun, such structures cannot be seen on other stars. But there are indirect evidences of the existence of stellar prominences. Above we described in detail the case of the fast rotator Gl 890, and BD+22°4409 and RE 1816 + 541. Later, the prominences in the atmospheres of rapidly rotating G dwarfs in the $\alpha$ Per cluster were suspected by Collier Cameron and Woods (1992) and Barnes et al. (1998), and in the binary systems OU Gem and BF Lyn by Montes et al. (2000). Collier Cameron (1999) thoroughly analyzed the structures, considering the problems of their experimental detection and the theory of their retention at considerable distances above the stellar surface. Jardine et al. (2001) suggested a general method for calculating support of prominences in the potential field of a rotating star and found that the stable minima of the gravitational potential exist both inside and outside the corotation radius. For the formation of prominences to occur at a significant distance from the equatorial stellar plane, like on AB Dor и PZ Tel, the field should have significant deviations from the dipole at a distance of $4R_*$ and more.

Using the advanced technique for detecting absorption details in the $H_\alpha$ line, Dunstone et al. (2006a) found for Speedy Mic (BO Mic) an active system of prominences with a dozen such formations on the stellar hemisphere per one revolution. The average distance for 25 prominences accounts for $(2.85 \pm 0.54)R*$, which is twice higher than the corotation radius. Next night some prominences revealed their "counterparts" with some evolution of height and phase, but the general structure of supporting prominences persisted up to 13 stellar revolutions. Later, Dunstone et al. (2006b) performed observations of this star with AAT and VLT, separated by two weeks, and detected that the largest prominences observed passing across the disk turned out to be at the same phases between observational epochs. This means the persistence throughout 2–3 weeks of the magnetic structures that support prominences. Owing to a high *S/N* ratio and a wide frequency range of VLT, the prominences were identified as absorption details in all Balmer lines toward $H_{10}$, in the $H_\alpha$ line they proved to be optically thick and their masses accounted for $(0.5–2.3) \cdot 10^{17}$ g, which is somewhat higher than that for the giant solar prominences. The rotationally modulated emission is seen outside $H_\alpha$ — these are prominences beyond the stellar disk, but one can associate them with prominences which were seen passing the stellar disk. It was concluded that the system of prominences is very flat and at low latitudes.

## 1.4.1. Soft X-ray Emission of Coronae: X-ray Photometry and Colorimetry

Contrary to the solar corona, information on stellar coronae can be directly obtained only from a few channels. The main source of information is the thermal X-ray radiation of the coronal plasma. The discovery of this radiation was not a surprise or accident: since the late 1950s the coronae of stars of different effective temperatures and luminosities have been calculated within the theoretical considerations on the generation of the flux of nonradiative energy in the subphotospheric layers and its dissipation in the atmosphere (see the review by de Jager, 1976). It was only a matter of time for space technologies to achieve the level sufficient for the detection of the expected weak point sources. Thus, when the X-ray radiation of red dwarfs was found, long-term monitoring or repeated observations were needed to confirm that this was weak radiation of quiescent coronae rather than that of strong sporadic flares.

Using the Astronomical Netherlands Satellite (ANS) with an aperture diameter of 10 cm, Mewe et al. (1975) recorded the X-ray radiation from Capella and Sirius within the range of 0.2–0.284 keV and established only the upper limit of the X-ray radiation of the K dwarf $\varepsilon$ Eri equal to $5 \cdot 10^{27}$ erg/s.

The first article that reported the direct detection of the X-ray radiation of the corona of the active dwarf was published by Nugent and Garmire (1978): on 19–25 August 1977 they recorded the radiation of the $\alpha$ Cen system within the A-2 experiment on the American satellite HEAO-1. The experiment was run in the scanning mode in the energy ranges of 0.18–0.44 and 0.44–2.8 keV. On scanning, the width of the FWHM band was about 1.5° along the scanning direction and about 3°in the perpendicular direction. Over 3 days they obtained about 50 scans of the region of $\alpha$ Cen. In the summary scan one could distinctly see the signal of the system, which in the band of 0.18–0.44 keV corresponded to a luminosity of $L_X \sim 3 \cdot 10^{27}$ erg/s. Using the same scans, Cash et al. (1979) studied the region of the 40 Eri triple system, composed of a K1 star, a white dwarf, and a dMe star. They concluded that the X-ray source found was most probably the K1 star. Later, Walter et al. (1980) used the HEAO-1 data to study four G0–K2 dwarfs — $\chi^1$ Ori, $\xi$ Boo, 70 Oph, and HR 6806 — with the same confidence of $3\sigma$. They suspected that the stars were coronal radiation sources. Using the HEAO-1 data, Ayres et al. (1979) studied the signals of 30 G-M stars, including 10 dMe objects, and, in addition to 40 Eri C, found the X-ray radiation from BY Dra and AD Leo of $L_X/L_{bol} = 8 \cdot 10^{-4}$ and $1 \cdot 10^{-3}$, respectively. During the A-1 experiment at HEAO-1, when the sky was scanned within the energy range of 0.5–20 keV with the rectangular field of view $1° \times 4°$, Ambruster et al. (1994b) recorded quiet X-ray radiation of EV Lac at the level of $4 \cdot 10^{28}$ erg/s between flares. Tsikoudi (1982) considered the vicinities of 70 known flare stars using the A-2 experiment observations, and in 13 cases found nonflare signals exceeding $3\sigma$. In the range of 2–20 keV these signals corresponded to the luminosity $L_X$ varying from $5 \cdot 10^{28}$ to $6 \cdot 10^{29}$ erg/s and the ratios $L_X/L_{bol}$ varying within $10^{-4}$–$10^{-2}$.

* * *

E i n s t e i n   O b s e r v a t o r y. A revolutionary advance in the X-ray studies was made by the second American High-Energy Astrophysics Observatory (HEAO-2), the Einstein Observatory with a 56-cm aperture diameter, launched in 1978. Among the devices installed on the apparatus, the most efficient was a position-sensitive imaging proportional counter (IPC) with a field of view of $1° \times 1°$ and an angular resolution of about 40″. The counter was a thousand times more sensitive than previous devices and was able to separate 32 bands in the energy range from 0.15 to 4.0 keV. Another device, the high-resolution imager (HRI), made it possible to reach a spatial resolution of 4″, but with a somewhat lower sensitivity and without spectral resolution. From December 1978 to August 1979 the Einstein Observatory carried out the Stellar survey, during which the X-ray radiation of 143 stars was recorded, including 32 G-M dwarfs (Vaiana and 15 coauthors, 1981). The detectors of the Einstein Observatory made it possible to reveal surface fluxes of soft X-ray radiation from stars within the range of $10^3$–$10^8$ erg/cm$^2 \cdot$ s, while the lower limit of this range was more than an order of magnitude lower than the flux from solar coronal holes and the upper limit was more than 10 times greater than the flux from solar-active regions. The analysis of these data formed a basis for the modern stellar X-ray astronomy: it was established that the X-ray radiation was intrinsic not only in previously known binary systems with accretion disks and the RS CVn-type systems, but also in the stars of all spectral types and luminosities, except for late giants and supergiants. In each spectral type and luminosity class the range of X-ray luminosity was 2–3 orders of magnitude, but average X-ray luminosities of G–K dwarfs amounted to $10^{27}$–$10^{28}$ erg/s (see Fig. 22), whereas bolometric luminosities within this spectral interval decreased considerably.

Such a high X-ray luminosity of medium- and low-mass dwarfs weakly depending on the effective temperature contradicts the traditional theoretical calculations and thus proves the inconsistency of the underlying conception of acoustic heating of stellar coronae. On the other hand, the large range of $L_X$ in every spectral type showed that classical stellar parameters, effective temperature and luminosity, were insufficient to explain X-ray radiation of the stars. Thus, the magnetic field related to rotation started playing the most important role in understanding of the solar-type stellar activity.

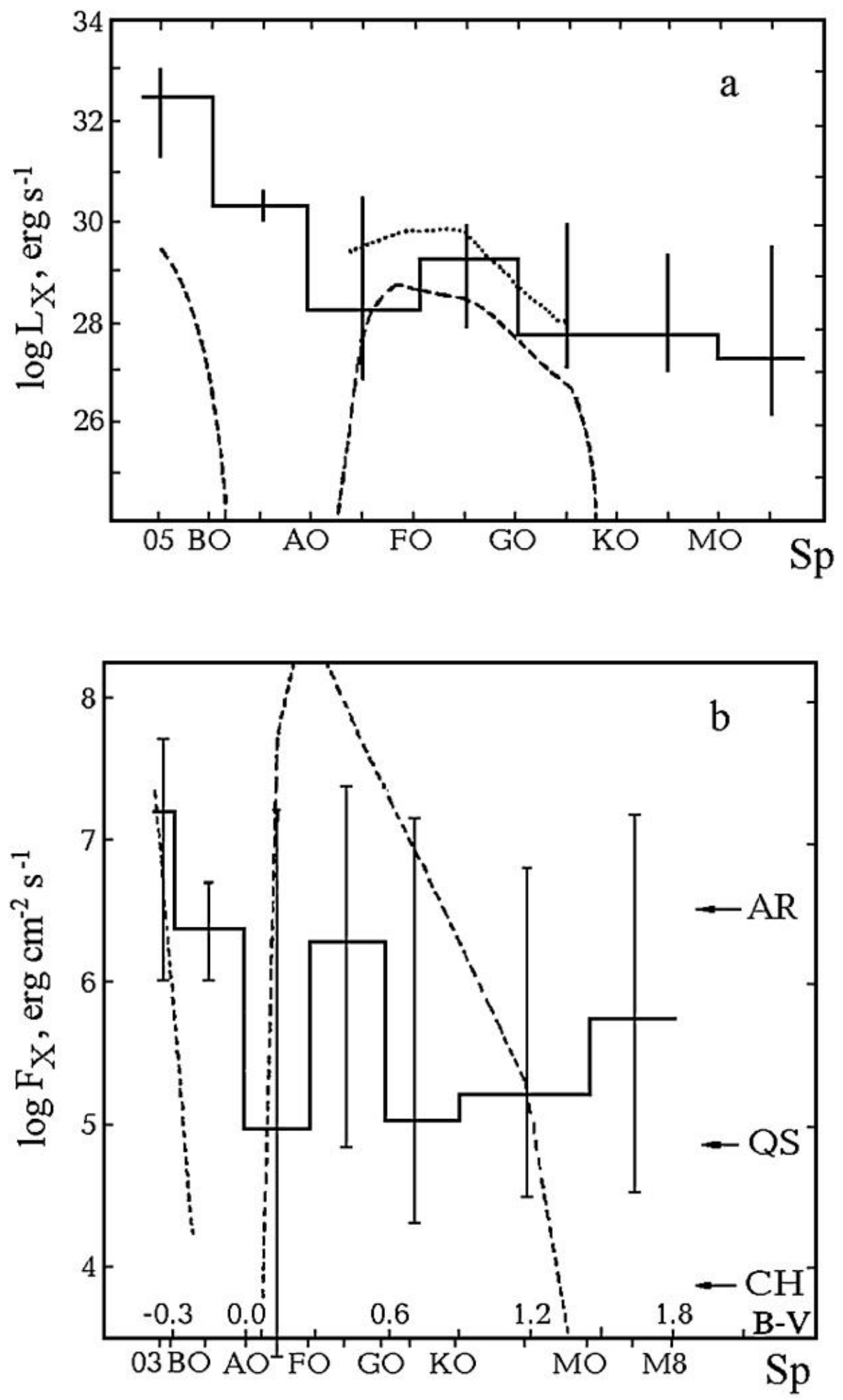

Fig. 22. Average luminosities (a) and average surface fluxes (b) in the soft X-ray range detected within the Stellar survey at the Einstein Observatory. AR is the level of a solar active region, QS is the level of the quiet Sun, and CH is the level of coronal holes; dashed and dotted curves show expected values based on the results of theoretical calculations of acoustic heating of stellar coronae (Vaiana, 1980)

Observations within the Stellar survey and some other programs of the Einstein Observatory enabled important conclusions on the properties of stellar X-ray radiation to be made, supplementing the above results of Vaiana and 15 coauthors.

Schmitt et al. (1985) constructed histograms of the distribution of luminosities $L_X$ of more than 120 A5–F7 stars and found that coronal emission started abruptly at about F0, where the external convective zone appeared in stars. In the sample under consideration, there is no correlation between $L_X$ and v sin*i*, which also evidences the decisive role of convection in the appearance of coronal emission. On the other hand, Golub (1983) and then Bookbinder (1985) found a noticeable decrease of $L_X$ in stars later than M5, on which convection involved the whole star. Earlier, the correlation of X-ray luminosity and the rotational velocity was found from different stellar samples within the above limits: $L_X/L_V \sim v_{rot}^{2.8}$(Katsova, 1981a), $L_X \sim (v\sin i)^2$ (Pallavicini et al., 1981), $L_X \sim \Omega$ (Walter, 1981). This provided an impetus for including the dynamo theory in the concept of stellar X-ray activity.

To find a parameter governing X-ray luminosity, Johnson (1981) considered photometric and kinematic characteristics of 23 closest stars (up to 6.5 pc to the Sun), from which the Einstein Observatory recorded X-ray radiation. The parameter was not found, but he suspected the correlations of X-ray with optical luminosities and of X-ray luminosity with the intensity of the CaII emission. Later, using the data observed by the Einstein Observatory, Johnson (1986) selected about 200 stars close to the Sun (up to 25 pc) that were not program objects and constructed the function of X-ray luminosity up to $3 \cdot 10^{26}$ erg/s for this sample independent of $L_X$. He revealed for the brightest young disk stars $L_X = 1 \cdot 10^{29}$ erg/s, while for the brightest old-disk stars the luminosity was only $2 \cdot 10^{28}$ erg/s.

Walter (1982) showed that the exponential dependence $L_X/L_{bol}$ on the period and the power law with a break of about 12 days equally well described the observational data. Marilli and Catalano (1984) compared the luminosities of $L_K$, $L_{CIV}$, and $L_X$ of late F–M stars with their rotational periods and found exponential dependences of the luminosity on the period, which yielded the following relations independent of the spectrum

$$L_{CIV} \sim L_K^{1.2} \text{ and } L_X \sim L_K^{2.6}. \quad (8)$$

The latter relation is close to the correlation obtained by Ayres et al. (1981a)

$$F_X/F_{bol} \sim (F_{MgII}/F_{bol})^3 \quad (9)$$

and the correlation found by Pallavicini et al. (1982)

$$F_X \sim F_{CaII}^{2-3}. \quad (10)$$

For a spatially limited sample of K dwarfs, Neff et al. (1987) found the correlation

$$F_X/F_{bol} \sim (F_{MgII}/F_{bol})^{2.7}. \quad (11)$$

Mewe et al. (1981) considered the sample of 20 cool stars with CaII emission and found a correlation of the surface fluxes $F_X$ and $F_{HK}$, which was amplified up to $r = 0.93$, when $F_{HK}$ was substituted for $\Delta F_{HK}$, the excess of this flux above the basal level of CaII emission. Schrijver (1983) completed a multifactor analysis of 66 F–G–K stars including flux densities in soft X-rays $F_X$, flux densities $F_{HK}$, and general stellar parameters — radius, mass, and luminosity — and found the close relation between $F_X$ and $\Delta F_{HK}$

$$\langle F_X \rangle = 3.4 \cdot 10^5 \langle \Delta F_{HK} \rangle^{1.67}. \quad (12)$$

The relation occurs in the range of 4–5 orders of magnitudes of $F_X$ and does not depend on the spectral type, luminosity class, and the multiplicity of a star. Considering only close in time observations and using EXOSAT observations, Schrijver et al. (1992) determined more precisely the exponent of $\Delta F_{HK}$ up to $1.50 \pm 0.20$. The nonlinearity of the relation between the characteristics of the chromosphere and corona radiation suggests that this relation is stipulated not only by the filling factor: the averaged structure of radiative coronal regions changes with the averaged activity level over the stellar surface. (Later, this conclusion was specified within the two-temperature models, the relation between the components of which systematically changed with $L_X$ — see below.) But Rutten et al. (1989) found that for M dwarfs the values of $\Delta F_{HK}$ were lower at least by an order of magnitude than those obtained from (12) for F–G–K stars. They associated this effect with heating by the flux of mechanical energy from deeper atmospheric layers up to the temperature minimum.

Teplitskaya and Skochilov (1990) suggested the existence of a basal level of X-ray radiation of the F–K main-sequence stars, that is, as for the quiet Sun in the minimum, due to the emission of bright X-ray points. They constructed the curve of the basal level that resembled a similar curve for the emission from the transition zone and found a rather close correlation between $\Delta F_{HK}$ and $\Delta F_X$:

$$\log\Delta F_X = -6.03 + (1.86 \pm 0.11)\log\Delta F_{HK} \qquad (r = 0.92)\,. \qquad (13)$$

The X-ray radiation of emission K–M dwarfs is consistent with the general tendency found by Linsky et al. (1982): toward more active stars the contribution of high-temperature components in the radiation of their atmospheres increased (see Fig. 18). Indeed, both the surface fluxes in the transition-zone lines and the soft X-ray radiation of dMe stars increase considerably as compared to the flux from the quiet Sun, and the luminosities $L_X$ of such stars substantially exceed the luminosity of the emission lines of the chromosphere and the transition zone, whereas the solar corona is much weaker than the chromosphere. The X-ray luminosity of dMe stars exceeds that of the Sun by a factor of 100–1000, while $L_X/L_{bol}$ is approximately $10^{-2}$–$10^{-3}$ for the stars and $10^{-6}$ for the Sun (Giampapa, 1987). For 13 M dwarfs Byrne and Doyle (1989) found a practically linear correlation of the coronal radiation and the emission of the upper chromosphere in the lines HeII $\lambda$ 1640 Å and CIV $\lambda$ 1550 Å:

$$\begin{aligned} \log F_X &= 0.87 + 1.04\log F_{HeII} \qquad (r = 0.96)\,, \\ \log F_X &= -1.97 + 1.13\log F_{CIV} \qquad (r = 0.97)\,, \end{aligned} \qquad (14)$$

which is valid in the range of about three orders of magnitude.

Mathioudakis and Doyle (1989a) compared the surface fluxes $F_X$ and $F_{MgII}$ in the group of dMe, dKe, dK, and dM stars and found that for the same values of $F_{MgII}$ the $F_X$ fluxes of emission dwarfs were greater by approximately an order of magnitude than in nonemission K–M stars. The relation is close to $F_X \sim F_{MgII}$, which differs significantly from the relation obtained by Ayres et al. (1981b) for F–G–K stars, which can be due to the saturation of the considered fluxes in the emission K–M dwarfs.

Young et al. (1984) proposed to estimate the emission of $H_\alpha$ as the difference between the observed equivalent width and the maximum equivalent width of the line in absorption in the stars of the same spectral type. They compared the values of $L_{H\alpha}$ with $L_X$ and found that the correlation of these values was close to linear: $L_{H\alpha} \sim 0.2L_X$. Commenting on this result, Young et al. (1989) noticed that in addition to the Cram hypothesis on the heating of the chromosphere by coronal radiation, a similar correlation occurs if the whole stellar atmosphere is heated by the same source of nonradiative energy. Following the method of Young et al. (1984), Doyle (1989b) calculated the equivalent widths of the $H_\alpha$ emission in the spectra of about 50 M dwarfs, determined the surface fluxes $F_{H\alpha}$ and luminosities $L_{H\alpha}$, and compared these values with the X-ray radiation. He found the correlation

$$\log L_X = 1.11\log L_{H\alpha} - 2.56 \qquad (r = 0.93). \qquad (15)$$

According to Cram, though this correlation is rather close to $L_{H\alpha} \sim L_X$ and suggests noticeable heating of the chromosphere of M dwarfs by their hot coronae, direct energy estimates for the active star YZ CMi showed that the heating of the corona accounted for not more than 50% of the radiative losses of the chromosphere.

In the low-resolution Einstein observations, the temperatures of stellar coronae were determined using the theoretical models of radiation of optically thin very hot low-density plasma. The lines of highly ionized atoms excited by an electron collision dominated in this radiation at temperatures up to $10^7$ K, while the contribution of continuous, free–free, recombination, and two-photon radiations was relatively small. At higher temperatures, a free–free radiation of hydrogen and helium started prevailing. Such calculations for the plasma of the given chemical content and temperature were convolved with the response curves of X-ray detectors. Comparison of the calculation results with the observations enables the estimate of coronal temperatures. (Initially the calculations involved the results for normal chemical content calculated by Kato (1976) and Raymond et al. (1976). In the 1970s–1980s, the RS model based on the results of Raymond and Smith (1977) and Raymond (1988) was applied. Today, the most frequently used is the MEKA model based on the results of Mewe et al. (1985) and Kaastra (1992) and its modifications.)

Within the isothermal model, Schrijver et al. (1984) determined the temperatures and column emission measures of stellar coronae, $\zeta = \mathrm{EM}/4\pi R_*^2$ and found that for dwarf stars these temperatures were, on average, equal to 2 MK and decreased for later spectral classes. Having considered the X-ray radiation within the model of static magnetic loops, Schrijver et al. concluded that the size of the structures varied considerably. Multifactor analysis revealed a close relation between the column emission measure, the temperature of coronae, and the rotational period

$$\zeta = 10^{28.6\pm0.2} T^{1.15\pm0.16}\, \mathrm{P_{rot}}^{-0.88\pm0.14}, \tag{16}$$

The relation covers all the sample stars and overlaps three orders of magnitudes of $\zeta$ and periods from one day to three months. Ayres et al. (1981b) found that the region occupied by dwarfs — X-ray sources in the Hertzsprung–Russell diagram coincided with the region of the emission sources in the line CIV $\lambda$ 1548/51 Å, but there were no X-ray sources — the stars with a strong cold wind.

Helfand and Caillault (1982) analyzed 270 fields, about 240 square degrees recorded by the Einstein Observatory to identify the additional X-ray sources with 1720 stars brighter than $10^m$ found in these fields. They identified 70 sources and most of them were F–G–K type stars.

Summing up the X-ray observations of M dwarfs made by mid 1982, Golub (1983) listed 35 objects and noted that their luminosities, as in F–G dwarfs, differed by up to 3 orders of magnitudes, but there was a systematic decrease of $L_X$ on redder stars with $B - V > 1.7^m$. All the determined coronal temperatures fell into a narrow range from 2 to $5 \cdot 10^6$ K. There was a tendency of growing temperature with increasing $L_X$ and decreasing axial rotation period.

Fleming et al. (1988) studied the complete sample of early M dwarfs found as X-ray sources in the fields of program stars. They found that 42% of luminosity of such dwarfs was $L_X > 6 \cdot 10^{27}$ erg/s, which corresponded to the sample of such stars based on optical properties, and confirmed the relation of $L_X$ with stellar rotation and mass. Comparison of $F_X$, $F_{H\alpha}$, and $F_K$ revealed the correlation of these values, but chromospheric emission was less dependent on rotation than the X-ray one, and in energy could be maintained by this hard radiation.

Mangeney and Praderie (1984) compared the X-ray radiation of 44 main-sequence stars measured by the Einstein Observatory with their Rossby numbers. They used the effective

Rossby numbers $Ro^* = v_m/2\Omega L_C$, where $v_m$ is the maximum velocity of convective flows and $L_C$ is the depth of the convective zone calculated for such zero-age stars, and obtained

$$L_X \sim (Ro^*)^{-1.2\pm0.1} . \qquad (17)$$

The relation covers stars of 0.5–20 solar masses.

Maggio et al. (1987) analyzed the observations of a spatially limited sample of about sixty F–G dwarfs by the Einstein Observatory and found that the stars involved in the binary systems had the same X-ray luminosities as single stars of the appropriate spectral types.

Fleming et al. (1989) compiled a sample of 128 F–M stars selected according to the value of $F_X$ among nonprogram X-ray sources. The optical study of the sample yielded spectral types, luminosities, distances, $L_X$, $v\sin i$, and $v_{rot}$. From this most complete sample of X-ray sources limited by the flux they found the correlation $L_X \sim (v\sin i)^{1.05\pm0.08}$ for single stars. Since $L_X$ did not correlate with $\Omega \sin i$, they concluded that the correlation was due to the radii of the stars involved in both values. This result differed from the earlier conclusions based on the analysis of the samples of X-ray sources selected on the basis of optical properties. Apparently, the samples selected on the basis of $F_X$ were enriched at the expense of the objects with saturated X-ray luminosities, while in the samples based on optical properties unsaturated objects dominated. Later, similar results were obtained by Schachter et al. (1996), who considered optical and X-ray characteristics of the sources detected by the Einstein Observatory during the Slew Survey when repointing the apparatus from one program object to another. The sample of 809 thus-recorded sources included objects of rather high luminosity. Single F7–K5 stars selected from the sample demonstrated a linear correlation of $L_X$ with $v\sin i$ and stellar size and a correlation with the Rossby number as $L_X \sim Ro^{-0.4}$, but there was no correlation of $L_X$ with the angular rotation rate of the stars.

Barbera et al. (1993) analyzed statistically the spatially limited sample of K–M dwarfs in which 257 stars observed with IPC were identified as X-ray sources. They found that these sources formed a representative subsample of the initial sample of about 1700 late stars from three optical catalogs of near stars. The statistical significance of the decline in $F_X$ after the spectral type M5 and a decrease in $L_X$ with age was proved: for K–M dwarfs inhabiting the young disk population the average luminosity was three times higher than in analogous old-disk stars. However, later Fleming et al. (1995) concluded that kinematic classes should not be used as indicators of age of the nearest stars.

Vilhu et al. (1986) estimated the upper limit of the coronal radiation in soft X-rays: for G0 stars and later spectral types $F_X/F_{bol} < 10^{-3}$. The solar active region was far below the saturation level and the filling factor on the Sun was well below unity.

Favata et al. (1995) compared the levels of X-ray radiation $L_X/L_V$, the abundance of lithium, and rotation in three stellar samples of G–K stars: from the Einstein sky survey, from the Pleiades cluster, and for active binary systems of the RS CVn type. They found that in the Pleiades all the three parameters were in good correlation with each other, whereas for the Einstein survey stars the abundance of lithium and rotation were in good correlation, but they did not correlate with $L_X/L_V$. In the sample of binary systems, no correlation between these parameters was found.

* * *

The following results were obtained by the Einstein Observatory for specific red dwarfs.

In March 1979, during the IPC observations of Proxima Cen, Haisch and Linsky (1980) first recorded the X-ray radiation of a quiescent corona of a flare star, and against this background the radiation of strong flares was found. A thorough analysis of these observations by Haisch et al. (1981) yielded the estimates of the X-ray luminosity of the star $L_X$ = $1.5 \cdot 10^{27}$ erg/s, the temperature of the coronal plasma of $4 \cdot 10^6$ K and the ratio $L_X/L_{bol}$ = $2.2 \cdot 10^{-4}$. It should be noted that for the Sun this ratio is $1.3 \cdot 10^{-6}$, whereas for the Sun completely covered by active regions it would amount to $5 \cdot 10^{-5}$; but the surface flux $F_X$ for Proxima Cen is equal to only 0.22 of the appropriate value in the solar-active region. The volume emission measure of the Proxima Cen corona is estimated as $2 \cdot 10^{49}$ cm$^{-3}$, whereas for the Sun it is $5 \cdot 10^{49}$ cm$^{-3}$. The measured flux from the star was presented by Rosner et al. (1978) within the plasma-loop model developed for the solar corona. Hence, it follows that magnetic flux tubes link sunspots of different polarity and form stable discrete arches, plasma loops, in which plasma can move along but not across the axes. Such an arch structure spreading up to the heights of the order of the solar radius determines to a considerable extent the properties of the quiescent corona. Theoretical analysis of the arch-shaped structures in the solar atmosphere (Rosner et al., 1978) showed that the basic equation of plasma loops was the equation of stationary thermal balance. For not too large loops, whose size is less than the height scale, and consequently, the gas pressure along the arch can be considered constant, the following similarity law is valid:

$$T_{max} = 1400(ph)^{1/3}. \tag{18}$$

Here, $T$ is the temperature at the top of the arch, $p$ is the pressure in it, and $h$ is the height of the loop. The numerical coefficient depends weakly on temperature. Haisch et al. (1981) concluded that the X-ray radiation recorded from Proxima Cen could be caused by the corona composed of plasma loops of length over $10^{10}$ cm and covering the whole star. During the repeated parallel Einstein Observatory and IUE observations of Proxima Cen in August 1980 Haisch et al. (1983) noticed the X-ray radiation of the quiescent stellar corona that corresponded to the same temperature as a year ago and a 4-times lower luminosity. Analysis of the data led to the conclusion that during this season the coronal loops occupied 6% of the stellar surface, while the necessary surface flux achieved $10^7$ erg/(cm$^2 \cdot$ s).

Golub et al. (1982) found that both components of the system $\alpha$ Cen were X-ray sources: $L_X(G2) = 1.2 \cdot 10^{27}$ and $L_X(K1) = 2.8 \cdot 10^{27}$ erg/s. Although the K1 star was weaker by $\Delta V$ = $1.3^m$ than the G2 component, it had a stronger corona. The observations were analyzed using the plasma loop model developed by Rosner et al. (1978) and modified by Serio et al. (1981). The modification made it possible to avoid constraints on the size of loops; for A and B components of $\alpha$ Cen the model yielded loops with a filling factor of 0.05 and 0.15 and the surface fluxes $F_X \sim 3 \cdot 10^5$ and $4 \cdot 10^5$ erg/(cm$^2 \cdot$ s).

Using the HRI Einstein Observatory, Cash et al. (1980) repeated the observations of the 40 Eri system and found that the sources of X-ray radiation were K1 and dMe stars having parameters: $L_X(K1) = 5 \cdot 10^{27}$ erg/s and $L_X/L_{bol} = 4 \cdot 10^{-6}$, $L_X(dMe) = 2 \cdot 10^{28}$ erg/s and $L_X/L_{bol}$ = $1.5 \cdot 10^{-3}$, that is, the optically weakest component of the system was the brightest in X-rays. Since the activity cycle of 40 Eri is estimated as 10–13 years, the star is fairly close to the Sun in all parameters.

* * *

Using the Einstein Observatory, Caillault (1982) studied five spotted stars: AU Mic, HD 218738, YY Gem, CC Eri, and HD 216803. As X-ray sources, these dKe-dMe stars were weaker by one–two orders of magnitude than the RS CVn-type systems, but three of these dwarfs complied with the relation ($L_X/L_{bol}$, rotational period) found for RS CVn variables, which suggested that the high luminosity in X-rays was due to fast rotation rather than binarity.

Agrawal et al. (1986) observed seven K–M dwarfs with the Einstein Observatory. The average X-ray luminosities of the stars out of flares were close to $2 \cdot 10^{28}$ erg/s. Multicolor analysis of three brightest late dMe stars – Gl 729, Gl 735, and Gl 791.2 – yielded the estimate of the coronal temperature as $\sim 3 \cdot 10^6$ K and the emission measure as $1.2–2.9 \cdot 10^{50}$ cm$^{-3}$. They analyzed all observations of flare stars performed using this equipment and found that red dwarfs and the RS CVn-type systems satisfied the correlation $L_X = 10^{-3.23\pm0.22} L_{bol}$ with a correlation coefficient of 0.94, but nonflare dM stars noticeably deviated from it. For the sample of late dwarfs, Agrawal et al. revealed neither the correlation between $L_X/L_{bol}$ and the axial rotation period, nor the correlation of $L_X$ with $v_{rot}$. These results differ significantly from the results obtained before for earlier dwarfs. They got a new explanation within the above concept of the saturation of the X-ray luminosity for the most active dMe stars completely covered by active regions. This concept made it possible to understand the results obtained by Fleming et al. (1989) and Schachter et al. (1996) for much greater stellar samples.

The system Wolf 630 AB was studied by Swank and Johnson (1982) using the solid state spectrometer (SSS) onboard the Einstein Observatory. SSS had the field of view of 6′, sensitivity to plasma radiation of $T > 2 \cdot 10^6$ K, and a spectral resolution of 5–10. The obtained energy distribution within 0.5–4.0 keV could not be presented as the radiation of isothermal plasma, thus a satisfactory presentation was achieved for two-component plasma with temperatures of $6.5 \cdot 10^6$ and $4 \cdot 10^7$ K, and on the assumption of the reduced iron abundance to 60% of the solar abundance. This assumption is concerned with the absence of highly ionized iron lines in the range of 0.7–1.0 keV seen in the spectrum of RS CVn-type systems. In the X-ray radiation of Wolf 630, EM = $1.8 \cdot 10^{51}$ cm$^{-3}$ and $L_X = 3.0 \cdot 10^{28}$ erg/s were estimated for the low-temperature component, while for the high-temperature component EM = $0.7 \cdot 10^{51}$ cm$^{-3}$ and $L_X = 0.9 \cdot 10^{28}$ erg/s, respectively.

In analyzing the flare star Gl 867 A over an interval of three days, Agrawal (1988) found a decrease in the level of its X-ray radiation by 40% and assumed that it could be due to the surface inhomogeneity of the corona. The multicolor X-ray photometry of the star was presented within the two-temperature model: $T_1 = (1.4-2.4) \cdot 10^6$ K and $T_2 = (1.2-1.7) \cdot 10^7$ K, $L_{X1} = (3-8) \cdot 10^{28}$ and $L_{X2} = (1-2) \cdot 10^{29}$ erg/s.

Giampapa et al. (1985) considered jointly the X-ray and ultraviolet observations of eight solar-type stars to determine the applicability of the hydrostatic loop model of the solar corona to them. For the Sun and $\alpha$ Cen B, they obtained a reasonable correspondence of the one-component loop model and the positions of the coronae on the "pressure-filling factor" plane. In particular, the Sun at the activity minimum has long loops ($h \sim 7 \cdot 10^{10}$ cm) with low pressure $p \sim 0.1$ dyn/cm$^2$ and a filling factor close to unity, whereas at the activity maximum the Sun has compact loops ($h \sim 7 \cdot 10^9$ cm) with high pressure $p \sim 1.5$ dyn/cm$^2$ and a filling factor of about 0.03. But in a general case the use of nonsynchronous observations of stellar transition zones and coronae did not provide a confident localization of stellar coronae on the mentioned plane. Probably, it would be more expedient to consider a multitemperature corona and add low-temperature loops contributing to the ultraviolet emission rather than to X-rays. As Golub et al. (1982) proved earlier, from the position of a corona on the "pressure-filling factor" plane one can find the filling factor, if all coronal loops are considered identical and their heights are equal to the scale of pressure heights.

Stern et al. (1986) analyzed the observations of several brightest active cool stars in the Hyades. They concluded that the isothermal model could not represent the observations. The ensemble of loops with a certain temperature distribution and equal maximum temperature or the two-temperature model would be more suitable for this purpose, though less physical. The models with significant variations of the transverse section of loops would not do either. The most acceptable model contains an ensemble of small loops ($h < 10^{10}$ cm) with high pressure (above 400 dyn/cm$^2$) with a maximum temperature of $10–15 \cdot 10^6$ K and a filling factor of up to 0.1–0.2; on the Sun such loops are typical of flares.

Ambruster et al. (1987) studied the observations of 19 active late dwarfs carried out with the Einstein Observatory to detect the variations of their X-ray emission on time scales from minutes to hours. They used the $\chi^2$ criterion modified for the detection of nonperiodic variability of weak sources. Such variability was revealed on 16 stars, for 40 Eri C its characteristic time was close to 150 s, while for the others it was more than 1000 s and the amplitude was up to 30%. The events of this variability did not fit the general energy spectrum of flares constructed using individually recorded flares, and the observation gaps due to the operation mode of the satellite did not allow them to determine the nature of this variability.

* * *

Wide-angle systems of the Einstein Observatory made it possible to study not only the X-ray radiation of individual stars but also that of stellar clusters. Below, we consider the results of investigations for the nearest clusters, in which F and later dwarfs are available.

Stern et al. (1981) studied the X-ray radiation of stars in 27 fields with the size of a square degree in the center of the Hyades. It was found that for a half of the 85 considered members of the cluster $L_X > 4 \cdot 10^{28}$ erg/s, more than 80% of F and G dwarfs were X-ray sources with an average luminosity of $\langle L_X \rangle \sim 10^{29}$ erg/s and this value varied over an order of magnitude. The strongest X-ray source was the F0V star 71 Tau with v sin*i* ~ 200 km/s, and on the whole one can suggest the correlation $L_X \sim v_{rot}^2$. G0–G8 dwarfs found in the Hyades in X-rays on average are 30 times as bright as the active Sun, while the G1 dwarf HD 27836 is brighter by a factor of 300. Combining the results of these X-ray observations with the ultraviolet observations at IUE, Zolcinski et al. (1982) constructed the differential emission measure curves for temperatures within $3 \cdot 10^4–10^7$ K for the model of the transition zone and static coronal loops. The resulting curves were similar to each other and to the appropriate solar curve. They enabled estimating the parameters of the stellar coronae of four F5–G1 dwarfs: temperatures within 1–46MK, the filling factors of the stellar surface within 0.002–1, the density at the base of the corona within $2 \cdot 10^9–3 \cdot 10^{10}$ cm$^{-3}$, and heights within $4 \cdot 10^9–3 \cdot 10^{11}$ cm. Micela et al. (1988) carried out a more detailed analysis of the Hyades: they used an improved algorithm for the detection of sources, considered the complete series of IPC observations of the cluster, 63 fields. Some 66 of 121 cluster members in the studied fields were identified as X-ray sources. The comparison of the resulting luminosity function for the Hyades with those of other stellar groups of the same age led to the conclusion that the dependence of $L_X$ on the stellar age was a function of the spectral type: $L_X$ of young solar-type stars in the Hyades was higher than in the field stars, $L_X$ of K–M stars in the Hyades corresponded to the young-disk stars, whereas $L_X$ of K–M old-disk stars was much lower.

At 4 square degrees, Caillault and Helfand (1985) found 61 X-ray sources in the Pleiades, of which 44 were cluster members. They found that X-ray luminosities of F stars in the Pleiades were close to the $L_X$ of the stars of this spectral type in the Hyades and the field stars, which evidenced slow evolution of $L_X$ of these stars. From the uniform sample of about 30 F stars the authors did not confirm that the dependence of $L_X$ on rotation started abruptly from F6. For G stars of the cluster the average luminosity is $\langle L_X \rangle = 4 \cdot 10^{29}$ erg/s, which exceeded by more than two orders of magnitudes that of the Sun and was higher by only 60% than for the solar-type stars in the Hyades. The Sun completely covered by active regions would have ten times lower $L_X$ than the brightest G star in the Pleiades Hz II 253 with $L_X = 2 \cdot 10^{30}$ erg/s. Therefore, Caillault and Helfand (1985) assumed that the structure of coronae of young stars differed from the solar corona by a greater number of loops. The absence of the rotational modulation evidenced a large number of uniformly distributed active regions.

Micela et al. (1990) analyzed all 14 IPC images of the Pleiades region and, using an updated algorithm for data processing, found that for slowly rotating stars $L_X \sim (v\sin i)^2$, while fast-rotating K dwarfs did not fit this relation, but they are brighter and thus are preferable for detection. In the Pleiades, K dwarfs are brighter by an order of magnitude in soft X-rays than similar stars in the Hyades and brighter by two orders of magnitude than K field dwarfs.

Gagné and Caillault (1994) analyzed the observations of the region of the stellar cluster of the Orion nebula, during which 65 of 157 F6–M5 cluster members were identified as X-ray sources. The comparison of $v\sin i$ of 29 stars and the rotation periods of 8 stars with $L_X/L_{bol}$ showed that in these samples fast rotators had lower ratios. This means that for the youngest stars the rotation is not the only factor that governs the activity level and suggests that the accretion disk can be such an independent factor.

Concluding the survey of the Einstein Observatory results it is worth noting that IPC observations covered only 10% of the sky, which enabled measuring of X-ray fluxes from more than 35000 stars in about 4000 fields. The methods of data analysis were brought to a state of near perfection by the team of the Harvard-Smithsonian Center for Astrophysics and the Palermo Astronomical Observatory. The data obtained and processing results are available on CD-ROM. This experience was then widely used in space experiments.

E X O S A T   O b s e r v a t o r y. Two years after the Einstein Observatory completed its operation, in May 1983 the European X-ray satellite EXOSAT with an aperture diameter of 28 cm was launched. Unlike the Einstein Observatory, the satellite was put on a highly eccentric orbit, which made it possible to observe selected objects continuously for up to three days. EXOSAT was equipped with instruments designed for the investigation for soft (LE) and medium (ME) X-ray emission. Soft X-rays were recorded by two telescopes of grazing incidence optics with four changeable filters, which to a certain extent narrowed the band of 0.05–2 keV; multichannel plates were used as detectors. At a field of view of 2 arcmin the angular resolution was 20 arcsec. The instrument for medium X-rays included two blocks, each composed of 4 proportional counters designed for the range of 1–20 keV; the axes of the blocks were slightly misaligned for simultaneous measurement of a source and an adjacent background. This instrument did not produce images; the angular resolution was determined by FWHM = 45′; spectral resolution in the region of 3 keV was equal to 3. Although EXOSAT was less sensitive than the Einstein Observatory, its advantage was a wider spectral band.

During the investigation of different characteristics of stellar activity of the UV Cet-type stars carried out in 1983 in the Crimea, a satisfactory linear correlation between X-ray luminosities $L_X$ of such stars measured by that time and time-averaged luminosities of their optical flares $\langle L^{fl}_{opt}\rangle$ was established (Gershberg and Shakhovskaya, 1983). This result was practically unnoticed, but then between October 1984 and January 1985 three independent studies with identical results were published by Whitehouse (1985) and Skumanich (1985), each for 9 flare stars, and by Doyle and Butler (1985) for 18 flare stars. The latter authors found a close correlation between $L_X$ and time-averaged flaring luminosity in the U band. Combining the Crimean data and those of Doyle and Butler, Shakhovskaya (1989) obtained the following relation from 23 stars

$$L_X = (4.4 \pm 1.9)\,\langle L^{fl}_{opt}\rangle \qquad (19)$$

Apparently, the linear relation of the optical radiation of sporadic flares and X-ray emission of quiescent coronae can be caused either by a common energy source or direct heating of coronae by flares. This issue will be considered in more detail in the following chapter.

Smale et al. (1986) studied the dM4.5e star 1E0419.2+1908 found occasionally in the course of EXOSAT X-ray observations of the region near T Tau and then found in the archive of the Einstein Observatory. Analysis of the data from both observatories made it possible to estimate the parameters of the quiescent corona: the temperature of $4 \cdot 10^6$ K, $L_X$(0.2– 4.0 keV) = $6 \cdot 10^{27}$ erg/s (IPC), $L_X$(0.02–2.5 keV) = $1.5 \cdot 10^{28}$ erg/s (LE), EM = $5 \cdot 10^{51}$ cm$^{-3}$. Within the spherically symmetrical homogeneous model these values yielded the height scale of the isothermal corona of $4.5 \cdot 10^9$ cm and mean-square density of $1.3 \cdot 10^{10}$ cm$^{-3}$, which does not differ much from the parameters of the solar corona, whereas the application of the correlations of Rosner et al. (1978) to the loop structure resulted in nonrealistic estimates of characteristic coronal loops.

As it was mentioned, Jordan et al. (1987) constructed spherically symmetrical models of hydrostatically equilibrium upper atmospheres of five G0–K2 stars on the basis of joint consideration of their ultraviolet spectra and X-ray luminosities. The models covered the range of structures from those similar to the quiescent solar atmosphere to the structure corresponding to a well-developed solar-active region.

Within the spherically symmetrical models of stellar coronae, Katsova et al. (1987) developed the method for estimating $n_0$, electron density, on the basis of the models. Using Einstein Observatory and EXOSAT X-ray observations, they estimated $n_0$ for 42 F8–M6 dwarfs (see Fig. 23). For this purpose, they took the coronal emission measures and temperatures, so that the resulting values of $n_0$ were averaged over the whole stellar surface. The upper envelope curve of the points corresponds to maximum values of $n_0$ for dwarf stars that belong to different spectral types. As to the found range of $n_0$, it should be noted that for the region of large polar coronal holes on the Sun $n_0 \sim 6 \cdot 10^7$ cm$^{-3}$, for quiet and active solar regions it was $3 \cdot 10^8$ cm$^{-3}$ and up to $10^9$ cm$^{-3}$, respectively; in the most dense stationary condensations above sunspots $n_0 \sim 10^{10}$ cm$^{-3}$ and up to $10^{11}$ cm$^{-3}$ in the flare loops. Thus, within the developed formalism, the objects placed in the upper part of Fig. 23 should have coronae with mean characteristics close to the densest condensations above sunspots. The upper envelope curve of the points grows monotonically for G–K–M3 stars and then has a clear decline. Katsova et al. (1987) explained the character of the envelope with the systematic growth of the filling factor at the ascending branch and the transfer to smaller loops and the mechanism of coronal heating by microflares in late M dwarfs. Later, analyzing the Balmer decrement of red dwarfs in the quiescent state Katsova (1990) suspected that the ratio $F_{H8}/F_{H\gamma}$ correlated with $L_X$: for the stars with higher X-ray luminosity the ratio was higher, i.e., the decrement was flatter.

Observing 7 M dwarfs close to the Sun, Schmitt and Rosso (1988) found that EXOSAT values of $L_X$ were close to the appropriate estimates for such stars based on the Einstein Observatory data and confirmed the systematic decrease of $L_X$ to late spectral types and the presence of structures with temperatures of about $10^7$ K.

Analyzing the X-ray spectra of the stars with increased luminosity — Capella and $\sigma^2$ CrB — within the range of 10–150 Å, Mewe et al. (1987) proved that the allowance for the expansion of coronal magnetic flux tubes with height considerably improved the presentation of observations. In this case, the tubes with a temperature of about 5 MK expanded by 30–50 times, whereas the tubes with hotter plasma (up to 30 MK), increased only by a factor of 2.5–4. An indirect confirmation of the validity of this idea is the fact that coronal condensations above bipolar magnetic fields on the Sun are greater by an order of magnitude than the underlying photospheric faculae. A more complete theory of static plasma loops in stellar coronae was developed by Ciaravella et al. (1993). They studied the influence of pressure at the loop base and its length on thermal conductivity, which governed the loop energy balance, temperature distribution, density, and EM. In particular, they found that for a given loop length the temperature at its top was minimum when the pressure at the base was maximum.

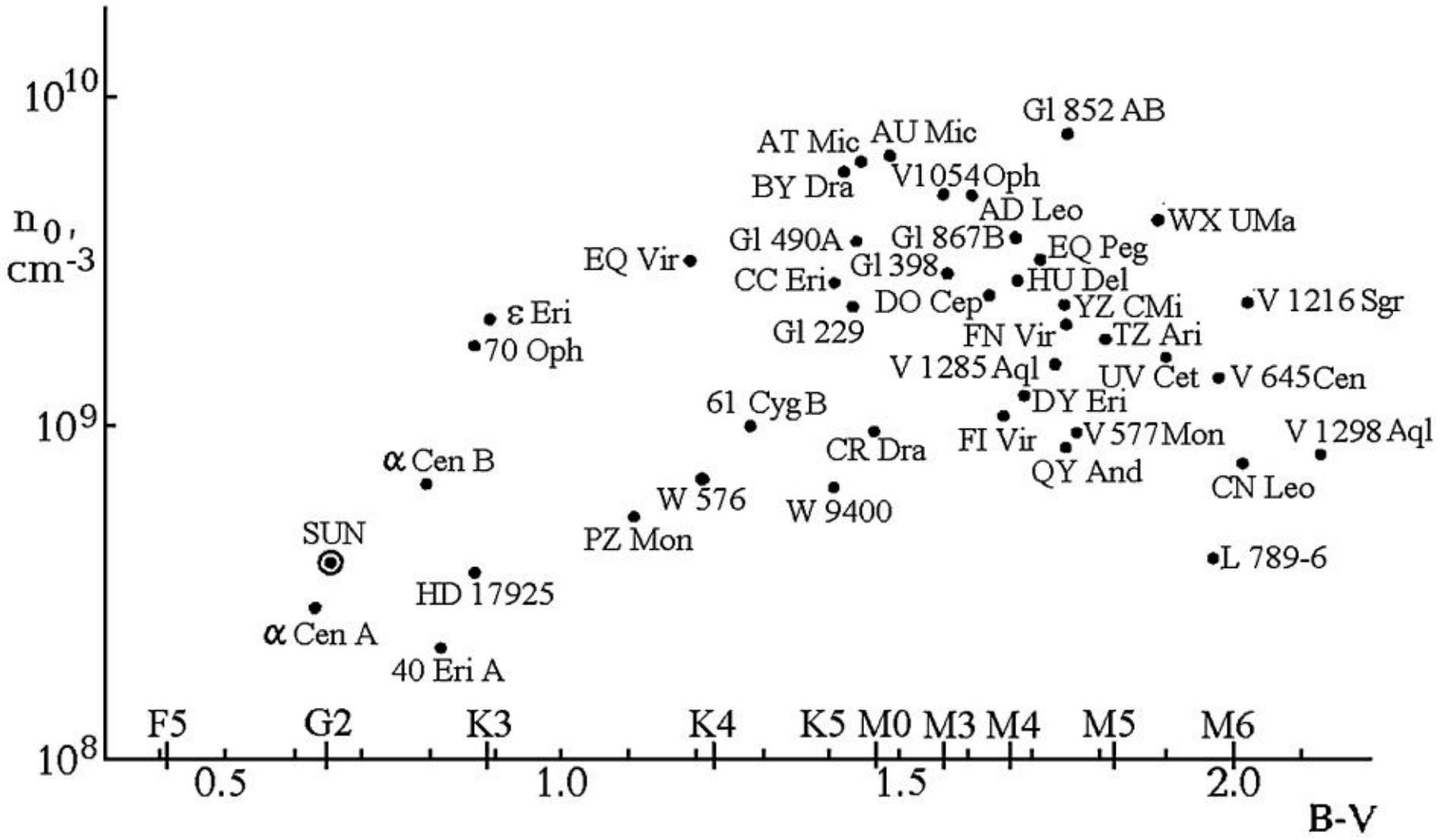

Fig. 23. Electron densities $n_0$ at the bases of red dwarf coronae calculated from X-ray emission within spherically symmetrical models (Katsova et al., 1987)

Johnson continued the studies of the Wolf 630 system at EXOSAT and found a cool stellar corona on one of the weakest flare stars: the coronal temperature of VB 8 was close to $6 \cdot 10^5$ K. For VB 8 Tagliaferri et al. (1990) estimated $L_X$ as $6 \cdot 10^{26}$ erg/s within the range of 0.1–3.5 keV and found that the observed ratio $L_X/L_{bol}$ corresponded to that on earlier M dwarfs, having stratified convective zone, whereas VB 8, whose mass is $0.12M_{\odot}$, should be completely convective.

Based on EXOSAT LE and ME X-ray observations of one of the fastest rotators Gl 890, Rao and Singh (1990) found that the estimated temperature and luminosity were the same as for slowly rotating stars, which could be due to the saturation of the dynamo mechanism on Gl 890.

Haisch et al. (1990a) considered ultraviolet and X-ray observations of the YY Gem system and concluded that each of the components was completely covered by active regions, where MgII emission was formed, since the latter was rather stable. Transition-zone lines demonstrated a certain rotational modulation, while the X-ray light curve had a deep minimum (to 50%) due to low-lying coronal loops or the X-ray emission concentrated on the eclipsed component.

Rao et al. (1990) studied the red dwarf BD+48°1958 A by the EXOSAT LE and ME detectors and estimated the parameters of its quiescent corona as $T = 18 \cdot 10^6$ K, $L_X(0.1-4.0 \text{ keV}) = 3 \cdot 10^{29}$ erg/s and $EM = 1.4 \cdot 10^{52}$ cm$^{-3}$.

Using EXOSAT observations, Pollock et al. (1991) revised the observational data of the Einstein Observatory for the Gl 867 AB system. They concluded that the components of the system contributed to the total radiation as 3:1 and that the change of the X-ray emission of the corona found by Agrawal (1988) from the Einstein Observatory was due to flares occurring on both components rather than to the spottedness of the corona. It is worth noting that between 1980 and 1984 X-ray luminosity of one of the components changed by a factor of 3.

Above, we mentioned the principle of estimating the coronal temperature through theoretical calculations of the radiation of optically thin hot plasma of the given chemical composition. However, the limited range of examined wavelengths and low spectral resolution of the X-ray Einstein IPC usually allowed only a one-temperature model of the corona to be elaborated. Nevertheless, the higher-resolution SSS observations by Swank and Johnson (1982) at the Einstein Observatory yielded the two-temperature model of the Wolf 630 corona, in which one component had a temperature of several MK and the other, an order of magnitude higher value. Then, Majer et al. (1986) noticed that the same temperatures were obtained for different stars observed with the same instrument, while the observations of the same star with different instruments systematically yielded different temperatures. They suggested that the estimated temperatures were governed not so much by the corona physics as by the different sensitivity curves of the detectors, since the shape of the curves was rather complicated. Further studies confirmed these suggestions (Schmitt et al., 1987). Schrijver and Mewe (1986) undertook the first attempt to simulate EXOSAT data within the concept of the differential emission measure (DEM) and compared them with the calculated spectra of the loop structure. Jordan et al. (1987) constructed DEM from separate well-determined spectral lines, whereas Schrijver and Mewe presented the whole EUV spectrum at once. Pallavicini et al. (1988) examined the possibility of wide-band EXOSAT observations to estimate temperatures, emission measures, and luminosities of stellar coronae. They established that there was a continuous temperature distribution up to $10^7$ K in the coronae of late stars in the quiescent state, but the measurements of EXOSAT filters prohibited construction of the DEM distribution and the selection of a multitemperature solution among different versions. Schmitt et al. (1990) studied the validity of two-temperature models of stellar coronae obtained from the wide-band IPC/Einstein data. To this end, they considered the observations with rather high *S/N* of 130 stars later than A0 and found that this ratio greatly affected the resulting parameters of the models and the observational data could also be well presented within a physically more reliable model on a continuous EM distribution with temperature. The low-flux objects can be presented only by the isothermal model, while the objects with higher fluxes can be presented within two-temperature models, with the relations of the obtained components being different for different stellar groups: in M dwarfs the hot component ($T > 10^7$ K) was found in combination with the component with $T \sim 3 \cdot 10^6$ K and EM of the hot component exceeded that of the cooler one, whereas in F and G dwarfs, as well as on the Sun, the high-temperature component was very weak or absent. Probably, the two-temperature models can account for the real situation when the radiative losses of the coronal plasma are predominant cooling mechanisms and over certain temperature intervals, as these losses grow with increasing temperature, the loops become stable to temperature disturbances. On the other hand, the two-peak DEM model suggests that the corona consists of two different ensembles of quasistatic magnetic loops with different maximum temperatures.

EXOSAT was operational from May 1983 until April 1986. Over this period, during 45 sessions with a total duration of about 300 h 25 UV Cet-type stars were examined. The results of these studies were summed up by Pallavicini et al. (1990a). The following results were obtained for quiescent coronae. In spite of the relatively narrow spectral range of considered objects, the X-ray luminosities of quiescent coronae $L_X$ (0.05–2.0 keV) calculated from the observations varied within a wide range from $1.4 \cdot 10^{27}$ erg/s (Proxima Cen) to $6.7 \cdot 10^{29}$ erg/s (YY Gem). These values did not display any correlation with axial rotation periods, while the correlation coefficient between $L_X$ and $v_{rot}$ was rather low, 0.37, but the dependence $L_X = 10^{27} v_{rot}^2$ found by Pallavicini et al. (1981) from the observations at the Einstein Observatory matched well the points of this comparison. A reliable dependence between $L_X$ and $L_{bol}$ with the correlation coefficient $r = 0.95$ was found by Pallavicini et al. (1990a) from EXOSAT data

$$\log L_X = -9.83 + 1.21 \log L_{bol} . \tag{20}$$

The correlation is close to $L_X \sim L_{bol}$ and is probably due to the saturation of stellar corona by active regions, since such a correlation is not valid for dM stars. During the long-term monitoring of red dwarfs slow intensity variations were observed: at times of the order of tens of minutes and hours. When different observational sessions of the same objects separated by several months were compared, the distinctions in the level of their intensity were usually equal to 10–20% and never exceeded 100%. To test the hypothesis of Butler and Rodonò (1985) suggesting that quiet radiation of stellar coronae is a superposition of numerous microflares, all EXOSAT data were analyzed using the dispersion analysis, the autocorrelation function, and the power spectrum. For this purpose, 26 runs of observations were divided into intervals of 2 and 5 min, and the dispersion of the X-ray intensity was equal to 10 to 20% in most cases and to about 40% only in some cases. In the resulting autocorrelation functions, the *e*-fold decay time was 30–60 min and never shorter than 10 min. Thus, though EXOSAT detected independently the variability of dMe stars at all time intervals from several minutes to hours, none of the stars provided definite evidence of the variability at times shorter than several hundred seconds, which could be associated with microflares. Pallavicini et al. (1990a) mentioned that this conclusion complied with the rigorous data analysis by Butler and Rodonò (1985) and by Ambruster et al. (1987). Earlier, conclusions practically coinciding with the results of Pallavicini et al. (1990a) were obtained by Collura et al. (1988), who analyzed the EXOSAT observations of 13 flare stars using the technique developed by Ambruster et al. (1987) for analyzing the Einstein observations of active red dwarfs.

The research results for time variations of the X-ray emission of active stars noticeably differ from the solar data. On the whole, the Sun as an X-ray source is to a greater extent more variable than red dwarf stars. The long-term observations for $F_X$ fluxes within the ranges of 0.5–4 Å and 1–8 Å with a time resolution of 5 min revealed the variations from fractions of a percent to 2–3 orders of magnitude. The greatest of them are due to flares, but the variations of lower amplitudes result from the development of the 11-year cycle, birth of new and death of old active regions, rotation of the Sun nonuniformly covered by such regions, the processes occurring in active regions, the fluctuations of coronal heating, and the exit of new magnetic fluxes. For the Sun, $L_X(1-8\ \text{Å})/L_{bol} = 10^{-9}$, therefore the variations of $F_X$ are so significant. The amplitudes of the variations due to the solar rotation are up to 10, while the changes between the extremes of the solar cycle are 100–200. The amplitudes of fluxes from quiescent stellar coronae do not exceed 2–3 times because the sample includes the most active stars with the surface filling factors for active regions that are higher by orders of magnitudes than the appropriate magnitudes on the Sun (Pallavicini, 1993).

* * *

R O S A T  O b s e r v a t o r y was orbited in June 1990 and operated for about 9 years. The Observatory was equipped with a German Wolter-type I X-ray telescope with an aperture diameter of 83 cm and the above-described English Wide-Field Camera (WFC). Two position-sensitive proportional counters (PSPC) and High Resolution Imager (HRI), very similar to the Einstein Observatory imager, were mounted on a carousel at the telescope focal plane. The PSPC efficiently recorded the radiation in two bands: 0.1–0.28 and 0.5–1.5 keV. The sensitivity of this system was much higher than in the previous experiments, and the main noise in the instrument was due to the diffuse X-ray background. The field of view of the device was about 2°, and the resolution was 15–25 arcsec on the axis and about 160 arcsec at a distance of 50 arcmin from the axis.

The first PSRC ROSAT all-sky survey, the RASS program, was undertaken between August 1990 and January 1991. In sky scanning with the orientation of the satellite to the Sun each object was accessible for at least two days and was measured for 20 s each 96 min. The RASS program made it possible to discover about 60000 X-ray sources, 1/3 of them being the stars with hot coronae. The important advantage of the RASS program was that it enabled direct statistical studies of X-ray sources based on their X-ray characteristics and not from earlier known manifestations of the activity in other wavelength ranges. Upon completion of RASS the studies were run in the mode of individual pointings of the apparatus with exposures of up to 40 min.

In considering the ROSAT observations of red dwarfs later than M5, Fleming et al. (1993) found that the upper border of the $L_X(M_V)$ curve continued the saturation curve outlined from earlier stars, while the $L_X/L_{bol}$ ratio remained constant, at about $2 \cdot 10^{-3}$, with $M_V$ ranging from $8^m$ to $20^m$. They concluded that there were no grounds to believe that M dwarfs were less efficient in creating coronae than hotter stars. The RASS program and individual pointings of ROSAT made it possible to observe an almost complete spatially limited sample of 114 K and M dwarfs in the vicinity of up to 7 pc from the Sun. $F_X$ fluxes were measured for 87% of sample stars, the upper limits of the fluxes were determined for 13% (Schmitt et al., 1995; Fleming et al., 1995). From the calculated values of $L_X$ a luminosity function of such stars was constructed, which covered almost 4 orders of magnitude. Comparison of $L_X$ and the hardness of X-ray emission suggested that the stars that were brighter in X-rays had hotter coronae. Comparison of ROSAT and Einstein Observatory data did not reveal the cases when $L_X$ varied more than by a factor of 2. The analysis of the sample did not provide any evidence of a decrease in the $L_X/L_{bol}$ ratio for the lowest-mass stars that were completely convective but suggested that the coronae of dM stars were systematically cooler and weaker than those of dMe dwarfs. The correlation of $L_X/L_{bol}$ and the radiation hardness with metallicity, i.e., with age, was established, but no correlation of these values with kinematic classes, which required cautious consideration of these classes as indicators of the age for close stars, was revealed.

Marino et al. (1999) studied the variability of M dwarfs in the X-ray range of 0.1–2.4 keV from the ROSAT PSPC data. They considered all M dwarfs from the CNS3 catalog found in the course of individual pointings of ROSAT and recorded not further than 48 arcmin from the center of the fields. The total number of such stars was 55 and they were observed during 86 sessions. The duration of exposures was shorter than in EXOSAT observations, but the sensitivity was much higher. For 29 of 32 sources, for which more than 1000 pulses were recorded, the variability was found at a significance level of over 99%. This result suggests that all M dwarfs are variable in X-rays and the detection of this variability is limited only by the statistics of recorded quanta. For the stars with $M_V \sim 13^m$, which is typical of completely convective red dwarfs, there was no noticeable decrease in the fraction of variable sources and the variability was independent of the average level of X-ray luminosity. Comparison of normalized distributions of $L_X$ of the considered M stars yielded results similar to those for young variable stars in the molecular cloud of $\rho$ Oph and in solar flares, which suggests that stellar flares are responsible for the discovered variability of M dwarfs.

Then, in a similar way, Marino et al. (2002) studied the X-ray variability of F7–K2 dwarfs. Analyzing the statistics of photons from 40 objects in 70 sessions, they found 8 variable sources in 10 sessions and suspected long-term changes, presumably of a cyclic character, for 9 stars observed at the beginning and at the end of the ROSAT operation. The amplitudes of the brightness variations were lower on F–K stars than those on M dwarfs.

Ambruster et al. (1994a, 1998) compared the radiation of the chromospheres (MgII), the transition regions (CIV), and the coronae (ROSAT PSPC) of 6 close K dwarfs of close age, whose rotational periods were within the range from 8 h to 6.6 days. The comparison of their surface fluxes with the rotational periods showed that $F_{\rm CIV}$ saturated at $P_{\rm rot} < 3$ days, and the saturation of $F_{\rm X}$ occurred at even shorter periods, while saturation of $F_{\rm MgII}$ required periods of about 5 or even 11 days.

Later, Micela et al. (1997) analyzed the observations of 12 close K4–M6 halo and old-disk stars at PSPC and WFC ROSAT and found that a half of the considered stars displayed variations of $L_{\rm X}$ to 50% on time scales from hours to days, for GJ 191 such changes occurred over six months and, as compared to the data of Schmitt et al. (1995), such variations were found at even greater time intervals. For the three brightest stars of the sample — GJ 845, GJ 866, and GJ 887 — two-temperature models were constructed, which demonstrated rather low temperatures of the components, $1.7 \cdot 10^6$ and $5.8 \cdot 10^6$ K, and as opposed to more active dwarfs, the $EM_2$ of high-temperature component was lower than $EM_1$ of the low-temperature one. In comparing the models obtained from nonsimultaneous data, the temperatures of the components were found to be constant, while $L_{\rm X}$ varied with the change of emission measures, and there was a close correlation of these values: $EM_2 \sim (EM_1)^3$.

Pan and Jordan (1995) found that the temperature and electron pressure in the source of quiescent X-ray emission of CC Eri corresponded to the well-developed active region on the Sun.

Pan et al. (1995) considered the IPC (Einstein Observatory) and PSPC (ROSAT) observations of the flare star EQ 1839.6+8002 discovered during a strong flare by the Ginga apparatus. They presented its X-ray emission in the quiescent state by the two-temperature model: $T_1 = 1.5 \cdot 10^6$ and $T_2 = 7.6 \cdot 10^7$ K, $EM_1 = 4 \cdot 10^{50}$ and $EM_2 = 6 \cdot 10^{50}$ cm$^{-3}$.

Giampapa et al. (1996) carried out long individual observations of 11 M dwarfs at ROSAT PSPC. Their program included dMe stars with $L_{\rm X}/L_{\rm bol} \sim 10^{-4}$ and dM objects with $L_{\rm X}/L_{\rm bol} \sim 10^{-7} - 10^{-6}$. In particular, they recorded the X-ray emission from Gl 411 (see Fig. 24) and Gl 754, which earlier were considered as stars without coronae. On the basis of the analysis of the obtained data they concluded that the coronae of such stars, probably except for Barnard's star corona only, contained two different components of the X-ray emission: the soft component with a temperature of 2–4 MK and the hard component with $T \sim 10$ MK. The temperature estimates are more reliable in calculations with reduced abundance of heavy elements (up to an order of magnitude) in the coronae as compared to that on the Sun. The hard coronal component on dMe stars makes a systematically greater contribution to the total X-ray emission than on dM stars and is responsible for the major part of variations observed in X-rays. Giampapa et al. (1996) modeled hydrostatic coronal loops for such components and found that the soft component could be presented by small-size loops ($h < R_*$) and high pressure, whereas the hard-component loops required either a small filling factor (< 0.1), large loops ($h > R_*$), and high pressure, or a very small filling factor (<< 1), small loops ($h < R_*$) and very high pressure. From the calculations, they concluded that the soft component of the coronal radiation of dMe stars was formed in the quiet active regions, while the hard component occurred in compact nonstable flare formations. The compact loops of the soft component should give a natural explanation of the observed evolution of angular momentum of late main-sequence stars.

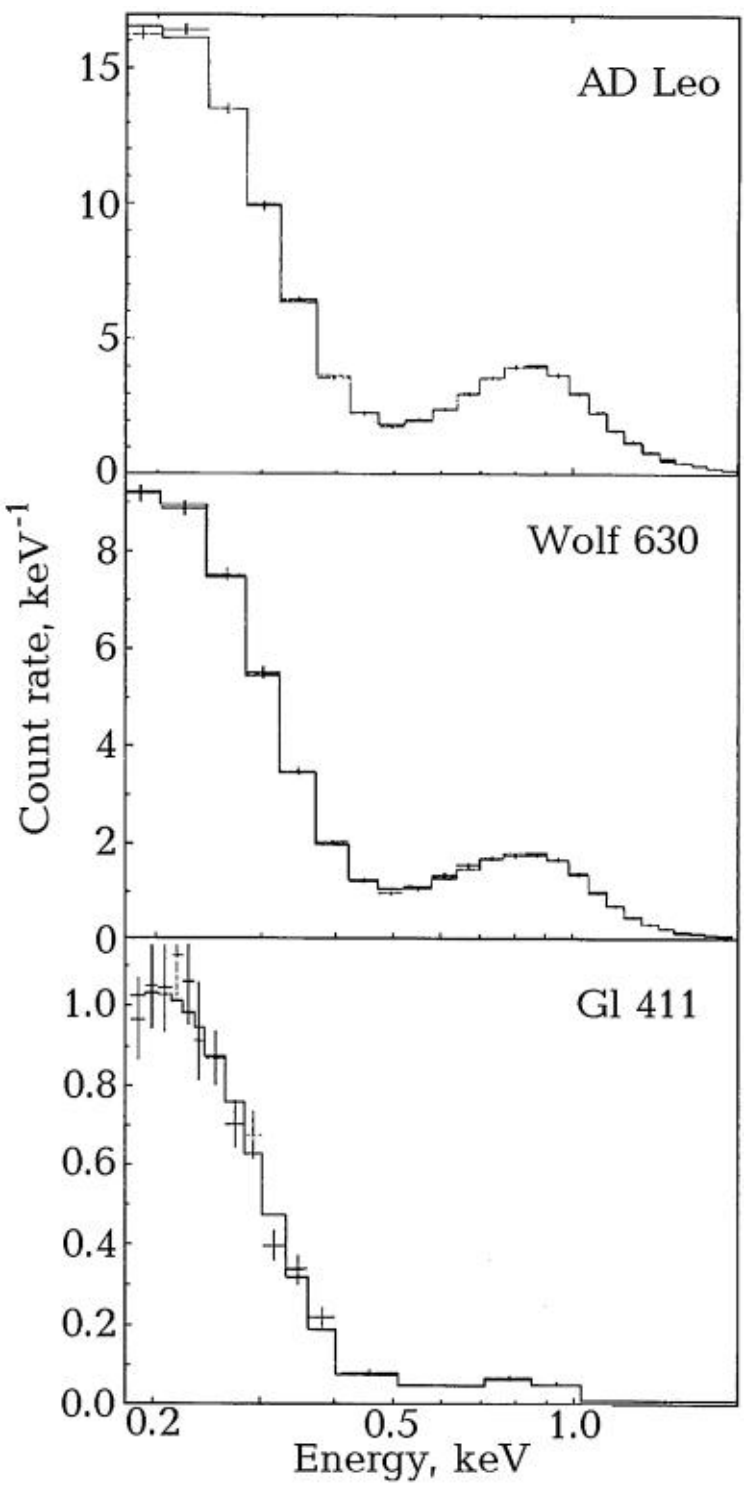

Fig. 24. Spectra of three M dwarfs recorded by ROSAT PSPC (Giampapa et al., 1996)

Using ROSAT PSPC observations of solar-type stars in young clusters, Maggio et al. (1997) thoroughly simulated the loop structure of coronae. They used the hydrostatic loop model modified by Serio et al. (1981) for arbitrary size that was described by the system of two equations linking four loop parameters: the temperature at the top, pressure at the base, the height of a loop, and the heating of a unit volume. Specifying the distribution of the plasma temperature and density along the loop height, one could calculate its integral optically thin X-ray emission. Then, upon normalizing to the sensitivity curve of a detector, the obtained spectrum could be compared with the observed one. Applying the developed algorithm to seven G–K dwarfs from several young clusters, they considered the models with identical loops and with two kinds of loops. In the resulting models, in all cases a hot component with a maximum temperature of $(0.6–4) \cdot 10^7$ K was available, which never occurs on the Sun, and three oldest stars from the Hyades had a cool component with $T_{max} = (1 - 6) \cdot 10^6$ K, typical of the quiet Sun. However, in the low-temperature loops, the pressure exceeds typical solar values by two orders of magnitude, and these loops contribute 30–70% to the total X-ray luminosity of these stars. The above algorithm yielded the following characteristics of the corona of the F9 dwarf HR 3625: low-temperature (1–2 MK) short ($< 10^8$ cm) low-pressure ($p_0 > 6$ dyn/cm$^2$) loops covering a large part of the stellar surface and high-temperature (over 7 MK) loops related to active regions. The characteristics of the latter loops could vary from $h < 10^9$ cm at $p_0 > 10^2$ dyn/cm$^2$ and a filling factor less than 0.5% of the surface to $h \sim 10^{10}$ cm at $p_0 \sim 5^{-10}$ dyn/cm$^2$ and f $\sim$ 0.1 (Maggio and Peres, 1997). Upon similar analysis of X-ray data for eight G dwarfs in the solar vicinity, Ventura et al. (1998) concluded that the most suitable presentation of these data was provided by the models with two types of loops: "cool" loops with $T \sim (1.5-5) \cdot 10^6$ K and $p_0$ varying from 2 to 100 dyn/cm$^2$ and "hot" loops with $T \sim (1-3) \cdot 10^7$ K and $p_0$ varying from $10^2$ to $2 \cdot 10^4$ dyn/cm$^2$. Ciaravella et al. (1997) proved the advantages of the two-temperature modeling using the SIS ASCA data (see below).

The fact that the hottest components of the quiescent stellar coronae have the temperatures close to the plasma temperature in solar flares suggests that the unresolved stellar flares can be responsible for these hot components. Within the Kopp–Poletto model (1992) (see Chapter 2.5), Güdel et al. (1998) calculated the summary effect of a great number of such flares. Using the observed solar-power distribution of flares with respect to energy and the correlation of the flare duration and its full energy, they found that at the simultaneous existence of several hundreds of magnetic loops with flares occurring in them the summary light curve of such loops in soft X-rays should display smooth oscillations, similar to the observed ones, while the numerically obtained bimodal distributions of EM($T$) reminded the appropriate distributions recorded on the four solar-type stars.

Additional data on calculated models of some quiescent coronae with flares developed on their background are presented in Sect. 2.4.1.

The Catalog of Chromospherically Active Binaries by Strassmeier et al. (1988) includes about 200 systems, 163 of them were found within the RASS program. Dempsey et al. (1997) studied the radiation of 35 binary stars of the BY Dra type and found that for these dwarf systems the dependence of $F_X$ on the rotational period was much weaker than in RS CVn systems: $FX \sim P_{rot}^{-0.2\pm0.3}$ and $P_{rot}^{-0.6\pm0.1}$, respectively. They did not find evidence of a basal level of X-ray emission of these stars. The temperatures of both components in the systems of both types found within two-temperature corona models were practically identical – 2 and 15 MK, but EM in RS CVn systems were higher by almost an order of magnitude and the dependences of temperatures and emission measures on rotation were identical for the systems containing the components of different luminosities.

Using the simplest geometrical model of the inhomogeneous stellar corona, Güdel and Schmitt (1996) estimated the electron density of the EK Dra corona as $3 \cdot 10^{10}$ cm$^{-3}$ from the measured modulation of its X-ray emission recorded within the RASS experiment. For another star with X-ray rotational modulation, 47 Cas, they failed to construct an unambiguous coronal model and obtained two density estimates: $2 \cdot 10^{10}$ and $2 \cdot 10^{11}$ cm$^{-3}$.

Piters et al. (1997) compared the values of $\Delta F_{HK}$ for 215 F–G–K stars calculated from the Mount-Wilson data and the values of $F_X$ calculated from the RASS data obtained within several days and found the relation

$$F_X/L_{bol} \sim (\Delta F_{HK}/L_{bol})^2, \qquad (21)$$

which is independent of the spectral type and luminosity, while the scatter of points near the dependence was explained by the measurement accuracy; the hardness of the X-ray spectrum of dwarf stars grew systematically with $F_X$.

Using new instruments, Schmitt (1997) repeated the study carried out in 1985: he considered the X-ray emission of a practically complete sample of A– F–G stars in the solar vicinity up to 13 pc and even with the most sensitive observations with long exposures did not find radiation from stars earlier than A7. This can be considered as an argument in favor of the universal character of corona formation on stars with outer convective zones. He found that $L_X$ correlated with the kinematic age of stars and with the hardness of their X-ray spectra, while the range of $F_X$ measured on the stars was close to the appropriate range of different structures on the Sun, from coronal holes to active regions.

Hünsch et al. (1998, 1999) published the catalog of RASS observations of 980 optically bright dwarfs and subgiants of A–K spectral types and the catalog of RASS observations of 1252 nearest stars.

Fuhrmeister and Schmitt (2003) analyzed about 30000 X-ray light curves obtained within the RASS program and discovered 1207 variable sources. Among the sources identified to optical objects there were about 450 F–M stars. In addition, many flares of the sources invisible in a quiescent state were recorded.

Caillault et al. (1992) performed preliminary observations of the Pleiades and found 47 X-ray sources, of which 36 were cluster members. A half of these sources were not revealed by the Einstein Observatory, on the other hand, a half of the sources discovered by the Einstein Observatory were not detected by ROSAT, which evidenced noticeable variability of the sources.

In studying the central nucleus of the Pleiades, Schmitt et al. (1993b) found that in the range of 0.5–1.8 keV the radiation density was $4 \cdot 10^{29}$ erg/(s · pc$^3$), which was higher by two orders of magnitude than in the solar vicinity.

Using long-exposure images, Stauffer et al. (1994) studied 3 regions in the Pleiades and identified 317 X-ray sources, of which 171 were cluster members. In a radius of 25 arcmin from the centers of considered fields practically all cluster members of G and later types were identified. Thus, for the first time the X-ray luminosity function of G–K–M stars was constructed without statistical estimates for the objects with determined upper limits of luminosity only. Apparently, there were low-mass cluster members among the found sources, which were not identified as such in the optics. The constructed dependence $L_X/L_{bol}(v_{rot})$ revealed the correlation of saturated type for the objects with B−V > $0.6^m$: in each of the considered B−V intervals there was a fast growth of $L_X$ for the rotation velocities of up to 15 km/s and then saturation for the velocities of about 100 km/s; the dispersion of rotational velocities for low-mass cluster members was a dominating factor that governed the dispersion of $L_X$. About 35% of B, A, and early F stars in the Pleiades were identified as X-ray sources, but the properties of their coronal radiation supported the hypothesis that invisible low-mass satellites were responsible for this radiation. Continuing the studies of Stauffer et al. (1994), based on the PSPC data, Gagné et al. (1994a, 1995) constructed one- and two-temperature models of a dozen of the brightest sources related to the stars of late spectral types. They found that the coronae of K and rapidly rotating G dwarfs were hotter than the coronae of F and slowly rotating G dwarfs. For F–G–K stars they found a dependence of the coronal temperature on $L_X/L_{bol}$.

Sciortino et al. (1994) studied the variability of the X-ray emission of stars in the Pleiades using the ROSAT PSPC images obtained within six months and found variability due to flares only. The comparison of the Einstein Observatory and ROSAT data showed that over 10 years the variations of the X-ray brightness were only slightly greater than over 6 months. Thus, Sciortino et al. concluded that this fact was stipulated by permanent high surface filling of young stars by active regions.

Barbera et al. (1993) and Fleming et al. (1993) came to different conclusions concerning the sensitivity of the X-ray radiation of stars to the transition from objects with the radiative core to fully convective structures. These differences could be explained by a different age composition of the stellar samples considered in the solar vicinity. To avoid this, Hodgkin et al. (1995) analyzed the indicators of chromospheric and coronal activity – $L_{H\alpha}/L_{bol}$ and $L_X/L_{bol}$, respectively – using a large sample of low- and very-low-mass stars, slightly more or less than $0.3M_\odot$, in the Pleiades. The $H_\alpha$ emission was found in all cluster members with $M_I > 7.5^m$, and the ratio $L_{H\alpha}/L_{bol}$ reached its maximum within $0.3–0.4M_\odot$, then rapidly decreased, i.e., the chromospheric activity decreased on fully convective stars. The situation with X-ray activity is more complicated. Since for weak X-ray sources, the members of the Pleiades cluster, ROSAT PSPC measured only the upper limits of $L_X$, the determined change of the ratio $L_X/L_{bol}$ depends on the specific consideration of these sources: if they are disregarded and only really measured sources are taken into account, $L_X/L_{bol}$ reaches its maximum at somewhat higher masses than $L_{H\alpha}/L_{bol}$, then it also decreases, according to Barbera et al. (1993). If we take into account the upper limits of $L_X$ as well, the conclusions of Fleming et al. (1993) are confirmed.

The analysis of long-term observations of the central square degree of the region in the Pleiades allowed Micela et al. (1996) to identify the sources with $L_X > (2-3) \cdot 10^{28}$ erg/s, to measure the fluxes from 99 of 214 cluster members, and to estimate the upper limits of $F_X$ for the rest. Inside the central 20 region they found all early M dwarfs, confirmed the dependence of $L_X$ on the age for G dwarfs revealed by the Einstein Observatory, and extrapolated it to K and M stars. G dwarfs in the Pleiades displayed an almost linear dependence of $L_X$ on the rotational velocity up to 100 km/s, but dK members of the Pleiades showed a weak dependence on the velocity, which required a certain additional parameter, in addition to rotation and age, to understand the X-ray emission of the stars. As opposed to the Hyades, the X-ray luminosities of dG and dK stars in the binary systems are of the same order of magnitude as the appropriate values of single stars. This means that the evolution of $L_X$ in single and binary stars to the age of the Pleiades is independent of their involvement in binary systems. Up to 15% of the members of the Pleiades displayed a variation of $F_X$ up to twice on an interval of 10 years.

Then, Micela et al. (1999a) analyzed the ROSAT HRI data for the Pleiades. This more sensitive instrument with higher spatial resolution made it possible to supplement the list of X-ray sources of the cluster by 34 objects, restrict the estimates of the upper limits of $L_X$ of weak sources, and continue the dependence $L_X(v_{rot})$ toward slow rotators.

Marino et al. (2003) analyzed the variability of stars of late spectral types in the Pleiades based on PSPC observations and divided them into two groups: 42 F7–K2 and 61 K3–M stars. Application of the Kolmogorov–Smirnov statistics to the time series of photon fluxes revealed an equal percentage of significant variations over short times – hours – in both groups. The analysis of amplitude-distribution functions showed that over short times K3–M stars were variable to a greater extent than F7–K2 dwarfs. For such stars in the Pleiades and the field the variability over short times depended on X-ray luminosity, whereas for later stars this dependence was not valid. The lowest-mass stars are characterized by fast variations, but not rotational modulation or cycles, whereas variations associated with rotational modulation and cyclic variability are more typical of F7–K2.

Prosser et al. (1995) studied the open stellar cluster NGC 6475, whose age is about $2 \cdot 10^8$ years, and identified 120 early F–M dwarfs — new weak members of the cluster — from X-ray emission with $L_X > 10^{29}$ erg/s. The upper envelope of the points on the plot of $L_X$ (Sp) is analogous to that in the Pleiades, while the scatter of the values of $L_X$ is apparently due to binary systems and fast rotators.

The X-ray properties of the IC 4665 cluster similar to those of the members of the Pleiades, a cluster of close age, were studied by Giampapa et al. (1998).

The central part of the Orion nebula was studied by Gagné et al. (1994b) using ROSAT HRI. They identified 361 sources with $L_X(0.1-2.0\text{ keV}) > 10^{30}$ erg/s and 55 of 87 F6–M5 optical cluster members as X-ray sources. The relation between $L_X/L_{bol}$ of young stars and rotation was not found from this sample.

Pye et al. (1994) observed 11 fields in the region of the Hyades: at a general coverage of 18 square degrees at an area of 3.5 square degrees their sensitivity increased by a factor of 5 as compared to the RASS program, and the X-ray emission of 75% of 70 stars of the cluster was found. This made it possible to construct the luminosity function up to $L_X(0.1-2.4\,\text{keV}) = 5 \cdot 10^{27}$ erg/s. They found that K dwarfs involved in the binary systems were systematically brighter than single K stars, but since these were all wide pairs with periods of more than a year, the cause of the higher brightness of the binary system components was not clear. For three dozen stars the two-component models of coronae were constructed, since the isothermal models did not provide a satisfactory presentation of the observations. Stern et al. (1995) studied the Hyades cluster from the RASS data covering the region of 30° × 30°. Some 187 of the 460 known cluster members were identified as X-ray sources with $L_X > (1-2) \cdot 10^{28}$ erg/s. In comparing the results with the measurements of the Einstein Observatory, they found the brightness changes by no more than a factor of 2, i.e., there were no manifestations of cyclicity, when one could expect variations up to an order of magnitude. Apparently, the point is that in these young stars the long-term dynamo mechanism has not been established as yet, as on the Sun, and a small-scale mechanism without long cycles operates there. Stern et al. measured $F_X$ of 90% G dwarfs and from the constructed X-ray luminosity function estimated the average luminosity of G dwarfs in the Hyades as $\langle L_X \rangle = 10^{29}$ erg/s.

Randich and Schmitt (1995) obtained 42 images of the Praesepe cluster that covered the area of 4° × 4°. On the composite map of the cluster they achieved the limit of $L_X \sim 2 \cdot 10^{28}$ erg/s. Only 40 of 255 G–K–M members of the cluster were identified as X-ray sources, i.e., a percentage much lower than in the Hyades. This distinction is particularly explicit for G dwarfs: in the Hyades they all are identified as X-ray sources, whereas in the Praesepe — only 28%. This contradiction was removed by Franciosini et al. (2003), who observed the Praesepe with the EPIC mounted on XMM-Newton and detected all F–G stars of the cluster, about 90% of K dwarfs and more than 70% of M dwarfs. As a result, the X-ray luminosity function of solar-type stars appeared to be the same, as in the Hyades.

Prosser et al. (1996) studied X-ray emission of the members of the $\alpha$ Per cluster of the age of about 50 million years. About 80 of the 222 detected sources were identified with the known cluster members, whereas in the central part of the cluster all K and the majority of M dwarfs were identified as X-ray sources. In the sources with $v\sin i < 15$ km/s $L_X$ grows together with the velocity in the range $-4.3 < \log L_X/L_{\text{bol}} < -2.9$ and at about 50 km/s reaches saturation. A noticeable decrease of this ratio for G–K at v > 50 km/s found in this and some other young clusters was called supersaturation by Randich (1997). Probably this effect is associated with a change of the structure of coronal loops on the fastest rotators and with a slight shift of the emission beyond the sensitivity limits of the detector (Stern, 1999; Jeffries, 1999). F and G dwarfs of the $\alpha$ Per cluster have greater $L_X$ than similar stars in the Pleiades, for K dwarfs the distinctions are less and for M dwarfs of both clusters the distribution functions of the X-ray luminosities are very close.

Using the ROSAT observations, Randich et al. (1996b) studied the Coma cluster and found that X-ray properties of the cluster members were similar to those in the Hyades and differed noticeably from those in the Praesepe, although all the three clusters are of almost the same age: in the Coma, as in the Hyades, almost all late F and G stars were recorded, whereas in the Praesepe such dwarfs had lower luminosity. The Coma cluster is distinguished by the deficit of K and later stars. The X-ray observations revealed 12 new members of the cluster.

Jeffries et al. (1997) carried out a ROSAT PSPC survey of the field containing the young open cluster NGC 2516, whose age is $1.1 \cdot 10^8$ years. Only 6 of 159 sources found in the range of 0.5–2.0 keV were identified within 0.1–0.4 keV. As many as 65 sources were reliably or hypothetically identified with the known members of the cluster, whereas most of the others were apparently low-mass cluster members. The maximum of the ratio $L_X/L_{bol} \sim 10^{-3}$ falls on the late G and early K dwarfs. The reduced metallicity of the stars of this cluster results in qualitative distinctions in the X-ray luminosity distribution functions from other clusters. Then, Micela et al. (2000) studied this cluster from the ROSAT HRI observations. They found 12 new X-ray sources and identified 12 sources as K and M dwarfs, probable cluster members. Comparing the X-ray luminosities of red dwarfs of this cluster and the Pleiades, they concluded that the distinctions in metallicity did not influence within a factor of 2 the saturation level of X-ray emission.

Jeffries and Tolley (1998) identified most of the 102 X-ray sources with weak stars, members of the young but relatively far open cluster NGC 2547 and mentioned that X-ray observations had certain advantages for detecting weak cluster members.

Using the ROSAT PSPC data, Hünsch et al. (2003) investigated the open clusters NGC 2451 A and B and in each of them identified tens of active stars.

Micela et al. (1999b) used the ROSAT HRI data to study the X-ray emission of stars in the region of the open cluster Blanco 1 placed high above the Galaxy plane. One could expect that it differed from the other stellar clusters. From the number of X-ray sources identified with the cluster members they determined that the cluster was rather young, and from the $L_X$ of G–K dwarfs its age seemed to be closer to that of the Pleiades rather than to the $\alpha$ Per cluster. Many low-mass cluster members were found among the X-ray sources as well. Pillitteri et al. (2003) continued this study up to M dwarfs. They specified the affiliation of X-ray sources to this cluster, constructed X-ray luminosity functions, compared them with the appropriate functions of the Pleiades, NGC 2516, and $\alpha$ Per cluster, and proved the increased metallicity of Blanco 1 resulting in the increased X-ray luminosity, probably at the expense of coronal emission lines.

Franciosini et al. (2000) used ROSAT HRI to study the open cluster NGC 3532, whose age is intermediate between the ages of the Pleiades and the Hyades. They found about 50 X-ray sources with a luminosity of above $4 \cdot 10^{28}$ erg/s, 15 of them were identified optically.

From the ROSAT PSPC/HRI data Barbera et al. (2002) investigated the open cluster NGC 2422, whose age is close to that of the Pleiades. They found 78 sources, 62 of them were identified with optical objects, of which 80% were late stars. The X-ray luminosity function for F–K stars of this cluster was indistinguishable from that of the Pleiades.

Fleming (1998) noted that X-ray observations could reveal unknown nearest stars. As a rule, the lists of the nearest stars were compiled on the basis of observations of proper motions; but young dwarfs have low velocities, thus, these objects are underrepresented in the lists. From the RASS data Fleming identified 54 M dwarfs within 25 pc. Earlier, near the north pole of the Galaxy 238 X-ray sources were found within the RASS program. Using wide-angle optical images, Richter et al. (1995) discovered five flare stars.

Using the ROSAT PSPC data, Pizzolato et al. (2002) revised the “activity–rotation” relation established from the Einstein Observatory measurements. They considered 115 field stars and 136 stars in the Pleiades, the Hyades, IC 2602, and $\alpha$ Per clusters in the range of $0.5^m < B-V < 2.0^m$ with the known rotational periods and confirmed the existence of two X-ray radiation regimes: in one of them X-ray luminosity $L_X \sim P_{rot}^{-2}$, in the other the saturation regime is achieved when $L_X/L_{bol} \sim 10^{-3}$ and $L_X$ varies from $5 \cdot 10^{30}$ erg/s for $B-V = 0.65^m$ to $10^{29}$ erg/s for $B-V = 1.5^m$.

Golub (1983) extrapolated the relations between the X-ray emission intensity of individual loops and the strength of the appropriate magnetic field established for the Sun to M dwarfs. Already from the first X-ray observations of red dwarfs he concluded that the observed X-ray emission of such stars should correspond to magnetic fields with the strength analogous to that in active regions of the Sun, and covering almost the whole stellar surface, or to the stronger fields covering a smaller area of the stellar surface. On the basis of the subsequent observations Saar and Schrijver (1987) found a close correlation between the average magnetic fluxes $\langle fB \rangle$ from G–K–M dwarfs and soft X-ray fluxes from them. The Sun satisfied this correlation. With allowance for Saar's refinement (2001), this correlation is close to linear:

$$\langle F_X \rangle = 6100 \langle fB \rangle^{0.95 \pm 0.05}. \quad (22)$$

Combining 22 and 12 refined by Schrijver et al. (1992), we obtain

$$\langle \Delta F_{HK} \rangle = 0.06 \langle fB \rangle^{0.62 \pm 0.14}, \quad (23)$$

which is valid up to $\langle fB \rangle \sim 300$ G.

* * *

Schmitt and Liefke (2004) published NEXXUS (Nearby X-ray and XUV-emitting Star database) — the catalog of ROSAT observations (PSPC, HRI, and WFC) of close stars with X-ray emission, full for F–G stars up to 14 pc, K stars — up to 12 pc, and M stars up to 6 pc. All these stars have indistinguishable distributions of surface fluxes up to $F_X = 10^4$ erg/cm$^2$, which corresponds to solar coronal holes. High amplitudes of X-ray flux variations are untypical of solar-type stars, but more typical of lower main-sequence stars.

* * *

On the border of centuries, NASA and ESA launched into the orbit the new generation X-ray telescopes Chandra and XMM-Newton, which due to their high sensitivity and high spectral resolution have opened a new page in the X-ray astronomy.

Chandra contains four grazing-incidence X-ray telescopes with the first-type mirrors Wolter with outer diameters up to 123 cm. This optics feeds the field CCD-spectrometer ASIC (Advance CCD Imaging Spectrometer), low- and high-energy spectrometers with transparent diffraction gratings LETGS and HETGS (Low and High Energy Transmission Grating Spectrometers) and high-resolution cameras (HRC). ACIS makes it possible to derive the high-resolution images and middle-resolution spectra simultaneously. HETGS operates in the range of 0.4–10 keV, LETGS — in the range of 0.08–0.2 keV. Both devices have a spectral resolution of about 1000. HRC is a successor of the HRI devices mounted on the Einstein Observatory and ROSAT, its field of view is 30′ × 30′. Chandra has a limiting magnitude that is by $2^m$ better than those of ROSAT and ASCA (see below) and an unprecedented spatial resolution of 0.5″. LETGS provides an opportunity to record simultaneously the triplets of OVII of about 22 Å, NVI of about 29 Å, and CV of about 41 Å, which are widely applied during the diagnostics of coronal plasma; in the region of > 100 Å the FeXXII lines may be used for this purpose.

The X-ray multimirror apparatus XMM-Newton is intended for observations in the range of 0.2–15 keV. The Observatory consists of three modules; each of them comprises an X-ray telescope made of 58 coaxial gold-covered mirrors with a maximum diameter of 70 cm, with a resolution of 4″ in the field center. Next to the two mirror modules the identical spectrometers (RGS) are mounted, which capture about a half of the rays from converging beams and plot spectra in the range of 0.3–2.1 keV with a resolution of 100–500. The rest half of beam rays, passing by the gratings of spectrometers, plot the direct images in the EPIC MOS devices (European Photon Imaging Camera) on the mosaics from seven MOS CCD, which overlap a field with a diameter of 0.5°. In the third module, the detector system pn-CCD with a size of 6 × 6 cm$^2$ with fast readout, sensitive to 15 keV, is mounted at the telescope focus; its resolution is about 4″. In addition to X-ray telescopes, XMM-Newton carries the 30-cm Ritchey-Chretien system intended for simultaneous observations of objects in the range of 1700–6500 Å in the central 17 square minutes of the field of view of X-ray devices. This XMM-OM system is equipped with a high-speed photometer with broadband filters and a prism to derive low-resolution spectra.

Each of these devices is essentially a modern X-ray observatory, and the high-class data on stellar flares were obtained with them. These results will be presented in Part 2. As to the X-ray radiation of the considered stars in the quiescent state, the acquired data are comparatively sparse and as follows.

Using the Chandra/LETGS observations, Audard et al. (2003) split components of the L 726–8 system in the X-ray range, and with the HRC-S camera acquired in the zero order an image of the system in a perfect agreement with the optical ephemeris.

During a 25 ks Chandra observation, Stelzer (2004) for the first time separated the binary brown dwarf Gl 569 Bab in X-rays, based on Gl 569 A estimated the age of the pair as 300–800 million years, and first measured the quiet X-rays, starting from an old brown dwarf. About a half of the exposure was occupied by a large flare.

Wood and Linsky (2006) first separated 70 Oph (K0+K5) and 36 Oph (K1+K1) into components and performed their cooperative analysis with $\varepsilon$ Eri. Densities and temperatures determined from the ratios of OVII and DEM lines proved to be very close for all five K dwarfs, but the conspicuous and unexpected distinctions were revealed in the abundance of elements. If the low FIP elements were enhanced with respect to the high FIP in the solar corona, and the increased solar-type FIP effect was detected for 70 Oph A, then the effect was absent or there was a weak inverse effect for 70 Oph B. This distinction, at all other parameters being similar, remains enigmatic. Wood et al. (2012) studied the Chandra X-ray spectra of two most inactive M0 dwarfs — components of the GJ 338 system, in both cases they found a medium inverse FIP effect.

Briggs and Pye (2003) derived an image of the Pleiades with a 40-ks XMM–Newton EPIC exposure and succeeded in studying objects by an order of magnitude weaker than those available with ROSAT/PSPC. They considered two groups of stars: solar-type F5–K8 and intermediate-type B4–F4. All eight objects of the former group had $L_X >\sim 10^{29}$ erg/s, which corresponds to the known rotation–activity correlation, and flares were detected on four of them. At the onset of a flare, there occurred an eclipse of the flare loop in HII 1100 by either a small satellite or a high-density prominence-type structure. Within the hydrodynamic model, they estimated a half-length of loops in two strongest flares of HII 1032 and HII 1100 as $< 0.5\ R_*$. This group revealed tendencies for growing coronal temperature and low metallicity from F and slowly rotating G stars to K and rapidly rotating stars.

Based on the XMM-Newton observations Judge et al. (2004) found that the surface X-ray flux of the extremely low-active dwarf G8 $\tau$ Cet in the range of 0.1–2.4 keV was several times lower than that of the Sun.

Pillitteri et al. (2004) carried out the XMM-Newton/ EPIC observations of the members of the high-metallicity open cluster Blanco 1, recorded all known F and G dwarfs in it, 80% and 90% of K and M dwarfs, respectively. Models of coronae with temperatures varying from 0.3 to 1 keV explained the quiescent X-ray radiation of the brightest members of the cluster, which underwent spectral and colorimetric X-ray analysis.

Continuing the described above studies of the NGC 2516 cluster by Jeffries et al. (1997), Pillitteri et al. (2006) studied this cluster with XMM-Newton. Within the EPIC data, they constructed a one-temperature (1T) and two-temperature (2T) models of coronae of the brightest stars. There were recorded 431 X-ray sources, 234 of them were identified with stars distributed over the main sequence with luminosity $\log L_X$ varying from 28.4 to 30.8 and coronal temperatures 3.5–8 MK of the cool and 12–23 MK of the hot components. At the absence of conspicuous distinctions in spectra, the solar-type stars in NGC 2516 definitely had lower luminosity than that in the Pleiades cluster, which is close in age. A comparison with the ROSAT data revealed a lack of amplitudes more than 2 for solar-type stars on the 11-year time scale, which implies the absence of such a cycle in young stars.

From the XMM-Newton observations, Scelsi et al. (2005) compared the coronal properties of three G stars of different age: WTT stars HD 283572, EK Dra, and the giant 31 Com in the Hertzsprung gap. All three have high X-ray luminosity: $10^{31}$ erg/s for the first and third stars and $10^{30}$ erg/s for the second one. Observations of the EM distribution were interpreted as an ensemble of loop structures with different apex temperatures. Coronae of the first and third stars were very similar in dominating coronal magnetic structures, despite distinctions in evolutionary phases, surface gravity, and metallicity. The EM distribution was flatter at the same apex temperature for EK Dra.

Using the XMM-Newton and Chandra observations, Grosso et al. (2007) studied the young brown dwarfs of the spectral type M at the age of about 3 million years in the molecular cloud in Taurus. Out of 17 program objects, 9 were detected, and a flare was recorded on one of them. They confirmed the mean ratio $\langle \log(L_X/L_{bol}) \rangle = 3.5$ but found no correlation of this ratio with the equivalent width of the $H_\alpha$ line. Accreting and non-accreting brown dwarfs revealed similar ratios $L_X/L_{bol}$, while the median value of this ratio was four times lower than the average saturation level of the fast rotators of the low-mass field stars. Brown dwarfs in Taurus detected higher ratios $L_X/L_{bol}$ than those in Orion.

Jardine et al. (2004) summarized investigations on the coronal structure of the solar-type stars, combining the results of constructing surface magnetograms within the ZDI technique and the results of their Chandra and XMM-Newton X-ray observations. They found that at high rotation rates the corona structure of such stars differed significantly from that of the Sun. Such stars are more spotted, and spots appear at high latitudes up to the poles. Their coronae are far brighter in X-rays and contain denser and hotter plasma than that on the Sun. Their coronae can maintain massive prominences at heights up to many stellar radii. Coronae contain the complex loop structures of different sizes, and a substantial fraction of X-rays emanate from high latitudes. High densities of the corona and EM in X-rays are an essential consequence of high density of magnetic fluxes on the stellar surface. A release of the corona matter by the centrifugal force due to high rotation rates can interpret the saturation and supersaturation of X-rays and, consequently, the rotational modulation of supersaturated stars.

Veronig et al. (2021) suggested estimating stellar CMEs on the dimming of EUV fluxes.

* * *

In the conclusion of this section, it is worth noting that one of the definitions of the onset of solar activity from the side of hot stars belongs to X-ray observations. Using the XMM-Newton observations of the A7 star Altair, Robrade and Schmitt (2009b) found that its $L_X/L_{bol} = -7.4$ and low-density "cold" plasma prevailed with a temperature of 1–4 MK and the solar-type FIP effect. The X-ray radiation level of the star varied by 30% due to rotational modulation and low activity without strong flares, it was close to saturation and by approximately four orders of magnitude lower than that for later stars.

## 1.4.2. EUV and X-ray Spectroscopy

In Subsect. 1.3.1.4 devoted to the ultraviolet spectra of the chromosphere and transition zone, the EUVE apparatus was briefly described, which enabled direct images and the first spectra in the intermediate region between UV and X-rays to be obtained and gave the results of EUVE multicolor photometry of the objects of the considered activity type. According to Mathioudakis et al. (1995a), radiative losses in EUV are comparable with the losses in the X-ray region, while a considerable part of the radiative losses of low-activity dwarfs falls exactly in the EUV range. Below, we present the results of EUVE spectral observations of such objects: the observation program included the objects discovered in the course of the earlier WFC and EUVE sky surveys in EUV. The Atlas by Craig et al. (1997) contains the EUV spectra of four G dwarfs ($\kappa$ Cet, $\chi^1$ Ori, $\alpha$ Cen AB, and $\xi$ Boo), six K dwarfs (GJ 117, $\varepsilon$ Eri, BF Lyn, LQ Hya, GJ 702 AB, and VW Cep), and ten M dwarfs (YY Gem, YZ CMi, AD Leo, Prox Cen, GJ 644, AT Mic, AU Mic, FK Aqr, EV Lac, and EQ Peg).

The system $\alpha$ Cen (G2V+K1V) was one of the first observed in all four photometric EUVE bands, these measurements led to the conclusion on the radiation of two components: $T_1 = 8.5 \cdot 10^5$ K, $EM_1 = 1.5 \cdot 10^{50}$ cm$^{-3}$, and $T_2 = 10^5$ K, $EM_2 = 5 \cdot 10^{49}$ cm (Vedder et al., 1993). However, the spectral lines in the EUV range give more reliable estimates of the temperature of radiating plasma, and Mewe et al. (1995a) carried out spectral observations of the system by EUVE. The recorded spectrum of $\alpha$ Cen was presented by a linear combination of isothermal plasma structures: the hot component with a temperature of about 3 MK, less hot with a temperature of 0.1 MK, and probably a very hot – with a temperature of several tens of megakelvins. The spectral lines of the FeX, XII, XIII, and XIV ions sensitive to the electron density yielded estimates of $(2-20) \cdot 10^8$ cm$^{-3}$ for a temperature of 1–2MK. Later, Drake et al. (1997) found that the DEM of this system had a minimum at $\log T = 5.5$ and a maximum at $\log T = 6.3$. Analyzing the abundances of the elements in the corona of $\alpha$ Cen AB, they established that the chemical composition of the corona differed from that of the photosphere, and, as on the Sun, the elements with low FIP were enriched by almost a half as compared to the elements with high FIP.

Monsignori Fossi and Landini developed a numerical method for analyzing differential emission measure (DEM). Using the method, they presented EXOSAT and IUE observations of AU Mic. Using the DEM model, they calculated the expected spectra of the star for the wide range of short-wavelength (70–190 Å) and medium-wavelength (140–380 Å) EUVE spectrographs and compared them with the observations. The lines of high-ionized ions of iron FeXIII–FeXXVI, OIV and HeII Ly$_\alpha$ dominated in these spectra. Using EUVE observations, they revised the DEM model and obtained the following parameters of the AU Mic corona in the quiescent state: $L_{EUV} = 4 \cdot 10^{29}$ erg/s, $EM = 6 \cdot 10^{51}$ cm$^{-3}$, and from two pairs of FeXXI and FeXXII lines sensitive to the electron density, the upper limit of the value was estimated as $n_e < 10^{12}$ cm$^{-3}$ (Monsignori Fossi and Landini, 1994; 1996; Monsignori Fossi et al. 1998). Monsignori Fossi et al. (1995a) similarly analyzed the flare star AT Mic using the EUVE and IUE data. In the quiescent state, the lines of high-ionized iron ions prevailed, while the ions of FeXVIII and lower ionization stages were presented more poorly. The DEM function had a minimum near $10^6$ K and a maximum near $10^7$ K. Similar results were obtained in analyzing another binary flare star EQ Peg: 13 iron ion lines up to FeXXIV were identified in the wavelength range of 94–360 Å and the same extremes of the DEM function were found (Monsignori et al., 1995b).

Schmitt et al. (1996b) analyzed the EUV spectrum of the K dwarf $\varepsilon$ Eri and found the emission lines of iron from FeIX to FeXXI with a maximum intensity in FeXV–FeXVI. From these iron lines they constructed DEM, which displayed a wide maximum; such a DEM was presented within the coronal model with two kinds of magnetic loops with maximum temperatures about 3 and 7 MK. From the FeXIII and FeXIV lines they estimated the electron densities, which were similar to those in the active regions of the Sun. With account of the found densities and emission measures they estimated the size of coronal loops as $3 \cdot 10^9$ cm and the filling factor as $f \sim 0.9$.

Landi et al. (1997) analyzed the EUV spectrum of $\kappa^1$ Cet, identified the iron lines from FeIX to FeXVIII, and, incorporating the IUE data, constructed DEM, whose maximum was about 3 MK, as in solar-active regions.

Analyzing the constraints on the stellar corona parameters observed using EUVE, van den Oord et al. (1997) found that the minimum expansion of magnetic loops from the base to the top was from 2 to 5, the coronae of $\alpha$ Cen should be composed of the loops of two different kinds, whereas for $\chi^1$ Ori loops of one kind were sufficient.

Summing up the results of three years of EUVE operation, Drake (1996) noted that the studies from this apparatus yielded the following conclusions on the general character of DEM: within $T \sim 10^4$–$10^5$ K there was a decline of the function, from $10^5$ to $10^6$ K there was a poorly determined minimum, then a rise started, which achieved a maximum in the range of $2 \cdot 10^6$–$10^7$ K depending on the general activity level of the star (see Fig. 25). Thus, the two-temperature corona model is the approximation of the structure with a continuous but not monotonic DEM($T$) distribution. Further, of principal significance were direct spectroscopic estimates of electron density in stellar coronae: the iron ion lines from FeXII to FeXXII in the wavelength range from 91 to 327 Å allowed an estimate of $n_e$ as $10^9$–$10^{15}$ cm$^{-3}$ at $2 \cdot 10^6$–$10^7$ K (Brown, 1994). The estimates of $n_e$ provided a considerable advance in the analysis of the inner structure of coronae, since in combination with the measured EM they allowed for the estimates of the characteristic volume of luminous matter, which, if a certain plausible geometry is specified, makes it possible to estimate the filling factor by the loop structure and the characteristic strength of magnetic fields. Thus, if from the analysis of FeXIX–XXII lines one assumes $n_e \sim 10^{13}$ cm$^{-3}$, compact stationary loops with $h \sim 100-1000$ km and $B \sim 1$ kG appear at $T \sim 10^7$ K, which substantially differ from the solar situation, where $n_e \sim 10^9-10^{11}$ cm$^{-3}$, $T \sim 2 \cdot 10^6$ K, $h \sim 20000$ km, and $B \sim 10$ G. For intermediate-activity stars, $\varepsilon$ Eri and $\xi$ Boo A, $n_e$ estimates from FeXIII and XIV lines yielded $10^{10}$ cm$^{-3}$. For the even less active pair $\alpha$ Cen AB from the FeIX–XIV lines $n_e \sim 10^9$ cm$^{-3}$, as in active solar regions. Probably the highest density estimates correspond to filaments inside the coronal loops. Thus, the EUVE results suggested that the growth of the X-ray flux for more active stars was due to the increase of the characteristic density rather than the volume of the luminous matter.

The first studies of the abundances of heavy elements in stellar coronae were started at EUVE to find an analogy with the solar so-called FIP effect. The effect consists in the increased abundance of the elements with first ionization potential (FIP) below 10 eV with respect to that of the elements with higher first ionization potentials. Laming et al. (1996) suspected this effect for $\varepsilon$ Eri; but although the star is noticeably more active than the Sun, the value of the FIP effect in it definitely did not exceed the solar value. In the coronae of $\alpha$ Cen AB system, Drake et al. (1997) for the first time reliably detected an analog of the solar FIP effect with a twice higher abundance of elements with low FIP than the elements with high FIP. Later, a distinct solar-type FIP effect was found by Laming and Drake (1999) and Drake and Kashyap (2001) in the corona of $\xi$ Boo A, whose activity level is higher by an order of magnitude than that of the Sun. Earlier, Kashyap, Giampapa, and Drake (2000) analyzed the EUVE/SW data of two inactive M dwarfs GJ 205 and GJ 411; from the DEM distribution they found the temperature peak to be the same as that for the quiet Sun and obtained data on the deficiency of metals in the corona of GJ 205 as compared to its photosphere.

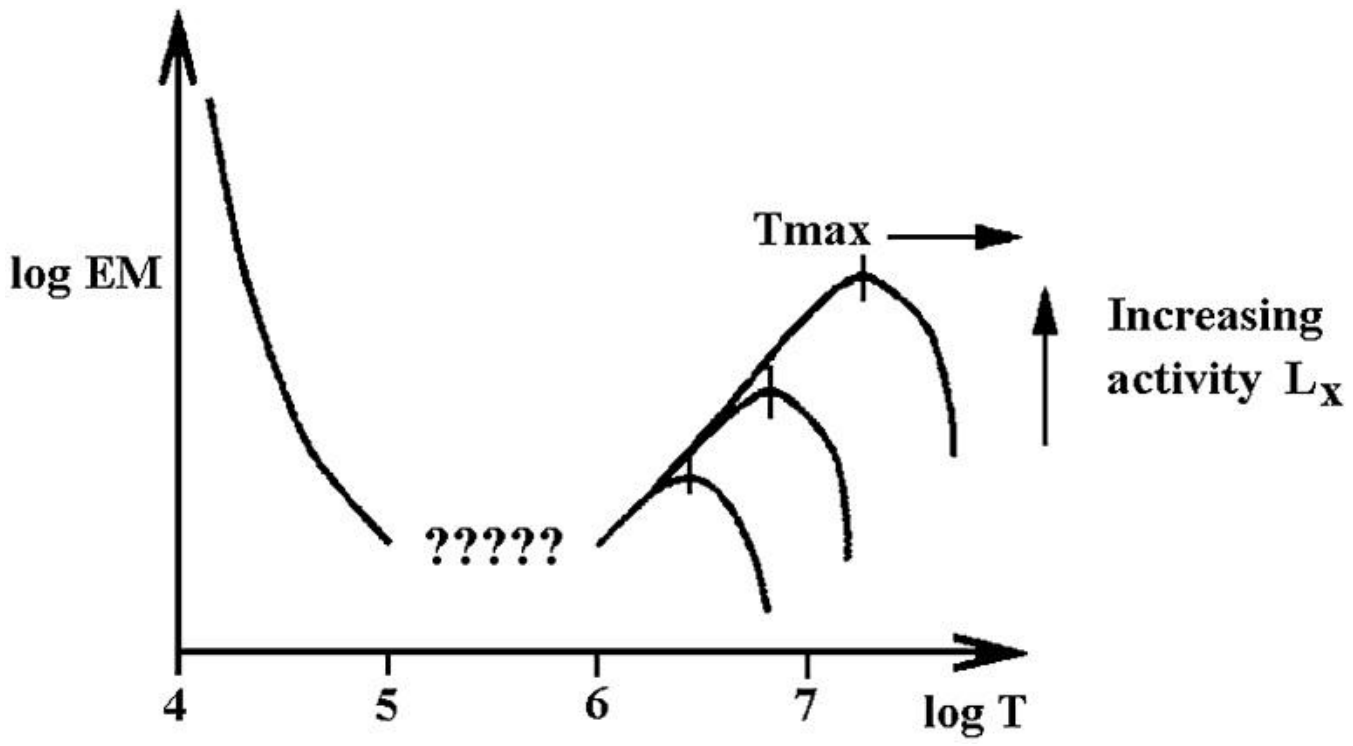

Fig. 25. Schematic dependence of the differential emission measure of stellar coronae upon the temperature and activity level (Drake, 1996)

Drake et al. (2000) compared EUVE spectral observations of $\varepsilon$ Eri (K2V) and $\xi$ Boo A (G8V) with Yohkoh solar observations near the maximum of its cycle on 6 January 1992 to find the solar coronal structures with EM($T$) corresponding to the coronae of these intermediate-activity stars. The intermediate status is illustrated by the following figures: solar $L_X$ is $2 \cdot 10^{26}$ erg/s at the activity minimum and $5 \cdot 10^{29}$ erg/s at the maximum, for the most active solar-type dwarfs $L_X > 10^{30}$ erg/s, whereas $L_X(\varepsilon$ Eri$) = 3 \cdot 10^{28}$ and $L_X(\xi$ Boo A$) = 9 \cdot 10^{28}$ erg/s, respectively. Drake et al. constructed the EM($T$) dependences for the whole solar disk and for the sections limited by isophotes 1/100 and 1/10 from the maximum brightness on the disk and found that these curves differed by maximum EM values, the positions of the maxima on the temperature scale, and the slopes of the curves beyond the maxima: if one approximates EM $\sim T^{\beta}$, then $\beta$ = 1.1, 1.6, and 3.8 for the listed EM($T$) dependences. The EM($T$) curve of the brightest solar regions is the most similar to the curves of the considered stars, but its maximum is approximately 70 times lower. Since the electron density in the coronae of these stars and the brightest solar-active regions was estimated by close values, Drake et al. concluded that to achieve the activity level corresponding to that of $\varepsilon$ Eri and $\xi$ Boo A, a filling factor for solar-active regions approaching unity should be sufficient. But an even higher activity level requires either an increase in the density of the coronal plasma, or an increase of its radial extension, i.e., the formation of structures differing from solar-active regions. The known increase in the coronal temperature with increasing $L_X$ argues in favor of such qualitative changes. Probably at $f \sim 1$ a new mechanism of flaring activity is launched due to the interaction of neighboring active regions, which engages stronger and higher-temperature processes than the flare activity in separate active regions on the Sun.

For complete quantitative analysis of the EUVE data, Drake (1999) performed detailed computations of the sensitivity of all systems of the apparatus to the radiation of optically thin astrophysical plasma within the temperature range of $10^4$–$10^8$ K.

* * *

The Japanese–American X-ray satellite ASCA (Advanced Satellite for Cosmology and Astrophysics) was orbited in February 1993 with a period of 96 min. The satellite carried 4 identical X-ray grazing-incidence telescopes designed to operate within the range of 0.5–10 keV and had an angular resolution of 1–3; at the focus of the telescopes 2 solid-state spectrometers (SIS) were installed with a spectral resolution of 50 and a field of view $20' \times 20'$ and two gas imaging spectrometers (GIS) with a spectral resolution of 13 and a diameter of the field of view of 50. The main advantage of ASCA over its predecessors was the possibility of studying the region of higher energies and higher effective area.

Drake S.A. et al. (1994) observed the active G1.5 dwarf $\pi^1$ UMa with SIS ASCA and found stable radiation at the level of $\log L_X = 28.7$. The analysis of the obtained spectra showed that isothermal RS and MEKA models could not represent the observations even with varying chemical composition but the two-temperature MEKA model with $T_1 = 4.0$ and $T_2 = 7.4$ MK and $EM_1 = EM_2 = 2 \cdot 10^{51}$ cm$^{-3}$ with noticeably reduced abundance of N, O, Ne, Si, S, and Fe yielded a good representation of observations.

Gotthelf et al. (1994) observed the $\alpha$ Gem system using SIS ASCA. The developed algorithm of image restoration made it possible to separate the radiation of YY Gem from that of Castor AB, then the quiescent radiation of YY Gem was presented as a four-temperature coronal model; the two-temperature model could also yield a good representation of observations, but only if the abundances of heavy elements were decreased by a factor of 5–10 as compared to their content in the solar photosphere.

The ASCA X-ray spectrum of $\alpha$ Cen AB was analyzed by Mewe et al. (1998a) within the concept of multitemperature corona and using DEM. It appeared that the corona was close to isothermal with $T \sim 3.5 \cdot 10^6$ K. The abundance of neon, silicon, and iron in it was close to that in the solar atmosphere, but there was a deficiency of oxygen and a several-fold excess of magnesium. When the system was observed by the BeppoSAX/LECS apparatus, which enabled simultaneous investigation of the range from 0.1 to 10 keV with rather high spectral resolution, Mewe et al. (1998b) found a two-temperature corona, whose cold component was recorded before by EUVE and ROSAT, and the hot one by the Einstein Observatory, EUVE, and ASCA. The established abundance of iron corresponded to the solar photosphere, but differed from the $\alpha$ Cen photosphere enriched by metals. However, there is no confidence that the hot component was not due to flare activity.

Using ASCA, Singh et al. (1999) observed two very fast rotators, HD 197890 and Gl 890. Analysis of these data and the observations of YY Gem showed that at least two temperature components should be available at each of these sources. In the YY Gem coronae the content of metals was definitely reduced, in two others it was preferential. Considering ROSAT and ASCA results in the general sample of 17 G–K–M dwarfs, the authors confirmed the clear correlation of $L_X/L_{bol}$ with the rotational period and the Rossby number. In the subsample of 10 stars observed from ASCA, they found that the temperature of the hot component correlated with the ratio. For all active dwarfs with $\log(L_X/L_{bol}) > -3.7$ the iron abundance was noticeably lower than the solar value, whereas for less-active stars it differed by no more than a factor of 2.

Using EUVE and ASCA, Gagné et al. (1999) studied the active corona of the F8 dwarf HD 35850. In the short-wavelength and medium-wavelength ranges of EUVE they identified 28 iron lines from FeIX to FeXXIV. From FeXXI lines they estimated the upper limit of the electron density $\log n_e < 11.6$. In the short-wavelength range they confidently discovered a continuum. In the obtained DEM curve one can see two distinct maxima at $\log T = 6.8$ and 7.4. During one-week observations no strong and long flares were recorded on the star, but there were traces of weak flares and probably rotational modulation. In comparing the nonsimultaneous EUVE and SIS ASCA spectra they found a similarity of the obtained DEM curves, but the ASCA spectra required stellar corona noticeably depleted in heavy elements. The surface X-ray fluxes from HD 35850 are comparable with those from EK Dra, and at an age of about 100 million years, vsin$i$ ~ 50 km/s, and high $L_X$ this star can be considered as an extremely active main-sequence F dwarf.

Favata et al. (2000a) analyzed X-ray observations of the flare star AD Leo with Einstein IPC, ROSAT PSPC, and ASCA SIS, and concluded that in six strong X-ray flares recorded by these instruments magnetic loops, where the flares were developed, had similar parameters. Hence, they accepted that such loops were typical of the quiescent corona as well. This suggests that the parameters of these magnetic loops corresponded to the compact solar loops with a 10-fold increase of pressure, 3-times higher loop diameter-to-height ratio, and a low surface filling factor. This conclusion is an alternative to the suggestion that an increase in $L_X$ of active stars occurs due to an increase of the filling factor. However, in this case one can expect an increase in $L_X$ only by 1.5–2 orders of magnitude, whereas for the most active dwarfs it should be increased by 3–4 orders. Considering the above results of Ventura et al. (1998) on two different magnetic-field structures, one can assume that the growth of the filling factor has a decisive role for the stars with low and medium activity, whereas for the most active stars the most important is the growth of the loop density. This conclusion coincides with that by Drake et al. (2000) on the structure of coronae of $\varepsilon$ Eri and $\xi$ Boo A obtained from EUVE data. According to the estimates of Favata et al. (2000a), thick-section magnetic loops of the AD Leo corona are characterized by a density of $5 \cdot 10^{10}$ cm$^{-3}$, a pressure of 70 dyn/cm$^2$, and a height $h \sim 0.3R_*$, and at $f \sim 0.06$ they yield EM ~ $2 \cdot 10^{51}$ cm$^{-3}$. From the spectra obtained by ASCA SIS Favata et al. constructed two-temperature models of the corona of AD Leo both for the solar abundance of elements and for varying compositions; in the second case $T_1 = 5.6 \cdot 10^6$ and $T_2 = 10^7$ K, EM$_1 = 1.6 \cdot 10^{51}$ and EM$_2 = 2.3 \cdot 10^{51}$ cm$^{-3}$ at the deficiency of [Fe/H] = −0.7 and [O/H] = [Si/H] = +0.3.

Covino et al. (2001) analyzed the corona of the young K2 star LQ Hya from the ROSAT PSPC (November 1992) and ASCA (May 1993) observations. The former were presented by the 1T MEKAL model with $T \sim 8$ MK, EM ~ $1.2 \cdot 10^{53}$ cm$^{-3}$, and metallicity $z/z_0 \sim 0.09$. The latter were presented by the 2T MEKAL model with $T$ = 9 and 15 MK, EM = $4.4 \cdot 10^{52}$ cm$^{-3}$ and $3.2 \cdot 10^{52}$ cm$^{-3}$, and $z/z_0 \sim 0.13$; whereas in the first half of these observations the star was very stable and in the second half showed fluctuations up to 50% and slow brightness variations with high amplitudes were associated with EM variations at practically constant temperature. (The strong flare recorded with PSPC on November 5, 1992 will be considered in Part 2.) In December 2000, the observations of this star were carried out at the BeppoSAX satellite (see below), they continued for 280 ks and allowed short variations to be considered, but flares were not recorded. These observations were presented within the 3T model with significantly reduced metallicity, as at previous epochs. Metallicity of the photosphere of this star was specially determined from the high-dispersion optical spectrum derived with the 1.5-meter telescope at ESO, and it proved to be somewhat higher than that of the Sun.

As to the above studies dealing with chemical anomalies of stellar coronae, it should be noted that Drake (1998) assumed that the observed effects could be due to the enrichment of coronae by helium. However, in 2001, in the beginning of the Chandra and XMM-Newton era, while surveying the chemical composition of stellar coronae, Drake (2002) did not mention this concept.

* * *

In April 1996, the Italian–Dutch X-ray Astronomy Satellite BeppoSAX was orbited. The satellite had a uniquely wide spectral coverage ranging from 0.1 to 300 keV and the wide-field cameras designed for studying the variability at long time intervals and detecting nonstationary phenomena. There were four X-ray concentrator spectrometers with apertures of 124 cm$^2$ each and proportional counters in the focal planes: a low-energy concentrator spectrometer (LECS) designed for the range of 0.1–10 keV and three medium-energy concentrator spectrometers (MECS) for the range of 1.3–10 keV, a high-pressure gas scintillation proportional counter for the range of 4–20 keV, and a proswich detection system (PDS) for the range of 15–300 keV that contained four scintillators with a total area of about 800 cm$^2$. The wide-field cameras were designed for the range of 1.8–28 keV, had a fields of view of 20°, angular resolution of 5′ and spectral resolution of 5 near 6 keV.

LECS and MECS instruments mounted on the BeppoSAX satellite recorded the spectra of the active binary system VY Ari and the flare star AD Leo within the range from 0.1 to 10 keV (Favata, 1998). The spectrum of VY Ari was presented by the two-temperature model with $T_1 = 1 \cdot 10^7$ and $T_2 = 2.4 \cdot 10^7$ K, the ratio $EM_1/EM_2 = 0.44$, and the iron abundance lower by a factor of 2.5 than that on the Sun. The preliminary analysis of the spectra of AD Leo yielded the two-temperature model of the quiescent stellar corona that is quite close to the model constructed using the ASCA data (see above).

Sciortino et al. (1999) observed AD Leo and EV Lac with BeppoSAX and compared the results with the ROSAT PSPC data. The ROSAT spectrum of EV Lac and AD Leo were presented by the one-component isothermal MEKAL model and by the two-component MEKAL model, respectively, with the metallicity of both stars lower by an order of magnitude than the solar value. For the BeppoSAX data, the two-component model was insufficient for any metallicity value, but the three-component model provided a satisfactory presentation of observations of AD Leo at noticeably reduced metallicity and the observations of EV Lac for a slightly reduced content of metals. The models with uniform magnetic loops were unsuitable, and the best correspondence was achieved in the model of two loop types with dominating hundreds of compact loops with relatively low maximum temperature, heights of about $0.1R_*$ and a filling factor of ~0.01 and tens of heavily elongated loops with very small filling factor that were responsible for the high-temperature radiation.

During 5 days in November 1998, Tagliaferri et al. (2001) observed the YY Gem system within the range of 0.1–10 keV. At two intervals out of flares they obtained the spectra of the quiescent state that were presented within the two-temperature MEKAL models with $T_1 = 7.7$ and 3.8 MK and $T_2 = 23$ and 16 MK at metallicities $z/z_0 = 0.2$ and 0.4, respectively. In addition, they suspected radiation within the range of 20–30 keV.

The total results of the BeppoSAX observations were reported by Pallavicini et al. (2000).

* * *

Fleming et al. (2003) observed M8 dwarf VB10 using Chandra ACIS and found a quiescent corona with an X-ray luminosity of $2.4 \cdot 10^{25}$ erg/s and the ratio $L_X/L_{bol} = -4.9$. These values comply with the fact that the star was discovered in X-rays only during the flare. The obtained luminosity is lower by two orders of magnitude than that of earlier M dwarfs. Nevertheless, the existence of corona on VB 10 contradicts the hypothesis on the absence of permanent coronae on such cool stars.

Ness et al. (2002) analyzed Chandra LETGS observations of 10 cool stars, whose spectra displayed clear qualitative distinctions: there is a strong continuum for the active $\varepsilon$ Eri and AD Leo in the range of 10–20 Å, which is absent on inactive $\alpha$ Cyg A and B. On the latter stars, the intensity of OVII lines (21.6 Å) exceeded that of OVIII lines (18.97 Å), whereas for $\varepsilon$ Eri and AD Leo the opposite relations were observed. Applying the theory of relative intensities to $Ly_\alpha$ and He-like triplets for various highly ionized atoms, Ness et al. carried out a systematic analysis of all appropriate emission lines. The following results were obtained:

- OVII lines: $\varepsilon$ Eri $n_e \sim 10^9$–$10^{10}$ cm$^{-3}$, $T = 2.2$–3.0 MK; AD Leo $n_e = 8 \cdot 10^9$ cm$^{-3}$, $T = 2.2$–3.5 MK; YY Gem $n_e = 2 \cdot 10^{10}$ cm$^{-3}$, $T = 2.2$– 3.6 MK;

- NVI lines: $\varepsilon$ Eri $n_e = 2 \cdot 10^{10}$ cm$^{-3}$, $T = 2.0$–$2.5$ MK; AD Leo $n_e = 2 \cdot 10^{10}$ cm$^{-3}$, $T = 2.0$–$2.9$ MK;
- CV lines: $\varepsilon$ Eri and AD Leo $T = 1.4$ MK;
- SiXIII lines: $\varepsilon$ Eri $n_e < 10^{14}$ cm$^{-3}$, $T = 8$ MK; AD Leo $n_e = 4 \cdot 10^{13}$ cm$^{-3}$, $T = 9$–$11$ MK, YY Gem $T = 12.5$ MK;
- MgXI lines: $\varepsilon$ Eri $n_e = 4 \cdot 10^{12}$ cm$^{-3}$, $T = 6$–$7$ MK; YY Gem $T = 8.7$ MK;
- NeIX lines: $\varepsilon$ Eri $n_e = 10^{11}$ cm$^{-3}$, $T = 3.3$–$4.4$ MK; AD Leo $n_e < 4 \cdot 10^{10}$ cm$^{-3}$, $T = 4.5$–$5.3$ MK; YY Gem $n_e < 2{\cdot}10^{11}$ cm$^{-3}$, $T = 4.7$–$5.6$ MK.

Apparently, the parameters for different ions of the same star were scattered mostly due to various temperatures of formation of the appropriate ions and coronal heterogeneity. Comparison of the ratios of intensities of the OVIII/OVII, NVII/NVI, and NeX/NeIX resonance lines with X-ray luminosity revealed a positive correlation: $L_X$ grows with an increase in these ratios.

Using Chandra ACIS, Tsuboi et al. (2002) discovered X-ray emission of brown medium-age dwarf TWA 5B, a companion of a T Tau-type star, at the level of $4 \cdot 10^{27}$ erg/s within 0.1–10 keV. They concluded that up to the age of $10^7$ years the saturation of X-ray emission occurred at the level of $L_X/L_{bol} \sim 10^{-3} - 10^{-4}$, then this ratio decreased to $10^{-5}$, and the coronal temperature started decreasing even earlier. The MEKAL model of this dwarf yielded $T$ = 3.5 MK, EM = $4 \cdot 10^{50}$ cm$^{-3}$, and $z/z_0 = 0.3$.

Hussain et al. (2005) carried out Chandra/LETG observations of AB Dor in the course of its two revolutions and recorded rotational modulation with three peaks in the X-ray light curve. Taking into account peculiarities of the OVIII, FeXVII, and FeXVIII line profiles, they constructed a two-component model of the quiescent corona: one component radiating 80% of the X-ray flux is a polar or uniformly distributed corona, the second component comprises two or three compact regions yielding the light-curve modulation and those located at different latitudes at a height of less than $0.3R_*$. Later, performing simultaneous AAT and Chandra observations of this star, from the ground-based spectra in the circularly polarized light Hussain et al. (2007) plotted a map of the surface magnetic field and, extrapolating it, constructed a model of the quiescent corona; the X-ray observations served as a test for feasibility of such a coronal model. Again, the obtained small height of the corona is determined by its high density and complex multipolar magnetic field on the stellar surface. The constructed model has a significant correlation of positions of surface and coronal active longitudes and the X-ray light curve. Contrary to earlier maps of the surface, in December 2002, AB Dor revealed one very large region of active longitudes overlapping almost a half of the star and manifesting in enhanced activity in the form of large spots, strong magnetic fields, and chromospheric emission.

With Chandra LETGS, Maggio et al. (2002) observed AD Leo in the range of 6–180 Å during 15-h observations out of strong flares. They identified most of the 110 detected emissions, including iron lines from XVI up to XXIII, C, N, O, Ne, Mg, Si, S, and Ni with formation temperatures varying from 0.6 to 15 MK. They constructed DEM overlapping this temperature range and from He-like triplets determined the electron density: $2 \cdot 10^{10}$ cm$^{-3}$ for $T = 1.5$ MK in OVII, $< 3 \cdot 10^{11}$ cm$^{-3}$ for $T = 4$ MK in NeIX and $1 \cdot 10^{14}$ cm$^{-3}$ for $T = 6$ MK in SiXIII.

Using Chandra ACIS, Preibisch and Zinnecker (2002) investigated the very young cluster IC 318, found 215 X-ray sources with masses varying from 0.15 to 2 solar masses, and identified them with 80% of the cluster members.

By means of Chandra HETGS and VLA, Brown et al. (2002) observed on 29/30 March 2001 active coronae in the short-period binary system ER Vul, composed of G0 and G5 dwarfs. The X-ray monitoring was run within 1.8–40 Å, radio monitoring was performed at 3.6 and 20 cm. The observations revealed a wide range of coronal temperatures from 2 (OVII) up to 30 (FeXXIV) MK with smooth variations of the level up to $A_X \sim 2$.

Raassen et al. (2003) were the first to obtain and analyze the spectrograms of each of the components of the $\alpha$ Cen system recorded by Chandra LETGS within the range 10–180 Å. From the ratios of the intensities of spectral lines they found that the corona of the K1 component was slightly hotter than that of the G2 dwarf. Then, using the 2T MEKAL program, they constructed the two-component models of the coronae from DEM of each component of the system and obtained the following parameters: for K1 star $T_1 = 1.2$ and $T_2 =$ 2.2 MK, $EM_1 = 1.0 \cdot 10^{49}$ cm$^{-3}$ and $EM_2 = 1.9 \cdot 10^{49}$ cm$^{-3}$ and for G2 star $T_1 = 1.1$ and $T_2 =$ 2.0 MK and $EM_1 = EM_2 = 1.1$ MK. On both components of the system, which is somewhat older than the Sun, they found the FIP effect, which apparently was slightly stronger on the K1 star.

Drake and Sarma (2003) observed the V 471 Tau system composed of white and red dwarfs by Chandra LETGS. In the range $\lambda < 50$ Å the radiation of the corona of the red dwarf prevailed, and from the carbon and nitrogen resonance lines they found [C/N] = −0.38. This value is intermediate between the non-evolved stellar matter and that characteristic of red giant. In the opinion of Drake and Sarma, this is due to the contamination of the atmosphere of the red dwarf when its companion was in the stage of a red giant that preceded the appearance of the white dwarf.

Sanz-Forcada et al. (2004) considered four F–K stars with different luminosity and activity levels, including K dwarf $\varepsilon$ Eri to study the FIP effect. Using Chandra LETGS-HRC they determined the abundance of 11 elements in the corona from the emission lines with the formation temperature varying from 50000 up to 300000 K, found the enhanced abundance of calcium and nickel, and suspected the tendency to a reduction of the abundance of elements with an increase of atomic weight.

One of the first objects studied with XMM-Newton was the Castor ABC system (Güdel et al., 2001b). During the observations, the components A and B were resolved in X-rays for the first time. A somewhat stronger luminosity of the A component was measured. Using EPIC data and the VMEKAL algorithm, three-component models of coronal plasma were constructed for YY Gem and Castor AB from low-dispersion spectra: for YY Gem $T_1 = 3.5$, $T_2 = 7.8$ and $T_3 = 16.5$ MK, $EM_1 = 0.9 \cdot 10^{52}$ cm$^{-3}$, $EM_2 = 1.3 \cdot 10^{52}$ cm$^{-3}$, and $EM_3 =$ $0.5 \cdot 10^{52}$ cm$^{-3}$, for Castor AB $T_1 = 2.0$, $T_2 = 7.8$, and $T_3 = 20.4$ MK, $EM_1 = 0.6 \cdot 10^{51}$ cm$^{-3}$, $EM_2 = 5.6 \cdot 10^{51}$ cm$^{-3}$, and $EM_3 = 1.0 \cdot 10^{51}$ cm$^{-3}$ with a distinct FIP effect. From the OVII lines the coronal density of cool components of YY Gem and Castor AB coronae were estimated as a few $10^{10}$ cm$^{-3}$. The recorded light curve of YY Gem with three eclipses enabled the construction of atmospheric models of the components to find the change of density with height and the prevailing localization of active regions at middle latitudes that is consistent with the Doppler imaging (Hatzes, 1995).

During simultaneous Chandra and XMM-Newton observations of YY Gem on 29/30 September 2000, Stelzer et al. (2002) found a good agreement between the calibrations of these instruments for the wavelengths and intensities. The high-resolution spectra obtained by the spacecrafts and averaged over the whole observation session encompassing about 75% of the orbital period duration demonstrated in the range of 5–133 Å a rich emission spectrum of highly ionized atoms: SiXII-XIV, NiIX, FeXVII-XXII, OVII and VIII, CVI, NeIX and X, NVII, MgXII, SX, and XII. The strong triplet OVII containing resonance, intercombination, and forbidden lines ensured estimation of the parameters of the "cold" component of the coronal plasma from their relative intensities: temperature of 2–3 MK and $n_e < 2 \cdot 10^{10}$ cm$^{-3}$. Even in the case of the strongest emission no self-absorption effect was revealed. From the low-resolution spectrum recorded with EPIC XMM-Newton in the beginning of the session, when the YY Gem system was the quietest, using the 3T VMEKAL algorithm, the following parameters of the coronal plasma were found: $T_1$ = 2.4, $T_2$ = 7.4, and $T_3$ = 20.8 MK, $EM_1$ = 2.2 · $10^{51}$ cm$^{-3}$, $EM_2$ = 14 · $10^{51}$ cm$^{-3}$, and $EM_3$ = 2.9 · $10^{51}$ cm$^{-3}$ with an appreciable deficiency of iron and some other elements. $EM_2$ and $EM_3$ decreased during the eclipse in the system.

Stelzer and Burwitz (2003) carried out simultaneous spectral and photometric observations of the Castor system using Chandra and higher-sensitivity XMM-Newton instruments. The resulting medium-resolution spectrum of Castor AB recorded with XMM-Newton EPIC was presented by the 3T VMEKAL model with temperatures of 3.1, 9.0, and 20 MK and close emission measures (1.7–3.1) · $10^{51}$ cm$^{-3}$; the high-resolution spectrum obtained on XMM-Newton RGS enabled the estimate of the ratio of the emission intensities as $T \sim 2$ MK and $n_e \sim$ (0.5−1.0) · $10^{10}$ cm$^{-3}$. The high-resolution spectra of each component, A and B, obtained using Chandra LETGS displayed the domination of OVIII, OVII, FeXVII, and NeIX emissions.

Inhomogeneity of the active corona was found during the XMM-Newton RGS observations of the eclipse in the $\alpha$ CrB system: according to Güdel et al. (2003a), the corona of the G component of this system is essentially asymmetrical and there are areas with a density of $10^9$–3 · $10^{10}$ cm$^{-3}$.

Krishnamurthi et al. (2001) observed the central area of the Pleiades with Chandra within a range of 0.1–10 eV and discovered 57 sources; some of them could be low-mass stars.

Harnden et al. (2001) studied the central part of the NGC 2516 cluster using Chandra and found more than 150 new sources. Probably due to low metallicity the X-ray luminosities of G and K stars in this cluster were lower than those in the Pleiades.

Using XMM-Newton RGS and Chandra LETGS and HETGS, Ness et al (2003) investigated the ratios of the intensities of FeXVII, OVII, and NeIX lines but failed to find the effect of optical thickness.

The spectral analysis of the ROSAT, ASCA, and BeppoSAX data on the young star LQ Hya performed by Covino et al. (2001) reveals a significant deficiency of metals in its corona: $z/z_\odot$ = 0.03–0.14. An analogous FIP effect in the corona of the LOPeg dwarf was detected by Pandey et al. (2005). The results concerning LQ Hya were confirmed then by the XMM-Newton observations. Meanwhile, the abundance of elements on the optical spectrum of this star displayed their closeness to solar abundances. The most informative BeppSAX observations made it possible to calculate the 3T model of the corona for LQ Hya of 2000 with $T_1$ = 0.25–0.45, $T_2$ = 0.9–1.1, and $T_3$ = 2.3–3.5 keV, and $z/z_\odot$ = 0.2–0.3.

Using XMM-Newton/EPIC, Briggs and Pye (2004) first recorded X-ray radiation of the middle-age brown M7 dwarf Roque 14 in the Pleiades. This seemed to be a constant emission, although one could not exclude a slow, about an hour, decay of the flare. At the time-averaged exposure, the luminosity $L_X = 3.3 \cdot 10^{27}$ erg/s and $L_X/L_{bol} \sim 10^{-3.05}$, and $L_X/L_{H\alpha} \sim 4.0$, which was close to the activity level of the old-disk M7 dwarf VB 8.

Using Chandra/LETG, Maggio et al. (2004) carried out a spectroscopic study of the quiescent corona of AD Leo in X-rays. The OVII–VIII, NeIX–X, and FeXVII–XIX lines dominated in the high-resolution spectrum, and these allowed them to construct the EM distribution and estimate the abundance of elements. The EM distribution proved to be the same as it was detected with EUVE, confirming the long-term stability of the corona, but differed from other analogous distributions by the fact that it corresponds to static models of isobaric loops with the constant section and homogeneous heating, as well as to the models with constant heating by flares. The ratios of lines to the continuum are consistent with the solar abundance, whereas other results were detected earlier.

Preibisch et al. (2005) carried out Chandra observations of 34 spectroscopically identified brown M6–M9 dwarfs in the nucleus of the Orion cluster and recorded X-rays from 9 of them; but if one considered only objects with $A_V < 5^m$, then the X-rays would be recorded from 7 of 16. The ten-day light curves in X-rays displayed constant radiation and its strong variability, including numerous flares. The X-ray properties of these young brown dwarfs — spectra, $L_X/L_{bol}$, frequency of flares — were similar to the characteristics of low-mass stars in the Orion cluster, and there was no evidence for variation of the magnetic component on the boundary of a star/substar, which in Orion falls onto M6. Since the X-ray properties of the young brown dwarfs are similar to those for M6–M9 field stars, then the key of magnetic activity of very young stars is a temperature that determines the ionization degree of the atmosphere.

During the 2-year regular XMM-Newton observations of the $\alpha$ Cen system, Robrade et al. (2005) found a constant predominance in X-ray radiation of the optically weaker component $\alpha$ Cen B and a weakening of X-ray radiation of the component $\alpha$ Cen A over this time at least by an order of magnitude, which could be apparently referred to the manifestation of an activity cycle in the corona.

Czesla et al. (2008) studied the eclipsed binary system of brown dwarfs 2MASS J05352184–0546085, in which a temperature inversion was detected: the primary massive component was cooler than the secondary one. Chandra and XMM-Newton recorded X-ray radiation of both components but without flares. The ratio $L_X/L_{bol}$ proved to be close to the saturation level $10^{-3}$. Applying the evolutionary concept of appearing convection in the presence of the magnetic field, MacDonald and Mullan (2009) suggested the following model of this pair: due to its youth there have not been achieved a synchronization of component rotations; the magnetic field with a strength of 120–320 G on the surface has already emerged in the primary component, it suppresses convection, increases the radius, and decreases the effective temperature, whereas on the secondary component there is no field and its effective temperature is higher.

Analyzing the XMM-Newton archive, for the M9 dwarf LHS 2065 Robrade and Schmitt (2008) detected soft X-ray radiation as quasi-quiescent emission without flares at the level $L_X = 2.2 \cdot 10^{26}$ erg/s and $\log(L_X/L_{bol}) = -3.7$, which corresponds to a fairly active star. Later, using Chandra/EPIC, Robrade and Schmitt (2009a) carried out observations of another M9 dwarf 1RXS J115928.5–524717 and detected weak variability in the soft X-ray region, which was attributed to quasi-permanent activity. Coronal emission of this star was represented by a two-temperature model with components of 2 and 6 MK with the total luminosity $L_X = 1.0 \cdot 10^{26}$ erg/s and $\log(L_X/L_{bol}) = -4.1$. Both M9 dwarfs turned out to be comparatively active in X-rays, i.e., they involved the effective dynamo producing their magnetic activities and coronal X-ray emission.

Using Chandra/HETG, Liefke et al. (2008) studied the M3.5 and M4.5 components of the binary system EQ Peg and found that the primary component was 6–10 times brighter than the secondary one, and the FIP effect was expressed weaker on EQ Peg B. This pair showed that toward later M dwarfs their X-ray luminosity weakened, and the temperature and frequency of flares decreased.

Continuing Chandra/ACIS observations, from an extensive sample of very cool dwarfs Berger et al. (2010) concluded that up to the level $L_X/L_{bol} \sim 10^{-5}$ about 50% of stars later than M7 had X-ray radiation, whereas for M7–M9 $L_X/L_{bol} \sim 10^{-4}$, for L dwarfs $L_X/L_{bol} \leq 10^{-5}$, and only 15% of L dwarfs displayed X-ray radiation.

Patel et al. (2013) determined temperatures of the 2T corona of the K5 star V 1147 Tau as $T_1$ = 0.8 and $T_2$ = 8 MK, whereas the light curve in X-rays revealed flares, rotational modulation and was in anticorrelation with the optical light curve. Optical observations indicated two active longitudes with recorded changes and the total coverage of spots up to 9–22%.

From the short-lasting stellar brightness dimming after the flare maximum in EUV and X-rays Veronig et al. (2021) suggested estimating parameters of coronal mass ejections that are associated with a flare.

## 1.4.3. Microwave and Shortwave Emissions

In addition to thermal X-ray emission and optical emission of the forbidden lines of highly ionized ions of iron, calcium, and some other metals excited by electron collisions, the solar corona radiates electromagnetic waves arising as a result of plasma oscillations and gyroresonance and gyrosynchrotron interactions of fast electrons and local magnetic fields. The coherent emission induced in the cyclotron maser and under plasma oscillations has a many orders of magnitude higher brightness temperature and a high degree of polarization from the incoherent emission. In principle, by analyzing such an emission one can estimate the density of the emitting plasma and the characteristics of the relevant magnetic fields. But the significant diversity of the mechanisms of radio emission of hot plasma in magnetic fields and a large number of independent parameters governing these emissions often makes ambiguous not only parameter estimates but even the identification of the specific emission mechanism.

According to Dulk (1985), at wavelengths of 3 cm and shorter the Sun looks like a homogeneous disk with a brightness temperature of about 15000 K that has bright active regions with a complex polarization pattern. By 30 cm the brightness of the disk slowly increases to $T_b \sim$ 50000 K and that of active regions to $(1-2) \cdot 10^6$ K. Dark coronal holes with $T_b \sim$ 30000 K emerge there. At meter wavelengths the temperature of the whole disk achieves $T_b \sim 10^6$ K, the active regions become indistinct, and the size of the solar disk becomes 1.5 times larger than that in the optical range. But large magnetic loops containing the main mass of the luminous matter at great heights are usually localized at low latitudes, which results in an asymmetry: the radio-Sun is greater in the equatorial direction.

As in studying the X-ray emission of red dwarfs, initially microwave flares simultaneous with the optical flares were found, and only launching of the Very Large Antenna (VLA) in New Mexico allowed the microwave emission of quiescent flare stars to be detected.

After a number of failures in the detection of such a quiescent radiation, during which the estimates of its upper limit subsequently decreased from 50 to 1–2 mJy (see the survey by Bastian, 1990) the quiescent microwave emission from red dwarfs was first successfully revealed by Gary and Linsky (1981). With 25 VLA antennas they observed 6 stars that were not included in close pairs and had high X-ray luminosity. At 6 cm they found emission of $\chi^1$ Ori (G0) at the level of 0.6 mJy and UV Cet (dM5.5e) at the level of 1.6 mJy. For $\pi^1$ UMa, $\xi$ Boo, 70 Oph, and $\varepsilon$ Eri they found only the upper emission limits. Each program star was monitored for several hours, which made it possible to consider the recorded emission as the quiescent corona emission rather than that of stellar flares. The flux measured from UV Cet corresponded to the brightness temperature of the radio emission source $T_b = 10^8(R_*/r_s)^2$ K, where $r_s$ is the characteristic size of the source. Due to the small optical thickness at 6 cm, the stellar corona that was responsible for the observed X-ray emission did not allow the recorded microwave emission to be explained by the free–free emission. Thus, Gary and Linsky used the mechanism of gyroresonance emission of thermal electrons that provided a greater optical thickness in relatively strong magnetic fields. They concluded that the mechanism occurred at the 6th or lower harmonic in the magnetic fields of at least 300 G. Then, Topka and Marsh (1982), using all 27 acting 25-meter VLA antennas, over 20 minutes of observations found microwave emission of each component of the EQ Peg system (dM3.5e+dM5.5e): at 6 cm the density of the recorded fluxes was 0.7 and 0.4 mJy (see Fig. 26). Simultaneous detection of the effect at both components justifies the conclusion that these short-term observations dealt with quiet coronae rather than flares. Topka and Marsh also decided that the discovered emission could not be due to thermal emission of the X-ray corona, but could be presented within the gyroresonance emission on the condition that the size of the radiating region was several times greater than the optical size of the star. Fisher and Gibson (1982) confirmed the results of Gary and Linsky (1981) and Topka and Marsh (1982) and determined quiescent radiation from the flare star YZ CMi at 6 cm at the level of 0.5 mJy and radiation of UV Cet at 21 cm at the level of 1.1 mJy.

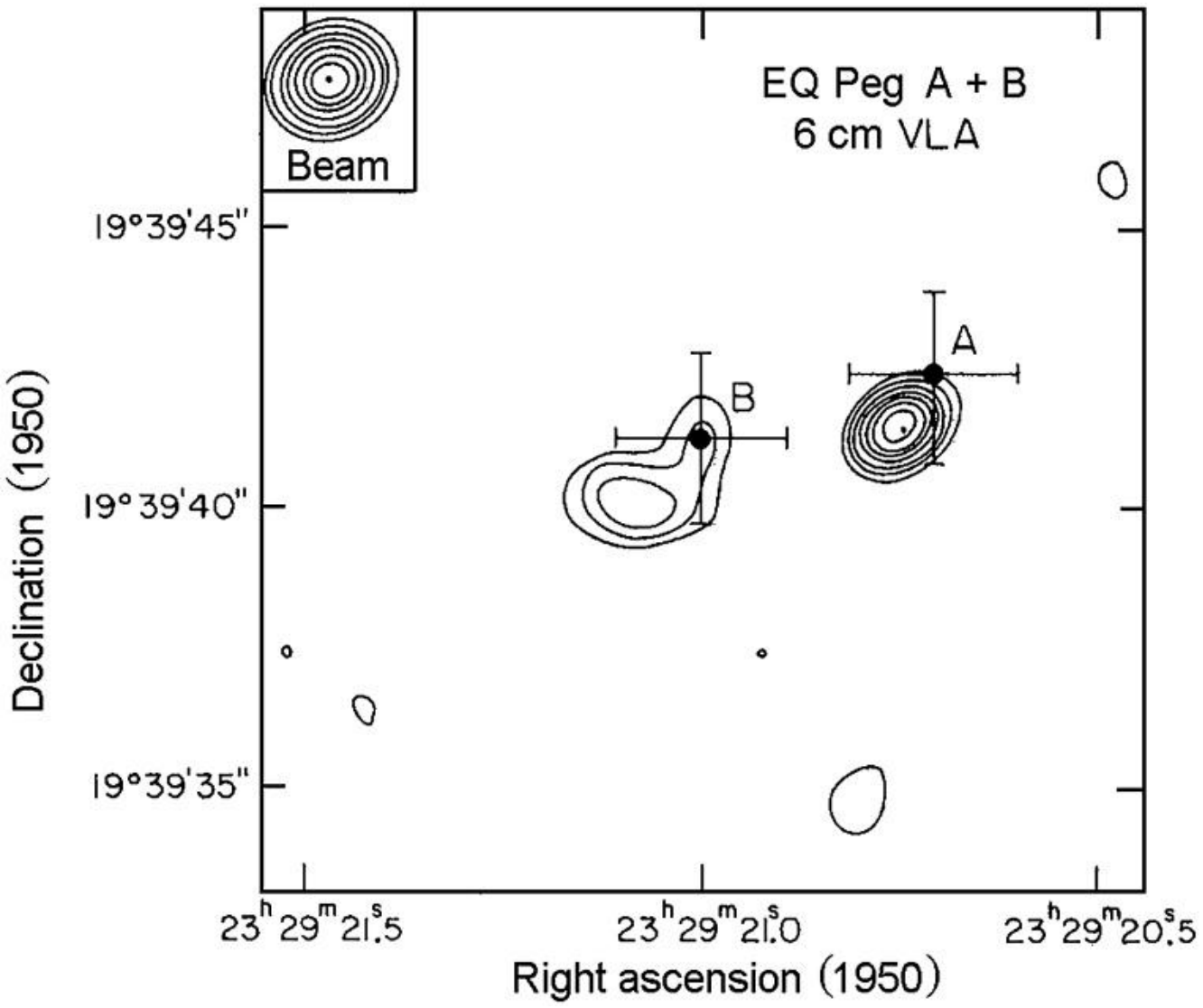

Fig. 26. Map of microwave emission of the EQ Peg region obtained with VLA; top left corner: the VLA beam (Topka and Marsh, 1982)

To continue the program, Linsky and Gary (1983) observed 14 other late stars. At 6 cm in the course of 2-to-4-h sessions they found slowly varying radiation of YY Gem and Wolf 630 AB binary systems, while the upper limit of F–K dwarfs was estimated as two orders of magnitude lower than $L_R$ for dMe stars. The YY Gem system was recorded during five of six sessions, the Wolf 630 AB system during four of five sessions, and all the dMe stars found by that time in the microwave range had the luminosity $L_R \sim (1–50) \cdot 10^{13}$ erg/(s · Hz). The measured fluxes from YY Gem and Wolf 630 AB could also be presented within the gyroresonance emission of thermal electrons. However, the required source size exceeding the stellar size by 3–4 times and the magnetic fields of hundreds of gauss that should correspond to photospheric fields of $10^3$–$10^4$ G raised doubts as to the validity of the model. Thus, Linsky and Gary (1983) concluded that the fluxes from UV Cet and YY Gem measured at 6 cm could be better explained by gyrosynchrotron emission of fast electrons whose required density was by 3 orders of magnitude lower than that of thermal plasma electrons responsible for the X-ray emission of the stars. The advantage of the gyrosynchrotron (GS) model is that with the growing electron energy the efficiency of their microwave emission increases rapidly, and consequently the number of particles necessary for the presentation of the observed flux decreases. Therefore, the model requires smaller sources of lower density than the gyroresonance (GR) model. Later, Gary (1986) concluded that both the power distribution of nonthermal electrons and the small addition of plasma with $T \sim 4 \cdot 10^8$ K to plasma with $T \sim 10^7$ K were suitable to form the recorded microwave emission. This small addition should be considered as an independent component of the stellar corona. In the course of six observational sessions, Linsky and Gary (1983) suspected the correlation of the radio flux with spottedness of YY Gem, which can be considered as a direct evidence of the relation between the microwave emission and magnetic fields of the star.

To explain the polarization features of the quiescent microwave emission from active red dwarfs, whose surfaces are usually spotted to a much greater extent than the Sun, Gibson (1983) proposed the model of a pair of spots: at equal magnetic fluxes the size of a leading sunspot is much less than that of an appropriate tail spot, so that the magnetic-field strength above the leading spot is much higher and consequently the emission above the leading spot is formed at a certain fixed frequency at a much higher level than above the tail spot. This asymmetry can result in noticeable polarization of the resulting microwave emission. However, the emission can be polarized both in the sources and along the propagation trajectory in the stellar atmosphere.

During five 3-hour intervals at different phases of the axial rotation period, Cox and Gibson (1984) observed the AU Mic star with VLA. They found rotational modulation at all wavelengths. At 2 cm, the modulation amplitude was minimum, probably because the filling factor was close to unity. Apparently, at 2 cm the emission was thermal, at two other wavelengths it was nonthermal. Gary (1985) performed 14-h observations of the YY Gem system with an orbital period of about 20 h and found an eclipse at all three wavelengths.

Using VLA, Kundu and Shevgaonkar (1985) observed UV Cet and YZ CMi at 6 and 20 cm. They interpreted the data obtained as gyrosynchrotron emission of nonthermal electrons with a power distribution with respect to energy occurring in the sources whose size was equal to 2–3 $R_*$ for UV Cet and 4–6 $R_*$ for YZ CMi and L 726–8 A. The proposed nonthermal electrons can fill stellar coronae as a result of flares or events similar to solar noise storms, so that the microwave emission should be considered as a time-averaged response of the quiescent corona to the sporadic occurrences of energy particles in it emerging during flare processes. The magnetic field on the photospheric level is estimated as several kilogauss from the lifetime of nonthermal particles of the order of an hour.

Using VLA, Pallavicini et al. (1985) observed 5 known flare stars and about a dozen chromospherically active G–K dwarfs and found the radiation of UV Cet, EQ Peg and YZ CMi at 6 cm: at about 1 mJy with low circular polarization for the first two stars and rather unstable radiation at the level of 2–8 mJy with the circular polarization of 50–90% for YZ CMi. For YY Gem and EQ Vir and all observed G–K dwarfs, including $\xi^1$ Ori, they obtained only the upper limits of microwave emission. The upper limit of the flux from $\xi^1$ Ori appeared to be 3 times lower than the value estimated by Gary and Linsky (1981). To explain this contradiction, they suggested that during the observations of Gary and Linsky a flare of the recently found satellite of the M dwarf was recorded (Gary, 1985).

Lang and Willson (1986a) continued the studies of the decimeter radiation of YZ CMi at two wavelengths close to 21 cm. They confirmed the variability of the radiation on the hour scale, but they also found its narrow-band feature: at $\Delta\nu/\nu \sim 0.1$ the variations of intensities occurred at each frequency band independently. Considering the measured level of X-ray emission of the star, one could expect in the decimeter region a thermal radiation flux two orders of magnitude lower than the recorded one. Within the gyroresonance model, the size of the source should be $\sim 200R_*$ and the magnetic field on this scale should achieve hundreds of G, which would require absolutely unrealistic photospheric fields. Thus, one might expect that the solution could be found in the gyrosynchrotron model. But neither the model of thermal free–free emission, nor those of gyroresonance emission of thermal electrons or gyrosynchrotron emission of nonthermal electrons suit the narrow-band emission of the whole stars. However, long-term narrow-band emission with high brightness temperature and strong circular polarization, which have no close analogs on the Sun, can be interpreted within the model of quasicontinuous bursts of coherent maser emission or plasma oscillations. It should be noted that Lang and Willson (1986a) estimated that in both cases the electron density should be about $10^{10}$ cm$^{-3}$, and the magnetic field strength at this level should be of several hundred gauss.

During the observations with the Molonglo interferometer in 1985, Vaughan and Large (1986a, b) found nonflare radiation at 35 cm on the CC Eri and AT Mic flare stars. During one session, the flux from CC Eri was at the level of 1 mJy and disappeared in a day, while the level of the flux from AT Mic was within 4–8 mJy during the four sessions within six weeks and in one of them meaningful variations at an interval of about an hour could be suspected.
Observing UV Cet in 1985, Kundu et al. (1987) recorded a flux of 2.8 mJy at 6 cm, which was twice that observed previously at the same frequency with VLA (Gary and Linsky, 1981; Fisher and Gibson, 1982; Kundu and Shevgaonkar, 1985). Thus, either there was a very slow weak flare or this was due to the flux variations on the time scale of years. The radiation of UV Cet at 6 cm was not polarized; the radiation at 20 cm was noticeably more variable. Usually, $F_6/F_{20} > 1$, which corresponds to an optically thick source. But when $F_6/F_{20} < 1$, there probably was a flare that is more clearly seen at 20 cm. In the observations of AU Mic a flux of 0.8 mJy without considerable polarization was recorded at 6 cm, and a flux of 4 mJy with a polarization of 70–90% was recorded at 20 cm. It is noteworthy that shortly before a flare occurred at 20 cm that was invisible at 6 cm. Developing the model of optically thick gyrosynchrotron radiation of nonthermal electrons for UV Cet quiet radiation at 6 cm, Kundu et al. (1987) estimated the necessary density of such particles as a few $10^7$ cm$^{-3}$ in magnetic fields of hundreds of gauss. Such a small addition of electrons should not be noticeable in the total X-ray emission of the star. Further, they showed that, as opposed to the solar corona where the height scale with respect to pressure is some hundredths of the solar radius, in the gyroresonance (GR) and gyrosynchrotron (GS) models of stellar coronae, where the height scale is comparable with $R_*$, the process of precipitation of fast particles down from the corona must play an important role: it should be the main mechanism of loss of particles determining their lifetime. In the GR model these losses are much higher than in the GS model, which imparts certain advantages to the latter. Jackson et al. (1989) observed several other flare stars. Each star was observed for several 5-min intervals with 10-s time resolution at 6 and 20 cm. Near each wavelength the radiation at two frequencies separated by 50 MHz was measured. During 2 h of monitoring at 6 cm UV Cet displayed smooth changes of nonpolarized radiation from 2.5 to 4.0 mJy, whereas L 726-8 A showed much greater variability and at 20 cm both components were more variable than at 6 cm. No evidence of narrow-band emission was found. A weakly varying flux from Wolf 630 was recorded at 6 cm, it contained the constant part at the level of 0.7 mJy and two 10–15-min flares with the amplitudes of $0.6^m$ and $0.8^m$. A weakly varying nonpolarized flux from YZ CMi was recorded at the level of 0.45 mJy. On the whole, the quiet radiation of dMe stars, as a rule, smoothly changing over hours and days, in some respects was closer to solar flares than to the quiescent radiation from the Sun. If one takes $r$s ~ $R$∗, then for flare stars at 6 cm $T$b = $10^8$–$10^{10}$ K and at 20 cm, $10^9$–$10^{11}$ K, which is much higher than for the quiet Sun and even for nonthermal solar flares. Apparently, in both cases the mechanism of gyrosynchrotron emission of nonthermal electrons is realized in magnetic fields above active regions, but it is not clear why the electrons live so long or are permanently generated in stellar atmospheres. Thus, the Sun, as a much weaker source of radio emission than active dwarfs, cannot always be a model for interpreting the radio emission of stars.

After several years of observations of flare stars in the microwave range and the discussion on the nature of the radiation recorded in them Dulk (1987) compiled the table of main characteristics of different radiation mechanisms, which was further elaborated by Linsky (1988), see Table 12.

Table 12. Characteristics of radio emission mechanisms

| Radio emission mechanism | Source size | Effective or brightness temperature | Circular polarization | Time variability |
|---|---|---|---|---|
| Thermal | $\gg R_*$ | ~ $10^4$ K | about zero | years |
| Gyroresonance | ≥ $R_*$ | ~ $10^7$ K | low | low? |
| Gyrosynchrotron and synchrotron | ≤ $R_*$ | $10^8 \div 10^{10}$ K | ≤ 30% | from minutes to hours |
| Cyclotron maser (λ ≥ 30 cm) | $\ll R_*$ | to $10^{20}$ K | to 100% | from milliseconds to hours |
| Plasma oscillations ($\lambda \leq 30$ cm) | $\ll R_*$ | to $10^{17}$ K | 10–90% | from milliseconds to days |

In 1985, Kundu et al. (1988a) conducted the first simultaneous observations of four flare stars (UV Cet, EQ Peg, YZ CMi, and AD Leo) with EXOSAT and VLA. Radio observations were run continuously for 7–10 h at two close frequencies near 6 and 20 cm with a time resolution of about 7 s. X-ray monitoring was executed in the range of 0.04–2 keV (LE) with a time resolution from 60 to 600 s. All the stars were found in the quiescent state in both wavelength ranges, at 6 cm the flux from UV Cet was about 1 mJy; from EQ Peg, 7–13 mJy; from YZ CMi and AD Leo, about 0.3 mJy. These values of the fluxes are 40, 230, 38, and 17 times higher than the expected values of free–free coronal radiation estimated from their X-ray emission measures at $T$ ~ $10^7$ K, which can be explained either by 2–3 orders of magnitude higher brightness temperature or by the source sizes exceeding $R_*$ by 2–20 times. This directly points to the difference of the physical mechanisms of X-ray and radio emissions. This conclusion is supported by the absence of any correlation between the time variations of these fluxes.

In 1986, Jackson et al. (1987b) observed 27 dMe stars (that were not observed before) with VLA and for nine of them detected emission at 20 cm at the level of 0.4–5.2 mJy and for 7 stars at 6 cm within 0.2–12.8 mJy. Each object was observed for about an hour. An extended observational program during the second epoch did not reveal new radio-emitting flare stars (White et al., 1989a). According to their statistics for 83 objects, the microwave emission was found for 42% of flare stars, 42% of dKe-dMe stars, 65% of BY Dra-type stars and for the only of 22 nonemission stars, Barnard's star. They considered various global characteristics of these stars and found that starting from dM5.5 the number of radio emission sources diminished noticeably, the mean fraction of radiating objects grew with increasing $L_X$ of red dwarfs, and according to their kinematic properties such objects mainly belonged to the young disk.

Willson et al. (1988) observed 16 K–M dwarfs at 6 cm with VLA. One-hour monitoring revealed a flux at the level of 0.45 mJy only from the dM2e star Gl 735, the strongest X-ray source in the sample. Similarly to the estimates of the mechanism of microwave radio emission for the other dwarf stars, the GR model in this case requires a size of the source of 20–30 $R_*$ with a magnetic field strength of hundreds of gauss, whereas for the GS model the density of nonthermal electrons of $10^7$ cm$^{-3}$ is sufficient, a linear size of the source of $10^8$ cm and the magnetic fields are weaker by an order of magnitude. Then, Lang and Willson (1988) observed YZ CMi with VLA near 20 cm in 15 narrow bands of 3.125 MHz with a time resolution of 10 s. The responses of the left- and right-circular polarization were recorded in turn at time intervals of 10 min. They recorded a slowly changing radiation with a characteristic change time of minutes and with 100% left-circular polarization. During several 10-min intervals in the band with a total width of 50 MHz there was a noticeable change of $F_\nu$, i.e., the radiation was narrow-band with $\Delta\nu/\nu \sim 0.02$, at other time intervals $F_\nu$ did not change noticeably. Lang and Willson interpreted the polarized emission as the coherent emission of an electron maser in a magnetic field of about 300 G.

During two seasons, Caillault et al. (1988) observed seven rapidly rotating BY Dra-type stars with known rotational periods with VLA at 6 cm and did not discover a constant flux. In 1985, the fluxes of 0.6–1.0 mJy were recorded from CC Eri, HD 218738, BY Dra, and EV Lac. During 3 nights in 1986, the radiation was found only from BY Dra and the measured flux varied from 0.8 to 1.5 mJy. For 22 dMe stars no correlation was found for the luminosities $L_X$ and $L_6$. For 14 stars no correlation between $L_6$ and $L_6/L_{bol}$ and the rotational period was revealed. Thus, Caillault et al. concluded that separate active events rather than quiescent corona were responsible for the recorded microwave emission. Then 16 K–M fast rotators were added to the research program and each program star was observed from two to four times over the interval from several days to a month (Drake and Caillault, 1991). Eight sample stars were identified reliably and four stars were detected supposedly as radio-emission sources. $F_6$ of four stars displayed variations at least by a factor of 3, the flux from Gl 867 B varied within 14–0.2 mJy on the monthly interval. Drake and Caillault found that the radio emission of M dwarfs was independent of whether the star was single or a member of a close binary system, while for M dwarfs the following correlation was valid

$$L_6 \sim L_X^{0.65\pm0.1}. \quad (24)$$

Later, Güdel (1992) formulated the following correlation for K dwarfs with a widely ranging rotational period

$$L_R \sim P_{rot}^{-1.9}. \quad (25)$$

Slee et al. (1988) carried out simultaneous observations of 24 chromospherically active stars, including six active dwarfs, at 6, 3.6, and 2 cm with VLA and the 64-meter telescope of the Parkes Observatory. A flux from YZ CMi was found with $F_6 = 0.5$ mJy, from AT Mic $F_6 = 3.2$ and $F_2 = 1.6$ mJy, from V 1054 Oph $F_2 = 1.4$ mJy with the left-circular polarization of 28%, from Gl 890 $F_2 = 1.4$ mJy. No microwave emission was recorded from AU Mic and Gl 182. The mean brightness temperature of measured emissions was $1.2 \cdot 10^8$ K. Apparently, all measured fluxes corresponded to the quiescent radiation of coronae or weak flares, since stronger fluxes from these objects were recorded before.

The VLA observations of 12 M dwarfs in a wide range of $L_X/L_{bol}$ ratios by Kundu et al. (1988b) made it possible to detect noticeable microwave emission from four stars. Three of them had a maximum $L_X/L_{bol}$ ratio, and on the fourth star, Gl 461, no flares and $H_\alpha$ emission were recorded before.

Güdel and Benz (1989) observed UV Cet with VLA at 90, 20, 6, 3.6, 2, and 1.3 cm. During two 5-h sessions no flares exceeding $3\sigma$ of the noise were found, and the order of magnitude of the fluxes at 20 and 6 cm was close to the earlier estimates, at higher frequencies they were obtained for the first time. No meaningful polarization was recorded at any frequency. During the first session the spectrum of microwave emission was measured within the range from 20 to 2 cm, in the second, from 6 to 1.3 cm. The obtained spectra revealed the statistically reliable U-like shape with a minimum at 3.6 cm and with spectral indices of −0.3 and +0.4 for low- and high-frequency branches, respectively. During the second session, for the main component of the L 726-8 A system at 6, 3.6, and 2 cm the fluxes of 0.8, 0.4, and 0.55 mJy were recorded, i.e., the minimum was also at 3.6 cm. The brightness temperature of UV Cet at 1.3 cm was close to the temperature of the X-ray emission for the microwave source $1{-}2R_*$ in size. Güdel and Benz presented the observed spectrum as a sum of two power spectra and obtained the spectral indices of −0.42 for $n < 10$ MHz and +2 for $n > 10$ MHz. They concluded that the high-frequency branch could not be due to the free–free plasma radiation responsible for the X-ray emission of the corona, but could be explained by the optically thick cyclotron emission at 4–6 harmonics of the plasma at a magnetic field strength at the photospheric level of 600–2100 G and a density of $3 \cdot 10^{9-10}$ cm$^{-3}$. This mechanism is not suitable for interpreting the low-frequency branch of the recorded spectrum, in which gyrosynchrotron or synchrotron radiation should be taken into account.

In the course of cooperative observations of YY Gem in 1988, the eclipses were simultaneously observed in optics, ultraviolet, and radio range, and a significant weakening of emission at 6 cm during the primary and secondary eclipses recorded with VLA indicated a relatively compact radio-emitting source between components of the system (Butler et al., 2015).

In 1992, Rucinski (1994) observed YZ CMi with VLA. At wavelengths of 3.6, 2, and 1.6 cm during two sessions he recorded the fluxes of 4–8, 2–7, and 3 mJy, respectively. The first observational session, during which maximum fluxes were found, lasted only 11 and 13 min; thus one cannot conclude with confidence that no flares occurred over that time.

White et al. (1989b) proposed the model for the generation of microwave radiation on the stars with separate active regions and high-strength photospheric magnetic fields. The model suggests that the nonthermal gyrosynchrotron radiation is emitted from these regions and the expected radiation spectrum depends on the vertical gradients of the magnetic-field strength and the density of matter. In particular, the negative spectral index for optically thick radiation is achieved at low values of these gradients and for low harmonics of 5–10. If the model is valid, constant sources of electrons with energies of 20–200 keV should exist on a star and there should be some evidence of precipitation of these electrons into the chromosphere. White et al. (1994) considered the constraints imposed on the models of the dMe coronae by the radio data collected by that time. According to their estimates, if the hot element of the corona with a temperature of about $2 \cdot 10^7$ K, which is usually found on such stars, was in the lower corona, where the strength of magnetic fields due to the proximity of the photosphere should be more than 1 kG, then due to higher optical thickness at 15 GHz it would be revealed by radio observations. Since observations for the considerable sample of dMe stars did not justify the expectations, White et al. concluded that the hot coronal component was localized at a height of the order of $R_*$ and was the cooling flare plasma.

Fomalont and Sanders (1989) surveyed all stars with $\delta > -28°$ at a distance of not more than 5 pc from the Sun with VLA at 6 cm. Among 241 objects they found microwave emission at the level of $F_6 > 1$ mJy only for five known flare stars: UV Cet, YZ CMi, AD Leo, Wolf 630, and EQ Peg.

In 1988, Caillault and Drake (1991) carried out 21 VLA observational sessions of BY Dra at 6 cm to reveal the rotational modulation of the quiescent microwave emission, but failed to reveal it, in their opinion, because of the small angle between the rotational axis of the star and the line of sight, about 30°, or because the magnetic loops, in which radio emission occurred, were very long.

In 1988, Spencer et al. (1993) observed YZ CMi at 6 cm with the broadband interferometer at the Jodrell Bank Observatory composed of the 76-meter Lovell antenna and the 25-meter telescope Mark II. They revealed stellar radiation varying at hourly intervals with an average level of 1.4 mJy.

The next step in the development of radio-astronomical techniques after VLA was the intercontinental Very Long Baseline Interferometer (VLBI). The first successful observations of dMe stars with VLBI were carried out by Phillips et al. (1989) within the framework of an astrometric program. Their VLBI consisted of two 70-meter antennas (Madrid and Goldstone), 100-meter Effelsberg antenna, and VLA. In observations at 3.6 cm, three of six program stars were found: YZ CMi, Wolf 630, and EV Lac. The recorded flux was at the level of 2–5 mJy and the size of sources was at least $R_*$. Using VLBI composed of VLA and the largest radio telescopes in Arecibo (Puerto Rico), Green Bank (West Virginia) and Effelsberg (Germany), Benz and Alef (1991) carried out four 13-min observation sessions of YZ CMi and recorded a smoothly varying correlated flux with 80% circular polarization at the level of 1.3 mJy. The upper limit of the source size was equal to 3.4 stellar diameter and $T_b > 4 \cdot 10^8$ K, which exceeded the temperature of slowly varying radio emission of coronal condensations on the Sun by orders of magnitude and required the nonthermal emission mechanism. It is still unclear how the necessary energy particles are retained for hours in the stellar atmosphere. Then, Benz et al. (1995) at VLBI composed of the above four telescopes, the 76-meter Lovell telescope and the 40-meter Owens Valley dish (California) observed AD Leo and the EQ Peg binary system at 18 cm. In one of the two sessions they estimated the properties of quiescent radiation of AD Leo: a flux of 0.75 mJy, a source size of less than 3.7 stellar diameters, and $T_b > 2 \cdot 10^9$ K, which was higher by an order of magnitude than that in the strongest solar flares and much higher than the appropriate temperature of YZ CMi. They suggested that the quiescent radiation of active dwarfs was physically related to IV-type solar stationary bursts, which continue for many hours after strong flares, or to noise storms, in which these bursts transform and then last for days at characteristic $T_b \sim 10^8$–$10^{10}$ K. If the suggestion is correct, the radio-emitting volume of dMe stars should be determined by the sizes of closed coronal magnetic loops. Using the VLBI composed of the above six radio telescopes, Alef et al. (1997) studied the binary YY Gem system at 18 cm during the phase of the main eclipse. They estimated the upper limit of the size of the radio source as 2.1 photospheric diameters of the star and $T_b = 1.1 \cdot 10^9$ K. The high brightness temperature and low circular polarization corresponded to gyrosynchrotron emission. The rather symmetrical radio image of the star made it possible to claim that the magnetic coronal loops trapping radio-emitting relativistic electrons achieved heights comparable with the stellar radius and were distributed rather isotropically. Benz et al. (1998) observed the system L 726-8 AB at 3.6 cm and found both components, while the UV Cet component, brighter in the radio range, was resolved at least into two components remoted by 4–5$R_*$ and with a noticeable variation of relative brightness during the 6-h session. Within the gyrosynchrotron model the recorded radiation requires coronal loops of several stellar radii and field strength between 20 and 130 G. In VLBI observations at 3.6 cm, Pestallozzi et al. (2000) found a level of the flux from YZ CMi of about 3 mJy and the left-circularly polarization up to 60%, for AD Leo, 0.5 mJy without noticeable polarization. They concluded that the size of the radio image of YZ CMi was 1.7±0.3 stellar diameters, that of AD Leo was less than 1.8 stellar diameter, the brightness temperature of YZ CMi and AD Leo was $7 \cdot 10^7$ K and more than $5 \cdot 10^7$ K, respectively.

Güdel et al. (1993) observed simultaneously 12 M dwarfs with ROSAT (0.1– 2.4 keV) and VLA (3.6 and 6 cm) and found a close correlation between the X-ray and microwave emission

$$\log L_X = (1.06\pm0.10)\log L_R + (14.5\pm1.3)\ (r = 0.95)\,. \tag{26}$$

This correlation became slightly weaker when nonsimultaneous observations were considered as well. The revealed linear correlation valid for the interval of three orders of magnitude continued with a slight change of slope toward stronger stellar coronae for the objects of the type of RS CVn, FK Com, and Algol. Then, using VLA, Güdel (1992) studied 12 active K0-K8 dwarfs, among them 5 single stars, 5 components of spectrally binary systems, and 2 apparent binary objects. All the program objects were known as strong X-ray sources with fast rotation and/or strong calcium emission. For 7 of them, nonflare radiation was found at 3.6 cm, while their luminosities were comparable with $L_R$ of M dwarfs. $L_R$ of studied K dwarfs was in good correspondence with the correlation $L_R \sim L_X$ previously found for M dwarfs. Faster-rotating K dwarfs were found to be stronger radio sources. The Sun, as a too-weak microwave source, does not satisfy the correlation $L_R \sim L_X$ that is common for active stars of different type and expands to 5 orders of magnitude. However, the solar flares satisfy the correlation, in particular in the maximum phase (Güdel and Benz, 1993; Benz and Güdel, 1994). This correlation, common for active stars, suggests that either the coronal heating and particle acceleration are of common origin or continuously accelerated particles are thermalized and heat the corona.

During observations at 6 and 3 cm with the Australian Telescope Compact Array (ATCA) Lim (1993) found a flux from the dM4e star Rossiter 137B close to that recorded from UV Cet. Since the former is ten times further from the Sun than the latter, the radio luminosity of Rossiter 137B is higher by 2 orders of magnitude than that of UV Cet. Rossiter 137B is very young, less than $10^8$ years, a fast rotator with $P_{rot}$ not exceeding 9 h; and has apparently not reached the main sequence and is in the post-T Tau stage.

Reasoning from the $L_R \sim L_X$ correlation for coronally active stars, Güdel et al. (1994a, b) selected 15 nearby solar-type stars with a high level of X-ray emission from the Catalog of Bright Stars and from the ROSAT data and observed them at 3.6 cm with VLA. For Gl 97, Gl 755, EK Dra, and HD 225239 during 45-min sessions the quiescent microwave emission at the level of 0.18– 0.33 mJy was found, which differs from the values expected from the correlation ($L_R$,$L_X$) for later active dwarfs by no more than 60%. The age and rotational rate of Gl 97 are close to those of the Sun. EK Dra is a "young Sun", its age is $\sim 7 \cdot 10^7$ years, the emission lines of its chromosphere and transition zone are very strong, the X-ray flux well satisfies the correlation with the axial rotation period. HD 225239, which based on the metal deficiency and the spatial velocity is related to the old-disk population, unexpectedly displayed radio luminosity that was several times higher than that of the strongest solar flares. The cause of the strong radio luminosity of these active stars is unclear.

* * *

Güdel (1994) summarized the results of the first decade of studying the microwave emission of the stars of late spectral types, for which the discovery of the slowly changing component was a surprise. This component, whose intensity exceeds by orders of magnitude the free–free radiation of the transition zone of the Sun and slowly changing thermal gyroresonance radiation of the regions with strong magnetic fields cannot be presented within the stationary corona. A concept of continuous or recurrent filling of a corona by the high-energy electrons, whose emergence could be related to flare phenomena, should be introduced. The characteristic times of variation of the radiation are hours and days, and there is no evidence of such variations over minutes. The low level of circular polarization of the radiation could be stipulated by relativistic electrons at a not very hard energy distribution in strong magnetic fields and small optical thickness of the source. On the other hand, mutual compensation of the polarization in the total radiation of many separate sources is possible. Thus, the investigation of the microwave emission of stars detected magnetic nature of the activity, close physical connections of the quiescent radiation with flare processes, and the hot component of the corona with the acceleration of particles to relativistic energies.

* * *

Using ACTA, Robinson et al. (1994) discovered at 6 cm the emission of 2 mJy from the young rapidly rotating K0V star HD 197890 with strong nonstationary EUV radiation; however, there is no confidence that the star already reached the main sequence.

Pallavicini et al. (1990b) analyzed the Einstein Observatory and EXOSAT observations of Castor ($\alpha$ Gem) and concluded that the quiescent X-ray emission of this system was due to the invisible red dwarf, a component of one of the A stars forming the system. HRI and PSPC ROSAT observations confirmed the X-ray emission of the Castor A+B system, but the microwave studies performed at the Very Large Antenna at 3.6, 6, and 20 cm with an order of magnitude higher spatial resolution showed that such a radiation was emitted only from the A component (Schmitt et al., 1994). Apparently, the X-ray emission of the system is also emitted by this component. The radiation at 20 cm changed from 0.3 to 1.2 mJy on one day and from 0.3 to 0.7 mJy on another with a measurement error of 0.1 mJy. X-ray and microwave studies led to the conclusion that the optically invisible low-mass component was responsible for the observed activity of $\alpha$ Gem, but its parameters have not been estimated unambiguously yet.

Güdel et al. (1995b) discovered a strong corona on the F0V star 47 Cas: strong X-ray emission corresponding to $L_X \sim 3 \cdot 10^{30}$ erg/s and strong microwave emission at a level of $10^{15}$ erg/(s $\cdot$ Hz) were recorded. The X-ray emission was presented by the two-temperature corona model with temperatures of 2 and 10 MK and emission measures of $4 \cdot 10^{52}$ cm$^{-3}$ and $10 \cdot 10^{52}$ cm$^{-3}$, with the “cool” component displaying rotational modulation, while the X-ray flares were recorded only in the “hot” component. The microwave emission of the star at 3.6 cm was recorded with VLA at a level of 0.63 mJy, it corresponded to $T_b = 8 \cdot 10^7$ K and was in good agreement with the correlation ($L_R$,$L_X$) for later active dwarfs. Apparently, 47 Cas is the youngest F0 single star with an extremely high level of magnetic activity, which agrees with fast rotation, vsin*i* = 95 km/s, and an age not exceeding that of the Pleiades. However, 20 years later this statement was shaken when the optical flares of late A stars were detected with Kepler (Balona, 2015).

Güdel et al. (1995c) studied the sample of 24 brightest, according to the RASS data, F and G stars with VLA at 3.6 cm. They looked for possible duplicity among the sample stars and found that at least for the F0V star HD 12230, one of the nine sources of microwave emission found, except for EK Dra, was definitely the star itself rather than the weak invisible component. The star belongs to the Pleiades moving cluster, and its age is 50–70 million years. Its X-ray emission is modulated with a period of about a day, with the cool component of the corona modulated to a greater extent than the hot one, as in EK Dra.

Güdel and Benz (1996) collected 33 VLA measurements for 20 dKe-dMe stars in the range from 20 to 2 cm and found that in most cases the intensity of radio emission decreased with increasing frequency, as one would expect in the gyrosynchrotron model, but in several measurements a more complicated change of $F_\nu$ suggested a contribution of thermal gyroresonance emission occurring at the plasma temperature of 10 MK and magnetic fields of about 1 kG.

In 1996, Leto et al. (2000) carried out the VLA observations of flare stars in the wavelength range from 3.5 cm to 7 mm to check the existence of a minimum in the radio spectrum at 3.5 cm detected earlier by Güdel and Benz and trying to associate an increase of the radio emission at higher frequencies with the known excesses of active stars at millimeter wavelengths and in the IR range (Mathioudakis and Doyle, 1991b, 1993). But in the course of these observations they measured the fluxes from UV Cet, V 1054 Oph, and EV Lac only at 3.5 cm, in other cases only the upper limits were estimated.

Lim and White (1995) observed four fast-rotating G–K dwarfs in the Pleiades at 3.6 cm with VLA. For three stars the microwave emission was detected. On the fastest G dwarf rotator HII 1136, a flare was recorded and after its decay a nonflare radiation was determined at a level of 0.16 mJy. From the fastest K dwarf rotator HII 1883, the quiescent radiation was recorded in two sessions separated by three months at a level of 0.10 and 0.05 mJy. The K0 dwarf HII 625 displayed a slowly changing emission at a level of 0.16 mJy. Such fluxes correspond to the luminosity $L_R \sim (1–3) \cdot 10^{15}$ erg/(s $\cdot$ Hz). The discovery of microwave emission of the open cluster members showed that solar-type stars that recently reached the main sequence were equally strong radio sources, as the nearby field stars with equally fast rotation. The radio fluxes of the stars averaged over the surface are comparable with the appropriate values of the T Tau-type stars, which suggests that the saturation mechanism is realized in both cases.

Between 1990 and 1994, Lim et al. (1996) carried out four ACTA observational sessions of Proxima Cen at 3.5, 6, 13, and 20 cm, but did not record quiescent radio emission. However, using the least estimates of the upper limits of the fluxes at 3.5 and 6 cm, 0.12 and 0.11 mJy, respectively, they obtained for the two temperatures of X-ray plasma, 3 and 20 MK, the lower estimates of the magnetic-field strength in the coronal loops from 400 to 1040 G and the upper estimates of the filling factor. These observations fit into two different models of radio emission sources: the filling factor is low at $B > 400$ G, or it is high at low $B$. Using the data of X-ray observations and (24), Lim et al. estimated the parameters of the photospheric magnetic field of the star: $B \sim 1$ kG and $f \sim 0.1 - 0.2$.

Güdel and Zucker (2001) presented the microwave emission of about a dozen stellar coronae recorded during 20 observational sessions within the framework of the gyrosynchrotron model. They found that the index of the power distribution of the energy of radio-emitting electrons was mainly within 2–3.5. Such a hard distribution of energy particles on the Sun is observed only in strong flares. In their survey of cool stars in the centimeter wavelength range, Güdel and Audard (2001) stated that the X-ray emission of almost all stars found in this range was close to $L_X/L_{bol} \sim 10^{-3}$, while the level of radio-emission saturation was $L_R/L_{bol} \sim 10^{-18}$ Hz$^{-1}$.

Using VLA, at 3.6 cm Krishnamurthi et al. (1999) sought for the microwave emission from four M7–M9 dwarfs and two brown dwarfs: Gl 229 B, LHS 2065, Kelu 1, Gl 569 B, VB 8, and VB 10. The accumulation time varied from 2.5 to 3.5 h, but no emission was recorded in either case. If the objects were removed at the distance of UV Cet, the upper limits would vary from 0.2 to 0.8 of the flux from UV Cet.

In late December 2001– early January 2002, Slee et al. (2004) carried out radio and optical observations of CC Eri, using ATCA and the Anglo-Australian Telescope, and detected a low level of radio emission with a spectral index of 0.26 and polarization up to 20%. Parameters of the emission turned out to be consistent with the GS model, in which radio emission becomes optically thick during a strong flare. Cross-correlation of the quiescent radio emission indicated the presence of microflares.

Burgasser and Putman (2005) performed the ATCA observations of radio emission from seven close M–L dwarfs. The M7 dwarf LHS 3003 and the M8 dwarf DENIS 1048–3956 revealed emission at 4.80 GHz, the observed emission conformed with optically thin GS emission of relativistic electrons with energy of 1–10 keV and density of $10^9$ cm$^{-3}$ in the field of $B >\sim 10$ G.

Berger et al. (2005) carried out simultaneous observations of the L3.5 dwarf 2MASS J00361617+1821104 with VLA, Chandra, and the 4-meter telescope of the Kitt Peak Observatory but recorded only strong variable radio emission, circularly polarized by 60% and with a periodicity of about 3 hours. They interpreted the radio emission as a gyrosynchrotron in the large-scale magnetic field with a strength of 175 G, which preserved stability over three years. The periodicity of emission was attributed due to the motion of the component at the distance of 5 radii from the star or due to its rotation with a velocity of 37 km/s, or due to weak flares. But a fully convective star with strong radio emission and without X-rays and $H_\alpha$ emission should have the atmosphere significantly different from those of earlier stars, even M dwarfs.

Osten et al. (2006b) first detected the radio emission of EV Lac at 2 cm, which excluded the spatial coincidence of the high-temperature coronal plasma and the source of radio emission. Assuming that this radio emission was due to accelerated electrons in the dipole configuration, they estimated the field strength and the total number of accelerated electrons: a rather surface equatorial field of several hundreds of gauss.

Berger (2006) carried out radio observations of 90 M5–T8 dwarfs, added 3 more radioactive dwarfs to the known 6, and found that 10% of stars of later than M7 were radio-emitting. In the quiescent state, the typical field strength was $\sim$ 100 G, in flares — 1 kG, i.e. lower than that for early M dwarfs. About M7 there occurred a fast jump from $\log(L_R/L_{bol}) \sim -15.5$ to $-12$ and this ratio increased toward later spectral types. A fraction of active stars dropped from 30% for M dwarfs to 5% for L dwarfs. This drop of activity reflects a variation of the chromosphere structure and/or the transition to the turbulent dynamo.

Audard et al. (2007) carried out Chandra and VLA observations of the binary close system of L dwarfs Kelu-1 and recorded it at a level of $L_X = 3 \cdot 10^{25}$ erg/s but did not detect it in the radio range.

Using VLA, Osten et al. (2007, 2009) and Phan-Bao et al. (2007) detected the radio emission at 8.5 GHz from the close binary system LP 349-25 consisting of M8 components; this radio emission was constant at the intervals from 10 s to 11 hours and from 0.6 to 1.6 years. This time frames resulted in estimating the electron density $n_e < 10^5$ cm$^{-3}$ and the field $B < 130$ G.

Throughout nine hours, Berger et al. (2008a) carried out the simultaneous radio, X-ray, ultraviolet, and optical spectral observations of the M8.5 dwarf TVLM 513–46546 and detected constant quiescent radio emission at a level of 0.2 mJy with a number of minute high-polarized bursts at a level of 2–5.5 mJy but found no previously reported periodic bursts, which points to their nonconstancy. The minimum level of $L_X/L_{bol} \sim 10^{-5.1}$ was detected for stars later than M5. The sinusoidal curves of $W_{H\alpha}$ and $W_{H\beta}$ show a period of about 2 hours coinciding with the rotation period of the star. The filling factor of the Balmer lines approaches 0.5, and their luminosity is several times higher than the X-ray luminosity, which excludes heating of the chromosphere by the corona. From the quiescent radio data, the large-scale magnetic field is with $B \sim 100$ G, from radio bursts — a field of $\sim 3$ kG of the multipolar configuration, but these do not significantly affect the chromosphere and corona.

Continuing these observations, Berger et al. (2008b) carried out the 9-hour multiwavelength observations of the M8.5 dwarf LSR 1835+32 and the M8 dwarf VB 10. Being investigated with VLA, the former dwarf revealed the constant radio emission and fairly variable $H_\alpha$ emission with characteristic times of 0.5–2 hours, but without UV excess, while Chandra did not detect X-rays, therefore the ratio $L_R/L_X$ exceeds by $2 \cdot 10^4$ times this value for F–M6 stars. The ratio $L_{H\alpha}/L_X > 10$ is by an order of magnitude higher than that for M0–M6 dwarfs and excludes heating of the quiescent chromosphere by coronal radiation. The absence of variability in the radio range during four stellar revolutions requires a homogeneous extensive field of 10 G, and variability in $H_\alpha$ — a microstructure source. Closeness of radio observation data to the previous results allows one to imply the field stability over years and to estimate the turnover time of convective cells as years. VB 10 revealed the correlated quiescent and flare emission in X-rays and UV, as for early M dwarfs. Optical emission lines displayed a similarity in amplitude and times with X-rays and ultraviolet and two different types of flare amplification — smooth and impulse; the mildly sloping Balmer decrement and helium emission in the impulse flare require significantly denser and hotter plasma. Taking into account the previous results on TVLM 513–46546, one can conclude that the late M dwarfs, contrary to F–M6 stars, have a combination of activity characteristics indicating the transition in the atmosphere structure and its heating by the large-scale magnetic fields, but coronae are generally weaker than those in early M dwarfs. A decrease of the chromospheric activity occurs not so rapidly and the ratio $L_{H\alpha}/L_{bol} \sim 10^{-5}$–$10^{-4.5}$ is by an order of magnitude lower than the saturation level of middle M dwarfs.

Then Berger et al. (2009) carried out the 8.5-hour simultaneous radio, X-ray, ultraviolet, and $H_\alpha$ observations of the binary system of L dwarfs 2MASSW J0746425+200032 and detected a strong radio emission with dominating pulsations at 4.86 GHz and with a period of 124.32 ± 0.11 min. Stability of the impulse profiles and moments of signal arrivals proved that they were due to rotational modulation of the magnetic field of 1.7 kG. The quiescent constant component was detected, which seemed to be due to the emission of the large-scale field. The sinusoidal variable $H_\alpha$ emission was found with the same period that radio emission had, but with a shift by 1/4 of a phase; such a shift eliminates the symmetric topology of the dipole field but presumes either the domination of the quadrupolar field or the absence of orientation on the field of radio and $H_\alpha$ emission. From the measured period and known orbit parameters the first size of the L dwarf was determined to be $(0.078 \pm 0.010)R_\odot$ and its mass was $(0.085\pm0/010)M_\odot$.

According to Berger et al. (2010), the value of $L_R$ remains practically constant in the range of M0–L4; hence, that ratio $L_R/L_{bol}$ increases from $10^{-9}$ to $10^{-6.5}$. Figure 27 shows a comparison of X-ray and microwave radiation from cool dwarfs following Berger et al. (2008a), the linear correlation of these values does not basically differ from (26).

Using ATCA, Burgasser et al. (2013) detected the radio emission of the superactive binary system L5e+T7 2MASS J13153094-2649513 AB at 5.5 GHz: $\log(L_{rad}/L_{bol}) = -5.44 \pm 0.22$, but detected no radio emission at 9.0 GHz. The recorded emission was quiescent, without traces of variability and bursts throughout three hours of observations, without noticeable polarization. This object is one of the radio-brightest, its luminosity in the quiescent state is comparable with luminosities of other cool sources during bursts; it shows that for middle L dwarfs there is no drop in radio luminosity, but powerful $H_\alpha$ emission is present.

Using VLA, Suresh et al. (2020) detected stable radiation of ε Eri in the range of 2−4 GHz at a level of 29 μJy. The absence of noticeable brightness variations and small circular polarization of radiation led the authors to its identification with thermal optical thick gyroresonance emission of the stellar corona.

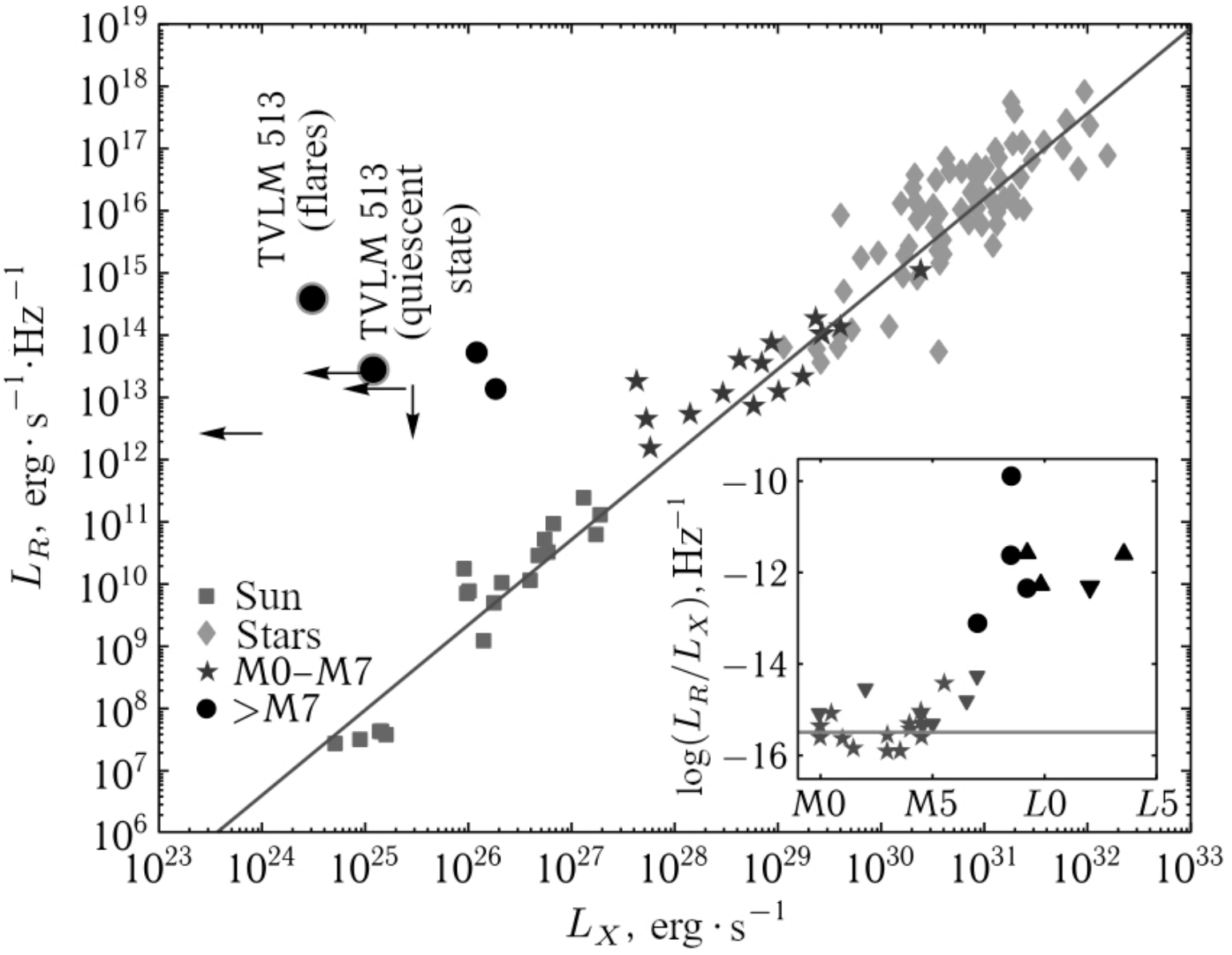

Fig. 27. Ratios of radio and X-ray luminosities of stars with coronal activity. Close correlation of $L_R$ and $L_X$ is evident, but it starts disrupting near M7 (Berger et al., 2008a)

* * *

Antonova et al. (2013) performed a radio review of M7–L3.5 dwarfs at 4.9 GHz, estimated a fraction of radioactive objects in this range as 9%, and detected a drop in this fraction toward later types.

In the case when it seemed that the gyrosynchrotron model for the interpretation of non-thermal microwave radiation of stars became widely accepted, Stepanov et al. (1999) and Stepanov (2003) showed that the plasma radiation near the second harmonic frequency excited by the instability of beams, which were captured into magnetic flux tubes of energetic electrons, is a strong alternative of the GS radiation of stellar coronae in the quiescent state. To excite this radiation, the density of energetic electrons by 6–7 orders of magnitude lower than the thermal plasma density is sufficient, the brightness temperature can reach $10^{14-16}$ K, noticeable and variable spectral polarization is typical of it, and within the conditions of high-temperature stellar coronae this radiation is excited with far greater effectiveness than on the Sun.

A direct evidence of the Stepanov model is periodic and 100% polarized radio emission from the M9 dwarf TVLM 513–46546 detected with VLA by Hallinan et al. (2006, 2007), its bursts are definitely coherent and due to electron cyclotron maser instability, which yields narrow radiation beams passing approximately every 2 hours across our field of view because of stellar rotation, as in pulsars. The model of thus-radiating region includes the low-density region above the magnetic pole of the stable dipole or multipole with a strength of up to 3 kG. Later, Hallinan et al. (2008) confirmed the Stepanov model by radio observations of the M8.5 dwarf LSR J1835+3259 and the L3.5 dwarf 2MASS J00361617+1821104. Similar to the TVLM 513–46546 star, the periodic bursts for these two dwarfs were circularly polarized by 100% and their periods were 2.84 ± 0.01 h and 3.08 ± 0.05 h, respectively. For the dwarf 2MASS J00361617+1821104, the detected nonpolarized periodic component had the brightness temperature which excluded the GS radio emission, and this nonpolarized component could emerge due to the polarization during the passage of the magnetosphere and scattering.

Osten et al. (2006a) constructed an analytical model of the corona for TVLM 513–46546 with a magnetic field less than 500 G and optically thin radiation at frequencies higher than 5 GHz. The results of observations of this object for years with VLA and Arecibo were published by Antonova et al. (2010). Doyle et al. (2010) established the regularity of radio bursts on the interval of several weeks, which means the stability of the electric field structure generated and maintained in the stellar magnetosphere and stability of the large-scale magnetic field during more than a year. Analogous properties of radio pulsations were detected for the fast M4 rotator V374 Peg by Hallinan et al. (2009).

Kuznetsov et al. (2011) considered two scenarios of the formation of dynamic radio emission spectra of the very cool dwarfs within the concept of electron cyclotron maser instability. Modeling the stellar magnetic field by the inclined dipole, they examined its interaction with the satellite and the narrow section of active longitudes and came to the conclusion that the radio light curve of TVLM 513–46546 could be represented by the second scenario. You et al. (2011) performed the modeling of radio bursts of this very cool dwarf TVLM 513–46546 within the same concept of electron cyclotron maser instability and interpreted high brightness variability and different impulse profiles with the model of the large-scale hot active region with the extended magnetic structure with a characteristic electron density of $(1–5) \cdot 10^5$ cm$^{-3}$, a temperature of $(1–5) \cdot 10^7$ K, a surface field at the level of 7000 G, a size of the active region of lower than the Jupiter radius and a zone of the impulse occurrence of an order of 0.007 Jupiter radii.

The radio emission from the M7 dwarf 2MASS J13142039+1320011 fits within the electron cyclotron maser mechanism. For this dwarf, at the level of ~ 1 mJy the radiation was recorded at four frequencies in the range from 1.43 to 22.5 GHz (McLean et al. (2011)). The 10-hour VLA observation detected a sinusoidal variation of radiation with a period of 3.89 hours with the 30% amplitude at 4.86 GHz and 20% — at 8.46 GHz; the periodicity was also seen in the circularly polarized radiation with a change of the sign in the phase with total intensity. Optical observations detected a period of 3.79 hours and $v\sin i$ = 45 km/s, hence, the size of the star was $> 0.13R_*$. The long-term stability of radio emission can be explained by the model of the inclined rotator with a field of at least 8 kG and a change of the sign — by the rotation of the magnetic pole with respect to the line-of-sight.

McLean et al. (2012) performed a VLA survey of about 100 M and L dwarfs to ascertain the role of rotation in the radio activity of ultracool stars; involving the literature data, they compiled a sample of 167 stars with the radio observations and measured rotation rates. Within M0–M6 they found that the radio activity–rotation relation had saturation at a level of $L_{rad}/L_{bol} \sim 10^{-7.5}$ for vsin$i$ > 5 km/s. But for M7 and later dwarfs, this luminosity ratio increased independently of the rotation rate, the divergence increased as well; at vsin$i$ > 20 km/s in X-rays and $H_\alpha$ there appeared a saturation, which was absent in radio range, and at such rates the fraction of radio detection of stars was three times higher than at vsin$i$ < 10 km/s. If the radio activity–rotation dependence was constructed using the Rossby number, then there would be no saturation of the radio emission up to Ro $\sim 10^{-3}$ for the range of stars from G to L, whereas for the X-rays and $H_\alpha$ it would occur at Ro $\sim$ 0.1 and saturation after M7. All this implies that the rotation effects are important in the regularization of topology or magnetic field strength at least for fully convective dwarfs. The fact that not all fast rotators are detected in the radio range provides additional support of the idea concerning two dynamos that results from spectropolarimetric observations.

## 1.4.4. Models of Stellar Coronae

Physical principles of the existence of the solar and stellar coronae are identical, but noticeable differences in values of the global stellar characteristics result in meaningful differences in the morphology of their coronae and the magnetic field structure. In this section, the started in three previous ones discussion of the physical characteristics of stellar coronae will be continued with particular attention to the models of stellar coronae following from these characteristics.

Throughout three years, Sanz-Forcada et al. (2003) carried out 16 XMM-Newton and Chandra observations of AB Dor with high resolution of the X-ray spectra. The EM distribution within log$T$ = 6.1–7.6 and the abundance of chemical elements in the corona yielded the agreed results from both instruments. As well as from the EUVE data, the EM distribution detected a steep rise to the maximum of about log$T \sim$ 6.9 and a substantial amount of plasma in the range of log$T \sim$ 6.9–7.3. The coronal abundance displayed a pronounced FIP effect as compared to the Sun. From the ratio of intensities of the FeXXI and FeXXII lines the following estimates were acquired: log$n_e \sim$ 10.8 at log$T$ = 6.3 and log$n_e$ = 12.5 at log$T \sim$ 7, i.e., a fast growth of density with the temperature. On the whole, it was concluded on the corona consisted of various families of magnetic loops that were smaller than the stellar radius and the filling factor $f \sim 10^{-4}$–$10^{-6}$.

Ness et al. (2004) contributed significantly to the knowledge on the stellar coronae. Based on the diffraction spectra derived with Chandra and XMM-Newton, they analyzed 42 such structures. From the ratios of intensities of the OVII lines they estimated densities of the cool (1–6 MK) components of coronae as log$n_e \sim$ 9.5–11 and from the NeIX lines log$n_e \sim$ 10.5–12, although the latter results may be distorted by the FeXIX and FeXXI lines. For the low-activity stars, the density did not exceed a few $10^{10}$ cm$^{-3}$, but for more active objects the certain trends were not found. A study of hotter coronal components from the FeXXI lines did not detect the presence of very-high-density regions $n_e > 10^{12}$ cm$^{-3}$. Then Ness et al. found that the plasma with a temperature of 1–4 MK dominated in the inactive coronae and it did not cover a large fraction of the surface. In active stars, such a plasma occupies a large fraction of the surface and a hot component appears. Its temperature correlates with the activity level and reaches the temperatures, which are known on the Sun only during the flares. The loops of the hot plasma seem to fill the space between cooler loops until the hot plasma starts dominating. At all the activity stages of such stars the filling factors are close or reach 0.1.

A study of Testa et al. (2004) was practically simultaneously published. Based on the observations of 22 active stars with Chandra/HETGS, from the OVII, MgXI and SiXIII lines they estimated the densities of coronae. These data do not confirm the EUVE results derived with lower resolution on the existence of structures with a density of more than $10^{13}$ cm$^{-3}$. The MgXI lines resulted in the density estimates of a few $10^{12}$ cm$^{-3}$ for the most sources with high X-ray luminosity (> $10^{30}$ erg/s), and stars with high ratios of $L_X/L_{bol}$ had an increased density at high temperatures. The ratios of the intensities of the OVII lines yielded a far lower density in units of $10^{10}$ cm$^{-3}$, which pointed to a different localization of hot (~ 7 MK) and cool (~ 2 MK) coronal structures. The compact hot coronal regions detected by the MgXI lines were characterized by the filling factor varying from $10^{-4}$ to $10^{-1}$, whereas the cooler regions with $f_{OVII}$ from $10^{-3}$ to ~ 1. Testa et al. found that $f_{OVII}$ approached unity at the X-ray flux level that is characteristic of solar active regions, i.e., when the stars are entirely covered by active regions. At the same surface flux $f_{MgXI}$ grew dramatically with growing flux. This result seems to confirm the suggestion that the plasma with $10^7$ K in active coronae emerges during the flare activity, and this activity grows significantly when the star becomes entirely covered by active regions.

Later, using Chandra/HETGS, Testa et al. (2007) studied the optical thicknesses of coronae of active stars in X-rays and found that in most cases they were small, as it was earlier established from the Fe XVII lines. A self-absorption was detected from the OVIII lines in the spectrum of EV Lac, which allowed one to estimate a length of quantum runs as several percent of the stellar radius and a very low filling factor. Analysis of the correlations with the basic stellar parameters showed that the objects with discernible optical thickness were at the upper activity level with the ratio of $L_X/L_{bol}$ at the saturation level and a high density of the hot plasma.

From the ratios of the intensities of the SiX lines recorded with Chandra/LETGS in the spectra of Procion, $\alpha$ Cen A and B, Capella and $\varepsilon$ Eri Liang et al. (2006) estimated the values of $\log n_e$ varying from 8.61 to 9.11, which are close to estimates on the analogs of the helium triplet, while Liang and Zhao (2006) estimated the temperature of coronae of these stars as 1.6 MK from the SiXI lines.

Peres et al. (2004) compared a large sample of data on the coronae of stars of different spectral types and activity levels derived with ROSAT and the X-ray solar data acquired with Yohkoh at different solar cycle phases and attributed to different objects from flares to the most quiescent regions. Whereas they detected a general close correlation of the surface X-ray fluxes with spectrum hardness, which covered all the targets, and concluded that the coronae of later spectral type stars were produced by the X-ray sources similar to the solar ones. But a spatial localization of these structures may differ significantly from that on the Sun. Thus, a fraction of the stellar surface covered by active regions, as well as their brightness, increase with growing activity. But the most active stars are brighter and hotter than if they would be fully covered by solar active regions. This may be possible if, apart from them, in each moment of time several flares are present.

One of the first models of stellar coronae was constructed by Güdel et al. (2005) for the eclipsed short-period binary system CM Dra, which based on many parameters resembles the YY Gem system but consists of later components of dM4.5e, whereas YY Gem — from the components of dM1e. From the X-ray and U bands of light curves derived with XMM-Newton in the vicinity of the primary and secondary minima, they found the three-dimensional distributions of the coronal matter of each component. The basic part of the coronal plasma was between latitudes −50° and +50° at the fairly inhomogeneous longitudinal distribution. The average radial profile detected the height scale of about $10^{10}$ cm, which corresponds to the isothermal corona with a temperature of about 8 MK. The density of the corona was $10^9$– $10^{10}$ cm$^{-3}$, similar to that in solar active regions. Being constructed from 23 emission lines in the range from 12.1 to 34.5 Å, the EM distribution yielded a maximum of this value of 10 MK and the FIP effect of [Ne/Fe] ~ 0.9 and [O/Fe] ~ 1.0.

Using the Chandra and FUSE LETG and HETG observations, Hussain et al. (2005) constructed a model of the corona of the single young star AB Dor. Their model consists of two components: one of them is responsible for 80% of the X-ray flux and forms a polar structure or a homogeneously distributed corona, the second one comprises two or three compact regions yielding the light curve modulation and shifts of the OVIII lines; these compact regions yield 16% of the total X-ray flux and are located at a level of less than $0.3R*$. In the FUSE spectrum of this star, Taylor et al. (2008) detected significant variations of the OVI and CIII lines between 1999 and 2003 and shifts of narrow cores of these lines with respect to the broad wings.

From the map of the radial magnetic field of the same young rotator AB Dor obtained with the ZDI technique, Cohen et al. (2010) calculated the MHD model of the corona and found that the strong entanglement of flux lines of the azimuthal field dominated in its general structure. From the detected solution they constructed the Alfvén surface and without excessive theoretical simplifications estimated the rates of mass losses and angular momentum; the former value was 10–500 times higher than the solar one, the second — by $3 \cdot 10^4$ times. They note that the MHD approach yields more information on the physical system than the used extrapolation of the potential field. Later, based on the dynamo model, Cohen et al. (2017) acquired the photospheric magnetic field, used it as initial for modeling the corona of the fully convective M dwarf, and found that the high-latitude giant dipole coronal loops of several kilogauss dominated in the quiescent X-ray radiation from the rapidly rotating stars.

Arzoumanian et al. (2011) considered a possible influence of starspots on the structure of the corona and concluded that an important factor of such an influence could be not only the magnetic flux of the spot but its position on the disk. For instance, on the strongly spotted star AB Dor, spots can double a fraction of the area with the open field, which shortens the braking time of the star and increases by two orders of magnitude the X-ray luminosity and rotational modulation of X-ray radiation.

The direct solar observations and the analysis of stellar X-ray studies show that the fundamental element of their coronae is numerous loops of the magnetized plasma, in which there occurred flares, plasma heating, and acceleration of particles. According to Stepanov (2004), the parameters of these loops from the numerous radio, optical, ultraviolet, and X-ray observations are listed in Table 13.

Within the survey report, Stepanov (2004) considered three possible mechanisms of the formation of these structures: the loop as a MHD resonator, the loop as an electric chain, and the loop as a magnetic mirror trap. Elaborating each one in detail, he associated them with the concrete observed phenomena in the coronae.

In the framework of the mentioned above multiwavelength studies of the quiescent upper atmosphere of EV Lac, Osten et al. (2006b) found that there was a practical constancy of the electron pressure in the transition zone and in the low-temperature corona in the temperature range $\log T = 5.2–6.4$, but there was detected a drastic jump of density up to $n_e = 10^{13}$ cm$^{-3}$ in the high temperature region $\log T = 6.9$; the situation is not in agreement with the hypothesis of the hydrostatic equilibrium for maintaining the quiescent state: the heat-conduction losses should be greater here, as well as a large energy inflow for their compensation. To retain the high-temperature structures, the strong magnetic fields are required, and the detected high frequencies of radio events require the high-temperature X-ray coronal plasma to be spatially separated with a radio-emitting source. Linear scales in the low-temperature corona are significantly greater than the structures of the high-temperature corona, which results in the conclusion on the inhomogeneous mixture of the thermal and nonthermal coronal plasma. The spectral analysis detected a decreased metallicity of the corona as compared to the Sun.

Table13. Parameters of the coronal loops of the Sun and red dwarfs (Stepanov, 2004)

| Parameters | on the Sun | on the red dwarfs |
|---|---|---|
| Loop length, cm | $(1–10) \cdot 10^9$ | $2 \cdot 10^9–3 \cdot 10^{11}$ |
| Loop section, cm | $(1–5) \cdot 10^8$ | $10^8–3 \cdot 10^9$ |
| Plasma density, $cm^{-3}$ | $10^9–10^{12}$ | $10^{10}–10^{13}$ |
| Plasma temperature, K | $10^6–10^7$ | $3 \cdot 10^6–10^8$ |
| Magnetic field, G | $10^1–10^3$ | $3 \cdot 10^2–10^4$ |
| Emission measure, $cm^{-3}$ | $10^{47}–10^{50}$ | $10^{50}–10^{53}$ |

Ness and Jordan (2008) analyzed the observations of $\varepsilon$ Eri with several instruments: Chandra LETG, EUVE, HST STIS, FUSE, and XMM-Newton RGS. From the measured emission lines they found the relative abundances of elements, the estimates of electron density and pressure, and the emission measure distribution. In their final coronal model, the temperature was 3.4 MK, the electron pressure $1.3 \cdot 10^{16}$ $cm^{-3} \cdot$ K at the temperature level of $2 \cdot 10^5$ K, the filling factor 0.14 at a level of $3 \cdot 10^5$ K, and no evidence of the FIP effect.

Lang et al. (2012) studied variations of the stellar magnetic topology at the intersection of the boundary between partially and fully convective stars and their effect on the properties of the corona. They considered both the open flux values that affect the angular momentum losses in the stellar wind and the X-ray EM. They extrapolated the constructed maps of radial magnetic fields on the surface into the three-dimensional coronal magnetic fields for making a sample from the early and middle M dwarfs. In the course of the magnetic reconstruction, it is possible to choose the field geometry that is symmetric or antisymmetric with respect to the equator, and they found that the dipole field component governing the large-scale structure strengthened as the stellar mass decreased, whereas the value of the open magnetic flux associated with the wind was proportional to the reconstructed flux value. Within the assumption of the hydrostatic and isothermal corona, they calculated the X-ray EM and rotational modulation of coronae and reproduced the observed saturation at $Ro < 0.1$. In their opinion, the rotational modulation of X-ray emission is an unreliable indicator of the magnetic structure, since it does not depend on the Rossby number but may be useful in the selection from different field geometries.

* * *

Thus, the X-ray luminosity of the coronae of active stars exceeds up to four orders of magnitude the solar corona luminosity, and this means the imperfection of the initial simple scaling of these structures. Favata (2001) performed a detailed comparison of structures of the solar and stellar coronae and concluded that contrary to the solar corona, where the long and close to the equator formations prevail and thin filaments are localized at low and middle latitudes, the coronae of active stars have a prevailing number of structures that are far smaller than the stars and located at high and near-polar latitudes with the temperatures of several and several tens of megakelvins; these are denser structures of the flare plasma which are constantly absent on the Sun. However, the basic result of studying the stellar coronae is in establishing the universality of these structures practically on all the F and later stars: such stars without coronae do not exist or they are exceptionally rare. The study of stellar coronae is one of the rapidly developing branches of the solar-stellar physics, which — as many other branches — has been raised conceptually and instrumentally from the solar physics. The peculiarities of this branch are determined by a remarkable subject for study — stellar coronae, whose structure is practically not associated with visible stellar surfaces but depends on such global parameters as rotation, the depth of the convective zone and the intensity of the matter mixing in it. This situation, though paradoxical at first sight, when the outermost atmospheric layers are determined by the internal structure of the star, can be easily explained: the physical state and spatial structure of stellar coronae are determined by magnetic fields, which are generated in the subphotospheric layers. This concept, common for the Sun and stars with convective shells, first made it possible to understand the phenomena, for which it sufficed to simply scale solar processes to stronger and hotter stellar coronae. Further extrapolation enabled qualitative interpretation of the phenomena nonexistent on the Sun: the very hot component of the coronae and strong microwave emission.

## 1.5. Heating Mechanisms for Stellar Atmospheres

In the previous sections, we presented experimental data on different atmospheric layers of active stars as thoroughly as possible. Below, we summarize the data analysis results that yielded the conclusions on possible heating mechanisms only as qualitative descriptions. This circumstance is due to both the absence of unambiguous and generally accepted solution so far concerning the problem of heating of stellar atmospheres and the mentioned in Preface general character of the current monograph.

The idea of heating of stellar atmospheres by nonradiative energy fluxes released from subphotospheric layers, which explains the existence of hot chromospheres and coronae, has been under development for many decades. The fluxes can be associated with different hydrodynamic and magnetohydrodynamic processes observed in the solar atmosphere and expected in stellar atmospheres, while the considerable diversity of the processes is due to the mobility and appreciable three-dimensional stratification of all physical parameters of the medium. First, the heating of the solar chromosphere was explained by acoustic waves generated by convection, then, by the field of velocities of 5-min oscillations, Alfvén and other magnetohydrodynamic waves, the Joule heating by electric currents, the quasistationary annihilation of magnetic fields, and fast annihilation of the fields both in the nonstationary low-energy processes, microflares, and in high-energy flares. But for the most active dMe stars the radiation of quiescent chromospheres and coronae is about $10^{-3}$ of the bolometric luminosity, i.e., higher by orders of magnitude than on the Sun, and it was not clear if the solar mechanisms were sufficient for the heating of such strong stellar atmospheres.

To generally orient in this issue, one should take into account the result on activity of several hundreds of dM1 stars obtained by Houdebine (2011): $L_X \propto (L_{CaII})^{3/2}$, i.e., the coronal emission grows faster than the chromospheric one, and the X-ray flux $F_X \propto (P/\sin i)^{-3.7}$, i.e., the coronal emission far stronger depends on the rotation period than on the chromospheric emission.

At the All-Russia Astronomical Conference in 2010, Zaitsev (2010) made a comprehensive report on the sources of heating of magnetic loops in the solar corona, but it did not contain a definite conclusion on the main source of such heating. Zaitsev and Kislyakova (2010) considered in more detail the parametric resonance mechanism of coronal magnetic loops with 5-min oscillations of the photosphere.

Zaitsev and Stepanov (2018) identified a slowly varying radio emission component of the coolest stars with gyrosynchrotron radiation of the system of magnetic loops that were quasi-uniformly distributed over the surface.

* * *

Fosbury (1973) concluded that nonthermal components in the widths of the $H_\alpha$, CaII H and K and MgII h and k lines in the stellar spectra were due to acoustic waves. However, Blanco et al. (1974) from the measurements of absolute luminosity of the CaII K emission line concluded that the acoustic waves were sufficient to maintain the chromosphere only on the stars with an effective temperature of above 5000 K, whereas for cooler stars this flux calculated from the Lighthill–Proudman theory was insufficient: for the coolest stars it is lower by two orders of magnitude than the required value. When Linsky and Ayres (1978) showed that the losses for the radiation of ultraviolet MgII h and k lines exceeded approximately by a factor of 3 those of the violet calcium lines, the insufficiency of the heating mechanism by acoustic waves became even more convincing.

The fact that the solar chromosphere, transition zone, and corona radiate mostly above the magnetic regions and there are clear correlations between the appropriate radiation fluxes and magnetic fluxes of active regions resulted in the conclusion on the domination of magnetic heating mechanisms. These mechanisms can be reduced to immediate dissipation of magnetohydrodynamic waves or to the generation of the beams of fast particles in the course of development of different instabilities and subsequent thermalization of the particles. As stated above, Mullan (1975a) proposed a concept of dMe stars as magnetic red dwarfs, in which strong magnetic fields enabled the heating of high-density chromospheres by magnetohydrodynamic waves. According to his estimate, such a heating can provide the radiative losses of chromospheres estimated by Blanco et al. (1974).

The estimates of radiative losses with regard to ultraviolet lines led Basri and Linsky (1979) to the conclusion of the existence of chromospheres with minimum nonradiative heating due to acoustic waves, which is compensated by radiative losses in the emission lines and $H^-$. In the solar flocculi, the chromospheric losses for the radiation of the MgII and CaII resonance doublets are 10 times higher, and since the intensity of the CaII K line correlates with the magnetic-field strength, the necessity of involving the magnetic heating mechanism became obvious. The insufficiency of the acoustic heating of the atmospheres of active red dwarfs was demonstrated by Haisch and Linsky (1980), who established that the losses of the corona of the dK5e star EQ Vir for radiation, heat conductivity, and wind were higher by two orders of magnitude than the heating expected from this mechanism.

The most notable argument against the concept of purely acoustic heating of stellar atmospheres was the detection of X-ray emission of the coronae of different spectral type stars and the weak dependence of this radiation on the effective temperature (see Fig. 22).

Stein (1981) considered the dependence of the flux of acoustic and various hydromagnetic waves leaving the subphotospheric convective layers toward the atmosphere on the general stellar parameters and showed that the observed very weak dependence of radiative losses of stellar chromospheres and coronae on the effective temperature corresponded to Alfvén or slow magnetohydrodynamic waves.

Comparing the absolute fluxes of ultraviolet emission lines formed in the chromospheres and in the transition zones and the X-ray emissions of G–K dwarfs, Ayres et al. (1981a) found that the ratios $R_{hk}$, $R_{lines}$ and $R_X$ correlated, but between $R_{hk}$ and $R_{chrom_lines}$ (chromospheric lines) the correlations were linear, whereas between $R_{hk}$ and $R_{trans_lines}$ (transition-region lines) and $R_X$ they were nonlinear. The difference between the dependences of ratios of $R_{hk}$ and $R_{CIV}$ and (5) on the effective stellar temperatures noted by Linsky et al. (1982) evidence certain differences in the heating mechanisms for the chromospheres, transition zones, and coronae. Later, Schrijver (1990) found that the exponent $\kappa$ in the relation $F_{CIV} \sim B^{\kappa}$ was between the exponents in the analogous expressions relating $F_{HK}$ and $F_X$ with $B$. This fact made it possible to admit the existence of two heating mechanisms for the transition zone: one analogous to the heating of the chromosphere and the heat conductivity from the corona. Along with this, as stated above, the widths of emission lines measured by Ayres et al. (1983b) led to the conclusion on the commonness of the heating mechanisms for the transition zones in the different luminosity stars from the solar-active regions to M dwarfs.

Cram (1982) calculated the model of such a dMe star chromosphere, which provided the relations of the equivalent widths of Balmer lines corresponding to the observations, and showed that in this case the coronal X-ray emission was sufficient to heat a stellar chromosphere. This hypothesis was supported by the closeness of the total losses for radiation of ultraviolet lines of the chromosphere and transition zone and for the X-ray emission of the corona found by Giampapa et al. (1981b). However, in this case the Balmer lines should appear after or simultaneously with X-rays but not before, as it was shown by observations. Further, upon calculating the atmospheric temperature distribution of the emission measure from the observations of ultraviolet lines in the spectra of three inactive G–K dwarfs $\tau$ Cet, $\delta$ Pav, and 61 Cyg A and constructing the transition-zone models, Fernandez-Figueroa et al. (1983) obtained an ambiguous result: the thermal conductivity of $\tau$ Cet was insufficient to compensate for the radiative losses, whereas that of $\delta$ Pav and 61 Cyg A could compensate for these losses, and the models of the transition zones of these two stars corresponded to the X-ray fluxes measured from them.

Marcy (1983) found that for G–K stars the dependence of the calcium emission on the magnetic field and the effective stellar temperature corresponded to the heating of chromospheres by magnetohydrodynamic waves, while the correlation of the field and soft X-ray emission from coronae evidenced that coronae were heated by Alfvén waves.

An important advance in studying the heating mechanisms for stellar atmospheres was achieved owing to the concept of differential emission measure (DEM) that enabled the first calculations of the whole energy in the atmospheric layers, where optically thin radiators prevailed. The models elaborated within this concept by Jordan et al. (1987) made it possible to calculate radiative losses at different levels, heat conductivity from the corona to underlying layers, required strength and height distribution of heating sources. It was shown that the distribution of the emission measure between $2 \cdot 10^4$ and $10^5$ K could be explained if the nonthermal broadening of the lines was due to the energy release caused by the passage of Alfvén or slow magnetohydrodynamic waves equilibrating the local radiative losses. But the outflow of matter, the diffusion and geometry of the corona make the situation somewhat ambiguous.

Continuing the studies of Basri and Linsky (1979) and Mewe et al. (1981), Schrijver (1987) found the minimum level of surface fluxes in the main emission lines of the chromosphere and the transition zone that was essentially dependent on the effective stellar temperature, the so-called basal level of chromospheric activity that has no correlation with the X-ray luminosity and is observed in old and slowly rotating stars and in the centers of solar supergranules far from noticeable magnetic fields. This level was associated with the acoustic heating whose presence determined the weak chromospheric emission in low-activity stars.

To check if the acoustic waves were sufficient to maintain the basal chromospheric activity level of M dwarfs, Mullan and Cheng (1993) thoroughly studied the plane-parallel atmospheric model with effective temperatures of 3000 and 4000 K. They found that the equilibrium state of stellar atmospheres could indeed be realized at equal dissipation of weak shock waves emerging under the upward propagation of acoustic waves emitted by subphotospheric convection, and the radiative losses of the atmosphere in the MgII and $Ly_\alpha$ lines. Later, Mullan et al. (1995) showed that in the sample of over 80 nearby red dwarfs practically all dM stars had surface X-ray fluxes of less than $10^{5.2}$ erg/(cm$^2 \cdot$ s), and practically all dMe stars had greater fluxes, while the maximum fluxes on acoustic heating of the coronae, according to the calculations of Mullan and Cheng (1994), were $10^{5.0-5.1}$ erg/(cm$^2 \cdot$ s).

Mathioudakis and Doyle (1992) considered the surface fluxes in the MgII lines of about 160 K–M dwarfs and found that the strong dependence of the basal level $F_{\rm MgII}{}^{\rm min}$ on the effective stellar temperature confirmed its acoustic nature.

Using the detailed calculations within the model of acoustic heating, Buchholz et al. (1998) presented the basal levels of calcium and magnesium emission of F0–M0 dwarfs within a range of two orders of magnitude to the accuracy of a factor of 2.

The excessive emission above the basal level correlates with rotation and is of magnetic origin, but it is not clear if the additional heating occurs under the dissipation of turbulence or magnetohydrodynamic waves, or the fast annihilation of magnetic fields generates the fluxes of accelerated particles, which cause the further heating of the ambient plasma.

Musielak et al. (1990) calculated the energy fluxes generated as transverse waves of magnetic tubes coming out of the convective zone and found that the fluxes were sufficient to maintain the observed X-ray emission of F–G–K dwarfs.

Schrijver and Aschwanden (2002) considered the heating of solar and stellar coronae by an ensemble of a great number of coronal loops. They found an expression for the energy flux spent for corona heating as a function of the magnetic-field strength at the loop base, its extension, and the velocity of gas motion at the base, and satisfactorily presented the observed X-ray emission of active stars with rotational periods of more than 5 days.

For the most active fast rotators, the very young stars and the components of close binary systems, Vilhu and Rucinski (1983) found the upper limit of surface fluxes in the transition zone lines — the saturation level. Then Vilhu et al. (1986) and Vilhu (1987) extrapolated the saturation concept to the maximum surface fluxes in the MgII and CIV lines and in soft X-rays. They advanced a hypothesis that maximum fluxes corresponded to the saturation due to the fact that the filling factor of magnetic structures on the stellar surface reached values close to unity. In this case, the reverse relation of the magnetic field to convection and differential rotation is switched on, which restrains a further growth of nonradiative energy and thus prevents further nonradiative heating of the stellar atmosphere. On the whole, the saturated fluxes in the X-rays and spectral lines forming in the chromosphere and transition zone are only 10 times lower than the estimates of the mechanical energy flux in the subphotospheric convective zone. Thus, the saturated fluxes can indicate the maximum efficiency of the conversion of the mechanical energy into the heating of outer layers of stellar atmospheres. Since the heating is maximum where the magnetic fields are the strongest, the saturation is realized when the star is completely covered by the strongest magnetic fields, which occurs when the magnetic and gas pressures in the photosphere are equal. But there are other points of view on the saturation nature. Thus, Doyle (1996b) concluded that this effect in the ultraviolet could be caused by the disregard of the radiative losses by hydrogen, which should be great on rapidly rotating dwarfs. Jardine and Unruh (1999) suggested that the saturation of X-ray emission of fast rotators was due to the effect of centrifugal forces, which at the level of corotation essentially disturbed the corona and initiated the above-mentioned large prominences retained by the magnetic tension at loop tops.

Summing up the results of the Conference on Mechanisms of Chromospheric and Coronal Heating held in Heidelberg in 1990, Linsky (1991) noted that the problem of heating of quiescent stellar chromospheres could be directly related to the conclusion of solar studies: the quiescent solar chromosphere can be heated by the dissipation of the energy of magnetohydrodynamic shock waves in the so-called bright points related to the elements of the chromospheric network. Such points are observed only in the places of localization of magnetic structures, their lifetime is 100–200 s, magnetic fields in them are of 10–20 G, and they repeatedly occur in the same magnetic elements. Only two mechanisms of Linsky's extensive survey on heating mechanisms for stellar atmospheres can be applied to the objects considered in this book: short-period acoustic waves that guarantee a weak basal level of the atmospheres of late stars and slow magnetohydrodynamic waves propagating from the convective zone to the chromosphere, heating the lower and middle chromosphere under shock wave dissipation, and easily excited under the interaction of other waves.

As it was mentioned, over a number of years, the insufficiency of the acoustic heating of the atmospheres of late stars was emphasized, and M stars occupied the region in the Hertzsprung–Russell diagram, where the insufficiency was the most evident. (After the revision by Bohn (1984) the theory of acoustic heating became much more efficient, but he apparently disregarded the weakening of the acoustic wave flux in the photosphere. Thus, it is not clear which fraction of the large flux of $10^6$–$10^7$ erg/(cm$^2\cdot$ s) ascertained by him can be spent for the heating of the chromosphere and corona.) However, it should be noted that the predictions of the acoustic theory were always compared with the most active dMe stars. But the acoustic heating can also play an important role in the atmospheres of dM stars, where the flux from the chromosphere is only $10^4$–$10^5$ erg/(cm$^2\cdot$ s) (Giampapa et al., 1989).

Apparently, the consideration of acoustic and magnetic heating as mutually exclusive mechanisms was, to a certain extent, exhausted by Cuntz et al. (1999), who calculated the chromospheric models of single K0–K3 dwarfs with rotational periods from 10 to 40 days, in which the surface magnetic component and the nonmagnetic component were heated by the waves in the longitudinal flux tubes and by acoustic waves, respectively. They found that the heating and maximum chromospheric emission were largest in rapidly rotating stars, while the stars with very slow rotation had the basal level of chromospheric emission. This can be illustrated by the example of the G8 V star $\tau$ Cet (Rammacher and Cuntz, 2003; Cuntz et al., 2007).

Thus, the chromospheric emission may generally have three different heating components: the basal level caused by acoustic heating and emission caused by magnetic heating owing to the large-scale and/or turbulent magnetic fields, and the latter component remains effective in low-mass stars where there are no regular solar-type fields (Durney et al., 1993).

De Pontieu et al. (2007), Jess et al. (2009), and McIntosh et al. (2011) recently discovered Alfvén waves with an amplitude of about 20 km/s and a period of 100–500 s in the lower solar atmosphere. This allowed them to conclude on the predominant heating of the atmosphere by this mechanism. The analogous results were recently obtained by Grant et al. (2018) for the spotted region on the Sun.

Van Ballergooijen et al. (2011) constructed the 3D MHD model of the propagation and dissipation of Alfvén waves in a solar coronal loop. The model involves the lower atmosphere at two loop bases. Waves arise in the photosphere in kilogauss elements with a radius of an order of 100 km. The model describes the nonlinear interaction of Alfvén waves in the magnetohydrodynamic approximation, the increase of the wave velocity with height in the chromosphere and transition zone produces a strong wave reflection resulting in their collisions and turbulence in the photospheric and chromospheric parts of the flux tube. A part of the wave energy passes into the transition zone and causes turbulence in the corona. The heating of the coronal loops and chromosphere was found to be provided by the Alfvén turbulence, if the small-scale motions at the loop base have velocities of 1–2 km/s and time scales of 60–200 s, and the heating of the corona increases with the field strength in the corona and decreases with the loop length.

Schröder et al. (2012) considered in detail an unusually deep and prolonged solar activity minimum in 2009, when the Sun achieved the minimum value of $S$ as compared to earlier observed minima, at some short intervals this index was the same as for inactive solar analogs, while its surface was free of flocculi and weak remnants of active regions, and the flux in calcium lines was identical with the basal threshold level. It was concluded that the chromosphere of the quiet Sun is a universal phenomenon typical of inactive stars, and their magnetic field generated by the turbulent dynamo provides a natural explanation of the minimum level of X-ray emission of inactive stars.

* * *

Different variants of the heating of stellar atmospheres by flares were discussed repeatedly, those by an ensemble of individually recorded events, and numerous microflares inaccessible for such a registration with characteristic energies of $10^{30}$ erg (Butler et al., 1986) and nanoflares with energies of the order of $10^{24}$ erg (Parker, 1988). Of course, heating by flares is not an alternative to magnetic heating, but one of its realizations.

As stated above, the concept of coronal heating by flares was advanced after the discovery of the linear correlation of X-ray luminosity of flare stars with the time-averaged optical luminosity of flares (see (19) in Chapter 1.4). At the same time, numerous short-lived bursts of hard X-rays were found on the Sun and the possibility of heating of the solar corona by the bursts was actively discussed. Therefore, when Butler and Rodonò (1985) found the fine structure in the EXOSAT records of soft X-ray emission of UV Cet, EQ Peg, and Proxima Cen out of bursts with characteristic times of about 20s and $L_X \sim 2 \cdot 10^{30}$ erg, and when the temporal correlation of X-ray bursts and bursts in the $H_\gamma$ line was found during parallel optical and X-ray observations of UV Cet, they concluded that these were microflares and that in X-rays they observed were not stellar coronae heated by microflares but microflares themselves. (It should be noted that since the energy of bursts is about $2 \cdot 10^{30}$ erg, they most probably correspond to compact solar flares rather than solar microflares with a typical energy of $10^{27}$ erg or Parker's nanoflares with an energy of $10^{24}$ erg.) Then, Butler et al. (1988) found the linear correlation of total energies $E_{H\alpha}$ and $E_X$, common for stellar and solar flares and covering a range of four orders of magnitude. Butler (1992) found that solar flares recorded in the range of 8–12 Å also satisfied this correlation, while the flare on the RS CVn-type star II Peg continued it for two orders of magnitude toward higher energies. Later, the arguments both in favor and against the substantial role of microflares in the energy of stellar atmospheres were obtained. Thus, Butler et al. (1986) believed that the established variations of $F_X$ at times of 100–1000 s in the newly processed observations of dKe and dMe obtained at the Einstein Observatory supported the concept. Mathioudakis and Doyle (1990) continued the correlation of $E_X$ and $E_{H\alpha}$ of flares to the quiescent state of active M dwarfs. Considering the results of studying the quiescent-state spectrum of EV Lac that proved a considerable contribution of microflares to the radiation of Balmer lines out of flares (Alekseev et al., 2003), the conclusion of Mathioudakis and Doyle seems to be a natural continuation of the conclusion of Butler et al. (1988). Young et al. (1989) found a rather close correlation between the luminosity of the $H_\alpha$ emission and quiescent X-ray luminosity of the coronae: $L_{H\alpha} \sim 0.2 L_X$, which may suggest the heating of the entire upper atmosphere from a single source or the coronal heating of the chromosphere. But, Haisch (1989) noticed that the X-ray emission in a very broad range of intensities was recorded in solar flares of the same $H_\alpha$ importance. Tsikoudi and Kellett (1997) detected slow and small-amplitude variations of the brightness of active stars in the EUV range and concluded that the processes, which they called milliflares, were important for coronal heating. But the detailed analysis of the EXOSAT X-ray observations carried out by Collura et al. (1988) and then more thoroughly by Pallavicini et al. (1990a) did not confirm the hypothesis on the decisive contribution of microflares. Besides, Schmitt (1993) did not find the expected effects of microflares in ROSAT observations.

When the power energy spectra of flares are extrapolated to the region of supposed microflares with the same spectral index, which is determined from individually recorded optical and X-ray flares and is usually equal to 0.7–0.9, the contribution of microflares to the total flare radiation appears to be low (see Chapter 2.3). However, even for the Sun there are observational and theoretical evidences that in the microflare range the spectrum of frequency energy distribution should be softer than within the range of stronger events. Therefore, even on the Sun the contribution of microflares should be higher than in the estimate with a constant spectral index for the whole range of flare energies. Finally, recently data confirming the significant role of microflares on active late stars were obtained. Based on the EUVE observations of 28 flares on active solar-type stars, for 47 Cas and EK Dra $\beta \sim 1.2 \pm 0.2$ within the energy range of $3 \cdot 10^{33}$–$6 \cdot 10^{34}$ erg (Audard et al., 1999). Considering EUVE data for F–M stars, Audard et al. (2000) concluded that for flares on F–G dwarfs $\beta > 1$, as for K–M stars it is more probable that $\beta < 1$. At $\beta > 1$ $E_{\rm min} = 10^{29}$–$10^{31}$ erg is sufficient for the time-averaged luminosity of flares to be comparable with the total luminosity of a quiescent corona, and such $E_{\rm min}$ are typical of flares on the Sun. This conclusion was later confirmed by Güdel et al. (2003b). From the 37-day EUVE monitoring of AD Leo and overlapping BeppoSAX monitoring they found that $1.0 < \beta < 1.4$ for EUV and X-ray flares of this star. These results were obtained by two independent statistical methods and under independent consideration of the EUVE, LECS, and MECS BeppoSAX data. Simultaneously, they developed the calculation method for the expected DEM for the stochastic flare ensemble, whose success provided another argument in favor of the entire statistical model of flare activity. Later, Arzner and Güdel (2004) developed a refined analytical approach to the analysis of light curves as a result of the flare superposition associated with coronal heating. Applying a developed formalism to the EUVE/DS observations of AD Leo, they found that the distribution of amplitudes could be represented by the reduced power law with an index of $1.3 \pm 0.1$.

Studying the active corona of the F8 dwarf HD 35850 with EUVE and ASCA, Gagné et al. (1999) concluded that the variability of its EUV radiation could be presented by the model of continuous flares, having integral energy spectrum with the spectral index 0.8 and capable of heating the corona.

If corona conductivity contributes to the heating of the quiescent chromosphere, during flares coronal heating occurs through the evaporation of the chromosphere initiated by energy particle beams from the corona. This conclusion was made by Güdel et al. (1996) on the basis of simultaneous X-ray and microwave observations of UV Cet and confirmed by the analysis of the flare on Proxima Cen of 12 August 2001 cited below (Güdel et al., 2002).

Discussing permanent microwave emission of stellar coronae, Kundu et al. (1987) noticed that fast particles emerging in flare processes were responsible for the radiation and should efficiently precipitate, and this could be an important mechanism of coronal heating for the chromosphere.

Güdel (1997) considered the EM($T$) distributions in the solar-type stars and showed that the observed two-peak distribution could be due to the sum of two components: the hot component stipulated by the cooling plasma of strong flares in large loops and the cold component that is due to a large number of microflares. Referring again to the EM($T$) distribution, Cargill and Klimchuk (2006) analyzed the strong peaks in this distribution at $10^7$ K at an electron density of $10^{13}$ cm$^{-3}$ and came to the conclusion that they appeared in numerous small, up to $10^8$ cm, loops with a field of about 1 kG, and in one such an event there released $10^{26}$–$10^{28}$ erg, as in a solar microflare.

One of the explanations of the revealed close correlation of the thermal X-ray and gyrosynchrotron microwave emissions of stellar coronae and solar flares (Benz and Güdel, 1994) consists in the common origin of the corona heating and particle acceleration. Already in 1985, Holman (1986) formulated a hypothesis that nonthermal electrons necessary for the gyrosynchrotron emission could appear in the current sheets, in which the Joule losses are the heating mechanism of stellar coronae. Airapetian and Holman (1998) developed this idea: they considered two mechanisms related to electric currents, within which the Joule heating of plasma to millions of K and the acceleration of electrons to subrelativistic energies occurred simultaneously. In the first model, electrons are accelerated by electric currents within classical current sheets in stellar coronae. The second model deals with MHD turbulence: in the small-scale regions in the presence of current sheets it excites ion sound waves, which intensify the heating and accelerate particles in the transition zones of stellar atmospheres. In the first model, one should expect a linear correlation of $L_X$ and $L_R$, in the second, a power correlation with an exponent of 0.6–0.8. Recently, Podlazov and Osokin (2002) concluded that the coronal heating and generation of fast particles occurred simultaneously within the "avalanche" concept (see below).

Stelzer et al. (2012) carried out the XMM-Newton observations of the nearby M9 dwarf DENIS-P J104814.7-395606 and found that its indicators of X-ray and chromospheric activity expanded the range of applicability of the flux–flux relation to the ultracool dwarfs. The approximated agreement of this and other ultracool dwarfs with the flux–flux relations for early M dwarfs allows one to suppose the same heating mechanism for their atmospheres. The observed Balmer decrement of this dwarf was consistent with the optically thick LTE plasma at the photospheric temperature or the optically thin LTE plasma at 20000 K. Difficulties with the interpretation of a high $H_\alpha/H_\beta$ flux ratio for this and two other cool dwarfs, VB 10 and LHS 2065, stimulated the idea of two types of very cool stars.

* * *

Let us consider the results of the studies dealing with particular aspects of the general problem of heating of stellar atmospheres. The analysis of several hundreds of ultraviolet spectra in the region of the CII line 1335 Å showed that strong emission of the high-temperature line was common for early F dwarfs and its intensity was comparable with the value recorded for the most active solar-type stars (Simon and Landsman, 1991). The emission maximum is achieved near F0V and decreases toward late A dwarfs. For stars earlier than F5, the CII emission does not correlate with the rotation rate, which confirms that the chromospheres of A–F stars are heated under shock dissipation of sound waves.

Garcia López et al. (1993) studied the He $D_3$ line in the spectra of 145 F– G stars and found that near the spectral class F0 the depth of the convective shell was sufficient to maintain the chromosphere by acoustic heating; the lack of a dependence on rotation and the Rossby number in such stars evidences the absence of magnetic heating in them, which appears only in F5 stars.

Wood et al. (1997) analyzed the SiIV and CIV line profiles obtained with HST in the spectra of stars of late spectral types and presented these profiles by the sum of two Gaussians with substantially different widths. They made the following conclusions on the heating mechanisms for the transition zone. The widths of narrow components correspond to subsonic thermal velocities and can be caused by the dissipation of turbulent motions or the Alfvén waves, whereas the wide components corresponding to supersonic motions can arise at microflares. The wide component of the SiIV $\lambda$ 1394 Å line profile with FWHM = 109 km/s was observed in the spectrum of $\alpha$ Cen A, which is practically identical to the Sun and in which analogous wide components are seen in the spectra of the so-called explosive phenomena. But on the star this wide component contains up to 25% of the total energy of the line, whereas on the Sun it contains not more than 5%. The connection of wide components of the profiles and microflares is supported by high velocities of plasma eruptions from the regions of annihilation of magnetic fields, as well as the established correlation of the fraction of energy in this component of the energy of the whole line profile with the stellar activity level and increased temperatures of coronae of the most active stars.

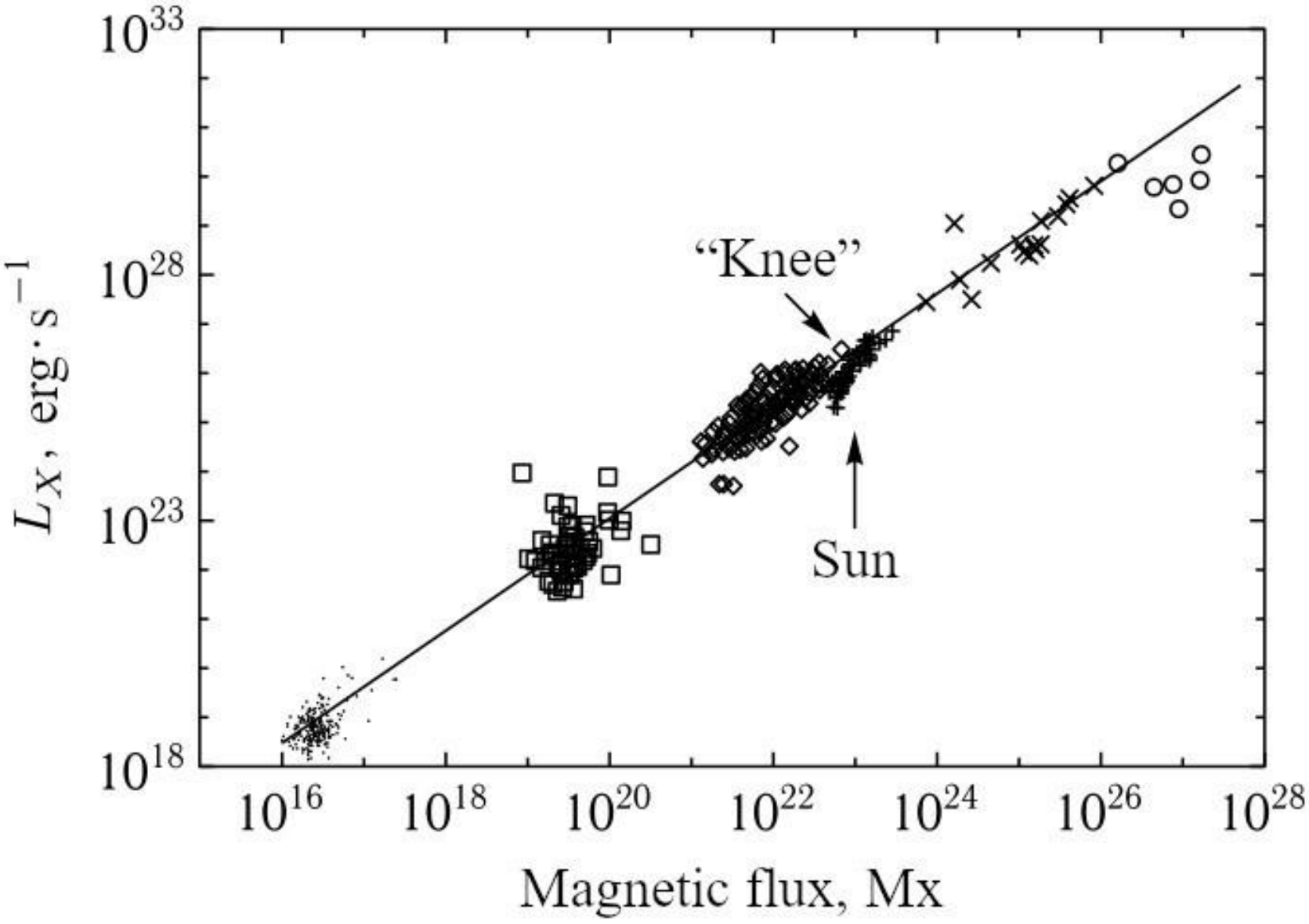

Fig. 28. Correlation of the X-ray emission $L_X$ and the full in absolute values magnetic flux for the solar and stellar objects: dots denote the quiet Sun, squares — X-ray bright points, diamonds — solar active regions, crosses — the average solar disk, times signs — G, K, and M dwarfs, circles — T Tau stars, solid line — the power approximation $L_X \sim \Phi^{1.15}$ of data ensemble (Pevtsov et al., 2003)

Saar et al. (1994c) found that in the spectrum of AD Leo from 10 to 40% of the emission in the SiIV lines were caused by low-amplitude flares, while Saar and Bookbinder (1998a) found that for fast rotators G0V EK Dra and K2V LQ Hya the radiation of the CIV and SiIV emissions was approximately for 10% determined by such flares. In the appropriate relations, according to Saar and Bookbinder, the transition zone is heated by microflares and permanent magnetohydrodynamic waves. Ayres (1999) obtained similar results from the observations of three G dwarfs. But the width of the coronal line FeXXI $\lambda$ 1354 Å in the spectrum of AD Leo was rather narrow, which points to the low contribution of shock waves to the corona heating (Pagano et al., 2000). As it was noted by Benz (2000), for the stellar coronae a continuous replacement of their matter by the ejected matter of chromospheres heated by flares may be substantial.

Studying the ultraviolet spectra of $\varepsilon$ Eri obtained with HST/STIS and the energy balance in the upper atmosphere, Sim and Jordan (2002) concluded that the surface filling factor was ~0.2 in the transition zone and approached unity in the corona. Then, from HST STIS and FUSE data they considered the heating of the upper atmosphere of $\varepsilon$ Eri (Sim and Jordan, 2003). They measured the widths of more than 50 emission lines in the range of 1030–2300 Å, calculated for each of them the width of the nonthermal component and found that the value grew from 7.5 km/s at a level of the formation temperature of 5000 K to 21.3 km/s at a level of 30000 K and then remained practically constant at this level up to 14 MK. The profiles of strong CIV, SiIII, SiIV, NV, and OIV lines were presented by two Gaussians. It appeared that with increasing temperature the contribution of the wide component systematically decreased. Sim and Jordan concluded that the narrow component was due to Alfvén waves reaching the corona, but tending toward the idea of microflares by Wood et al. (1997) they left the question of the nature of the wide component open.

* * *

Within the whole diversity of the considered heating mechanisms for stellar atmospheres, the detected close correlation of X-ray emission and magnetic flux (Fig. 28) is fundamental; the practically linear correlation between these parameters of the Sun and active stars expands up to 12 orders of magnitude.

Toriumi and Airapetian (2022) analyzed 10-year multi-wavelength observations of the Sun-as-a-star and young sun-like stars and significantly extended the study by Pevtsov et al. (see Fig. 28), revealed a close power-law scaling between the chromospheric, transition region and coronal Far UV, Extreme UV and X-ray emission fluxes, and the total unsigned magnetic flux. They found that the power-law exponent coefficient is the smallest at activity maximum and increases during solar maximum and concluded that the mechanism of atmospheric heating is universal for the Sun and Sun-like stars, regardless of age and activity level. Recently, Toriumi et al. (2022) published a catalog of power-law indices between solar activity proxies and various spectral line fluxes. Compared to previous studies, they expanded the number of proxies, which now includes the total magnetic flux, total sunspot number, total sunspot area, and the F10.7 cm radio flux, and further enhances the number of spectral lines by a factor of two. This provides the data to study in detail the flux-flux scaling laws from the regions specified by the temperatures of the corona to those of the chromosphere, as well as the reconstruction of various spectral line fluxes of the Sun in the past, F-, G-, and K-type dwarfs, and the modeled stars.

* * *

Stelzer et al. (2013) performed an all-wavelength survey of M dwarfs in the 10 pc vicinity of the Sun, and it may be treated as to some extent resultant for a series of sections of Part 1. They found that the surface fluxes in $H_\alpha$, near and far ultraviolet, and in X-rays were related by the power dependence. If one considers the ratio of a flux in some band to the bolometric flux as an activity index, then for a given spectral type such activity indices reveal a divergence up to three orders of magnitude and they are maximal near M4. The average activity index of fast rotators related to the saturated emitters decreases from the X-rays through the ultraviolet to $H_\alpha$. For the ultraviolet and X-rays, at the transition from young objects of an age of 10 million years to the objects of several billion years the indices decrease up to three orders of magnitude. Several young dwarfs of the sample show that this activity weakening with age occurs constantly, and in the X-rays it is steeper than in the ultraviolet.

Analyzing the starspots and the most powerful flares on A stars recently discovered with Kepler (see Subsect. 2.4.4.1), Balona (2016) concluded that these stars had magnetic fields but no convection; this fact excludes the emergence of coronae, which is evidenced by the X-ray observations.

## 1.6. Stellar Winds

The solar wind was discovered "at a tip of a pencil" by Parker (1958), although there were less known predecessors (Ponomarev and Rubo, 1965). The main conclusion of the Parker theory was that the solar corona could not be a hydrostatic structure continuously transforming into the interstellar medium but that it could be described by a particular solution of the nonlinear differential hydrodynamics equation for the continuously expanding atmosphere. The theory became widely accepted almost immediately because of the elegant derivation and because the solution provided a simple and natural explanation for a number of observed facts concerned with the structure of comets, the brightness of the zodiacal light, the variation of cosmic rays, and various geomagnetic and geophysical phenomena. Further, Parker's model was revised on the basis of direct measurements of physical parameters of interplanetary medium at different distances from the Sun and with the transition from a spherically symmetrical purely hydrodynamics scheme to the concept of solar magnetosphere containing open and closed magnetic flux tubes.

Higher densities and temperatures of the coronae and higher strengths of photospheric magnetic fields of active dwarfs suggest that they should have solar-wind analogs, which are stronger than the prototype. As stated above, the first quantitative model of the stellar wind of an M dwarf was constructed by Kahn (1969). He estimated the secular loss of stellar matter at the expense of the expanding corona as $3 \cdot 10^{-12} M_{\odot}$/year, which exceeded the appropriate solar value by a factor of 300.

Mullan et al. (1989a) were the first to find the stellar wind of a late dwarf based on optical data: during the eclipse phase of the eclipse system V 471 Tau composed of red and white dwarfs they found absorption lines of neutral and ionized metals in the ultraviolet spectrum, i.e., the spectral details occurring when the white dwarf was examined through the atmosphere of the red one. The discovered absorption components were shifted by –700–800 km/s, which definitely exceeds the parabolic velocity of the K2 dwarf, and their common character was preserved for at least 4 months. The multiplicity of the absorption components evidences the nonstationarity and/or nonsphericity of the stellar wind. From Mg lines, Mullan et al. estimated the secular mass loss as $3 \cdot 10^{-11} M_{\odot}$/year, which exceeds the appropriate solar value by three orders of magnitude.

Then, winds were sought in the far-IR and millimeter wavelength regions primarily using the observational results from the Infrared Astronomical Satellite (IRAS). The spacecraft carried a telescope with an aperture of 57 cm and 62 infrared detectors. Between January and November 1983, IRAS scanned 98% of the sky in the regions of 12, 25, 60, and 100 μm, while 72% of the sky was scanned more than three times and 24% twice. IRAS recorded about 280000 point sources with a characteristic accuracy of coordinates $30'' \times 7''$ and a photometric accuracy of 5–15%. Initially IRAS data were analyzed irrespective of stellar winds. Mullan et al. (1989b) compared the IRAS data for the samples of 78 flare red dwarfs and 38 dM stars. For the stars with identical stellar magnitudes in the *K* band they found that the brightness of dMe stars at 12 μm was systematically and noticeably higher, by 70%, than that of dM stars. At 12 μm they recorded the overwhelming majority of the stars of both samples, but only 15 dMe stars and probably 2 dM stars were identified as IR sources at 100 μm; 9 dMe and no dM stars were recorded in all IRAS bands. Thus, dMe stars are definitely stronger IR sources than dM stars. The fact that the V−K color indices of dMe stars also differ systematically from analogous values of dM stars suggests that IR excesses of dMe stars spread to 2.2 μm. For those flare stars, from which fluxes were recorded at 100 and 60 μm, a noticeable increase in intensity toward the long-wavelength range was observed. The authors suggested that since the total duration of IRAS observations of each object did not exceed 100 s, then it was due to stationary emission of dMe stars, rather than flares on them. They concluded that the IR excesses could not be due to the circumstellar dust. If the reason was synchrotron radiation, one should expect noticeable fluxes of about 1 mm. Based on the IRAS data for the BY Dra-type stars, Chugainov and Lovkaya (1989) suspected that dust circumstellar structures were responsible for the discovered IR excesses for 11 of 22 considered stars. Tsikoudi (1989, 1990) compared the IRAS data for 36 active F–K dwarfs and 30 inactive stars of the same spectral types and found that 27 active stars were recorded at 12 μm, half of them at 25 μm and only 2 stars at 60 and 100 μm. However, no systematic distinctions between these two samples were revealed: with the correlation coefficient 0.96 for active stars she obtained the relation $L_{12} = 10^{-2.6\pm0.3}L_{\rm bol}$ and $L_{12} = 10^{-2.7\pm0.2}L_{\rm bol}$ for inactive stars. Only for the active stars Gl 113.1 and Gl 735 were noticeable excesses at 12 μm found, for $\varepsilon$ Eri and 70 Oph probable excesses were found at 60 and 100 μm. Then, Mathioudakis and Doyle (1991b) considered the radiation of seven active M dwarfs in the range from the ultraviolet to the IR region, including the IRAS data. All objects were recorded at 12 μm, 5 at 25 μm, 2 at 60 and one at 100 μm. Clear IR excesses were discovered on Gl 65 AB, Gl 644 AB, and Gl 873. Mathioudakis and Doyle noted that these excesses could result from the radiation of relict dust or synchrotron emission of relativistic particles emerging in microflares, but did not select one of the mechanisms.

By the 1990s, the concept of gaseous shells expanding around hot stars had been developed and confirmed by observations: at certain frequencies they become opaque for the free–free radiation, and the fluxes going from them at these frequencies can be used for estimating their sizes and kinematics. Within this concept Mullan et al. (1992a) calculated the expected radiation spectrum of stellar wind from cool stars and determined that the frequency range within which it should be opaque and radiate as an absolutely blackbody was within tens of micrometers, the "photospheric radius" of the wind should vary as $\nu^{0.6-1.2}$ and the flux expected from YZ CMi was estimated as 14 mJy. Since, within the Rayleigh–Jeans approximation the blackbody radiation is proportional to $\nu^2$, the optically thick wind should provide a certain addition to the radiation of the stars in the far IR region. Such an excessive radiation of flare stars had indeed been discovered already. To check these calculations, Mullan et al. (1992a) performed special observations at 0.8 and 1.1 mm. Only for six of 13 program stars were rather unreliable results obtained. However, combination with the IRAS and VLA data for YZ CMi, Wolf 630, and EV Lac yielded fluxes that were in rather good correspondence with the calculation predictions. Paradoxically, the microwave emission successfully interpreted within the gyrosynchrotron emission of stellar coronae matched the single power spectrum with the expected index with the IRAS data for the far-IR range and the measurements by Mullan et al. in the millimeter range.

To find stellar winds of G dwarfs younger than 250 million years, Vikram Singh (1995) observed HD 39587, HD 72905, HD 115383, and HD 206860 with VLA at 6, 3.6, and 2 cm. At 6 cm meaningful signals were recorded from the first two stars, at 3.6 cm only from HD 39587. Within the thermal radiation of spherically symmetrical expanding shells the obtained estimates correspond to the secular mass losses of (2–3) · $10^{10}M_\odot$/year for the two former stars and of < $10^{-11}M_\odot$/year for two others.

Another attempt of interpreting the excessive radiation of active late dwarfs in the far-IR region was associated with direct confrontation of this radiation with the manifestations of magnetic activity of these stars. First, for 14 K–M dwarfs, for which the measurements of magnetic fields and IR fluxes were available, Katsova and Tsikoudi (1992) found a correlation of magnetic fluxes $fB$ with the radiation in the 12-μm band that exceeded the values expected for blackbody radiation at the temperature estimated from the B−V color index. They also suspected the correlation between the excessive IR radiation and the rotation rate. Then they compared the excessive IR radiation of G–K–M dwarfs with the fluxes in the soft X-ray range and found that in the region of G–K dwarfs, where $\log L_X/L_{bol} = -5.5$–$-3$, there was a correlation of these radiations, while at high $\log L_X/L_{bol}$, typical of flare M dwarfs, there was a saturation in X-rays (Katsova and Tsikoudi, 1993). Considering the found dependences, Katsova and Tsikoudi justified why IR excesses could not be caused by interstellar dust or synchrotron radiation and advanced the hypothesis on the responsibility of starspots covering a considerable part of the stellar surface for the generation of radiation. However, later they accepted the model of cool ($T \sim 10^4$ K) stellar wind (Katsova et al., 1993).

Probably, the solution to this complicated problem was found by van den Oord and Doyle (1997), who noticed that the optical thickness of the wind for cool stars was much lower than for the hot ones, and on cool stars the wind was hotter than the photosphere, whereas for hot stars this relation was reverse. These differences introduce significant changes in the theory of expanding shells as radio sources, and, according to the estimates of van den Oord and Doyle, the mass loss by dwarf stars due to the stellar wind did not exceed $10^{-12} M_\odot$/year. But on the other hand, several years earlier, Badalyan and Livshits (1992) considered possible stellar analogs of solar streamers — coronal beams in which the effective outflow of solar coronal plasma occurred. They found that with allowance for the X-ray observations of the densities at the base of stellar coronae (Katsova et al., 1987) on five considered F8–M0 dwarfs one could expect mass loss at the expense of these structured stellar winds at a level of $(1–2) \cdot 10^{-11} M_\odot$/year.

Cohen et al. (2009) considered another variant of the interaction of the structured stellar wind and starspots. Within the Michigan MHD model of the solar corona, they considered the influence of latitudes of starspots, strength and topology of their magnetic fields on the stellar wind structure and mass and angular momentum loss of a star and found that when spots are at low latitudes, the angular momentum loss is controlled by the mass flow, but at the high-latitude spots these losses are controlled by increasing plasma density between the stellar and Alfvén surface. Cohen et al. suggested that there should exist a feedback mechanism between the magnetic field distribution, wind distribution, and motions in the convective zone that generate the magnetic field.

Solar studies and gas-dynamic calculations showed that the expanding stellar wind is separated from the colliding interstellar gas by two shock waves isolated by a "hydrogen wall" of heated and compressed neutral gas. Such a "hydrogen wall" was initially found by Linsky and Wood (1996) during the HST observations of the $\mathrm{Ly}_\alpha$ line in the spectra of $\alpha$ Cen A and $\alpha$ Cen B. Then, they found analogous structures for 61 Cyg A, 40 Eri A, $\varepsilon$ Ind, and the giant $\lambda$ And (Wood and Linsky, 1998). From the parameters of the "walls" found they estimated the pressure in stellar winds $p_w$ and found its correlation with the surface X-ray flux

$$p_w \sim F_X^{-1/2} \tag{27}$$

This correlation is valid for the Sun at different phases of the activity cycle. Hence, it follows that as stellar activity attenuates, one should expect a strengthening of the wind, and consequently an increase of the secular mass loss. Further UV observations of several more solar-type stars allowed Wood et al. (2001) to estimate the rates of mass loss for the stellar wind as $2\dot{M}_\odot$ for $\alpha$ Cen and less than $2\dot{M}_\odot$ for Proxima Cen, to find a correlation between the mass-loss rate and the X-ray emission level as $\dot{M} \propto F_X^{1.15\pm0.20}$ and to suspect an evolutionary weakening of the wind as $\dot{M} \propto t^{-2.00\pm0.52}$ (Wood et al., 2002). Then Wood et al. (2005a) clarified this conclusion: owing to the stellar wind, the stellar mass loss increases with the activity level, but there was detected evidence for an abrupt weakening of the wind at some activity levels.

Independent considerations on the generation of strong stellar winds as a result of considerable coronal ejections of matter were developed by Houdebine et al. (1990) in analyzing the flare on AD Leo of March 28, 1984, whose spectrum contained a short-lived blue wing of the $H_\gamma$ emission line expanded up to 80 Å (see below).

Later, Wood et al. (2005b) analyzed 33 $Ly_\alpha$ spectra of stars within 100 pc from the HST/ STIS archive and found four new heliospheric absorptions for 70 Oph, $\xi$ Boo, 61 Vir, and HD 65185 and seven new astrospheric absorptions for EV Lac, 70 Oph, $\xi$ Boo, 61 Vir, $\delta$ Eri, HD 128987, and DK UMa, doubling the earlier known number of such phenomena. A high fraction of the detected astrospheric absorptions — 10 of 17 lines-of-sight — means that in most cases within 10 pc the interstellar medium should be partially neutral, but beyond 10 pc this fraction drops lower than 10%, i.e., the medium is more ionized and hotter.

For a limited number of nearby stars, Linsky et al. (2013) performed calculations of the intrinsic line of $Ly_\alpha$, which could not be observed due to the strong absorption in the interstellar medium, constructed the correlations of fluxes in this line with fluxes in other lines forming in the chromosphere, found that these correlations were rather weakly dependent on the fluxes in other lines but closely related to the stellar flows in X-rays for F5–K5 dwarfs and with higher dispersion for M dwarfs. They showed that based on the stellar effective temperature and rotation period one might estimate with sufficient accuracy the flux in $Ly_\alpha$ for G and K stars, but less reliably for cooler stars.

To analyze Chandra observations of individual components of the $\xi$ Boo G8V + K4V system, Wood and Linsky (2010) found that the component of $\xi$ Boo B yielding only 11.5% of the X-ray flux dominated in the whole wind of the system, whereas $\xi$ Boo A had a weak wind despite the strong corona. Furthermore, they found that the FIP effect did not depend on activity but depended on the spectral type.

Vidotto et al. (2011) performed a three-dimensional MHD calculation of the strong stellar wind of the fast M4 rotator V374 Peg, taking into account a self-consistent interaction of the wind and magnetic field. Considering this interaction, the field configuration and the coronal wind structure were detected, which allowed estimating the stellar angular momentum loss rate. They found that the wind of this star significantly differs from the low-velocity solar wind with a low mass loss rate: its threshold velocity for the whole range of latitudes accounts for $(1500–2300)(n_{12})^{-1/2}$ km/s, where $n_{12}$ is the density at the base of the coronal wind in units of $10^{12}$ cm$^{-3}$, the rotation braking time is $28(n_{12})^{-1/2}$ million years, and the wind pressure on V374 Peg exceeds the solar value by 5 orders of magnitude.

To interpret the prominences at a height of $(2–5)R_*$ of a rapidly rotating solar-type star, Jardine and van Ballegooijen (2005) proposed a model of prominences that was alternative of the traditional one: the long thin magnetic loops that forms them are filled with the stellar wind gas rather than the hot coronal plasma radiating in X-rays. The height maximum of such systems drops with increasing rotation rate and grows with decreasing temperature. For a solar-mass star with a rotation period of 1/2 days, whose X-ray corona extends up to $1R_*$ above the stellar surface, the loops at the temperature $10^4$ K reach $5R_*$.

Leitzinger et al. (2022) considered an event on the young Me dwarf V374 Peg in the course of which the extra emission appeared on the blue side of the Balmer lines. From 1D NLTE modeling, they presented distributions of physical parameters in the structure responsible for this emission and found that except for temperature and area all parameters are at the upper range of typical solar prominence parameters. The temperature and the area of the event were found to be higher than for typical solar prominences. But there are more solutions for the filament than for the prominence geometry.

Suzuki in cooperation with colleagues constructed a MHD model of the solar wind, in which, at the photospheric level, the transverse fluctuations excite Alfvén waves propagating upward through the open magnetic tubes; the calculations showed that velocities of transverse fluctuations of about 1 km/s are sufficient for deriving a pattern observed on the Sun. Alfvén waves are effectively reflected, but 10% of the energy flux penetrates into the corona and causes heating and mass acceleration. Recently, Suzuki (2013) expanded this model up to the general evolutionary scheme of winds of the solar-type stars. Since the young solar-type stars are notably active, then Suzuki considered the case when the magnetic field strength and fluctuation velocities on the stellar surface exceed the current solar values by a factor of 2 and obtained a 20-fold increase in the mass-loss rate. This nonlinearity is due to the fact that the mass loss rate is very sensitive to the energy inflow because of the global instability associated with reflection and nonlinear dissipation of Alfvén waves, since the most energy is spent for radiative losses rather than the kinetic energy of the wind. Moreover, the beginning of the transition zone was shifted by $h = 0.1R_*$ as compared to the solar value $0.003R_*$. At the end of the main-sequence phase, when the stellar radius increases approximately by a factor of 10, the stationary corona with the temperature of million degrees rapidly disappears, while the chromospheric matter with hot inclusions arising due to the temperature instabilities emanates outward. Therefore, the wind of a giant is nonstationary and structured.

Drake et al. (2013) analyzed data on the solar coronal ejections and associated flares throughout a period of 1996–2007 and found confident power relations between the flare X-ray fluxes in the range of 1–8 Å and masses and kinetic energy of CME. But one failed to extend these relations to stars probably because of saturation of the stellar X-rays.

Vidotto et al. (2014a) performed a three-dimensional numerical calculation of the stellar wind from an early M dwarf. Despite a significant diversity of the magnetic field topology in the considered sample they found that the stellar wind blows near the equatorial plane carrying most part of the stellar angular momentum, but there are no predominant latitudes for the mass loss, and a more axinonsymmetrical magnetic field results in a more asymmetrical flux and total pressure in the wind.

Fionnagain and Vidotto (2018) considered the solar wind evolution caused by the angular momentum loss and solar magnetism variations. From the X-ray observations of solar-type stars they estimated constraints on the thermal acceleration of the wind for such stars at an age of 120–7000 million years and found that the drop of X-ray properties results in a drop and more drastic weakening of the mass-loss rate after 2 billion years. The latter may be the cause of the fact that older stars are less effective in the angular momentum loss, which may be due to the anomalously fast rotation observed for older stars.

From the COS HTS observations of close pairs of red and white dwarfs Wilson et al. (2018) estimated a wealth of C, O, Si, and some metals in the atmospheres of white dwarfs, and from these data — the stellar wind from red dwarfs as a function of their spectral types and rotation periods.

Garraffo et al. (2018) showed that, according to the MHD model of the stellar wind, the topology on the surface magnetic field of a star strongly affects the angular momentum losses through the wind, and, following observations, the rapidly rotating stars contain the most part of the magnetic flux in the magnetic field components of the highest orders. The transition from the complex structure of the magnetic field to a simpler configuration seems to be associated with the stellar rotation braking, which may be interpreted by the bimodal distribution of fast and slow rotators in open stellar clusters.

Sakaue and Shibata (2021) calculated a nonlinear model of Alfvén waves in coronae and winds of M dwarfs and concluded that the coronae of these dwarfs should be colder than that of the Sun, and the stellar winds are characterized by higher velocities and far lower matter fluxes as compared to the solar wind.

From the Hubble spectra Wood et al. (2021) studied the absorption profiles of the $Ly_\alpha$ line for nine M dwarfs, arising during the interaction of the stellar wind and interstellar medium. They estimated the mass-loss rates, constraints on the stellar winds, and their dependence on the coronal activity. Taking into consideration the previous data for 13 of 15 M dwarfs, the winds turned out to be weaker or comparable with the solar wind, but for YZ CMi and GJ 15AB the mass losses were 30 and 10 times higher, respectively, than those for the Sun. Wood et al. concluded that not only coronal activity and spectral type determine the wind properties but, possibly, strong winds are determined to a great extent by CME.

Airapetian et al. (2021) developed a three-dimensional MHD model of the stellar corona + wind system and, using the X-ray, EUV, Hubble, and TESS system observations, constructed models of the heated corona and wind by the Alfvén wave flux for the young star $\kappa^1$ Cet at two epochs separated by 11 months. They found that over this time the global magnetic field structure experienced a restructuring from the simple dipole to the inclined and weaker dipole with developed multipole components; the mass flux in the wind dropped by 40%.

To investigate the space weather of the Sun at an age of 0.6 Gyr, Evensberget et al. (2021) calculated the three-dimensional models of winds for five young solar-type stars in the Hyades cluster. Using the magnetograms of these stars and the Alfvén wave driven wind modeling, they found a mass loss of 2–4 times higher and an angular momentum loss of 2–10 times higher than those for the Sun for the five-fold range of the magnetic field strength.

Alvarado-Gómez et al. (2022) performed 3D numerical modeling of the stellar wind from AU Mic, as well as simulations describing the evolution of highly energetic coronal mass ejection. Observational constraints on the stellar magnetic field and properties of the eruption are incorporated in the models.

Using spectropolarimetry at the Large Binocular Telescope, literature data on Zeeman Doppler images, as well as the X-ray, Gaia and astroseismology data, Metcalfe et al. (2022) compared the rotation rates of the components in pairs of the solar-type stars of different age, estimated the rate of angular momentum loss, and found that it drops from 2.6 to 3.7 Gyr by more than an order of magnitude and continues to decrease modestly to 7 Gyr.

## 1.7. Cold Dust Disks

At the beginning of the century, Zuckerman et al. (2001) carried out infrared observations of the young A5 star $\beta$ Pic and detected a circumstellar dust disk. Soon, such disks were detected for a number of different spectral type stars. The morphology of these structures — a rather abrupt outer boundary and central cavity in many cases — did not allow one to treat them as formed by stellar winds, and now they are considered as the fifth element of stellar electromagnetic radiation in addition to the photosphere, chromosphere, corona, and stellar wind.

Kalas et al. (2004) detected such a disk for one of the most active and youngest flare dwarfs AU Mic that belong to the nearby moving group $\beta$ Pic. Within a search for the cold dust disks near M dwarfs Lestrade et al. (2006) carried out observations of 32 such stars at wavelengths 0.85 and 1.2 mm and estimated the detection probability of such disks for them as 13%. Since for A–K stars such a probability was estimated as 9–23%, then it was concluded on a wide occurrence of these structures for stars of all spectral types.

Kalas et al. (2004) detected the AU Mic disk in the scattered light at a distance from 50 to 210 AU from the star, and at such a distance the lifetime of dust particles exceeds the lifetime of the AU Mic star, which is about 10 million years; consequently, this means the discovery of solid matter required for the formation of planets.

Soon, using the adaptive optics of the Keck telescope, Liu (2004) studied the inner disk of AU Mic from 15 to 80 AU and found asymmetry of the disk and a drastic variation of its structure by 35 AU: in the range from 25 to 40 AU clumps and gaps are seen between them, which may be affected by invisible third bodies and resemble the structures expected at the recent formation of the planet. The similarity of $\beta$ Pic and AU Mic disks points to the synchronous development of these formations for such different stars at least during the first 12 million years of their development.

Strictly speaking, we are not confident that circumstellar dust disks should be considered in the book devoted to solar activity in the realm of stars: these structures are sooner important elements in the proper evolution of stars and their planets, and considering the physics and evolution of such disks, the basic place is occupied by the processes of formation and dissipation of dust, which are extremely weakly dependent on the global characteristics of stars and, moreover, on the phenomena of their nonstationarity (see, for example, the review of Wyatt (2008) and the comprehensive work of Gaspar et al. (2013)). A number of works have been published so far devoted to the emergence and dynamics of dust during the collisions with larger bodies, the excitation of disks by planets, the size distribution of dust particles at different distances from a star, the formation mechanism of the central disk hole, its thickness, the thickness dispersion of dust particles of different sizes, etc. But here we do not imply to provide a comprehensive review of these investigations on the evolution of planetesimales and planet genesis, but present only knowledge on a cold dust disk near the stars which definitely or supposedly have solar-type activity.

* * *

AU Mic is still a main target for studying cold disks of M dwarfs, and the obtained basic data are as follows.

From the FUSE observations in the far ultraviolet Roberge et al. (2004, 2005) estimated the upper threshold of the gas $H_2$ in the disk: $1.7 \cdot 10^{19}$ cm$^{-2}$ and the $H_2$/dust ratio proved to be lower than 6:1, similar to $\beta$ Pic. This means that the initial gas generally dissipates over the time of less than 12 million years from the both disks despite significant differences in masses and luminosities of stars.

Krist et al. (2005) carried out the Hubble multicolor observations in the coronagraphic mode of AU Mic and recorded a dust disk at the distance from 0.75″ to 15″ (7.5–150 AU); its thickness accounts for 2.5–3.5 AU at the distance less than 50 AU and to 6.5–9 AU — about 75 AU. The radial brightness profile of the disk is flat at a radius of up to 15 AU from the center, after which it drops as $r^{-1.8}$ to $r \sim 43$ AU and then as $r^{-4.7}$. The disk plane is inclined to the line of sight by 1–3°. The three-dimensional model of the disk points to the absence of dust up to 12 AU, which corresponds to a spectral IR estimate of 17 AU. This disk is bluer than the similar disks, which may be associated with a great number of very small particles.

Metchev et al. (2005) studied this star with the adaptive optics at the Keck telescope. They considered the disk in the H band in the range of 17–60 AU from the star with a resolution of 0.4 AU, identified the dust clumps and gaps between them, which could be caused by planet orbits. At a distance of more than 20 AU there were no planets with masses of about Jupiter mass, but could be smaller planets. At a radius of 10 AU from the center, the disk was empty, up to 50 AU there was a poor number of small particles. About 33 AU the spectral index of the power distribution of dust was changed, which might be due to both the collision of particles and the Poynting–Robertson effect.

Using the Spitzer satellite, Chen et al. (2005) observed 39 A–M dwarfs and for AU Mic detected excessive radiation at 70 μm twice exceeding the predicted photospheric radiation. The simultaneous absence of excess at 24 μm means the low temperature of dust particles. Later, Chen et al. (2006; 2008) enlarged a list of the studied objects with excessive radiation in the 60 μm region up to 59; from the spectra in the range of 5.5–35 μm they estimated the blackbody temperature of the continuum as 80–200 K, the temperature of silicates as 290–600 K and concluded that the model of disks in the form of narrow rings around stars is preferential.

Augereau and Beust (2006) analyzed the photometric properties of the dust disk around AU Mic and found that, independently of the scattering properties of particles, the most part of emission occurred in the asymmetric shock dominated region near the maximum of the disk surface brightness at a distance of 35 AU. Elementary scatters of the visible light are submicron particles, which are transported into the high-eccentric orbits by the pressure force dependent on a size of particles, it is due to the stellar wind, and the color gradient is a result of the dust dynamics.

Strubbe and Chiang (2006) determined the photometric disk profile of AU Mic as $b^{-\alpha}$, where $b = 43$ AU and $\alpha \sim 1$–2 inside 43 AU, and $\alpha \sim 4$–5 outside 43 AU. At a distance of 43 AU "the parent bodies" are located, which collide and generate micron dust particles. Inside this boundary the dust particles drop out on the star, and large particles are preserved outside it. The simultaneous modeling of surface brightness and thermal spectrum led to the conclusion on the domination of collisions in the disk and that the narrow parent ring contains decimeter particles of the total mass as that of the Moon. Mieda et al. (2010) carried out the HST colorimetric and polarization observations of this star, resolved its disk at the distance from 9 to 200 AU and after 75 AU detected irregularities as in the $\beta$ Pic disk. With increasing distance from the star, the blue light scatter increases, and under rational assumptions on the composition and porosity of dust particles, the obtained observations confirm the dynamic model by Strubbe and Chiang (2006).

Later, Riaz et al. (2006) carried out the Spitzer IR observations in the bands 24, 70, and 110 μm of eight M stars of different activity levels — dMe, dM, and sdMe — and came to the conclusion that throughout more than 10 million years there occurred a dissipation of inner parts of the disk up to several astronomical units, the V–K anomalies could be due to metallicity, and the chromosphere of AU Mic yielded not more than 2% of the flux in the submillimeter range.

Using the Hubble telescope, Graham et al. (2007) constructed a map of the polarization of the AU Mic disk and found that it increased from 5% at a distance of 20 AU to 40% at 80 AU. The polarization is perpendicular to the disk, which corresponds to the scatter of micron particles in the optically thin layer. Up to 40–50 AU there is a hole in the disk, where the density of micron particles is increased by more than a factor of 300. Strongly scattering and yielding high polarization, such spherical particles cannot be those of usual composition, but only very porous particles as dust in comets, where the sublimation of ice leaves the refracting material. Porous particles of AU Mic may be relic and, consequently, AU Mic may detect traces of the relic process of agglomeration, where the interstellar particles clump and form microscopic bodies.

Using the Keck II telescope, Fitzgerald et al. (2007) in the JHK bands resolved the circumstellar disk of AU Mic in the range of 8–60 AU. From the measured light they confirmed the existence of the substructure in the inner disk, and the measured polarization indicates the existence of porous dust particles, which strongly affects the obtained disk structure. The belt of parent bodies at a distance of 35 AU produces dust that is thrown out then by the wind and radiation. Very small particles up to 0.05 μm are available in the outer regions.

Using FUSE, France et al. (2007) detected a molecular gas component of the circumstellar disk of AU Mic: they recorded the fluorescent emission line of $H_2$ $\lambda$ 1031.87 Å that was excited by the resonance line of OVI $\lambda$ 1032 Å. The gas temperature was estimated within 800–2000 K, the ratio of the $H_2$ abundance to the dust $<\sim 1/30 : 1$, and the total gas mass was in the range from $4 \cdot 10^{-4}$ to $6 \cdot 10^{-6} M_\odot$.

Using the ALMA complex of radio telescopes, Wilner et al. (2012) at 1.3 mm resolved the disk around AU Mic by the same geometry with the central radius 35 AU, which was found in optics.

Cranmer et al. (2013) considered constraints on the MHD heating of the corona for AU Mic, which were put by X-ray, radio, and millimeter observations. The obtained self-consistent model of the coronal loop competed with the model of the inner asteroid belt at a distance 3 AU from the star constructed from the ALMA observations at 1.3 mm.

Observations of the circumstellar disk of AU Mic in 2010–2011 detected the moving of its structures from the star with velocities from 4 to 10 km/s and some of them exceeded the escape velocity. According to the HST observations, the disk acquires a more diffuse form (Grady et al., 2018).

* * *

The knowledge regarding dust disks near other active dwarfs are rather sparse.

Using the 15-meter radio telescope SEST in Chile, Schuetz et al. (2004) detected at 1.2 mm the presence of an extended dust circumstellar disk for $\varepsilon$ Eri and estimated its size as 88 AU. Greaves et al. (2005) recorded the submillimeter images of the extended dust disk around $\varepsilon$ Eri throughout 5 years. The obtained images correspond to the dust distribution, which reflects its emergence during the collision of comets, “cleaning”, and excitation of dust by planets. The previous observations at 850 and 450 μm were confirmed. The disk is inclined by 25°, the total disk size is 105 AU, the emission maximum is at 65 AU, the inner cavity is twice weaker. The disk structure allows one to suspect the excitation from the planet at a distance of tens of astronomical units, two clumps and an arc were detected, rotating with a velocity of 1° per year in the anti-clockwise direction. Comparing the disk images in different years, Poulton et al. (2008) estimated its rotation with a velocity of 2.75° per year in the anti-clockwise direction. This value is significantly higher than the Kepler velocity, which allows one to suspect the influence of a planet in the disk.

By means of the interferometry in the near IR range, Di Folco et al. (2007) undertook a search for warm dust in the inner regions of disks around $\tau$ Cet and $\varepsilon$ Eri and found a sought-for excess at a level of 1% above the photosphere in the former case and its upper limit in the latter one.

Using Spitzer and instruments at the Caltech Submillimeter Observatory, Backman et al. (2009) derived images and spectrophotometry of $\varepsilon$ Eri in the range from 3.5 to 350 μm and detected new bright details. At 350 μm the presence of a ring was confirmed from 35 to 90 AU, in the middle and far IR the ring was traced to 110 AU. The energy distribution evidences the complicated structure of the system with particles varying from 135 to 15 microns. The innermost belt with a radius of 3 AU consists of silicon dust particles, and the stability of the system requires the presence of three planets.

Brogi et al. (2009) considered a dynamic stability of the inner belt around $\varepsilon$ Eri and found that at the excentricity of $\varepsilon$ Eri B 0.15 and higher the belt source noticeably lost its initial mass and decreased the width.

Reidemeister et al. (2011) proposed a model for the interpretation of the observed energy distribution in the spectrum of $\varepsilon$ Eri from the middle IR range to submillimeter wavelengths, which involves ice and silicates in comparable amounts. Warm dust is transported from the outer Kuiper belt inward by the action of the Poynting–Robertson effect and the stellar wind.

Williams et al. (2004) detected radiation at 450 μm of the G2 dwarf HD 107146 with the age varying from 80 to 200 million years, emanating from the region of 300 AU × 210 AU The spectral energy distribution corresponds to a dust temperature of 51 K with a mass of 0.10 Earth mass. Observations at 18 μm show a very small amount of warm dust and the presence of a large inner hole in the disk, at least up to 31 AU. Later, Ertel et al. (2011) constructed a model of this stellar disk in the form of a ring with a peak of the surface density at a distance of 131 AU.

Liu et al. (2004) found at 850 μm a dust disk near GJ 182, due to which there is an excess at 25 μm, i.e., contains warm dust at several AU of the inner disk.

Greaves et al. (2004) discovered the excessive radiation in the far IR region of the G8 dwarf $\tau$ Cet and ascribed it to the dust particles revolving near the star, which emerge during the collision of planetesimales. At 850 μm the dust disk was seen up to 55 AU, the dust mass was estimated as exceeding the Kuiper belt mass in the Solar system by an order of magnitude, the mass in the colliding bodies of a size up to 10 km — about 1.2 Earth masses, whereas in the Kuiper belt — 0.1 Earth mass.

Using HST, in the near IR region Kalas (2005) discovered a circumstellar dust cloud around the second after $\beta$ Pic A star HD 32297. The scattered light of the disk was recorded at the distance from 560 to 1680 AU from the star. He assumed that an abrupt asymmetry and blue color of the disk were due to the collision of the star of an age of 30 million years or less with the interstellar matter. Then Kalas et al. (2006) reported on the discovery of dust disks near the K star HD 53143 and the F star HD 139664; at an age of 0.3–1 billion years these are the oldest optically detected disks; the width of the former is more than 55 AU, the latter has the form of a belt with the maximum near 83 AU and the outer boundary at 109 AU. These two morphological forms of disks seem not to be associated with the spectral type and age of the star.

In the submillimeter and middle IR range, Wyatt et al. (2005) detected a dust disk near the F2 star $\eta$ Corvi with a size up to 100 AU at the 850 μm wavelength; at 450 μm the disk is significantly asymmetric, two maxima are seen on the large axis, and this pattern may be presented by a ring of a radius of 150 AU inclined by 45°. The central 100 AU of the disk are free of submillimeter dust radiation and the age of the system is close to 1 billion years.

At the 0.85 mm wavelength, Lestrade et al. (2006) detected a dust disk near the M0.5 dwarf GJ 842.2 at the distance of up to 300 AU.

Schneider et al. (2006) discovered a dust ring of a radius of 86 AU around the F6 star HD 181327. At 1.1 μm it radiates 0.17% of the stellar light, the particle scattering function corresponds to the Henyey-Greenstein theory and a weak diffuse halo is detected outside the ring in the range from 4″ to 9″. Besla and Wu (2007) proposed a mechanism of forming such rings without planets.

In the middle IR range, Fitzgerald et al. (2007) found a ring of warm dust around the star HD 32297, which seems to be later than A0: the recorded flux at 11.2 μm is twice higher than the purely photospheric one. The inner boundary of the ring is at a distance of about 65 AU from the star. The structure thickness of about 1% of the size means the significant role of mutual collisions of dust particles in their dynamics and evolution. Redfield (2007) detected a strong absorption by the sodium doublet in this structure and estimated the upper limit of its mass as 0.3 Earth mass.

Using HST, Kalas et al. (2007a) detected a scattering dust disk around the F2 dwarf HD 15745 at a distance from 128 to 480 AU. The disk is asymmetric with respect to the star, resembles a fan, and the total dust mass comprising particles of 1–10 μm accounts for $10^{-7}M_{\odot}$. A more asymmetric dust disk was detected with HST and the Keck telescope near another dwarf HD 15115 (Kalas et al. (2007b)).

By means of two MIPS (Multiband Imaging Photometer for Spitzer), Forbrich et al. (2008) detected nine new dust disks of the M dwarfs in the NGC 2547 cluster with excessive radiation at 24 μm.

Löhne et al. (2008) developed a theory of the long-term collisional evolution of dust disks, applied it to the disk near the solar-type star G2 V, and proposed a general scaling rule. They obtained that the characteristic time of the collisional evolution is proportional to the initial mass of the disk and varies with the radial distance as $r^{4.3}$ and with the eccentricity of planetesimales as $e^{-2.3}$, for the ages from 10 million to 10 billion years the decay laws of the dust mass and the total disk mass are different since the time between collisions of planetesimales depends on their size; in each moment there is a distance separating large objects which maintain the initial distribution in the growth phase, from those of smaller sizes, whose distribution is determined by collisional destructions. The evolution occurs in such a way that this distance captures larger sizes. Under some standard assumptions, the dust mass, its luminosity, and thermal fluxes decrease with time as $t^{\xi}$, where $\xi$ is within −0.3 and −0.4.

The evolutionary pattern depends on several model parameters — the energy fragmentation as a function of size, initial distribution of the greatest planetesimales, eccentricity, and orbit inclination. This theory represents well the observed distribution of fluxes at 24 and 70 μm and color as a function of age.

Using the IR spectrograph of the Spitzer Space Telescope, Morrow et al. (2008) derived spectra at 0.7 to 40 μm of three young brown dwarfs in the TW Hya association and for two of them detected significant excesses at $\lambda > 5$ μm. These excesses were represented within their model of radiating accretion disks with dust particles of more than 5 μm. The absence of silicon emission at 10 and 20 μm is in agreement with the previous conclusions on the weakening of such emission for older lower-mass stars. This fact suggests that either the growth of dust particles and, possibly, planetesimales near the brown dwarfs occurs faster than that for high-mass stars or the sizes of IR-radiating regions on the brown dwarfs smaller than those on such stars, and dust particles grow faster at lower radii of the disk.

In the middle IR range, Rhee et al. (2008) detected a strong silicon emission on the F6 star HD 23514 in the Pleiades, which indicates the presence of warm small dust.

Phan-Bao et al. (2008) suspected that the dust disks around stars could be formed from the molecular fluxes ejected by them.

Corder et al. (2009) presented an image of the dust disk around HD 107146 at 1.3 mm and 350 μm; in both cases they estimated a diameter of images of more than 10″, while the morphology of dust emission suggests a resonance of dust motion with the planet at a distance of 45–75 AU from the star.

Roccatagliata et al. (2009) detected a correlation of dust disk mass and age, and all the revealed disks had central regions free of small micron particles.

Using HST and MIPS, Krist et al. (2010) studied a ring around HD 207129 in the scattered rays and its thermal emission at 70 μm, respectively. The disk has a width of about 30 AU and is at the distance from the star at 163 AU. At the average surface brightness $V = 23.7^m$ from a square second, this is the weakest disk recorded in visible rays. The size of dust particles is estimated as 2.8 μm, their albedo is 5% and the distribution of dust particles on the power law is with an index of −3.9.

Within 115 and 250 AU, Eiroa et al. (2011) detected the dust disks with a temperature of about or lower than 22 K and luminosity $L_{dust}/L_* \sim 10^{-6}$, which is close to the solar Kuiper belt. These coolest and weakest disks cannot be represented within the models of such structures elaborated earlier.

From the 70, 100, and 160 micron observations at the Herschel telescope Lestrade et al. (2012) resolved a dust disk near the M3 dwarf GJ 581 with several planets. Contrary to the very young star AU Mic (12 million years), whose disk was resolved earlier, GJ 581 is an old star with an age of 2–8 billion years and quiescent in X-rays. Its disk extends from 25 to 60 AU and resembles the Kuiper belt with luminosity $L_{dust}/L_* \sim 10^{-4}$. The submicron-size particles of collisional origin prevail in it, and they are not thrown out of the system by radiation or pressure of the wind.

Golimowski et al. (2011) performed the HST and SST observations of the close K dwarf HD 92945 and detected an inclined axisymmetric disk composed of the inner ring in the range within 43–65 AU and the extended outer ring with slowly decreasing brightness and albedo 0.1 at a distance of 65–110 AU from the star. In the constructed disk model, in the range of 24–350 μm the minimum size of particles is 4.5 μm and their distribution is described by the function $r^{-3.7}$ at a total mass of 0.001 Earth mass.

Fujiwara et al. (2012) studied a dust disk around the F3 dwarf HD 15407A with an anomalously high amount of warm dust of $10^{-7}$ Earth mass. At a temperature of 500–600 K the dust is located at a distance of 0.6–1.0 AU, and its luminosity $L_{dust}/L_* \sim 0.005$, which is higher than the predictions of the stationary model of dust production by collisions of planetesimales.

Using WISE, Schneider et al. (2012) studied four new members of the TW Hya association, which due to excesses at 12 and 22 μm revealed noticeable dust disks around these young (8 million years) stars.

Simon et al. (2012) performed the WISE observations of stars in the four bands from 3.4 to 22.1 μm in five moving groups. At 11.6 and 22.1 μm all the known disks were confirmed, excluding the coolest one around AU Mic, and a disk was detected for the pre-main-sequence star 2M J0820-8003 with episodic accretion; the inner radius of this disk turned out to be at a distance of about 0.02 AU, and its luminosity accounted for about 0.1 stellar luminosity.

Kains et al. (2011) considered the evolution of dust disks near the solar-type stars and found that planetesimales near such stars should be somewhat larger than those near A stars, but the mass of disks was higher for A stars.

Boccaletti et al. (2012) performed a modeling of observations of the HD 32297 star at 1.6 and 2.2 μm and detected a strongly inclined thin disk with an inner cavity of 110 AU at the distance 0.5–0.6″ around the star.

Kirchschlager and Wolf (2013) considered the influence of porosity of particles on the properties of their absorption, radiation pressure, and balanced temperature and found that the temperature of such particles was lower than that of compact spheres.

During the first observations with the ALMA system with the resolution 0.6″ of the dust disk of AU Mic, MacGregor et al. (2013) confirmed the already known dust belt extending up to 40 AU and detected the unresolved central peak. The cold dust with a mass of about that of the Moon reveals the profile with growing emission, which is abruptly interrupted at the outer boundary. The central peak is 6 times brighter than the stellar photosphere, which points to the additional emission process in inner regions of the system. The emission may emanate from the corona or activity, but observations did not show time variations or flares in the radio region.

* * *

The following investigations provide notions on the occurrence of circumstellar cold dust disks.

In the range of 14–35 μm, Jura et al. (2004) carried out the Spitzer spectral observations of 19 close main-sequence stars with IR excesses, in 6 cases the spectral peculiarities proved to be so that the radiating particles had diameters of more than 10 μm.

Bryden et al. (2006) searched for the IR excesses in a sample of 69 F–G–K dwarfs of the field selected without taking their age and metallicity into account. At a wavelength of 70 μm they detected excesses for seven stars. This radiation emanates from the cold ($T$ < 100 K) matter localized at 10 AU and corresponds to the solar Kuiper belt with the 100-fold increased surface. The disks with $L_{\rm dust}/L_* > 10^{-3}$ are rare near old F–G–K stars and their frequency increases from 2% ± 2% for $L_{\rm dust}/L_* > 10^{-4}$ to 12% ± 5% for $L_{\rm dust}/L_* > 10^{-5}$. The IR excesses do not correlate with metallicity and spectral type, but there is a weak correlation with age: the stars younger than 1 billion years have excesses with higher probability.

Using the Spitzer telescope, Gorlova et al. (2006) performed a survey of the region 2°× 1° in the Pleiades at 24 μm and detected excesses, which correspond to dust disks, on five early stars and four solar-type stars; this means the existence of such disks on 25% of the B–A stars and on 10% of the F–K3 stars.

Moor et al. (2006) compiled a list of 60 dust disks at the distance up to 120 pc from the Sun with the dust luminosity $L_{\rm dust}/L_{\rm bol} > 10^{-4}$ and found that such structures often belong to young kinematic groups, practically all disks with $L_{\rm dust}/L_{\rm bol} > 5 \cdot 10^{-4}$ are younger than 100 million years. But there exist many disks with middle dust luminosities. Later, Moor et al. (2009) revealed warm excesses for four F stars, three of them were in agreement with predictions of the theory of stationary disk evolution, for the fourth, the oldest system HD 169666, the dust luminosity was so high that it made one assume the recently started process of its formation ongoing at least for several years.

During the Spitzer observations of 18 supposedly young systems at a radius of up to 65 pc from the Sun, Smith et al. (2006) discovered for 15 of them the radiation excess at 70 μm from 2 to 30 times and for only one of them detected the radiation excess at 24 μm; this confirms a rather low temperature of radiating particles — less than 150 K.

Using the Spitzer telescope and literature data, Trilling et al. (2008) observed about 200 F–G–K stars at wavelengths of 24 and 70 μm and compiled a sample of more than 350 A–F–G–K–M main-sequence stars. There are 4.2% with excesses at 24 μm among F–G stars, 16.4% — with excesses at 70 μm; the distributions of excesses for A, F, G, and K stars turn out to be indistinguishable but with decreasing toward cooler stars, which may be the age effect.

Greaves et al. (2009) studied 13 G dwarfs within 12–15 pc from the Sun and at 850 μm confirmed a disk near HD 30495 with a mass of 0.008 Earth mass and lower than 0.0025 Earth mass for other stars under investigation.

Lestrade et al. (2009) studied at 1.2 mm 50 M dwarfs involved into the moving groups and close objects of unknown age and confidently determined only one dust disk around GJ 842.2. Taking into account the submillimeter surveys of objects with the ages of 10 and 190 million years, they found an occurrence of disks for M dwarfs to be 5.3%, 15% for F–G–K stars and 22% for A stars; however, the uncertainties of these estimates are comparable or even more than the estimates themselves. Thus, for such age groups one may suggest that the frequency of disks probably decreases for cooler stars.

Plavchan et al. (2009) carried out the MIPS observations at 24 and 70 μm of 70 A–M stars at an age of 8 million to 1.1 billion years and detected at 70 μm excesses for five objects, whereas such excesses are rarer for stars with temperatures of less than 5000 K.

Using the Spitzer Infrared Spectrograph, Dodson-Robinson et al. (2011) studied 111 solar-type stars, involving 105 stars that comprise planets. They recorded 11 disks, involving two new ones, and estimated the dust temperature within 60–100 K. The dust rings rotate around stars at the distances from 15 to 240 AU, depending on the particle sizes.

Using the TESS data, Rebull et al. (2022) studied two regions in the Centaurus constellation with objects of an age of 17 and 16 million years. From the brightness modulation due to spots, they estimated rotation periods for 90% of these objects and for 13% of them found infrared excesses, suggesting circumstellar disks. They also revealed a strong concentration of disk-free M stars at a period of about 2 days, hinting that perhaps these stars have recently freed themselves from their disks.

Using archive and own HST/STIS observations of the star AU Mic, Flagg et al. (2022) studied the molecular hydrogen lines in the wavelength range 1144–1710 Å with a resolution of 46000 in the quiet state of the star and during a flare. Based on the contours of these lines, they estimated the temperature of the gas at 1000 to 2000 K. Four possible sources of $H_2$ emission were considered: a background source independent of the star, a circumstellar cool disk, an exoplanet or a star. The authors concluded that the most probable is the AU Mic star itself with anomalously cool spots or areas of the cool photosphere of this young star.

# Part 2

# Flares

Short-term flares are the most easily accessible for observations manifestations of the activity of red dwarfs. Being the last link in the chain of magnetohydrodynamic processes occurring on the variable stars under consideration, flares are a challenge for those who examine their mechanism to reconstruct the whole chain of magnetohydrodynamic processes.

The general problem of studying flare activity of UV Cet-type stars can be split into two particular problems: elaboration of a physical model of an individual flare and presentation of an ensemble of flares within a certain statistical model. The first problem should be solved by analyzing the relatively few flares, for which the most detailed observations are available. The second problem can be solved using as long and uniform observational series as possible. This approach determined the structure of this part of the book. In the first chapter, we describe several strong flares to provide a general idea of the phenomena of stellar nonstationarity. In the two next chapters, we consider the principal statistical properties of flare activity, temporal characteristics of flares and their energy. Then we survey and analyze the observational data underlying the physical modeling of flares and finally present simulation results.

Before proceeding to the description of stellar flares let us recall the principal properties of solar flares.

Solar flares are one of the most impressive phenomena of the solar activity. They were discovered in the middle 19th century but have been studied intensely only since the 1940s. In visual observations, a solar flare is a rapid increase in the brightness of small regions, a few tenths of a percent of the solar disk, near the groups of sunspots. If observations are carried out through a specially selected color filter picking out the radiation of the red $H_\alpha$ hydrogen line, the contrast of flares against the background of the quiescent solar surface multiplies and the frequency of recorded flares consequently increases. The fast burning is usually followed by a phase of maximum brightness of comparable duration, which gradually turns into the decay phase, which is several times longer.

Most solar flares develop near spots, but for about 10% of flares the relation to spots is not evident, and all flares occur in the facular areas.

Visual observations and filming reveal the extraordinary diversity of solar flares: visible patterns of the development of such flares are unique and their temporal and energetic characteristics are very diverse. Thus, a typical flare lasts one hour, but the fastest manage to burst and die out in several minutes, while the strongest flares last for many hours. The total energy of optical emission of the weakest flares recorded on the Sun is $10^{26}$– $10^{27}$ erg and that of the strongest flares is $10^{30}$–$10^{31}$ erg. For a long time it was supposed that the most powerful flares were rare flares of white light, in which apart from emission the flare continuum was detected. However, recently, Kretzschmar (2011) put forward a confirmation regarding a wide distribution of the “white light” identified with blackbody luminosity at 9000 K in the total radiation of solar flares.

During solar flares, the heated middle layers of the atmosphere, the chromosphere, intensely radiates at optical wavelengths. Therefore, as long as flares were studied only within the optical range, they were called chromospheric flares. But radio-astronomical and X-ray observations showed that optical flares accounted only for the secondary effects of this complicated phenomenon. It was established that solar flares are caused by magnetohydrodynamic and plasma processes related to strong magnetic fields of sunspots. Spots of opposite polarity are coupled in the solar atmosphere by magnetic flux tubes, and the onset of a flare is usually recorded as an appearance of a large number of charged high-energy particles in the bases of the arches formed by flux tubes. The particles are found from the strong burst of nonthermal X-ray emission. Moving along the flux tubes, these particles heat plasma confined in the tubes to high temperatures and induce a longer thermal glow of flares in the X-ray range. The excitation of deeper and denser layers by these energetic particles leads to optical glow of the lower chromosphere. In some cases, the particles in the flares have such a high energy that nuclear reactions occur, which is evidenced by the recorded gamma quantum spectral line corresponding to the annihilation of electron–positron pairs.

The high-energy particles emerging at the onset of a solar flare cause disturbances propagating upward along the solar corona. These disturbances are detected in the form of nonthermal radiation in a wide range from centimeter and decimeter wavelengths to tens of meters, whereas several different types of radiation are valid, which differ by a mechanism, place, and time of generation, polarization and frequency characteristics, their variations with time.

The diverse motions of matter occur during solar flares. Ten minutes before the onset of many flares, prominences start moving high above an active region. These are fantastical filament structures of relatively cold plasma, "hanging" in the hot corona; these motions of prominences are one of the most reliable precursors of flares. In the films of flares made in the light of the red hydrogen line, the fast gas motions are well seen in the lower chromosphere. But ejections of matter associated with flares seem to be the most substantial, which is traced from the photospheric level to the interplanetary space.

The disturbances extend up to the middle chromosphere in the strongest solar flares. Thus, during the solar flares, the short-term processes cover a huge space from the visible solar surface up to the nearest planet orbits. In the whole flare energy, the contribution of radio emission is minimum; the total energies of emission in the optical and X-ray ranges are comparable in an order of magnitude, but the kinetic energy of the involved by a flare matter exceeds the energy of electromagnetic radiation by an order of magnitude.

From the secular and millennial terrestrial archives of $^{14}C$ and $^{10}Be$ isotopes Usoskin and Kovaltsov (2012) searched for proton ejections from the Sun with energy of higher than 30 MeV and found four candidates with fluxes of (1–1.5) $\cdot 10^{10}$ cm$^{-2}$ over past 600 years at the secular data resolution, but none with a flux of more than $2 \cdot 10^{10}$ cm$^{-2}$; at a more rough time resolution over past 11400 years they found 19 candidates with fluxes of (1–3) $\cdot 10^{10}$ cm$^{-2}$ but none with a flux of more than $5 \cdot 10^{10}$ cm$^{-2}$. These ejections of high-energy particles contain about 10% of energy of all ejections.

Solar physics in general, and the physics of solar flares in particular, are the major impetus for the development of cosmic electrodynamics and magnetohydrodynamics. Two types of stable magnetic topological structures with a limited energy reserve in a limited volume were discovered experimentally and theoretically on the Sun: magnetic flux tubes and arcades. Solar flares of two types – compact and two-ribbon – are associated with these structures. Compact flares occur in one or two loops with fast energy release at which the plasma temperature achieves $10^7$ K and nonthermal particles emerge. Two-ribbon flares occur in active regions with dark filaments, which evidence a definite structure of the magnetic field. It is believed that in this case a neutral layer is formed, and the motion of the filament provokes the reconnection of the magnetic flux tubes. Strong electric fields are induced and the arcades of loops are formed, whose behavior is analogous to the behavior of loops in compact flares. The energy of two-ribbon flares exceeds that of compact flares by a factor of 10–1000.

The dominant point of view on the immediate energy sources of solar flares supposes that the sources are nonpotential magnetic fields arranged in loop structures, whose bases are in the photospheric regions of different polarity and whose tops are in the corona.

Zirin and Ferland (1980) thoroughly compared various characteristics of solar and stellar flares. Two aspects should be considered in comparing these structures.

First, we cannot see stellar disks and, consequently, individual structures on the surface of any star, as on the Sun. Therefore, there are only few indirect data on the correlation of stellar flares with active regions. Thus, Busko and Torres (1978) noted a weak correlation of the visibility of spots with the flare frequency, but from the observations of EV Lac and BY Dra in the B band Mavridis and Avgoloupis (1993) suspected that the season of high flare activity is followed by the high-spottedness season, and Contadakis (1995) found the alternation of 1–2-month intervals during which either the flare activity and the rotational modulation of the brightness of EV Lac intensified in turn, or both manifestations of stellar activity disappeared or substantially weakened. Later Contadakis (1997) suspected the situation on BY Dra and CE Boo was similar. Leto et al. (1997) analyzed the flare activity of EV Lac in 1967–77 and found that in 1970 flares were concentrated at a certain longitude, in the following year this longitude was the place of maximum spottedness. Doyle (1987) found that in 1973–76 there were definitely more flares on one hemisphere of EV Lac than on the other. Mavridis et al. (1995) found that flares on BY Dra occurred more often within a certain range of axial rotational phases. Doyle and Mathioudakis (1990) and then Butler et al. (1994) concluded that in the YY Gem system the best visibility of spots was in the quadrature phase and flares were drawn to this phase.

Secondly, even on Proxima Cen, the star nearest to the Sun, the number of optical quanta per unit time is $10^{15}$ times lower than the appropriate number of solar quanta.

These two circumstances make practically impossible the application of the same technique and the same tools to the studies of the Sun and flare stars. Therefore, the determination of the substantial effects of the observational selection should be taken into account in comparative studies.

## 2.1. General Description of Stellar Flares

During a strong flare, the optical luminosity of a star increases by several fold over tens of seconds. The character of stellar radiation changes markedly. Continuous emission fills in the absorption details of the short-wavelength region of the optical spectrum. The intensity of emission lines of the chromospheres and the transition zone from the chromosphere to the corona observed in the optical and ultraviolet parts of the spectrum sharply increases. Many emission lines observed in the optical range broaden noticeably. The intensity of thermal coronal radiation increases, so does the temperature of the coronal plasma. Over a wide frequency range, nonthermal radiation emerges and propagates in the corona. Flares on red dwarfs are distinguished by extreme diversity: the light curves of the flares are unique, even those occurring on the same star successively during one night (see Fig. 29). The burning time of a flare varies from fractions of a second to several minutes, and the time of decay varies from seconds to many hours. Absolute luminosities in the optical range of recorded flares range from $10^{26}$ to $10^{32}$ erg/s, as the total energy of the optical emission of flares cover at least the range of $10^{28}$–$10^{34}$ erg. Flares differ substantially in the relative intensities of optical, radio, and X-ray emissions, in the relative contribution to the optical emission of continuous and line emission, in the broadening and symmetricity of emission lines, in the temperature and emission measure of hot coronal plasma, in the temporal sequence of the occurrence of flares in the optical and radio ranges, and in the frequency and polarization characteristics of radiation. Such a diversity of recorded flares, the relatively short time of investigation of the phenomena, which can be insufficient to detect the strongest and rarest events, and the inaccessibility of the weakest flares for observations due to the permanent radiation of flare stars – all this makes describing a typical flare a rather ambiguous task. Therefore, instead of artificial construction of a certain average flare let us consider several real most studied flares.

T h e f l a r e o n A D L e o o f 1 8 M a y 1 9 6 5 was recorded by Gershberg and Chugainov (1966) in the Crimea during simultaneous photometric and spectral observations at the 64-cm meniscus telescope and at the 2.6-meter Shajn reflector, respectively. Figure 30 shows the light curve of the flare and indicates the time intervals of spectral exposures. At the phase of flaring up, the brightness of the star in the blue band increased by more than 5 times, approximately over a minute. At the maximum brightness of about 20 s the luminosity of the flare in the B band was about $3.3 \cdot 10^{30}$ erg/s. Several minutes after the main maximum a fast decrease was followed by a slower decline. Then 12 min after the main flare maximum a secondary, smoother and lower-amplitude brightness peak occurred. Figure 30 shows the changes of the spectrum of AD Leo during the flare. The first spectrogram of the flare was obtained in the photographic region of 18 s during a single run of the star along the spectrograph slit. One can see the moment of the flare maximum brightness in the spectrum. On the whole, the spectrum of AD Leo in the blue region at this time changed beyond recognition: the absorption of CaI $\lambda$ 4227 Å became indistinguishable from the fluctuations of the continuous spectrum intensity, all hydrogen emission lines strengthened abruptly, the half-widths of the $H_\gamma$ and $H_\delta$ emission lines grew from 5–6 Å to 10–11 Å, the emission line of neutral helium $\lambda$ 4471 Å appeared, which is invisible in the spectra of quiet stars. In the second spectrogram, further strengthening of the hydrogen and calcium lines was found and quite measurable helium lines of $\lambda$ 4026 Å and $\lambda$ 4471 Å were seen; the absorption of CaI $\lambda$ 4227 Å was also rather low but reliably recordable. In the third spectrum, the hydrogen lines strengthened, the absorption of CaI $\lambda$ 4227 Å increased, the helium line $\lambda$ 4026 Å disappeared, and the line $\lambda$ 4471 Å was difficult to distinguish. The fourth and fifth spectrograms were rather similar, the equivalent widths of the hydrogen lines achieved the maximum values of $W_{H\gamma} = 65$ Å and $W_{H\delta} = 50$ Å, the widths of the lines slightly decreased as compared to the first image. If we neglect the continued narrowing of hydrogen lines, the sixth spectrum would repeat the third one in the intensity of hydrogen and calcium lines, in the absorption of CaI $\lambda$ 4227 Å, in the difficult-to-measure traces of the $\lambda$ 4471 Å line. The seventh image was obtained in the blue region of the spectrum during a four-fold run of the star along the spectrograph slit and fell on the rise of the brightness to the secondary flare maximum. Here, the $H_\gamma$ line was overexposed, though the emissions to $H_{11}$ were measurable, the absorption depth of CaI $\lambda$ 4227 Å slightly decreased and the $\lambda$ 4471 Å line became more distinct. The equivalent width of the $H_\alpha$ line measured in spectra 8 and 11–14 varied from 20 to 14 Å at 8 Å in the quiet star, while $W_{H\beta}$ in spectra 9–10 was about 50 Å as compared to 7 Å out of the flare; the HeII $\lambda$ 4686 Å line was not seen in images 9–10. The last spectrum 15 obtained in the blue region shows that even one and half hours after the brightness maximum the equivalent width of $H_\gamma$ line noticeably exceeded the normal value.

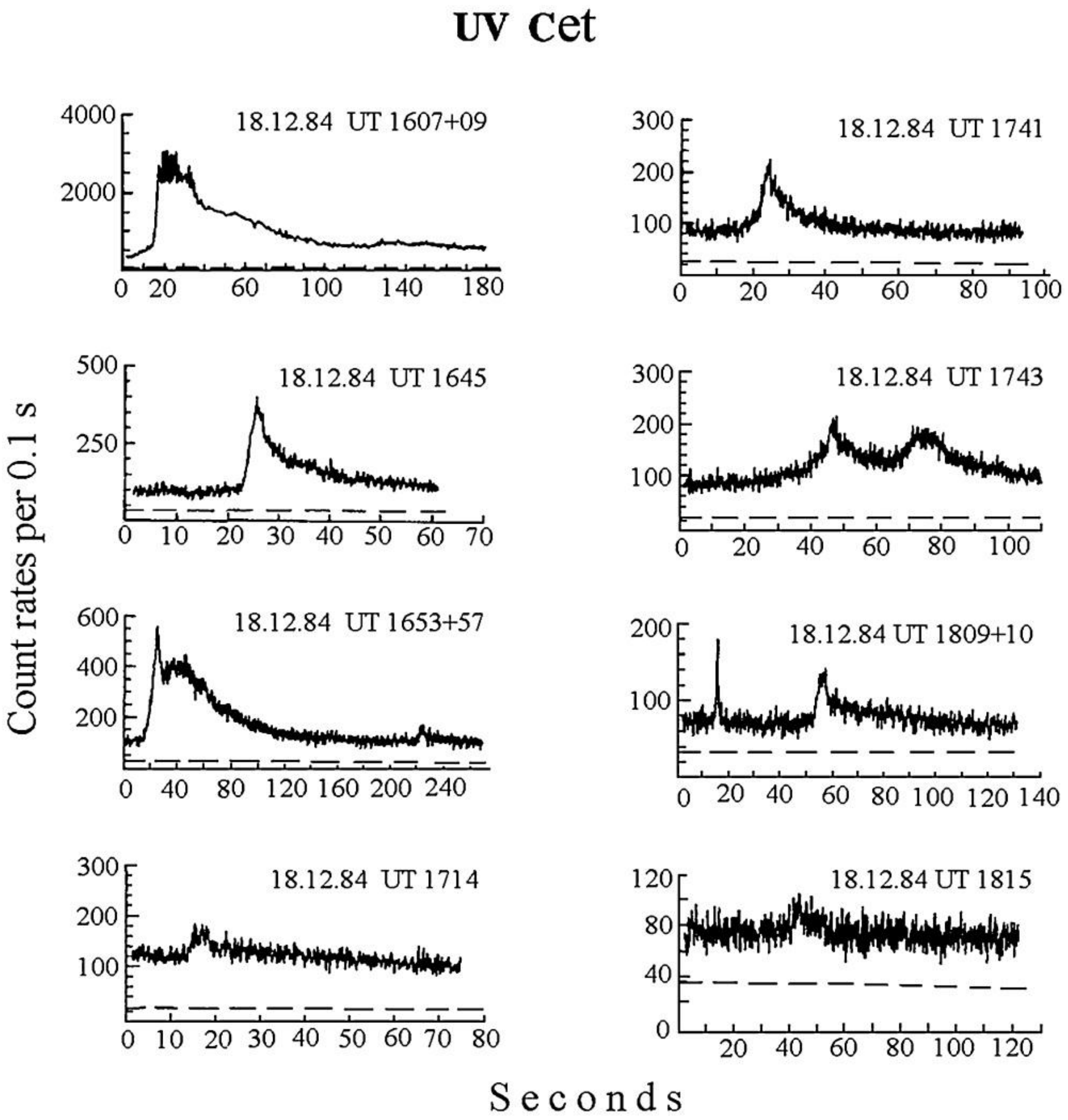

Fig. 29. Light curves of the UV Cet flares recorded with the 6-meter telescope of the Special Astrophysical Observatory, Caucasus on 18 December 1984 during a 2.5-h photoelectric monitoring in the U band (Beskin et al., 1988)

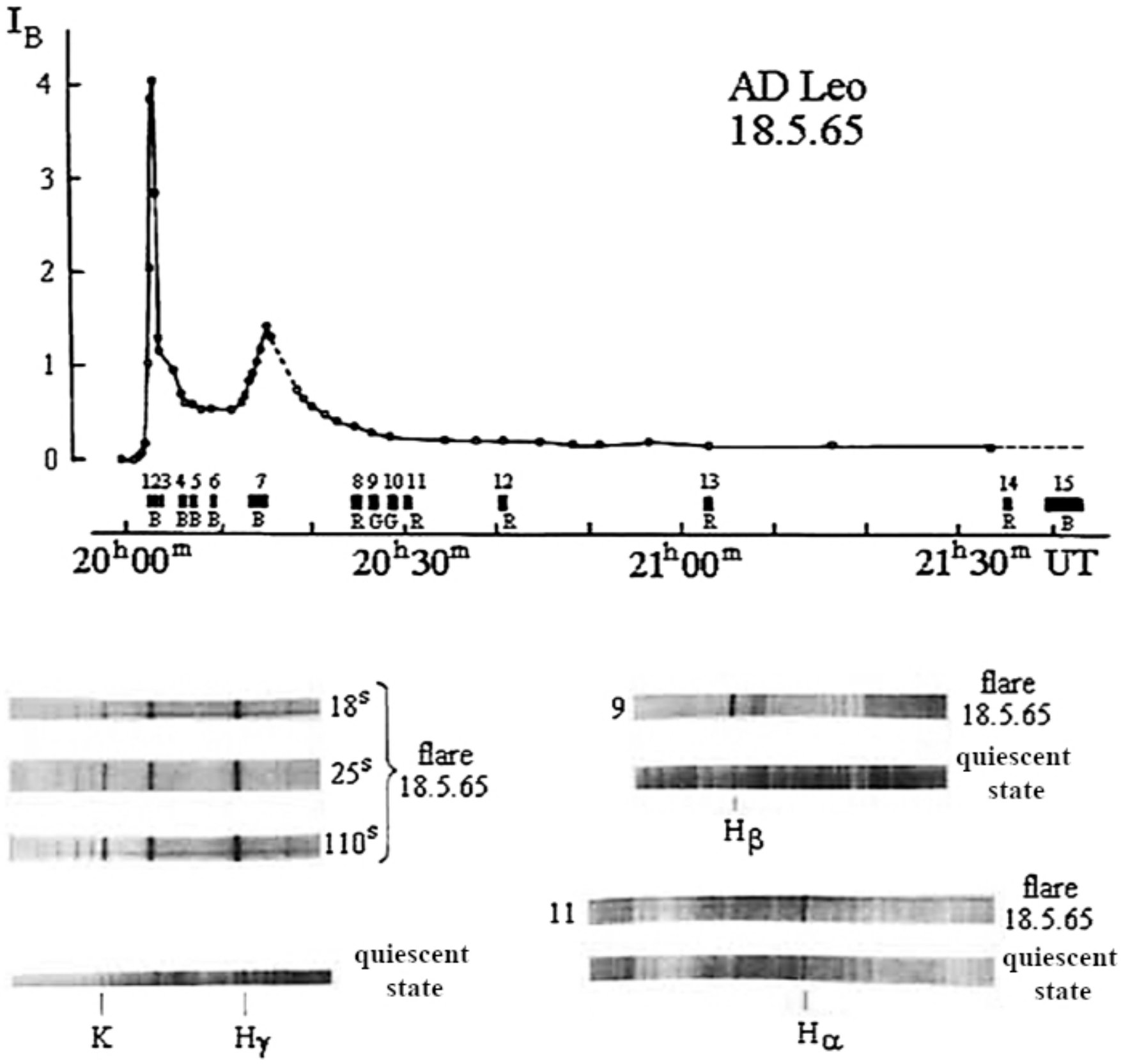

Fig. 30. Flare on AD Leo of 18 May 1965. *Top*: the light curve in the B band. Numbered rectangles mark the time intervals when the spectrograms were obtained in the blue (B), red (R), and green (G) regions of the spectrum. *Bottom*: spectra of the flare, numbers on the left of the spectra correspond to spectrogram number (Gershberg and Chugainov, 1966)

The flare on EV Lac of 11 December 1965 (Kunkel, 1967) was observed by the 90-cm reflector of the McDonald Observatory in Texas. The moving cassette of the spectrograph made it possible to obtain a clear picture of the flare spectrum (Fig. 31). From the onset of the flare the Balmer lines were sharply strengthened, while the continuous emission, well seen only during the first three minutes after the maximum brightness, almost completely filled in the absorption line of CaI $\lambda$ 4227 Å and noticeably weakened the absorption details of the normal spectrum in the region of larger wavelengths. The Balmer lines were traced in the spectrogram to $H_{11}$, while from $\lambda$ 3750 Å, where the emission lines merged due to low resolution, to $\lambda$ 3500 Å, which was the sensitivity threshold of the equipment, and the level of continuous radiation was practically constant. The increased radiation was observed in the lines even one hour after the maximum brightness, the decay in the CaII K line was the slowest. From the spectrum of the flare, Kunkel obtained one of the most reliable photographic estimates of the emission Balmer jump in the intrinsic radiation of the flare and the Balmer decrement at different stages of the flare.

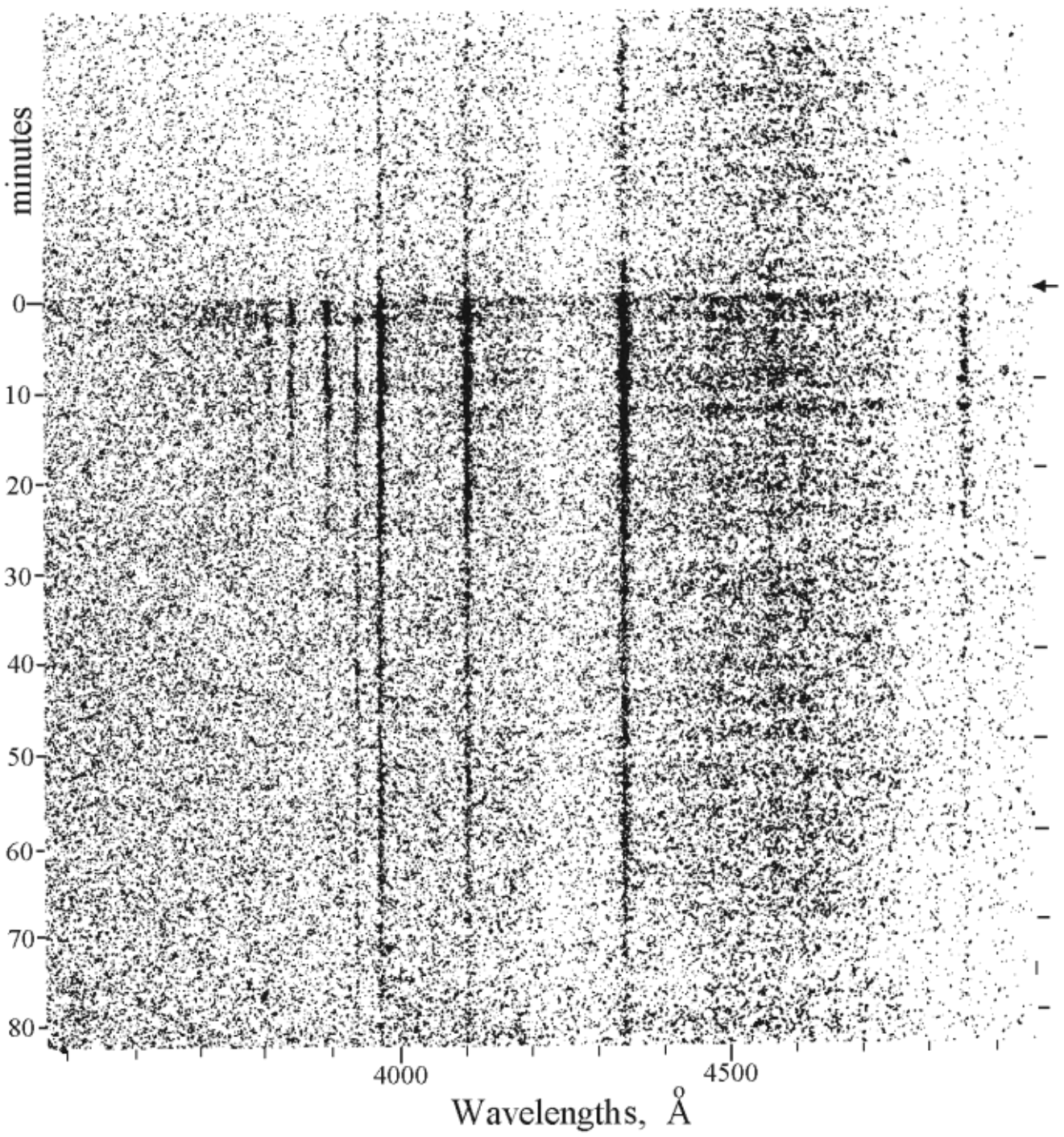

Fig. 31. Spectrogram of the flare on EV Lac of 11 December 1965, figures on the left show the time since the flare onset (Kunkel, 1967)

Two flares on UV Cet (Bopp and Moffett, 1973) were recorded on 14 October 1972 during simultaneous photometric and spectral observations with the 76-cm reflector and the 2.1-meter Struve telescope at the McDonald Observatory in Texas. As one can see from Fig. 32, the initial slow burning of the first flare was due to the strengthening of the line emission, then the continuum had a fast rise to the sharp brightness maximum, during which the luminosity of the flare in the U band achieved $2 \cdot 10^{28}$ erg/s. The second flare was 15 times stronger and started from an abrupt rise of continuum. During the flares, detailed data on the time of appearance and vanishing of the flare spectral features were obtained. In the spectrum of the first flare the effect of asymmetry of the emission line profiles was localized on the light curve with the highest time resolution: the red wings of hydrogen lines were approximately twice as long as the appropriate blue wings; less pronounced asymmetry was recorded also in the CaII K line. No asymmetry was observed in the spectrum of the second flare.

The flare on UV Cet of 11 October 1972 (Lovell et al., 1974) was observed simultaneously at 408 MHz at the Jodrell Bank radio telescope (England) and in the B band with the 76-cm reflector of the Stephanion Observatory at Peloponnese (Greece). Figure 33 shows the light curves of the flare in the different wavelength ranges. The radio flare started practically simultaneously with the optical flare, but the abrupt peak of radio emission occurred 8 min after the optical maximum, when the optical flare decayed. The radio flare faded almost 5 times slower but the total energy of optical emission exceeded the total energy of radio emission of the flare by several orders of magnitude.

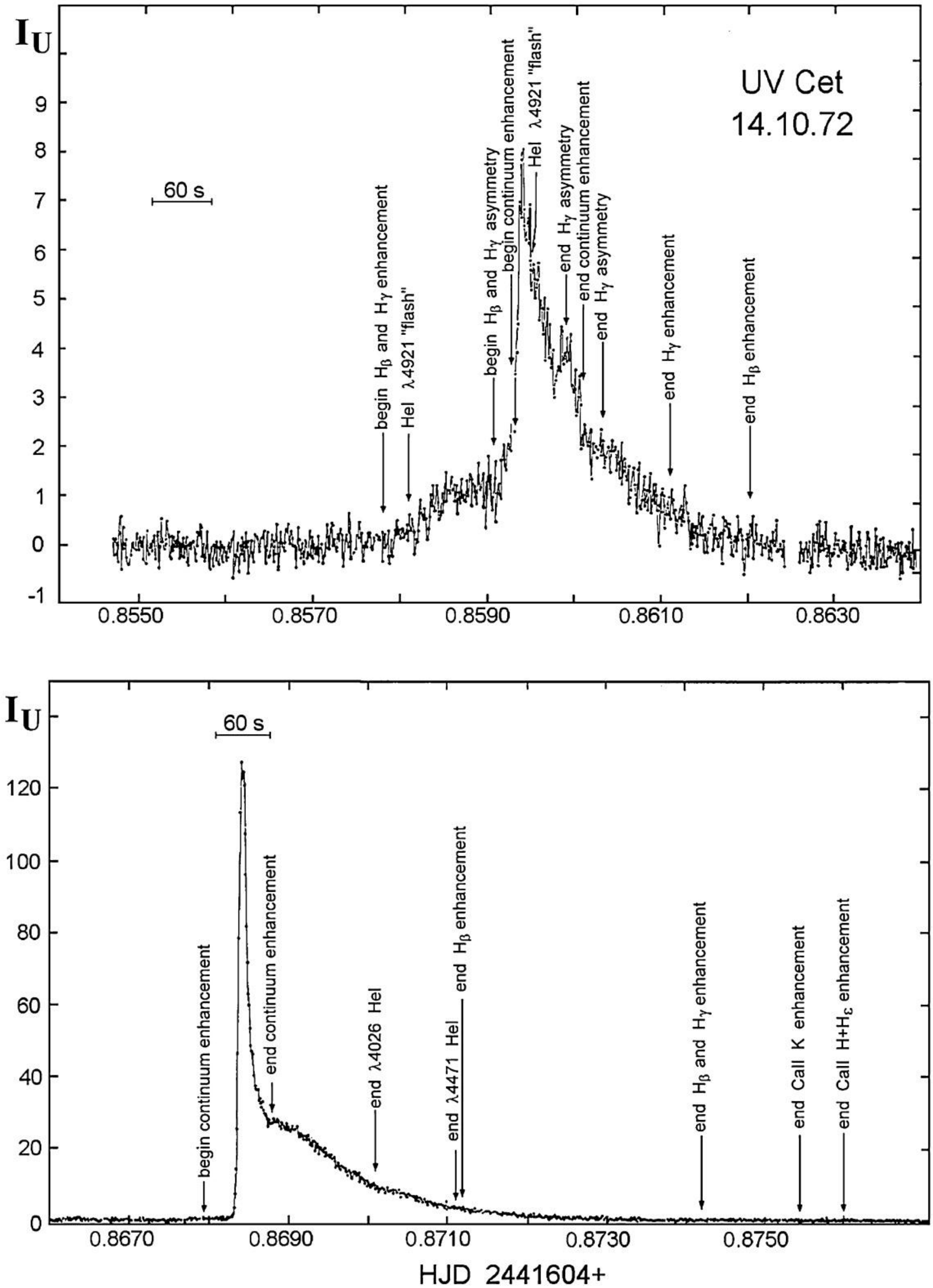

Fig. 32. Flares on UV Cet of 14 October 1972. The sequences of the development of spectral features of flares are specified on the light curves recorded in the U band
(Bopp and Moffett, 1973)

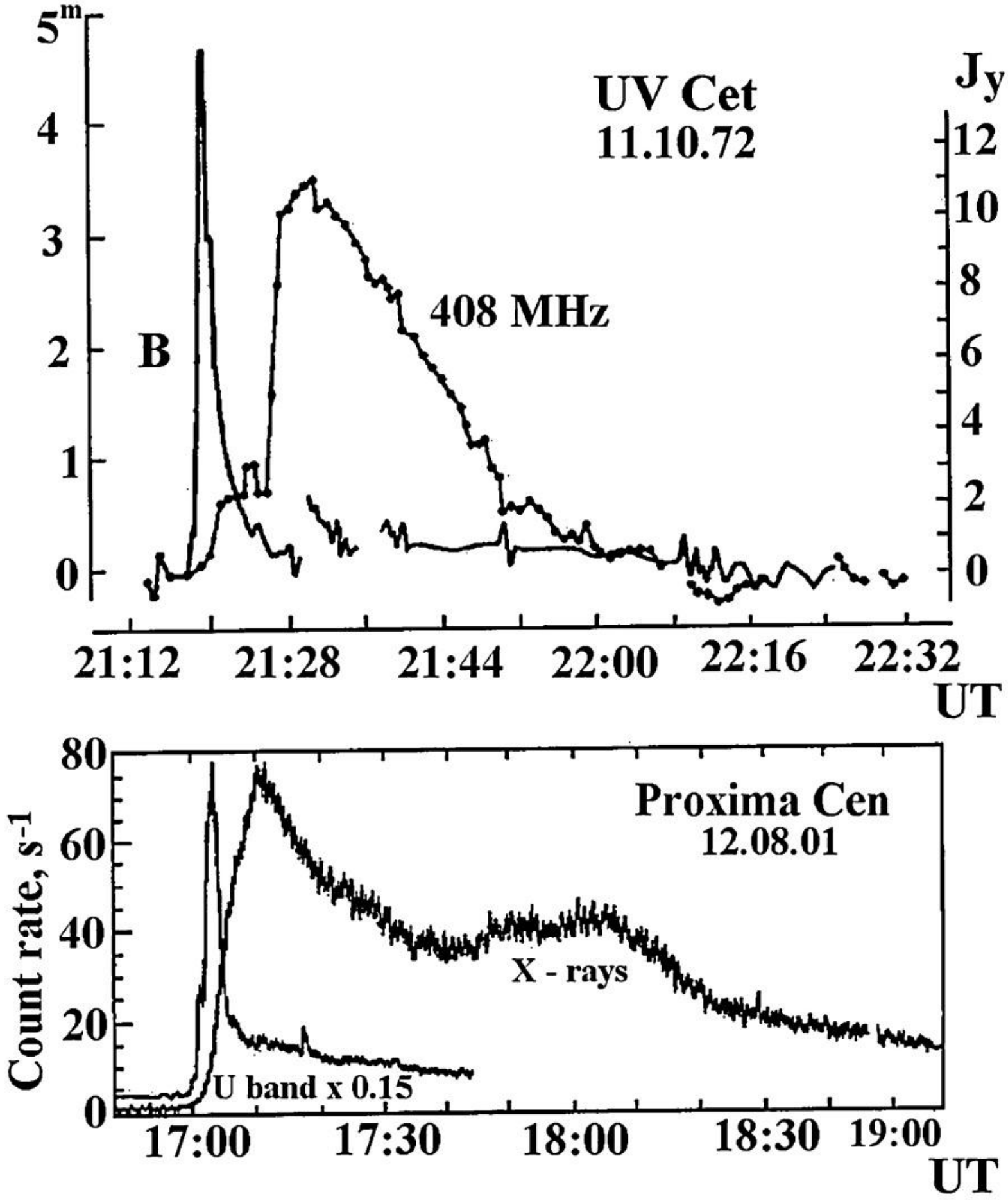

Fig. 33. Responses of stellar coronae to the disturbances resulting in impulsive optical flares. *Top*: flare on UV Cet of 11 October 1972 recorded in the B band and in the radio range (Lovell et al., 1974). *Bottom*: flare on Proxima Cen of 12 August 2001 recorded in the U band and X-rays (Güdel et al., 2002)

The flare on Proxima Cen of 12 August 2001 (Güdel et al., 2002) was detected by XMM-Newton simultaneously in X-rays and the optical range. Figure 33 illustrates the similarity of the coronal response to the optical burst or to its cause in the radio and X-ray ranges.

The flare on Wolf 424 of 30 January 1974 (Spangler and Moffett, 1976) was recorded at 196 and 319 MHz by the radio telescope in Arecibo (Puerto Rico) and in the U band by the McDonald Observatory. Figure 34 shows the light curves of the flare for the three wavelength regions. One can see that practically simultaneously with the abrupt peak of the optical radiation, which at maximum achieved $1 \cdot 10^{28}$ erg/s, a smooth rise of radio emission started at a frequency of 196 MHz, two minutes later a fast and short radiation peak occurred at both radio frequencies. (Over 20 hours of patrol observations of this binary system Pettersen (2006) recorded 57 flares with energy in the U band from $2 \cdot 10^{28}$ to $2 \cdot 10^{31}$ erg. Presumably, a variation of the flare activity level was detected over a year and a dependence of this level on the stellar position on the orbit was suspected.)

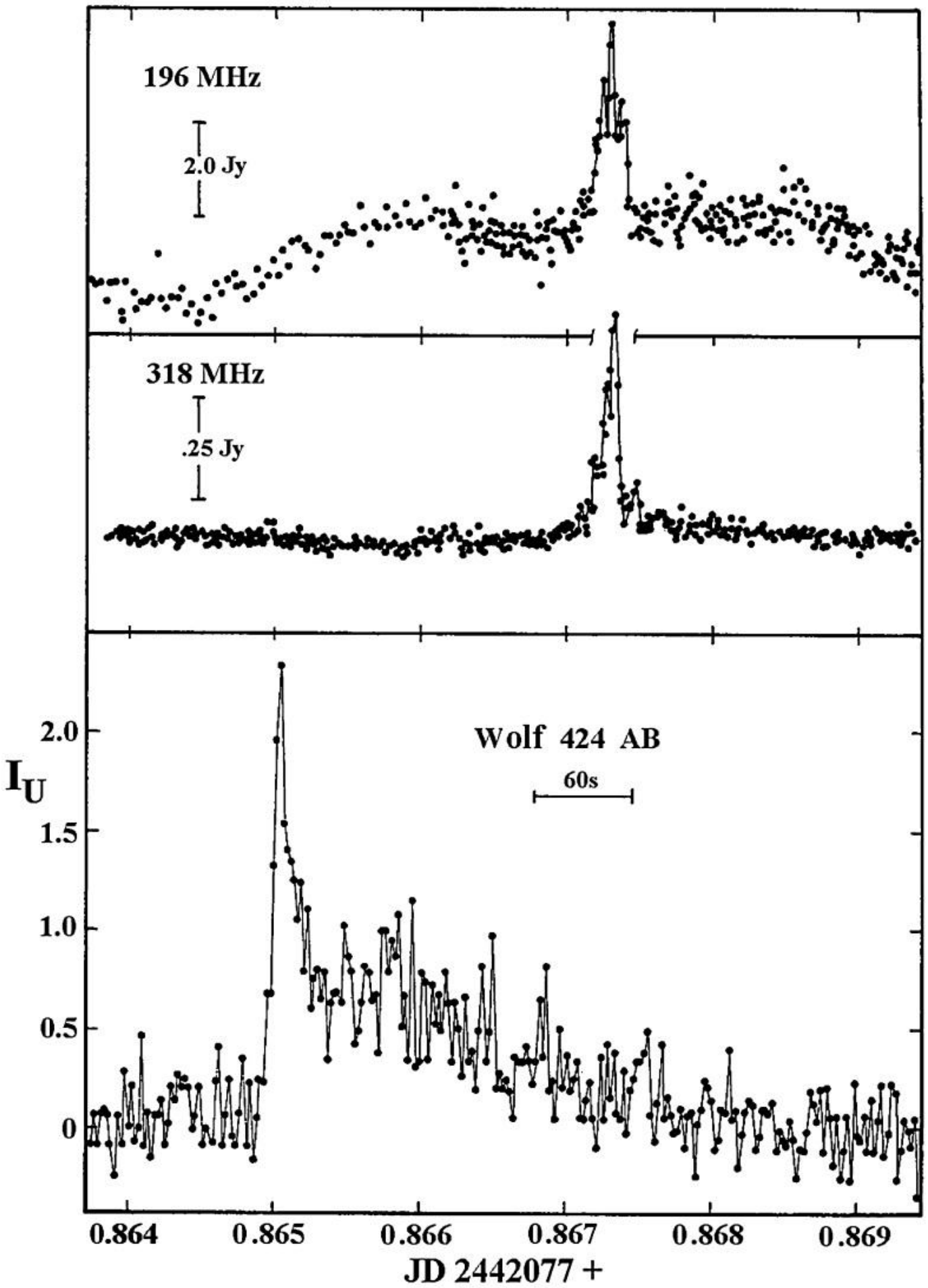

Fig. 34. Flare on Wolf 424 of 30 January 1974 recorded at two radio frequencies and in the U band (Spangler and Moffett, 1976)

T h e  f l a r e  o n  Y Z  C M i  o f  2 5  O c t o b e r  1 9 7 9 (Kahler et al., 1982) was recorded in the course of cooperative program that involved the X-ray Einstein Observatory, seven optical telescopes with diameters of 61–230 cm, and seven radio telescopes with diameters ranging from 26 to 300 m. Figure 35 shows the light curves of the flare in different wavelength ranges and the temperatures and emission measures of coronal plasma calculated from the X-ray emission. Over 35 s the luminosity of the flare in the U band achieved a maximum of $2 \cdot 10^{29}$ erg/s and after two additional short peaks faded relatively rapidly; a similar behavior was observed in the radiation of the continuum near $\lambda$ 4680 Å. Radiation in the $H_\beta$ and $H_\gamma$ lines and the X-ray emission achieved the maximum a little later than the optical continuous emission and weakened much more slowly. The maximum flare luminosity in soft X-rays (0.2–4 keV) was $8 \cdot 10^{28}$ erg/s, in this case the temperature of the coronal plasma achieved 20 MK, and the emission measure was $4 \cdot 10^{51}$ cm$^{-3}$. The total flare radiation in the soft X-ray region was close to the total radiation in the optical range. The radio emission at a frequency of 408 MHz was recorded only 17 min after the optical maximum. At higher radio frequencies the flare was not reliably detected.

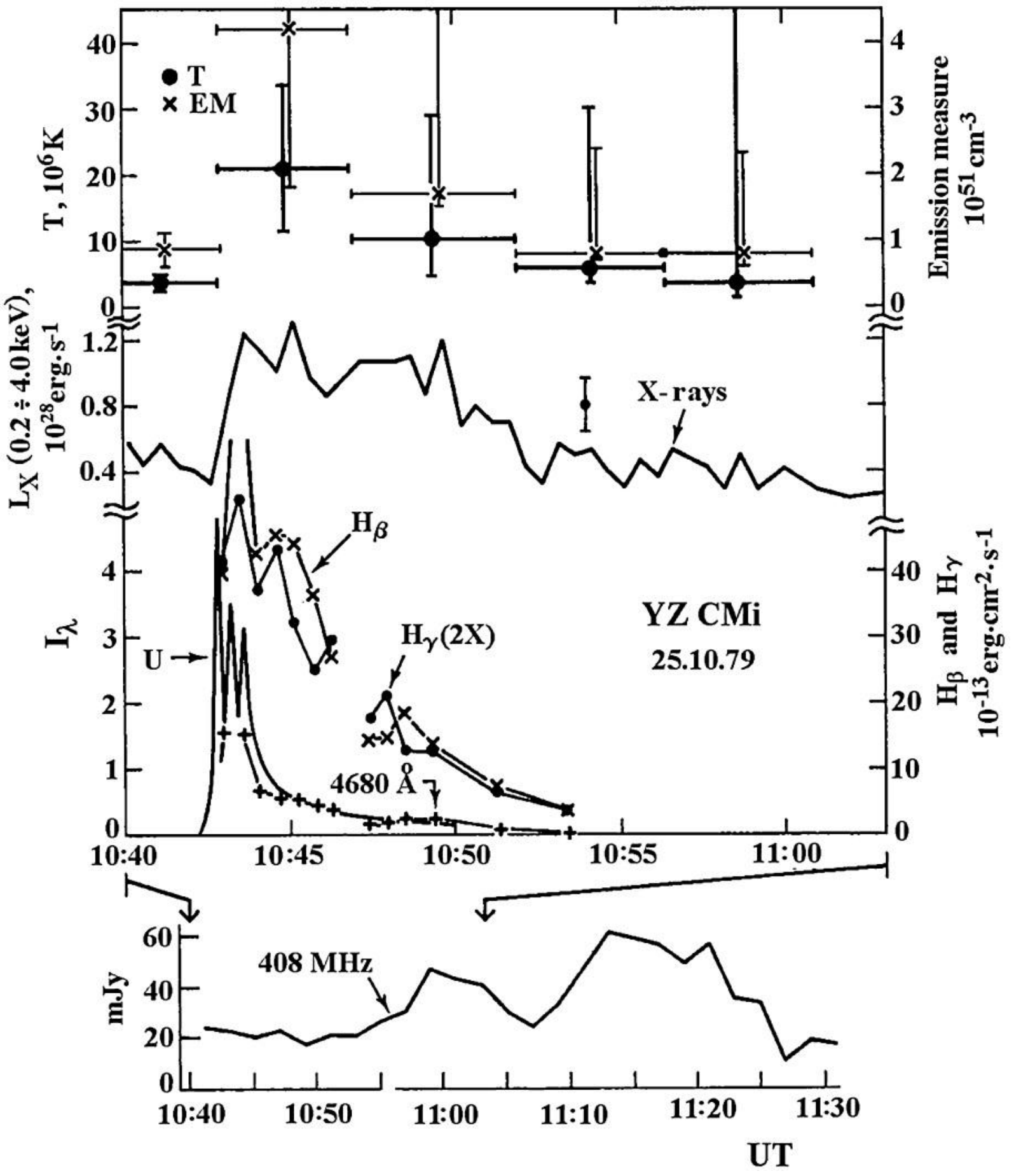

Fig. 35. Flare on YZ CMi of 25 October 1979. *Bottom*: light curves in the radio region, in optical continuum near $\lambda$ 4680 Å, in the U band, in Balmer lines $H_\gamma$ and $H_\beta$, and in soft X-rays; *Top*: the time changes of temperature and emission measure of coronal plasma (Kahler et al., 1982)

The flare on Proxima Cen of 20 August 1980 (Haisch et al., 1983) was recorded in the course of coordinated extraterrestrial observations. The upper panel of Fig. 36 shows the light curve of the flare in the soft X-ray region obtained at the Einstein Observatory, and the time intervals of the spectrography in the ultraviolet region performed from the IUE satellite. In the lower panel the records of four spectra in the range of 1100–2000 Å are presented. The analysis of the data showed that the maximum luminosity of the flare in X-rays achieved $2 \cdot 10^{28}$ erg/s; the total radiation in this range was about $4 \cdot 10^{31}$ erg, the radiative losses for the line ultraviolet emission – to the $Ly_\alpha$ line and the transition zone lines from the chromosphere to the corona – were lower by an order of magnitude than the X-ray emission of the flare. The temperature of the coronal plasma in the flare was 6–7 times higher than in the quiescent corona and reached a maximum value of 27 MK immediately before the maximum of X-ray luminosity of the flare.

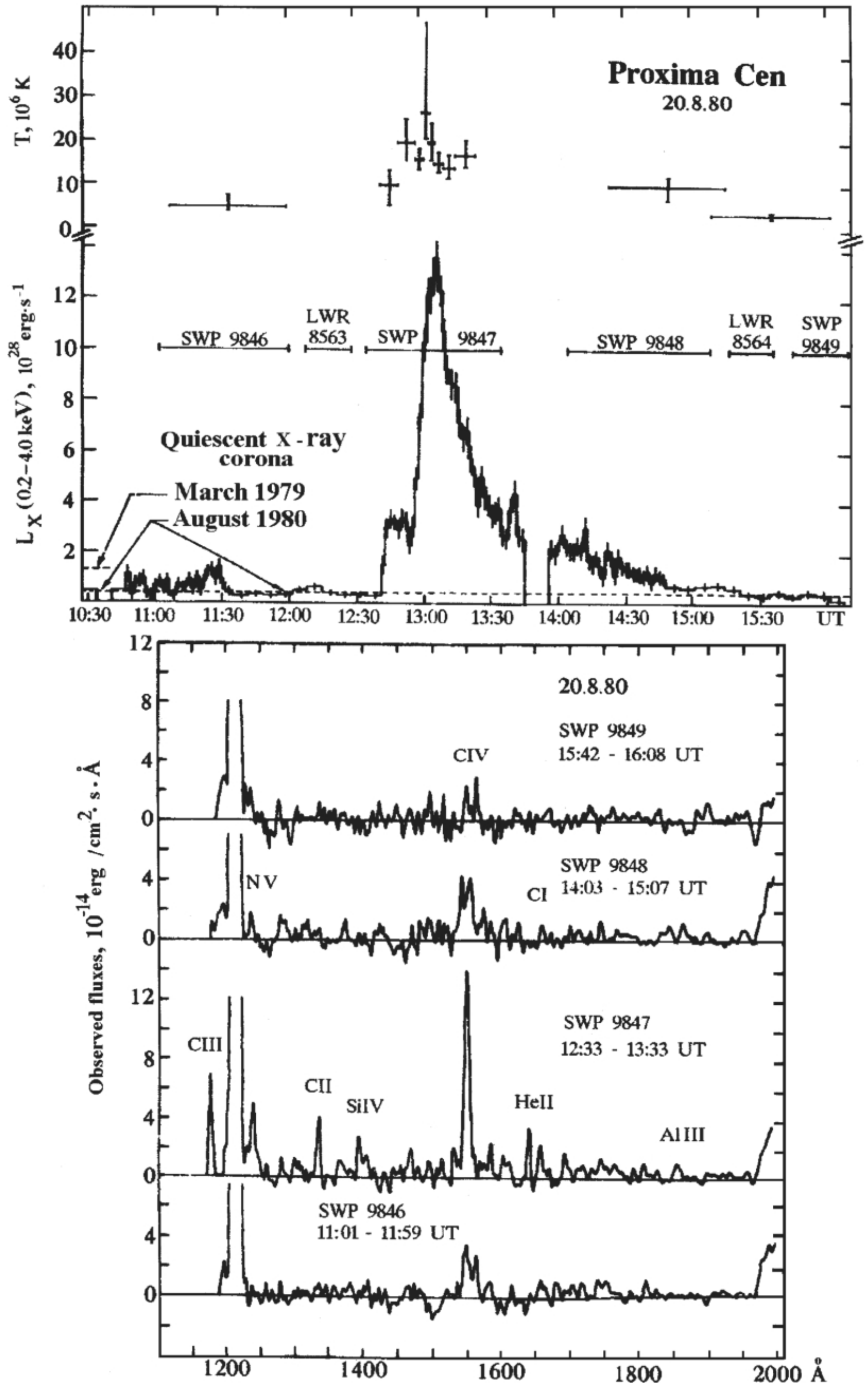

Fig. 36. Flare on Proxima Cen of 20 August 1980. *Top panel*: the light curve in the soft X-ray region derived with the Einstein space observatory and time intervals of the spectrography of the star in the UV region taken with IUE. *Bottom panel*: tracings of ultraviolet spectra (Haisch et al., 1983)

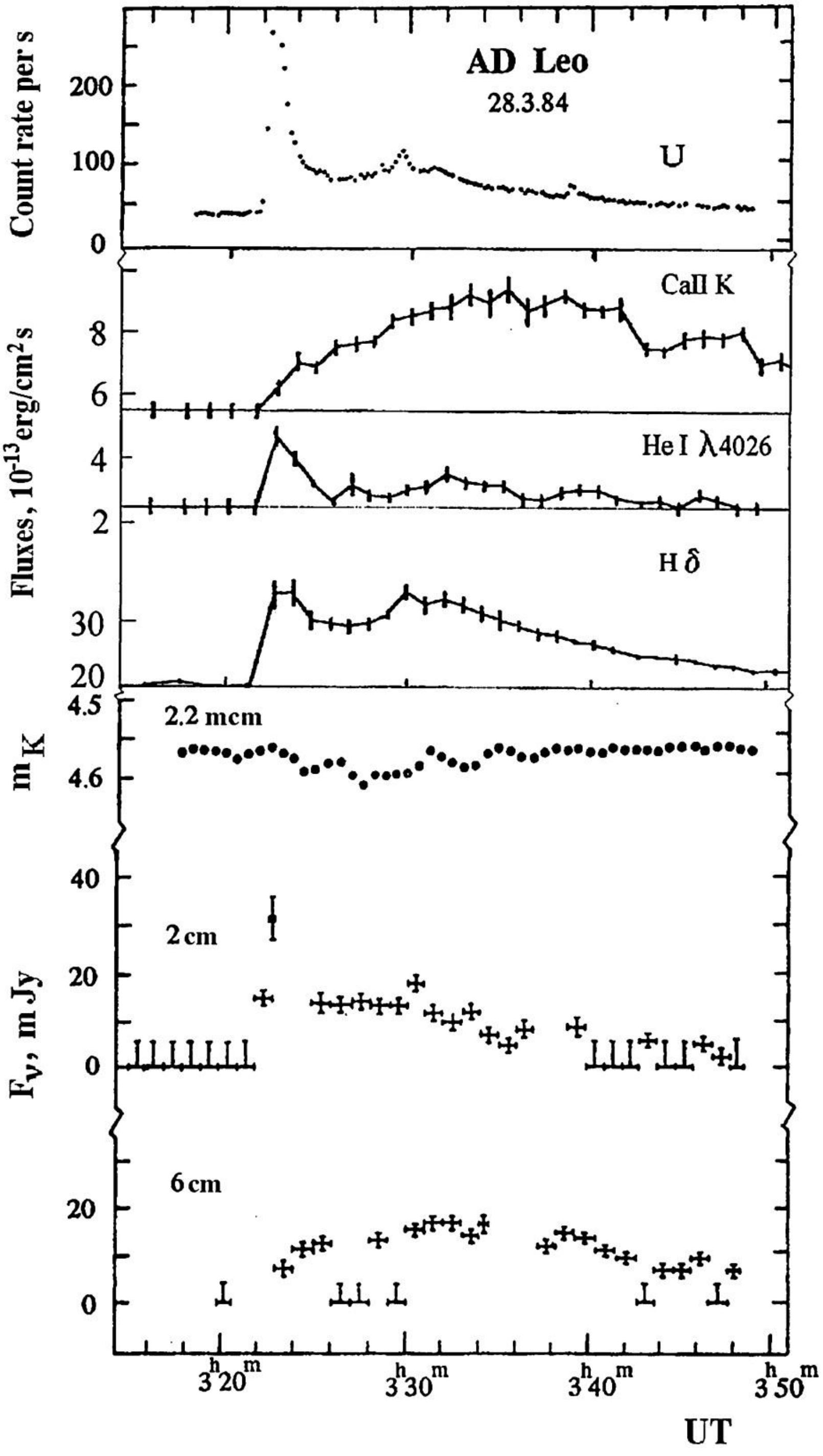

Fig. 37. Light curves of the flare on AD Leo of 28 March 1984 in the U band, in calcium, helium, and hydrogen emission lines, in the infrared K band and at 2 and 6 cm in the radio range (Rodonò et al., 1989)

The flare on AD Leo of 28 March 1984 (Rodonò et al., 1984) was observed within the cooperative program including four telescopes of the European Southern Observatory in Chile, the 91-cm reflector of the McDonald Observatory, the Very Large Antenna in New Mexico, and the IUE satellite. Figure 37 shows the light curves of the flare in different wavelength ranges. The optical light curve of the flare in many respects resembles that of 18 May 1965 on the same star (Fig. 30): both had fast main maxima, lower-amplitude secondary peaks, and the same characteristic times of all details of the light curves. The luminosity of the flare of 28 March 1984 in the U band achieved the maximum value of $1 \cdot 10^{30}$ erg/s 30 s after the burning, the fluxes in the Balmer lines $H_\gamma - H_{11}$ increased approximately two-fold, fluxes in the CaII K line by a factor of 1.4. The emission at 2 and 6 cm was recorded in the radio range, and the weakening of stellar brightness in the K band (2.2 μm) approximately by 3% was found between the main and secondary maxima of the optical flare. 15–25 min after the main flare maximum the ultraviolet observations showed that the MgII λ 2795 Å line was still enhanced, almost 2-fold, the blend of FeII λ 2600 Å, by 9-fold, and the continuum near λ 2500 Å, was 10 times stronger than the appropriate emissions of the quiet star.

The flares on AU Mic of 15 and 16 July 1992 (Cully et al., 1993) were discovered during a four-day monitoring with the Extreme Ultraviolet Explorer (EUVE): in the 65–190 Å band with a time resolution of 100 s. The derived curve is shown in Fig. 38.

Burning of the strong flare on 15 July 1992 lasted for an hour and a half, then the maximum luminosity in this band remained at the level of $10^{30}$ erg/s for about 2 h, the process of decay lasted for more than a day: first there was an exponential decay with a characteristic time of 1.3 h, then a slower decrease of brightness, and the whole decay phase lasted about one and a half days. The total radiative energy of the flare was $3 \cdot 10^{34}$ erg and the volume emission measure was estimated as $6 \cdot 10^{53}$ cm$^{-3}$, assuming that the temperature of the radiating plasma was 30 MK. A day after on the descending branch of the strong flare, there was a slightly weaker flare with a maximum luminosity of $6 \cdot 10^{29}$ erg/s, a duration of about 3 h, a total energy of EUV radiation of $2 \cdot 10^{33}$ erg, and a volume emission measure of $3 \cdot 10^{53}$ cm$^{-3}$. As to the energy, these EUV flares belong to the strongest events recorded on the UV Cet-type stars. In addition to the strong flares, weaker peaks are seen in Fig. 38.

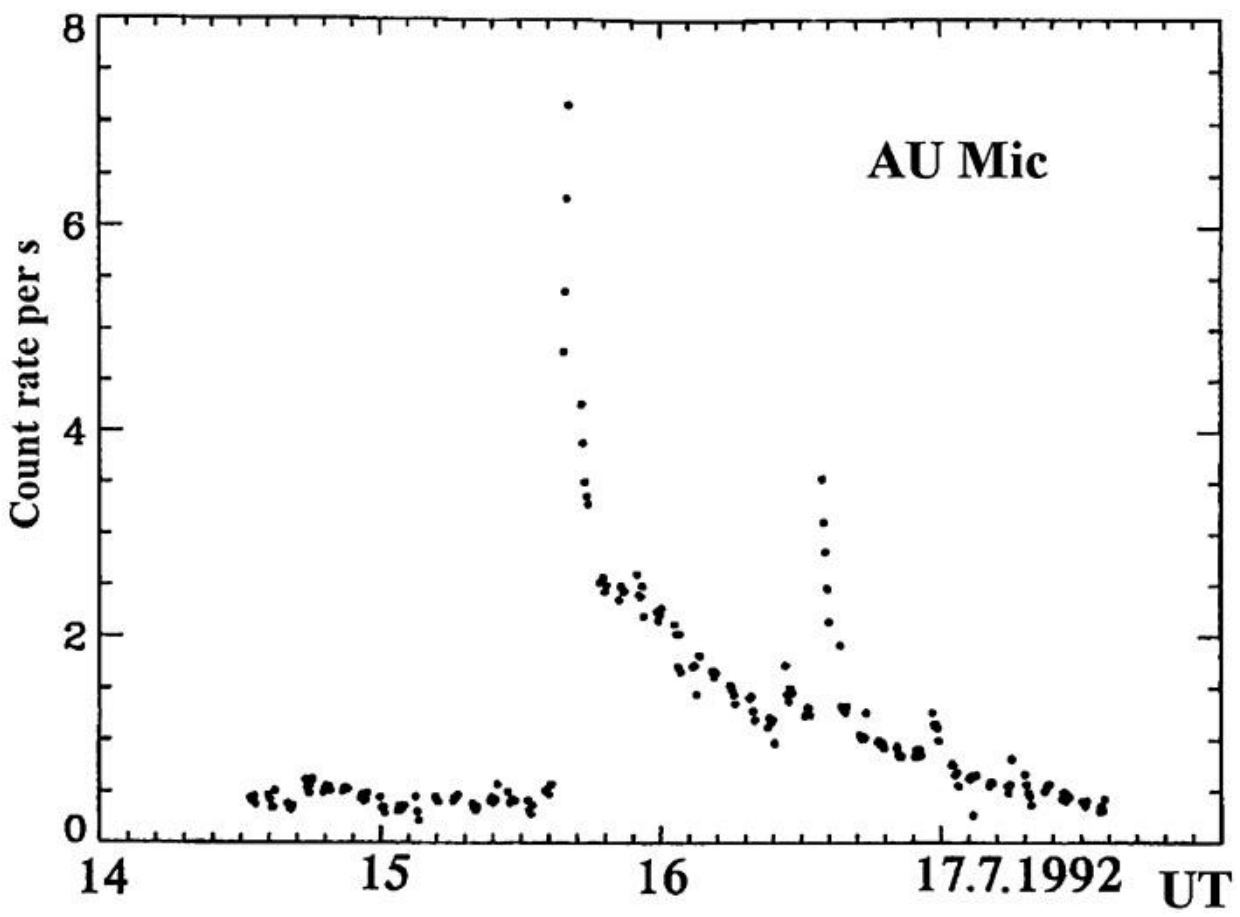

Fig. 38. The light curve of AU Mic in the far ultraviolet (65–190 Å) recorded with EUVE on 14–18 July, 1992 (Cully et al., 1993)

In recent years, hundreds of flares with higher energy — up to many units of $10^{35}$ erg were recorded with the space apparatus Kepler. They are of particular interest for researchers and were named as superflares.

* * *

The above flares belong to the most thoroughly studied flares on the UV Cet-type stars and are the strongest and longest processes of the considered type. The vast majority of the recorded flares on red dwarfs are shorter and weaker. However, there are no grounds to think that the weaker phenomena are more uniform: a certain uniqueness and diversity of solar flares, whose energies approximate to weak stellar flares, suggests that weak flares are very diverse as well.

The next chapter will consider the two most general temporal characteristics of stellar flares: their durations and temporal distribution and energy of flares. The character of flare development, i.e., the inner time structures and physical processes in different wavelength ranges are to be discussed in Chapter 2.4. The last chapters of this part are devoted to the models and theory of stellar flares.

## 2.2. Temporal Characteristics of Flares

### 2.2.1. Time Scales of Flares

Available data on the duration of stellar flares are determined both by the intrinsic properties of flares and the capacities of recording equipment. During the first visual observations brightness peaks lasting from fractions of a minute to tens of minutes were recorded. When the visual monitoring was replaced by photographic observations, the characteristic duration of the exposures limited the possibilities of detection of fast flares, and the recorded durations of many hundreds of bursts revealed mostly in several nearest stellar clusters varied from 5–10 min to many hours.

Continuous photoelectric monitoring provides the most complete data on the duration of optical flares. The data for several series of long-term observations are presented in Table 14.

Table 14. Distribution of durations of optical flares

| | | Flare duration, min | | | | | | |
|---|---|---|---|---|---|---|---|---|
| | | < 1/3 | 1/3–1 | 1–3 | 3–9 | 9–27 | 27–81 | > 81 |
| 1 | EV Lac | 1 | 23 | 42 | 69 | 71 | 13 | 3 |
| 2 | CN Leo | 5 | 16 | 35 | 38 | 17 | | |
| 3 | UV Cet | 5 | 21 | 38 | 33 | 11 | | |
| 4 | YZ CMi | | 5 | 16 | 31 | 8 | 2 | |
| 5 | AD Leo | 10 | 32 | 45 | 5 | 1 | | |
| 6 | 8 stars | 8 | 24 | 28 | 12 | 1 | | |

The first row of Table 14 shows the results for the EV Lac flare star monitored in the Crimea: in 1986–95 the star was regularly examined within international cooperative campaigns. 227 flares were recorded over 307 h of photoelectric UBVRI monitoring with a resolution of 12 s (Alekseev and Gershberg, 1997a). The table presents the distribution of duration of 222 flares. The data show that almost 2/3 of flares on this star lasted from 3 to 27 min. Usually, these observations were performed throughout the night from evening to morning, thus the decrease in the number of events longer than half an hour is a real fact rather than the result of observational selection.

The next three rows of the table show the data on the flares on CN Leo, UV Cet, and YZ CMi obtained from the data of Moffett (1974), who in 1971–72 monitored the stars with a resolution of 1–4 s with four telescopes at the McDonald Observatory (USA). The duration of continuous monitoring of one star rarely exceeded 1.5–2 h, thus the small number of long-term flares recorded by Moffett on these stars is largely due to the observational selection.

According to the statistics of flares on these three stars, 2/3 and more flares lasted from 1 to 9 min.

The fifth row of the table lists the data on the flares on AD Leo based on the observations of Pettersen et al. (1984a) with two telescopes at the McDonald Observatory with a time resolution varying from 1 to 10 s (1974–1979). Contrary to the publications describing the above mentioned studies and involving the total durations of flares Pettersen et al. present the durations of the descending branches of flares, from brightness maxima to their half-brightness level. The full time of flare decay is usually several times longer. Thus, to compare the data on AD Leo with previous data, one should shift them to the right by one column at least. Then, it will appear that the distribution of duration of flares on AD Leo is practically the same as for the three stars observed by Moffett.

Finally, the sixth row of the table shows the observational results for 8 flare stars – YZ CMi, HU Del, GQ And, V 577 Mon, EQ Peg, Wolf 424, UV Cet, and CN Leo – obtained in 1982–85 with the 6-meter telescope of the Special Astrophysical Observatory, Caucasus, using the measuring complex MANIA (Beskin et al., 1988). These observations were performed with a time resolution of $3 \cdot 10^{-7}$ s during relatively short time intervals. Over 35 h of monitoring 118 stellar flares were recorded and for 73 of them the light curves with a resolution of 0.1 s were constructed to estimate the duration of flares. Table 14 shows that more than 2/3 of flares of this sample were 1/3 to 3 min long. The small number of long-duration flares is due to the short observational time for each star, and an increase in the fraction of fast events is due to the higher resolution of MANIA as compared to standard photometers and the higher efficiency of recording of weak bursts with the larger telescope.

Having analyzed the first observations of flare stars in clusters and in the solar vicinity, Haro and Chavira (1955) suspected that flares on brighter stars were longer, on average. This conclusion was supported by the photoelectric observation of five stars by Kunkel (1969b), Gershberg and Chugainov (1969), and Chugainov (1974), the number of flares on stars in the solar vicinity was greater. Later, using the duration of flares at the level of a half-maximum brightness as the characteristics of their duration, Kunkel (1974, 1975a) considered about 600 flares he had detected on the three stars and concluded that statistically these values depended on the absolute luminosity of the star

$$\langle \log T_{0.5} \rangle = \text{const} - 0.12 M_V. \tag{28}$$

Then, the relation was confirmed by the observation of about fifty flares on four stars by Busko and Torres (1978). On average, solar flares develop slower than flares on dKe–dMe stars, which qualitatively supports the conclusion. Table 14 also contains an indirect confirmation of the Kunkel conclusion. However, it should be noted that the range of the durations of flares on each star substantially exceeds possible systematic shifts of their average values from one star to another at the expense of different luminosity of the stars. Thus, Kunkel (1973) found that in the sample of 140 flares the duration of the longest flare was 10 times longer than the average duration of flares in the sample.

There are two points of view on the weak dependence of the average duration of flares on the luminosity of a flare star. On the one hand, for higher luminosity stars the density of matter in the atmosphere is systematically lower, which can stipulate slower development of flares. On the other hand, weak flares, as a rule, are shorter, and the contrast with the background of a quiet photosphere results in rather bright and consequently longer flares recorded on higher-luminosity stars. Certainly, one cannot exclude a joint action of the effects, but the influence of the observational selection apparently prevails (Pettersen, 1989a).

To understand the physical sense of these processes, one should know not only the characteristic durations of stellar flares, but also the threshold values of the durations, i.e., the durations of the shortest and longest stellar flares.

Figure 39 shows the light curves of the shortest flares recorded with MANIA at the 6-meter telescope of the Special Astrophysical Observatory (Beskin et al., 1988). It leaves no doubts about the reality of the few second long stellar flares. With high confidence, an optical flare on EV Lac of 2.4 s was recorded on 24 February 1984 at the Astron space station (Gershberg and Petrov, 1986). There were doubts about the stellar origin of the flares of one second and shorter recorded by standard one-channel photometers. Some confidence was gained when monitoring with high-speed two-channel photometers produced in the early 1990s in Kiev and Byurakan was started, which enabled simultaneous recording of stellar brightness in two spectral bands. The flare on EV Lac of 26 August 1990 recorded in Terskol with a time resolution of 0.05 s (Zhilyaev and Verlyuk, 1995) and two flares on EV Lac of 7 August 1994 23:43 and 23:59 UT (see Fig. 40) and the flare on V 577 Mon of 10 January 1997 06:12 UT recorded with a resolution of 0.1 s in Byurakan and Pueblo, respectively (Tovmassian et al., 1997), detected stellar flares with a total duration of not more than 0.3 s. To find ultimately weak stellar flares with HST, Robinson et al. (1995) tried to observe one of the weakest flare stars CN Leo in the range of 2400 Å, where the maximum contrast of the flare and the stellar photosphere was expected. Using the method of statistical photometry, they singled out 32 several-seconds long flares with details to 0.1 s from the two-hour monitoring of stellar brightness. The energy spectrum of the flares (see below) was in agreement with that of the flare spectrum of the star established on the basis of observations in the optical range.

The listed results of the past century were confirmed with the recently built 1.3-meter telescope at the Skinakas Observatory on the island of Crete: during the monitoring of UV Cet 3 hours each on 6–7 October 2008 with a resolution of 4 ms Schmitt et al. (2016) recorded two flares with characteristic times of burning and fast decay of about 2 s.

The data on the duration of the longest stellar flares are rather fragmentary. The longest flare (of the above 227 flares) of 29 August 1990 on EV Lac lasted for more than 1.5 h. On this star, Rojzman and Shevchenko (1982) and Roizman (1983) recorded flares of 4.5 and 5 h, respectively. A very strong X-ray flare on this star was more than 4 h long (Favata, 1998). The onset of the flare on YZ CMi of 19 January 1969 was recorded in Armagh (Northern Ireland), and its end, in Chile, the total duration of the event exceeded 4 h (Kunkel, 1969a). The flare on EQ Peg recorded at ROSAT on 6 August 1985 lasted more than 2.5 h (Poletto, 1989). The flare on BY Dra of 1 October 1990 was recorded during the all-sky survey in the far ultraviolet using the ROSAT WFC. It was recorded during three successive passages of the star in the camera's field of view, so that the full duration of the flare was about 4.5 h (Barstow et al., 1991). On 19 October 1990, ROSAT recorded a flare on AZ Cnc in soft X-rays during six revolutions, i.e., it lasted at least eight hours (Fleming et al., 1993). The flare on the T177 star in the Orion Nebula cluster recorded photographically in Tonantzintla (Haro and Parsamian, 1969) lasted about 20 h, while the flare on T48 star in the same cluster recorded during spectral monitoring lasted for several hours (Carter et al., 1988). The longest known flare on the UV Cet-type stars that lasted more than one and a half days was the flare on AU Mic 15 July 1992 (Cully et al., 1993) (see Fig. 38).

Thus, the durations of stellar flares cover a wide range from fractions of a second to tens of hours, i.e., more than 5 orders of magnitude.

All the above long-duration events had the characteristic light curves with an abrupt brightness increase and long decay. In addition, many-hour substantial increases of brightness

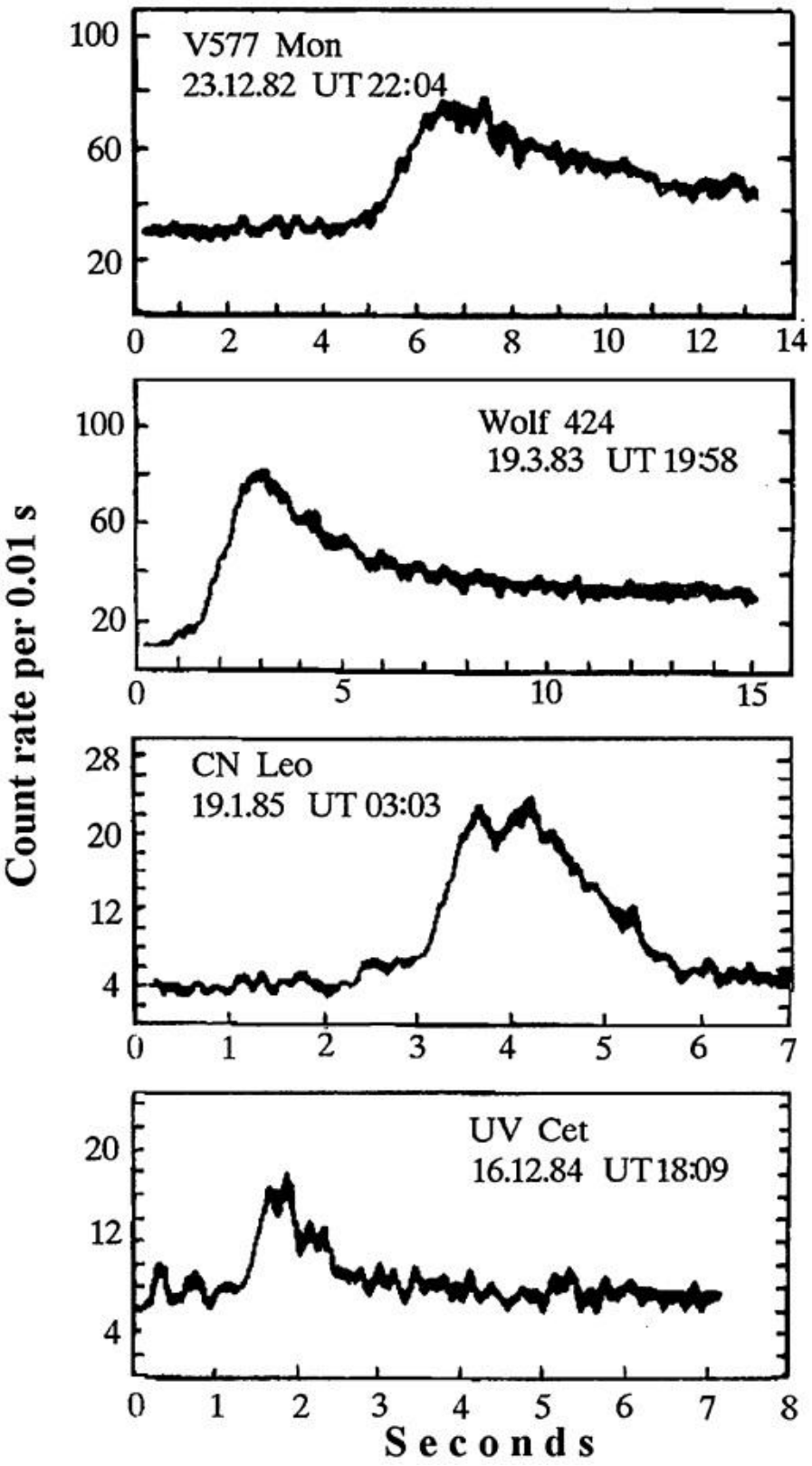

Fig. 39. Fast flares on four UV Cet-type stars recorded with the MANIA complex at the 6-meter telescope of the Special Astrophysical Observatory (Beskin et al., 1988)

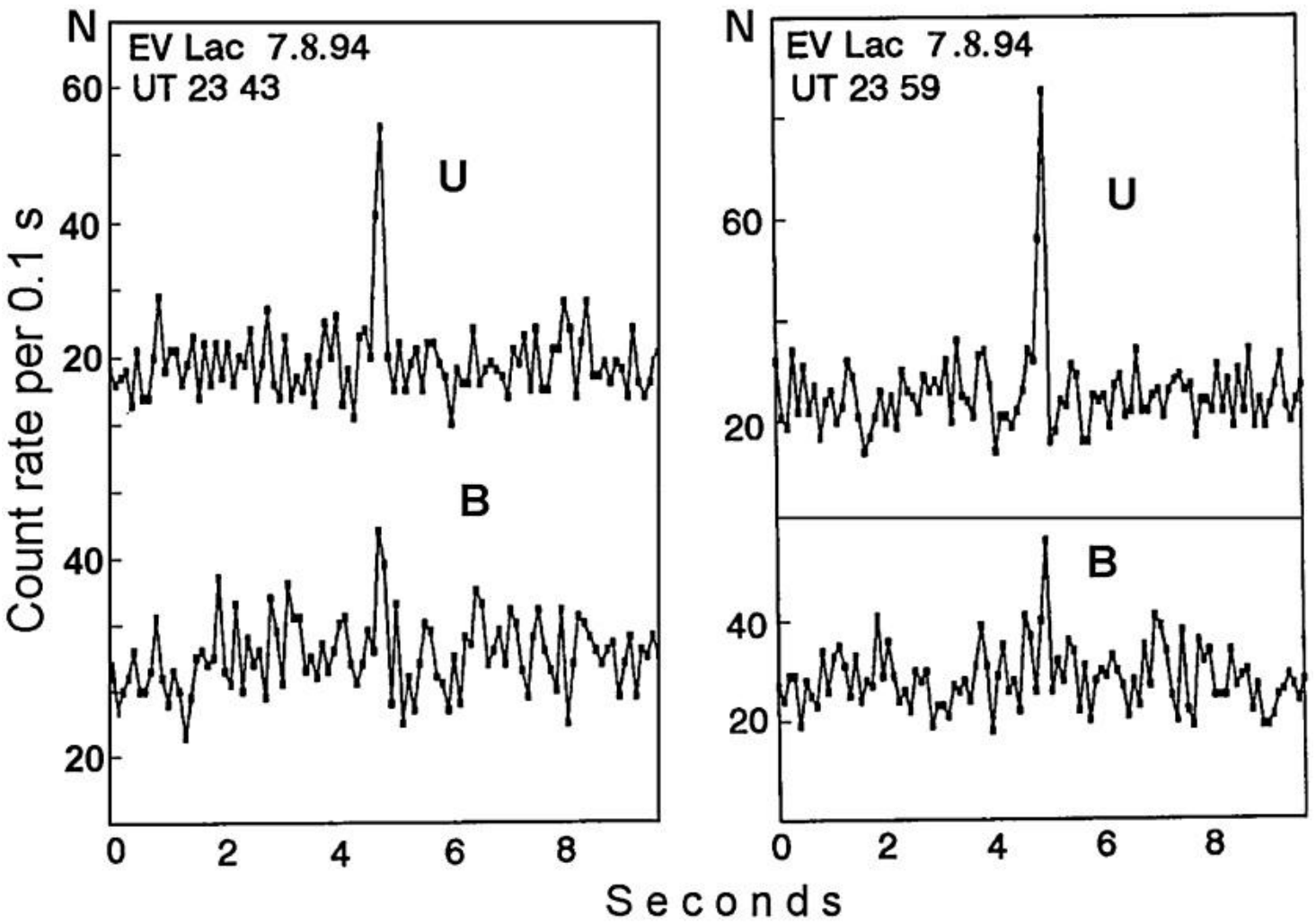

Fig. 40. Two very fast flares on EV Lac recorded with 0.1 s time resolution using a two-channel photometer of the Byurakan Astrophysical Observatory (Tovmassian et al., 1997)

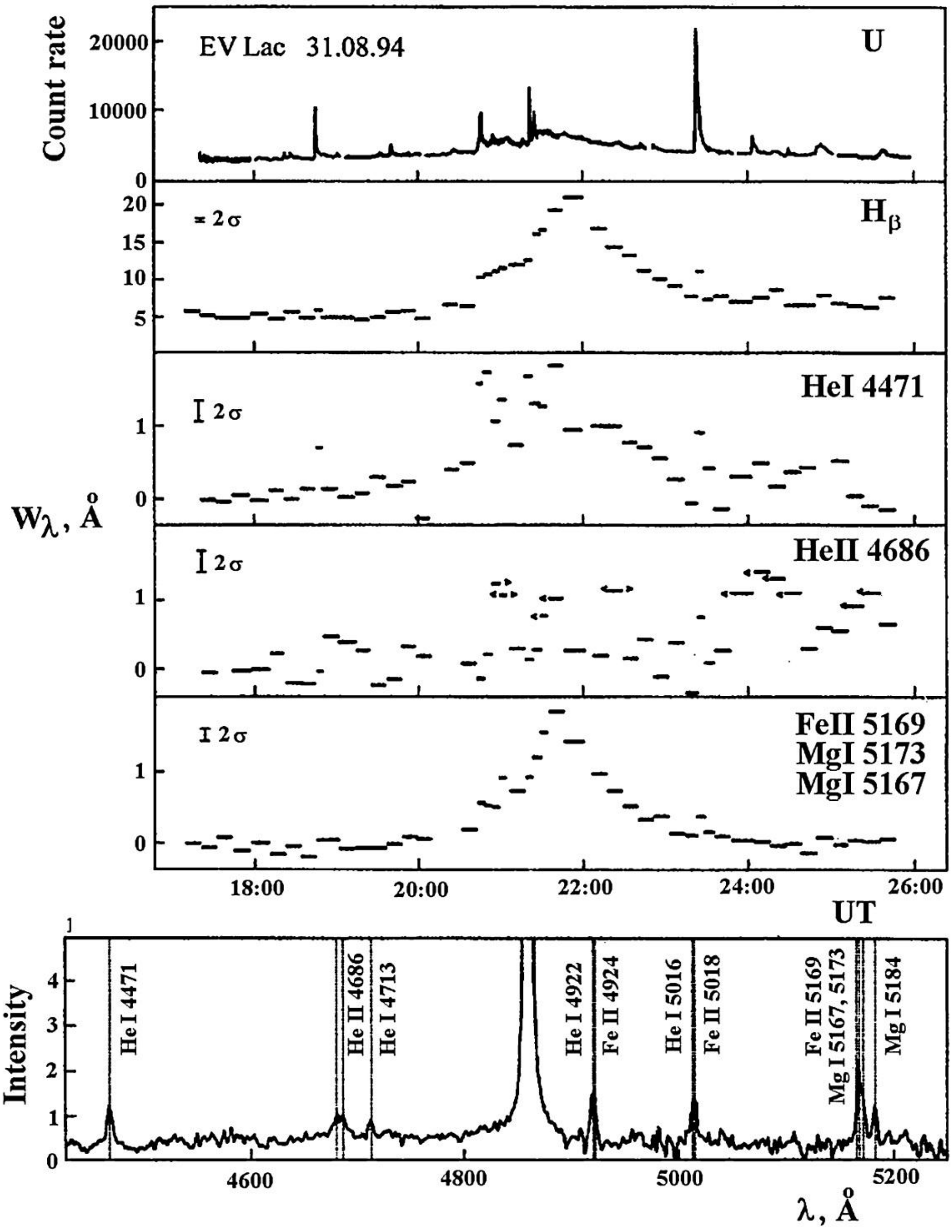

Fig. 41. Results of photometric and spectral monitoring of the flare on the red dwarf EV Lac on the night between 31 August and 1 September 1994 obtained in the Crimea: the light curve in the U band, the equivalent widths of the emission lines $H_\beta$, HeI $\lambda$ 4471 Å, HeII $\lambda$ 4686 Å, the blends of FeII $\lambda$ 5169–5173 Å, and the resulting spectrum of active states of the star (Abranin et al., 1998b)

with almost symmetrical ascending and descending branches of the light curves are sometimes recorded on red dwarfs (see EV Lac light curve in Fig. 41 (Abranin et al., 1998b)). Probably, some slow flares recorded photographically in stellar clusters (see, e.g., Fig. 1 in Parsamian, 1980) are just the same events. Another probability — a many-hour passage of the facular field in the field of view.

Gurzadian (1986) developed a theory of the light curves of such slow flares, assuming that not only direct but also scattered glow of flares occurring behind the stellar limb was observed in these cases. However, Dal and Evren (2011a) specially studied several hundred observed flares and did not confirm the idea of Gurzadian.

Studying 38000 spectra of M dwarfs obtained by West et al. (2008) within the SDSS project, Hilton et al. (2010) detected 63 flares and estimated that for the early M dwarfs flares took 0.02% of time, whereas for the late ones — 3%.

From the statistics of 321 flares recorded in the U band on AD Leo, EV Lac, EQ Peg, and V1054 Oph, Dal and Evren (2010) proposed to separate the stellar flares into fast and slow, depending on the ratio of the decay time to the burning time, taking as a boundary ratio 3.5. Moreover, they found significant differences in the amplitudes and burning times of these flare groups and differences in their duration. Later, they analyzed 222 flares in the B band for the same stars, confirmed their separation into fast and slow and found that the mean equivalent durations of the former exceeded such values for the latter by a factor of 16; the minimum total duration, the maximum burning time, and the total duration of flares increased toward later spectral types (Dal and Evren, 2012).

In another statistical investigation of flares, Dal and Evren (2011b) found that the minimum duration of a flare, the maximum of the total duration, and the maximum of the burning time increased toward later spectral types. Finally, Dal and Evren (2011c) considered their observations of the flare star V 1285 Aql, in the course of which they recorded 83 flares. From this sample they found that the differences in the mean equivalent durations of fast and slow flares were 73 seconds. Dal (2011, 2012) summed the analogous observational data on DO Cep and CR Dra.

### 2.2.2. Time Distribution of Flares

The time distribution of flares is one of the major characteristics of the flare activity. Prior to astrophysical analysis, the consideration of the time distribution makes it possible to distinguish the type of stellar variability, which presumably can determine the physical nature of such activity. Thus, if the distribution is strictly periodical with rather short periods, the activity of red dwarfs could be associated with global phenomena of the type of stellar pulsations. A strict Poisson distribution would evidence the physical independence of individual flares in the observed sequence of such events, the local character of the processes, and apparently the dependence of the flare-triggering mechanism and the course of its development on several independent physical parameters. Intermediate situations can be stipulated by different factors: physical relation of contiguous flares, the modulating effect of stellar rotation on the visibility conditions of a low number of active regions on its surface, and the randomizing effect of the weather on the fulfillment of ground-based observations.

* * *

The first researchers of flares on UV Cet-type stars, who spent tens of hours to record one flare, unanimously marked a random character of their time distribution. But, with the beginning of regular photoelectrical observations of red dwarf stars that increased the frequency of recorded flares by an order of magnitude, some quasiperiodicity of the distribution was called into question (Andrews, 1966a, b). The character of the time distribution of flares became one of the most topical issues. To solve the problem, a working group was set up in 1967 by the Commission on Variable Stars of the International Astronomical Union. Over a number of years, the Group had been organizing the campaigns of round-the-clock observations of such objects. Annually, the Working Group selected four to seven UV Cet-type stars and recommended certain 10–20 days for monitoring. For ease of comparison of the data obtained by various observers, a standard format for publishing the observational results was agreed (Andrews et al., 1969). By the end of 1975, the Working Group arranged 36 cooperative campaigns that involved astronomers from more than 20 observatories from Australia, Great Britain, Hungary, Greece, India, Italy, New Zealand, Norway, Poland, Southern Africa, the USSR, the USA, Yugoslavia, and Japan. The summary results of the cooperative campaigns, the analysis of uniformity and completeness of the obtained data were published by Gershberg (1972a) and then supplemented by Shakhovskaya (1979).

The analysis of the results obtained during the first cooperative observations again led to the conclusion about the existence of quasiperiods in the time distribution of flares (Osawa et al., 1968; Higgins et al., 1968; Chugainov, 1969b; Jarrett and Eksteen, 1972; see also Gershberg, 1969 and Jarrett and Grabner, 1976). However, it was questionable whether all suspected characteristic intervals between flares – from an hour to tens of hours – on all considered flare stars ( YZ CMi, V 1216 Sgr, and UV Cet) was either divisible by a day, or kept an integer number of times within an interval divisible by a day. Apparently, the small number of considered flares combined with the daily variation of the observation schedule led to false quasiperiods. This assumption is rather plausible in the context of Evans's report (1975) on the detection of the period of 29.53 days in flare distribution, which suggests an obvious effect of Moon phases in the observation schedule of flare stars. The subsequent autocorrelation analysis of flare activity of UV Cet, YZ CMi, and Wolf 630 (Kunkel, 1971, 1973; Lukatskaya, 1976) did not reveal any periodicity in the sequence of flares on the stars. However, it should be noted that there are various opinions on the technique and efficiency of such an analysis as applied to the set of discrete brightness bursts (Oskanian and Terebizh, 1971a; Kunkel, 1975a; Lukatskaya, 1976).

Oskanian and Terebizh (1971b, c) were the first to perform statistically strict and full consideration of the time distribution of flares based on cooperative observations: they considered more than 300 flares on UV Cet and about 100 flares on YZ CMi recorded during 1967–70. The authors analyzed the distributions of small identical time intervals along a number of recorded flares within them, the distributions of time intervals since the start of observations till the first flare, between consecutive flares, and in continuous observations, with respect to the number of flares recorded within the intervals. As a result, they concluded that all these distributions in general did not contradict the assumption about the Poisson distribution of the time sequence of flares, though the number of very close flares exceeded their expected number for such a distribution. (This conclusion was confirmed by Haupt and Schlosser (1974) in observations of 94 flares on UV Cet.) Since the excess of close flares was concerned with the events separated from each other by tens of seconds, Oskanian and Terebizh assumed that this excess was caused by flares with several fast bursts near the phase of maximum, i.e., the flares whose light curves were similar to those shown in Fig. 35.

Later, similar statistical studies were carried out using the data of more photometrically uniform observations. Thus, Melikian and Grandpierre (1984) in studying the distribution of small time intervals with respect to the number of flares – about a hundred bursts on UV Cet recorded during 1978–1982 at Maidanak – found a similar excess of the number of close flares over the number expected from the Poisson distribution. Using the result of Moffett (1974), Lacy et al. (1976) analyzed the distributions of intervals between flares during continuous monitoring and concluded that these data did not contradict the Poisson distribution of flares. Pettersen et al. (1984a) came to a similar conclusion upon analyzing 115 flares on AD Leo recorded at the McDonald Observatory: as in the previous case, the criterion $\chi^2$ did not disprove the random time distribution of flares at the 0.05 significance level. This conclusion was valid for the joint consideration of observations of AD Leo from the McDonald Observatory and the Bulgarian National Astronomical Observatory (Pettersen et al., 1986a).

Different results were obtained by Pazzani and Rodonò (1981) who considered the time distribution of 424 flares on UV Cet, 123 flares on EQ Peg, and 80 flares on YZ CMi recorded in Catania, Sicily. They analyzed the distributions of time intervals from the beginning of observations till the first flare, the time intervals between successive flares, the number of flares on small intervals of specified duration and the ratios of recorded and expected (for the Poisson distribution) number of flares on the intervals of continuous monitoring and concluded that the fraction of contiguous flares essentially exceeded the expected value. Thus, a purely random distribution of flares was of small probability. This conclusion was drawn on the basis of independent consideration of flares as a whole and of individual peaks on light curves, therefore the excess of close bursts was definitely not provided by the secondary peaks. The conclusion on the nonrandom distribution of flares on UV Cet and EQ Peg is the more unexpected, since photometric observations record the total brightness of two components, both are flare red dwarfs, which definitely should randomize the observed total time distributions of flares.

It is still not clear why Texas and Sicily observations with identical photometric uniformity and distribution of monitoring over the day yielded different conclusions, whereas Byurakan and Texas observations with differing photometric uniformity and distribution of monitoring yielded identical conclusions. Nevertheless, one can conclude that the observed time distribution of flares on UV Cet-type stars definitely differs from strictly periodic and is random to some extent. The principal cause of deviations of the observed distribution from the Poisson one toward the increased frequencies of contiguous flares is not clear yet. Probably, it is due to weak precursor flares that provoke subsequent stronger flares (Moffett, 1974). Other possible explanations are that the percent of sympathetic flares induced by the general cause at close time moments but at different sites of the stellar surface is very high or the effect of the type of active longitudes leading to nonrandom distribution of flares on the stellar surface is very strong (Pazzani and Rodonò, 1981).

Andrews (1982) considered the season of 1975, when many flares were recorded on EV Lac: 50 flares were detected over 208 h of monitoring at six observatories. He found an alternation of 5–6 day intervals with high and low level of flare activity and attributed this effect of flare grouping to the coexistence of 4 or 5 active regions. Probably, this also explains the longitudinal asymmetry of flares on EV Lac found in 1970 by Leto et al. (1997).

Examining the time variations of flare frequencies from photographic monitoring, Szecsnyi-Nagy (1990) considered 1414 flares on 448 stars recorded in the Pleiades by seven observatories over 3300 observation hours. He selected 17 stars with not less than 10 flares detected on each and found that flares were distributed randomly on five stars, whereas there were time variations of flare frequencies occurred on the other 12: hyperactivity periods were 4–20 times shorter than low-activity periods, but flares during the former were 5–25 times more frequent than during the latter.

The high number of binary systems among flare stars repeatedly raised the question as to the effect of this factor on the level of flare activity. Kunkel (1975b) suspected the modulation of $U_O$ (see below) in the statistical data for flares in the system V 1054 Oph. But Lacy et al. (1978) did not find regular variations of flare level in the system L 726-8 AB, though due to its large orbit eccentricity the distance between components varied by a factor of 4.2. Rodonò (1978) made a similar conclusion in studying the system EQ Peg AB.

Finally, it should be noted that the trains of four flares recorded over 12 min in the system Wolf 424 (Moffett, 1973) and over 2.5 h in the system YY Gem (Doyle et al., 1990c, Gao et al., 2008) hardly match any of the considered statistical models.

* * *

From the observations of 50000 M dwarfs within the SDSS project and 1321 M dwarfs in the 2MASS survey Davenport et al. (2012) found that the strong flares were detected not only in the blue region of the spectrum but in the gri bands of SDSS, while the strongest ones — in the z band as well. Their frequency estimates in the optical range are in agreement with the previously obtained values, but the frequencies of flares in the near IR range are by two orders of magnitude lower than those in the optics. Studying the flares in eight photometric bands — u, g, r, i, z, J, H, and $K_s$, — they found that for events with amplitudes $\Delta J > 0.01^m$ the frequency of flares for M0 dwarfs was 0.003 flares per hour, whereas $\Delta u = 1.4^m$, for M2 dwarfs — 0.001 flares per hour, while $\Delta u = 2.0^m$, and for M6 dwarfs — 0.1 flares per hour, while $\Delta u = 4.5^m$.

Summing up this section, one should note an interesting result obtained by Hunt-Walker et al. (2012): over 5.8 days of observations of AD Leo with MOST they recorded 19 flares, detected the 2.23-day rotational modulation of brightness, and came to the conclusion that the energy of flares was accumulated not at the same place and this supply was not released during each flare.

## 2.3. Flare Energy

The energy of flares is another major quantitative characteristic of the flare activity of stars. Strictly speaking, the most general qualitative definition of a flare as "fast release of a noticeable amount of energy that disturbs the steady state of the star on the whole or a part of it" is directly concerned with this parameter. As the time distribution of flares, the characteristic energy gives a hint on the set of physical processes associated with the flares under consideration. Thus, the total radiative energy of supernovae and the kinetic energy of the shell removed under a burst require a thermonuclear explosion that should involve a considerable part of the stellar mass. The energy of flares of novae is lower by several orders of magnitude, and the flares are caused by thermonuclear explosions that involve only a small part of the stellar mass.

As it was noted in Chapter 2.1, flares on the UV Cet-type stars are accompanied by energy release practically in all ranges of electromagnetic spectrum. By analogy with solar flares, one should expect that strong gas motions should also accompany these stellar flares and generate fluxes of fast particles. Based on the present-day notions, solar flares and relative flares on the lower main-sequence stars are associated with the development of local magnetic structures, which will be discussed at the end of this part.

### 2.3.1. Energy of Optical Flare Emission

All energy parameters of optical emission of flares on the UV Cet-type stars – the absolute luminosity at maximum brightness and during other phases, the total radiative energy, the rate of luminosity increase under burning, etc. – are determined directly by photometric observations. For absolute calibration of observational results, stellar luminosity out of the flare is used, which is determined in absolute units from the known stellar magnitude and distance to the star.

As a result of several thousand hours of photoelectric monitoring from observatories placed in different parts of the world, thus far more than 3000 flares have been recorded on the stars of the considered type. The main contribution to the databank was made by the international cooperative programs of 1967–75 and by intensive observations of red dwarfs beyond the framework of these programs carried out at the observatories in Sicily, the Crimea, Chile, Texas, Southern Africa, Greece, Uzbekistan, Armenia, and Japan. The data on many tens of flares were published by Kunkel (1968, 1973), Cristaldi and Rodonò (1970, 1973), Moffett (1974, 1975), Bruevich et al. (1980), Pettersen et al. (1984a). Rather complete lists of publications about recorded flares can be found in statistical studies by Crimean researchers (Gershberg and Chugainov, 1969; Gershberg, 1972a; Shakhovskaya, 1979; Gershberg and Shakhovskaya, 1983) and in reports of the IAU Commission 27 on Variable Stars (Chugainov, 1979; Gershberg, 1982; Gershberg and Shakhovskaya, 1985, 1988, 1991).

Photographic observations of flare stars in clusters were started in Mexico, continued in Italy, and then run intensively in Armenia, Hungary, Georgia, and Bulgaria. Mainly, they resulted in the identification of these objects and determination of their number and spatial distribution. The most complete summary of the observational results is presented in the monograph by Mirzoyan (1981), the Catalog of Flare Stars in the Pleiades (Haro et al., 1982), and the reviews by Mirzoyan (1986, 1990) and Tsvetkova (2012). According to the statistics of the latter, as a result of the photographic monitoring of stellar clusters and associations of a total duration of 9551 hours, 1519 flare stars were recorded with 3024 flares on them. Careful analysis of these observations allows one to obtain certain parameters of flare energy on flare stars in clusters (Krasnobabtsev and Gershberg, 1975; Kunkel, 1975a; Korotin and Krasnobabtsev, 1985).

Under observations in the standard photoelectric system UBV, the intrinsic flare radiation is rather "blue", while flare stars are the reddest objects. The contrast leads to a fast increase in the amplitude of flares when passing to observations in a band with shorter effective wavelength. The overwhelming majority of flares were recorded in U and B bands and in the appropriate photographic systems. However, this does not mean that U-band observations are always more informative. Because flare stars are red, photometric measurements in violet rays conflict with the quantum noise of the recorded radiation flux earlier than those in the range of longer wavelengths. For weaker objects, the greater amplitude of the useful signal, flare, does not compensate for the increasing noise in the record of the photometric standard, stars out of flares. Therefore, for each telescope equipped with a photometer with a rather low level of intrinsic noise, there is such a critical visual stellar magnitude determined by the telescope size, radiation colors of stars and flares, and viewing conditions that photoelectric observations aimed at recording of flares on stars brighter than this critical magnitude should be carried out in the U band, whereas on weaker stars, the B band should be used.

Started in the current century, photometric observations of flare stars with spacecrafts by the panorama photometry method led to an abrupt growth of the number of known objects of this type and made weaker and more distant stars to be available for studying (see Subsect. 2.4.4.1).

Considerable diversity of flare light curves complicates comparison of the energy of various events of this kind. The problem of quantitative determination of some average flare activity required to compare the activity of a star at different periods or the activity of various flare stars is even more ambiguous. Originally, such values as average frequencies, average amplitudes, average flare energies and total flare energy attributed to a full duration of monitoring were used for this purpose (Gershberg and Chugainov, 1969; Gershberg, 1969; Gurzadian, 1971a; Sinvhal and Sanwal, 1977). However, these average values are heavily influenced by observational selection.

On the basis of consideration of photoelectric observations of about 90 light curves of flares on the UV Cet-type stars recorded photoelectrically before the beginning of cooperative observations, Gershberg and Chugainov (1969) estimated absolute flare energy, burning and decay rates, and the duration of flares. They made an attempt to find a correlation among these values and between them and absolute luminosities of flare stars. This paper considerably affected the ideology of subsequent cooperative campaigns.

The first strict statistical research of the energy of flare activity of red dwarfs on the basis of cooperative observations of four of the most active stars – YZ CMi, AD Leo, EV Lac, and UV Cet – was carried out by Oskanian and Terebizh (1971a). They estimated the luminosity range and the total energy of optical flares, distributions of their amplitudes and energy, and constructed the frequency functions of these parameters. They noticed that the functions of the first three brighter stars differed essentially from those of the absolutely weaker UV Cet, on which even weaker flares could be observed. They also estimated the possible contribution to total flare emission by the events that were beyond the detection threshold.

The effects of the observational selection can be revealed, if, instead of average flare characteristics, the distributions of these characteristics are considered. Reasoning from this, Kunkel (1968) introduced into examination the distributions of flares over absolute stellar magnitudes of flare radiation at maximum brightness. Chugainov (1972b) proposed to consider distributions of flares over equivalent durations

$$P = \int [(I_{\text{flare}} - I_0) / I_0]\, dt, \quad (29)$$

where $I_{\text{flare}}$ is the radiation flux from a star recorded during a flare and $I_0$ is the flux recorded from a quiet star. If, in the distribution of flares with respect to equivalent durations $n(P)$, we proceed from the flare number to their average frequency $n = n/T$, where $T$ is the time of monitoring, and from equivalent duration to total flare energy $E = PL$ , where $L$ is the stellar luminosity out of the flare in the appropriate photometric band, we obtain the observed energy spectrum of flares $n(E)$ (Gershberg, 1972a). Analysis of these two distributions determines the modern concepts of statistical properties of the energy of flare activity. It should be noted that considering 386 flares detected by Moffett (1974) on eight red dwarfs, Lacy et al. (1976) found a practically linear correlation between average flare energies in the U band and stellar luminosity in this band over five orders of magnitude

$$\langle \log E_U \rangle = \log L_U^* + 2.0 \pm 0.7. \quad (30)$$

They obtained similar relations for the total energy of flares. Later, these relations became widely used, but one should be careful, because in extrapolations to the Sun they yield an error of several orders of magnitude.

From the extensive photoelectric observations of five flare stars with MOST, Dal and Evren (2011b) came to the conclusion that for each star the flare energy could not be higher than some level irrespective of its duration, and this level was a saturation level in the U band; the time maximum of flare burning and their total duration decreased toward cooler stars.

**2.3.1.1. Spectrum of Maximum Flare Brightness.** It is obvious that the flare distribution over the absolute stellar magnitude at maximum brightness in physical terms corresponds to the spectrum of maximum flare radiation. Figure 42, reproduced from the paper by Kunkel (1968), shows the spectrum of the maximum brightness of flares on UV Cet recorded over 78 h of monitoring in 1967 at Cerro Tololo (Chile). The logarithm of the cumulative frequencies $\tilde{\nu}(U < U_0)$, i.e., the logarithm of frequencies of flares, whose brightness in the U band at maximum brightness exceeded the brightness corresponding to the value $U_0$, is plotted along the $x$-axis. Experimental points fit well the straight line corresponding to the equation

$$\tilde{\nu}(U - U_0) = \exp [1.04\, (U - 13.58)]\ h^{-1}. \quad (31)$$

Hence, it follows that flares on UV Cet occur on average every hour, at maximum brightness of the visual magnitude in the U band is at least $13.58^m$.

Kunkel (1973) proved that the relation

$$\tilde{\nu}(m) = \exp [a(m - m_0)] \quad (32)$$

represented all the spectra of maximum flare brightness that he had constructed. From observations of 12 flare stars he calculated $a$ and $m_0$ and found that all the calculated values of $a$ were within a rather narrow range 0.76–1.24. The average value of $a$ appeared to be very close to unity, and differences of individual values of this parameter from unity, as a rule, were not greater than the determination error of these values (Kunkel, 1975a). Therefore in considering stars with a few detected flares one could postulate $a = 1.0$ (or 0.9) and determine the only free parameter $m_0$ from observations. The results of the analysis of flare activity of 27 flare stars are summarized in Table 15 reproduced from Kunkel's review (1975a).

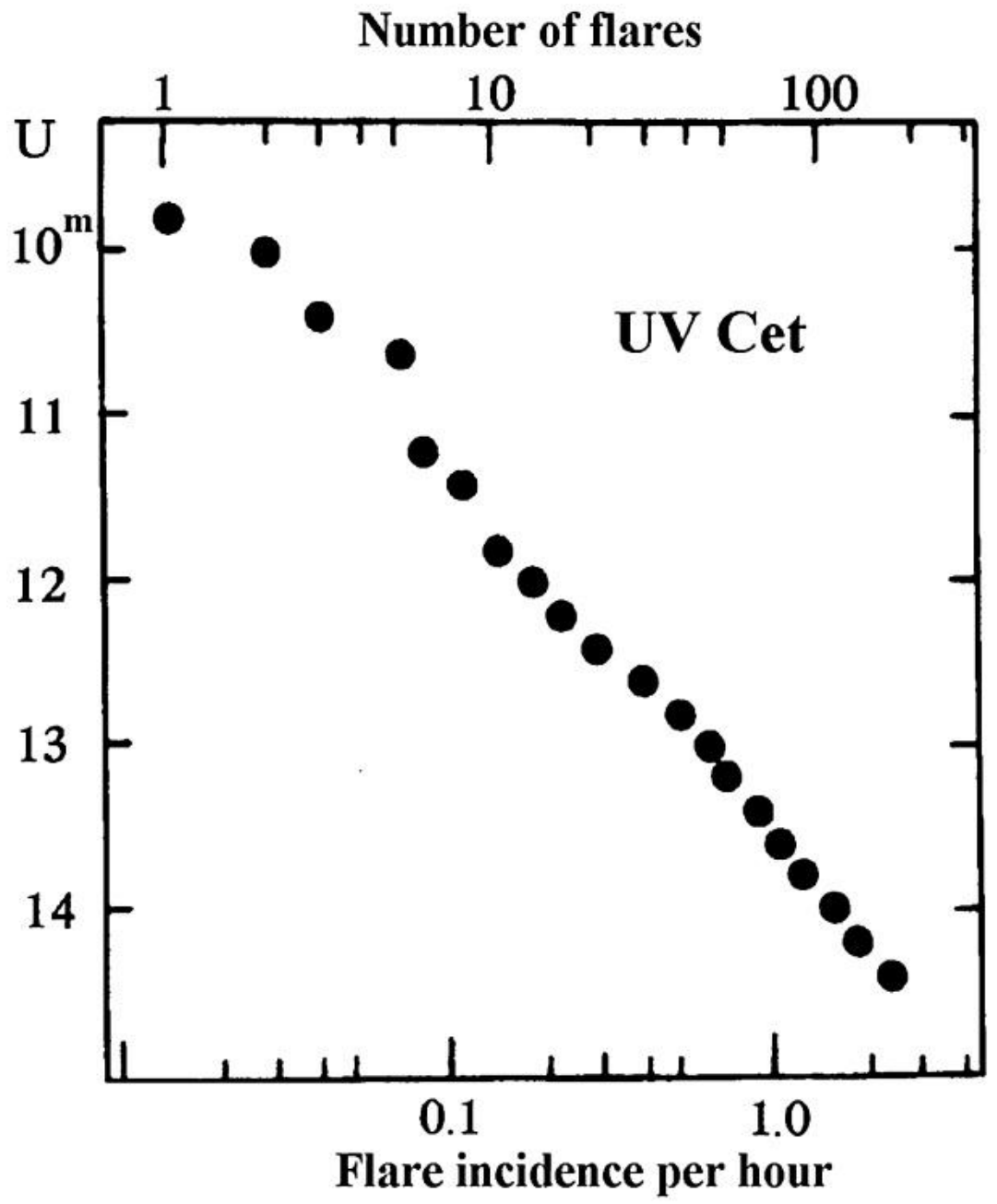

Fig. 42. Distribution of flares on UV Cet over maximum brightness (Kunkel, 1968)

The parameter $m_0$ evidently represents the observed stellar flare activity and allows one to predict the expected number of flares, which should be recorded over the set duration of observations with the equipment that is able to detect flares with certain minimum amplitude $\Delta m_{\min}$. It is obvious that in (32) the difference of apparent stellar magnitude can be replaced by the difference of absolute stellar magnitudes, but the absolute magnitude $M_0$ corresponding to $m_0$ characterizes the physical properties of flare activity rather than the observed pattern. Figure 43, based on Kunkel's data (1975a), shows the flare luminosity at maximum brightness obtained by transition from the absolute stellar magnitudes $M_0$ to luminosity energy units $L_0$. In addition to flare stars in the solar vicinity listed in Table 15, there are two signs $\Pi$ corresponding to flare stars in the Pleiades. These data were obtained in analyzing the observations of flare stars in this cluster: there are two groups of flare stars of similar brightness on which flares with a duration close to the characteristic time of photographic exposures were recorded. Then, the flare frequencies on these stars were estimated on the assumption that the samples of flare stars satisfied the ergodicity principle: a set of flares recorded over a time $T$ on $n$ stars was close enough to the set of flares recorded on each of them over the time $nT$. As it follows from the figure, such an estimate of $L_0$ for flare stars in the Pleiades is in good agreement with individual values determined for the brightest stars in the solar vicinity.

Using the points in Fig. 43 one can plot the envelope from above with confidence. Obviously, it shows the maximum of flaring activity of flare stars, if for the measure of this activity we accept the luminosity of flares occurring with a frequency of $\tilde{\nu} = 1\ \mathrm{h}^{-1}$. Most likely, there are flare stars that fill in the plane of the figure below the envelope; however, observers prefer to examine the most active flare stars, thus the data on low-activity flare stars are few in number and to a great extent accidental.

Table 15. Parameters of spectra of maximum flare brightness (Kunkel, 1975a)

| Designation of stars | Observational epoch | Photometric region | Number of recorded flares | *a* | $m_0$ |
|---|---|---|---|---|---|
| 1 | 2 | 3 | 4 | 5 | 6 |
| Gl 15 B = GQ And | 1969.3 | B | 15 | 0.83 | $16.2^m$ |
| Gl 54.1 = YZ Cet | 1969.7 | U | 7 | 0.9* | 17.5 |
| Gl 65 = UV Cet | 1966–71 | U | 802 | 1.06 | 14.0 |
| Gl 166 | 1968.0 | U | 38 | 0.98 | 15.0 |
| Gl 206 | 1967.9 | U | 3 | 1.0* | 15.3 |
| Gl 229 | 1970.0 | U | 2 | 1.0* | 17.2 |
| Gl 234 = V577 Mon | 1969.1 | U | 35 | 1.24 | 14.8 |
| Gl 278 C = YY Gem | 1971.2 | U | 10 | 1.0* | 14.3 |
| Gl 285 = YZ CMi | 1969–70 | U | 85 | 1.10 | 15.0 |
| Gl 388 = AD Leo | 1965.3 | U | 27 | 0.91 | 14.5 |
| Gl 406 = CN Leo | 1969.1 | U | 38 | 1.19 | 15.1 |
| Gl 473 | 1972.2 | U | 10 | 1.0* | 14.7 |
| Gl 493.1 | 1969.1 | U | 3 | 1.0* | 16.1 |
| Gl 494 = DT Vir | 1970.3 | B | 2 | 1.0* | 16.0 |
| Gl 540.2 | 1969.4 | U | 3 | 1.0* | 17.6 |
| Gl 551 = V645 Cen | 1969.2 | U | 28 | 0.83 | 14.8 |
| Gl 616.2 | 1970.4 | B | 5 | 1.0* | 16.7 |
| Gl 644 = V1054 Oph | 1968–70 | U | 125 | 1.14 | 14.5 |
| Gl 719 = BY Dra | 1970.5 | U | 9 | 1.0* | 15.1 |
| Gl 729 = V1216 Sgr | 1970.5 | B | 10 | 1.0* | 16.6 |
| Gl 735 = V1285 Aql | 1970.6 | B | 5 | 1.0* | 16.8 |
| Gl 799 = AT Mic | 1967–70 | U | 80 | 0.76 | 14.5 |
| Gl 803 = AU Mic | 1970.6 | U | 31 | 1.03 | 13.8 |
| Gl 815 | 1969 | B | 7 | 1.0* | 16.4 |
| Gl 860 B = DO Cep | 1969.7 | U | 10 | 1.0* | 17.2 |
| Gl 866 | 1971.7 | U | 7 | 1.0* | 15.9 |
| Gl 873 = EV Lac | 1970.7 | U | 67 | 1.00 | 14.9 |

* Postulated values.

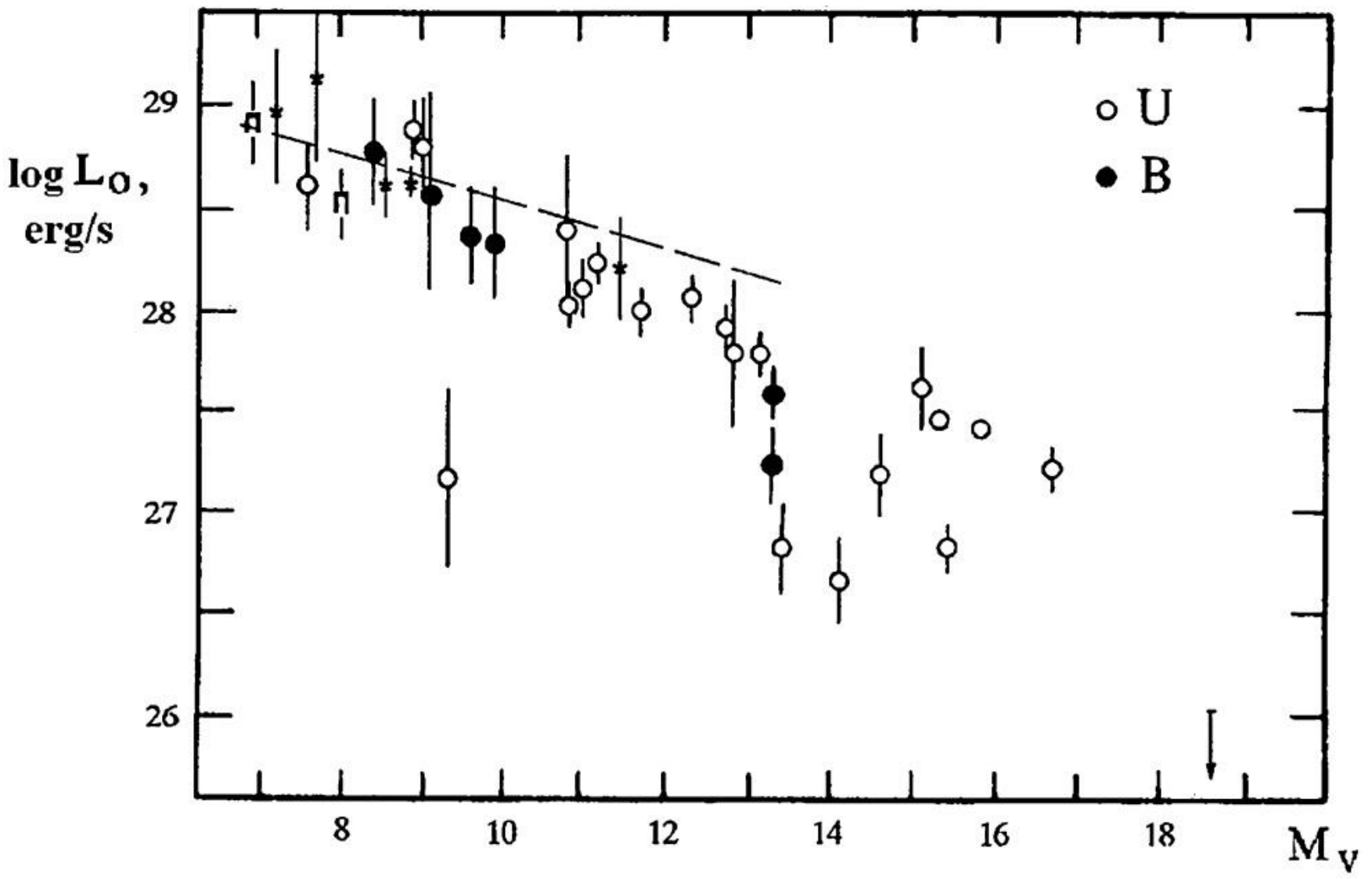

Fig. 43. Absolute luminosities at maximum brightness of flares with a frequency $\tilde{\nu} = 1$ h$^{-1}$ on flare stars of different luminosity (Kunkel, 1975a); asterisks mark the data by Busko and Torres (1978); the dashed line shows 1% of bolometric luminosity

The main conclusion from Fig. 43 is that on the most active UV Cet-type stars the maximum absolute luminosity of flares occurring with an average frequency of $\tilde{\nu} = 1$ h$^{-1}$ in the U and B bands reaches $10^{29}$ erg/s for K stars of middle subtypes and systematically decreases for late M stars. On the other hand, according to the calculations by Kurochka and Rossada (1981), when the Sun approached the epoch of maximum activity in 1980, on average each hour there occurred a flare, whose maximum luminosity in all lines and continua of the hydrogen spectrum amounted to a few $10^{25}$ erg/s, in the B band the radiation of such a flare should be at least an order of magnitude lower. Thus, stellar flare activity similar to the analogous solar activity covers the range of at least 4–5 orders of magnitude of $L_0$.

**2.3.1.2. Energy Spectrum of Flares.** Analysis of the dependence of the average frequency of flares on their total energy $\nu(E)$, i.e., the energy spectrum of flares, is today the most widespread method of statistical consideration of the energy of stellar flares (Gershberg, 1972a, 1985; Krasnobabtsev and Gershberg, 1975; Lacy et al., 1976; Shakhovskaya, 1979; Walker, 1981; Byrne, 1983; Gershberg and Shakhovskaya, 1983; Pettersen et al., 1984a; Korotin and Krasnobabtsev, 1985; Mavridis and Avgoloupis, 1987; Beskin et al., 1988; Ishida et al., 1991; Hilton, 2011).

Figure 44 shows the energy spectra of the flare stars YZ CMi and BY Dra constructed in the double-logarithmic scales from the B band observations (Gershberg and Shakhovskaya, 1983). To reduce the influence of random spread, cumulative frequencies $\tilde{\nu}(E)$ are used instead of average frequencies of flares with the energy $E$, values of $\nu(E)$

$$\tilde{\nu}(E) = \int_E^{E_{\max}} \nu(E)\,\mathrm{d}E\,. \tag{33}$$

The figure proves that for rather strong flares the relation between log$E$ and log$\tilde{\nu}$ is close to linear, and weak flares demonstrate a sharp slope.

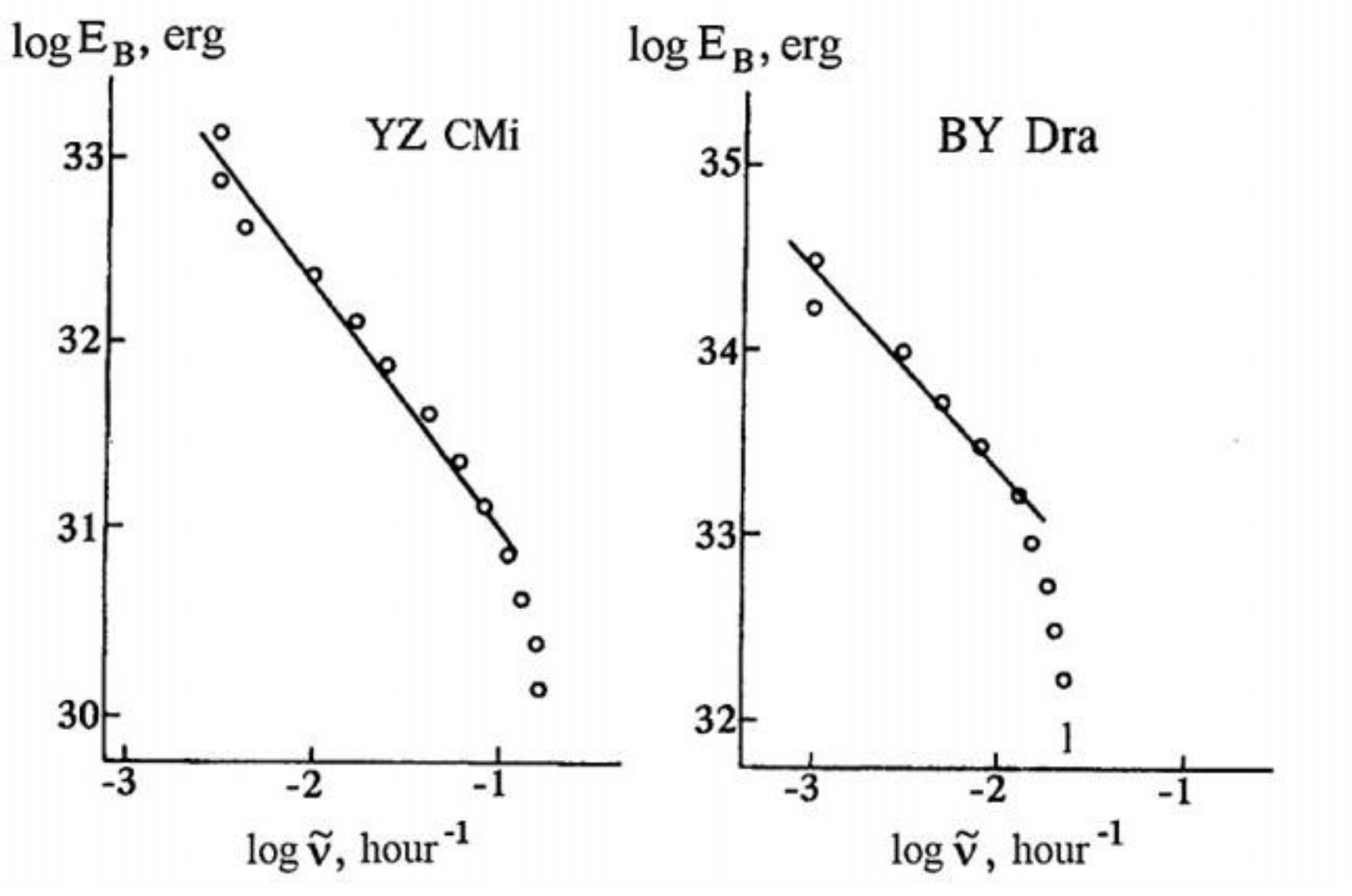

Fig. 44. Distribution of flares on YZ CMi and BY Dra over the total radiative energy in the B band, the integral energy spectra of flares (Gershberg and Shakhovskaya, 1983)

Contrary to the flare stars in the solar vicinity, where over several tens of hours of photoelectric monitoring a number of flares can be recorded, which is sufficient for the construction of the energy spectrum of flares, the frequency of photographically recorded flares is lower by 1–3 orders of magnitude. This fact prevents construction of individual energy spectra of flares on flare stars in clusters. But, as was mentioned, it seems plausible that flare stars of practically identical age and initial chemical composition, identical brightness and, hence, identical mass in clusters should have similar flare activity, that is, an identical energy spectrum of flares. In this case, flares recorded over the time $T$ in any cluster of stars of similar brightness can be considered as the flares that occurred on the same flare star over the time $Tn$, where $n$ is the total number of flare stars of similar brightness in this cluster. The energy spectra of flares constructed within the framework of this ergodicity hypothesis are called group spectra.

Figure 45, reproduced from the paper of Korotin and Krasnobabtsev (1985), presents the group energy spectra of bursts on flare stars in the Pleiades. Each group spectrum is constructed for flare stars in the range of apparent brightness $\Delta m = 2^m$, flare stars in the neighboring groups are overlapped by half, and all together they cover all flare stars in the Pleiades. Group spectra are given as dependences $\tilde{N}$ ($L_{max}$), where $\tilde{N}$ is the cumulative number of flares and $L_{max}$ is the absolute flare luminosity at the maximum. Since in photographic observations of flare stars in clusters the exposures $\Delta t$ take many minutes, they are comparable with the duration of flares and are identical for all flares on all stars of a cluster, the values of $L_{max}\Delta t$ differ from $E$ only by a multiplier of about unity, which can be estimated from observations (Krasnobabtsev and Gershberg, 1975). Hence, the values plotted in Figs. 44 and 45 differ only in constant shifts. Thus, the two figures can be considered together and one can suggest that the linear relation of log$E$ and log $\tilde{\nu}$ for strong flares and the abrupt break of the energy spectrum in the range of weak flares are characteristic of all flare stars. Let us discuss

the thus-found general structure of the observed energy spectra of flares and consider some formal consequences of the dependence $\tilde{\nu}$ (*E*) and physical meaning of the established power character of this dependence.

A. The structure of observed energy spectra of flares. First, one should make sure that the break of the spectrum in the region of weak flares is the result of observational selection. Indeed, the criterion of reality of a flare is usually a certain excess of the amplitude of burst over the width of the noise path. At low *E* the fastest flares satisfy this criterion, whereas smoother flares with equal total energy *E* are attributed to noise. The closeness of the break of energy-spectrum to the detection threshold for flares was established in observations (Chugainov, 1972b). (We note that trying to connect the frequency of the energy-spectrum break with intrinsic properties of flare activity Rosner and Vaiana (1978) erroneously took published cumulative frequencies for average frequencies.)

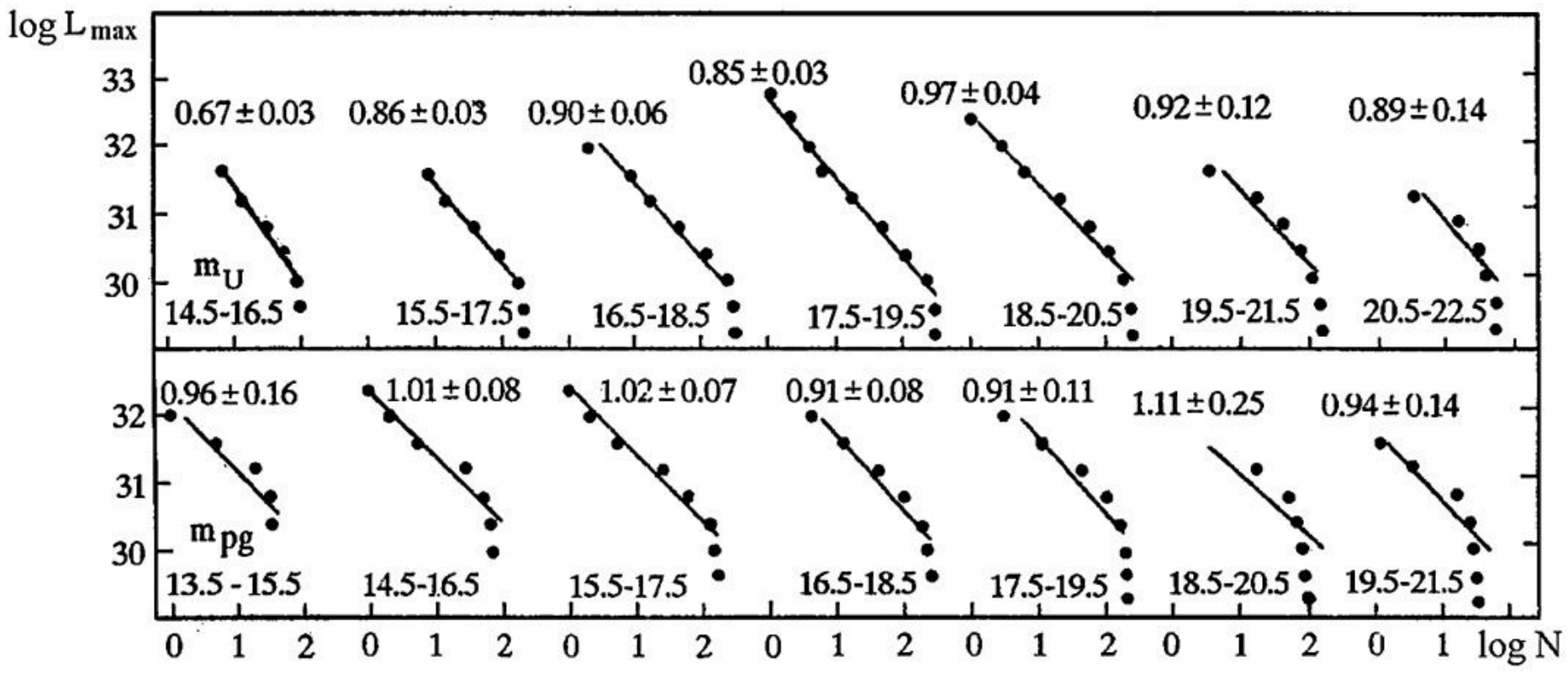

Fig. 45. The group energy spectra of flares on the stars of different luminosity in the Pleiades (Korotin and Krasnobabtsev, 1985). Above each spectrum the value of the spectral index $\beta$ is specified, below is the range of brightness of the stars included in this group spectrum

It is obvious that $\tilde{\nu}$ of the last point ahead of the break in the linear part of the spectrum in Figs. 44 and 45 corresponds to the mean frequency of flares that are recorded independently of the detection threshold. In other words, only above the break does the energy spectrum of flares contain physically significant information on the properties of stellar flare activity. On the other hand, $\nu_{break}$ should depend on the absolute luminosity, since on brighter stars only absolutely stronger flares can be noticed against the background of permanent radiation of the photosphere. The flare/star contrast depends on the used effective wavelength of photometric band, and on the recording technique, sensitivity, time resolution, intrinsic noise and other properties of the equipment. However, according to the calculations by Shakhovskaya (1979), when using a sample of uniform data the break of the observed energy spectra of flares, as a rule, is rather abrupt, thus the break frequency and the appropriate energy of flares are determined with sufficient confidence.

In the region of strong flares the basic uncertainty of the considered energy spectra of flares is due to the limited scope of the sample: the points corresponding to the strongest flares, are statistically least reliable, since by definition they are based on the minimum number of flares. However, as for flares of any physical nature, there should be a constraint on the maximum energy of an individual flare, in the region of high energy in the axes in Figs. 44 and 45 the integral energy spectrum of flares $\tilde{\nu}(E)$ should necessarily turn into a horizontal line. If observations do not definitely show such a transition, maximum-energy flares have not been recorded as yet on the star under consideration.

Results of the most numerous constructs of integral energy spectra of flares on the UV Cet-type stars from photoelectric monitoring are presented in the publication by Gershberg and Shakhovskaya (1983). Table 16, reproduced from the paper, shows the values of initial data: times of monitoring and numbers of recorded flares in the U and B bands separately — and the coefficients of power presentation of the energy spectrum of flares

$$\log\tilde{\nu} = \alpha - \beta \log E, \tag{34}$$

obtained for each of the considered flare stars above the break of an energy spectrum. Korotin and Krasnobabtsev (1985) published the most complete results on the group energy spectra of flares on flare stars in clusters. In both studies, the observations of flares in blue and violet rays were analyzed separately, but the energy spectra obtained were close enough. Figure 46 combines the results of the studies of stellar flares in blue rays: the individual spectra of flares on flare stars in the solar vicinity and group spectra of flares on flare stars in the Pleiades and Orion. The plot shows only the significant parts of the energy spectra located above the breaks of spectra.

It should be noted that to construct the group energy spectrum providing a value of the spectral index $\beta$, it suffices to detect twenty to thirty flares in the considered group of flare stars. But to localize the spectrum in Fig. 46, to transfer from the cumulative number of recorded flares on all stars of considered brightness to the number of flares on a star, one should know the number $n$ of flare stars of similar brightness in a stellar cluster. This is determined following the Ambartsumian method (1969) assuming that the sequence of flares on a number of similar flare stars can be presented by the Poisson law. If this hypothesis is valid and the average frequency of flares on all flare stars is the same, the number of stars on which $k$ flares occurred over the time $t$ is determined by the relation

$$n_k = n(\nu t)^k \frac{\exp(-\nu t)}{k!}. \tag{35}$$

Hence, it follows that the number of flare stars on which over this time no flares occurred can be calculated using the formula

$$n_0 = n_1^2/2n_2. \tag{36}$$

In other words, from the number of stars on which one or more flares occurred one can determine the full number of flare stars in a cluster as $n = \sum n_k$. However, to get not too coarse determination, the denominator of (36) should be rather reliable, say, more than 5. Not all considered groups of flare stars satisfy this condition; therefore the number of such spectra in Fig. 46 is rather small.

Figure 46 shows that for flare stars in general the power dependence of the integral spectrum of flares $\tilde{\nu}(E)$ can be traced in the energy range of about seven orders of magnitude, but on the spectra of flares on individual stars and groups of stars of identical luminosity the dependence can be traced only in the energy range not exceeding 2 orders of magnitude, because of the effect of observational selection in the region of weak flares and short observational time for recording the strongest flares. It is obvious that the upper limit of the band occupied in Fig. 46 by the energy spectra of stellar flares corresponds to the peak efficiency of optical emission of flares of the considered type. The lower limit of the band, as in Fig. 43, is due to the inclination of observers to study the most active stars.

The energy spectra of flares on three flare stars shown in Fig. 46 demonstrate appreciable deviations from the power dependence $\tilde{\nu}$ ($E$). The curved lower part of flare spectrum (shown by the dot-and-dash line) in the binary system EQ Peg is caused apparently by the heterogeneity of the used observational data: as noticed first by Byrne and McFarland (1980), the integral energy spectrum of flares constructed on the basis of independent series with different $\nu_{break}$ can yield more than one break in the integral spectrum. In this case it is difficult to separate the physically significant part of the observed spectrum of flares. Curved upper parts of the energy spectra of flares on EQ Peg, UV Cet, and AD Leo suggest that we are approaching the strongest flares on these stars. Any more resolute statements are premature, because some similar curvatures of energy spectra of flares considered earlier as indications of $E_{max}$ "straightened" later for increasing number of recorded flares. Another distorting effect of the energy spectra of flares was suspected by Doyle and Mathioudakis (1990) in examining the YY Gem system. They found that the energy spectra of flares during the eclipse phase and out of it differed. They explained this by stronger flares between the system components when their interaction was intense.

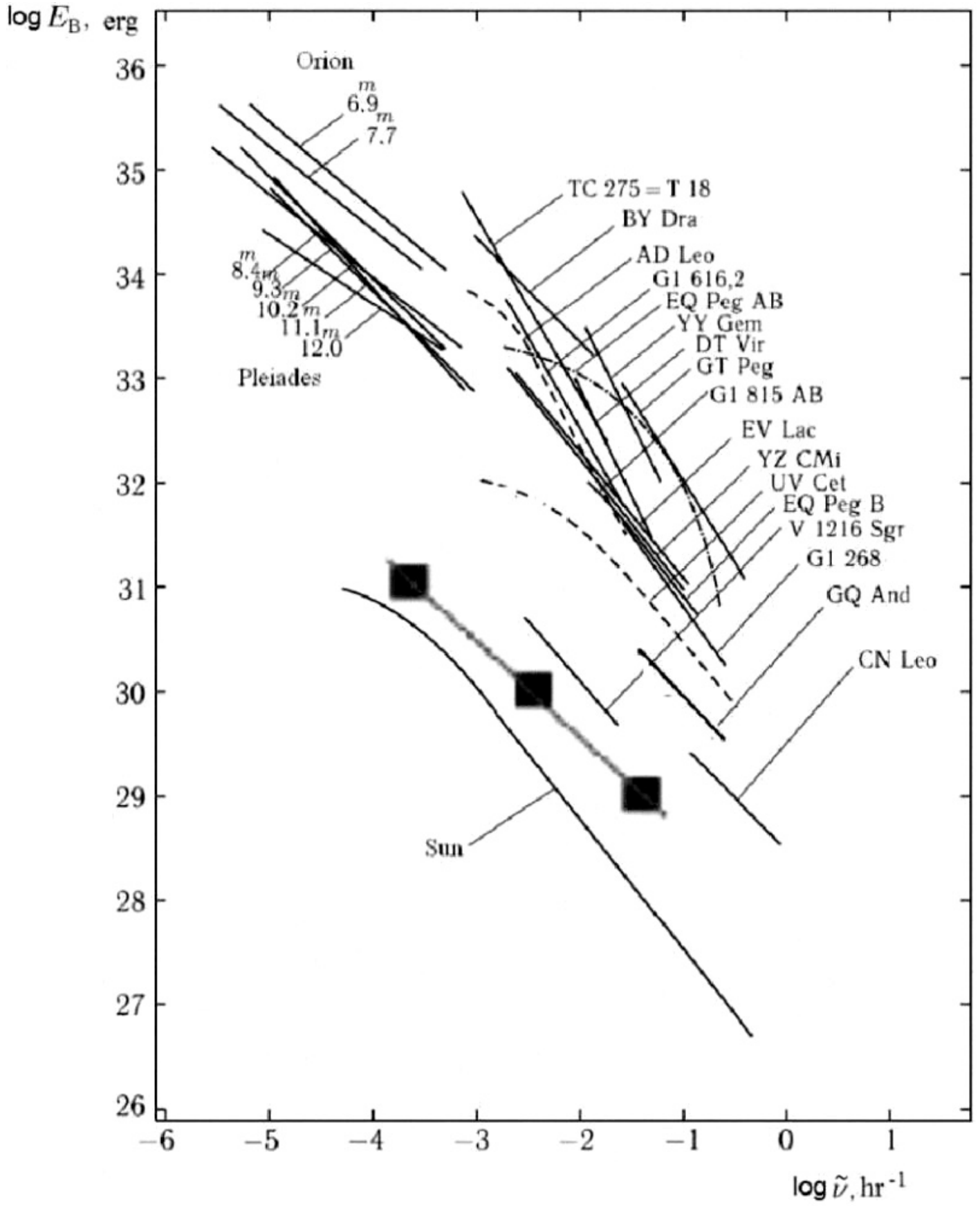

Fig. 46. Energy spectra of flares on stars in the solar vicinity, in the Pleiades and Orion clusters, and solar flares — physically notable parts of integral spectra above the breaks due to the observational selection (Gershberg et al., 1987) and complemented by Katsova et al. (2022) with squares and the thick line that are plotted according to the solar X-ray data

Table 16. Parameters of integral energy spectra of flares (Gershberg and Shakhovskaya, 1983)

| Designation of stars | $M_V$ | B | | | | U | | | |
|---|---|---|---|---|---|---|---|---|---|
| | | $T$, h | $N$ | α | β | $T$, h | $N$ | α | β |
| Gl 15B = GQ And | $13.29^m$ | 46 | 16 | 28.38 ± 0.03 | 0.98 ± 0.11 | | | | |
| Gl 65(A + B) = UV Cet | 15.35 + 15.89 | 849 | 320 | 24.64 ± 0.07 | 0.84 ± 0.06 | 95 | 205 | 26.04 ± 0.05 | 0.89 ± 0.06 |
| Gl 166C = 40 Eri C | 12.73 | | | | | 16 | 30 | 26.73 ± 0.18 | 0.90 ± 0.14 |
| Gl 234AB = V577 Mon | 13.08 | | | | | 8.8 | 28 | 19.96 ± 0.10 | 0.67 ± 0.12 |
| Gl 268 = Ross 986 | 12.62 | 37 | 10 | 20.15 ± 0.04 | 0.68 ± 0.08 | | | | |
| Gl 278C = YY Gem | 8.26 | 92 | 6 | 14.54 ± 0.07 | 0.49 ± 0.05 | 89 | 6 | 8.98 ± 0.03 | 0.32 ± 0.04 |
| Gl 285 = YZ CMi | 12.29 | 999 | 170 | 22.61 ± 0.03 | 0.76 ± 0.04 | 117 | 89 | 26.65 ± 0.05 | 0.89 ± 0.09 |
| Gl 388 = AD Leo | 10.98 | 1010 | 54 | 13.55 ± 0.08 | 0.48 ± 0.06 | 198 | 53 | 16.67 ± 0.05 | 0.57 ± 0.05 |
| Gl 406 = CN Leo | 16.68 | 21 | 59 | 31.57 ± 0.05 | 1.10 ± 0.09 | 42 | 70 | 41.12 ± 0.08 | 1.43 ± 0.11 |
| Gl 473(A + B) = Wolf 424(A + B) | 14.98 +15.2 | | | | | 2.6 | 11 | 27.43 ± 0.09 | 0.93 ± 0.16 |
| Gl 494 = DT Vir | 9.4 | 77 | 6 | 13.50 ± 0.05 | 0.47 ± 0.09 | | | | |
| Gl 616.2 = BD + 55°1823 | 8.9 | 411 | 20 | 18.83 ± 0.03 | 0.64 ± 0.04 | | | | |
| Gl 644(A + B) = V1054 Oph | 10.79 + 10.80 | | | | | 11.4 | 11 | 14.20 ± 0.16 | 0.49 ± 0.19 |
| Gl 719 = BY Dra | 7.9 | 1064 | 35 | 27.58 ± 0.07 | 0.89 ± 0.08 | 234 | 19 | 20.38 ± 0.03 | 0.68 ± 0.04 |
| Gl 729 = V1216 Sgr | 13.3 | 227 | 16 | 25.24 ± 0.04 | 0.90 ± 0.06 | | | | |
| Gl 799(A + B) = AT Mic | 11.09 +11.2 | | | | | 11.5 | 22 | 20.44 ± 0.08 | 0.66 ± 0.09 |
| Gl 815(A + B) | 9.8 + 11.8 | 69 | 8 | 27.21 ± 0.05 | 0.91 ± 0.07 | | | | |
| Gl 860B = DO Cep | 13.3 | | | | | 54 | 22 | 20.74 ± 0.08 | 0.72 ± 0.09 |
| Gl 867B = L 717−22 | 11.8 | | | | | 6.6 | 21 | 42.26 | 1.4 ± 0.4 |
| Gl 873 = EV Lac | 11.65 | 1348 | 126 | 29.48 ± 0.05 | 0.97 ± 0.05 | 380 | 117 | 26.76 ± 0.05 | 0.89 ± 0.06 |
| Gl 875.1 = GT Peg | 10.6 | 76 | 30 | 17.30 ± 0.02 | 0.60 ± 0.04 | | | | |
| Gl 896(A + B) = BD + 19°5116 | 11.33 + 13.4 | 485 | 142 | 20.56 | 0.68 | 165 | 70 | 17.6 | 0.59 |
| Gl 896B = EQ Peg | 13.4 | 58 | 9 | 21.01 ± 0.06 | 0.71 ± 0.06 | | | | |

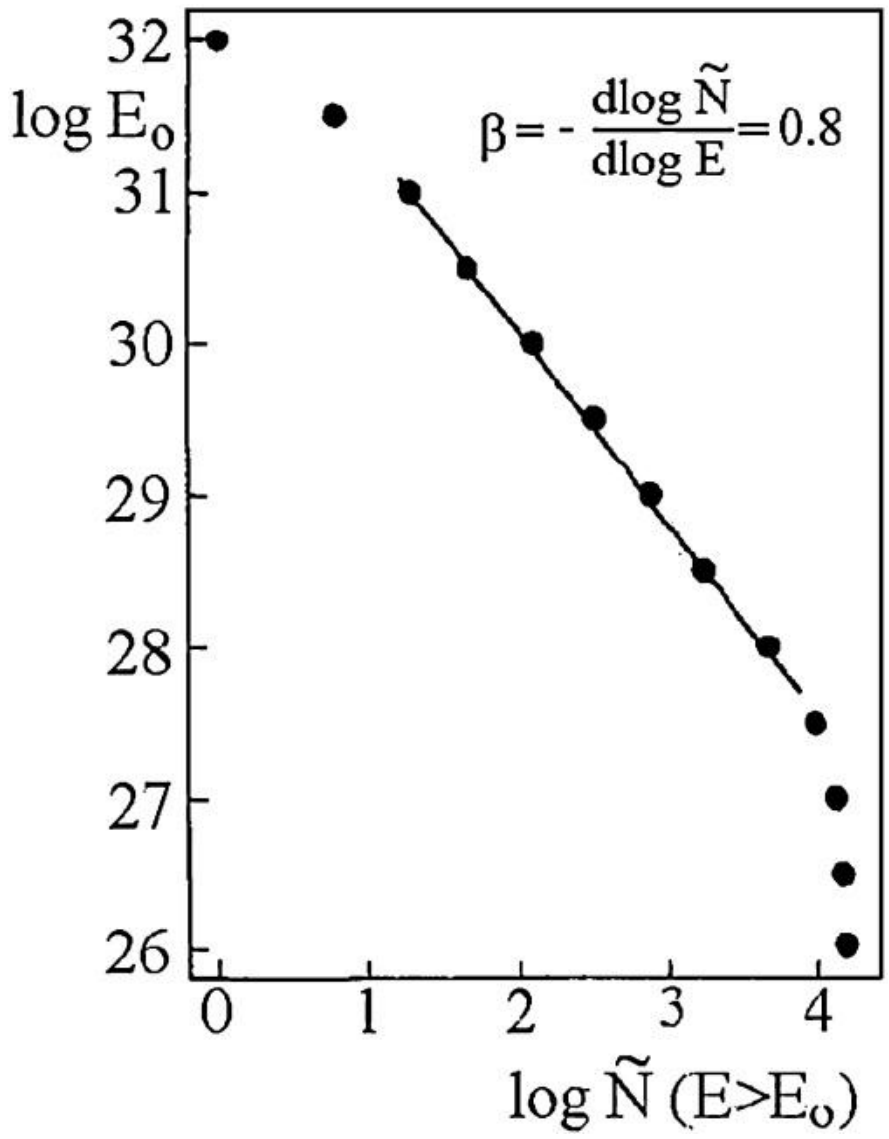

Fig. 47. Integral energy spectrum of solar flares (Kurochka, 1987)

Figure 47 shows the energy spectrum of solar flares reproduced from the paper of Kurochka (1987), who on the basis of observations of more than 15500 flares in $H_\alpha$ rays calculated the energy of their complete radiation in all lines and continua of hydrogen series starting from the Lyman series. Three of the above-discussed elements of the energy spectra of stellar flares are seen: the break in the region of weak flares due to observational selection, the linear section corresponding to the power dependence $\tilde{\nu}(E)$ and the beginning saturation section. The physically significant part of this energy spectrum of solar flares is transferred to Fig. 46. The transition from $\tilde{N}$ $(E)$ in Fig. 47 to $\tilde{N}$ $(E)$ in Fig. 46 is obvious, because the duration of monitoring of the Sun is known, while the transition from the energies of the total hydrogen radiation calculated by Kurochka to the values of $E_B$ is less univocal. Using the relations of total hydrogen radiation of solar flares to the radiation in the $H_\alpha$ line provided by Kurochka and Stasyuk (1981) and estimates of the ratio $E_B/E_{H\alpha}$ ~10–20 for the number of stellar flares (Gershberg and Chugainov, 1966; Gershberg and Shakhovskaya, 1972; Kulapova and Shakhovskaya, 1973; Worden et al., 1984; Rodonò 1986a), one can accept that, considering the weak continuous emission of solar flares, the total hydrogen radiation of such flares exceeds their radiation in the B band by an order of magnitude. Despite such a rough estimate in which an error of a factor of 3–5 cannot be excluded, the integral energy spectrum of solar flares in Fig. 46 matches the general picture of the energy spectra of flares. Similarly to Fig. 43, Fig. 46 evidences that the level of flare activity of the Sun is approximately 4 orders of magnitude lower than that of the most active red dwarfs. Thus, the observations show that the power character of the energy spectrum of flares is a universal feature of stellar flares of the considered type. A number of interesting consequences follow from this experimental fact.

B. Spectral indices of flare energy spectra. Let us assume that the differential energy spectrum of flares $v(E)$ in the whole energy range $E_{min}– E_{max}$ is presented by a uniform power dependence. In this case, the total radiation of all flares over the time $T$ sufficient to realize the whole energy spectrum of flares can be written as

$$\mathbf{E}=T\int_{E_{min}}^{E_{max}} E\,\nu(E)dE = -T\int_{E_{max}}^{E_{min}} E(d\tilde{\nu}/dE)dE = \begin{cases} T\cdot 10^{\alpha}\beta(E_{max}^{1-\beta}-E_{min}^{1-\beta})/(1-\beta) \\ T\cdot 10^{\alpha}\ln(E_{max}/E_{min}) \end{cases} \quad (37)$$

The bottom relation is valid for $\beta = 1$, the upper for the other cases.

Hence, it follows that the value of the spectral index $\beta$ determines which flares make the major contribution to **E**: at $\beta < 1$ these are rare but strong flares, at $\beta > 1$ frequent weak flares.

Figure 48 presents the values of spectral indices of energy spectra of flares depending on the absolute luminosity of flare stars. The values of $\beta$ for stars in the solar vicinity are taken from Table 16, for cluster stars — from the paper by Korotin and Krasnobabtsev (1985). Since the energy spectra of flares recorded in blue and violet rays are rather similar, the average values of $\beta$ are plotted for the stars for which these parameters are available in both photometric systems.

Figure 48 shows that $\beta$ = 0.4–1.4. Since the errors in determining $\beta$ usually do not exceed 0.10–0.15, then it is clear that there are flare stars with spectral indices of integral energy spectra of flares above and less than unity. However, in Table 16 only 2 of the 23 objects have $\beta > 1$, and for one of them the estimate of this value is obtained with the greatest error.

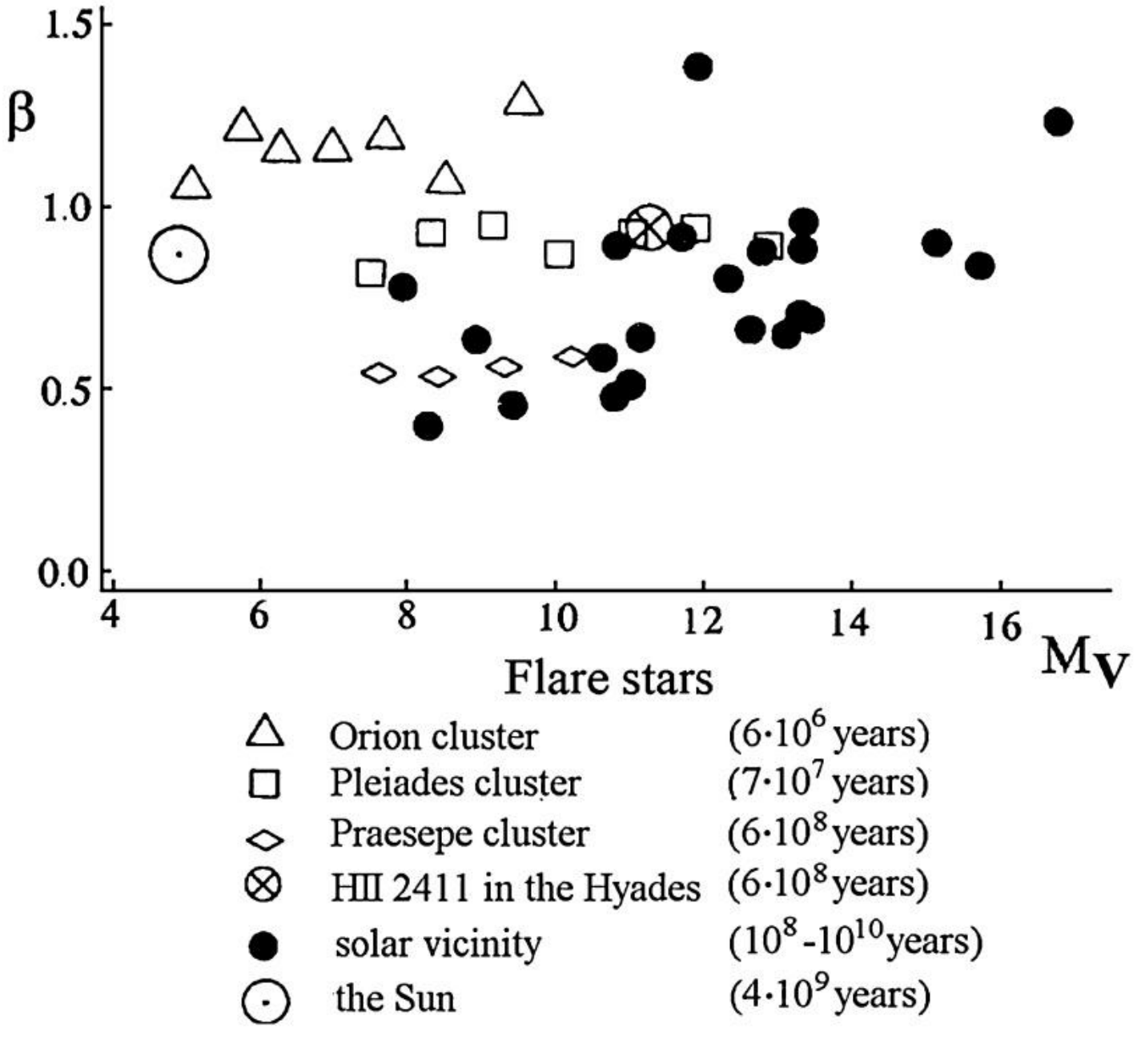

Fig. 48. Spectral indices of flare energy spectra (Gershberg, 1989)

For flare stars in the solar vicinity between spectral indices $\beta$ and absolute stellar magnitudes $M_V$ there is a weak positive correlation: $r(\beta, M_V) = 0.60 \pm 0.14$, which causes low confidence in the coefficient of linear regression between the values

$$\beta = (0.047 \pm 0.015)M_V + 0.18 \pm 0.19 \quad (38)$$

Finally, Figure 48 suggests the existence of the dependence of $\beta$ on the age of the flare star. Indeed, the age of each of the three investigated stellar clusters — Orion, the Pleiades, and Praesepe — exceeds the age of the previous cluster by an order of magnitude. In Fig. 48, the spectral indices of group energy spectra of flares in these clusters form sequences distinctly spaced for $\beta$: the older the cluster, the less is the value of $\beta$. Further, among flare stars in the solar vicinity there are objects of various age, including those older than flare stars in Praesepe. Confirming the dependence of $\beta$ on age, spectral indices of flare stars in the solar vicinity demonstrate the greatest scatter in Fig. 48 and expand to the region of values lower than $\beta$ values characteristic of Praesepe. However, it should be noted that the $\beta$ invariance of flare stars in each of the considered clusters in the luminosity range up to $4^m - 5^m$ is a rather unexpected result, because it means that the character of flare activity, in general, is insensitive to the changes in the internal structure of stars. At the same time, for example, such an important parameter of the internal structure as the thickness of the convective zone in this luminosity range should definitely vary noticeably, and one could try to associate it with the correlation of $\beta$ and $M_V$ observed on flare stars in the solar vicinity. In the framework of a purely observational approximation the existence of this correlation and the correlation of $\beta$ with the cluster age suggest that, on average, less bright flare stars in the solar vicinity are younger objects than brighter stars.

The correlation of $\beta$ and the age of the revealed flare stars in clusters is violated in Fig. 48 by spectral indices of two individual energy spectra of flares: the star HII 2411 in the Hyades and the Sun, whose spectral indices in this figure correspond to much younger stars. The cause of this distortion can be interpreted using the research results for the energy spectra of solar flares in the region of soft X-ray emission obtained by Kasinsky and Sotnikova (1989, 1997). They analyzed such energy spectra from 1972 to 1993 and revealed a clear change of some characteristics of flare activity of the Sun with the 11-year cycle phase. In particular, it follows from their data that as the phase shifts for one year there is a close correlation of $\beta$ with the Wolf numbers: $r(\beta_{+1}, W) = 0.84$. There is quite good agreement with the results obtained by Kurochka (1987) for optical flares on the Sun in 1978–79 $\beta = 0.80$, Kasinsky and Sotnikova found for X-ray flares $\beta = 0.66$ and 0.68 in 1978 and 1979, respectively. But throughout the two cycles $\beta_X$ covers the range from 0.47 near the epoch of minimum solar activity to 0.93 near maximum activity. This means that during the epoch of minimum activity, when rare and strong flares dominate in the flare ensemble, the value of $\beta$ of the Sun is older than the oldest cluster Praesepe in Fig. 48 and is much younger, as in the Pleiades, during the epoch of the maximum, when $\beta$ approaches unity and the contribution of frequent low-energy events sharply increases. Based on photographic observations of HII 2411, Szecsnyi-Nagy (1986) suspected the cyclicity of its activity with a two-fold increase of the number of flares during the maximum phase.

For the active flare dwarf EV Lac there are no sufficiently consistent data. Mavridis and Avgoloupis (1987) from observations in 1974–79 in the B band found a significant decrease of the spectral index $\beta$ during low flare-activity seasons when the total radiative energy decreased two-fold. Alekseev and Gershberg (1997a) analyzed the observations carried out in 1986–95 in the U band and found cyclicity in the changes of $\beta$ with a characteristic amplitude of ±0.1, but did not detect any changes of the level of flare activity. On UV Cet variations of $\beta$ on short time intervals were suspected by Lacy et al. (1978).

Thus, the data on individual energy spectra of flares in Fig. 48 can be influenced by appreciable effects of physical variability of these spectra.

The mentioned statistical study of X-ray flares on the Sun (Kasinsky and Sotnikova, 1989) is based on the consideration of more than 23000 events. The total number of X-ray flares recorded on dwarf stars by now does not reach a hundred and fifty and there are inconsistent data again. According to Pallavicini et al. (1990a), there is a power-energy spectrum of flares with $\beta \sim 0.7$ in the sample of about thirty X-ray stellar flares on red dwarfs detected by EXOSAT. But, as stated above, based on EUVE observations, Audard et al. (2000) concluded that for flares with $E_{EUV} > 10^{32}$ erg $\beta > 1$ for F–G dwarfs and for K–M stars $\beta < 1$ was more probable.

C. Mean flare energies. The power differential spectrum follows from the power integral energy spectrum of flares

$$dN = \text{const}\; dE/E^{1+\beta}, \tag{39}$$

which allows the average energy of recorded flares to be estimated as

$$\bar{E} = \int_{E_{\min}}^{E_{\max}} E\mathrm{d}N \bigg/ \int_{E_{\min}}^{E_{\max}} \mathrm{d}N = E_{\min} \frac{\beta\left[(E_{\max}/E_{\min})^{1-\beta} - 1\right]}{(1-\beta)\left[1 - E_{\max}/E_{\min})^{\beta}\right]}. \tag{40}$$

This relation is especially evident at $\beta = 1$

$$\bar{E} = E_{\min} \frac{\ln\left(E_{\max}/E_{\min}\right)}{1 - E_{\min}/E_{\max}}. \tag{40$'$}$$

In other words, the average energy of flares recorded on a star is equal to the product of the minimum energy of flare detected on the star using a certain observation technique, and a slowly varying function of the ratio $E_{max}/E_{min}$. For the most intensely studied flare stars, on which many tens or even more than one hundred flares were recorded, this ratio is close to one hundred, and for absolutely brighter stars it is apparently a little higher. Assuming for weak flare stars $\beta = 1.4$ and $E_{max}/E_{min} = 50$, and for bright stars, $\beta = 0.4$ and $E_{max}/E_{min} = 200$, we obtain

$$\langle E \rangle = (2.8 - 17)E_{\min}. \tag{41}$$

Consideration of extensive observational series of AD Leo carried out by Pettersen et al. (1984a) and flare stars in the Pleiades by Haro et al. (1982) demonstrated that the average amplitudes of recorded flares fell into the range of values expected from (41) (Gershberg, 1985).

Thus, the mean energy of flares in a rather large sample is determined first of all by the energy of an extremely weak flare detected on the given star. The extremely low energy is determined in turn by the observational techniques, thus the average energy (or the average amplitude) – as opposed to the conventional concept – cannot be the measure of flare activity of stars.

D. Dependence of the frequency of observed flares on stellar luminosity. Analyzing photoelectric observations of eight flare stars by Moffett (1974), Mirzoyan (1981) discovered an increasing average frequency of flares with an increase in absolute stellar magnitudes. Let us consider this question in more detail on the basis of more complete data.

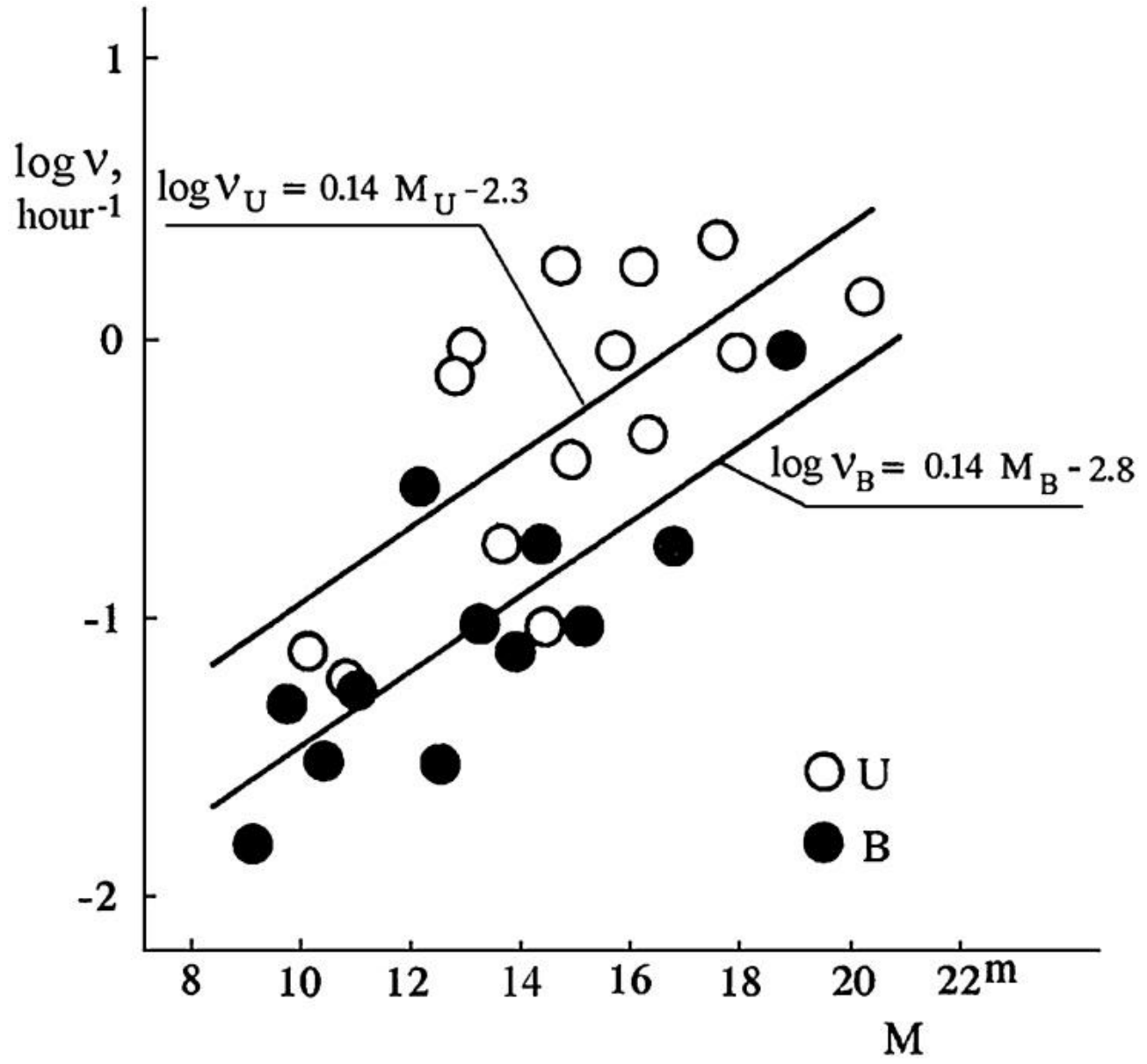

Fig. 49. The average frequency of recorded flares vs. absolute luminosity of flare stars (Gershberg, 1985)

On the *x*-axis in Fig. 46 one can see the cumulative frequencies of flares, thus the abscissa of the rightmost point before the break of each spectrum — to an accuracy of an insignificant extent in the logarithmic scale of frequencies of the observed energy spectrum after the break — corresponds to the average frequency of all recorded flares. There is a reliable positive correlation between the average frequencies and absolute stellar magnitudes: $r(\log\nu, M_B) = 0.81 \pm 0.10$. Figure 49 presents the results of a comparison of these values and those obtained from the observations in violet rays (Gershberg, 1985). The equations of straight lines of the linear regressions are as follows:

$$\log\nu_U = (0.14 \pm 0.04)M_U - 2.3 \pm 0.6\ ,$$
$$\log\nu_B = (0.14 \pm 0.03)M_B - 2.8 \pm 0.4\ . \qquad (42)$$

Therefore, the statistical dependence of the average frequency of all flares observed on flare stars on their absolute luminosity can be presented by the relation

$$\nu \propto L^{-0.35\pm0.09}\ . \qquad (43)$$

The observed dependence is doubtless, and the appreciable dispersion of points in Fig. 49 and appreciable probable errors of the numerical factors corresponding to it in (43) may be caused by the heterogeneity of the initial observational data providing a nonuniform set of frequencies of the breaks of the flare-energy spectra, and by physical nonuniformity, for example, with respect to age. Finally, the dispersion of points in Fig. 49 may be due to the fact that flare stars are often components of binary systems, which can contain stars of different brightness and different levels of flare activity, therefore the dispersion in the observed dependence logν on M should manifest itself even if there is a functional relation between these parameters for individual flare stars.

The decrease of average frequency of recorded flares with the increase in stellar absolute luminosity is due to observational selection: on brighter stars the flare detection threshold is higher. However, using the observed energy spectra of flares, one can show that for the same sample of flare stars the effect of observational selection should lead to the relation

$$\nu \propto L^{-0.96\pm0.13} \tag{44}$$

(Gershberg, 1985). A significant difference between (43) and (44) means that at the transition from absolutely weak to absolutely brighter flare stars the average frequency of flares decreases not as fast as expected because of an increasing detection threshold. Apparently, this effect can be connected with the increase in the surface area of brighter stars and the availability of a greater number of active regions producing flares. Indeed, using the data of Pettersen (1976, 1980) on absolute luminosities and the sizes of single flare stars, one can obtain the statistical relation between the values

$$L \propto R^{5.0\pm0.06}. \tag{45}$$

Then, for the multiplier that distinguishes (43) from (44), we obtain

$$L^{0.6\pm0.2} \propto R^{3.0\pm1.4} \propto 4\pi R^2 \cdot R^{1.0\pm1.4}. \tag{46}$$

The last factor $R^{1.0\pm1.4}$ is too uncertain for any conclusions to be made on the level of "specific" flare activity per unit surface of a flare star; but the influence of the size of stellar surface on the frequency of flares is rather probable.

Thus, the dependence of average frequency of recorded flares on absolute stellar luminosity is determined both by the dependence of the detection threshold on stellar luminosity, and real distinctions of average frequencies of flares on the stars of various luminosity, and — contrary to the widespread belief — the true average frequency of flares is higher on brighter stars. The erroneous opinion arose because the higher frequency of recorded flares found observationally on low-luminosity stars was not considered from the point of view of inevitable effects of observational selection (Petit, 1970; Gurzadian, 1971a). This essential effect of observational selection was not taken into account by Ambartsumian (1978) in calculations of the distribution of average frequencies of flares from the chronology of opening of flare stars in the Pleiades cluster.

Thus, the red dwarf stars are supposed to be the most pronounced carriers of the considered type of activity not because such activity is the most advanced, but only because it is the most noticeable on such low-luminosity cool stars.

Recently, from the direct observations with Kepler, Balona (2012, 2013, 2015) clarified these 30-year-old statistical conclusions. He found that the frequency of flares on A–F stars was four times lower than that on K–M dwarfs. The detailed studies of the power energy spectra of flares on the solar-type stars were recently carried out with Kepler by Wu et al. (2015), Namekata et al. (2018), and with 5 telescopes by Paudel et al. (2021).

E. The energy spectrum of flares and the SDSS project. Within the SDSS project, Hilton (2011) performed about a thousand hours of photometric observations of 39 M0–M8 dwarfs, involving active and inactive stars, and recorded 239 flares. Selecting out of them 137 events, which were called as classic flares with the impulse burning, fast initial and subsequent exponential decay, he constructed their energy spectra, found correlations between the total energy of flares and times of their burning and decay, and detected that the secondary bursts were over 40% of the total duration of events. Using the original quantitative model of flare light curves, their energy spectra and the model of stellar population of the Galfast Galaxy, Hilton constructed a model of the flare mode of the Galaxy, which allowed one to perform the qualitative predictions of frequencies and flare energies of M dwarfs over the whole Galaxy and, particularly, to obtain from observations a dependence of activity of such stars on the height above the Galactic plane (see Fig. 4). This model predicts several flares with $\Delta U > 0.1^m$ on each 9.6 square degree of the full frame of the exposure at galactic latitude $b = 20°$ and $\sim 0.1$ of a flare at $b = 80°$. The invisible in the quiescent state but recorded in a flare stars produce about 1 % of radiation in the U band at a latitude of 20° and about 0.01 % at a latitude of 80°.

F. On the physical sense of the power dependence $\tilde{\nu}(E)$. As stated above, the formal analysis of the observed energy spectrum of flares leads to physically significant conclusions on some statistical properties of the flare activity of stars. It is obvious that interpretation of the discovered power character of the dependence $\tilde{\nu}(E)$ can provide additional information on the nature of flares. The first attempt to analyze energy spectra of flares was undertaken by Pustil'nik (1988). He concluded that if a stellar flare was a final result of dissipation of nonpotential excesses of magnetic field in the active region and optical emission made up a certain fraction of the total energy released during the dissipation, the flare radiation of the flare should be described by the relation

$$E(l) \propto \int_V (\Delta H^2/8\pi)\, dV \propto \begin{cases} H^2 l^3 & \text{at } l < z \\ H^2 l^2 & \text{at } l > z \end{cases} \tag{47}$$

Here, $l$ is the characteristic size of the region of magnetic-field dissipation and $z$ is the characteristic scale of heterogeneity of the field with height; $l < z$ and $l > z$ suggest volumetric and layered dissipation of nonpotential excesses of magnetic field, respectively.

Simple dimensional considerations make it possible to obtain the following expression from (47)

$$\tilde{\nu}(E > E_0) \propto \begin{cases} E_0^{(s-2)/3} & \text{- for volumetric dissipation} \\ E_0^{(s-2)/2} & \text{- for layered dissipation} \end{cases} \tag{48}$$

Here, $s$ determines the dependence of turbulent velocity on the size of the turbulent element $l$

$$v_{\text{turb}} \propto l^s\,. \tag{49}$$

In the existing theories of stellar convection, the value of $s$ varies in a rather narrow range: $s = 1/2$ in the case of magnetic-field convection, $s = 1/3$ for Kolmogorov's turbulence, and $s = 1/4$ for acoustic and Alfvén turbulence. If the strength of the dissipating magnetic field $H$ is assumed independent of $l$, for the given values of $s$ (48) yields the sought power dependence $\tilde{\nu}(E)$ with spectral indices within 0.5–0.9. A weak negative correlation $H \propto l^{-1/3}$ would allow this range of spectral indices to be shifted to (0.85; 1.4).

Kasinsky and Sotnikova (1988) considered the model with a dissipating kinetic part of the energy of turbulent plasma motions rather than a magnetic one. The following expressions were valid instead of (47)

$$E(l) \propto v_{\rm turb}^2 l^3 \propto l^{2s+3} \qquad (50)$$

and

$$\tilde{\nu}(E) \propto E^{(s-2)/(2s+3)} \qquad (51)$$

Thus, this model also yields the power dependence $\tilde{\nu}(E)$, but with systematically lower spectral indices, from 0.4 up to 0.5.

However, it should be noted that both proposed schemes include nonobvious assumptions: the structure of the considered size lives until the first flare and that its lifetime is $t(l) = l/v_{\rm turb}$. However, the existence of homologous flares on the Sun shows that in the structures that generate flares such processes can develop repeatedly. On the other hand, if the lifetime of these structures is not determined by turbulence, but rather by the diffusion of magnetic field, i.e., the lifetime of the structure $t(l) = l^2/n_{\rm m}$, where $n_{\rm m}$ is the magnetic viscosity, the following expression is valid

$$\tilde{\nu}_{\rm flares}(l > l_0) \propto l_0^{-3}\,, \qquad (52)$$

and we obtain the power spectrum with $\beta = 1$ or 3/2 for volumetric or layered dissipation of the magnetic field, respectively.

Lu and Hamilton (1991) showed that in the case of a general approach to coronal magnetic fields on the Sun as to the self-organizing complex system close to the critical state, and within the context of the notion on observed solar flares as a superposition of numerous elementary processes of reconnection in these fields, one should expect the energy distribution of flares with a spectral index of 0.4. But within the framework of the notion the spectral index of the energy spectrum should not depend on the level of solar activity, which contradicts the results of Kasinsky and Sotnikova (1988, 1989). Recently, Podlazov and Osokin (2002) simulated numerically flares within the framework of the updated avalanche concept proposed by Lu and Hamilton and found $\beta = 0.37$, which is still much lower than the appropriate values in the energy spectra of flare stars.

Litvinenko (1994) showed that within the framework of the traditional scheme of origin of flares from reconnection of magnetic flux tubes, one could obtain the distribution of probabilities of the occurrence of flares of certain energy that for high-energy flares – more than $10^{26}$ erg – yields the differential energy spectrum of power type $dN \propto E^{-7/4}dE$, i.e., $\beta = 0.75$ – and a softer spectrum for low-energy flares. Probably, an important confirmation of the situation in the region of sufficiently strong stellar flares is the above results of Audard et al. (1999, 2000).

The notions of current sheets, developed recently, led to the concept of essentially nonuniform dynamic structures of turbulent layers containing numerous clusters of the regions with switching over normal and sharply reduced conductivity. The passage of current through such a nonstationary medium is described by the percolation process, that is "filtering" through a stochastic network. In the process, the dissipation of current differs quantitatively from the case of electron–ion collisions. Based on the concept of fractals, the percolation theory also leads to a statistical power dependence of global properties of the system on the properties of its separate parts with exponents of 0.6–0.9 for two-dimensional systems and 1.5 for three-dimensional systems (Pustil'nik, 1997, 1999).

Recently, Mullan and Paudel (2018) proposed a new model of flares leading to the observed values of spectral indices of their energy spectra. They suppose that the convective flows yielding granules make the field lines at the bases of coronal magnetic loops frozen into the photospheric gas produce random wanderings leading to their twisting. At a critical value of such a twisting the loops become unstable and flares occur in them. Mullan and Paudel thoroughly considered the described above energy spectra of flares of tens of stars, tens of thousands of solar X-ray flares and more than 850000 flares on 4041 stars recorded with Kepler (Davenport, 2016). As a result, they obtained the rational estimates of characteristic frequencies of flares, the fraction of energy of the local magnetic fields was found to be sufficient to supply the energy of flares and, according to observations, they detected a quite abrupt power energy spectrum of relatively weak flares and a flatter one in the region of strong flares.

Thus, the power character of the dependence $\tilde{\nu}(E)$ follows from the consideration of stellar flares as processes proceeding in separate structures of turbulent magnetized plasma. This statement remains valid despite the significant progress in the actual content of these ideas. The present-day state of these notions is given in Section 2.6.6. of Chapter 2.6. "Physical nature of flares".

**2.3.1.3. Total Energy of Flare Emission.** Simultaneous observations of stellar flares in three bands of the UBV system (Lacy et al., 1976) showed that there was a distinct statistical relation between total emission of flares in these bands

$$E_{\mathrm{U}} = (1.20 \pm 0.08)E_{\mathrm{B}} = (1.79 \pm 0.15)E_{\mathrm{V}}, \tag{53}$$

which is valid in the energy range $10^{27}$–$10^{34}$ erg. It follows from (53) that

$$E_{\mathrm{UBV}} = 2.4E_{\mathrm{U}} = 2.9E_{\mathrm{B}} = 4.3E_{\mathrm{V}}. \tag{54}$$

For a wider UBVRI system covering the whole optical range, similar relations can be obtained from long-term observations of one of the brightest and most active flare stars EV Lac (Alekseev and Gershberg, 1997a). As stated above, in 1986–95, 227 flares on this star were recorded in the Crimea in the UBVRI system. Among them only those flares were selected for which all color parameters of flare emission at maximum brightness were determined with an error not exceeding $0.15^m$. Eight flares were selected and for all of them the brightness amplitude $\Delta U$ at maximum exceeded $1.8^m$, and the average values of color indices of flare emission at maximum brightness were

$$\begin{aligned} &\langle U - B\rangle = -0.87^m \pm 0.15^m, \quad \langle B - V\rangle = 0.04^m \pm 0.1^m,\\ &\langle V - R\rangle = 0.8^m \pm 0.06^m, \quad \langle V - I\rangle = 0.81^m \pm 0.21^m \end{aligned} \tag{55}$$

(Earlier, from extensive UBV observations of eight flare stars for flares of different amplitudes, Moffett (1974) found average color indices of intrinsic flare radiation $\langle U - B\rangle = -0.88^m \pm 0.31^m$ from 153 flares and $\langle B - V\rangle = +0.34^m \pm 0.44^m$ from 77 flares. Ishida et al. (1991) from 127 flares on YZ CMi, AD Leo, and EV Lac estimated $\langle U - B\rangle = -0.98^m \pm 0.36^m$ and $\langle B - V\rangle = +0.24^m \pm 0.30^m$, from 59 flares with amplitudes of $\Delta U > 1.5^m$ – $\langle U - B\rangle = -1.03^m \pm 0.23^m$ and $\langle B - V\rangle = +0.14^m \pm 0.20^m$, and from 17 flares with $\Delta U > 2.5^m$ – $\langle U - B\rangle = -0.98^m \pm 0.17^m$ and $\langle B - V\rangle = +0.05^m \pm 0.13^m$. Combining the data from different publications, Grandpierre and Melikian (1985) estimated from 276 flares $\langle U - B\rangle \sim -1.0^m$ and from 174 flares $\langle B - V\rangle \sim +0.4^m$.) Using the absolute calibration of the scale of stellar magnitudes, from (55) one can obtain

$$L_{\rm UBVRI} = 1.6L_{\rm UBV} = 4.2L_{\rm U} = 3.8L_{\rm B} = 7.6L_{\rm V} \qquad (56)$$

Differences in the appropriate coefficients in (54) and (56) by 10–20% are caused both by the real dispersion of colors of flare emission at maximum brightness and small systematic changes of these colors during the development of flares. However, such distinctions are negligible for the estimates of the total flare energy.

It is natural to use (54) or (56) after the estimate of total flare emission in one of the photometric bands with the help of the energy spectrum of flares found for this band. The total energy of flare emission calculated from (37) depends essentially on the accepted limiting values of flare energy: on $E_{\rm max}$ for the energy spectra with $\beta < 1$ and on $E_{\rm min}$ for $\beta > 1$. The calculations of Shakhovskaya (1979) showed that if first the total energy of optical emission of flares was calculated on the assumption that on each star $E_{\rm max} = 3 \cdot 10^{35}$ erg, the value corresponded to the strongest flares recorded in clusters, and $E_{\rm min} = 2 \cdot 10^{27}$ erg corresponded to the weakest flares on the faintest flare star in the solar vicinity CN Leo, and secondly the values $E_{\rm max}$ and $E_{\rm min}$ really recorded on each flare star were then used, the first estimate of total flare energy would exceed the second estimate of the same value by 2 to 40 times. Thus, even using an average of these two estimates of total energy of flare emission, one cannot exclude the error of the average by an order of magnitude.

The total energy of flare emission is usually specified as the ratio of time-averaged observations of the intensity of flare emissions to a certain constant intensity related to the stellar radiation rather than in absolute units. Thus, initially the sum of equivalent durations of flares to the time of monitoring $\Sigma P/T$ was often used, which apparently was equal to the ratio of the total flare energy in a certain photometric band to permanent stellar radiation in the same band. For the most active flare stars this dimensionless ratio varies from several thousandths to 3–5 hundredths. However, in the U and B bands regularly used in flare recording, flare stars emit only an insignificant portion of energy, therefore the ratio $\Sigma P/T$ determines only the visible pattern of flare activity. More suitable, from a physical viewpoint, is the ratio of total energy of all optical emission of flares to bolometric luminosity. The ratios for 23 flare stars presented in Fig. 50 are reproduced from the paper by Gershberg and Shakhovskaya (1983). In plotting the diagram, the values of limiting flare energies really recorded on each flare star were used as $E_{\rm max}$ and $E_{\rm min}$. According to these data, the ratio of the time-averaged emission of flares to bolometric luminosity of flare stars is within $10^{-5}$–$10^{-3}$. No dependence of this ratio on the absolute luminosity of stars was found.

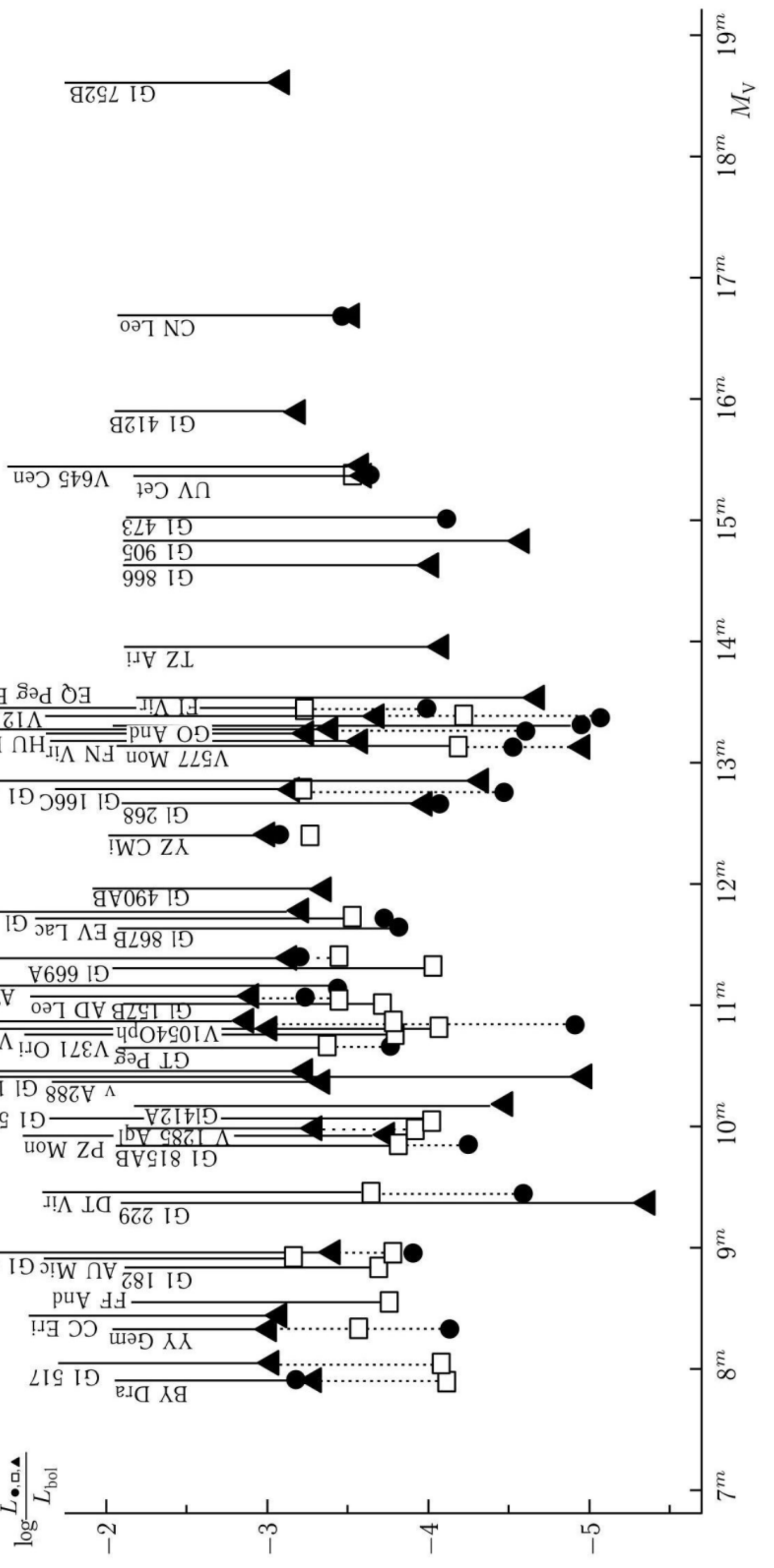

Fig. 50. Energy losses of flare stars for optical flare emission (*filled circles*), emission of quiet chromospheres (*squares*) and coronae (*filled triangles*) vs. bolometric luminosity of flare stars of different luminosity (Gershberg and Shakhovskaya, 1983)

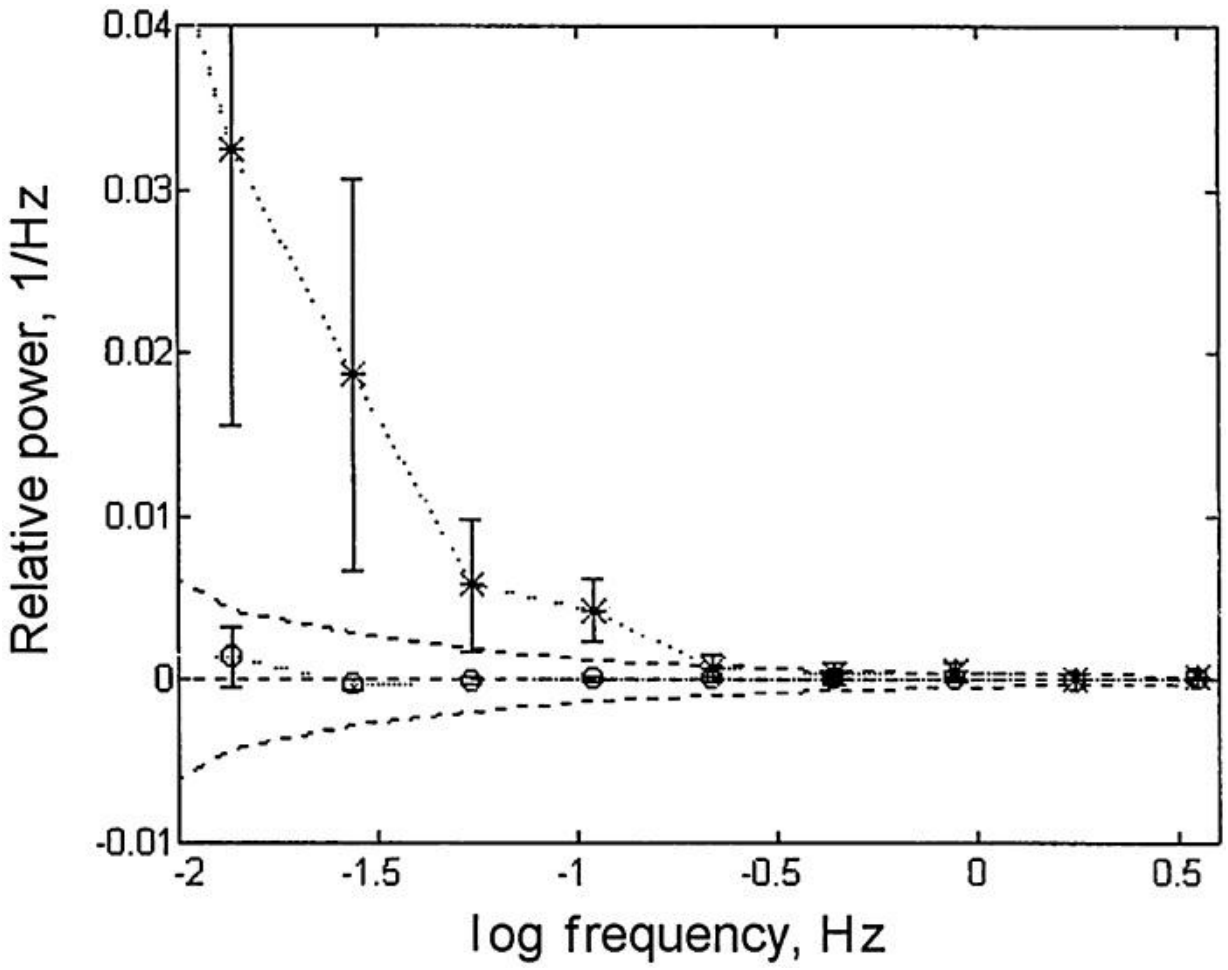

Fig. 51. Relative intensity of the variable brightness component in the U band within 0.05–3.5 Hz: quiescent state of EV Lac (*asterisks*) and check stars (*circles*) (Zhilyaev et al., 2001)

As to the above concept of heating of stellar coronae by numerous microflares, it should be noted that the energy spectrum method does not provide confident estimates of the contribution of low-energy flares to the general flare emission: at $\beta < 1$ when formally their contribution is insignificant, there is no confidence that the spectral index found for strong flares is valid for low-energy flares, and at $\beta > 1$ the lack of knowledge of the value of $E_{min}$ brings essential uncertainty. The recent results obtained by Zhilyaev et al. (2001) with statistical photometry should be cited (Fig. 51): within the frequency range of 0.05–3.5 Hz the relative brightness of the variable component of EV Lac reaches 0.25% of the stellar brightness in the U band, which is almost an order of magnitude lower than the average intensity of individually recorded flares. This estimate is much stricter than that obtained earlier by Beskin et al. (1995).

## 2.3.2. Estimates of Total Energy of Flares

The data on energy losses of stellar flares, which differ from the losses for optical radiation, are rather poor and sketchy. It is known that the energy of radio emission of flares is much lower than the energy of their optical radiation $E_{radio}/E_{opt} = 10^{-5}–10^{-2}$ (Lovell, 1971).

The ratio $E_X/E_{opt}$ differs from unity apparently by not more than one order of magnitude in both directions (Katsova and Livshits, 1986); probably the span of the estimate is partly explained by its essential variability during the flare (Kahler et al., 1982; Haisch, 1983). According to Haisch (1983), $L_X/L_{trans.z.} \sim 20$.

The contribution of radiation in the far ultraviolet to the total flare emission is rather uncertain. Based on ROSAT WFC, IUE, and optical observations of the flare on BY Dra of 1 October 1990, Barstow et al. (1991) concluded that EUV radiation contained an essential part of the total flare emission. Audard et al. (1999, 2000) carried out a statistical study of EUV flares. To carry out photometric observations of two young G dwarfs, 47 Cas and EK Dra, in the deep-photometry mode by EUVE, they applied an algorithm of statistical identification of flares from the fidelity test of every readout and identified 28 flares, which occupied about 60% of the total duration of observations. They showed that the power function represented well the energy spectrum of these flares and found that in the energy range of $3 \cdot 10^{33}$– $6 \cdot 10^{34}$ erg the spectra had spectral indices of 1.2 ± 0.2, i.e., numerous weak flares made a decisive contribution (up to 90%) to the total EUV radiation of flares on these stars. Then, Audard et al. (2000) similarly investigated eight other stars, and in joint consideration of the whole sample of 10 F2–M6 objects found a certain change of spectral index of the energy spectrum from 1.3 for F–G through 0.9 for K to 0.8 for M dwarfs.

In March 2000, a wide cooperative program was arranged to monitor AD Leo: photometrical observations at three observatories, spectral studies at two observatories, ultraviolet observations with HST STIS and EUVE, and radio observations at Jodrell Bank (Hawley et al., 2003). The program made it possible to estimate the contribution of different components to the total flare emission. In analyzing four flares of different amplitude and duration in all cases the main contribution was attributed to continuous emission, at the impulsive phase the contribution of ultraviolet lines achieved several percent of the total radiation and exceeded the contribution of optical lines, whereas at the stage of smooth decay the optical lines provided up to 10–20% of the total radiation and exceeded the contribution of ultraviolet lines.

If $E_{\rm opt} \sim E_{\rm UV} \sim E_{\rm X}$ and $< E_{\rm EUV}$, as a rough approximation one can accept $E_{\rm bol} \sim (5-6)E_{\rm opt}$.

Thus far, no direct data on the energy of stellar flares spent for particle acceleration have been obtained. As to the kinetic energy of the matter of flares, some preliminary estimates can be made on the basis of available spectral data.

It is known that in the spectra of flares on red dwarf stars some shifts of emission lines for several tens of km/s and appreciable broadening of their profiles of the lines — up to 10–15 Å at the level of half-intensity — were found (Greenstein and Arp, 1969; Gershberg, 1978). The fact that in the initial stages of flares, emission lines often have asymmetric profiles with a red wing extended noticeably more than the blue one (Gershberg and Shakhovskaya, 1972; Kulapova and Shakhovskaya, 1973; Bopp and Moffett, 1973; Bopp, 1974b; Byrne, 1989; Houdebine et al., 1993a) indicates that the motion of matter plays an essential, maybe even dominating, role in the observed variations of profiles. Thus far, only one detailed study of flare kinematics has been undertaken for the flare on AD Leo of 28 March 1984 (Houdebine et al., 1993a). However, the upper and lower estimates of kinetic energy of this event differ by two orders of magnitude. From the analysis of ultraviolet spectra, Byrne (1993a) concluded that the kinetic energy of matter in flares exceeded their radiation, whereas Gunn et al. (1994b) estimated the excess of emission energy over the kinetic energy from the spectra in the optical range of the flare on AT Mic as two orders of magnitude. In ten other flares on red dwarf stars the analysis of the observed profiles of emission lines near maximum brightness made it possible to present them as a superposition of different Gaussians (Byrne, 1989; Robinson, 1989; Eason et al., 1992; Abdul-Aziz et al., 1995): one narrow profile of instrumental width and one wide profile or the sum of several narrow Gaussians. This testifies to the kinematic heterogeneity of flares: apparently the narrow unshifted components are kinematically similar to the quiet chromosphere, whereas other components — widened and shifted — correspond to the motion of matter with velocities of several hundreds of kilometers per second. The removal of matter from a stellar surface at about 400 km/s was found from shifted components of the emission line HeII $\lambda$ 4686 Å in the spectrum of active EV Lac (Abranin et al., 1998b). These emissions were noticeably delayed with respect to flare maxima; thus, they were interpreted as the analogs of solar coronal ejections of matter into the interplanetary space.

Now let us focus on solar flares.

It is known that many characteristics of recorded stellar flares are very close to those of relatively rare solar white-light flares. Neidig and Kane (1993) considered the data on eight flares of this type and found that in these events the ratio of the power of nonthermal electrons with an energy of more than 50 keV, which was responsible for hard X-ray emission, to that of optical emission varied within 0.5–22, and the ratio of intensities of optical emission and soft X-rays in the range of 1–8 Å varied within 4–100.

According to the results of Kurochka (1987), the maximum energy of the total optical and ultraviolet emission of hydrogen plasma of solar flares in 1978–79 calculated from $H_{\alpha}$ emission was equal to a few $10^{32}$ erg. According to Lin and Hudson (1976), in the strongest solar flares in August 1972 the recorded optical and ultraviolet radiation reached $5 \cdot 10^{31}$ erg that made up from 8 to 33% of the total energy losses including the radiation in the whole electromagnetic spectrum, losses for acceleration of fast particles and motion of matter. Based on these estimates, we conclude that the total energy of the strongest solar flares apparently does not exceed $10^{33}$ erg. We note that, according to Hudson (1978), the energy spectra of radiation of solar flares in soft X-rays, near 20 keV and in the microwave range have a power character with spectral indices of 0.80–0.85, which practically coincides with the optical and ultraviolet data. For proton fluxes from solar flares, the spectral index noticeably differs: $\beta$ = 0.15, but the contribution of these losses to the total energy of flares is rather insignificant. The same is valid for hard X-ray emission of solar flares for which, according to Dennis (1985), $\beta$ = 0.40. On the other hand, as searches for solar "superflares" showed (Lingenfelter and Hudson, 1980), during 19 and 20 cycles of solar activity, during the last 7000 and even $10^7$ years there were no flares on the Sun in which the energy of hard particles exceeded the flare of 4 August 1972, whose total energy was of $2 \cdot 10^{32}$ erg, by more than an order of magnitude.

Let us turn back to the flare stars.

Figure 46 shows that on the brightest flare stars in the solar vicinity — BY Dra and AD Leo — in the strongest flares in the B band more than $10^{34}$ are emitted, maybe even $10^{35}$ erg. On flare stars in the Pleiades and in Orion where the technique of simultaneous observations of several hundred flare stars enables recording of very rare events on brighter stars, the maximum value of $E_B$ approaches $3 \cdot 10^{35}$ erg. Taking into account (56) and assuming that in stellar flares the same ratio between radiative and nonradiative losses of energy as for the strongest solar flares is valid, the total energy of the strongest stellar flares of the considered type should reach $3 \cdot 10^{36}$ erg. The flare of such energy was recorded by Batyrshinova and Ibragimov (2001) during the photometric monitoring of the X-ray object IRXS J220111+281849 in 1999.

This estimate has recently been confirmed in space investigations. Maehara et al. (2012) analyzed 365 strong flares with energy of more than $10^{33}$ erg recorded on 148 G dwarfs during observations of 83 000 stars within the Kepler experiment throughout 120 days in 2009. The typical duration of such flares accounted for several hours, the amplitude — 0.1–1% of the stellar bolometric luminosity; at the energy estimates the flares were assumed to radiate as absolute blackbodies at a temperature of 10000 K. The bolometric luminosity estimates of such flares are within $9 \cdot 10^{29}$ and $4 \cdot 10^{32}$ erg/s and the total energy estimates are within $10^{33}$ and $10^{36}$ erg; the uncertainty of these estimates approaches 60 %. The quasiperiodicity detected by Maehara et al. provides evidence for the higher number of spots than that on the Sun. The maximum energy of flares does not correlate with rotation periods, but strong flares occur more often on stars with fast rotation.

From the inferred estimates Maehara et al. obtained that so powerful flares should occur on a G star once every 350 years. On slowly rotating stars, like the Sun, such flares should occur once every 800 years, and flares with energy of $10^{35}$ erg — once every 5000 years. These high-energy estimates of rare flares do not contradict the results outlined above acquired from the statistics of ground-based observations of the most active flare stars in clusters (Gershberg et al., 1987; Gershberg, 2015). If it is true, then the Earth has already undergone million (!) such tremendous flares on the Sun.

Later, Shibayama et al. (2013) repeated the studies by Maehara et al. (2012) with more comprehensive data: as a result of the Kepler 500-day monitoring, 1547 superflares were recorded on 279 G stars. Following these data, the energy spectrum of superflares has the power form $dN/dE \propto E^{-\alpha}$ with $\alpha \sim 2$ and the flares with energies of $10^{34}$–$10^{35}$ erg occur once every 800–5000 years on such slowly rotating stars; but on some G stars 57 superflares were recorded throughout 500 days; apparently, huge spots and their groups exist on them.

Paudel et al. (2019) recorded 11 flares with a duration varying from one to 18 hours on three late and slowly rotating M dwarfs: a flare with energy of $1.3 \cdot 10^{35}$ erg on the M7 dwarf 2MASS J08315742 + 2042213 (the age of 700 million years) with a rotation period of 0.556 days, involved into the Praesepe cluster, five flares with energies from $5.2 \cdot 10^{33}$ to $3.5 \cdot 10^{35}$erg on the M8 dwarf 2MASS J0837832 + 2050349 with a rotation period of 0.193 days, and five flares with energies from $1.4 \cdot 10^{33}$ to $6.1 \cdot 10^{34}$ erg on the M9 dwarf 2 MASS J08312608 + 2244586 with a rotation period of 0.292 days, being a probable member of Praesepe. The maximum energy of the recorded flares on the mentioned M8 dwarf is 2.7 orders of magnitude higher than that on the M8 dwarf TRAPPIST-1 of an age of 7.6 billion years.

* * *

Thus, to interpret the observations, the theory of stellar flares of the considered type should explain the energy release up to $10^{33}$ erg, but not more than $6 \cdot 10^{33}$ (Aulanier et al., 2013), during one event on the Sun and at least up to $3 \cdot 10^{36}$ erg on the brightest UV Cet-type stars. Before the start of Kepler observations, there were reports concerning exceptionally rare flares with energy of two orders of magnitude higher than the indicated limits (Schaefer et al., 2000). But numerous superflares with energies of more than $10^{34}$ erg detected with Kepler raise the question otherwise regarding the maximum energies of flares of the considered type. On the other hand, the interpretation of stellar flares with such energy encounters some difficulties (Livshits et al., 2015).

## 2.4. Dynamics and Radiation Mechanisms of Flares at Different Wavelengths

Though flares recorded on red dwarf stars of the UV Cet-type are extremely diverse, available observational data make it possible to present a preliminary physical picture of these short-term phenomena. To sort the data logically, let us use the data on solar flares.

It is known that the so-called precursors of flares are observed on the Sun. One of the indisputable precursors is the activation of prominences: shortly before a flare, formerly quiet filament structures, which are seen as hanging bright structures in the corona and are found as dark filaments in rays of the red line of hydrogen on the solar disk, come into motion with velocities of tens of km/s in the place where the flare then occurs. These motions are connected with incipient reorganization of the magnetic field of the appropriate active region as prominences “hang” on magnetic force lines. There are observations in which the reorganization is recorded directly. Unfortunately, neither prominences in an individual active region on red dwarf stars nor its magnetic field have been recorded so far.

A confidently recorded onset of a solar flare is concerned with the occurrence of nonthermal X-ray emission. The spectrum of this radiation and its polarization features unambiguously evidence that this radiation is nonthermal and is caused by fast particle fluxes in the lower corona. The question on the mechanism of formation of the fluxes and the source of their energy was a disputed one for a long time. In the past decade, there was an appreciable progress, which is outlined in Chapter 2.6. However, most existing phenomenological models of these phenomena start from the observed phase, since strong fluxes of fast particles allow one to understand the subsequent thermal X-ray, ultraviolet, optical, and radio emission of various natures that appears further as the flare develops. Thus, it is reasonable to start the consideration of observations of stellar flares from the discussion of X-ray and radio emission from the corona. Then, we consider the results of ultraviolet observations that provide data on the luminescence of the transition zone and the upper chromosphere, and the most numerous data on optical emission of flares, which is formed in the stellar chromosphere and probably encompasses the upper layers of the photosphere.

Widely using the research results for solar activity as guiding the studies of stellar activity, one should not forget about the essential quantitative distinctions between the phenomena. For example, the average flux $F_X$ from a unit surface of dMe stars in the quiescent state is higher than the appropriate flux from the Sun at least by 3 orders of magnitude and by hundreds of times at the maximum of the 11-year solar cycle; the time-averaged optical emission of flares differs by 3–4 orders of magnitude. The qualitative difference of solar and stellar flares is that radiative losses in the optical, ultraviolet and X-ray regions in the latter are of the same order of magnitude, whereas in the former they differ essentially. Therefore, starting research of stellar activity from simple scaling of solar phenomena, one should be ready to discover qualitatively new processes on stars.

### 2.4.1. X-ray Emission of Flares

If optical flares on red dwarf stars were discovered occasionally, the accompanying bursts in hard radiation were purposefully sought for a long time.

Already in 1967, the data on the level of cosmic noises recorded from the Crimea and Gorky were considered in studying the geophysical effects of solar activity. These records were used to find sudden absorption of the noises during recorded optical flares on red dwarfs, the absorptions occur under additional ionization of the earth's atmosphere by sporadic hard radiation (Gershberg et al., 1969). This effect was not found; therefore, the upper limit of the ratio of luminosities of stellar flares in two wavelength ranges was estimated very roughly as $L_X/L_B < 10^5$.

Within the framework of certain physical notions, X-ray emission of stellar flares for the first time was precalculated by Grindlay (1970). He considered a number of models, which explained radiation of the bursts recorded by then on the UV Cet at meter wavelengths by synchrotron radiation of monoenergetic relativistic electrons or electrons with a power-energy spectrum. As a result, he estimated the efficiency of the relativistic particles in the generation of X-ray emission due to the inverse Compton effect and nonthermal bremsstrahlung emission. He found that the latter mechanism was more effective for radiation in the range of energies $E > 10$ keV. Then, Gurzadian (1971b) estimated the expected X-ray emission of flares, reasoning from his own concept (Gurzadian, 1965) of stellar flares as a result of the occurrence of a large number of subrelativistic (~ 1 MeV) electrons above the surface of a cool star and their interactions with photospheric radiation. Gurzadian also concluded that bremsstrahlung emission of fast electrons was most effective in the generation of X-ray emission. But, according to Edwards (1971), the level of X-ray emission of stellar flares predicted by Grindlay should lead to such a high total radiation of flare stars in the Galaxy that it would essentially exceed the observed X-ray background. Tsikoudi and Hudson (1975) analyzed the data obtained by the OSO-3 X-ray telescope during 82 optical flares on four red dwarf stars, but did find the effects of flares. Thus, they established another rigid constraint on X-ray emission of stellar flares $L_X/L_U < 300$, which also contradicted Grindlay's model and to a lesser degree, the Gurzadian model. However, according to Kahler and Shulman (1972) and Crannell et al. (1974), the contradiction is completely eliminated if the nonthermal models of Grindlay and Gurzadian are rejected and the expected X-ray emission of stellar flares is estimated by scaling the known data on solar flares.

For the first time, X-ray emission of a flare on a red dwarf was recorded by the Astronomical Netherlands Satellite (ANS) during monitoring of YZ CMi on 19 October 1974 (Heise et al., 1975). The satellite was equipped with two instruments produced by the Utrecht Laboratory for Space Research for recording soft (0.2–0.28 keV) and medium (1–7 keV) X-rays. Another instrument produced in the Cambridge Center for Astrophysics (USA) was designed for detecting hard X-rays (1–30 keV). In soft and medium X-rays, the emission burst on YZ CMi occurred practically simultaneously, but in soft X-rays the flare was observed for six minutes, whereas in medium X-rays — only for one and a half minutes. The Cambridge instrument did not detect the flare. The luminosity at flare maximum $L_X^{max}$ in the specified energy ranges reached $2.5 \cdot 10^{29}$ and $3.6 \cdot 10^{30}$ erg/s, and full radiation $E_X$ was $4 \cdot 10^{31}$ and $2 \cdot 10^{32}$ erg, respectively. The maximum recorded flux in soft X-rays was 0.14 quanta/(cm$^2 \cdot$ s). (Since flare stars are the objects nearest to the Sun, distances to them are determined rather precisely from trigonometrical parallaxes and interstellar absorption that depends on the number of neutral atoms on the line of sight can be neglected. Thus, in calculating the absolute luminosity $L_X$ from the observed flux $F_X$ an error can be caused by absolute calibration of the equipment and the deviation of radiation from isotropicity, the latter is practically excluded for thermal emission.) During the X-ray flare the star was not monitored in other wavelength ranges, and this flare was only recorded for 5 h, 5–12 min at each revolution.

The flare on UV Cet of 8 January 1975 was recorded in soft X-rays with ANS and in the optical range from the Earth (Heise et al., 1975). The optical flare was rather strong — $\Delta U \sim 6^m$ — and lasted more than six minutes, but X-ray emission of the flare was recorded from optical maximum only for 48 s. At the moment of maximum the recorded flux was 0.21

quanta/(cm$^2$· s), which corresponded to the flare luminosity of $6 \cdot 10^{28}$ erg/s. According to the observations, the ratio $L$(0.2−0.28 keV)/$L_U \leq 0.03$ was valid for this flare. These values appeared to be less than all these precalculated, the closest estimate to the real value was that of Kahler and Shulman. On the whole, ANS observations of UV Cet lasted for more than five hours, over this time three other weak optical flares were recorded, but they did not have appreciable X-ray emission.

The first successful observations of stellar flares in X-rays stimulated Mullan (1976a) to develop a purely thermal quantitative model. According to the model, hot coronal gas arising in the stellar atmosphere during a flare determines its further development: flare X-ray emission is caused by thermal radiation of gas, optical emission arises due to heating of underlying layers through heat conductivity. In the model, the ratio $L_X/L_{opt}$ near maximum brightness should be determined only by the ratio of radiative losses and losses for heat conductivity of hot coronal gas. The expected value of this ratio in flares on red dwarfs should be noticeably lower than on the Sun and make a few thousandths or hundredths.

The concept of thermal X-ray emission of stellar flares put forward by Mullan has now become conventional. During the joint USSR–USA manned space flight of the Soyuz–Apollo satellites the flare star Proxima Cen was observed in soft X-rays in the range of 0.065–0.28 keV for 1100 s on 21 July 1975 and for 78 s on 22 July 1975 using the telescope produced at the University of California (Haisch et al., 1977). During the second observational session, the signal, which noticeably exceeded both the background and the signal recorded the day before, was registered. Unfortunately, the X-ray observations were not accompanied by optical monitoring. American researchers attributed the signal to a flare and postulated its purely thermal nature. In the context of this hypothesis, quantitative characteristics of high-temperature components of the stellar flare were estimated for the first time. These estimates used a simple relation connecting specific luminosity of optically thin plasma of unit density and determined chemical composition $l_\lambda$(T) with the luminosity of the whole plasma body $L_\lambda = \int l_\lambda n_e^2 \, dV$; for homogeneous medium the relation is as follows

$$L_\lambda = l_\lambda\,(T) \int n_e^2 \, dV = l_\lambda(T) \mathrm{EM}, \quad (57)$$

where $EM$ is the volume emission measure of the source. To use (57), one should determine the plasma temperature from independent considerations. But if it is somehow fixed, from the found values $L_\lambda(T)$ and $EM$ one can easily estimate the total X-ray emission in the range of sensitivity of the detector

$$L_X = \mathrm{EM} \int l_\lambda (\mathrm{T}) \, d\lambda \quad (58)$$

According to Haisch et al. (1977), if on 22 July 1975 thermal emission of the optically thin plasma was recorded, at 2.5–100 MK the emission measure of the radiating volume was $3 \cdot 10^{52}$–$1 \cdot 10^{53}$ cm$^{-3}$, and its full luminosity throughout the X-ray range varied from $6 \cdot 10^{29}$ to $3 \cdot 10^{30}$ erg/s. Similarly, they considered X-ray flares on YZ CMi and UV Cet and estimated the full X-ray luminosity as $8 \cdot 10^{30}$ and $2 \cdot 10^{30}$ erg/s and emission measures as $5 \cdot 10^{53}$ and $1 \cdot 10^{53}$ cm$^{-3}$, respectively, for a temperature of radiating gas in the flares of $10^7$ K. (In the described calculations the values of $l_\lambda(T)$ were used for the standard chemical composition, found by Kato (1976) and Raymond et al. (1976); today, usually the Raymond-Smith and MEKA models and modifications of the latter are used.)

Quantitative parameters of the above stellar flares and those recorded later in different X-ray experiments are summed up in Table 17. The data are not quite uniform, since they were taken from original publications that appeared at different times and were not adjusted to the

Table 17. Characteristic of X-ray flares

| | Star | Date | Experiment | Range, keV | $T_b$, min | $T_a$, min | $L_X^{max}$, erg/s | $E_X$, erg | $T$, MK | EM, $cm^{-3}$ | $h$, cm | $n_e$ $cm^{-3}$ | Description and analysis of experiment |
|---|---|---|---|---|---|---|---|---|---|---|---|---|---|
| 1 | 2 | 3 | 4 | 5 | 6 | 7 | 8 | 9 | 10 | 11 | 12 | 13 | 14 |
| 1 | YZ CMi | 19.10.74 | ANS | 0.2–0.28<br>1–7 | 6 | 1.5 | $2.5 \cdot 10^{29}$<br>$3.6 \cdot 10^{30}$ | $4.2 \cdot 10^{31}$<br>$1.9 \cdot 10^{32}$ | $10^*$ | $5 \cdot 10^{53}$ | | | Heise et al., 1975 |
| 2 | UV Cet | 8.1.75 | ANS | 0.2–0.28 | | 0.8 | $6 \cdot 10^{28}$ | $3 \cdot 10^{30}$ | $10^*$ | $1 \cdot 10^{53}$ | | | Heise et al., 1975 |
| 3 | Prox Cen | 22.7.75 | Soyuz–Apollo | 0.065–0.29 | | >1.3 | $1.5 \cdot 10^{29}$ | | 2.5–100* | $5 \cdot 10^{52}$ | | | Heise et al., 1977 |
| 4 | YZ CMi | | SAS-3 | 0.15–50 | | | | | | | | | Karpen et al., 1977 |
| 5 | Prox Cen | | SAS-3 | 0.15–50 | | | | | $10^*$ | $<5 \cdot 10^{51}$ | | | Heise et al., 1978 |
| 6 | AT Mic | 25.10.77 | HEAO-1 | 0.15–18 | | >0.5 | $7 \cdot 10^{30}$ | $>2 \cdot 10^{32}$ | 30 | $1.4 \cdot 10^{54}$ | | | Kahn et al., 1979 |
| 7 | | II | | | | | | | 40 | | | | Connors et al., 1986 |
| 8 | AT Mic | 27.10.77 | HEAO-1 | 0.15–18 | | | $2 \cdot 10^{30}$ | | $30^*$ | $4.0 \cdot 10^{53}$ | | | Kahn et al., 1979 |
| 9 | AD Leo | 22.11.77 | HEAO-1 | 0.15–2.5 | | | $5 \cdot 10^{29}$ | | $30^*$ | $1.1 \cdot 10^{53}$ | | | Kahn et al., 1979 |
| 10 | | II | | | | | $7 \cdot 10^{29}$ | | $30^*$ | $1.4 \cdot 10^{53}$ | | | Kahn et al., 1979 |
| 11 | EV Lac | 25.12.77 | HEAO-1 | 0.5–20 | <300 | ~22 | $2.9 \cdot 10^{29}$ | | 10–30* | $3 \cdot 10^{52}$ | $5 \cdot 10^{9}$ | $(4–10) \cdot 10^{11}$ | Ambruster et al.,1984 |
| 12 | EV Lac | 30.12.77 | HEAO-1 | 0.5–20 | | <78 | $2.2 \cdot 10^{30}$ | $4 \cdot 10^{33}$ | 10–30* | $2 \cdot 10^{52}$ | | | Ambruster et al.,1984 |
| 13 | EQ Vir | 13.1.78 | HEAO-1 | 0.5–20 | | | $7.4 \cdot 10^{30}$ | | 32 | $8 \cdot 10^{51}$ | | | Ambruster and Wood, 1984, 1986 |
| 14 | Wolf 630 | 28.2.79 | IPC | 0.5–4.8 | 11 | >6 | $4 \cdot 10^{29}$ | | | | | | Jonhson, 1981 |
| 15 | WX UMa | 24.5.79 | IPC | | 8 | | $1 \cdot 10^{28}$ | | | | | | Jonhson, 1981, 1983b |

*Table 17 (continued)*

| 1 | 2 | 3 | 4 | 5 | 6 | 7 | 8 | 9 | 10 | 11 | 12 | 13 | 14 |
|---|---|---|---|---|---|---|---|---|---|---|---|---|---|
| 16 | Prox Cen | 6.3.79 | IPC | 0.2–4 | 5 | 25 | $7.4 \cdot 10^{27}$ | $>6 \cdot 10^{30}$ | 12–17 | $1 \cdot 10^{51}$ | $10^{10}$ | $10^{11}$ | Jonhson, 1981, 1983b |
| 17 | Prox Cen | 7.3.79 | IPC | | | 50 | | | | | | | Haisch et al., 1980, 1981 |
| 18 | YZ CMi | 25.10.79 | IPC | 0.2–4 | 1 | 10 | $3 \cdot 10^{28}$ | $3.6 \cdot 10^{31}$ | 20 | $2 \cdot 10^{51}$ | $10^{10}$ | $3 \cdot 10^{11}$ | Haisch et al., 1980, 1981 |
| 19 | YZ CMi | 27.10.79 | IPC | | 7 | | | $4 \cdot 10^{28}$ | | | | | Kahler et al., 1982 |
| 20 | AD Leo | 13.5.80 | IPC | 0.2–3.5 | | 160 | $3 \cdot 10^{29}$ | $2 \cdot 10^{33}$ | 12 | $1 \cdot 10^{52}$ | $1.3 \cdot 10^{10}$ | | Kahler et al., 1982 |
| 21 | TZ Ari | 14.7.80 | IPC | | 5 | >7 | $3 \cdot 10^{28}$ | | | $2 \cdot 10^{51}$ | | | Favata et al., 2000a |
| 22 | Prox Cen | 20.8.80 | IPC | 0.2–4 | 34 | 130 | $1.3 \cdot 10^{28}$ | $2 \cdot 10^{31}$ | 27 | | | | Haisch et al., 1983 |
| 23 | HD 27130 | 19.9.80 | IPC+ MPC | 0.2–10 | | >200 | $10^{31}$ | $2 \cdot 10^{34}$ | 50 | $6 \cdot 10^{53}$ | $2 \cdot 10^{10}$ | $4 \cdot 10^{11}$ | Stern et al., 1983 |
| 24 | BD + 14°690 | | IPC | 0.2–4 | 16 | 30 | $2 \cdot 10^{30}$ | | | | | | Stern and Zolcinski, 1983 |
| 25 | vA 500 | | IPC | 0.2–4 | 16 | 30 | | | | | | | Stern and Zolcinski, 1983 |
| 26 | Gl 669 | 27.3.79 | IPC | 0.2–4 | <3 | 10 | $>10^{29}$ | $10^{31}$ | 24 | | | | Harris and Johnson, 1985 |
| 27 | Gl 669 B | 19.8.80 | HRI | 0.2–4 | <8 | <8 | | | | | | | Harris and Johnson, 1985 |
| 28 | Gl 34 B | 4.2.80 | HRI | 0.2–4 | | | | | | | | | Harris and Johnson, 1985 |
| 29 | Gl 338 A | 27.4.80 | HRI | 0.2–4 | <12 | | | | | | | | Harris and Johnson, 1985 |

*Table 17 (continued)*

| 1 | 2 | 3 | 4 | 5 | 6 | 7 | 8 | 9 | 10 | 11 | 12 | 13 | 14 |
|---|---|---|---|---|---|---|---|---|---|---|---|---|---|
| 30 | V 1216 Sgr | 24.3.81 | IPC | 0.2–4 | <2 | 5 | $5 \cdot 10^{27}$ | $10^{30}$ | | | | | Agrawal et al., 1986 |
| 31 | HZ 1733 | 8.2.81 | IPC | 0.2–4 | | 17 | $6 \cdot 10^{28}$ | $6 \cdot 10^{31}$ | 20 | $3 \cdot 10^{51}$ | | | Caillault and Helfand, 1985 |
| 32 | 1E0419.2 + 1908 | 21.8.83 | LE | 0.02–2.5 | 40 | 80 | $6 \cdot 10^{29}$ | $6 \cdot 10^{32}$ | 20* | $3 \cdot 10^{53}$ | | | Smale et al., 1986 |
| 33 | $\pi^1$ UMa | 31.1.84 | LE + ME | 0.1–10 | 8 | 24 | | $2 \cdot 10^{33}$ | 30 | $7 \cdot 10^{52}$ | $2 \cdot 10^{10}$ | $7 \cdot 10^{11}$ | Landini et al., 1986 |
| 34 | Wolf 1561 | | LE | 0.05–2 | 10 | 30 | $5 \cdot 10^{29}$ | $5 \cdot 10^{32}$ | | | | | Pallavichini et al., 1990a |
| 35 | BY Dra | 24.9.84 | LE | | 15 | 30 | $1 \cdot 10^{30}$ | $6 \cdot 10^{32}$ | 10* | $1.2 \cdot 10^{52}$ | | | de Jager et al., 1986 |
| 36 | EXO 040830-7134.7 | 11.10.84 | LE + ME | 0.05–2 | | | $1.2 \cdot 10^{30}$ | $7 \cdot 10^{33}$ | | | | | van der Woerd et al., 1989 |
| 37<br>38 | YY Gem | 14.11.84<br>II | LE + ME | 0.05–6 | 35<br>10 | 65<br>25 | | $4.7 \cdot 10^{30}$<br>$1.3 \cdot 10^{30}$ | $6 \cdot 10^{33}$<br>$4 \cdot 10^{32}$ | from 64 to 24 | $6 \cdot 10^{52}$ | | Pallavichini et al., 1990ab |
| 39<br>40 | Gl 867 A<br>Gl 867AB | 18.11.84<br>II | LE<br>LE | 0.05–2<br>0.05–2 | 8<br>10 | 8<br>15 | $4.4 \cdot 10^{29}$<br>$5.1 \cdot 10^{29}$ | $6 \cdot 10^{31}$<br>$1 \cdot 10^{32}$ | | | | | Pollock et al., 1991 |
| 41<br>42<br>43<br>44 | UV Cet | 6.12.84 I<br>II<br>III<br>IV | LE<br>LE<br>LE<br>LE | 0.05–2 | <2<br><2<br>5<br><3 | 4<br>5<br>10<br>10 | $2.0 \cdot 10^{28}$<br>$1.9 \cdot 10^{28}$<br>$2.5 \cdot 10^{28}$<br>$3.0 \cdot 10^{28}$ | $2 \cdot 10^{30}$<br>$4 \cdot 10^{30}$<br>$5 \cdot 10^{30}$<br>$5 \cdot 10^{30}$ | | | | | Pallavichini et al., 1990a, Haisch et al., 1987 |
| 45<br>46<br>47<br>48 | EQ Peg | 7.12.84 I<br>II<br><br>III<br>IV | LE<br>LE<br>ME<br>LE<br>LE | 0.04–2<br><br>1.5–5.5 | 5<br>45<br>20<br>5<br>10 | 15<br>20<br>40<br>10<br>15 | $2 \cdot 10^{29}$<br>$4 \cdot 10^{29}$<br>$7 \cdot 10^{28}$<br>$1.5 \cdot 10^{29}$<br>$3.0 \cdot 10^{29}$ | $2 \cdot 10^{31}$<br>$2 \cdot 10^{32}$<br><br>$2 \cdot 10^{31}$<br>$4 \cdot 10^{31}$ | 26 | $1 \cdot 10^{52}$<br>$5 \cdot 10^{51}$ | $1 \cdot 10^{10}$<br>$6 \cdot 10^{9}$ | $2 \cdot 10^{11}$ | Haisch et al., 1987 |
| 49 | Prox Cen | 2.3.85 | LE | | 10 | 25 | $4 \cdot 10^{27}$ | $2 \cdot 10^{30}$ | | | | | Pallavichini et al., 1990a |

*Table 17 (continued)*

| 1 | 2 | 3 | 4 | 5 | 6 | 7 | 8 | 9 | 10 | 11 | 12 | 13 | 14 |
|---|---|---|---|---|---|---|---|---|---|---|---|---|---|
| 50<br>51 | YZ CMi | 4.3.85 I<br>II | LE<br>LE | 0.02–2<br>0.02–2 | 25<br>10 | 30<br>40 | $1 \cdot 10^{29}$<br>$1 \cdot 10^{29}$ | $7.8 \cdot 10^{31}$<br>$1.4 \cdot 10^{32}$ | | | | | Doyle et al., 1986 |
| 52 | Wolf 630 | 8.3.85 | LE<br>ME | 0.05–3<br>2–7 | 1<br>1 | 30<br>30 | $1.4 \cdot 10^{29}$ | | 22 | $8 \cdot 10^{51}$ | | | Tagliaferri et al., 1987 |
| 53 | VB 8 AB | 8.3.85 | LE | | 4 | 60 | $7.3 \cdot 10^{28}$ | $8 \cdot 10^{31}$ | | $6 \cdot 10^{50}$ | | $1 \cdot 10^{11}$ | Jonhson, 1987, Tagliaferri et al., 1987, 1990 |
| 54 | AT Mic | 24.5.85 | LE | | 3 | 30 | $6.5 \cdot 10^{29}$ | $1.3 \cdot 10^{33}$ | 39 | $5 \cdot 10^{52}$ | | | Pallavichini et al., 1990a |
| 55<br>56<br>57 | UV Cet | 4.8.85 I<br>II<br>III | LE | | <5<br>5<br>5 | 10<br>15<br>10 | $2 \cdot 10^{28}$<br>$2 \cdot 10^{28}$<br>$2 \cdot 10^{28}$ | $5 \cdot 10^{30}$<br>$5 \cdot 10^{30}$<br>$9 \cdot 10^{30}$ | | | | | Pallavichini et al., 1990a |
| 58<br>59 | EQ Peg | 6.8.85 I<br>II | LE<br>ME<br>LE | 0.04–2<br>120 | 10 | 120<br>60 | $2 \cdot 10^{30}$<br>$3.4 \cdot 10^{29}$ | $4 \cdot 10^{33}$<br>$3 \cdot 10^{32}$ | from 42 to 18 | from $2 \cdot 10^{53}$ to $2 \cdot 10^{53}$ | | | Pallavichini et al.,1986, 1990a |
| 60 | Wolf 630 | 25.8.85 | LE<br>ME | 0.05–2<br>2–7 | 5<br>5 | 15<br>25 | $4.6 \cdot 10^{29}$ | $9 \cdot 10^{32}$<br>$4 \cdot 10^{32}$ | from 48 to 29 | $3 \cdot 10^{53}$ | | $1 \cdot 10^{12}$ | Doyle et al., 1988a, Tagliaferri et al., 1987 |
| 61 | EV Lac | 13.10.85 | LE | | 120 | | | | | | | | Ambruster et al., 1989a |
| 62 | EV Lac | 15.10.85 | LE | | 10 | 75 | $1.3 \cdot 10^{32}$ | | | | | | Ambruster et al., 1989a |
| 63 | YZ CMi | 19.11.85 | LE | | 8 | 20 | $8.2 \cdot 10^{29}$ | $7 \cdot 10^{31}$ | | | | | Pallavichini et al., 1990a |
| 64 | AD Leo | 15.12.85 | LE | | 10 | 60 | | $4.8 \cdot 10^{29}$ | $1.5 \cdot 10^{33}$ | | | | Pallavichini et al., 1990a |
| 65 | BD + 48°1958A | 16.12.85 | LE+<br>ME | | | | $6 \cdot 10^{29}$ | $2 \cdot 10^{32}$ | | $3 \cdot 10^{51}$ | | $2 \cdot 10^{11}$ | Rao et al., 1990 |

*Table 17 (continued)*

| 1 | 2 | 3 | 4 | 5 | 6 | 7 | 8 | 9 | 10 | 11 | 12 | 13 | 14 |
|---|---|---|---|---|---|---|---|---|---|---|---|---|---|
| 66<br><br>67 | UV Cet | 23.12.85<br><br>II | LE<br>ME<br>LE | | <3<br><3<br><3 | 20<br>30<br>10 | $1.0 \cdot 10^{29}$<br><br>$1.0 \cdot 10^{29}$ | $6 \cdot 10^{31}$<br><br>$3 \cdot 10^{30}$ | 10*<br>40 | | | $5 \cdot 10^{11}$<br>$2 \cdot 10^{11}$ | de Jager et al., 1989, Pallavichini et al. |
| 68 | CC Eri | 10.7.90 | PSPC | 0.2–2 | 60 | 220 | $7 \cdot 10^{29}$ | | 28–12 | | | | Pan and Jordan, 1995 |
| 69 | HD 147365 | 30.7.90 | PSPC | | 360 | 360 | $4 \cdot 10^{29}$ | $1 \cdot 10^{34}$ | | | | | Güdel et al., 1995b |
| 70 | 47 Cas | 19.8.90 | PSPC | 0.1–2.0 | | <90 | $4.3 \cdot 10^{30}$ | | 25* | $4 \cdot 10^{53}$ | $4 \cdot 10^{10}$ | $5 \cdot 10^{11}$ | Güdel et al., 1995b |
| 71 | 47 Cas | 21.8.90 | PSPC | 0.1–2.0 | | 90 | $1.2 \cdot 10^{31}$ | $5 \cdot 10^{34}$ | 25* | | | | Güdel et al., 1995b |
| 72 | AZ Cnc | 19.10.90 | PSPC | 0.1–2.4 | | 500 | | | | | | | Fleming et al., 1993 |
| 73 | Prox Cen | 8.10.90 | PSPC | 0.1–2.4 | | 300 | | | | | | | Fleming et al., 1993 |
| 74 | CN Leo | 25.11.90 | PSPC | 0.1–2.4 | | 200 | | | | | | | Fleming et al., 1993 |
| 75 | EK Dra | 11.90 | PSPC | 0.4–2.4 | 60 | | $1 \cdot 10^{30}$ | $4 \cdot 10^{33}$ | | | $(1–2) \cdot 10^{10}$ | $2 \cdot 10^{11}$ | Güdel et al., 1995a |
| 76 | EQ 1839.6 +8002 | 14.2.91 | LAC GINGA | 1–37 | 4 | 20 | $10^{31}$ | $10^{34}$ | | | | | Pan et al., 1997 |
| 77 | VB 8 | 25–27.2.91 | PSPC | | | >100 | | | 2 and 10 | $5 \cdot 10^{49}$ | | | Reale and Micela, 1998 |

*Table 17 (continued)*

| 1 | 2 | 3 | 4 | 5 | 6 | 7 | 8 | 9 | 10 | 11 | 12 | 13 | 14 |
|---|---|---|---|---|---|---|---|---|---|---|---|---|---|
| 78<br>79 | AD Leo | 8.5.91 I<br>II | PSPC<br>PSPC | | 60 | 160 | $2 \cdot 10^{29}$ | $2 \cdot 10^{32}$ | 12 | $1 \cdot 10^{51}$ | $<1.7 \cdot 10^{10}$ | | Springfellow, 1996, Favata et al., 2000a |
| 80 | UV Cet | 31.12.91 | PSPC | | | | | | 10 | | | $10^{11}$ | Stepanov et al., 1995 |
| 81<br>82 | UV Cet | 2.1.92 I<br>II | PSPC<br>PSPC | | 0.1<br>0.1 | 0.1<br>5 | | | | | | | Schmitt et al., 1993a |
| 83 | EV Lac | 13.7.92 | PSPC | | | | | | 8 and 35 | 8 and $23 \cdot 10^{51}$ | | | Sciortino et al., 1999 |
| 84 | LQ Hya | 5.11.92 | PSPC + ASCA +SAX | 0.5–10 | | 240 | $A_X > 10$ | | 7–10 | $(0.9–21) \cdot 10^{53}$ | | | Covino et al., 2001a |
| 85 | G 201-21 | 23.9.93 | PSPC | 0.2–2.4 | | 16 | $5 \cdot 10^{29}$ | $>5 \cdot 10^{32}$ | 5 and 16 | $(8–50) \cdot 10^{51}$ | | $10^{12}$ | Micela et al., 1995 |
| 86 | YY Gem | 26.10.93 | ASCA | 0.4–7 | 1.6 | 70 | | | 11–3 | | | | Gotthelf et al., 1994 |
| 87 | Castor AB | 26.10.93 | ASCA | 0.4–7 | | | | | | | | | Gotthelf et al., 1994 |
| 88 | H II 2034 | 11.8.90 | PSPC | 0.5–1.8 | | | $3 \cdot 10^{31}$ | | 10 | | | | Schmitt et al., 1993b |
| 89 | H II 2147 | | PSPC | | <18<br>0 | <1000 | $2.5 \cdot 10^{31}$ | | 4 and 14 | 2 and $11 \cdot 10^{53}$ | | | Gagné et al., 1994a, 1995 |
| 90 | H II 1516 | | PSPC | | <80 | 80 | $1.6 \cdot 10^{31}$ | | 13 | $2 \cdot 10^{54}$ | | $>1.3 \cdot 10^{11}$ | Gagné et al., 1995 |
| 91 | HCG 97 | | PSPC | | <60 | 80 | $1.6 \cdot 10^{30}$ | | | | | | Gagné et al., 1995 |

*Table 17 (continued)*

| 1 | 2 | 3 | 4 | 5 | 6 | 7 | 8 | 9 | 10 | 11 | 12 | 13 | 14 |
|---|---|---|---|---|---|---|---|---|---|---|---|---|---|
| 92 | H II 174 | | PSPC | | 90 | 220 | $3.2 \cdot 10^{30}$ | | | | | | Gagné et al., 1995 |
| 93 | H II 191 | | PSPC | | 16 | 80 | $3.2 \cdot 10^{30}$ | | | | | | Gagné et al., 1995 |
| 94 | H II 212 | | PSPC | | <70 | 100 | $2 \cdot 10^{30}$ | | | | | | Gagné et al., 1995 |
| 95 | HCG 143 | | PSPC | | 10 | 25 | $2.5 \cdot 10^{30}$ | | | | | | Gagné et al., 1995 |
| 96 | H II 345 | | PSPC | | <70 | 110 | $5 \cdot 10^{30}$ | | | | | | Gagné et al., 1995 |
| 97 | HCG 181 | | PSPC | | 16 | 80 | $3 \cdot 10^{30}$ | | | | | | Gagné et al., 1995 |
| 98 | H II 1100 | | PSPC | | 50 | 160 | $3 \cdot 10^{30}$ | | | | | | Gagné et al., 1995 |
| 99 | XR 191 | | PSPC | | <75 | 180 | $8 \cdot 10^{30}$ | | | | | | Gagné et al., 1995 |
| 100 | H II 2244 | | PSPC | | <80 | 130 | $4 \cdot 10^{30}$ | | | | | | Gagné et al., 1995 |
| 101 | HE 421 | | PSPC | | <70 | <750 | $6 \cdot 10^{30}$ | $>6 \cdot 10^{33}$ | | | | | Prosser et al., 1996 |
| 102 | AP 20 | | PSPC | | <150 | <840 | $1 \cdot 10^{31}$ | $>2 \cdot 10^{34}$ | | | | | Prosser et al., 1996 |
| 103 | AP 108 | | PSPC | | <70 | <840 | $4 \cdot 10^{31}$ | $>3 \cdot 10^{34}$ | | | | | Prosser et al., 1996 |
| 104 | Prox Cen | 10–20.3.94 | ASCA | 0.5–12 | | | | | 44 and 7 | | | | Haisch et al., 1995 |
| 105 | EQ Peg | 23.6.94 | ROSAT HRI | 0.1–2.4 | 2 | 10 | $1 \cdot 10^{30}$ | $2 \cdot 10^{32}$ | 27 | $3 \cdot 10^{52}$ | | $4 \cdot 10^{10}$ | Katsova et al., 2002 |

*Table 17 (continued)*

| 1 | 2 | 3 | 4 | 5 | 6 | 7 | 8 | 9 | 10 | 11 | 12 | 13 | 14 |
|---|---|---|---|---|---|---|---|---|---|---|---|---|---|
| 106 | UV Cet | 5/6.01.95 | ROSAT + ASCA | 0.1–2.4 | | 80 | $3 \cdot 10^{27}$ | $2 \cdot 10^{31}$ | | | | | Güdel et al., 1996 |
| 107 | UV Cet | 6/7.01.95 | ROSAT + ASCA | 0.5–10 | | 70 | $6 \cdot 10^{27}$ | $3 \cdot 10^{31}$ | | | | | Güdel et al., 1996 |
| 108 | HD 197890 | 20.4.95 | ASCA | 0.4–10 | | | $2 \cdot 10^{30}$ | | 9 and 37 | 6 and $12 \cdot 10^{52}$ | | | Singh et al., 1999 |
| 109 | Gl 890 | 19.11.95 | ASCA | 0.4–10 | | 40 | $4 \cdot 10^{29}$ | | 9 and > 23 | 2 and $0.2 \cdot 10^{52}$ | | | Singh et al., 1999 |
| 110<br>111<br>112 | AD Leo | 4.5.96 I<br>II<br>III | ASCA<br>ASCA<br>ASCA | 0.5–10<br>0.5–10<br>0.5–10 | | 10<br>20<br>60 | $9 \cdot 10^{28}$<br>$3 \cdot 10^{29}$<br>$7 \cdot 10^{28}$ | $1 \cdot 10^{32}$<br>$8 \cdot 10^{31}$<br>$2 \cdot 10^{32}$ | 20<br>48<br>38 | $1 \cdot 10^{52}$<br>$5 \cdot 10^{52}$<br>$2 \cdot 10^{52}$ | $<10^{10}$<br>$4 \cdot 10^{9}$<br>$7 \cdot 10^{9}$ | | Favata et al., 2000a |
| 113 | AU Mic | 13.6.96 | RXTE | 2–15 | <13 | 1000 | | | 20 and 93 | | | | Gagné et al., 1998 |
| 114 | EQ Peg | 2.10.96 | RXTE | 2–10 | | | | | | | | | Gagné et al., 1998 |
| 115 | AD Leo | 23-24.4.97 I, II, III | SAX | 0.1–7 | | 13, 67, 73 | $A_X \sim 4–5$ | | 4, 11 and >100 | 2, 3, and $0.3 \cdot 10^{51}$ | | | Sciortino et al., 1998, 1999 |
| 116 | VB 10 | 19.10.97 | PSPC | | | <19 | $3 \cdot 10^{27}$ | | | | | | Fleming et al., 2000 |
| 117 | AB Dor | 9.11.97 | SAX | 0.1–8 | | | | | 9–109 | $6 \cdot 10^{54}$ | | $(5–16) \cdot 10^{11}$ | Maggio et al., 2000 |
| 118 | AB Dor | 29.11.97 | SAX | 0.1–8 | | | | | 8–110 | $4 \cdot 10^{54}$ | | $(3–9) \cdot 10^{11}$ | Maggio et al., 2000 |
| 119 | EQ Peg | 3.12.97 | SAX | 0.1–5 | 300 | 330 | | | | | | | Landi et al., 2001 |

*Table 17 (continued)*

| 1 | 2 | 3 | 4 | 5 | 6 | 7 | 8 | 9 | 10 | 11 | 12 | 13 | 14 |
|---|---|---|---|---|---|---|---|---|---|---|---|---|---|
| 120 | EV Lac | 7–8.12.97 I, II | SAX | 0.1–7 | | 75 and 23 | | | | 3, 2, 8, 3 and 25 | 5, 4 and $8 \cdot 10^{51}$ | | Sciortino et al., 1999 |
| 121 | EV Lac | 13.7.98 | ASCA | 0.5–10 | ~240 | | $A_X$~300 | $1.5 \cdot 10^{34}$ | 73 | $7.4 \cdot 10^{53}$ | $1.3 \cdot 10^{10}$ | $(2\text{-}0.2) \cdot 10^{12}$ | Favata et al., 2000b |
| 122 | YY Gem | 4–5.11.98 | SAX | 0.1–10 | | | | | | | | | Tagliaferri et al., 2001 |
| 123 | AD Leo | 2.4–16.5.99 | SAX | | | | | | | $10^{51}$ | | | Güdel et al., 2001 |
| 124 | YY Gem | 25.4.00 | XMM-Newton | | | | | | 9–37 | $(0.4–3.4) \cdot 10^{52}$ | | $3 \cdot 10^{10}$ | Güdel et al., 2000b |
| 125 | LP 944-20 | 15.12.99 | Chandra | 0.1–4.0 | | 90 | $1 \cdot 10^{26}$ | $2 \cdot 10^{29}$ | 3 | | | | Rutledge et al., 2000 |
| 126 | AB Dor | 30.4/1.5.00 | XMM-Newton | 0.3–10 | | | | | 1–3 | $(2–6) \cdot 10^{52}$ | | $3 \cdot 10^{10}$ | Güdel et al., 2000c |
| 127<br>128 | YY Gem | 29/30.9.00 I<br>29/30.9.00 II | XMM-Newton + Chandra | | | 30<br>45 | $8 \cdot 10^{29}$<br>$5 \cdot 10^{29}$ | | 10–40<br>10–46 | $3 \cdot 10^{51}$<br>$2 \cdot 10^{51-52}$ | $10^{9}$<br>$10^{9}$ | | Stelzer et al., 2002 |
| 129 | Castor A | 29/30.9.00 I-VI | XMM-Newton+ Chandra | | 12 | 36 | | $1 \cdot 10^{32}$ | 2–18 | $(1.7–3.1) \cdot 10^{51}$ | | $(0.5–1) \cdot 10^{10}$ | Stelzer and Burwitz, 2003 |
| 130 | Castor B | 29/30.9.00 I-VI | XMM-Newton+ Chandra | | 9 | 16 | | $6 \cdot 10^{31}$ | 3–18 | $(1.7–3.1) \cdot 10^{51}$ | | $(0.5–1) \cdot 10^{10}$ | Stelzer and Burwitz, 2003 |
| 131 | YZ CMi | 9.10.00 | XMM-Newton | 0.3–10 | | | $(19–27) \cdot 10^{27}$ | | 1.3–42 | $(14–22) \cdot 10^{50}$ | $(5–13) \cdot 10^{9}$ | | Raassen, 2005; Raassen et al., 2007 |

*Table 17 (continued)*

| 1 | 2 | 3 | 4 | 5 | 6 | 7 | 8 | 9 | 10 | 11 | 12 | 13 | 14 |
|---|---|---|---|---|---|---|---|---|---|---|---|---|---|
| 132 | AT Mic | 16.10.00 | XMM-Newton | 1–40Å | | | | | 1–60 | $2\cdot 10^{52}$ | | $4\cdot 10^{10}$ | Raassen et al., 2003b |
| 133 | AD Leo | 14.3.01 I-VI | XMM-Newton + Chandra | 0.3–10 | | | $4.5\cdot 10^{28}$ | | 2.7–20 | $(0.8–2)\cdot 10^{51}$ | | $(3–7)\cdot 10^{10}$ | van den Besselaar et al., 2003 |
| 134 | ER Vul | 29/30.3.01 | Chandra | | | 30 | | | 2–30 | | | | Brown et al., 2002; Hussain et al., 2012 |
| 135 | UV Cet AB | 26/27.11.01 | Chandra | 0.8–10 | | | $2\cdot 10^{26}$ | | | | | | Audard et al., 2003 |
| 136 | Prox Cen | 12.8.01I | XMM-Newton | 0.15–2.5 | | 10 | $(27–14)\cdot 10^{28}$ | $2\cdot 10^{32}$ | 2–5 | $10^{50}$ | $1.1–0.2\cdot 10^{11}$ | | Güdel et al., 2002a; Reale et al., 2004 |
| 137 | | II | | | | 10 | $4\cdot 10^{28}$ | $1\cdot 10^{32}$ | 2–4 | | $10^{10}–4\cdot 10^{11}$ | | |
| 138 | EV Lac | 20.9.01 I-IX | Chandra | 0.3–10 | 7.7 236 | | $(2.3–29)\cdot 10^{28}$ | $(0.21–43)\cdot 10^{31}$ | 3.6–30 | $(1.0–6.8)\cdot 10^{51}$ | | | Osten et al., 2005 |
| 139 | CC Eri | 8.8.03 | XX-Newton | 0.5–10 | | | | | 3–20 | | $1.4\cdot 10^{10}$ | $(1–2)\cdot 10^{11}$ | Crespo-Chacon et al., 2007 |
| 140 | CN Leo | 19.5.06 | XX-Newton | 0.6–1.1 | 0.1 25 | | | | | | | $>10^{12}$ | Liefke et al., 2007 |
| 141 | DG CVn | 23.4.14 | Swift | 15–50 | | | | | | | | | Caballero-Garcia et al., 2015 |
| 142 | EQ Peg | 21-23.10.19 I-III | AstroSat | 0.3-7 | 3.4–11 | 1.6–24 | $(0.5–1.0)\cdot 10^{31}$ | $(0.2–1.3)\cdot 10^{35}$ | 3–9 | $(3.9–7.1)\cdot 10^{53}$ | $(2.0–2.5)\cdot 10^{11}$ | $(2.2–4.2)\cdot 10^{10}$ | Karmakar et al., 2022 |

*Coronal temperatures are taken from independent concerns

development of stricter analysis methods. Moreover, for burning $T_b$ and decay $T_a$ times of flares some publications present the full duration of these phases, and the others, the times of an *e*-fold increase of brightness to the maximum level and an *e*-fold decrease after the maximum. In some case, the values $L_X^{max}$ and $E_X$ were calculated for Table 17 from the published values of luminosity and total energy in the whole X-ray range. Therefore, Table 17 is not suitable for strict statistical consideration but is convenient for approximate estimates.

Table 17 has sufficiently been supplemented with data from the work recently published by Pye et al. (2015), which contains information on the X-ray flares occasionally detected with XMM-Newton on cool, according to Hipparcos and Tycho-2, stars. This work has information as regards 130 flares on 70 stars; the duration of these flares is within $10^3$–$10^4$ s with the maximum luminosity $10^{32}$–$10^{35}$ erg. The given variations of X-ray radiation hardness during the flares provide evidence for significant temperature variations of the radiating plasma.

In late 1975, the first simultaneous monitoring in optical, radio, and X-ray wavelength ranges was carried out (Karpen et al., 1977). Observations in the range of 0.15–50 keV were executed from SAS-3 using the X-ray telescope produced at the Massachusetts Institute of Technology. The X-ray monitoring of YZ CMi continued during 15 optical and 7 radio flares but did not reveal hard emission. The three strongest optical flares yielded the ratio $L$(0.15–0.8 keV)/$L_B$ < 0.3. This value is between the values predicted by Kahler and Shulman (1972) and Mullan (1976a).

In May 1977, Proxima Cen was observed within the framework of similar cooperative radio–optical–X-ray studies from SAS-3 (Haisch et al., 1978). During the X-ray observations 22 optical flares were recorded from the Earth, but no reliable X-ray flares were revealed. Like before, for the strongest optical flare the ratio $L_X/L_{opt}$ < 0.08 was obtained.

Important results on hard flare emission from the UV Cet-type stars were obtained at the American High-Energy Astrophysics Observatory HEAO-1. Between August 1977 and January 1979 the satellite carried out three all-sky surveys in soft (0.15–3 keV), medium (2–20 keV), and hard (2–60 keV) X-rays. Each source was in the field of view of the HEAO-1 instruments for 10 s at each revolution, every 30 min within a week. In October 1977, two X-ray flares on AT Mic were recorded. The flare on 25 October 1977 was so strong that the for first time an X-ray spectrum of the flare was recorded (see Fig. 52) (Kahn et al., 1979). In the spectrum within the energy range of 0.2– 20 keV thermal radiation of an optically thin plasma was confidently found at 31 ± 7 MK and the iron emission line $K_\alpha$ with excitation energy of 6.7 keV, which at the particular temperature really had to be excited effectively. The total X-ray luminosity of this flare was $1.6 \cdot 10^{31}$ erg/s, the emission measure of $1.4 \cdot 10^{54}$ cm$^{-3}$, and total X-ray emission of at least $5 \cdot 10^{32}$ erg. If one assumes that the spectrum of the second, weaker flare on AT Mic of 27 October 1977 was the same, the following parameters are valid for it: total X-ray luminosity of $4.6 \cdot 10^{30}$ erg/s and EM = $4.0 \cdot 10^{53}$ cm$^{-3}$.

During similar HEAO-1 observations of the flare star AD Leo on 22 November 1977 two rather weak X-ray flares were revealed. Their parameters cited in Table 17, on the assumption that they had the same spectrum as the flare on AT Mic of 25 October 1977, are rather close to the appropriate values of the flare on AT Mic.

HEAO-1 observations were not accompanied by optical monitoring. But using the statistical properties of optical flares obtained for the stars observed by Kunkel (1975a), Kahn et al. concluded that in these flares $L_X/L_{opt}$ > 1. This contradicted Mullan's thermal model, in which heat conductivity from the corona downward was considered as a determinative of excitation of an optical flare, but was in agreement with the observations of the strong solar flare of 4 August 1972, which yielded much more complete and authentic data than previous estimates from many weaker solar flares.

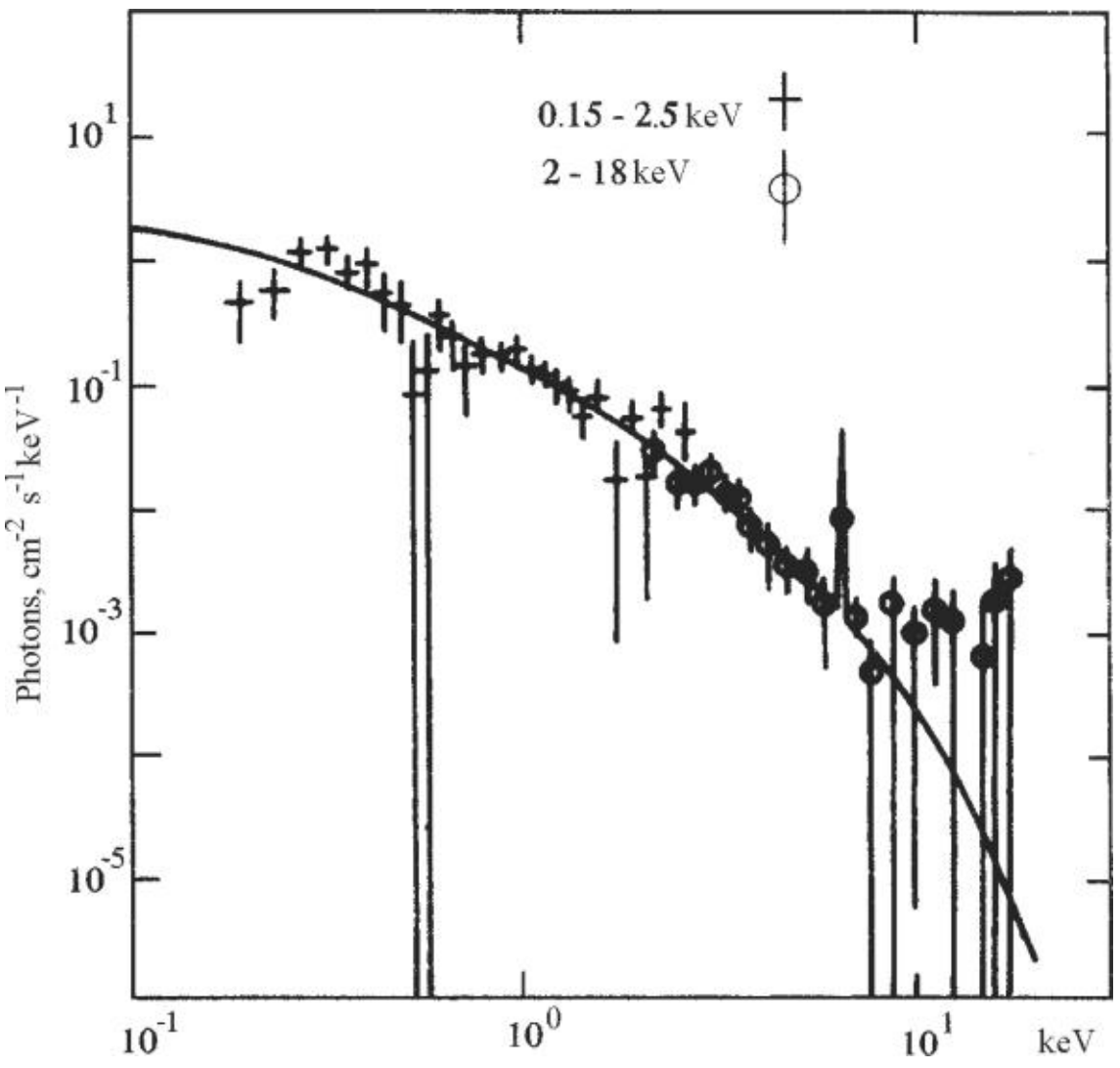

Fig. 52. X-ray spectrum of the flare on AU Mic of 25 October 1977 (Kahn et al., 1979)

Ambruster et al. (1984) found from the HEAO-1 data two flares on EV Lac (25 and 30 December 1977) with X-ray luminosities of $3 \cdot 10^{29}$ and $2 \cdot 10^{30}$ erg/s, respectively. There are suspicions that on 25 December 1977 it was the decay stage of a strong flare that was recorded, and three and four hours after the burst on 30 December 1977 there was still increased X-ray luminosity. Since the time interval between these bursts was equal to 1.25 rotational periods of the star it is rather probable that both flares took place in the same active region. The amplitude of the X-ray flare of 30 December 1977 was very high: $A_X \sim 54$, higher by an order of magnitude than that of earlier recorded X-ray flares on red dwarfs. At 10–30 MK the measured flux corresponded to EM = $3 \cdot 10^{52}$ and $2 \cdot 10^{53}$ cm$^{-3}$ for the first and second flares, respectively. The authors estimated for X-rays $(\Sigma P/T)_X = 0.5$, which was higher by one to one and a half orders of magnitude than the appropriate ratio for the U band. But one should take into account the qualitative distinctions of these ratios in the two considered intervals of the spectrum: in the optical range the stellar photospheric emission is used as a photometric standard, which physically is not connected to flares, whereas in X-rays physically homogeneous emissions of hot gas of the disturbed and quiet stellar corona are compared. Though two flares are not sufficient for statistical conclusions, we note that the total energy of X-ray emission recorded on EV Lac bursts by one order of magnitude surpassed the total energy of flare emission in the U band, which was estimated from the available statistics for optical flares on this star.

Similar analysis of the HEAO-1 data allowed Ambruster and Wood (1984) to find a flare on EQ Vir on 13 January 1978.

Finally, Connors et al. (1986) thoroughly analyzed the entire databank of about 40000 HEAO-1 scans obtained over one and a half years to find fast phenomena (1–5 s). They analyzed the records in medium X-rays, which are less subject to the effect of interstellar absorption than soft X-rays, and found five new bursts. On the basis of statistical consideration of the data they concluded that annually about 20000 flares of duration from 1 to 30 min occurred in the whole sky, which provided fluxes within 2–20 keV of more than $10^{-10}$ erg/cm$^{-2}$ s on the Earth. Most probably, the majority of bursts are concerned with flares on red dwarf stars.

* * *

The next important step in the studies of X-ray emission of stellar flares was the launch of the Einstein Observatory. By this time, an up-to-date concept on the upper solar atmosphere and the governing role of local magnetic fields in its formation were formed. According to these concepts, the magnetic flux tubes connecting solar spots of various polarity form steady discrete arches, plasma loops, in which plasma can move along but not across the axes (see Fig. 21). The arches reach extreme heights of the order of the solar radius and to a considerable extent determine both the properties of the quiet corona and the development of solar flares.

Already during the epoch of pure optical observations two characteristic types were distinguished from the visible picture in $H_\alpha$ rays in an immense variety of solar flares: rather fast and less strong compact flares and longer and stronger two-ribbon flares. Further studies showed that the basic distinction of these types was related to the topology of local magnetic fields: compact flares developed in individual closed magnetic loops, which remained invariant throughout the flare process, whereas two-ribbon flares involved the whole arcades of loops, a strong initial disturbance in them led to breakage of initial loops; the flux tubes of a new open magnetic configuration then closed again, the field energy reduced, and the portion of energy released provided long additional heating of flare matter. In most solar flares, impulsive and gradual phases can be distinguished. The impulsive phase is characterized by a short-lived burst of hard X-ray emission, which comes from a few small sites at the base of coronal arches placed close on both sides of the neutral line of magnetic field. This suggests bombardment and heating of the chromosphere by fluxes of energetic particles. The gradual phase of flares begins simultaneously with the impulsive one, but lasts longer; it has a smooth light curve, radiation occurs in the range of soft X-rays and proceeds from the rather extended diffuse region. The radiation is attributed to hot plasma, which rises from the arch base, propagates over a large area, and slowly spends its energy for radiation and heat conductivity. Finally, one should bear in mind that the described types of flares are seldom realized in a pure form in active solar regions: as a rule, in real flares the characteristics of one or the other kind dominate.

Table 18 lists typical parameters of the arches of weak and strong solar flares obtained in observations.

Let us return to stellar flares.

Rewriting (18) in the following way

$$T_{\max} = 6.3 \cdot 10^{-3} \, (n_e h)^{1/2}, \qquad (59)$$

where $n_e$ is the density at $T_{\max}$, Natanzon (1981) showed that flares on YZ CMi and UV Cet recorded by ANS, the flare on Proxima Cen detected by Soyuz–Apollo, and two flares on AD Leo recorded by HEAO-1 satisfied this relation as well. It should be noted, however, that the estimates of the parameters of stellar flares involved in (59) from the first rather limited X-ray observations required additional assumptions: the decay times of flares were determined for the heat conductivity of hot gas, cross sections of arches with flares covered the same part of stellar surface as was estimated for the total area of active regions on two flare stars — AD Leo and EQ Peg — from optical observations during different seasons, the vertical extension of flares was equal to the scale of heights of the appropriate stellar chromosphere. These assumptions are not obvious, but, nevertheless, it is clear that a detailed discussion of the local processes occurring in the atmospheres of flare stars is impossible without taking into account the structure of the atmospheres.

Table 18. Parameters of solar flare arches (Pallavicini et al., 1990a)

| Parameters | Weak flares (compact) | Strong flares (two-ribbon) |
|---|---|---|
| Luminosity at maximum, erg/s | $10^{26}$–$10^{27}$ | $10^{27}$–$10^{28}$ |
| Temperature, MK | 10–30 | 10–30 |
| Emission measure, cm$^{-3}$ | $10^{48}$–$10^{49}$ | $10^{49}$–$10^{50}$ |
| Electron density, cm$^{-3}$ | $10^{11}$–$10^{12}$ | $10^{10}$–$10^{11}$ |
| Loop height, cm | $10^{9}$ | $10^{10}$ |
| Field strength, G | 100–300 | 50–100 |

On 28 February 1979, the Einstein Observatory recorded the first stellar flare in the Wolf 630 AB system (Johnson, 1981). This flare and those of 24 May 1979 in the BD+44°2051 AB system and of 14 July 1980 on TZ Ari stars were detected as a byproduct of the survey undertaken by Johnson in X-rays for a large sample of objects near the Sun (Johnson, 1981, 1983a). Since each object was observed only for about half an hour and there was no optical support, the data on these flares are rather limited. The flare on Wolf 630 AB was recorded from the beginning of burning till the beginning of decay for 11 min and, since the duration of 12 recorded optical flares on this star varied from 8 to 250 s, either X-ray emission was found during a flare that was very slow for this star or it lasted noticeably longer than the optical one. The rise and decay of the flare were synchronous in the soft and harder regions of the spectrum, but the flare amplitude was higher in harder X-rays. Spectral analysis of these observations of Wolf 630 AB yielded, at a flare temperature of 32 MK, X-ray luminosity at maximum reached $4 \cdot 10^{29}$ erg/s, and EM = $8 \cdot 10^{51}$ cm$^{-3}$. The X-ray flare on BD+44°2051 AB lasted about eight minutes and its luminosity at maximum was about $1 \cdot 10^{28}$ erg/s. The flare on TZ Ari lasted about 10 min, but till the end of observations of this star that lasted for about 20 min its X-ray emission remained at the level that was noticeably higher than that of the quiescent state; the luminosity of the flare in a maximum was $3 \cdot 10^{28}$ erg/s and EM = $2 \cdot 10^{51}$ cm$^{-3}$.

On 6 March 1979, the Einstein Observatory recorded the first stellar flare within special monitoring of Proxima Cen (Haisch et al., 1980). On time averaging of 5 min at the stage of burning one readout was obtained, at luminosity maximum — two, and in the decay phase — three; an *e*-fold decrease of brightness took 20 min. As to the time characteristics, the flare was rather close to typical strong solar flares. During this event, for the first time the change of spectral structure of X-ray emission was confidently recorded, and the analysis of this radiation resulted in the estimates presented in Table 19.

Thus, the maximum temperature was achieved at the phase of burning of the flare, i.e., prior to maximum X-ray luminosity, and the value of maximum temperature and the time shift between temperature and luminosity maxima were close to the appropriate values typical of solar flares.

Table 19. Parameters of the X-ray flare on Proxima Cen of 6 March 1979 (Haisch et al., 1980, 1981)

| Parameters | $L_X$(0.2–4 keV), erg/s | $T$, MK | EM, cm$^{-3}$ |
|---|---|---|---|
| Preflare state | $1.5 \cdot 10^{27}$ | 4 | |
| Burning | $6.0 \cdot 10^{27}$ | 17 | $0.8 \cdot 10^{51}$ |
| Brightness maximum | $7.4 \cdot 10^{27}$ | 12 | $1.3 \cdot 10^{51}$ |
| 70 min after the maximum | $1.8\ 10^{27}$ | 6 | |

During the flare on Proxima Cen simultaneous observations of the star were arranged for the first time: in X-rays, in the ultraviolet from the IUE satellite, in the optical range at four telescopes of three Australian observatories, and in the radio range at the 64-meter Parkes radio telescope. But the flare was recorded only in X-rays. Haisch et al. (1981) concluded that the thermal radio emission, which should be expected from hot plasma responsible for the revealed hard emission, should have intensity below the detection of the equipment used, while the rather probable nonthermal radio emission of the flare could be directed and thus did not reach the Earth. The absence of accompanying phenomena in optical and ultraviolet ranges can be explained by spatial separation of the regions of formation of flare radiations, which is typical of the loop model and shielding of the bottom parts of arches, where optical and ultraviolet flare radiation is formed, with stellar limbs. However, another interpretation is possible. Estimating the upper limit of optical emission from observations, we get $L_X/L_{opt} > 1$. On the other hand, the recorded X-ray emission can be provided by 3–4 loops similar to those in which strong solar flares of importance 3 are observed: loops with a height of $10^{10}$ cm, cross-sectional radius of $10^9$ cm, and an average density of $10^{11}$ cm$^{-3}$. For such characteristics of hot gas the ratio of radiative losses to those for heat conductivity definitely exceeds unity. Thus, we should consider an alternative to the model proposed by Mullan (1976a): hot coronal structure relaxing at the expense of radiative cooling. In this case, it is natural to expect $L_X/L_{opt} > 1$, and the time of radiative cooling of such gas for the obtained estimates of its temperature and density should coincide with the characteristic time of flare decay. The parameters $h$ and $n_e$ of the EV Lac flare of 25 December 1977 (Ambruster et al., 1984) summarized in Table 17 were also obtained on the assumption that the flare decay time was determined by the time of radiative cooling of relaxed coronal plasma.

The successful observations of the flare on YZ CMi of 25 October 1979 (see Fig. 35) at the Einstein Observatory were mentioned above. This event lasted for about 10 min and for the first time the development of a stellar flare was traced in optical, radio, and X-ray ranges. The burst in optical and X-ray emission started almost simultaneously, but the maximum of optical luminosity was achieved much earlier than that of X-ray emission. Figure 35 shows the flare light curve in X-rays and calculated the temperature and emission measure of hot gas: at maximum, the temperature reached 20 MK and EM = $4 \cdot 10^{51}$ cm$^{-3}$. As before, calculations were based on the adjustment of the distribution along the energy of flare quanta to known distributions $l_\lambda(T)$ for thermal radiation of homogeneous hot plasma at various temperatures and standard chemical composition.

In the YZ CMi flare of 25 October 1979, no delays of the emission measure maximum with respect to the temperature maximum was noticed, as in the Proxima Cen flare of 6 March 1979, this is also typical of solar flares. However, one cannot exclude that the lack of delay was due to fast burning of the YZ CMi flare and averaging of the X-ray data over appreciable time intervals. If the cross section of a flare plasma loop is assumed equal to the area of this flare measured in optical observations, $10^{19}$ cm$^2$, and the loop height is taken equal to $10^{10}$ cm, an electron density of a few $10^{11}$ cm$^{-3}$ follows from the found estimate for the emission measure. In this case, the time of radiative cooling of hot gas, as in the Proxima Cen flare of 6 March 1979, is close to the time of the observed flare decay, which agrees with the ratio found for total emission energies $E_X/E_U \sim 2$. On the other hand, Kahler et al. (1982) found that the ratio $L_X/L_{opt}$, which was widely used earlier for the diagnostics of flare radiation, underwent essential changes during the YZ CMi flare of 25 October 1979: from 0.1 near optical maximum to 1 and more at the decay phase of an optical flare, and the average value of this ratio of about 1.5.

During the 3-day cooperative observations of YZ CMi on 27 October 1979 one more weak X-ray flare was recorded. An essentially quantitative conclusion following from the data on this event confirms the estimate of $E_X/E_U$ found for the flare of 25 October 1979 (Kahler et al., 1982).

On 20 August 1980, during cooperative observations with IUE, a rather strong flare that continued for more than 2 h was recorded on Proxima Cen at the Einstein Observatory (Haisch et al., 1983). Due to slow development and rather high luminosity, the light curve of the flare was confidently constructed and the temperature changes of radiating gas were found (see Fig. 36). The maximum temperature of 27 MK was achieved 2–3 min prior to the maximum luminosity, as during the flare of 6 March 1979, and the maximum luminosity within 0.2–4 keV was $1.4 \cdot 10^{28}$ erg/s. For the estimated temperature of hot gas this value corresponds to full luminosity at a maximum of $2 \cdot 10^{28}$ erg/s and a total energy of X-ray emission of $4 \cdot 10^{31}$ erg. An hour and a half before the strong flare the X-ray activity of the star, several overlapping weak bursts, was observed for half an hour. The bursts were accompanied by a slightly increased temperature of radiating gas. Immediately after the maximum of X-ray luminosity the *e*-fold decay lasted for about 20 min, then the rate of brightness decay and gas cooling slowed down, and secondary bursts of similar amplitude and duration to those that occurred prior to the strong flare were observed on the light curve.

Near the maximum brightness of the Proxima Cen flare of 20 August 1980 a distortion of the X-ray spectrum was observed for several minutes in the low-energy region, which corresponds to the occurrence of cold neutral matter with the number of particles of about $10^{20}$ cm$^{-2}$ in the line of sight. It is natural to attribute this effect to the passage of a strong prominence above the flare, and, as Haisch et al. (1983) noted, the long decay phase and luminosity of this stellar flare resembled two-ribbon solar flares, which are characterized by prominences in the active regions. Finally, the ratios of total emission energies in the transition-zone emission lines and in the $Ly_\alpha$ to $E_X$ in this stellar flare were estimated as 0.05–0.06, which was lower by an order of magnitude than similar ratios in the well-examined solar flare of 5 September 1973. However, later Byrne and McKay (1989) analyzed this flare again using the method of differential emission measure, which takes into account the contribution of many weak lines to the general radiative losses, and the ratios of transition-zone emission lines sensitive to electron density. As a result, they removed the discrepancy and found that volumes and luminosities of the regions of corona radiating in the flare and the transition zone were comparable.

On 19 September 1980, Stern et al. (1983) recorded a strong X-ray flare on the star HD 27130 (= BD+16°577) in the Hyades at the Einstein Observatory using IPC and MPC sensitive to higher-energy emissions. The earlier stellar survey of this cluster ranked this star as a strong source of X-ray emission. The star was observed during five revolutions, and only during the first of them was it in a quiescent state. Flare burning was missed: when the star came out of the Earth shadow its radiation was already at a level of $10^{31}$ erg/s, which is 35 times higher than its quiet-state luminosity. During the initial stage of observations the flare emission corresponded to a temperature of about 50 MK, EM = $4 \cdot 10^{53}$–$10^{54}$ cm$^{-3}$, and the decay rate, *e*-fold decrease, was about 40 min. Full X-ray emission of the flare within the range of 0.2–10 keV exceeded $3 \cdot 10^{34}$ erg. By that time, extensive X-ray observations of solar flares at Skylab had already been analyzed (Moore et al., 1980) and led to the conclusion that near the maxima of such flares the time of an *e*-fold decay $\tau_d$, the time of radiative cooling $\tau_{rad}$, and the time of cooling due to heat conductivity $\tau_{cond}$ were approximately equal

$$\tau_d \sim \tau_{cond} = 3 \cdot 10^6 k n_e h^2 / T^{5/2} \sim \tau_{rad} = 3kT/n_e \int l_\lambda(T)\, d\lambda. \qquad (60)$$

As stated above, at known (or given) temperature the emission measure of a flare is determined from the measured luminosity $L_X$; then the condition (60) combined with EM allows one to estimate the average density of flare plasma, the height of magnetic loops, the product of their number and the squared loop thickness to height ratio. Assuming that (60) is valid for the HD 27130 flare, Stern et al. (1983) from found values of EM and temperatures estimated the characteristic electron density and volume of the flare: $n_e \sim 4 \cdot 10^{11}$ cm$^{-3}$ and $V = 4 \cdot 10^{30}$ cm$^3$. Thus, in the decay rate, temperature, and density of this event have close analogs among X-ray flares on the Sun and on UV Cet-type stars, but in total energy and, especially, volume it stood out. Probably, since HD 27130 is a binary system of G and K dwarfs with a period of 5.6 days, the recorded flare can be related to X-ray bursts on RS CVn-type stars, which, as is well known, involve much greater volumes than flares on UV Cet-type stars, and their physical nature can differ from that of solar flares. It should be noted that the X-ray luminosity of HD 27130 in the quiescent state is close to the lower limit of those values corresponding to RS CVn-types stars, but unlike the latter, HD 27130 does not contain the components that left the main sequence. It is unknown whether another important characteristic of RS CVn-types stars, fast and synchronous rotation and revolution of one or both components, holds in this system. In addition to this huge flare on HD 27130, during the survey of the Hyades in soft X-rays, Stern and Zolcinski (1983) recorded X-ray bursts in the spectral binary G0 V system BD+14°690 and on the K dwarf vA 500, as well as slow — about eight hours — flare decay on the flare dMe star vA 288. During analogous observations of the Pleiades Caillault and Helfand (1985) detected flares on HZ 1136 and HZ 1733, the former was similar to the flare on HD 27130, the latter — to many X-ray flares on M dwarfs.

Using the high-resolution imager (HRI) installed at the Einstein Observatory, whose sensitivity was lower than that of IPC and that did not provide information about the spectrum but had a spatial resolution of 4 arcsec, Harris and Johnson (1985) observed four binary systems, including one or two dwarf M stars, from February till August 1980, one session each. During the observations they found flares on Gl 34B and Gl 338A at constant X-ray luminosity of Gl 338B and the flare on Gl 669B at constant luminosity of Gl 669A. The temporal characteristics of these bursts are summarized in Table 17. Harris and Johnson analyzed IPC data on Gl 669 obtained occasionally in 1979 during two observational sessions of another nearby source, and in one session they found a flare with a burning time of less than 3 min and longer decay: the maximum luminosity of the flare was over $10^{29}$ erg/s, the total energy $10^{35}$ erg, and temperature 24 MK.

On 24 March 1981, Agrawal et al. (1986) recorded at the Einstein Observatory an X-ray flare on the star V 1216 Sgr with a time of burning < 2 min and decay time ~ 5 min. Its luminosity at maximum was $5 \cdot 10^{27}$ erg/s and total energy $E_X(0.2-4$ keV$) \sim 10^{30}$ erg. The development of the flare was similar to characteristic optical flares on this star.

Studies of X-ray flares on the UV Cet-type stars at the Einstein Observatory were reviewed by Haisch (1983). Supplementing (60) by the approximation

$$\int l_\lambda(T)\mathrm{d}\lambda \sim 10^{-26.2}T^{1/2} \tag{61}$$

and the estimate for volume $V = h^3/100$, i.e., accepting the solar ratio of loop thickness to its height as 1/10, Haisch estimated anew the temperature, density, and linear size of loops in eight stellar flares. He obtained temperatures within 30–350 MK and systematically higher — by a factor of 1.5 to 5 — than those found directly in X-ray spectra of flares. The density varied within $8 \cdot 10^{11}$–$10^{13}$ cm$^{-3}$ and the linear sizes — from $2 \cdot 10^9$ to $>6 \cdot 10^{10}$ cm, the latter estimate is for the flare on HD 27130. For four of eight flares — on Proxima Cen of 6 March 1979 and 20 August 1980, on TZ Ari of 14 July 1980, and on V 1216 Sgr of 24 March 1981 — the direct determination of temperature from the spectrum and from (60) and (61) differ by no more than a factor of 2–2.5. They are similar to solar flares of moderate power, but their electron density is higher by approximately one order of magnitude, which results in large EM and $L_X$ and a shorter decay time. Simple scaling of the remaining three flares — on YZ CMi of 19 October 1974 and 25 October 1979 and on HD 27130 of 19 September 1980 — did not yield a "scaled" solar flare. The status of the flare on Wolf 630 of 28 February 1979 in this sample remained uncertain.

Assuming that flare loops persist due to magnetic field pressure, which should exceed the gas pressure, Haisch estimated the lower limit of field strength from the found values of plasma density and temperatures: it was over 400–1600 G in the flares, comparable with solar flares, and over 700– 9000 G in the others. For such fields, the characteristic time of development of instability of magnetic structures for scales $h$ appeared to be of close order of magnitude to the times of burning of the appropriate flares.

Further, upon comparing the full stock of thermal energy at the maximum of flares $E = 2n_e(3kT/2)h^3/100$ and the total radiative energy, which for plasma at temperatures of 10–100 MK observed within 0.2–4 keV differed from directly measured $E_X$ only by some tens of percent, Haisch (1983) found that for flares on Proxima Cen of 6 March 1979 and 20 August 1980 and probably the flare on Wolf 630 of 28 February 1979 the value of $E$ was higher by one and a half or two times than $E_X$. This means that in these two or three events, in addition to the initial pulsed energy release, one can suspect on-going heating during the decay, as in strong two-ribbon flares on the Sun.

* * *

As many as 22 flare stars were observed by EXOSAT, about 30 flares were recorded on 11 of them. Furthermore, flares were found on three stars observed within other programs, which occasionally came into the field of view of the satellite instruments. Figure 53 shows the light curves of four flares recorded by EXOSAT.

The first X-ray flare on a red dwarf star was recorded by EXOSAT by Smale et al. (1986). In studying the field in the vicinity of T Tau on 21 August 1983 they detected the event on the object that was later named 1E0419.2+1908 and was a typical UV Cet star. The flare was observed with both soft X-ray telescopes, its burning and decay time were about 40 and more than 80 min, respectively. Based on the estimate of the distance to the star, assuming that the temperature of the coronal plasma was 4 and 20 MK in quiescent and flare states, from the measured flux they estimated the amplitude of the X-ray flare as $A_X$ ~ 40, maximum luminosity as $L_X \sim 6 \cdot 10^{29}$ erg/s, the emission measure as $3 \cdot 10^{53}$ cm$^{-3}$ and the total energy — to a factor of 3 — as $E_X \sim 10^{33}$ erg. This object, as an X-ray source, was found also in two images from the archive of the Einstein Observatory, which showed the star in quiescent and active states. The active state of the star was manifested in a slightly increased level of X-ray emission and slightly increased hardness. Quantitative parameters of this and successive flares recorded by EXOSAT are cited in Table 17.

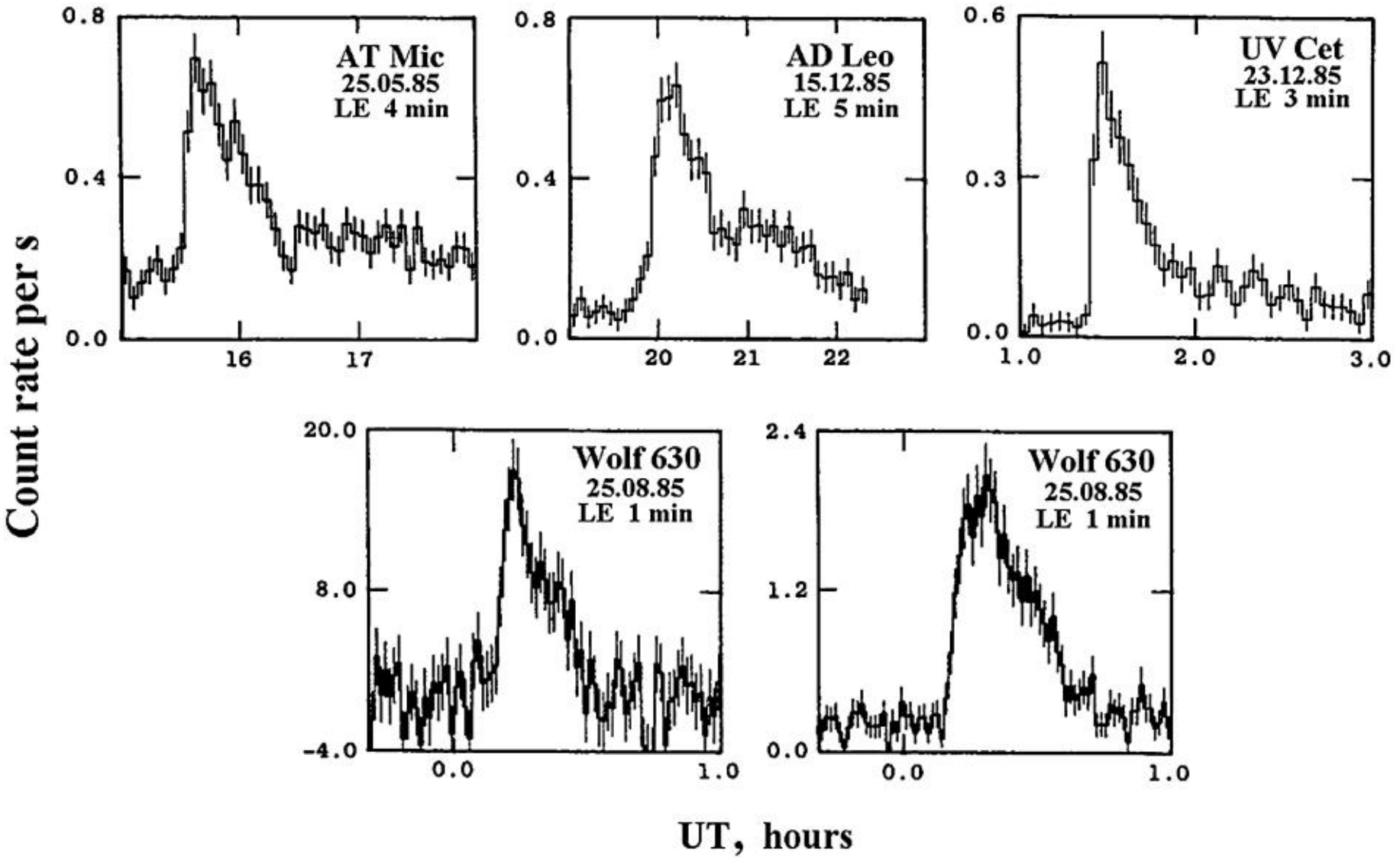

Fig. 53. EXOSAT light curves of X-ray flares on four dMe stars; the observation range and the time of averaging of readouts are specified below the date (Cheng and Pallavicini, 1991)

On 31 January 1984, both EXOSAT instruments recorded a strong flare on the single G0 V star $\pi^1$ UMa, which is an order of magnitude older than the Sun, has several times higher rotational rate, and displays strong emission of the chromosphere, a transition zone and corona in the quiescent state (Landini et al., 1986). At burning over about 8 min the maximum of flare luminosity in ME was achieved approximately 12 min earlier than in LE, and the maximum hardness of this range was recorded at the burning phase. The spectrum averaged over the whole flare in the band of 1–8 keV was presented within the framework of the model of optically thin plasma by Landini and Monsignori Fossi (1984) as thermal radiation with $T \sim 30$ MK and EM $\sim 7 \cdot 10^{52}$ cm$^{-3}$. These estimates did not contradict LE flare luminosity. If the flare decay was determined only by radiative losses, $n_e \sim 7 \cdot 10^{11}$ cm$^{-3}$ and $h \sim 7 \cdot 10^9$ cm $\sim 0.1R_*$, the area of this flare did not exceed 0.1% of the stellar surface. The total thermal energy of the flare plasma was close to $E_X$, which suggested a lack of supporting heating in the decay phase. From the energy properties this flare is close to the strongest events on dMe stars and exceeds the strongest solar flares by two orders of magnitude. Though its volume is close to that of solar two-ribbon flares, optical monitoring did not reveal this flare on $\pi^1$ UMa because of insignificant contrast with the photospheric radiation.

The flare on Wolf 1561 was recorded occasionally in studying another X-ray source (Pallavicini et al., 1990a).

The flare in the BY Dra system on 24 September 1984 was recorded during joint X-ray, optical, and radio monitoring observations (de Jager et al., 1986). The development of the stellar X-ray flare was similar to the gradual phase of a solar flare. The lack of a signal in medium X-rays made it possible to accept $T <\sim 10$ MK and estimate the emission measure as $1.2 \cdot 10^{52}$ cm$^{-3}$. De Jager et al. believed that a rather short optical flare that occurred at the beginning of the long X-ray flare corresponded to an impulsive phase during which photospheric layers of the star were burnt out and evaporated. Within the model, they estimated the coronal plasma density as $n_e \sim 2.3 \cdot 10^{11}$ cm$^{-3}$.

The X-ray source EXO 040830-7134.7 that was later identified with a dMe star was found in soft X-rays in the field of a dwarf nova during 30 sessions and seven times higher emission was recorded from it on 11 October 1984 (van der Woerd et al., 1989).

EXOSAT observations of the $\alpha$ Gem system made it possible to distinguish the red dwarf system YY Gem from the brighter pair $\alpha$ Gem AB and record two flares in the YY Gem system in the LE and ME regions on 14 November 1984 (Pallavicini et al., 1990b). The peak of the first, stronger and longer, flare in ME occurred somewhat earlier than in LE, and the flare emission in ME was almost half as long. On the basis of the analysis of this radiation the temperature of the flare plasma was estimated as 64–24 MK and EM as $5 \cdot 10^{53}$–$4 \cdot 10^{52}$ cm$^{-3}$. This was the strongest X-ray flare recorded by EXOSAT on red dwarfs.

Another visual pair resolved by EXOSAT was Gl 867 AB. The distance between its components FK and FH Aqr is only 25 arcsec, thus only the methods of maximum likelihood used to analyze X-ray images of the system made it possible to distinguish quiet radiation from the coronae of the components and conclude that X-ray flares occurred on both of them: the first X-ray burst recorded during four-hour monitoring on 18 November 1984 was due to the flare on FK Aqr, and the second due to the superposition of two flares on each of the components at a 13-min interval (Pollock et al., 1991).

On 6 December 1984, over less than four hours, four flares of similar power were recorded on UV Cet: the time of burning and the decay time varied from < 2 to 5 min and from 4 to 10 min, respectively. The luminosity at maxima was $(2-3) \cdot 10^{28}$ erg/s and the total emission in X-ray range was $(3-7) \cdot 10^{32}$ erg (Pallavicini et al., 1990a).

On 7 December 1984, Haisch et al. (1987) detected four flares in the EQ Peg system by EXOSAT. The second flare that lasted for more than an hour and was confidently recorded in the LE and ME regions is the most interesting: though the light curve in ME looked as usual with a fast rise and smoother decay, burning in LE was very slow and took approximately twice as long as in ME, and even longer than the decay in LE. Pallavicini et al. (1990a) believed that the unusual form of this light curve was caused by the superposition of two events in LE. But Haisch et al. explained this feature by temperature heterogeneity of the flare plasma from the data about the temperature gradient along arch axes known from the observations of solar flares and considerations on the radiation of soft X-rays at different stages of such flares. However, the observations of stellar flares did not allow a detailed model to be elaborated, and quantitative estimates of plasma parameters in the flare on EQ Peg were carried out using the traditional homogeneous model. The hardness of radiation in ME determined as the ratio of fluxes $F$(4–7 keV)/$F$(1.5–4 keV) was maximum at the onset of the flare and then smoothly decreased. A similar conclusion follows from spectral analysis of the emission in ME: the temperature of the flare plasma reached 26 MK at the very beginning of burning and went down to 14 MK during the decay phase. The total emission in the LE region was $5 \cdot 10^{32}$ erg. If the decay was caused by radiative cooling, then $n_e \sim 2 \cdot 10^{11}$ cm$^{-3}$. LE and ME measurements yielded a practically identical emission measure of $5 \cdot 10^{51}$–$10^{52}$ cm$^{-3}$. If this flare occurred in the arch structure, its height, density, and emission measure were closer to the appropriate parameters of solar two-ribbon flares rather than compact flares. Three other flares recorded during the 8-h observational session were shorter and their energy release in the X-ray range was lower by 5–10 times (Pallavicini et al., 1990a).

In observations of YZ CMi on 4/5 March 1985, two rather slow X-ray flares were recorded by Doyle et al. (1986). During these events, the optical monitoring was incomplete. But between the detected X-ray flares there was a rather strong optical burst with $\Delta U \sim 1.2^m$ and a total energy of optical emission of about $6 \cdot 10^{30}$ erg, which did not manifest itself in X-rays. This result contradicts the earlier revealed close relation between the variations of UV Cet emission in the $H_\gamma$ line and in soft X-rays (Butler et al., 1986). Doyle et al. (1988b) considered three probabilities for optical flares without X-ray emission: reconnection in low loops where, due to a high density of matter, all released energy quickly relaxed and no hot coronal plasma occurred; the absorption of X-ray emission by overlying cold matter, as on 20 August 1980 (Haisch et al., 1983) near the maximum of the flare on Proxima Cen; deeper heating of the chromosphere by penetrating proton rather than electron beams.

The Wolf 630 system including at least seven members was repeatedly observed by EXOSAT. In processing the 13-h session of 8 March 1985 using the initial standard program, Johnson (1987) found a flare in soft X-rays with $A_X \sim 30$, a duration of about 10 min and increased luminosity during the next 40 min in the binary star VB 8 involved in the Wolf 630 system and containing an M7 dwarf. It is essential that both components of VB 8 AB have too low masses for maintaining thermonuclear burn of hydrogen, and optical flares in it were not recorded before. Table 17 lists the parameters of this flare obtained later by Tagliaferri et al. (1987, 1990) with the improved program for analyzing EXOSAT data.

Using this program, Tagliaferri et al. (1987) reprocessed the data of the session of 8 March 1985 and, in addition to the flare on VB 8, found a flare on Wolf 630 in LE and ME. But the most complete data were obtained during the session of 25 August 1985 from a stronger flare on Wolf 630 itself, which was also found in LE and ME (Doyle et al., 1988a; Tagliaferri et al., 1987). This event occurred after the star was in the quiescent state for about seven hours. The maximum of this flare in ME took place 30 s before the maximum in LE. The temperature of the flare plasma reached 48 MK during the burning phase and reduced to 29 MK during the decay phase. The maximum hardness of the spectrum took place two minutes before the maximum of X-ray brightness. If one assumes that the flare plasma did not obtain additional energy from outside after the maximum and its cooling was due to radiative losses only, the rate of flare decay would correspond to the density $n_e \sim 10^{12}$ cm$^{-3}$ and its volume would be $3 \cdot 10^{28}$ cm$^3$. If the flare was related to $N_1$ loops and the loop radii amounted to the $\alpha$ share of their height, the found volume corresponds to $h \sim 2 \cdot 10^9(\alpha^2 N_1)^{-1/3}$ cm. If cooling losses dominate over heat conductivity in the flare, $h \sim 2 \cdot 10^{10}(\alpha^2 N_1)$ cm, and at $\alpha = 1/10$, as on the Sun, this value is lower by an order of magnitude than the previous one. Since the intermediate mode of cooling is the most probable, the heights of flare loops should make a few $10^9$ cm, i.e., about 10% of the stellar radius. Parallel optical observations revealed a rather good correlation of soft X-ray emission with flare emission in the $H_\alpha$ line: though stellar spectra were obtained with a time resolution of 14 min, they proved the coincidence of the moments of maxima and the increase of equivalent width of the line from 2.1 Å in the quiescent state to 3.2 Å at flare maximum. Excessive radiation in the line lasted approximately twice as long as the X-ray emission, but its total energy amounted to about 4% of the flare emission within the range of 0.05–7 keV. The same $E_{H\alpha}/E_X$ ratio is observed in compact solar flares.

Observations of UV Cet on 4 August 1985, 6 December 1984, and 23 December 1985 revealed several individual flares of similar power on the background of rather unstable radiation of the star (Pallavicini et al., 1990a).

On 6 August 1985, EXOSAT LE and ME detected one of the strongest and longest X-ray stellar flares in the EQ Peg system (Pallavicini et al., 1986, 1990a). Its duration was similar to that of two-ribbon solar flares, about 2 h, but its energy was two orders of magnitude higher than that of solar flares. The flare spectrum at different phases is well represented by free–free radiation of the coronal plasma plus the emission in the iron line of 6.7 keV. Reduced hardness of the spectrum was revealed during the decay and cooling of the radiating plasma from 42 to 18 MK. The decay rate of the flare in the case of the prevailing radiative losses and the absence of additional heating corresponded to the density $n_e \sim 6 \cdot 10^{11}$ cm$^{-3}$, which is typical of solar flares.

The nine-day cooperative program aimed at studying EV Lac in October 1985 resulted in two successful sessions of EXOSAT observations (Ambruster et al., 1989a). On 13 October 1985, noticeably increased X-ray emission was observed for more than two hours. Comparison of X-ray light curve and optical photometry suggests that there were two long — an hour or an hour and a half — X-ray flares that, however, were not accompanied by significant changes in stellar brightness in optical wide bands, and two fast X-ray bursts, one of which was identified as an optical flare with amplitude $\Delta U = 1.8^m$, which was accompanied by appreciable but short amplification of $H_\beta$, CaII, and MgII emissions. Upon termination of this two-hour X-ray activity within the next three hours X-ray emission was similar to that of a constant source. During a 12-hour session on 15 October 1985 on the background of obvious but weak variability of X-ray emission with a characteristic time of more than one hour a rather strong X-ray flare was recorded, which had an abrupt start, but an optical burst with an amplitude of only $\Delta U \sim 0.4^m$ corresponded to it. As the flare decayed, there was one more optical burst of equally low amplitude. There was also a flare with $\Delta U = 1.5^m$ with a simultaneous fast X-ray burst. As on the night before, these rather fast events were accompanied by enhanced emission in $H_\alpha$ and CaII but without appreciable changes of the intensity of CIV $\lambda$ 1550 Å.

In the course of cooperative investigations of the binary system YY Gem in 1988, when the X-ray observations were being carried with GINGA, two flares were recorded with the ratio $L_X/L_{opt}$, which corresponded to the closed magnetic structures (Butler, 2015).

Considering the results of EV Lac observations, Ambruster et al. (1989a) noted the lack of close correlation of optical and X-ray emission of flares: rather weak bursts in X-rays corresponded to appreciable optical flares and weak optical flares to strong X-ray activity. At the same time, there is a close relation between the events found photometrically in the U band and chromospheric emission. Recalling a similar situation with YZ CMi observations on 4 March 1985 (Doyle et al., 1988b) and the X-ray flare on Proxima Cen of 3 August 1979, which were not accompanied by strengthened optical emission (Haisch et al., 1981), Ambruster et al. advanced the idea on isolated magnetic regions in stellar atmospheres and cited some solar data supporting the idea. Later evidence of such structures was provided by Malaschuk and Stepanian (2013).

Kundu et al. (1988a) carried out simultaneous EXOSAT and VLA observations: flares recorded on UV Cet, EQ Peg, YZ CMi, and AD Leo had a rather weak correlation of activity in X-rays and in the microwave range.

Rao et al. (1990) analyzed EXOSAT data for BD+48°1958A, which was supposed to be a flare star because of the variability of the $H_\alpha$ emission profile, and found evidence of a long — about 2 h — weak X-ray flare with rather hot flare plasma.

During one of the most successful cooperative observations of UV Cet on 23 December 1985 a flare of very high optical amplitude, $\Delta B \sim 5^m$, was recorded with LE and ME EXOSAT (de Jager et al., 1989). The preflare LE flux corresponded to a temperature of 4 MK and an emission measure of $3.3 \cdot 10^{50}$ cm$^{-3}$. At maximum, which was achieved in 140 s, the LE radiation increased by a factor of 20 and 20 min after the maximum was still twice as high as the preflare level. Radiation in ME increased at maximum to a level of $3\sigma$ and was observed for about 30 min. Quantitative analysis of the data resulted in a conclusion about a significant heterogeneity of flare plasma being responsible for X-ray emission in this event. If LE and ME radiations were caused by the same plasma body, for the recorded radiation in soft X-rays its temperature should be 1 MK, i.e., lower than the coronal temperature in the quiet star, while for the radiation recorded in medium X-rays, it should be 40MK. If one assumes that emission in both ranges was caused by plasma at 40 MK, the emission measure for LE should be an order of magnitude higher than for ME.

The results of EXOSAT X-ray observations of flare stars were reviewed by Pallavicini et al. (1990a). About 30 X-ray flares recorded on some fifteen stars allowed for statistical conclusions. First, a rather high frequency of the recorded flares in the considered sample of stars should be noted — on average, one event every 8–10 h. Secondly, there is a clear positive correlation of this frequency with the level of X-ray emission of quiet stellar coronae. Thirdly, the range of characteristic times varied from minutes to hours, and total energy of X-ray emission of flares, from $3 \cdot 10^{30}$ to $10^{34}$ erg. Thus, the temporal characteristics of stellar and solar flares are identical; the energy of X-ray flares on red dwarfs is higher by orders of magnitude than the energy of solar flares. In spite of the significant difference, two types of flares occur on stars: pulsed, similar to solar compact flares, with burning in minutes and decay in ten minutes, and with a long decay — up to an hour and more, similar to solar two-ribbon flares. Stellar flares in soft X-rays, similar to solar compact flares, are usually smoother and longer than optical bursts accompanying them. Though flares of different energy of both specified types can occur on the same star, as on the Sun, there is a tendency: flares with higher $L_X^{max}$ and $E_X$ are recorded on stars with higher quiescent X-ray emission. The energy spectrum of X-ray flares constructed by Pallavicini et al. (1990a) has power characteristics with the spectral index of 0.7, as in the optical range. Flare plasma is quickly heated to 40–50 MK and cools down to 20–10 MK. Emission measures of the recorded stellar flares are equal to $10^{51}$–$10^{53}$ cm$^{-3}$ and essentially exceed solar values. The ME emission reaches a maximum somewhat earlier than LE and is noticeably shorter. Pallavicini et al. did not confirm the concept of microflares and gave another interpretation to the X-ray observations that underlies the concept. Earlier, Collura et al. (1988) obtained similar results in analyzing EXOSAT observations of 13 dMe stars: quiescent X-ray emission of the stars could not be presented by the superposition of microflares with characteristic times of hundreds of seconds, and the spectral index of the energy spectrum of individually recorded flares was close to 0.52. They concluded that if microflares played an essential role in the heating coronae, their distribution should have been noticeably different from the spectral index.

* * *

In the late 1980s–early 1990s, when the results of EXOSAT that had completed its operation were intensely studied together with the first data from ROSAT, new approaches to the analysis of X-ray stellar flares were advanced.

The relations (60) underlying some of the above estimates of the parameters of stellar flares are approximately fulfilled in many solar flares, but do not have a distinct physical substantiation. This shortcoming was removed by van den Oord and Mewe (1989), who proposed for flare decay a model of quasistatic cooling of flare loops as a series of equilibrium states of a magnetic structure. They showed that if there was power dependence of $\int l_\lambda(\sim T)d\lambda$ on temperature, a physically exact model was realized at $\tau_{rad}/\tau_{cond} = 0.18$ and $\int l_\lambda (T)d\lambda\ T^{0.25}$. In this case, radiative losses prevailed, but one should not neglect heat conductivity, while simple power relations described the change of temperature and emission energy with time. Comparing the relations with observations, one can estimate the density of matter in flares, loop heights and all other parameters, as in analyzing (60). Basically, the model of quasistatic cooling can also include additional heating at the flare decay stage, but it is not clear whether the estimate of flare parameters remains unambiguous. It is obvious that ideologically this model is close to compact flares, but originally van den Oord and Mewe applied it to a very strong flare on Algol recorded by EXOSAT, which does not match the simple morphological classification of solar flares.

* * *

Considering the results of the ROSAT investigation of X-ray stellar flares, one should remember that the satellite had two operating modes. During sky surveys, each object was recorded for 20 s every 96 min during five days, which enabled recording of long and strong flares. At long individual pointings with considerably increased sensitivity, weak and fast flares could be recorded.

ROSAT was used to continue checking the concept of microflares: within two weeks Schmitt (1993) collected about twenty half-hour records of UV Cet, but their consideration did not reveal the expected effects of microflares. Some of these X-ray observations were accompanied by monitoring in the U band. Two contiguous optical flares on 2 January 1992 with $\Delta U \sim 1.2^m$ were accompanied by X-ray bursts. The first purely impulsive optical flare was about 12 s long and was accompanied by a short X-ray burst. The second optical flare had a very fast impulsive phase, but it was followed by a gradual five-minute phase, and in X-rays there was a fast single burst, which was followed by long X-ray luminosity three minutes after. A similar delay of soft X-rays after an impulsive beginning of optical flare was recorded in the flares on the Sun, UV Cet, and BY Dra (de Jager et al., 1986, 1989). Both X-ray bursts on UV Cet took place 30 s after optical pulses, and they could be low-energy tails of hard X-ray emission concerned with initial beams of energetic particles (Schmitt et al., 1993a), but the reason for the 30-second delay remains unclear.

In studying late Me dwarfs with ROSAT, Fleming et al. (1993) recorded several X-ray flares on three stars. On 19 October 1990, the already mentioned strong flare with $A_X \sim 160$ was recorded on AZ Cnc during six revolutions around of the Earth. Within two days, when Proxima Cen was accessible within the RASS program, they recorded a flare during four revolutions and two flares, with individual readouts and apparently the decay of a slow flare, whose maximum was missed. On 25 November 1990, a flare on CN Leo was recorded during three consecutive revolutions.

Pan and Jordan (1995) carried out two cycles of ROSAT PSPC observations of the flare star CC Eri within the range of 0.1–2.4 keV with a spectral resolution of 2.1. They found its variability at times from several minutes to several hours. On 10 July 1990, they recorded a rather slow flare with a burning time of about an hour and a decay of about 3.7 h, an amplitude $A_X \sim 2$ and maximum luminosity $L_X(0.2{-}2\ \text{keV}) \sim 7 \cdot 10^{29}$ erg/s. There was an additional burst on the descending branch of the flare an hour and a half after the maximum. The spectra of CC Eri in the quiescent state and during the flare were presented within two-temperature models, and in both states various algorithms inevitably suggested the existence of a coronal component with a temperature of about $10^7$ K. The time of development of the flare of 10 July 1990 was similar to two-ribbon solar flares, but its energy was an order of magnitude higher compared to the latter. To analyze the "pure flare spectra" obtained by subtraction of the stellar spectrum out of the flare from the observed spectra of active star, Pan and Jordan applied the analytical theory of Kopp and Poletto (1984) and Poletto et al. (1988) on magnetic reconnection in two-ribbon flares. They presented the descending branch of the light curve of the flare on CC Eri within the framework of this model and found that during the decay phase the temperature of the plasma decreased from 28 to 12 MK, and EM — from $42 \cdot 10^{51}$ to $1 \cdot 10^{51}$ cm$^{-3}$. Depending on the specified extension of the flare on the stellar latitude from 33° to 5°, its area varied within 8–0.2% of the stellar hemisphere, the magnetic-field strength within 250–1500 G, an initial rise speed of the reconnection point within 14–2 km/s, the maximum height of rise of this point within $2 \cdot 10^7$–$4 \cdot 10^9$ cm, and ranges of electron density within 13–220 in units of $10^{10}$cm$^{-3}$. Thus, the model of the flare on CC Eri with continuous energy release should have either rather high density, or much greater volume than large two-ribbon solar flares. Within the framework of the flare model with only pulsed energy release, parameters obtained by Pan and Jordan are close to those in the previous models where the total flare area was close to 8% of the stellar hemisphere.

Within the RASS program the star F0V 47 Cas, which probably belongs to the Pleiades moving group, was observed every 96 min during 3.5 days. On the background of clear flux oscillations with a period of about a day that were visible only in the soft part of the recorded range and were apparently caused by the rotation of the star with an asymmetrical corona, Güdel et al. (1995b) detected, on 19 and 21 August 1990, flares at the maxima of periodic oscillations. The first was visible only in the hard part of the recorded range. The estimates showed that the observed pattern could be related to the large active region on the star with coronal temperature of about 2 MK. At maximum brightness the luminosity of flares was $4.3 \cdot 10^{30}$ and $1.2 \cdot 10^{31}$ erg/s. If the temperature of the flare plasma is assumed to be 25 MK, the total energy of X-ray emission of the second flare would be about $7 \cdot 10^{34}$ erg. This event is one of the strongest among stellar flares. If the duration of the first flare is assumed to be due to radiative relaxation, at EM $\sim 4 \cdot 10^{53}$ cm$^{-3}$ and $n_e \sim (2 - 4) \cdot 10^{10}$ cm, and the size of the flare loops is about the size of a loop in the cool component of the preflare corona.

Springfellow (1996) published light curves of AD Leo on 8/9 May 1991 and of VB 8 on 25–27 February 1991. Over four hours of AD Leo observations one strong flare was found with $A_X \sim 3$ and the decay to an FWHM level of eight minutes and two weaker flares. On the light curve of VB 8 there was one long flare of about 100 min in total.

Under spectral analysis of a strong flare on EV Lac recorded with ROSAT PSPC on 13 July 1992, Sciortino et al. (1999) found that a hot component with $T \sim 36$ MK should be introduced in addition to isothermal MEKAL models that enabled presentation of the spectrum of the quiet star; MEKAL models are modifications of MEKA (Mewe et al., 1995b).

A byproduct of ROSAT PSPC observations of Gl 213 stars was the detection of a high activity level of the M3 dwarf G 102-21 (Micela et al., 1995). Later, this X-ray source was found in the archives of the Einstein Observatory. The ROSAT data for G 102-21, eight sessions within 3.35 days, were analyzed within the range of 0.20–2.0 keV. In the records of 23 September 1993, a strong flare with $A_X \sim 8$ and $L_X^{max} \sim 5 \cdot 10^{29}$ erg/s was found. Representation of the flare spectrum within the framework of the two-temperature model yielded $T_1 = 4.8$ and $T_2 = 16$ MK, $EM_1 = 8 \cdot 10^{51}$ and $EM_2 = 51 \cdot 10^{51}$ cm$^{-3}$. The high-temperature coronal component appeared to be necessary for the models of the quiescent corona as well. The value of $T_1$ was practically equal for the flare and quiescent corona, whereas the $EM_1$ of flares was twice that in the quiescent state. The values of physical parameters of this flare found resembled those of compact solar flares and, if it was located in one loop, the field strength in it should have been 630 G.

The young G star EK Dra that has just reached the main sequence was in the field of view of ROSAT PSPC for five days during the sky survey. A flare invisible in softer rays of 0.1–0.4 keV was recorded on it on 23 November 1990 in the range of 0.4–2.4 keV. The characteristic time scale of the flare was about an hour, $L_X^{max} \sim 1 \cdot 10^{30}$ erg/s and $E_X \sim 4 \cdot 10^{33}$ erg. If the decay of the flare was caused by radiative cooling, then $n_e \sim 2 \cdot 10^{11}$ cm$^{-3}$ and $h \sim (1{-}2) \cdot 10^{10}$ cm (Güdel et al., 1995a). Similarly, on 30 July 1990 a slow flare whose burning and decay took about six hours was recorded on the F star HD 147365 (Güdel et al., 1995c).

Of particular interest for the general picture of the considered activity are completely convective objects with masses lower than 0.3 solar masses. Fleming et al. (2000) carried out additional long-term observations of the M8 star VB 10 (= Gl 752 B), which had not been found in ROSAT surveys or two-hour individual monitoring. Upon six-hour observations they found an extremely weak image of the star in the image of the field obtained throughout the monitoring. However, detailed analysis proved that all of the 10 photons composing this image were recorded on 19 October 1997 during a 19-min session. Assuming that the star was recorded during the flare, Fleming et al. estimated its luminosity at maximum as $L_X = 3 \cdot 10^{27}$ erg/s and the stellar luminosity out of the flare as less than $2 \cdot 10^{25}$ erg/s. In the full duration and the part of total monitoring the X-ray flare took, this event was close to the UV flare recorded on this star on 12 October 1994 by HST (Linsky et al., 1995). Another important similarity of the two flares is that the level of quiescent radiation of the star in both wavelength intervals was lower by 1–2 orders of magnitude than the radiation in flares. Based on this fact, Fleming et al. concluded that on this star the solar-type corona with $T \sim 10^6$ K could be absent because of an essential difference of thermodynamic conditions in the photosphere.

During the ROSAT sky survey, Schmitt et al. (1993b) found a very strong flare on HII 2034, quickly rotating K2 dwarf in the Pleiades with v sin$i$ > 50 km/s and $P_{rot} \sim 8.3$ h: at least a ten-fold increase of the flux was recorded in one of the scans, whereas the whole duration of the event was no more than 3 h, and the greater part of the recorded photons was in the hard region of the recorded wavelength range. Sciortino et al. (1994) found flares with $A_X \sim 10$ on three stars in the Pleiades: a flare with fast burning and decay in tens of minutes on the very weak star Hz 892; an event on HCG 307 similar to a compact solar flare, and a flare with slow burning and slow decay on HCG 144 similar to a two-ribbon solar flare.

In individual pointing images, Gagné et al. (1994a) recorded a flare on the fast G8 rotator HII 2147 in the Pleiades and estimated the parameters of the two-temperature model of the flare: $T_1 = 3.6$ and $T_2 = 14$ MK; $EM_1 = 2 \cdot 10^{53}$ and $EM_2 = 11 \cdot 10^{53}$ cm$^{-3}$. In a more detailed analysis of these data, Gagné et al. (1995) found 11 other flares with $L_X > 10^{30}$ erg/s, which were among the strongest X-ray flares. These results confirmed the conclusion of Schmitt (1994): strong flares occurred on stars of all late types from G to M and in a wide range of rotational rates v sin$i$ = 9–45 km/s. The data on a flare on the K star HII 1516 also enabled the parameters of this event to be estimated within the framework of the simple model of quasistationary cooling: $n_e > 1.3 \cdot 10^{11}$ cm$^{-3}$, $T \sim 13$ MK, and EM $\sim 2 \cdot 10^{54}$ cm$^{-3}$.

Pye et al. (1994) noted at least 10 members of the Hyades with variable X-ray brightness whose flare spectra were recorded with PSPC. X-ray flares occurred with the greatest probability during observations of VB 141 and VB 190.

In studying three regions of the $\alpha$ Per cluster in the mode of individual pointing, three slow flares with $L_X^{max} \sim 10^{31}$ erg/s and $E_X > 10^{34}$ erg with burning from 1 to 2.5 h and decay within many hours were detected on three stars of the cluster (Prosser et al., 1996).

During the ROSAT sky survey, Güdel et al. (1994b) recorded a strong flare on the G star Gl 97, which at v sin$i$ ~ 4 km/s had high luminosity $L_X \sim 10^{29}$ erg/s. If the flare light curve is approximated by a triangle whose vertex is at maximum brightness, the duration of this event in FWHM is about 100 min and $E_X \sim 2 \cdot 10^{33}$ erg, which is higher than the strongest solar flares by an order of magnitude.

During five-day ROSAT HRI observations of AT Mic, McGale et al. (1994) found fast oscillations of X-ray flux with a characteristic time of about 20 s and an amplitude of about 10% imposed on smooth variations with a period of about 8 h and $A_X \sim 2$.

Schmitt (1994) summed up the ROSAT studies of stellar flares completed by early 1993 both during the sky survey and in individual pointings on selected objects. The basic conclusion from the observations within the RASS program is that it enabled X-ray flares on late stars of all spectral types to be revealed. In addition to the above flares on three Me dwarfs (Fleming et al., 1993) and the fast rotator K0 V HD 197890 (Bromage et al., 1992), Schmitt cited the 40-day light curve of the F5 dwarf 36 Dra, on which among about 700 measurements two flares with $A_X \sim 4$ and 10 are clearly seen, and the 3-day light curve of B star in $\pi$ Lup binary system with a strong flare at the end of the observation period. A stronger flare on 36 Dra had $L_X \sim 2 \cdot 10^{30}$ erg/s and $E_X \sim 10^{33}$ erg, whereas the flare on $\pi$ Lup — $3 \cdot 10^{31}$ and $10^{35}$, respectively. On the three-day light curve of EV Lac obtained during the first half of the monitoring — on 18/19 December 1990 — several fast bursts are seen, all the second half (19–20 December 1990) took a long time — about a day — increased brightness and one very strong ($A_X \sim 25$) and fast flare. The hardness of the spectrum calculated from PSPC data during this state was definitely higher than that before it. Spectral analysis of these observations resulted in the two-temperature model of the quiescent corona with temperatures of 1.8 and 5 MK and the one-temperature flare model with $T = 25$ MK. Schmitt simulated this flare on EV Lac within two different models: quasistatic cooling by van den Oord and Mewe (1989) and a two-ribbon flare by Kopp and Poletto (1984). Within the framework of the first model, the light curve of the flare was presented at $T = 30$ MK, EM = $1.5 \cdot 10^{52}$ cm$^{-3}$, $n_e \sim 3 \cdot 10^{10}$ cm$^{-3}$ and the length of the coronal loop of $6 \cdot 10^{11}$ cm or $10R_*$, i.e., a structure of rather low density and large extension. Within the two-ribbon model, an appreciable dependence of the ascending branch of the flare on the chosen parameters was kept, whereas during the decay phase this dependence was very weak, and observations could be presented by theoretical curves with $n_e = 3 \cdot 10^{12}$–$5 \cdot 10^{10}$ cm$^{-3}$ and the magnetic field strength was within 2–5 kG. Comparison of the calculations did not allow selection of one of the models, though the fact that the loop size was ten times greater than the stellar radius raised suspicion. Consideration of the resulting seven-hour ROSAT monitoring of the UV Cet star in the mode of individual pointing confirmed the flickering in the spectrum of brightness intensity found by Pallavicini et al. (1990a) with EXOSAT and advanced its region to 10 MHz, white noise occurred at higher frequencies.

* * *

Apparently, the first stellar flares in the range of medium X-rays from 2 to 18 keV were recorded from the Ariel V satellite: Rao and Vahia (1987) identified eight flare stars as fast variable objects found by the satellite, at maximum brightness the luminosity of these stars varied from $2 \cdot 10^{30}$ to $3 \cdot 10^{31}$ erg/s.

During GINGA LAC observations of the radio galaxy 3C390.3, within 2–36 keV on 14 February 1991, a flare was recorded on the dM4e star EQ 1839.6+8002 (Inda et al., 1994). The flare fell onto the last 37 min of the 14-h monitoring of the radio galaxy field, over four minutes it reached the maximum with $A_X \sim 10$ and then smoothly decayed approximately over 20 min. The large-angle counter (LAC) onboard GINGA, performed spectrophotometry in 48 energy channels with a 16-s time resolution. Pan et al. (1997) analyzed in detail the unique observations. The flare flux at $E > 20$ keV was negligible, the rate of development of the flare

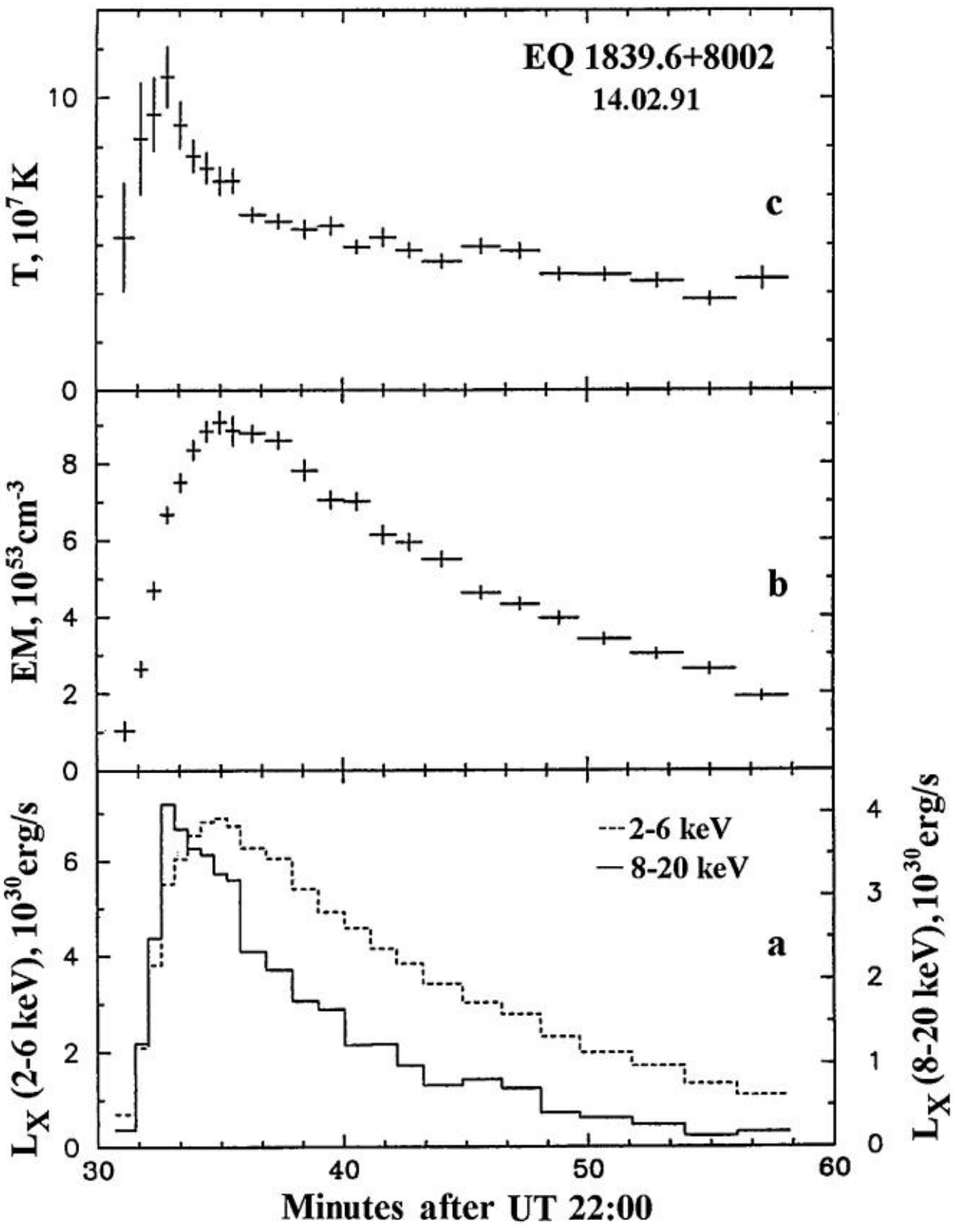

Fig. 54. Flare on EQ 1839.6+002 of 14 February 1991 recorded by GINGA in medium and soft X-rays: (a) light curves within 2–6 and 8–20 keV; (b) development of the emission measure; (c) dynamics of flare plasma temperature (Pan et al., 1997)

within 8–20 keV was noticeably higher than within 2–6 keV. Pan et al. (1997) divided the whole recorded interval into 24 small intervals and adjusted the model of isothermal optically thin plasma, 1T RS model, the “pure spectra of the flare” of each of them was adjusted by varying the temperature and emission measure of the flare plasma. Results of the analysis are presented in Fig. 54. The plots show a 2–3-minute delay of $EM^{max}$ with respect to $T^{max}$, similar dynamics of the temperature and a flare intensity within 8–20 keV and similar dynamics of EM and intensity in the softer range. The total energy of the X-ray emission of this flare was about $10^{34}$ erg, while its luminosity at maximum was $10^{31}$ erg/s, which corresponds to $L_X/L_{bol} \sim 0.25$. Using the method proposed by Cheng and Pallavicini (1991) to analyze flares on the ($T$,EM) plane, Pan et al. (1997) distinguished three phases of the flare: a burning phase when the temperature and emission measure grew simultaneously, the phase of continued growth of EM at reducing temperature, and the decay phase when both parameters decreased. They carried out preliminary analysis of each phase on the assumption of constant mass or size of radiating plasma and found that in both cases the beginning of the second phase corresponded to an essential decrease of heating after which EM grew due to evaporation, but slight heating took place at the decay phase as well. The key parameters of the flare are as follows: the length and section of coronal loops (1–4) $\cdot$ $10^{10}$ cm and 2 $\cdot$ $10^{19}$ cm$^2$, respectively, and $n_e \sim (0.4–1.7) \cdot 10^{12}$ cm$^{-3}$. Comparison of the parameters of this flare with other flares on M dwarfs showed that here $L_X$, $E_X$, and EM were higher by 1–2 orders of magnitude, $T^{max} \sim 10^8$ K was several times higher, while $n_e$, loop length, and magnetic flux tube strength were the same as in typical X-ray flares on M stars. In the archives of the Einstein Observatory, Pan et al. found flares on EQ 1839.6+8002 of 10 April 1980 with $A_X \sim 23$ and on 7 November 1985 with lower amplitude.

* * *

Advancement of observations of stellar flares in medium X-rays opened qualitatively new opportunities: since many heavy elements play an essential role in this range, the adjustment of theoretical models to the observed spectral distributions enables one to assess the chemical composition of radiating plasma. These qualitatively new results were obtained by ASCA and BeppoSAX in X-rays and by EUVE in extreme ultraviolet.

The system $\alpha$ Gem was monitored for one day within 0.4–7 keV with SIS ASCA. On 26 October 1993, a flare on YY Gem with $A_X \sim 4$, following a weaker burst that occurred approximately 10 h before, and a strong flare on Castor AB with $A_X > 10$ were recorded (Gotthelf et al., 1994). The resulting quiescent-state spectrum and the spectra of the flare on YY Gem were analyzed using the RS and MEKA models of radiation of hot equilibrium plasma. (It should be noted that representation of an X-ray spectrum in the framework of the MEKA model yields higher temperatures, a much higher $EM_1/EM_2$ ratio and much lower values of $EM_1$ and $EM_2$ than the RS model for the same spectrum (Sciortino et al., 1999).) The spectrum of flare maximum on YY Gem could be presented by means of rising temperature of the hottest component of the quiet corona, and the flare spectrum of Castor AB, by means of the two-temperature RS model with $T_1 = 3$ and $T_2 = 11$ MK.

The star Proxima Cen was monitored for 50000 s with ASCA SIS and GIS on 18–20 March 1994. Over this time, two flares took place on it in the range of 0.5–12 keV with amplitudes of about 5 and 10 (Haisch et al., 1995). Within the framework of the two-temperature model the active stellar corona was composed of components with temperatures of 7 and 44 MK, whereas no hot element was found in the quiet corona.

Using ASCA, Singh et al. (1999) carried out X-ray spectroscopy of flares on two rapidly rotating late dwarfs HD 197890 (Speedy Mic) and Gl 890 and revised the observational results obtained for YY Gem system by Gotthelf et al. (1994). The flare on HD 197980 of 20 April 1995 was recorded at the end of a 13-h observational session: its luminosity increased approximately twice over 33 min and reached $2 \cdot 10^{30}$ erg/s. The flare on Gl 890 was found at the end of a 27-h observational session on 19 November 1995: at $A_X \sim 2$ the exponential decay lasted for about 40 min. The spectra were analyzed using MEKAL and VMEKAL models accounting for variations of chemical composition of the radiating plasma. Isothermal and two-temperature coronae and coronae with a continuous dependence of EM on temperature (CEM) set by Chebyshev's polynomials were considered. The spectrum of the flare on HD 197890 in the range of 0.4–6 keV was better represented by the two-component corona with $T_1 = 8.6$ and $T_2 = 37$ MK, $EM_1 = 6 \cdot 10^{52}$ and $EM_2 = 12 \cdot 10^{52}$ cm$^{-3}$. As compared to the quiescent state, the greatest — two-fold — amplification in the flare was noticed in $EM_2$. Attempts to present the quiet corona and flare with identical abundance of heavy elements within the 2T models failed, but the solar abundance was presented within the CEM model and in the 2T model with a five-fold depletion of heavy elements. The spectrum of Gl 890 during the flare was presented by 2T or CEM models of coronae for the solar abundance of elements, but this representation noticeably improved at a multiple of 3–16 decrease of the abundance of heavy elements. Gotthelf et al. (1994) analyzed the flare on YY Gem of 26 October 1993 within the MEKA model using the SIS data only. Singh et al. (1999) repeated the analysis using the MEKAL model and the data of all ASCA instruments and found that 2T models for the solar abundance of heavy elements well presented the observations, but introduction of the third component noticeably improved the representation, and a decrease of the abundance of heavy elements by a factor of 2 yielded the best result. Singh et al. found that emission measures of hot components of HD 197890 and YY Gem coronae and the temperature of the hot component of Gl 890 corona increased considerably during flares.

The results of the analysis of the quiet corona of AD Leo obtained by Favata et al. (2000a) in observations from three X-ray observatories were reported above. Now let us consider their conclusions about flares on this star. They examined the flare of 13 May 1980 with $A_X \sim 7$ recorded with IPC of the Einstein Observatory, two flares on 8 May 1991 at an interval of 12 h with $A_X \sim 5$ and 3 found by ROSAT and three flares with $A_X \sim 3$– 10 recorded with ASCA SIS. ASCA observations were run on 2–4 May 1996: during the first day the star was in the quiescent state, then with intervals of 5.6 and 1.9 h rather strong flares occurred, then till the end of observations the star was active with appreciable oscillations of X-ray luminosity. All the flares were uniformly analyzed using the hydrodynamics approach with obvious allowance for the supporting heating of radiating flare plasma. An empirical relation of the characteristic decay time with the inclination of the flare track on the $\log n_e - \log T$ plane was established for each instrument separately, since this relation depended on the used range of energies. The fullest results were obtained for the second flare of 1996: its descending branch was split into seven intervals; each of them was analyzed independently. Within the framework of the model this flare had essential heating during the decay phase, thus its light curve at this phase was determined by the temporal dependence of heating, rather than by self-decay of the loop. The estimate $h \sim 4 \cdot 10^9$ cm is seven times lower than $R_*$ and tens of times lower than the pressure scales of heights. At a maximum luminosity of about $2 \cdot 10^{29}$ erg/s and total energy $E_X = 3 \cdot 10^{31}$ erg a field of 0.6 kG was sufficient to maintain such a loop and 1.4 kG were sufficient for its complete emission. The decay phase of the weaker third flare involved supporting heating as well, and the size of its loop was similar. Decay heating was also found in the first flare recorded with ROSAT PSPC. The ROSAT light curve of the longest second flare on AD Leo of 8 May 1991 was divided into five intervals 1.7 to 2.3 h long, which were analyzed independently. The decay of this flare was completely determined by the time dependence of additional heating of the radiating plasma. Thus, only the upper limit of the loop size could be estimated: $h < 17 \cdot 10^9$ cm. Analogously, the light curve of the flare recorded by the Einstein Observatory on 13 May 1980 was divided into intervals of 10–26 min. Their analysis suggested essential supporting heating and a largest loop of $13 \cdot 10^9$ cm. Thus, in all 6 flares the loop sizes fell in a narrow interval and were appreciably less than the stellar radius. The maximum temperature in loops was within 10–50 MK, but $E_X$ covered a wide range of values due to differences in the times of flare decay. Comparison of these results with the conclusions of Cully et al. (1997) for the EUV flare on AD Leo obtained on the assumption of an absence of supporting heating revealed an essential difference in the estimates of the loop sizes, which were overestimated by an order of magnitude. An important conclusion was made by Favata et al. for the loop thickness: the thickness-to-height ratio is 0.3, not 0.1, as on the Sun. Favata et al. (2000a) concluded that all considered flares developed in similar compact coronal loops and their decay was determined by the process of supporting heating rather than self-decay of flare loops. Otherwise, loops would be five times larger. For the standard small size of loops the transition from the level of solar activity to that of dMe stars should not be determined by the filling factor (which is limited to values 1.5–2 orders of magnitude higher than the solar level) but rather by pressure in loops, i.e., densities of flare loops. This pattern is in agreement with the models of the quiet corona of AD Leo with f << 1 and with thick loops obtained earlier by Giampapa et al. (1996) and Sciortino et al. (1999).

During many-hour monitoring of EV Lac with ASCA GIS on 13–15 July 1998, Favata et al. (2000b) recorded three flares, and the second flare with a maximum close to 13 July 1998 at 20:20 UT was one of the strongest stellar X-ray flares: its amplitude reached 300 at full duration of about four hours and a luminosity at maximum of $L_X/L_{bol} \sim 0.25$. For several minutes its X-ray luminosity was comparable with photospheric luminosity. Figure 55 shows the light curve of the flare with a time resolution of one minute and the spectra of the flare within each of the time intervals: (1) preflare state, (2) and (3) ascending flare branch, (4) maximum, and (5–9) decay. The results of quantitative analysis of spectra at separate intervals are presented in the bottom diagrams. Within the two-temperature model, quiescent X-ray emission with luminosity $L_X = 3 \cdot 10^{28}$ erg/s was presented by $T_{1,2}$ = 9 and 22 MK and $EM_{1,2}$ = $2.4 \cdot 10^{51}$ and $0.8 \cdot 10^{51}$ cm$^{-3}$. Flare radiation was presented within the one-temperature model: a maximum temperature of 73 MK was achieved during the pre-maximum time and the greatest EM = $7.4 \cdot 10^{53}$ cm$^{-3}$ occurred at the moment of flare maximum. On intervals 3–6 a 2–3-fold increase of the abundance of heavy elements and the subsequent return to the preflare level were found. Physical parameters of plasma radiating in the flare were estimated by Favata et al. from their decay character using both models: in the approximation of quasistatic cooling and in the hydrodynamic scheme of decay of an arcade of flare loops with long heating. In the first model, the character of flare decay corresponded to a flare loop of $5 \cdot 10^{10}$ cm ($\sim 2R_*$) and matter density of $6 \cdot 10^{11}$ cm$^{-3}$. In the second analysis scheme, the drift of the flare was considered during its decay on the $\log n_e - \log T$ plane, which, as stated above, made it possible to find additional heating. In this case, as for the above flare on AD Leo, the light curve of the flare was determined not by the self-cooling rate but by the time variations of additional heating. This approximation yielded the following parameters for the flare on EV Lac: the true temperature of the flare at maximum was 150 MK, a loop height of $1.3 \cdot 10^{10}$ cm ($\sim 0.5R_*$), a plasma density in the loop of $(2–0.2) \cdot 10^{12}$ cm$^{-3}$, and pressure $(8–0.9) \cdot 10^4$ dyn/cm$^2$. The obtained loop size is large, but not too large, similar loops occur even on the Sun.

Numerous conclusions within the framework of the hydrodynamic model of flares about the rather small size of coronal loops $h < R_*$ led Favata et al. (2000a) to the conclusion on the nonrealistic nature of the free relaxation model that regularly yielded $h > R_*$.

Covino et al. (2001) analyzed ROSAT and ASCA X-ray observations of LQ Hya. On 5 November 1992, ROSAT recorded a flare with $A_X > 10$ and a decay of about four hours, the maximum hardness of its X-ray spectrum was achieved approximately 2–3 min after the maximum brightness. The spectrum of this flare was presented by 2T model at metallicity of $z/z_\odot \sim 0.1$. During the flare, the cool component of coronal radiation practically did not change, and the size of the flare loops was estimated as $1.5R_*$. The data on LQ Hya flare of 7 May 1993 recorded by ASCA were presented by the 2T model with $z/z_\odot$ = 0.2–0.6 for different elements. The data from all X-ray instruments lead to significantly lower abundance of elements in the corona, whereas in optics it is close to solar one. In addition to the strong flare, two- and three-fold amplifications of X-rays were recorded generally due to EM at maintaining temperature. The ASCA data require the second component and more wide-wave BeppoSAT — the third component.

Liefke et al. (2009) estimated the frequencies and amplitude distributions of flares on the M dwarf CN Leo in the course of simultaneous observations in the coronal X-rays, chromospheric lines, and optics and found that most appreciable events were seen at all atmospheric levels; several weaker events were seen only in chromospheric lines, which corresponded to the $H_\alpha$ flares on the Sun. One strong event in the X-rays proved to be very weak in chromospheric lines and was not seen in the photospheric continuum, which could be attributed to a large size of the flare loop. No correlations of flare amplitudes and decay times were detected at different atmospheric levels. Later, by means of these observations, they studied temporal variations of the corona during a giant flare. Despite its very fast decay, about 5 min, the flare had $A_X \sim 100$, an average electron density of more than $5 \cdot 10^{11}$ cm$^{-3}$, twice enhanced iron at a rise and in a maximum and a size of < 9000 km. They assumed that the flare occurred not in a separate loop but in a compact arcade.

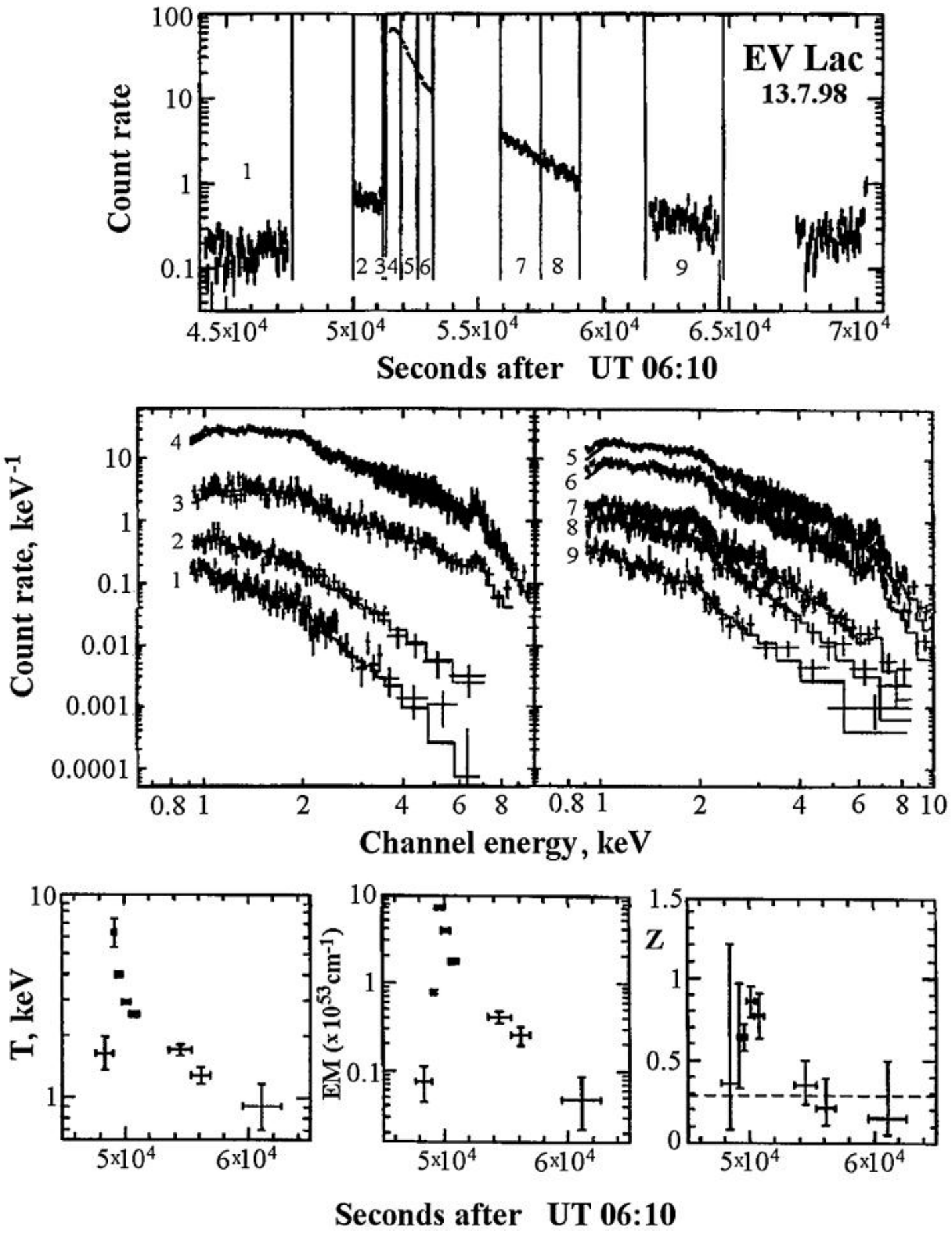

Fig. 55. The flare on EV Lac of 13 July 1998 recorded with ASCA. *Top down*: the light curve; the spectra on the separated sections of the light curve; results of spectral analysis: the dynamics of temperature, emission measures, and the abundance of heavy elements (Favata et al., 2000b)

* * *

BeppoSAX LECS and MECS observations with a spectral resolution of 5 and 12 in the ranges of 0.1–7 keV and 1.5–7 keV, respectively, yielded additional data on flares in medium X-rays.

Landi et al. (2001) observed EQ Peg for 19 h and on 3 December 1997 recorded a flare in the range of 0.1–5 keV that lasted for many hours. Within the 2T MEKAL model the recorded radiation was presented as a sum of components with temperatures of 6.7 and 15 MK.

Katsova et al. (2002) analyzed a strong flare on this star, detecting six abrupt bursts in the B band and somewhat late total X-ray maximum. The gas-dynamic modeling of bursts in the range from 0.4 to 10 s with an exciting flux $F_0 = 3 \cdot 10^{11}$ erg · cm$^{-2}$ · s$^{-1}$ led to a density of the downward condensation of somewhat higher than $10^{16}$ cm$^{-3}$, a temperature of about $10^4$ K and a condensation size within 1–5 km, an optical flare size of $6.6 \cdot 10^{18}$ cm$^2$; the temperature of the flare loop was about 27 MK and its density $10^{11}$ –$2 \cdot 10^{10}$ cm$^{-3}$ from the base to the top. Arguments were provided in favor of similarity of this event to <superhot> solar flares, when hot coronal loops are formed but they rapidly cool down and disappear.

Within a day, Sciortino et al. (1999) recorded three flares with amplitudes of 4–5 and a time of exponential decay of 13, 67, and 73 min on AD Leo. The star was active for 30–50% of the monitoring time. Two strong flares with a decay time of 75 and 23 min and several weaker bursts were revealed over 9 h on EV Lac. Spectral analysis of these flares showed that 2T models were insufficient irrespective of the accepted chemical composition of the radiating plasma. The flare spectra were successfully presented by 3T MEKAL models with metallicity of the flare plasma decreased by 3–5 times on AD Leo and by a factor of 2 on EV Lac. Evidence of the existence of at least two types of loops was obtained: hundreds of dominating low-temperature compact loops with a size of less than $0.1R_*$ and a total area of bases of less than 1% of the stellar surface and tens of extended loops responsible for the hottest radiation with very small total area of bases. In 1999, AD Leo was observed throughout 45 days within the cooperative program using BeppoSAX and EUVE, the VLA radio system, and optical telescopes in Pennsylvania and the Crimea (Güdel et al., 2001). During about seven days the star maintained high activity with almost continuous brightness variations with an amplitude of 1.5–2, and about a dozen EUV flares with a decay time up to 2–3 were recorded in the deep-photometry mode. About a dozen X-ray flares, partially coincided with EUV events, were recorded with LECS and MECS at a shorter time interval. On the other days, activity in EUV coincided with microwave radiation bursts and a strong optical flare on 30 April 1999.

During five-day observations of YY Gem in November, 1998 Tagliaferri et al. (2001) recorded three flares. The hardness of X-ray emission was the greatest during the weakest flare. Spectral analysis of the third flare using the 2T MEKAL model yielded the following parameters of flare plasma: temperatures of components of 9 and 40 MK and increase of metallicity from $z/z_\odot = 0.2$–$0.4$ in the quiescent state to 0.8.

On 9 and 29 November 1997, Maggio et al. (2000) carried out BeppoSAX observations of two flares on AB Dor and first detected hard X-rays. Using LECS and MECS, they performed an analysis of flare spectra and the subsequent quiescent state within the framework of 2T models. The first flare lasted for about 12 hours at $A_X \sim 200$, at a peak temperature of about $10^8$ K and $2.5 \cdot 10^7$ K at the decay phase, whereas the temperature of the low-temperature component all the time remained at a level of $10^7$ K. The peak EM $\sim 3 \cdot 10^{54}$ decreased up to $8 \cdot 10^{52}$ cm$^{-3}$. The composition of elements in the corona turned out to be at a half photospheric level without noticeable variations during the flares. The exponential decay of X-ray brightness, evolution time of the temperature plasma and EM were similar to these values in the compact solar flares. A maintaining of plasma heating was detected at the decay phase and the maximum height of loops was less or comparable with $R_*$. One flare was observed during the full stellar revolution without a visible self-eclipsing effect; this is possible at its location in the near-polar region. No association with the magnetic field maintaining prominences at several stellar radii above the surface was found. In the quiescent state, after the second flare lasting for two revolutions, a low-amplitude variability was detected but without rotational modulation.

As it was already mentioned, the total results of BeppoSAX observations were inferred by Pallavicini et al. (2000).

* * *

The X-ray satellite RXTE with a sensitivity of 2–15 keV was used for multiwavelength observations of AU Mic and EQ Peg on 12–15 June and 2–6 October 1996, respectively (Gagné et al., 1998). At the onset of the flare on AU Mic of 13 June 1996 that was thoroughly tracked in EUV over a day, with a time resolution of eight minutes, there was an individual X-ray burst with $A_X \sim 20$. The spectrum of this flare was presented within the 2T MEKAL model with $T_1 = 20$ and $T_2 = 93$ MK, $EM_1 = 5.0 \cdot 10^{52}$ and $EM_2 = 0.6 \cdot 10^{52}$ cm$^{-3}$, whereas for the quiet star the 1T MEKAL model with $T = 18$ MK appeared to be sufficient. But the flare spectrum could be equally well presented also by the thermal MEKAL model + power spectrum corresponding to nonthermal energetic electrons. Thus, the observations did not reveal the nonthermal component of X-ray emission of the AU Mic flare, though this could be expected from the light curve. Further, at least five strong bursts on EQ Peg were recorded within the range of 2–10 keV, and their spectra could be presented by purely thermal models and the combination of thermal and nonthermal components. Significant similarity of light curves in the optical U band and in medium X-rays was found in the EQ Peg flare of 2 October 1996 recorded in all four ranges of the cooperative observations.

* * *

To date, the studies of stellar X-ray flares being carried out with XMM-Newton and Chandra launched on the border of centuries are the most effective.

In Chandra ACIS observations, Rutledge et al. (2000) recorded on 15 December 1999 a flare on the rapidly rotating brown dwarf LP 944-20, which was not detected in X-rays out of the flare. The maximum flare luminosity was $10^{26}$ erg/s, which is lower by an order of magnitude than in the flare on the M8 dwarf VB 10 (Fleming et al., 2000). Within the RS model of the flare plasma they estimated the flare temperature as about 3 MK.

In the course of acceptance tests of XMM-Newton/RGS, Güdel et al. (2001a) carried out observations of AB Dor with the aim of studying variations of temperature structure, density, and composition of elements in bright flares. They detected variations of fluxes in lines of highly ionized iron and continuum. In the quiescent state the composition of elements was lower than that in the solar photosphere and increased at the onset of flares up to a factor of 3, i.e. an inverse FIP effect was present. The average electron density $n_e \sim 3 \cdot 10^{10}$ cm$^{-3}$ did not vary during the flares. The half-length of magnetic loops is $2.5 \cdot 10^{10}$ cm $\sim 0.3R_*$ in the constructed model.

Matranga et al. (2005) carried out observations of AB Dor with XMM-Newton and analyzed the region of Fe XVII $\lambda$ 15–17 Å lines. The ratio of $\lambda$ 16.78/$\lambda$ 15.01 lines increased significantly in hotter flare plasma. This variation contradicts theoretical predictions and was attributed to the scattering of $\lambda$ 15.01 Å line at an optical thickness of 0.4. In the flare, $n_e$ was estimated as $4.4 \cdot 10^{10}$ cm$^{-3}$ at a loop length of $8 \cdot 10^8$ cm and the iron abundance as 0.5 of the solar one.

In the course of the mentioned observations of the $\alpha$ Cen system, when Güdel et al. (2001b) with XMM-Newton first resolved components A and B in X-rays, high flare activity of both components was detected; on component B the frequency of flares proved to be somewhat higher than that on component A, but a slower decay of flares was on component A. Using the 2T VMEKAL algorithm, the modeling of flare coronal plasma was performed for the strong flare of 25 April 2000 on YY Gem, whereas the 3T model derived for the quiescent state (see above) was used as initial parameters. The detected parameters of flare plasma are as follows: $T_1$ = 9 and $T_2$ = 37 MK, $ME_1$ = 0.4 and $ME_2$ = 3.4 in units of $10^{52}$ $cm^{-3}$ at a significant iron enrichment of matter.

On 12 August 2001, in the course of XMM-Newton observations with EPIC of Proxima Cen, Güdel et al. (2002a, 2004) recorded X-ray radiation of the star, a spectrum with a resolution of 300 within the range of 0.35–25 keV was recorded with RGS in the region of OVII emission about 22 Å, and the XMM-OM instrument recorded the stellar brightness in the U band. They revealed a strong variability of coronal X-rays with flares of amplitudes up to $A_X \sim 100$ and variations of low-level emission on three time scales: slow decay over several hours, modulation for about an hour, and weak flares for several minutes. Whereas the weakest recorded flare had $L_X$(0.15–10 keV) = $2 \cdot 10^{26}$ erg/s, which corresponds to the average solar flare. Some flares were preceded by short-lived bursts in the U band, and Güdel et al. assumed that these ultraviolet bursts were consistent with untraceable nonthermal bursts of hard X-rays implied by the analogy with solar observations. The strongest flare also preceded by the powerful burst in the U band (see Fig. 33) lasted for more than five hours, its maximum luminosity $L_X$ (0.15–10 keV) = $4 \cdot 10^{28}$ erg/s and $E_X = 2 \cdot 10^{32}$ erg; it first made it possible to measure significant variations of electron density from the OVII and NeIX triplets with a maximum of $4 \cdot 10^{11}$ $cm^{-3}$ at the temperature 1–5 MK. In convolving the optical light curve with a decay constant of 200 s, its correlation with the X-ray one noticeably increased. From the density estimates in combination with the measured emission measures they obtained the following mass estimates: $5 \cdot 10^{14}$, $2 \cdot 10^{15}$, $3 \cdot 10^{14}$, and $2 \cdot 10^{15}$ g in moments of the primary flare maximum, its decay, secondary maximum and its decay (see Fig. 33); whereas the volume of hot plasma amounted to $7 \cdot 10^{26}$, $2 \cdot 10^{28}$, $4 \cdot 10^{26}$, and $5 \cdot 10^{28}$ $cm^3$. The density and mass increase at decay phases of the relatively cold component of the flare plasma was interpreted as a result of cooling down of the hotter component arising at the flare maximum. Based on the estimates, Güdel et al. concluded that the evaporation of the chromosphere played the main role in the development of the flare. They estimated the characteristic size of flare loops in two independent ways and obtained close values: $10^{10}$ cm $\sim R_*$. The relative abundances of elements revealed weak variations with an increase of FIP as compared to solar values. The strong flare model did not lead to an appreciable optical thickness of the FeXVII line. Comparing this large flare with other analogous flares occurring far often in more active stars, they suggested that the X-ray properties of active stars were the result of overlapping flares with hotter plasma.

Reale et al. (2004) continued the analysis of the strong flare on Prox Cen of 12 August 2001 within the framework of the nonstationary hydrodynamic modeling of burning, flare maximum, and the larger part of the decay phase, including the inclination change and secondary maximum. Splitting the light curve into six segments— burning, decay, secondary maximum and, four regions of decay with different rates, they derived EM and temperatures within the 2T models, which allowed the morphology of loops and heating mechanisms to be constrained and made it possible to show that a flare could be represented by two components: the first large excited by an intense heating impulse in one flare loop, having a half-length of $10^{10}$ cm, $\log T = 7.6$, and log EM = 51.3, and a less intensive heating impulse in half an hour, probably, in the loop arcade of the same length. The heating functions in both cases are rather similar: impulses are localized at the bases of loops followed by a smooth decay in the coronal part of the loops.

During the mentioned simultaneous Chandra and XMM-Newton observations of YY Gem on 29/30 September 2000 throughout 11.5 hours Stelzer et al. (2002) observed two X-ray flares with luminosities in maxima of 2 and $8 \cdot 10^{29}$ erg/s. When comparing the high-dispersion spectrum of a more stronger flare of them with the appropriate spectrum of the quiescent state of the star before the flare, a significant strengthening of emission lines was observed in the region of 16 Å. Constructing the coronal plasma models of these flares from the XMM-Newton EPIC spectra, two more components with enhanced iron abundance were added to the described three-component model of the quiescent corona. As in previous investigations, the temperature of the flare plasma approached its maximum during the burning of flares and the emission measure — at the beginning of decay.

In the course of simultaneous Chandra and XMM-Newton observations, Stelzer and Burwitz (2003) resolved the spectroscopic binary systems Castor A and Castor B with primary components of early spectral types. At that time only Chandra could resolve radiation of components A and B separated by 4″; XMM-Newton was not capable of doing it but due to high sensitivity it allowed a quantitative analysis of the total high-resolution spectrum to be performed. The oxygen triplet yielded $n_e = (0.5–1) \cdot 10^{10}$ cm$^{-3}$, the neon triplet yielded less confidently a higher density and $T \sim 2$ MK. Strong flare activity was recorded — six flares of both components with a burning time of about 10 min and 2–3 times more prolonged decay, which is typical of X-ray sources among late-type stars.

In a close pair of ER Vul consisting of two G dwarfs with an orbital period of 0.69 days Brown et al. (2002) recorded an X-ray flare with a duration of 30 min. The same system was observed by Hussain et al. (2012) throughout 140 ks with Chandra. These observations overlapped three strong flares, and an increased brightness level was recorded between the first and second of them. The X-ray and $H_\alpha$ spectra allowed one to suggest a high activity of both components of the system. All the flares revealed a hot component with a temperature of higher than 10 MK. The lengths of flare loops were estimated as $0.7R_*$, $1.5R_*$, and $1.8R_*$, the two latter were close to a half-distance between components of the system, which points to the magnetospheric interaction of coronae of the components.

In the course of the mentioned Chandra/LETGS observations of the system L 726-8, Audard et al. (2003) detected a significantly higher flare activity of component B than that of component A and distinguished on it, i.e. on UV Cet, flares with two types of light curves: symmetric with slow burning and slow decay, which are on component A as well, and flares with fast burning (about 80 s) and decay of about 40 min. This seems to be associated with appreciable difference of flare activity of these stars in the radio range as well. Furthermore, the spectra of the quiescent state of components are rather close; their analysis leads to the conclusion on the noticeable emission measures at the temperatures of 3–6 MK.

Van den Besselaar et al. (2003) carried out XMM-Newton and Chandra observations of AD Leo and recorded five flares with the former instrument and one with the latter. The spectra of flares and quasiquiescent state were analyzed separately, but several weak flares hampered consideration of the stellar quiescent state. The temperature structure was considered within the framework of the multitemperature model and EM distribution. In the 3T models, EM of a hot component out of flares is significantly lower than that in flares; the temperatures span the range from 1 to 40 MK in flares and from 1 to 20 MK in the quasiquiescent state. There is no difference in the abundances of elements, although one may speak on a weak inverse FIP effect on AD Leo. From the triplet lines of OVII and NeIX $n_e$ and $T_e$ were determined. Both instruments revealed an increased EM at a high temperature in the flare state.

Maggio et al. (2003) carried out XMM-Newton observations of AU Mic and recorded six flares and numerous weak events. The coronal density was determined from the triplet lines of OVII for all observations and for the most recent four flares. The average density over all the observations proved to be at the limits of this triplet, $1 \cdot 10^9$ cm$^{-3}$. But in the fourth and fifth flares $n_e$ increased up to $2.0 \cdot 10^{10}$ and $(1.0–1.6) \cdot 10^{10}$ cm$^{-3}$.

Raassen et al. (2003b) carried out XMM-Newton RGS and EPIC-MOS observations of AT Mic in the range of 1–40 Å and on 16 October 2000 recorded one strong and three weak flares on the star. They represented the stellar quiescent state and the flare within the 3T and DEM models: the temperature range is within 1–60 MK, EM = 12 and 20 in units of $10^{51}$ cm$^{-3}$. The H- and He-like C, N, O, Ne, and Fe XVII transitions within the temperature range 1–10 MK dominate in the high-resolution spectra from 8 to 37 Å at $\Delta\lambda \sim 0.07$ Å. The Ne, Si lines and the iron K-shells arising in the hot component at $T$ = 30 MK are seen lower than 10 Å in the EPIC-MOS spectrum; the K-shell transitions are the most noticeable during a flare. The spectral lines involved into EM generally arise at $T_1$ = 3 MK and $T_2$ = 8 MK, whereas at $T_3$ > 20 MK there forms a continuum lower than 10 Å and the iron K-shell lines. From the OVII lines the electron densities are estimated as $2 \cdot 10^{10}$ cm$^{-3}$ in the quiescent state and $4 \cdot 10^{10}$ cm$^{-3}$ — in the flare. A depletion of FIP effects, i.e. the inverse FIP effect, is detected in the quiescent state of AT Mic, but it weakens in the flare.

Observing the EQ Peg system with XMM-Newton/EPIC, Robrade et al. (2004) first distinguished the components of this close pair, established the X-ray flare activity of each of them, estimated a high brightness of component A and more frequent flares on component B.

Robrade and Schmitt (2005) carried out XMM-Newton observations of two single active dwarfs EV Lac and AD Leo and two nonseparated pairs AT Mic and EQ Peg. Numerous flares were detected in all the cases. The multitemperature models with variable abundances derived with EPIC and RGS yielded the consistent results. The inverse FIP effect was detected in all the cases: the metallicity is lower than that in the solar photosphere but seems to agree with photospheres of M dwarfs. During the flares, prominences significantly increase in the hot component, but the cold component reacts weakly with flares.

Later, using XMM-Newton, Robrade et al. (2010) carried out observations of the binary system M8.5/T5.5 SCR 1845-6357. The system was not resolved, the recorded X-rays were attributed to the M dwarf, and the latter was analyzed as an analog of VB 10. The quasiquiescent radiation and a strong flare with $A_X \sim 30$ and duration of 10 min were detected in the range of 0.2–2.0 keV. In the quasiquiescent state, $T \sim 5$ MK and $\log(L_X/L_{bol}) = -3.8$, which corresponds to rather high activity. A fairly noticeable brightening occurred in the flare in the near ultraviolet, the temperature increased up to 25–30 MK and luminosity $L_X$ up to $8 \cdot 10^{27}$ erg/s. A high luminosity is present despite the age of the system of several billion years.

Stelzer (2004) recorded a strong flare on the old brown dwarf Gl 569 Bab, whose age of 300–800 million years was determined from a bright component of the pair M2e Gl 569 A. This means that relatively evolved brown dwarfs show that coronae of very low-mass objects can maintain their strength beyond the earliest age as well. Later, Stelzer et al. (2006) conducted simultaneous XMM-Newton observations of the M8 dwarf LP 412-31 in X-rays and in the V band and recorded a giant flare with the amplitude $A_V = 6^m$ and a total radiative energy of $3 \cdot 10^{32}$ erg in both channels. The flare was fully recorded in both channels with a resolution of 20 s. The observations were in agreement with the impulsive energy release and subsequent radiative cooling without additional feeding at the decay phase. It was concluded that the optical flare emerged on a smaller part of the stellar surface, whereas the size of the flare X-ray plasma was of about the stellar radius. The absence of small-scale variations in the light curve makes one suspect a nonstandard energy distribution of flares.

In the described cooperative observations of EV Lac in September 2001 (Osten et al., 2005), the central moment was a continuous X-ray monitoring of the star with Chandra/HETGS of a total duration of about 100 ks; the first 40 ks the star was quiet in X-rays, during the next 60 ks nine X-ray flares were recorded with amplitudes of more than 1.6, which accounts for, on average, 0.32 flares per hour (see Fig. 56 and Table 17). The energy distributions in quiescent X-rays of the star, for six flares both the ascending and descending branches of the three most prolonged flares were analyzed from the low-resolution spectra within the 2T models. Parameters of the obtained models are given in Table 17. The energy distributions in ascending branches of flares differ significantly from the energy distributions in flares as a whole, and a hot component appeared in the most prolonged eighth flare. Metallicity everywhere was reduced, but variations of this value were not noticeable. Then Osten et al. analyzed intensities of the emission lines of highly ionized ions in the range of 6.18 to 21.6 Å and from the ratios of temperature dependent pairs of lines in spectra of the quiescent state of six weak and three strong flares obtained an independent confirmation of the increased plasma temperature in strong flares. In the course of monitoring of EV Lac in X-rays the star was observed in the U band with the 2.1- and 2.7-meter telescopes of the McDonald Observatory. A small optical flare at 3:04 UT, a middle one at 4:04 UT, and a strong flare at 8:30 UT did not cause any discernible variations in X-rays, but seven or eight optical flares of different amplitudes were recorded during a remarkable X-ray flare at 7:02 UT. Unfortunately, these optical observations were carried out at a cloudy sky and one could not make any conclusions on the correlation of optics and X-rays. The behavior of EV Lac in this campaign in the radio and ultraviolet ranges will be discussed in the subsequent sections of this chapter.

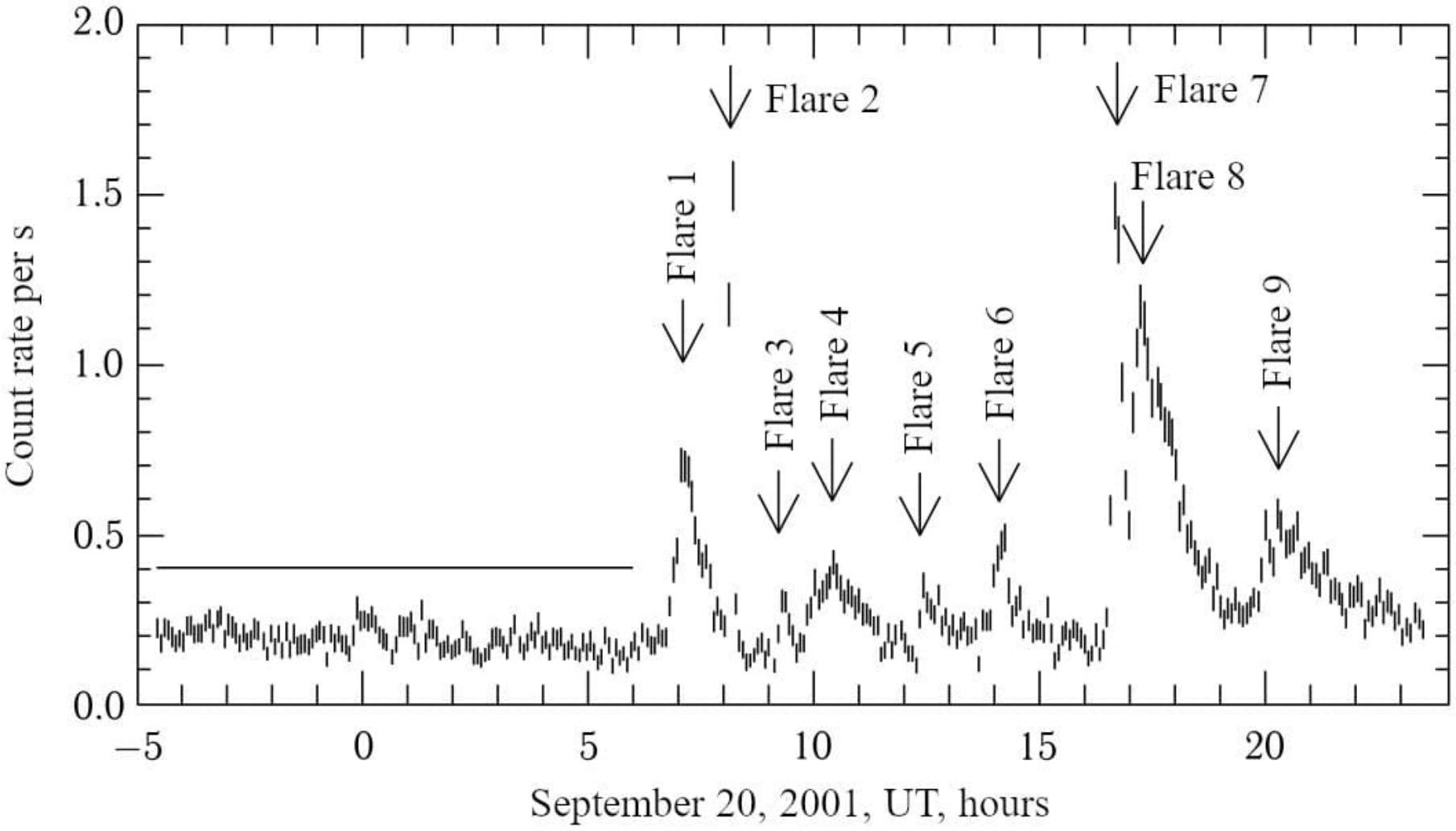

Fig. 56. The light curve of EV Lac in X-rays recorded with Chandra on 19/20 September 2001 (Osten et al., 2005)

Telleschi et al. (2004) carried out XMM-Newton and Chandra X-ray observations of five solar-type stars of different age and with different axial rotation periods. From the OVII lines the electron density was estimated as a few $10^{10}$ $cm^{-3}$ for all the stars. The EM distribution was constructed in the temperature range log$T$ = 5.5–8.0. DEM extended up to 20–30 MK for the most active stars 47 Cas and EK Dra, but this was absent for older stars. The abundances of elements revealed a weak tendency for a growth of high FIP for the most active stars — an inverse FIP effect (IFIP effect).

Sanz-Forcada et al. (2004) compared the coronal abundance of elements for four stars of different activity levels with the well-known photospheric abundances, but did not make any straightforward conclusions. Since the contents of coronal holes, active regions, and flares are different on the Sun as well, one fails to unambiguously associate the FIP effect and age, activity level, and other parameters of active stars.

Laming (2004) proposed a universal layout for explaining the FIP and IFIP effects based on the interaction of waves with partially ionized gas at the base of the loops. Instead of consideration of the ion heating in this place, he studied the effect of waves of ponderomotive force on the upper chromosphere plasma. These may arise when the upward waves from the chromosphere intersect or reflect from the boundary chromosphere–corona and to a great extent are determined by the properties of the overlying coronal loop. Advantage of this scenario is in the fact that both positive and negative variations of the abundance of ionized particles, i.e., the FIP and IFIP effects, may occur for the realistic wave energy densities.

Comparing the results of three measurements of the coronal density of AD Leo from the density-sensitive lines derived with Chandra, Maggio and Ness (2005) found a variability of this value apparently caused by flare activity of the star.

Using XMM-Newton, Mitra-Kraev et al. (2005a) carried out simultaneous observations of AT Mic, AU Mic, EV Lac, UV Cet, and YZ CMi in X-rays, 0.2–12 keV, and in the ultraviolet, 1800–3200 Å, throughout 40 hours they recorded 13 flares in both ranges, detected a systematic delay of X-rays by hundreds of seconds with respect to the ultraviolet, and in individual flares first found a power correlation between these emissions with a spectral index between 1 and 2. They constructed 3T models with temperatures of 2–3 MK, 7–8 MK, and > 20 MK. The lines in the range of 0.2–1 keV and an exponential decay of 1 keV dominated in X-rays. During a flare, the decay occurred slower, pointing to a higher temperature. The range of 0.2–1 keV comprised 70–75 % of energy, 1–2 keV — 15–20 %, and 2–12 кэВ — 10 %; during a flare the contribution of the former range decreased by 5 % and the latter one increased. These observations confirmed the presence of the Neupert effect and the process of chromospheric evaporation.

Ness and Jordan (2008) analyzed the observations of $\varepsilon$ Eri derived with several instruments: Chandra/ LETG, EUVE, HST/STIS, FUSE, and XMM-Newton RGS. From the measured emission lines they found relative abundances of elements, estimates of electron density and pressure, and the emission measure distribution. In their resulting coronal models, the temperature was 3.4 MK, electron pressure — $1.3 \cdot 10^{16}$ cm$^{-3} \cdot$ K at a temperature level of $2 \cdot 10^5$ K, the filling factor 0.14 at a level of $3 \cdot 10^5$ K and there was no evidence of the FIP effect.

During the strong X-ray flare on YZ CMi of 9 October 2000, which was observed by Raassen (2005) with EPIC-pn, EPIC-MOS, and RGS onboard XMM-Newton in the range of 0.3–10 keV, the total EM increased from 11 to 17 in units of $10^{50}$ cm$^{-3}$, $L_X$ — from 19 to 27 in units of $10^{27}$ erg/s and this growth was caused by the hot component of the corona with a temperature of about 2 keV. A small inverse FIP effect was noted. A more comprehensive analysis of this flare performed by Raassen et al. (2007) in the range of 1–40 Å with XMM-Newton yielded the total range of coronal temperatures from 1.3 to 42 MK, EM — from 14 to 22 in units of $10^{50}$ cm$^{-3}$ and the inverse FIP effect in the quiescent state but without difference in the active state. The length of flare loops was estimated as (5–13) $\cdot$ $10^9$ cm and the coronal field strength as 50–100 G.

Smith et al. (2005) carried out observations of five active M dwarfs with XMM-Newton/ EPIC and VLA and recorded a series of flares. A comparison of radiation in these channels led to the conclusion on different flare mechanisms. In the cases when a distinct correlation of radio and X-rays is present then the heating and X-ray radiation are produced by the same particles which radiate in radio. When radio waves are not accompanied by X-rays, the radio emission seems to be excited by coherent processes involving a relatively few particles. There was definitely a case when the radio emission was polarized. The case of X-rays without radio is difficult for the interpretation, possibly, the heating particles accelerate up to very high energies and radio synchrotron leaves the field of view.

In the course of XMM-Newton observations of AT Mic in the range of 0.2–12 keV Mitra-Kraev et al. (2005b) first detected oscillations of a flare in X-rays. The flare was rather prolonged, the oscillation started at its maximum with a period of about 750 s, an amplitude of 15 %, and with an exponential decay time of about 2000 s. They assumed that the oscillations were a standing magneto-acoustic wave tied to the flare loop and found that the most possible interpretation was a longitudinal slow-mode wave at a fundamental frequency of the loop of a length of $2.5 \cdot 10^{10}$ cm with a local magnetic field of $105 \pm 50$ G. Stepanov et al. (2006) showed that the decay of this flare was due to electron heat conductivity in the medium with a density of $(3–10) \cdot 10^{10}$ $cm^{-3}$ at the minimum magnetic field in a region of energy release of 100 G.

Throughout three nights Liefke et al. (2007) carried out simultaneous XMM-Newton and VLT/UVES observations of CN Leo and on 19 May 2006 recorded a strong flare with an amplitude of 500 in optics and about 100 in X-rays with a total duration of about 25 min. In the blue and red rays and X-rays, the onset of the flare was impulsive and practically simultaneous, but immediately after the maximum a fast decay of optical brightness started in these rays, whereas the maximum radiation persisted for about 5 min in X-rays. From the X-ray lines of OVII the hot plasma density was estimated at a level of $10^{10}$ $cm^{-3}$ in the stellar quiescent state and $> 10^{12}$ $cm^{-3}$ in the flare.

Later, they returned to these observations (Liefke et al., 2009), estimated frequencies and amplitude distributions of flares on the M dwarf CN Leo in the course of these simultaneous observations in the coronal X-rays, chromospheric lines, and optics and found that most discernible events were seen at all atmospheric levels; several weaker events were seen only in chromospheric lines, which corresponded to the $H_\alpha$ flares on the Sun. One strong event proved to be very weak in chromospheric lines and was not seen in the photospheric continuum which could be attributed to a large size of the flare loop. But no correlations of flare amplitudes and decay times were found at different atmospheric levels. From these observations they studied time variations of the corona during a giant flare on 19 May 2006, estimated an average electron density as more than $5 \cdot 10^{11}$ $cm^{-3}$, twice enhanced iron at a rise and in maximum, and a size of < 9000 km. They assumed that the flare occurred not in a separate loop but in a compact arcade.

Crespo-Chacon et al. (2007) studied two relatively weak flares on CC Eri and found that all regions of their light curves were well represented by 3T models with variable EM but surprisingly stable temperatures of 3, 10, and 22 MK. With 3.4 and 7.1 ks exposure time and amplitudes of 1.5–1.9 they covered arcades with tens of similar coronal loops, which were less than the radii of each component of the system. The derived results are in agreement with the notions that the X-ray light curve of CC Eri may be a result of superposition of many low-energy flares and the very flares may be scaled by solar flares.

Using Chandra, Wargelin et al. (2008) studied the M dwarf Ross 154. During the 61 ks of the exposure a strong flare occurred with $A_X > 100$, and signatures of the Neupert effect appeared in the initial phase. Microflares seemed to make a contribution into the quiescent radiation, but their distribution failed to be represented by a spectral index. The ratio Ne/O in the quiescent state was strengthened by a factor of 2.5 with respect to the solar one, but during the flare this ratio increased as compared to the quiescent state.

Nordon and Behar (2007) studied the strongest six X-ray flares recorded with Chandra/ HETG on OU And, CC Eri, and three RS CVn-type stars and found that a high-temperature component appeared in flares in all the cases, and bursts increased the temperature of the cold component. The FIP effect with respect to the quiescent state was recorded in five cases. Adding eight flares recorded with XMM-Newton, Nordon and Behar (2008) suspected a tendency for the FIP effect in the coronae, whereas the IFIP effect was present in the quiescent state. The situation was inverse in the coronae with the FIP effect in the quiescent state. This tendency is in agreement with the hypothesis of chromospheric evaporation as a mechanism of the effect occurrence.

Schmitt et al. (2008) carried out optical and X-ray observations of CN Leo with a high time and spectral resolution, which overlapped a flare with $A_X \sim 10^3$. The flare revealed a fast rise and a slow decay in optics and a delay of the X-ray peak by approximately 200 s. The two-second burst of soft X-rays was likely of thermal nature. Although the nonthermal bursts of hard X-rays usually accompany the onset of solar flares, the bursts of soft X-rays on the second scale were observed neither on the Sun nor on stars. The one-dimensional hydrodynamic model of this event requires the energy release over several seconds, i.e., this is a coronal explosion making it possible to study microflares as a coronal heating mechanism.

Pandey and Singh (2008) analyzed 17 X-ray flares recorded with XMM-Newton on six active G–K stars. The strongest of them was on $\chi$ Boo with a decay time of 10 ks and $A_X \sim 2$. They found that the size of flare loops was smaller than the stellar radius, and the hydrodynamic modeling led to the presence of the maintaining heating during the decay of the majority of flares.

In the course of XMM-Newton/EPIC/MOS observations of $\xi$ Boo Pandey and Srivastava (2009) detected variations of X-ray radiation of a flare, which were attributed to the helical instability of magneto-acoustic waves; the wavelet and periodogram analysis yielded an oscillation period of about 10–19 s.

Wolter et al. (2008) carried out simultaneous XMM-Newton and VLT observations and performed an analysis of photospheric, chromospheric, and coronal structures during a moderate flare, which radiated in X-rays and ultraviolet but not in the white light, and was located at a middle latitude between the large group of spots and the weakly spotted pole. The flare differed in longitude and height (about $0.4R_*$) from the prominences which were observed at the distance $2R_*$ from the stellar surface. Contrary to the photosphere, neither chromospheric nor coronal emission detected rotational modulation.

Huenemoerder et al. (2010) performed an analysis of 25 X-ray flares on EV Lac recorded with Chandra/HETGS over 200 ks in 2001 and 2009. Among the considered events there were very short-term (about 100 s) and long-term (up to 10 ks) with different light curves and amplitudes. Within the three-parameter Weibull distribution they represented the light curves of flares and performed a statistics of their parameters; from 139 spectra they traced their evolution on the density–temperature plane and found noticeable heating of flare matter. Short-term flares turned out to be significantly hotter than the long-term ones, and they noticeably differed in a loop size, therefore the simple scaling was unsuitable. The estimated range of the loop sizes amounted to $(0.1–1)R_*$. The detected upper limit of the flux in the fluorescent iron K line proved to be close to the expected one for the compact loops.

Fuhrmeister et al. (2011) carried out XMM-Newton and VLT/UVES observations of Prox Cen during the quiescent state of the star and during a middle-strength flare. From the X-ray data they estimated the density variability of the coronal plasma and abundance of elements, parameters of loops in the flare. The magnetic field and its variability were estimated from optical data. The chromospheric model of the star was constructed from emission lines. During its active phase the forbidden Fe XIII $\lambda$ 3388 Å line was detected, as well as the asymmetry of the hydrogen, helium, and calcium lines. A model of the flare chromosphere was constructed within the PHOENIX program.

Using XMM-Newton, Lopez-Santiago et al. (2010) recorded a flare of the member of the young association (8 million years) of the TWA 11B star with anomalously slow burning: over 35 ks the burning was recorded and, possibly, a maximum was covered. They supposed that initially the burning of the first group occurred, which was followed by burning of the subsequent two-ribbon flare. The half-length turned out to be very large, about $4R_*$; earlier, such structures were observed in very young stars of the age varying up to 1–4 million years. The iron fluorescence at 6.4 keV was recorded during the two-ribbon phase.

Gupta et al. (2011) recorded with XMM-Newton a strong X-ray flare with the amplitude $A_X$ = 52 on the M8–M8.5 dwarf 2MASS J04352724-1443017 at a distance of 67 pc; in its maximum the flare radiated about $8 \cdot 10^{28}$ erg/s, and over the whole flare — $2.3 \cdot 10^{32}$ erg. The flare radiation was represented by the two-temperature model with 10 and 46 MK with a dominating hot component.

Using XMM-Newton, Tsang et al. (2012) recorded a strong X-ray and ultraviolet flare on the late K or early M dwarf in the region of the Large Magellanic Cloud LMC 335 = 2MASS J 05414534-6921512, in the course of which the X-rays increased by two orders of magnitude, and ultraviolet — by one order, and the total energy of the flare amounted to $(0.4–2.9) \cdot 10^{35}$ erg. The flare was traced for about 10 h from the precursor to the late decay, it is one of the most extensively studied ones; its decay was represented by two exponents with characteristic times of 28 min and 4 h, and X-ray spectra — by two components with temperatures of 40–60 MK and about 10 MK and a length of the flare loop of $(1.1–1.3)R_*$.

From XMM-Newton observations Srivastava et al. (2013) detected slow acoustic oscillations in the post-flare loop of the Prox Cen corona. A wavelet analysis distinguished periods of 1261 and 687 s, and these oscillations lasted for 90 and 50 min, i.e. more than three periods. The 20-minute period can be a fundamental mode of slow magnetoacoustic waves with a phase speed of 119 km/s in the loop with a length of $7.5 \cdot 10^9$ cm, which was heated in the maximum up to 33 MK and cooled down to 7.2 MK at a decay. The second period could be an overtone of the first one. The oscillations of the fundamental mode decayed with a characteristic time of 47 min, while the ratio of periods indicated the excitement of the stellar loop with longitudinally stratified density.

Over 8 years, Pillitteri et al. (2022) carried out 25 XMM-Newton observations of HD 189733. They detected an increase in the mean coronal temperature during the flares from 0.4 to 0.9 keV at constant fluxes and rigidity of radiation outside flares and found a power-law energy distribution of flares.

* * *

Favata et al. (2005) studied the brightest X-ray flares on young stars in Orion recorded throughout 13 days with Chandra. From the maximum temperature and EM at the decay phase they estimated the size, maximum density, pressure, and minimum of the limiting magnetic field. Consideration of 32 strongest flares revealed a wide range of decay times varying from 10 to 400 ks. The temperature was higher than 100 MK for a half of these flares, for the majority of flares there was heating at the decay stage. In many cases, there were long structures up to $10^{12}$ cm with a magnetic field of hundreds of gauss, which extended up to a comparable distance from the surface. Such very large flare structures with $l \gg R_*$ were not observed in older stars where, as a rule, $l < R_*$. Since most young stars of the Orion cluster are surrounded by disks, Favata et al. (2005) suppose that the large-scale magnetic structures confining the flaring plasma are the same structures which canalize plasma in the magnetospheric accretion, associating the stellar photosphere with the accretion disk.

Caramazza et al. (2007) studied X-ray flares of 165 low-mass stars in Orion and found that the energy distribution of their flares was described well by the power law with an index of about 2.2 and light curves could be represented by the superposition of flares with the uniform power law. A comparison of low-mass stars with stars in the range of 0.9–1.1 solar masses showed that at an age of 1 million years stars with similar X-ray luminosity had similar frequencies of flares.

Using the XMM-Newton/EPIC camera, Pillitteri et al. (2005) studied the X-ray variability in the young open cluster Blanco 1. Observations with the EPIC camera made it possible to detect variability on a time scale of hours, while the comparison of these data with ROSAT results — on a time scale of 6 years. As a result, they found that on a short scale M dwarfs were more variable than the solar-mass stars; the flare-type events with typical durations of an hour were characteristic of M dwarfs, whereas the smoother variations of X-ray emission were observed for dF–dK stars. Two strong, possibly, associated flares were recorded on one of the stars; analysis of their light curves and high-speed spectroscopy led to the conclusion on the flares in arcades consisting of loops of $10^{10}$ cm in size. Variations of X-ray emission on a time scale of 6 years had a twice smaller amplitude, whereas on the Sun this coefficient is close to 10. Thus, the processes, as the 11-year cycle, are less pronounced in the young cluster than on the older Sun.

* * *

In the course of observations at the Suzaku observatory Laming and Hwang (2009) recorded a flare with an amplitude of about 50 on EV Lac; its quiescent spectrum confirmed the known for this star IFIP effect with significant depletion of FIP elements, whereas during the flare the abundance of elements in the corona approached the photospheric one. To explain this fact, they attracted ponderomotive forces of Alfvén waves: FIP and IFIP

[1] Swift is a multiwave special space observatory for studying gamma-ray bursts developed by the USA, England, and Italy and launched into orbit 20 November 2004. It consists of three telescopes: BAT, XRT, and UVOT. BAT (Burst Alert Telescope) is a wide-angle system of an area of 5200 $cm^2$ capable of examining totally of more than a steradian and partially three steradians; over 15 s it surveys the selected field and reports to the Earth the detected X-ray sources at an interval of 15–150 keV and their coordinates with an accuracy of up to 1–4 min; an encrypted mask with several tens of thousands of apertures is mounted in front of the system focus. The X-ray telescope of the narrow field of view XRT of the Woltier 1 system operates in the range of 0.2–10 keV, consists of 12 mirrors as EPIC-MOS/XMM-Newton, specifies coordinates up to 2″, and performs a long-term monitoring over days and weeks. The 30-cm Ritchey–Chrétien telescope UVOT is based on XMM-Newton/OM, specifies coordinates up to fractions of seconds, operates with optical and ultraviolet prisms in the range of 170–650 nm. One of the first successes of the Swift observatory is a record of flares on CC Eri (Evans et al., (2008).

Konus-Wind is a Russian spectrometer Konus onboard the American spacecraft Wind mounted on the point L1 of the Sun–Earth system, rotates with a period of 20 s around its axis that is perpendicular to the ecliptic; it surveys the whole sky in hard X-rays and gamma-rays, patrols gamma-ray bursts, solar flares and other nonstationary phenomena with the average energy resolution that is available for scintillation spectrometers; it operates in three channels in the range of 10–770 keV with a time resolution of 64 ms, and 2 ms for bursts.

effects can arise due to different directions of these forces in the evaporating chromospheric plasma of the flare and heat conductivity downward from the corona.

* * *

Osten et al. (2010) carried out the first successive observations with spacecrafts Swift and Konus-Wind [1] applying the Liverpool Telescope. Swift was a hard X-ray trigger of the flare on EV Lac, which increased by a factor of 7000 in the range of 0.3–100 keV and by $4.7^m$ at the Liverpool Telescope, and in the flare maximum $L_X/L_{bol} \sim -3.1$. Within the hydrodynamic modeling the half-length of a loop amounted to $0.37R_*$ and the iron K line was detected in emission; it yielded the loop height $0.1R_*$ and variability on a time scale of 200 s. Osten et al. showed that this emission could arise due to collisional ionization, and the found parameters of spectra of ionizing particles accounted for a flux in the $K_\alpha$ line and the absence of nonthermal emission in the range of 20–50 keV.

Caballero-Garcia et al. (2016) reported on two flares recorded on the known radio and X-ray source in the binary dM4e system DG CVn. A strong X-ray flare on 23 April 2014 of one of the components of the system switched on the search system BAT of the Swift satellite, which in two minutes pointed its small field-of-view telescope to the target. The optical ground-based observations were carried out at the wide-angle telescope with high time-resolution Pi of the Sky and at two small telescopes of the created then in Spain the BOOTES system (Burst Optical Observer and Transient Exploring System) equipped with Compact Low-Resolution Spectrographs (COLORES). The duration of the first optical burst accounted for 60 s, an amplitude of about 4 stellar magnitudes, and it finished prior the X-ray maximum; This delay was attributed due to the Neupert effect first detected in hard X-rays. The luminosities of flares in maxima amounted to $L$(0.3–10 keV) = $1.4 \cdot 10^{33}$ and $3.1 \cdot 10^{32}$ erg/s; such values were encountered for other red dwarfs, but they require a much stronger dynamo than on the Sun.

Based on the X-ray observations of flares on Prox Cen with ASCA, Swift, Chandra, and XMM-Newton, Kashyap et al. (2018) considered a spectral index of their power energy spectrum and found its constancy during the activity cycle. If one attributes the onset of the flare due to the self-organization of critical processes, the filling factor of the magnetic field and the plasma heating mechanism do not vary during the activity cycle.

Fuhrmeister et al. (2022) carried out simultaneous observations of Prox Cen in X-rays with Chandra/LETGS and in FUV with HST/STIS. Based on 18 optically thin emission lines in both spectral ranges, the authors determined the temperature structure and the differential emission measure of the transition region and corona for both the quiescence and flaring state. The amplitudes of flares reach $A_X = 30$ and $A_{FUV} = 20$. DEM is represented with Chebyshev polynomials in the range of logT from 4.25 up to 8. The constructed synthetic spectra at 1-1700 Å may be considered as representative for the high-energy irradiation of Prox Cen b during flare periods.

In spring 2019, using their own equipment, a large team of Japanese researchers carried out monitoring of AD Leo in X-rays (0.2–12 keV), in the $H_\alpha$ line and optical continuum throughout 8.5 nights (Namekata et al., 2020b). They recorded 12 flares, involving one superflare with total radiative energy of more than $10^{33}$ erg. The optical continuum was not detected in the flares with a small ratio of intensities $H_\alpha$/X, while the cophased rotational modulation of emission was detected in the quiescent state of the star in X-rays and in the optical range with a period of 2.24 days.

In October 2019, using the Indian satellite AstroSat, three long-duration strong flares were recorded from the binary system EQ Peg in the 0.3-7 keV range (Karmakar et al., 2022). The peak luminosities of flares are found to be within (5-10) · $10^{30}$ erg/s; the rise- and decay times are derived to be up to 11 and 24 ks; the flare temperatures peaked at 26, 16, and 17 MK; the peak emission measures are (4-7) · $10^{53}$ $cm^{-3}$, the coronal loop lengths are about 2 · $10^{11}$ cm, and their density is unities of $10^{10}$ $cm^{-3}$ at a magnetic field of less than 100 G and full energy of up to $10^{34\text{-}35}$ erg.

In continuation of the described above long-term TESS observations (Ioannidis and Schmitt, 2020) of the young fast rotator AB Dor, Schmitt et al. (2021) compared the optical photometry of this very active star with its X-ray observations derived with eROSITA of the joint Russian-German mission Spectrum-Roentgen-Gamma during about 20 days. Continuous X-ray emission of the star proved to be very stable throughout one and a half years without hints of rotational modulation, whereas during a very strong flare with emission not less than 4 · $10^{36}$ erg the X-ray emission was at least by an order of magnitude less.

Núñez et al. (2022) compiled extensive lists of Praesepe and Hyades members and found 326 and 462 X-ray emission sources, of which 273 and 164, respectively, have rotation periods. At an age of these clusters of 700 billion years, only M dwarfs remain saturated in X-rays, with only tentative evidence for supersaturation. They found a tight relation between the Rossby number and value $L_X/L_{bol}$ in unsaturated single members, suggesting a power-law index between -3.2 and -3.9. Lastly, they found no difference in the coronal parameters between binary and single members.

Using Chandra observations of EV Lac, Chen et al. (2022) studied variations of emission profiles in several flares. Based on variations in OVIII, FeXVII, MgXII, and SiXIV lines, they detected coronal plasma flows with velocity up to 130 km/s, whereas upflow velocity generally increases with temperature. Variable line ratios of the SiXIII triplet reveal that these plasma flows in most flares are accompanied by an increase of the coronal plasma density and temperature. Chen et al. associate these results with explosive chromospheric evaporation in flares.

* * *

Let us briefly sum up the results.

X-ray flares on active dwarfs are nonstationary relaxation of optically thin gas at an initial temperature of tens of megakelvins. They occur on stars of F–M spectral types with rather strong coronae. Gas results from primary energy release in a flare, and its relaxation is due to heat conductivity, radiative losses, long heating at the decay phase, or combinations of these processes. In the strongest flares this hot gas fills magnetic loops of sizes up to the stellar radius, its total radiative energy in X-rays reaches $10^{34}$ erg, while the weakest flares depend on the opportunity of detection of such events with a low *S/N* ratio, since, as a rule, stellar flares are recorded on the background of quiet luminosity of stellar coronae. Some stellar flares are considered as analogs to solar compact flares, others to solar two-ribbon flares. However, this classification is not always certain and simple. Feldman et al. (1995) found that the correlation between temperature and emission measure of solar flares, which covers the ranges of $T$ = 5–35 MK and EM = $10^{47}$–$10^{51}$ $cm^{-3}$, toward higher temperatures and EM was continued by the active stars AU Mic, Algol, $\pi$ Peg, and UX Ari.

The relatively simple mechanism of thermal radiation of optically thin gas made it possible to determine a number of its characteristics in stellar flares: electron density, dimensions of magnetic loops, strength of magnetic fields, and even the variation of chemical composition of matter during flares.

## 2.4.2. Radio Emission of Flares

Radio emission of flares in the solar atmosphere is the field of solar physics that has plenty of data is due to the fact that many magnetohydrodynamic disturbances arise and propagate in low-density plasma sited in magnetic fields. On the other hand, various electromagnetic radiations induced by these disturbances can be observed in a wide wavelength range from millimeters to tens of meters. By the late 1950s, numerous radio data on solar flares were accumulated. Fast growth of the interest in flare red dwarfs naturally encouraged attempts to find similar phenomena on these stars.

After successful detection of stellar flares their research was stimulated by the progress of radioastronomical equipment and in its turn stimulated the development of plasma astrophysics.

* * *

On 28 September 1958, Lovell started observations of flare stars at the then largest 250-ft radio telescope in Jodrell Bank. During the first night two radio bursts were recorded. Over one and a half years data on 474 h of radio monitoring of UV Cet, YZ CMi, AD Leo, EV Lac, and BD+19°5116 AB at 100, 158, and 240 MHz were gathered. Radio observations were conducted with additional detectors displaced from the axis to control the atmospheric noise. Over this time, 13 bursts that were not in the comparison channel were recorded, but the absence of parallel optical control left doubts as to the stellar nature of these events. In autumn 1960, simultaneous radio and optical observations were started with a network of five Baker–Nunn cameras mounted by NASA to monitor artificial satellites. The time resolution of photographic observations was 2 min. In superimposing the epochs of the obtained radio emission records at the moment of maxima of 23 recorded rather weak optical flares a distinctly increased radio emission was found in the interval from 2 to 8 min after the summed up optical maximum (Lovell et al.,1963).

On 25 October 1963, the Jodrell Bank reflector simultaneously recorded a flare on UV Cet at 240 and 408 MHz. The brightness maximum at greater frequency took place two minutes after the optical maximum, while the maximum brightness at lower frequency — a minute later. Comparison of these features with the characteristics of solar radio flares showed that, based on the emission delay with respect to the optical maximum, the total duration of the event, frequency drift rate, and spectral index, this radio flare on UV Cet could be ranked between solar radio bursts of II and III types (Lovell et al., 1964).

A more distinct time relation of optical and radio flares was established using the results of photoelectric monitoring. According to Lovell and Chugainov (1964), in three flares on UV Cet, one flare on YZ CMi, and one flare on EV Lac the events in both ranges began almost simultaneously, but the maximum of radio brightness was reached 6–8 min after the optical maximum. This conclusion was confirmed by the most successful cooperative observations in October 1963, when the data of 132-h radio monitoring, 55-h observations with Baker–Nunn cameras, and 61-h photoelectric and visual observations by Soviet astronomers were collected (Lovell and Solomon, 1966).

For one and a half years since autumn 1960, UV Cet, Proxima Cen, V 371 Ori, and V 1216 Sgr were observed from Australia with optical support from amateur astronomers (Slee et al., 1963a). For more than 1000 h observations were conducted on the north–south baseline of the interferometer near Sydney at 15 and 3.5 m and for 52 h with the 64-meter reflector in Parkes at 75 and 20 cm. As a result, a flare on UV Cet and two flares on Proxima Cen were recorded in the radio and optical ranges and five flares on UV Cet, Proxima Cen, and V 371 Ori were recorded without optical support. Optical monitoring was visual, the duration of the optical flares accompanied by radio bursts was 40–60 min at $\Delta m_{\rm vis} = 0.3-1^m$, the flux density of radio signals was up to 16 Jy, though several of the brightest optical flares were not accompanied by radio bursts.

The strongest was the flare on V 371 Ori observed on 30 November 1963 at both frequencies at Parkes and at 15 m with an interferometer. In the latter case, a half baseline observed the star, while the other, the control region of 3° from the star (Slee et al., 1963b). A flare at 75 cm lasted for about 15 min, its rather symmetric light curve had a maximum three minutes after the optical peak, but on the descending branch one could see numerous 6–20 s weakenings of radio brightness. The optical monitoring was executed by the Baker– Nunn camera and visually by a team of amateurs: $\Delta m_{\rm vis} \sim 0.6^m$. At 20 cm and 15 m radio emission was recorded several minutes earlier, just before the beginning of the optical burst. The main feature of this stellar flare was enormous radio-emission energy of $10^{31}$ erg, which was higher than that of earlier recorded stellar flares by two orders of magnitude and that of strong solar flares in this range by 6 orders.

Slish (1964) and then Zheleznjakov (1967) proved that recorded essentially nonthermal radiation of the flare on V 371 Ori with a brightness temperature to $6 \cdot 10^{21}$ K in decameters could not be due to noncoherent synchrotron radiation of relativistic electrons, but probably was a result of coherent synchrotron emission under conditions close to the upper layers of the solar corona.

In May–August 1965, UV Cet, V 371 Ori, V 1054 Oph, and EQ Peg were monitored by the Parkes antenna at 153, 408, and 1410 MHz (Moisseev et al., 1975). Optical monitoring was executed photoelectrically in the Crimea, by the Baker–Nunn camera, and by a team of amateurs in Australia. Over 60 h of joint observations seven optical flares were recorded and for three of them radio emission was recorded at 153 MHz; no bursts were observed at other radio frequencies. The optical flare on UV Cet of 28 August 1965 had $\Delta m_{pg} = 1.5^m$, and burning and decay times of three and four minutes, respectively. Five minutes after optical maximum a seven-minute radio flare started with a flux density of 5 Jy at maximum. The flare on V 371 Ori of 22 August 1965 with $\Delta B \sim 0.15^m$ lasted for eight minutes and a radio burst of 3.5 Jy began simultaneously with the optical flare. The optical flare on V 1054 Oph on 9 May 1965 had $\Delta m_{\rm vis} \sim 0.3^m$ and radio flux of 7 Jy at maximum.

In October 1967, during cooperative observations, UV Cet was monitored in Australia by the Parkes radio telescope at 150 and 2650 MHz and by the Culgoora radio heliograph at 80 MHz (Higgins et al., 1968). On 3 October 1967, with an interval of two hours two strong optical flares were recorded, both had high activity at 150 MHz. During a stronger flare that started at 14:36 UT, the maximum flux at 150 MHz was 25 Jy and at 80 MHz, 37 Jy. At 150 MHz, 4 more radio flares simultaneous with optical bursts were found, but they were not confirmed at 80 and 2650 MHz.

During three nights in December 1963 and in one night in November 1968 Slee et al. (1969) carried out simultaneous radio and optical observations of the Orion nebula containing a large number of flare stars. Radio observations were carried out with the Parkes antenna at 136/150 and 408 MHz, optical monitoring, using the chain method by a Schmitt camera with a time resolution of eight minutes. Over 34.1 h of joint observations nine flares with $\Delta B > 0.6^m$ and duration from 22 to 70 min were recorded. The maximum radio emission delayed for 1–19 min with respect to the optical peak, but in two flares it preceded the optical maximum for 4 and 15 min. The duration of radio flares varied from 1.5 to 50 min. Slee et al. assumed the existence of two different types of radio flares with respect to their duration. In the strongest radio flare that occurred at 13:16 UT on 20 December 1963 and corresponded to two practically simultaneous optical flares in different parts of the nebula, the maximum flux at 136 MHz reached 230 Jy, while one of the longest flares at 16:24 UT on 21 December 1963 had smooth burning and decay and lasted for more than 40 min; close to maximum brightness the spectral index was equal to −1.65. For the strongest flare, Slee et al. estimated the full electromagnetic radiation as $5 \cdot 10^{35}$ erg, radio emission amounted to 1% of this value.

One of the most interesting radio flares was recorded on 19 January 1969 on YZ CMi during 12-h radio monitoring at Jodrell Bank (Lovell, 1969). At 1:59 UT a fast optical burst with $\Delta m_{vis} = 1.7^m$ was recorded by Andrews in Armagh; five minutes after the sharp maximum a fast decrease was replaced by a smooth decay, which proceeded until the star disappeared from sight in Northern Ireland. But at 2:49 UT Kunkel in Chile began photoelectric monitoring and tracked the slow decay of the flare in the U band during the following 3.5 h. The 4.5-h optical flare was recorded successfully at Jodrell Bank at 240 and 408 MHz. Records at different frequencies differed noticeably due to the frequency drift of radiation, the time change of the spectral index, and some technical reasons. The flare intensity at maximum reached 30 Jy, and in the middle part, where the estimates of the spectral index were the most authentic, it varied from −1 to −2.5, which in general agreed with similar estimates for the values recorded in Australia (Higgins et al., 1968). Assuming that the frequencies of the recorded radio emission corresponded to the frequencies of plasma oscillations, Lovell estimated the density of matter in the corona as $7 \cdot 10^8$ and $2 \cdot 10^9$ $cm^{-3}$ for 240 and 408 MHz, respectively. He concluded that the long luminosity of plasma required an input of additional energy during flare decay and estimated the total energy of radio emission of the flare as $3 \cdot 10^{29}$ erg, which is lower than its optical radiation by five orders of magnitude. Lovell noted the closeness of the frequency drift to the corresponding value in solar events of type II. The flare brightness temperature was $10^{15}$ K at meter wavelengths, but this estimate could be increased by four orders of magnitude if the flare was a local rather than global event in the stellar atmosphere. In total, up to 1969, within the program of radio-optical observations of flare stars at the Jodrell Bank radio telescope 36 campaigns were run and radio monitoring data for 4000 h were accumulated (Lovell, 1964, 1971). Shortly thereafter the telescope stopped operating because of modernization.

On 11 October 1972, the upgraded Jodrell Bank radio telescope, at a frequency of 408 MHz, and the 30-inch telescope of the Stephanion Observatory in Greece recorded a strong flare on UV Cet in the B band (Lovell et al., 1974). Figure 33 shows the light curves of the flare in both wavelength ranges. One can see a considerable delay of radio emission and its noticeably long duration as compared to the optical flare. The flare intensity at maximum reached 12 Jy. As the strong flare on YZ CMi of 19 January 1969, it was used by Kahn (1974) to construct a model of the disturbed atmosphere (see below).

In autumn 1972, the 37-meter radio telescope of the Vermilion River Observatory was used to observe EV Lac at 170 MHz over three nights (Webber et al., 1973). Optical support was provided by the 102-cm reflector of the Prairie Observatory (Illinois) and the 51-cm Palomar telescope. On 2 October 1972, both optical telescopes recorded a flare with $\Delta U \sim 0.14^m$ and duration of about 25 s, two and three minutes later there were radio bursts of about 25 Jy that decayed completely in nine minutes. At the same time using the same radio telescope, Tovmassian et al. (1974) observed the Pleiades with support from optical telescopes in Tonantzintla and Palomar. Over eight hours of joint observations one radio flare with an intensity of about 35 Jy was found, this result was confirmed by optical observations. But an attempt to find radio emission of flare stars in the Pleiades with the help of the Ooty radio telescope (India) failed (Sanamian et al., 1978).

Between October 1973 and January 1974, Spangler et al. (1974b) monitored YZ CMi, AD Leo, Wolf 424, and V 371 Ori with the 305-meter Arecibo radio telescope at 196, 318, and 430 MHz for more than 70 h. The sensitivity of the telescope was an order of magnitude higher than that of its predecessors. On the first three stars, 26 flares were recorded with amplitudes of tenths of Jy and duration of tens of seconds, the average frequency of the events was comparable with that of optical flares — from several hours to ten hours for one flare. For the flares on YZ CMi and Wolf 424 recorded at 196 and 318 MHz, the spectral index was estimated as −2.5 and −2.9, respectively. Then during 15 nights in January–February 1974 Spangler and Moffett (1976) carried out simultaneous observations of YZ CMi, AD Leo, and Wolf 424 with the Arecibo radio telescope and the 91-cm and 76-cm telescopes of the McDonald Observatory. Over 57.8 h of joint monitoring the radio fluxes of 13 radio events were detected directly from the records; one of the events definitely was not accompanied by an optical flare. In analyzing the moments of 62 optical flares, 15 more radio flares were found. Figure 34 presents the light curves of a flare on Wolf 424 recorded during this campaign. On the whole, the observations revealed a considerable diversity of light curves at two radio frequencies and distinctions of their light curves from the light curves of relevant optical flares. Most probable was a delay of the radio burst with respect to the optical one for 0–5 min. Out of 10 radio flares recorded at both frequencies four started simultaneously, in three flares a frequency of 31 8MHz and in three of 196 MHz were in the lead. The recorded radio flares did not show a preference to strong optical flares, and one of 3 strong optical flares on AD Leo definitely was not accompanied by a radio event. All this evidences the coherent mechanism of directed radio emission of stellar flares. By the end of 1974, the list of thus studied flare stars was supplemented by BD+16°2708, Ross 867, G 25–16, EQ Peg, and G 3–33, the total duration of radio monitoring exceeded 400 h and the number of recorded events exceeded 70, and the range of spectral indices extended from −2 to −10 (Spangler et al., 1975).

In observations of AD Leo with the Arecibo radio telescope on 1 April 1974 Spangler et al. (1974b) obtained the first radiopolarimetric data: all four Stokes parameters were measured at 430 MHz. A burst was recorded with an intensity of 0.5 Jy and the duration at the levels of 1/2 and 1/10 of the maximum intensity was 12 and 40 s. The degree of circular polarization at maximum was 56% and increased to 92% by the end of the flare, and the linear polarization was 21% at the maximum. By analogy with decimeter bursts on the Sun, Spangler et al. used the gyrosynchrotron mechanism of emission of moderately relativistic electrons in magnetoactive plasma to interpret the observations.

During a three-day cooperative campaign of observations of YZ CMi in November–December 1975 radio observations were carried out with eight radio telescopes at 12 frequencies in the range from 38 to 6300 MHz. For the first time radio–optical–X-ray coverage was arranged over 1/3 of the campaign duration (Karpen et al., 1977). In total, 11 radio events were recorded. During three of the longest events lasting from half an hour to an hour and a half no optical monitoring was arranged, the duration of all other radio bursts did not exceed one minute. The greatest number of events was recorded by the Ooty radio telescope at 327 MHz. The burst at 16:12 UT on 1 December 1975 was recorded with the Culgoora telescope at 160 MHz and with Parkes at 5000 MHz. Some 100 s after the radio flare at 20:44 UT on 1 December 1975 there was an optical flare with $\Delta B \sim 0.4^m$. The flare at 7:05 UT on 3 December 1975 was recorded simultaneously at 196 and 318 MHz and in the optical range.

Nelson et al. (1979) summarized the data of radio–optical observations of 15 flare stars obtained in 1972–76 with the Culgoora radio heliograph at 80 and 160 MHz in the mode that detected only long events, but not fast bursts. Over 110 h of observations supported by amateur astronomers 21 optical flare and 19 radio events were recorded independently, variability was found on 11 objects. The greatest activity was displayed by AD Leo: 30 and 58 radio flares were recorded over 34.5 h of observations at 80 MHz and 67.7 h at 160 MHz, respectively. The least active were AT and AU Mic, on which over 10.8 and 24 h of observations at 160 MHz no radio events were recorded. At the same time, no optical flares were found on AD Leo and AU Mic, while on AT Mic three flares were recorded. Only four pairs of radio and optical flares were contiguous: a five-minute flare on UV Cet of 21 July 1975 with $\Delta m_{pg} = 0.3^m$ and $F_R = 1.4$ Jy, two flares on BD+16°2708 on 18 May and 17 June 1975 with $\Delta m = 0.3^m$ and $0.8^m$ and duration of 70 s and >35 min with $F_R = 1.1$ and 1.9 Jy, and a 15-s flare on V 1054 Oph of 10 August 1975 with $F_R = 0.8$ Jy. In all cases, radio observations were executed at 160 MHz and $F_R$ was determined by integration over half an hour. Three of the four optical flares were short-pulsed events, after which long, up to 33 min, radio flares were observed, which resembled solar bursts of types II and IV and noise storms after strong solar flares. Thus, most optical flares did not cause detectable radio emission and radio flares were frequently recorded in the absence of optical flares. Nelson et al. assumed that this could be explained by the fact that both were recorded at the detection threshold or, as on the Sun, optical flares were not always accompanied by radio flares at meter wavelengths. For eight radio flares, significant circular polarization was found from 40 to 61% and more. If the size of the radiation source was $\sim 0.1R_*$, then the brightness temperature was $\sim 10^{15}$ K. Robinson, one of the authors of this publication, showed that coherent synchrotron or coherent cyclotron radiation matched such a temperature, but both yielded heavily collimated beams, which could not be observed for more than two hours. To maintain such a radiation, a continuous supply of relativistic electrons to the medium is required.

Davis et al. (1978) carried out interferometric observations of YZ CMi at 408 MHz with the Jodrell Bank radio telescope and the 25-meter Defford antenna. Over 48 h two flares were found: on 15 December 1977 with an amplitude of > 0.04 Jy and on 18 December 1977 with an amplitude of 0.12 Jy. The energy of the latter was lower by two orders of magnitude than that of the flare of 19 January 1969.

* * *

In January 1975, Moffett et al. (1978) carried out a 33.9-h radio–optical monitoring of YZ CMi, AD Leo, and Wolf 424 to detect centimeter flare emission. Radio observations at 1420 MHz were conducted with the Arecibo radio telescope. The optical monitoring was carried out at the McDonald Observatory and with the 152-cm telescope of the Mount Hopkins Observatory. Radio flares were suspected near two of the 41 optical flares, but 12 of the 14 radio events were not accompanied by optical flares.

In May 1977, during 3-day cooperative studies of Proxima Cen radio observations were carried out at 6 cm with the 64-meter Parkes antenna and at 13 cm with the 25-meter antenna in Johannesburg. Optical monitoring was carried out with the 76-cm telescope of the South African Observatory and the 41-cm telescope in Cerro Tololo, Chile. Observations at Parkes were carried out during four optical flares, one of them was strong. Radio observations at 13 cm covered the time of 26 optical flares, but no events in the radio range were detected. Hence, it was concluded that $L_R/L_{\rm opt} < 2 \cdot 10^{-5}$ (Haisch et al., 1978).

During cooperative observations on 6–8 March 1979 Proxima Cen was radio monitored at 6 cm at Parkes (Haisch et al., 1981). Over 24 h six flares were recorded with $\Delta U$ varying from $0.6^m$ to $4.5^m$; independently 12 radio flares with $F_R$ varying from 7 to 12 mJy were found; but only three pairs appeared to be close in time. Radio emission was not revealed during the strong X-ray flare. The most confident correlation was between flares on 8 March 1979 at 17:30 UT: they started and reached the maximum of 8 mJy and $\Delta U \sim 4.5^m$ simultaneously. Three radio flares were not accompanied by optical events five minutes before and after and seven radio bursts were not accompanied by optical flares. On the basis of the rather modest sample Haisch et al. suspected that there was a tendency of a five-minute delay of radio bursts and a lack of correlation between the amplitudes of the events.

A cooperative campaign of multiwavelength observations of YZ CMi arranged in October 1979 suggested radio monitoring from six radio telescopes at 12 frequencies from 275 to 7875 MHz (Kahler et al., 1982) over 40 h. One strong flare was recorded on 25 October 1979 (see Fig. 34). A confident signal from the flare was obtained with the Jodrell Bank radio interferometer at 408 MHz: radio emission began 17 min after the impulsive onset of the optical and X-ray flare, reached a maximum of 60 mJy, and then lasted for more than half an hour, while at other frequencies it decayed. Green Bank observations at 515 MHz detected one more radio event, probably a narrow-band one, since it was not recorded at 430 and 1428 MHz. If the radiation source was of the order of the stellar diameter, the brightness temperature of the flare of 25 October 1975 was $2 \cdot 10^{12}$ K. The temperature and the delay with respect to the optical flare are similar to solar bursts of type II caused by plasma oscillations and bursts of type IV in a continuum caused by gyrosynchrotron emission of quasirelativistic electrons. But on the Sun, the delay of bursts of type II usually takes about two minutes, which makes the gyrosynchrotron nature of radio emission of the flare on YZ CMi more probable.

Slee et al. (1981) described radio–optical observations of AT Mic and a flare that occurred on 25 October 1980. Radio observations were carried out at 6 cm in Parkes and photographic monitoring with the 26-inch refractor of the Mount Stromlo Observatory with a five-minute resolution. During serial measurements of the star and check region with an integration time of 100 s an accuracy of about 3 mJy was achieved, and the flux at maximum was $F_R$ = 20 mJy. Burning of the radio flare took less than 100 s, its decay lasted for about one hour. The optical flare started almost simultaneously with the radio burst and was of almost the same duration. This is a unique case of a very close correlation of radio and optical events. Three years later in Australia, new radio-optical observations of AT Mic (Nelson et al., 1986) were organized. During 3 nights the radio monitoring was carried out at 5 GHz in Parkes and at 843 MHz in Molonglo, optical observations were conducted in U, B, and J bands by the telescopes in Mount Stromlo and Perth. Radio observations covered the time of 10 optical flares, but the signal $> 3\sigma$ was recorded only during one of them at 15:30 UT on 3 August 1983 with $\Delta B = 0.4^m$. On the other hand, no optical flares were found during two radio bursts. But both flares on AT Mic — of 25 October 1980 and 3 August 1983 — were recorded in two ranges and had appreciable durations, whereas two flares without optical events were short radio bursts.

* * *

In the early 1980s, a new stage in studying stellar flares started with the launch of the Very Large Antenna (VLA) designed for recording waves of centimeter lengths.

In spring 1980, YZ CMi, Wolf 424, and BD+16°2708 were observed at 20 cm with VLA. Over 22 h the only event on YZ CMi was detected with a maximum luminosity of $2 \cdot 10^{14}$ erg/(s · Hz) and duration of about two hours, and strong circular polarization (Gibson and Fisher, 1981). The obtained light curve can be interpreted as three bursts separated by 45- and 41-min intervals, or as one event with 30-min burning and 12-min decay. The observations were continued in autumn 1980 with the optical support from the 200-inch Palomar reflector and the 60-inch Mount Wilson telescope. On 19 December 1980, at 6 cm Fisher and Gibson (1982) recorded a flare on UV Cet with a fast rise and an exponential decay, immediately after which another flare began at 21cm. It is not clear whether it was an analog to solar drift bursts or an independent event. During three nights they monitored UV Cet, YZ CMi, and CN Leo for 4 h and recorded 10 or 11 radio events. All the events lasted for about 20 min and did not have an impulsive beginning. At the same time 36 optical flares were recorded but they had no correlation with radio events, though there were rather strong optical flares: $\Delta B = 0.88^m$ on UV Cet, $1.43^m$ on YZ CMi, and $1.04^m$ on CN Leo. Based on the lack of correlation of radio and optical emission of flares and high circular polarization of radio emission of some flares, Fisher and Gibson concluded that gyrosynchrotron or another coherent mechanism was responsible for the centimeter radiation of stellar flares.

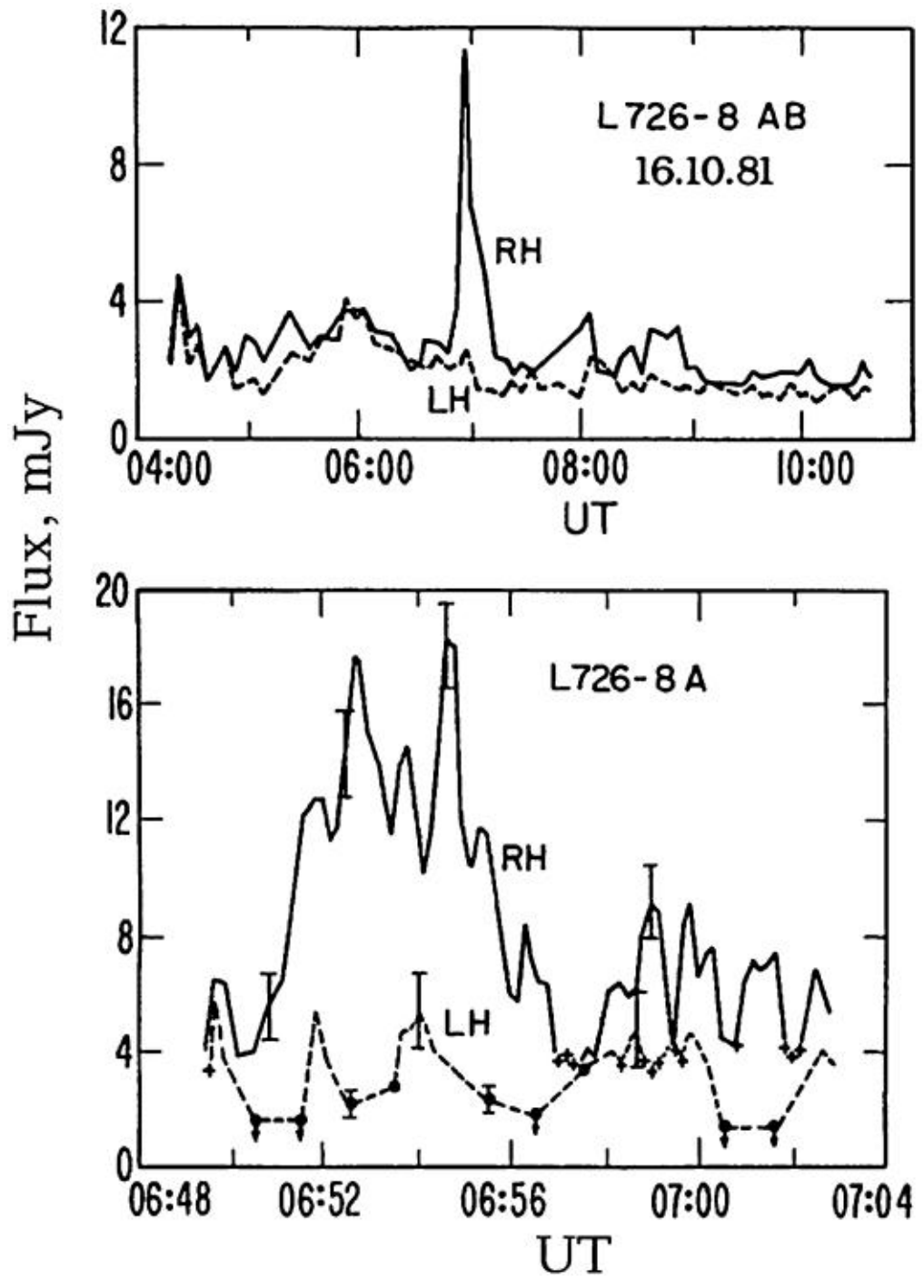

Fig. 57. The flare on L 726-8 A of 16 October 1981 recorded in right and left circularly polarized radiation at 6 cm, RH and LH, respectively (Gary et al., 1982)

Six-hour VLA observations of the system L 726-8 on 16 October 1981 at 6 cm were fulfilled with separate recording of circular polarization components (Gary et al., 1982). At 6:52 UT the right-polarized component of L 726-8 increased from 4 to 16 mJy, whereas the left-polarized component remained invariant. The light curve of the flare constructed with a resolution of 10 s demonstrated five or six quasiperiodic oscillations with a characteristic time of about 56 s (Fig. 57), over which the degree of circular polarization varied from 40 to 82%. Gary et al. considered various opportunities for the occurrence of radiation: modulation of a radio source by an external agent and variations of the mechanism of energy release. In the first case, these could be the oscillations of magnetic flux tubes, but they modulate incoherent radiation and consequently could not account for high brightness temperature and high circular polarization. The second variant can be realized in electron-cyclotron maser, assuming that the conditions required for this mechanism – sufficient strength of the magnetic field, availability of free energy for electrons and free emission at the second harmonic of the gyrofrequency – are met in the atmosphere of dMe stars. According to Gary et al., this mechanism operates at $B \sim 900$ G, brightness temperature of $10^{14}$ K, and $f \sim 10^{-4}$; but the mechanism of 56-s oscillations remained obscure.

In the course of observations of AD Leo on 1 February 1983 at 1400 MHz at Arecibo with a higher time resolution, Lang et al. (1983) recorded a very interesting flare. Its burning phase lasted for 18 min, and the decay phase was six minutes, whereas the degree of circular polarization did not exceed 15%. But on the ascending branch numerous fast bursts with 100% left-circular polarization occurred within three minutes, the burning time of these bursts did not exceed the time resolution of the instrument — 200 ms, the flux density achieved 130 mJy, and the brightness temperature was $10^{13}$ K. The event was interpreted as the emission of an electron-cyclotron maser at the second gyrofrequency in a longitudinal field of $B \sim 250$ G.

During the next days the cooperative program recorded significant variation of AD Leo activity (Gary et al., 1987). On 2 February 1983, simultaneously with an optical impulsive flare with $\Delta U \sim 0.7^m$ and duration of about three minutes, a radio flare with essential and variable polarization that lasted for about an hour and a half was recorded at 6 cm with VLA. On 5 February 1983, numerous independent flares were recorded at 20 and 6 cm over 7 h, and one of the flares, seen only at 6 cm, was 100% left polarized. According to Gary et al., a weak correlation of flare activity at 20 and 6 cm can mean various mechanisms of flare generation or various regions of their exit from the corona.

On 15 July 1985, Lang and Willson (1986b) carried out 1.5-h observations of AD Leo at 1415 MHz using the Arecibo telescope with a maximum time resolution for the instrument of 5 ms. They detected two events 50 and 25 s long with intensities of 30 and 10 mJy and a high degree of left-circular polarization. During the second half of the 50-s event quasiperiodic brightness oscillations occurred with a characteristic time of 3.2 s and the strongest component of these oscillations, in its turn, was composed of five separate bursts with a characteristic time of 32 ms, a flux density of 70 to 400 mJy, a rise time of not more than 5 ms and 100% circular polarization. Components next to the strongest one did not display such a structure. At such a fast burning the size of the radiation sources could not exceed $1.5 \cdot 10^8$ cm, the filling factor was estimated as $2.5 \cdot 10^{-5}$ and the brightness temperature as $\sim 10^{16}$ K. The radiation source could be caused by electron-cyclotron maser or plasma oscillations. In the first case, the longitudinal field should be 250 G and $n_e \sim 6 \cdot 10^9$ cm$^{-3}$. In the second case, $B << 500$ and 250 G and $n_e \sim 2 \cdot 10^{10}$ and $6 \cdot 10^9$ cm$^{-3}$ for the first and second harmonics of plasma oscillations, respectively. Quasiperiodic oscillations and fast bursts, according to Lang and Willson, could be caused by radial pulsations of coronal loops excited by trapped energetic particles or their pulsed source.

On 3 February 1983, VLA microwave observations of YZ CMi were carried out within a wide cooperative program (Rodonò, 1986a; van den Oord et al., 1996). At 6:08 UT an optical flare occurred with $\Delta U = 0.84^m$ and before it completely decayed at 6:12 UT there was an optical burst with $\Delta U = 3.83^m$. This optical maximum coincided with a burst at 6 cm, which reached a maximum of 5 mJy in seven minutes and lasted for about half an hour; significant circular polarization was absent. At 6 cm, at a time resolution of 10 s, the fine structure of the light curve was not seen, and at 2 and 20 cm no radio emission was recorded. If the radiation source was equal to the stellar size, its brightness temperature was about $10^9$ K. The analysis of the data sets by van den Oord et al. resulted in the conclusion that there was optically thick synchrotron radiation of electrons with energy of megaelectron-volts in the magnetic field of about 100 G.

On 5 October 1983, VLA recorded the activity of Gl 182 at 2 cm (Rodonò, 1986a). Quiescent radio emission of this star was not revealed, thus the light curve of the flare, constructed with a minute time resolution had three symmetric peaks at 10 to 6 mJy with rise and fall times of about one minute. The flare was not visible at 6 and 20 cm.

On 25 October 1983, a 3.5-h VLA recording of the radiation of YZ CMi at 6 cm revealed interesting phenomena (Pallavicini et al., 1985). At first there was a fast radio burst with an amplitude of about 6 mJy and duration of about 30 s on the background of quiescent radiation of 2.5 mJy. Then 83 and 75 min after there were gradual 20- and 30-min flares. All these brightness variations occurred on the background of growing circular polarization. Pallavicini et al. found analogs of all these events in solar centimeter-range radio emission. Observing this star on 10 December 1984, Lang and Willson (1986a) found two slow one-hour flares with an intensity of ~20 mJy, the events at 1515 MHz preceded for about an hour similar events at 1415 MHz. Identification of the appropriate events at these frequencies means narrow-band radiation $\Delta\nu/\nu << 0.1$ with slow frequency drift.

The flare on AD Leo of 28 March 1984 was studied in the wide wavelength range from ultraviolet to microwaves (see Fig. 37). At 3:22 UT there was an optical burst with $\Delta U = 2.1^m$, and after several additional rises the flare faded 22 min after the onset. Near the optical maximum there was a radio burst of up to 32 mJy at 2 cm, the radio emission at 6 cm started simultaneously, but had a smooth rise and reached a maximum of 10 mJy 10 min later (Rodonò, 1986a). The microwave radiation decayed simultaneously with the optical radiation. This flare was not found at 20 cm, which probably suggests its localization in the lower corona.

* * *

In early 1985, Kundu et al. (1987) observed UV Cet, AT and AU Mic with VLA at 20 and 6 cm and recorded microwave flares on all stars. In the binary systems AT Mic they found flare activity on both components at both frequencies. During one-hour observations on 5 February 1985 over 13 min an increase of radio emission from 3.6 to 5.6 mJy from the northern star of the pair was detected at 20 cm. Over the next 12 min the value decreased to the initial level. On 22 March 1985, the southern star displayed variations of weakly polarized radiation at 20 cm from 10 to 7 mJy. On 6 February 1985, during one-hour VLA observations of UV Cet, Kundu et al. recorded a smooth rise of radio emission from 1.7 to 3.2 mJy at 20 cm that lasted for 10 min. Twenty minutes later there was a sharp jump to 10 mJy. Circular polarization increased to 40%, while radio emission at 6 cm was invariant at the level of 2.5 mJy. There was a weak 100% polarized flare on the main component of the system L 726-8 A. Two-hour VLA observations of AU Mic on 22 March 1985 recorded the moment of maximum at 26 mJy and subsequent decay of a strong radio flare at 20 cm: over 85 min its brightness halved and circular polarization was kept throughout the observations at the level of 70–90%. But this strong flare was not visible at 6 cm (Kundu et al., 1987).

On 22 March 1985, Jackson et al. (1989) carried out one-day VLA observations of AD Leo, EQ Peg, UV Cet, Wolf 630, YY Gem, and YZ CMi at 20 and 6 cm. In most cases they found some kind of flare activity. The most interesting was the following result: at a time resolution of 10 s at the burning phase of a flare on AD Leo the sign of circular polarization changed from $V = +4$ mJy at 1465 and 1515 MHz to −13 and −7 mJy, respectively; at the maximum the flare reached 18 mJy. There was a single increase of intensity from 0.6 to 1.4 mJy at 6 cm with a subsequent decrease to 1.0 mJy, but no peculiarities in polarization were found. Another important conclusion drawn from these observations was that optically identical components of binary systems could essentially differ in the radio range.

As it was mentioned, in August–December 1985, Kundu et al. (1988a) carried out simultaneous VLA and EXOSAT radio and X-ray observations of UV Cet, EQ Peg, YZ CMi, and AD Leo to detect a correlation between flare radiation in these two wavelength ranges. Each star was observed continuously for 7– 11 h with a time resolution of 6.7 s. AD Leo was monitored continuously in each microwave region by a half of VLA-antennas, other stars by all antennas that were switched every 5 or 10 min from 20 cm to 6 cm and back. During nine-hour observations of the system L 726-8 on 4 August 1985 two right-polarized flares with intensities of 10 and 35 mJy were found at 20 cm on UV Cet and L 726-8 A, respectively. At the same time, bursts with an intensity of several mJy were recorded at 6 cm, but their duration was longer than that of flares at 20 cm. EXOSAT simultaneously revealed increased activity, but there was no correlation between the peaks of radiation in the two ranges. It is noteworthy that in the frequency range of 1390–1450 MHz the flux from the flare on L 726-8 A was approximately twice that at 1490–1540 MHz, and the general character of the light curves differed appreciably: at low frequencies a maximum was clearly recorded, whereas at higher frequencies it was absent. This could be due to the drift of radio-burst frequencies, but then one should consider the flare radiation as a narrow-band one with a width of about 100 MHz (White et al., 1986).

In observations of EQ Peg on 6 August 1985 at the beginning of the session a strong X-ray flare was recorded, then at 20 cm there was a many-hour activity at a level of 3–5 mJy and at 6 cm at a level of 10–12 mJy. But it is possible that this microwave activity started already on 4 August 1985, when the system was monitored in the radio range for half an hour and over this time a strong flare with an intensity of ~50 mJy was recorded on EQ Peg B at 20 cm (Kundu et al., 1988a).

During the observations of YZ CMi on 19 November 1985 from 10:00 to 13:30 UT at 20 cm a 100% left-polarized flare was found with an intensity of 14 mJy; at 6 cm a flare started at 9:24 UT and decayed by 13:00. An X-ray flare started at 8:30 UT and faded in an hour and a half. There was no increased X-ray emission during the strong radio flare at 20 cm.

During 11-hour observations of AD Leo on 15 December 1985 a 100% left-polarized 2.5-h flare with an intensity of 80 mJy was recorded, but it was not noticed at 6 cm or in X-rays. This flare was visible at 1415 MHz, but did not manifest at 1515 MHz, which could not result from the frequency drift, since the star was monitored long before the flare and after it. Details of less than a minute were not visible in the light curve of the flare. Such a long and strong polarized radiation has no analogs in solar phenomena (White et al., 1986; Kundu et al., 1988a).

Thus, no distinct correlation between the flare radiation in the microwave range and in soft X-rays was found, at least distinct maxima in one range have no analogs in another one.

On 23 June 1986, Willson et al. (1988) carried out three-hour sessions of VLA observations of AD Leo and YZ CMi. Two flares were recorded on AD Leo at 6 cm, while no activity was recorded at 20 cm. The first flare had a pulsed light curve with burning shorter than 10 s, a full duration of about a minute, and 100% right-circular polarization. Two hours later there was a flare with a complex light curve and duration of about 20 min and 100% left-circular polarization. On YZ CMi, a radio flare at 20 cm with a rather symmetric light curve was recorded, its total duration was about 20 min, the polarization was low and there were no signatures of narrow-band emission, but immediately after its decay a slow radio flare started at 6 cm with strong left-circular polarization.

* * *

Bastian and Bookbinder (1987) obtained with VLA the first dynamic spectra of stellar flares in the microwave range. The working band of 50 MHz centered at 1415 MHz was divided into 15 narrow bands, and recording was conducted in each of them. VLA antennas were divided into two groups for independent recording of right- and left-polarized components. The whole system worked with a resolution of 5 s. On 28 June 1986, five-hour monitoring of UV Cet yielded unique data. On the background of nonpolarized radiation smoothly changing within 2–18 mJy, which was higher than the regular level of 1 mJy, for an hour and a half Bastian and Bookbinder recorded two flares that had completely different parameters. The first event was 100% left polarized. Its light curve was rather simple, with 10-s burning and a twice longer decay. The maximum intensity was about 220 mJy. The event was rather broadband, since light curves in all 15 narrow frequency bands were identical. The second event consisted of two bursts separated by several minutes with a 3-min burning phase and a 1.5-min decay phase. It had a time structure up to 5 s. The radiation was 70% right polarized. The first and second bursts achieved 80 mJy and 100 mJy, respectively. Essential differences in the structure of the event at close frequencies were found within the 41-MHz band. If the source was about stellar size, the brightness temperatures were estimated as 3 and $2 \cdot 10^{11}$ K. Bastian and Bookbinder analyzed the conditions at which both cyclotron maser instability and plasma radiation, which originally arise as monochromatic effects, can provide narrow-band emission.

On 3 July 1986, a dynamic spectrum of a flare on UV Cet was obtained by Jackson et al. (1987a). Observations at about 20 cm were carried out at 1385 and 1502 MHz with 13 VLA antennas and at 1435 and 1652 MHz with other 14 VLA antennas that were switched every 10 min from 20 to 6 cm and back. Near 6 cm, emission at four close frequencies was studied. The session continued for six hours with a time resolution of 6.7 s. Within this period there was a weak (up to 2 mJy) 100% right polarized flare with a duration of about 80 min on the star L 726-8 A. On UV Cet, at 20 cm a 100% polarized flare was detected at about 11:00 UT with an intensity of ~10 mJy and a nonpolarized flare at about 14:00 UT with an intensity of ~ 4 mJy. At 6 cm there were weak flares with an intensity of about 2 mJy at 11:45 and 14:20 UT. If the events were considered as uniform bursts at these two frequencies, one would suggest that there was a frequency drift toward high frequencies, i.e., the motion of a disturbing agent toward the stellar surface. The dynamic spectrum displayed a rather complicated pattern of frequency and time changes of microwave radiation.

In November 1987, the studies were continued at Arecibo by Bastian et al. (1990). Since the Arecibo radio telescope is six times more sensitive than VLA, it enabled separation of bursts of terrestrial origin with lower reliability, but in the dynamic spectra the noise was efficiently rejected based on the spectrum type. During three-hour monitoring of AD Leo near 1415 MHz with a resolution of 20 ms a 4.5-min burst with a distinct peak of about one second was recorded. The whole burst had 100% right-circular polarization, and a one-second peak and its nearest environment displayed a rather complex structure with a lifetime of separate details to ten of milliseconds and an appreciable change of light curves in 2.5-MHz bands from 1395 to 1435 MHz, corresponding to a spectral index of about 12. The one-second peak at maximum achieved 940 mJy, the maximum level for stellar flares at this frequency, and with regard to the time structure the temperature brightness exceeded $6 \cdot 10^{15}$ K. Six minutes after the decay of this burst another 100% right polarized burst with duration of about 6 min took place on the star, the burst peak was several seconds long. But unlike the previous burst, this peak consisted of ten oscillations with a quasiperiod of 0.7 s and coherent frequencies within the whole frequency band of 40 MHz. The oscillations started 20 s before the maximum of the peak, and its decline was much more abrupt than burning. Bastian et al. observed YZ CMi for two hours with a resolution of 20 ms at 430 MHz and found a single 100% left polarized burst with a duration of 6.5 s and intensity up to 1.6 Jy. As in the case of AD Leo, the light curve of the burst had short — up to 50 ms — decays with a frequency drift of −250 MHz/s. Other details evidence both negative and positive frequency drift. If the size of the source was determined by the time scale of 50 ms, its brightness temperature was $\sim 3 \cdot 10^{16}$ K. But this value can be only the lower limit: first, the propagation velocity of disturbances can be only 0.1 of the speed of light and, hence, the linear size of the source is 10 times smaller. Secondly, 50 ms is the resolution of the instrument rather than the real burning time, which leads to an overestimation of the size of the source. Bastian et al. concluded that the decisive role in the selection of the nature of the observed coherent radiation was the ratio of the frequency of plasma oscillations to the electron-cyclotron frequency: if this ratio was less than unity or close to it, instability of the electron-cyclotron maser could occur in the medium, but its appreciable broadband character required a heterogeneous medium. If the ratio was over or close to 3, the conditions for plasma oscillations were created.

In November 1987, Güdel et al. (1989) carried out simultaneous observations of AD Leo, UV Cet, YZ CMi, and Wolf 630 with Effelsberg, Jodrell Bank, and Arecibo radio telescopes at 1665, 1666, and 1415 MHz, respectively. On 7 November 1987, all three telescopes recorded a sharp burst of right-polarized microwave radiation, all telescopes recorded identical time, brightness, and polarization characteristics. The total duration of the radio event was 40 s, but its light curve consisted of numerous pulsations of different amplitude with duration up to 125 ms and intensity up to 670 mJy. Thus, the event is one of the brightest observed at 18 cm. The obtained data allowed one to suspect narrow-band radiation $\Delta\nu/\nu < 0.1$ and frequency drift with a velocity of 240 MHz/s. The value is close to similar values in solar decimeter pulsations, the time division of pulses is analogous as well, but solar events are weaker than the burst on AD Leo by four orders of magnitude.

White et al. (1989a) carried out two observational series of all known flare stars accessible to VLA and located at a distance of up to 10 pc from the Sun, which earlier were not identified as microwave-radiation sources. In the first series, each star was monitored for 20 and 15 min at 20 and 6 cm, respectively. In the second series, monitoring was slightly shorter. As a result, they found microwave radiation from Wolf 47, Gl 234 AB, Wolf 424, Wolf 461, DT Vir, Ross 867, DO Cep, Gl 867 AB, and EV Lac. Together with earlier known sources they made about 40% of the known nearby flare stars and, naturally, this percentage would increase if the exposures were longer, because in all cases apparently a flare rather than quiescent microwave radiation was recorded. An extremely strong 95% polarized flare was recorded on DO Cep at 6 cm: it was almost invisible at 20 cm and was essentially narrow-band, since the light curves differed markedly at frequencies with $\Delta\nu = 450$ MHz. Comparing the fact of detection of microwave radiation with other global stellar parameters, White et al. noted the maximum for M4e– M5.5e stars and the reduced occurrence of this radiation before and after the maximum. The data of this survey suggest that the sources of microwave radiation coincide with the strongest X-ray sources among flare stars and that flares are more often found at 20 cm, while quiescent radiation is found at 6 cm.

In January 1987, Lang and Willson (1988) carried out four-hour VLA observations of YZ CMi. Like Bastian and Bookbinder (1987), they divided the 50-MHz band into 15 narrow bands and with a resolution of 10 s conducted serial monitoring of the right- and left-polarized components of radiation. On the background of practically constant radiation of the right-polarized component, the left-polarized one displayed significant variations of brightness with an intensity up to 18 mJy and with a characteristic time of minutes. Detailed consideration of eight local maxima in the light curve showed that four of them contained significant differences between radiations in narrow bands, i.e., microwave emission was narrow-band with $\Delta\nu/\nu \sim 0.02$, whereas in four other local minima the effect of narrow-bandness was not noticed: $\Delta\nu/\nu > 0.03$. Interpreting the recorded radiation in the context of an electroncyclotron maser, Lang and Willson estimated the magnetic field as $B \sim 260$ G and noted that since radiation was limited to a narrow cone with the opening angle from 70° to 85° with respect to the field flux tubes, one should expect its narrow-bandness to be from 0.01 to 0.1, and the sharp field gradient results in the broadening of the band, while the small size of the source results to its narrowing.

In the course of cooperative program of observations in 1988 the binary system YY Gem was observed with VLA, and eclipses on it were recorded in optical, ultraviolet, and radio ranges (Butler et al., 2015). A significant weakening of emission at 6 cm at phases of the primary and secondary eclipses pointed to a compact radio source, which could lie between components of the system.

During the strongly polarized flare of 29 March 2001 in the close binary system ER Vul that lasted for about an hour and was recorded with VLA at 20 cm, Brown et al. (2002) found only a very weak response with Chandra HETGS.

* * *

On 15 July 1989, using the upgraded Arecibo radio telescope Lecacheux et al. (1993) recorded a four-second burst on AD Leo at 6 cm. At a resolution of 20 ms and a maximum intensity of the burst of 400 mJy the brightness temperature of the source was $10^{10}(R_*/r_s)^2$ K, where $r_s$ is the size of the radio source. The burst radiation was 20% polarized and concentrated mainly within the band of 135 MHz.

On 24 November 1990, in the course of a 14-h session of observations of the young star AB Dor with the Australian Telescope Compact Array (ATCA) Lim (1993) recorded two very strong circularly polarized flares on the nearby star Rositter 137 B — a very young fast rotator — and their polarization was opposite to that of quiescent stellar emission.

On 6 October 1988, Spencer et al. (1993) observed YZ CMi at 6 cm using the broadband interferometer (BBI) at Jodrell Bank. They recorded flare activity as a set of short-lived bursts with a frequency of 5–6 per hour and an intensity of ∼15 mJy on the background of slowly varying emission of about 1.4 mJy. There was a burst with an intensity of up to 42 mJy and total duration of several minutes that consisted of a number of fast pulsations with a duration of less than 20 s. The total energy of the burst was higher than that of similar events on the Sun by four orders of magnitude.

In 1990–93, Abada-Simon et al. (1994, 1997) for more than 81 h observed nine known flare stars at 1.4 and 5 GHz with the Arecibo telescope. Using certain criteria to reject bursts of nonstellar origin, they selected only 12 bursts on AD Leo detected after 40 h of radio monitoring from the initial list of more than fifty radio events at 1.4 GHz. The duration of bursts varied from several seconds to about one and a half minutes, four bursts lasted for 3.5 min, the maximum intensity achieved 70 mJy, half of the bursts were 100% circularly polarized, the others were not polarized. One of the bursts consisted of one right-polarized peak and two left-polarized peaks. Over five hours of monitoring at 5 GHz no bursts were found. Upon more detailed analysis of the strongest and 100% right-circularly polarized burst on AD Leo of 13 February 1993 Abada-Simon et al. found that during this flare the emission frequency repeatedly increased and decreased with the drift module of 1–5 MHz/s and formed arches on the dynamic spectrum with a duration of 10–20 s. The limitation of the detection band did not allow the determination of extreme values of emission frequencies, but sometimes it filled the whole 50-MHz band. In considering the central 5 s of the burst with 20-ms resolution, clear differences in the adjacent bands with a width of 10 MHz with separate peaks up to 350 mJy were ascertained, they increased "from zero" during the next 20-ms intervals. Such peaks corresponded to brightness temperature of more than $10^{15}$ K. These characteristics, according to Abada-Simon et al., conform to the maser mechanism with a magnetic field of about 500 G.

On 31 December 1991, the 100-meter Effelsberg telescope at 4.75 GHz recorded a radio flare on UV Cet. The flare started five minutes after the maximum of optical flare with $\Delta U \sim 1^m$ that lasted in the B band for about six minutes. Optical observations were performed at the 80-cm telescope of the Wendelstein Observatory using the five-channel UBVRI photometer with a resolution of 20 ms (Stepanov et al., 1995). The radio burst was more than 75% left-circularly polarized. When 12 narrow 3.1-MHz bands were separated from the full 50-MHz band, significant differences were found in the light curves at the radio flare maximum and in its nearest vicinities: at a frequency lower than 4751 MHz the flare practically was not visible, but it definitely went below the detection limits of 4725 MHz. In the range of 4725–4744 MHz the light curve of the flare constructed with a resolution of 0.125 s had a total duration of about one minute, a maximum intensity of 250 mJy and numerous narrow peaks with an intensity of up to 150 mJy. The characteristic brightness temperature reached $3 \cdot 10^{12}$ K. Comparing the mechanisms of the electron-cyclotron maser and plasma oscillations, Stepanov et al. concluded that in the case of the maser, the magnetic field should be $\sim$ 850 G and the electron density less than $7 \cdot 10^{10}$ cm$^{-3}$, but above 10 MK there would be problems with an exit of radiation from the medium. In the case of plasma oscillations, $n_e = 3 \cdot 10^{11}$ cm$^{-3}$ and B > 200 G. This mechanism explains the fine structure of the dynamic spectra. In both cases, the radiation source size should be about $10^9$ cm.

Using the same instruments, Stepanov et al. (2001) recorded a microwave burst on AD Leo of 19 May 1997 with a flux of about 300 mJy, a duration of about one minute, and 100% right polarization. In the decay phase they found quasiperiodic oscillations with a characteristic period of about 2 s, which most probably were due to magnetohydrodynamic oscillations of the flux tube containing the radiation source. Stepanov et al. identified the burst with $T_{br} > 5 \cdot 10^{10}$ K with plasma oscillations at the fundamental frequency of the source with $n_e \sim 2 \cdot 10^{11}$ cm$^{-3}$ and a field of about 800 G. Only rather nonuniform corona with average density that was 100 times lower than that in the source could pass this radiation through without appreciable absorption. Acceleration of electrons by an electric current at the loop base could initiate plasma oscillations. Continuing the analysis of this flare, Zaitsev et al. (2004) detected two components in the low-frequency radio emission spectrum: a harmonic narrow-band component with slowly decreasing frequency during the flare from 2 to 0.2 Hz and a periodic series of impulses with iteration every 2 Hz during the flare. They considered possible combinations of effects of MHD and LCR oscillations of the radio source through the accelerated particles and obtained estimates of the source parameters: a thickness of coronal loops of $2 \cdot 10^8$ cm, their length of $4 \cdot 10^{10}$ cm, electric current in the loop of $4.5 \cdot 10^{12}$ A, its energy of $5.5 \cdot 10^{33}$ erg. Thus, this flare on AD Leo was similar to microwave flares on the Sun: magnetic loops carry electric current, which is a source of free energy of flares. Free oscillations of magnetic loops as MHD systems and electric chains modulate radio emission. The energy supply and rate of its release in the flare on AD Leo exceeds the solar flares by 2-3 orders of magnitude.

Güdel et al. (1995a) investigated microwave radiation of the very young solar-type star EK Dra at 6 and 3.6 cm. Analysis of 36 one-hour sessions of VLA observations (September–October 1993) revealed the variability of radio emission at 3.6 cm. From its average intensities the intervals of a “strong flux” were established when, on intervals of 10–180 min the intensity achieved 0.4– 0.5 mJy that was 5–6 times higher than the quiescent level, but polarization in this case was low. In the case of a “weak flux”, such flare-like events did not occur. If the size of the radiation source is equal to the solar radius, the brightness temperature should be $2 \cdot 10^7$ K. Measurements at 6 cm conducted during high-intensity radiation at 3.6 cm did not reveal such events.

As stated above, Lim and White (1995) observed four rapidly rotating G–K dwarfs in the Pleiades at 3.6 cm with VLA. On HII 1136, the fastest G rotator in the cluster, they found a low-polarized radio flare, whose emergence was accompanied by a one-hour increase of intensity up to 1 mJy. The flare maintained this state for 2 h and then exponentially decayed with a characteristic time of 1.4 h. In the descending curve there was a short 100% circularly polarized burst.

Summarizing the properties of the not numerous microwave flares on F–G– K stars, Benz (1995) noted their regular distinction from flares on M dwarfs: the former were low-polarized and could be caused by gyrosynchrotron emission; they were longer and less frequent; their radiation was apparently incoherent and required more energy for the same intensity, whereas highly polarized mostly narrow-band radio flares on M dwarfs were caused by the coherent mechanism.

To determine the sizes of radiation sources, Benz et al. (1995) observed EQ Peg B and AD Leo at 18 cm using the intercontinental interferometer system VLBI. The system included antennas of VLA, Arecibo, Effelsberg, Jodrell Bank, Green Bank, and Owens Valley observatories. On 15 March 1990, a radio flare was recorded on EQ Peg B with an intensity of 8 mJy at maximum, duration of about one hour, and right-circular polarization to 75%. The size of the radiation source was estimated as less than 1.8 stellar photospheres. During two observational sessions of AD Leo in September 1991 slow oscillations of the intensity of radio emissions at a level of 0.6 and 1.8 mJy were recorded, as well as 50% right-circular polarization. Based on the analysis of the obtained data, the upper limits of the sizes of the radiation sources were estimated as 1.9 and 3.7 stellar photospheres. The estimates of the sizes correspond to the lower limits of brightness temperatures of $4 \cdot 10^{10}$ K on EQ Peg B and $2 \cdot 10^{9-10}$ K on AD Leo.

One of the critical moments of the phenomenological model of solar flares consists in the consideration of the fluxes of accelerated particles that appear during the initial energy release — the causal process is not clear as yet — quickly heat up the chromosphere, which leads to its evaporation and an increase in thermal radiation from the corona. One of the most forcible experimental confirmations of the idea is the so-called Neupert effect: the intensity of flare thermal radiation from the corona at each moment of time is proportional to the integral from the nonthermal radiation initiated directly by accelerated particles since the flare onset till this moment. For the Sun, the measure of thermal coronal radiation is thermal X-ray emission. Hard nonthermal X-ray emission of electrons with energies of hundreds of kiloelectron-volts in the form of gyrosynchrotron emission is usually considered as a measure of nonthermal emission, though originally nonthermal radio emission was studied as such a measure. To detect the stellar analog of the Neupert effect, Güdel et al. (1996) observed the system L 726-8 AB in January 1995 at 6 and 3.6 cm with VLA and in X-rays with ROSAT and ASCA. During two nine-hour sessions they recorded a half dozen weakly polarized radio flares: one on L 726-8 A and the rest on UV Cet. Upon detailed consideration of their radio and X-ray emissions the authors found that three or four pairs of such radio and X-ray events to some extent satisfied the criterion of the Neupert effect: radiation in both ranges began practically simultaneously, fast growth of X-ray emission occurred at the moment of maximum radio emission, while maximum X-ray emission took place when radio emission decreased to the preflare level. Güdel et al. showed that one could expect regular distinctions between the Neupert effect on stars and on the Sun. They called the events "candidates for the Neupert effect".

During four runs in 1990–94 Lim et al. (1996) studied microwave radiation of Proxima Cen at 20, 13, 6 and 3.5 cm using ATCA. They recorded flare radiation only during one run on 31 August 1991 at 20 cm. The flare lasted for about six minutes, reached a maximum of 20 mJy, and its radiation was 100% left-circularly polarized. The absence of signal at other wavelengths testified to the narrow-bandness of the flare.

In October 1996, EQ Peg was observed within the framework of the wide cooperative program that involved VLA, EUVE, and RXTE, and the McDonald Observatory for studies in the optical range (Gagné et al., 1998). On 2 October 1996, for nine hours EQ Peg activity was recorded in four wavelength ranges, six or eight radio flares at 3.5 cm with an intensity of 3–4 mJy and duration of up to 30–40 min were recorded. After one of the strongest optical flares with $\Delta U \sim 2.5^m$, simultaneously recorded with EUVE and RXTE, a radio burst started with an appreciable delay, and the strongest radio flare simultaneous with the flare in medium X-rays was recorded with RXTE. Based on the results of these and previous observations, Gagné et al. concluded that the strongly polarized radio flares occurred irrespective of activity in other ranges, whereas low-polarized radio events were often correlated with flares at other wavelengths.

On 29 April 1999, during 10-hour VLA monitoring of AD Leo a radio flare was recorded at maximum EUV brightness of the flare (Güdel et al., 2001a).

During the described multiwave observations of EV Lac on 20 September 2001 Osten et al. (2005) carried out monitoring of this star with VLA at 3.6 and 6 cm. The radio observations continued for about a half day and a strong flare was recorded in both ranges about 8:51 UT. This seemed to be the strongest flare among such events on dMe stars — 61 mJy, whose interpretation within the GS model covered the volume comparable with the star and involved the trapped electrons, whose lifetime was accounted for hours. At a delay of maximum at 6 cm with respect to the maximum at 3.6 cm amounted to several seconds, the maximum of the strong optical flare with $\Delta U \sim 3.3^m$ with a duration of 7 min was prior to the maximum at 3.6 cm by 54 s. The flux at 6 cm was almost 100% polarized; furthermore, to interpret a flare, a coherent mechanism was required. Osten et al. traced variations of the spectral index at an interval of 3.6–6 cm and came to the conclusion that at the brightness maximum the flare was optically thick at both frequencies and became optically thin due to Coulomb collisions at decay. Magnetic fields were estimated as 60 and 110 G in the luminosity regions at 6 and 3.6 cm, while the maximum effective temperature achieved $3.4 \cdot 10^9$ K. In addition to this strong flare, in the course of radio monitoring of EV Lac weak events were recorded at a wavelength of 6 cm, which were significantly circularly polarized.

On 13–14 June 2003, Osten and Bastian (2006b) carried out broadband observations of this star in the range of 1120–1620 MHz with Arecibo; the monitoring was conducted with a resolution of 0.78 MHz and a time resolution of 10 ms. A radio burst of 13 June was characterized by numerous short-term, about 30 ms, rapidly moving subbursts of high brightness temperature of more than $10^{14}$ K and width of $\Delta \nu/\nu \sim 5$ %; these subbursts emerged with an average frequency of about 13 per second and showed no variability, generally resembling solar decimeter bursts. A strong high-polarized burst of 14 June differed noticeably from the previous one by a slowly varying intensity profile, slow frequency drift, and brightness temperature varying from $6 \cdot 10^{11}$ to $6 \cdot 10^{13}$ K.

Osten and Jayawardhana (2006) carried out VLA observations of three young brown dwarfs 2MASSW J1207334-393254, TWA 5B, and SSSPM J1102-3431, which are presumably related to the young association TW Hyades; an upper limit at 3.6 cm was estimated for all the targets. They came to the conclusion that radio emission of these cool stars was not determined by age.

In May 2000, using AAT and Australian telescope, Slee et al. (2003) carried out optical and radio observations of Prox Cen. They recorded eight optical flares throughout two days, but no radio event at frequencies of 1.38 and 2.50 GHz. Throughout the whole time of observations a slow decay at 1.38 GHz was 100 % polarized and constrained in intensity by a rather narrow band and by significant variations of polarization inside the receiver band of 104 MHz. This is the first case of existence throughout several days of high-polarized narrow-band emission of a flare star which is attributed to coherent radio emission of the stellar corona. But the obtained observations did not allow one to distinguish the fundamental plasma emission and electron-cyclotron maser emission.

In December 2001–January 2002, Slee et al. (2004) carried out radio and optical observations of the binary system CC Eri at 3.5 and 6.25 cm during three observing runs of 12 hours each and half-hour optical observations with AAT. These resulted in the detection of weak radio emission that was 20 % circularly polarized and two flares with positive spectral indices without polarization. In a stronger flare of 30 December 2001, emission at 3.5 cm was ahead of emission at 6.25 cm by 5 minutes, which was due to the distribution of perturbations through the corona upward and acceleration of electrons up to near-relativistic velocities. Parameters of emission throughout three nights were comparable with GS, in which the radio source becomes optically thick during a strong flare. Within the framework of the simple developed model, which supposes the number of electrons to be a function of the magnetic field in the source, the values of the field, the source size, and the number of radiating electrons were obtained. Spectropolarimetry revealed the presence of the strong surface magnetic field, and optical photometry detected a spotted region with a radius of about 14° on the brighter component.

The M8 dwarf DENIS 1048-3956 showed two fast flares: one at 4.80 GHz and one that was 10 minutes later at 8.64 GHz with fluxes of 6.0 and 30 mJy, respectively (Burgasser and Putman, 2005). A high brightness temperature of more than $10^{13}$ K, duration of 45 min, almost 100 % circular polarization, and, apparently, a narrow spectral band mean the coherent emission in the region with a high electron density of $10^{11}$–$10^{12}$ cm$^{-3}$ and magnetic field of about 1 kG. If these two flares are associated, then the frequency drift of emission means that the source moved to the region with higher density or with higher density of the magnetic flux or was compressed by the twisting of field lines or gas motion.

During the multiwave observations from the X-rays to the radio range of the binary system CC Eri Budding et al. (2006), operating at ATCA, detected constant low-level radio emission and flares with maxima of several mJy on its background.

Performing a reanalysis of observations of the flare source GCRT J1745-3009 carried out on 28 September 2003 at 325 MHz, Roy et al. (2010) detected strong circularly polarized emission. They established an upper limit of the source size of $8R_{\odot}$ and coherent character of emission — electron cyclotron and plasma radiation.

* * *

In 1984–85, Jackson et al. (1990) studied flare activity of known dMe stars at decameters at the Clark Lake Radio Observatory: 12 objects were monitored at seven frequencies from 30.9 to 110.6 MHz over 143 h. Over this time meaningful signals were suspected only from AD Leo at the onset and the end of the flare of 15 December 1985, which was recorded at 1415 MHz with VLA. According to Jackson et al., this means that flares, as a rule, do not reach the decameter wavelength range.

On 1–6 September 1992, UTR-2[1], a radio telescope near Kharkov, with optical support from Crimea, Sicily, and Greece monitored EV Lac in the decameter wavelength range (Abranin et al., 1994; Abdul-Aziz et al., 1995). Simultaneous observations at 25 and 20 MHz with a time resolution of 0.2 s were carried out for 5 h about midnight. Some special methods were used to suppress noise. Over 30 h more than 30 single and group radio bursts from 1–3 to 10 seconds were detected; one burst lasted about four minutes. During radio monitoring numerous optical flares were detected. Eight radio bursts could with great probability be related to optical events. It should be noted that optical flare activity varied considerably from one night to another, stellar activity in the radio range varied in parallel. For example, groups of radio bursts were observed only during the night between the 1st and 2nd of September, when optical activity was also maximum. Radio bursts were more often recorded at 25 than at 20 MHz, and the effect of frequency drift was not found. The bursts with the greatest probability identified with optical events had an intensity of 150–900 Jy at 25 MHz and 200–535 Jy at 20 MHz. These values correspond to brightness temperatures within $5 \cdot 10^{17}$–$3 \cdot 10^{18}$ K. In a number of characteristics the recorded radio bursts were similar to type III solar bursts but their intensity was higher by 3–4 orders of magnitude. Melnik (1994) thoroughly considered the mechanism of an electron-cyclotron maser as applied to dMe stars and concluded that it could explain all features observed on EV Lac — narrow-band radiation, an absence of frequency drift, and the duration of bursts — and admitted brightness temperatures up to $10^{21}$ K.

[1] UTR-2 is the world largest radio telescope for this range designed as a flat fixed system of dipoles with an electrically controlled orientation diagram.

After a number of technical and methodical updatings in campaigns of 1993 and 1994 searches for decameter radiation of EV Lac were carried out again. But in 1993, throughout 25 hours of radio observations, no event was recorded which could be confidently compared to the optical flare of the star. In 1994, throughout 33 hours, 18 radio flares were detected, but after a hard selection only one left which most likely had stellar origin: a radio flare of 26 August 1994 at 22:08 UT at 20 MHz with a duration of about 10 s, an intensity of ~ 150 Jy and with an onset up to 3 s coincided with the onset of the optical flare with $\Delta U \sim 0.5^m$ (Abranin et al., 1998a, b). In December 1994, using URT-2, with support in the Crimea and Greece, the radio-optical observations of YZ CMi were carried out (Abranin et al., 1997). During 33-h radio monitoring of the star 12 radio flares were recorded with a duration of 5 to 20 s. 8 of them were detected during the photometric monitoring of the star in the U band in the Crimea, but none of them coincided with the optical flare. The only flare recorded in the B band in Greece occurred 160 s after the radio burst, its brightness temperature was $10^{16}(R_*/r_s)^2$ K.

The total results of decameter observations with UTR-2 are presented in the survey of Abranin et al. (2001).

On 4 April 2005, during observations of AD Leo at Arecibo in the decimeter range with a record time resolution of 1 ms and broad-bandness $\Delta\nu/\nu = 0.36$ Osten and Bastian (2008) recorded a burst of 100 % circularly polarized emission with a maximum of 500 mJy. The dynamic spectrum of the burst showed a wide diversity of structures: nonhomogeneous emission, diffuse bands, narrow rapidly drifting bands; dispersion plays a significant role in the formation of the latter. Rather specific conditions are required for such a manifestation of cyclotron maser instability.

During 9-h monitoring with VLA Antonova et al. (2007) detected both pulsations and long-term sporadic variability of the quiescent radiation level at centimeter wavelengths on the very cool dwarf 2MASS J05233822-1403022.

Antonova et al. (2008) performed a special search among very cool dwarfs — fast rotators known as nonradiating at 8.5 MHz, radiators at 4.9 MHz. Out of 8 studied objects, the emission was detected only for 2MASS J07464251+2000321 at a level of 286 mJy. By the end of observations, a burst was detected in the light curve up to the level of 2.4 mJy with 100 % circular polarization. An hour before the burst the average level increased up to 160 mJy, and 40 minutes prior to the burst — up to 400 mJy.

Using the 305-meter Arecibo radio telescope, Route and Wolszczan (2013) carried out observations of 33 brown dwarfs at 4.75 GHz with the aim of determining magnetic activity of objects cooler than L3.5. They confirmed flare activity of the T6.5 dwarf 2MASS J10475385+2124234, extending the range of effective temperatures of active in the radio range objects from 1900 to 900 K.

In the course of observations of UV Cet with the Murchison wide-field antenna at 154 MHz Lynch et al. (2017) recorded four polarized flares at a level of 1065 mJy, which is one hundred times weaker than the known flares at this frequency. The circular polarization was at a level of > 28%, the linear polarization was > 18% for the strongest flare. The brightness temperature $10^{13}$–$10^{14}$ K indicated the coherent electron-cyclotron emission mechanism.

MacGregor et al. (2018) analyzed observations of Prox Cen at 1.3 mm, which were carried out throughout 20 hours in January–April 2017 at two multi-element radio telescopes ALMA and ACA (Atacama Compact Array), and on 24 March 2017 detected a flare with a 1-min duration and a maximum flux of 100 ± 4 mJy, which corresponds to the thousand-fold amplitude. In the brightness maximum, continuous emission corresponded to $F_\nu \sim \nu^\alpha$ c $\alpha = -1.77 \pm 0.45$ and a lower limit of linear polarization of 0.19 ± 0.02. The excess of stellar luminosity by 101 ± 9 mJy over the estimate of the expected luminosity of the photosphere at a level of 74 ± 4 mJy was attributed due to the corona heated by the continuous weak flares, which allows one to refuse from the warm dust up to 0.4 AU discussed earlier.

Later, using the ALMA complex, MacGregor et al. (2020) carried out monitoring of AU Mic at 1.35 mm and recorded two flares with radiation at maxima of 15 and 5 mJy, with a maximum luminosity of 196 and $69 \cdot 10^{13}$ erg/(s · Hz), duration of 35 and 9 s, and with a linear polarization minimum of the stronger flare of 12 %. The frequencies of such flares were estimated as 20 and 4 events per day on AU Mic and Prox Cen, respectively, i.e., as usual phenomena in flare activity of these stars. According to the analysis of authors, the spectral index of the distribution of megaelectron-volt electrons and the absence of exponential decay are interpreted as their formation at spilling in a single magnetic loop, while radiation is due to the synchrotron or gyrosynchrotron mechanism.

In April-July 2019, during 40-h monitoring of Proc Cen in the range from radio to X-rays, MacGregor et al. (2021) recorded a short-lived flare of enormous amplitude on 1 May in the range from mm to FUV: up to 1000 with ALMA and 14000 with HST. These bursts recorded with 1-s resolution were simultaneous, whereas optical radiation recorded with the TESS system with 2-min resolution had an amplitude of less than 2 and a delay with respect to the mentioned bursts of about a minute. The flare started as a 5-second burst at mm and FUV, after which there was an approximately just as fast decay, therefore, the light curve was represented by a symmetric Gaussian without a noticeable exponential decay phase. Radiation at mm at the burst maximum achieved 2.14 ± 0.15 in units of $10^{14}$ erg/(s · Hz), meanwhile, a variation of the radiation spectral index occurred from +2, which corresponds to the blackbody radiation of the quiescent photosphere, up to -2.5, which corresponds to the synchrotron or gyrosynchrotron, and a significant variation of radiation linear polarization. This may be the third case of recording synchrotron in a stellar flare after observations of Beskin et al. (2017) and MacGregor et al. (2018).

Using VLA, throughout 58 hours Villadsen and Hallinan (2019) carried out a survey of radio sources on five active M dwarfs in three bands in the ranges of 224–482 MHz and 1–6 GHz and detected 22 radio bursts in 13 epochs, first recording a large sample of broadband dynamic spectra of coherent stellar radio bursts. The observed bursts had different morphology, durations varying from seconds to hours and strong circular polarization 40–100%. All the bursts of hour duration were polarized by the large-scale magnetic field, which suggests the responsibility of cyclotron maser radiation by accelerated electrons. Peaks of the bursts were within 1–1.4 GHz with a weakening at lower or higher frequencies, which points to the localization of sources in the lower corona. Active M dwarfs should be the most widespread galactic variable sources at these frequencies.

In the course of test observations of UV Cet with ASKAP (Australian Square Kilometer Array Pathfinder) at 888 MHz Zic et al. (2019) detected high-polarized coherent bursts with a rotation period of the star of 5.447 ± 0.008 days and attributed them due to the beams of electron-cyclotron maser instability with a density that is by seven orders of magnitude less than the average coronal density, which regularly intersect the field of view of an observer. Solar-type activity seems to joint with auroral activity. In 2019, with the same instrument, 14-h monitoring of Prox Cen was carried out and ~50% circular polarization of emission was detected (MacGregor et al., 2021). But during the mentioned above burst of 1 May no microwave event was recorded at millimeter and ultraviolet wavelengths. On the next day, microwave observations of this star were maintained photometrically and spectroscopically by two optical telescopes and 42 s prior to the start of the 1-h strong optical flare with an amplitude of more than 1.5 and energy of $1.6 \cdot 10^{32}$ erg the first strong burst of coherent radio emission. According to Zic et al. (2020) and following polarization and temporal characteristics, this emission is analogous to the solar decimeter radio emission of type IV caused by fluxes of accelerated electrons.

In October 2021, using the MeerKAT[1] radio telescope, Bastian et al. (2022) carried out observations of UV Cet in the frequency range of 886–1682 MHz. They recorded a radio outburst with a duration of about 2 hours with a time resolution of 8 s and a frequency resolution of 0.84 MHz, enabling a dynamic spectrum to be considered. It shows three peaks and a variety of broadband arcs. The arcs are highly right-hand circularly polarized. During the end of the third peak, brief bursts occur that are significantly elliptically polarized. Bastian et al. interpreted this event by the model with the dipole magnetic field and the radiation mechanism associated with the cyclotron maser instability, whereas the elliptically polarized radiation may be the result of reflection on an over-dense plasma structure at some distance from the source.

* * *

Thus, observations of stellar flares in the radio range have provided extremely important notions on nonthermal radiation and, consequently, on nonequilibrium processes arising in stellar coronae that could be responsible for this emission. Although thermal X-ray emission and nonthermal radiation of flares occur in the same coronal regions, their roles in flares are totally incomparable: the basic fraction of energy is relaxed in X-rays that is acquired by the coronal gas, in essence, from the initial energy release and possible additional heating during a flare, whereas the radio range is primarily an information channel carrying a negligibly small fraction of flare energy but allowing one to investigate the independent component in the corona — accelerated particles. It is natural that in a wide wavelength range from millimeter to tens of meters fairly different facilities were used, therefore, to conclude on these studies, one should do it separately for several subranges.

[1] MeerKAT is a set of 64 Gregorian antennas located in South Africa each with an effective diameter of 13.5 m, that sample antenna baselines ranging from 29 m to 7.7 km. It currently supports observations in bands 856–1712 MHz and 544–1088 MHz.

The meter wavelength range is associated with both the discovery of radio emission of stellar flares and the first decades of their studying, which resulted in the established high brightness temperature of sources, high spectral indices, and a high degree of circular polarization of radiation. These properties led to the conclusion on the coherent and directed radio emission of flares. As to the concrete mechanisms, there initially prevailed a concept of plasma oscillations excited, on the whole, by intensive motions of flare plasma. In combination with regular delays with respect to the optical events, the detected properties of stellar flares were compared with much weaker solar bursts of types II and IV. But from the experimental results of the studies of the initial period only the strongest flares are currently of importance, which, as a rule, had distinct optical confirmations, since there were doubts with reliability of accounting terrestrial noise in those years and mess with near radio sources.

The studies of stellar radio flares in the centimeter wavelength range proved to be the most effective. Radio events revealed a considerable diversity of time, brightness, and polarimetric characteristics, broad-bandness and a degree of correlation with optical flares. It is worth noting that the time change of microwave flares expands from that typical of optical light curves with a fast rise, slower decay, and a total duration of tens of minutes up to separate bursts and even sharp peaks lasting for tens of milliseconds. If, in the 1960s–1970s, many hundreds of hours of monitoring were needed to detect a radio flare in the meter wavelength range, modern telescopes record microwave flares in hours and more often. As to physical models of such events, two coherent mechanisms are actively discussed: plasma oscillations and instabilities of an electron-cyclotron maser. The main point of the discussion is the fact that the parameters of the radiating medium obtained within both models — plasma density and strength of magnetic fields — are rather similar. Thus, it is not clear whether these models are alternative or additive.

The richest data on radio flares in centimeter waves yielded dynamic spectra with considerable diversity of spectra and time characteristics. Frequency drift of both signs, fast decays, and quasiperiodic pulsations of these radio bursts resemble solar decimeter bursts.

Except for the unique flare on V 371 Ori of 30 November 1963 recorded simultaneously in decameter and meter waves and in the optical range, no decameter range flares were unambiguously associated with optical events: about twenty radio events with different reliability are compared to contiguous optical flares, both individually and at the level of general activity on certain nights. As to the physical model of such radio events, there are competing maser and plasma mechanisms, which basically can provide brightness temperatures up to $10^{20}$–$10^{21}$ K.

Finally, the available pattern of stellar radio flares differs from the solar one not only by the orders of magnitude higher intensity of processes, but also by such qualitatively new components as a high-temperature plasma with $T > 10^7$ K and constantly replenished resources of relativistic particles that are absent on the Sun. Recently, Kavanagh et al. (2021) put forward an idea of the excitement of radio emission of AU Mic by a planet moving in its corona.

### 2.4.3. Ultraviolet Emission of Flares

Information about ultraviolet emission of flares on active dwarfs is less abundant than on X-ray flares. Ultraviolet emission, as X-ray emission, is accessible only from space vehicles, but the flare activity of stellar coronae was investigated by wide-angle instruments during sky surveys and in studying the vicinities of some objects within other programs, whereas flares in the chromospheres and transition zones require long monitoring with individual pointing.

As in a number of the above cases, ultraviolet spectra of solar flares serve as initial data in considering the appropriate stellar spectra, however, essential differences between the spectral resolution of stellar and solar spectra require certain caution (Doyle and Cook, 1992).

* * *

In monitoring of Proxima Cen on 6 March 1979 with IUE and the Einstein Observatory, an X-ray flare with appreciable amplitude was recorded. Simultaneously, a spectrogram was obtained in the range of 1150–1950 Å. However, no changes as compared to the spectrum of the quiet star were revealed (Haisch et al., 1981).

IUE observations of Gl 867 A without photometric support detected an additional continuum in the range of 1200–1950 Å, approximately twice as large an intensity of the emission lines of CII $\lambda$ 1335 Å, SiIV $\lambda$ 1396 Å, and CIV $\lambda$ 1550 Å, and less strengthened other lines in one of three spectrograms obtained with hour exposures (Butler et al., 1981).

In 1980, simultaneous observations of flare stars and RS CVn-type variables in ultraviolet and other wavelength ranges were carried out by researchers from Armagh, Boulder, Palo Alto, Socorro, Caltech, and Catania. Participants of this cooperation performed most of the further observations of active dwarfs in the ultraviolet (Rodonò, 1986a).

In monitoring of Proxima Cen on 20 August 1980, IUE and the Einstein Observatory recorded a strong X-ray flare with a duration of about one hour. Spectrograms of the flare in the short-wavelength range recorded with IUE are shown in Fig. 36. The CIV, NV, HeII, CII, and CI lines were many times stronger near the maximum brightness and SiIV, CIII, AlIII, and CI emissions appeared, which were not seen in the spectra of the quiet star (Haisch et al., 1983). The total emission energy of the transition zone in the flare was estimated as $10^{30}$ erg, 20 times lower than the X-ray emission of this flare. However, later, when Byrne et al. (1987) developed the technique of differential emission measures and estimating electron density from the ratio of line intensities to analyze IUE spectra, Byrne and McKay (1989) revised the observations of this flare and found that the characteristic density in the transition zone was $4 \cdot 10^{10}$ cm$^{-3}$, the volumes of flare X-ray and ultraviolet emissions were comparable, radiative losses for $Ly_\alpha$ and the transition-zone lines were $1.5 \cdot 10^{31}$ erg and were close to $E_X = 3.5 \cdot 10^{31}$ erg.

Bromage et al. (1983) during 17.5-h IUE monitoring detected strong flares on UV Cet, AT Mic, EV Lac, and EQ Peg, while usually, on average, one ultraviolet flare is detected over a day of monitoring. Flares on AT Mic and EQ Peg with maximum enhancement of the CIV emission were monitored without photometric support, and rather strong optical flares on UV Cet and EV Lac demonstrated rather weak enhancement of the emission. This fact confirmed that there was no simple dependence between the optical continuum and ultraviolet line flare emission, which could be connected with the occurrence of these radiations in spatially different regions. The strong continuum in the range of 1700–1900 Å in the flare on AT Mic could be presented as blackbody radiation with a temperature of 13000 K. In the same wavelength range a weaker continuum was found in a flare on EQ Peg; the availability of CIII $\lambda$ 1176 Å emission in the absence of CIII $\lambda$ 1909 Å and SiIII $\lambda$ 1892 Å testified to a density of matter of above $10^{10}$ cm$^{-3}$.

Baliunas and Raymond (1984) recorded a flare in the system EQ Peg AB on 2 September 1981: IUE provided the spectrum of the system as a whole, while the Whipple Observatory recorded the spectrum in the visible range of EQ Peg B, in which strengthening of $H_\alpha$ and $H_\beta$ emissions was found during the flare. The intensity of CIV emission increased approximately by a factor of 3, the SiIV, NV, HeII, and CII lines appreciably enhanced, in the range of 1700–1900 Å the continuum grew above the detection threshold. In the long-wavelength range obtained for the decay phase, MgII emission was amplified approximately by 10%, and FeII blend within 2600–2650 Å, by 20%. A maximum of helium emission in this flare took place earlier than in other transition-zone lines close to the beginning of strengthening of hydrogen lines, whereas the flare on AU Mic of 4 August 1980 observed without photometric support but with shifting of the stellar image along the slit during three hours had another time change of emission: HeII and SiII lines remained bright enough after appreciable fading of CIV, and maximum intensity of CI, CII, SiII, and HeII took place approximately 20 min after the CIV maximum (Butler et al., 1983).

Using an improved technique to extract data from IUE telemetry and their absolute calibration, Bromage et al. (1986) thoroughly analyzed the observations of a flare on AT Mic of 19 September 1980 and reprocessed the archives of observations of flares on Gl 867 A of 11 September 1979, Proxima Cen of 20 August 1980, and EQ Peg B of 2 September 1981. They found that in both spectrograms obtained during one-hour observations of AT Mic on 19 September 1981, emission lines were appreciably stronger, and in the first spectrogram the strengthening was more intense: fluxes in CIV, SiIV, CIII, OI, and CII lines increased from 7.5 to 2 times as compared to the quiescent state. Absolute fluxes in these lines, averaged over 30-min exposures, were 4–5 orders of magnitude higher than those in the strong solar flare of 15 June 1973 recorded by Skylab, and the increase was even higher for high-temperature lines. Considering the data for the four stellar flares, Bromage et al. noted that in a solar flare the strengthening of the $Ly_\alpha$ line was much greater than that in stellar flares, which could be a consequence of the large optical thickness of the line in stellar atmospheres. In the first spectrogram for the flare on AT Mic of 19 September 1980, a flare continuum was seen in the whole range of 1150–1970 Å, in the second, only in the range $\lambda > 1500$ Å and about 1900 Å its intensity was four times lower than on the previous spectrogram. The continuum in the first spectrogram of this flare was rather flat and similar to that recorded in the flare on Gl 867 A. The continuum in the second spectrogram was close to that of the flare on EQ Peg, and both were much higher than the continuum of the flare on Proxima Cen of 20 August 1980. The continuum of the solar flare was 4–5 orders of magnitude lower in the region of 1900 Å and lower by one more order of magnitude in the region of 1300 Å.

Using the same technique, Phillips et al. (1988) analyzed the spectra of UV Cet obtained on 17 September 1980. In a 55-min spectrogram obtained in the range of 1150–1950 Å, during which a five-minute optical flare occurred, they found strengthening of the CIV line by 30%, while in the spectrogram of the range of 1950–3200 Å obtained almost half an hour later, the MgII emission was strengthened by the same extent.

Upon more detailed analysis of AU Mic observations on 3–6 August 1980, during which in the range of 1150–1950 Å a dozen flares with $\Delta U < 1^m$ were recorded, Butler et al. (1987) noticed that as the energy of flares $E_U$ varied within two orders of magnitude, the fluxes of the strongest ultraviolet emissions of CIV, CII, and HeII changed by not more than a factor of 2. In different flares these emissions behaved in a different way: CIV and HeII emissions changed nonsynchronously, some flares with appreciable amplitude $\Delta U$ were not accompanied by significant changes of ultraviolet emissions and vice versa. Of a dozen BY Dra spectra obtained in October 1981 in the range of 1150–1950 Å, two were presumably obtained during stellar flares: CIV, SiIV, CII, and NV lines were stronger by some tens of percent as compared to the other spectra; there was a tendency of growth of this increase as the formation temperature of the lines increased (Butler et al., 1987).

Based on the observations of BY Dra and AU Mic, Butler et al. (1987) constructed the curve of column emission measures for these stars in quiescent and active states. They found that the EM of these stars in the active state was higher by 1.5 than the appropriate EM for a quiet star, but all the EM of the stars were an order of magnitude higher than the appropriate value of the quiet Sun at about $3 \cdot 10^4$ K and by two orders of magnitude than that of about $2 \cdot 10^5$ K, at which they are much higher than even the EM of a solar flare.

In the spectrogram of YZ CMi obtained with a 40-min exposure that covered almost the whole strong flare of 3 March 1983 with $\Delta U = 3.8^m$, the strengthened lines are well seen: CIV by a factor of 5.6, SiIV by a factor of 10, HeII by a factor of 5, and NV, CII, and SiII by a factor of 3–4. The total luminosity of the lines in the flare achieved $2 \cdot 10^{29}$ erg/s and $3 \cdot 10^{28}$ erg/s in the continuum in the range of 1150–1950 Å. Comparison of the curves of radiative losses in a quiet star and during the flare showed a shift of the transition zone to the depth of the atmosphere during the flare (Rodonò, 1986a; van den Oord et al., 1996).

On 4–5 October 1983, two ultraviolet flares were recorded on Gl 182. In the first and second one-hour exposures of 4 October 1983, CIV lines were four and two times stronger. The spectrum obtained later in the range of 1950–3200 Å had a 30% stronger MgII emission. The unique spectrogram in the range of 1150–1950 Å of the flare of 5 October 1983 revealed a 20-fold amplification of CIV and SiIV lines, the highest recorded on dMe stars, other emission lines increased by 7–14 times, the line $Ly_\alpha$ by one and a half times (Mathioudakis et al., 1991). From the ratios of intensities of the lines CIII $\lambda$ 1176 Å and SiIV $\lambda$ 1396 Å to the intensity of the line CIII $\lambda$ 1908 Å electron densities were estimated during two exposures of the flare of 4 October 1983 and one of the flare of 5 October 1983 as $1.5 \cdot 10^{10}$, $1.0 \cdot 10^{10}$, and $1.7 \cdot 10^{10}$ cm$^{-3}$. Using the DEM technique, radiative losses of flares on 4 and 5 October were estimated as $8 \cdot 10^{32}$ and $3 \cdot 10^{33}$ erg within $4.3 < \log T < 5.4$ and $1 \cdot 10^{34}$ and $6 \cdot 10^{34}$ erg within $4.0 < \log T < 8.0$. In the flare of 5 October 1983, within 1150–1950 Å, a strong continuum was found with a total energy of $2 \cdot 10^{33}$ erg, the ratio of this energy to that of CIV emission was close to the values found by Phillips et al. (1992) in flares on other stars. From absolute fluxes and estimates of $n_e$ Mathioudakis et al. found the volumes of flare plasma in both flares on Gl 182 and then the radii of curvature of magnetic flare loops: $9 \cdot 10^9$ and $2 \cdot 10^{10}$ cm in the events on 4 and 5 October, respectively.

In cooperative observations of AD Leo in early February 1983 the star was monitored with IUE and VLA, and in the optical range. In total, nine flares were recorded. The following conclusions were made from this small sample: a) CIV fluxes correlate with brightness in the U band; b) flares in the U band are much shorter than the duration of strengthening of HeII and radio emissions; c) radio emission correlates with HeII, but does not correlate with CIV. Thus, it was concluded that HeII in flares was excited by coronal radiation (Gary et al., 1987). In the spectrogram obtained eight minutes after the fast flare on 2 February 1983 at 11:37 UT, which lasted in the U band for about three minutes, a two-fold strengthened line of HeII $\lambda$ 1640 Å was found without any traces of strengthening of other transition zone lines; the strengthened helium line was observed 20 min after and the maximum $Ly_\alpha$ took place 37 min after the optical flare (Byrne and Gary, 1989).

In the decay phase of the strong flare on AD Leo of 28 March 1984, the spectrum within 1950–3200 Å was obtained. In the spectrogram one can see a many-fold strengthened MgII doublet and FeII $\lambda$ 2600 Å blend. Since during this exposure the amplitude of the optical flare was already low, probably the emissions were due to the slow decay of the strong flare with $\Delta U = 2.1^m$, which occurred 15 min before the beginning of the IUE exposure (Rodonò, 1986a).

During the very strong flare on AD Leo on 12 April 1985, whose amplitude reached $4.5^m$, and the total duration, according to ground-based photometric observations, was more than two hours, Pettersen et al. (1986b) obtained a spectrogram within 1150–1950 Å. The total exposure was 56 min, the first half of the impulsive phase, including its maximum, lasted 15 min, so the contribution of the quiescent state during the first 41 min was negligible. In this spectrum, the lines of $Ly_{\alpha}$, CII, CIV, and SiIV, and the continuum in the region $\lambda > 1780$ Å were overexposed, but rough estimates by means of completion of profiles from the wings showed that in this spectrum the $Ly_{\alpha}$ line was 20 times stronger, CIV lines — stronger by 45 times, HeII and NV — stronger by 14 and 9 times, respectively. In the spectrogram obtained within 1950–3200 Å, numerous unidentified details were seen, while the intensities at the blend limits of FeII $\lambda$ 2630 Å and $\lambda$ 2750 Å were an order of magnitude stronger than in the quiet star. It was concluded that the observed flare continuum resulted from imposing many metal lines and recombination continua. In any case, UBVR measurements could be equally well presented by free–free radiation at 10 MK and blackbody radiation at 10000 K, but none of these models matched the ultraviolet flare continuum (Hawley and Pettersen, 1991).

During the flare on EQ Peg of 7 December 1984, EXOSAT provided light curves in LE and ME, and IUE recorded three 10-min spectrograms within 1950–3200 Å that encompassed a slow LE ascending branch and flare maximum phase (Haisch et al., 1987). In all three spectrograms the emission of MgII was practically identical, 70% higher than the level of the quiet star. The emission of FeII blend increased from the first to the third spectrograms from two to four times as compared to the level of the quiet star. The continuum near 2950 Å behaved as FeII blend, whereas near 3100 Å the continuum regularly decreased from an 8- to 4-fold level of the quiescent state. This could be due to the fact that many faint lines of iron and other ions fell into the band near 2950 Å, whereas the band of 3100 Å is a purer continuum that behaves as one developing in a flare in the U band.

Mathioudakis and Doyle (1989b) analyzed two IUE spectrograms of the system Wolf 630. In one of them obtained on 12 June 1981 within 1150– 1950 Å, a flare was diagnosed from strengthened lines and the continuum. From the ratio of intensities of the lines of CIII $\lambda$ 1176 Å and SiIV $\lambda$ 1396 Å to CIII $\lambda$ 1908 Å they estimated the electron density at the level of $\log T = 4.8$ as $1.5 \cdot 10^{10}$ and $4.0 \cdot 10^{10}$ cm$^{-3}$ in the quiet star and during the flare, respectively. As in flares on BY Dra and AU Mic analyzed by Butler et al. (1987), the DEM curve in the flare on Wolf 630 was shifted toward higher values without any change in the inclination. From absolute fluxes in the lines and the estimate of $n_e$ Mathioudakis and Doyle determined the radius of the magnetic flare loop as about $5 \cdot 10^{10}$ cm, if a single loop was responsible for the flare. If the flare involved $n$ loops, their characteristic radius should be $n^{1/3}$ times less. The spectrophotometric gradient of the flare continuum corresponded to free–free radiation at 25000 K, and its total emission within 1250–1950 Å was $8 \cdot 10^{31}$ erg, which is only half that of the total radiation of emission lines in the temperature range $4.3 < \log T < 5.4$.

We mentioned already that on 13–15 October 1985 EV Lac was monitored for 17 h by IUE and EXOSAT with photometric and spectral ground support. However, only a weak correlation of activity was found at different levels of the atmosphere (Ambruster et al., 1989a). The flare on 15 October 1985 at 6:25 UT with $\Delta U = 1.5^m$ was not accompanied by appreciable strengthening of the CIV lines, though the flux in $H_{\beta}$ increased by $6\sigma$, and in the CaII K line, by $2.3\sigma$. The stellar chromosphere better correlated with variations in the U band than with coronal flares, and long X-ray activity on 13 October 1985 was not accompanied by essential changes in other atmospheric layers. This suggests magnetic isolation of the loops of the chromosphere and transition zone. Another example of isolation is the X-ray flare on Prox Cen of 6 March 1979, which was not accompanied by variations in the ultraviolet, optical, and radio ranges. However, such independence is quite rare.

Phillips et al. (1992) compared the ultraviolet spectra of about a dozen stellar flares, in which a flare continuum was found within 1150–1950 Å, with solar flares. They found a linear correlation between flare emission in the CIV lines and in the continuum and concluded that the observed flare continuum was the recombinant luminosity of neutral silicon at the level of the temperature minimum of the stellar atmosphere ionized by the radiation of the transition-zone lines.

In the ultraviolet flare of 26 December 1986 on the very fast rotating dM2e star Gl 890, Byrne and McKay (1990) found a several-times strengthened emission of CIV and a strong continuum, whose total energy within 1250–1950 Å was estimated at an order of magnitude higher than in CIV lines: $4 \cdot 10^{32}$ and $3 \cdot 10^{31}$ erg, respectively. Comparable energy of flare emission in the continuum was found in the flares on Gl 687 A and AT Mic (Bromage et al., 1986). Half an hour before this exposure the spectrum of Gl 890 within 1950–3200 Å was recorded, in which the MgII doublet and the blend of FeII $\lambda$ 2600 Å were strengthened almost by a factor of 1.5, their total energies were $2 \cdot 10^{31}$ and $1 \cdot 10^{31}$ ergs, but a flare continuum was not found in this spectrum. In the spectrum of the same region obtained after the above shortwave spectrogram, the iron blend was still strengthened.

On the ascending branch of the strong X-ray flare of 14 November 1984 in the system YY Gem that lasted for more than three hours, the spectrogram within 1950–3200 Å displayed overexposed MgII emission. In the spectra of the 1150–1950 Å region obtained before and after the X-ray flare, one can see strengthened CIV, HeII, CII, and CI lines, but eclipses in the system hampered unambiguous interpretation of these data (Haisch et al., 1990a).

Enhanced emission of MgII found in the ascending branch and at the maximum of a moderate X-ray flare on Proxima Cen of 2 March 1985 lasted for about an hour and a half after the termination of the X-ray flare (Haisch et al., 1990b).

During a five-day monitoring of the young K2 dwarf HD 82558 (= LQ Hya) an ultraviolet flare was recorded on 30 October 1988 with three-fold amplification of the emission of CIV and a considerable continuum within 1250–1850 Å. Increased luminosity of CIV lasted for at least two hours (Ambruster and Fekel, 1990).

CC Eri was monitored by IUE on 2–4 November 1989. Over 48 h in two of twenty spectra obtained within 1150–1950 Å 2–3-times strengthened CIV emission and smaller strengthening of emissions of CII and SiII were revealed (Byrne et al., 1992a; Amado et al., 2000). $E_{\mathrm{CIV}}$ in these flares was a few $10^{31}$ erg, which is several orders of magnitude higher than the values typical of solar two-ribbon flares.

On 11 June 1991, the dM5.5e star Gl 866 was observed by IUE with good spectral and photometric ground support (Jevremovic et al., 1998a). The energy of two of five optical flares was comparable with the strongest solar flares and MgII emission was strengthened by a factor of 1.5–2.

A strong flare with a duration of more than three hours was recorded on LQ Hya on 22 December 1993 by IUE and ground-based telescopes (Montes et al., 1999). In the spectrogram within 1150–1950 Å obtained immediately before the optical flare, the lines of the transition zone and continuum were stronger by a factor of 1.5, but chromospheric lines remained at the level typical of the quiet star. CIV lines were shifted by −250 km/s, SiIV and NV lines also displayed a blue shift but to a lesser extent, while chromospheric lines remained stationary. The spectrograms obtained for the range of 1950–3200 Å during the flare maximum showed that MgII emission was 2.5 times stronger, shifted by −40 km/s, and its FWHM was widened to 250 km/s. In another spectrogram obtained during smooth decay for the range of 1150–1950 Å, transition-zone lines were 20 times stronger, while chromospheric lines were only 4 times stronger. The continuum recorded near 1500 Å displayed up to a 10-fold increase. Analysis of this radiation does not contradict the hypothesis that it is conditioned by SiI recombinations.

Table 20 presents stellar flares recorded on active dwarfs with IUE and in the subsequent ultraviolet experiments.

Summing up the IUE studies of stellar flares, one should note that the instrument recorded less than 30 flares on red dwarfs and the conditions of their observations, as a rule, were unique.

Table 20. Flares on active dwarfs in UV and EUV

| № | Star | Date | Experiment | Strengthening | | Other characteristics of flares | Description and analysis of observations |
|---|---|---|---|---|---|---|---|
| | | | | CIV | MgII | | |
| 1 | Gl 867 A | 11.9.79 | IUE | ×2 | | | Butler et al., 1981; Bromage et al., 1986 |
| 2 | AU Mic | 4.8.80 | IUE | ×2 | | | Butler et al., 1983 Butler et al., 1987 |
| 3 | Prox Cen | 20.8.80 | IUE | ×7 | | ×35 in X-rays | Haisch et al., 1983; Bromage et al., 1986; Byrne and McKay, 1989 |
| 4 | UV Cet | 17.9.80 | IUE | × <1.5 | | $\Delta U = 2.4^m$ | Bromage et al., 1983; Phillips et al., 1988 |
| 5 | AT Mic | 19.9.80 | IUE | ×3.7 | | | Bromage et al., 1983, 1986 |
| 6 | Wolf 630 | 12.6.81 | IUE | ×2.8 | | | Mathioudakis and Doyle, 1989b |
| 7 | EV Lac | 3.9.81 | IUE | × <1.5 | | $\Delta B = 0.7^m$ | Bromage et al., 1983 |
| 8 | EQ Peg B | 3.9.81 | IUE | ×9 | | | Bromage et al., 1983 |
| 9 | EQ Peg B | 2.9.81 | IUE | ×3 | | ×2 in the $H_\beta$ line | Baliunas and Raymond, 1984; Bromage et al., 1986 |
| 10 | BY Dra | 2.10.81 | IUE | ×1.5 | | | Butler et al., 1987 |
| 11 | BY Dra | 4.10.81 | IUE | ×1.3 | | | Butler et al., 1987 |
| 12 | YZ CMi | 3.2.83 | IUE | ×5.6 | | $\Delta U = 3.8^m$ | Rodonò, 1986a; van den Oord et al., 1996 |
| 13 | AD Leo | 2–5.2.83 | IUE | | | | Gary et al., 1987; Byrne and Gary, 1989 |
| 14 | Gl 182 | 4.10.83 | IUE | ×4 | | | Mathioudakis et al., 1991 |
| 15 | Gl 182 | 5.10.83 | IUE | ×20 | | | Mathioudakis et al., 1991 |
| 16 | AD Leo | 28.3.84 | IUE | | ×7.5 | $\Delta U = 2.1^m$ | Rodonò et al., 1989 |

*Table 20 (continued)*

| № | Star | Date | Experiment | Strengthening | | Other characteristics of flares | Description and analysis of observations |
|---|---|---|---|---|---|---|---|
| | | | | CIV | MgII | | |
| 17 | YY Gem | 14.11.84 | IUE | | overexposure | ×4 in X-rays | Haisch et al., 1990a |
| 18 | EQ Peg | 7.12.84 | IUE | | | | Haisch et al., 1987 |
| 19 | Prox Cen | 2.3.85 | IUE | | ×2.5 | | Haisch et al., 1990b |
| 20 | AD Leo | 12.4.85 | IUE | | | $\Delta U = 4.5^m$ | Pettersen et al., 1986b; Hawley and Pettersen, 1991 |
| 21 | EV Lac | 13.10.85 | IUE | | overexposure | $\Delta U = 1.8^m$ | Ambruster et al., 1989a |
| 22 | EV Lac | 6.2.86 | Astron | | | | Burnasheva et al., 1989 |
| 23 | Gl 890 | 26.12.86 | IUE | ×4.5 | | | Byrne and McKay, 1990 |
| 24 | LQ Hya | 30.10.88 | IUE | ×3 | | | Ambruster and Fekel, 1990 |
| 25 | CC Eri | 3.11.89 | IUE | ×3 | | | Byrn et al., 1992a; Amado et al., 2000 |
| 26 | BY Dra | 1.10.90 | IUE + WFC | ×2.4 | | $\Delta U \sim 0.7^m$ flare in EUV and soft X-rays | Barstow et al. 1991; Phillips et al., 1992 |
| 27 | BY Dra | 2.10.90 | IUE | ×1.9 | | | Phillips et al., 1992 |
| 28 | YY Gem | 3.10.90 | WFC | | | | Bromage, 1992 |
| 29 | HD 197890 | 17.10.90 | WFC | | | EUV burst | Matthews et al., 1994 |
| 30 | AU Mic | 18.10.90 | WFC | | | | Bromage, 1992 |
| 31 | EV Lac | 17.12.90 | WFC + IUE | | | | Bromage, 1992 |

*Table 20 (continued)*

| № | Star | Date | Experiment | Strengthening | | Other characteristics of flares | Description and analysis of observations |
|---|---|---|---|---|---|---|---|
| | | | | CIV | MgII | | |
| 32 | EV Lac | 20.12.90 | WFC + IUE | | | | Bromage, 1992 |
| 33 | AT Mic | 3.5.91 | | | | Flare in X-rays | McGale et al., 1994 |
| 34 | AD Leo | 9.5.91 | HST | ×90 | | | Bookbinder et al., 1992; Byrne et al., 1993a |
| 35 | Gl 866 | 11.6.91 | IUE | | ×2 | | Jevremovic et al., 1998a |
| 36 | 35 EUV flares recorded by WFC within RASS | | | | | | Tsikoudi and Kellett, 1997 |
| 37 | AU Mic | 3.9.91 | HST | | | flare in SiIII λ 1206 E and in the $Ly_\alpha$ wing | Woodgate et al. 1992 |
| 38 | AT Mic | 1.7.92 | EUVE | | | | Vedder et al., 1994; Monsignori Fossi et al., 1994a; Brown, 1994 |
| 39 | AU Mic | 15.7.92 | EUVE | | | the strongest flare in EUV: $A_{EUV} \sim 20$ | Cully et al., 1993, 1994; Monsignori Fossi et al., 1994b, 1996; Drake J.J. et al., 1994; Brown, 1994, 1996; Schrijver et al., 1996; Katsova et al., 1999a |
| 40 | Proc Cen | 20.7.92 | EUVE | | | | Vedder et al., 1994 |
| 41 | EUVE 2056-17.1 | 3.8.92 | EUVE | | | $A_{EUV} \sim 10$ | Mathioudakis et al., 1995b |
| 42 | HII 314 | 3.9.92 | HST | ×3 | | | Ayres et al., 1994 |
| 43 | AU Mic | 9.9.92 | IUE + HST | | | flare in SiIV doublet lines | Linsky and Wood, 1994 |
| 44 | YZ CMi | 25.2.93 | EUVE | | | | Vedder et al., 1994 |

*Table 20 (continued)*

| № | Star | Date | Experiment | Strengthening | | Other characteristics of flares | Description and analysis of observations |
|---|---|---|---|---|---|---|---|
| | | | | CIV | MgII | | |
| 45 | AD Leo | 2.3.93 | EUVE | | | $\Delta U \sim 0.6^m$<br>$A_{EUV} \sim 4$ | Hawley et al., 1995; Cully et al., 1997 |
| 46 | AD Leo | 3.3.93 | EUVE | | | $\Delta U \sim 0.6^m$<br>$A_{EUV} \sim 2$ | Hawley et al., 1995 |
| 47 | EQ Peg | 30.8.93III | EUVE | | | ×3 in EUV<br>×5 in EUV | Monsignori Fossi et al., 1995b |
| 48 | EV Lac | 10.9.93I | EUVE+ IUE | ×6 | × > 3 | | Ambruster, 1995; Brown, 1996 |
| 49 | | II | EUVE+ IUE | ×4.4 | ×1.6 | | Ambruster, 1995 |
| 50 | LQ Hya | 22.12.96 | IUE | | | | Montes et al., 1999 |
| 51 | Gl 752 B | 12.10.94 | HST | | | | Linsky et al., 1995 |
| 52 | YZ CMi | 21.12.94 | HST | | | ×25 in the SiIV line | Robinson et al., 1996; Mathioudakis et al., 1999 |
| 53 | YZ CMi | 22.12.94 | EUVE | | | ×3 in EUV + radio + optical ranges | Robinson et al., 2001 |
| 54 | YZ CMi | 4.96 | HST | | | | Mathioudakis et al., 1999 |
| 55 | AU Mic | 13.6.96 | EUVE | | | ×5 in EUV+ X-rays | Gagné et al., 1998 |
| 56 | EXO 2041.9-3129 | 14 and 15.6.96 | EUVE | | | | Gagné et al., 1998 |
| 57 | EQ Peg | 2.10.96 | EUVE | | | $\Delta U \sim 2.5^m$ | Gagné et al., 1998 |
| 58 | EUVE J1438-432 | 1.5.97 | EUVE | | | ×16 in EUV | Christian and Vennes, 1999 |
| 59 | AU Mic | 6.9.98 | HST | | | 4 flares and numerous microflares | Robinson et al., 2001b |

*Table 20 (continued)*

| № | Star | Date | Experiment | Strengthening | | Other characteristics of flares | Description and analysis of observations |
|---|---|---|---|---|---|---|---|
| | | | | CIV | MgII | | |
| 60 | 12 flares recorded by EUVE within RAP | | | | | | Christian et al., 1999 |
| 61 | AD Leo | 2.4–16.5.99 | EUVE | | | | Güdel et al., 2001 |
| 62 | AU Mic | 26.08.00 | FUSE | | | | Redfield et al., 2002 |
| 63 | AU Mic | 10.10.01 | FUSE | | | | Redfield et al., 2002 |
| 64 | EUVE J0613-23.9B | 22.10.00 | EUVE | | | ×200 DS/S | Christian et al., 2003 |
| 65 | EV Lac | 20.9.01 | HST | 4 and 5 min | | $A_{UV}$ = 1.5 and 1.8 | Osten et al., 2005 |
| 66 | AB Dor | | FUSE | | | Downward flux in OVI with a velocity of 600 km/s | Ate et al., 2000 |
| 67 | LHS 2076 | | HST | | | Strong forbidden coronal line of FeXIII λ 3388 Å | Fuhrmeister et al., 2004 |
| 68 | EK Dra | 22.4.10 | HST | | | SIV, coronal rain | Ayres and France, 2010 |
| 69 | EK Dra | 28.3.12 | HST | | | SIV, CIV, coronal rain | Ayres, 2015a |
| 70 | GJ 1243, two flares | 31.8/1.09.14 | COS/HST and others | | | | Kowalski et al., 2019 |
| 71 | 2MASS J02365171-5203036 | 9.08.17 | COS/HST | | | FUV, absolute blackbody, strengthening of CII, SiIII, CIII, SiIV, NV emissions with a red shift of 50–80 km/s | Loyd et al., 2018b |

*Table 20 (continued)*

| № | Star | Date | Experiment | Strengthening | | Other characteristics of flares | Description and analysis of observations |
|---|---|---|---|---|---|---|---|
| | | | | CIV | MgII | | |
| 72 | 8 early dwarfs | 17 flares | COS/ HST | | | | Loyd et al., 2018b |
| 73 | GJ 674 | 2–3.04.18 | COS/ HST | | | | Froning et al., 2019 |
| 74 | Proc Cen | 1.05.19 | HST, ALMA and others | | | $A_{ALMA}$~1000 $A_{HST}$~14000 | MacGregor et al., 2021 |

Further, the stars could be observed by IUE with a time resolution of not less than 10–20 min, which appreciably exceeded the characteristic duration of most stellar flares. Therefore, these observations provide only the most general pattern of UV flares.

CIV emission is the most appreciable on quiet stars and in flares on active dwarfs, where its strengthening reaches one and a half orders of magnitude. As a whole, the level of strengthening of emissions in flares increases with their formation temperature. This law is satisfied for solar flares, where the transition zone goes down to large densities and with growing EM its dependence on formation temperature becomes steeper.

There is no unequivocal relation between the amplitudes of optical flares and the effects occurring in the ultraviolet region, which may be due to spatial separation of these radiation sources.

Ultraviolet emission of flares usually lasts longer than that in an optical continuum.

During UV flares practically all emission lines strengthen, but to different extents, and their light curves do not coincide.

The HeII $\lambda$ 1640 Å line falls out of the general law: it correlates poorly with CIV, and there are arguments in favor of coronal excitation of this line in flares.

In strong flares, the ultraviolet continuum occurs down to 1200 Å, but its physical nature has not been determined unequivocally. The most developed is the hypothesis about its recombination on silicon. When the continuum was recorded, its energy was comparable with ultraviolet lines.

In one of the most thoroughly studied flares that occurred on AU Mic on 19 September 1981, absolute fluxes in CIV, SiIV, CIII, OI, and CII lines averaged over a 30-min exposure were higher by 4–5 orders of magnitudes than those in the strong solar flare of 15 June 1973 recorded by Skylab. The excess was greater for high-temperature lines.

In stellar flares, strengthening of the $Ly_\alpha$ line was lower than that in solar flares, which can be a consequence of the large optical thickness of the line in stellar atmospheres.

The continuum of a solar flare is 4–5 orders of magnitude lower near 1900 Å and still an order of magnitude lower near 1300 Å as compared to several stellar flares for which it was measured.

The total energy of flares in transition-zone lines achieves $10^{31}$–$10^{33}$ erg and is comparable with the appropriate values of $E_X$.

Smooth variations of the intensities of ultraviolet lines out of flares suggest that they are caused by numerous low-amplitude flares.

* * *

All spectra of stellar flares were recorded by IUE with a spectral resolution of 350, whereas HST GHRS enabled a resolution of 2000, 25000 and 80000, covering the wavelength ranges of 285, 30, and 8 Å, respectively. Therefore, HST provided more refined but less systematic data about the activity of considered stars in the ultraviolet.

According to high time resolution HST observations aimed at estimating the contribution of numerous weak flares to the observed emission spectrum from individually recorded events — Saar et al. (1994c) on AD Leo and Saar and Bookbinder (1998a) on HD 129333 and LQ Hya — the contribution reaches 10 or even 20%. Similar results were obtained by Ayres (1999) in observations of three G dwarfs in the $\alpha$ Per cluster and the Pleiades.

On 9 May 1991, observations of AD Leo were carried out with a spectral resolution of 2000 and a time resolution of 1 s alternately in the ranges of 1170–1450 Å and 1390–1670 Å with five-minute exposures. On each revolution four exposures were made and the whole experiment was run on four revolutions of the instrument. At the maximum of a strong flare the emission of CIV for 25 s was strengthened by a factor of 90 and that of SiIV for 15 s by a factor of 60; amplification of the HeII line was halved and in CI $\lambda$ 1561 Å and $\lambda$ 1656 Å lines it was not appreciable. Many previously unknown emissions were revealed in this spectrum, CIV, SiIV, and HeII lines displayed components shifted by +1800 km/s, but 25 s later the shift decreased to 600 km/s (Bookbinder et al., 1992; Byrne, 1993a). According to Byrne, the kinetic energy of this motion is 25 times higher than the CIV radiative energy.

In two-hour observations of AU Mic on 3 September 1991 with a spectral resolution of 10000 and a time resolution of 0.4 s during a flare in the SiII $\lambda$ 1206 Å line the only three-second emission increase was recorded in the red wing of the $Ly_\alpha$ line. In duration and shift from the line center the increase complied with the pattern expected from the descending flux of fast protons accelerated during the impulsive phase. The energy of protons was sufficient to excite flare luminosity of the transition zone. But, in repeated 3.5-h observations carried out one year later the effect was not found (Woodgate et al., 1992; Robinson et al., 1993).

Three rapidly rotating G dwarfs in the Pleiades were observed on 3 September 1992 during three consecutive revolutions of HST around the Earth. Activity of HII 314 was discovered with confidence in the CIV lines at characteristic times of several minutes (Ayres et al., 1994). Over 89 min of general spectral monitoring on each revolution seven spectra were obtained in the range of 1150–1606 Å: in the spectra of the first revolution the line intensity was scattered, which obviously surpassed regular scattering, one of the spectra of the second revolution suggested a fast burst, and the spectra of the third revolution evidenced flare decay in the lines. The activity was not found on two other G dwarfs. In this connection, only smooth brightness variations with characteristic times of several hours and an amplitude of about 2 were found in the X-ray range on HII 314. Ayres et al. assumed that these variations in X-rays resulted from superimposing of numerous bursts with characteristic times corresponding to the found flares in CIV.

In 17 spectra in the region of the CIV doublet and 22 spectra in the region of the SiIV doublet obtained by Linsky and Wood (1994) on 9 September 1992 in high-spectral-resolution HST observations of AU Mic (162-s exposures), a fast flare was found on two successive SiIV spectra: in the first spectrogram one of the doublet lines was shifted by 40 km/s at FWHM = 260 km/s, in the second, by 20 km/s at FWHM = 430 km/s, and the flux increased by 3–4 times. Linsky and Wood attributed the initial red shift to the disturbance of the atmosphere by primary energy release in the corona and the subsequent blue shift to evaporation of the chromosphere.

To find very fast brightness variations related to possible microflares, Robinson et al. (1995) carried out a high-speed photometric study of one of the weakest flare stars CN Leo (dM8e) by HST on 30 May 1993. The observations were run in the 645 Å broad band centered at 2400 Å, where the greatest contrast of the flare and quiet star was expected. During four half-hour monitoring sessions with a time resolution of 0.01 s 32 flares with time structures of up to 0.1 s were recorded. The strongest of them had an amplitude of 18 and was about two minutes long. The weakest flares had $E_{UV} \sim 10^{27}$ erg and seldom lasted longer than several seconds. Strong flares resembled close groups of weaker events. The energy spectrum in the region of weak flares constructed from the 32 flares was appreciably below the spectra constructed earlier for the star from recorded optical flares.

On 12 October 1994, within one hour 11 spectrograms with a resolution of 22000 were obtained for the star Gl 752 B (= VB 10, spectral type M8Ve), which is intermediate between red and brown dwarfs. No emission lines were seen in the first 10 spectrograms. The eleventh spectrogram showed a strong spectrum of the transition zone (Fig. 58). This fact proves the existence of high-level chromospheric activity even on low-mass stars (Linsky et al., 1995).

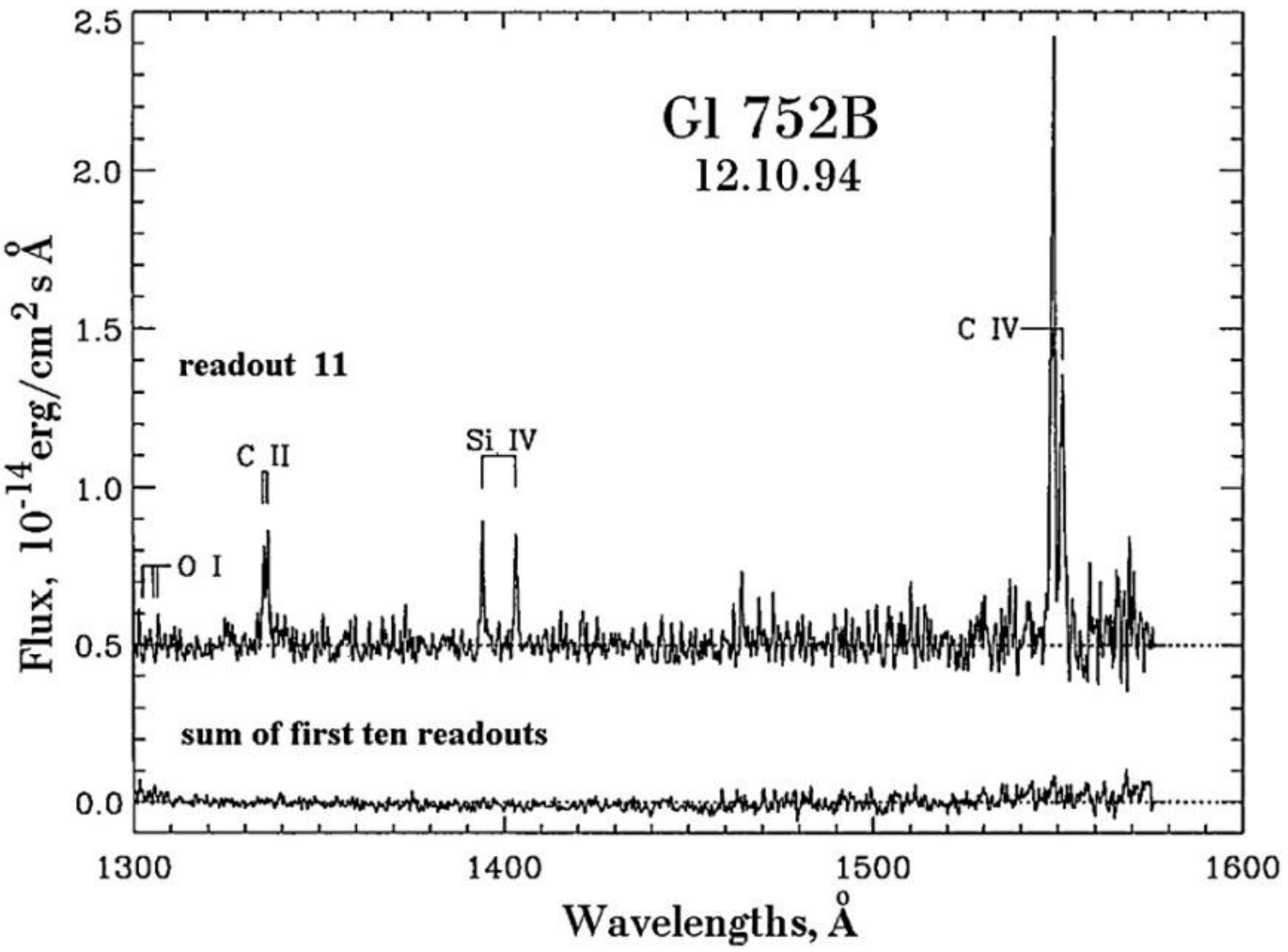

Fig. 58. HST ultraviolet spectra of Gl 752 B recorded on 12 October 1994: 10 spectra of the quiet star and a spectrum of the flare (Linsky et al., 1995)

The 7.5-h monitoring of YZ CMi on 21 December 1994 with a spectral resolution of 1000 and a time resolution of 0.4 s over the range of 1150–1440 Å revealed 29 bursts with durations of 15–200 s. During one moderate-intensity burst the dependence of the line-emission amplification on the formation temperature was clear. But NV did not correspond to this dependence, while the coronal line of FeXXI disappeared during the flare. The flare continuum, as a rule, followed the line emission, but in some cases it grew sharply without appreciable amplification of lines (Robinson et al., 1996). Later, in the spectrum of the strongest flare, Mathioudakis et al. (1999) found significant deviations of the intensity ratios of the components of the SiIV doublet ($\lambda$ 1394/$\lambda$ 1403) from 2:1, which suggests appreciable optical thickness in the $\lambda$ 1394 Å line. A similar effect was found in the strong flare on this star that occurred in April 1996, simultaneously, a similar effect was found in the CIV doublet and consistent estimates of optical thicknesses were obtained; based on these, the geometrical thickness of the transition zone was estimated as several kilometers.

Observing the eclipse system CM Dra composed of dM4e stars with a time resolution of 5 s, Saar and Bookbinder (1998b) found numerous weak flares in the MgII line and in the adjacent continuum. Flares in the lines were weaker, less impulsive and longer than in the continuum. The total energy of flares in lines was at a level of 1% of the total radiation in these lines, at the same time they took up to 20–30% of the time.

Using HST/STIS, Robinson et al. (2001b) recorded four flares and numerous microflares on AU Mic over 10 ks in the 1170–1730 Å range. The flares lasted from 10 s up to 3 min and were identified from the continuum and resonance lines of CIV and SiIV, variations of other lines were less noticeable. No enhancement of the red wing of the $Ly_\alpha$ profile was detected, which would indicate a flux of energetic protons.

From the optical and ultraviolet observations of flares on AD Leo in March 2000 Hawley et al. (2003) constructed the light curves of the nine strongest ultraviolet lines of carbon, silicon, oxygen, helium, and nitrogen, and the six optical lines of hydrogen, helium, and the component of IR CaII triplet. In the strongest flare they found close to linear correlations of the intensities of CIV, SiIV, and NV lines with U-band radiation.

In studying 15 M dwarfs of different activity levels in close ultraviolet, Fuhrmeister et al. (2004) found a strong forbidden coronal line of FeXIII $\lambda$ 3388 Å during a flare on LHS 2076 and confirmed the existence and variability of this emission for CN Leo.

During the described above X-ray and radio observations of EV Lac on 20 September 2001 Osten et al. (2005) carried out a monitoring of this star in the ultraviolet range with HST/STIS in the course of four revolutions of the satellite around the Earth. The observations were conducted with a spectral resolution of 45800, 44 echelle orders were recorded with a time resolution of 60 s. Two small brightenings at 16:53 and 19:48 UT were associated with X-ray flares, the duration of these UV bursts was 4 and 5 min, their amplitudes 1.5 and 1.8, and the total energy in the 1270–1750 Å range amounted to $1.4 \cdot 10^{29}$ and $5.0 \cdot 10^{29}$ erg. Small amplitudes and short duration of these flares made it possible to estimate only total spectral variations of the six strongest lines from SiIII $\lambda$ 1206 Å to CIV $\lambda$ 1550 Å. Turbulent broadenings of lines in the two bursts were different.

From the UV spectra of $\varepsilon$ Eri derived with HST/STIS and FUSE Sim and Jordan (2005) newly determined the emission measures and constructed a semi-empirical model of the chromosphere and transition zone of the star. Calculating the ionization equilibrium, they took into account the dielectronic recombination of C, N, O, and Si and obtained constraints on the FIP effect. Continuing this work, Ness and Jordan (2008) analyzed the observations of $\varepsilon$ Eri with several instruments: Chandra/LETG, EUVE, HST/STIS, FUSE, and XMM-Newton RGS. From the measured emission lines they found relative abundances of elements, estimates of electron density and pressure, and the emission measure distribution. In their resulting coronal model the temperature accounts for 3.4 MK, electron pressure $1.3 \cdot 10^{16}$ cm$^{-3}$ K at a temperature level of $2 \cdot 10^5$ K, the filling factor 0.14 at a level of $3 \cdot 10^5$ K, and no evidence of the FIP effect.

Hawley et al. (2007) performed an analysis of the spectra of flares on YZ CMi in the near ultraviolet derived with HST/STIS and detected a similar behavior of MgII and FeII lines. In one of the flares they found broad MgII lines corresponded to FWHM $\sim$ 250 km/s, whereas the kinetic energy should exceed significantly the radiative one.

Ayres and France (2010) carried out HST/COS observations of EK Dra in the region of 1290–1430 Å and detected noticeable activity in the different-temperature lines of SiIV $\lambda$ 1400 Å ($6 \cdot 10^4$ K), CII $\lambda$ 1335 Å ($3 \cdot 10^4$ K), FeXXI $\lambda$ 1354 Å (10 MK). This peculiarity indicated the substantial dynamicity of subcoronal plasma and the warm "coronal rain" carrying SiIV in the lower atmosphere. Such a "coronal rain" with high velocity was recorded again by Ayres (2012) with the same instrument during a many-hour strong flare of this star in March 2012. Upon more detailed consideration of this event, Ayres (2015a) put forward an idea that since the red shift of the transition-zone lines of SIV and CIV by 10–20 km/s is observed practically all the time and the estimates of density from OIV] in the flare and out of it are close, then the structure of the stellar atmosphere is constantly close to that of the flare, i.e., there is a constant active process of the flare type on the star that was called a flare-ona.

The relatively weak dM4e star GJ 1243 ($V = 12.8^m$) was observed on 1 September 2014 with HST/COS during its visibility at eight revolutions of the telescope around the Earth and with nine ground-based telescopes (Kowalski et al., 2019). Whereas at the second and sixth revolutions made by HST in the near ultraviolet range at 2444–2841 Å two flares were recorded with $\Delta U \sim 1.5^m$ and with the accurate calibration of fluxes within U and I by ground-based instruments. In these flares, strong Balmer lines were recorded, as well as a large emission Balmer jump and a significant growth of emission in NUV. At a resolution of about 1 min, the decay with respect to the NUV-flux maxima occurred faster than that of the fluxes in lines. Within a radiative-hydrodynamic model elaborated by them, these flares corresponded to the heating over several seconds, whereas their impulsive phases lasted for several minutes. At a 5-s time resolution, the light curves in NUV and U did not coincide, and such a resolution allowed the fast and slow decay phases of flares to be separated. The flare NUV flux revealed true flare emission based on a significant variation of the ratio of fluxes lines/continuum. The blackbody radiation with $T \sim 9000$ K underestimated by a factor of 2 the enhancement of the continuum of flares in NUV, though represented well the distribution of broadband optical colors. Within the framework of the elaborated radiative-hydrodynamic model, the measured Balmer jumps — the largest ones among the flares on M dwarfs — excluded the possibilities of the excitement of flares by F11 fluxes, the heating by the F13 flux proved to be more appropriate. Thus, this model suits for the observed Balmer jump of 3–4 in the emission of flares on dMe stars and corresponds to the ratio in the flare blue-red continua of 1–1.4.

Loyd et al. (2018b) observed with HST/COS throughout 10 hours 12 young (younger than 40 million years) M0–M2 dwarfs and recorded 18 flares on 9 of them with amplitudes $A_{FUV}$ varying from 2.5 to 63 and duration from 5 s to 13 min. On 9 August 2017, during the strongest among the recorded flares on the M2 dwarf 2MASS J02365171-5203036 with bolometric energy exceeding $10^{33}$ erg, blackbody radiation with a temperature of 15500 ± 400 K was detected in the far ultraviolet (1170–1430 Å), which yielded 18% of the flux in this range and a significant strengthening of emission lines of the transition zone from the chromosphere to the corona with a noticeable redshift.

During the Hubble observations in the far ultraviolet Froning et al. (2019) recorded several weak and one strong flare on the M2.5 dwarf GJ 674 on 2–3 April 2018. The energy of the latter in FUV was $10^{30.75}$ erg and it lasted over one revolution of the telescope — one of the greatest events in this wavelength range. The spectrum of the flare revealed a strengthening of emission lines at all the levels of the stellar atmosphere and a strong FUV continuum with a temperature of up to 40000 K, which could be interpreted within the framework of the radiative-hydrodynamic model of the atmosphere with an additional dense component.

We have already described above observations of Proc Cen on 1 May 2019 in a wide wavelength range in the course of which a fast high-amplitude flare was recorded at mm and ultraviolet wavelengths, whose bolometric energy accounted for $10^{31.2}$ erg (MacGregor et al., 2021). Based on the energy spectrum of flares of this star constructed from observations by the TESS system, such flares and much stronger ones occurred on it in the optical range, on average, one per day. Emissions of hydrogen, helium, sodium, and calcium were recorded in the flare spectrum with different maximum moments. 510 seconds after this flare a new burst occurred with the symmetric light curve; it had lower amplitude but was more prolonged with a slow decay phase. A strong continuum was recorded in the ultraviolet and visible ranges during both bursts. The obtained multiwavelength observations of the flare were interpreted by a combination of the blackbody and synchrotron radiation mechanisms.

* * *

On spectral monitoring of EV Lac in the CIV $\lambda$ 1550 Å line with a time resolution of 0.61 s from the space station Astron, a strong optical flare was recorded on 6 February 1986. In the smooth light curve of the flare in these lines 50 s after the onset a very short burst was found (Burnasheva et al., 1989). The analysis of these observations led to the conclusion about a probable analogy of this burst with explosive phenomena observed in active regions of the Sun (Katsova and Livshits, 1989).

* * *

The wavelength range accessible for FUSE observations adjoins the short wavelength range border accessible for HST.

In the course of calibrating and test observations with FUSE Ate et al. (2000) derived spectra of AB Dor with a resolution of 12000–15000. The downward flows with a velocity of up to 600 km/s were detected in the OVI line during the flare.

Using FUSE, Redfield et al. (2002) recorded on 26 August 2000 and 10 October 2001 two six-minute flares on AU Mic. The first flare displayed a continuum increasing to short wavelengths, which was interpreted as free–free radiation of hot plasma, and the emission lines of CIII and OVI shifted up to +200 km/s to the long-wavelength range. The second flare displayed an enhancement of the wings of these lines without the Doppler shift. The flares burned in the lines and in the continuum almost simultaneously. The amplitudes of flares in different lines were practically comparable, except for the emission of CIII $\lambda$ 1176 Å: its amplitude in the first flare was almost six times higher than that in the second one.

Christian et al. (2004) found a noticeable optical thickness in the CIII $\lambda$ 977 Å line both in the quiescent state and during the flare on Prox Cen, which resulted in losing up to 70% of the flux in this line. This casted some doubt on the estimates of electron density acquired for this line but did not affect its observed widths. The combination $n_e \sim 6 \cdot 10^{10}$ cm$^{-3}$ derived for the transition zone from the OIV lines of about 1400 Å and the found optical thickness yielded a length of the free passage of about 10 km, which corresponded to the small-scale structures in the solar transition zone. Later, from the FUSE spectra of AD Leo Christian et al. (2006) detected a noticeable optical thickness of the transition zone in the CIII lines.

* * *

HST enabled investigation of stellar flares with a much higher spectral resolution than IUE, and a time resolution of HST observations was similar to the time variations of flares, but spectral ranges of these observations were quite similar. ROSAT and HST were put into orbit almost simultaneously. ROSAT was equipped with a wide-field camera (WFC) that could record flares in the extreme ultraviolet from 60 to 1000 Å or 0.012–0.2 keV, i.e., the region adjacent to soft X-rays. EUVE was put into orbit later and operated in the same range of the extreme ultraviolet. These space observatories provided qualitatively new data on stellar flares. Observations in the extreme ultraviolet are important because this wavelength range contains numerous emission lines formed in the temperature range $10^5$–$10^7$ K that enable the analysis of the state of radiating matter but are not observed in other regions of the spectrum (Monsignori Fossi et al., 1995a, b, 1996).

The first stellar flare recorded by WFC was the flare of 1 October 1990 on BY Dra, which was observed simultaneously by ROSAT and IUE, and in the optical range. The flare was recorded in all four wavelength regions. Using WFC data, Barstow et al. (1991) found that during the flare EM of hot plasma increased by almost a factor of 2 with respect to the level of the quiet star and reached $10^{52}$ cm$^{-3}$. If the observations of the flare at three revolutions suggest that its decay was about 4.5 h long and was caused by radiative relaxation, the density of the flare plasma was $10^{10}$ cm$^{-3}$, but in X-rays the decay lasted for about 10 h. The total energy of the flare in the range of 0.08–0.18 keV reached $7 \cdot 10^{32}$ erg, which made up an appreciable fraction of the total emission of the flare.

On 17 October 1990, WFC revealed a flare on the single K0 dwarf HD 197890 within 90–210 Å. The duration of the flare was more than three hours with subsequent activity for 20 h. Further optical observations showed that this star had vsin*i* ~ 240 km/s, thus, it is one of the fastest and youngest rotators that probably has not yet reached the main sequence and is similar to AB Dor (Matthews et al., 1994).

From the WFC sky survey data, Tsikoudi and Kellett (1997) studied the regions of 58 flare stars from the Pettersen list (1976) and found 28 EUV sources. In studying the fields of 67 stars with CaII emission they found 21 EUV sources. On 23 stars they detected 35 flares. Strong flares were found only on the most active flare stars: BY Dra, AU Mic, and EV Lac. Moderate-intensity flares were recorded on the well-known flare stars: YZ Cet, V 577 Mon, L 1113-55, CN Leo, Prox Cen, V 1258 Aql, and Gl 867. Weak flares were found on the known flare stars and stars for which the optical flare activity had not been established. As many as 19 of the 35 flares took place on 13 dMe stars. Except for these distinct several-hour-long flares, on a half of the examined objects Tsikoudi and Kellett found low-amplitude brightness variations with characteristic durations from 1–2 h to several days. Apparently, these slow variations did not correlate with individually recorded flares and were not caused by rotational or orbital modulation. Tsikoudi and Kellett called them milliflares. Comparing flare activity of late dwarfs from WFC data with other parameters of these stars, they established a positive correlation between the EUV number of flares and bolometric luminosity and negative correlation of $L_{\rm EUV}$ with the rotational period. The dependence of flare activity on spectral class was similar to that found by Kunkel for flare luminosity in the U band (see Fig. 43): the maximum of this distribution falls on early K dwarfs. Finally, Kellett and Tsikoudi (1997) concluded that the flare activity of F–K dwarfs was more appreciable in EUV than in the optical range and that low-amplitude EUV "milliflares" were essential for heating of stellar coronae.

The star AT Mic was used for calibration of EUVE. On 1 July 1992, all three spectrometers and a deep-survey photometer recorded burning of a strong flare. Based on these data, the DEM curve was plotted (Monsignori Fossi et al., 1994a).

During four-day EUVE monitoring of the star AU Mic on 15 July 1992 within 65–190 Å a strong flare was recorded by Cully et al. (1993) (Fig. 38). Its total energy in the extreme ultraviolet was $3 \cdot 10^{34}$ erg and EM ~ $6 \cdot 10^{53}$ cm$^{-3}$. The following weaker flare had $E_{\rm EUV}$ ~ $2 \cdot 10^{33}$ erg and EM ~ $3 \cdot 10^{53}$ cm$^{-3}$. These flares were 1–2 orders of magnitude stronger than the flare on BY Dra of 1 October 1990 recorded by Barstow et al. (1991) with WFC. Assuming that the decay of the strong flare of 15 July 1992 was due to radiative losses in the rapidly expanding plasma structure, from the light curve of the flare Cully et al. (1994) estimated the characteristic density of matter in it as (4−6) $\cdot 10^{11}$ $cm^{-3}$ and the size as $5 \cdot 10^{10}$ cm. With these parameters, the flare luminosity should be high due to the large volume of luminous matter rather than its high density. Then, the flare would resemble more the processes on RS CVn-type stars rather than those on dMe. Monsignori Fossi et al. (1994b, 1996) analyzed independently simultaneous spectral data of three EUVE spectrometers and obtained different results: from the ratios of intensities of spectral lines of FeXXI they estimated the density of the quiet corona and near-flare maxima from $3 \cdot 10^{12}$ to $2 \cdot 10^{13}$ $cm^{-3}$. Using theoretical curves by Monsignori Fossi and Landini, from ten emission lines of FeIX–XXIV and the HeII $\lambda$ 304 Å line Monsignori Fossi et al. (1996) constructed DEM curves for each of 7 intervals into which the whole monitoring interval of AU Mic was split. In the first and the last intervals, corresponding to the quietest state, the DEM curve reached maximum at about 8 MK. Then the curves were broken, whereas on the intervals with flares a significant amount of radiating plasma was found with a temperature above 50 MK. Synthetic spectra SW and MW EUV constructed using these parameters well presented the observations. According to these calculations, $L_{SW}^{max} = 5 \cdot 10^{29}$ erg/s, and this radiation is equal to 1/30 of the total radiative losses in the range of 1– 2000 Å. The characteristic size of coronal loops should be about $3 \cdot 10^9$ cm and the filling factor is about $10^{-4}$. Brown (1996) concluded that the maximum density occurred during the decay phase rather than at the maximum of the strong flare. According to Schrijver et al. (1996), the high density of matter can even lead to appreciable optical thickness of the corona in resonance lines and scattering in these lines will lower the ratio of intensities of the lines and the continuum. Drake J.J. et al. (1994) constructed the light curves of AU Mic in the lines of FeXXIV, XXIII, XVIII, and HeII $\lambda$ 304 Å and found that the first two high-temperature lines of iron and the helium line sharply increased during the maximum of the flare of 15 July 1992, whereas the amplification of the FeXVIII line was appreciably weaker. Katsova et al. (1999a) assumed that the many-hour luminosity of high-temperature lines could be caused by additional heating of the flare plasma in the extended vertical current sheet.

The object EUVE J2056-17.1 was the brightest among the sources discovered during the sky survey. With the help of ground-based observations Mathioudakis et al. (1995b) identified it with an active dK7e-dM0e star. On 3 August 1992, within the range of 60–200 Å a strong flare was recorded with burning longer than an hour, an amplitude of 10, a luminosity at maximum of $1.3 \cdot 10^{30}$ erg/s, a total duration of about one day, and a total energy of more than $10^{35}$ erg. Thus, the parameters of this flare were similar to the flare on AU Mic of 15 July 1992. An extremely strong line of lithium absorption corresponding to the abundance of this element in the stellar atmosphere $\log N(\mathrm{Li}) = 2.4 \pm 0.4$ suggests that this star can be post T Tau or lithium is formed in it as a result of spallation during strong flares.

AD Leo was monitored on 1–3 March 1993 by EUVE with a good ground-based support. On 2–3 March, a high activity level was recorded (Hawley et al., 1995). First, on 2 March 1993, a flare with $\Delta U \sim 0.7^m$ was found, which practically was not seen in EUV, but three hours later a flare with $\Delta U \sim 0.8^m$ marked the onset of a EUV flare with a total duration of about eight hours. The total energy of this flare was $E_U \sim 5 \cdot 10^{32}$ erg and $E_{EUV} \sim 4 \cdot 10^{32}$ erg. Approximately a day later, a weaker EUV flare occurred on the star ($E_U \sim 3 \cdot 10^{32}$ and $E_{EUV} \sim 8 \cdot 10^{31}$ erg) that lasted for about four hours during which a rather fast burst with $\Delta U \sim 0.6^m$ and subsequent long photometric activity were recorded. Comparison of light curves in different bands revealed a delay at the onset of the flare in EUV with respect to U, while comparison of $L_{EUV}$ and $L_U$ during the active state supported the validity of the relation

$$L_{\rm EUV}(t) \propto \int_0^t \mathrm{I_U}\, dt, \qquad (62)$$

which corresponds to the Neupert effect, if the emission in the optical U band is considered proportional to the flux of accelerated particles that emerged in the flare, and the EUV emission is proportional to the thermal radiation flare plasma heated by the particles. (Later, the relation (62) was confirmed in analyzing the flares on AD Leo of 10 March 2000 (Hawley et al., 2003).) Analysis of the EUV data within the theory proposed by Hawley et al. (1995) for burning and fast decay of flares in static coronal loops yielded the estimates for the flare of 2 March 1993: a loop length of $4 \cdot 10^{10}$ cm, which exceeds the stellar radius, a loop section of $9 \cdot 10^{19}$ cm$^2$, an average electron density of $3 \cdot 10^{10}$ cm$^{-3}$, a maximum pressure of 180 dyn/cm$^2$, and an emission measure of $8 \cdot 10^{51}$ cm$^{-3}$. Optical observations yielded the size of the flare $1 \cdot 10^{18}$ cm$^2$ or 0.01% of the stellar surface. In the flare of 3 March 1993, the loop length was $1.5 \cdot 10^{10}$, its section was $2 \cdot 10^{19}$ cm$^2$, the pressure was 280 dyn/cm$^2$, and EM $\sim 2 \cdot 10^{51}$ cm$^{-3}$. Later, Cully et al. (1997) thoroughly analyzed the spectra of AD Leo obtained by three EUVE spectrometers during the same observations averaged over four time intervals: during the first and second flares, during the decay of the first flare and in the quiet star before the first and after the second flare. Domination of emission lines of multiply ionized iron and the HeII $\lambda$ 304 Å line was found in all averaged spectra, but low *S/N* ratio obtained in these observations did not enable consideration of “pure flare spectra”. Cully et al. calculated DEM in two different ways and obtained consistent results. Further, DEM were calculated for the standard chemical composition of coronal plasma of AD Leo and for a ten-fold depletion of heavy elements $z/z_\odot = 0.1$. The estimated sizes of coronal loops are $\propto z^{1/2}$ and the average electron density in them is $n_e \propto z^{-1}$. If in a quiet star DEM has a wide maximum about $10^{7\pm0.2}$ K, which is higher by an order of magnitude than DEM ($10^{6.2}$ K) and can be caused by a set of coronal loops with different temperatures at the tops, then the DEM of flares has a maximum at $T > 10^7$ K, and in the decay phase this maximum goes down and shifts to lower temperatures, which is naturally associated with cooling of the coronal plasma and condensation of matter of the evaporated chromosphere. Spectral analysis suggests that during the flares mainly the high-temperature component of the corona changes, it yielded a somewhat lower temperature of the flare plasma as compared to the previous photometric consideration that underestimated the pressure and overestimated the filling factor of coronal loops. Thus, the conclusion on the decisive contribution of long loops to total coronal radiation and the range of density $10^9$–$10^{11}$ cm$^{-3}$ was confirmed.

The system EQ Peg was observed by EUVE during 20 revolutions of the satellite in the DS/S mode: on 30 August 1993 two strong flares with amplitudes of about 3 and 5 and total durations of more than six and nine hours, respectively, were recorded during photometric and spectral observations (Monsignori Fossi et al., 1995b).

On 9–13 September 1993, EUVE monitoring of EV Lac was conducted in the spectral range and in the photometric range of 65–190 Å. On 10 September 1993 at 6:10 UT, a rather strong flare was recorded almost from the very beginning. Burning of the flare lasted for several minutes, its EUV amplitude achieved 10, an initial decline was almost symmetrical to the rise to the maximum, then a smoother decay followed with an appreciable secondary burst approximately eight minutes after the main maximum. Using an exponential approximation of the descending branch of the flare the time of the *e*-fold decay was 19 min. The amplitude and duration of this flare are similar to those of two flares on this star recorded earlier with WFC. During the decay phase the star passed out of sight of the satellite, and at the next revolution there were no traces of the flare, though weak variations occurred in the EUV light curve.

During the flare, approximately ten-fold-amplified lines of FeXVI–XXIV and HeII were found in the spectra in SW and MW ranges, the ratio of intensities of the FeXXII line sensitive to electron density yielded $n_e \sim 10^{11}$ cm$^{-3}$ for the flare and $10^{13}$ cm$^{-3}$ for the quiet star. A slightly weaker flare that took place on the same day at 22:30 UT was studied in detail by ground-based telescopes (Abranin et al., 1998a). EUVE observations of EV Lac partly overlapped IUE monitoring. Comparison of these data led to the conclusion that in the flare of 10 September 1993 at 6:10 UT CIV and MgII emissions were amplified by a factor of 6 and at least 3, respectively, whereas in the flare at 22:30 UT by a factor of 4.4 and 1.6, respectively. The total losses for radiation in the earlier flare were $8 \cdot 10^{31}$, $5 \cdot 10^{30}$, and $> 4 \cdot 10^{30}$ erg in the EUV range, CIV and MgII lines, whereas in the flare at 22:30 UT they were $7 \cdot 10^{30}$, $3 \cdot 10^{30}$, and $> 1 \cdot 10^{30}$ erg, respectively. After the flare at 6:10 UT the activity in MgII lines continued for two days, which can be associated with the visibility of a strong active region on the star, whose axial rotation period was about four days (Ambruster, 1995; Brown, 1996).

During EUVE observations in the DS/S mode, with individual pointings to a selected object using deep-survey photometers and spectrometers, other objects placed at a right angle to the object could be observed photometrically with a 20-times higher sensitivity than during the all-sky survey. In considering the observations of two dozen G–M dwarfs in October 1993–November 1994, EUV flares were found in the binary system of K dwarfs BD+22°669, on the dK3e star V 834 Tau, and the M0.5 star Melotte 25 VA 334 (Christian et al., 1998). During the observations within the Right Angle Program on 1 May 1997 a strong flare was found on EUVE J1438-432 identified with one of dMe stars with a 16-fold increase of brightness within 2.7 h, a subsequent *e*-fold decrease over 2.2 h, and a full decay over 11 h. The total energy of flare was about $5 \cdot 10^{33}$ erg, which is two orders of magnitude lower than in the strongest flare on AU Mic but close to the other events in EUV (Christian and Vennes, 1999). The results of the Right Angle Program were summarized by Christian et al. (1999): 45% of EUV sources were identified with the stars of late spectral types and during these observations flares were recorded on approximately a dozen stars, including EUVE J0202+105, EUVE J0008+208, G 32-6, EUVE J0213+368, V 837 Tau, EUVE J0725-004, EUVE J1147050, EUVE J1148-374, EQ Vir, WT 486/487, EUVE J1808+297, and G 208-45. Later, the number of objects with flares recorded using this technique increased to 16, most of them were M dwarfs. The amplitudes of these flares varied within 1.5–16, burning times within 1–11 h, decay times within 3–18 h, luminosity at maximum within $8 \cdot 10^{28}$–$4 \cdot 10^{32}$ erg/s, and full energies within $3–500 \cdot 10^{32}$ erg (Christian, 2001).

During cooperative observations in X-ray, UV, optical and radio ranges, AU Mic was monitored by EUVE on 12–15 June 1996: on 13 June 1996 a long (of about a day) flare was recorded with 13-min burning and 5-fold amplification of brightness in the range of 70–160 Å. In SW and MW spectra EUV radiation was presented by the two-temperature model with $T_1$ = 20 and $T_2$ = 80 MK (Gagné et al., 1998).

In the course of similar multiwavelength observations of EQ Peg with EUVE on 2–6 October 1996, a flare was detected on 2 October 1996 in EUV with $\Delta U = 2.5^m$ (Gagné et al., 1998).

From EUVE data, Christian et al. (2003) discovered a strong flare on the dM3.5e star EUVE J0613-23.9B: a 200-fold brightness increase within the range 60–200 Å was primarily due to emission lines with a formation temperature above 10 MK. A strong Ly continuum in the range of 320–650 Å lasted less than 500 s, whereas high-temperature radiation continued for about eight hours and the total flare energy was $E \sim 3 \cdot 10^{34}$ erg. Within the semi-empirical model the coronal density was estimated as $10^{14} - 10^{15}$ cm$^{-3}$.

On the basis of statistical analysis of EUVE observations of 10 F2–M6 dwarfs, Audard et al. (2000) concluded that the distribution of flares with respect to EUVE energies, as in the optical range, could be presented by a power function with spectral indices from 1.3 to 0.8. Values of these indices do not correlate with the period and velocity of axial rotation and Rossby numbers; but if we constrain consideration to only strong flares with $E > 10^{32}$ erg, a correlation between flare frequency and stellar luminosity $L_X$ is found.

* * *

So, both stellar X-ray and ultraviolet flares have first of all higher intensity and total energy and in many cases longer duration than the appropriate solar flares. Unlike stellar radio flares, stellar ultraviolet flares did not show qualitatively new phenomena and processes. Among the features that differentiate stellar ultraviolet flares from solar ones, one should mention a less close correlation with optical and apparently X-ray flares. However, this can be due to different observational conditions. Certainly, quantitative analysis of these strong stellar phenomena requires cautious assumptions about small optical thickness of sources in the emission lines, about equal duration of luminosity of different lines, etc.

## 2.4.4. Optical Emission of Flares

In the optical wavelength range, solar flares are recorded first of all in spectral lines and only some rare and the strongest, the so-called white-light flares, display flare continuum. The sizes of such flares usually do not exceed 3 arcsec, which is 2000 km, their luminosity reaches $2 \cdot 10^{29}$ erg/s and total radiative energies are $10^{30}$–$3 \cdot 10^{31}$ erg, which corresponds to a flare on a dMe star with $\Delta U \sim 1^m$. But recently, there appeared reports that the statement concerning the rarity of solar white-light flares is somewhat overestimated and the improved technique allows one to detect continuous radiation in many flares. In strong flares, the Balmer series is traced up to higher members, in the strongest, to $H_{16}$, the lines of neutral and ionized metals are excited, which testifies to the heating of deeper atmospheric layers. Absorption lines of neutral helium appear in strong flares at greater height, the lines in even stronger flares are replaced by emission. In the strongest events, the HeII $\lambda$ 4686 Å line is recorded. Narrow emission cores of 1–2 Å and wide wings (to 10 Å) of hydrogen lines confirm a decisive role of the Stark broadening of these lines in solar flares. Often, the short-lived “blue asymmetry”, amplification of the short-wavelength wing of the line, occurs. Usually after 1–2 min it is replaced by “red asymmetry” that lasts somewhat longer. CaII lines of solar flares lose the central absorption cores, but on the whole these lines change to a lesser degree than hydrogen lines. As a rule, the lines of metals in solar flares are rather narrow and only the strongest of them display self-absorption. In the rich variety of solar flares one can distinguish the events with limiting spectroscopic characteristics: flares with merging Balmer lines in the range of $\lambda < 4000$ Å, broad Stark wings of hydrogen lines, and strong emission of HeI and HeII and metals, and flares without a Balmer quasicontinuum with narrow hydrogen lines, weak emission of HeI and metals and without HeII. The former group is associated with the impulsive phase and is related only to the small bright knots representing the bases of magnetic loops. The second group is associated with the gradual phase and large flare regions surrounding the bases of magnetic loops.

Considering solar optical flares as an element of the general flare process in a wide range of the electromagnetic spectrum, one will find that a continuum and the lines of extreme ultraviolet that reflect the events occurring in the transition zone appear simultaneously with hard X-rays, microwave bursts and white-light flare during the impulsive phase and several minutes prior to the maximum in $H_\alpha$ and in soft X-rays. An EUV burst lasts not more than several minutes and is much shorter than flares in $H_\alpha$ and soft X-rays, but longer than bursts in hard X-rays. Radiation from the chromosphere and transition zone dominates in the early flare stage, while coronal radiation dominates in the late stage.

It is often stated that flares on red dwarfs, on average, are an order of magnitude shorter than solar flares. However, one should bear in mind that the duration of solar flares is usually estimated from the time of $H_\alpha$ luminosity, while that of stellar flares is from the luminosity of continuum. Since the latter, with the greatest probability, corresponds to solar white-light flares, whose duration is an order of magnitude shorter than the luminosity of $H_\alpha$, one should carefully use this statement in physical estimates.

Discovery of optical flares on red dwarfs and the first steps in their investigation are described in the Introduction and in detail in the small book by Gershberg (1970a). Only the most general properties — the time and energy characteristics — of stellar flares in the optical range were considered in Chapters 2.2 and 2.3. Below, the features of the light curves of optical flares, their colorimetric and polarimetric properties, and the results of spectral studies are considered in more detail.

**2.4.4.1. Light Curves.** Light curves are among the most important characteristics of stellar optical flares, but they practically have no analogs in solar flares: if the light curve of a stellar flare is obtained directly during photometric observations, for each point of a similar curve of a solar flare one should carry out a labor-consuming procedure of two-dimensional integration over the solar disk. Certainly, now this can be easily done by computers, but 30–40 years ago, when solar-activity studies were mainly optical, such facilities were not available, solar researchers did not believe that the curves were necessary and preferred filming of the development of flares.

Several hundred flare light curves on red dwarfs have been published since 1968 in almost three hundred issues of the Information Bulletin on Variable Stars (IBVS) of Commission 27 of the International Astronomical Union, and also in numerous publications by observers from Catania, Crimea, Byurakan, Armagh, Okayama, McDonald, South African observatories, etc. Examples of the curves are presented in Figs. 29, 30, 32–35, 37, and 39–41.

Light curves of flares on UV Cet-type stars, as a rule, are sharply asymmetric: a fast brightness burst turns into a smooth decay. This feature was found already in visual and photographic observations. The overwhelming majority of flares have a narrow and sharp maximum, but approximately 15% of flares have fast irregular oscillations near maximum. The descending branch of the light curve usually consists of two parts: an initial fast decrease, whose rate in absolute values is lower than that of the rise to maximum by a factor of 1.5–3, and the subsequent slow decay during which stellar brightness smoothly approaches the preflare level. The transition from the fast decrease to the slow decay is usually rather sharp and occurs at the level of 0.2–0.3 maximum flare brightness. Often, secondary brightness maxima are found on the descending branch of the light curve.

Despite a number of common properties, one can hardly find two flares with identical light curves, especially in the case of high time resolution observations. In observations with a low noise level a small preflare is sometimes found before the sharp rise to maximum: about a one-minute long smooth increase of brightness to the level of 20–35% of the amplitude of the future flare. There are reliable data about a slight brightness decrease below the level of the quiet star directly before the rise to maximum.

As flare light curves were accumulated, repeated attempts were undertaken to find a correlation between various parameters of the curves. MacConnell (1968) considered 21 flares on AD Leo and 13 flares on BD-8°4352, recorded in the U band and suspected statistical dependences between amplitudes and flare burning and decay times. But Pettersen et al. (1984a, 1986a) did not establish such a correlation in 241 flares on AD Leo. Cristaldi et al. (1969) found a weak correlation of amplitude and the ratio of burning and decay times in flares on six red dwarfs. Gershberg and Chugainov (1969) from 90 light curves of flares on seven flare stars concluded that slower flares occurred on brighter stars. Using a greater number of initial data, Kunkel (1975a) expressed this conclusion in (28). Kiljachkov et al. (1979) found a weak correlation between the burning time and total durations of flares, and Sanwal (1995) suspected a faster burning of flares on lower-luminosity stars.

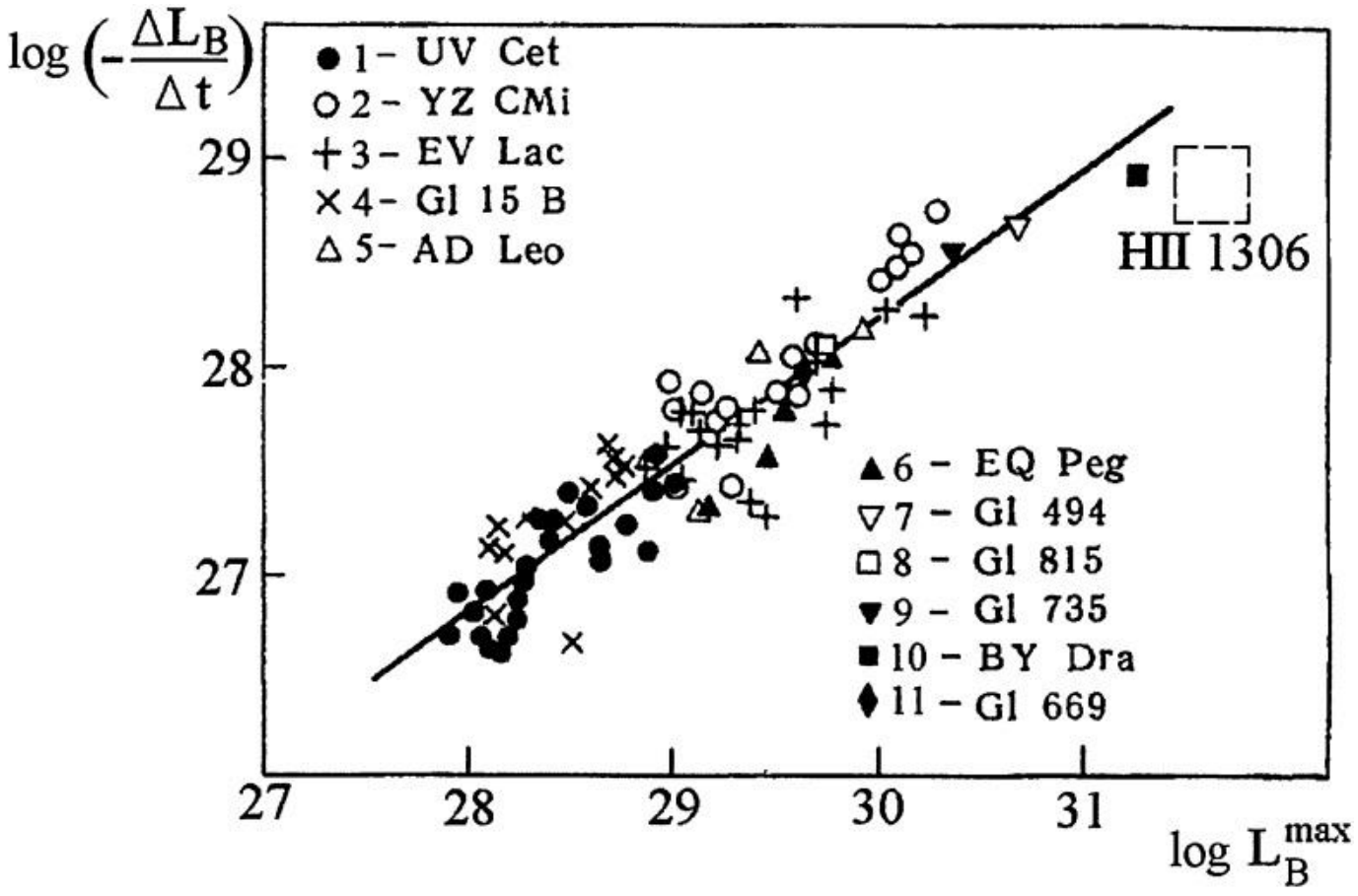

Fig. 59. The rate of initial decay of absolute flare luminosity after maximum brightness vs. absolute flare luminosity at maximum (Shakhovskaya, 1974a)

Correlations between various parameters of flare light curves were studied in detail by Shakhovskaya (1974a). In particular, she found a distinct correlation between the rate of initial fast decay of flares $-\Delta L_B/\Delta t$ and their luminosity at maximum $L_{max}$, where $\Delta L_B$ is the decrease of flare luminosity from the maximum to the start of slow decay and $\Delta t$ is the duration of fast decay. More than 80 flares on 13 stars in the range of absolute luminosities of $8^m$ (Fig. 59) correspond to the found statistical relation with a correlation coefficient of 0.90. A similar but a less confident correlation was found for $-\Delta L_U/\Delta t$. The correlations are close to the relation

$$-dL/dt \sim L_{\max}^{3/4} \tag{63}$$

And only recently, in the course of high-speed colorimetry of a number of flares a simple physical sense of this relation was found (see Subsect. 2.4.4.2).

A slow and smooth decay of most flares encouraged attempts to present the descending branch of the light curve analytically. Abell (1959), Roques (1961) and Chugainov (1961, 1962) used 1–2 exponents for this purpose. Later, Coluzzi et al. (1978) developed an interactive method for analyzing the light curves of flares in which the curves were split into elementary components; each of them had a linear growth to the maximum and an exponential drop after maximum. The first physical model of flare decay suggested relaxation of a hot photospheric spot but did not provide satisfactory representation of the observations on appreciable extent of flares (Roques, 1961). An alternative model of relaxation of optically thin gas was proposed by Gershberg (1964): based on recombinations of hot gas and expansion into some dimensions, six of the 10 considered light curves of flares on EV Lac were presented. Then, this model was supplemented by the effect of gas cooling (Gershberg, 1967). However, the resulting characteristic rates of expansion were unrealistically high. A similar model was considered independently by Andrews (1965): assuming that the initial decrease after the maximum was caused by free–free emission of gas cooling down in a constant volume and recombination relaxation was responsible for the slow decay, he presented well the light curves of 15 flares. But Kunkel (1967) could use this scheme only for four of the 13 flares. Korovyakovskaya (1972) presented several flare light curves within the model of gas relaxation behind the shock-wave front.

Within his model of "fast electrons" (see Chapter 2.5) Gurzadian (1969) presented the light curves of about 20 flares.

However, further spectral observations of flares on UV Cet by Bopp and Bopp and Moffett (1973) (see Fig. 32) and on some other flare stars by Moffett and Bopp (1976) showed that the content of optical emission from flares essentially varied during their development. When this fact was established, the attempts to construct the models of the whole light curves mainly terminated. (The only exception is the work by Kolesov and Sobolev (1990) in which the problem of construction of a light curve in a one-dimensional homogeneous stellar atmosphere was stated and solved analytically, but the quantitative comparison of the curves obtained with the observed flare light curves was not fulfilled.) However, clarification of the physical sense of individual characteristic details of light curves is still an urgent task. The most interesting results are considered below.

In the context of the idea of the decay of stellar flares through the relaxation of hot gas, Pettersen et al. (1984a) considered probable mechanisms of such relaxation and found that at the stage of initial fast decay the main role should be played by heat conductivity. Since the heat conductivity rate is proportional to electron density, flares on lower-mass stars with denser atmospheres, where one can expect a higher density of matter, should decay faster. These reasons were physical substantiation of the statistical relation (28), which is valid for the absolute values $M_V = 8–17^m$.

Analyzing about a hundred flares he recorded, Kunkel (1967) suggested considering their light curves as combinations of fast and slow components in different ratios. Independently, reasoning from the rate of flare burning and the character of descending branches of their light curves, Osawa et al. (1968) proposed to classify flares as fast bursts and flares with smooth decay. Oskanian (1969) independently proposed a more detailed classification of four morphological types of light curves. In the late 1960s, this partition seemed quite a formal procedure, its physical sense became clear later: according to Moffett and Bopp (1976), continuous radiation prevails in fast bursts, whereas in slow flares an essential contribution is made by emission lines. On the other hand, these could be analogs of solar impulsive and two-ribbon flares. Oskanian and Terebizh (1971a) noted an appreciable increase of the number of fast bursts in the total number of flares on UV Cet, the weakest among the stars considered by them. Avgoloupis (1986) analyzed statistically the light curves of 183 flares from the full sample and from each of Oskanian's morphological types separately and found that in the subsamples of separate types there were correlations between burning and decay times to the level of half-maximum and amplitudes, which in the whole sample were not significant.

In recent years, with the appearance of new technologies of observing flares, new models of light curves emerged, using different variants of linear burning and two decay exponents (Hilton, 2011; Davenport et al., 2014; Jackman et al., 2018).

A. Preflare brightness decrease and the "emission peak in the absorption saucer". In the late 1960s, during photoelectric monitoring of the brightness of flare red dwarfs in blue and green rays in the B and V photometric systems, respectively, Cristaldi et al. (1969) noticed that often stellar brightness slightly decreased prior to a flare and all flares on EV Lac took place in minima of slow oscillations of quiet brightness up to about $0.1^m$. Earlier, Roques (1961) revealed a similar effect in observations of YZ CMi. On 18 October 1968, a rather strong flare was recorded on UV Cet at the minimum of slow oscillations of stellar brightness (Chugainov et al., 1969). The flare on EV Lac on 9 October 1973 was observed in violet, blue, and red rays: a 15-s preflare brightness decrease by $0.1^m$ was recorded in red rays but was not noticed in violet and blue rays (Flesh and Oliver, 1974) (Fig. 60). Andersen (1976) recorded a similar situation on EV Lac on 7 October 1975: a preflare brightness decrease in the red region of the spectrum and less distinct decrease in blue rays.

Moffett et al. (1977) recorded a preflare decrease of the $H_\beta$ line in the flare on UV Cet of 6 January 1975. Rodonò et al. (1979) recorded very distinct preflare brightness decreases in the flare on YZ CMi of 5 January 1978 in blue rays using a two-channel photometer. This effect is seen extremely distinctly in the light curve of the flare on EQ Peg of 19 July 1980 recorded by Giampapa et al. (1982b) in violet rays, the preflare decrease lasted almost three minutes. The greatest preflare brightness decreases were recorded on BD+22°3406 by Mahmoud and Soliman (1980) on 25 and 28 May 1980. On 15 June 1994, Ventura et al. (1995) recorded the longest preflare brightness decrease on V 1054 Oph that lasted for about 36 min and was visible in all UBV bands.

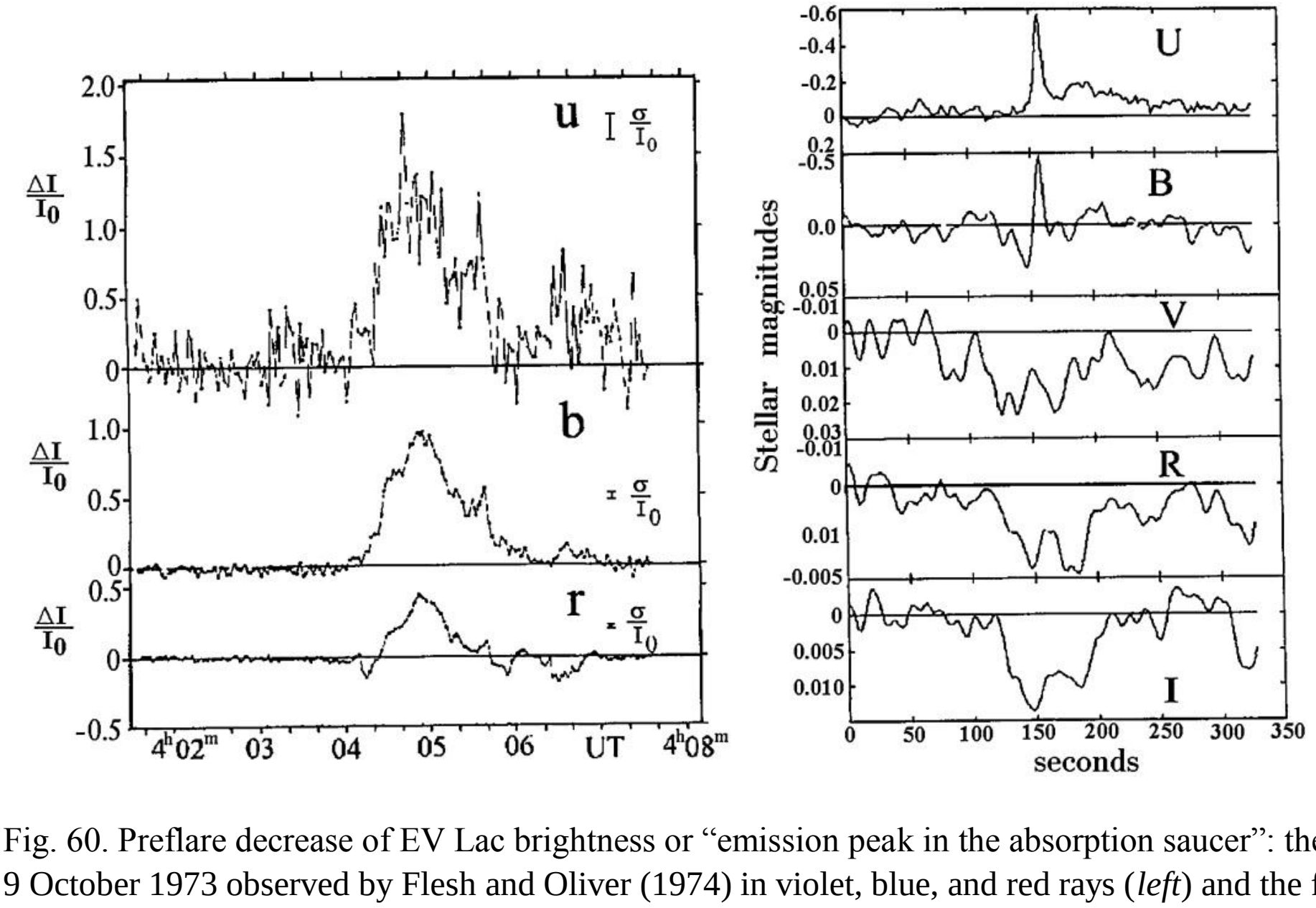

Fig. 60. Preflare decrease of EV Lac brightness or "emission peak in the absorption saucer": the flare of 9 October 1973 observed by Flesh and Oliver (1974) in violet, blue, and red rays (*left*) and the flare of 5 October 1996 (Zhilyaev et al., 1998) in UBVRI bands (*right*)

Shevchenko (1973) carried out the first statistical research of preflare brightness decreases of flare stars. He selected among published data 144 confidently recorded light curves of flares on four UV Cet-type stars. He concluded that half of the light curves did not demonstrate any preflare variations of brightness. In the other half, preflare brightening, i.e., a smooth increase of stellar brightness immediately before the sharp rise in the flare and a preflare brightness decrease were almost equally frequent. One could suspect that preflare brightness decreases occur mainly in the flares that are weaker than preflare brightness amplification. Cristaldi et al. (1980) analyzed the light curves of 277 flares on seven red dwarf stars recorded in blue rays in Catania in 1968–76 and found that 61% of the flares did not have preflare brightness variations, 30% had preflare brightening, and 9%, preflare brightness decreases. Simultaneously, it was shown that during preflare brightening the stars turned bluer, while during the preflare brightness decrease they reddened.

The observations of Flesh and Oliver (1974) initiated the first theoretical model of preflare brightness decrease: Mullan (1975b) assumed that the observed brightness decrease in red rays was connected with a fast transition of strong emission of $H_{\alpha}$ into absorption. However, further spectral observations of flares did not support this model.

An alternative model proposed by Grinin (1973, 1976) was based on the fact that, according to calculations, on fast heating the atmosphere of cool stars should become less transparent; therefore, the flux of radiative energy should decrease at the same time. The source of fast heating could be the impulsive phase, which by analogy with solar flares should begin with the appearance of short-lived hard radiation from the chromosphere. Developing this idea, Henoux et al. (1990) assumed that “black flares” should precede white-light flares on the Sun by 20s.

In addition to this thoroughly developed model of abnormal relaxation of the atmospheres of cool stars after a temperature disturbance, Grinin (1980a) put forward an idea of suppression of radiation of active regions in the initial phase of stellar flare. This concept matches the interpretation of observations proposed by Rojzman and Kabichev (1985): almost always there is a smooth and slow brightening of stars before flares, which cannot always be recorded photometrically, while the fast brightness decrease recorded directly before the fast rise is concerned with the short-time disappearance of an additional source responsible for the slow preflare rise. This scheme does not foresee the absolute decrease of stellar brightness below its normal level. This scheme has not been developed further.

As Grinin’s model of abnormal relaxation of stellar atmospheres predicted the most appreciable preflare brightness decrease in the range of about one micrometer, since the mid-1970s this phenomenon has been studied mainly in the near-IR region.

Bruevich et al. (1980) analyzed about 150 light curves of flares on three red dwarfs recorded in optical and near IR regions and found that practically all strong optical flares were accompanied by synchronous IR flares. A preflare decrease of stellar brightness was observed in almost 70% of cases. The study confirmed that the decrease was predominantly detected in rather weak optical flares. These conclusions were supplemented and refined by the observations of Tashkent researchers (Kiljachkov and Shevchenko, 1980).

During the strong flare on AD Leo of 28 March 1984 Rodonò et al. (1989) found a long brightness decrease in the K IR band at a rather insignificant preflare brightness decrease in this band. To explain the decrease of IR radiation of the star during this optical flare and the flare on UV Cet of 24 November 1985, Rodonò and Cutispoto (1988) used Gurzadian’s concept (1977) according to which optical emission of flares is a side effect of eruption of a strong flux of relativistic electrons from the stellar interior: at the expense of Compton scattering they transformed a part of the infrared quanta radiated by the stellar photosphere into ultraviolet and optical ranges. This model predicted only a synchronous changes in the brightness antiphase in IR and in the optical range, but it did not explain the preflare decrease of brightness and there are serious doubts as to its physical validity on the whole (see Chapter 2.5).

During cooperative studies of EV Lac in 1991–93 its brightness was monitored simultaneously in the near K IR band from the Canary Islands and in all standard UBVRI bands from the Crimea (Alekseev et al., 1994; Abdul-Aziz et al., 1995; Abranin et al., 1998a). From six rather weak flares recorded simultaneously at both observatories in 1991–92, only during one burst in violet rays was the brightness decrease visible in the K band at the level of $2\sigma$. In 1993, both observatories recorded a strong optical flare on September 10, but at the moment of optical maximum only a decrease at the level of $3\sigma$ was visible in the K band. However, several similar decreases were visible before the flare as well. In 1994–95, in parallel with observations in UBVRI bands from the Crimea, the K IR band was monitored from Catania: in one of six flares recorded in optical rays a slight smooth decrease of IR brightness could be suspected 20–30 min after the optical maximum (Abranin et al., 1998b).

Decades later, during 47 hours of simultaneous high-accuracy optical and IR observations of YZ CMi, EV Lac, and AD Leo Tofflemire et al. (2012) recorded four strong flares with energies in the U band varying from $7.8 \cdot 10^{30}$ to $1.3 \cdot 10^{32}$ erg but only upper limits of variations in the J, H, and K bands.

Hence, it follows that rather numerous photometric observations in the long-wavelength range did not yield a distinct and unambiguous picture of the preflare brightness decrease of flare stars. Of particular interest is a rather weak flare recorded on EV Lac during high time resolution observations by the UBVRI system at the Crimean Observatory on 5 October 1996 by Shakhovskoy.

Figure 60 shows the light curves of this flare obtained by Zhilyaev et al. (1998) after processing the initial data by smoothing a Gaussian window with a width of 5 to 12.5 s. One can see two components of the flare: the traditionally recorded emission component, best seen in the U band, and the wide absorption component that is better detected in the long-wavelength range of the spectrum. This distinct two-component model, “emission peak in the absorption saucer”, makes it possible to consider from a common point of view a number of photometric characteristics of stellar flares that seemed independent.

1. Since the late 1960s, observers have noted that fast flares took place during the slow minima of stellar brightness because observations were originally carried out mainly in blue and green rays, B and V bands. Later, observers started monitoring in violet rays, U band, which increased the sensitivity to the emission component, but simultaneously reduced the probability of detection of the “absorption saucer”.

2. The effect of stellar bluing during preflare brightening and reddening during preflare brightness decrease discovered by Cristaldi et al. (1980) suggested that these photometric variations are caused by various mechanisms and are substantially independent. Since the amplitudes of the absorption component are small, in strong flares the emission component with greater probability fills in the absorption one, which results in predominant detection of the preflare brightness decrease in less bright flares noted by Shevchenko (1973), Cristaldi et al. (1980) and Bruevich et al. (1980).

3. The preflare brightness decrease on EV Lac of 9 October 1973 found by Flesh and Oliver (1974) was recorded with a high signal-to-noise ratio, had high amplitude, and is usually cited as one of the most illustrative examples of such phenomena (see Fig. 60). But on close consideration of the whole light curve of the flare one can suspect that after the decay of the emission component there was a final fragment of the absorption component, whose depth was comparable with the preflare brightness decrease. Thus, this flare with a “demonstration” preflare brightness decrease also matches the scheme of "emission peak in the absorption saucer".

4. Coming back to the observed behavior of EV Lac in the K band during optical flares, one can suspect that in the two low-amplitude flares recorded in 1993 simultaneously in the Crimea and on the Canary Islands there were shallow "saucers" in the K IR band.

It is essential that there was detected traditional flare activity in the R band: Qian et al. (2012) carried out monitoring observations of the eclipsed binary system CU Cnc in the R band whose components were on the upper boundary of the fully convective stars, and over 73 min detected its flare activity with the maximum amplitude $A_{\mathrm{R}} \sim 0.52^m$ in this band. This system was totally observed over 80 hours, and the frequency of its flares accounted for about 0.05 flares per hour. To explain the relatively low brightness of the system, an idea of the substantial spottedness of components was invoked.

The above facts suggest that the "emission peak in the absorption saucer" scheme reflects essential properties of stellar flares and deserves special theoretical consideration. In other words, this scheme substantially changes the statement of the problem of the theory of the light curves of stellar flares: one should determine the decrease mechanism of photospheric radiation whose duration is comparable with or longer than the flare observed in violet rays rather than explain only short brightness decrease directly before the flare.

As it was repeatedly mentioned, abundant and diverse phenomena of solar activity provide prompting for interpreting many manifestations of the activity of UV Cet-type stars. But one can hardly find the analogs of rather long-living "absorption saucers" on the Sun. The disappearance of filaments recorded on the Sun at the very beginning of the development of flares proposed by Giampapa et al. (1982a) as a cause for the preflare decrease of ultraviolet stellar emission can hardly have an impact in the range of longer wavelengths. Though abundant solar observation data are available, there are no adequate observations to the broadband stellar observations during which the fluxes from the whole hemisphere containing the active region with spots and flares are recorded with high accuracy. But observers still have to check the hypothesis of Henoux et al. (1990) about "black" flares as precursors of white-light flares.

According to Tovmassian et al. (2003), the slow part of the flare decay is reradiation of the initial fast phase of a flare by the photosphere, and the ratio of a flare and a preflare brightness decrease is determined by the position of the flare on the stellar disk.

One should consider carefully a probable effect of coincidence of the flare localization on the stellar disk with dark spots and facular fields, which can be responsible for both the preflare decreases and stellar brightenings.

B. Fast burning. A characteristic feature of many stellar flares is extremely fast burning. Within the hydrodynamic model of stellar flares, Katsova and Livshits (1986) suggested that the duration of this phase is ultimately determined by the time interval required to shift the shock-wave front to the stellar photosphere at about one scale height of the stellar atmosphere. This suggestion was confirmed by the analysis of fast flares detected by the 6-meter telescope (Beskin et al., 1988).

C. Secondary brightness maxima. Figures 30 and 37 show the light curves of flares on AD Leo of 18 May 1965 and 28 March 1984, where secondary maxima of brightness are distinctly seen. This feature is often observed in the light curves of strong flares. Andrews (1966a) noted that all secondary maxima found in four of the nine flares on YZ CMi took place approximately six minutes after the main maximum. The time interval between the maxima of light curves was rather typical of different flares on different stars with a total duration of tens of minutes (Gershberg and Chugainov, 1969). Apparently, small humps on descending branches and even practically horizontal sections of the light curves, called decay halts, are physically related to secondary maxima.

Mullan (1977) assumed that these details of light curves could be caused by cooling of expanding flare plasma owing to simultaneous action of heat conductivity and radiative losses, whose dependence on temperature was different.

Katsova and Livshits (1991) believed that secondary maxima were a response of the base of the second leg of the magnetic loop to the initial disturbance, which reached it through the whole loop from the first leg base, where the events that caused the primary flare maximum occurred.

However, in the flare on AD Leo of 2 March 1993 with $\Delta U \sim 0.6^m$ brightness remained at a noticeably high level for about three hours after the beginning of the decrease, thus the main flare energy was released in this state rather than, as usual, at about maximum (Hawley et al., 1995).

D. Short-period oscillations. On 28 November 1972, during monitoring of the star HII 2411 with the 207-cm Struve telescope of the McDonald Observatory Rodonò (1974) recorded a flare with an unusual light curve: its smooth brightness changes were superposed by high-frequency pulsations (Fig. 61). Recording was stopped twice to check whether the oscillations were due to technical problems, in so doing, the entrance diaphragm and the time of integration were changed. But the cause was not found. Shortly before this, in high-speed photometry studies with the same telescope, Moffett (1972) suspected a thin structure of the light curves of stellar flares, but upon thorough examination of his results this structure was attributed to regular statistics of quanta (Gershberg and Shakhovskaya, 1973). As opposed to Moffett's data, where the light curve was composed of a set of symmetric parts of different duration, on the light curve of the flare on HII 2411 one could see regular details with systematically faster rise than decay, and with an average duration of 13.08 ± 0.06 s, which slowly increased by the end of the flare. Mullan (1976c) interpreted these oscillations as cyclotron waves propagating between spots on different sides of the equator. Zaitsev et al. (1994) considered them as magnetohydrodynamic oscillations of magnetic loops with a moderate quality factor.

Though there were telling arguments in favor of the reality of the oscillations discovered by Rodonò, his results remained unique for more than a quarter of a century. The data obtained by Rojzman and Kabichev (1985) somewhat resemble them. Andrews (1989a,b, 1990a–c, 1991) tried to find short-period variations of brightness on a number of flare stars using various statistical methods. In analyzing the data for the quiescent state of V 1285 Aql in the R and I bands, he suspected quasiperiodic brightness variations with a characteristic time of about four minutes. Then he found a brightness variation in the U band of quiet stars with the following characteristic times: AU Mic — 25.4 s, V 1285 Aql — 39.3 s, V 645 Cen — 31.3 s, and V 1054 Oph — 24.7 s. In the records of brightness of AT Mic in the U band soon after the flare Andrews suspected quasiperiods of 13.2 and 7.9 s, whereas the latter apparently took place during the flare decay phase as well. The data of U photometry of YZ CMi provided several quasiperiods in the range from 7.1 to 92.9 s, which were supported at different time intervals. The analysis of brightness records of the star V 1054 Oph in the U band at eight time intervals before a strong X-ray flare revealed quasiperiods from 7.2 to 86.4 s, and many of them repeated at different intervals. During a strong flare on Gl 182, Andrews found quasiperiods at all the phases, the maximum characteristic time of variations was 59 s. However, all these results could not be compared in reliability with the data obtained by Rodonò, which was fully "retrieved" only recently.

In the late XXth century, Zhilyaev et al. (2011a) elaborated and created the Synchronous Network of distant Telescopes in Ukraine, Russia, Bulgaria, and Greece, which made it possible to carry out simultaneous observations based on GPS receivers with an accuracy of not worse than one millisecond. The valid observational system at Peak Terskol (Caucasus), in the Crimea, at the Stephanion Observatory (Greece), and in Belogradchik (Bulgaria) in combination with the implemented procedure of digital filtration of the derived data dramatically increased the effectiveness of photometric observations of flare stars and confidently confirmed the existence of low-amplitude high-frequency stellar brightness variations during flares. Thus, the lower panel of Fig. 61 shows the initial light curve of the flare on EV Lac of 11 September 1998 and the high-frequency components derived by subtracting a smoothed component from the initial curves; the solid line marks Crimean observations, the dashed line denotes the observations in Greece. Individual impulses acquired during observations at different observatories are in good agreement, which raises no doubts in their validity.

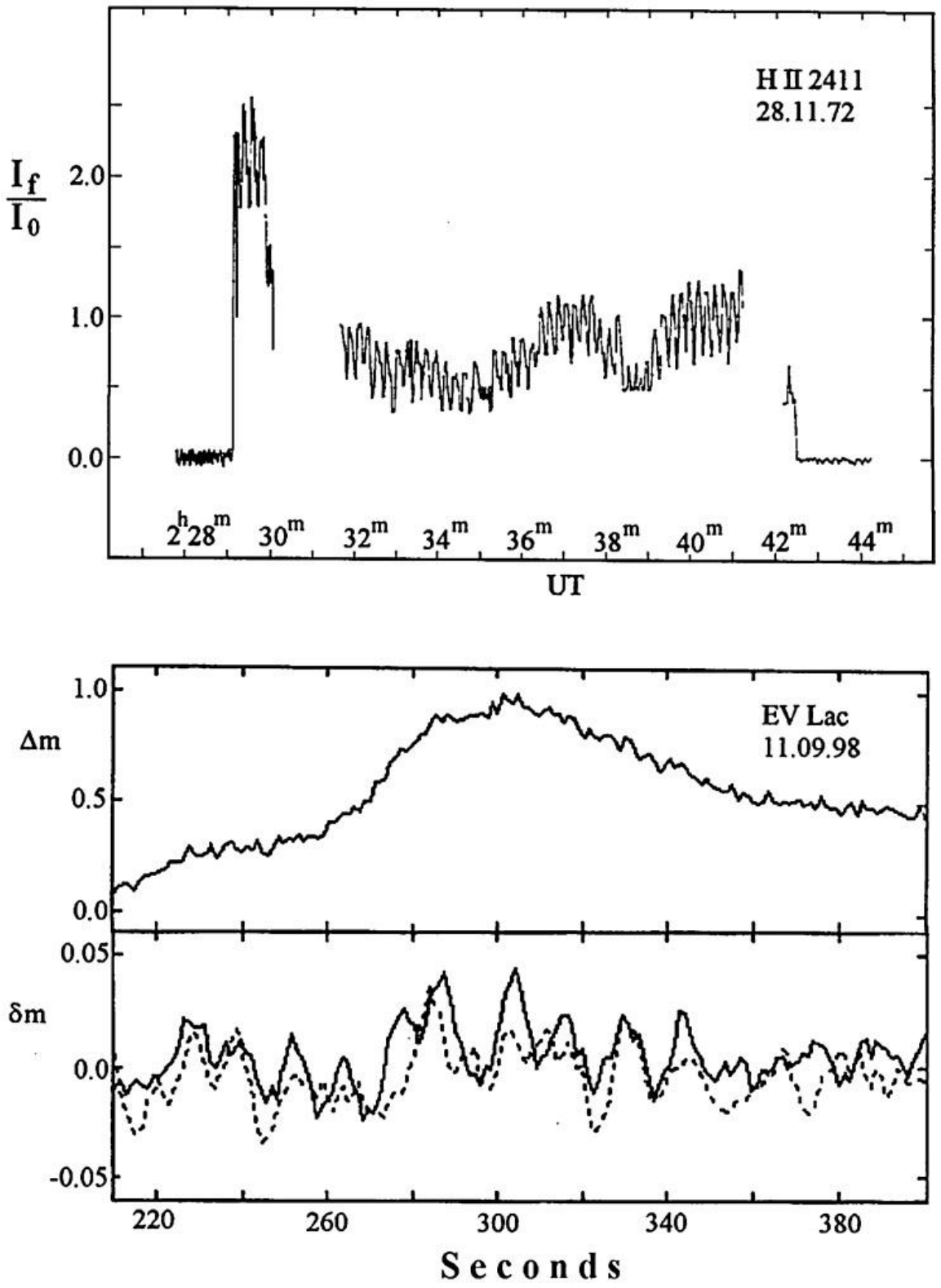

Fig. 61. Quasiperiodic brightness oscillations of stellar flares: *upper diagram*, the light curve of the flare on HII 2411 (Rodonò, 1974); *bottom diagram*, the initial light curve of the flare on EV Lac of 11 September 1998 at 21:55 UT and the residual light curve of this flare after the subtraction from the initial curve of the smoothed component; observations from the Crimea (*solid line*) and Greece (*dashed line*) (Zhilyaev et al., 2000)

One of the most studied by the Synchronous Network of distant Telescopes was a flare on EV Lac of 14 September 2004 at 20:31 UT recorded simultaneously with the 2-meter telescope at Peak Terskol, the 1.25-meter AZT-11, and the 122-inch telescope in the Crimea (Lovkaya, 2014). After the digital filtration of data all three curves revealed the well-agreed high-frequency brightness oscillations, and the time series from the telescope at Peak Terskol and the 122-cm Crimean telescope were used to construct the power wavelet spectra. Both procedures detected high-frequency brightness oscillations with a period of about 4.5 s, an amplitude in the U band of a few hundredths of a stellar magnitude at an interval of about 20 s in the vicinity of the flare maximum. Analogous analysis of a flare on EV Lac of 12 September 2004 recorded in the Crimea, Bulgaria, and Greece detected the high-frequency oscillations with a period of about 17 s, which appeared about an onset of the flare and existed for about a minute. These observations resulted in recording the high-frequency brightness oscillations with periods varying from several seconds to several tens of seconds; these were recorded in all the UBVRI filters and at different phases of the flares.

Investigations of these high-frequency brightness pulsations were actively continued in the subsequent years. Thus, Contadakis et al. (2006) analyzed a light curve of the flare on EV Lac of September 2004 recorded in the U band with the 30-inch telescope of the Stephanion Observatory. They found brightness oscillations with periods of about 1.2 min, 11 s, 7.5 s, and 5.5 s in the preflare state and during the flare, oscillations with periods of 20 to 4 s grew after the flare maximum, and the oscillations with only preflare periods persisted at the end of the flare. These results do not contradict the notions that the evolution of the fast mode of the magnetoactive wave arising at the impulsive phase of the flare propagates through the magnetic loop. From the observations of AD Leo Contadakis et al. (2010) detected brightness oscillations with a period of 2 min to 3 s in the quiescent and flaring states of the star. In the course of observations of EV Lac in the B band in 2001 Contadakis (2012a) detected nonstationary pulsations at the frequencies in the range of 3.5 to 36 s. During the same season Contadakis (2012b) conducted observations of UV Cet and suspected oscillations in the quiescent state and during a weak flare in the range of periods of 4.3 to 108 s. Analogous results were obtained by Contadakis et al. (2012) from observations of YZ CMi: its oscillations were detected in the range of periods of 3 s to 2 min, while during the flare there were added oscillations with periods of 2 to 1.5 min and 60, 11, 7.5, and 4 s.

Using ULTRACAM, in a flare on EQ Peg B Mathioudakis et al. (2006) recorded variations of intensity with a period of about 10 s and with an increasing period as the flare developed. As an interpretation, the fast MHD wave was suggested with emission modulation due to the magnetic field or a series of periodic reconnections of the magnetic field.

Using GALEX (Galaxy Evolution Explorer) launched in 2003 for studying ultraviolet radiation of galaxies, Welsh et al. (2006) performed high-speed photometry in the near (1750–2800 Å) and far (1350–1750 Å) ultraviolet for the four dMe stars: GJ 3685 A, CR Dra, AF Psc, SDSS J084425.9+513830.5. Statistical analysis of the 700-s preflare registration of CR Dra and SDSS J084 detected no microflare activity on the times of 0.2, 1, and 10 s, but by selecting electron density, it allowed one to represent ratios of the measured fluxes in two bands. In all four recorded flares, the first significant oscillations were detected in the range of periods of 30–40 s. These were interpreted as acoustic waves in coronal loops of a length of $10^9$ cm, which is less than 10% of the stellar radius, and at a temperature of 5–20 MK. Earlier, with the same instrument Robinson et al. (2005) recorded with high time resolution a strong ultraviolet flare on the M4 dwarf GJ 368 A: in the range of 1350–1750 Å the star became brighter by a factor of 1000 over 200 s.

To interpret the detected fast brightness oscillations, different physical ideas were invoked: effects of the terrestrial atmosphere, falling of the remnants of the dust envelope on stars, astroseismology, and the Ionson–Mullan hypothesis (Mullan, 1984) on the interaction of convective motions with coronal magnetic loops at their base. The similarity of the found statistical quasiperiods with the above periods of brightness variations on HII 2411 and EV Lac is intriguing.

Later, Mullan et al. (1992b) found quasiperiodic variations of brightness with a characteristic time of about six minutes on AD Leo and with characteristic times from three to six minutes on Gl 549. To explain them, they developed the model of oscillating coronal loops.

Peres et al. (1993) suspected similar short brightness variations with a quasiperiod of two and four minutes shortly before the flare on FF And and variations with longer quasiperiods on V 1054 Oph and BY Dra. But, according to their quantitative estimates, the model of oscillating coronal loops by Mullan et al. (1992b) does not fit the energy considerations.

In photometric observations of YY Gem on 6 March 1988 Doyle et al. (1990c) within one night recorded four flares with time intervals of 48 ± 3 min. They tried to explain this series of flares within the context of the events initiated by oscillating filament located above the active region. According to their calculations, if the filament is placed at small height, less than half the size of the active region, oscillations with a frequency of up to 1 Hz are possible.

From the early 2000s, the MHD concepts have prevailed in the interpretation of high-frequency brightness oscillations. Thus, prior to the mentioned above MHD models, Kouprianova et al. (2004), Stepanov et al. (2005), and Zhilyaev (2011b) considered such brightness oscillations of the flare stars within MHD framework and came to the conclusion that they were localized at the bases of magnetic loops, and pulsations were determined by modulation of downward fluxes of energetic particles along a loop by magnetoacoustic oscillations.

Searching for the quasiperiodic pulsations of flares from the data of panorama photometry within the Kepler experiment (see below Item E), Balona et al. (2015) considered 257 strong flares on 75 stars in frames derived with minute exposures and in 18% of cases various “humps” were detected at the descending stages of flares. In seven flares on five stars oscillations overlapped several cycles, but their periods did not correlate with any stellar parameters; this suggests that they are associated with loop oscillations. None of the two mechanisms proposed for the Sun (magnetohydrodynamic processes and activation of flare loops observed in the lines of highly ionized atoms) can explain the quasiperiodic pulsations of strong stellar flares recorded in the continuum.

Recently, from TESS observations with a 2-min cadence Ramsay et al. (2021) for 7 M dwarfs found quasiperiodic pulsations of flares with periods of 10 to 72 minutes and associated these phenomena with magnetoacoustic waves in flare coronal loops.

E. Mechanisms of flare decay will be discussed below, but here, in the context of the discussion regarding the flare light curves, it is worth noting that Mullan et al. (2006) from the light curves of 134 flares recorded with EUVE DS considered lengths of the magnetic loops on stars within F2–M6 and found that the loops that were less than the stellar half-radius took place on stars of all these spectral types, whereas the loops up to two stellar radii were on the stars later than K5 or M0.

F. Panorama photometry — flares. The first panorama photometric studies of flare stars were described above considering the results of photographic observations of several open stellar clusters and calculating the group energy spectra of flares. The panorama photometry from spacecrafts allows one to derive fairly prolonged photometric series of many tens, hundreds, and thousands of objects simultaneously with an accuracy by two orders of magnitude higher than photographic observations and thus to record extremely rare events.

The first results of studying stellar activity by the GALEX spacecraft were reported by Welsh et al. (2005): they published the first catalog including 84 variable objects in the ultraviolet range in which the most appreciable were flares on M dwarfs. The strongest flare on the dM4e star GJ 3685 lasted for less than 200 s, had an amplitude of $10^m$ and energy release of $10^{34}$ erg.

Later, Welsh et al. (2007) analyzed 1802 images derived over 630 hours with the aim of searching for their fast variations and found 49 such objects, their average amplitudes accounted for $2.7^m$. For 32 of them they determined photometric parallaxes, which placed sources at the distances of 25 to 1000 pc; the majority of them were active M dwarfs, and continuous emission dominated in their flares. The average energy of the UV flare amounted to $2.5 \cdot 10^{30}$ erg, which is comparable with flares on M dwarfs in the U band, X-rays, and EUV. Flares of M0–M5 stars radiated approximately 5 times more energy than the flares of M5–M8 stars. Earlier, using the same instrument, Robinson et al. (2005) recorded with high time resolution a strong ultraviolet flare on the M4 dwarf GJ 368 A: within 1350–1750 Å the star became brighter by a factor of 1000 over 200 s.

Brown et al. (2009) carried out GALEX observations of two regions in the Hyades and Pleiades: searching for the fast variability, during about 4 hours they recorded about 400 objects in each cluster. These resulted in the detection of 16 variable objects: 13 M dwarfs and two G stars. From the seven light curves of six M dwarfs the flare energies were estimated within $2 \cdot 10^{27} – 5 \cdot 10^{29}$ erg. One of the flaring stars was definitely a member of the Hyades. A number of M dwarfs revealed periods of increased brightness, which can be interpreted as an effect of many weak flares and this may serve as an argument for supporting the idea that all emission of M dwarfs is an ensemble of weak flares.

Throughout 227.6 hours of observations Rockenfeller et al. (2006b) recorded five flares on four late M dwarfs during the monitoring of 19 such dwarfs with the four-channel CCD photometer BUSCA at the 2.2-meter telescope of the Calar Alto Observatory with participation of the 1.5-meter telescope of the Maidanak Observatory. The BUSCA photometer is equipped with dichroic filters to separate a beam into four wavelength ranges, and each one is recorded with its own CCD camera. Of particular interest is a flare on the M9 dwarf 2MASS J1707183+643933 whose amplitude was more than $6^m$ in the ultraviolet range.

The most abundant results of the panorama photometry of flare stars were derived with Kepler that allowed one to accumulate photometric series of thousands of hours duration simultaneously for about 105 thousand objects.

Over 33.5-hour monitoring with Kepler Walkowicz et al. (2011) studied about 23000 cool dwarfs and identified 373 flare stars; some of them flared repeatedly during the time of observations. Having determined energies, frequencies, and durations of flares, they found that frequencies of flares on M dwarfs were higher but their durations were shorter than those for K dwarfs. High photometric accuracy of Kepler made it possible to detect stellar variability out of flares and it was found that stars with such variability in the quiescent state radiated more energy, but this variability only weakly correlated with frequencies of flares. The flares with more duration occurred on stars with lower frequency of flares.

Analyzing several thousand A–G stars from the Kepler archive, Balona (2012) discovered flares on 25 G, 27 F, and 19 A stars. Later, Balona et al. (2013) found 13 more flare A stars. The amplitudes of flares were from 0.001 to 0.1, their durations from several minutes to several hours, typical energies of flares on F stars were $10^{35}$ erg and on A stars $10^{36}$ erg.

Almost all the considered stars showed low-amplitude brightness variations with probable rotation periods. The recorded flares on A stars were not definitely due to their cold components but might arise in magnetic fields between components. Then Balona et al. (2015, 2016) found a certain tendency for the whole range of spectral types A–M: the strongest flares took place on the fastest rotators. Differential rotation seems to play an important role in increasing energy of flares, and the stars with flares have significantly larger spots than nonflare stars, which points to the correlation between flare energy and size of the active region. Based on the estimates by Balona (2015), 1–2% main-sequence stars of all spectral types from M to A are flaring. But Svanda and Karlicky (2016) studied with Kepler K, G, F, and A stars and found that for A stars the total energy flux of flares at least four times less than that for G stars, which, in their opinion, explains the absence of hot coronae in such stars. They detected a regular shift in the histogram of flare frequencies for A stars toward higher energies and a significant role of nanoflares in the heating and formation of coronae.

In continuation of the outlined in Sect. 2.3.2 results by Maehara et al. (2012) and Shibayama et al. (2013), who from Kepler data detected superflares with energies up to $10^{36}$ erg, Notsu et al. (2013) found that on stars with relatively slow rotation there might occur flares with energies similar to those on fast rotators, but the average frequency of flares on slow rotators was lower, and the energy of superflares correlated with the total area of stellar spottedness.

As stated above, Roettenbacher et al. (2013) detected 17 flares on the star KIC 511407 over 172 revolutions.

Based on Kepler observations with 1-min exposures, Ramsay et al. (2013) recorded two flares with energies $\sim 7 \cdot 10^{32}$ erg and a duration of about 10 min and several weaker bursts on the M4 dwarf KIC 5474065, which is a fast rotator with a rotation period of 2.47 days. The obtained results were compared with data on the faster rotator M4 V KIC 9726699 with a rotation period of 0.60 days.

From 117661 stars observed with Kepler Candelaresi et al. (2014) considered 380 G, K, and M stars and recorded 1690 superflares with energies of more than $5 \cdot 10^{34}$ erg, among them only two turned out to be binary systems. They studied whether the effective temperature and rotation rate had effect on the frequency of such flares and found that with growing temperature the frequency of superflares decreased; earlier, there was found a decrease of dynamo activity with increasing temperature. For the slowly rotating stars they found a square growth of the average frequency of events with increasing rotation rate up to some level above which frequencies linearly decreased. For the rapidly rotating stars, their fraction with higher spottedness increased, which leads to more frequent flares. Turbulent dynamo was used for studying ohmic dissipation as a source of flare energy at the differential rotation. The obtained statistics of energy dissipation as a function of dynamo number is analogous to the observed statistics of the inverse Rossby number with strong fluctuations.

Maehara et al. (2015) studied the results of Kepler observations with 1-min exposures. They found 187 superflares on 23 solar-type stars with bolometric energy of $10^{32}$ to $10^{36}$ erg. Having combined these results with data on half-hour exposures, they detected a power energy spectrum of superflares with the spectral index −1.5. The average frequency of flares with energy of $10^{33}$ erg was estimated as one event over 500–600 years. The upper energy limit of superflares was comparable with magnetic energy of starspots. The duration of superflares, which was determined as $e$-fold weakening after the maximum, increased with their energy as $E^{0.39 \pm 0.03}$, which corresponds to the development of the flare with Alfvén time.

As stated above, from the Kepler data Wu et al. (2015) and Namekata et al. (2018) recently performed comprehensive investigations of power energy spectra of superflares on solar-type stars.

Savanov and Dmitrienko (2016) studied 279 G stars on which Shibayama et al. (2013) detected 1547 superflares with energies of $10^{33}$–$10^{36}$ erg, estimated a fraction of the stellar surface occupied by flares under the assumption of their blackbody radiation at maxima and found that the areas of flares were about a half of the areas of spots.

From the Kepler and GALEX data Dmitrienko and Savanov (2017) analyzed the spottedness and activity of 1570 M dwarfs.

Comprehensive studies from the Kepler data were carried out by the team of Susan Hawley.

First, in the paper "Kepler flares. I", from the two-month observations with short exposures Hawley et al. (2014) considered six M dwarfs, separating them into active (M4e Gl 1243, M5e Gl 1245 A, and M5e Gl 1245 B) and inactive (M1 Gl 4099, M2 Gl 4113, and M3 Gl 4083) based on the presence or absence of $H_\alpha$ emission in the quiescent state; data on this emission were derived with the ground-based 3.5-meter telescope of the Apache Point Observatory. The recorded flare in the U band allowed one to bound energy estimates from the Earth with those from Kepler. Active M dwarfs showed numerous flares and well-defined rotational modulation of brightness due to spots. Inactive M dwarfs displayed a less noticeable effect of spots and a less number of flares. The scaling of Kepler data from the ground-based photometry in the U band allowed one to compare these data with the previously found frequency distributions of flares. From the observations of the GJ 1243 star a confident correlation was found between the energy of flares, their amplitudes, duration, and the decay time, but only weak dependence on the burning time. (We have mentioned above the high time resolution observations of this star (Kowalski et al. (2019).) The flares with more than one brightness peak proved to be more prolonged, they radiated more energy at the same amplitude. The flares with energy more than $10^{31}$ erg had a power-law energy distribution and concerning this distribution there was a discernible deficiency of less strong flares. There was no correlation between the frequency and energy of flares with phases of spots, the wait times of a subsequent flare corresponded to their random distribution in time and energy. All this corresponds to the scenario with numerous independent active regions on the stellar surface in agreement with magnetometric results by Morin et al. (2008b). The energy spectra of flares on active and inactive M dwarfs showed a continuous change of frequencies and energies of flares, which means that the separation in $H_\alpha$ emission in the quiescent state does not fully reflect the complexity of magnetic activity. A fraction of complex flares grew with duration, which suggests such flares as a superposition of independent events in different stellar regions.

Then in the paper "Kepler flares. II", based on the Kepler data derived with 1-min exposures over 11 months, Davenport et al. (2014) reported on the detection of 6107 individual flares with energies of $10^{29}$ to $10^{33}$ erg on the single star M4 GJ 1243. They developed a method of automatic searching for flares from the short exposures of Kepler, based on 885 events constructed a template of light curves with the burning phase, which is described by the fourth-order polynomial, and two exponential decay phases, and found that only 15% of flares did not fit this template. Using it, they separated flares with complex light curves into individual components and detected that during 80% of events with a duration of 50 min and longer the light curves had not one maximum, and for the flares with a duration of more than 10 min the distribution of durations had a two-component power-law form. In the paper "Kepler flares. IV" Silverberg et al. (2016) continued the study of this star and, using a Monte Carlo Markov Chain sampling, analyzed the light curve over the years. The obtained results on stellar spottedness are outlined above. They attributed the star to the Argus association, which determines its age as 30–50 million years, considered the flare frequency distribution and found variations in the spectral index of the flare energy spectrum, depending on the completeness of the sample and a small number of the strongest flares. The ground-based spectral observations of three weak flares derived simultaneously with Kepler observations allowed one to put constraints on the radiative-hydrodynamic model in the region of such stellar flares.

In the paper "Kepler flares. III" from the 9-month observations with short exposures and four-year monitoring with half-hour exposures, Lurie et al. (2015) considered flare activity of the binary system GJ 1245 AB consisting of two M5e dwarfs. While separating components by 7″ and at a plate scale of 4″/pixel, the analysis of the recorded flux was performed applying the statistics of pixels. According to their estimates, the average flare frequency of the A component was 3.0 and B component 2.6 flares per day. The long-tern effects of spots were recorded on both components and the spin-down of the B component was consistent with the differential rotation. The components of GJ 1245, having rotation periods that are differentiated by almost a factor of 3, confirm the known statement that the differential rotation is higher for slowly rotating stars. The four-year observations led to the conclusion on a decrease in the separation of components, which could be caused by the motion of a very weak component of the system M8 GJ 1245 C near the A component. As in GJ 1243, the spectral index of energy spectra of flares GJ 1245 A and B was close to 2. Here the analysis of spottedness leads to a pattern of numerous small active regions on both components.

Davenport (2016) published the Catalog, including more than 850000 candidates for flares recorded with Kepler on 4041 stars and comprising 1.9% of the database of this instrument. In this sample of stars, a fraction of flare ones grew toward lower masses, and flare activity decreased based on the power law with increasing Rossby number (as well as the chromospheric $H_\alpha$ emission) and saturated at Ro ~ 0.03. However, Yang and Liu (2019) claimed that 60% of the events marked by Davenport were erroneously attributed to flares.

From the Kepler data, Gao et al. (2016) studied 1049 close binary systems and on 234 of them detected 6818 flares. They investigated the dependence of the flare activity level on the orbital period and rotation period and found that this level increased with decreasing orbital period up to 3 days or rotational period up to 1.5 days; then the activity level decreased independently of the fact how rapidly stars rotated. For the two eclipsing systems with the greatest number of recorded flares, the frequencies of flares were considered as functions of the orbital phase, but correlations of these values were not detected. From 203 flares on 20 noneclipsing ellipsoidal systems a bimodal distribution of amplitudes was found with maxima at orbital phases 0.25 and 0.75.

Then a team of Chinese colleagues (Yang et al., 2017) studied activity of 540 M dwarfs from long exposures in the Kepler field. This resulted in the detection of 103187 flares. According to their estimates, on average, several flares per day occurred on M0–M4 dwarfs. Using the luminosity ratio $L_{\text{flare}}/L_{\text{bol}}$ as a characteristic of the flare activity level, they found, as for X-ray radiation, the modes of supersaturation, saturation, and exponential decay for ultrashort, short, and long rotation periods. The activity level and the fraction of flare stars grew rapidly starting from M4, which is predicted by the turbulent dynamo theory. The fraction of flare stars in early M dwarfs was higher than the fraction of emission dwarfs; these fractions were comparable for intermediate subtypes. The size of starspots correlated with the flare activity level, and the ratio $L_{\text{flare}}/L_{\text{bol}}$ was associated with the chromospheric activity level by a power-law way: $L_{\text{H}\alpha}/L_{\text{bol}} \sim (L_{\text{flare}}/L_{\text{bol}})^a$, $a \sim 0.44$, therefore, small increases of chromospheric activity could cause a significant increase of flare activity. Thus, Yang et al.

concluded that superflares did not require any excitation mechanism. The rotation period 10 days or the flare activity level $6 \cdot 10^{-6}$ separated active and inactive stars, higher than this border all the flare stars had $H_\alpha$ emission. The spectral index of the flare energy spectrum was $2.07 \pm 0.35$.

Yang and Liu (2019) compiled a catalog of the Kepler data derived with long-term exposures and detected numerous errors in previous publications. Their catalog comprises 3420 flares stars on which 162262 flares were recorded. There was revealed an increase of the fraction of flare stars from 0.69% among F stars to 9.74% among M objects. This distinction was attributed due to growing thickness of the convective shell. The flare energy spectrum on stars of spectral types ranging from F to M was power-law with the same spectral index, which implies the same $\alpha$-$\omega$ dynamo mechanism of generation of flares, but this index was different by unity for A stars and the fraction of flare ones among them was 1.16%, which may mean another dynamo mechanism (see Chapter 4.3). The relation activity–rotation acquired by Yang and Liu for the cool stars is similar to what obtained earlier, but with increasing temperature it became more disperse, which could be caused by the combined effect of two different dynamos: envelope and convective. For 70% of flare stars, the rotation periods were shorter than 10 days and for 95% these were shorter than 30 days. At such estimates, the flares on the Sun with energy of $10^{34}$ erg occur once per 5500 years.

Gizis et al. (2017) studied the flares of young brown dwarfs. From the Kepler data with 1-min exposures they detected 22 flares on 2MASS J035350208+2342356 and constructed their power-law energy spectrum ranging from $10^{31}$ to $10^{33}$ erg with the index $-1.8 \pm 0.2$; taking the data on other M6–M8 dwarfs into account, the flare energy spectrum was followed within 4.3 orders of magnitude. On the other object of the L5 type from 2MASS they found no flares. Moreover, from the Kepler data derived with half-hour exposures Gizis et al. presumably detected a superflare with energy of $2.6 \cdot 10^{34}$ erg on the other late M star CFHT-PL-17.

From the Kepler data, Katsova and Livshits (2015) considered the superflares on G stars of early age and found that on such young stars of about 1 billion years, when the activity cyclicity was formed, the maximum total energy of flares was $10^{34}$ erg and the basic source of continuous radiation in the optical range was low-temperature condensations forming at the impulsive heating of the chromosphere. At a given flux of heating accelerated electrons of $3 \cdot 10^{11}$ erg $\cdot$ cm$^{-2} \cdot$ s$^{-1}$ they estimated flare sizes to $10^{19}$ cm$^2$, and if the spectrum of electrons was rather rigid, then determined its expected microwave emission.

Then Katsova et al. (2016) considered G and K stars on which superflares were detected and found that most of them had saturated activity and rotation periods of 0.5–7 days; the second peak of bimodal distribution fell into 11–15 days. The transition from basically saturated activity of G stars to solar-type activity took place when their rotation period was approximately 1.4 days, while for K stars this period was 4–5 days. The stars with superflares were young and characterized by fast rotation, their activity level definitely correlated with enhanced abundance of lithium.

From the Kepler observations with short exposures Doyle et al. (2018) considered 40 M0–L1 dwarfs, revealing flares with energy up to $3 \cdot 10^{34}$ erg, and found that flare activity decreased for stars with rotation periods of more than 10 days and there was no preferential phase in the distribution of flares in rotation period phases.

As it was mentioned, from the Kepler data Maehara et al. (2018) considered an association of spottedness with flare activity for G, K, and M stars with superflares and found that there is a good correlation between bolometric energy of the largest superflare on a star and magnetic energy near spots. The average frequencies of flares of the specified bolometric power are approximately proportional to the area of spots. These results suggest that the flare activity level is determined by the area of spots.

From the Kepler data derived with short exposures, Paudel et al. (2018) studied activity of 10 cool dwarfs: two M6 stars, three M7, three M8, and two L0; on all the targets they detected flares of the total number of 283 with energies in the range of $10^{29–33.5}$ erg and with power-law energy spectra, which were more mildly sloping for cooler stars. On one L0 and one M7 dwarfs the superflares were recorded with a duration of about two minutes, amplitudes of 144 and 60, and bolometric energies of 3.6 and $8.9 \cdot 10^{33}$ erg, whose achieving required magnetic fields up to 5–10 kG.

In the course of simultaneous observations with Kepler in the optical range and with XMM-Newton in X-rays, Guarcello et al. (2019) analyzed 12 strong flares in the Pleiades: on 10 K–M dwarfs, on one F9, and one G8 star. The total energy of optical flares was in the range of $10^{32.9}$ to $10^{34.7}$ erg, energy in X-rays from $8 \cdot 10^{32}$ to $10^{34}$ erg, and in 10 flares the optical radiation energy exceeded the energy in X-rays, as in most solar flares. But radiations in these two ranges were weakly correlated. The durations of flares in these ranges turned out to be comparable, and these flares occurred mainly on stars with an axial rotation period of less than a half-day. Guarcello et al. concluded that the considered events were tied by single coronal loops and there was no any additional heating in the decay phase.

Notsu et al. (2019) considered short-exposure Kepler observations of the solar-type stars with superflares in combination with the results of spectroscopic observations of 64 such stars with the 3.5-meter telescope APO within the SDSS project, the Japanese telescope Subaru, and GAIA-DR2 data. Consideration of values of $v\sin i$ and chromospheric CaII H, K, and IR lines showed that brightness variations of stars with superflares were produced by the rotation of objects with large spots, and the energy maximum of superflares steadily decreased with increasing axial rotation period. Having estimated the age of stars from the lithium line, Notsu et al. found that the superflares with energies of $< 5 \cdot 10^{34}$ erg occurred on old slowly rotating solar-type stars once per 2000–3000 years, whereas the young — several hundred million years — fast rotators with rotation periods of a few days showed the flares with energy up to $10^{36}$ erg and one hundred times more frequent. The spottedness maximum of young stars does not depend on the rotation period and amounts to $A_{\text{spot}} \sim 0.05–0.1$ of the hemisphere, but it starts decreasing when at the age of 1.4 billion years the rotation slows down, starting from the period of 12 days, and the filling factor drops to 1%. More than 40% of stars with superflares classified earlier as the solar-type objects proved to be subgiants.

From a substantially later sample including 2341 superflares on 265 solar-type stars, Okamoto et al. (2021) refined the results of Notsu et al. (2019): the maximum energy of superflares on solar-like stars amounts to $4 \cdot 10^{34}$ erg and on the Sun the flares with energy of $7 \cdot 10^{33}$ and $1 \cdot 10^{34}$ erg can occur once per 3000 and 6000 years, respectively. Furthermore, they associated a decrease of the maximum energy of flares and increase of the rotation period with age with a decrease of the total area of spots from 10 to a few percent and found that the frequencies of superflares on young stars with rotation periods of 1-3 days were one hundred times higher than those on old slow rotators, but the spectral index of the power-law flare energy distribution was approximately the same. No exoplanets were found on the considered stars with superflares, which means that there is no need in them for such strong flares.

From the Kepler high time resolution data, Jackman et al. (2021) detected 4430 flares with energies up to $1.5 \cdot 10^{35}$ erg on 403 stars. 515 flares were found in the proximity to other sources or in binary systems, and the frequencies of flares turned out to be regularly higher than those for weak components.

In the early 2010s, the principally new instrument of panorama photometry Evryscope[1] was put into operation. Over 1344 hours of observations of Prox Cen between January 2016 and March 2018 23 flares were recorded with bolometric energies from $10^{30.6}$ to $10^{33.5}$ erg. The flare energy spectrum was represented by a power-law function with the spectral index −1.22, and the strongest flare of 18 March 2016 at 8:32 UT, which is $2.5^m$ stronger than others, was noted by Howard et al. (2018) as the first superflare of this star.

Later, Howard et al. (2019) reported on the Evryscope observations of 285 southern sky stars which were identified as flare ones within TESS. Furthermore, 576 flares were recorded with the average energy $10^{34}$ erg, 8 of them with the amplitude more than $3^m$, and a noticeable increase of flare activity was detected near the spectral type M4, i.e., on the boundary of the fully convective structures. Howard et al. found a decrease in the average energy of flares for later spectral types, represented the average energy index of flares as a function of the spectral type of the star, and measured the frequency of superflares for each spectral type, which turned out to be maximal for late K and early M dwarfs. The flares of maximum amplitude were found to increase toward later types and the frequency of superflares decreased for older stars at high galactic latitudes. The strongest from the recorded flares with the amplitude $5.6^m$ and energy $10^{36.2}$ erg occurred on the star of the age of 40 million years in the Tuc-Hor cluster.

In the late 2015, the ground-based panorama photometric system NGTS[2] was put into operation. Its first scientific result was a record of two superflares of 17 December 2015 and 3 January 2016 on the bright G8 star NGTS J030834.9-211322 with bolometric energies of 5.4 and $2.6 \cdot 10^{34}$ erg, an estimate of a rotation period of 59 hours from photometric modulation, and acquisition of evidence for its differential rotation (Jackman et al., 2018); the fast stellar rotation causes the saturation of its X-ray emission measured by ROSAT.

On 13 August 2017, using the NGTS system, Jackman et al. (2019) recorded a flare of the L2.5 dwarf ULAS J224940.13-011236.9 with the effective temperature 1930 ± 100 K, which based on its proper motions was attributed to the thin disk of the Galaxy. The flare with the amplitude $\Delta V \sim 10^m$ and bolometric energy of $2 \cdot 10^{33}$ erg lasted for 9.5 minutes. This is the sixth flare recorded on L dwarfs.

In April 2018, the TESS satellite was launched into orbit with the aim of recording the passage of planets less than Neptune across the disk of bright not large stars. Günther et al. (2020) studied the observations derived with it throughout two months with 2-min exposures and from the light curves of more than 24800 stars, using special software, detected 3247 flares with bolometric energies from $10^{31}$ to $10^{38.7}$ erg on 763 stars, from which 453 were early and 179 late M dwarfs. They considered frequencies and energies of flares as functions of the spectral type and rotation period and found that 60% of rapidly rotating M dwarfs were flare ones, whereas among M dwarfs with unknown rotation periods only 10% flared; stars with high frequency of flares had increased amplitudes and durations of flares, but amplitudes were not dependent on the rotation period. The flares were recorded on 30% of middle and late M dwarfs, on 5% of early dwarfs and less than 1% on F, G, and K stars.

---

[1] Evryscope was designed as follows: on the 1.8-m hemisphere made of reinforced alunimium installed on the German equatorial mount there are twenty-seven 61-mm individual telescopes with a scaling in the focal plane of 13.3″/pixel. 50% of light is accumulated by the telescopes in a circle of 2 pixels in diameter in inner 2/3 of the field of view and of 2–4 pixels in outer 1/3 of the field of view. The high-aperture correcting lenses and filter wheels are mounted on 5 positions in the telescopes. The total field of view is 8660 deg$^2$, i.e., about 42% of the celestial hemisphere, recording simultaneously about 15 million sources. Due to the high-sensitive receivers at 2-min exposures the photometry accuracy of stars brighter than $12^m$ is 1%, the limiting magnitude in the V band is $16.4^m$, $18.2^m$ at the 1-h exposure, and $19.0^m$ at the 6-h exposure (Law et al., 2015).

[2] The ground-based system Next Generation Transit Survey consists of twelve 20-cm telescopes (F/2.8) with a total field of view of 96 deg$^2$, operates in the range of 5200–8900 Å with a time resolution of 10 s and records all the stars brighter than $16^m$.

Later, within the TESS experiment Zhan et al. (2019) studied the short-period variability of stars with an effective temperature of lower than 4000 K and among 371 M dwarfs detected 10 rapidly rotating objects with rotation periods of less than 1 day and with fairly structured light curves; the structured details existed for days on the background of periodicity of light curves for week. They considered several models of this structuring and the most probable was the rotation of a spotted star inside the dust ring surrounding the star at a distance of tenths of its radius. 17 M dwarfs showed the axial rotation periods of less than 4 hours and the fastest rotator was with a period of 1.63 hours. Zhan et al. recorded 32 flares with amplitudes > 2 and just after the strongest of them on TIC 206544316 with $A$ = 4 and $E_{\mathrm{bol}} = 1.7 \cdot 10^{34}$ erg throughout several subsequent revolutions the maximum increased and one of the short structures changed.

Tu et al. (2020) analyzed the TESS observations over the first year of operation with 2-min exposures of 25734 solar-type stars, confirmed 1216 superflares on 400 stars and detected their increased frequency as compared to the Kepler results.

Then from the data of the second observational year within TESS Tu et al. (2021) studied more than 22500 solar-type stars and detected 1272 superflares on 311 of them. The spectral index of the power-law flare energy distribution was -1.76 ± 0.11 at the dependence of the duration of superflares on energy as $E^{0.42 \pm 0.01}$, which coincides with the situation in solar flares. The chromospheric activity parameter S was derived with the LAMOST telescope for 7454 stars and indicated that the Sun was less active than these stars with superflares. The hotter stars of the sample flared rarer than the less hot ones. Presumably, the superflare energy saturation was at a level of $10^{36}$ erg, while on the star with the most energetic superflare TIC93277807, exceeding this limit more than by an order of magnitude, another mechanism could work.

The spectral index of the power-law flare energy distribution accounted for 2.16 ± 0.10, which is somewhat higher than on the Sun but in agreement with the Kepler data. The star TIC43472154 showed about 200 superflares per year. The duration of flares correlated with their power-law energy of 0.42 ± 0.01, which is somewhat higher than 1/3 from the predictions of the magnetic reconnection theory.

From the Kepler data Ilin et al. (2021a) studied flares based on 3435 80-day light curves of 2111 members of the open clusters Pleiades, Hyades, Praesepe, Ruprecht 147, and M67. In these clusters of fairly different ages, they confirmed 3844 flares on G-M stars, whose energy distributions had a power-law form with spectral indices of 1.84-2.39. They found that flare activity decreased from middle M stars toward G stars and from ZAMS stars toward the solar-age stars, confirmed a decrease in flare frequencies with age and it was higher for more massive stars. They also found values of mass and rotation rate, higher than which the flare activity was no longer saturated.

Analyzing TESS observations of four strong and prolonged flares on fully convective dwarfs, from their brightness modulation by a fast stellar rotation Ilin et al. (2021b) estimated their latitudes between 55 and 81°, which, in their opinion, makes the influence of such stellar activity on exoplanets unsubstantial.

Howard and MacGregor (2022) considered 3792 flares on 226 low-mass stars recorded with TESS with 20-s resolution and, analyzing 440 of them with energy of more than 10>33 erg, concluded on the absence of flares with light curves without a fine structure and quasiperiodic pulsations.

Using the 1-meter Kiso Schmidt telescope of the University of Tokyo equipped with the Tomo-e Gozen camera[1], Aizawa et al. (2022) over 40 hours of observations detected 22 flares from M3-M5 dwarfs with a 0.5 s cadence. The amplitudes of flares are in the range from 0.5 to 20, the rise time is from 5 to 100 seconds, the bolometric energies are estimated to be $10^{31}$–$10^{34}$ erg. Ninety percent of stars with detected flares show Hα emission, and the occurrence rate of flares from these stars is 0.7 events per day.

Pietras et al. (2022) performed a statistics of the stellar flares from three-year observations with the TESS satellite (the first 39 sectors) with a 2-min cadence. Using the elaborated software to make an automatic search for flares and faculae from the light curves, they considered 330,000 stars and detected over 25,000 stars showing flare activity with the total number of more than 140,000 flares with energies in the range between $10^{31}$ and $10^{36}$ erg, i.e. the majority of the recorded events are superflares. About 7.7% of the analyzed objects are flaring stars, but among M dwarfs this fraction reaches 50%. The maximum of the flares duration distribution is 50 minutes, the mean ignition time is less than 10 minutes, and the most prolonged flares lasted for a few hours. The secondary peaks of the light curves have been referred to by the authors due to the heating of the photosphere by nonthermal electrons. The maximum in the size distribution of flares, according to various estimates, amounts to 0.2–0.3% of the stellar surface. The spectral index in the energy distribution of flares is estimated to be 1.7 and 1.5. From the energy and duration of flares, the magnetic field strength is estimated to be from 10 to 200 G, and the lengths of flaring coronal loops are between $10^{10}$ and $2 \cdot 10^{11}$ cm.

G. Out-of-flare brightness variations. Many observers noted low-amplitude brightness variations in flare stars out of flares (Roques, 1958; Oskanian, 1964). Some of them were mentioned above. The most detailed research of these phenomena was done by Roizman (1983, 1984) and Rojzman and Kabichev (1985). During photoelectric monitoring of EV Lac in the U band with regular measurements for comparison and check stars, they found small variations of brightness on EV Lac at time intervals of hours and with amplitudes of up to $0.3^m$. With a lower amplitude this effect was observed in the B band. In many cases, such an increase in brightness preceded a flare or occurred on the nights when no flares were recorded (see Fig. 62), which enabled a conclusion on the similarity of these stellar phenomena to preflare slow brightening on the Sun. However, this analogy does not shed light on the physical aspects of the brightness variations. The hypothesis of the passage of facular fields across the disk seems to be the most natural.

Later, similar results were obtained by Mahmoud (1993b) in observations of EV Lac in the photometric B band.

From the power spectra of radiation of BY Dra, Chugainov and Lovkaya (1988) singled out oscillations with periods of about 188 and 100 min with the amplitude within $0.002^m$–$0.005^m$ and periods of 10–59 min with twice less amplitude, while from the direct brightness measurements of V 1285 Aql in the U band oscillations with periods from 10 min to 1–2 h with amplitudes up to $0.1^m$ (Chugainov and Lovkaya, 1992). Finally, owing to high-accuracy radial velocity measurements, Bouchy and Carrier (2002) detected p-modes of oscillations of the $\alpha$ Cen A star at frequencies on the interval of 1.8–2.9 mHz, while Carrier and Bourban

[1] The 1.05 m Kiso Schmidt telescope F/3 with the Tomo-e Gozen camera has a field of view of 20.7 deg$^2$, one pixel of the matrix corresponds to 1.19″. During the observations in the range of 3000–10000 Å, the limiting magnitude is $17^m$.

(2003) found oscillations of $\alpha$ Cen B at frequencies on the interval of 3–4.6 mHz, i.e. analogs of solar 5-min oscillations.

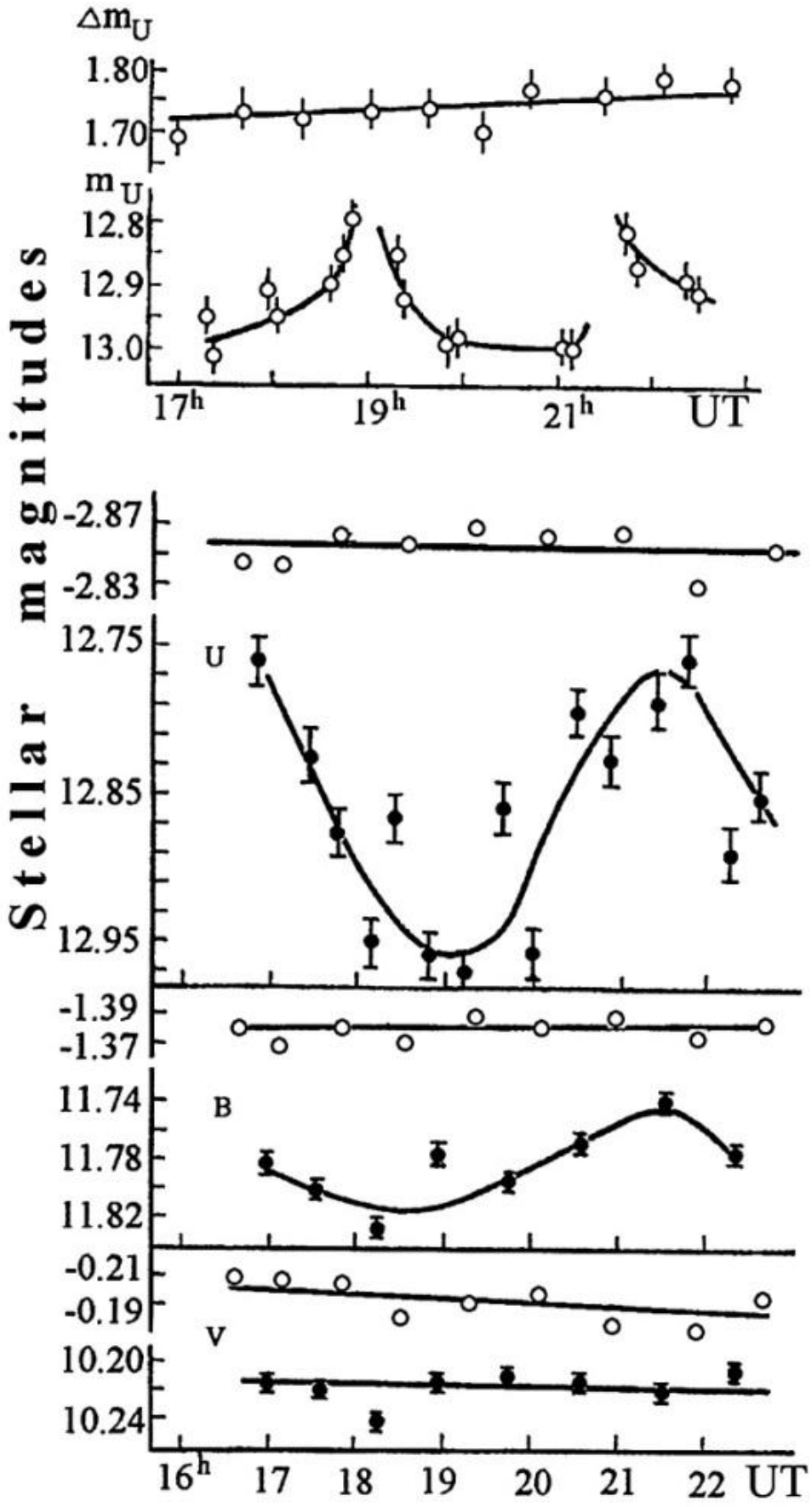

Fig. 62. Out-of-flare brightness variations on EV Lac: *top diagram* shows preflare increase and two descending branches of flares of 22 August 1980 in the U band (Roizman, 1983); *bottom diagram* illustrates the behavior of the star on 14 August 1981 in the UBV bands (Rojzman, 1984); *empty circles* above the light curves show the differences of stellar brightness of comparison and check stars in the appropriate band

E. Flare energy spectrum and the SDSS project. Within the framework of the SDSS project, Hilton (2011) carried out about one thousand hours of photometric observations of 39 M0–M8 dwarfs including active and inactive stars and recorded 239 flares. Having selected 137 events of them, which were called by him as classic flares with impulsive burning, fast initial and subsequent exponential decay, he constructed their energy spectra, found correlations between the total energy of flares and times of their burning and decay, and detected that the secondary bursts took place during the first 40% of the total duration of the event. Using the original quantitative model of flare light curves, their energy flares, and the Galactic stellar population model Galfast, Hilton constructed the Galactic flare rate model, which allows one to make qualitative predictions of frequencies and energies of flares on M dwarfs over the whole Galaxy and, particularly, to obtain the found from observations dependence of activity of such stars on the height above the Galactic plane (see Fig. 4). This model predicts several flares of $\Delta U > 0.1^m$ in every 9.6 $deg^2$ full-frame exposure at the galactic latitude $b = 20°$ and ~ 0.1 of a flare at $b = 80°$.

**2.4.4.2. Colorimetry.** Sharp bluing of red dwarfs during strong flares was revealed in visual observations (Oskanjan, 1953). The rather "blue" color of the flare on the background of "red" photospheres results in a fast growth of the flare amplitude toward shorter wavelengths and, hence, to a noticeable increase of frequencies of recorded flares. The first photoelectric colorimetric results were obtained by Johnson and Mitchell (1958) in observations of the flare on the star HII 1306 in the Pleiades — $\Delta U > 3.5^m$, $\Delta B \sim 1.6^m$, and $\Delta V \sim 0.7^m$ — and by Abell (1959) during the flare on AD Leo of 9 March 1959 — $\Delta U \sim 1.5^m$, $\Delta B \sim 0.3^m$, and $\Delta V \sim 0.1^m$. Noticeable differences in the amplitudes stimulated the use of colorimetric analysis in flare diagnostics. Basically, the analysis can be performed within two alternative models: one assumes that during the flare process the radiating body changes on the whole, the other assumes that a new source is added to the initial one that remains invariable. The first model corresponds, for example, to pulsations of cepheids. On the other hand, spectral observations of flares showed that even during the strongest flares molecular bands were kept in the red region of the spectrum, i.e., flares are local phenomena, not covering the whole stellar surface, and the additive model can be used to analyze them. Apparently, the color characteristics of intrinsic flare radiation were determined first by Abell (1959): $U{-}B \sim -1.4^m$ and $B{-}V \sim 0.1^m$. A month later from the flare on AD Leo of 13 April 1959 Engelkemeir (1959) determined intrinsic flare radiation in the B and V bands. Assuming that the flare was caused by a hot spot on the stellar surface, from the color indices of main-sequence stars he estimated the effective temperature of the spot at the flare maximum as 11000 K and its area as 0.11% of the stellar disk. Chugainov (1965) found a systematic growth of the color index of B−V flare radiation from $-0.2^m$ to $0.3^m$ as the flare decayed. Kunkel (1967) tracked the drift of flares in the two-color diagram (U−B, B−V) and associated it with the increasing contribution of the "burnt" photosphere to the total radiation of the flare. (The theory of such a "burn" in the approximation of small disturbance was developed by Grinin (1973)). The first numerous colorimetry data on flares on red dwarfs were published by Kunkel (1970), Cristaldi and Rodonò (1973), Moffett (1974), Lacy et al. (1976). Later results are summed up in the relations (55). One should bear in mind that a number of radiations of various nature are characterized by rather "blue" U−B color indices, sharply distinct from this parameter of red dwarfs. Therefore, flare tracks on two-color (U−B, B−V) diagrams, describing the drift of stars during flares, weakly depend on the specific mechanism of flare radiation, and colorimetric analysis based on the tracks is extremely uncertain. In other words, the analysis should be based of the intrinsic flare radiation rather than on the position of the star in the two-color diagram during the flare, otherwise erroneous results can be obtained. Thus, Gurzadian (1970), Moffett (1973), and Cristaldi and Rodonò (1975) compared observations with theoretical tracks "star + inverse Compton effect" and obtained satisfactory agreement. They thought it was an argument for the validity of the model. However, comparison of intrinsic colors of flares with the inverse Compton effect clearly proved the inconsistency of this model (Gershberg, 1978).

Colorimetric analysis disproved some other hypotheses on the nature of luminosity of stellar flares. Thus, Arakelian (1959) showed that proper colors of B−V and U−B flares on HII 1306 were much more "blue" than expected under synchrotron radiation in the wide range of the index of energy spectrum of relativistic electrons. Thus, synchrotron radiation cannot be considered as a basic mechanism of optical emission of flares on UV Cet-type stars. Later, this conclusion was made independently by Kunkel (1967). Klimishin (1969) estimated the color indices of a relaxed cloud of molecular hydrogen: U−B $\sim -1.0^m$ and B−V $\sim -0.5^m$, and this point on the two-color diagram is also rather far from the most probable position of stellar flares.

The first colorimetric models within the notions on fast occurrence and further relaxation of hot hydrogen structure above the photosphere of a cool star (Gershberg, 1964; Kunkel, 1967) did not yield a satisfactory presentation of observations (Cristaldi and Rodonò, 1975) and later were repeatedly revised. Strictly speaking, Kunkel (1967) made the first refinement by adding radiation of the heated photospheric section "burnt" by the flare to the luminosity of the hydrogen plasma.

Chugainov (1972a) observed three flares on EV Lac using a three-channel spectrocolorimeter of the Crimean Astrophysical Observatory with simultaneous recording of radiation in the ranges 3350–3650 Å, 4155–4280 Å, and 5120–5320 Å. Analysis of these data, which are analogous to UBV photometry with noticeably narrowed pass bands, suggests that flares at the brightness maxima do not lay on the lines of an absolutely blackbody or in the region of optically thin hydrogen plasma in the appropriate two-color diagram.

On 3 August 1975, EV Lac was monitored using the 188-cm Okayama reflector; observations were carried out with a five-channel spectrocolorimeter in the wavelength range of 3300–6005 Å in 400 and 800 Å wide bands. Analysis of two strong flares recorded by Kodaira et al. (1976) made it possible to conclude that over about half an hour during the stronger second flare with $\Delta m_{\mathrm{UV}} \sim 5.9^m$ and during several tens of seconds at the maximum of the first weaker flare with $\Delta m_{\mathrm{UV}} \sim 1.9^m$ the energy distribution in the flare radiation was very flat, as, approaching decay, it became "redder". Kodaira et al. noted that their results were consistent with the data of Chugainov (1972a) and identified a flat spectrum with $m_\lambda \sim$ const with the radiation of hot hydrogen plasma at temperatures above $10^5$ K.

Shmeleva and Syrovatsky (1973) showed that, depending on the rate of initial energy release, one of two temperature structures occurred in solar flares: with constant density (CDR) at very fast energy release or with constant pressure (CPR) at smoother burning. In this connection, developing the above concept of plasma heated to coronal temperatures at initial energy release as the basic component of stellar flares, Mullan (1976b) calculated U−B and B−V color indices for the structures. He concluded that in both cases the designed parameters were close enough to the observed parameters at brightness maxima of flares, though in CDR the basic contribution to optical emission was made by plasma with $T \sim 10^7$ K, whereas in CPR, it was by a plasma with $T \sim 20000$ K. Mullan attributed the sharp transition from fast decay to smooth decay with the reorganization of radiating plasma from CDR into CPR. He believed that the proposed model, as a whole, developed Andrews' scheme (1965) with prevailing continuous radiation at the beginning of the flare decay and line radiation, during slow decay, and explained two components of flares following Kunkel (1967) and Moffett and Bopp (1976) (see below).

On the basis of the above observations of flares on EV Lac of 3 August 1975, Kodaira (1977) developed the Andrews–Mullan idea about the decisive role of nonstationary high-temperature plasma in flares and proposed the following model: as a result of primary energy release, a structure of stellar size emerges with $n_e \sim 5 \cdot 10^{10}$ cm$^{-3}$ and $T_e \sim 10^8$ K, at the base of the structure there is a smaller volume of denser and cooler plasma with $n_e \sim 5 \cdot 10^{13}$ cm$^{-3}$ and $T_e \sim 10^5$ K; the latter is the source of optical and ultraviolet emission and the former is the energy pool, which relaxes in X-rays and heats up the latter due to the heat conductivity.

Further, Mullan and Tarter (1977) considered the influence of flare X-ray emission on its position in the two-color (U−B, B−V) diagram and found that the X-ray quanta falling on a star degraded as a result of multiple scattering and thus up to 10% of such quanta leaving the star fell into the UBVR bands. This resulted in a shift of the flare in the diagram to the side corresponding to the observed drift. Thus, instead of Kunkel's model of "hydrogen recombination emission + burn of photosphere" Mullan and Tarter proposed "luminosity of Shmeleva–Syrovatsky hydrogen structure + reflection of flare X-ray flux". Within this model, Schneeberger et al. (1979) presented the light curves of two weak flares on AD Leo recorded by them, and Worden et al. (1984) adjusted the initial decay of the flare on YZ CMi of 9 February 1979 with $\Delta U = 1.5^m$ for parameters close to those of a solar flare: $n_e \sim 10^{13-14}$ cm$^{-3}$ and $T_e = 20000$ K.

Using UBV observations, Lukatskaya (1972, 1977) suggested studying stellar flares based on their positions on the plane ($\Delta U/\Delta B, \Delta V/\Delta B$); however, she detected no color differences of flares at the brightness maximum and decay phase, and this approach has not been widespread.

* * *

Using a five-channel UBVRI photometer-polarimeter designed by Piirola (1975), Panov et al. (1988) in observations of flares on EV Lac established systematic variations of U−B and B−V color indices of flare radiation during the development of flares. But the most complete colorimetric analysis of stellar flares was executed using a similar photometer in the Crimea during a long-term research of EV Lac. Figure 63 illustrates this basically comprehensive colorimetric analysis.

The left panel of Fig. 63 shows the light curve of the flare on EV Lac of 10 September 1993; by the numbered vertical straight lines the points of this curve are denoted, as well as appropriate points on the color index curves located lower, which were involved in further analysis. In four two-color diagrams rightward, which exhaust the colorimetric information from the UBVRI data, these points are marked with rectangles whose sizes correspond to the accuracy of determination of color indices at the appropriate moments of time. Independent color indices of the UBVRI system were previously calculated for a set of known sources, and the results of these calculations are displayed in these two diagrams by the lines indicated by Roman numerals: curve I — absolute blackbody, curves II and III — emission of optically thin (in the Balmer continuum) hydrogen plasma at the temperature 10000 K, electron density of $10^{12}$ to $10^{14}$ cm$^{-3}$, and probabilities of the escape of Ly$_\alpha$ quanta from the medium within $10^{-5}$ and $10^{-8}$; curves IV and V — emission of optically thick (in the Balmer continuum) hydrogen plasma at the temperatures of 10000 to 15000 K; curves VI, VII, and VIII — radiation of the upper layers of red dwarfs heated by the fluxes of fast particles (Abranin et al., 1998a). Figure shows that in the (U–B, B–V) diagram flare radiation at the moment of maximum (rectangle 3) is close enough to the blackbody radiation curve, but in other diagrams rectangles 3 are noticeably shifted from this curve toward hydrogen plasma radiation. Preflare radiation (rectangles 1 and 2) in the (U–B, B–V) diagram is also located on the blackbody curve but at higher temperature, and is also shifted toward hydrogen plasma localization in other diagrams. One can see the selected four points on the descending branch of the flare (4,5,6, and 7) to be already shifted from the blackbody curve in the (U–B, B–V) diagram and, as the (B–V, V–R) and (U–B, V–I) diagrams show, the hydrogen plasma emission dominates at that time. Thus, this analysis shows that at none of the stages of flare development none of the considered radiation mechanisms by itself can explain the observed colorimetric characteristics of flares in the whole range of the UBVRI system, and it is required to invoke some combinations of these mechanisms: the most probable seems to be a combination of short-lived blackbody radiation near the maximum and more prolonged hydrogen plasma luminosity.

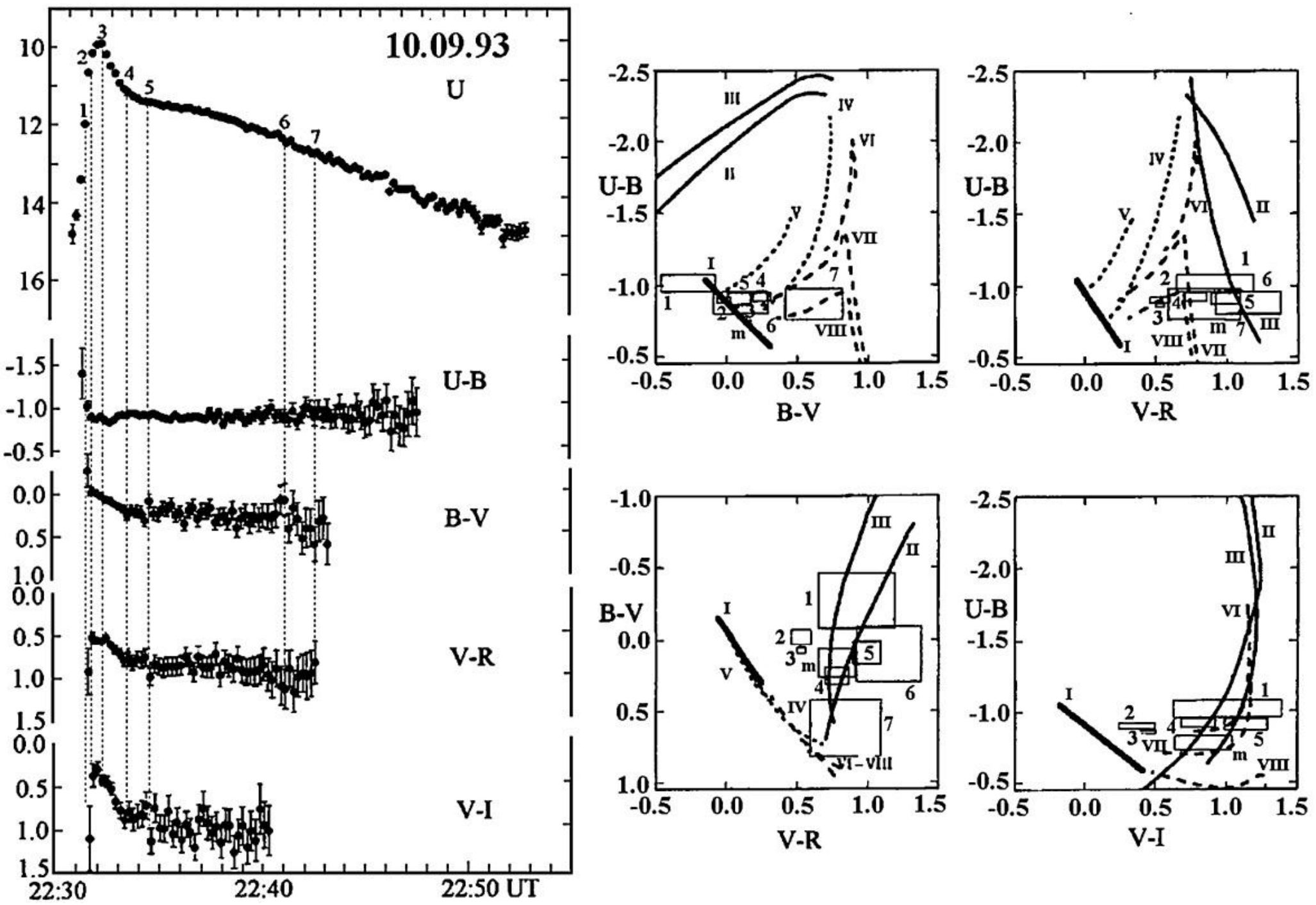

Fig. 63. The light curve of the flare on EV Lac of 10 September 1993, curves of color indices in the UBVRI system *(left)* and two-color diagrams for studying the nature of flare radiation *(right)* (Abranin et al., 1998a)

Figures similar to Fig. 63 were based on the Crimean UBVRI observations of flares on EV Lac, which yielded similar results (Gershberg et al., 1991a, 1993; Alekseev et al., 1994; Abdul-Aziz et al., 1995; Abranin et al., 1998b). Here it is worth making two remarks. First, the conclusion about insufficiency of one of the mechanisms for modeling optical flares was made by Bopp and Moffett (1973) based on spectral and photometric observations of flares on UV Cet. Second, the fact that one of the strongest stellar flares on G 102-21 was located, according to Pagano et al. (1995), in the region of radiation of hydrogen plasma in the (U−B, B−V) diagram supports the validity of the method of colorimetric analysis of stellar flares.

Taking flare radiation in the brightness maximum as blackbody one, from the color indices U−B and B−V of flare radiation Alekseev and Gershberg (1997a) estimated the temperatures of nine strong flares on EV Lac observed in the Crimea with amplitudes $\Delta U > 1.8^m$ at maxima to be within 10000 and 25000 K. Furthermore, based on the known stellar luminosity and found blackbody temperatures $T_{bb}$ of flares, their sizes were estimated from the relation

$$S = L_U (10^{0.4\Delta U} - 1) / \pi B_U (T_{bb})\ cm^2 = \qquad (64)$$

$$=1.7\cdot 10^{33}(10^{0.4\Delta U}-1)/\int_U d\lambda/\lambda\cdot[\exp(1.4388/\lambda T_{bb})-1]cm^2$$

where $B(T)$ is the Planck function; these turned out to be in the range of 1.1 to 25 in units of $10^{18}$ cm$^2$, i.e., from 0.06 to 1.3% of the stellar disk. Table 21 lists all the numerical characteristics of these nine flares. Note that, according to Neidig (1989), the typical size of a solar white-light flare is $2\cdot 10^{17}$ cm$^2$.

* * *

Application of the digital filtering technique for photometric data to the flare on EV Lac of 11 September 1998 recorded in the course of operation of the Synchronous Network of distant Telescopes made it possible to detect the rapidly varying color characteristics of flares (Zhilyaev et al., 2003) and resulted in appreciable clarification of the results described above.

(U−B) – (B−V) diagram of EV Lac of 15 October 1996, AZT-11, CrAO

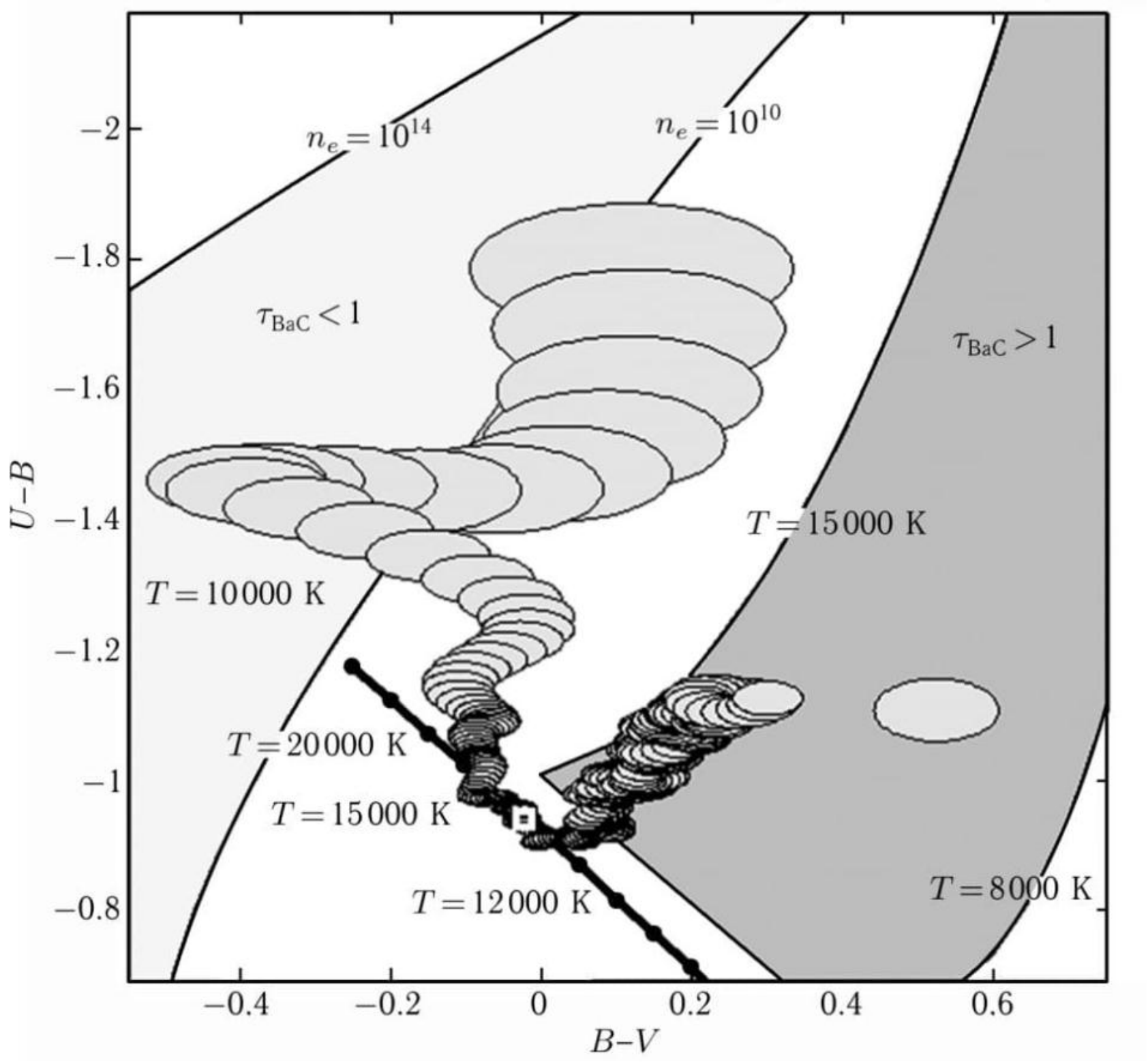

Fig. 64. A track of the flare on EV Lac of 15 October 1996 in the (U–B, B–V) diagram. The white square denotes the position of the flare maximum. The separately located ellipse rightward corresponds to the end of observations (Lovkaya, 2014)

Thus, Fig. 64 represents a track of the flare on EV Lac of 15 October 1996 recorded with a five-channel photometer-polarimeter in the Crimea in the (U–B, B–V) diagram, which is identical to the appropriate diagram in Fig. 63. The flare track is formed by a sequence of ellipses within the uncertainties of ±2$\sigma$ with a time resolution of 1 s. The figure shows that at the initial phase the flare radiates as the optically thin (in the Balmer continuum) hydrogen plasma, then at the moment of brightness maximum and over a minute about this moment it radiates as an absolute blackbody, cooling down approximately from 20000 to 12000 K, and then drifts in the region of the optically thick plasma in the Balmer continuum. The observations of about a dozen strong flares on EV Lac и AD Leo have been reduced by this technique so far (Zhilyaev et al., 2007; Lovkaya, 2012, 2013, 2014), the results on the six of them are given in Table 22. All the flare tracks start in the region $\tau_{BaC} < 1$ and pass the region of absolute blackbody at the moment of maximum. A flare, as a rule, falls to the region $\tau_{BaC} > 1$ at the decay phase, but in some cases there is a significant divergence and a significant increase in the ellipses of uncertainties, presumably, due to nonmonotonicity of the decay branch of the flare. As to the acquired values of temperatures and sizes of flares at maxima, then the data of Tables 21 and 22 are in agreement. Here it is worth noting that the model of the blackbody luminosity of flares was repeatedly used in spectral investigations (see, for example, publications by Mochnacki and Zirin (1980), Kahler et al., (1982), Giampapa (1983), Pettersen et al. (1986b), de Jager et al. (1989), Katsova and Livshits (1991), Hawley and Fisher (1992), Montes et al. (1999), Hawley et al. (2003), Kowalski et al. (2010)). However, the technique of observations with the Synchronous Network of distant Telescopes, using the digital data filtering of the derived data, made it possible to reliably and quite precisely determine the temperature of flares at brightness maxima and to first obtain estimates of their cooling rate near maxima. This cooling process seems to result in the initial fast decay of the optical flare and after the decrease of the optical thickness of flare plasma and disappearance of conditions for the blackbody luminosity the decisive role is taken on by its recombination radiation.

Table 21. Characteristics of nine strong flares on EV Lac (Alekseev and Gershberg, 1997a)

| Date | Moments of brightness maxima | Duration of rise/decay, min | $P_U$, min | $\Delta U$, $^m$ | Color indices of flare radiation at the moment of brightness maximum | | | | $T_{abb}$, K | $S$, $10^{18}$ cm$^2$ |
|---|---|---|---|---|---|---|---|---|---|---|
| | | | | | U–B, *m* | B–V, *m* | V–R, *m* | V–I, *m* | | |
| 11.09.86 | $19^h18^m53^s$ | 10.5/55.0 | 28.4 | 2.3 | –0.78±0.02 | 0.06±0.04 | 0.49±0.06 | 0.37±0.11 | 11000 | 6.9 |
| 13.09.86 | 25 50 49 | 5.3/5.5 | 13.6 | 2.8 | –0.69±0.02 | –0.10±0.04 | 0.26±0.06 | 1.87±0.13 | 10000 | 15 |
| 26.08.90 | 18 22 14 | 1.1/33 | 23 | 2.4 | –0.85±0.02 | 0.00±0.03 | 0.31±0.06 | 0.36±0.14 | 12500 | 5.6 |
| 28.08.90 | 24 23 26 | 0.6/48 | 14 | 2.0 | –1.11±0.03 | 0.07±0.06 | 0.75±0.09 | 1.25±0.13 | 25000 | 1.1 |
| 29.08.90 | 23 39 21 | 2/90 | 41 | 2.0 | –1.07±0.02 | 0.09±0.05 | 0.65±0.08 | 0.92±0.13 | 20000 | 1.6 |
| 14.09.91 | 25 29 06 | 2.9/29 | 154 | 3.7 | –0.87±0.01 | 0.07±0.01 | 0.37±0.02 | 0.97±0.02 | 11500 | 25 |
| 10.09.93 | 22 32 30 | 3.0/22 | 53.3 | 3.1 | –0.86±0.01 | 0.07±0.02 | 0.53±0.03 | 0.45±0.06 | 11500 | 14 |
| 31.08.94 | 23 23 37 | 0.9/2.2 | 10.5 | 1.8 | –0.72±0.02 | 0.06±0.05 | 0.50±0.09 | 0.32±0.15 | 10500 | 4.6 |
| 03.09.95 | 18 48 59 | 0.9/9.0 | 9.2 | 2.1 | –0.91±0.03 | –0.02±0.06 | 0.40±0.10 | 0.49±0.15 | 14000 | 3.1 |
| | | Average | | 2.5 | –0.87±0.05 | 0.03±0.02 | 0.47±0.05 | 0.78±0.18 | 14000 | 8.5 |

Table 22. Parameters of flares on AD Leo and EV Lac (Lovkaya, 2014)

| | Date | $t_{max}$, UT | $\Delta t$, min | ΔU, $^m$ | U–B, $^m$ | B–V, $^m$ | $T_{abb}$, K | $S$, cm$^2$ |
|---|---|---|---|---|---|---|---|---|
| AD Leo | 04.02.2003 | 23:17:58 | 5 | 1.65 | –0.95 ± 0.03 | –0.01 ± ± 0.04 | 14000 | $2.1 \cdot 10^{18}$ |
| | | 23:29:41 | 5 | ~ 0.2 | | | 14000* | $0.1 \cdot 10^{18}$* |
| | | 23:41:20 | > 8 | 1.76 | –0.84 ± ± 0.03 | –0.03 ± ± 0.04 | 13000 | $3.0 \cdot 10^{18}$ |
| EV Lac | 12.09.2004 | 22:53 | 2.5 | 1.49 | –1.11 ± ± 0.04 | –0.09 ± ± 0.07 | 18500 | 1.1% |
| | 14.09.2004 | 20:31 | 0.7 | 2.1 | –1.06 ± ± 0.06 | –0.10 ± ± 0.10 | 18500 | 1.3% |
| | 15.10.1996 | | 25 | 3.73 | –0.93 ± ± 0.01 | –0.02 ± ± 0.01 | 15000 | $5.1 \cdot 10^{18}$ |
| | 10.10.1998 | | > 20 | 2.72 | –0.97 ± ± 0.01 | –0.07 ± ± 0.01 | 16000 | $1.6 \cdot 10^{18}$ |

* The temperature and area of the weak flare on AD Leo, which occurred between two flares of appreciable amplitude, were estimated from the following consideration: if one supposes that the flare thermodynamics depends weakly on their power and the temperature regime of the weak flare on AD Leo is the same as in strong flares, then the area of this weak flare is approximately 20 times less than that of strong flares.

The results acquired by Lovkaya (2014) have recently been confirmed by Loyd et al. (2018). In the course of spectroscopic HST/COS observations of the M2 dwarf 2MASS J02365171-5203036 they recorded a strong three-peak flare on 9 August 2017, whose maxima showed blackbody radiation at a temperature of 16000–15500 K and a filling factor of about 0.01%; 3 minutes later the radiation temperature decreased up to 14000 K.

* * *

Let us return to Fig. 59 and relation (63).

The presented above data on the flare blackbody radiation at the moments of their brightness maxima allow one to write for their bolometric luminosity

$$L \sim T^4$$

and

$$dL/dt \sim 4T^3(dT/dt) \sim L^{3/4}(dT/dt). \quad (65)$$

Since at the considered temperatures the U and B bands cover a significant part of blackbody radiation, then relation (65) should be valid enough for these bands. Furthermore, since in the right side of (65), the first multiplier is basic, then (65) practically coincides with (63) (Gershberg, 2014).

Later, the colorimetric observations of flares were carried out with more improved equipment by the team of Susan Hawley, observing with the high-speed triple-beam camera ULTRACAM mounted on the 4.2-meter William Herschel Telescope in La Palma and on the 3.6-meter NTT at La Silla (Kowalski et al., 2011, 2013, 2016). The channels of the camera recorded radiation with a subsecond resolution in the near ultraviolet ($\lambda$ 3500 Å) and in the blue ($\lambda$ 4100 Å) and red ($\lambda$ 6010 Å) regions of the optical spectrum in the bands of 50 to 120 Å. On five red dwarfs YZ CMi, Prox Cen, EQ Peg A, Gl 644 AB, and AD Leo in 2008–2012 they recorded more than one hundred flares and selected 20 of them with a duration at the FWHM level of 11 s to 5.5 min for qualitative analysis. Two middle-strength flares on YZ CMi of 14 January 2012 at 2:59 and 4:32 UT were recorded simultaneously with the 3.5-meter telescope at the Apache Point Observatory and served for confident reference of the photometric data derived with ULTRACAM to the low-dispersion spectra. The ratios of the flux in the ultraviolet range to the flux in the blue region were used to estimate the Balmer jump, and the ratios of two fluxes in the optical region — to estimate the color temperature of the optical continuum. The two-color diagrams constructed from these ratios detected an inverse correlation between the ratios, which corresponds to the known property of the recombination hydrogen spectrum: the Balmer jump decreases with increasing temperature. And divergence in “colors” of different flares was attributed by Kowalski et al. due to the different heat rates. Two indicated flares — of 14 January 2012 and the strongest of the recorded on this star of 13 January 2012 at 22:44 UT with $\Delta U > 5^m$ — were analyzed in more detail. To interpret the acquired measurements, Kowalski et al. developed an one-dimensional radiative-hydrodynamic model within which they represented the obtained data on flares as on the luminosity of dense chromospheric condensations at the temperatures of 12000–13000 and at variable optical thickness in the Balmer continuum, which were heated by fluxes of nonthermal high-energy electrons with variable hardness and power of an order of $10^{13}$ erg · cm$^{-2}$ · s$^{-1}$. The results of further analysis of this strongest flare on YZ CMi were reported by Kowalski et al. (2018) at the Cool Stars 20 conference. Supposing that the ratio of fluxes in blue and red rays corresponds to the blackbody radiation, they estimated the color temperature of the flare in the optical range at different phases of its development. A rise of the flare brightness and its fast decay after maximum were without a visible fine structure, whereas the secondary brightening on 50–180 s after maximum showed such a structure.

**2.4.4.3. Polarimetry.** Polarimetric observations of flares can provide meaningful astrophysical information only if the full cycle of measurements of polarization parameters is much shorter than the characteristic time of brightness variation and the telescope used is so large for the statistical errors of the measured flux to be much less than the measured values (Efimov, 1970). Unfortunately, many initial polarimetric observations conducted in the 1960s–1970s do not meet these requirements or do not contain exhaustive analysis of the achieved accuracy.

The first confident polarimetric results on stellar flares were obtained by Efimov and Shakhovskoy (1972) in observations with the Shajn telescope in the Crimea: they measured polarization parameters during the flare on EV Lac on 17 August 1969 in three points on the ascending branch of the light curve, at maximum and in 21 points on the descending branch. They found that flare radiation in the B band was not polarized to the accuracy of the measurement errors of ±0.5% caused mainly by the statistics of quanta of the recorded flux. In other words, these measurements suggested that the polarization of stellar radiation during the flare was the same as during its absence. Similar conclusions were made by Karpen et al. (1977) based on the Crimean observations of the flare on YZ CMi of 2 December 1975 in the V band, by Pettersen and Hsu (1981) from polarimetric observations of the flare on AD Leo of 29 October 1979 in the U band using the 76-cm telescope of the McDonald Observatory, Eritsian (1978) from observations of several flares on EV Lac and AD Leo from Byurakan, and Tuominen et al. (1989) from the Crimean observations of the YY Gem system on 6 March 1988. But de Jager et al. (1986), based on the U band observations from the Crimea, suspected that the level of polarization of radiation of the flare on BY Dra of 24 September 1984 was $3\sigma$. The most reliable polarimetric results for stellar flares were obtained by Shakhovskoy using the Shajn reflector in the Crimea in the course of the cooperative program of EV Lac examination in 1989 and 1991 (Alekseev et al., 1994; Berdugin et al., 1995). Figure 65 shows the light curve of the strong flare of 14 September 1991 in the U band, the Stokes parameters of stellar radiation during the flares, and the ratios of the values of these parameters to their root-mean-square errors. The analysis of these data suggests that the degree of polarization does not exceed 2% for a time resolution of 10 s and 1% for a time resolution of 50 s. In weaker flares, these limits are accordingly higher.

In observations of the flare star BD+26°730 with the 1.25-meter telescope of the Crimean Astrophysical Observatory with a UBVRI polarimeter Saar et al. (1994b) detected some changes in the degree and angle of polarization plane over about three hours in the U and B bands. Upon considering several models of this phenomenon, they decided that the most probable was the model concerned with the flare during which a flux of directed particles of $10^{9-10}$ erg/(cm$^2 \cdot$ s) was generated.

Slower variations of polarization of active dwarfs are usually associated with the nonuniform magnetic structure of the stellar surface (see, for example, the results of Alekseev (2000) for MS Ser).

Beskin et al. (2017) recently reported on a qualitatively new and principally important polarimetric result.

In the course of monitoring with the 6-meter telescope at the Special Astrophysical Observatory RAS of the binary system L 726-8 AB on 28 December 2008 a strong flare was recorded with a duration of about 25 min with $\Delta U \sim 3^m$ (see Fig. 66). This event seemed to occur on the secondary component, i.e., on UV Cet itself, which is half-magnitude weaker than the primary component, hence, the real amplitude of the flare star could be one stellar magnitude higher. Observations were carried out with the MANIA complex, which consists of the panorama photometer-polarimeter with a time resolution of 1 ms equipped with the double Wollaston prism and the multichannel system for recording photons Quantochron. Near the flare maximum, throughout about one minute, a dozen bursts with a duration of 0.6–1.2 s were recorded on the light curve, most of them had linear polarization exceeding 35%. Out of bursts, both in the flare and prior to it, the polarization was constantly absent. Comprehensive analysis of the possible physics of the phenomenon led to the conclusion that there seemed to be first detected short impulses of synchrotron radiation generating in the magnetic field with a strength of about 1.4 kG by the beams of accelerated electrons with energy of hundreds of megaelectron volts.

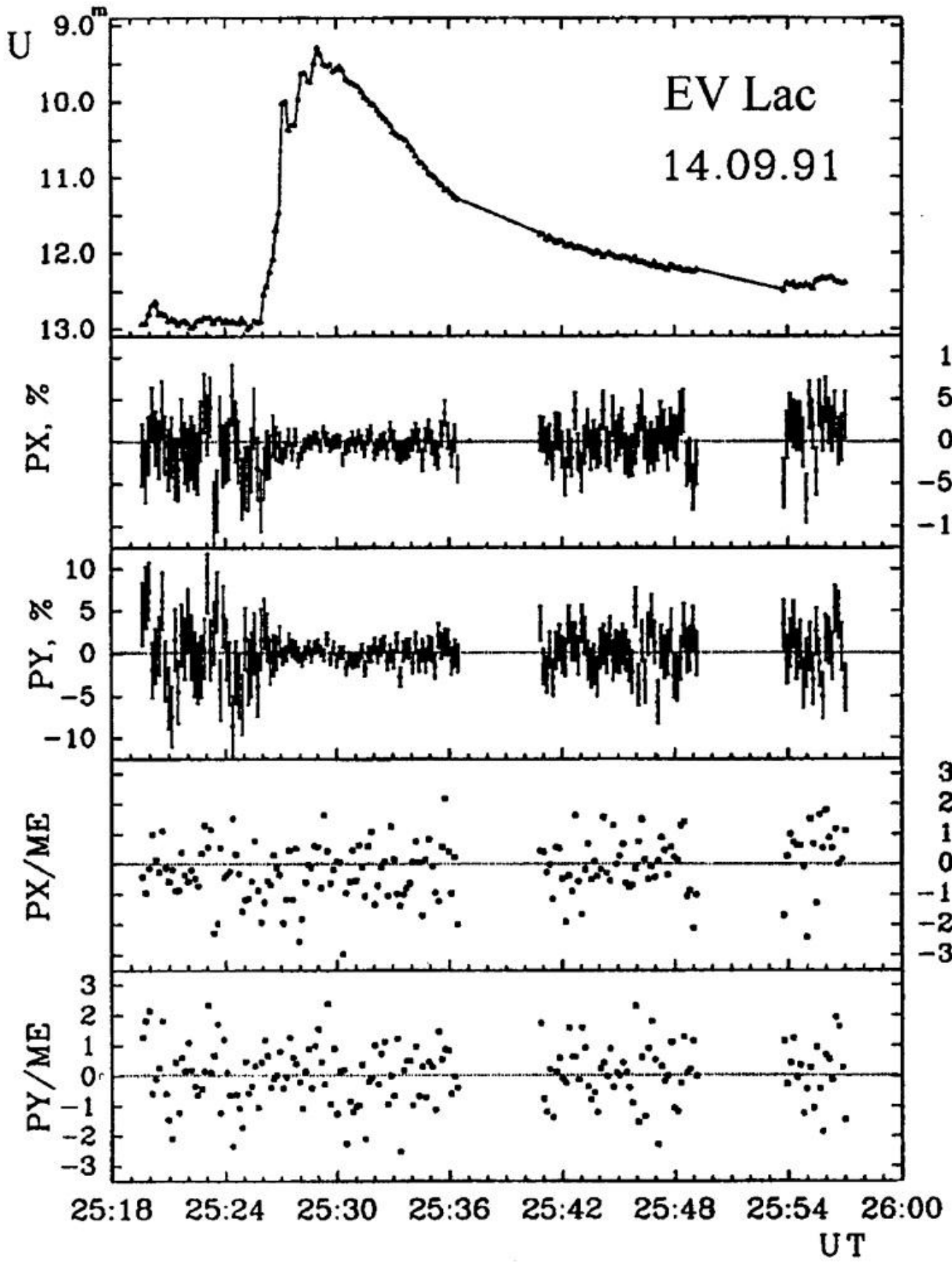

Fig. 65. Polarimetric observations of the strong flare on EV Lac of 14 September 1991: the flare light curve in the U band (*top panel*); the Stokes parameters (*the second and third panels*); the ratios of the Stokes parameters to their errors (*two bottom panels*) (Alekseev et al., 1994)

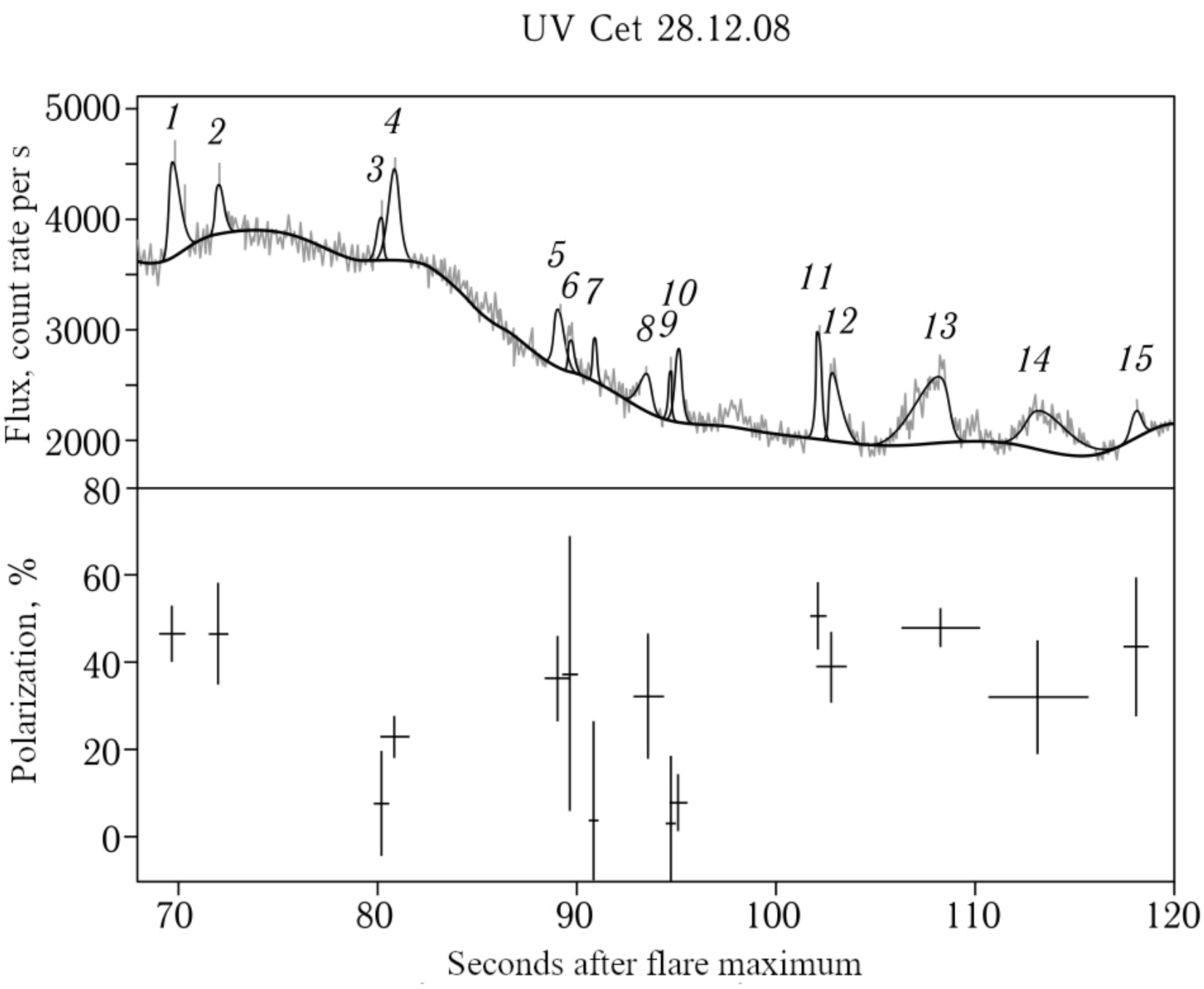

Fig. 66. Section of the brightness curve of the flare on UV Cet of 28 December 2008 from 70 to 120 s after brightness maximum with a dozen strongly polarized subsecond bursts (*top*) and lower estimates of linear polarization of their radiation (*bottom*) (Beskin et al., 2017)

Section 2.4.2 reports on the recorded in the millimeter wavelength flare on Prox Cen of 24 March 2017 with a duration of 1 min with a thousand-fold amplitude, a rapidly varying power-law spectrum and a lower limit of linear polarization of 0.19 ± 0.02. Analyzing these observations, MacGregor et al. (2018) assumed that synchrotron radiation was one of possible mechanisms of this event. The described above flare of the same Proc Cen of 1 May 2019 recorded by MacGregor et al. (2021) for the third time yielded arguments for the existence of synchrotron radiation of stellar bursts: an abrupt drop of the spectral index up to a negative value in the power-law representation of emission and variation of the linear polarization degree, a symmetric light curve without the slow decay phase, relative rapidity and large amplitudes in the mm and FUV ranges.

**2.4.4.4. Spectral Studies.** The optical spectrum of a flare on a UV Cet-type star differs heavily from the spectrum of its quiescent state and changes rapidly in the process of development of the flare. The main distinctive features of flares are strong continuous short-lived emission in the short-wavelength region of the spectrum and intensive emission lines with slower development. Figures 2, 30–32, and 36 present various spectral characteristics of optical stellar flares. Let us consider these characteristics in more detail with quantitative estimates and physical conclusions, whenever possible.

A. Continuous radiation, hydrogen emission, and Balmer decrement of flares. Strong continuous emission and strong emission lines of hydrogen were discovered as characteristic features on the very first spectra of stellar flares.

During the first high time resolution observations with an image intensifier tube spectrograph at the 2.6-m Shajn reflector in the Crimea, the degrees of filling by continuous emission of some absorption details — jumps of intensity at the heads of TiO bands and depths of the CaI $\lambda$ 4227 Å line — were measured in the spectra of flares on AD Leo and UV Cet (Gershberg and Chugainov, 1966, 1967). However, the low accuracy of these measurements and small range of simultaneously recorded wavelengths prohibited reconstruction of the energy distribution in the continuous flare spectrum. But quantitative analysis of the equivalent widths of Balmer lines in the spectra of these flares suggested that the lines could be optically thin if they emerged at temperatures above 80000 K or optically thick if they were formed at chromospheric temperatures (Gershberg, 1968). Further observations unambiguously resolved this dilemma in favor of the second variant.

Near the maximum of the flare on EV Lac of 11 December 1965 (Fig. 31) Kunkel (1967) recorded practically constant emission near the Balmer continuum from 3750 Å to 3500 Å. Together with the measured Balmer emission jump it was the first direct evidence of the involvement of hydrogen recombination emission of not too high temperature in this emission.

Kunkel (1967) traced the Balmer series to $H_{11}$ in the spectrum of the flare on EV Lac of 11 December 1965, to $H_{14}$ in the flare on YZ CMi of 7 December 1965 and in the spectrum of EV Lac obtained during long-duration increased brightness of the star on 5 December 1965, and to $H_{10}$ at the maximum of the flare on AD Leo of 11 December 1965. The latter corresponds to the density of radiating plasma of about $10^{15}$ cm$^{-3}$. In the spectra of flares on EV Lac and AD Leo of 11 December 1965 Kunkel (1967) estimated Balmer jumps as 4.6 and 5.7, which corresponds to temperatures of 23000 and 19000 K in a purely recombination spectrum of hydrogen plasma.

From the spectrum of the flare on YZ CMi of 5 December 1965 Kunkel (1967) obtained distinct evidence of faster decay of the continuum, brightness in the U band, as compared to the brightness decay of emission lines. The same conclusion followed from a systematic growth of the equivalent width of the Balmer lines in the process of development of flares on AD Leo on 18 May 1965 (Gershberg and Chugainov, 1966) and from photoelectric measurements of equivalent widths of the $H_\beta$ line in the flares on UV Cet and EV Lac executed by Chugainov (1968, 1969a). Special observations of the strong flare on UV Cet of 6 January 1975 carried out by Moffett et al. (1977) using a photometer and a scanner showed that the delay of the maximum emission in the $H_\beta$ line was a real effect, which was connected not only with the growth of equivalent widths of the line because of a decrease of the continuum after the flare maximum. The delay was later confirmed many times (Kahler et al., 1982; Worden et al., 1984; Hawley and Pettersen, 1991; McMillan and Herbst, 1991; Melikian et al., 2006; Montes et al., 1999), but Pettersen and Sundland (1991) suspected that it took place only during sufficiently strong flares.

The first Balmer decrement measured by Kunkel (1967) in five phases of the flare on EV Lac of 11 December 1965 at maximum brightness looked as follows: $H_\beta$:$H_\gamma$:$H_\delta$:$H_8$:$H_9$:$H_{10}$:$H_{11}$ = 1:1.24:1.48:1.22:1.17:0.94:0.80, and the upper members of the series weakened faster when the flare decayed. Kunkel presented the decrement within the framework of the chromospheric model with coherent scattering in lines at $n_e = 3 \cdot 10^{13}$ cm$^{-3}$, $T$ = 20000 or 25000 K and optical thickness at the center of the $H_\alpha$ line of about $10^3$; as the flare decayed the latter had to be replaced by $10^2$.

The characteristics of hydrogen emission of the strong flare on AD Leo of 2 March 1970 recorded in the Crimea by Gershberg and Shakhovskaya (1972) were slightly different: the Balmer jump was close to 1.4 and the $H_\alpha$ and $H_\beta$ lines were apparently always more intensive than the following members of the series, i.e., the inverse decrement was not observed. Small-emission Balmer jumps of close values were estimated by Chugainov (1972a) near maxima of three flares on EV Lac, and the inverse Balmer decrement was suspected in flares on YZ CMi (Gershberg, 1972b). Spectrocolorimetric observations by Kodaira et al. (1976) admit a small Balmer jump in the flare on EV Lac of 3 August 1975. According to the observations by Hambarian (1982), the Balmer jump in the spectrum of the flare in the binary system YY Gem on 5 February 1981 can be estimated as 1.5–2.

As it was mentioned, in observations of two flares on UV Cet of 14 October 1972 separated by 15 min, Bopp and Moffett (1973) obtained direct evidence of the decisive contribution of continuous emission to the optical flare radiation in maximum brightness and during the initial fast decay (Fig. 32). According to Moffett and Bopp (1976), the contribution of emission lines to the radiation of the UV Cet flare recorded in the B band on 11 November 1971 was 11% at the maximum and 16%, during the decay stage, to the flare on UV Cet of 12 November 1971 — 11 and 17%, and to the flare on EV Lac of 5 November 1971 — 5 and 28%, respectively. According to Giampapa (1983), within the range of 3600–4600 Å the contribution of emission lines to the total radiation of the quiescent state of UV Cet was 6%, at the onset of the strong flare of 8 September 1979 — 18%, at maximum — 3%, during the decay phase it was about 30%. According to Doyle et al. (1988b), in the flare on YZ CMi of 4 March 1985 the contribution of emission lines to the total radiation in the U band was 10%. For the very strong flare on AD Leo of 12 April 1985 with $\Delta U = 4.5^m$ and a duration of about four hours Hawley and Pettersen (1991) found that within the range of 1200–8000 Å the continuum prevailed at all stages: the contribution of lines to the total radiation of the flare was 9%, about 4% during the impulsive phase and 17% during the gradual phase.

At maximum brightness of the stronger of two flares on UV Cet of 14 October 1972 Bopp and Moffett (1973) recorded the inverse Balmer decrement similar to that of Kunkel's model at $\tau_{H\alpha} \sim 10^2$, but its changes in the process of flare decay and the decrement of the previous weaker flare on UV Cet on the same day did not comply with Kunkel's calculations.

The data on the emission Balmer decrements of flares on UV Cet-type stars accumulated by the early 1970s were analyzed within the theory of radiation of optically thick hydrogen plasma with the velocity gradient of internal motions. Assuming the temperature of plasma to be equal to 15000 or 20000 K, all observational data were presented by models with an electron density from $10^{12}$ to $10^{14}$ cm$^{-3}$ and the escape probability of Ly$_\alpha$ quantum of $10^{-5}$–$10^{-4}$ (Gershberg, 1974b). Later, Bruevich et al. (1990) elaborated the theory of the Balmer decrement without using the hypothesis about the velocity gradient of internal motions, but with allowance for multiple scatterings of quanta that allowed them to leave the medium in the line wings. Within this concept, Katsova (1990) presented the Balmer decrement of the flare on EV Lac of 11 December 1965 both at maximum and 15 min later, when the flat decrement was replaced by a steep one, as well as the flares on AD Leo of 4 March 1970, UV Cet of 17 September 1980, and YZ CMi of 4 March 1985. The observations were presented with an electron density of about $10^{14}$ cm$^{-3}$, somewhat lower temperatures and greater optical thickness $\tau_{Ly\alpha} \sim 10^6$ than in calculations within the concept of moving media. The transition from flat or inverse decrement at the flare maximum to a steep decrement at the decay phase was associated with the decrease of electron density of the radiating plasma. Within the framework of this model, Katsova estimated the area of the flare on YZ CMi of 4 March 1985 in the Balmer lines as $5 \cdot 10^{18}$ cm$^2$, which exceeds by an order of magnitude the estimate of the area of the source of continuous radiation (see below). In this flare, an appreciable decay of $H_8$ and $H_9$ at practical invariance of the ratio $H_\delta/H_\gamma$ was recorded (Butler, 1991). Grinin (1980b) showed that under the formation of the Balmer emission lines in the medium with considerable gradients of physical conditions combined with radiating interaction between high- and low-density zones an abnormally high inverse ratio $I_{H\gamma}/I_{H\beta}$ could occur.

Drake and Ulrich (1980) calculated the luminosity of flat hydrogen layer using the probability method for $n_e = 10^8$–$10^{15}$ cm$^{-3}$, $T_e$ = 5000–40000 K, and $\tau_{Ly\alpha} = 10^4$–$10^6$. Using their calculations, from the Balmer decrement at the 15th minute of the flare on AD Leo of 28 March 1984 Butler (1991) estimated its electron density as $10^{14}$–$10^{15}$ cm$^{-3}$ assuming $T_e$ = 15000 K and $\tau_{Ly\alpha} = 10^4$. Within the Drake and Ulrich model, Jevremovic et al. (1998b) analyzed the Balmer decrement of three flares on the dM5.5e star Gl 866 recorded on 11 June 1991 during four-hour spectral and photometric observations at the South African Astronomical Observatory. To enable the best representation of relative intensities of the lines from $H_\beta$ to $H_{10}$, the analysis involved the selection of four free parameters of the model: electron temperature, electron density, optical thickness in $Ly_\alpha$, and temperatures of underlying surface. As a result, for the flares with rather different light curves the appreciably varying parameters were obtained: $\log n_e$ = 12.5–14.9, $T_e$ = 8000–20600 K, $\log\tau_{Ly\alpha}$ = 6.7–4.2, and temperatures of the underlying surface within 3800–7200 K. The models yielded the areas of flares from 0.1 to 5% of the stellar surface, or $8 \cdot 10^{17}$–$6 \cdot 10^{19}$ cm$^{-2}$ and thickness from 0.5 to 1100 km.

In the strong but short flare on YZ CMi of 5 March 1985 Doyle et al. (1988b) found a nonmonotonous evolution of the Balmer decrement: in the range from $H_\delta$ to $H_9$ first it became flatter than at maximum brightness, then it steepened.

The Balmer decrement in the strong flare on AD Leo of 12 April 1985 was traced by Hawley and Pettersen (1991): it was rather flat at the flare maximum $H_\beta : H_\gamma : H_\delta : H_8 : H_9$ = 1.20 : 1 : 0.89 : 0.66 : 0.53 and slightly steeper 1.30 : 1 : 0.72 : 0.50 : 0.36 8000 s later, but it did not reach the normal state.

On 26 March 1986, Doyle et al. (1990b) recorded a rather flat Balmer decrement in the spectrum of the flare on Gl 375. However, later Montes et al. (2006) detected a binarity of the system Gl 375 consisting of two M3.5e dwarfs, and it is not clear whether this decrement concerns a flare of any of its components. A flat or even inverse decrement was found in the flare on YZ CMi of 18 May 1992 (Gunn et al., 1994a) and in flares on EV Lac of 10 and 11 September 1993 (Abranin et al., 1998a). But the Balmer decrement found by Phillips et al. (1988) in the flare on UV Cet of 17 September 1980 $H_\beta$:$H_\gamma$:$H_\delta$:$H_8$:$H_9$ = 2.00:1:0.69:0.47:0.42 approached the nebular decrement. In the flare on LQ Hya of 22 December 1993 the ratio $H_\alpha/H_\beta$ was close to 1.5 (Montes et al., 1999).

Petrov et al. (1984) examined the $H_\beta$ emission in the spectra of several flare stars with a resolution 0.7–1.0 Å, using the 2.6-m Shajn telescope. During joint photometric monitoring they recorded two flares on UV Cet, three flares on AD Leo, and three flares on YZ CMi. Analysis of the data confirmed the conclusions about independent changes in the profiles and intensities of the Balmer lines in flares and the decisive contribution of the linear emission to preflare brightening. It also revealed significant variations in the central intensity of $H_\beta$ an hour before the flares.

Mochnacki and Zirin (1980) carried out spectral observations of five flare stars using a 32-channel spectrometer at the 5-meter telescope. In the blue and visual regions of the spectrum, the resolution was 160 Å, in red 20 or 80 Å, whereas in the region $\lambda > 5700$ Å only separate sections of the spectrum were measured, thus the lower members of the Balmer series were measured separately, and starting from $H_\delta$ higher members were not separated. On 9 October 1979, they recorded a strong flare on YZ CMi with a time resolution of 10 and 30 s and on 10 October 1979, two fast flares on UV Cet. Analyzing the obtained data within the two-component model of Kunkel (1967), Mochnacki and Zirin found that at a similar Balmer decrement the Balmer jump in these flares was much smaller than in Kunkel's observations. They concluded that the emission from the heated photosphere dominated in these events even at maximum brightness. Assuming the radiation to be blackbody, from the measurements in the range of 4200–6900 Å they estimated the temperature of such radiation at maximum of the flare on YZ CMi as a value close to 9000 K, on UV Cet 9500 and 7400 K; at measured absolute luminosities such temperatures correspond to the flares of $38 \cdot 10^{17}$, $2.9 \cdot 10^{17}$, and $3.3 \cdot 10^{17}$ cm$^2$. According to Mochnacki and Zirin, fast burning of flares is connected with the increase of their areas, at maximum brightness the maximum of the blackbody temperature is reached and fast decay is caused by the decrease of this temperature.

At maximum brightness of the strong flare on YZ CMi of 25 October 1979 the energy distribution in the range of 4200–5900 Å was also represented by blackbody radiation at a temperature of 8500 K. In this case, the area of the flare was estimated as $1 \cdot 10^{19}$ cm$^2$ or 0.5% of the stellar disk (Kahler et al., 1982). Katsova et al. (1991) approximated the optical continuum in the range of 3600–4600 Å at maximum brightness of the same star on 4 March 1985 by the blackbody radiation at 10000 K and estimated the size of this six-minute flare with $\Delta U = 1.2^m$ as $5 \cdot 10^{17}$ cm$^2$.

Giampapa (1983) compared the energy distribution in the continuous spectrum of the strong flare on UV Cet of 8 September 1979 with $\Delta U = 5^m$ in the range of 3700–4600 Å (Fig. 67) with several theoretical distributions: in the radiation of an absolutely blackbody and $H^-$ at 7500 K, in the recombination spectrum of hydrogen at $T_e$ = 20000 K and $n_e = 10^{12}$ cm$^{-3}$, and in free–free radiation of the hydrogen plasma at 50 and 100 kK. Comparison showed that both models of free–free radiation were suitable for representation of observations and there was a substantial divergence between them only in the range of $\lambda < 1800$ Å. Purely recombination luminosity occurs in the long-wavelength region of the range, but is insufficient for its "blue" part, similar situations with the luminosity of $H^-$ and blackbody radiation exist, all of them require additional short-wavelength emission. The ion of $H^-$ was used because it is considered as one of the probable radiators in solar white-light flares with a luminosity of $2 \cdot 10^{29}$ erg/s and a total energy of $3 \cdot 10^{31}$ erg. In these flares within 4000–6000 Å the spectrum is flat, the Balmer jump is 2–3, 90% of total radiation makes up the continuum and only 10% the lines. On the other hand, Grinin (1976) drew a conclusion on the strengthening of $H^-$ emission in the field of the temperature minimum of stellar atmospheres under the effect of the flare impulsive phase. The last member of the Balmer series in the spectrum of the flare on UV Cet of 8 September 1979 was the $H_{15}$ line, which corresponds to a plasma density of $10^{13}$ cm$^{-3}$, while the ratios of intensities $H_{11}$–$H_{15}$ corresponded to the Boltzmann distribution for the population of levels, which means thermal excitation of this emission.

Using the 3.9-meter Anglo–Australian telescope, Robinson (1989) recorded a flare on Wolf 424 of 4 April 1987 with a resolution of 1.5 Å with 30-s exposures. He found a flat spectrum near the maximum in the range of $\lambda > 3900$ Å and a smooth quasicontinuum in the short-wavelength region due to the superimposing of the Balmer lines. On the background of an almost monotonic decay of the flare continuum, Robinson found three appreciable bursts in hydrogen and calcium lines.

Zarro and Zirin (1985) obtained the spectra of YZ CMi in the range of 3600–4000 Å with a resolution of 3 Å using the 5-meter telescope for the quiet star and during the flare of 19 February 1984. They showed that the flare continuum in the region $\lambda < 3800$ Å was formed as a result of superimposing of the upper members of the Balmer series.

The mentioned strong flare on AD Leo of 12 April 1985 with $\Delta U = 4.5^m$ recorded by Pettersen et al. (1986b) with IUE in the ultraviolet and UBVR bands with a high-speed photometer of the 90-cm telescope of the McDonald Observatory allowed an important conclusion to be made on the nature of flare continuum in the optical range.

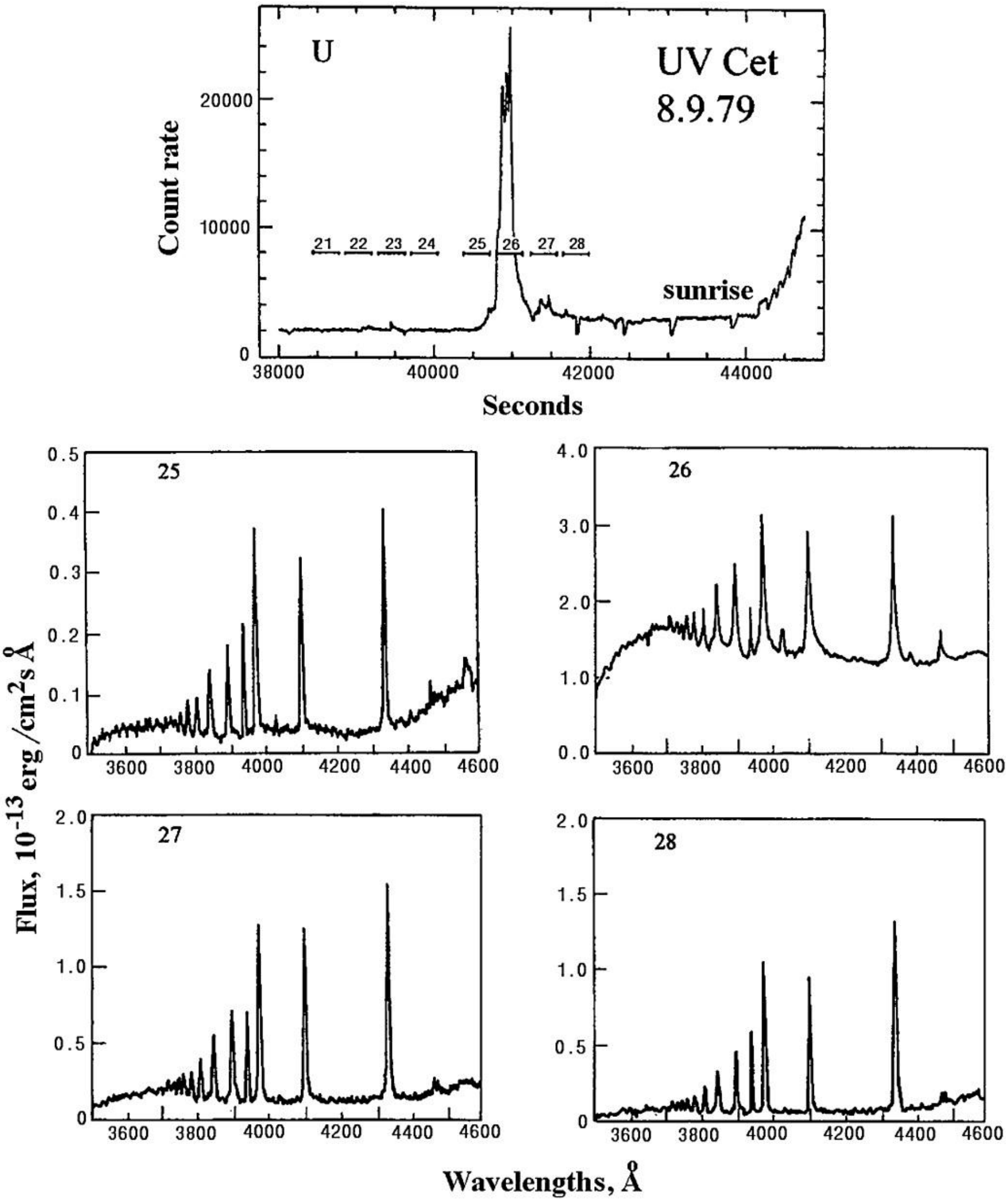

Fig. 67. The light curve of the flare on UV Cet of 8 September 1979 in the U band: seconds after 00:00 UT and energy distribution in its spectrum over the time intervals shown by horizontal sections (Eason et al., 1992)

In Fig. 68 one can see that in the optical range both distributions well represent the observations, but the ultraviolet points exclude the luminosity of plasma at coronal temperatures as the basic mechanism of optical emission of the flare. At least a two-component model is needed to describe the flare continuum. Apparently, this is the only strong flare of the dMe star recorded with IUE and in UBVR bands. Flare spectra obtained simultaneously in the optical range allowed the last member of the Balmer series $H_{15}$ to be recorded.

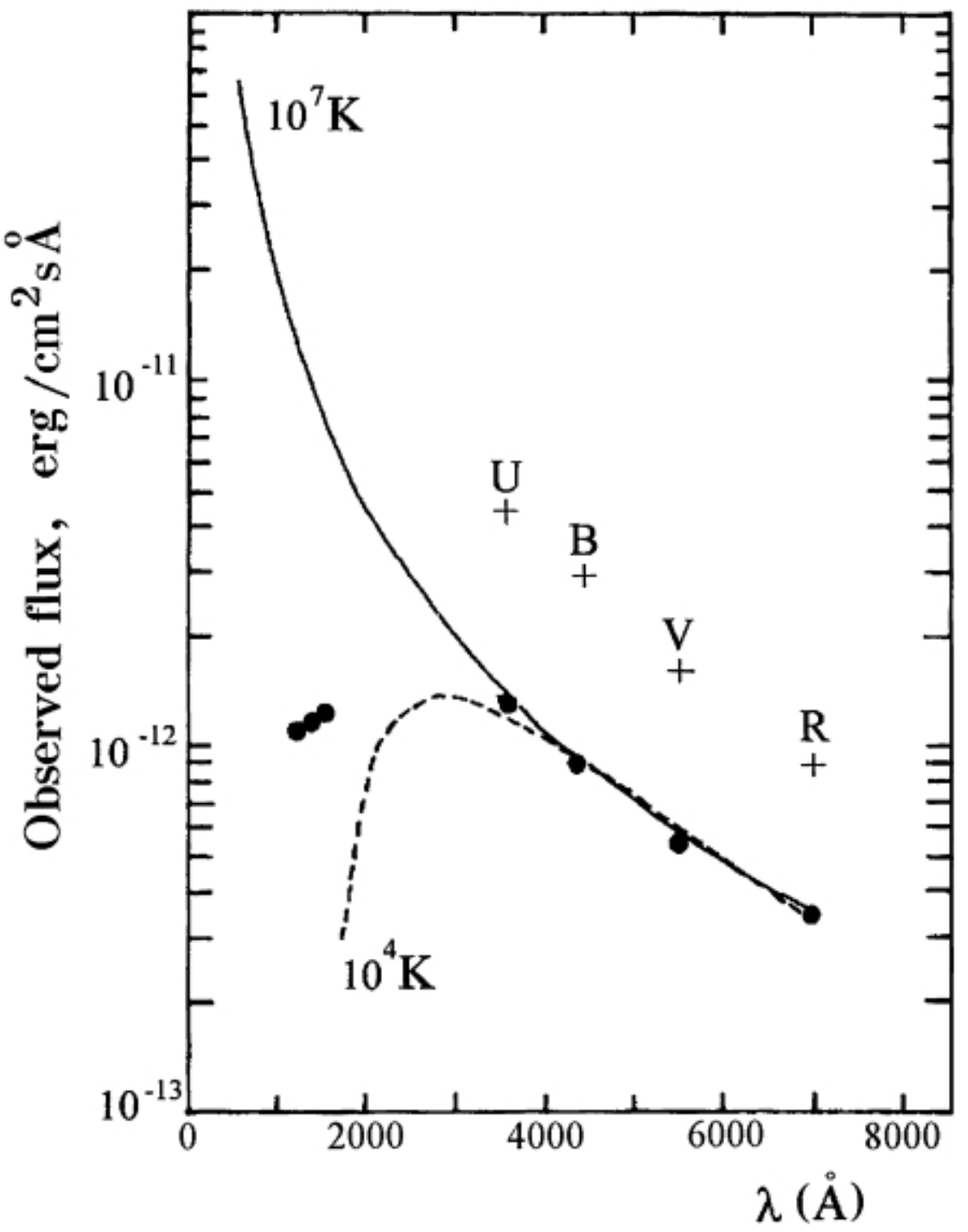

Fig. 68. Absolute fluxes in the UBVR bands at maximum of the flare on AD Leo of 12 April 1985 (*crosses*) and fluxes averaged over a duration of deriving an IUE spectrogram of 15 min (*circles*), the *solid curve* shows the energy distribution in free–free radiation of plasma at $10^7$ K, the *dashed line* shows blackbody radiation at $10^4$ K (Pettersen et al., 1986b)

De Jager et al. (1989) showed that at the maximum of the fast and strong flare on UV Cet of 23 December 1985 with $\Delta U > 6^m$, whose spectra were obtained at the 1.5-meter telescope of the European Southern Observatory, the energy distribution in the range of 3500–7000 Å was well described by blackbody radiation at 16000 K, which yields a flare size of $3 \cdot 10^{18}$ cm$^2$, but as the flare decayed the blackbody distribution represented its continuum even less well.

Doyle et al. (1989) found that near the maximum of the flare on V 577 Mon of 28 February 1985 with $\Delta U = 3.8^m$ during several tens of seconds continuous emission in BVRI bands was well presented by synchrotron, free–free, and recombinant radiation. But to represent the observed luminosity, the synchrotron needs too many luminous electrons, free–free radiation should exit from the volume comparable to the stellar one, whereas the recombination luminosity is 4000 times more effective.

McMillan and Herbst (1991) found a correlation between the equivalent widths of the emission line $H_\alpha$ and the degree of its variability and some evidence of the connection of strong flares on EV Lac with localization of spots. Similar results were obtained by Dal and Evren (2011) from the observations of EV Lac, V1005 Ori, and EQ Peg.

During observations of the young fast rotator K2 LQ Hya at the 4.2-meter Herschel telescope with the Utrecht echelle spectrograph in the range of 4840–7720 Å Montes et al.

(1999) recorded on 22 December 1993 a strong flare that lasted for more than three hours. The first spectrogram revealed a strong continuum, which filled all the photospheric lines and was identified as blackbody emission at 7500 K from the degree of filling.

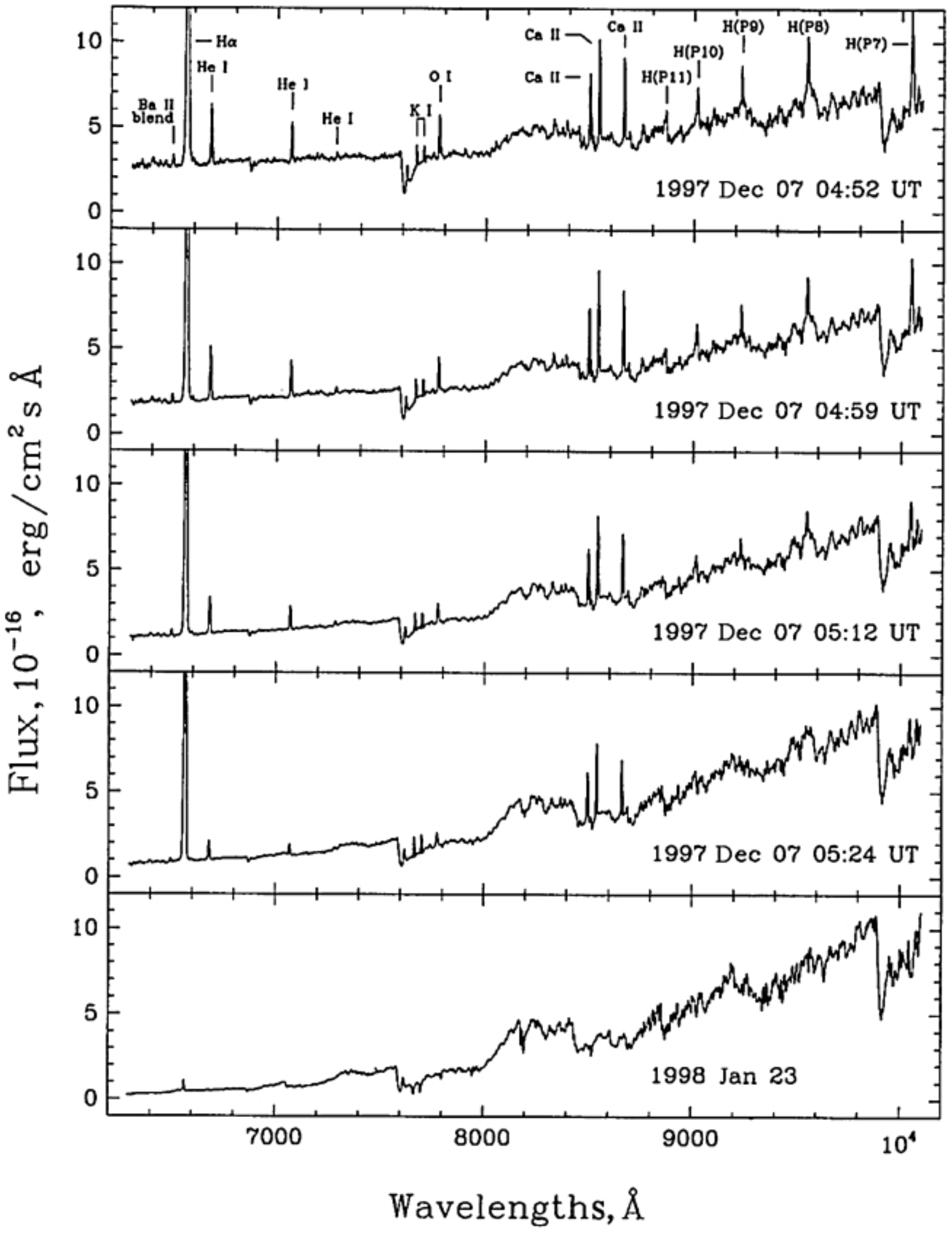

Fig. 69. Spectrograms of the flare on the dM9.5e star 2MASSW J0149090+2951613 of 7 December 1993 in the range of 6150–10100 Å; the *bottom panel* shows the spectrum of the quiet star (Liebert et al., 1999)

Liebert et al. (1999) recorded a unique flare on the dM9.5e star 2MASSW J0149090+2951613 on 7 December 1993 in observations with the Keck telescope in the range of 6150–10100 Å with a spectral resolution of 9 Å. They found an abnormally strong emission of $H_\alpha$ in the five-minute spectrum, which weakened in three following 10-min spectra (Fig. 69). $W_{H\alpha}$ achieved 300 Å, and its FWHM did not exceed the instrument width, whereas in the quiet star $W_{H\alpha} = 8–14$ Å. In the short-wavelength part of the spectrum, the continuum was enhanced by several times. The estimate of $L_{H\alpha}$ showed that at the impulsive phase it was identical to or even higher than the bolometric luminosity of the star.

Fuhrmeister and Schmitt (2004) recorded the high-resolution spectra of a strong flare on the old M9 dwarf DENIS 104814.7-395606.1. The flare lasted for 1.3 hours and the mass motion in it was recorded from the blue shift of $H_\alpha$-, $H_\beta$-, and NaI D lines. The last spectrogram showed an increase of the blue wing of two Balmer lines, which was interpreted as a matter rise with a velocity of 100 km/s, which could result in significant mass ejection.

During the spectrally detected flare on DX Cnc the equivalent width of $H_\alpha$ increased from 3 to 87 Å (Meusinger et al., 2007).

As it was mentioned, Liefke et al. (2007) throughout three nights carried out simultaneous XMM-Newton and VLT/UVES observations of CN Leo and on 19 May 2006 recorded a strong flare with an amplitude of about 500 in the optical range and about 100 in X-rays with a total duration of about 25 minutes. The optical monitor of XMM-Newton recorded a preflare in the U band, a later maximum, and a secondary burst 20 minutes later. The optical spectra from UVES showed a significant increase of the continuum at the flare onset, then the enhancement and broadening of Balmer lines, and appearance of emission in the Paschen series, in the CaII IR triplet, and HeI lines.

The mentioned above 63 flares found within the SDSS project were identified by Hilton et al. (2010) from the abnormally strong emissions in the $H_\alpha$ and $H_\beta$ lines or from the significant variability of these lines during the exposure.

On 16 January 2009, Kowalski et al. (2010a,b, 2012) recorded an extremely strong flare on YZ CMi at the Apache Point Observatory: at an amplitude ΔU of about $6^m$ the flare lasted for more than 8 hours, its radiation energy in the U band was more than $1.7 \cdot 10^{34}$ erg. Several secondary bursts were detected on the descending branch of the flare, and at the beginning of the decay the star during 1.3 hours was spectrographed with the 3.5-meter telescope in blue rays and in the near ultraviolet range with a time resolution of 28 s. Analyzing in detail the obtained spectra near one of the secondary bursts, Kowalski et al. came to the conclusion that the ultraviolet continuum was formed by two components: blackbody radiation at $T \sim 10000$ K with the filling factor 0.22% and the Balmer continuum. These components had different time development up to anticorrelation at some time intervals, and the blackbody radiation region was 3–16 times less than the radiation region of the Balmer recombination spectrum. Subtracting the spectrum obtained at the onset of the burst from the spectrum derived during almost two-fold stellar brightening in the course of this secondary burst, they detected a typical spectrum of the A star with absorption in the Balmer lines and in the continuum. (The detected strong absorption in the red wing of $H_\alpha$ during the flare on EV Lac of 15 August 2015 may be physically related to this phenomenon (Honda et al., 2018).) It is obviously that this two-component representation of the blue and near UV also fits for a series of the listed above observations. In particular, it explains a small Balmer jump in a number of flares. Anfinogentov et al. (2013) studied the descending branch of this huge flare and found that the light curve in the U band at the decay phase could be represented by an exponent with a characteristic time of 46 min, which was superimposed by oscillations with a period of 32 min. Such flares are typical of solar flares in EUV and radio range and are associated with standing slow magnetoacoustic waves.

Furthermore, from the SDSS data Kowalski et al. (2013) studied the spectra and photometry of 20 flares on M dwarfs in the near ultraviolet and optical range with the aim of disentangling the white-light continuum on either side of the Balmer jump and came to the following conclusions. The emission Balmer continuum is present in all the flares and contributes in the wide range of values toward the blue side from the jump. The blue continuum at the flare maximum, decreasing linearly from 4000 to 4800 Å, indicates blackbody radiation at 9000–14000 K. Toward the red side from $H_\beta$ the continuum becomes relatively more important in the energy budget at the late decay stage. The blackbody component, which is described by the one-temperature Planck function, includes absorption details and resembles a spectrum of the A star. The continuation of these studies was described above when discussing a flare on GJ 1243 in the near ultraviolet.

Within the MUSICOS campaign, using the Isaac Newton Telescope at the Roque de los Muchachos Observatory, Montes et al. (2004) carried out spectral monitoring in the range of 3560–5170 AA of AD Leo and V 1054 Oph with exposures of 15–300 s and breaks of less than a minute to readout the CCD matrix. During four nights they recorded 459 spectra of AD Leo and 233 spectra of V 1054 Oph, and in the presented graphs of $EW(H_\beta)$ a number of flares were clearly seen. The Balmer lines, CaII H and K lines, and the neutral helium $\lambda$ 4026 Å line could be clearly seen in most spectra. The durations of flares were 14–31 min on AD Leo and 21–96 min on V 1054 Oph; besides flares, the short-term small-amplitude variations were seen in the lines. The calcium emission has a less amplitude, slower development as compared to the hydrogen one, and its maximum is achieved later. Red asymmetry of the $H_\beta$ line was noticeable in the quiet star as well, and the maximum broadening of lines took place in the flare maxima.

Using the Isaac Newton Telescope, Crespo-Chacon et al. (2004) carried out spectral observations of V 1054 Oph and with high time and intermediate spectral resolution derived spectra of the quiescent state, four strong, and several weak flares. Very strong emission lines were detected even in the quiescent state. Flares were determined within the development of the Balmer and CaII H and K lines, which lasted for 25–95 min, $EW(H_\beta)$ increased by a factor of 2.3, broad wings and red asymmetry were observed.

Later, following Jevremovic, Crespo-Chacon et al. (2006) analyzed Balmer decrements of about 600 spectra of AD Leo, which were derived throughout four nights at the Isaac Newton Telescope with exposures ranging from 15 to 300 s with a resolution of about 1 Å. Analysis was performed for 14 moderate-intensity flares. When reproducing the Balmer decrement, the following ranges of values of the flare physical parameters were obtained: $6 \cdot 10^{13} < n_e < 2 \cdot 10^{14}$ cm$^{-3}$, $12000 < T_e < 24000$ K, $3.60 < \log\tau_{Ly\alpha} < 4.62$, the temperatures of the underlying surface varying from 8000 to 13500 K, and areas from 0.12 to 2.3% of the stellar disk. These estimates are in agreement with the previously found values. Crespo-Chacon et al. found no correlation between the area, temperature, and duration of flares and the dependence of these values on the temperature of the underlying surface.

Using the 2.6-meter telescope of the Byurakan Astrophysical Observatory, on 18 May 2012 Melikian et al. (2012) recorded a strong flare on WX UMa — one of the first stars with detected flare activity (see Fig. 1). During the flare of this M6 dwarf with $\Delta B \sim 5^m$ there was an abrupt increase in the continuum and hydrogen emission lines.

Using a slitless spectrograph mounted on the 600-mm telescope at Peak Terskol, in May and August 2010 Zhilyaev et al. (2012) carried out monitoring of EV Lac and in the flare maximum on 10 August 2010 with amplitudes in the UBVR band from $2.83^m$ to $0.28^m$ they detected blackbody radiation with a temperature of 13400 ± 500 K, and in the flare brightness maximum on 30 May 2010 with amplitudes from $0.65^m$ to $0.10^m$ at 5700 ± 100 K; the size of the May flare accounted for 7% of the stellar radius, the size of the August flare was of about 4%.

With the 3.5-meter telescope of the Apache Point Observatory, Schmidt et al. (2012b) carried out infrared spectral observations with photometric support of active M dwarfs

EV Lac, AD Leo, YZ CMi, and VB 8 and during 50 hours recorded 16 flares; they first recorded emissions of $P_\beta$, $P_\gamma$, $P_\delta$, HeI $\lambda$ 10830 Å, and $Br_\gamma$ in the strongest of them. The strongest of the recorded flares — EV Lac with $\Delta U = 4.02^m$ — was analyzed using the nonLTE program of the radiative transfer of the one-dimensional static atmosphere, and within this program there was obtained a good agreement with the measured ratios of lines.

During five summer nights in 2015, using the Japanese 2-meter NAYUTA Telescope, Honda et al. (2018) carried out spectral observations of EV Lac in the $H_\alpha$ region with a resolution of 10000 and exposures of 3 and 5 min. During the observations throughout four nights the star was in the quiescent state with $W_{H\alpha} = 3$–5 Å, and on 15 August throughout 2.5 hours 30 spectra were acquired with exposures of 5 min each with $W_{H\alpha} > 7.9$ Å, and there was a flare with burning of about 20 min and decay of about 1.5 h. During this flare $W_{H\alpha}$ increased up to 10.5 Å without a shift of the profile maximum wavelength, blue asymmetry of the profile took place over the whole flare. In some spectra, absorption was suspected in the red wing, and ratios of intensity of the blue and red components varied in the profile peak. The HeI $\lambda$ 6678 Å emission, absent during other nights, was seen in all the spectra.

Recently, Kowalski et al. (2017) published calculations of the Balmer decrement for solar and stellar flares in which the broadening of hydrogen lines by the fluxes of nonthermal particles was taken into account. In combination with the decay phase of the giant flare on YZ CMi of 16 January 2009, these calculations within the multiloop model with occurring at different times development of individual elements lead to the estimate of electron density in the wide range from $10^{14}$ to $5 \cdot 10^{15}$ cm$^{-3}$ and to the concept of formation of chromospheric condensations. Later, these calculations were successfully applied to the analysis of ultraviolet and optical continua of two flares on the M4e dwarf GJ 1243 (see Sect. 2.4.3).

From the spectra of more than 600,000 all F–M main-sequence stars derived within SDSS Koller et al. (2021) searched for the connection between stellar flares and coronal mass ejections, the latter was determined from the asymmetry of $H_\alpha$ and $H_\beta$ emission profiles. Through the automatic analysis of spectra they identified 281 flares on K3-M9 stars and 6 possible coronal mass ejections with masses of $6 \cdot 10^{16}$–$6 \cdot 10^{18}$ g and velocities of 300–700 km/s. Thus, optical spectroscopy provides rather sparse knowledge on coronal mass ejections.

B. Calcium emission in the H and K lines enhances in stellar flares later and to a lesser extent than hydrogen lines, and fades more slowly. Monitoring of YZ CMi revealed a many-hour increased luminosity of this emission after strong flares (Gershberg, 1972b). According to the estimate of Moffett and Bopp (1976), the delay of the CaII emission maximum with respect to the flare maximum varied from 4 to 52 min. Thus, they suspected that a greater delay is typical of stars of higher luminosity.

At the onset of the strong flare on UV Cet of 8 September 1979 the flux in the CaII K line decreased almost 3-fold, then increased to a 2–3-fold level of the quiescent state of the star, which could be due to the evaporation of the low chromosphere or further ionization of calcium by X-ray emission of the flare (Eason et al., 1992).

Houdebine et al. (1993b) found that flare CaII emission, to a considerable extent, was stipulated by radiative pumping by the $H_\varepsilon$ emission.

In the flare on the dM9.5e star 2MASSW J0149090+2951613 of 7 December 1993 the CaII IR triplet was recorded in emission (Fig. 69). In the flare on AD Leo on 13 March 2000 Hawley et al. (2003) discovered a substantial delay of this triplet lines with respect to emission in the U band.

Analogous results were obtained by Schmidt et al. (2007) in a flare on 2MASS J1028404-143843 of 25 January 2002: in its spectrum $EW_{H\alpha} = 97$ Å, whereas in the quiescent state of the star it accounted to 12–23 Å, and within 7000–10000 Å the CaII IR, HeI, and Paschen hydrogen emission lines dominated, the strong flare continuum filled in the deep NaI IR line and heads of many molecular bands.

We mentioned already that in observations of the active K2 star $\varepsilon$ Eri with a four-slit spectrometer in the focus of the 2.5-meter telescope of the Mount Wilson Observatory, Baliunas et al. (1981) recorded correlated variations of brightness of the H and K lines with amplitudes up to 7% and with characteristic duration of up to 15 min. These variations correspond to the flares with an energy of $2 \cdot 10^{30}$ erg.

C. Lines of helium and metals. Emission of neutral and ionized helium was found by Joy and Humason (1949) in the first slit spectrogram of the flare on UV Cet of 25 September 1948 obtained with a 144-min exposure. The emission lines of neutral helium $\lambda$ 4026 Å and $\lambda$ 4471 Å were recorded near the main maximum of the flare on AD Leo of 18 May 1965. By the secondary maximum phase (Fig. 30) they weakened, but during further flare decay, which was spectrographed in the green and red regions of the spectrum, other HeI lines were recorded (Gershberg and Chugainov, 1966). Alongside the neutral helium lines, at maximum brightness and during the fast decay phase of flares on UV Cet of 24 September 1965 the HeII $\lambda$ 4686 Å line was found in its spectrum (Gershberg and Chugainov, 1967). But this is not a rule: in the flare on Wolf 359 of 20 March 1968 HeI lines were appreciably strengthened and the HeII line was absent (Greenstein and Arp, 1969). A similar situation with respect to helium occurred during the flares on UV Cet of 14 October 1972, from the stronger flare the emission of the MgI b line was detected for the first time (Bopp and Moffett, 1973). Moffett and Bopp (1976) found a vast diversity of helium emission in flares that were recorded with high time resolution, but never detected HeII emission.

In the weak flare on AD Leo of 30 March 1977 with $\Delta U = 0.4^m$ Schneeberger et al. (1979) recorded the behavior of the HeI line $\lambda$ 5876 Å identical with $H_\alpha$: an increase of the central intensity by 40% at constant width. During the decay stage in the spectrum of the strong flare on UV Cet of 8 September 1979 Eason et al. (1992) found strong lines of neutral helium $\lambda$ 5876 Å with equivalent widths from 16 to 4 Å and $\lambda$ 6678 Å with equivalent widths from 4 to 1 Å and numerous lines of FeI and FeII in the wavelength range of 5200–5400 Å; equivalent widths of D components of the sodium doublet varied from 3.8 to 1.5 Å.

On 2 September 1981, Baliunas and Raymond (1984) recorded about an hour-long flare on EQ Peg B simultaneously with IUE in the ultraviolet and in the range of 4700–7000 Å with a resolution of about 6 Å using the Reticon spectrograph. The intensity of $H_\beta$ emission increased in the flare by a factor of 2, $H_\alpha$ by 30%, CIV lines almost by a factor of 4, HeII $\lambda$ 1640 Å lines approximately 3-fold, with an enhancement of ultraviolet emission being measured in the resulting spectrum of the system. It is essential that the intensity maximum of helium lines took place approximately 15 min prior to the maximum of $H_\alpha$ and $H_\beta$. Assuming that recombinations in coronal plasma dominated in the emission of the HeII $\lambda$ 1640 Å line, Baliunas and Raymond suggested the X-ray energy of the flare as $6 \cdot 10^{32}$ erg and interpreted the flare as an analog of the gradual phase of a solar two-ribbon flare or as quickly cooling plasma of coronal temperature.

During the mentioned spectral and photometric observations of several flare stars using the 2.6-meter Shajn telescope, Petrov et al. (1984) looked for the HeII $\lambda$ 4686 Å emission in the spectra with a resolution of 0.7–1.0 Å, but did not find it in any of the six flares on AD Leo and YZ CMi recorded during this campaign.

Investigating radial velocities of K–M dwarfs in spring 1978, Mochnacki and Schommer (1979) carried out high-dispersion observations of YZ CMi with an image intensifier tube multichannel coude spectrometer of the 2.5-meter telescope of the Mount Wilson Observatory. On 29 March 1978, over two hours they obtained 11 spectra with emission lines of MgI b, FeI (15, 37) and FeII (42, 48, 49) in the range of 5090–5350 Å. On other dates these emissions were not visible. Radial velocities of the emissions did not show significant differences from the absorption lines, and in the course of observations the intensity of emissions weakened by approximately 30%. The components of MgI b of similar intensity suggested an appreciable optical thickness in these lines. Apparently, during this night a flare took place on the star, and the found emissions could have even longer lifetime than CaII lines.

In the maximum of the flare on AD Leo of 12 April 1985 Hawley and Pettersen (1991) identified seven HeI emission lines and ten FeI lines within the range of 3700–4433 Å, the emission peak in the center of the absorption line of CaI $\lambda$ 4227 Å was observed for about an hour and a half after the flare onset.

In the spectrum of the flare on Wolf 424 of 4 April 1987 Robinson (1989) noticed numerous weak emission lines of FeI and FeII, the MgII $\lambda$ 4481 Å line, and inverse emission of normally absorption lines of CaI $\lambda$ 4227 Å and Mn I $\lambda$ 4030 Å.

In the spectra of active EV Lac obtained during the cooperative observations in 1992 (Abdul-Aziz et al., 1995), one can easily see the lines of neutral and ionized helium $\lambda$ 4471 Å and $\lambda$ 4686 Å, respectively, three strong emissions of FeII $\lambda$ 4924, 5018, and 5169 Å and several weaker lines of iron and magnesium, some of which were mentioned earlier by Mochnacki and Schommer (1979).

On 24 January 1986, during observations of the young G5 dwarf $\kappa^1$ Cet with $P_{\rm rot}$ = 9.4 days, using an echelle spectrograph at the 1-m telescope of the Ritter Observatory, Robinson and Bopp (1987) recorded strong emission of the HeI $D_3$ line in the 40-min spectrogram; this emission was absent in the subsequent spectrogram derived a half an hour later. Previously, this line was detected only in absorption.

The occurrence of the HeI lines was noticed in the flares on Gl 866 of 11 June 1991 (Jevremovic et al., 1998b), on AT Mic of 15 May 1992 (Gunn et al., 1994b), on Proxima Cen of 9 May 1994 (Patten, 1994), on LQ Hya of 22 December 1993 (Montes et al., 1999), on 2MASSW J0149090+2951613 of 7 December 1993 (Liebert et al., 1999), and on LHS 2065 of 12 December 1998 (Martin and Ardila, 2001).

Of particular interest are the continuous spectral observations of EV Lac with the Shajn telescope in the Crimea on the night of 31 August–1 September 1994 (Abranin et al., 1998b). The top panel in Fig. 41 shows the light curve of the star in the U band. One can see half a dozen characteristic fast flares of different amplitudes with durations of several minutes and long increased luminosity of the star, which started after the flare at 20:46 UT and lasted until the strongest flare on this night began at 23:24 UT. As it was noted, this prolonged stellar brightening can be essentially associated with an emergence of extended facular fields in the field of view. The next panel in Fig. 41 presents the equivalent width of the $H_\beta$ line. Close examination of the diagrams reveals the response of $H_\beta$ emission to individual fast bursts of stellar brightness. But the main feature of the time course $W_{H\beta}$ is the dominating maximum concerned with an almost three-hour increased brightness of the star. The time course of $W_{H\beta}$ is rather similar to that of an equivalent-width blend of emission lines of FeII $\lambda$ 5169 Å and MgI $\lambda$ 5167/73 Å presented on the fifth diagram: the correlation coefficient $r$ ($W_{H\beta}$, $W_{\rm blend}$) = 0.94. The third panel shows the equivalent width of the HeI $\lambda$ 4471 Å line. Comparison of this diagram with the first two shows that this line reacts to individual fast flares more explicitly than $H_\beta$, though as a whole the two lines correlate well with each other: $r(W_{H\beta}, W\lambda_{4471}) = 0.76$. (Earlier, Hawley and Pettersen (1991) noted that HeI $\lambda$ 4026 Å emission in the flare on AD Leo of 12 April 1985 preceded the Balmer lines.) The most interesting and unexpected results were obtained in the monitoring of the emission line of ionized helium $\lambda$ 4686 Å (see fourth panel). Formally, the line does not correlate with the above emissions: the appropriate correlation coefficients vary within 0.2– 0.3. But thorough comparison of $W\lambda_{4686}$ with the light curve of the star detects that almost all spectra with increased emission of ionized helium were obtained 15–30 min after fast flares. The second feature of emission of the HeII $\lambda$ 4686 Å line is shown in the bottom panel, where the resulting spectrum of all active states of the star recorded during the campaign of 1994 is presented. One can see the splitting of HeII emission: the long-wavelength component has a normal wavelength, whereas the short-wavelength one is shifted by −400 km/s. Among the spectra obtained on the night of 31 August –1 September, eight spectra had only a short-wavelength component, 2 spectra had both components, and one spectrum had only a long-wavelength component; these features are marked by <,<>, and >, respectively. The meaning of the observations is not clear. Probably, similar enhancement of HeII lines took place in the flare on AD Leo of 2 February 1984, when in the spectrum obtained by Byrne and Gary (1989) with IUE with a 20-min exposure started eight minutes after the maximum of the fast flare, doubled intensity was recorded in the HeII $\lambda$ 1640 Å line arising in the recombination spectrum of HeII at the subsequent cascade transition after the radiation of the quantum of the $\lambda$ 4686 Å line. Apparently, the short-wavelength component of this line can arise in the transition region from the chromosphere to the corona or in low corona under the formation of moving structures, whose analogs on the Sun yield transients into the interplanetary space.

Recently, Moschou et al. (2019) performed a survey of observations of flare stars, which similar to the described above observations of EV Lac in the Crimea could be interpreted as analogs of solar coronal mass ejections (CME). Using a simple model of these events, they estimated masses of the ejected matter within $10^{15-22}$ g for 12 stellar CME candidates, energy of the associated X-ray flares in the range of $10^{31-37}$ erg and found that energy of stellar CME followed the average ratio between masses of solar CME and X-ray energy of flares. But stellar CME showed 200 times less kinetic energy than it was predicted by extrapolation of solar data. There seem to be constraints on velocities of ejections caused by braking in strong magnetic fields and by the stellar wind for CME associated with the strongest stellar flares.

Using the Utrecht echelle spectrograph with a resolution of 110000 mounted on the 4.2-meter Herschel telescope, Martin (1999) obtained on 6 June 1995 four spectra of the dM7e star VB 8 within 5400–10600 Å. In one of them he found the evidence of a flare: an almost 2-fold increased $W_{H\alpha}$ and the increased line width, helium $D_3$ line emission, and the enhanced emission cores of Na D lines and resonance K $\lambda$ 7699 Å line. In the flare on LQ Hya of 22 December 1993 recorded with the same instruments, in addition to helium lines, emissions of Na D lines, the triplet MgI b and numerous lines of ionized iron multiplets 42, 48, and 49 were found (Montes et al., 1999).

Considering the general properties of relaxing plasma with regard to variations of its temperature, expansion, ionization by soft X-rays, ionization and excitation of ultraviolet and optical continua, Houdebine et al. (1991) concluded that in spite of complicated geometry, expansion and radiative transfer, collision processes dominated in optically thin lines of flares with nonthermal heating only at the initial phase with not too high amplitude of ΔU and at $n_e = 10^{11} - 10^{14}$ cm$^{-3}$. For such models they calculated the functions of luminosity of optically thin lines of helium, hydrogen, and ionized calcium depending only on temperature and correlated the recorded luminosity maxima of the Balmer lines and CaII K with temperatures at the maxima of the appropriate luminosity functions. They constructed the cooling curves of flares on Proxima Cen of 24 March 1984, on UV Cet of 8 December 1984, and on AD Leo of 28 March 1984. The thus-obtained experimental cooling curves of flare plasma appeared to be rather close to theoretical Gurzadian curves (1984) for $n_e = 10^{11}$ cm$^{-3}$, but differed significantly from his curves for $n_e = 10^{10}$ cm$^{-3}$.

D. Broadening and shifts of spectral lines. Doubling of the width of hydrogen emission lines in the spectrum of the flare on UV Cet of 7 October 1957 was recorded by Joy (1958) in one of the first spectrograms of flares of this type. The line widths in flare spectra in separate phases of their development were first measured in observations of AD Leo in 1965 in the Crimea (Gershberg and Chugainov, 1966). During the flare of 13 May 1965 with $\Delta B = 1.5^m$ the FWHM of the $H_\alpha$ line grew from 5 to 7–8 Å at an instrumental width of 4.6 Å. During a stronger flare on 18 May 1965 (Fig. 30) $H_\alpha$ was observed only at the late decay stage, but the near-maximum brightness FWHM of $H_\gamma$ and $H_\delta$ lines increased to 10–11 Å from 5–6 Å in the quiescent state. Several minutes after the maximum and termination of fast decay (see spectrograms 4 and 5 in Fig. 30) these lines started narrowing. During weak flares on AD Leo of 24 March and 16 May 1965 with $\Delta B \sim 0.2^m$ no changes in the $H_\alpha$ profile were noticed. Later that year the same equipment was used to observe UV Cet: six flares of different amplitude were recorded (Gershberg and Chugainov, 1967). The FWHM of $H_\beta$ lines was 10–11 Å at the decay phase of the flare of 20 September 1965 with $\Delta V \sim 1.0^m$ and 12 Å in the weaker flare started 20 min later at a width of the instrumental contour of 7.8 Å. The strongest flares recorded during this campaign demonstrated the broadening of lines: the flare of 24 September 1965 with $\Delta V > 1.8^m$ near the FWHM maximum the $H_\beta$ line was 8.4 Å and the $H_\gamma$ line was 7.6 Å at the quiescent state width less than 4.5 Å; in the flare of 26 September 1965 with $\Delta V = 1.4^m$ FWHM = 5.4 Å of the $H_\alpha$ line at the same width in the quiescent state.

On the spectrogram of the flare on Wolf 359 of 20 March 1968, Greenstein and Arp (1969) traced with the strong Balmer continuum the Balmer series to $H_{13}$ and found an increase in FWHM of $H_\beta$ and higher members of the series from 8 to 15 Å. Gershberg and Shakhovskaya (1972) in the flare on AD Leo of 2 March 1969 found that the FWHM of $H_\gamma$,$H_\delta$ and H + $H_\varepsilon$ were 10–13 Å at the width of the instrumental contour of 6.2 Å and near the flare maximum there was “red asymmetry”, i.e., the long-wavelength wing was longer than the short-wavelength one. During the decay phase this asymmetry disappeared, profile tops flattened, and a decrease of intensity in the center was outlined in some of them. Bopp and Moffett (1973) found practically the same width of $H_\beta$, $H_\gamma$, $H_\delta$, and $H$ +$H_\varepsilon$ lines in the flares on UV Cet of 14 October 1972, but strong red asymmetry of structures was found only in the first of the two flares.

Kulapova and Shakhovskaya (1973) recorded FWHM of $H_\alpha$ lines corrected for the instrumental profile at maximum of the flare on AD Leo of 18 February 1971 with $\Delta B \sim 1.2^m$ equal to about 15 Å. The line profile had red asymmetry and during the flare decay phase the profiles of $H_\beta$,$H_\gamma$, and $H_\delta$ had flat tops with a small decrease in intensity in the center.

Joint consideration of observations of more than twenty flares from the Crimea and the McDonald Observatory in the late 1960s–early 1970s showed that there was no correlation between photometric and spectral characteristics of flares: between the flare amplitude and the width of $H_\alpha$ emission line, between the intensity and width of the line, the flare light curve and occurrence of helium lines, the relative contribution of the lines and the continuum.

Using an image intensifier tube echelle spectrograph of the 4-meter reflector and a three-channel photometer of the 90-cm telescope of the Kitt Peak Observatory, Schneeberger et al.

(1979) observed AD Leo: on 30 March 1977 two flares with $\Delta U = 0.40^m$ and $0.35^m$ and total durations of one and three minutes were recorded. In observations with a spectral resolution of 0.22 Å and a time resolution of 2 min the central intensity of the $H_\alpha$ line in the second flare increased by 40%, but the line, whose width in the quiet star was 1.4 Å, remained invariant. Using the same instruments, Worden et al. (1984) recorded six short flares on YZ CMi with amplitudes up to $\Delta U = 1.5^m$, but did not find significant broadening of $H_\alpha$ or $H_\beta$ in any of them. In the strongest flare of 9 February 1979 the time of the two-fold decay in the U band was 60 s, the total duration of the event was about 4 min, but the smoothly decaying afterglow of $H_\alpha$ lasted for about 2.5 h.

Eason et al. (1992) failed to present the profile of $H_\alpha$ emission in the spectrum of the flare on UV Cet of 8 September 1979 by the Stark curve and presented it as a superposition of two Gaussians in which the wide Gaussian had a "blue" shift by 70 km/s at a Doppler half-width of 150 km/s. Eason et al. considered this result as a direct evidence of evaporation of plasma of chromospheric density.

In observations of the flare on AD Leo of 31 May 1981 at maximum ($\Delta B = 0.62^m$) with a spectral resolution of 0.7–1.0 Å Petrov et al. (1984) found the wings of the $H_\beta$ line up to ±15 Å with red asymmetry. Near the maxima of flares on YZ CMi of 30 January 1981 and 2 March 1981 with $\Delta B = 0.5^m$ and $0.7^m$, respectively, they found the wings of this line up to ±10 Å. Comparing the measured structures with the Stark theoretical profiles, Petrov et al. estimated the electron density as $10^{14}$–$10^{15}$ cm$^{-3}$ in the mentioned flare on AD Leo of 31 May 1981 and $10^{14}$ cm$^{-3}$ at the maximum of the second flare on YZ CMi.

In the spectrogram of the flare on AD Leo of 28 March 1984 with $\Delta U \sim 2^m$ recorded with the 3.5-meter telescope of the European Southern Observatory, one can see that all emission lines are considerably broadened (Rodonò et al., 1986). In the spectrum of the flare on AD Leo of 12 April 1985 obtained at the 2.1-meter telescope of the McDonald Observatory with a resolution of 3.5 Å, the width of the hydrogen wings at the base exceeded 20 Å at maximum and its broadening lasted for more than two hours, as long as the flare continuum lasted. $H_{16}$ was the last distinct member of the Balmer series, and the broadening of calcium lines was very slight (Hawley and Pettersen, 1991).

In observations of the $H_\beta$ line in the flare on YY Gem of 18 February 1986 Baliunas et al. (1986) found a doubling of the line width and a four-fold increase of its equivalent width in the spectrum of the B component of the system.

In the flare on YZ CMi of 4 March 1985 with $\Delta U = 1.2^m$ Doyle et al. (1988b) found broad wings at the maximum and especially during the fast decay of the flare in all the recorded Balmer lines. The lines $H_\gamma$ and $H_\delta$ were symmetric, whereas $H_6$ and $H_7$ manifested red asymmetry. Before the flare, $H_\gamma$ and $H_\delta$ were presented by a Gaussian with FWHM = 1.5 Å, but flare profiles could not be presented either by one Gaussian or by one or two Stark profiles. The profiles were presented by the sums of two Gaussians, which corresponds to the directed motions at ± 250–300 km/s or to turbulence with a velocity of 500–600 km/s in the broad component and of 55 km/s in the narrow component. At these velocities, over the time of one exposure, the motion should throw an appreciable mass toward the corona and produce an X-ray flare. The event was not observed in the flare, thus Doyle et al. proposed to replace one long ejection by a chain of successive ejections in the loops of several adjacent flare nuclei, as is observed on the Sun.

During a 6-min flare on UV Cet of 17 September 1980 with $\Delta U = 2^m$, whose spectra were registered in the range of 3800–5100 Å with a resolution of 1 Å at the 1.9-meter telescope in Southern Africa, Phillips et al. (1988) found broadening of the $H_\beta$ and $H_\delta$ lines for about 1 Å, but with noticeable red asymmetry corresponding to directed motions with a velocity of about 100 km/s, which can be attributed to downward motion of chromospheric condensation. Similar red asymmetry of the $H_\alpha$ profile was observed in the spectra of flares on EV Lac of 11 September 1986, 14 September 1986, and 10 September 1987 recorded in the Crimea (Gershberg et al., 1991a).

Robinson (1989) failed to present the wings of the Balmer lines near the maximum of the flare on Wolf 424 of 4 April 1987 by the Stark profiles.

Falchi et al. (1990) found broad wings of Balmer line profiles in the flare on V 1054 Oph of 15 June 1987 recorded spectroscopically with the 1.5-m telescope of the European Southern Observatory with a resolution of 6 Å.

Kinematics and dynamics of radiating gas in the flare on AD Leo of 28 March 1984 with $\Delta U = 2.1^m$ and duration of about 50 min (Fig. 37) were studied in detail from the spectra obtained at the 3.6-meter telescope of the European Southern Observatory in the range of 3600–4400 Å with a resolution of 1.7 Å and an exposure of 60 s (Houdebine et al., 1990, 1993a, b). Before the flare, faint P Cyg components were found in the structure of CaII K lines, which were associated with dark filaments, typical precursors of solar flares. In the impulsive phase a blue wing extended to 80 Å was revealed in the $H_\gamma$ line, which corresponds to the emission of matter at about 5800 km/s. This lasted for several minutes as long as the flare continuum weakened and practically disappeared. At the same time, a red shift was observed in the cores of the Balmer lines, in the lines of CaII K and HeI $\lambda$ 4026 Å, which corresponds to the downward motion of chromospheric condensations that could be initiated by the flux of energetic particles with a power of $9 \cdot 10^{10}$ erg/(cm$^2 \cdot$ s). Later, oscillations of the line centroids with a period of 2.7 min and an amplitude of about 95 km/s were found, which Houdebine et al. attributed to a $2 \cdot 10^9$ cm long prominence with a magnetic field of 20 G. The estimates of Houdebine et al. showed that the kinetic energy was $5 \cdot 10^{34}$ erg, which was 500 times higher than in solar coronal ejections (CME), and the amount of ejected matter was 40 times greater. It is possible that such events can lead to the loss of matter that has evolutionary significance for the star.

For several well-studied flares, Houdebine (1992) summed up the behavior of emission line widths. The FWHM of high members of the Balmer series during the impulsive phase achieve 20 Å and then monotonically decrease. In the gradual phase even at significant fluxes the broadening of the Balmer lines is insignificant. $H_\alpha$ and $H_\beta$ display broadening less often than high members of the series. The intensity of lines correlates with the width directly neither during impulsive nor gradual phases, but broadening was most appreciable near the maximum brightness in the U band, and in both phases the correlation of its broadening with the absolute flux in the U line is outlined. Helium lines are usually much narrower than hydrogen lines by 2–3 Å. A slightly broadened CaII K line, about 1 Å, does not react to brightness variations during the flare. On the basis of a quantitative analysis of these data, Houdebine concluded that broadening of upper members of the Balmer series was caused by the Stark effect in the medium, whose density essentially exceeded the density in solar flares, while the broadening of the lower members was due to self-absorption. Under the decomposition of the Balmer line profiles from $H_\gamma$ to $H_9$ in the spectrum of the flare on AD Leo of 28 March 1984 into two Gaussians, Houdebine obtained FWHM of the broad component from 22 to 17 Å and that of the narrow one, from 6.6 to 5.0 Å. He concluded that the narrow component was caused by a radiatively excited chromosphere around the flare, and the broad component by the flare core emission arising on deep penetration of particles into the layers with a density of up to $10^{15}$ cm$^{-3}$ during the first 30 s of the flare. The size of such a core was about 10 cm$^2$, and its luminosity was detected only in the first flare spectrum. Broadening of the helium line $\lambda$ 4026 Å could be caused by the Stark effect at a density of $4 \cdot 10^{15}$ cm$^{-3}$, and the broadening of the CaII K line to 1.5 Å by macroscopical motions. Later, comparative analysis of twenty flares on M dwarfs in the range of 3600–4500 Å by Houdebine (2003) confirmed the conclusions and revealed various correlations between the parameters of line and continuous emission during the impulsive and gradual phases of flares of different intensity.

The data on broadening and shift of emission lines in the spectrum of EV Lac were obtained at the 2.6-meter Shajn telescope in the Crimea and the 6-meter telescope in North Caucasus during a cooperative observation series. In the Crimea, observations of the $H_\alpha$ region with a resolution of 24 km/s near the flare maxima on 28 and 29 August 1990 a red asymmetry to 130 km/s was found, and in the flare of 30 August 1990 a blue asymmetry to −100 km/s. The 6-meter telescope observations showed that the flare decay on 29 August 1990 in the $H_\delta$ line occurred much faster than in $H_\gamma$ and $H_\beta$ (Gershberg et al., 1993). On 1 and 3 September 1992 spectral monitoring in the Crimea in the range of 4450–5500 Å with a resolution of 2.2 Å/pixel and with exposures of about 10 min detected small flares with appreciable amplitudes on EV Lac. During these events the FWHM of $H_\beta$ practically did not change, but full widths of the line at the level of about 20% and 10% of the maximum noticeably increased: up to 12–15 Å wide wings appeared (Abdul-Aziz et al., 1995). Seven profiles of this line in an active star were presented by pairs of Gaussians of various widths: with the FWHM of narrow components from 4.0 to 5.7 Å and the FWHM of broad components up to 21 Å, and the centers of these components in pairs were spaced on the wavelength axis from +1.3 to −2.2 Å. Near the brightness maxima the broad component contributed more than half of the total flux of the line and during the decay this share vanished. Then, the same recorded profiles were presented by superposition of three to seven Gaussians with instrumental width and with various shifts along the wavelength axis. In this representation, the root-mean-square divergence of the velocities of components varied from 400 to 700 km/s. Similar results were obtained in observations of the $H_\beta$ line in the spectrum of EV Lac in 1994 (Abranin et al., 1998b). Thus, both representations of the observed profiles evidenced kinematic heterogeneity of the matter radiated in flares. Observations of the $H_\alpha$ line in the spectrum of EV Lac during the campaign of 1995 in the Crimea did not reveal a straightforward correlation between stellar brightness in the U band and $W_{H\alpha}$. In many cases, the profile was appreciably asymmetric: the profiles were presented by pairs of Gaussians with the widths of narrow components within 1.2–3.1 Å and broad components within 4.2– 5.8 Å and with a relative shift of their centers up to 60 km/s; broad components contained up to 79% of the total flux in the line (Abranin et al., 1998b).

During the strong flare on AT Mic of 15 May 1992, which was observed by the 1.9-meter telescope of the South African Astronomical Observatory with a resolution of 0.9 Å and exposures of 2–3 min, strong blue asymmetry in the $H_\delta$ and $H_8$ profiles was found and for the first time in CaII H and K lines (Gunn et al. 1994b). Radiation in the blue wings of the lines was comparable with that of the central components, and the shift along the wavelength axis corresponded to velocities from 200 to 700 km/s. The preliminary estimate of the density of evaporating matter as $10^{14}$ cm$^{-3}$, made with a certain assumption, led to the conclusion that the kinetic energy of the rising flux was much less than the flare radiative losses in the range of 3600–4200 Å. Several days later, the same equipment was used to observe YZ CMi with exposures from two to six minutes (Gunn et al., 1994a). As opposed to the described flare on AT Mic, at the maximum of the flare on YZ CMi of 18 May 1992 broad wings appeared only in the Balmer lines, in the CaII lines this effect was not observed and broadenings of hydrogen lines were symmetric. On decomposition of these profiles into the sums of two Gaussians the widths of the broad component corresponded to velocities of up to 250 km/s. Thus, Gunn et al. assumed that such components were formed by superimposing many differently oriented ejections during the exposure. At maximum brightness of the flare on AT Mic of 21 August 1985 with $\Delta U \sim 4^m$ the widths of the $H_\gamma$–$H_9$ lines near the base increased to 25 Å. This broadening persisted as long as an enhanced continuum was seen, whereas CaII lines did not broaden in this case either (Garcia-Alvarez, 2002).

During spectral and photometric observations of the dM4e star RE J0241-53N from South Africa, Ball and Bromage (1996) recorded a strong flare with $\Delta U = 4^m$ that lasted for more than 2.5 h. Appreciable broadening of the Balmer lines increased in high members of the series and was equal to 8 Å in the $H_\delta$ profile. Using the profiles of $H_8$ and $H_9$ they estimated $n_e$ as $\sim 2 \cdot 10^{15}$ cm$^{-3}$. The lines $H_\delta$, $H_8$, $H_9$, and CaII K showed simultaneous Doppler shifts at the phase of slight preflare brightening of the star in the U band, and similar simultaneous shifts took place at slight brightenings during the gradual phase of the flare.

When purely flare line profiles of the flare on LQ Hya of 22 December 1993 were represented by two Gaussians, the FWHM of the narrow and broad components of $H_\alpha$ emission was 59–69 km/s and 190–293 km/s, respectively. The broad component at the impulsive phase provided about 80% of the total flux in the line. The broad component of the $H_\beta$ profile at the maximum contained 86% of the total flux in the line, and this emission decayed slightly faster than $H_\alpha$; at the maximum brightness both Balmer lines manifested a slight shift of the broad component toward short wavelengths, which was then changed to the shift of another sign (Montes et al. 1999). Simultaneous emission was also recorded for the helium $D_3$ and $\lambda$ 6678 Å with broad wings.

In two flares on AD Leo recorded with HST/STIS in March 2000, a shift of SiIV $\lambda$ 1403 Å and C IV $\lambda$ 1548 Å up to 30–50 km/s toward long wavelengths was detected, which Hawley et al. (2003) interpreted as a result of condensations.

Fuhrmeister et al. (2005) detected a strong asymmetry of lines in the spectrum of a long duration flare on the M6 dwarf LHS 2034. The asymmetry was shown by hydrogen and helium lines but not by metal lines. The PHOENIX model chromospheres elucidated this effect by the formation of lines in depth or by downward condensations, which brake in the medium with increasing density in the lower chromosphere. An analogous effect but less pronounced was observed for LHS 2397a.

Then Fuhrmeister et al. (2007, 2008) carried out simultaneous XMM-Newton and VLT/UVES observations of CN Leo, which overlapped both the quiescent state of the star and a strong flare. From the FeH lines the magnetic field variations were detected from night to night. They found that different activity levels could be followed from both the X-rays and the forbidden FeXIII $\lambda$ 3388 Å line, and the latter proved to be in good agreement with predictions of the differential EM constructed from X-rays. During the flare the two HeII lines from the transition zone were detected, the estimate of $n_e > 10^{12}$ cм$^{-3}$ was acquired from the OVII triplet lines, a significant flux increase in chromospheric lines was accompanied by a significant enhancement of the continuum, and the Balmer lines showed strong broadening. 1143 emission lines were identified in the 3000–10000 Å range, an asymmetry of the lines of hydrogen, helium, and ionized calcium was detected. The blue wing of the profiles was enhanced at the onset of the flare, and the red one during the decay phase.

Reiners (2009) studied 181 UVES spectra of CN Leo obtained during three nights, and during a strong flare he detected a shift of radial velocities by several hundred meters per second, which was correlated with intensity of $H_\alpha$.

Schmidt et al. (2013) recorded a flare with an amplitude of ΔV ~ $9^m$ on the M8 dwarf SDSS J022116.84+194020.4, a star of $22^m$. Based on the optical and IR spectra and SDSS ASASSN photometry, the spectral type of the star was determined, strong constant $H_\alpha$ emission was detected, the age was estimated to be more than 200 million years, and the total energy $E_U = 10^{31}$ erg. The area of the flare was by an order of magnitude greater than that for most flares on middle M dwarfs. The flare was grandiose in amplitude and filling factor but not extraordinary in total energy. Later, Schmidt et al. (2016) recorded a flare with ΔV < $11^m$ and a total energy of more than $3.7 \cdot 10^{34}$ erg on the L0 dwarf SDSS0533 with strong $H_\alpha$ emission and the NIR continuum. Following the kinematic characteristics, the star is old and attributed to the thick disk.

Aschwanden et al. (2008) collected numerous definitions of different parameters of stellar and solar flares and considered scaling relations between them. Taking into account the fractal relation between the linear size and volume $V \propto L^{2.4}$ instead of Euclidean $V \propto L^3$, they found the following statistical relation between the emission measure and the temperature in the flare peak: $EM_p \propto T_p^{4.3}$ and $L(T_p) \propto T^{0.9}$.

With the aim of studying coronal ejections of matter from stars Vida et al. (2019) considered more than 5500 spectra of 25 single M dwarfs and in 478 spectra detected asymmetry of the Balmer lines, most of these spectra were attributed to the moments of maximum radiation in lines. In most cases the estimated velocities were within 100–300 km/s and did not achieve the escape velocity on the stellar surface, while masses of ejections accounted for $10^{15}$–$10^{18}$ g. These events took place more often on cooler and faster rotating stars with high chromospheric activity.

Wichmann et al. (2014) undertook a high-dispersion study of 11 stars on which the superflares were detected with Kepler, with the aim of searching for their common peculiarities. A part of them proved to be very young and fast rotators, which can explain a high activity level, whereas for other stars with slow rotation the common peculiarities resulting in flares were not found.

The primary mechanism of hydrogen line broadening in plasmas is the Stark effect. According to the classical notions, it is due to microfields of plasma particles. However, Oks (1978, 1981, 2006) proposed and developed a concept of the Stark broadening by low-frequency electrostatic plasma turbulence (LEPT); in the framework of this concept, this broadening can be the dominating factor. At first, Koval and Oks (1983) considered the profiles of hydrogen emissions between $H_5$ and $H_{11}$ from two strong proton flares recorded in Crimea and found that the Stark broadening in them is caused by LEPT. Then, Oks and Gershberg (2016) analyzed three spectrograms of two flares in AD Leo (recorded in Crimea), as well as one flare of EV Lac. From the widths of $H_\alpha$, $H_\beta$, $H_\gamma$, and $H_\delta$ the authors concluded that when interpreting these data in the framework of the classical Stark effect, the electron density of radiating plasma should be very high, up to $10^{15}$ cm$^{-3}$, whereas within the LEPT concept it turns out to be close enough to the electron density derived by Katsova (1990) from the Balmer decrement in flare spectra.

From the high-dispersion spectra of Prox Cen derived with HARPS during more than 13 years Pavlenko et al. (2019) studied the time variations of emission lines of hydrogen, calcium, sodium, and helium. They revealed that all the lines showed variations at least with a characteristic time of up to 10 min and intensities of all lines, except for HeI $\lambda$ 4026 Å, correlated with $H_\alpha$; another helium line $\lambda$ 5876 Å better correlated with $H_\alpha$, but at the activity maximum both helium lines were not seen. During the strong flares the $H_\alpha$ line showed a blue asymmetry at > −100 km/s, and the helium lines $\lambda$ 4026 and $\lambda$ 5876 Å appeared, at the activity minimum the CaII lines almost disappeared and sodium lines weakened.

Using the high-speed spectrograph ULTRASPEC mounted on the 2.4-meter Thai National Telescope, Doyle et al. (2022) recorded two flares from YZ CMi with a total energy close to $10^{34}$ erg with subsecond cadence. Both cases reveal quasi-period pulsations of a few minutes and the absence of such oscillations at higher frequencies. These peculiarities are interpreted as a dynamics of magnetohydrodynamic waves in coronal loops having a length of 0.2–0.7 $R_*$.

E. Magnetometry of flares. The measurement of magnetic field variations during solar flares is one of the most complicated experiments of the solar physics. According to Babin and Koval' (2016), during solar flares there occur variations of both weak and strong magnetic fields of the active region on different time and spatial scales and in different places of the active region, having an irreversible and impulsive character. In one of the studied flares, they found that the impulse amplification in the radio range and in hard X-rays was accompanied by impulse variations of the magnetic field strength in cores of the $\delta$ spot.

As to the stellar flares, one should mention the study of Liefke et al. (2007) who in May 2006 carried out simultaneous three-day monitoring of CN Leo with the XMM-Newton X-ray telescope and VLT/UVES in optical wavelengths. On 19 May 2006 they recorded a strong flare in both wavelength ranges. The results of observations in X-rays and in the red region were described above, and the molecular magnetometry of FeH showed an increase of the magnetic flux of 22 May with respect to that of 20 May by 100 G and the flux of 24 May with respect to that of 22 May by 25 G more.

A set of the listed data leads to the conclusion that there is a real and significant diversity in the behavior of emission line widths in different flares, and one should compare this diversity with the known properties of solar flares in which the half-widths of $H_\alpha$ in bright compact cores achieve 5–10 Å, whereas out of these structures they do not exceed 1 Å. In the cases when the profile asymmetry is observed, the blue anomaly is usually recorded in the pre-maximum phase, whereas the red one achieves its maximum value at the flare maximum, although recently Honda et al. (2018) detected a blue asymmetry of $H_\alpha$ in a flare on EV Lac, which persisted practically during the whole flare.

In the conclusion of this chapter on the observed characteristics of stellar flares in different electromagnetic radiation ranges it is worth noting that their basic difference from the solar flares lies in the fact that among them there are events by 2–3 orders stronger than those on the Sun. Furthermore, in the X-ray and radio ranges there are phenomena qualitatively different from the solar ones, whereas in the ultraviolet and visible regions such qualitative differences are not detected.

## 2.5. Models of Stellar Flares

The secular anniversary of the discovery of stellar flares of the considered type is approaching: as it was mentioned, in 1924 Ejnar Hertzsprung recorded a short-term flare on the weak star in the Carina constellation and attributed this event to the splashes of stellar matter during an asteroid falling on a star. In the previous sections, we outline in detail the collected over this century experimental data on such stellar flares and short reference is made to their theoretical interpretation. In this chapter, we comprehensively consider the phenomenological models which answer the question regarding what radiates in flares and what is the structure of radiating matter with inevitable repetition of some statements from the previous sections. Although the mechanisms of stellar flare radiation in the entire electromagnetic spectrum wavelength ranges have to a certain extent been understood, the self-consistent theory of the ensemble and interaction of such processes on stars is still being developed. The following chapter exposes the state of this theory of flares with consideration of dynamic model explaining the physics of powerful energy release and the general understanding of the fact that the magnetic field energy is a finite source of their energy, and the initial phase of a flare consists in the formation of beams of high-energy particles.

In the 1960s, when the data on stellar flares were accessible only in optical range photometry and from the first results of radioastronomical studies, about ten models of flares were proposed. They were surveyed by Gershberg (1970a). Even for the appearance of ionized plasma above the photosphere of a cool star various schemes were discussed (Gershberg, 1968; Korovyakovskaya, 1972). Here, we shall briefly consider only those initial schemes that are somehow involved in further modeling of the flare activity of red dwarfs.

* * *

In the model of an asteroid falling onto a star proposed by Hertzsprung in 1924 (Oskanjan, 1964) and in the model of accretion of magnetized plasma proposed a quarter of a century later by Greenstein (1950), flare stars were considered as a gravitating center only. However, the data available now on the dependence of the flare activity level on the age of dwarf stars removes all doubts as to the decisive role of stars themselves, their rotation first of all, in the nature of this activity.

The concept of a flare as a hot spot was proposed by Gordon and Kron (1949) to interpret the first photoelectric observations. Over half a century it evolved essentially from a photospheric region cooling down due to blackbody radiation to a short-lived dynamic structure in the atmosphere and is now a natural element of the general picture of optical flares.

An alternative to the hot-spot model, the nebular model, was originally only phenomenological, but it was developed to determine the place and physical conditions of the appearance of chromospheric emission in flares. Today, its confrontation with the hot-spot model has been removed, since both components are included in the complex dynamic picture of optical flares.

Ambartsumian's concept (1954) about the eruption of prestellar matter with a large energy potential from the stellar depth initiated the hypothesis of Gurzadian (1965) about fast electrons and Papas' idea (1977, 1990) about electromagnetic solitons. But, if Papas derived only general equations of such solitons, Gurzadian elaborated the hypothesis: assuming that during a flare above the surface of a cool star a cloud of nonthermal electrons appears, he considered the expected consequences, and compared them to the observed properties of stellar flares. This hypothesis was repeatedly published by Gurzadian and drew the attention of many observers. However, it was subject to serious and all-round criticism, for example, by Haruthyunian et al. (1979), Gershberg (1980), Fomin and Chugainov (1980), and Mullan (1990). Leaving aside their forcible critical arguments, we will make only two remarks. First, the efficiency of fast electrons in the birth of quanta seen in the flare is $10^{-5}$–$10^{-6}$; so this concept can hardly be considered as a solution, if it postulates the existence of an unknown energy source exceeding the energy of flares by many orders of magnitude to explain stellar flares. Secondly, if in the mid-1960s one could still argue about fast electrons that were ejected so far from stars that ionization losses were insignificant for them, which is the approximation underlying Gurzadian's theory, after the discovery and research of stellar coronae such reasonings make no sense.

Having considered a number of models in which radiation of meter wavelength bursts recorded by the late 1960s on UV Cet could be explained by synchrotron radiation of relativistic electrons, Grindlay (1970) estimated the efficiency of these relativistic particles in the generation of X-ray emission due to the inverse Compton effect and nonthermal bremsstrahlung and found that the latter mechanism was more effective for radiation in the energy range $E > 10$ keV. However, as stated above, the level of X-ray emission of stellar flares predicted by Grindlay should lead to such a high total radiation of flare stars of the Galaxy that it would considerably surpass the observed X-ray background.

On the basis of simultaneous optical and radio observations of the flare on UV Cet of 11 October 1972 (Fig. 33) during which the radiation in these ranges was appreciably divided in time, Kahn (1974) proposed the model, which for the first time combined different aspects of the event. From the luminosity, mass, and size of the star he estimated thermodynamic characteristics of its photosphere: at convection up to a depth less than $\tau = 2/3$ an effectively radiating layer should be one kilometer thick, have a mass of 16 g/cm$^2$, and an energy of convective motions capable of generating surface magnetic fields up to 900 G. A recorded optical flare on UV Cet had luminosity of about $2 \cdot 10^{30}$ erg/s and a duration of about 10 s. Kahn estimated the total number and energy of relativistic electrons that could provide such luminosity in the lower corona through the synchrotron mechanism. Kahn simulated further development of the event with the radio emission at 408 MHz within the framework of the hypothesis about the emersion of a plasma bubble in the corona that was formed during reconnection of magnetic flux tubes. Assuming that the stellar corona was in the state of a stationary isothermal wind, Kahn estimated the size of the flare as about $2 \cdot 10^9$ cm, the magnetic-field strength in it as about 600 G, the total number of "optical" relativistic electrons as $10^{34}$ and the rise of the plasma bubble in the corona up to $3R_*$. Within the model, up to 40% of the energy produced by convective motions in the photosphere was spent for the maintenance of the corona. However, polarimetric and colorimetric observations showed that the contribution of synchrotron radiation to the optical flare was negligible.

Grandpierre (1986, 1988, 1991) formulated and developed the convective model of stellar flares. According to this model, ascending convective cells acquire the speed of sound in subphotopheric layers, shock waves formed ahead of them generate all the observed effects of flares: shock waves cause heating and turbulence in the chromosphere, which are observed as its evaporation, and the fast particles generated by such waves cooperate with the overlying magnetic field tubes, generate electric fields, and produce bremsstrahlung, recorded as flare radiation. The distribution of accelerated particles along the flux tubes to the bases of loops yields the effects considered within the magnetic models of flares. But since traditional convective motions are subsonic, to obtain fast convective cells, Grandpierre postulated the existence of additional fast cells caused by the heterogeneity of thermonuclear burning in the stellar nucleus. However, solar-type activity was observed also in M6–M9 dwarfs in which no thermonuclear burning occurs. One should note that starting the development of his hypothesis, Grandpierre strictly criticized all the magnetic models, but in the version of 1991 he included practically all the elements of such models.

Apparently, all models of flares on red dwarfs that to various extents do not correlate with modern models of solar flares have lost their value now.

* * *

When the initial version of the nebular model with the relaxation of homogeneous plasma was found to be unable to present the observed colorimetric characteristics of optical flares, several attempts were undertaken to present these characteristics within the framework of the two-component model.

Mullan (1976a) constructed the model in which the basic component of a stellar flare was a significant volume of plasma of coronal temperature arising at the onset of a flare. Heat conductivity was considered as the basic heating mechanism of underlying layers of the stellar atmosphere. However, such a structure could not provide fast development of flares during the impulsive phase and better corresponded to the gradual phase only. But, with the help of the model Mullan estimated the expected ratios of luminosities $L_X/L_{opt}$.

Kodaira (1977) made calculations within the two-component model to present the characteristics of the strong flare on EV Lac of 3 August 1975 (Kodaira et al., 1976). According to his estimates, the observed radiation could emanate from such a structure: the hot component with $T \sim 10^8$ K, $n_e \sim 5 \cdot 10^{10}$ cm$^{-3}$ at its sizes of about a stellar one and its "cold legs" with $T \sim 10^5$ K, $n_e \sim 5 \cdot 10^{13}$ cm$^{-3}$. It was assumed that the hot component contained the main energy store of the flare, which due to heat conductivity was transported down and radiated in X-rays, whereas the cold component cools down due to the effective radiation in the saturated ultraviolet lines and the continuum, and in the flat optical continuum recorded on EV Lac. The system is able to retain at a magnetic field of several gauss. However, further observations did not reveal the strong X-ray emission predicted by Kodaira corresponding to $T \sim 10^8$ K and the expected ultraviolet emission in the region of $\lambda < 1500$ Å.

* * *

Grinin and Sobolev (1977, 1988, 1989) stated and developed the idea about the higher energy of fast particles in stellar flares as compared to solar flares. They found that electrons with energies higher than 100 keV or protons with energies of 5 MeV in the total flux of $10^{12}$ erg/cm$^2$ can reach the layers with density to $10^{15} - 10^{17}$ cm$^{-3}$ and directly heat the stellar photosphere and initiate continuous flare emission. In their model, deep atmospheric layers were heated simultaneously with fast particles reaching these layers and with optical and ultraviolet quanta generated in higher layers under the passage of fast particles. Later, Sobolev and Grinin (1995) calculated the Stark profiles of $H_\alpha$, $H_\beta$, and $H_\gamma$ within this model and found that profile wings could expand for 20 Å from the centers of lines. They also well presented the wings of the $H_\beta$ line beyond 6 Å from its center in the spectrum of the flare on EV Lac of 1 September 1992 recorded in the Crimea. But, according to their calculations, $H_\alpha$ should be more broadened than $H_\beta$ and $H_\gamma$, which apparently does not fit the observations.

Van den Oord (1988) compared the efficiency of excitation of flares by fluxes of electrons and protons and concluded that the efficiency of the latter was higher, since in this case fewer fast particles were required. Furthermore, one should not expect a correlation between optical and microwave bursts under excitation of optical flares by protons, which is supported by the observations.

Later, Grinin (1991) made a detailed review of the role of accelerated particles in solar and stellar flares. Having analyzed the observations, he concluded that in both cases the role of particles was important, but the sections of particle beams and their energy per unit surface in strong stellar flares were an order of magnitude higher than in solar flares. According to his estimates, in stellar flares protons play a considerable or dominating role in the fluxes of accelerated particles, whereas in solar flares, electrons dominate. The lower limit of the energy spectrum of particles in stellar flares is several times higher than the appropriate value in solar flares. Extrapolating variations of the energy spectrum of accelerated particles in solar flares of different intensity, one can expect that in strong stellar flares this spectrum is harder than in solar white-light flares, which shifts the maximum of energy release into the deeper layers of the stellar atmosphere and on stars one can expect a higher energy ratio $(L_{opt} + E_{UV})/L_X$ than on the Sun.

The most detailed analysis of heating and ionization of the chromospheres of red dwarfs by fluxes of protons with regard to the observed shift of $Ly_\alpha$ in the flare spectrum was made by Brosius et al. (1995): according to their estimates, the shift should proceed over 0.1–14 s.

Finally, Grinin et al. (1993) calculated the heating of fast electrons in stellar flares and estimated the expected color indices in the UBVRI system depending on the characteristics of electron beams and the position of the flare on the stellar disk.

Recently, Nizamov (2019) put forward an idea that the "photospheric burn", yielding the optical continuum of a flare, occurs under the action of X-ray flare radiation of coronal loops.

* * *

Following the pioneering work by Kostjuk and Pikel'ner (1974), who were the first to show that the flux of fast electrons arising at the onset of a solar flare resulted in fast formation of a high-temperature plasma sheet at some depth of the atmosphere from which two hydrodynamic disturbances propagate upward and downward. Katsova et al. (1980) and Livshits et al. (1981) found that in the denser stellar atmosphere at the total power of the flux of particles to $10^{12}$ erg/(cm$^2$ · s) a downward disturbance should consist of a jump of the temperature and a shock wave (Fig. 70).

The shock wave goes ahead of the temperature jump slowly moving downward, and due to strong radiative relaxation behind the wave front a jump of density by two orders of magnitude is formed. At thicknesses from 1 to 10 km, which is less by an order of magnitude than that on the Sun, this compression can reach an appreciable optical thickness in the continuum, which will provide appreciable continuous radiation of a stellar flare. In this downward disturbance one can expect a density of matter up to $10^{15}$ cm$^{-3}$ and a temperature up to $10^4$ K, in this case the profiles of Balmer lines should be broadened by the Stark effect and have a red asymmetry, as is observed in many stellar flares. Physical parameters of this source of continuous radiation appeared to be close to those of the stationary model calculated by Grinin and Sobolev (1977), but now it is formed in an essentially nonstationary situation caused by short injection of the flux of fast particles and is displaced from the region of the temperature minimum in the stellar chromosphere.

The gasdynamic model was repeatedly and successfully used to interpret various observed facts. Thus, Katsova and Livshits (1986) showed that this model provided a natural explanation for some features of the light curves of impulsive stellar flares, in particular the fact that all structures on them were longer than hundreds of milliseconds. Analyzing the burst of CIV $\lambda$ 1548/51 Å emission during the flare on EV Lac of 6 February 1986 recorded by Astron (Burnasheva et al., 1989), Katsova and Livshits (1989) interpreted it as a direct observation of the process of shock-wave formation with radiation accompanied by explosive evaporation of the chromosphere. The analysis of spectral and photometric observations of the impulsive flare on YZ CMi of 4 March 1985 within the framework of this model allowed Katsova et al. (1991) to estimate the area of the source of continuous optical radiation as $> 5 \cdot 10^{17}$ cm$^2$ and to explain the rather fast evolution of the emission Balmer decrement.

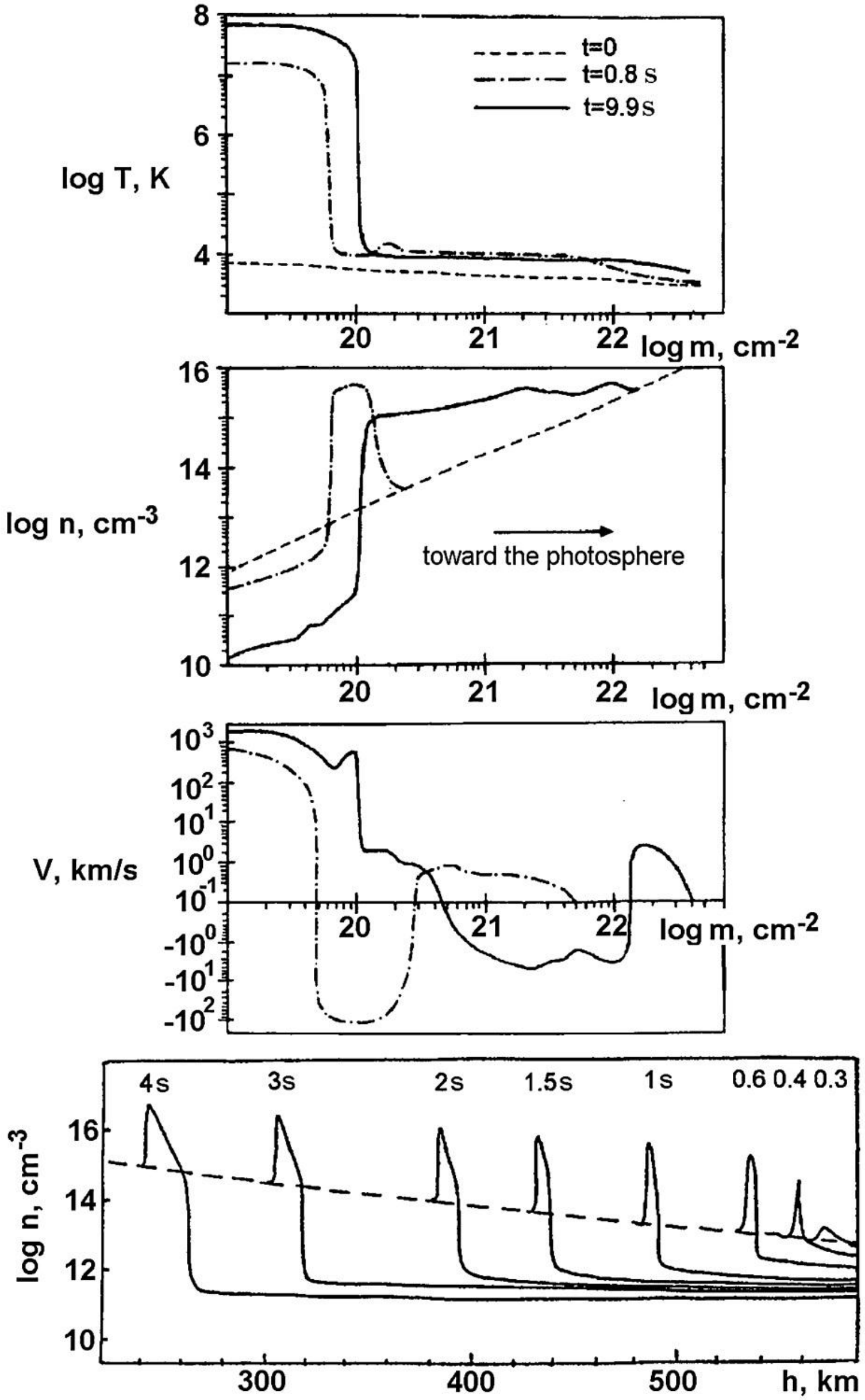

Fig. 70. Distributions of temperature, velocity, and density of matter in the gasdynamic model of a flare (Katsova, 1981b) and shock-wave propagation in the atmosphere of AD Leo (Katsova et al., 1997)

Later, Katsova et al. (1987) redesigned their initial calculation algorithm for the gasdynamic model by including separate consideration of electron and ion temperatures of plasma and recalculated the consequences of injection of a separate 10-s beam of fast electrons with an energy of $3 \cdot 10^{11}$ erg/(cm$^2 \cdot$ s) on the magnetic tube throughout its height from the photosphere to the corona. Calculation showed that during the first 0.1–0.2 s the plasma was considerably heated at the level of the upper chromosphere and, as in previous calculations, two gasdynamic disturbances were formed and propagated up and down from the region of initial heating. In the disturbance going down, the resulting compression produces optical radiation. The compression has a nonuniform temperature structure that varies appreciably with time. The obtained model is applicable to flares that are not too strong. It agrees with the results of Cheng and Pallavicini (1991) for the top part of the tube and refines their results on the chromospheric level. The bottom panel in Fig. 70 presents the calculation of the density profile of the downward disturbance in the atmosphere of AD Leo (Katsova et al., 1997), the velocity of this disturbance smoothly varies from 160 km/s 0.4 s after the onset of the flare to 18 km/s 8 s later. The density of matter in this compression achieves $10^{16}$ cm$^{-3}$ and conditions are close enough to LTE, the optical thickness becomes close to unity 2 s after the onset of the flare and is kept at this level for several seconds. In the upward disturbance, the ion temperature at first considerably lags behind the electron temperature and only after 6 s do they equalize at the level of 30 MK, and the velocity of this disturbance changes from 200 km/s 0.28 s after the onset of the flare to 1000 km/s 8.5 s later. The evaporating chromosphere fills the loop at a speed of $5 \cdot 10^{18}$ particle/(cm$^2 \cdot$ s). If a flare is modeled as a sequence of fast injections, the proposed algorithm enables simultaneous calculation of optical, EUV, and soft X-ray emission of the heated tube. Katsova et al. found quite good agreement of the calculations with characteristic properties of flares in these wavelength ranges. With the help of this algorithm, Livshits and Katsova (1996) considered the conditions of occurrence of extreme ultraviolet emission during pulsed heating of the magnetic tube and found three different regions of formation of such radiation with different time characteristics. Then, Katsova et al. (1999b) considered the influence of stationary X-ray emission of the atmosphere on the condensation responsible for optical emission of the flare on AD Leo. They found that the allowance for this radiation noticeably improves agreement with observations regarding the intensity of $H_\alpha$ and CaII K lines and continuum, since because of X-ray absorption a background source of continuous and line radiation is formed. After evaporation of the chromosphere the efficiency of fast particles falls in this place, so the pulsed heating is needed in a new place to restore such structures.

Katsova et al. (2002) analyzed the flare on EQ Peg of 23 June 1994 observed in X-ray and optical ranges: at its maximum several fast optical bursts took place, which were followed by a gradual decay. An X-ray maximum started four minutes after the optical one. The ratio of luminosities $L_X/L_{opt}$ grew during the flare from 0.24 in the first peak to 15 in the transition to a slow decay. Gasdynamic modeling of bursts enabled estimation of the size of the optical continuum source as $7 \cdot 10^{18}$ cm$^2$ and the amount of evaporated plasma in the corona. The calculated mass of coronal plasma and the observed rather slow decay of flares suggest that this event combined the properties of compact and two-ribbon flares.

* * *

Using the formalism of Cram and Mullan (1979) proposed for calculations of the chromospheres of quiet stars, Cram and Woods (1982) calculated the structure of the chromosphere of a stellar flare, assuming that a quiet stellar atmosphere with $T$ = 3500 K, $\log g$ = 4.75, and with a solar abundance of elements was influenced by some heating mechanisms that changed the temperature variation with height. Calculations were executed for a plane-parallel geometry, hydrostatic equilibrium, non-LTE ionization of hydrogen, and LTE ionization of other elements. From six constructed models, i.e., from the dependences of temperature and density of matter on column density, Cram and Woods calculated the equivalent widths of $H_\alpha$, $H_\beta$, and $H_\gamma$ and the flare continuum in the range from 500 to 10000 Å. They obtained different anticipated variations of the initial spectrum of the star. As the transition zone shifts downward, which can be due to the influence of heat conductivity or in the gasdynamic model, a strong and narrow $H_\alpha$ emission with a weak central inversion should appear. A temperature increase in the upper chromosphere can occur because of the action of X-ray or ultraviolet emission or low-energy electron fluxes and lead to strong $H_\alpha$ emission with deep central inversion and a large Balmer jump. A temperature growth in the lower chromosphere and upper photosphere can arise under the effect of fluxes of high-energy electrons or protons, or hard X-rays, or magnetic energy dissipation in the depth of the atmosphere. At the very large energy release in the chromosphere the obtained model is localized in the two-color diagram near to observed flares.

* * *

Natanzon (1981) seemed to be the first to apply the concept of stationary coronal loops proposed by Rosner et al. (1978) for the Sun to the analysis of stellar flares developing in such structures: from the then known X-ray observations of five flares on four stars he estimated a 5–6-fold temperature rise in stellar flares as compared to flares in solar loops, an excess of density of matter by many tens of times, slightly higher luminosity $L_X$, and an emission measure higher by 4–5 orders of magnitude.

Gasdynamic models of X-ray stellar flares also evolved from numerous models of solar X-ray flares, which from the early 1980s considered the response of coronal loops to various temperature variations. The coronal models considered the chromosphere as a boundary layer and the mass holder, and all attention was focused on the events occurring in the optically thin top part of the loop.

If we ignore the differences in the suggested structures of local magnetic fields, the basic qualitative difference in the energy of solar compact and two-ribbon flares will be that in the first case only initial pulse heating is suggested, while in long events of the second sort there is additional heating in the decay stage. A source of such heating, according to the solar model by Kopp and Pneumann (1976) and Kopp and Poletto (1984), is the process of reconnecting of magnetic flux tubes in the vertical current sheet formed at the break of initially closed loops during the primary strong energy release. The reconnection results in restoration of closed structures. Even for the one-dimensional case this model is described by a system of nonlinear equations with partial derivatives and supposes only numerical solutions. They yield the rate of energy release at each moment of time as a result of reconnection, density and temperature of the flare plasma, position and the speed of rise of the point of reconnection of the magnetic flux tubes. The resulting model depends on the product of the flare size and magnetic-field strength, thus absolute values of flare parameters can be obtained only by an independent estimate of one of the factors.

Poletto et al. (1988) applied the model of magnetic reconnection to analyze the decay phases of two long stellar flares: the light curves of the flare on EQ Peg of 6 August 1985 recorded by EXOSAT in the ME region and those of the flare on Proxima Cen recorded by the Einstein Observatory on 20 August 1980. Within the framework of this model the total energy of each flare and descending branches of their light curves were presented. Adding the Saar and Linsky (1985) estimates of the strength of photospheric magnetic fields on dMe stars, Poletto et al. estimated the area of the flare on EQ Peg as about 1% of the stellar surface at a loop height of $0.18R_*$ and a density of flare matter from $6 \cdot 10^{12}$ to $4 \cdot 10^{11}$ cm$^{-3}$ and on Proxima Cen as about 0.2% of the stellar surface at the loop height not higher than $0.09R_*$ and a density of about $1 \cdot 10^{12}$ cm$^{-3}$.

For the preliminary analysis of stellar flares attributed to the compact type, Kopp and Poletto (1990, 1992) proposed point-wise (zero-dimensional) analytical models that provided the time change of thermodynamic parameters of plasma averaged over the whole magnetic loop of the flare. Within the framework of these models they concluded that fitting of input conditions and loop geometry enabled flares with EM = $10^{51}$–$10^{53}$ cm$^{-3}$ to be obtained, and that strong flares could arise only in large loops of up to $2 \cdot 10^{10}$ cm.

Simultaneously with Poletto et al. (1988), Reale et al. (1988) carried out independent analysis of the flare on Proxima Cen of 20 August 1980 within the framework of the hydrodynamic model of solar compact flares modified for stellar events using the one-dimensional and one-liquid Palermo–Harvard program. As opposed to the previous model with reconnection of magnetic flux tubes, here pulsed heating of an originally static coronal magnetic loop from inside was suggested. By fitting model parameters, Reale et al. reproduced the observed flare light curve, the distribution of energy in its spectrum and the maximum temperature under energy dissipation over 700 s of $6 \cdot 10^{31}$ erg in a semicircular rigid loop with $h = 5 \cdot 10^9$ cm, which is slightly less than the stellar radius and $n_e \sim 10^{11}$ cm$^{-3}$. Introduction of supporting heating enabled further reduction of the loop size. It is necessary to remember that Haisch et al. (1983) considered a rather slow decay of this flare as evidence of its similarity to solar two-ribbon flares.

To reveal the common properties of hydrodynamic models as applied to stellar analogs of solar compact flares irrespective of any actual events, Cheng and Pallavicini (1991) executed extensive numerical calculations of the models for dMe stars with the help of one-dimensional two-liquid programs aimed at determining the real area of allowable variations of the parameters involved in the calculations. At first, they constructed the equilibrium models of magnetic loops of $2 \cdot 10^9$ and $8 \cdot 10^9$ cm in size expanding from the chromospheres to coronae under doubled solar acceleration of gravity, energy absorption providing loop balance, and a chromospheric temperature of 9000 K. Then, assuming five-minute energy release of different power at the loop top, they traced its further evolution. It appeared that the energy release resulted in fast heating of gas to 40–100 MK, a fast — over several seconds — shift of the temperature front downward, and heating of the chromosphere that caused its evaporation with a velocity of 400 to 2700 km/s, to lowering of the transition zone and the compression of matter at the base of the loop. In the decay phase gradual or catastrophic cooling of gas occurs. If the energy release stops at the flare maximum, conditions for the formation of condensations are formed in cooling down plasma, the condensations with a velocity of several hundred kilometers per second reach the loop base. If the energy release proceeds after the flare maximum, the plasma cools down slowly and condensations are not formed. These calculations reproduce the observed X-ray flares, in general their intensity, light curves, and average coronal temperature. In particular, a flare with EM up to $10^{52}$ cm$^{-3}$ can be obtained in a rather small loop with $h \sim 2 \cdot 10^9$, and the loops with $h \sim 8 \cdot 10^9$ cm are enough for the flare with EM $\sim 10^{53}$ cm$^{-3}$. Cheng and Pallavicini managed to reproduce the observed correlation of the calculated total energy during the development of flare $E_X$ and EM in the range of $10^{50}$–$10^{53}$ cm$^{-3}$. As stated above, Katsova et al. (1997) refined these calculations for chromospheres.

Within the hydrodynamic model of compact flares, Reale et al. (1993, 1997) developed the technique of analyzing the decay phase for various gravity accelerations, loop sizes, maximum temperatures, and characteristics of additional heating. On the basis of extensive calculations of such models with the obvious account of supporting heating they found the empirical relation between the time of an *e*-fold decay of the flare brightness and the inclination track of the flare on the plane ($\log n_e$, T), where $n_e$ was estimated from EM$^{1/2}$; the value of inclination also provided information on the presence or absence of supporting heating during the flare: at a slope of 1.5 there was no heating and at a slope of 0.5 this heating determined the rate of flare decay. In particular, their analysis showed that there was no such heating in the flare on Proxima Cen of 20 August 1980. In the Yohkoh observations of the Sun, Reale et al. (1997) tested the technique of quantitative analysis of stellar flares from their decay phase:

comparison of direct measurements of the sizes of solar flares and the estimates with the help of the (EM,$T$) diagram yielded quite good agreement.

Continuing the studies of Reale et al. (1997), Reale and Micela (1998) elaborated the algorithm for direct estimate of the sizes of coronal loops from ROSAT PSPC data. The algorithm, without modeling of an actual flare, used earlier calculations of many hydrodynamic models of compact flares with the obvious account of supporting heating. They found that errors of such estimates were ~20% for 40000 readouts and ~70% for 1000 readouts. Application of the algorithm was illustrated by the example of flares on AD Leo and CN Leo. Using the parameters of the above flare on YY Gem of 29/30 September 2000, on the basis of the (EM, $T$) diagram, Stelzer et al. (2002) constructed the model of a flare loop with a length of $2 \cdot 10^9$ cm, the density increased to $5 \cdot 10^{11}$ cm$^{-3}$, and noticeable heating during the decay phase. The short length of the loop — of about $0.6R_*$ — points to the affiliation of the flare to one of the components of this binary system rather than to the intercomponent medium.

Observations of a strong flare on Algol during an eclipse allowed consideration of two competing models of the decay of stellar flares — quasistatic cooling in long loops and heating in the decay phase — and established that the latter yielded more exact estimates of loop sizes. Therefore, an algorithm for analyzing flares with supporting heating was applied to a strong flare on EV Lac recorded with ASCA and to five flares on AD Leo recorded with different instruments. In all cases, rather compact loops with $h \sim (0.1 - 0.5)R_*$ and the essential effect of supporting heating were obtained. In this case, the light curve should be determined by the evolution of additional heating rather than by free cooling of flare plasma (Favata et al., 2001).

Reale (2002) compared all the above modeling schemes of flare-decay curves and provided the list of their applications to real events.

Within the hydrodynamic approach, Reale et al. (2004) thoroughly modeled the strong flare on Proxima Cen of 12 August 2001 (Fig. 33). They found that its burning, maximum brightness, initial decay and secondary maximum could be presented by the two-component model: a single pulse-heated loop responsible for the main maximum and an arcade of fewer hot loops of similar large size that flared half an hour later and were responsible for the secondary maximum; pulsed heating occurred at the level of loop bases, with smooth decay in their coronal parts. Then Reale et al. (2005) and Testa et al. (2005) presented the calculations of structures and evolution of coronal loops in different regimes of their heating, an account of additional heating results in a noticeable decrease in estimates of loop sizes. Later, Reale (2007) supplemented the known analysis of flares at the stage of their decay with the analysis of burning and brightness maximum of flares. From the assumption concerning the equilibrium of the flare loop at maximum and using the relations of equilibrium and hydrodynamic calculations he acquired diagnostic formulae for the loop length, density, inclination angle, and time.

Allred et al. (2006) performed the detailed calculations of radiative hydrodynamic models of optical and ultraviolet emission of flares on M dwarfs. The flares were modeled as a hydrodynamic response of the M dwarf model atmosphere to the flux of nonthermal electrons. Equations of radiative transfer and statistical equilibrium were considered in the nonLTE approximation for many transitions of hydrogen, helium, and ionized calcium; the calculations were performed for medium and strong heating beams. In both cases the flare dynamics is divided into two phases: the initial soft phase in which hydrogen and helium radiate most part of the beam energy and the explosive phase characterized by strong hydrodynamic waves. At the initial phase the matter of the lower chromosphere evaporates into the upper atmospheric regions, producing a dramatic increase of many lines and continua, the HeII $\lambda$ 304 Å line becomes the strongest in the flare spectrum. The Balmer lines become brighter as well and show broad wings. Comparison of the calculated and observed Balmer decrements during several flares revealed general agreement. During the explosive phase there appeared waves of condensation and evaporation, in mid and strong flares the separation velocities reached ~ 130 km/s, the rates of condensation decrease ~ 30 km/s. The optical continuum with a maximum of about 6000 Å increased substantially — up to 130% — in the strong flare. Within the performed calculations, an attempt was undertaken to explain the preflare decrease of the stellar brightness. Whereas, the long-term effect of this decrease in red rays detected by Zhilyaev has not been considered.

Vaeaenaenen et al. (2009) analyzed 27 solar flares in two bands of the X-ray range and in $H_\alpha$, found a correlation between the diameter of the X-ray radiating region and the time of radiation decay in $H_\alpha$, used this correlation to estimate a half-length of the flare loop, and then applied it to estimate the sizes of flare loops in flares on solar-type young stars. Without taking into account the heating of loops, their size was $10^{10}$ cm, and taking this heating into account, it was 2–10 times less, i.e., a half-length of flare loops is comparable with the stellar radius for very young solar-type stars.

* * *

Among flares that were the most successfully observed using different methods at different wavelengths, several events enabled rather detailed schemes of these processes to be obtained. Let us consider some of them.

The flare on BY Dra of 24 September 1984 with $\Delta U = 0.22^m$ was observed in the visible range and in soft X-rays by de Jager et al. (1986). Data analysis showed that they could be interpreted as a "burning out" in the stellar atmosphere during the 5-minute impulsive phase by a beam of fast particles of a "well" of $2 \cdot 10^{17}$ cm$^2$ with a bottom temperature above 25000 K. Gas heated up to $10^7$ K with an average density of $2 \cdot 10^{11}$ cm$^{-3}$ goes up covering the area of up to $10^{20}$ cm$^2$ and slowly relaxes radiatively during the gradual phase.

In spring 2019, this radiative hydrodynamic model was used to interpret the described observational results regarding flares on AD Leo (Namekata et al., 2020b). The performed analysis involved consideration of two variants of heating of the flare matter with the aim of representing the obtained widths of the $H_\alpha$ line and ratios of intensities of this line and continuum: the variant of nonthermal heating by the flux of energetic electrons and that of heating by thermal conductivity. The calculations led to the conclusion that the $H_\alpha$ profile was broadened by the Stark effect and self-absorption, and this resulted in the observed nonlinear ratio of intensities in this line and continuum.

De Jager et al. (1989) analyzed a strong impulsive flare on UV Cet of 23 December 1985 with an amplitude of more than $5^m$ in ultraviolet, which was observed photometrically, spectrally within the whole optical range, and in microwave ranges, and in X-rays. They came to the conclusion that the model of this event could be constructed analogously to solar impulsive flares with a vertically extended "cool" component — an optically thick source of visual radiation — about 700 km, which is higher by 1.5–2 orders of magnitude than that on the Sun, a temperature of about 16000 K and electron density of $10^{15}$ cm$^{-3}$, whereas the structures with temperatures of 40 and 10 MK and densities of (2–5) $\cdot$ $10^{11}$ cm$^{-3}$ were found in the hot component. On the whole, this supports the above model of "burnt wells".

The flare on YZ CMi of 3 February 1983 with $\Delta U = 3.8^m$ described earlier by Doyle et al. (1988b) was observed in optical, ultraviolet, and radio ranges. The analysis of the data by van den Oord et al. (1996) led to the conclusion on a downward shift of the transition zone, an effective blackbody temperature of optical radiation of about 9000 K, and an increase of the emission measure in the whole range of temperatures from $10^4$ to $10^7$ K. Radio emission at 6 cm that appeared seven minutes after the optical maximum was interpreted as an optically thick synchrotron in the arcade of loops growing under the current sheet. The absence of emission at 20 cm was attributed to the absorption in the upper corona layers. In general, the flare was compared with a two-ribbon solar white-light flare with reconnection of magnetic flux tubes.

Fisher and Hawley (1990) considered the evolution of the parameters of coronal plasma averaged over a magnetic loop subjected to heating. Confining themselves to heating times longer than the time of passage of sound through the loop, they reduced the system of nonstationary nonlinear equations with partial derivatives to an ordinary differential equation in which the average rate of cooling of a homogeneous-thickness coronal loop was described by its global parameters — size and column density. Without allowance for the role of fast particles, this equation enabled calculations for the flare on AD Leo of 12 April 1985: the change of average temperatures and emission measures of a loop at evaporation, self-similarity, and condensation stages distinguished by Fisher and Hawley at the gradual phase based on the value of the ratio of plasma heating and cooling rates. Then, Hawley and Fisher (1992) calculated five models of magnetic loops from the level of the photosphere to the transition zone under strong flare energy release in the corona. The basic variable parameter of the designed models is coronal temperature at the loop top. At higher temperature, the transition zone determined by the balance of heat conductivity from the corona and radiative losses of optically thin ultraviolet lines is narrowed and displaced downward to higher column densities. But the structure in the region of the chromosphere and temperature minimum is determined by the X-ray and ultraviolet emission field from the top of the loop rather than by heat conductivity. In these calculations of plasma temperatures, non-LTE functions of cooling on atoms of hydrogen and ions of calcium and magnesium were taken into account. The intensities of chromospheric lines calculated using the equations of transfer as functions of the same temperature at the loop top enabled estimation of this temperature variation from the observed change of chromospheric lines and their comparison with the prediction of coronal models of flares. Comparing these calculations with observations for flares on AD Leo of 12 April 1985, Hawley and Fisher (1992) found quite good agreement of the change of coronal temperatures found from the $H_\gamma$ line and the model of evolution of a coronal loop, but there was regular divergence in the analysis of the CaII K line, indicating the necessity of applying additional heating of flare matter one and a half hours after the beginning of its decay. This suggests that the event was similar to a two-ribbon rather than a compact solar flare.

Hawley and Pettersen (1991) compared in detail the characteristics of two flares on AD Leo: of 28 March 1984 with $\Delta U = 2.1^m$ and of 12 April 1985 with $\Delta U = 4.5^m$. In spite of essential differences in luminosity and duration of these events, they found close similarity in the ratio of line and continuous emissions at different phases of these flares, in the character of development of many spectral details. Thus, they concluded that the structures of flares were substantially similar, and the basic difference was concerned with the sizes and duration of the process of energy release. This conclusion was confirmed in analyzing cooperative observations of AD Leo in March 2000, when eight flares of different intensity were registered: they had strongly different amplitudes and durations but equal blackbody temperatures and close ratios of continuous and line radiation (Hawley et al., 2003).

* * *

To interpret the long and strong flare on AU Mic of 15 July 1992 recorded in the extreme ultraviolet with EUVE (Fig. 38), Cully et al. (1994) used the model of quickly extending structure that was earlier used to represent the light curves of optical flares. Fast expansion in the model sharply lowered the density of matter, keeping its temperature rather high for EUV radiation. But later, Katsova et al. (1995) proposed an alternative model of such an event: heating of ejected plasma in the vertical current sheet that provided luminosity at sufficiently high density.

Then Katsova and Livshits (2001) proposed a generalized concept of flares on stars of late spectral types in which their duration was a regulating parameter of the set of such phenomena: from the very fast bursts of 0.1–10 s, through the impulsive events of 100–1000 s to the flares with long-term decay of hours-days; to the latter one there was attributed the already noted extreme for red dwarfs flare on AU Mic of 15 July 1992 and most flares on RS CVn-type stars. Katsova and Livshits (2001) assumed that the first two groups that are characteristic of red dwarfs evolved in low-lying loops of local magnetic fields, while the long-fading flares that are more characteristic of late subgiants were the result of post-eruptive energy release in giant coronal loops and their energy reached $10^{35}$–$10^{36}$ erg, which could be provided by only large-scale magnetic fields.

Robinson et al. (2001) observed the flare on YZ CMi on 22 December 1994 in optical, microwave, and EUV wavelength ranges and found rather different light curves. At 16:20 UT a sharp maximum of a strong optical flare with $\Delta m \sim 3^m$ and a total duration of about 40 min occurred. Simultaneously, a radio flare started at 3.5 and 6 cm, whereas the increased radiation at 13 cm was observed for one and a half hours before and after the optical maximum and reached a maximum level after the decay of higher-frequency radio emission. The flare in EUV was even longer: from 13 to at least 24 UT with a maximum two hours after the optical flare. To interpret these observations, Robinson et al. used the concept of expansion of a magnetic loop, excitation of Alfvén waves in it in resonance with convective motions. Dissipation of Alfvén waves generates intensive turbulence, which causes heating of the plasma with radiation in the EUV range. Turbulence also generates Langmuir waves, which transform into electromagnetic radiation and accelerate particles. But, in further expansion of the loop it gets out of resonance with convective motions and its heating terminates. When a critical height is achieved, a current sheet is formed in the loop, which corresponds to an optical flare. Instabilities developing in this phase result in the formation of a plasmoid from the loop top. The plasmoid moves away from the star: it slowly cools down and produces long luminosity in EUV. High-frequency microwave radiation comes from the dense part of the loop connected to the stellar surface, whereas low-frequency radiation comes from the extending plasmoid.

From the EUVE/DS observations of 134 flares on 44 F2–M6 stars Mullan et al. (2006) carried out a comparative study of flare loops, using the approach of Haisch (1983). The basis for the latter is the obtained from solar observations conclusion on the approximate equality of the time of e-fold flare decay and times of its radiative and thermally conductive cooling at the flare maximum. They found that the flare loops of less than a half of the stellar radius in size occurred for stars of all the considered types — spectral types from F to M and luminosity from V to III, whereas the large loops — up to two stellar radii — only for M dwarfs. The transition from one type of loops to another occurred between K2 and M0 or K5 dwarfs and it was associated with a possible substitution of the envelope dynamo by turbulent one, which can take place not during the transition to a fully convective star but when the convective envelope would cover an outer half of the stellar radius.

* * *

In calculations of the above semiempirical model for the quiet chromosphere of AD Leo, Mauas and Falchi (1996) constructed the semiempirical hydrostatic models for two moments of the strong flare on AD Leo of 12 April 1985 described by Hawley and Pettersen (1991). Certainly, there are no a priori proofs of the applicability of the hydrostatic models to stellar flares. This can be justified only by numerous constructs of such models for solar flares that provided consistent results for various events.

Mauas and Falchi (1996) analyzed the data for the moments 15 and 20 min after the onset of this strong flare: they considered fluxes in UBVR bands and in the $H_\beta$, $H_\gamma$, $H_\delta$, $H_8$, $\lambda$ 4227 Å CaI and CaII K lines. Calculations were carried out for independently chosen values of the filling factor of the stellar surface by the flare. The results are presented in Fig. 71 by curves B and C that correspond to filling factors of 5% and 1%. Spectral features arising in the upper layers of the chromosphere are, in general, better represented in calculations within the model of the flare occupying 5% of the stellar surface, whereas the model of flare covering 1% of the stellar surface better represents the features arising at greater depths. In the 1%-model a temperature plateau is distinctly found at a level of about 8200 K that extends from $\log m$ = −1.1 to −2.3. This model ensures the best consistency of observed and calculated continuum of the flare. Observations carried out 20 min after the onset of the flare, when the continuum appreciably weakened, were presented by the model with even smaller area.

Baranovskii et al. (2001b) calculated semiempirical models for six moments of active state of EV Lac: from the Crimean observations described by Abranin et al. (1998b) two fast flares and the phase of slow decay of the flare of 29 and 31 August 1994, and three moments of decay of the slow flare of 4 September 1995 were selected. The algorithm of Baranovskii et al. (2001a) used for the elaboration of the semiempirical models of the quiet chromosphere was supplemented by an additional unknown, the flare size. The models that were the best in representing the amplitudes of flares ΔU, ΔB, and ΔV , and equivalent widths of emission lines $H_\beta$ and $H_\gamma$ of 1994, and amplitudes in the same bands and equivalent widths, and line profiles of $H_\alpha$ of 1995 were sought in the calculations. Assuming that the temperature plateaux known in the solar chromosphere and in the quiet chromosphere of EV Lac (Baranovskii et al., 2001a) are formed only in a sufficiently stationary environment, the first Crimean calculations of flare models did not involve the plateau. However, the models did not yield positive results: in adjusting the profile of $H_\alpha$ the intensity of the flare continuum was too small, the temperature rise increased the continuum, but the line was too broad. Moreover, the models yielded too high ΔU/ΔB ratios due to a large Balmer jump. Thus, it was concluded that one of the necessary conditions of successful representation of the observations was the inclusion of an extended temperature plateau in the sought flare model. It should be noted that the temperature plateau in the chromosphere appears in the gasdynamic models of solar flares as well (Abbett and Hawley, 1999). The quantitative characteristics of the models obtained by Baranovskii et al. (2001b) are as follows: at increased stellar brightness ΔU = $0.11^m$ –$0.70^m$ optical depths are (2–6) · $10^9$ in the center of $Ly_\alpha$ and 200–2000 in $H_\alpha$, the bottom and upper borders of the plateau are $\log m$ = −1.0–−0.7 and −3–−2, respectively, the electron density on them is $2 \cdot 10^{12-13}$ cm$^{-3}$, the temperature gradient on the plateau varies from 140 to 400 K at absolute temperatures of 5500–6900 K, and the size of flares is 1.3–4.4% of the stellar surface. Thus, the structures responsible for hydrogen emission of flares on EV Lac are less extended in depth, but have greater electron density than in the quiescent chromosphere. The upper limit of the temperature plateau in flares is at a height of 200–300 km, whereas in the quiescent chromosphere it is equal to 700 km and 1800 km in the quiescent solar chromosphere.

On the whole, this picture of hydrostatic flares is consistent with the concept of Grinin and Sobolev (1977) about the localization of a source of optical luminosity of flares in the depth of the stellar atmosphere. Comparison in Fig. 71 of the Crimean models and the B and C curves of Mauas and Falchi reveals a substantial divergence in the lower chromosphere, since Mauas and Falchi considered a strong flare with excitation reaching the upper photospheric layers, and a significant difference in electron densities at the unexpected similarity of models regarding the characteristic sizes of flares, the fact of existence of temperature plateau and the depth of their occurrence, though this refers to the flares of different stars, different initial data, and independent calculation programs.

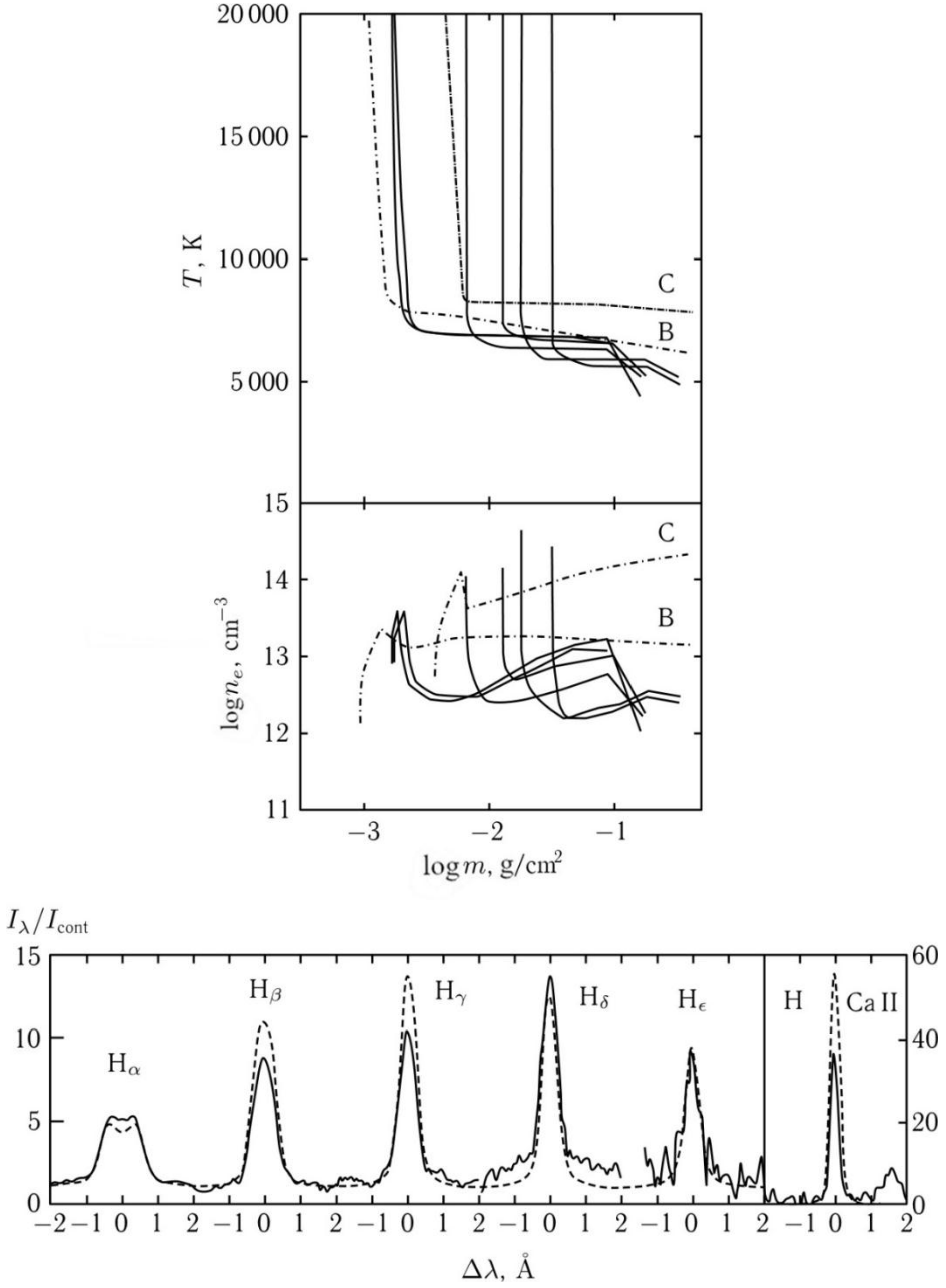

Fig. 71. *Top panel*: semiempirical models of flares on AD Leo (Mauas and Falchi, 1996) and EV Lac (Baranovsky et al., 2001b; see the text); *bottom panel*: representation of emission line profiles by resultant emission of active regions, flares, and microflares on EV Lac of 30 August 1994 at 23:19 UT (Alekseev et al., 2003)

Then, the Crimean algorithm was used for quantitative analysis of the spectrum of the impulsive flare on EV Lac of 30 August 1994 at 23:19 UT and stellar spectra at 15-min intervals before and after the flare recorded with the echelle spectrograph of the Nordic Optical Telescope. It was mentioned that the range of wavelengths was recorded from $H_\alpha$ to CaII H. Owing to the high resolution, broad-profile wings were discovered out of the photometrically recorded flare. The results of multicomponent modeling are presented in the bottom panel of Fig. 71. The top diagrams show the models of three flares by solid curves, which in sum represent the line profiles of the flare on EV Lac of 30 August 1994, curves B and C are the mentioned models of two moments of the strong flare on AD Leo constructed by Mauas and Falchi (1996), dashed lines represent the models of flares on EV Lac from the Crimean observations (Baranovskii et al., 2001b). In this paper it was first showed that flares and microflares overlapped in the depths of occurrence, microflares had slightly lower densities and temperatures. On the other hand, on average, the densities of microflares were two orders of magnitudes higher than those of active regions. They occurred at depths that were hundreds of times deeper in the parameter *m* than the depths of occurrence of active regions. Comparison of the observed and calculated line profiles in the bottom panel of Fig. 71 with Fig. 20 shows that the stellar spectrum during a flare is represented less successfully than for the quiescent state, probably because the hydrostatic stationary model is insufficient for describing impulsive flares.

Jevremovic et al. (1998b) tried to construct the semiempirical hydrostatic models of three weak flares on the dM5.5e star Gl 866 using the similarity of the calculated and observed Balmer decrement as the consistency criterion. They used a spectra in the range of 3400–5400 Å with a spectral resolution of 3 Å obtained at the South African Astronomical Observatory, the model of the photosphere for $T_{\mathrm{eff}}$ = 2900 K with the solar abundance of elements and the condition $dT/d\log m$ = const in the chromosphere. The model of the quiet chromosphere was successfully constructed with the help of the accepted algorithm. Then, the response of the calculated chromosphere to the bombardment by electron beams with energies from 20 to 200 keV was calculated, which resulted in the change of its height, electron density, and the degree of hydrogen ionization. However, the observed Balmer decrements could not be reproduced in such a way.

Later, Garcia-Alvarez et al. (2002) used this approach to analyze a flare on AT Mic of 21 August 1985 with $\Delta U \sim 4^m$ and a total energy of about $4 \cdot 10^{33}$ erg within 3600–4500 Å. They estimated the flare-plasma temperature as 20000–40000 K and the density as $10^{13}$–$10^{14}$ $cm^{-3}$ and its geometric thickness as 200–500 km.

Within the PHOENIX program Fuhrmeister et al. (2010) calculated the model chromosphere in a strong flare on the M5.5 dwarf CN Leo, including the photosphere and a part of the transition zone. In the obtained model, the temperature minimum was deep in the atmosphere, which increased the electron pressure in the flare chromosphere. The filling factor of the flare was 3% and decreased as the flare developed, and in the photosphere it was 0.3%.

As it was already noted, to interpret the observations of flares derived with the high-speed triple-beam camera ULTRACAM, Kowalski et al. (2016, 2018) developed the one-dimensional radiative hydrodynamic model within the framework of which they represented the acquired data on flares as on emission of dense chromospheric condensations at temperatures of about 10000 K and variable optical thickness in the Balmer continuum heated by the fluxes of nonthermal high-energy electrons with variable hardness and a power of an order of $10^{13}$ erg $\cdot$ $cm^{-2}$ $\cdot$ $s^{-1}$. Supposing that the ratio of fluxes in the blue and red rays corresponds to blackbody radiation, they estimated a color temperature of the flare in the optical range at the beginning of burning at a level of about $A_{4170A} \sim 2$ by the value 8500 K, at the flare brightness maximum lasting for 1.1 s — 10000 K, and 250 s after the maximum — 6000 K. An increase of the flare brightness and its fast decay after the maximum without a visible fine structure was used by them to refine the spatial model. But Morchenko et al. (2015), Morchenko (2016), Belova and Bychkov (2019), and Kupryakov et al., (2021) argue against the concept of chromospheric condensation by Kowalski (2015), asserting that a great optical thickness is not achieved in a shock wave in the stellar atmosphere and the Kunkel "photospheric burn" is not responsible for the observed blackbody radiation.

* * *

This may sound strange, but the first paper concerning MHD of space objects by Larmor (1919) was published five years before the discovery of flare red dwarfs by Hertzsprung (1924)! The real beginning of the MHD modeling of flares was attributed to the paper of Giovanelli (1939). His and subsequent MHD studies are outlined in the next chapter of the monograph.

By the mid 1960s, the magnetohydrodynamic concept of flares was actively involved into the theory of solar flares: a floating of the magnetic field tubes from the subphotospheric layers and the reconnection of their field lines. This process is believed to be inevitable at the approaching plasma motions with opposite signs of magnetic fields. This concept has undergone a significant development over half a century.

The electrodynamic model of stellar flares goes back to the paper of Alfvén and Carlqvist (1967), who suggested considering a solar flare as a break in the chain of the strong electric current; such a break can be caused by a local increase of resistance by many orders of magnitude.

Contrary to the Sun, on the stars the hard X-rays were not directly recorded, and only details of spectral line profiles in the high resolution spectra of several flares were an evidence of fast particle fluxes. Other significant difficulties for the construction of the theory of stellar flares are the absence of notions on nonradiative components of their total energy and the absence of unambiguous relations between flare manifestations in different wavelength ranges. The latter circumstance should be considered in detail, although some comparisons of such a kind have already been made above.

We have already described the long-term discussion concerning the role of microflares in the heating of stellar atmospheres associated with a detection of the correlation between the quiescent X-ray emission and the averaged radiation of flares and Balmer emission out of the strong flares. Here it is reasonable to add that X-ray emission emanates from the optically thin plasma of very high temperature, whereas the emission of hydrogen lines from the one thousand times less hot medium and optically thick in the considered lines, i.e. the X-ray emission emanates from the volume, while the Balmer emission from the surface. Balmer lines are detected at the very beginning or even in preflare brightenings, while soft X-rays at the gradual thermal stage of a flare. Thus, the practical proportionality of these emissions is required to be explained.

Not all the individual flares satisfy the found statistical relation between the optical and X-ray emissions.

Thus, an optical flare on YZ CMi of 4 March 1985 with $\Delta U = 1.2^m$ recorded with the 75-cm and 1.9-meter telescopes of the South African Astronomical Observatory and with EXOSAT was not accompanied by a simultaneous flare in X-rays, there was no correlation of the low-amplitude variations in the U band and in the X-ray stellar emission (Doyle et al, 1988b). We have already outlined above the description of observations of EV Lac in October 1985 when rather complicated ratios of activity of this star were detected in X-ray and optical ranges (Ambruster et al., 1989a). An X-ray flare on Proxima Cen of 6 March 1979 was not accompanied by ultraviolet activity. In the course of simultaneous ultraviolet observations with IUE and ground-based photometric and spectral observations of UV Cet Phillips et al. (1988) on 17 September 1980 recorded a flare with $\Delta U = 2^m$ but with a rather weak, by 30%, increase of CIV emission.

There are other similar examples in the previous chapter in which we provide descriptions of simultaneous observations in different wavelength ranges, and for their interpretation there was attracted an idea of anisotropic radio emission of flares, the localization of lower parts of the flares beyond the stellar limb, the concept of development of these flares in the lower atmospheric layers not associated by magnetic fields with the corona — such magnetoizolated regions are known on the Sun (Malaschuk and Stepanian, 2013), the hypothesis on the absorption of hard emission in the upper layers of the atmosphere.

* * *

Thus, from the phenomenological point of view, the existing concept of stellar flares has made it possible to unambiguously identify the flare emission mechanisms almost in all the wavelength ranges and at different stages of their development. However, a physical model of flares is required which would yield the sequence of occurrence and evolution of radiating structures. The next chapter is devoted to these physical models of flares.

## 2.6. Physical Nature of Flares

Since the mid XXth century, the term "solar flare", similarly to "stellar flares" on UV Cet-type red dwarfs, has steadily been associated with the process of extremely fast (100–1000 s) conversion of the magnetic field energy in the solar atmosphere occurring above an active region (Svestka, 1976). In the course of this conversion, the energy of the magnetic field (of an order of $10^{29}$–$10^{32.5}$ erg)[1] and that of electric currents forming it are spent for:

1. the local heating of appreciable gas masses in the solar atmosphere (hereafter the thermal channel $E_{th}$) accompanied by a drastic increase of the electromagnetic radiation intensity[2] in a wide spectral range covering from radio wavelengths and optical radiation to X- and gamma-rays;
2. the acceleration of particles (electrons and protons) up to relativistic energies (hereafter the nonthermal channel $E_{nth}$) accompanied, in particular, by the formation of solar cosmic rays (hereafter SCRs);
3. the formation of ejections of the huge masses of matter from the solar corona (Coronal Mass Ejection — CME) with supersonic and superalfvénic (locally) speeds (hereafter the kinematic channel $E_{CME}$).

The relative fraction of magnetic field energy that is transformed into different release channels differs drastically from flare to flare (in particular, one can observe flares dominating in radiation and flares which are weak in radiation and nonthermal manifestations but with strong coronal ejections (stealth CME), and vice versa, flares accompanied by the generation of a large number of accelerated particles and simultaneously by weak radiation and absence of CME). A wide diversity of flare activity manifestations seem to be caused by differences in configurations of the coronal magnetic field above the flare activity regions determining a type of the flare energy release, by high sensitivity of the "escape" of accelerated particles from the localization of the energy release region toward the magnetic structure, and by its distance from the photospheric surface.

However, one should stress that the maximum energy value that can be released in a flare in each of the channels listed above (heating, acceleration, ejections) turns out to be of one order: $E_{hot} \sim E_{nth} \sim E_{CME}$ (Miroshnichenko, 2015; Ramati et al., 1995; Lin and Hudson, 1976; Aschwanden et al., 2014, 2015, 2016a, 2016b, 2017a, 2017b).

Of particular note is a series of six papers of Aschwanden and co-authors known as Global Energetics of Solar Flares in which authors studied the basic energy release channels in 399 M and X flares (i.e., the strongest ones) observed over the first three years (2010–2012) of the Solar Dynamics Observatory mission (SDO) and in 860 flares associated with CMEs during the first 7 years (2010–2016) of the SDO mission. In this series of papers, including the first performed statistical analysis of the basic manifestations of flare energy release and accumulated magnetic energy for each event (2014 — I. Global Energetics of Solar Flares, 2015 — II. Thermal Energies, 2016a — III. Nonthermal Energies, 2016b — IV. Coronal Mass Ejection Energetics, 2017a — V. Energy Closure in Flares and Coronal Mass Ejections. 2017b — VI. Refined Energetics of Coronal Mass Ejections), Aschwanden could estimate the typical relative effectiveness of each of the main flare energy release channels and the role of secondary conversion effects of flare energy as follows.

---

[1] The flares with energy release of $10^{33}$–$10^{36}$ erg were recorded for the flares on UV Cet-type red dwarfs having the same kind of activity as the Sun (Maehara et al., 2012).

[2] Solar flares are indebted for their name to this "flare" manifestation detected earlier than others.

1. The sum of the mean nonthermal energy of flare-accelerated particles $E_{nt}$, the energy of direct heating $E_{dir}$, and the energy in CMEs $E_{CM}$, which are the primary energy dissipation processes in a flare, is associated with the dissipated magnetic energy as $(E_{nt} + E_{dir} + E_{CME})/E_{magn} = 0.87 \pm 0.18 \approx 1$. Thus, the observed flare energy output in basic channels is comparable with free energy of the magnetic field in an active region, which unambiguously corroborates the magnetic origin of flares and CMEs.
2. The energy partition of the dissipated magnetic field outflowing into the nonthermal channel is: 0.51 ± 0.17 for the electrons with energy of 6 keV, 0.17 ± 0.14 for the ions with energy of 1 MeV, 0.07 ± 0.14 for the kinetic energy of CMEs, and 0.07 ± 0.17 for the direct heating of coronal and chromospheric plasma.
3. The flare energy output into the thermal channel is almost always less than that into the nonthermal channel, which is consistent with the thick-target model.
4. The bolometric luminosity in white-light flares (strong flares visible in the optical continuum similar to the Carrington flare) is comparable to the thermal luminosity in soft X-rays.
5. The energy that is released into the channel of accelerated solar energetic particles (SEP) represents a small fraction of the energy released into the kinetics of CME, which is in agreement with the assumption concerning the acceleration of these particles in a shock wave produced by coronal ejection.
6. The "warm-target" model predicts a lower limit of the low-energy cutoff of the nonthermal electron spectrum at ≈ 6 keV based on the mean differential emission measure (DEM) and the average temperature of electrons $T = 8.6$ MK during flares.

The observed fact of approximate equipartition of the flare energy outputs into the acceleration of particles (electrons and protons), into the anomalous heating and the kinetic energy of ejections is particularly important for further discussion of the suggested mechanisms of flare energy release, as well as the fact of identity of physical processes of energy release occurring in flares, being observed on the Sun and red dwarfs, which was established in the second half of the past century (Gershberg and Pikel'ner, 1972).

Note that the overwhelming majority of external manifestations of solar and stellar flares directly observed by researchers are secondary and often even tertiary manifestations of the primary energy release due to magnetic field dissipation. These secondary-tertiary manifestations were formed as a result of complex superposition of a number of processes:

- the secondary heating of lower dense layers of the chromosphere and photosphere by the thermal conductivity from the superhot region of direct energy release combined with the bombardment of underlying dense layers by accelerated electrons;
- the evaporation of the overheated dense matter of the chromosphere with its subsequent expansion into the corona along with reconnected magnetic fields with the filling of coronal magnetic tubes with the evaporated plasma;
- the relaxation of the heated secondary plasma;
- the fragmentation of coronal magnetic structures by thermal and plasma instabilities with resultant cooling and outflow of the earlier ejected matter downward to the chromosphere;
- simultaneous ejection of the other part of coronal plasma to the solar wind with the formation of CME and a shock wave at the ejection front;
- the escape of accelerated particles and protons from the energy release region to the corona and solar wind;

- the formation of shock waves in the corona and chromosphere with the acceleration of charged protons and electrons at its fronts, the formation of solar energetic particles (SEP), etc.

These and many other manifestations of solar flares are causally connected forming a complex and picturesque pattern of the observed manifestations of flares and determining their dynamics.

## 2.6.1. Historical Introduction

The first recorded observations of a solar flare were made accidentally by the amateur astronomer Carrington on September 1, 1859 during five minutes (11:18–11:23 GT), who observed visually the development of a giant group of spots (Carrington, 1859). At the periphery of the spot groups, he detected an abrupt and short-term (5-min) brightening of four compact regions (Fig. 72), and this proved to be the first widely accepted evidence for the presence of solar flares[1]. On this day, Carrington shared his discovery with his colleagues but initially got a skeptical response, which was dispersed through the confirmation of this fact by observations of a "white-light flare" by the English astronomer Hodgson (1859) who observed these brightenings in the same active region and at the same time independently from Carrington. The strongest aurorae followed just after the flare (18 hours later) and illuminated the whole night sky of the Earth up to equatorial Caribbean islands, as well as the strongest magnetic storm exciting extracurrents that burnt then the system of cable transoceanic-telegraph line, were the first demonstration of both the fact of the presence of solar-terrestrial relations and the huge destructive potential of solar activity for the Earth technology.

Since the white-light flares (the formation of a hot layer of plasma in the solar atmosphere, which is optically opaque in the continuum) require a huge energy release and take place only in very strong and rare flares (once per hundreds of years for flares of the Carrington kind), one failed to detect the new events of this class in spite of many-year attempts of many astronomers-visual observers. A principal breakthrough was made 36 years later (July 15,

---

[1] The first observation of a solar flare in the history of mankind seems to belong to another English amateur astronomer Stephen Gray who 300 years ago (150 years before Carrington) in the course of visual observations of solar flares on December 27, 1705 noticed a short-term bright flash near the spot. Gray determined it in his unpublished then observer's diary and in the letter to his colleague-observer John Flamsteed as a "lightning flash near the spot". Unfortunately, contrary to Carrington, Gray had no independent confirmations of his discovery and did not make up his mind to publish it for public access, hoping to confirm the phenomenon of flashes later in further observations. From "Sunspot Observations during the Maunder Minimum from the Correspondence of John Flamsteed" of V.M.S. Carrasco, J.M. Vaquero, (article in: "Letter 1062 (1705 December 27): [...] I am in Persute [pursuit] of a new Phenomenon of the suns Spots [sunspots.] I say Persute [pursuit] because though I suspect that I have seen it more than once yet I have often looked for it without success tis [sic] this there seems sometimes to Proceed from the West side the Spot as it were flash of lightening which moves round the spot by the north to the East and is there extinguished generaly [generally] but sometimes it arives [arrives] to the south before extinction this is soon after followed by an other [another] such like Phenomenon they succeed each other in about a second of time. the Tremulation of the Atmosphear [atmosphere] I cannot think to be the cause of this Phenomenon but however shall suspend my judgment till I have confirmed the apearance [appearance] by more observations."

1895) by the outstanding astronomer Hale who designed the first[1] spectroheliograph (a device for deriving an image of the solar surface in narrow spectral bands, preferentially in chromospheric lines of hydrogen and calcium) (Hale and Ellerman 1904). In the first observations, Hale derived an image of the solar flare in the chromosphere and its dynamics.

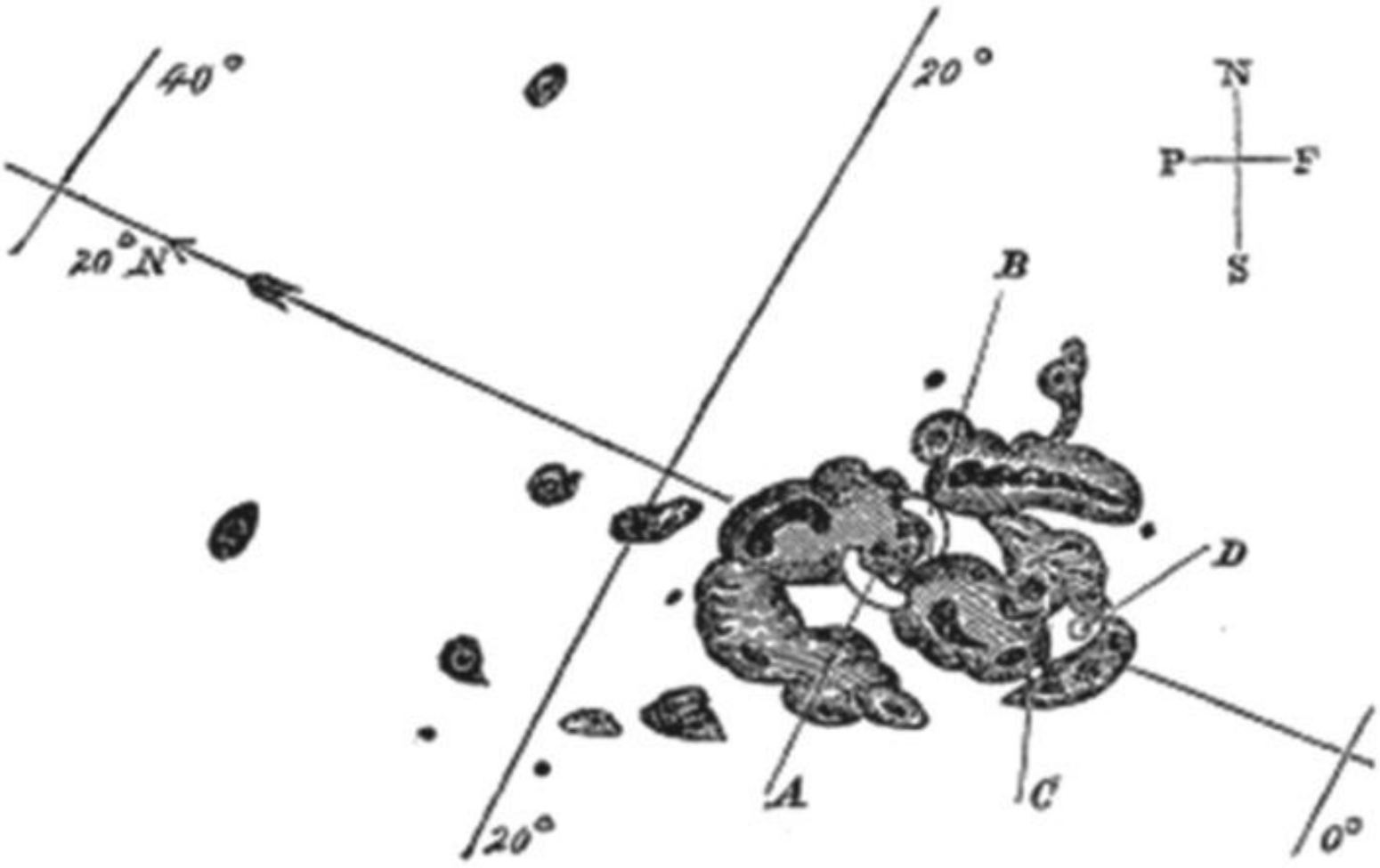

Fig. 72. A sketch of the giant group of sunspots and four knots A, B, C, D of the observed "white-light flare" (Carrington, 1859)

A great contribution to understanding a place of solar flares was made by the global solar patrol organized by Hale in 1930 (35 years later), applying the Hale spectrographs distributed across the globe. This program allowed one to carry out the 24-hour observations of chromospheric activity and showed that flares in chromospheric radiation took place very often and were an ordinary event. Thus, for a long time this phenomenon was named as "chromospheric flares". Just after the observations in ultraviolet and X-rays of flare manifestations not only in the chromosphere but in the corona and recognizing the unified source of both chromospheric and coronal manifestations of flare energy release, there was established the widely accepted today name "solar flare"[2]. Moreover, the numerous observations of flares on red dwarfs showed a very close similarity of the physics of flare processes on the Sun and on the low-mass UV-Cet-type red dwarfs, inducing the term "solar-stellar spot-flare activity".

[1] If to formulate it more precisely, then the first spectroheliograph of the Hale design was constructed in 1893 by Toepfer from Potsdam (see in the article of Annibale Ricco (1895, page 31)) who tried to observe the extrasolar corona at the observatory on the Etna Mountain (unfortunately, the level of images did not allow the structure on the solar disk to be seen).

[2] The history of evolution of the terminology and knowledge on solar flares ("eruption" vs "flare") is described by Cliver (1995).

## 2.6.2. Problem Statement

In this section, we will try to survey the attempts to describe physical processes providing the realization of the observed stellar and solar activity. The corresponding theoretical and model description should answer the following questions:

1. *What is the source of flare activity?* Since it is practically widely accepted now that this source is a magnetic field generated in the convective stellar interior, the question concerning the source should be formulated otherwise: how is the stellar magnetic field generated? This "dynamo-problem" is considered in detail in Chapter 4.3 of the monograph.

Are the generated magnetic fields maintained at the static equilibrium state when the field generation process is balanced by the opposing processes of its dissipation and diffusion outward, or the equilibrium is dynamic with the constant cycle of global field oscillations in a star around the equilibrium value (an attractor in the phase plane)? How stable is the state of such dynamic equilibrium and are the prolonged turnings-off the dynamo process possible, as well as its transition from one state of the attractor to another one with different periods and oscillation (cycle) amplitudes around the equilibrium state? Answers to these questions will allow one to understand the physics of long-term magnetic activity of the Sun and stars, the nature of the observed cycles of their activity and stability of the observed cyclicality.

2. *How do the generated magnetic fields penetrate from the convective zone into the atmosphere of the Sun and stars?* How do the magnetic fields emerging from the convective zone form the observed spots, faculae, active zones — the places of the future flare energy release?

Is this process threshold requiring the achievement of some limit field values in the convective zone or its torsion driving instabilities of the magnetic flux ropes, their emergence on the surface of the Sun (stars) and the subsequent concentration into the magnetic dark spots and brighter faculae around them, or there is a permanent stationary ejection of the excess magnetic flux upward into the photosphere? What is the equilibrium state of magnetic fields in the atmosphere above an active region and in what way does the slow accumulation of free magnetic energy of a future flare occur in the course of preflare evolution? What is the reason of the preflare equilibrium break ("threshold state") and what can perform functions of the triggering mechanism that initiates the flare energy release?

3. *How does the flare dissipation of magnetic fields occur in the atmosphere of a star (the Sun)?* What is the physical mechanism providing an ultrafast conversion of magnetic field energy and currents generated by them in a hot (hundreds of thousands and millions of degrees) and almost ideally conducting plasma of the stellar (solar) corona and the chromosphere? What does the magnetic field energy evolve into in the course of the primary process of its dissipation: into thermal energy of the superhot plasma or into the fluxes of nonthermal and above-thermal accelerated particles, or into kinetic energy of the chromospheric and coronal plasma ejected outward during a flare?

4. *Finally, how are these primary components of the flare energy release transformed into secondary manifestations that one can observe*: into the evaporation and ejection upward of the overheated dense atmospheric layers with emission in optical chromospheric lines and sometimes in the optical continuum, ultraviolet, and X-rays; into the complex dynamics and kinematics of relaxing gas flows partially returning back in the chromosphere, partially ejected to the corona and solar wind, in solar cosmic rays, etc.?

## 2.6.3. Magnetic Nature of Flare Energy Release on the Sun and Flare Red Dwarfs

The understanding by astrophysicists of the magnetic nature of solar and stellar flares was quite dramatic. In the previous chapter, it was referred to the nonmagnetic models of Hertzsprung (1924) — "falling of a body similar to a minor planet on a star", Greenstein (1950) — an accretion of the circumstellar envelope with the magnetic field onto the magnetized stellar chromosphere, Ambartsumian (1954, 1957) — the energy release in the course of decomposition of superdense prestellar states of matter ejected into the outer layers, Gurzadyan (1965) — the appearance in the stellar atmosphere of a large amount of ultrarelativistic electrons reradiating by inverse Compton scattering the background radiation of the red photosphere into the blue and UV continuum was suggested to be a flare source, Grandpierre (1981) — a shock wave formed in the stellar atmosphere by subphotospheric convective motions was suggested to be a flare energy source. The other variant of the flare model that refers to the energy of subphotospheric motions was proposed by Mogilevskii (Krishan and Mogilevskii, 1990) within the developed concept of MHD solitons. He showed that the wave motions in the nonlinear and nonstationary magnetized plasma of the convective zone could cause the formation of MHD solitons which contrary to MHD waves propagate with superalfvénic speeds practically without dissipation and therefore are able to carry out rapidly the significant discrete portions of energy and matter from the subphotospheric layers. Under the conditions of flare stars, this mechanism can provide with energy the strongest stellar flares of the considered type (Gershberg et al., 1987). To date, all these models are only of historical interest for a reader.

The first paper indicating a close relation between solar flares (named then as eruptions) and magnetic fields of spots, in the proximity of which they occurred, was a publication of the young Australian post-graduate student Ronald Giovanelli (1939). Using the statistics of observations of spots and flares, he showed that:

a) the frequency of flares is proportional to the area of spots;

b) the frequency of flares is proportional to the growth rate of the spot area for the positive growth rate and exceeding some critical value (generally $10^{-4}$ of the solar hemisphere area per day);

c) the more complex and tangled the spot configuration ($\gamma$ configuration), the higher the probability of flares;

d) the frequency of flares does not depend on the maximum value of the magnetic field observed in a spot (active region) but only on the rate of the change of a spot (a magnetic flux producing it).

As can be seen in Fig. 73a from the paper of Giovanelli (1939), the probability of flares is proportional to the area of spots (and, consequently, to the magnetic flux), while in Fig. 73b — the probability of flares grows with increasing rate of change in the spot area (whereas only at the spot growth stage). Thus, these diagrams are the first evidence for the detected by Giovanelli close relation of flare activity with area and dynamics of sunspots.

It is noteworthy that despite evident proofs of the univocal relation of flare activity with dynamics of sunspots (and dynamics of the magnetic flux generating them) Giovanelli said that: a) "the more the area of the spot group, the more often the occurrence of flares" and b) the faster the growth of spot area, the more often the occurrence of flares", but he did not dare to formulate in his article an evident conclusion on the magnetic nature of flare energy release.

The main reason of his hesitation was in total impossibility within the existing then notions on the physics of plasma and magnetic hydrodynamics to explain fast magnetic field dissipation under conditions of the solar atmosphere with ideally conducting and fully ionized plasma of the hot chromosphere with a temperature of $10^4$–$10^5$ K (nobody suspected then the existence of the corona with a temperature of millions of degrees). As it was mentioned by Cowling (1953) in his discussion about the magnetic hypothesis of the origin of flares, the time of magnetic dissipation/diffusion in the active region volume should exceed many days and months, which is by many orders of magnitude superior to the observed times of flares. The only possibility to ensure the observed flare energy release is to create a very thin transition sheet between different magnetic fluxes above an active region and to compress it to the thickness of several meters (for the variant of the Coulomb plasma conductivity in this sheet) with a huge density of the current in it, to pump all the plasma through this current sheet, and to dissipate in it all the free fields of the active region. Cowling estimated such an assumption as absolutely unreal.

Thus, a dramatic and deadlock situation arose: the observations clearly pointed to the magnetic nature of the flare energy source, but the existing physical pattern of conditions and processes in the solar atmosphere entirely ruled out such a possibility.

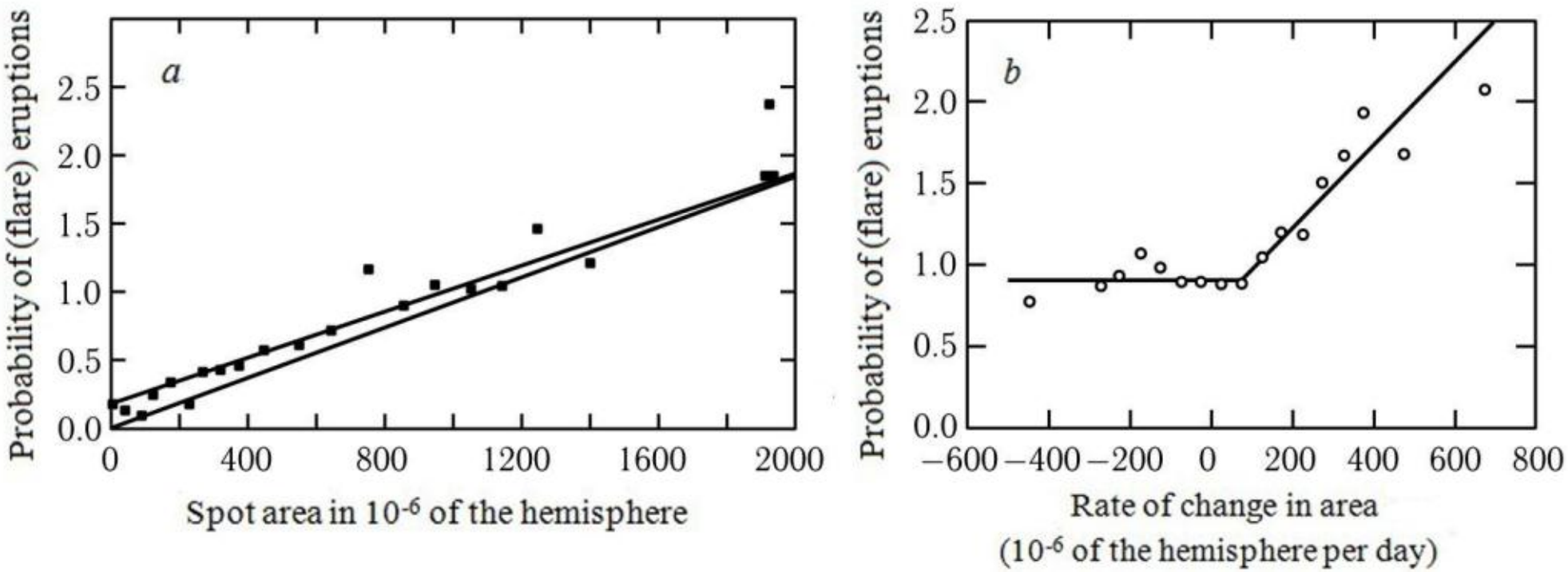

Fig. 73. Diagrams from the paper of Giovanelli (1939)

To resolve the arisen contradiction, Giovanelli (1947, 1948) considered a possible explanation of the flare energy release as a result of the acceleration of charged particles in the atmosphere above an active region under conditions of a rapidly changing magnetic flux of a spot in the particular singular (neutral) point of magnetic topology. He indicated that, first, in the general volume the electric fields produced by this variation of the magnetic flux are orthogonal to the magnetic fields and, consequently, cannot accelerate particles; second, in the case of the presence of several magnetic fluxes with opposite directions of magnetic fields, a neutral point of the magnetic field can arise in which regular magnetic fields are absent, and the induced electric field will be able to effectively accelerate particles in a small neighborhood of this neutral point.

Despite the absolutely incorrect assumptions made by Giovanelli to get a possibility of the analytical description of the proposed scheme (neglecting plasma conductivity in the solar atmosphere in total contrast to the observed almost ideal conductivity in the chromosphere and corona which short circuits electric fields in the plasma (but not currents), an assumption concerning the homogeneous distribution of conductivity, which is absolutely incorrect within the conditions of very high temperature and density gradients in the chromosphere and transition layer, etc.), a suggestion itself to treat the neutral points and currents in them as the only place where can be ensured a fast dissipation of magnetic fields and accelerations of particles in induced electric fields proved to be fruitful and is currently predominant in the astrophysical community.

The other variant of solar flare models based on the assumption about the magnetic nature of flare energy release is the model of a break in the current circuit forming magnetic fields (Alfvén and Carlqvist, 1966) with effective dissipation of the field (current) energy in the region of the break and with acceleration of particles in the double electrostatic layer forming in it.

## 2.6.4. Preflare Equilibrium of the Magnetic Field Above an Active Region

There exist two standard approaches to the description of preflare state of the magnetic field above an active region. The first of them refers to the description of preflare configuration as a superposition of potential fields of several sources[1] comprising one or several singular (null) points at which, in the authors' opinion, the energy for a future flare is accumulated (Syrovatski, 1966a, 1966b, 1968; Somov and Syrovatskii, 1975; Priest and Raadu, 1975). A detailed description of this approach is given in the monographs of Priest and and Forbse (2000) and Somov (2000).

As a mechanism of accumulation of nonpotential magnetic energy at the null point of the magnetic field[2], two processes were considered: a) collapse of the external plasma with the magnetic field with formation of a self-pinching rope (pinch) in which the external magnetic pressure is compensated by the internal gas pressure (Dungey, 1958; Severny, 1958, 1961; Imshennik and Syrovatskii, 1967); b) splitting of the null point into two and more with formation of the current sheet as a result of slow deformations (shifts and rotations) by the convective motions of magnetic fields frozen into the photospheric plasma that is involved into the convective motions (Syrovatskii, 1966).

The widespread appreciation of this approach was stimulated by the possibility of strictly analytical calculation of the two-dimensional preflare configuration formed by the combination of potential magnetic fields of $H_x$, $H_y$, whose sources are located in the $y = 0$ plane (photosphere), and by particular (null) points of the interaction between them.

At the same time, it is this requirement of the two-dimensionality of the task that comprises the basic shortcoming of this approach. This requirement suggests the absence of the longitudinal magnetic field of $H_z$ directed perpendicular to the $x, y$ plane. The introduction into consideration of the longitudinal component makes the point of contact between oppositely directed magnetic fluxes to be revoked of the status of null or singular. The resultant field geometry is reduced to force-free with the magnetic field shear (a turn of the field line grid in the $x$, $y$ plane in the course of variations at the $z$ axis (see Fig. 1 at http://femto.com.ua/articles/part_2/3846.html) — the situation typical of problems of plasma retention. The arisen additional back pressure from the longitudinal magnetic field, $H_z^2/8\pi$, changes radically the problem situation, blocking the development of collapse and formation of current sheets or linear currents (see the review of Pontin (2011)).

---

[1] It should be recalled that the combination of several potential magnetic fields, generally speaking, is not potential and comprises particular points on the boundary between them where the potential approach does not work.

[2] The places where the magnetic back pressure, by definition, is absent.

In the place of the former null point, there appears a region with the magnetic field shear (a shift with turning), and the problem of preflare equilibrium of the magnetic configuration refers to the well-studied problem of plasma confinement in tokamaks with magnetic configuration including the magnetic field shear (Kadomtsev, 1987). Such a description is reduced to the equilibrium of force-free configuration in which the magnetic field, electric current, and the associated electric field are directed at each spatial point strictly parallel to each other and form a force-free extended current sheet on the boundary between different magnetic fluxes with various topologies. Thus, the singularity appearing in the two-dimensional description of preflare magnetic configuration as a combination of potential magnetic fields is a purely mathematical result and it disappears at the transition from the degenerated two-dimensional case to the more general three-dimensional one.

Nonetheless, as shown by Schindler et al. (1988) within the developed concept of general magnetic reconnection, in the three-dimensional case the presence of null points and separatrices may be a particular degenerated case generating reconnections on a ("spike"), plane, and separator. More generally, the reconnection takes place in a "particular" magnetic field line where the effects of local nonideality cause the appearance of the longitudinal component $E$ of the electric field along this line, and the planes perpendicular to the particular line have X-type topology. Whereas a number of particular force lines represent, generally speaking, a continuum[1].

The second approach to the description of preflare equilibrium is based on a force-free character of the magnetic field maintaining it (Lust and Schluter, 1954; Chandrasekhar, 1961; Parker, 1979; Molodensky, 1974). It considers configurations with the nonpotential field generated by current flowing parallel to the magnetic field and having no impact on it: $\mathrm{rot}\mathbf{H} = \alpha\mathbf{H}$, where $\alpha$ determines the level of the local magnetic field twisting. Due to the high complexity of analytical calculations of such force-free configurations, this approach was not popular for a long time. Analytical solutions were obtained only for several simplest variants of geometry, such as a linear force-free rope, flat fields with a shear, toroidal configuration (Priest, 1985).

However, in recent decades, the effective numerical methods were developed allowing one to quite successfully calculate the expected global geometry from the data on the observed photospheric magnetic fields, having a series of simplifying assumptions. The selection of appropriate hidden parameters made it possible to fit model field structures with the observed ones, solving the inverse problem. In such numerical simulations, one succeeds in describing the state of preflare equilibrium in interaction with magnetic fluxes of different twisting and in localizing the places of current concentration (current sheets) on the boundaries between different magnetic fluxes. The force-free description of preflare equilibrium and phase of energy accumulation seems to be the most adequate and widely accepted as compared to the combination of potential fields including singular points or lines.

But this "force-free" approach to the preflare field modeling is not perfect. In particular, the observed photospheric fields used as boundary conditions, generally speaking, are not strictly force-free, since at the photospheric level the contribution of gas pressure and kinetic motion energy in equilibrium of magnetic fields is not negligibly small. Moreover, the general problem of reproducing the force-free structure from the given boundary conditions and fitting to the observed pattern of the distribution of plasma structures in the corona is not exactly defined — absolutely different distributions of the helicity parameter $\alpha(\mathbf{r})$ in space can produce very close resultant field structures. Therefore, in the majority of approaches the modeling is constrained by the case of the variable parameter in space $\alpha(\mathbf{r})$ (nonlinear force-free approximation) found by fitting a model within the framework of which one obtains a pattern that is the closest to the observed one. Progress in computational capabilities achieved in recent years allowed several groups to perform modeling in more general case of the inhomogeneous helicity parameter of the field $\alpha$, which showed a significant difference with the results of earlier and simplified descriptions (Tadesse et al., 2014).

[1] A detailed description of possible variants of the reconnection in the three-dimensional geometry is given by Priest and Forbse (2005).

However, in recent years, there appeared an understanding of irrelevance of the standard force-free approach that is based on the description of preflare equilibrium fields as a structure distributed over the entire active region. The ground for revising the concept of the distributed field was the observations of solar coronal structures with high angular and time resolution carried out in recent years with space telescopes TRACE, Hinodo[1], SDO, at the ground-based 1.6-meter solar telescope with adaptive optics of the Big Bear Observatory (Sharykin and Kosovichev, 2014) and at the stratospheric balloon-borne solar telescope SUNRISE (Requerey et al., 2014). As a result of a series of observations with high and superhigh spatial resolution, it was detected that the base element of magnetoplasma configurations in the solar corona was tens and hundreds of isolated (!) thin and superthin magneto-current ropes up to 150 km in diameter. The observations showed that the equilibrium of magnetic fields in the solar atmosphere was determined by the "magnetic skeleton"— an ensemble of tens and even hundreds of thin magnetic ropes of constant section (Klimchuk et al., 2007; Lypez Fuentes et al., 2008). Such ropes, by definition, cannot be described within the framework of the purely force-free approach but obviously extensively interact with each other, with surrounding background fields and the photosphere in which their bases are frozen-in. Due to the persistence of the magnetic flux in such thin current-magnetic threads, the longitudinal magnetic field in them should be constant all over the thread from the photosphere to the corona, while their magnetic energy in upper atmospheric layers should dominate in the total energy balance and determine the flare energy release in the process of both the accumulation of free magnetic energy and the global equilibrium loss and the triggering of anomalously fast flare energy release.

The idea of self-compression of distributed magnetic fields and currents maintaining them into thin filaments was discussed by Pikel'ner and Gershberg (1959) in the context of structures in the solar chromosphere. These basic elements of the magnetic structure of the solar atmosphere became observable from space in the far ultraviolet (see Fig. 21 and the review of Reale (2014)). Analysis of physics of such twisted isolated magnetic ropes and the problem of their equilibrium are presented in the review by Solov'ev (2008).

[1] The galleries of magnetic skeletons in the solar atmosphere above different active regions are exhibited on the websites of TRACE (http://soi.stanford.edu/results/SolPhys200/Schrijver/TRACEpodarchive.html) and Hinode https://www.nasa.gov/mission_pages/hinode/gallery.html#lowerAccordion-set1-slide1

Global equilibrium of such an ensemble of strongly interacting threads is not reduced to the sum of force-free equilibriums of individual tubes, but it is dynamic equilibrium including the constant redistribution of magnetic tensions between adjacent threads, the surrounding background field, and the photospheric plasma in which bases of current-magnetic threads are frozen-in. On the boundaries between current-magnetic threads there arise tangential field discontinuities — current sheets having numerous nanoflares (Parker, 1988; Rappazzo and Parker, 2013) that are responsible for coronal heating. Equilibrium of such isolated magnetic threads was considered by Solov'ev (2008), while statistical properties of an ensemble of interacting current-magnetic threads were studied by Vlahos et al. (1995, 2002). Observations with the subsecond angular resolution 0.1″–0.2″ (100–200 km) carried out by Sharykin and Kosovichev (2014) in the optical wavelength range ($H_\alpha$), and analogous results derived in the ultraviolet with the AIA instrument of the solar space observatory SDO univocally provided evidence for the presence of magnetic bunches with a diameter of individual threads of $\sim$ 150 km with signatures of nanoflares — spontaneous heating of individual regions up to 4–6 MK for tens of minutes.

## 2.6.5. Modern Description of Physics of Flare Energy Release

To date, it is widely accepted that the physical basis for energy release in both stellar and solar flares is the same process (Gershberg, Pikel'ner, 1972) — energy dissipation of the nonpotential component of the magnetic field and electric currents $\mathbf{J}_1$ which are its sources[1]. As stated above, applying to the solar flares, this breakthrough was made in the pioneer paper by Giovanelli[2] (1946) who not only provided direct observational arguments in favor of the magnetic nature of flare energy release but directly indicated the most important role of null points (in particular, X points) of magnetic configuration as places where this energy release can principally occur under conditions of highly conducting solar and stellar atmospheres.

### 2.6.5.1. Necessity for the Thin Turbulent Current Sheet with Reconnection from the "First Principles" as a Basis for Flare Energy Release of Solar and Stellar Flares.

A. Necessity for the thin current sheet (TCS). The simplest and obvious explanation of the mechanism of magnetic energy conversion during solar flares at the anomalous plasma heating, acceleration of particles, and formation of supersonic ejections of the large gas masses is proposed by the model of thin turbulent current sheet (TTCS) with reconnection of dissipating magnetic fields. The rate of magnetic energy dissipation in such a sheet (energy release power) is described by the Joule-Lenz law $Q = j^2/(4\pi\sigma)$, where $j = neu$ is the electric current density, $e$ — the electron charge, $u$ — the electron current velocity, $n$ — their concentration, and $\sigma$ — the effective plasma conductivity in the region of magnetic energy dissipation. It directly follows from the Joule-Lenz law that to ensure abnormally high observed power of flare energy release, it is required either to ensure the abnormally high current density $j$ in the dissipation region or to create there abnormally low conductivity $\sigma$ (or both former and latter). Let us remember that current is determined by the magnetic field vortex $j \sim \Delta H/(4\pi d)$, where $\Delta H$ is the value of the magnetic field jump in the transition layer between two magnetic fluxes, $d$ is the thickness of this transition layer. It is obviously that the requirement of high current density $j$ to ensure a flare definitely leads to the conclusion on a small thickness $d$ of such a transition layer, i.e., to the conclusion that the flare energy release zone should take the form of a thin current sheet. For the case of the Coulomb plasma with the Cowling conductivity in conditions of the solar or stellar chromosphere, or the corona, this thickness proves to be microscopic (several hundred meters) with respect to the observed scales of magnetic fields and currents in the atmosphere above an active region — tens and hundreds of thousands of kilometers. And a spontaneous compression of currents above the active region into such a superthin sheet would not be possible (Cowling, 1953).

[1] Let us recall that the potential part of the magnetic field $\mathbf{H}_0$ with rot$\mathbf{H}_0 = 0$ corresponds to the state of minimum magnetic energy and, by definition, cannot be a source of energy release.

[2] According to Dungey (1958), Giovanelli first came to this idea in 1938 when he did his practical work at the Mount Stromlo Observatory in Australia. Giovanelli (1939) then showed a close correlation of flare activity with the area of a group of spots (the closer it was, the more complicated the magnetic configuration of the group was) and with the rate of change in the area of spots, especially at the stage of birth and early development of an active region.

B. Necessity for plasma turbulence in TCS and formation of a thin turbulent current sheet (TTCS). The situation is somewhat easier if one takes into account that the growth of the current density in a sheet is up to the critical value $j_{cr}$ at which the current velocity of electrons $u = j_{cr}/ne$ becomes higher than the phase velocity of plasma waves in it, $u > V_{ph}$, which leads to an intensive generation of plasma waves, and such a current sheet automatically transforms into the turbulent state[1]. In this state, in the process of scattering current electrons in the electric microfields of plasma waves, a very low "abnormal" turbulent conductivity is formed (Sagdeev and Galeev, 1973; Bychenkov et al., 1988).

$$\sigma_* = \frac{\omega_{0e}^2}{4\pi\nu_{eff}} \sim k\omega_{pl} \, . \quad (66)$$

Here $\omega_{pl} = \omega_{0e}$ or $\Omega_{Hi}$ is the frequency of basic plasma oscillations responsible for abnormal conductivity (Langmuir's ionic or Larmor's ionic), $\nu_{eff} = \beta\omega_{pl}$ is the effective frequency of scatterings of current electrons in microfields of plasma waves, and $V_{ph} \equiv \omega_{pl}/k_{pl}$ is the phase velocity of plasma waves excited by current electrons (ion-acoustic, magneto-acoustic, Bernstein modes, etc.). $k$ and $\beta$ are the coefficients reflecting specific properties of plasma turbulence and dependent on the level of supercriticality, anisotropy, a spectrum of plasma waves, and similar details (at strongly developed turbulence $k \sim \beta \sim 10^{-1}$–$10^{-2}$). Furthermore, achieving the transition of TTCS into a qualitatively new state is possible by a variation of the only one parameter — its thickness $d$, since at its decreasing the current density $j$ caused by the disturbance of the magnetic field $\Delta H$ increases as $j \propto \Delta H/d^2$. To create such turbulent

---

[1] A basic description of different aspects of plasma turbulence in the solar atmosphere and in solar flares is given by S.A. Kaplan, S.B. Pikel'ner, and V.N. Tsytovich in the monograph *Physics of Plasma of the Solar Atmosphere* (Kaplan et al., 1977).

[2] The conclusion on the formation of "thin turbulent current sheet" as an indispensable condition of a flare generates a new and very "awkward" question: how did currents, initially distributed, according to standard approaches, in the whole volume of an active region with the typical size $D \approx 10^{9-10}$ cm self-compress up to a thickness of five orders less than $d \approx 10^5$ cm, being in equilibrium over the entire preflare and compression periods? By analogy to the daily life, let us note that the example of such compression can be a situation when the slight breeze from air conditioners, being distributed over the

plasma in the current sheet located in the solar/stellar atmosphere, the required sheet thickness should not exceed several kilometers, which also looks microscopic for the scale of an object (active region) of tens and hundreds of thousands of kilometers. With all exoticism of the required geometry and without understanding of how it can naturally self-form in the distributed fields above the active region, neither the physics of plasma nor MHD can propose better for solving the problem of the observable superstrong flare energy release.

C. Necessity for the process of magnetic field reconnection in TTCS to ensure the flare process. Dissipation of the magnetic field and electric current, by definition, occurs as diffusion of the magnetic field smoothing its inhomogeneity and blurring the magnetic field jump that generates current. In the process of blurring the field jump by diffusion the rate of diffusion broadening (as well as the rate of the magnetic energy dissipation produced by it) should, by definition, drop with time as $V_D = \sqrt{D/t}$, where $D$ is the diffusion coefficient and $t$ is the time of the diffusion process. In other words, even if one was initially able, after creating an extremely thin turbulent current sheet, to ensure a field jump from $H_0$ to $H_0 + \Delta H$ and start a fast field dissipation in it, this rate of dissipation would not be further maintained, since a fast blurring of thickness of the current sheet in the course of diffusion dissipation would result in a fast slowing-down of dissipation.

One needs thus to find a way to constantly maintain a small thickness of the sheet and a high field gradient (current density) in the dissipation region. The only possibility for this is yielded by the process of reconnection of lines of force forming a new topology with unbalanced magnetic tension of new magnetic loops that ejects out the already “processed” magnetic fluxes and hot flare plasma from the dissipation zone (current sheet) by a slingshot mechanism. As a result of such ejection and redistribution of pressure of the field and plasma caused by it, the current sheet self-compresses again and has an opportunity to further retain a small thickness. Let us stress that the process of field dissipation with magnetic-field reconnection in the current sheet and their ejection from the sheet consequently becomes dynamic and involves the plasma flow (slow — into the sheet, and fast — out of the sheet) as an indispensable component. The use of the assumption concerning the reconnection of lines of force in the current sheet is a fundamental condition for ensuring the process of flare energy release. Here we will not review different models of reconnection; the number of publications on them over the past 10 years exceeds ten thousand and is increasing by many hundreds each year (Cassak et al., 2008)[1]. We will limit us by a description of three basal approaches including opposed plasma flows with antiparallel magnetic fields reduced to either slow Sweet-Parker reconnection in the extended current sheet or fast Petschek reconnection in the small vicinity of the null (particular) point, and the accumulation of external magnetic disturbance in the null-point region (Syrovatskii, 1966a, 1966b) on the basis of which the overwhelming majority of published models are constructed.

**2.6.5.2. Variants of Models of Current Sheets with Reconnection.** As shown in the early 60s of the past century, the reconnection regimes can be summarized in two limiting cases: the slow Sweet-Parker reconnection in an extended current sheet (Sweet, 1958; Parker, 1957) or the fast reconnection in a very short sheet in the vicinity of a particular point in the course of the Petschek flow (1964).

---

conference hall of tens of meters, would spontaneously shrink into a layer thinner than a sheet of paper with accelerating to supersonic speed.

[1] See also the review of Cassak in Parker’s lecture of 2008 “The Theory of Magnetic Reconnection: Past, Present, and Future” (Cassak, 2008).

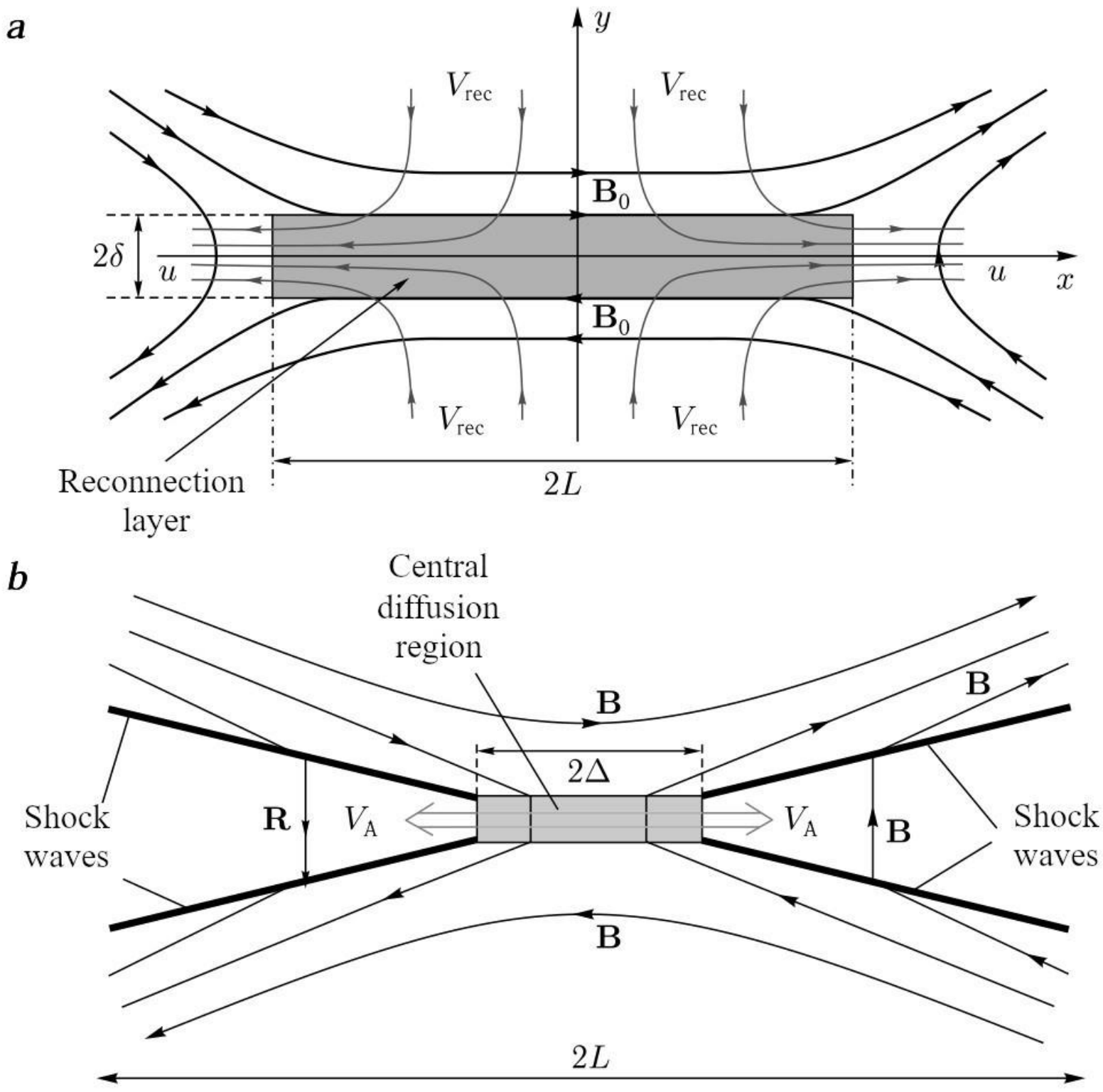

Fig. 74. Two variants of reconnection in the current sheet: *a* — slow Sweet-Parker reconnection in the extended current sheet; *b* — fast Petschek reconnection in the short current sheet between two colliding plasma flows with the field (http://www.cmso.info/cmsopdf/general_jul08/Talks/Uzdensky.pdf)

For the Sweet-Parker variant (Fig. 74, *a*), two magnetic fluxes with plasma and oppositely directed magnetic fields $\pm H_x$ are compressed slowly by magnetic pressure toward each other along the axis $Y$, forming an extended along the axis $X$ current sheet with the zero magnetic field (and respectively, by zero back pressure). It is in this sheet where the magnetic field dissipates and reconnects; the reconnected magnetic flux is ejected afterward by the tension of lines of force along the axis $X$ with the Alfvén speed. The speed of inflowing field into the sheet is determined by the magnetic Reynolds number (the ratio of the Alfvén speed to the diffusion speed, $R_m = V_A/v_d = L/(V_A/D_m)$, where $L$ is the spatial scale of the problem (current sheet width), $D_m = c^2/4\pi\sigma$ is the magnetic diffusion coefficient determined by the conductivity $\sigma$. As a result, the time of flare field dissipation for the Sweet-Parker scheme turned out to be hybrid, $\tau_{fl} = \sqrt{\tau_A \tau_d}$, i.e., the geometric mean between diffusion and Alfvén. For the high-temperature solar corona with giant Spitzer conductivity, the reconnection in the Sweet-Parker current sheet requires tens of hours and is too slow[1].

[1] Note that the account of plasma turbulence in the current sheet with a decrease in effective conductivity by 4–6 orders of magnitude ensures the time of flare dissipation in spatial scales of the active region in the Sweet-Parker scheme of an order of hundreds of seconds, which is fairly consistent with the observed times of flares.

An alternative variant of the fast dissipation of magnetic field energy in the current sheet with reconnection is the Petschek solution (Fig. 74, *b*) in which the objective is reformulated from the creation of a transition layer between magnetic fluxes to the objective on the collision of oppositely directed plasma flows initially approaching each other and containing antiparallel magnetic fluxes inside the colliding plasma. Within the Petschek scheme, these two converging plasma flows with the field collide in a very compact (within a particular point) central diffusion region, and in this region (and only in it) there occurs a reconnection of colliding magnetic fields and conversion of the magnetic field energy into heating and acceleration of particles. This central region is a source reflecting the incoming plasma flow, as a result the flow turns perpendicularly to the initial direction of colliding flows and ejects together with reconnected magnetic lines. In addition to the turn of the flux in the incoming plasma, there forms a departed standing slow magnetoacoustic weak shock wave on the front of which the turn of the plasma motion vector occurs.

The dissipation of a part of the incoming magnetic field with its turn occurs in this wave, and its energy converts into kinetic energy of the orthogonal ejection of plasma out of the front end. The process runs extremely fast on times comparable with Alfvén, $\tau_{fl} = \tau_A \left(\frac{\pi}{8 \cdot log\,(Re_m)}\right)$, which is quite sufficient to explain the observed rate of energy release (here $\mathrm{Re}_m$ is the magnetic Reynolds number).

Despite the seeming effectiveness of reconnection in the Petschek scheme, it has a series of principal shortcomings not allowing one to use it as a basis for the theory of flare energy release. First, by definition, it requires the incoming from outside plasma flows in the diffusion zone and does not work if the plasma is in equilibrium, as it takes place in the preflare state of fields and plasma above the active region. Within the Petschek scheme, the external disturbance at the photospheric level is immediately processed by reconnection in the vicinity of the null point into the motion of ejected plasma, not allowing the energy for a flare to be accumulated. Second, as indicated by Pikelner and Tsytovich (1976), the Petschek scheme allowing one to explain the conversion of magnetic energy into kinetic energy of ejection and partially into the heat is not organically capable of explaining the widely observed fact of effective energy conversion of the magnetic field into the acceleration of particles up to high nonthermal energies with total energetics that is comparable with an exit into thermal and kinetic energy.

The reason for such a constraint is a small size of the diffusion region where the acceleration of particles can actually occur, as compared to the sheet width (see Fig. 74, b): $\Delta \ll L$.

**2.6.5.3. Breakdown of Preflare Equilibrium and Transition into the Flare State.** Currently, there is no widely accepted mechanism for the transition of preflare equilibrium configuration with force-free fields and currents distributed in the atmosphere above an active region on scales of 1000–10,000 km into the flare state with a thickness of the current sheet of hundreds of meters–few kilometers. The following possibilities are suggested:

a) interrupting instabilities of the tearing mode type, redistributing local currents and plasma density up to the level of currents that are sufficient for generation of plasma waves and plasma turbulence (Furth et al., 1963; Sturrock and Coppi, 1966);

b) thermal instabilities of the dense plasma of the current sheet with relaxation, resulting in an abrupt drop in pressure in local regions of the current sheet with subsequent compression of the sheet and transition into the turbulent state (Somov and Syrovatskii, 1982);

c) self-compression of the initially thick transition layer (shear zone) into the thin current sheet, which enhances the current density and current velocity of electrons up to critical values with subsequent generation of plasma turbulence (Cassak et al., 2006).

As a trigger mechanism, there were also considered variants of external perturbation of the current sheet located near the instability threshold. The ballooning modes of flute instabilities of the quiet prominences and coronal condensations were suggested to play a role of such triggers (Pustil'nik, 1974, 1977, 1978; Zaitsev and Stepanov, 1992). External perturbations in the form of a shock wave coming from a flare in the nearby active region were treated as a reason of "sympathetic flares" (Fritzova-Svestkova et al., 1976). Stepanov and Zaitsev (1992) show that the ejection of cool dense plasma of the quiet prominence with a high number of neutrals into the current sheet abruptly increases resistivity in it (by 6–8 orders of magnitude) due to the collision of neutral atoms with magnetized ions and capable of producing the onset of flare energy release.

**2.6.5.4. Postflare Redistribution of Flare Energy and External Observational Manifestations of Flares.** The basic products of flare energy release in the turbulent current sheet are high-temperature plasma and particles accelerated up to relativistic and ultrarelativistic energies. The temperature of the primary flare plasma can be estimated according to the order of magnitude as $T \approx H^2/8\pi nk$, where $H$ is the dissipated field, $n$ is the plasma density, $k$ is the Boltzmann constant. For the typical coronal values, the effective plasma temperature in the sheet $T_* \approx (1\text{–}3) \cdot 10^9$ K corresponds to the mean energy of particles $\varepsilon_* \approx (30\text{–}100)$ keV, whereas the energies of particles accelerated by electric fields in the turbulent current sheet cover the range from 100 keV to 100 GeV (Miroshnichenko, 2015). This primary superhot plasma forms a powerful flux of nonthermal particles (primarily electrons) and heat (due to thermal conductivity) in lower cool chromosphere layers, heating them up to temperatures of tens of millions of degrees. This pre-heating of lower dense layers results in intensive secondary evaporation of the overheated chromosphere plasma upward, the filling of coronal magnetic arches with it, and further relaxation in soft X-rays and hard ultraviolet. The thermal energy flux from the primary plasma approaches lower dense layers and forms a downward shock wave with relaxation behind the front where, consequently, the temperature drops down to tens of thousands of degrees, whereas the density, accordingly, grows by many orders of magnitude. As shown by Kostiuk and Pikelner (1975) and Livshits et al. (1981), it is this zone behind the shock wave front that is responsible for the formation of a chromosphere brightening in optical lines and for the record megaflares — in the optical continuum as well.

The subsequent thermal instabilities in the plasma evaporated from the chromosphere, which fills the magnetic loops in the corona and forms coronal X-ray arcades, decrease the temperature in the part of loops down to tens-hundreds of thousands of degrees. These arch structures are observed simultaneously in both X-ray and ultraviolet ranges, and in lines of the optical part of the spectrum. High magnetic isolation prevents arch structures of different temperatures from transverse thermal conductivity, allowing them to exist a long time in close proximity to each other. Another part of the primary superhot plasma, expanding upward, fills the upper parts of the magnetic structure above the active region. In those places where magnetic retention of the hot plasma proves to be insufficiently effective, there may occur ejections of a part of coronal structures in the solar wind, forming eruptive prominences and coronal mass ejections. We stress that all this wide diversity of manifestations of solar flares is a secondary (and sometimes even tertiary) consequence of flare energy release in the current sheet, whereas the primary superhot plasma itself and accelerated cosmic rays directly in the region of their formation are inaccessible for direct observations with current technologies even on the Sun.

The basic regions of emission of solar and stellar flares are presented in Fig. 75 from the paper of Dennis and Schwartz (1989). Within this scheme, the primary energy release in the region of tops of magnetic arches involved in flare reconnection produces a primary volume of the hot plasma radiating in hard X-rays, whose cooling by conductivity and escaping of primary particles downward into the dense chromosphere layers generates their evaporation upward and filling of the magnetic rope with the dense plasma with a temperature of (1–2) · $10^7$ K, which radiates in soft X-rays and cools down with time up to ultraviolet temperatures of (1–3) · $10^5$ K.

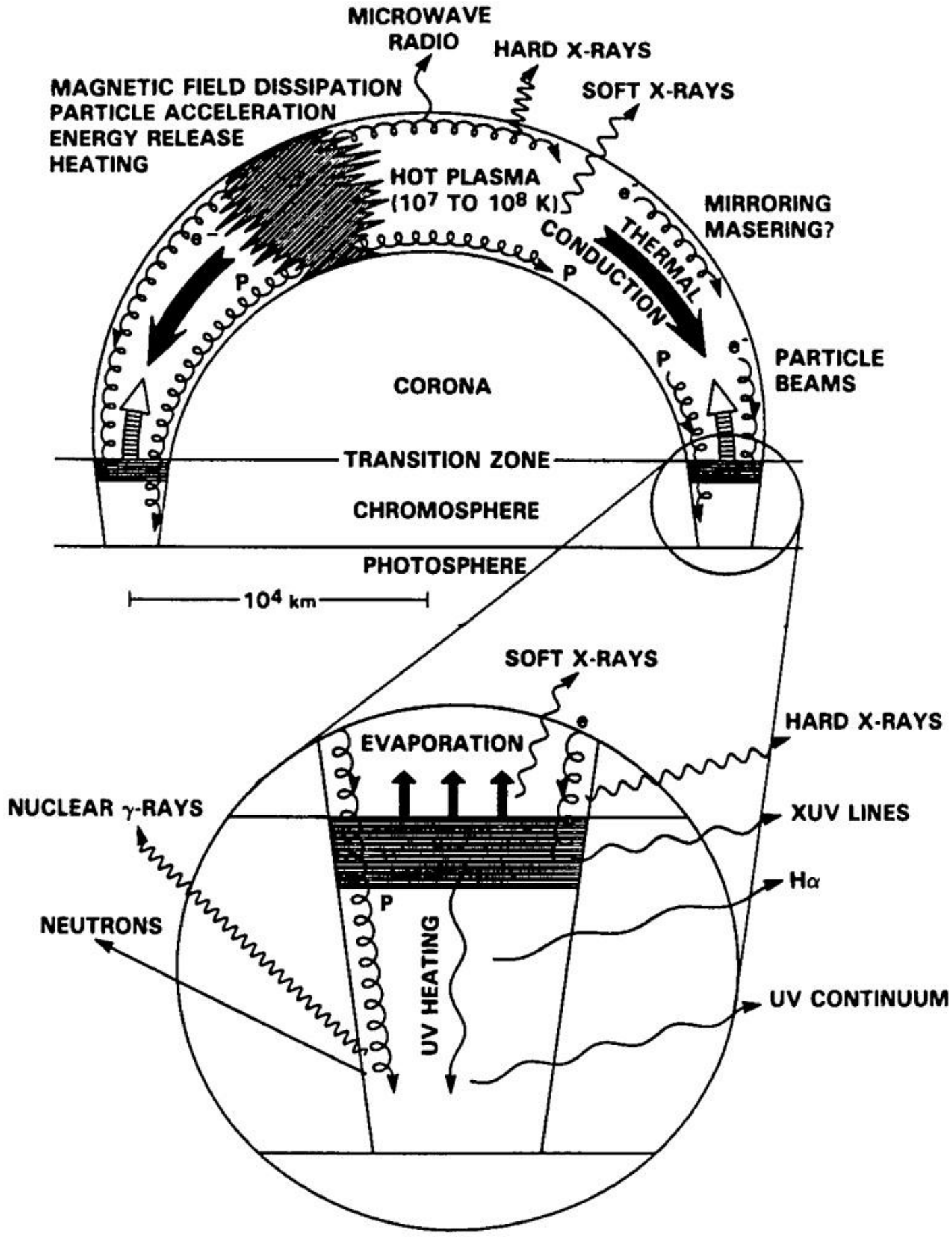

Fig. 75. A scheme of the elementary impulse of a solar flare with subsequent redistribution of the released energy from secondary (conductivity and particle fluxes) and tertiary (evaporation, emission of the primary and secondary plasma) (Dennis and Schwartz, 1989)

As the dense plasma cools down, it cannot be retained by its pressure and flows through the magnetic tubes/ropes downward to the chromosphere, illuminating and manifesting the magnetic field structure (picturesque coronal rains at

(https://apod.nasa.gov/apod/ap180527.html)). The electrons accelerated in a flare radiate flare radio emission at millimeter to meter wavelengths by nonthermal mechanisms (maser, plasma, and gyroresonance).

During the bombardment of dense layers of the lower chromosphere the energetic suprathermal protons generate hard gamma-ray emission (continuum in the nuclear lines as well), whereas suprathermal electrons form soft gamma-ray emission by the bremsstrahlung scattering on the chromosphere atoms. A part of the accelerated protons and electrons escape to the solar wind and reach the Earth, forming a proton flux of corpuscular radiation and filling the radiation belts. In the case when energy release is rather high and coronal magnetic fields are not capable of maintaining the hot plasma expanding outward, a part of the plasma leaves the energy release region and expands outward, through the corona into the solar wind, producing CMEs.

Although the radiation mechanisms for solar and stellar flares in each individual range have been understood to a certain extent, one failed to create a comprehensive theory of all this chain of processes. There are too many inner (hidden from observations) parameters of the field and plasma which affect drastically the characteristics of radiation required to be taken into account to solve such an objective (from the geometry and topology of the magnetic field above a particular active region of a future flare and the place where the current sheet is located in this structure up to the processes of hydrodynamic transfer, evaporation, relaxation, thermal instability of the heated plasma, the role of accelerated suprathermal particles, etc.). In stellar flares, the situation is even more complicated than that in solar flares, since the spatial resolution is absent and a significant part of flare emission cannot be recorded at such a distance (gamma-ray emission and hard X-rays, protons and CMEs). Nonetheless, the existing total understanding that the source of their energy is magnetic field energy and the initial phase of a flare lies in the formation of beams of high-energy particles allows one to interpret successfully the observed processes and determine parameters of the plasma and field in the stellar flare region[1].

We have already described above a long discussion on the role of microflares in heating stellar atmospheres associated with the detection of correlation of the quiescent X-ray emission with averaged flare emission and with Balmer emission out of strong flares. It should be added that X-ray emission emanates from the optically thin plasma of very high temperature, whereas emission of hydrogen lines — from a plasma that is thousand times less hot and optically thick in the discussed lines, i.e., X-ray emission emanates from the volume, whereas Balmer emission from the surface. Since Balmer lines are detected at the onset of flares or even in preflare brightenings, whereas soft X-rays — in the gradual thermal stage of a flare, then the practical proportionality of these emissions requires explanations.

[1] Observations of Beskin et al. (2018) in the optical range with the 6-meter telescope within the MANIA complex turned out to be direct evidence for manifestation of emission of primary suprathermal particles accelerated in stellar flares up to high energies. At the peak of a strong flare of 28 December 2008 in the binary system L726-8AB (comprising the active red dwarf) the observations showed the presence of short-lived (0.6–1.2 s) bursts of optical radiation with high (30%) linear polarization. As an explanation, the authors suggested the synchrotron radiation of primary ultrarelativistic electrons of the flare in regular magnetic fields of the active region of the red dwarf. Observational data can be treated as a first detection of direct emission of primary accelerated particles in stellar flares. For solar flares, the analog of such emission is hard X-rays from the bremsstrahlung emission of primary particles and radio bursts of type III at the plasma frequency of the background plasma radiated due to the conversion of plasma oscillations into electromagnetic in the beams of accelerated electrons of the primary suprathermal electrons.

### 2.6.6. Power Character of Amplitude and Energy Distributions in Flares as a Manifestation of "Self-Organizing Criticality" in an Open Multielement System with Strong Interaction

The universal power character of distribution of flares in energy release amplitudes is a fundamental observational fact detected in the course of studying solar and stellar flares. The conclusion on this universal property of flare activity was first formulated in the pioneer paper by Gershberg and Shachovskaya (1983) in the course of studying a large number of flare UV-Cet stars (see Subsect. 2.3.1.2 and references therein). A stimulated by their result study of statistical properties of solar flares in a wide spectrum of manifestations ($H_\alpha$ flares, X-ray and microwave bursts, bursts in UV and soft X-rays (Kurochka, 1987; Kasinsky and Sotnikova, 2003; Yashiro et al., 2006) showed that the power-law character of distribution of flares over amplitude is a common property of flare activity for both flare stars and the Sun[1]. The basic property of such a power distribution is scale invariance — the absence of a characteristic scale (scaling).

The attempts to explain this universal property of flares during the first stage were reduced to the association of flare energy with the value of magnetic or kinetic energy of turbulent cells rising from the convective zone to the photosphere and carrying the appropriate portions of free magnetic energy to the solar (stellar) atmosphere (Pustil'nik, 1988). The resultant frequency amplitude spectrum of flare energy release was considered as a direct consequence of the frequency spectrum from the scales of turbulent cells, which always has a power character for developed turbulence. Under different assumptions about the nature of turbulence in the convective zone (taking into account and without the impact of magnetic fields and dimension of the system), the index of the predicted power spectrum turned out to be in the interval 0.5–0.9 for the Kolmogorov spectrum of convective turbulence, 0.4–0.5 for the dissipation in the flare of kinetic energy of convective cells, 1–1.5 when taking into account the finite lifetime of the outflowed magnetic flux, and 1–1.5 to consider the process of redistribution of currents of both two- and three-dimensional percolation, which is close to the observed values.

However, in recent years, there has been a radical breakthrough in understanding the deep physics of formation of power frequency-energetic spectra of flares on the Sun and stars on the basis of the SOC paradigm (Self-Organized Criticality), i.e., the self-organizing criticality in open nonlinear systems involving a strong chaotic interaction of their constituent elements on the background of the flux through the system of energy and entropy (Bak et al., 1987; Mogilevskii et al., 2005; Mogilevsky and Shilova, 2006; Feinstein et al., 2022). The direct manifestation of the presence of self-organizing criticality is a formation of specific self-similar fractal structures consisting of smaller structures similar to the basic one; those, in their turn, consist of even smaller similar structures, and so on[2]. The property of self-organized criticality is currently found for a series of phenomena having the following common external signatures: the power law of distribution of events, fractality, and resistance to variations of parameters. To such phenomena one can attribute: dynamics of the Earth's crust (earthquakes, formation of the fractal coastline), traffic jams, spread of forest fires, evolution of living organisms, human economic activities, statistics of military conflicts, neurophysiological models, the emergence of cancer tumor, phenomena of plasma physics, radiation of quasars, and more. Lu and Hamilton (1991) paid attention to the role of self-organizing criticality in solar flares, having considered flare energy release as an avalanche-like self-organizing process. The most extensive review of fractal properties and self-organizing criticality in the physics of solar flares is given in Section "SOC and Solar Flares" by Paul Charbonneau (2015), in the monograph and review by Markus Aschwanden (2013, 2016c). Let us note some of them.

[1] Note that in the case when solar and stellar flares are random and independent events, the statistics of their distribution (in amplitudes, in interflare intervals, etc.) with necessity should be reduced to a normal distribution, with the presence of characteristic scale in space, time, and amplitudes.

[2] The beginning of such an approach is associated with the publication of the monograph by Mandelbrot (2002) in which there was demonstrated a wide distribution in nature of similar situations with the self-organization of complex multielement systems and with the inner interaction between elements that are different from the standard characteristics of random processes by the formation of a power spectrum of distributions of key characteristics (spatial, temporal, amplitude, frequency, and so on).

1. In a wide range of spatial scales (6 orders of magnitude), the observed structure of the photospheric magnetic field demonstrates just the power dependence on sizes.
2. Another important result of studying statistical properties of solar activity is a detection of the fractal character in the distribution of photospheric, chromospheric, and coronal structures (Abramenko, 2012; McAteer et al., 2005).
3. Although additional processes (heating, hydrodynamics, etc.), dependent on external factors (field geometry, physical conditions in the region of emission, etc.), are involved in the transformation of primary free magnetic energy into the observed flares of emission in the UV and X-ray ranges, nevertheless, solar and stellar flares are the most impressive example of scale invariance (power frequency amplitude spectrum) observed in astrophysics. It is notable that the slope index of the power spectrum of flare energy release does not vary during a solar cycle, although both the frequency of flares and their amplitude vary from the cycle maximum to minimum by many orders of magnitude (Lu et al., 1993; Aschwanden et al., 2011a; Aschwanden, 2011b; Aschwanden et al., 2012).

The basic idea used to interpret power distributions of flare properties on the Sun and stars appeals to ideas of Eugene Parker.

1. The magnetic field of coronal structures frozen into the convective elements in subphotospheric layers takes part in their local motions (rotations and shifts), leading to a tangling of individual current-magnetic threads around each other.
2. In the zones where individual current-magnetic threads contact with each other (tangential discontinuities in Parker's terminology), the local surface current sheets are formed in which upon favorable threshold conditions of high current concentration the plasma turbulence is generated and the reconnection process is triggered. This initiates the following process of flare energy release in the adjacent places of contacts of the tangled current-magnetic threads (local surface current sheets). This process has an avalanche-like character; its amplitude is defined by the spatial size of the cluster of local surface current sheets, and the spectrum of dependence on scales (and, accordingly, flare energy) is defined by the self-organization process and is power-law. This mechanism governs flare energy release on scales from nanoflares heating the corona up to megaflares capable of causing appreciable negative consequences (proton radiation fluency and magnetic storms) in the space weather (Parker, 1983a, 1983b, 1988; Morales et al., 2008, 2009).
3. As a result, the magnetic tension, formed by the motion of a magnetic rope base on the photosphere, from the primarily force–free configuration forms the tangled magnetic threads with tangential discontinuities (current sheets) between them, with the presence of magnetic field force component, the interrope interaction and formation of primary threads into an ensemble of strongly interacting current-magnetic threads ensuring the self-organization into fractal structures and formation of a power spectrum of magnetic energy accumulated in current-magnetic fractals. Achieving the threshold of plasma instabilities and plasma turbulence in the current sheet, there occurs a fast reconnection of tangled current-magnetic threads, a release of free magnetic energy of the current-magnetic fractal with simplification (untangling) of magnetic threads and relaxation to the state that is almost force-free and close to initial. Such a cycle of accumulation of magnetic tensions and their subsequent discharging in the current sheets of tangential discontinuities is the basis for supplying and releasing flare energy.

4. The power character of statistics is steadily manifested also in the distribution over energies of particles accelerated in solar flares. This fact was noted as a result of analysis of spectra derived during observations of flares in hard X-ray range (Dennis, 1988) and directly analyzing the solar cosmic ray spectrum (Dorman, Miroshnichenko, 1968; Miroshnichenko, 2015).

The observable domination of power-type statistics of distribution over spatial, temporal, and energetic scales of various flare activity manifestations on the Sun and stars and the presence of fractality in the observed structures and temporal dynamics can be explained only by suggesting in these objects a presence of the self-organization process in the open system with strongly interacting elements and leakage (percolation) of energy and entropy through it. It is self-organizing criticality as a universal property of such systems that naturally leads to the power character of distribution spectra and fractal type of dimensionality for a large number of applications for both terrestrial and astrophysical objects. We refer the reader to the recent edition of the collection of reviews edited by Aschwanden (2013) and also a later review by Aschwanden and coauthors (Aschwanden et al., 2016c) for more detailed consideration of this particularly interesting phenomenon.

## 2.6.7. Open Questions in the Physics of Flares

In this section, we would like to pay attention to a series of open questions producing inner contradictions in the standard approach to the modeling of solar and stellar flares and indicate possible ways of their solutions.

**A. Contradiction between the observed picture of coronal arcades and two-ribbon $H_\alpha$ structures and the expected within the standard approach three- or four-ribbon result of interaction of two magnetic fluxes.** The standard flare models appeal to the interaction of two magnetic fluxes: either new floating up from the photosphere with old overlapping their way out into the upper layers or in the form of "collision" of two magnetic fluxes from the already existing active regions (Ikhsanov et al., 2004). As a result of such an interaction, after a flare there should be observed two systems of flare arcades emerging due to the reconnection in the zone of contact and the filling of magnetic loops by the matter evaporated from the chromosphere. Furthermore, emission at the four bases of new loops should form four flare ribbons along the neutral line at the chromosphere level (Fig. 76 *a*, *b*). However, the overwhelming part of observations after a flare record only one system of flare arcades and only one two-ribbon structure of emission in the chromosphere, as, for example, in the case of the famous Bastille Day flare shown in Fig. 76*c*. This contradiction makes one suggest that the widely accepted pattern of collision of magnetic fluxes as the flare sources misses some important factor leading to the very systems with one arcade in the corona and to the two-ribbon structures in the chromosphere.

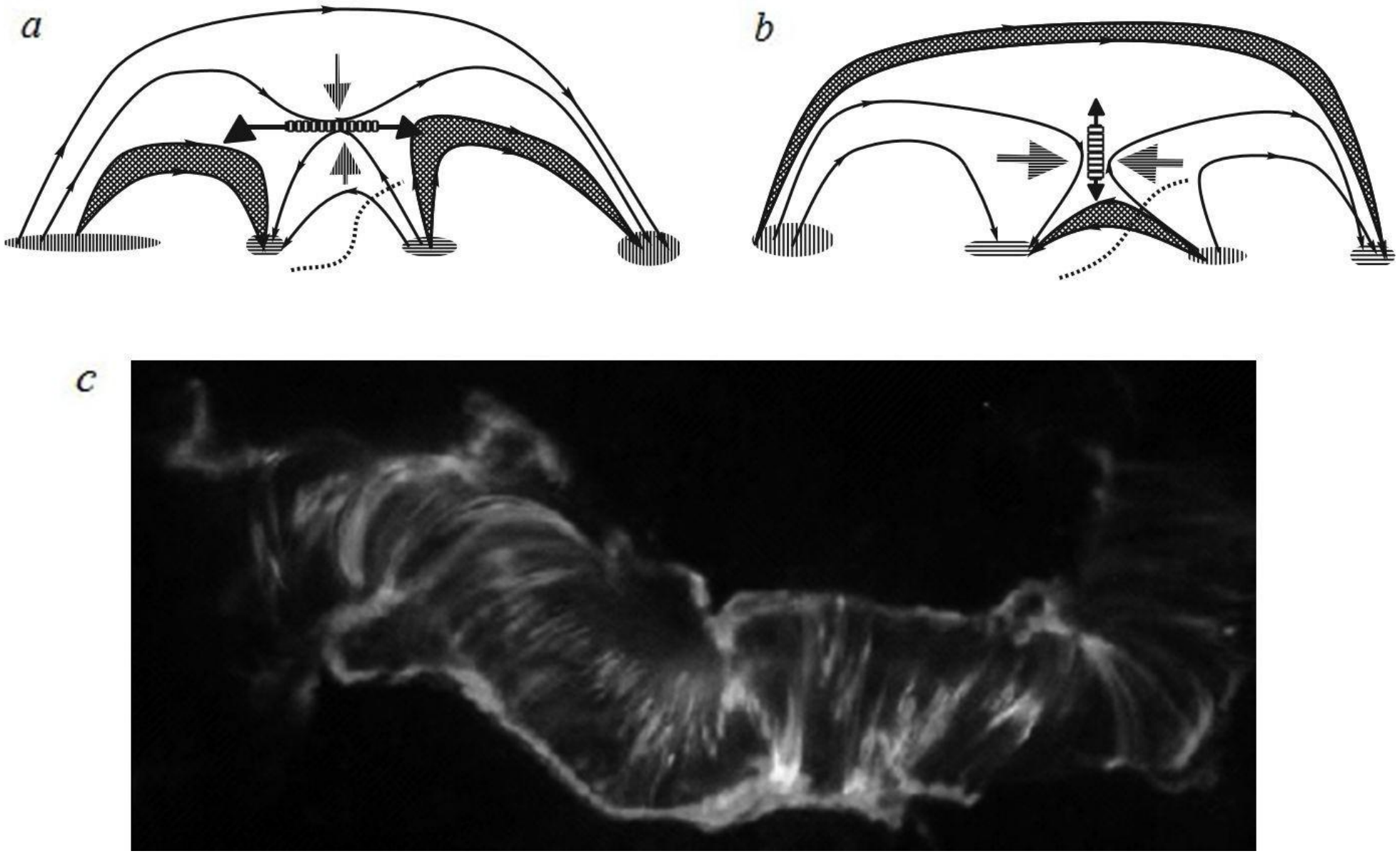

Fig. 76. Expected flare geometry in the form of two coronal arcades in X-rays and UV and four-flare ribbons in $H_\alpha$: *a* — for the case of “floating from below” of a new magnetic flux; *b* — the same for the case of “collision” of already existing magnetic fluxes; *c* — typical observed real geometry on the example of a flare recorded on 14 July 2000 known as the Bastille Day flare, which shows only one system of coronal arcades and only two flare ribbons in $H_\alpha$ at the bases of arcades (a more illustrative color version can be found at http://soi.stanford.edu/results/SolPhys200/Schrijver/images/BastilledaySlinky3band.jpg)

**B. Presence of the magnetic skeleton from thin magnetic threads of the constant section from the photosphere to the corona, determining the energetics and magnetic field equilibrium in upper layers of the solar atmosphere.** During recent observations with high angular resolution in the ultraviolet, the $H_\alpha$ line, and soft X-rays, there was detected a phenomenon of magnetic skeleton in the magnetic field structure above an active region — the presence of numerous current-magnetic ropes of the constant cross-section extending from the photosphere to the corona. The first indications of cross-section constancy of magneto-current coronal arches were obtained by Klimchuk (1992) from solar corona observations at the orbital X-ray telescope SXT. With increasing angular resolution up to fractions of angular seconds, it was ascertained that the bulk of the magnetic flux was involved in numerous (tens and hundreds) thinnest magnetic ropes with a thickness that was less than the resolution limit for the best instruments: 0.2″ (150 km) and with constant cross-sections $S_j$ along an arch loop, which practically did not change extending from the photosphere to the upper parts of the corona (see Fig. 21 in this book and Fig. 20 in Lypez Fuentes et al. (2008)). Conservation of the magnetic flux in such a tube, $H_iS_i$ = const, means that the magnetic field strength in it, $H_i$, should be constant, extending from the photosphere to the corona and be of hundreds or thousands of gauss.

On the other hand, the magnetic skeleton threads themselves diverge away from each other with height, similar to the pattern expected for lines of force of potential and force-free configurations, and the field strength between threads, accordingly, falls with height $h$ as $(h/R_{spot})^{-k}$, where $k \approx$ 1–3 (depending on the relative contribution of the force-free component). Since the magnetic field energy is proportional to $H^2$, then the respective fraction of magnetic skeleton energy in the total energetics of magnetic fields in the atmosphere above an active region grows with height and at coronal heights it should dominate in the total balance. The detected fine structure in the form of a set of isolated magnetic ropes in this case proves to be a basic element of both preflare equilibrium and flare energy release. The magnetic field configuration including the set of isolated magnetic ropes is not force-free. Magnetic ropes in such a configuration should strongly interact with surrounding background fields and through them with each other. The preflare state manifests itself as a magnetic carpet made of strongly interacting current-magnetic threads constantly redistributing among themselves new magnetic deformations and tensions coming from the photospheric layers (Fig. 77).

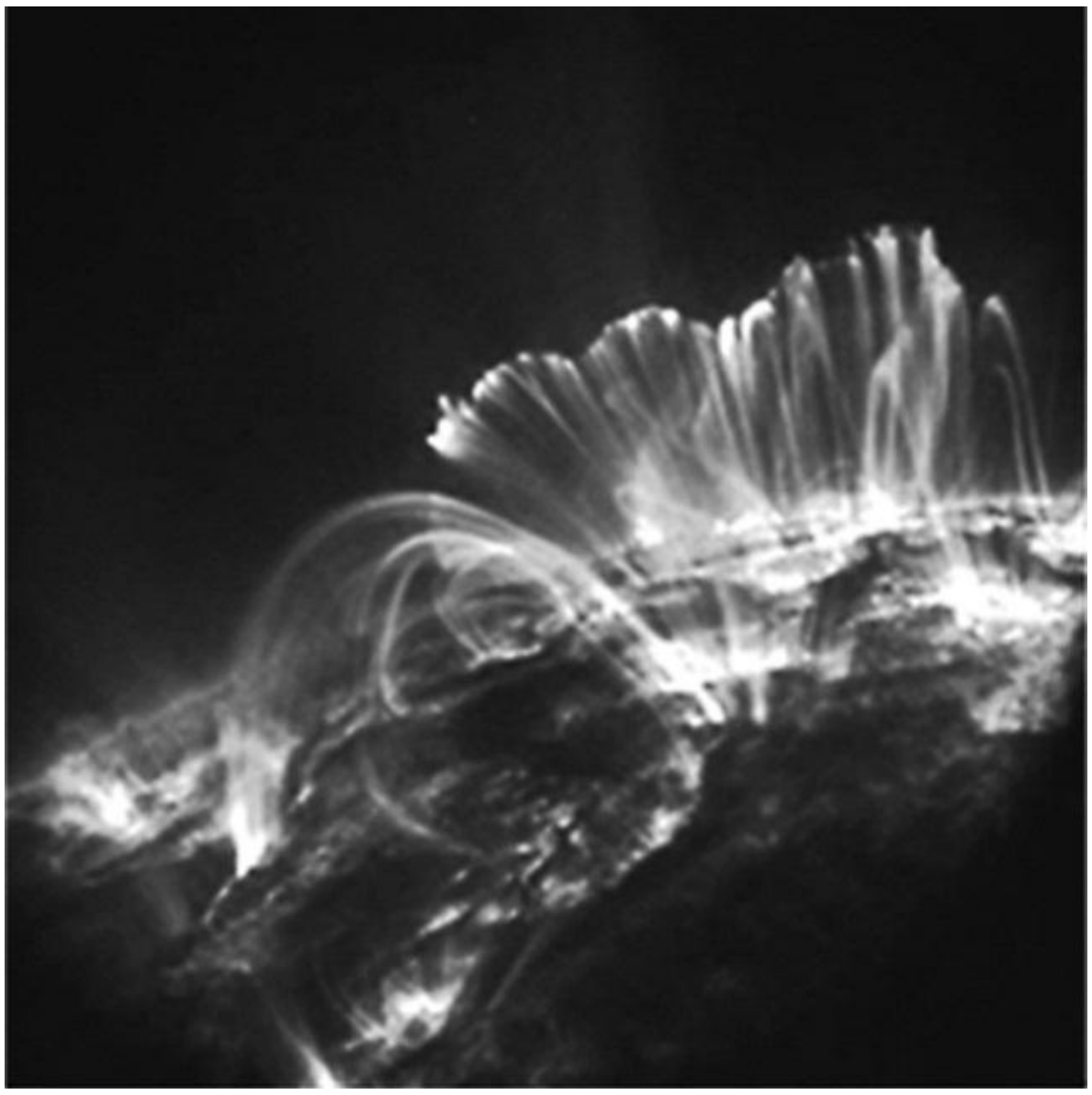

Fig. 77. Demonstration of the fine structure of magnetic fields in the atmosphere above an active region; the basis of this structure is numerous thin magnetic ropes/plasma threads and the magnetic field tubes of constant section. Source: SDO (https://youtu.be/_wPBmKzIyAo; https://youtu.be/aSIDelAAHiA)

Note that the equilibrium described by such a scheme is dynamic rather than static, and the process that controls its behavior is a percolation of magnetic tension energy through a network of numerous interacting current-magnetic ropes of the magnetic skeleton in the atmosphere above an active region (Rappazzo and Parker, 2013). The stability of such ropes is determined by their bases that are frozen into the photosphere; however, when the torsional stresses are excessive, there may emerge kink instabilities, which was observed in a series of cases (Török et al., 2014). The resultant equilibrium is dynamic rather than static; each individual element can significantly change its state, but the whole system remains in equilibrium until reaches the critical state of precatastrophe.

**C. Instability (tearing and pinch) of the thin turbulent current sheet.** The thin turbulent current sheet with a magnetic field shear is an extremely unstable formation and from the first moment of its generation it must be split by tearing instabilities into thin parallel current ropes with surrounding magnetic islands (Furth et al., 1963). This tearing occurs on hybrid times (for the tearing mode — on the geometric mean of Alfvén time and the time of field diffusion through the sheet, $\tau_{\rm fl}\sqrt{\tau_{\rm A}\tau_{\rm d}}$), which for solar flare conditions corresponds to seconds and fractions of a second. These thin threads represent a system of linear pinches and, in their turn, should undergo the pinch-type instabilities of direct currents (overstressings, helical modes) occurring on purely Alfvén time scales (for solar flares — less than milliseconds). The emerging current sheets intertwist and transform the distribution of the flat current into TTCS into percolation through a randomly spread network of currents.

Drake et al. (2006, 2007) used this pattern of disintegration of the current sheet into current threads and magnetic islands and suggested a mechanism for the acceleration of energetic particles at the initial phases of flares, which is based on their reflection from the ends of numerous small compressing magnetic islands. They showed that the resultant energy spectrum of such electrons should have a power-law form with a spectral index that is close to the observed one. According to Drake et al., the accelerated energetic particles should form narrow fluxes with characteristic width of an order of the dissipation region width on their way out from the current sheet. The outgoing flux represents not a wide front but a tangle of numerous channels-filaments. Through the computer calculations, Che et al. (2011) confirmed the idea of the current sheet disintegration into an ensemble of filaments at which the reconnection velocity abruptly increases.

**D. The problem of local overheating of the thin turbulent current sheet.** One of the reasons of the nonstationary energy release process in TTCS is a threshold dependence of instability that generates plasma waves and turbulence on the ratio of the current velocity to the thermal velocity of electrons (Pustil'nik, 1980). As already mentioned, for the generation of plasma waves by current in the sheet, which provide abnormally low turbulent conductivity, the current velocity $u = j/ne$ should not exceed the threshold value that is tied up in the phase velocity of these waves $v_{\rm pl}$, otherwise the Landau damping on thermal electrons or ions initially swallows these waves. These thresholds are different for various types of instabilities, and the most popular among them $u_{cr} = C_{Te} = \sqrt{T_e/m_e}$ — the thermal velocity of electrons for the Buneman instability and $u_{cr} = C_{si} = \sqrt{T_e/M_i}$ — the ion sound speed for ion-acoustic variants of instabilities. But then "abnormal" heating of the turbulent plasma with growing electron temperature $T_{\rm e}$ by three orders of magnitude over supershort times of plasma oscillations will extremely fast (during the first microseconds) increase the thermal threshold of current instabilities, locally turn off plasma turbulence and flaring process. Such self-quenching of a flare by its own heating makes the energy release process in TTCS ineffective.

## 2.6.8. Flares as Three-Level Energy Percolation Coming from the Deeper Layers of a Star

The arguments listed in this and in previous sections allow one to consider solar and stellar flares from the unified point of view as an implementation of the universal self-organizing process of energy percolation from the center of the Sun into its atmosphere with the formation of the fractal character of distributions that is typical of percolation processes. This process is implemented at three levels and leads to the formation of universal power spectra in the system.

**2.6.8.1. Three levels of percolation.**

1. Megapercolation of primary thermal energy from the central regions to the photosphere of a star (Sun) through the convective zone. This process contributes to the formation of the power spectrum of turbulent cells carrying magnetic loops of different scales to the photosphere during the dynamo process[1] and thus determines the character of the primary spectrum of magnetic energy that is subsequently released in flares.

2. Macropercolation of magnetic energy and tensions through the dynamic network of strongly interacting magnetic skeleton (carcase) elements of fields in the atmosphere above an active region. An ensemble of these elements forms a magnetic carpet with tangled current-magnetic threads of different scales (Meyer et al., 2011, 2012). This process is responsible for the formation of multiply connected current structures penetrating through the corona and generating the global turbulent current sheet (current system).

3. Micropercolation of currents through the current sheet emerged at the flare stage, forming a dynamic network of turbulent and normal domains (Pustil'nik, 1999, 2017). This process is responsible for properly flare energy release and acceleration of suprathermal particles up to relativistic energies.

Such a three-level percolation with the formation at each stage of different power spectra with different fractality determined by dimensionality of the object and the type of interaction between elements in the percolating network can explain the often observed multifractal character of organization of solar magnetic fields and solar activity (Lawrence et al., 1993; Abramenko, 2005), i.e., such a character of distribution when no unified power law exists, which holds for all possible scales of events, but the spectrum slope varies depending on scale[2]. An exceptionally interesting result was obtained by Abramenko and Yurchyshyn (2010): the higher the multifractality degree, the stronger the flare productivity of an active region. Moreover, analysis of the preflare development of multifractality in the photosphere, chromosphere, and corona (Abramenko et al., 2008) for a particular flare on X3.4 observed on 13 December 2006 showed that approximately a day before the flare there occurred a rise and decay of multifractality in the photosphere, and by the moment of the flare the multifractality in the chromosphere and corona gradually achieved maximal values. This means that the degree of multifractality reflecting the complexity and intermittency of the magnetic structure at the photospheric level may be an effective characteristic of preflare energy accumulation suitable for the prediction of flares.

**2.6.8.2. Acceleration of particles in the percolating turbulent current sheet.** Instabilities (tearing and pinch modes) and overheating of TTCS result in layering the initial current sheet into numerous local plasma domains with "normal" and "turbulent" conductivity. Whereas the turbulent domains, whose electric resistance exceeds by 3–5 orders of magnitude the background plasma resistance, are essentially electrostatic double layers (Pustil'nik, 1978, 1997) and form a random network of linear microaccelerators. The distribution of "escaping electrons" in this network of accelerators with fractal distribution of their number in the volume of acceleration (current sheet) automatically leads to the power dependence of the number of accelerated particles on their energy $N(\varepsilon) \sim \varepsilon^{-\gamma}$ observed in solar flares (Dorman, Miroshnichenko, 1968; Miroshnichenko, 2015).

---

[1] The role of percolation in modernizing the Babcock-Leighton dynamo model that describes the magnetic field generation on the Sun and stars is considered in Shatten (2007).

[2] In fact, all the diversity of existing in nature fractal structures are multifractals — superpositions of a number of monofractals, whereas each one with its own scaling, i.e., with its power law being implemented in its limited interval of scales (see, for example, Mandelbrot, 1982, Frish and Parisi, 1985, Maruyama, 2016). A degree of multifractality (as a degree of complexity of the structure) may be estimated from the parameters of multifractality spectra (Lawrence et al., 1993, Abramenko et al., 2002, Abramenko, 2005, 2012).

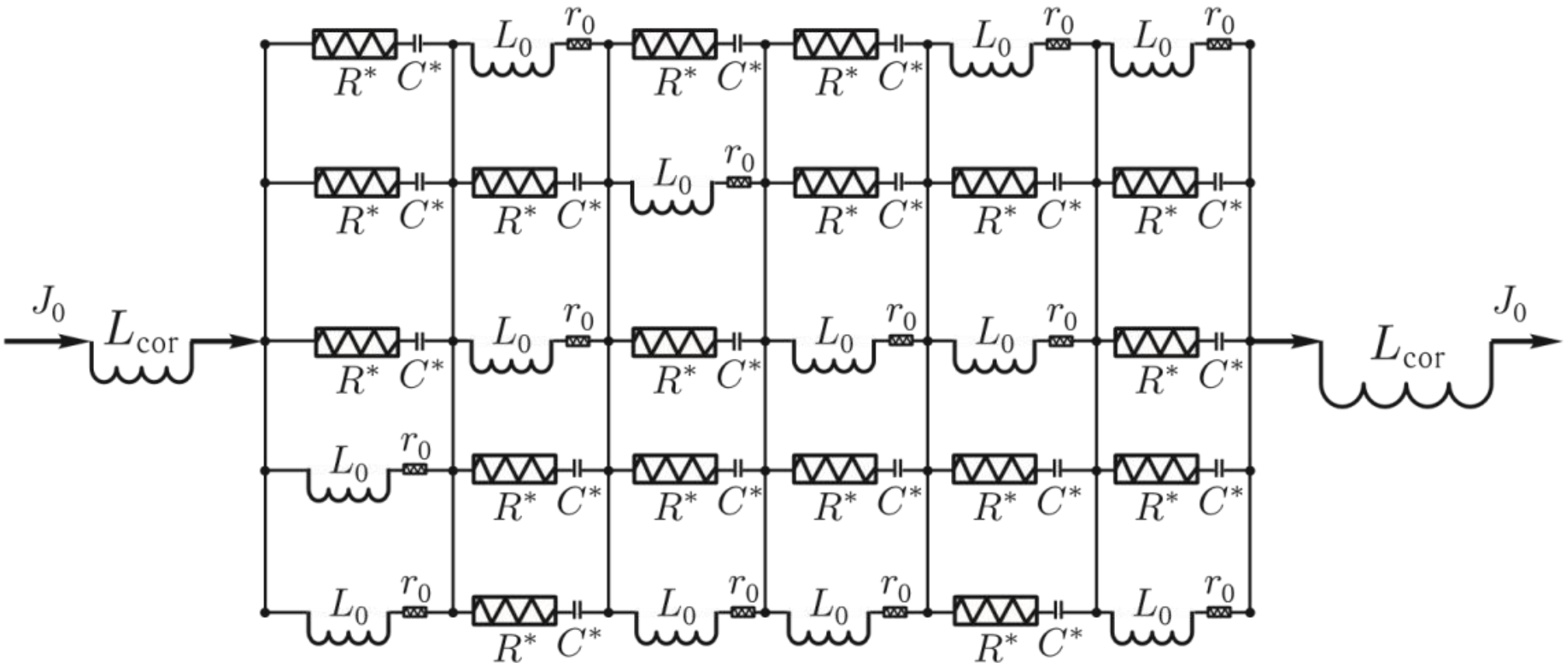

Fig. 78. Dynamic percolation of currents and electric fields through a random network of resistors of turbulent and normal domains with threshold sensitivity of network elements to the level of local current (Pustil'nik et al., 2013), turbulent elements $R^*$, normal highly conducting elements $r_0$ with high self-inductivity $L_0$ and double electrostatic layers $C^*$

The distribution of current through such a network of resistors manifests itself as a percolation with constant redistribution of current and electric fields determined by the requirement of preserving current in the system and by the Kirchhoff laws for the networks of resistors. We also stress that the equilibrium in such a system is dynamic with constantly occurring variations of local values of currents and resistances while maintaining global properties of the system (Fig. 78).

**2.6.8.3. Advantages of the percolation approach for understanding the physics of flares.** The common properties of systems with percolation through the networks of strongly interacting elements are possibilities of spontaneous switching of the state of network elements, depending on a value of the local percolating flux and the presence of feedback. Due to this, in such systems there occurs a formation of fractal complexes with universal power character of the network's fractal spectrum having sizes from mega- to micro-scales and with a spectrum index determined by the fractal dimensionality of the system through which percolation proceeds. Such a situation is typical of the self-organization process at the criticality threshold (Aschwanden, 2013). It is this fundamental property of percolating systems that makes them the most attractive tool when modeling the observed universal power character of the amplitude spectrum of solar and stellar flares and the power energetic spectrum of particles accelerated during solar flares.

The other important property of percolating systems is a possibility of phase transitions within which their global properties (conductivity, heat conductivity, elasticity, etc.) undergo drastic and significant variations (Feder, 1988). Such a phase transition may happen if the total density of "bad" elements of the network achieves a critical value at which the energy flux leakage accumulated by the current system through the remaining "good" elements results in their turbulence and "burning out". It is this process that describes the phase transitions

“conductor–superconductor” and “conductor–semiconductor”. In our case, such a phase transition brings the current system into the flaring state. This occurs at the moment when the density of turbulent domains (i.e., the domains with abnormally high resistance) achieves a critical level. Whereas the global conductivity of the current structure drops by several orders of magnitude, and the rate of magnetic energy dissipation in it achieves a level observed in solar flares. The strength of a flare is determined by the size of the fractal of abnormal resistors involved into the process of local current dissipation. The amplitude-frequency dependence of the frequency of flares on the amplitude is essentially power-law with an index of the spectrum that is determined by fractal dimensionality of the dynamic ensemble of the percolating resistive-inductive-capacitive network.

## 2.6.9. Conclusions

The current situation with understanding the physics of solar and stellar flares perfectly repeats the situation of 80 years old when Giovanelli (1939) in his pioneer paper unambiguously showed that the source of flares was the magnetic fields of spots and their dynamics, but he failed to formulate it explicitly, since all the known at that time mechanisms for magnetic field dissipation excluded such a possibility.

Eighty years later, despite several substantial breakthroughs (the role of thin turbulent current sheets and the reconnection process in them), we are in some sense at the same position that Giovanelli was 80 years ago — we see that flare dissipation of magnetic fields takes place, but we cannot explain the way how it works. Hence, it is premature to tell about some progress in understanding the physics of flares.

The main problem of the existing tens and hundreds of models which are based on idealized simplified assumptions on the magnetic field geometry and processes in the current sheet plasma lies in their instability arising when taking into account the real geometries and processes (three-dimensionality of the magnetic configuration and the role of the magnetic field shear, instability of the current sheet with respect to its splitting by dissipative current modes into thin current-magnetic ropes with subsequent discontinuity of the latter by fast MHD instabilities of overwinding and torsional types with the formation of double current sheets between them, local break of plasma turbulence due to overheating of turbulent domains with subsequent rearrangement of currents in the network triggered by the jump-like rearrangement of local resistances). The standard approach to the process of a flare as stationary and static equilibrium is not able to involve permanent inner instability of the key processes in the current sheet.

An essential way out is the transition from the description of a flare within the framework of the paradigm of static and stationary equilibrium to the description within the paradigm of dynamic equilibrium when each element of the system constantly rearranges, causing a chain of changes of adjacent elements and forming clusters of fractal dimensionalities. Percolation of the global current through such a dynamic network of clusters, whose elements’ properties critically depend on the value of the local current in it, integrates these changes from micro- to macro-scales, ensuring the fractal (or multifractal) behavior of flare energy release.

The detection over active regions of a thin and ultrathin magnetic field structure consisting of numerous thin current-magnetic isolated ropes sinking into the weak background magnetic field and strongly interacting with both each other and background fields also rules out the static MHD approach and requires consideration of preflare equilibrium as dynamic, with a constant percolation exchange of magnetic tensions, currents, and magnetic energy between adjacent elements of the ensemble and the background. A mathematical apparatus required for such a description is currently absent, and the most perspective is a numerical simulation of such equilibrium with the greatest possible consideration of the real geometry and currents, threshold dependence of local resistances and tensions on the local current values and the possibility of plasma instability generation, processes of turbulence and overheating in such an ensemble of current-magnetic threads and currents sheets on the boundaries between them.

## Appendix. Using a Wide-Field Survey for Studying Solar Flare-Like Activity in Flare Stars

We currently understand the basic scheme of the origin of sunspot-flare activity in solar flares. Its main elements are as follows.

1. Generation of the magnetic field in the solar convective zone by dynamo mechanisms, including differential rotation, anisotropic convection (helicity with alpha effect), the formation of magnetic bundles and threads in the generated field.

2. The Parker buoyancy of magnetic ropes that carries them out into the solar atmosphere and the Coriolis twist of floating magnetic fluxes that generates a shear of the magnetic fields of various floating magnetic fluxes.

3. Flash dissipation of emerging non-potential magnetic fields with the formation of turbulent current sheets at the intersections of oppositely directed magnetic fluxes and current-magnetic braids of isolated magnetic threads and ropes.

To verify this scheme of the origin of spot-flare activity, it is extremely important to check it under different conditions of various physical parameters that determine this activity (the masses of stars, the power and the spectrum of convection in the convection zone, and the values of convection anisotropy and helicity at different scales (alpha parameters), different rotation periods and differential rotation, different Coriolis effects, etc).

Unlike a physical experiment based on a direct impact on the object in the process of research, astronomy is based exclusively on observations and cannot influence directly the object of research and conditions in it. Astronomy is not able to change purposefully the properties of the object under study. We cannot change the mass of the Sun, its energy release and rotation, heat transfer in the inner parts of the Sun, and the formation of convection. We have to be content with the Sun provided by nature. This greatly limits the possibility of verifying the theoretical description of the physics that determines the origin of spot-flare activity.

Observations of the same physical processes of activity in other stars (first of all, in solar-like flare stars with different masses, rotations, evolutionary statuses (Main Sequence, Hayashi stage,...) open up the fundamental possibility to discover the dependence of activity parameters on the state of stars (mass, temperature, rotation, evolutionary status, etc.). This is a reason of our interest to observations of hundreds of flare stars as testers for models of spot-flare solar-like stars activity under conditions other than those of the Sun.

In recent years, tremendous progress has been made in this direction thanks to many-year all-sky surveys with the TESS (Transiting Exoplanet Survey Satellite) and KEPLER telescopes.

The recent review of Feinstein et al. (2020) presents very interesting results of using the convolutional neural network (CNN), “stella”, specifically trained to find flares in Transiting Exoplanet Survey Satellite (TESS) short-cadence data. The authors applied the network to 3200 young stars to evaluate flare rates as a function of age and spectral type. They also measured rotation periods for 1500 of our targets and find that flares of all amplitudes are present across all spot phases, suggesting high spot coverage across the entire surface. Additionally, flare rates and amplitudes decrease for stars with age > 50 Myr across all temperatures $T_{eff} \geq 4000$ K, while stars from $2300 \leq T_{eff} < 4000$ K show no evolution across 800 Myr. Stars of $T_{eff} \leq 4000$ K also show higher flare rates and amplitudes across all ages. Authors investigated the effects of high flare rates on photoevaporative atmospheric mass loss for young planets and showed that in the presence of flares, planets lose 4-7% more atmosphere over the first 1 Gyr.

Seligman et al. (2022) used the TESS data for more detailed analysis of the dependence of the power spectrum slope of the flare frequency and flare amplitude as indicator of Coriolis effects influence on dynamo process, stellar convection, and flare activity. The authors expanded the standard topological DC model of coronal heating (Parker, 1972) along with the model of magnetic braiding and reconnection of flare dissipation of the magnetic field to include the Coriolis effect, and demonstrated that Coriolis process produces a shallower distribution of flare energies in stars that rotate more rapidly (corresponding to a weaker decline in occurrence rates toward increasing flare energies).

Another interesting attempt to determine the maximum possible energy of solar and stellar flares was made by Cliver et al. (2022). On the basis of the modern observations, historical or long-term data, including the auroral and cosmogenic radionuclide records, and Kepler observations of Sun-like stars, the authors compiled a table of 100- and 1000-year events based on occurrence frequency distributions for the space weather events. Questions considered include the Sun-like nature of superflare stars and the existence of impactful but unpredictable solar “black swans” and extreme “dragon king” solar phenomena that can involve different physics from that operating in events which are merely large.

# Part 3

# Long-Term Variations in Stellar Activity

## 3.1. Activity Cycles

Cyclicity in solar spottedness was discovered in the mid-XIXth century by the German amateur astronomer Heinrich Schwabe. Today, the 11-year cycle of solar activity, along with sunspots and solar flares, is the most well-known phenomenon in the solar life. The body of data collected thus far on the cyclicity of solar activity is huge. It was found that all other characteristics of solar activity change in parallel with the spottedness of the Sun, which is characterized by Wolf's numbers, including the number of spots and their groups, and by the average latitude of spots. These characteristics are the size and number of active regions of the chromosphere, the frequency and intensity of flares, the structure of the solar corona and the intensity of its X-ray emissions, the characteristics of the solar wind including the parameters of the interplanetary magnetic field. All manifestations of solar–terrestrial relationships and most geophysical phenomena are synchronized with the solar-activity cycles. After detecting a close relation between the level of solar activity and the thickness of annual tree rings and the abundance of carbon isotope $^{14}C$ in them, the so-called dendrochronological scale was constructed following these terrestrial marks, which helps in tracking the variations of solar activity over up to 200 million years (Dmitriev et al., 2015). There are geological structures evidencing the existence of the same cyclicity up to 680 million years and, possibly, even up to 2 billion years ago (Tyasto et al., 2017, 2018).

An average solar cycle is 11.2 years, whereas the duration of individual cycles varies from 7 to 17 years. The depth of modulation of all quantitative characteristics of activity varies essentially from cycle to cycle. The change in sign of the total magnetic field and those of magnetic fields in the pairs of leading and tail spots on each solar hemisphere during neighboring cycles suggests 22-year solar magnetic cycles.

Two other important features of solar activity connected with spottedness are essential for the subsequent discussion. First, in addition to the 11-year cycle, there is the so-called Gleissberg secular cycle that lasts for 80–90 years. It is suspected that a longer cycle of several centuries exists. Second, in 1645–1716 the solar spottedness was tens of times lower than usual: spots could hardly be detected at maximum phases and only few aurorae, one of the major geophysical effects that allows determining the solar-cycle phase, were seen from the Earth. It was the so-called Maunder minimum of solar activity. Apparently, this happens once every 2–3 centuries and in total such periods cover 1/3 of the time.

All manifestations of cyclicity of solar activity are associated with various changes of the subphotospheric magnetic field of the Sun, and the solar 11-year cycle was a starting point for the elaboration of the solar dynamo model. Analogously, the cyclicity of stellar activity is one of the basic directions in studying the generation mechanisms of stellar magnetic fields. As the data on the cycles of stellar activity were accumulated, attempts were undertaken to select one of the models of stellar dynamo. The study of the cyclicity in stellar activity uses not only known characteristics of solar activity but in its turn supports solving this important problem of solar physics. A comprehensive review regarding the solar cycle models was published by Charbonneau (2010), a chapter in the monograph of Obridko and Nagovitsyn (2017) was devoted to the issues concerning the cyclicity of solar activity within the concept of multifractality of nonlinear dynamic systems.

Further, we consider available data on stellar cycles at different levels of stellar atmospheres.

* * *

The first experimental data on possible activity cycles of red dwarfs were obtained in interpreting their broadband photometry after ten years of intense photoelectric observations.

Based on the Crimean observations of BY Dra brightness in the B and V bands, Chugainov (1973) suspected a 8–9-year cycle of spot formation with minimum brightness in 1965. This conclusion had ambiguous continuation: Vogt (1975) doubted this period. The data of Oskanian et al. (1977) on the 20-year interval did not contradict Chugainov's conclusion. From photoelectric observations of the star from the Stephanion Observatory, Mavridis et al. (1982) concluded that in the late 1970s the star entered a new minimum stage with a depth of not less than $0.3^m$ and duration of not less than 14 years. Observations by Cutispoto (Rodonò, 1987) and Pettersen et al. (1992b) during the next decade confirmed the 14-year cycle. The data of Panov et al. (1995) evidenced the end of this cycle that, however, was completely dissimilar to the minimum of 1965–67. Further, Mavridis et al. (1982) suspected slow oscillations in quiet brightness of EV Lac in the B band with an amplitude of $0.3^m$ and characteristic time of about 5 years. Mahmoud (1993a) confirmed this period on the basis of expanded observations. From the annual average flare radiation in the U band Andrews and Marang (1989) estimated the cyclicity of the dMe star FL Aqr as 10–18 years. On the basis of long UBV observation series, Aslan et al. (1992) found an 8–14-year activity cycle of DH Leo. From a 10-year series of photometric observations of the rapidly rotating K2 dwarf LQ Hya Jetsu (1993) and Cutispoto (1993) estimated the duration of its activity cycle as 6.2 years. On the basis of observations of EK Dra in 1983–94 in the V band, Dorren et al. (1995) suspected the photometric variability cycle of 12–14 years. From UBV observations by two automated telescopes in 1989–98 Messina et al. (1999a) found the 3.9-year cyclicity period of the average brightness change for the K0 V star DX Leo of the Pleiades moving group. The phase of this cycle correlates with the duration of the photometrically determined axial rotation period. Based on 11-year VRI observations of more than 40 M dwarfs, Weis (1994) revealed 2.7- and 2.9-year photometric periods for Gl 213 and Gl 876, respectively, for which other manifestations of activity were not known. From the 22-year observational series of VY Ari Strassmeier et al. (1997) suspected a 14-year photometric cycle and from the 14-year observations of LQ Hya, a 7-year photometric cycle, which did not contradict the above results of Jetsu and Cutispoto. Continuing these studies, Olah et al. (2000, 2001) analyzed long-term photometric observations of 10 active stars, searching for multiperiodicity. For three program dwarfs the following cycles were obtained: LQ Hya – 11.4 and 6.8 years, V 833 Tau – 6.5 and 2.4 years, and BY Dra – 13.7 years. Later, based on 34-year observations, Olah and Strassmeier (2005) confirmed the earlier obtained durations of short cycles for LQ Hya, V833 Tau, and several RS CVn-type variables with an accuracy of up to 20–30 %. Taking these and other literature data into account, they marked on the plane ($\log(1/P_{\rm rot})$, $\log(P_{\rm cyc}/P_{\rm rot})$) three linear sequences with slopes 0.54, 0.72, and 0.69 and the correlation coefficients 0.96, 0.94, and 0.96 for short, middle, and long cycles, whereas Baliunas et al. (1996) found a slope of 0.74 for slowly rotating stars within the HK project described in detail in Subsect. 1.3.1.1.

Berdyugina et al. (2002) found 3 periods of spottedness for LQ Hya: a 5.2-year period of switching of active longitudes, a 7.7-year period of change of amplitudes of brightness oscillations, and a 15-year period of variations of average stellar brightness during which the differential rotation period was detected. Kővári et al. (2004) obtained slightly different values — 13.8, 6.9, and 3.7 years — from a 21-year observational series.

Continuing the studies of Messina et al. (1999a), Messina and Guinan (2002) analyzed 10-year observations of six young G-K dwarfs and established photometrical periods: 6.7 years for BE Cet, 5.9 years for $\kappa^1$ Cet, 13.1 years for $\pi^1$ UMa, 9.2 years for EK Dra, 5.5 years for HN Peg, and 3.2 years for DX Leo. In addition, a secondary period of 2.1 years was found for $\pi^1$ UMa. A confident correlation of the brightness amplitude during the cycle and the Rossby numbers was revealed.

However, using 1030 negatives from the collection of the Sonneberg Observatory, Fröhlich et al. (2002) studied the brightness of EK Dra since 1958 and detected since the mid-1970s its systematic decrease with a velocity of $0.0057^m \pm 0.0008^m$ per year. If it is not a manifestation of the very long — half a century— activity cycle, then there are no solar analogs of such a long process. Fröhlich et al. discussed it due to such a qualitative distinction in stellar dynamo, depending on the rotation rate, since the axial rotation period of EK Dra is ten times shorter than that of the Sun.

After enumerating the photometrically detected cases of cyclic activity of red dwarfs one should mention unexpected results obtained by Eaton et al. (1996) in the numerical experiment. Considering the cyclic activity of RS CVn-type stars, they showed that the numerous spots of moderate size and with a characteristic lifetime of about one year randomly distributed over the stellar surface could yield a light curve, which can be accepted as evidence of cyclic activity.

* * *

Other parameters considered in searching for activity cycles were the frequency and energy of stellar flares.

In studying EV Lac in the B band, Mavridis et al. (1982) suspected the changes of average frequency and total energy of flares, parallel to variations of the average stellar brightness. However, no periodicity in their distribution was revealed in a more extensive sample of flares on this star in the U band (Alekseev et al., 2000). Pettersen et al. analyzed 241 flares on AD Leo and suspected an 8-year period of flare frequency (Pettersen et al., 1986a), but in considering flares on AD Leo over two decades they did not find variations of the average annual flare amplitudes and energies by more than a factor of 2 (Pettersen et al., 1990a). Pooling together the observational results for UV Cet flares obtained in 1966–88 in Chile, the USA, and Bulgaria, Pettersen et al. (1990b) selected 808 events with an amplitude over 1.4 and considered their time distribution. Following the $\chi^2$ criterion, they rejected the assumption of the constancy of flare frequencies and suggested a 10–15-year cycle. From the observations in the B band, Mahmoud (1993a) suspected a 6.6-year cycle of UV Cet activity. Based on the 20-year observation series of BY Dra in the B band, Mavridis et al. (1995) noted the maximum of average flare energy in 1973–75 and subsequent increase of this value in 1987. As stated above, Alekseev and Gershberg (1997a) found periodic variations of the index of the energy spectrum of flares on EV Lac with a characteristic time of about 7.5 years without appreciable changes in the level of flare activity.

Mirzoyan and Ohanian (1977) found evidence of variability of the activity level of flare stars in the Pleiades, which they attributed to cyclic activity.

* * *

The cycles of stellar activity at the photospheric level can be studied using glass libraries. The American researchers invoked the Harvard collection of negatives comprising images of the sky since the late XIXth century.

Using the Harvard Collection, Phillips and Hartmann (1978) investigated the behavior of average annual brightness of BY Dra-type stars: BY Dra itself, CC Eri, YZ CMi, and AU Mic. Images of each star were measured on an iris photometer using more than a hundred negatives from the early XXth century to the 1960s. As a result, a distinct photometric wave was found on BY Dra with an amplitude of $0.3^m$ and duration of 50–60 years. Considering the binarity of BY Dra, the amplitude of the found maximum in the early 1930s was determined as $0.5^m$. Similar behavior was found for CC Eri, whereas two other stars did not show brightness cyclicity. Then, Hartmann et al. (1981) studied 225 negatives of the Harvard Collection to determine the long-term brightness dynamics of the dK5e star BD+26°730 (= Gl 171.2 = V 833 Tau) and for the late 1930s rather confidently found cyclic changes with an amplitude minimum of about $0.5^m$ with a cycle a duration of about 60 years. In addition, they suspected overlapping cycles of lower amplitude with a duration of about 10 years. Subsequent photoelectric observations by Olah and Pettersen (1991) yielded results that did not contradict the photographic light curve by Hartmann et al.

From a 20-year series of photographic observations of the flare star HII 2411 in the Hyades Szecsnyi-Nagy (1986) suspected that its activity cycle was 10–15 years.

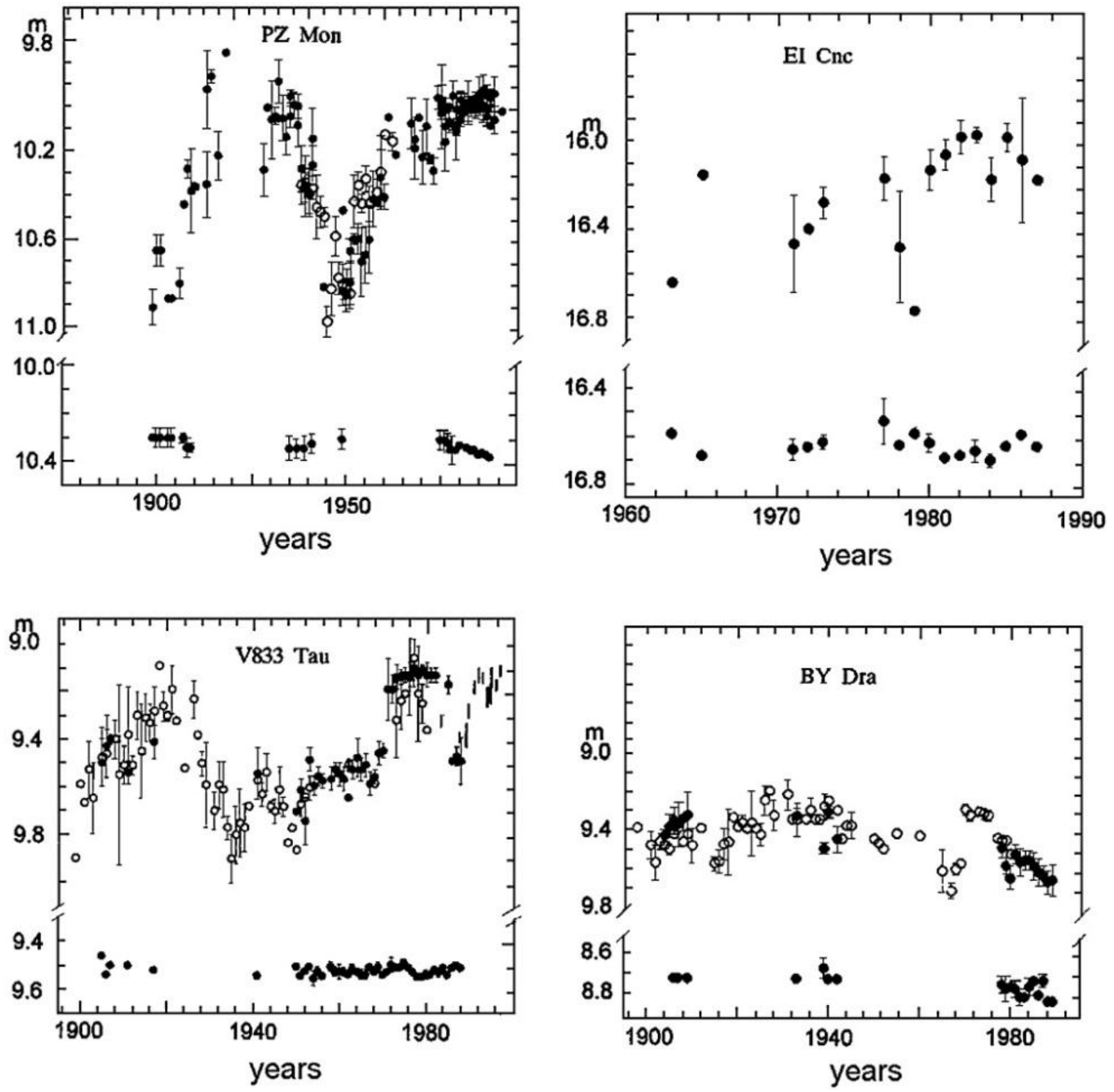

Fig. 79. Long-term brightness variations of red dwarfs: *dark circles* after Bondar (1996), *light circles* on PZ Mon diagram after Wachmann (1968), *light circles* on V 833 Tau diagram after Hartmann et al. (1981), *light circles* on BY Dra diagram after Phillips and Hartmann (1978); *vertical strokes* on PZ Mon diagram from photoelectric measurements by Alekseev and Bondar (1998) The light curves of comparison stars are given at the bottom of each diagram (Bondar, 2000)

Bondar (1995, 1996, 2000) undertook the most extensive research in this direction: using the photograph collections of Moscow University, and Odessa and Sonneberg observatories, she considered the behavior of average annual brightness levels of 40 dKe-dMe stars. In eye estimates and iris photometer measurements of about 5900 negatives obtained in 1896–1992, she found a variation of the average annual brightness with amplitudes from $0.3^m$ to $1.0^m$ and characteristic lifetimes from 3 to 60 years for 21 stars. On eight of them cyclic spottedness could be ascertained confidently, on the others, only presumably. Amplitudes higher than $0.5^m$ were found in four red dwarfs: V 833 Tau, PZ Mon, EI Cnc, and BY Dra (Fig. 79). The light curves constructed by Bondar for BY Dra and V 833 Tau are in good agreement with the results obtained from the Harvard glass library, and the light curve of PZ Mon, with the data from the Heidelberg collection. The variability amplitudes found by Bondar exceeded the values obtained earlier from shorter intervals and used for constructing the spottedness models of such stars. In Fig. 80, the cyclicity parameters estimated by Bondar are compared with other stellar characteristics. One can see that long activity cycles are typical of stars with a rotational

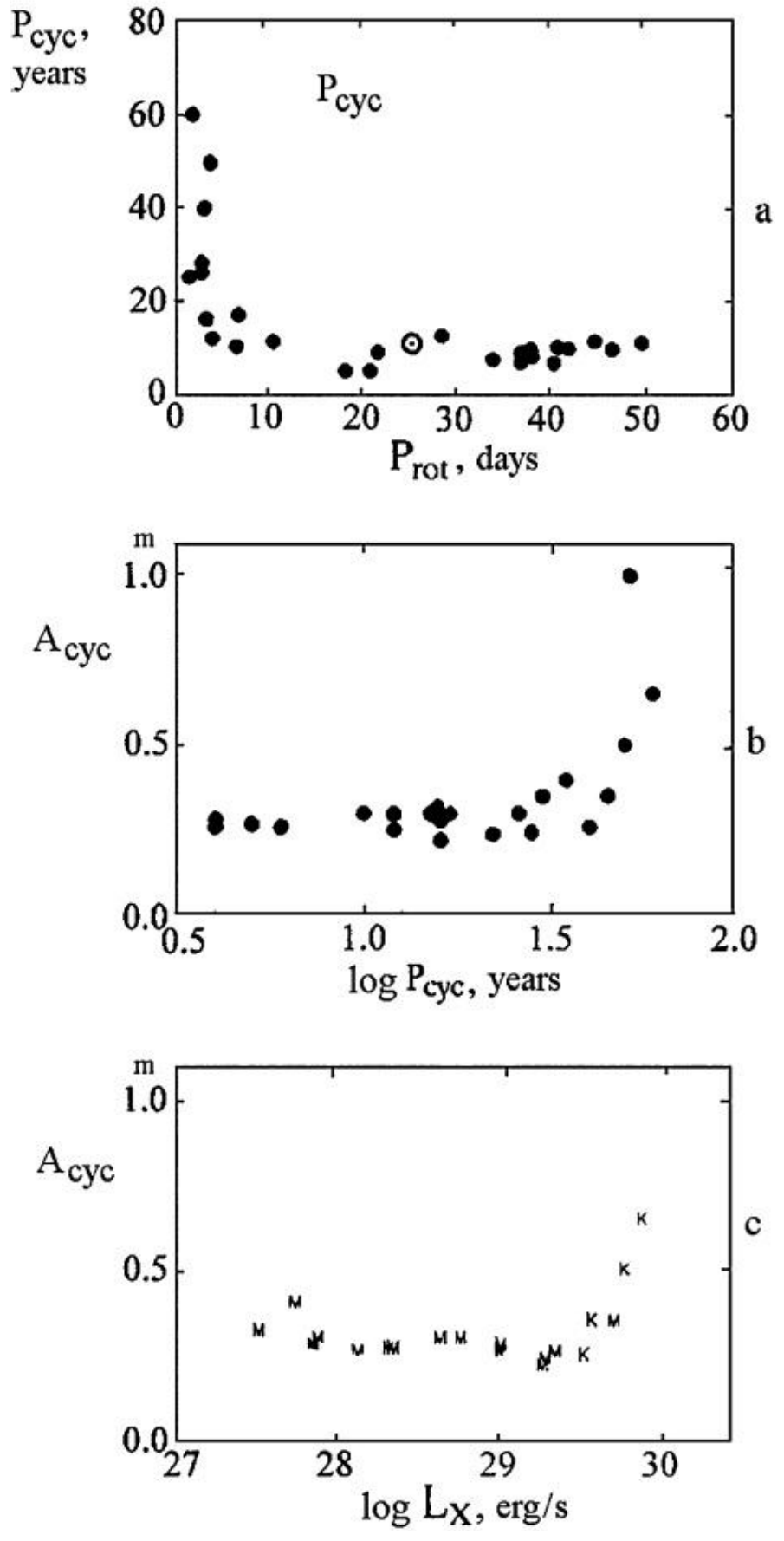

Fig. 80. Comparison of characteristics of long-term brightness variations and global properties of red dwarf stars: (**a**) axial rotation periods and duration of activity cycles; (**b**) duration of activity cycles and the amplitudes of optical brightness; (**c**) luminosities of K and M dwarfs in soft X-rays and the amplitudes of optical brightness (Bondar, 2000)

period less than five days (Fig. 80a). Later, this result was confirmed by Messina and Guinan (2002). From less homogeneous data a similar result was obtained earlier by Vogt (1983) and from the chromospheric data for a sample of earlier stars it was suspected by Saar and Baliunas (1992a). Figures 80b and 80c show that the amplitudes of average annual brightnesses above $0.5^m$ are characteristic of stars with cycles longer than 30 years and with luminosity in soft X-rays above $3 \cdot 10^{29}$ erg/s. High values of $L_X$ are mainly found for K dwarf stars.

A few years later after Bondar, Lorente and Montesinos (2005) constructed a dependence of $P_{cyc}$ on $P_{rot}$ using 25 stars with well-determined parameters from the HK project, and their results have much in common with Fig. 80a but also have some noticeable differences. As well as in the results of Bondar, at small $P_{rot}$ the values of $P_{cyc}$ are maximal, but small $P_{rot}$ extend here up to 20 days rather than 5, as it was found by Bondar, and maximal $P_{cyc}$ amount to just 20 years, whereas that of Bondar — 60. On the other hand, according to Bondar, at $P_{rot} > 10$ days, $P_{cyc}$ becomes stable at the level of about 10 years, whereas, following Lorente and Montesinos, in the region of $P_{rot} \sim 30$–50 days there takes place a rise of $P_{cyc}$ up to 15 years. Lorente and Montesinos did not construct the dynamo models for each star, but they somewhat modified the tachocline dynamo with the aim of representing the magnetic cycles of late stars, undertook an attempt to reproduce the total dependence of the duration of cycles on the rotation period, and showed that within this concept in tachocline a strong gradient of the angular velocity dominates and the toroidal field is generated, whereas the poloidal field is formed at the bottom of the convective zone. Furthermore, they found that a half of stars with the shortest rotation periods had two cycles — short and long. If one singles out only short cycles, then along with other considered stars a linear change of the function $P_{cyc}(P_{rot})$ is detected. By selecting a gradient of the angular velocity and tachocline thickness as a function of rotation they reproduce the linear dependence $P_{cyc}(P_{rot})$.

Recently, Bondar (2017) obtained one more direct confirmation of a significant duration of cycles for rapidly rotating stars: based on the 80-year photometric series for the red dwarf YZ CMi with the rotation period 2.77 days, she detected an activity cycle of 27.5 years. However, as it was noted above (Karmakar et al., 2016), for the superfast rotator LO Peg with an axial rotation period of 0.4231 days the activity period was 2.7 years.

A study of stellar activity cycles from the results of constructing models of their zonal spottedness was carried out by Alekseev (2005). From more than three tens of the considered objects for 10 of them, including 7 red dwarfs, he detected cyclicity of mean latitudes and total areas of starspots, the revealed cycles had durations from 4 to 15 years, whereas the durations of cycles found from mean latitudes and areas of spots were in good agreement. The durations of cycles showed no distinct dependence on the stellar spectral type, rotation rate, and Rossby number. For most stars, there is a rough analog of solar butterfly diagram — a decrease in the mean latitude of spots with their growing area. For a number of objects, particularly for the "young Sun" LQ Hya, there was revealed a flip-flop effect of switching active longitudes that occurred at the maximum epochs of mean latitudes, and also a decrease of the photometric stellar rotation period with a drift of spots toward the equator, i.e., an analog of the solar differential rotation. For V 833 Tau, BY Dra, EK Dra and VY Ari, short Schwabe periods coexist with the analogs of the long-term solar Gleissberg period at which the area of spots achieves a half of the total stellar surface. Figure 81 shows the drift rates of spots in latitude $\delta\phi$ and differential rotation coefficients $D_r$ obtained by Alekseev. The latitude drift rates of spots $\delta\phi$ quite strongly vary from cycle to cycle and from star to star, accounting for −0.8 to −2.6 degrees per year, whereas on the Sun this value is from −3 to −4. But three coolest stars V 833 Tau, BY Dra, and EV Lac show the picture that is contrary to that of the Sun — the drift of spots toward the pole as their area increases. For these stars, the differential rotation coefficient is from −0.01 to −0.04, whereas on the Sun it is +0.19.

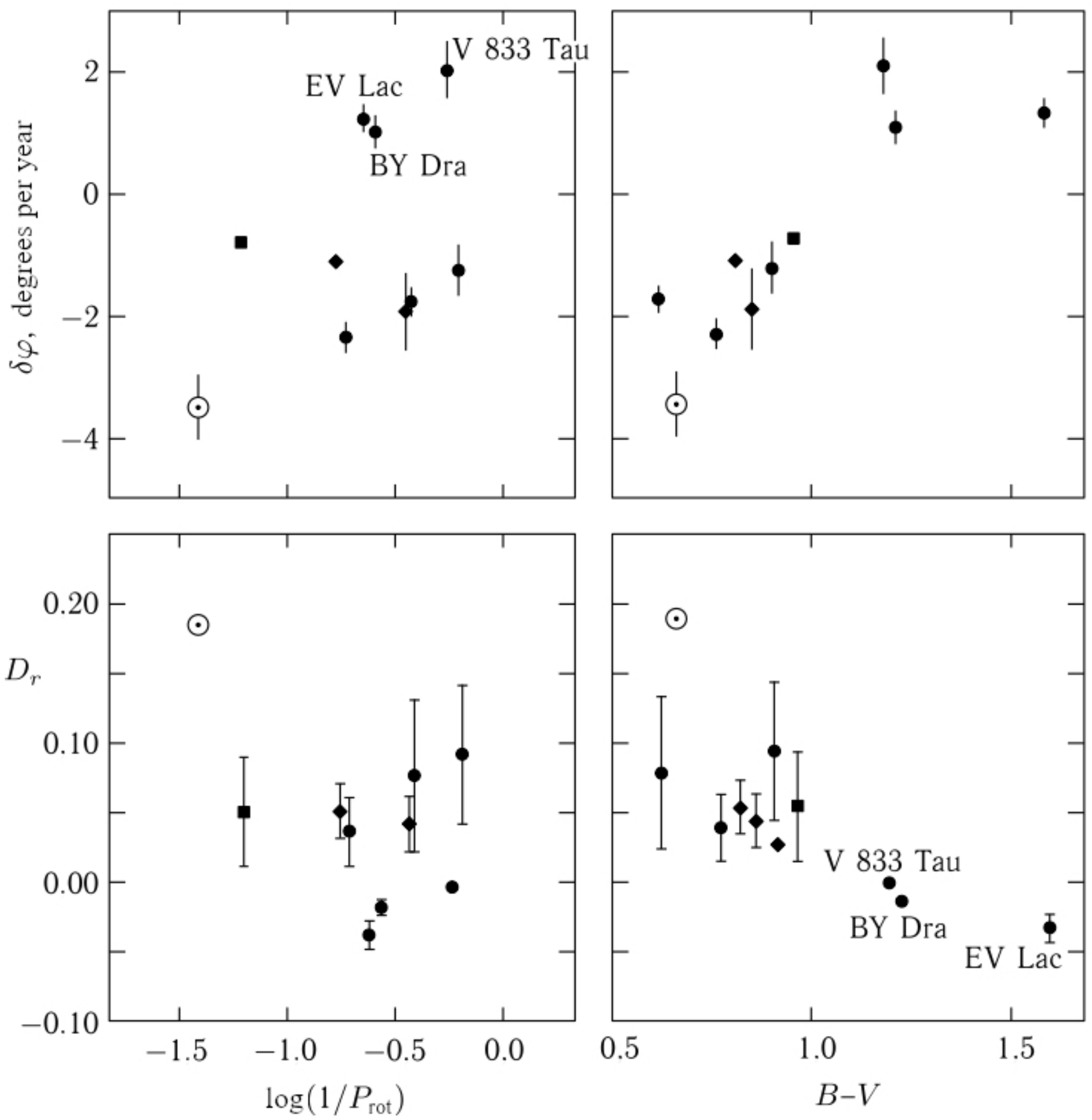

Fig. 81. Dependence of the latitude drift rate of spots $\delta\varphi$ and the differential rotation coefficient $D_r$ on the stellar angular velocity $1/P_{rot}$ and the color index B–V (Alekseev, 2005)

From observations of the CaII K for the closest solar twin 18 Sco, Hall and Lockwood (2000) recorded in July 2000 a fast rise of emission after the minimum of 1998 and attributed this due to a cycle with larger amplitude than that of the Sun.

Jaervinen et al. (2005b) analyzed the photometric observations of AB Dor for more than 20 years and detected a long-lived nonaxisymmetric distribution of spots — active longitudes in different stellar hemispheres and two activity cycles: the switching of active longitudes with a period of 5.5 years and the 20-year cycle in the levels of mean and maximal stellar brightness.

Akopian (2010) suggested a method for searching for the cycles in stellar flare activity and detected such cycles for the stars in the Pleiades Ton 91 and Ton 377 with a duration of 15.2 and 17.7 years, respectively.

Searching for the cycles from calcium emission for the southern solar-type stars, Metcalfe et al. (2010) detected the shortest period 1.6 years for $\iota$ Hor. Thereafter, Sanz-Forcada et al.

(2013) during about two years carried out XMM-Newton observations of this star, confirmed cyclicity of its activity in calcium lines, and found it in X-rays; this star proved to be the most active among those for which coronal activity was detected.

Savanov (2012) carried out a study of activity cycles for 31 M dwarfs from the photometric ASAS data using the amplitude power spectra and wavelet analysis. Most of the stars examined had a periodicity of light curves on the time scales of hundreds of days–years. In the diagram, in axes $\log(P_{cyc}/P_{rot})$ and $\log(1/P_{rot})$ all the data concerning the stars with convective envelopes and fully convective stars can be interpreted by the total linear dependence but do not fall into the dependences for relatively young and active stars and seem to form one more branch of low-mass stars having irregular magnetic activity that is generated and maintained either by the distributed or small-scale dynamo.

* * *

In a review paper, Rempel (2008) considered stellar cycles in the context of different dynamo models developed for the 11-year solar cycle; taking into account the differences of characteristics of stars with respect to those of the Sun in rotation rates and the depth of the convective zone, he concluded that two groups of stellar observations — the dependence of the cycle duration on rotation or, better, on Ro and the dependence of activity on rotation — represent the basic interest for the dynamo theory.

Outlining the general scheme of solar activity cyclicity, Lanza (2009) noted that activity observed on the surface may be a result of the contribution of different dynamo models that differ in localization, mechanisms for maintaining the azimuthal field, and its amplitude with respect to the mean field, and emphasized an important role of the meridional circulation for the basic characteristics of activity — the duration of a cycle, the direction and transfer rate of the azimuthal field, i.e., the dynamo waves.

For many years it was believed that during the global solar magnetic field reversal, i.e., at the beginning of a new cycle, the dipole momentum value passes through the zero. However, considering in detail variations of this magnetic momentum over three solar cycles, Livshits Jr. and Obridko (2006) found that at that time the zeroing of the dipole filed component did not occur, but there took place its complex drift. Supposing that its drift was associated with meridional flows, Baklanova and Plachinda (2014) considered the Sun and 27 F9−K7 dwarfs with known rotation periods and activity cycles and found the average velocity of such a dynamo-wave drift to be 5.4 m/s, whereas on the Sun it amounts to 6.29 m/s. This velocity value for the considered stars does not depend on the Rossby number, but the activity cycle duration is well represented analytically as a function of this velocity:

$$P_{cyc} = 5.74 + 35 \exp[-(\langle v \rangle - 2.1)/1.66] \text{ years.} \quad (67)$$

* * *

The most reliable and numerous initial data for the cycles of stellar activity are the results of long-term monitoring of CaII H and K emission in the spectra of single late main-sequence stars started in the 1960s by O. Wilson at the 100″ telescope of the Mount Wilson Observatory. In the first paper from the series of works published within this program Wilson (1978) presented the measurement results for calcium emission in the spectra of 91 F5–M2 dwarfs over 9–11 years during which a measurement accuracy of up to 1– 3% was achieved. For 13 program stars the cyclicity of changes of calcium emission was confidently found, for 12 it was established presumably. The duration of found cycles varied from 7 to 14 years. Further, 20% of the studied stars demonstrated the constancy of calcium emission, the others, significant nongradual variations. Then, Vaughan (1980) found regular differences in the variability of calcium emission among the stars considered by Wilson. On older stars with rather weak emission the changes were gradual enough, similar to the solar cycle, and the emission bursts were short-lived, as on the Sun, at maximum phases of these cycles. On younger stars with rather strong emission the variation proceeded rather chaotically and, as a rule, no cyclicity was revealed. Then, Vaughan and Preston (1980) added four or five stars with cyclic variations of calcium emission to the Wilson list. On the basis of these observations Vaughan et al. (1981) concluded that 10–12-year cycles were typical of stars whose axial rotation periods exceeded 20 days, and that the duration of these cycles did not correlate with rotation rate. However, later, Dorren and Guinan (1994) in observations of ultraviolet lines of the transition zone found a 12-year cycle on “the young Sun”, HD 129333, with an axial rotation period of 2.7 days.

Between the groups of young and old stars on the planes (log $S$, B−V), where $S$ is a value proportional to the sum of equivalent widths of CaII H and K lines (see Subsect. 1.3.1.1, Fig. 14), a certain deficiency of objects was found, which was called the Vaughan–Preston gap. Durney et al. (1981) associated it with a fast reduction of the dimensionless dynamo number $N_D$ at a certain stage of stellar evolution. The number is the ratio of generating and dissipative members in the magnetohydrodynamic equation. It determines the efficiency of the dynamo mechanism and is functionally connected with the Rossby number: $N_D \sim \text{Ro}^{-1/2}$. They showed that, according to theoretical expectations of Parker (1971) at high $N_D$ small-scale rapidly variable magnetic fields were generated, whereas at low $N_D$ global magnetic fields of solar type with slow cyclic variations were excited.

After the late 1970s the studies within the Wilson program were actively continued by O. Wilson’s followers at the 60″ telescope of the Mount Wilson Observatory and regularly yielded new results with an ever-increasing time interval. Simultaneously, the criteria for selecting cyclic variations of chromospheric emission were specified and the number of stars with such variations was determined more precisely.

When activity cycles were revealed on 27 program stars, Vaughan (1983) compared the distribution of their durations with those in solar cycles since 1740 and found a similarity of these distributions, but some stellar cycles were shorter than the shortest solar cycles. Analysis of the data obtained over 18 years using the power-spectrum formalism, in general, confirmed the difference in variability of calcium emission on active and inactive stars found by Vaughan (1980) But the difference was not very large: activity periods of 2.6, 3.8, and 12 years were found for rather active dwarfs. In the sample of stars with activity periods found from the power spectrum, no correlations of cycle durations with the Rossby number or the amplitude of changes in calcium emission were revealed. The periods shorter than five years were found only on F6-G1 stars, about seven years – only on K dwarfs, and periods of 10 years and more were found on stars of all spectral types (Baliunas and Vaughan, 1985).

* * *

After successful establishment of the correlation between calcium emission intensity and the Rossby number (Noyes et al., 1984a), Noyes et al. (1984b) on 13 slowly rotating late stars, including the Sun, found the correlation between the durations of activity cycles and Rossby numbers as

$$P_{\text{cyc}} \sim \text{Ro}^{1.3\pm0.5}. \tag{68}$$

Later, from 22 slowly rotating stars with well-determined periods of chromospheric activity Ossendrijver (1997) refined the exponent by 2.0 ± 0.3.

However, Maceroni et al. (1990) considered the distribution of durations of the cycles of 60 Wilson program stars and did not confirm the relation (68) but found an obviously asymmetrical distribution of cycle durations with maxima of about 6 and 11 years. As stated above, separate consideration of various spectral types of stars revealed a regular growth of cycles from F to G and further to K stars. Thus, the suspected bimodality of the general distribution can be a result of summing different distributions with noncoincident maxima.

On the basis of a 20-year observational series, Baliunas and Jastrow (1990) analyzed the distribution of magnetic activity on 74 G1–G8 dwarfs with mass and age close to those of the Sun. The histogram of *S* had a precise two-peaked structure: a broad component, whose gravity center practically coincided with the average value on the Sun during the solar cycle, and distinctly separated the narrow component, whose gravity center was appreciably less than the minimum value of *S* on the Sun during the cycle and very close to the value of *S* in solar regions with zero magnetic activity. The width of the broad component of the histogram slightly exceeds the range of solar values of *S* during the cycle. The stars of the broad component displayed activity cycles, whereas the stars of the narrow component did not show cyclicity. It is natural to consider the stars in the narrow component of the histogram as being at the stage of the Maunder minimum, whereas the clear split of the components of the histogram means that the transition from the stage of activity observed now on the Sun with well-expressed cyclicity to the stage of Maunder minimum occurs very quickly. This fast transition can be observed directly on HD 3651, where, after the cycle maximum of 1977–78, *S* decreased below the values recorded earlier, and in the further change of this value the traces of cyclicity disappeared. The ratio of the areas of the histogram components is close to 3:1, which corresponds to the ratio of times spent by the Sun in two stages of different activity. Soderblom and Clements (1987) arrived at a conclusion that the Maunder minima occurred very infrequently on young stars. Wright (2006) analyzed the values of *S* for about a thousand dwarfs and subdwarfs and detected their dependence on the evolutionary status of a star and its metallicity. In particular, for objects that are older than 6 billion years the values of *S* decrease with growing [Fe/H], and these stars can be erroneously attributed to those at the Maunder minimum. But when Judge and Saar (2007) from the value of *S* selected the most probable candidates for objects at the Maunder minimum — HD 10700 and HD 143761, then these stars measured at the Hubble telescope and within the ROSAT project showed chromospheres, transitions zones, and coronae that were close or not much weaker than in the current solar minimum.

Based on a nearly 50-year series of S values for HD 66620, which is somewhat older and less active than the Sun, Luhn et al. (2022) detected its activity cycle and confidently recorded its entering into the Maunder minimum.

For a more strict analysis of *S* measurements, Horne and Baliunas (1986) applied a search algorithm for periodicities with respect to nonuniform time series with the estimate of probability of false periodicity, formalized the selection of noise peaks on the power spectra from a useful signal, and estimated a number of independent frequencies, signal-to-noise ratios, and a number of measurements required to determine one and two periodicities in the time series. Gilliland and Baliunas (1987) showed that the basic source of noise in the determination of stellar cycles was rotational modulation of active regions stochastically distributed over the stellar surface.

In 1991, the Wilson program included 99 stars: 36 F, 38 G, and 25 K. Within a strict approach to the establishment of periodicity, according to Horne and Baliunas (1986), for 10–15% of sample stars periodic activity was not revealed, for 40% chaotic variations or rather probable false periodicity were found, for 15% — long-term trends or periods longer than 25 years. Only for 30 stars — 5 F, 10 G, and 15 K dwarfs — activity cycles were confidently determined. As a rule, these stars, as the Sun, are characterized by a fast rise to maximum and a slower decline to minimum. Saar and Baliunas (1992a) compared the periods of activity of the stars with B−V color indices, depths of convective zones, activity indicators $S$ and $R_{HK}$, axial rotation periods, and Rossby numbers. They did not find a precise correlation, though they noted a significant range of values $P_{cyc}$ at low $P_{rot}$ and narrowing of this range at $P_{rot} > 30$ days. Considering the amplitudes of cyclic variations, Saar and Baliunas found maximum values of $A_{cyc}$ near the K2 spectral type and a systematic decrease of the maximum amplitude with growing $R_{HK}$.

For the active G dwarf $\kappa$ Cet with a chromospheric cycle of about 5.6 years, Saar and Baliunas (1992b) compared measurements of the magnetic field and chromospheric emission over four seasons and revealed a weak correlation between these values, similar to that on the Sun.

Preliminary analysis of the results obtained over 25 years within the O. Wilson program for variations of calcium emission at time intervals more than a year were published by the team of 27 researchers, including the recently deceased initiator (Baliunas et al., 1995). Examples of the long series for $S$ are presented in Fig. 82. The full sample included 111 stars; for 52 of them, including the Sun, the cycles were found, for 31 stars — constant emission or linear trends, and for 29 — nonperiodic variability. The found periods varied within 2.5 to 25 years, but all "good" and "excellent" periods thus classified based on the probability of false periodicity exceeded seven years. Solar cycles shorter than seven years did not occur for 250 years. Probably, the point is that short periods are less steady and they do not persist even over 25 years. From 25-year statistical data, Donahue (Saar et al., 1994a) found that the average logarithm of stellar age for stars with irregular changes of $S$ was 9.03, for stars with multiperiodic changes — 9.22, for stars with cyclic oscillations — 9.46, and for stars with constant values — 9.86. On stars with cyclic variations of calcium emission $\log t < 9.5$ and on stars with constant emission $\log t > 9.6$. Cyclic activity appears on G and K dwarfs at $\log t \sim 8.8$ and on F stars at $\log t \sim 9.3$. According to the results of 30-year observations, 60% of the program stars had confident or assumed activity cycles, 25% had nonperiodic variability, and 15% had constant calcium fluxes (Baliunas et al., 1998).

Saar and Baliunas (1992) performed a search for relations between $P_{cyc}$ and stellar parameters within the framework of the $\alpha - \omega$ dynamo theory. Comparing the normalized frequencies of cyclic variations and dimensionless dynamo numbers, they found correlations between these values in two subsamples containing active and inactive stars: most stars considered by Noyes et al. (1984b) and the Sun were in the subsample of inactive older stars, whereas in the subsample of young active stars the ratios of $R_{HK}$ were twice as high, the dynamo number was greater, and the frequency of cyclic variations was lower. Considering the amplitudes of cyclic variations, Saar and Baliunas found that $A_{cyc}$ grew together with $P_{rot}$ and the frequency of cyclic variations.

Soon et al. (1994) considered together the 250-year observational series of the Sun and 25-year observations of the Wilson program stars and found the general regularity: a systematic decrease of the amplitude of activity $\Delta R'_{HK}$ with increasing ratio $P_{cyc}/P_{rot}$. From the same data Baliunas and Soon (1995) found an inverse correlation between the cycle duration $P_{cyc}$ and the average activity level $\langle R'_{HK}\rangle$ and a direct correlation between the amplitude of $S$ and photometric amplitude during the activity cycle. Baliunas et al. (1996) divided the Wilson program stars for which cyclic activity was ascertained, into groups of more and less active stars using $\langle R'_{HK}\rangle$ and $P_{rot}$ and found that for the second group of stars, which included the Sun, the following expression was valid:

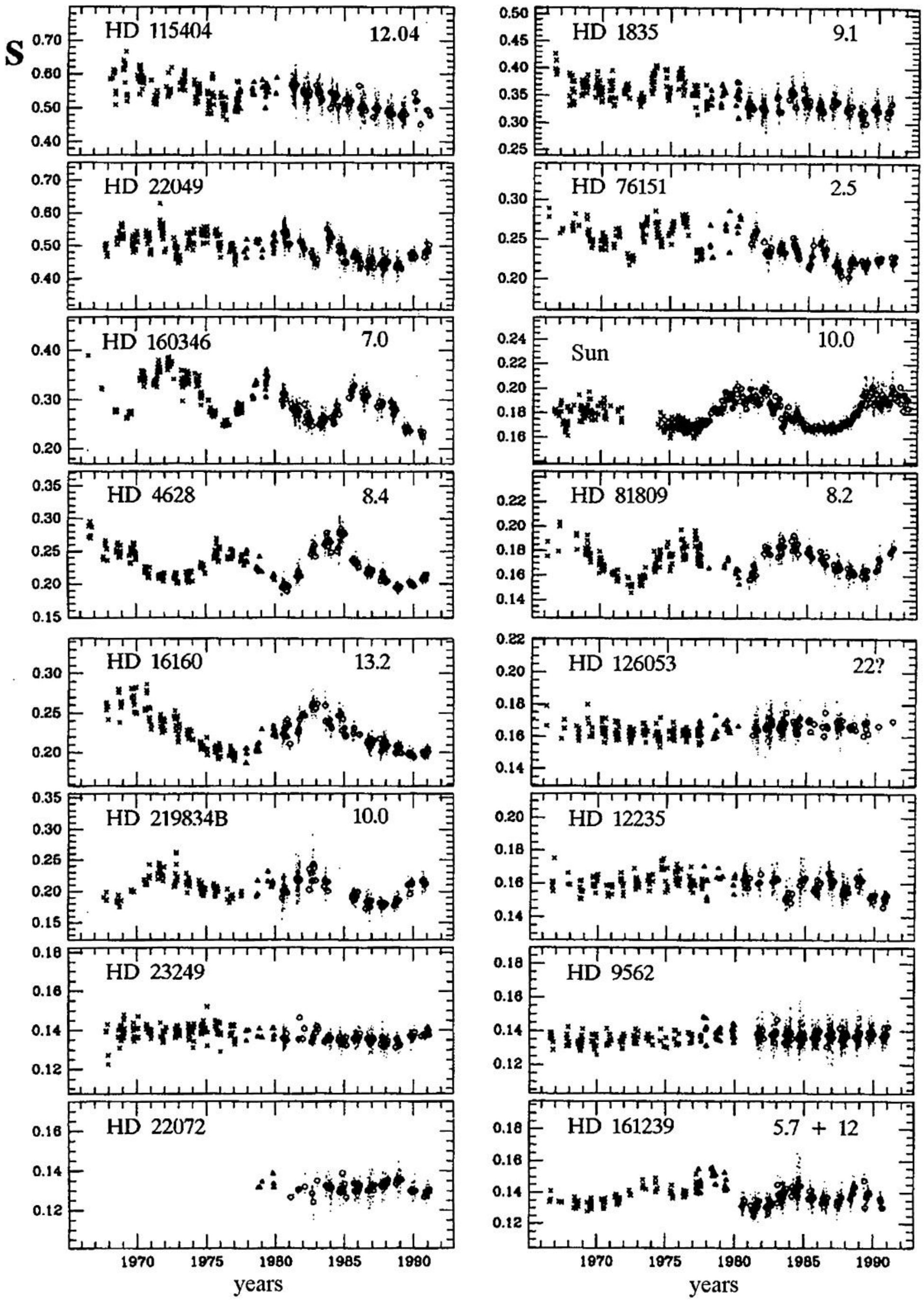

Fig. 82. Long-term variations of calcium emission from 25-year observations within the framework of O. Wilson program. Estimates of cycle durations in years are specified in top right corners of the diagrams (Baliunas et al., 1995)

$$\Delta R'_{HK} / \langle R'_{HK} \rangle \sim (P_{cyc}/P_{rot})^{-1.35+0.35/-0.65}, \quad (69)$$

which provides evidence in favor of the magnetic field generation by the nonlinear dynamo.

Splitting into the high- and low-activity groups was kept when plotting the stars on the plane (X-ray luminosity, dynamo number). Later, Brandenburg et al. (1998) and Saar and Brandenburg (1999) compared the ratios $P_{cyc}/P_{rot}$, Rossby numbers and the level of chromospheric activity $\langle R'_{HK} \rangle$ on 21 stars with well-determined values of $P_{cyc}$ and different levels of activity, including the Sun. They found that on the planes ($P_{cyc}/P_{rot}$,Ro) and ($P_{cyc}/P_{rot}$, $\langle R'_{HK} \rangle$) the stars were also divided into groups of young active and old inactive stars. For each group, Brandenburg et al. (1998) found clear power dependences between compared values and a power dependence between $\langle R'_{HK} \rangle$ and Ro that is common for all considered stars. The obtained relations were interpreted as initial growth of the ratio $P_{cyc}/P_{rot}$ with stellar age, then a sharp reduction of $P_{cyc}$ by about six times at the age of 2–3 billion years and the subsequent new increase of the ratio following a $t^{0.35}$ law. A sharp reduction of the cycle duration corresponds to the division into young and old stars on the basis of the level of calcium emission mentioned by Vaughan (1980). Some older stars have two periods, and Gleisberg's secular solar cycle satisfies this law.

Boehm-Vitense (2007) found that inside each of the sequences scheduled by Brandenburg et al. (1998) the number of rotation periods during the activity cycle was approximately the same, but these numbers were different for various sequences: the number of revolutions was within 300–500 for the sequence of active stars and less than 100 for inactive. She associated this circumstance with different dynamo mechanisms: caused by the latitude gradient of the rotation rate and short periods of cycles for active stars and caused by the radial gradient of the rotation rate and prolonged periods of cycles for inactive; the change of sequences occurred at a rotation period of 21 days, and for a series of stars with periods close to this value two periods were simultaneously observed. In the inactive sequence, the flux in CaII lines, effective temperature, and rotation period were linked by correlation in the form: $F$(CaII) ~ $T_{eff}^{4} P_{rot}^{-4/3}$, whereas in the active sequence the dependence of the calcium flux on rotation was stronger. According to Boehm-Vitense, the Sun, whose ratio of the indicated durations is close to 160, does not fall into a single one of the considered sequences and may be at the state of transition from one to another, showing characteristics of two dynamo mechanisms implemented at different depths and, consequently, of two cycles simultaneously.

Based on the long series of high-precision photometric observations of six young solar-type stars, Messina and Guinan (2003) found that all the considered stars showed variations of the duration of the rotation period, i.e., had differential rotation. These variations were periodic also at the phase with the spottedness cycle for BE Cet and DX Leo; presumably, the analogous situation was for $\pi^1$ UMa, EK Dra, and HN Peg; for BE Cet, $\pi^1$ Uma, and EK Dra the rotation period decreased as the cycle developed, then it by a bump returned to the initial value at the beginning of the next cycle, but DX Leo, $\kappa^1$ Cet, and HN Peg, as well as LQ Hya, showed an inverse change of the period duration. The duration of the cycle did not correlate with the dynamo number but showed a positive correlation with the differential rotation amplitude.

The next step in analyzing the results of the Wilson program was made by Frick et al. (1997) who used the methods of wavelet analysis. Strictly speaking, the standard Fourier method is not fairly adequate to the considered cyclic activity of stars with an irregular change of variables and can result in false periodicity. Frick et al. reduced the method of wavelet analysis to the form in which the finiteness of studied time series and missing data did not much influence the results. They applied the constructed algorithm to 25-year observational series of four stars: HD 10476 with normal cyclicity, HD 201091 with a changing period, HD 10700 at the state of the Maunder minimum, and HD 3651 entering the minimum. The calculations done with the updated wavelet analysis algorithm compared with earlier calculations by Baliunas et al. (1995) confirmed the basic conclusions, but revealed new important details: much lower noise in the obtained periodograms in all cases, a confident change of the activity period on HD 201091 over 25 years from 6.6 to 8 years, and less certain smooth oscillations of the period of HD 10470 within 10.2–9.7 years. Nothing was added to the characteristics of activity of HD 10700; the period of HD 3651 of about 14 years was confirmed, but the reliability of the value was rather low.

In the mid-1990s, spectral observations within an expanded O. Wilson program, the HK project, were recommenced at the upgraded 100″ telescope. In parallel, an extensive photometric program was started with automated telescopes providing an accuracy of better than $0.001^m$ (Baliunas et al., 1998).

As is known, during the solar cycle various indices of the activity level can reach their extreme values at different times. Gray et al. (1996) studied this aspect using the example of several G–K dwarfs. The level of calcium emission was considered as an indicator of magnetic activity, photometric data as indicators of temperatures and luminosity, the ratio of depths of two close absorption lines of vanadium and iron as a temperature indicator, and line bisectors as an indicator of stellar granulation. The 10-year observations of G8 $\xi$ Boo A proved that magnetic activity preceded variations of all other considered parameters: photometric brightness — by 1.4±0.4 years, color index — by 1.5±0.5 years, ratios of line depths — by 1.8±0.3 years, and line bisectors — by 2.1±0.3 years. Similar studies were run for several other stars. In all cases, the preempting of magnetic activity was revealed and the dependence of the delay of temperature changes on effective temperature was suspected: from 3 years for the G0 dwarf $\beta$ Com to 0.3 years for the K2 star $\varepsilon$ Eri.

Between 1984 and 1995, Radick et al. (1998) carried out high-precision photometric observations of 34 F5–K7 stars within the HK project, whose level of chromospheric activity varied from five-fold to a half of the solar activity level. Analysis of the obtained data showed that the amplitudes of variability of photospheric and chromospheric radiation — on short and long time scales — of the sample stars and the Sun were connected by the power ratios with the average level of chromospheric activity $\langle R'_{\rm HK}\rangle$. Young and more active stars were found to weaken with increasing chromospheric emission, i.e., their optical brightness changed due to dark spots, whereas low-activity stars, such as the Sun, brightened, i.e., their optical brightness varied due to bright flocculi. This result was later confirmed by Lockwood et al. (2007) from more prolonged observations up to 20 years.

Using the original approach, Olah et al. (2009) studied variations of cycles for 20 dwarfs, giants, and supergiants with a duration of photometric and spectroscopic observations lasting for decades. They found that 15 of them distinctly showed a multiplicity of periods, whereas for the rest the observations were not prolonged enough to detect the second period. As a rule, cycles show systematic variations. A positive correlation between rotational and cyclic periods was confirmed.

From 7-year observations of fully convective dM5.5e star Proxima Cen, Cincunegui et al. (2007) confidently found an activity cycle of 442 days based on spectra of the quiescent state, whereas Wargelin et al. (2010) were amazed to suspect a series of periods from 1.2 to 7 years.

During spectroscopic observations of the star $\tau$ Boo, Fares et al. (2009) detected two consecutive polarity switchings of its weak magnetic photospheric field over two years and suspected that such a short cycle could be caused by the influence of a close giant planet.

From 45-year observations Metcalfe et al. (2013) confirmed a simultaneous implementation of 2.95- and 12.7-year activity cycles for $\varepsilon$ Eri.

After the detection of an effect of switching active longitudes for a series of stars, Berdyugina (2006) proposed considering periodicity of these switchings as a new activity cycle.

The CaII H and K lines are obviously the most convenient but not the only chromospheric lines which can be used to study activity cycles. Thus, Larson et al. (1993) throughout 12 years carried out spectral observations of the K5 dwarf 61 Cyg A in the near IR range and from variations of equivalent widths of the CaII $\lambda$ 8662 Å line detected a distinct activity period of 7.22 years.

Robertson et al. (2012) performed an analysis of about 3000 spectra of 93 K5–M5 dwarfs in the $H_\alpha$ region acquired over a decade and detected periodicity varying from a day to more than 10 years for six stars and long-term trends for seven ones. Cycles with periods of more than a year were found for at least 5% of objects. The M0–M2 stars were more active than later dwarfs, and the increase of metallicity made the activity level higher.

From photometric observations, Amado et al. (2001) suspected a cycle of about 20 years for AB Dor. Using spectral and photometric data, Buccino et al. (2011) estimated a duration of the cycle for the M1/2 dwarf Gl 229 A of approximately 4 years and for the M2.5 dwarf Gl 752 A of about 7 years. Later, Buccino et al. (2013) analyzed calcium emission of the star AD Leo between 2001 and 2013 and its V photometry between 2004 and 2010 and confidently detected an activity cycle of 7 years and a less confident cycle of 2 years. In their opinion, the dynamo acting in subphotospheric layers was responsible for the 7-year cycle, whereas the dynamo in the depth of the convective zone — for the 2-year cycle.

As stated above, Vida et al. (2014) analyzed 4-year Kepler observations of 39 late-type rotators with rotation periods of less than a day and for 9 of them from variations of spot latitudes suspected the activity cycles of 300–900 days, which significantly supplements the region of poorly studied short cycles.

Ibanez Bustos et al. (2019) performed a search for the activity cycle of the young and one of the most active M dwarfs AU Mic independently from spectral observations in 2004–2016 and photometry of the star between 2000 and 2009. In both cases, they found a cycle duration of about 5 years at the rotation period 4.85 days.

Combining the results of Baliunas et al. (1995) and the data on calcium emission of several dozen solar-type stars over approximately 50 years, Egeland et al. (2018) found that most known stars with good cycles, similar to that of the Sun, were K dwarfs with large Rossby numbers.

Recently, from the 36-year series of data within the HK project, Olah et al. (2016) studied an association between the activity cycles and age for 29 G–K dwarfs. Cycles were detected for 28 objects. Twelve of them for stars with high rotation periods of 39.7 ± 6.0 days have simple gradual cycles with a duration of 9.7 ± 1.9 years, whereas for other stars with rotation periods of 18.1 ± 12.2 days there were detected complex, sometimes abruptly changing cycles with a duration of 7.6 ± 4.9 years. The age division between these groups of stars is at a level of 2–3 billion years.

* * *

Saar (1998) thoroughly studied the stars in the state of flat activity, i.e., whose $S$ was invariant for many years. He showed that not all of them could be considered as very old objects in which the dynamo mechanism had become ineffective, for example, because of too slow rotation. But there were stars even younger than the Sun among them, which should be considered as Maunder minimum stars. Saar found that chromospheric CaII and CII lines of these stars corresponded to the basal level presumably identified with acoustic heating of the chromosphere and amplified toward earlier stars, while radiation fluxes from the transition zones and coronae were lower than on the stars with cyclic activity, grew toward later spectral types, and were independent of rotation. Apparently, formation of these fluxes was a weak nonacoustic process, which did not display variability and was more likely dependent on the depth of the convective zone. This process could be the turbulent (distributed) dynamo in the convective zone, which is apparently effective to a certain extent on all stars and is noticeable under the weakening of a cyclic solar-type dynamo.

* * *

As stated above, the amplitude of oscillations of X-ray emission of the Sun during the solar cycle exceeds by one–two orders of magnitude the amplitude of calcium-emission oscillations, but the data on activity cycles of stars in X-rays are much poorer than the data on such cycles of chromospheric emission. This is explained by a much lower variability of active stars in X-rays as compared to the Sun. The second and the most important explanation is that no X-ray studies analogous to the O. Wilson program have been undertaken, since space vehicles operated for 2–3 years only and even the "long-liver" ROSAT did not provide required long-term observation series. Nevertheless, attempts to reveal cyclicity from X-ray observations were undertaken.

Haisch et al. (1990b) considered ultraviolet and X-ray observations of Proxima Cen during four epochs — in March and August 1979, in March 1984, and March 1985 — and found a synchronous decline and the subsequent rise of CIV, MgII, and soft X-ray emission, which they considered as the manifestation of the activity cycle.

The central region of the Pleiades was analyzed by Schmitt et al. (1993b) based on the ROSAT sky survey. They revealed 24 X-ray sources, 20 of them were already discovered by the Einstein Observatory. But Schmitt et al. found changes of X-ray luminosity up to one order of magnitude between the observations separated by a time interval of 10 years, which could not be explained by flares or rotation. Schmitt et al. (1993) associated them with activity cycles. However, later, Gagné et al. (1994a, 1995) considered the variability of X-ray sources in the Pleiades at intervals of 1 and 16 days, 12 months and 10 years. They revealed variability of 22 out of 44 bright sources, of which 12 were flares: 33% of the found variable sources were variable on the time scale of 16 days, 64% within 12 months, and 55% within 10 years. The closeness of the two last values put in question the conclusion of Schmitt et al. (1993b) on cyclic activity as a cause of long-term changes of the level of X-ray emission.

Hempelmann et al. (1996) considered the properties of X-ray emission from single F–K dwarfs divided into 3 groups: stars with constant, periodically changing, and chaotically varying calcium emission. They found that average X-ray luminosities in these groups regularly differed: $\langle \log F_X \rangle$ = 4, 5, and 6 in the first, second, and third group, respectively, i.e., the stars with irregular changes of $F_{\mathrm{HK}}$ were the most active in X-rays. To determine the cyclicity of X-ray emission, they calculated the phases of chromospheric cycles at the moment of X-ray observations and calculated the differences between the observed and expected values of $F_{\mathrm{X}}$, which were determined from the dependence $F_{\mathrm{X}}(\mathrm{Ro})$. These differences had a slightly asymmetric curve, similar to the change of Wolf's numbers during the solar cycle, which made it possible to assume the existence of coronal cycles synchronous with chromospheric ones.

Later, Hempelmann et al. (2003, 2006) studied two long-term series of X-ray observations with ROSAT HRI and HK monitoring of components of the system 61 Cyg carried out in 1993–1998. Although the interval of X-ray observations is noticeably shorter than the known periods of chromospheric activity of stars, 7 and 12 years, these observations overlap the epochs of maximum activity of each component. Analysis of the data led to the conclusion on the existence of a close correlation of chromospheric and coronal activity of each component and that both stars indeed show coronal cycles. Amplitudes of coronal variations prove to be significantly higher than amplitudes of variability of HK fluxes, and these cyclic variations are pivotal for the total variability of X-ray emission of the system 61 Cyg.

Marino et al. (1999) undertook a systematic search for the cycles of stellar activity in X-rays comparing the observations of 29 stars at the Einstein Observatory in 1978–1981 and with ROSAT PSPC in 1990–1994. They found that if long-term variations similar to solar cycles existed, their amplitudes were much lower than in variations at short time intervals, which dominated in ROSAT data and that were identified as stellar flares. This result confirmed the earlier conclusion by Schmitt et al. (1995) based on a smaller sample of K–M stars and the conclusion by Stern et al. (1995) obtained for the Hyades stars. This may be due to the fact that on stars of the age of the Hyades and younger and on all red and brown dwarfs the generation of small-scale turbulent magnetic fields dominates over the large-scale dynamo responsible for magnetic cycles on the Sun. It should be noted that Kitchatinov et al. (2000) and Donati et al. (2003b) interpreted long-term magnetometric observations of the young K dwarf LQ Hya through the Zeeman–Doppler imaging within the framework of the distributed dynamo model.

During 2.5 years Favata et al. (2004) carried out XMM-Newton observations of the G2 star HD 81809, whose chromospheric cycle is known to be of 8.2 years, and detected a 10-fold modulation of X-ray luminosity with its maximum in 2002, whereas its chromospheric maximum was in 2001. After passing these extrema, chromospheric and coronal activity synchronously decreased, which is evidence of coherence of cycles.

Using XMM-Newton observations, Marino et al. (2005) and, as stated above, Pillitteri et al. (2006) compared observations of the young clusters NGC 2516 and IC 2391 with ROSAT observations carried out 7, 9, and 11 years earlier and found no analog of the 11-year solar cycle. Apparently, the point is in a noticeable difference of ages of these clusters and the Sun.

Robrade et al. (2007, 2012) performed XMM-Newton observations of components of the 61 Cyg systems: K5V (A) and K7V (B), and $\alpha$ Cen AB (G2V and K1V) and confidently detected their long-term variability in X-rays. For 61 Cyg A, the coronal cycle reflected well the chromospheric 7.3-year activity cycle, agreed with the phase of ROSAT observations in the 1990s, and variations of X-ray brightness occurred basically in the hot component. For 61 Cyg B, coronal activity was less regular but also followed chromospheric activity. Long-term variability was also present on $\alpha$ Cen AB. In this system, the component B dominated in X-rays with a cycle of 8–9 years and an amplitude of 6–8; the component A — a star of the spectral type G2 — with an activity cycle of 12–15 years, very similar to the Sun, weakened in X-rays at least by a magnitude during observations, which had not been observed earlier. According to Ayres (2009), $\alpha$ Cen B is near the X-ray maximum in the mid-1990s, at its minimum in the late 1990s, and again at maximum in 2004–2005. Earlier, Ayres et al. (2008) reported on a long-term weakening of X-ray emission for $\alpha$ Cen A. From observations of the K1 dwarf $\alpha$ Cen B in X-rays, in near and far ultraviolet, DeWarf et al. (2010) detected an activity cycle of 8.84 ± 0.4 years at a rotation period of 36.2 days and age of 5.6 billion years. Finally, using Chandra, Ayres (2015b) studied variations of X-ray emission of the components of the system $\alpha$ Cen AB. In 2008–2013, he recorded a brightness increase, its maximum, and the beginning of luminosity weakening of the component B; taking earlier ROSAT and XMM-Newton observations into account, estimated a duration of the period to be 8.2 ± 0.2 years with an amplitude of 4.5. Since 2005, the component A was at the state of low luminosity and during these observations started to emerge from this state; this development could be represented by a cycle with a duration of 19.1 ± 0.7 years with the contrast 3.4. Observations in 2010–2014 with HST/STIS in the range of 1150–1700 Å confirmed an escape of this component from the prolonged minimum based on SIV, CIV lines, and the [FeXII] $\lambda$ 1242 Å coronal line.

Between 2001 and 2007, Favata et al. (2008) carried out regular XMM-Newton observations of the solar-type star HD 81809 and detected the cyclic variations of coronal luminosity with $A_X \sim 1$ dex and coronal temperature, and the average luminosity was by an order of magnitude higher than that of the Sun; the stellar corona can be modeled by solar active regions with greater coverage than that for the Sun.

Coffaro et al. (2022) added Kepler-63 with an age of 210 Myr and a duration of the photometric cycle of 1.27 years to seven solar-type dwarfs with the known activity cycles in X-rays, among which ε Eri and ι Hor are the youngest – 400 and 600 Myr, respectively. The authors also note that all these young objects have the shortest durations of cycles and small amplitudes. The latter circumstance is related to a high filling factor of coronal magnetic structures.

* * *

In almost a dozen papers of the first decades of the XXIst century, Livshits and Katsova with colleagues suggested and developed a general pattern of solar-type activity in stars of the lower main sequence in which activity cycles take a specific, physically caused place. They continued the analysis of original observations within the HK project described by Baliunas et al. (1995) and many hundreds of measurements of the $S$ value collected in subsequent years and suggested a physical sense of the diversity of solar-activity manifestations for the stars of different masses and ages.

Bruevich et al. (2001) performed a systematic analysis of X-ray emission of about 1500 late F, G, and K stars, including 78 objects studied within the framework of the HK project of chromospheric activity, and showed that stars with irregular variations of chromospheric radiation have more powerful coronae, whereas the stars with cyclic activity are characterized by relatively low X-ray luminosity and appropriate ratios of $L_X/L_{bol}$. They confirmed high X-ray luminosity of stars with fast rotation. The stars with well-pronounced cycles have these cycles in the range from 7 to 16 years and an approximately similar value of $\log(L_X/L_{bol}) = -5.7$, whereas the objects with irregular activity occupy a wide range of these values from $-4.2$ to $-6.0$, but they are far from saturation at $-3$. For a quantitative discussion of the collected data, Bruevich et al. (2001) invoked a concept of dynamic systems and, following Parker, calculated a dimensionless dynamo number determining the dynamo wave and effectiveness of the magnetic field generation. They found that the well-pronounced cycles correspond to the relatively low values of this number and a low value of the magnetic field; with increasing dynamo number, the magnetic field strength increases, but the degree of cycle organization decreases. Thus, the stars with well-pronounced cycles are inferior in the field to the stars with high and unstable activity but a weakly pronounced cycle. A discussion of the collected data within the Parker concept of cycles made it possible to apply the general properties of dynamic systems with different degrees of freedom, i.e., to the stars in a wider range of thickness of convective zones, excluding extremely narrow zones in hot and disappearing fully convective objects.

Katsova et al. (2003) analyzed the long-term data of 20 stars within the HK project using the modified wavelet algorithm, determined the rotation periods of all the considered stars, and for a few of them found variations of these periods from season to season. For two of them, more active than the Sun, the rotation slows down during high activity, and the ratio $\Delta\Omega/\Omega$ is 0.14 and −0.074. If one uses the solar analogy, then the epochs of slow rotation correspond to the reversal time of the dipole total field.

Katsova and Livshits (2006) considered coronal and chromospheric activity of late stars on the basis of recently revealed many hundreds of such objects and confidently confirmed the existence of three groups of stars distinguished by Baliunas et al. (1995) with different ratios to activity cyclicity. They ascertained that during the transition from the Sun to K dwarfs with cycles and to more rapidly rotating F and G stars there occurs an increase of the high-temperature component of coronae with $T > 10$ MK, and the level of X-ray emission is closely associated with the spottedness degree of the stellar surface. In the case of the well-pronounced cyclicity, the association between chromospheric and coronal radiation is weak, whereas at less regular activity it becomes distinct. It was suggested that for the stars with well-pronounced cycles the coronae are heated due to quasistationary processes, whereas for F and G stars with high irregular activity — at long-term nonstationary coronal phenomena.

Katsova, Bruevich, and Livshits (2007) devoted their paper to the stars with cycles becoming established. From the general notions concerning a place of cyclicity in the evolution of activity of late stars, it follows that to the stars with cycles becoming established one should attribute the objects whose level of chromospheric and coronal activity is higher than that for the Sun and other stars with the well-pronounced cycles and which rotate faster than the stars with cycles. This seems to be associated with their younger age; spots on such stars occupy a few percent of the surface, which is by an order of magnitude larger than on the Sun at the activity maximum. The conducted wavelet analysis of variability of chromospheric radiation shows that the axial rotation period of some of the stars from the HK project varies from year to year. This is the most notable for the stars HD 149661 and HD 115404, and for the star with more complex type of variability HD 101501. But this effect is absent for the stars with the most clearly pronounced activity cyclicity. These variations of the period are observed at the epochs of existence of the noticeable rotational modulation of chromospheric radiation fluxes apparently immediately after the maximum of the many-year wave. This seems to evidence on the existence of huge activity complexes in the chromospheres of these stars. Over the years their longitudes have remained almost unchanged, but they drift from high latitudes toward the equator with a velocity close to that of the Sun. Thus, the most likely cause of variations of periods is differential rotation coinciding in sign with solar one.

Katsova et al. (2010) continued consideration of long-term variations of chromospheric radiation for 20 stars from the HK project using the modified wavelet algorithm that takes nonuniformity of initial data into account and for several of them detected variations of rotational periods from year to year. With growing activity, the rotation slowed down and the epochs of rotation braking took place at the increased level of activity; this repeated in adjacent cycles of activity. The characteristic variations of rotational periods covered about three years and occurred when the amplitudes of rotational modulation were high. These variations can be represented by butterfly diagrams with minimum assumptions. From the comparison with solar data, it follows that the epochs of rotation braking for surface inhomogeneities are synchronous with spins of the global magnetic dipole.

From observations of 1334 objects, Katsova and Livshits (2011) considered the evolution of solar-type activity for low-mass stars and found that chromospheric activity of the Sun was higher than that for the overwhelming majority of stars in the solar vicinity. They detected an appreciable group of stars with low chromospheric activity whose coronal radiation was within wide limits and suggested that in the process of rotation braking both chromospheric and coronal activity weaken simultaneously, whereas in the other group the chromospheric activity decreases up to the solar level, but coronae remain much more powerful than that of the Sun. Presumably, the point is in different thickness of convective zones, and the spectral type G6 is a border between these groups. They formulated arguments in favor of the two-level dynamo and a different role of the large-scale and local magnetic fields in the formation and evolution of activity.

Continuing the previous study, Katsova et al. (2013) found a close correlation between the lithium abundance and the parameter of calcium emission $R_{HK}$ for two groups of G stars — hotter and cooler than the Sun. This association was best pronounced at high activity, notably in the group of G6–K3 stars with many BY Dra-type variables it was pronounced better than for F8–G5 stars. For the stars with high activity, they confirmed that both the enhanced lithium abundance and the activity level are determined by the rotation rate that is dependent on age. The enhanced lithium abundance is differently associated with chromospheric activity, depending on its level; the cooler stars with detected lithium and the chromosphere, similar to that of the Sun, have a markedly more powerful corona. A comparison of the indices of chromospheric and coronal activity showed that, except for the objects with good correlation of these indices, there is a noticeable fraction of stars definitely deviating from this correlation: at appreciable X-ray emission their chromospheres are relatively weak.

Having compared the processes on the present-day Sun and G stars HD 152391 and HD 1835 (BE Cet) with rotation periods of 11 and 8 days, Katsova and Livshits (2014) studied solar activity at the age of 1–2 billion years, when its quasistationary level was established and the rotation period was close to 10 days. Then the total area of spots was by 2–3 orders of magnitude larger than the present-day at the cycle maximum, the levels of chromospheric and coronal activity were significantly higher, as in Hyades, but did not reach the saturation level. The average value of the longitudinal magnetic field on active G stars was an order of magnitude higher than the average present-day field of the Sun as a star at the maximum of cycle 21. Moreover, a large-scale toroidal field was detected on these stars. Taking the obtained estimates of activity levels and magnetic fields into account, from the statistics of Kepler observations Katsova and Livshits estimated the frequency range of superflares with a total energy of $10^{34}$ erg: one per 5 and 500 years at the rotation periods of 3 and 12–15 days, respectively. The mass loss rate of the young Sun achieved $10^{-11}$ $M_{\odot}$/year, and the contribution of CME into this value was significantly higher than it is now.

Finally, as a continuation of the mentioned studies of stars at the epoch of activity cyclicity being established (Katsova et al., 2007), Katsova et al. (2015) estimated the age of the young Sun at the epoch of slightly more than 1 billion years at a cycle duration of 8.7 years and concluded that the cycle duration of such activity increases as the rotation slows down, i.e., with age.

Metcalfe and van Saders (2017) came back to the results of Brandenburg et al. (1998), namely to the two relations between the rotation rate and duration of the activity cycle for young fast rotators with short cycles and for older slowly rotating stars with more prolonged cycles. They complemented the results of Boehm-Vitense (2007) with the data of recent observations of the middle-age stars (see Fig. 83) and concluded that at the critical value Ro ~ 2 the surface rotation rate changes slower, whereas the cycles gradually grow up to their disappearance. Such a scenario, in their opinion, was confirmed by the stars of spectral types F, G, and K. Whereas the Sun, which does not satisfy any of the relations of Brandenburg et al. (1998), presumably has already been involved into the transition activity phase when its cycle grows on the evolutionary time scale and can disappear during the next 0.8–2.4 billion years.

Suarez Mascareno et al. (2016) studied 50000 photometric light curves, rotation periods, and activity cycles for 125 stars from late A to middle M stars recorded at southern latitudes; for 47 of them they found activity cycles, for 36 measured the rotation periods, and from the original and literature data for 44 estimated the activity level $\log R_{HK}$. Durations of the detected cycles proved to be in the range of 2.5–14 years at photometric amplitudes of 5–20 mmag, whereas these amplitudes for G–K dwarfs were lower than those for early and middle M stars, and for several objects more than one cycle were detected. The distributions of cycle durations were similar for different spectral types, but the average durations for F stars were 9.5 years, for G — 6.7, for K — 8.5, for early M — 6.0, and for middle M stars — 7.1. On the other hand, the detected rotation periods were within the range of fractions of a day to 150 days, and their distribution differed from approximate constancy of activity cycles: there was a distinct increase toward later spectral types from 8.6 days for F dwarfs to 85.4 days for middle M-types. The empirical relation between the level of chromospheric activity and rotation period was proceeded up to log 4. For F, G, and K stars, a correlation between the cycle duration and rotation period was detected; the faster the star rotates, the higher the amplitude of its cycle.

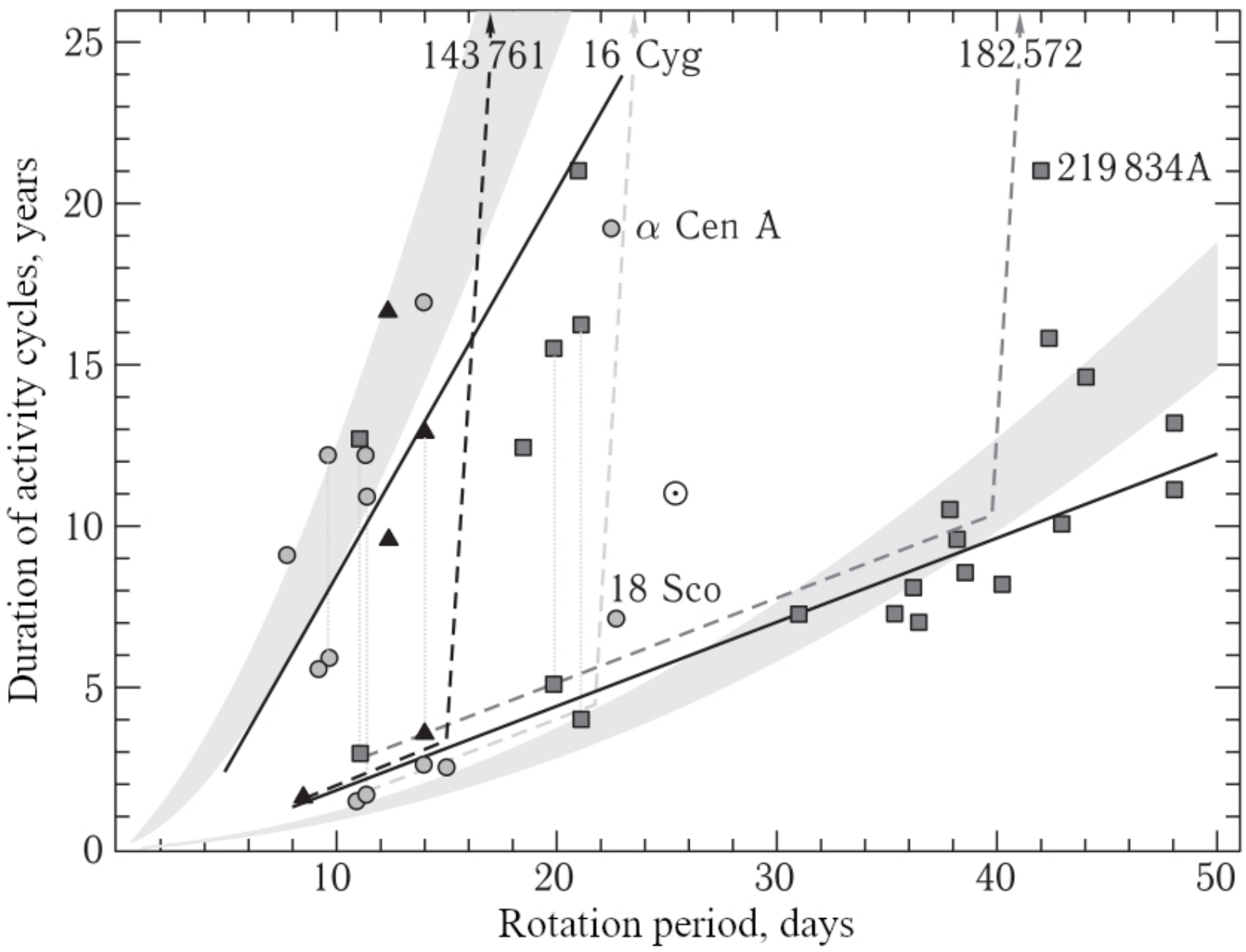

Fig. 83. An updated version of the figure by Boehm-Vitense (2007) demonstrating two different relations between the rotation period and the duration of activity cycle. The cycles acting simultaneously on the same star are connected by vertical dashed lines. The shaded areas correspond to the regions marked by Brandenburg et al. (1998) for G stars with a duration of rotation of convective vortices of 7–14 days and for K stars with a duration of rotation of such vortices of 17–23 days

From the Kepler data over four years of observations, Reinhold et al. (2017) studied 23601 stars and for 3203 of them detected the periodicity of activity in the range of $0.5 < P_{cyc} < 6$ years at the rotation periods of $1 < P_{rot} < 40$ days. These studies showed a weak dependence of the cycle period on the rotation rate, its increase for more prolonged rotation periods; no statistically significant dependence on the effective temperature of a star was found and no clear separation into two sequences of active and inactive objects was confirmed, although for the periods 5–25 days a sequence was envisioned for inactive stars.

Within the framework of the CARMENES project, from their own and literature data, Diez Alonso et al. (2019) analyzed 622 lights curves for 334 M dwarfs and found 142 rotation periods and 10 activity cycles. They detected no correlation between the rotation periods and durations of cycles, confirming the results of Suarez Mascareno et al. (2016).

A review of studies by Katsova and Livshits regarding the evolution of solar activity cyclicity was presented in Katsova (2020).

Bondar, Katsova, and Livshits (2018) analyzed the activity cycles for 65 solar-type stars and rapidly rotating cool dwarfs observed for several decades and found that the cycles with a duration of 7−18 years, which could be compared with solar 11-year cycles, were characteristic of more than 50% of the objects studied. For cooler dwarfs with the rotation periods of shorter than 5 days, the cycles achieved 80 years, but they had no distinct correlation between the rotation and the cycle duration. The most active K dwarfs had prolonged cycles with the highest amplitudes. Young and old solar-type stars showed a tendency for increasing cycle duration with decreasing rotation period.

Stepanov et al. (2020) reanalyzed photometry of the V 833 Tau star over 120 years using the latest version of wavelet analysis and concluded that its brightness variations in the range of 3 to 100 years had a power character and were similar to solar variability in the range of 0.1 to 10 years. Probably, the star has a period of cycle activity of 80–100 years, i.e., surpasses tenfold the solar cycle. However, possibly, the star has not reached yet the stage of pronounced activity cyclicity.

Kichatinov (2022) proposed an idea that the observed dependence of stellar activity cycles on rotation rate can be a manifestation of a stronger dependence on effective temperature. A combined model for differential rotation and dynamo is applied to stars of different masses. Computations show shorter dynamo cycles for hotter stars, which is related to the larger amplitude of the α-effect required for the dynamo.

## 3.2. Evolutionary Changes in the Activity of Mid- and Low-Mass Stars

Unlike all previous chapters, this chapter does not start with the description of the solar situation as initial, since in this case the Sun is the driven and not the driving object: it gives us only one point on the evolutionary tracks of activity, which should be constructed based on the characteristics of the activity of stars of different age and mass.

Since stars with identical effective temperatures and luminosities display a much varying activity level, i.e., the closeness in the Hertzsprung–Russell diagram does not guarantee the closeness of activity levels, one should find another essential parameter defining the activity. Originally, UV Cet-type flare stars were attributed to young stars, since they had certain similarities with T Tau-type stars (Ambartsumian, 1954). The concept of the youth of T Tau-type stars was advanced by Ambartsumian shortly before (1952) and got wide recognition. The idea of the youth of flare stars obtained independent support, when the analysis of kinematic characteristics of red dwarfs showed that dMe stars were much younger than dM stars (Herbig, 1962). Then, Kraft and Greenstein (1969) found that there were many more K5–M3 dwarfs with calcium emission in the Pleiades than in the Hyades. Robinson and Kraft (1974) found spotted K3–M0 stars in the Pleiades and did not find such objects in the Hyades. But, as the number of known UV Cet-type stars increased, in particular, in clusters of various age and among binary systems with flare components, rather old objects were found among them, even subdwarfs of up to 10 billion years of age. Old UV Cet-type stars were found in studying kinematic characteristics of such objects (Shakhovskaya, 1975; Poveda et al., 1995). This fact disproved short-lived relics of star formation and youth in itself as the main cause of activity. The sources of modern evolutionary concepts of this activity go back to the studies of stellar rotation, the parameter, which practically does not influence the position of a star in the Hertzsprung–Russell diagram, but determines its magnetism and all related processes in the photosphere and atmosphere, to the studies of evolution of rotation and chromospheric emission, as the most accessible manifestations of stellar magnetism.

In 1955, Parker (1955b) proposed the idea of a stellar dynamo: the scheme of amplification of the stellar magnetic field due to the interaction of convective motions and rotation. Soon, Schatzman (1959, 1962) explained the strong distinction of rotation rates of stars of early and late spectral types by magnetic braking of stars with convective zones, assuming that this process was the most effective when stars approached the main sequence.

In the mid-1950s, O. Wilson started systematic observations of calcium emission to reveal its relation to other parameters. In the early 1960s, he summarized preliminary results of studying G0–K2 dwarfs among field stars and in four stellar clusters. He showed that emission stars occurred in clusters more often than among field stars and that the level of emission from the stars of the young cluster Pleiades was higher than that in the older clusters Praesepe and the Hyades (Wilson, 1963). Being completed with the photometry of more than 100 main-sequence stars (Wilson and Skumanich, 1964) and the spectroscopy of more than 300 stars (Wilson, 1966), this study initiated an evolutionary approach to the phenomenon of calcium emission and chromospheric activity in general. Soon, Wilson and Wooley (1970) compared the intensity of calcium emission in the spectra of 325 late main-sequence stars close to the Sun with the parameters of their galactic orbits and found that stars with strong emission had orbits close to circular and with small inclination angles to the Galaxy plane, i.e., the orbits of young objects, whereas stars with weak emission had orbits typical of old objects. Thus, the kinematic characteristics of a large number of stars in the solar vicinity provided independent confirmation of the slow decay of calcium emission.

From the spectrograms with a dispersion of 6 Å/mm, Kraft (1967) found that F–G stars with calcium emission had systematically higher rotation rates than stars without the emission, and solar-type stars in the Pleiades rotated faster than the Hyades stars. On this basis, he suggested secular braking of rotation of main-sequence stars with convective envelopes and linked this effect to the Schatzman concept, assuming, however, that magnetic braking continued during the main-sequence life of the stars. According to Kraft, the velocity of a star whose mass was equal to 1.2 solar masses decreased by a factor of 2 over $4 \cdot 10^8$ years. Soon, the close relation between chromospheric emission and local magnetic fields was established in solar studies (Frazier, 1970), which made it possible to advance a hypothesis about secular decay of the stellar magnetic fields eventually caused by the braking of stellar rotation. The concept of braking of stellar rotation and simultaneous decay of magnetic activity on evolutionary time scales has become conventional and is enriched by the studies of evolution of various manifestations of this activity. The results of such studies comprise the contents of this chapter.

## 3.2.1. Evolution of Stellar Activity

It is appropriate to start outlining the evolution of stellar solar-type activity from the results of studying X-ray flares on the pre-main sequence stars in Orion. Such a research was performed by Getman et al. (2008a, b) using Chandra. They studied 216 flares on 161 stars and found that the flares on these young stars were among the strongest, the most prolonged and hottest events corresponding to the largest coronal structures. However, no statistically significant differences were detected in the distributions of peak luminosity or temperature, which provides evidence of a common mechanism of all flares. The superhot flares in Orion were brighter but shorter than the cooler flares, and most of bright flares in Orion could be treated as the enhanced analogs of rare solar "prolonged events". Comparing the flares in accreting systems and in systems that are free of disks, they found that the superhot flares with temperatures higher than 100 MK were predominantly in accreting systems, and such flares were shorter.

As stated above, due to the detection of high-latitude spots on rapidly rotating RS CVn-type stars, Schüssler and Solanki (1992) performed magnetohydrodynamic calculations of the rise of magnetic flux tubes to the stellar surface and showed that fast rotation displaced the rise of such tubes to high latitudes. This idea was advanced by Granzer et al. (2000) for stellar evolution: they calculated the dynamics of magnetic flux tubes for stars with masses of 0.4–1.7 solar masses and with angular rotation rate within 0.25–63 solar values, which covered the interval from the classical T Tau stars to the stars of the $\alpha$ Per cluster. Granzer et al. found that the latitudes of spots should grow quickly with an increase of the rotation rate, slightly decrease for stars with large masses, and quickly decrease with age. Later, Granzer (2002) concludes that the second most important parameter defining spot latitudes is a relative size of the radiative core of a star, whereas its mass and evolutionary status does not play a significant role.

The situation with solar-type activity at an even earlier stage is yielded by ZDI of T Tau stars (Gregory et al., 2012). The topology of large-scale fields on pre-main sequence stars proved to be strongly dependent on their inner structure and with a fast extinction of the dipole component of the multipolar magnetic field as the radiative core developed. They found that the general characteristics of the global field of such very young stars could be determined based on their position in the Hertzsprung-Russel diagram and singled out four possible variants: stars with a large radiative core — more than 40 % of the stellar mass — have a very complex magnetic field with a dominating nonaxisymmetric component, whereas at a less radiative core there dominates an axisymmetric field with higher orders of components than a dipole. The axisymmetric fields with kilogauss dipole components dominate in fully convective stars. The similarity between magnetic properties of pre-main sequence stars and M dwarfs with a similar inner structure allows one to discuss on the dual nature of dynamo in young low-mass stars.

The experimentally critical phase of emergence of a solar-type star on the main sequence was considered by Folsom et al. (2015, 2018). They observed stars in open clusters and associations of the known age and applied the Zeeman-Doppler imaging. They first studied 15 stars in 5 associations of an age from 20 to 250 million years, stars with masses from 0.7 to $1.2M_{\odot}$ and rotation periods of 0.4 to 6 days, filling the evolutionary gap between young T Tau stars and quite developed main-sequence objects. As a result, they detected a complex geometry of the large-scale fields with an average strength of 14 to 140 G with a distinct tendency for decreasing this strength with age and a close correlation with the Rossby number: before the main sequence, the magnetic field parameters are determined by the stellar inner structure, on this sequence — by rotation. Then they studied 15 young stars from four associations with masses varying from 0.8 to $0.95M_{\odot}$, rotation periods of 0.326 to 10.6 days, and age from 120 to 625 million years. On these objects, they detected large-scale fields with a strength of 8.5 to 195 G and with a significant diversity of geometries and a noticeable scattering of about 120 million years. Again, there was detected a tendency for weakening strength with increasing age and a power weakening of the field with growing Rossby number, and the field saturation at $\mathrm{Ro} < 0.1$ was suspected.

From HST/COS observations of 10 M dwarfs throughout 73 hours, Loyd et al. (2018a) compared flare activity in the far ultraviolet of active and inactive M dwarfs, attributing stars to these groups based on the values of equivalent widths of calcium emission: $W_{\mathrm{CaII\ K}} >$ 10 Å — active and $W_{\mathrm{CaII\ K}} < 2$ Å — inactive. Comparing the absolute energies of flares, the events in the former group of M dwarfs were an order of magnitude more energetic than the flares belonging to the latter group of stars, but when considering equivalent durations both groups showed an identical power energy spectrum of the accumulated flares with a spectral index of 0.76.

From HTS/COS observations in the far ultraviolet, Loyd et al. (2018b) compared the activity of early M dwarfs of the field and the young cluster Tuc-Hor of an age of 40 million years and found that they had similar spectral indices of power energy spectra of flares, but the absolute energetics of flares on young dwarfs was 2–3 orders of magnitude higher than that for dwarfs of the field.

* * *

Let us turn to main-sequence stars.

The most abundant data on the evolutionary change of the level of stellar activity were accumulated in the studies of stellar chromospheres.

After qualitative conclusions by Wilson and Kraft on the relation of the chromospheric calcium emission to the age and rotation of stars, based on the observations of stars in the Pleiades and the Hyades, the UMa moving group, and the Sun, Skumanich (1972) proposed the $t^{-1/2}$ law describing the weakening of calcium emission and braking of stellar rotation. (According to Smith (1979), this relation is valid after the transition to photoelectric recording of spectral-line profiles, which noticeably increased the estimates of rotational line broadening.) Braking of rotation was attributed to the constant loss of angular momentum in the coronal wind (Durney, 1972), which is the product of activity. The decay of chromospheric activity was associated with the weakening of the magnetic dynamo initiated by rotation (Parker, 1970).

Continuing the studies of calcium emissions started by Wilson, Vaughan, and Preston (1980) and Vaughan (1980) discovered a certain "Vaughan–Preston gap" dividing young stars with strong emission and less-active old (older than $10^9$ years) stars. The relation between more and less active stars was in agreement with the hypothesis about a secular weakening of this emission at a constant rate of star formation. From an extensive sample of 486 stars in the solar vicinity Soderblom (1985) calculated $R'_{HK}$ and found that these values satisfied $t^{-1/2}$. Later, Soderblom and Clements (1987) found that this relation started to be valid for an age close to that of the UMa moving group ($3 \cdot 10^8$ years) or the Hyades moving group ($6 \cdot 10^8$ years).

From the subsample of G0–K5 dwarfs of the sample of 111 F2–M2 stars investigated by the O. Wilson team, Baliunas et al. (1995) picked out young stars with a high average level of activity, fast rotation, an absence of the Maunder minima and rare smooth cycles, stars of intermediate age of 1–2 billion years with an average level of activity and rotation and more frequent cycles, and old stars, such as the Sun and older, with slower rotation, lower level of activity, smooth cycles, and the Maunder minima.

Based on the depths of absorption details of the calcium $H_1$ and $K_1$ lines, Barry et al. (1981) considered regular changes of the level of the chromospheric activity of solar-type stars in six clusters with ages from $10^7$ to $5 \cdot 10^9$ years. They concluded that these spectral characteristics were dependent on the intensity of emissions and could be used for age calibration, which for stars younger than the Hyades was determined to an accuracy of up to a factor of 2 and for stars older than $4 \cdot 10^9$ years to an accuracy of up to 25%.

Hartmann et al. (1984b) analyzed the "Vaughan–Preston gap". They concluded that the minimum of the measured *S* caused by the photospheric contribution and independent of chromospheric activity, on the one hand, and emission saturation on the youngest stars found in observations of the Pleiades, where $t^{-1/2}$ definitely was not satisfied, on the other hand, resulted in the concentration of the values of *S* at the band borders, which produced an impression of a "gap". Further, they showed that the observed distribution of chromospheric emission could be presented under the assumption of constant birth of stars provided that on young stars chromospheric emission decayed exponentially and then $t^{-1/2}$ took effect.

Jackson and Jeffries (2010) studied in detail the chromospheric emission of stars in the young open cluster NGC 2516. Having rotation rates, intensities of the infrared CaII triplet lines, and the confirmed membership of 210 K–M stars in their belonging to the cluster, they found a noticeable growth of the fraction of fast rotators for cooler objects: for 20% of M0−M1 dwarfs v sin*i* > 15 km/s, whereas for M4 this fraction increases up to 90%. The activity determined from calcium lines depends differently on the rotation period and mass on opposite sides of the transition to fully convective stars: for more massive K3–M2.5 dwarfs chromospheric activity increases with decreasing Rossby number and saturates at Ro < 0.1, whereas for cooler fully convective dwarfs, among which almost all have Ro < 0.1, the chromosphere weakens with growing V–I and drops by a factor of 2–3 between M2.5 and M4; however, this weakening is absent in X-rays. This weakening of chromospheric emission and a growth of rotation after M3 was associated by Jackson and Jeffries with variations of the magnetic field topology. Considering different models of solar-type stars with different level of chromospheric activity, Vieytes and Mauas (2004) conclude that the Vaughan–Preston gap is due to the nonlinearity of sensitivity of the CaII lines to chromospheric heating.

In 1964, Herbig (1965) discovered a secular decrease of the lithium abundance in main-sequence stars. Skumanich (1972) compared this effect with the weakening of calcium emission and braking of rotation, but did not find confident synchronism. Then, from the spectra of about a hundred F5–G5 dwarfs, Duncan (1981) estimated the intensity of calcium emission and the lithium abundance. He found that on the general background of parallel secular weakening of these values there were stars with weak emission but with high lithium abundance. To interpret the objects, Duncan assumed that at the age of 1–2 billion years a sharp decay of emission occurred, whereas the jump was absent in the burning out of lithium. Walter (1982) showed that at power representation of the ratio $L_X/L_{bol}$ as a function of the rotational period there was a break near the period of 12 days, this period corresponded to the Vaughan–Preston gap and the age of single G dwarfs of one billion years. On the other hand, the strongest lithium line in the spectrum of V 1005 Ori is not related to the extremely high activity. Probably, this is due to the fact that in the atmospheres of M dwarfs the lithium line $\lambda$ 6708 Å becomes dependent on the activity level, as the $H_\alpha$ line, and ceases to be an independent indicator of age (Houdebine and Doyle, 1994b).

Using the observations with the echelle spectrograph of the 3-m Lick telescope and his own calibration of the lithium abundance as the age characteristic, Soderblom (1983) found that for solar-type stars the relation $\langle v\sin i\rangle \sim t^{-1/2}$ was valid. However, it was not satisfied for the stars younger than the Pleiades and appreciably overstated their velocities as compared to the observed values. The intensity of calcium emission was proportional to the rotation rate, but the value $v\sin i$ did not display an analog to the Vaughan–Preston gap.

Observational evidence of the fact that rotation rather than the age in itself determines the level of stellar activity was obtained in studying the components of stellar pairs: as stated above, the dependences of the fluxes $F_{HK}$ and $F_X$ on rotation rate were common for the components of binary systems and single stars. This commonness is violated only in the systems with very short orbital periods – in semidetached and contact systems (Middlekoop, 1982; Walter, 1982; Rutten, 1986, 1987; Maggio et al. 1987). Therefore, all research results for activity depending on rotation listed above have an exact evolutionary sense.

From a dozen stars observed with IUE, Blanco et al. (1982) found a linear dependence between $\log F_{MgII}$ and the axial rotational period, common for the considered F0–K5 dwarfs. Then, Catalano and Marilli (1983) compared the luminosity of calcium emission and rotation of twenty F8-K7 stars and found that this luminosity decreases with growing axial rotation period as $\mathrm{dex}(-P/27^d)$. Comparison of luminosities of the Pleiades and the Hyades stars with their masses revealed a practically identical dependence $L_K \sim M^{5.1}$ for both clusters, but with a shift toward lower luminosities in older Hyades stars. If one excludes young G stars of the Pleiades, for the stars within the range of $10^8$–$5 \cdot 10^9$ years the relation is valid

$$L_K(M/M_\odot, t) = L_K(1.0)(M/M_\odot)^{5.1} \cdot 10^{-1.5 \cdot 10^{-5}\, t^{1/2}} \tag{70}$$

with a correlation coefficient of 0.98 and at $L_K(1.0) = 1.90 \cdot 10^{29}$. Further, Catalano and Marilli found that the relation $L_K \sim t^{-1/2}$ represented observations poorer than the exponent. But an exponential decay of emission yields the rotation deceleration rate as $\Omega \sim t^{-1/2}$ for stars older than $3 \cdot 10^8$ years. Finally, they found that the lithium abundance on the Sun and the appropriate stellar values determined by Duncan and Soderblom well suited the general exponential dependence for the rate of lithium burning out, and the exponent of the dependence included $t^{-1/2}$.

Using a more complete sample of single stars, Marilli et al. (1986) revised the results of Catalano and Marilli (1983). They revealed a dependence on B−V, i.e., the depths of the convective zone, coefficients of the linear relation between the logarithm of luminosity of calcium emission and the axial rotation period. Rutten (1986) arrived at a similar conclusion. Taking this dependence into account, Marilli et al. concluded that there was a relation between $L_{HK}$ and the Rossby number.

Using the lithium abundance after Duncan (1981) and the depths of $H_1$ and $K_1$ absorptions after Barry et al. (1984) as time scales, Cabestany and Vazquez (1983) considered the time change of magnesium emission and found that luminosity in the MgII k line could be presented by the relation

$$L_k(\mathrm{MgII}) = 2.8 \cdot 10^{29-1.82 \cdot 10^{-5} t^{1/2}}, \qquad (71)$$

which is rather similar to (70) for the CaII K line found by Catalano and Marilli (1983).

For the low-resolution spectra Barry et al. (1984) constructed an analog of the Mount Wilson value $S$ and determined $R_{HK}$ from the spectra of F-G dwarfs in seven open stellar clusters of various age and the Sun. The obtained values did not match the Skumanich relation $t^{-1/2}$, but could be presented by one or two exponents: in the first case, the exponent included $t^{-1/2}$, in the second, the first addend described the long-term evolution of chromospheric emission in the main sequence, and the second addend — fast evolution of emission for the youngest stars. Having added the data on the middle-aged cluster NGC 752, Barry et al. (1987) specified the chromospheric scale of the age of solar-type stars within $10^7$–$6 \cdot 10^9$ years and confirmed that the above two analytical representations equally well described the secular decay of chromospheric emission without any breaks as the Vaughan–Preston gap. According to Pace (2013), the Skumanich relation is valid up to the age of one and a half–two billion years.

According to Noyes et al. (1984a), Fig. 15 shows the dependence of $R'_{HK}$ — the ratio of fluxes in CaII H and K lines corrected for the contribution of the radiative atmosphere to the bolometric luminosity — on the Rossby number, which is the key parameter of the dynamo theory. As stated above, this dependence appeared to be so close that it was used to estimate the axial rotation period from the observed luminosity of calcium emission. Similar results were obtained in comparing the values of $R_{hk}$ for G–K stars with the axial rotation periods from 5 to 50 days and the Rossby numbers, but F and the earliest G dwarfs had some differences (Hartmann et al., 1984a). Doyle (1987) considered the most active emission dwarfs with the rotational periods from 0.8 to 6 days and confirmed the correlation between $R_{hk}$ and Rossby numbers, which again was so close that it allowed estimation of axial rotation periods. At $P_{rot}/\tau_c \sim 1/10$ ($\tau_c$ is the turnover time of the convective element, the denomination in the determination of the Rossby number), there is saturation analogous to that found by Vilhu (1984) for the radiation of the transition region. For late spectral types this border corresponds to four days. Later, the times of rise of convective elements involved in Rossby numbers were thoroughly studied by Stepien (1994) for the whole range of F–M stars: he included experimental corrections to theoretically calculated values and essentially decreased the scattering of points in "activity level–Rossby number" diagrams for dwarf stars.

Figure 84 shows a modern analog of the result of Noyes et al. (1984a) — the correlation between axial rotation periods and values of $R'_{HK}$ (Toledo-Padron et al., 2019).

Using the sample of 115 chromospherically active early F and G dwarfs, Barry (1988) estimated possible errors of chromospheric ages. He compared rotation rates and the chromospheric age of sample stars and found that the rotation periods $P_{rot}$ increased as $t^{0.37}$.

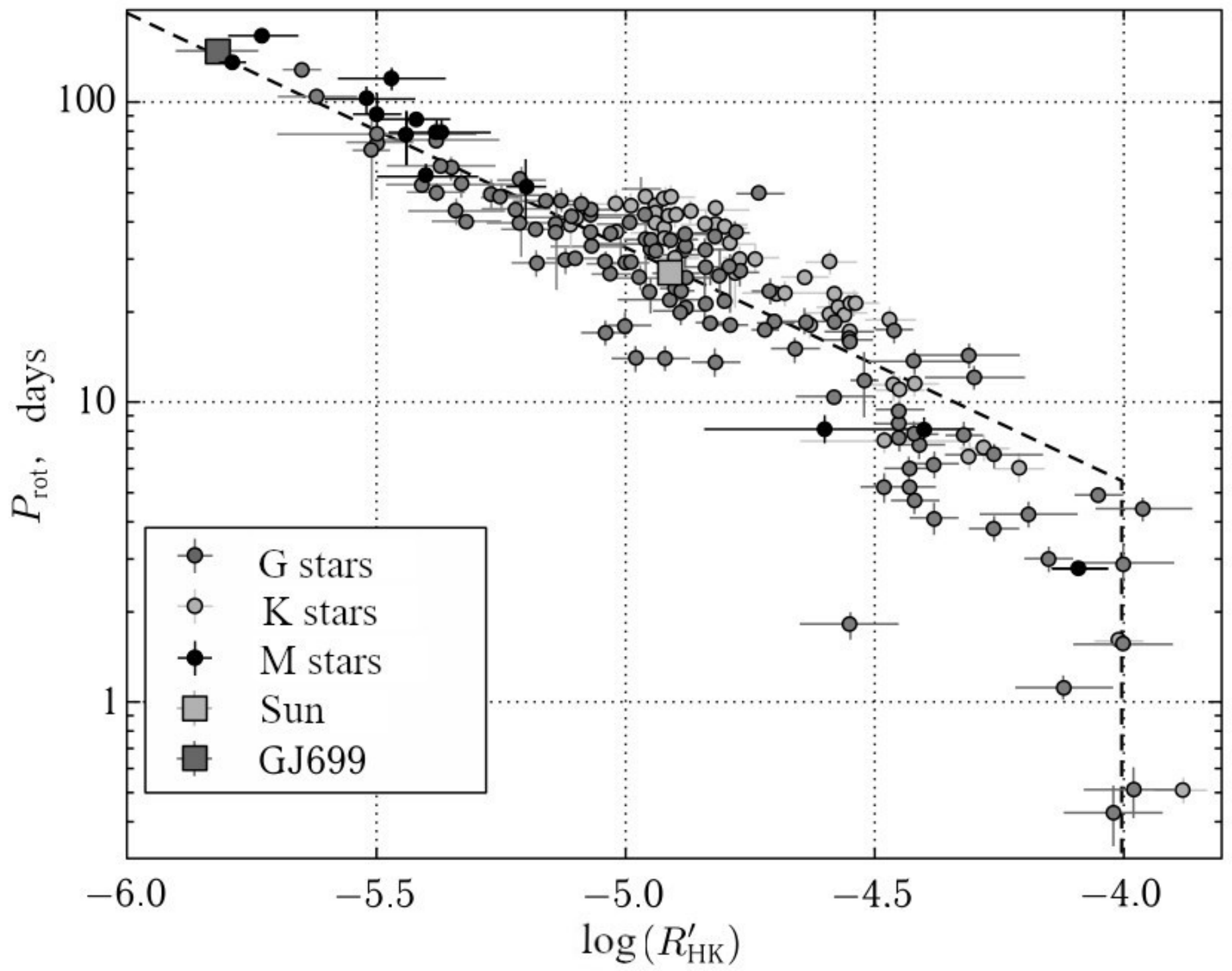

Fig. 84. Correlation between the axial rotation periods and values of $R'_{HK}$ (Toledo-Padron et al., 2019)

The appropriate change of angular velocity is shown in Fig. 85. In this case, braking is proportional to angular momentum and inversely proportional to age and rotation. Rotation depends only on the stellar mass and age. The entire activity decay is determined by an increase of the Rossby number. For a wider range of stellar masses from 0.5 to $1.1M_{\odot}$, Catalano et al. (1989) found that $P_{rot} \sim t^{-1/2}$ with a factor growing toward low masses. According to Barry, the Vaughan–Preston gap is caused by nonuniformity of star formation. The monotonous curve of Barry in Fig. 85 covers billions of years of stellar evolution on the main sequence. But during the first millions of years, the stellar rotation was determined by its active interaction with a gas-dust disk. Following Grankin (2013), the positions of about 50 young stars are shown on the "rotation period–age" plane in Fig. 86, as well as the constructed family of dashed curves that describe the initial evolution of stellar rotation periods depending on the moment of time when the magnetospheric relation of a star and a disk ceases. The figure shows that if this relation is disrupted rather early, then $P_{rot}$ decreases from a dozen days to a few hours.

As stated above, Herbig (1985) measured the intensity of the $H_{\alpha}$ emission line in the spectra of 40 F8–G3 dwarfs and, using the age scale of Duncan (1981), found a secular decay of $H_{\alpha}$ emission on the background of great divergence of data, which can be presented as $\sim t^{-0.4}$, or using the exponent with an *e*-fold decay over (3–5) · $10^9$ years. Mekkaden (1985) compared the intensity of $H_{\alpha}$ emission measured by Herbig for F8–G3 dwarfs with their rotation and found a close correlation between $R_{H\alpha}$ and axial rotation periods and a less close relation to Rossby numbers.

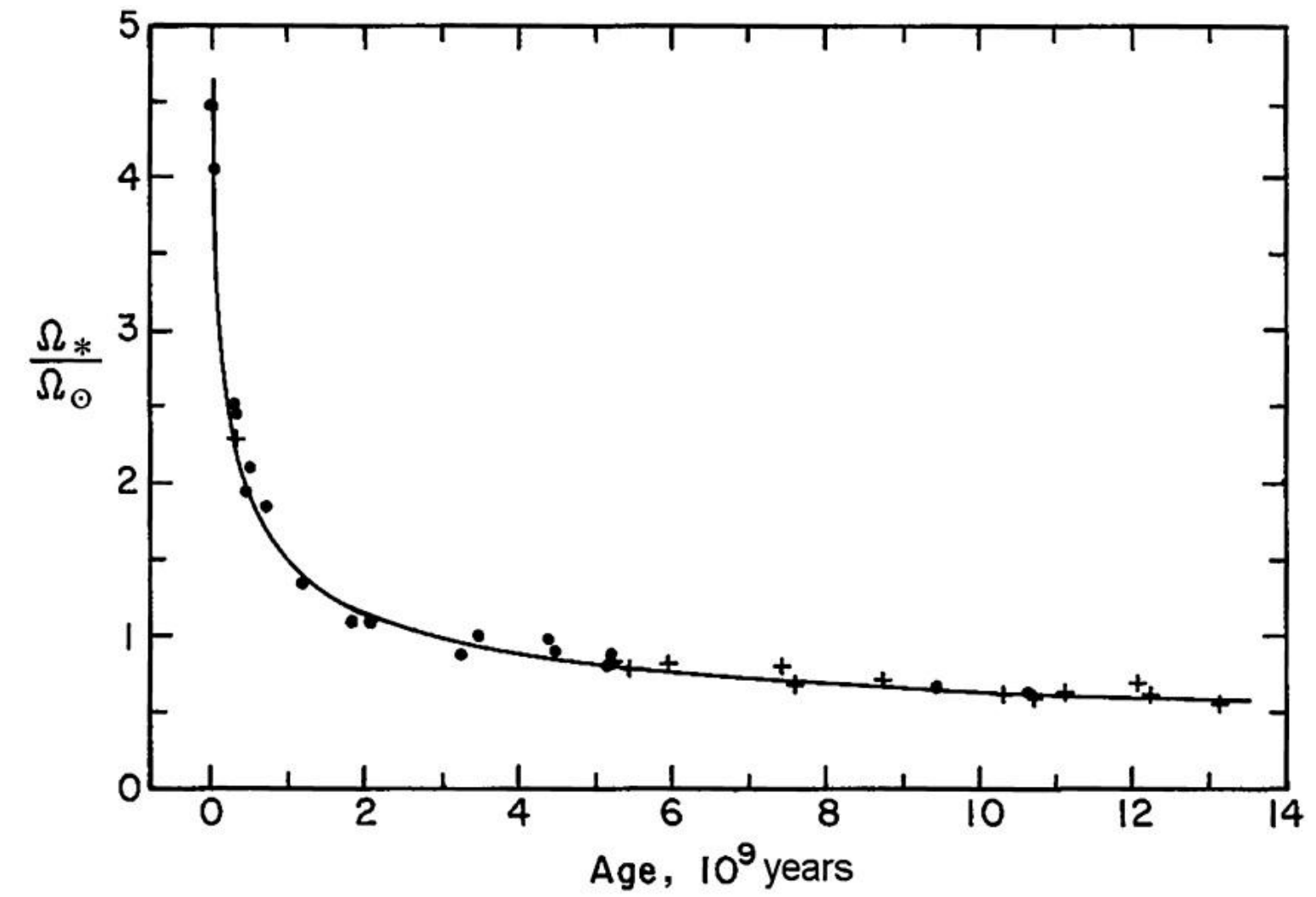

Fig. 85. Evolution of the angular velocity of F–G dwarfs (Barry, 1988)

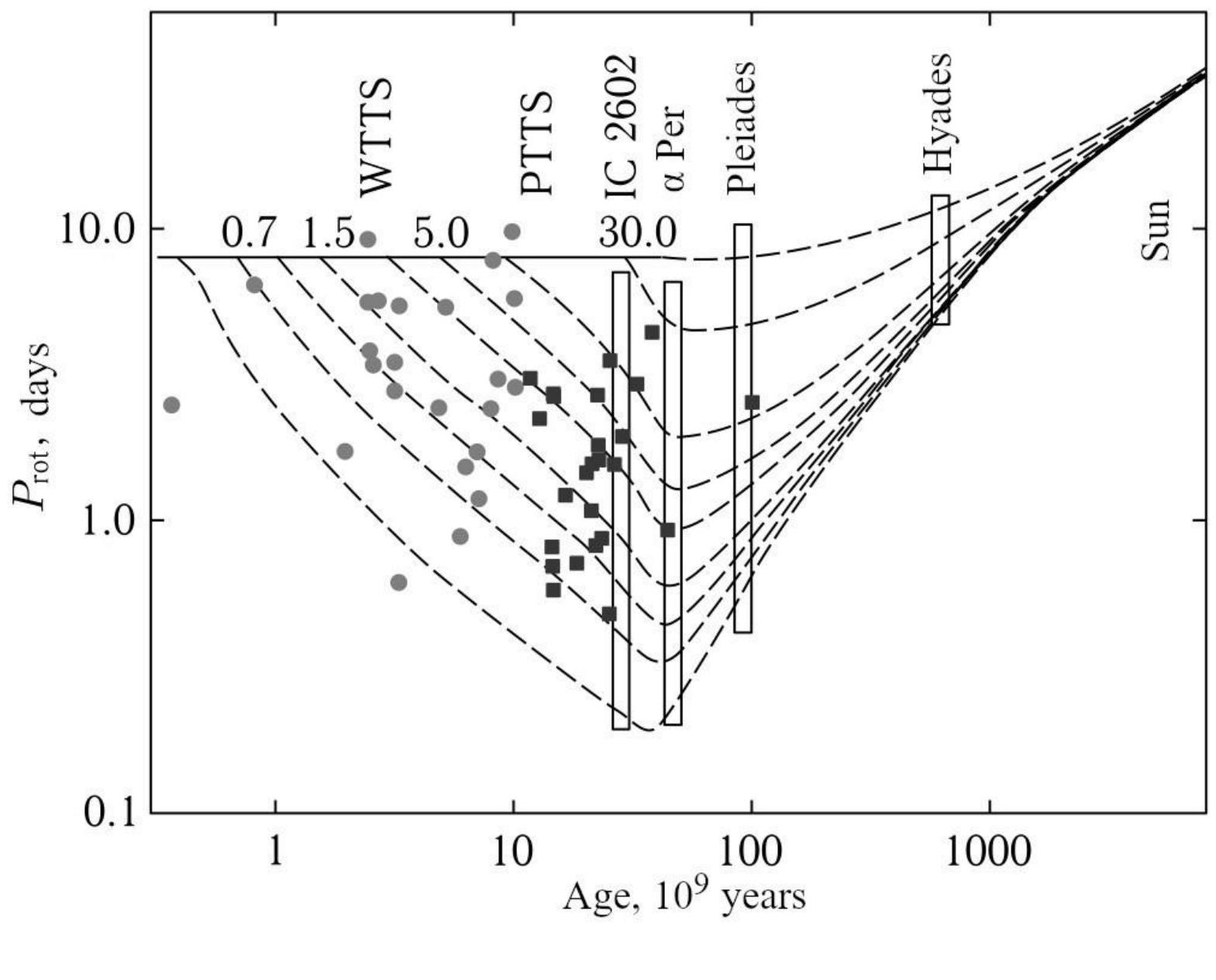

Fig. 86. Dependence of the rotation period on the age of young stars without disks in the Taurus-Auriga star-forming region. The stars younger than 10 million years are marked by circles, older — by squares. The family of dashed curves describes the evolution of rotation periods depending on the moments when the magnetospheric relation of a star and a disk ceases (Grankin, 2013)

Simon et al. (1985) thoroughly analyzed the UV spectra of the Sun and 31 F7–G2 dwarfs, for which lithium or other age estimates were available, and found that the sum of surface fluxes in four strongest UV lines of the chromosphere and CaII H and K lines weakened with time as $t^{-0.51\pm0.05}$, whereas the sum of surface fluxes in five strongest transition-region lines weakens as $t^{-1.00\pm0.09}$. But in both cases observations are better presented by an exponential decay with the time of an *e*-fold decay of 2.6 and 1.4 billion years for the chromosphere and the transition zone, respectively. Further, following Noyes et al. (1984a), Simon et al. constructed the dependences of flux ratios normalized to bolometric luminosity on Rossby numbers in all considered UV lines. They found that the decay of CaII and MgII emissions was slower than those of CIV and SiIV, and X-ray emission was faster than the emission of the transition zone. They presented arguments supporting the hypothesis of Hartmann et al. (1984b) and Barry et al. (1984) that stellar activity decayed at a different rate in the phase of T Tau and in the main sequence. They proposed various physical mechanisms of the phenomenon. In the first case, decay occurred due to a decrease of the depth of the convective zone, inherited from the evolution during the Hayashi phase, whereas lowered activity of main-sequence stars starting from a certain plateau inherent in the youngest stars of the sequence was caused by rotation braking. These two processes can be combined assuming the decisive role of Rossby numbers: reduced activity in the first case is connected with a growth of the Rossby number due to the reduction of the denominator, whereas in the second case the former grows together with the numerator. Changes in integrated intensity of chromospheric emission can occur due to changes of the brightness of the chromospheric network, which on the Sun determines this integrated intensity, or due to the changes of the filling factor, as during the solar-activity cycle. Based on detailed comparison with solar data, Simon et al. (1985) concluded that on young stars excessive emission could be caused by flares covering up to 4% of the surface and flocculi occupying up to 30% of the surface, whereas on old stars the filling factors were 5–10 times less, and in all cases radiation in the transition-zone lines occurred basically in flares, while in chromospheric lines it was from flocculi. Finally, they showed that the ratios of $R_{\mathrm{MgII}}$, $R_{\mathrm{CIV}}$, and $R_{\mathrm{X}}$ found by Ayres et al. (1981a) corresponded to a regular change of these values at a monotonic change of the Rossby number. Being based on the observations of solar-type stars, these correlations appeared to be fair for the whole range of F–M dwarfs.

As stated above, according to Stauffer and Hartmann (1986), an *e*-fold decay of $H_\alpha$ emission takes approximately one billion years. From independent observations of a number of moving stellar groups, Eggen (1990) estimated the rate of decay of equivalent widths of $H_\alpha$ emission as $t^{-0.5}$ and MgII k and h emissions as $t^{-0.3}$. Later, Stauffer and Hartmann (1987) examined the distribution of rotation rates of low-mass stars in the Pleiades. Detection of fast rotators among K and M dwarfs and minimum rotation rates of G stars led to the conclusion that low-mass stars before achieving the main sequence, appreciably accelerated and then quickly decelerated rotation. Thus, they assumed that the wide range of angular momenta found on K and M stars of the Pleiades could be caused by the considerable range of the initial angular momenta in combination with braking on the main sequence, regardless of rotation rates, rather than a combination of braking on the main sequence with the considerable age range of these stars. However, by the age of the Hyades the effect of the initial distribution of angular momentum practically disappears. As compared to the Pleiades, in the Hyades the beginning of $H_\alpha$ emission and fast axial rotation occur in later stars (Stauffer et al., 1991). According to Leggett et al. (1994) and Stauffer et al. (1995), all cluster members in the Pleiades and the Hyades with a mass less than 0.3 solar masses displayed $H_\alpha$ emission, and in the Pleiades $R_{\mathrm{H}\alpha}$ achieves its maximum at about $0.3M_\odot$.

Simon (1990) analyzed the emission of MgII lines in the spectra of G–M stars of several young clusters observed with IUE. He established that as the stars achieved the main sequence, there occurred a very fast braking of rotation and only then did the Skumanich ratio $t^{-1/2}$ come into force. But for low-mass stars the time of initial loss of angular momentum and activity decay was much longer than for solar-type stars. This conclusion was confirmed by Stauffer et al. (1991). Having found the closeness of radio luminosity of fast rotators in the Pleiades and T Tau stars, Lim and White (1995) concluded that the stars reached the main sequence in the mode of saturated activity. Soderblom et al. (1991) measured the calcium emission from solar-type stars in binary systems with known age and, upon combining the results with the data on open clusters, inferred that there was a functional rather than statistic relation between age and chromospheric emission. But they could not resolve the dilemma whether the decay of chromospheric emission was described by the power function of time or by variations of the star-formation rate over the past billion years.

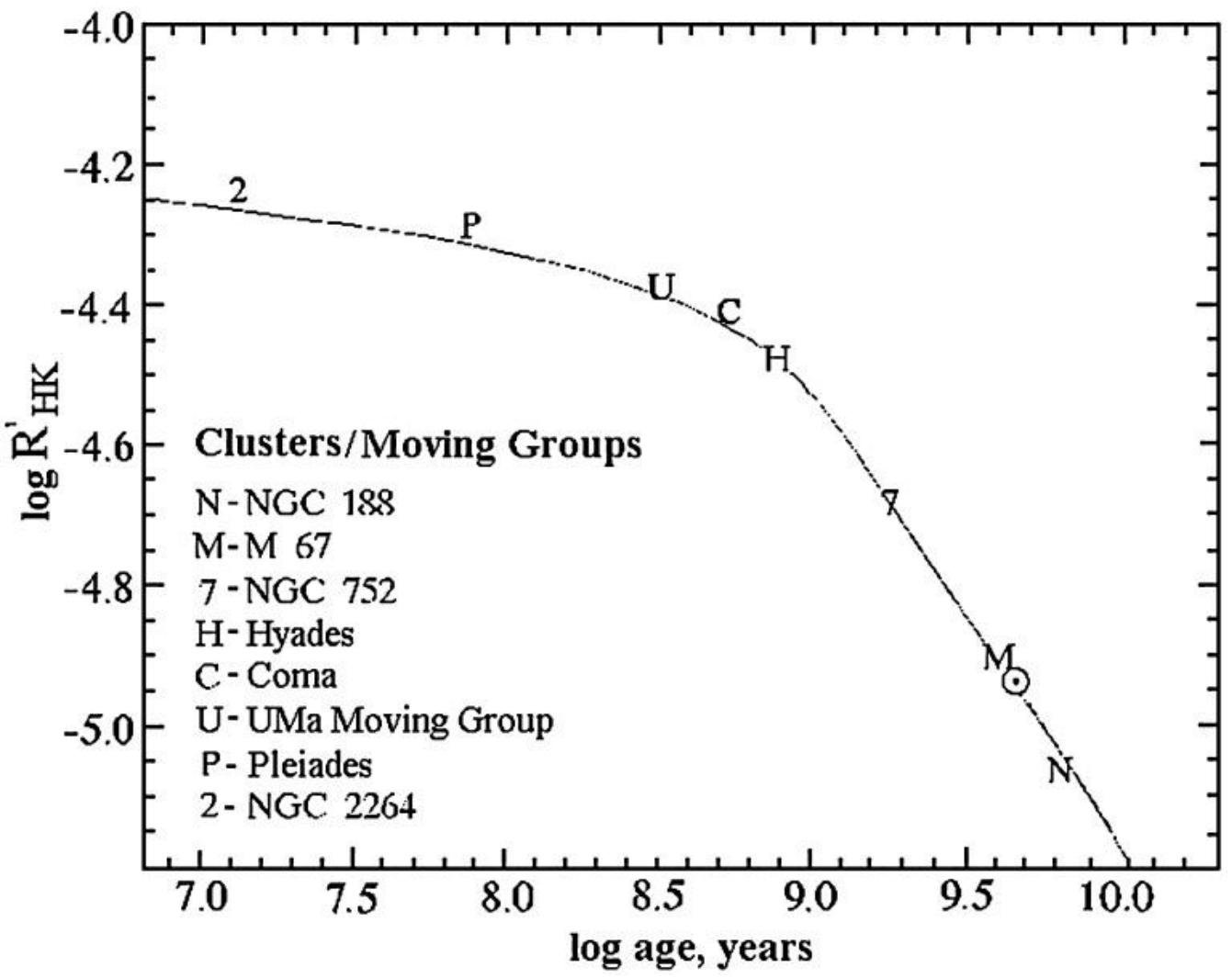

Fig. 87. The decay curve of calcium emission from observations of stars in clusters and moving groups of different age (Baliunas et al., 1998)

A secular weakening of calcium emission is given in Fig. 87 from the stellar assemblies of different age. Note that in this study the cyclic variability of such emission was found for the subdwarf Groombridge 1830, whose age is about 10 billion years. The recent age calibration for F7–K2 stars seemed to be acquired by Mamajek (2009) from the emission of CaII H and K lines and from the ratio $L_X/L_{bol}$.

* * *

Above, we mentioned the results of Simon et al. (1985) proving that the rate of evolutionary decay of emission of the transition zone was much greater than that in the chromosphere. A similar conclusion was made by Marilli and Catalano (1984) who revised the preliminary results and found the following ratios:

$$L_{HK} = 1.1 \cdot 10^{29-P_{rot}/27.0d},\ L_{CIV} = 5.5 \cdot 10^{27-P_{rot}/22.8d},\ \text{and}\ L_X = 5.7 \cdot 10^{29-P_{rot}/10.4d} \qquad (72)$$

determining absolute luminosities of the chromosphere, the transition zone, and corona as a function of the stellar rotation period. Simon and Fekel (1987) compared the values of $R_{CIV}$ and axial rotation periods and confirmed the importance of the Rossby number for unifying the relations of activity and rotation of dwarf stars. From the very precise dependence $R_{CIV}$ (Ro) Simon et al. (1985) estimated the deceleration of rotation rate of a star with a mass of $1.1M_{\odot}$ with increasing rotation period from 4 days for the age of $10^8$ to 22 days for the age of $4 \cdot 10^9$.

Using HST, Ayres et al. (1996) and Ayres (1999) observed ten early G dwarfs in three open clusters of different age: in $\alpha$ Per, the Pleiades, and the Hyades. Combining these results with those of Ayres et al. (1995) for F9–G2 field dwarfs, they found an essential decline in CIV emission from the youngest to the oldest stars well correlated with $v_{rot}$. Saturation of this and soft X-ray emission occurred at about $v_{rot} \sim 35$ km/s. The correlation of intensity of these carbon lines and soft X-rays is of the form $R_X \sim R_{CIV}{}^{1.7}$, and the Sun at different phases of its activity cycle matched the relation.

The character of decline of stellar coronal activity was determined by comparing the activity of individual stars of different age and by analyzing the objects in various stellar clusters.

Johnson (1983b) found that, as a rule, M dwarfs with maximum X-ray luminosity, according to their kinematic characteristics, could be ranked among young objects, but some of them belonged to the old-disk population. Thus, he suggested that these M dwarfs recently appeared in fast clouds of interstellar gas.

Dobson and Radick (1989) compared soft X-ray emission and rotation of 157 single main-sequence late stars in the solar vicinity and in different clusters and found that the best correlation was between normalized X-ray luminosity $R_X$ and Rossby numbers; this correlation includes the stars of all spectral types from early F and covers three orders of magnitude in Ro and four orders of magnitude in $R_X$.

On the basis of kinematic properties, Hawley and Feigelson (1994) selected five nearest and oldest M dwarfs of the halo population without $H_\alpha$ emission and observed them in X-rays. Four stars displayed the luminosity $L_X = (1–5) \cdot 10^{26}$ erg/s, the fifth $L_X < 6 \cdot 10^{25}$ erg/s. Thus, the activity can be maintained within the Hubble time. It should be noted that, according to Peres et al. (2000), the luminosity of the solar corona measured by ROSAT PSPC would make $3 \cdot 10^{26}$ erg/s at the activity minimum and $5 \cdot 10^{27}$ erg/s at the activity maximum.

Micela et al. (1997) analyzed the observations of 12 nearby K4–M6 halo and old-disk stars using ROSAT PSPC and WFC. They found that for halo stars the values of $L_X$ were systematically lower than for the old-disk stars: $(1–5) \cdot 10^{26}$ and $(1–9) \cdot 10^{27}$ erg/s, respectively, and extremely low $L_X \sim 3 \cdot 10^{25}$ erg/s for Barnard's star (GJ 699), which showed no signs of activity in optical range. The obtained values of $L_X$ of the halo stars correspond to the lowest luminosities of nearby dK and dM stars of a spatially full sample by Schmitt et al. (1995), whereas $L_X$ of the old-disk stars is within average luminosities.

Hempelmann et al. (1995) selected about 100 single stars of late spectral types with directly measured rotation periods and different age — both field stars and members of the Pleiades and the Hyades — and analyzed RASS data for them. They found that field stars and members of young clusters satisfied the same correlations $L_X/L_{bol}(P_{rot})$ and $L_X/L_{bol}$(Ro) that were valid over three orders of magnitude, which evidenced the decisive role of rotation for the level of X-ray activity, but not the age in itself. Another argument supporting this statement was recently obtained by James et al. (2000) who compared the features of X-ray emission for rapidly rotating M dwarfs — single stars and components of binary systems with $v_{rot} > 6$ km/s. They found that for the former, including objects with $0.2 < P_{rot} < 10.1$ days, $\langle \log L_X/L_{bol} \rangle = -3.21 \pm 0.04$, whereas for the latter, including objects with $0.8 < P_{rot} < 10.4$ days, $\langle \log L_X/L_{bol} \rangle = -3.19 \pm 0.10$. On looking for statistical dependences of $L_X$, $F_X$, and $L_X/L_{bol}$ on some power of the period or the Rossby number, Hempelmann et al. found that exponents in all cases were close to −1. But at Ro>1/3 $L_X$ and $F_X$ decreased with growing Ro much faster than at lower Rossby numbers.

Combining HST and ROSAT data, Ayres (1999) found for early G field dwarfs and from $\alpha$ Per, the Pleiades, and the Hyades clusters a distinct correlation $R_X \sim R_{CIV}^{2.0}$ with a certain saturation of the youngest $\alpha$ Per stars. Comparison of these photometric values with rotation rates revealed precise correlations $R_X \sim v_{rot}^{3.0\pm0.6}$ and $R_{CIV} \sim v_{rot}^{3.15\pm0.3}$ and with saturation at about 35 km/s and 100 km/s, respectively. During the magnetic activity cycle the Sun drifts along the correlation ($R_X$,$R_{CIV}$) constructed using the data on the considered stars. Another important conclusion is that young stars have a noticeable dispersion of high-temperature emission, whereas in the Hyades it is insignificant. Within the model of simple magnetic loops (Rosner et al., 1978) the nonlinear correlation of $R_{CIV}$ and $v_{rot}$ should be connected with the pressure rise at the feet of such loops as compared to the appropriate solar values, and the nonlinearity of the correlation of $R_X$ and $R_{CIV}$ indicates the rise of average temperatures at the loop tops. Thus, Ayres et al. concluded that on old stars, such as the Sun, loops of X-ray bright points and of active regions dominated, whereas on the youngest and most active stars large-scale structures or postflare loops prevailed.

The data on evolution of stellar magnetic activity obtained in X-ray observations of stellar clusters containing stars of different masses but the same age, identical chemical composition, and remote by the same distance, show that the decay of X-ray activity with age is typical of all stars of late spectral types and this process depends on the stellar spectrum.

Stern et al. (1981) found that the higher X-ray luminosity of the Hyades as compared to the field stars corresponds to the dependence $L_X \sim (v\sin i)^2$, while the Pleiades demonstrate a more complex dependence and greater divergence.

Caillault and Helfand (1985) found that $L_X$ in several clusters decreased not as $t^{-1/2}$, but first up to log$t$ = 9 very slowly and then much faster than according to Skumanich. Since $L_X$ of the Sun, if it were completely covered by active regions, would be 10 times lower than that of the brightest G stars in the Pleiades, Caillault and Helfand assumed that the coronae of young stars had bigger loops than the solar corona, which was later confirmed by Ayres (1999). The fact that $L_X$ on K0–M3 stars does not correlate with v sin$i$ found by Pallavicini et al. (1982), but fits the exponential relation of Walter (1982) suggests the existence of an additional factor that determines $L_X$ of young stars along with rotation.

Stern (1984) and Maggio et al. (1987) compared the X-ray luminosity of stars in Orion, the Pleiades, the Hyades, and the Sun and found a systematic decrease of maximum values of $L_X$ with age. Using the Einstein Observatory data, Micela et al. (1985) detected that the X-ray luminosity of G stars in the Pleiades was much higher than in the Hyades. Essentially, the greater number of X-ray bright G dwarfs as compared to K dwarfs, the absence of M dwarfs, and the availability of rapidly rotating K stars in the Pleiades led to the suspicion that in this young cluster the rotation of K dwarfs accelerated, whereas G stars already entered the phase of magnetic braking.

According to Caillault (1996), by the end of 1995 ROSAT provided data on 13 stellar clusters with ages varying from 20 to 600 million years separated by 45 to 400 pc, the number of identified X-ray sources varied from 15 to 185. Figure 88 shows normalized distribution functions of the X-ray luminosity of G dwarfs in seven clusters. One can see a distinct change of the total decay of X-ray emission with age, though significant divergence of these functions for the Hyades and Praesepe — clusters of the same age — requires additional explanations. One cause is the different contribution of binary systems to the total X-ray luminosity or different metallicity of stars in the clusters (Barrado y Navascues et al., 1997). On the whole,

X-ray data obtained for the clusters make it possible to conclude that the distribution function of $L_X$ depends on the age of clusters and the mass of the considered stars, which is caused by the dependence of the deceleration scale on stellar mass, which in turn results in the age dependence of the distributions of rotation rates of stars in a cluster. Comparison of coronal activity and rotation rates reveals the saturation of the ratio $L_X/L_{bol}$ at the level of $10^{-3}$ for rapidly rotating stars.

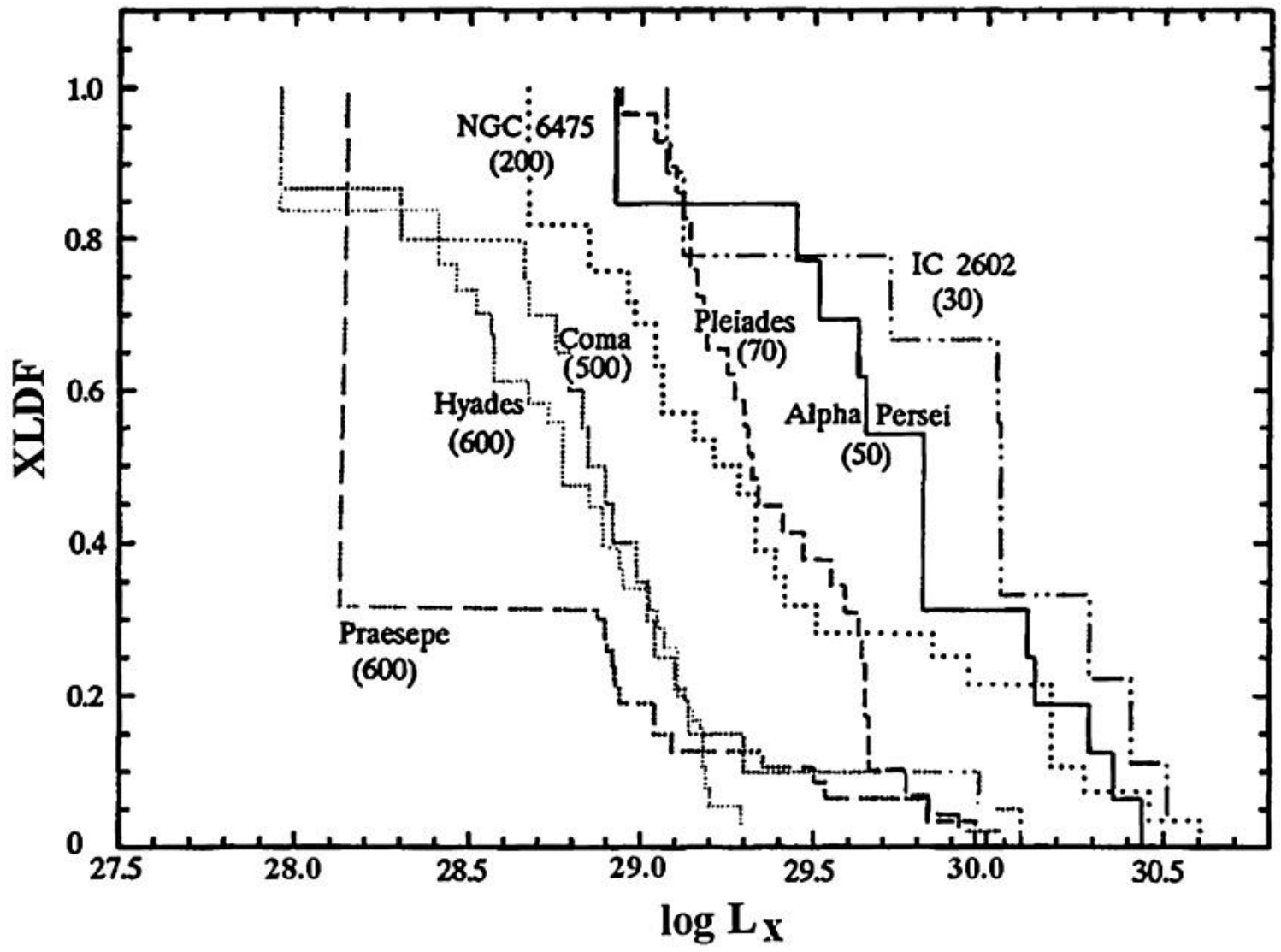

Fig. 88. Normalized distribution functions of X-ray luminosity of G dwarfs in seven stellar clusters of different age after Caillault (1996); numbers in brackets indicate the age in million years

Patten and Simon (1993, 1996) analyzed ROSAT PSPC observations of the very young — about 30 million years — cluster IC 2391, where at the center 76 X-ray sources with $L_X(0.2-2.0 \text{ keV}) > 2 \cdot 10^{28}$ erg/s were found, 19 of them were identified as cluster members. Using ground photometric and spectral observations of new members of the cluster, they concluded that for solar-type stars rotation periods differed by more than 20 times, thus stars of very different activity level reached the main sequence. In comparison with older clusters of the Pleiades and the Hyades, it was found that at the general decrease of $\langle v_{rot} \rangle$ and $\langle L_X \rangle$ with age, the scattering of $L_X$ grew because the stars left the saturation mode for the phase, where their X-ray luminosity was determined by rotation.

In a later review of the results of a study of 15 clusters younger than 1 billion years, Randich (1997) considered distribution functions of X-ray luminosities of stars with different masses. The distributions were determined by the number of fast rotators in the cluster, whose Rossby numbers were less than 0.16 and X-ray luminosity was at the saturation level, and by the distribution of rotation rates among slow rotators. Since the characteristic times of braking of fast and slow rotators differ and depend on stellar masses, in the age range from $3 \cdot 10^7$ to $6 \cdot 10^8$ years the values of $L_X$ are not described by the simple Skumanich or other relation common for different mass stars.

In ROSAT/PSPC studies of old and intermediate age open clusters, Belloni (1997) and Belloni and Tagliaferri (1997) confirmed the conclusion obtained for other old clusters: at an age of more than $10^9$ years, high X-ray luminosity was within $10^{29}$–$10^{30}$ erg/s only in binary systems.

The Chandra Deep Field-North survey carried out with high sensitivity was used by Feigelson et al. (2004) to identify and estimate the parameters of X-ray emission of old high-latitude main-sequence stars. They compiled a sample of 11 stars — two G, two K, and seven M stars — with an average value of $V$ of about $19^m$ at the distance 300 pc with luminosity $\log L_X \sim 27$ plus two objects with high luminosity; whereas their variability with a higher amplitude on time scales of hours made the emission of these objects to be attributed to flares rather than to quiescent coronae. The compiled sample was compared to calculations based on folding the X-ray luminosity function with the known distribution of old disk stars — these are the members of the sample. The calculation showed that without taking into account the decay of X-ray luminosity on the interval from 1 to 11 billion years there should be 39 such objects and the found significant distinction was caused by the omission of evolutionary decrease of X-ray luminosity. One can agree calculations and observations by an assumption that $L_X$ decreases as $t^{-2}$ rather than $t^{-1}$, which is a widely accepted expression for rotation braking and weakening of X-ray activity.

For slowly rotating M dwarfs it is difficult to determine the rotation period required for checking the rotation–activity relation due to its smallness. To overcome this difficulty, Reiners (2007a) derived spectra of 10 M dwarfs in the IR region, where the absorption lines of the FeH molecule were strong, and determined rotation rates of the sample stars. For one star, he detected an upper limit of velocity of 1 km/s and measured it for the remaining nine stars. For inactive objects, it was not more than 2 km/s, for AD Leo and YZ CMi — 40–50% lower than it was obtained in earlier studies. Thus, the activity and rotation rates in the sample were in agreement with the ratio found for earlier young stars.

Studying more than 38000 low-mass stars from SDSS spectra, West et al. (2008) confirmed a decrease in the fraction of active stars from $H_\alpha$ emission with growing distance from the galactic plane, whereas the value and rate of this decrease depended on the spectral type. Within the framework of the 1D model, they showed that this effect was triggered by dynamic heating of the galactic thin disk with a rapid decrease of magnetic activity. From the comparison of observational data and calculations, the age-activity relations were calibrated for each spectral type of M dwarfs. West et al. (2008) also provided evidence for a probable decrease of metallicity with distance from the galactic plane.

Studying 40 000 M dwarfs within the SDSS project, West et al. (2009) found that the age at which an abrupt decrease of surface magnetic activity occurs significantly grew for later spectral types and among them a fraction of active objects noticeably increased. But the fraction of active stars decreased with increasing distance from the galactic plane, and this decrease also depended on the spectral type. Within the simple dynamic model of vertical heating of the Galaxy thin disk, one can attribute the age of activity weakening to each spectral type and thus determine the activity lifetime of M dwarfs. In particular, the rapid decrease of magnetic activity for M7 dwarfs occurs at an age of 6–7 billion years (West et al., 2006).

With the same aim of studying the decay of stellar dynamo and X-ray activity, Wright et al. (2010) constructed a distribution of stellar X-ray luminosity, combining the model of galactic population synthesis with available theories of the decay of rotation and the rotation-activity ratio for the magnetic dynamo. Having included 60 new stellar sources from the C-COSMOS survey, identified as objects of the thin disk and Galaxy halo, from the comparison of calculations they revealed a good agreement in the region of the brightest sources and a small deficiency of medium-intensity sources. This divergence decreases if, following

Feigelson et al. (2004), one takes the mentioned more abrupt change of decay of X-ray emission with time.

Of considerable interest is the high-resolution spectroscopy carried out by Reiners el al. (2007), as well as the magnetometry of all the components of the system LHS 1070 comprising the A component of the middle M type and components B and C on the boundary of brown dwarfs. For all the components the magnetic fluxes are a few kilogauss, for B and C components vsin$i$ = 16 km/s, for the A component — twice less. This means that for late M dwarfs the rotation braking weakens and it is valid for M and L dwarfs. Further, the magnetic flux of the B component is twice higher than that for the C component, which also indicates that in fully convective objects the field topology also plays a significant role.

Browning et al. (2010) considered the broadening of lines by rotation and chromospheric activity based on a sample of 123 M dwarfs from the spectra derived with the Keck telescope. Rotation was represented more common for stars later than M3, and less than 10% of early M stars showed rotation. This supports the view that the braking by rotation becomes less effective for fully convective stars.

From the molecular absorption lines of FeH, Reiners et al. (2009b) performed magnetometry of four accreting young brown dwarfs and one inactive young low-mass star. They found no evidence of the kilogauss field on either of brown dwarfs, but the field of 2 kG was detected on the low-mass star. Thus, the magnetic field on a young brown dwarf is five or more times weaker than those on young solar-mass stars or on older brown dwarfs.

Delorme et al. (2011) determined the precise rotation periods for 120 confident members of the Hyades and Praesepe clusters and found a magnetic braking effect in the wide range of spectral types for the total age of these clusters of 600 million years. In both clusters, a close and almost linear association of the color index $J$–$K_s$ and the rotation period was detected, which means that the angular momentum loss was sufficient for the stars of such a mass, initially having fairly different rotation rates, to have the same period by 600 million years. In the case of the Hyades, the color–period sequence continued up to M dwarfs and showed an increase of the divergence of this sequence with the appearance of numerous fast rotators from $M \sim 0.5M_\odot$ up to a lower limit of the considered sample of about $0.25M_\odot$. This means the end of the time scale of rotation braking and the use of gyrochronology for a precise determination of age for the main-sequence stars. Barnes (2007) substantiated and developed the gyrochronology technique: the accuracy up to 15% was achieved for late F, G, K, and early M stars; for the sample of Pizzolato et al. (2003) the gyroage estimates were in the range from less than 100 million years up to a few billion years. The decisive test of the technique was performed by Barnes using the components of broad pairs of $\xi$ Boo AB, 61 Cyg AB, and $\alpha$ Cen AB: the age of components of each pair were basically similar. From the extensive sample of $R'_{\rm HK}$ values for F7–K2 stars ($0.5^m$ < B–V < $0.9^m$) Mamajek and Hillenbrand (2008) refined the gyrochronology scale ranging from 0.6 (Hyades) up to 4.5 billion years (Sun) and concluded that the accuracy of their age estimates was about 0.2 dex.

Osten et al. (2012) carried out HST/ACS observations of the objects in the Sagittarius Window to constrain the flare rate toward the older stellar population in the Galactic bulge. During seven days of monitoring about 230000 stars brighter than $V$ = 29.5 they found evidence for flaring activity in 105 stars between $20^m$ and $28^m$. Attributing the objects with large-scale details in their light curves to the variable stars, they estimated for them a frequency of flares by a factor of 700 higher than that for non-variable stars and found a significant correlation between the amount of stellar variability and peak flare amplitude. The flare energy loss rates are generally higher than those of nearby well-studied single dMe stars. The distribution of their proper motions is consistent with the distance and age of the Galactic bulge. If these are single dwarfs, then their masses are in the range of 1.0 to $0.25M_{\odot}$. A majority of flaring stars exhibited periodic photometric modulations with a period of less than 3 days.

Considering two variants of evolution of stellar magnetic activity on main sequence, namely its gradual weakening and abrupt drop by 1–2 billion years with subsequent relatively inactive constant level, Lorenzo-Oliveira et al. (2018) revised the chromospheric activity–age relation up to 6–7 billion years from the analysis of about 9000 HARPS spectra of 82 stars and found that the chromospheric activity index continues to decrease after the solar age, whereas the significant divergence in this relation is due to activity cyclicity.

In two open stellar clusters, $\alpha$ Per of 60 million years and M 37 of 500 million years in age, Núñez et al. (2018) determined the X-ray fluxes, $W_{\mathrm{H}\alpha}$, rotation periods, and Rossby numbers. They considered the way how two different criteria of magnetic activity — coronal and chromospheric — depend on the Rossby number, whether they are simultaneously in the unsaturated mode.

Shkolnik and colleagues (Shkolnik and Barman, 2014; Miles and Shkolnik, 2017; Schneider and Shkolnik, 2018) analyzed an archive of photometric observations of M dwarfs in the ultraviolet range with the panoramic camera HALEX within the HAZMAT project. From observations of 215 M1–M4 dwarfs in 6 young moving groups at an age from 10 to 650 million years and field stars (5 billion years) they found that the median ultraviolet flux remains saturated for a few hundred million years. In the Pleiades (650 million years), it drops by a factor of 2–3 and then decreases up to the level of field stars; up to 300 million years the ultraviolet weakening occurs as $t^{-1}$. Despite this distinct common pattern, at each age there is a divergence by 2–3 orders of magnitude. On average, a drop in the X-ray flux occurs 65 times, in the far ultraviolet ($\lambda$1344–1786 Å) — 30 times, and in the near ultraviolet ($\lambda$1771–2831 Å) — 20 times.

Expanding a sample of stars of spectral types K7–M7 up to 377 in the photometric series of 145 of them, there was detected an average decrease of fluxes in FUV by 16 % and in NUV — by 11 %; variability increased toward later types in the near ultraviolet, whereas this effect was nor seen in the far ultraviolet. Flares increased the average fluxes in FUV to a greater degree than in NUV. The ratios of these fluxes were within 0.008 and 46 and by 1–2 orders of magnitude higher than those for G dwarfs. As to evolutionary variations, then the late M dwarfs with masses of 0.08–0.35 solar masses do not follow the pattern found for early M dwarfs with masses more than 0.35 solar masses: ultraviolet activity in the late M dwarfs weakens in both ranges up to the age of field stars on average by a factor of 4.

* * *

Stellar wind is one of the corona properties that today cannot be measured in other dwarf stars but for the Sun. Though modern properties of the solar wind are investigated in detail, its history is practically unknown. Some indirect evidences are accessible from the observations of very young clusters $\alpha$ Per and IC 2391, which are characterized by the wide distribution of rotational periods of G dwarfs, whereas in older clusters, as the Hyades, this distribution is much narrower. Thus, it was suggested that at a rotational velocity below some critical value the braking mechanism did not depend on rotational velocity, whereas at velocities higher than the critical level it became a steep function of the velocity.

On the basis of the theory of magnetic braking of stars by the thermal wind advanced by Mestel (1984), Stepien (1989) calculated the evolution of stellar rotation in the range of color indices (B −V) from $0.5^m$ up to $1.3^m$ for the Hyades, assuming initial rotational periods of all stars as 0.7, 1.3, and 3 days, and for slowly rotating single stars in the solar vicinity with age from 5 to 13 billion years, assuming initial periods of their rotation of two and five days. In the first case, the calculated periods to the age of the Hyades reached 14 days, in the second — 48 days. These values and distributions $P_{rot}$(B−V) well represented the observations and were used by Stepien to calculate expected fluxes $\Delta F_{CaII}$ and $F_X$ of stars in the Pleiades, young cluster NGC 2264, and the Hyades. The calculations yielded observed fluxes of calcium emission to an accuracy up to ± 0.05 dex and X-ray emission up to ± 0.5 dex. Then, Stepien considered the evolution of solar-type stars assuming the initial rotational period of 1, 3, and 5 days (see Fig. 89). According to these diagrams, at the age of over $10^8$ years $\Delta F_{CaII}$ decreases as $t^{-0.5}$ and $F_X$ as $t^{-1.7}$. The agreement of the calculations with observations confirms the validity of the Mestel theory (1984).

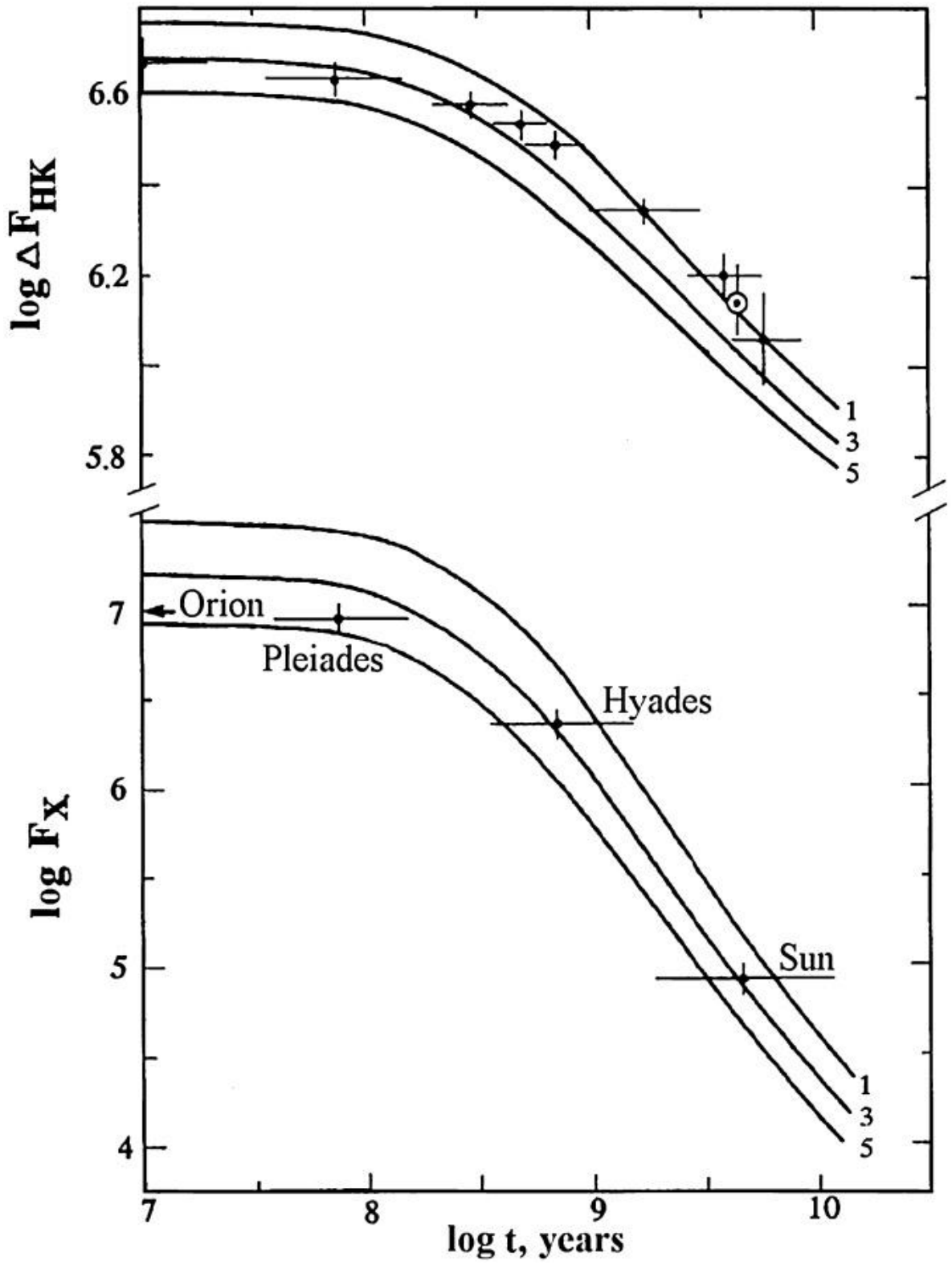

Fig. 89. Secular decay of calcium and X-ray emission from observations of cluster stars and the Sun and from calculations of the evolution of stellar rotation by Stepien (1989)

Here, it is appropriate to remember the conclusion made by Wood and Linsky (1998) in studying "a hydrogen wall" in circumstellar space: during the decay of a surface X-ray flux pressure in the stellar wind increases and one should expect a growing rate of secular loss of stellar mass. On the other hand, at reducing coronal temperature below a critical value a solar-type wind cannot be maintained on a star.

From the TESS observations, Yamashita et al. (2022) considered the light curves of 33 zero-age stars in young stellar clusters IC 2391 and IC 2602, estimated their brightness amplitudes ranging between 0.001 and 0.145 mag, similar to those of young stars in Pleiades, and found the filling factors of spots to be 0.1–21%, and a strong CaII emission, similar to that of stars with superflares, which is two orders of magnitude higher than that of the Sun. For 12 stars of these clusters with saturated chromospheric emission, they detected 21 flares with energies of $10^{33}$–$10^{35}$ erg.

* * *

The data on evolutionary changes of the level of flare activity of red dwarf stars are relatively few. In essence, they suggest the revision of the initial opinion about flare stars as extremely young objects. Affiliation of several flare stars to old objects was suspected by van de Kamp (1969), Chugainov (1972b), Lee and Hoxie (1972), and Kunkel (1972). Shakhovskaya (1975) considered the kinematic characteristics of 646 red dwarfs in the solar vicinity. She concluded that flare activity detected both from directly recorded flares and from hydrogen emission was observed among the objects of the young-disk population, whose age is less than $5 \cdot 10^8$ years, and among those of the old-disk population, whose age varies from $5 \cdot 10^8$ to $5 \cdot 10^9$ years, and even among halo objects, whose age exceeds $5 \cdot 10^9$ years. But on young stars, this activity was the strongest and it weakened toward older objects. From photographic observations of flare stars in four stellar clusters, Parsamian (1976) found a systematic decrease of the maximum luminosity of such stars with increasing age of clusters.

Based on the TESS, Gaia, and own high-resolution spectral observations, Medina et al. (2022) compiled a comprehensive sample of 219 single M dwarfs with masses between $0.1-0.3M_\odot$ within 15 parsecs from the Sun and performed an analysis of their kinematic characteristics and parameters of activity. They found that all stars are consistent with a common value of $\alpha = 1.984 \pm 0.019$ for the exponent of the flare frequency distribution. If we assume star formation has been constant in the thin disk of the Galaxy for the past 8 Gyr, then the transition from the saturated to the unsaturated flaring regime is $2.4 \pm 0.3$ Gyr. The stars with rotation periods less than 10 days have an age of $2.0 \pm 1.2$ Gyr, the stars with a rotation period between 10 and 90 days have an age of $5.6 \pm 2.7$ Gyr, and stars with rotation period over 90 days have an age of $12.9 \pm 3.5$ Gyr. The stars with rotation periods of less than 10 days and mass of $0.2-0.3M_\odot$ have an average age of $0.6 \pm 0.3$ Gyr, and those with mass $0.1-0.2M_\odot$ have an age of $2.3 \pm 1.3$ Gyr.

* * *

As stated above, Fig. 48 illustrates evolutionary changes of the spectral index of the energy spectrum of flares: this spectrum flattens with age and, hence, the contribution of rare but strong flares to the total flare radiation increases. But from the light curves of 347 stars recorded with Kepler Davenport et al. (2019) studied the evolution of stellar activity and confirmed that the activity of low-mass stars weakens as the rotation brakes, whereas the spectral index of the energy spectrum of flares does not change.

Skumanich (1986) imparted an evolutionary sense to the above correlation of total flare radiation and quiet X-ray emission from the corona. Based on the statistics on solar white-light flares, which are the most similar to stellar flares, and the luminosity of quiet X-ray emission from the Sun, he showed that the Sun also satisfied this correlation. Further, he found that dM dwarfs, on which optical flares and X-ray emission were recorded, satisfied the relation. Considering dM stars as evolved dMe dwarfs, Skumanich arrived at a conclusion that during stellar evolution the frequency of flares was preserved, while their power decreased, as did the X-ray luminosity of corona.

Using the GAPS program, Maldonado et al. (2022) analyzed the activity-rotation relationship and evolution of active regions for objects of different age. The age was estimated from the stellar kinematics, whereas the rotation and activity – from high-resolution spectra. They confirmed the decline of activity and rotation as the star ages. They also found that the rotation rate decays with age more slowly for cooler stars and that, for a given age, cooler stars show higher levels of activity. The authors concluded that the lifetime of active regions is longer on younger, cooler, and more active stars.

* * *

Schröder et al. (2013) analyzed in detail the positions of stars with activity levels measured from calcium lines in the Hertzsprung-Russel diagram, whereas their localization was carried out taking into account the Hipparcos parallaxes, values of B–V from the SIMBAD database, and all the data were attributed to metallicity $Z = 0.02$. In the HR diagram they also marked the evolutionary tracks with masses from 0.8 to $1.5M_\odot$, the main-sequence line of zero age and the line of half-life of these stars on the main sequence. Based on the activity level the stars were distributed into three groups: minimum activity stars or with calcium emission at the basic level ($S < 0.17$), moderate activity stars ($0.17 < S < 0.25$), and high activity stars ($S > 0.25$). Comparison of the listed stellar groups with the curves marked in the HR diagram led Schröder et al. to the following conclusions. The weakening of stellar activity depends not on absolute but on relative age of the staying of a star on the main sequence. The most active among incorrectly varying stars produce the largest distance near the zero-age main sequence. The moderate activity stars, both with distinct cycles, as on the Sun, and the stars without dominating periodicity in activity occupy the second quarter of main-sequence evolution. Almost inactive and stars with phases of the solar Maunder-type minimum are located in the third quarter of their main-sequence evolution. All quite inactive stars are in the last quarter on the main sequence. The solar Maunder minimum indicates the future stable (“flat”) state of staying on the main sequence, and about 70% of late main-sequence stars are in this state.

* * *

Thus, evolution of the activity of lower main-sequence stars depends on a set of processes. For the youngest stars, one can find the final phase of rotation acceleration related to the end of compression when a star reaches the main sequence, and stars arrive at the sequence with significant dispersion of angular momenta. When this phase is over, rotational velocities of most stars are rather high, and there is activity saturation with maximum values of $R_{HK}$, $R_{hk}$, $R_{H\alpha}$, $R_{CIV}$, $R_X$, and $R_R$. When stellar compression is over, magnetic braking of rotation starts, which first operates very efficiently, and when the rotational velocity decreases to some critical values, a more gradual monotonic activity decay begins. It is obvious that all critical points in the specified sequence of events are determined by both stellar mass and particularly the considered manifestation of its activity, which results in significant variety of activity indicators of various stars even inside one cluster. It should be noted that, according to D’Antona and Mazzitelli (1985), the time it takes for a star of 0.6 solar masses to reach the main sequence is $1.2 \cdot 10^8$ years, whereas for stars of $0.1M_\odot$ it is $1.5 \cdot 10^9$ years and $0.08M_\odot$ it is $1.3 \cdot 10^{10}$ years.

## 3.2.2. Evolution of Solar Activity

Above, we mentioned some properties of G dwarfs, which should be included in the general picture of the evolution of solar activity: the initial angular momentum of the Sun in the wide range of probable values; expected high-latitude spots on the quickly rotating young Sun; the change of speed of secular braking shown in Fig. 87; the Marilli–Catalano relationship (70) and the correlations of $R_{MgII}$, $R_{CIV}$, and $R_X$ and the ensuing conclusion on different rates of activity decay at various levels of the atmosphere; correlations of $R_X$, $R_{CIV}$, and $v_{rot}$ and the ensuing conclusion about evolutionary changes of parameters of coronal loops.

Solar-type stars achieve a universal dependence of activity on rotation at the age of 200–300 million years. Below, all these general conclusions are specified and elaborated upon.

There are two approaches to studying the evolution of solar activity: statistical consideration of various characteristics of the activity of G dwarfs of different age and detailed research of actual stars similar to the Sun using as many parameters as possible, except for the age. For the second problem, Gaidos et al. (2000) selected 38 G–K dwarfs as the analogs of the young Sun within 25 pc from the Sun; the less active among them turned out to be slow rotators of the greatest ages. Both approaches yielded certain results.

A stumbling block of former evolutionary schemes of the solar system was the distribution of angular momenta between the Sun and planets: though their mass is negligible, planets are responsible for the overwhelming part of the total angular momentum of the system. The concept of magnetic braking cuts the Gordian knot: soon after the achievement of the main sequence a star quickly dumps the major part of the initial angular momentum. In other words, magnetic braking is not only universal, but also an effective mechanism of fast dumping of the initial angular momentum resulting in further braking depending on stellar mass and age. These reasons enable discussions of the early stages of the development of the Sun with faster rotation and higher magnetic activity and predictions for its future. According to Soderblom (1983), the rotational velocity of the Sun corresponds to the average rotation rate of G dwarfs of solar age, which is justified by the statistical approach to its evolution.

Feigelson et al. (1991) estimated the expected properties of solar activity at the earliest stages of its development, assuming that at the age of $10^6$ years the Sun was a T Tau-type star with weak lines (WTT) for which, unlike for classical T Tau stars (CTT), the surrounding circumstellar medium was not a decisive factor in the observed activity. Spots occupy from 5 to 40% of the surface of WTT stars, the level of optical and ultraviolet activity of their chromospheres exceeds the solar one by approximately 50 times, the intensity of X-ray and radio emissions — by 3–5 orders of magnitude, and these fluxes vary on time scales of a day, hours and minutes and with the greatest probability are caused by strong flare activity of magnetic nature. Within this approach to the young Sun, at the age of 0.3 to 1.5 million years it should have a constant temperature of 4200 ± 500 K and spectral type K5 ± 2 IV. At the age of one million years its luminosity was 1.7 of the current luminosity, $L_X \sim 10^{30}$ erg/s, and radius — 2.5 of the current radius. It was compressed to the current size at the age of 10–20 million years. At an axial rotation rate of about 25 km/s the period $P_{rot}$ was about 5 days. Spots covered up to 25% of the solar surface and their temperature was 3300 ± 500 K.

Stepien and Geyer (1996) carried out photometric observations of 16 solar-type stars and for 9 of them detected appreciable brightness oscillations with amplitudes of several hundredths of the stellar magnitude, for most of them $P_{rot}$ < 10 days. Probably, as in the Hyades, small amplitudes of brightness oscillation are common characteristics of active solar-type field stars with rotation periods of about one week and less, some variables display strong modulation of variability amplitude within one year.

The ideas on the character of evolution of solar activity on the basis of the HK project were stated by Baliunas (1991). From the full sample of project stars she selected single objects with close to solar B−V color indices ($\sim 0.66^m$) and the range of *S* in this subsample was explained by the age difference. Table 23 presents statistical data for the stars of this subsample with axial rotation periods estimated from the observations of calcium emission. The second column of the table contains data for young stars, whose age is about one billion years, only the objects with variable *S*, both periodic and irregular, are cited. The objects with ages of several billions years (third column) display periodic changes of *S* or its constancy. Figure 90 illustrates *S* of two stars from the second column and two stars from the third column. The second from the bottom row presents the correlations of stellar brightness with the level of magnetic activity: for young stars there is an anticorrelation, i.e., large dark spots dominate in brightness variations, whereas on the Sun and old stars, brightening magnetic areas dominate (Radick et al., 1998). Figure 91 shows the MgII h and k line profiles in the spectra of G stars of various ages (Dorren and Guinan, 1994).

Table 23. Parameters of magnetic activity of solar-type stars of different age (Baliunas, 1991)

| Parameters | Young stars | Old stars | the Sun |
|---|---|---|---|
| Number of stars | 7 | 12 | 1 |
| Age, billion years | ~1 | several | 4.6 |
| $\langle S\rangle$ | 0.314 | 0.165 | 0.171 |
| $\langle P_{rot}\rangle$, days | 9.1 | 27 | 25 |
| Character of $S(t)$ | periodic or irregular without minima | periodic with minima taking up to 1/4 of time | 11-year cycle with minima taking up to 1/3 of time |
| Correlation of brightness and magnetic activity | inverse | direct | direct |
| Amplitude | <several percent | <0.4% | ~ 0.1% |

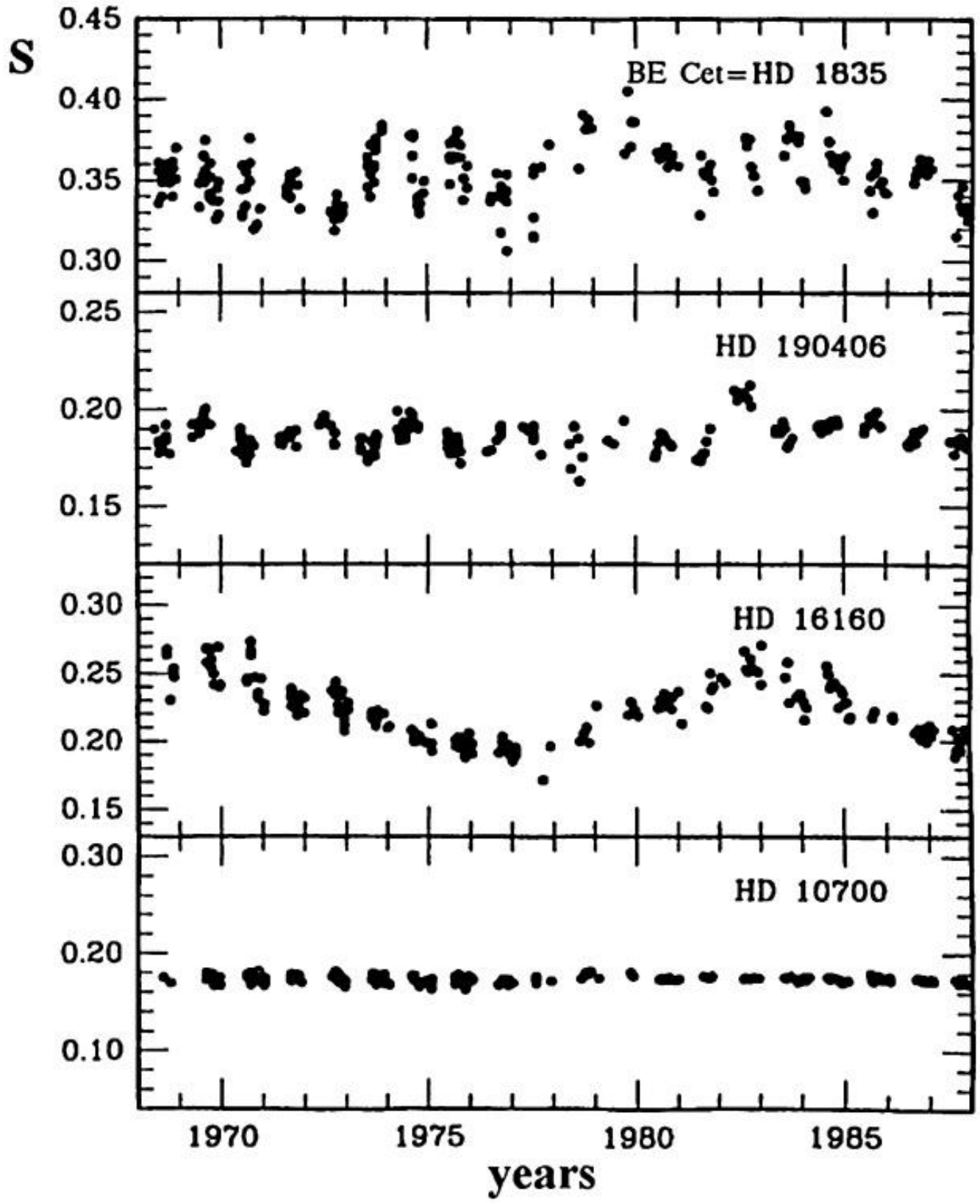

Fig. 90. Four characteristic time changes of S for solar-type stars (Baliunas, 1991)

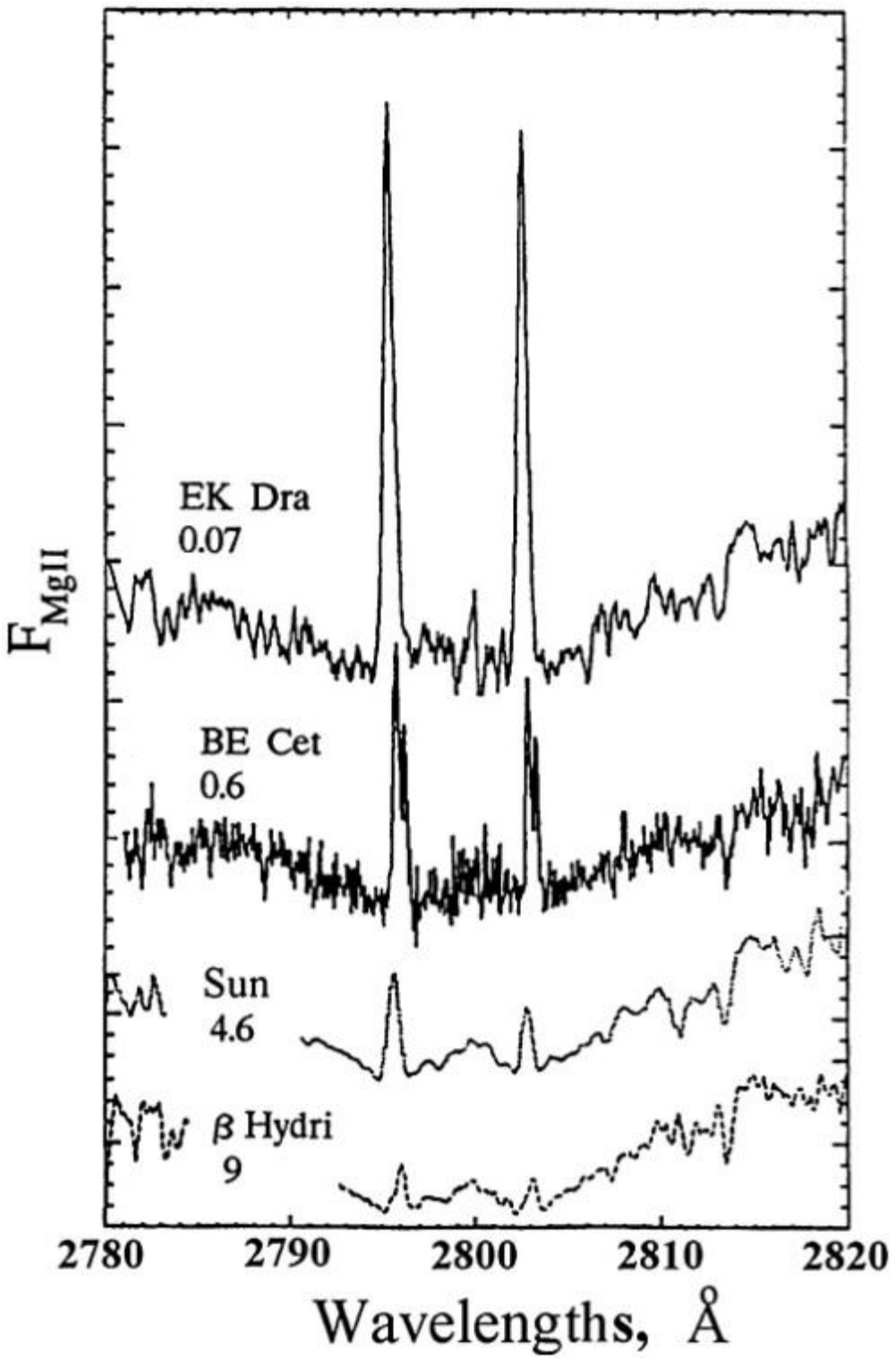

Fig. 91. Four characteristic MgII h and k line profiles in the spectra of G dwarfs of different age; the age in billion years is given below the names of stars (Dorren and Guinan, 1994)

Thus, the young Sun had several times higher *S*, which had considerably greater nonperiodic or periodic variations with a shorter period, and did not experience the Maunder-type minima. One can expect that in the future the current periodicity will be preserved until the Sun leaves the main sequence, because this pattern was observed for stars up to 10 billion years old, but one cannot specify the expected frequency of the Maunder-type minima.

To trace the expected changes of solar activity, Dorren et al. (1994) selected nine single G0–G5 dwarfs with masses of 0.9 to 1.1 solar masses, with known axial rotation periods of 2.7–45 days, and with ages of $7 \cdot 10^7$–$9 \cdot 10^9$ years (Fig. 92). Comparison of X-ray luminosity, luminosity of the CIV lines of the transition zone and chromospheric MgII lines revealed the power dependence of these emissions on the rotation period with exponents of −2.5,−1.6, and −0.76, respectively. In comparing the amplitudes of brightness oscillations in the V band due to stellar spottedness and the rotation periods a decrease of these amplitudes from $0.09^m$ at $P_{rot}$ = 1.5 days to $0.008^m$ at $P_{rot}$ = 14 days was revealed. For greater rotation periods, brightness oscillations were below the detection limit. (Later, Güdel and Gaidos (2001) suspected a similar dependence of $L_R/L_{bol}$ on $P_{rot}$ for G dwarfs.) Comparison of $P_{rot}$ with the age yielded the relation $P_{rot} = 0.21t^{0.57}$, where $P_{rot}$ was in days and age in billions of years. Later, Bochanski et al. (2000) carried out analogous work, expanding the range of considered values up to the radio range and including observations in EUV, FUV-NUV, and Strömgren photometry. The coolest stars of the sample showed rotational modulation with an amplitude from $0.03^m$ to $0.09^m$, whereas the stars with an age of more than 2 billion years — far less than $0.01^m$.

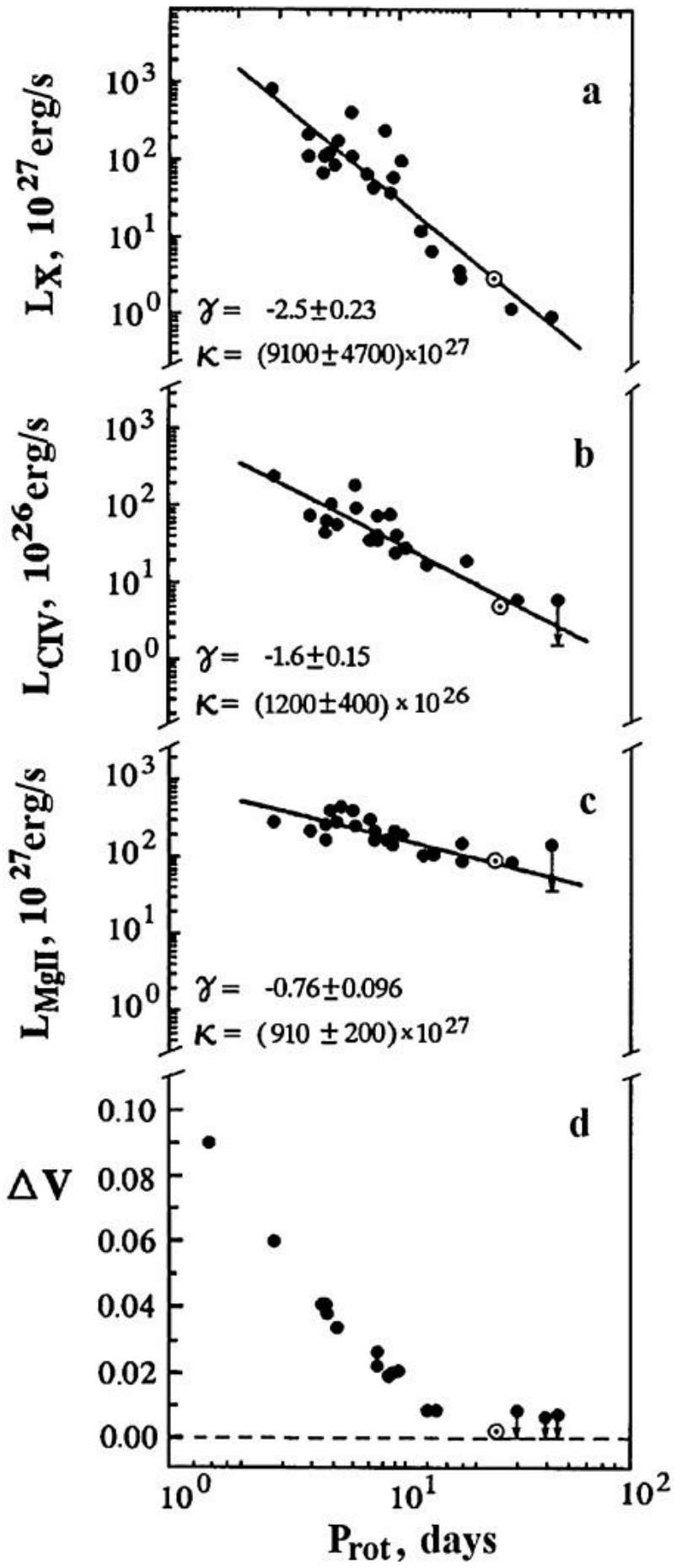

Fig. 92. X-ray luminosity, luminosities of CIV lines in the transition zone and chromospheric MgII lines, and amplitudes of apparent brightness of G dwarfs vs. rotation period; dependences $L(P_{rot})$ were approximated by the power functions of $L = \kappa P^{\gamma}$ type, the values of $k$ and $\gamma$ are presented in each diagram (Dorren et al., 1994)

Güdel et al. (1997b) analyzed ROSAT and ASCA spectra of 11 single G0–G5 dwarfs and the subgiant $\beta$ Hyi within the age range from 70 million to 9 billion years and with $L_X$ varying from 1 to 500 solar X-ray luminosities and relevant data for the Sun. They found that with increasing age the energy distribution changed systematically in the X-ray spectrum: both the temperature and emission measure of the high-temperature component of stellar corona decreased (Fig. 93). Thus, on the young star EK Dra, the temperature of the hot component of the corona reached 20–30 MK, which is equal to the temperature of solar flares. On the whole, the change of this temperature with age can be presented as $T_2 \sim t^{-0.3}$ or $P_{rot}^{-0.55}$, and the dependence of X-ray luminosity on this temperature as $L_X \sim T_2^4$. $EM_2$ also decreases rapidly with age and near 500 million years this corona component becomes insignificant. Radio observations revealed a similar correlation $L_R \sim T_2^4$ and fit into this evolutionary picture. This suggests that the existence of energetic electrons is connected with the hot component of the corona. But saturation of $L_R$ occurs at rotation periods less than two days, i.e., when rotation is faster than the saturation of $L_X$. Developing this scheme, Güdel et al. (1998) described the following evolutionary picture. On young stars, a dynamo generates high surface magnetic activity with large volume filling factor of the corona with magnetic loops. The frequent and strong flares arisen in these conditions spend a considerable part of energy for the acceleration of fast particles and plasma heating in loops up to 20–30 MK, which are recorded as microwave gyrosynchrotron radiation and a high-temperature thermal component in X-rays. On rapidly rotating stars, X-ray coronae are saturated, which may suggest filling of all loops with hot and dense plasma, whose temperature is maintained against cooling by frequent flares. With age, the filling factor of the corona with loops decreases, strong flares become less common, and after 500 million years such flares become isolated events, therefore quasipermanent microwave radiation disappears. It should be noted that similar evolution of flare activity follows from Fig. 24. Later, Argiroffi et al. (2008) put forward an idea that the hot component of coronae is a superposition of unresolved flares, whose number decreases as the star gets older.

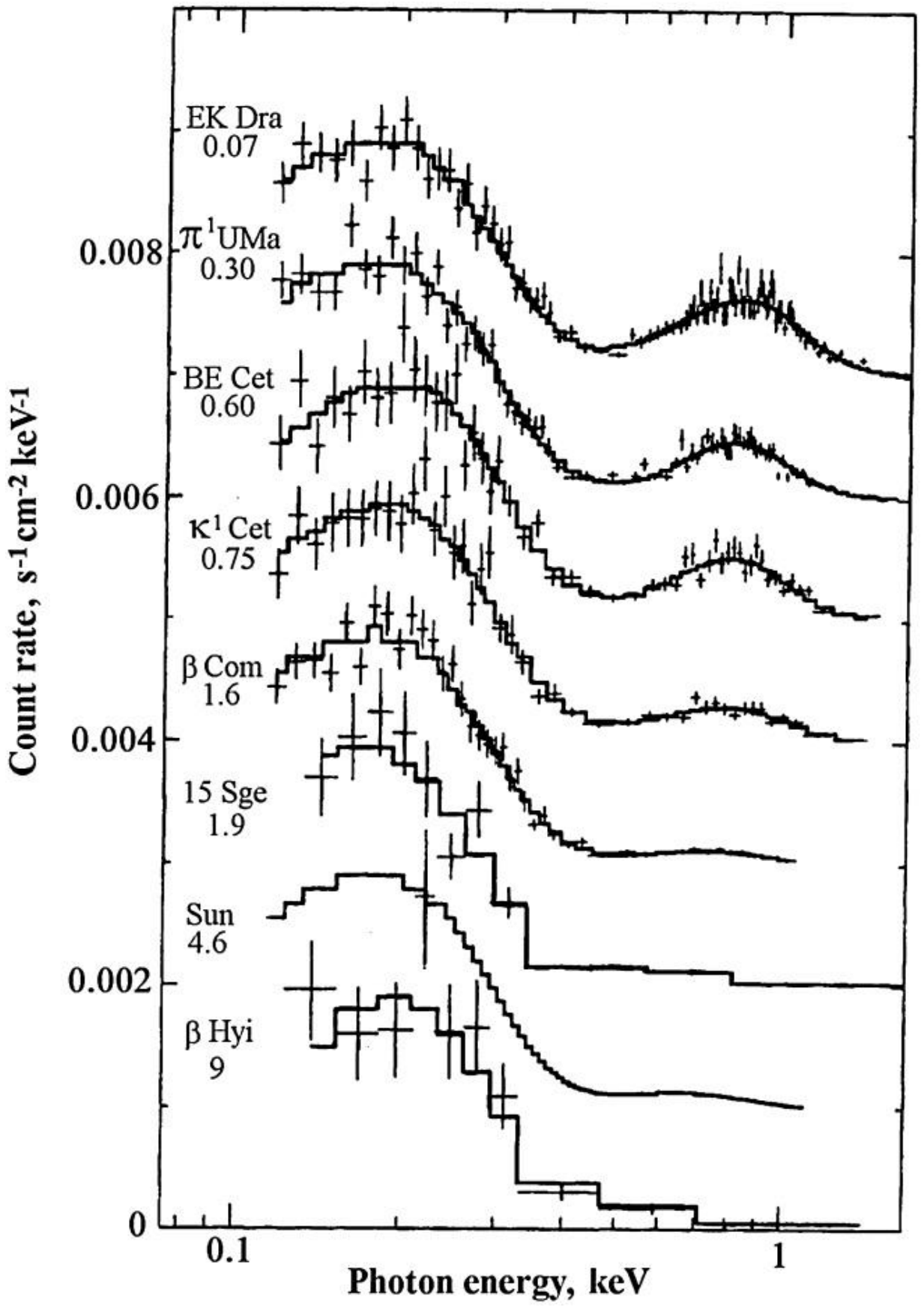

Fig. 93. The ROSAT and ASCA X-ray spectra of G stars of different age; the age is given in billion years (Güdel et al., 1997b)

* * *

Let us consider in detail actual stars to illustrate the evolution of solar activity.

Dorren and Guinan (1994) analyzed the activity of the G0 dwarf HD 129333 (= EK Dra), as an analog of the very young Sun that had passed the T Tau stage and recently achieved the main sequence. The star is close to the Sun and belongs to the Pleiades moving group. It is about 70 million years old, the planet-formation process being completed. The effective temperature of EK Dra is 150 K higher than the solar one, its mass is $\sim 1.05 M_{\odot}$ and radius $\sim 0.92 R_{\odot}$. Photometry of the star showed its spottedness up to 6%, the largest among G stars, and $\Delta T \sim 500$ K. According to DePasquale et al. (2001), the amplitude of stellar rotational modulation is from $0.05^m$ to $0.09^m$, whereas rotation periods are from 2.55 to 2.80 days.

As stated above, Scheible and Guinan (1994) presented the light curve of the star with the help of two spots with radii of 20°, while the Doppler imaging revealed cool spots at the equator, at middle latitudes and at the pole, though the polar spot was seen in only 9 of 12 spectral lines (Strassmeier and Rice, 1998). The emission of $H_\alpha$ is maximum in the hemisphere with the greatest spottedness. The luminosity of EK Dra in soft X-rays exceeds the solar luminosity by a factor of 300, in the ultraviolet lines of the transition zone by 20–100 times, in ultraviolet chromospheric lines by 3–20 times, CaII and MgII emission by 3–5 times, which is the strongest among G dwarfs that do not belong to binary systems. At $v\sin i \sim 18-20$ km/s the rotation period is 2.7–2.8 days, the shortest among G stars, it modulates the calcium emission with an amplitude of 5%. The activity level of EK Dra satisfies the Skumanich relation, though it is already close to saturation. As mentioned above, Saar and Bookbinder (1998a) found that low-amplitude flares contributed up to 8% to the flux in CIV and SiIV lines. X-ray emission of the star was presented by the two-temperature coronal model with $T_1 = 1.3 \cdot 10^6$ and $T_2 = 9.6 \cdot 10^6$ K, but later Güdel et al. (1997b) estimated the temperature of the hot component as 20–30 MK. Radio emission at 3.6 cm was 3000 times higher than the appropriate solar value. According to Dorren and Guinan (1994) and Dorren et al. (1995), variations of average brightness and fluxes in ultraviolet lines suggest an activity cycle of about 12 years. But, unlike the old Sun, the young EK Dra weakens at the activity maximum phase and brightens at its minimum. The amplitude of changes of CIV lines in the spectrum of EK Dra during the cycle is much higher than that of CII and MgII lines. Later studies of this star refined its characteristics: a mass of $0.9 \pm 0.1 M_{\odot}$, it has a satellite with a mass of $0.5 \pm 0.1 M_{\odot}$ with a very eccentric orbit ($e = 0.82$), the effective temperature of EK Dra is 5700 K, $\log g = 4.37$, [Fe/H] = −0.16, and the rotation period is 2.767 days (Koenig et al., 2005).

* * *

The X-ray luminosity of coronae of active stars exceeds by up to four orders of magnitude the luminosity of the solar corona, and this calls into question the correctness of the initial scaling of these structures. Favata (2001) performed a comprehensive comparison of the structure of solar and stellar coronae and came to the conclusion that, contrary to the solar corona where long and close to the equator formations prevail and fine filaments are located at low and middle latitudes, for the coronae of active stars there prevail structures which are far less than the stars located at high and circumpolar latitudes with temperatures of a few and a few tens of MK; these are denser structures of the flare plasma, which are constantly absent on the Sun.

* * *

Güdel et al. (1995a) investigated EK Dra in the microwave and X-ray ranges. In both ranges the star was brighter than single flare M dwarfs and G dwarfs in the Pleiades. For the brightness temperature of microwave radiation of about $10^8$ K, the most suitable mechanism of this radiation is optically thin gyrosynchrotron emission of relativistic electrons. According to RASS data, $L_X = 8 \cdot 10^{29}$ erg/s, in the two-temperature coronal model its components had temperatures of 1.9 and 10 MK and emission measures of $1.2 \cdot 10^{52}$ and $2.5 \cdot 10^{52}$ cm$^{-3}$, and the cooler component displayed much greater rotational modulation than the hotter one. Güdel and Schmitt (1996) from the modulation of X-ray radiation of EK Dra estimated an electron density of stellar corona as $3 \cdot 10^{10}$ cm$^{-3}$. From this modulation, the electron density of the cooler component was estimated as $4 \cdot 10^{10}$ cm$^{-3}$ and its size as not more than 0.1–0.2$R$*. Thus, high X-ray luminosity of the star is caused by high density rather than by a great volume of luminous matter, and for retaining such structures a magnetic field of about 240 G is required.

Using four ASCA instruments and two EUVE (SW and MW) spectra of EK Dra, Güdel et al. (1997a) analyzed the structure of DEM in the range of temperatures from 0.1 up to 100 MK. Five different analysis methods led to the coordinated conclusion that DEM had broad maxima in the ranges of 5–8 and 15–40 MK, a narrow deep minimum at about 8–20 MK, and a small amount of plasma at 1–3 MK. This coldest part of the EK Dra corona corresponds to the solar corona, though its EM is higher by an order of magnitude than the solar value. Apparently, the general view of DEM is determined by the structure of the cooling function of the optically thin plasma with regard to its thermodynamic stability, and the hottest corona component arises in flares.

Senavci et al. (2021) carried out the extensive studies of EK Dra in the optical wavelength range. They determined the abundance of 23 elements with [Fe/H] = 0.03 and substantial excesses of lithium and barium; elements of the s-process Sr, Y, and Ce showed a little excess, whereas Ni, Cu, and Zn displayed a little deficiency with respect to solar abundances. The excess of barium was probably due to the microturbulent velocity independent of the depth, the excess of lithium due to the youth of the object. The authors estimated the stellar mass as $1.04M_\odot$ and age as 27 million years, which corresponds to the pre-main sequence post T Tau phase. The Doppler imaging covering 15 days led to the conclusion on the existence of the circumpolar spot and spots at middle latitudes in their absence at low latitudes.

Messina et al. (1999b) compared the young Sun and the single G5 dwarf HD 134319, a probable member of the Hyades moving group with strong chromospheric emission. According to their photometric observations in 1991–95, the star has an axial rotational period of 4.45 days and stable spottedness with spots uniformly distributed along the equator plus two active longitudes of opposite longitude with a total area of dark spots from 7 up to 24%, depending on the accepted inclination angle of the rotational axis of the star and the algorithm selected for the inverse photometric problem.

Marino et al. (2003b) observed the young fast rotator VXR45 in IC 2391 with XMM-Newton/pnEPIC. This G9 dwarf has v$\sin i$ > 200 km/s and $P_{\text{rot}}$ = 0.223 days, i.e., it is a supersaturated star and its X-ray radiation is slightly lower than the maximum observed at rotation rates of several tens of kilometers per second. From the X-ray light curve they confidently established rotational modulation during two photometric periods. As opposed to the case of EK Dra observed by Güdel et al. (1995a), this modulation did not impact the hardness of radiation. From the spectrum within 0.3–7.8 keV they constructed the two-component coronal model: $T_1 = 7$, $T_2 = 14$ MK, and $EM_1/EM_2 = 1.4$ at an FIP effect of about 0.27. The revealed rotational modulation evidences structural inhomogeneities of the corona and disproves the hypothesis on supersaturated stars as the objects completely covered by active regions. Jardine (2004) proposed the model in which rotational modulation was directly related to the supersaturation phenomenon.

Another experimental support to reproduce the history of solar activity is the G1 star 15 Sge with an axial rotation period of 13.9 days and age of about 2 billion years. The Sun in the current state of this star was at the epoch when the primitive life had already existed on the Earth and Mars was warm, wet, and liveable (Dulude et al., 2004).

With the aim of ascertaining the role of acoustic and magnetic heating mechanisms for outer atmospheres of solar-type stars Judge et al. (2004) undertook a comparison of solar spectra near the activity minimum, the HST/STIS spectra of $\alpha$ Cen A and absolutely inactive star $\tau$ Cet; in the latter case, the rotational modulation sometimes occurred, but never long-term cyclic variations did, and possibly the star is at the Maunder minimum phase. Comparison of spectra showed that the line profiles of $\tau$ Cet systematically differ from the spectra of two other stars: the flux densities in the lines from the upper chromosphere and up to the middle transition zone are approximately twice less, lines are significantly narrower, the soft X-rays from this star are several times lower than those from the Sun.

According to the estimate of Ribas et al. (2005), X-ray and EUV emission of the young Sun on the main sequence is 100–1000 times higher than the present-day level, whereas emission of the transition zone and the chromospheric ultraviolet are 20–60 times and 10–20 times higher, respectively. Two and a half billion years ago radiation in the range 1–1200 Å exceeded by a factor of 2.5 the present-day level, and 3.5 billion years ago — by a factor of 6.

From the high-dispersion XMM-Newton observations Telleschi et al. (2005) studied the long-term evolution of X-ray coronae on six nearby G stars in the age range from $\sim$ 0.1 to 1.6 billion years with rotation periods from $\sim$ 1 to 12.4 days. They found that the inverse FIP effect for the stars with an age of about 0.1 billion years and transition to the solar FIP effect for the objects with an age of about 0.3 billion years, which coincides with a fast weakening of nonthermal radio emission. Qualitatively, this coincides with a simple model in which the flow of electrons in magnetic fields suppresses the diffusion of slight FIP ions from the chromosphere to the corona. The EM distribution reveals a power character on both sides of the maximum, which is at 10 MK for the fastest and young rotators and shifts to 4 MK for the oldest of the stars considered here. The spectral index of the cool side of EM distribution accounted for 1.5–3, which corresponds to coronal emission as a superposition of stochastic flares with their power energy distribution. From the spectral index of this distribution the required flare energy range was estimated and for more active stars this range was found to be wider than for less active. On the whole, flaring activity plays a larger role for more active stars. Thus, in the proposed model, the high frequency of flares is responsible for both higher temperature and higher X-ray luminosity of the corona.

Vardavas (2005) studied a dependence of the Rossby number and radiation fluxes in X-rays, in the far ultraviolet (1–1200 Å), and in Ly$_\alpha$ on the age of G stars in the range from 0.1 to 8.5 billion years and found that the Rossby number grew with time as $t^{0.5}$, rotation period as $t^{0.6}$, the flux in the far ultraviolet decreased as $t^{-1.25}$ and in Ly$\alpha$ as $t^{-0.78}$.

According to Gondoin et al. (2012), the high level of chromospheric activity, the rotation period 1.625 days, the absence of differential rotation, and the strong lithium line prove the youth of the solar-type star CoRoT 102899501. Throughout 35 days its filling factor varied from 5–14% to 13–29%. Such a level of magnetic activity was on the Sun at the epoch of formation of planets.

From the archive data derived with spectropolarimeters NARVAL, ESPaDOnS, and HARPSPol Rosen et al. (2016) performed a study of six solar-type stars in the age range from 100 to 650 million years. They detected a significant decrease of the magnetic field strength and its energy in the range of 100–250 million years, whereas there were no such variations on the interval of 250–650 million years. The meridional field component is less than radial and azimuthal in 15 maps out of 16, and 89–97% of magnetic field energy is maintained in components with $l$ = 1–3. In the sample, there is no clear dependence of the ratios of poloidal to toroidal and axisymmetric to nonaxisymmetric components on age. The numerical results of Rosen et al. are listed in Table 24.

Table 24. Magnetic fields of the solar-type stars of different age ($\langle B \rangle$ — average field strength, $\langle B_r \rangle$ , $\langle B_m \rangle$, and $\langle B_a \rangle$ — local strengths of radial, meridional, and azimuthal field components)

| Name of a star | HD number | Epoch of observations | $T_{eff}$, K | vsin*i*, km·s$^{-1}$ | $P_{rot}$, days | Cluster member | Age, million years | $\langle B \rangle$, G | $\langle B_r \rangle$, G | $\langle B_m \rangle$,G | $\langle B_a \rangle$, G |
|---|---|---|---|---|---|---|---|---|---|---|---|
| EK Dra | 129 333 | 2007.1 | 5845 | 16.8 | 2.6 | Pleiades | 100 | 66 | 15 | 12 | 61 |
| EK Dra | 129 333 | 2012.1 | 5845 | 16.8 | 2.6 | Pleiades | 100 | 89 | 29 | 22 | 74 |
| HN Peg | 206 860 | 2007.6 | 5974 | 10.6 | 4.6 | Hercules-Lyra | 250 | 22 | 14 | 5 | 14 |
| HN Peg | 206 860 | 2008.6 | 5974 | 10.6 | 4.6 | Hercules-Lyra | 250 | 13 | 6 | 3 | 9 |
| HN Peg | 206 860 | 2009.5 | 5974 | 10.6 | 4.6 | Hercules-Lyra | 250 | 15 | 10 | 7 | 5 |
| HN Peg | 206 860 | 2010.5 | 5974 | 10.6 | 4.6 | Hercules-Lyra | 250 | 20 | 14 | 5 | 8 |
| HN Peg | 206 860 | 2011.5 | 5974 | 10.6 | 4.6 | Hercules-Lyra | 250 | 25 | 14 | 8 | 15 |
| HN Peg | 206 860 | 2013.7 | 5974 | 10.6 | 4.6 | Hercules-Lyra | 250 | 25 | 15 | 5 | 17 |
| $\pi^1$ UMa | 72 905 | 2007.1 | 5873 | 11.2 | 4.9 | Ursa major | 300 | 24 | 9 | 4 | 21 |
| $\chi^1$ Ori | 39 587 | 2007.1 | 5882 | 9.8 | 5.08 | Ursa major | 300 | 15 | 7 | 5 | 10 |
| $\chi^1$ Ori | 39 587 | 2008.1 | 5882 | 9.8 | 5.08 | Ursa major | 300 | 13 | 5 | 2 | 11 |
| $\chi^1$ Ori | 39 587 | 2010.8 | 5882 | 9.8 | 5.08 | Ursa major | 300 | 20 | 6 | 5 | 17 |
| $\chi^1$ Ori | 39 587 | 2011.9 | 5882 | 9.8 | 5.08 | Ursa major | 300 | 16 | 6 | 4 | 13 |
| BE Cet | 1835 | 2013.7 | 5837 | 7.0 | 7.65 | Hyades | 600 | 19 | 11 | 6 | 11 |
| $\kappa^1$ Cet | 20 630 | 2012.8 | 5742 | 5.2 | 9.2 | – | 650 | 26 | 9 | 7 | 21 |
| $\kappa^1$ Cet | 20 630 | 2013.7 | 5742 | 5.2 | 9.2 | – | 650 | 21 | 7 | 5 | 17 |

From the data of spectral (Robinson and Bopp, 1987) and magnetometric (Rosen et al., 2016) observations of the young G5 dwarf $\kappa^1$ Cet with an axial rotation period of 9.4 days, Lynch et al. (2019) constructed a numerical three-dimensional MHD model which showed that the global shear concentrated near the polarity inversion line of the radial magnetic field can supply the magnetic field of the stellar corona with energy sufficient for CME and superflares with a strength of the Carrington flare of 1869 on the Sun and a flare on Prox Cen recently visible with the naked eye (Howard et al., 2018). Within the framework of this model, Lynch et al. represented a superflare on $\kappa^1$ Cet and CME recorded by Rosen et al.: the model shows a gradual self-consistent accumulation of free magnetic energy on the interval of 150 hours from the measured background magnetic field, and its fast release into CME over 10 hours and maximizes both the value of the free field energy and the spatial scale of CME, whereas the CME energetics amounts to 1/3 of the accumulated free field energy.

* * *

The early stages of solar activity are successfully analyzed by the example of numerous solar-type stars with a high activity level and confidently determined age. However, studying the future of solar activity is much more difficult because the activity level is much lower and old stars of known age are, as a rule, in remote globular clusters. To get an idea of the far future of solar activity, i.e., to determine its absolute minimum, Dravins et al. (1993a–c) studied the single G2 IV star $\beta$ Hyi, which recently left the main sequence. This nearest subgiant is at a distance of 6.5 pc from the Sun, its age determined from the position on the evolutionary track is 9.5 billion years, i.e., the star is twice as old as the Sun and brighter by $1^m$ than the Sun by absolute value. With the help of high-resolution spectroscopy and three-dimensional hydrodynamic calculations the parameters of the $\beta$ Hyi photosphere were determined: the temperature differs from the solar temperature by no more than 100 K, the acceleration of gravity is four times lower than on the Sun, and the stellar radius is $1.6R_\odot$. The photosphere contains several times larger and more contrast granules, the velocities of matter are higher by a factor of 1.5–2. Lower gas pressure can admit approximately twice weaker local magnetic fields. At $v\sin i = 2 \pm 1$ km/s the axial rotation period is about 45 days. A general small deficiency of metals [Fe/H] $\sim -0.2$, corresponding to old stars, the lithium abundance is higher by an order of magnitude than the solar abundance, which can be associated with deepening of the convective zone when a star leaves the main sequence. This mechanism for lithium enrichment of the atmospheres of subgiants has not been completely clarified, but the excess of this element observed on other subgiants does not raise doubts about the considerable age of $\beta$ Hyi.

Calcium emission in the spectrum of $\beta$ Hyi is half that in the solar spectrum, but it displays the same profile asymmetry. About a hundred IUE high-dispersion spectrograms of MgII emission obtained over 12 years show gradual and regular changes of emission without signs of fast appearance and growth of individual active regions. It is such a character of calcium emission that is determined in low-activity stars within the HK project. The recorded minimum and maximum of magnesium emission allows one to estimate the activity cycle of $\beta$ Hyi as 15–18 years. The measured amplitude of MgII emission of about 30% corresponds to variations of the Mount–Wilson index $S$ of 1%, which is the lowest recorded amplitude of the stellar activity cycle. However, the measured level of chromospheric activity and established cyclicity evidence that the observed chromosphere does not comply with the basal level but is due to the magnetic activity of the star.

The surface fluxes of $\beta$ Hyi measured with IUE in ultraviolet high-excitation lines are comparable with the appropriate solar values in the minimum phase, but different algorithms of data processing yielded essentially different results.

Observations in soft X-rays with different EXOSAT filters revealed that the flux and spectrum of $\beta$ Hyi radiation differed from those of the Sun. The analysis of similar solar observations within the isothermal coronal model yields $\log T = 6.5$ and logEM = 49.6, but the data on $\beta$ Hyi provided two solutions: $\log T = 5.7$ with logEM = 49.1 and $\log T = 6.6$ with logEM = 49.6. The "hot" model is rather close to the solar corona, but the "cold" variant falls into another steady thermodynamic minimum of optically thin high-temperature plasma. The temperature of "cold" $\beta$ Hyi corona approaches or has already reached the critical level, below which a stellar wind similar to the solar one cannot exist and the termination of such an outflow should lead to the termination of magnetic braking of stellar rotation.

# Part 4

# Magnetism of Stars with Solar-Type Activity

Semi-centennial intensive studies of UV Cet-type flare stars coincided with the period of formation and fast development of multiwavelength astrophysics. By now, observational data on the activity of such stars were accumulated practically in the whole range of electromagnetic radiation – from X-rays to decameter radio waves. As a result, the identity of the physical nature of the activity of such stars and of the Sun has been ascertained. It has been revealed that such activity is inherent in all lower main-sequence stars at a certain — frequently rather long — stage of their development, that the energy of this activity is eventually derived from stellar rotation, and the "driving belts" are hydrodynamic and magnetohydrodynamic processes.

In the previous chapters of the book, we show that for the considered activity, for nonstationary phenomena with characteristic times from milliseconds to decades, the crucial parameter is stellar rotation, and it has practically no influence on its position in the Hertzsprung–Russell diagram. Though the rotation of upper main-sequence stars is maximum, there are no manifestations of the discussed activity on them. Hence, there should be an important intermediate link including this energy source in the chain of considered activity. This is subphotospheric convection. Indeed, this activity occurs on stars with a mass of less than 1.5 solar masses, at this threshold stars acquire a convective envelope. How abrupt is the threshold at which the activity is triggered? The full range of masses of the main sequence stars is more than three orders of magnitude, but cardinal changes occur on the interval from two to one solar masses: stars of two solar masses have a characteristic rotational velocity of 120 km/s and display no magnetic activity, whereas the rotational velocity of the Sun is 2 km/s and it is characterized by intense magnetic activity. The decisive role of rotation is observed both for young single mid- and low-mass stars and for similar components of close binary systems of different age in which high rotational velocity is maintained for long periods due to the energy of their orbital motion.

The experimentally established close connection of sunspots and faculae, chromospheric network and flocculi with local magnetic fields, development of transient flares in the regions of complicated magnetic configuration, strong coronal radiation in the region of closed magnetic loops and coronal holes in the regions of open magnetic structures — all suggests the magnetic nature of the solar activity. According to Obridko and Nagovitsyn (2017), "solar activity is variations of solar magnetic fields on different spatial and temporal scales". The fundamental role of magnetic fields is stipulated by the fact that their nonpotential components can contain considerable energy reserves, which can rapidly convert into other forms accessible to direct observations, and that hydromagnetic waves governing the atmospheric energetics and its nonthermal heating are excited in magnetic flux tubes and efficiently transferred along them. The loop configuration of magnetic fields determines the geometry of plasma structures in chromospheres and coronae and the time character of the development of flares in them. During the life of a single main-sequence star the loss of angular momentum due to magnetic braking noticeably slows its rotation and lowers the level of magnetic activity.

Therefore, the exact definition formulated by Parker 20 years ago "solar activity is first of all the result of displacement of magnetic flux tubes by convective gas motions under the solar surface and the subsequent shift of rarefied gas by magnetic-field discontinuities and instabilities above the solar surface" is completely valid for stars with the solar-type activity considered in this monograph.

In the previous chapters, we outline in detail the observational facts that have been collected so far and related to such an activity. These facts primarily differ by their diversity; this is natural if one considers that the phenomena of solar-type magnetic activity were recorded on stars in the mass range of at least 1.2 to 0.06 solar masses with surface temperatures of 6000 to 2500 K, with axial rotation periods of 40 days to 4 hours, and with age of millions to ten billion years, whereas the quantitative characteristics of the phenomena themselves — the degree of surface spottedness, energetics of permanent chromospheres and coronae and sporadic flares — lie within the range of 2 to 7–8 orders of magnitude. Tens of monographs, extensive reviews, and many hundreds of individual publications are devoted to the theoretical studies of magnetism of all this stellar diversity. However, their detailed discussion is beyond the scope of this book.

In the final part of the monograph, following Parker, we first discuss two basic groups of problems of the considered stellar activity associated with their subphotospheric and atmospheric magnetic fields. The first of them basically determine different temporal characteristics of stellar activity, the second are associated with its energetics since the real energy release attributed to magnetic activity is observed in stellar atmospheres, where the dissipation in active phenomena occurs from the photosphere to the corona. While discussing these problems, it is inevitable to discuss additionally a series of previously outlined observational data. In the concluding chapter, we briefly outline the theoretical models of the most developed envelope stellar dynamo.

## 4.1. Photospheric and Subphotospheric Magnetic Fields

As stated above, studies of the 11-year solar cycle gave rise to observational and theoretical research of the solar dynamo, i.e., the interaction of convection and rotation resulting in the emergence of magnetic structures in the convective envelope and atmosphere, and stimulated a study of local magnetic structures on the stellar surface. However, a variety of phenomena found on stars was incomparably vaster than on the Sun. To interpret them, various dynamo models — models of generation of stellar magnetic fields — were advanced: the kinetic or linear model, which specifies the velocity field and disregards the back coupling of magnetic field and plasma motion; the hydrodynamic or nonlinear model that takes into account the back coupling and solves the complete system of magnetohydrodynamic equations; the cyclic envelope model of solar type in which the interface between the convective zone and radiative core is essential; and the model of a distributed or turbulent dynamo, where boundary conditions are insignificant. (Since in the envelope dynamo the maximum rotation gradient and, consequently, the field generation maximum occur on the boundary between the convective envelope and the radiative core, and this interface layer is called a tachocline, then sometimes such a dynamo is called an interface dynamo.) In the process of accumulation of data on stellar activity cycles researchers tried to give preference to one or another model of the stellar dynamo or to correlate them with various structures of stellar interiors and/or evolution phases. But solutions of magnetohydrodynamic equations in different approximations sometimes resulted in opposite conclusions. Thus, Belvedere et al. (1982) constructed a dynamo theory in which the duration of activity cycles should grow from F5 to M0 stars, whereas the theory of Robinson and Durney (1982) decreased it from G0 to M5 stars. Since the first work of Parker (1955a) on the mechanism of the solar dynamo the spiral character of plasma motion required for the magnetic field generation, the so-called $\alpha$-effect, has been attributed to convection, but Brandenburg (1998) put forward an alternative idea that this effect could be due to magnetic buoyancy and instability, whereas Getling (2001) completely rejected the concept of emerging magnetic tubes and explained the magnetic field generation only by convective motions. Discussing the reasons for a jump in the activity cycle period $P_{\rm cyc}$ for the stars with an age of 2–3 billion years, Brandenburg et al. (1998) suggested that it could be associated with rearrangement of the internal magnetic field — with excitement of high dynamo modes or a change of the dominating instability factor.

Considering the nonlinear dynamo of a rotating spherical envelope with three-dimensional convection to the Sun, whose full and differential rotation and surface convection parameters are well known, Gilman (1983) could not reproduce the observed duration of the solar cycle and observed migrating toroidal fields, but showed that in the nonlinear $\alpha - \omega$ dynamo the cyclic behavior was possible in the limited ranges of medium electroconductivity and rotational velocities: at very low velocities the $\alpha$-effect was very weak, at very high velocities the rotation suppressed convection. In the nonlinear dynamo, the induced magnetic field should suppress the differential rotation, and this feedback is strong enough: when a certain critical level is exceeded by a factor of 3–4, magnetic cycles should disappear. Solov'ev and Kirichek (2004) proposed a diffusion theory of the solar magnetic cycle based on precise solutions of the equation of the magnetic field diffusion in a spherical layer. Baliunas et al. (2006) put forward a hypothesis of the standing dynamo waves, which operate in a regime that is substantially supercritical, when the alpha effect is strongly suppressed over a significant part of the cycle. Cyclicity also takes place in the model of vortex dynamo that generates a magnetic field on the bottom of the convective envelope (Tlatov, 2013).

Olemskoy and Kitchatinov (2010) represented a nonlinear dynamo model that takes into account the dependence of turbulent diffusion on the magnetic field. Within a certain range of values of the dynamo number two solutions can be possible: decaying oscillations of weak fields and magnetic cycles with constant and large amplitude. Fluctuations of the $\alpha$ parameter characterizing the alpha effect value (see Chapter 4.2) cause the transitions between these regimes, and calculations show their intermittency. Such a behavior may serve as a model of the Maunder-type minimum events.

Moss et al. (2011) considered a dynamo wave directed to the pole and concluded that one can construct very different models of these waves, depending on the parameters of the convective zone. In particular, the sign of $\alpha(d\Omega/dr)$ to a large extent determines the direction of dynamo wave distribution, and for the wave motion to the pole the rotation inside a star should differ significantly from the solar one. They noted that this is a preliminary conclusion, and additional 2D calculations are required.

Analysis of observational data on the distribution and statistical properties of magnetic structures on the solar surface shows a high degree of multifractality of magnetic and plasma structures both in space and time (Abramenko et al., 2008), both for zones of weak fields (Abramenko et al., 2012) and for zones of strong spots (Abramenko and Yurchyshyn, 2010b; Abramenko, 2013). The level of flaring activity is to a large extent determined by the multifractal and turbulent state of the photospheric plasma (Abramenko and Yurchyshyn, 2010a,b). The presence of multifractality allows one to suppose that the magnetic field generation may be explained by the action of a unified dynamo mechanism operating — as a nonlinear dynamic dissipative system — simultaneously on all the scales: from the solar radius to the sizes of the smallest magnetic elements. In this context, the mentioned approaches to the nonlinear dynamo modeling seem to be a promising direction.

In studying stellar magnetism, the features of axial rotation, in particular surface differential rotation, were considered together with cyclic activity. Donahue et al. (1996) analyzed the rotation periods of 36 stars of the Wilson program on which the rotation periods were confidently determined using the algorithm of Horne and Baliunas (1986) at least over five seasons. They found the following correlation between the average rotation periods of stars $\langle P_{\rm rot}\rangle$ and their complete ranges $\Delta P$:

$$\Delta P \sim \langle P_{\rm rot}\rangle^{1.3\pm0.1}, \tag{73}$$

The correlation coefficient was 0.90. It matched the solar situation well. The nonlinear relation (73) contradicted the assumption on proportionality of the radial gradient of angular rotation velocity to the angular velocity used in theoretical mean-field dynamo models. It did not match the $\alpha - \omega$ dynamo model in which surface differential rotation should grow together with angular rotation velocities, whereas, according to (73), it should grow with increasing $\langle P_{\rm rot}\rangle$, i.e., with stellar age.

Since the sequence of Doppler maps defining the positions of spots on a star makes it possible to estimate stellar differential rotation, Lanza (2007) carried out the long-term investigations of two fast rotators, AB Dor and LQ Hya, and detected variability of their equatorial angular velocities and amplitudes of surface differential rotation. Within the mean-field theory, he associated this variability with azimuthal magnetic field strength in the convective zone. Summing the results of studying the surface differential rotation on F, G, K, and M stars, Collier Cameron (2007) noted its increase with growing effective temperature along the main sequence, while the tidal forces in binary systems seem to suppress it.

Relation (68) with specification of Ossendrijver within the framework of the linear mean-field dynamo corresponds to a decrease in differential rotation and a fast growth of the $\alpha$ effect with increasing stellar rotation rate.

Based on the long-term series of high-precision photometric observations of six young solar-type stars, Messina and Guinan (2003) found that all the considered stars show variations in the rotation period duration, i.e., they have differential rotation. These variations are periodic in phase with the cycle of spottedness for BE Cet and DX Leo; the analogous situation seems to be valid for $\pi^1$ UMa, EK Dra, and HN Peg; for BE Cet, $\pi^1$ UMa, and EK Dra the rotation period decreases as the cycle develops, then by a jump it returns to the initial value at the beginning of the next cycle, but DX Leo, $\kappa^1$ Cet, and HN Peg, as well as LQ Hya, show an inverse change of the period duration. The amplitude of variations in the rotation period shows a dependence that is close to (73). The cycle duration does not correlate with the dynamo number but reveals a positive correlation with differential rotation amplitude.

* * *

Donati et al. (2006) considered critically the existed then dynamo models of fully convective stars and concluded on their insufficiency. In one models, it is stated that in the very low-mass stars the dynamo mechanism generates a nonaxisymmetric and noncyclic oscillating field with negligible differential rotation, whereas other models predict that such stars can have significant differential rotation and mainly the axisymmetric oscillating field. With the aim of disentangling this confrontation of models, throughout three nights with the high-resolution spectropolarimeter CHFT Donati et al. (2006) derived 64 circularly polarized spectra of fully convective rapidly rotating M4 dwarf V374 Peg and simultaneously analyzed 5000 spectral lines. The Stokes I spectra provided information on cool spots, the Stokes V spectra — on the magnetic field. Using the maximum entropy technique and the resolution 10°, they determined potential fields and found mainly radial fields with a strength of up to 2 kG, the azimuthal and meridional fields that were 4 times weaker. Comparison of data over different nights led to the negligible estimate of differential rotation. Thus, such observations contradict both groups of the mentioned field models in fully convective stars. But following later 3D MHD calculations of Browning (2008), the fully convective stars can generate the magnetic fields with kilogauss strength even without a tachocline. The obtained model was all over unstably stratified, the amplitudes and character of convective cells varied strongly with radius, but in depth the situation was quieter. Despite the absence of strong differential rotation, the significant mean axisymmetric filed was present, which was partially attributed to the strong influence of rotation on slow convective flows. Wright and Drake (2016) evidenced in favor of the fact that the tachocline is not a decisive factor in the process of stellar magnetic field generation: for four fully convective stars they found a correlation of X-ray emission with rotation periods, similarly to solar-like stars, and concluded that in such stars the dynamo operates over the whole convective zone.

To interpret the coexistence of two different types of large-scale magnetic fields on the low-mass fully convective stars with similar masses and rotation rates but generating drastically different types of magnetic fields, Morin et al. (2011a, 2011b) put forward an idea of bistability “weak/strong dynamo field”: in their opinion, the point is either in other global parameters (for instance, age) or in switching between two states in this mass range. Dynamo bistability may lead to the hysteresis, thus the magnetic properties of a star may depend not only on its current parameters but on its backstory — strong accretion in younger years and the initial rotation period. Gastine et al. (2013) found that at high values of the Rossby number the multipole magnetic fields are detected, whereas at Ro < 0.1 there is a bistability region in which the large-scale dipole and multipole fields can coexist.

Reiners and Basri (2009) compared the measured magnetic fluxes in Stokes V and Stokes I spectra in a sample of four partially convective and two fully convective stars, supposing this boundary on the stars with a mass of $0.35M_{\odot}$. The Stokes V measurements are sensitive mainly to the large-scale fields, since the signals of opposite polarities are mutually destroyed, whereas the variations in Stokes I are sensitive to the whole average flux regardless of polarity and topology. As a result, they concluded that in M dwarfs more than 85% (96%) of the magnetic flux (energy) containing in the magnetic field were not seen in Stokes V; the fraction of the magnetic flux detected in Stokes V shows a significant jump on the boundary with fully convective stars, and in such convective stars the large-scale fields, visible in Stokes V, comprise a 2–3 times higher magnetic flux (5 times higher magnetic energy) than that in partially convective stars.

Brown (2012) performed the calculations of stellar convection and dynamo, where he first derived globally organized magnetic fields with cycles of polarity reversal. In particular, he carried out a 3D MHD calculation of rapidly rotating solar-type stars and found cyclic dynamo. These dynamos in the form of lighting chains fill the stellar convective zone even in the case when the stably stratified radiation zone is involved in the model. This means that the globally organized fields can exist in the stars with very deep convective zones.

Marsden et al. (2014) spectropolarimetrically considered 167 solar-type stars for the presence of the magnetic field, and it was detected for 67 stars older than 2 billion years. They found that the maximum absolute value of the longitudinal field grew with rotation rate, decreased with age, and correlated well with traditional activity indicators.

Using 104 maps of surface magnetic fields for 73 stars with masses of 0.1 to 2 solar mass and age of 1 million to 10 billion years, a big team of researchers considered how the observed large-scale magnetic fields of stars reconstructed by the ZDI technology varied with age, rotation period, Rossby number, and X-ray emission (Vidotto et al., 2014b). They found that the large-scale field that is mean in absolute magnitude varies with age as $t^{-0.655\pm0.045}$, which is close to the relation of Skumanich (1972), which laid the basis of gyrochronology and may be treated as a basis of the alternative method of magnetochronology. They revealed that this large-scale field decreases as $P_{\rm rot}^{-1.32}$ and $\mathrm{Ro}^{-1.38}$, supporting the notions on the operation of linear dynamo. The relations found by the ZDI method for the large-scale stellar fields are in agreement with the results derived during the analysis of Zeeman line broadening sensitive to the absolute values of large- and small-scale fields. This agreement of two technologies suggests that there is a relationship between processes of field generation of both scales.

Recently, Shulyak et al. (2017) studied the rapidly rotating fully convective M stars WX UMa, Wolf 47, UV Cet, and V374 Peg and found that if in the stars with a multipole magnetic field structure its strength was not higher than 4 kG, then in the stars with dipole fields it exceeded this saturation limit and achieved 7 kG. They associated this result with dynamo bistability, which was observed for stars with a mass that is equal or less than 0.2 solar mass. The magnetic fields of these different structures may contain different total magnetic energy. An alternative of the concept of dynamo bistability is the hypothesis of the magnetism cyclicity of such stars, whose checking requires long-term magnetometric observations.

Figure 94 displays the results of studying stellar magnetism derived over several years with two mentioned special spectrographs and collected by Shulyak et al. (2017).

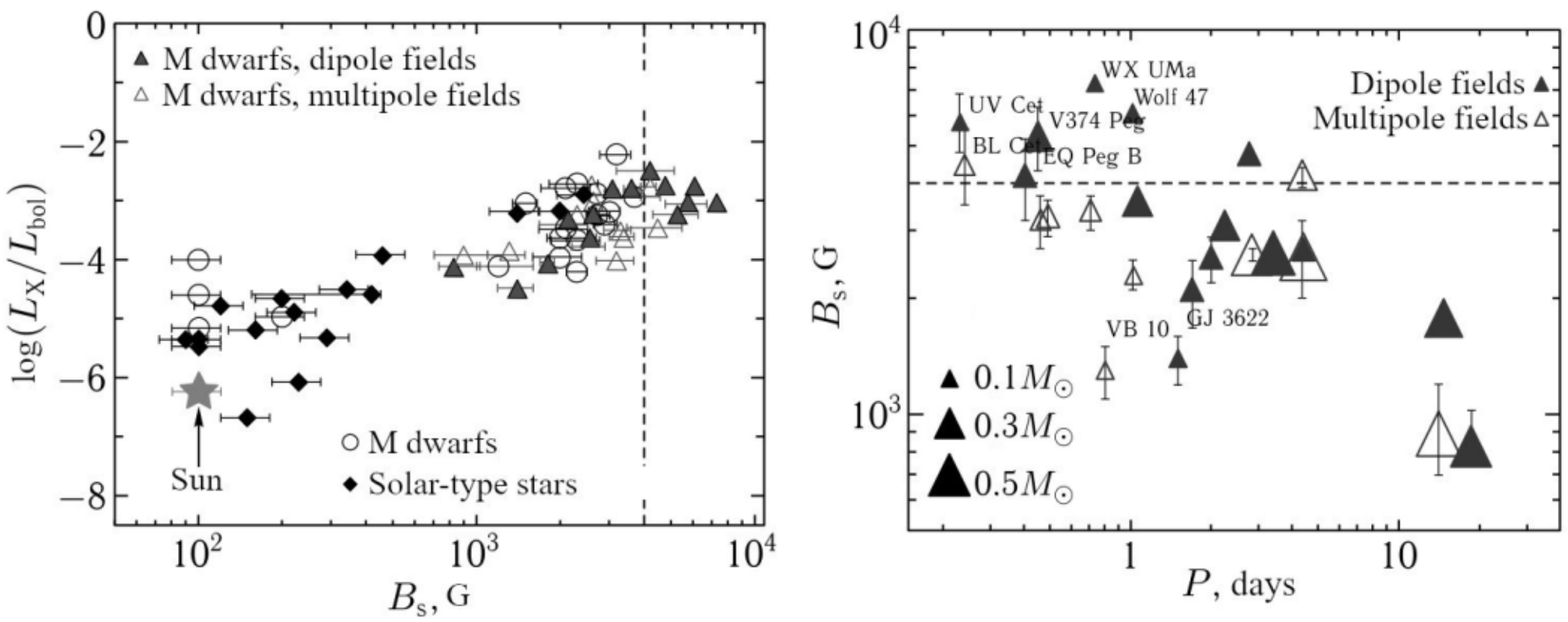

Fig. 94. Results of studying stellar magnetism: correlation of the mean surface magnetic field and stellar X-ray luminosity (*left*); correlation of the rotation period and mean surface magnetic field of a star (*right*). Dashed lines denote the 4 kG level (Shulyak et al., 2017)

During two years, using the special high-dispersion instrument CARMENES[1], Shulyak et al. (2019) performed the first magnetometric observations of 29 active M0–M8 dwarfs and on each of them detected strong magnetic fields with $G > 1$ kG, on 12 of them — fields with $G > 4$ kG. The observations were carried out in the near IR range, and magnetic fields were estimated from the Zeeman broadening of TiI and FeH lines. For 17 objects, the axial rotation period $P_{rot} < 1$ days and at $P_{rot} < 4$ days the field saturation effect was possible. The maximum filling factor was 0.6, the typical one — 0.2–0.4.

Using the same equipment, from more than 15000 spectra, Reiners et al. (2022) measured magnetic fields on 292 M dwarfs and detected a relation between average field strength and Rossby number, resembling the well-studied rotation-activity dependence. Among the slowly rotating stars, the magnetic flux is proportional to rotation period, and among the rapidly rotating stars the average fields do not grow significantly beyond the level set by the available kinetic energy. Furthermore, they find close relations between nonthermal coronal X-ray emission, chromospheric Hα and Ca H and K emission, and magnetic flux. Taken together, these relations make it possible to trace the rotation-activity relation to a dependence of the magnetic dynamo on rotation. The calcium emission saturates at average field strengths of 800 G while Hα and X-ray emission grow further with stronger fields in the more rapidly rotating stars.

* * *

The above survey shows that despite a significant wealth of experimental data theoretical studies are still far from being completed. Current general notions of subphotospheric magnetic fields in stars with solar-type activity are briefly as follows.

As stated above, Durney et al. (1981) associated the dispersion of chromospheric emission of stars of different age with the evolutionary reduction of the dynamo number, the ratio of generating and dissipative terms in the magnetohydrodynamic equation determining the efficiency of the magnetic-field generation. Knobloch et al. (1981) arrived at a conclusion that the Vaughan–Preston gap was caused by a nonmonotonic dependence of the strength of the magnetic field generated by the stellar dynamo on the Rossby number: at some critical value of the number depending on stellar mass, one should expect a fast decrease of the strength. Durney and Robinson (1982) estimated the expected magnetic-field strength and sizes of the regions occupied by such fields on the lower main-sequence stars assuming that field strength was determined by the condition of equal time of emerging of a magnetic flux tube and the time of an *e*-fold amplification of the field through the dynamo mechanism. As a result, the following conclusions were drawn: at a given axial rotation period the magnetic field should grow from several hundred gauss on G0 stars to many kilogauss on M5 stars; the fraction of the surface covered by magnetic fields should quickly grow with a decrease of the axial rotation period, and for periods close to solar period, M5 stars should be practically completely covered by magnetic fields. Based on the model by Durney and Robinson, Montesinos et al. (1987) found relations connecting the parameters of the internal structure of stars with the strength of the magnetic field generated in it and the surface filling factor with magnetic structures. Applying the relations to the sample of more than 30 G0–K3 main-sequence stars, they estimated the expected strength of magnetic fields and surface filling factors with them. The resulting filling factors displayed a clear dependence on Rossby numbers, and the measured total fluxes in CaII and MgII lines on the calculated filling factors. Further, analyzing Mount–Wilson monitoring measurements of calcium emission in the spectra of single F–G dwarfs, Hempelmann et al. (1996) divided all the examined stars into 3 groups with regard to the emission character: constant, regular, and irregular variables. They found that on regular variables the Rossby number, as a rule, was above unity and on irregular variables it was less than unity. Then, Hempelmann et al. found that stars belonging to different groups had different $\Delta F_{HK}$ and $F_X$: at the power form of specifying this ratio the exponent value was 0.99±0.13 on stars with periodic changes of $F_{HK}$ and 1.7±0.4 on stars with chaotic changes of these fluxes. These features can be related to the transition from the nonlinear dynamo operating on stars with irregular changes of $F_{HK}$ to a linear dynamo in stars with periodic variations of fluxes.

[1] CARMENES is a spectrograph with a resolution of 80000 mounted on the Calar Alto Observatory. It is intended for the high-accuracy, up to 10 m/s, measurements of line-of-sight velocities in the range of 5000–17100 Å, in the work of Shulyak et al. observations were carried out in the near IR range of 9600–10000 Å.

Reiners et al. (2014) studied the X-ray coronal emission and rotation periods of 821 stars with masses of less than $1.4M_\odot$ and found that stellar activity defined as $L_X/L_{bol}$ was closer associated with the rotation period rather than with the Rossby number.

As stated above, Bruevich et al. (2001) updated the concept of Durney et al. (1981) within the framework of the theory of dynamic systems. They found that young stars with high dynamo numbers should be compared with dynamic systems with a high degree of freedom, which are characterized by the modes of chaotic behavior and strange attractors. In the course of evolutionary rotation braking the dynamo number decreases, the dynamic system enters the limit cycle mode with auto-oscillations, which is manifested in the appearance of cyclicity in activity. In other words, the evolutionary transition from chaotic to cyclic activity is concerned with a decrease of the effective dimension of the dynamic system.

But the concept of evolution of dynamic systems through a cyclic dynamo is certainly insufficient for describing the whole variety of stars with solar-type activity, in particular, the youngest and lowest-mass stars.

* * *

For the youngest and, hence, the fastest rotators, the effect of saturation was found, when the activity level ceased to depend on rotation rate (Simon, 2001). Apparently, first, there is an approximation to the full coverage of the stellar surface by magnetic fields ($f \to 1$); second, the physical restrictions of the dynamo mechanism are triggered, for example, in suppressing differential rotation necessary for a solar-type dynamo, by a fast rotation of the whole star. Only the first factor $f \rightarrow 1$ is insufficient, since in this case the level of activity can hardly surpass the solar value by two orders of magnitude, whereas a much greater increase of this level has been observed. On the other hand, a growth of the degree of spottedness on active K–M dwarfs is an observationally established fact, and this cause of saturation should not be rejected.

Donati et al. (2003) estimated differential rotation on the surfaces of AB Dor and LQ Hya, revealed differences in the estimates based on cool spots and magnetic structures and changes of its amplitude over several years. They assumed that their results evidenced the effect of dynamo distributed over the whole convective zone rather than that concentrated on the bottom of the zone, as on the Sun.

The simple relation of the rotation rate with the activity level on late M dwarfs is violated: there are late M dwarfs — fast rotators without $H_\alpha$ emission, which contradicts the known correlations on G stars, and active M dwarfs within the wide range of rotation rates, whereas L dwarfs rotate quickly but are inactive (Basri, 2001). In other words, the activity weakens in achieving a completely convective structure and disappears on objects later than M9. Apparently, on low-mass stars the main role is played by a distributed or turbulent small-scale dynamo (Rosner, 1980; Durney et al., 1993). Remember the study of the very low-mass star VB 8 in the extreme ultraviolet, whose activity, according to Drake et al. (1996), is maintained by a turbulent dynamo. Its main difference from the large-scale envelope dynamo is a weak connection with rotation, slower transfer of the angular momentum. This circumstance can cause fast rotation of M dwarfs in the Pleiades and the Hyades and lack of X-ray variability in late stars in the Hyades, which is typical of the Pleiades (Stauffer et al., 1998; Gagné et al., 1995; Stern et al., 1995; Durney et al., 1993).

As stated above, Messina et al. (2003) found correlations between the maximum amplitude of rotational modulation of stellar brightness $A_{\max}$, which they considered as a value related to the total magnetic flux $fB$ and rotation period, and the Rossby number. These correlations distinctly show discontinuities near a period of 1.1 days, which were interpreted as evidence of the transition from one dynamo mechanism to another.

According to Mohanty et al. (2002), due to low temperature, high total density, and therefore very low degree of ionization of matter on the latest M dwarfs and on L dwarfs, one of the main features of cosmic plasma is violated – its freezing into magnetic fields. This results in a qualitatively new magnetohydrodynamic situation: noticeable nonpotential magnetic fields responsible for the stellar activity cannot exist in these layers of the stellar atmosphere.

From the spectra derived with the Keck telescope and within the SDSS project, West and Basri (2009) studied 14 M6–M7 dwarfs, most of which were inactive but with strong $H_\alpha$ emission. However, for three stars this emission was absent, though at their rotation rates both the strong fields and stellar activity had been observed earlier. This means that for the late M dwarfs rotation and activity are not always associated, which can be due to a possible dependence of stellar parameters on the Rossby number; at the latter the saturation of activity occurs for fully convective stars.

Reiners and Basri (2010) studied the spectra of 63 M7–M9.5 dwarfs derived with VLT and Keck telescope. They determined rotation rates, averaged magnetic field strengths and measured $H_\alpha$ emission, confirming that this emission was weaker at lower temperatures, and divergence of its values increased. The measured magnetic fields lie within 0 and 4.5 kG without a certain dependence on temperature, but at a fixed temperature the field strength correlates with $H_\alpha$ luminosity. For very slow rotators, a weak field and weak hydrogen emission occur, but all the stars with rotation faster than the detection limit reveals a field of at least a few hundred kilogauss. Contrary to earlier stars, the fields weaker than 1 kG are observed on the stars with rotation of more than 3 km/s but without a visible correlation between these values. To interpret this pattern, Reiners and Basri attracted the domination of turbulent dynamo for the coolest stars.

* * *

Walter (1990) found that the dependence of soft X-rays on rotation appeared at $B-V > 0.45^m$ as a threshold of emergence of the solar-type dynamo. Probably, the mechanism of the turbulent dynamo is effective not only on the objects where cyclic magnetic fields of solar type are absent. It is probable that the turbulent dynamo is responsible for the activity in the equatorial zone of the Sun with a low phase activity level independent of the cycle, for the ubiquitous "magnetic carpet" on the Sun, presumably responsible for the diffuse X-ray radiation and the residual activity of the Sun during the Maunder minimum. Stars with flat activity can also be candidates for turbulent dynamo objects: their coronae are weak and there is no dependence of coronal emission and radiation of transition zones on rotation (Saar, 1998, 2001).

Schmitt et al. (1998) considered the simultaneous action of two different mechanisms of magnetic field generation.

According to Kitchatinov et al. (2001), the interaction of magnetic fields in the radiative core and convective envelope can lead to the effect of active longitudes.

Within the envelope dynamo theory, Korhonen and Elstner (2005) calculated a flip-flop effect for different thicknesses of the convective zone, different values of differential rotation, and different inclination angles of the rotation axis to the line-of-sight, constructed synthetic light curves during several periods of this effect, and compared the calculated models with concrete observations. For a solar-size star, the differential rotation of 10% of the solar one is sufficient for the appearance of the effect, for a star twice larger — of 40%. Later, they paid attention to the circumstance that the starspots caused by the large-scale dynamo not necessarily followed stellar differential rotation, whereas this was better done by spots associated with fields of less scales (Korhonen and Elstner, 2011).

Based on three-dimensional MHD calculations, Chabrier and Kueker (2006) developed a model of the large-scale magnetic field generation in fully convective stars. This $\alpha^2$ dynamo model in which helicity is generated by the action of the Coriolis force on the convective motion in the rotating medium differs from the envelope dynamo of more massive stars: it generates a large-scale nonaxisymmetric stationary field, which is symmetric with respect to the equatorial plane. It saturates at a strength of a few kilogauss and yields a qualitative explanation for the saturation of late M dwarfs. For the brown dwarfs with a conducting core, which is typical of the most massive and old such stars, they considered the $\alpha^2\omega$ dynamo, i.e., the field generation in the medium with differential rotation; in this case, the toroidal axisymmetric and oscillating fields, similar to those on the Sun, are generated predominantly. The field topology of fully convective objects reveals the fields of high multiplicity, which differs them from the fields generated by the $\alpha$–$\omega$ dynamo. A significant decrease of the dipole component due to nonaxisymmetricity noticeably decreases the Alfvén radius and, consequently, magnetic braking, interpreting a decrease in the angular momentum loss rate. Despite this large-scale field, a decrease of conductivity in the predominantly neutral atmosphere may hamper the generation of currents required to maintain chromospheric activity. As an observational confirmation of this dynamo model there may be an asymmetry of activity, contrary to spatially homogeneous activity expected in the small-scale turbulent dynamo, and the absence of cycles in homogeneously rotating fully convective low-mass objects.

Discussing the conclusions, on the one hand, on the total maintenance of magnetic activity on the boundary of the transition to the fully convective stars and, on the other hand, on an abrupt decrease in the rotation braking of lower-mass stars, Reiners (2007b) put forward an idea that the reason of this is in varying magnetic field topology: from dipole in more massive objects to fine-structured in fully convective stars.

Feiden and Chaboyer (2012) elaborated a program to calculate stellar evolution with the global magnetic field, tried it on the solar-type components of the binary system EF Aqr, and found that the addition of magnetic perturbations corrects the observed in this system discrepancies in radii and effective temperatures, whereas the required magnetic field strength is twice higher than the value estimated from X-ray luminosity and predicted by CaII emission. Thus, these models confirm that the suppression of thermal convection by the magnetic field is sufficient to change significantly the structure of solar-type stars — increasing radius and lowering effective temperature.

Nelson et al. (2013) showed that the convective dynamo can yield the constant toroidal current structure — magnetic ring — inside the turbulent convective zone, whereas a high rotation rate contributes to the cyclic reversal of these rings. Magnetic cycles can be reached at a decrease in diffusion, increasing the magnetic Reynolds number. In such a more turbulent model, diffusion ceases to play a crucial role in the dynamic balance, which is established and maintained by differential rotation and magnetic rings. An enhancement of turbulence leads to larger intermittency in toroidal magnetic rings, contributing to buoyancy of magnetic loops that arise from the depths to outer regions.

The dependence of the turbulent diffusion coefficient on the scale was detected by Abramenko and coauthors (2011) on the solar surface. It turned out that for small scales (of an order of 30 km and 10 s) the turbulent diffusion accounts for $10^{11}$ $cm^2/s$, increasing up to $6 \cdot 10^{12}$ $cm^2/s$ on scales of an order of 2000–4000 km and 4 h. A growth of the diffusion coefficient with scale means the superdiffusion regime and helps to explain the uneven distribution of large spots over the stellar surface as opposed to the even distribution of small magnetic formations: the latter spend much more time in the convective zone due to lower diffusion.

Fluri and Berdyugina (2004) interpreted stellar cycles based on the decomposition of activity into a series according to dipole and quadrupole dynamo modes, whereas Moss et al. (2008) expressed a belief that neither the dynamo theory nor the observational data can provide serious support for the notions that stellar magnetic fields can have dipole rather than quadrupole symmetry with respect to the stellar equator. They showed that even the basic model of stellar magnetic activity — the Parker dynamo — yields many possibilities to excite large-scale magnetic fields of nondipole symmetry and possibilities of spontaneous transition from the excited magnetic field dynamo with one kind of symmetry to another one. Some observational data can be treated as evidence of the quadrupole magnetic field component.

Mullan and MacDonald (2010) showed that the localization of components of the system HD 130948 BC — brown dwarfs with masses of 0.0555 and $0.0530 M_\odot$ and luminosity $\log(L/L_{Sun}) = -3.82$ and $-3.90$ — in the effective temperature–age diagram and the Hertzsprung–Russell diagram can be explained within the magnetic convection model, taking into consideration the internal magnetic pressure which accounts for 0.007–0.038 of the gas pressure. Then, MacDonald and Mullan (2012, 2013b) showed that Mullan's magnetoconvective model somewhat raises the bottom of the convective zone, lowering by this way its maximum temperature, which should lead to a decrease in the lithium burning out. Taking this circumstance into account, they found that this model operated well in binary systems CM Dra and YY Gem but did not fit for the system CU Cnc. Considering the change of equivalent widths of the CaII K line in K5−M4 dwarfs, Mullan et al. (2015) proposed the interface dynamo model in which the surface magnetic field strength is linearly related to the intensity of calcium emission as in the case of local fields in solar active regions.

* * *

All the above considerations should be included in the not yet constructed general theory of magnetic fields of mid- and low-mass stars. An important step in this direction has been made by Barnes (2003a, b). He analyzed a large sample of stars with known rotation periods within $0.5^m$ < B−V < $1.5^m$ and found two sequences in the diagram $P_{rot}$,(B−V), which he compared with different structures of stellar magnetic fields. Clear localization of these sequences in the diagram depending on the age of stars within the range of 30 to 4500 million years made it possible to advance the concept of gyrochronology and estimate the characteristic time of the evolutionary transition of the global magnetic field from one structure to another in stars of different masses.

## 4.2. Stellar Magnetism and Stellar Atmospheres

Flux tubes of the photospheric magnetic field extend upward to the solar atmosphere, define its spatial structure and, to a great extent, its physical state. A close relation between the magnetic field and various manifestations of solar activity — dark spots and bright photospheric faculae, bright chromospheric flocculi, coronal condensations, and sporadic flares — is revealed when directly comparing magnetograms with localization of the listed structures on the solar disk. This relation is expressed in clear correlations between the local values of $fB$ and fluxes of $\Delta F_{HK}$ and $F_{CIV}$. As shown in previous chapters, a similar situation occurs on active red dwarfs (see correlations (22) and (23)), although the analog of solar magnetograms does not exist yet, and direct comparisons are impossible here. However, there is less detailed but equally confident evidence for the relation between photospheric magnetic fields and the activity level in the atmosphere. Thus, Fig. 95 displays the comparison of average magnetic fluxes from the stellar photospheres with X-ray luminosity of coronae $L_X/L_{bol}$. The figure reveals a distinct correlation between the compared values. Saar (2001) found a quantitative relation (22) for the stars of the same sense as the local relations for the Sun. Essentially, the same physical sense has a comparison of absolute coronal plasma radiation in the range of 2.8 to 36.6 Å, $L_X$, and magnetic fluxes Φ performed by Pevtsov et al. (2003) for a series of solar and stellar objects. The correlation acquired by them is close to linear

$$L_X \propto \Phi^{1.13 \pm 0.05} \tag{74}$$

(see Fig. 28), overlaps 12 orders of magnitude based on Φ, and combines the quiet Sun (dots), X-ray points on the Sun (squares), solar active regions (diamonds), the averaged solar disk (+), G, K, M dwarfs (×), and T Tau stars (circles). Jordan and Montesinos (1991) detected a correlation between the coronal temperature and the Rossby number, which also means the relation of the dynamo mechanism in the depth of the convective zone with physical processes in the upper atmosphere. The same sense has a correlation between the amplitude maxima of rotational modulations of brightness found by Messina and ratios $L_X/L_{bol}$, which will be discussed below.

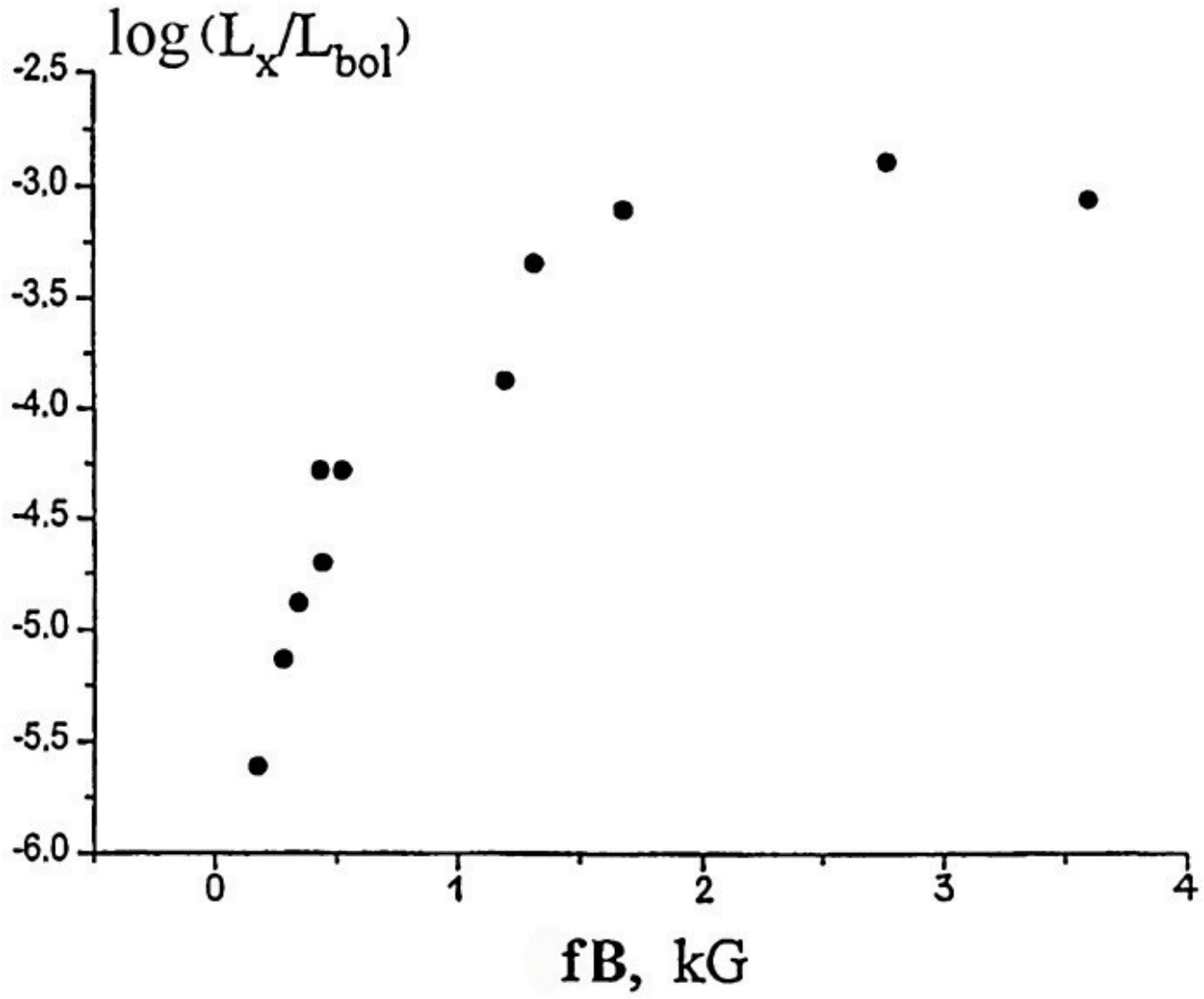

Fig. 95. Comparison of average magnetic fluxes *fB* on stellar surface and relative luminosity of coronal X-ray emission (Katsova, 1999)

From a sample of several hundred stars of solar and later types with known X-ray emission and rotation periods Wright et al. (2011) estimated again the turnover time of convective elements and found that $L_X/L_{bol} \propto \mathrm{Ro}^{-2/7}$. This confident difference from the canonical index −2 requires an additional member in the equation for the dynamo number and leads to the relation $\Delta\Omega/\Omega \propto \Omega^{0.7}$, i.e., to a gradual decrease of differential rotation as the stellar rotation brakes.

* * *

Let us present two more concrete facts.

The extensive campaign for the comprehensive study of radiation of the active G8 star $\xi$ Boo A was carried out in 1986: magnetometric observations based on the $\lambda$ 6173 Å line and spectral observations in the region of the helium $D_3$ line, multicolor broadband polarimetric observations, spectral observations in the ultraviolet and in the region of CaII H and K lines (Saar et al., 1988). Analysis of the collected data reveals distinct synchronous variations of the magnetic flux, intensity of CIV and CII ultraviolet emission lines and CaII emission, which is a direct indication of the connection between the magnetic flux of the photospheric field and emission of the outer stellar atmosphere. The absorption maximum in the helium $D_3$ line and the filling maximum of the sodium D line cores are also near the magnetic flux maximum. From the comparison of the derived data it was suggested the existence of four longitudinal sectors with enhanced magnetic fluxes on the star.

By analogy with the Sun, one can expect that monitoring of stars with asymmetric distribution of active regions on the disk will reveal a growth and decline of ultraviolet emission, X-ray emission, and magnetic flux at certain phases of the rotation period. This prediction was confirmed by observations of ε Eri (Saar et al., 1986b) and $\xi$ Boo A (Saar et al., 1986b, 1988).

* * *

The first two parts of the book consistently reported the data on the quiescent state of flare stars and flares on them. This was a natural separation of experimental data obtained for the objects with different scales of time changes, and their interpretation required different theoretical constructions: models of atmospheres of active stars and models of flares. Independent development of these models led to the notions on active processes — flares — in the passive medium of stellar atmospheres. However, this dichotomy should not be absolutized, since there is a deep physical commonness of these phenomena, which consists in possible relations of the heating mechanism of quiet atmosphere of an active star and the energy source of flares. In solar studies, the experimental fact stimulating the integration of the mentioned problems was a discovery of microflares in hard X-rays and bright ultraviolet points: both phenomena, as well as flares, are caused by fluxes of fast electrons interacting with the dense chromosphere at the feet of coronal loops. Their energy is apparently sufficient for heating of the solar atmosphere. This concept is supported by the data on the heating of the solar wind not only at the base, but also at great distance from the solar surface due to surges, transients, and coronal eruptions. For active stars, the above statistical correlations of X-ray luminosity of quiet coronae and the averaged power of optical flares and the very hot component of stellar coronae, whose temperature is close to the temperature of solar flares, can be considered as unifying facts.

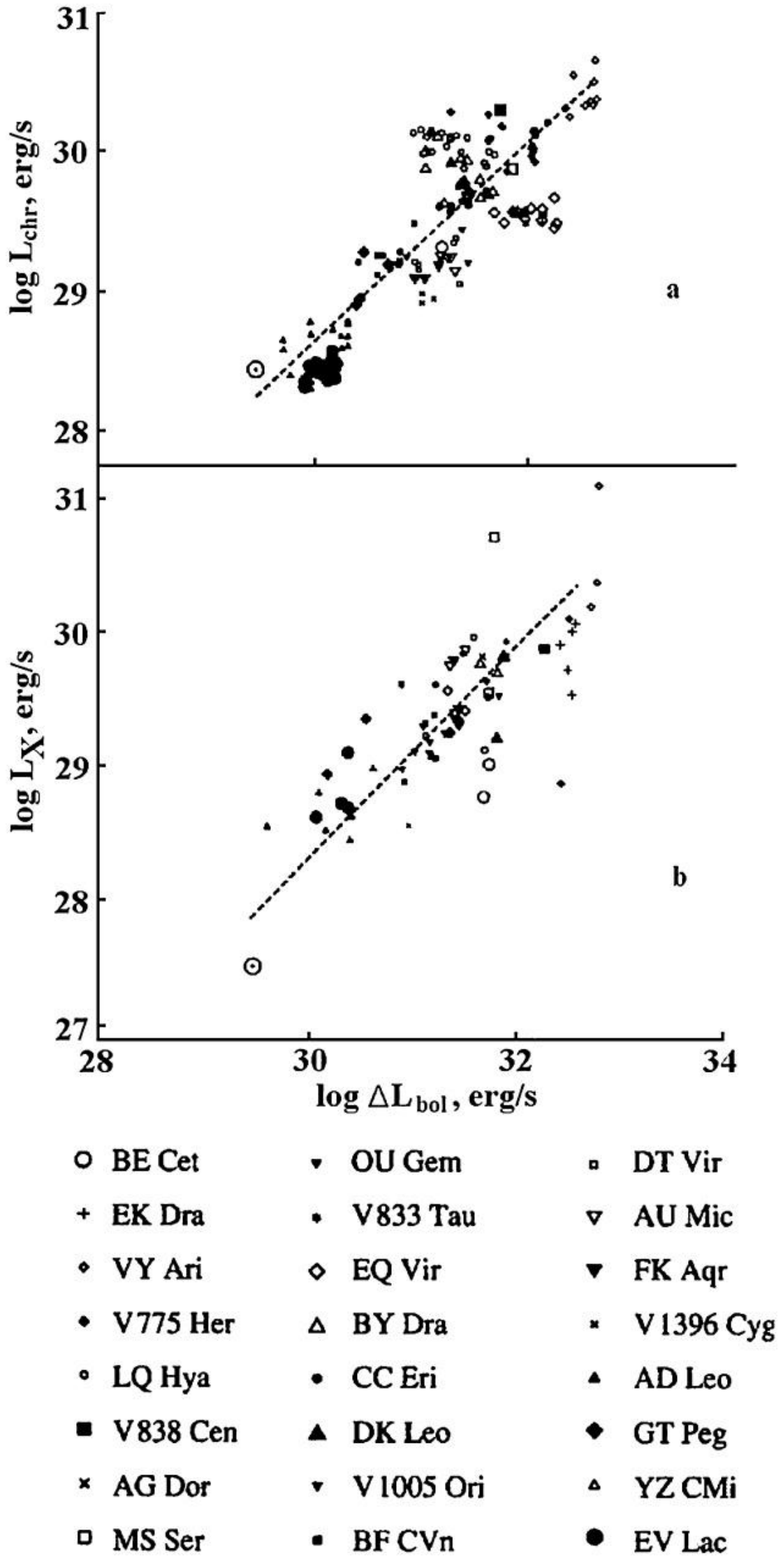

Fig. 96. Comparison of the deficits of bolometric luminosity of photospheric radiation with radiative losses of chromospheres (**a**) and coronae (**b**) (Alekseev et al., 2001)

Using this unifying approach, we shall reconsider the problem of the deficiency of photospheric radiation of spotted stars mentioned in Subsect. 1.2.3.4 assuming that the total photometric effect of stellar spottedness is, in a way, connected with the integral magnetic flux. In Fig. 96 the bolometric deficits of photospheric radiation of two dozen spotted stars are compared with the estimates of radiative losses by the chromospheres and coronae during the epochs close to optical observations (Alekseev et al., 2001). The diagrams show certain correlations of the compared values, which eventually, as in Fig. 28, correspond to correlations of the photospheric magnetic flux and the flux in X-rays. On the basis of quantitative analysis of the found correlations, Alekseev et al. (2001) concluded that the bolometric deficit of photospheric radiation of the most active spotted stars definitely exceeded their radiative losses for permanent radiation from all atmospheric layers and for the radiation of sporadic flares. In this connection, Katsova and Livshits formulated the idea about the transfer of deficient energy of photospheric radiation into a global rearrangement of the structures in the upper atmospheres of such stars analogous to the local reorganization of atmospheres on shorter times during flares (Alekseev et al., 2001).

On the Sun and on other less-active dwarfs, the deficit of photospheric radiation can be converted into the spread of this energy over the near photospheric regions, formation of facular areas and related chromospheric flocculi. Indeed, exact measurements of the solar constant showed that the deficit of photospheric radiation of the Sun was substantially balanced by additional radiation of the atmosphere: the measurements detected a decrease of this constant during the passage of a large group of spots, but during the maximum of solar activity at the greatest spottedness this constant grew at the expense of numerous active regions. High-precision photometric measurements of some stars within the Wilson program showed that the total brightness of the most active stars slightly decreased during the epoch of maximum chromospheric activity, i.e., the brightness variations were determined by large spotted regions with decreased temperature. As to the less-active stars, during the maxima of chromospheric activity their brightness increases, i.e., as on the Sun, the additional radiation of faculae prevails (Radick et al., 1998; Alekseev et al., 2001). It should be noted that domination of faculae above spots during the solar maximum suggests the increasing contribution of large magnetic structures to their total spectrum. Abramenko (2002) found a local manifestation of this process in considering some active solar regions: the closer the region to the realization of flares, the flatter is the structural function of photospheric magnetic fields, i.e., the contribution of large-scale components increases.

* * *

Analogous conclusion on the physical relation between photospheric and coronal activity was acquired by Messina et al. (2003). From the field stars and stars of different-age clusters IC 2602, IC 4665, IC 2391, $\alpha$ Per, the Pleiades, and Hyades they compared the amplitude maxima $A_{max}$ — variability that follows from the long-term photometry of spotted stars and the ratio $L_X/L_{bol}$. These values apart characterize the levels of magnetic activity of the photosphere and corona. First, based on the constructed distributions of F, G, K, and M stars on the rotation period–activity plane they found that the upper envelopes of distributions $A_{max}(P_{rot})$ grow monotonously with decreasing rotation period but for G and K stars show a breaking-off about 1.1 days; for shorter periods a saturation mode was revealed; analogous distributions were constructed for $A_{max}$(Ro). In the same way the ratios $L_X/L_{bol}$ depend on $P_{rot}$ and Ro. A direct comparison of $A_{max}$ and $L_X/L_{bol}$ (see Fig. 97) reveals their correlations, which means the dependence of coronal activity on the photospheric magnetic field. The distribution $L_X/L_{bol}(A_{max})$ for M dwarfs has an upper envelope at the saturation level $\log(L_X/L_{bol}) =\sim -3$, whereas the envelopes of earlier stars — a power form with steepness decreasing from spectral types F–G to M. The average values of these activity indices for each cluster grow monotonously with age, showing that the levels of photospheric and coronal activity evolve in time according to the unified power law up to the age of the Sun.

Based on spectropolarimetric data, Brown et al. (2022) compared stellar chromospheric $\log R'_{HK}$ activity and the surface-averaged longitudinal magnetic field for 954 F-M stars and found a positive correlation. Chromospheric activity, activity variability, and toroidal field strength decrease on the main sequence as rotation slows. For G stars, the disappearance of dominant toroidal fields occurs at the same chromospheric activity level as the change in the relationships between chromospheric activity and mean field strength.

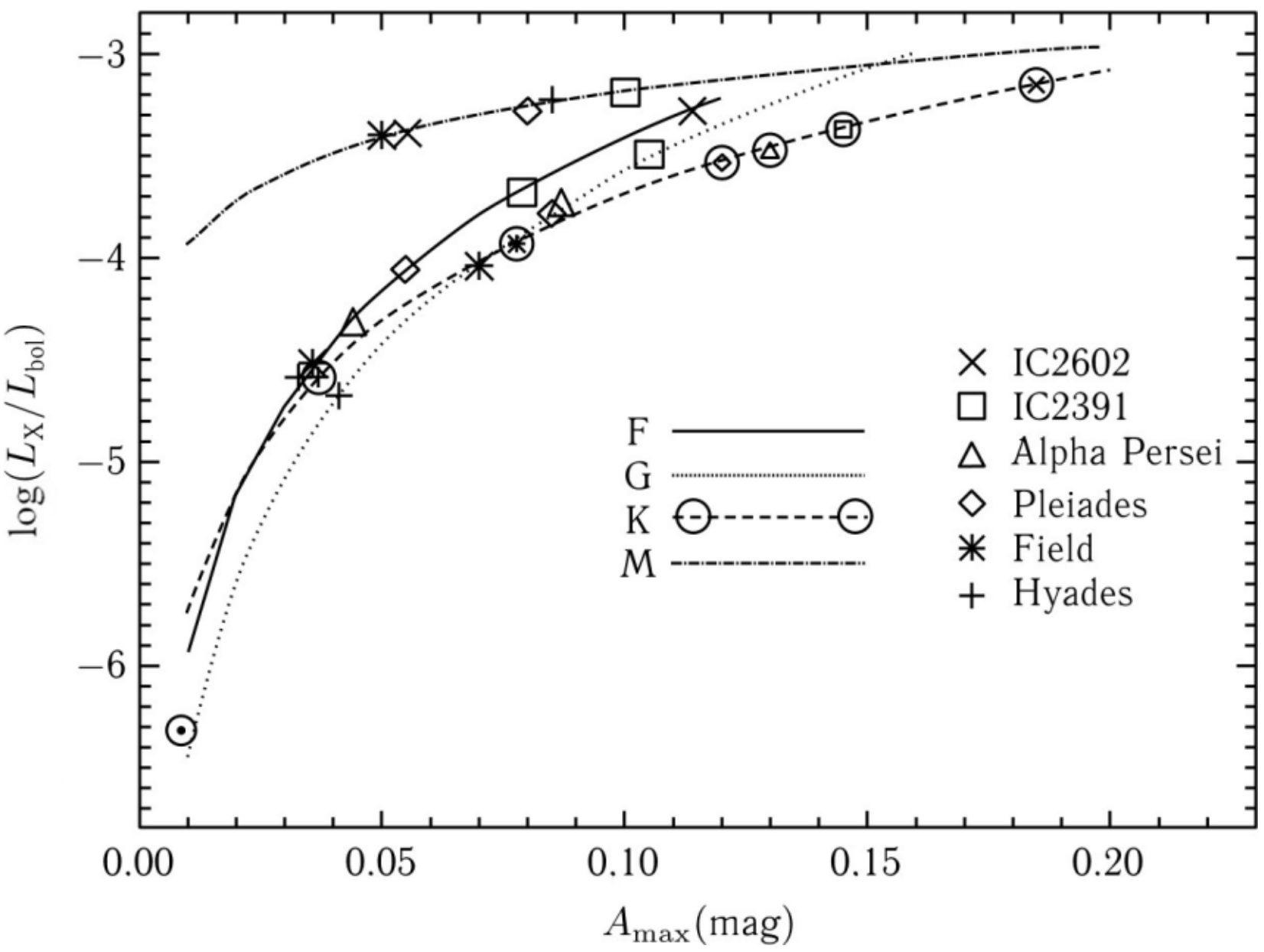

Fig. 97. Dependencies of $L_X/L_{bol}(A_{max})$ for the stars of different spectral types from the stellar clusters of different ages and field stars (Messina et al., 2003)

* * *

The atmospheres of the most active red dwarfs, according to a series of characteristics, differ systematically from the solar atmosphere. Roughly speaking, on the Sun, where spots occupy the fractions of a percent of the total surface, the deficiency of photospheric radiation can spread over the active regions surrounding the spots. But on the most active stars, where spots occupy tens of percent of the surface, there is no place enough for the required extended active regions, and the deficiency energy of photospheric radiation seems to be channeled into the upper layers of atmospheres. Chapters 1.3 and 1.4 have already listed the qualitative differences in the degrees of excitement in the atmospheres of the most active stars and the Sun: for such stars, the electron density of atmospheres is higher, the total radiative losses are higher, and most fraction of such losses are caused by the corona; in their corona there is constantly a hot component that may arise during flare processes, and their flare activity is higher. As noted in Subsect. 1.2.3.4, the absolute values of bolometric deficits of line-of-sight energy proved to be very high in the most active stars: up to $5 \cdot 10^{32}$ erg/s, and they definitely exceed the radiative losses of atmospheres and flares, which even in the most active stars do not exceed $10^{-2}L_{bol}$ (Pettersen, 1988). In the Katsova–Livshits concept, the governing parameter of stellar activity is the degree of stellar spottedness, which varies both during the evolution of an individual star due to secular braking of its rotation and decay of its magnetic activity and in the spectral sequences of stars due to systematic changes of the structure of their convective zones along the main sequence.

Phenomenological notions on the decisive role of the deficit of photospheric radiation in the energy of atmospheres of active stars brings us back to the idea proposed by Mullan (1975b) on Alfven waves as a source of heating of the atmospheres of flare stars and the energy of flares. (But already in 1968 de Jager stated that the deficit of radiation of a large sunspot can save energy sufficient for a large solar flare up to $10^{32}$ erg.) Mogilevsky (1980, 1986) developed the concept of self-organization of plasma in the convective zone into MHD solitons, which directly connect subphotospheric magnetism with the activity phenomena in the stellar atmosphere and have certain advantages over MHD waves in a higher propagation velocity and negligible dissipation. Not giving preference to any specific form of energy flux coming from subphotospheric layers, one can cite some arguments for this scheme of energy supply for the whole atmospheric activity: "All forms of observed radiative losses — quiet atmospheres and flares — can have a common cause, and this cause should be sought inside a star" (Pettersen, 1988).

Passage of the flux of nonradiative energy through a rather stratified stellar atmosphere in different configurations of magnetic flux tubes can include more than one change of the actual energy carrier: various waves, including slow MHD waves and MHD solitons, fluxes of fast particles, reverse — top down — heat conductivity, radiative heating of the upper chromosphere by the Lyman series from the transition zone and heating of the temperature minimum region by the Balmer series from the chromosphere. The character of the energy flux should have threshold properties and significantly depend on its power: at low energy density the flux proceeds in a "laminar" way, heating up the atmosphere, whereas at high energy density there should be breaks of the flux, which can be identified with flares. These phase transitions are essential in the present-day models of flares (Pustil'nik, 1997, 1999). A strong energy flux through a nonlinear medium is one of the basic requirements of thermodynamics of open systems needed for self-organization of the medium. The formation of current sheets, which are often considered as direct energy sources of flares and heating of stellar atmospheres, can be one of the results of the self-organization, while the actual channeling of their energy can depend on the amount of free energy contained in them (Mullan, 1989). Probably, the slowly varying deficit of photospheric radiation during a flare can be a source of additional heating of plasma at the phase of flare decay.

Turning back from the most energetic phenomena on active stars to one of the least energetic manifestations of activity — basal emission — it is worth noting the paper of Bercik et al. (2005), who, with the aim of ascertaining whether the convective dynamo can supply the observed lower limit of the X-ray flux on solar-type stars, carried out the numerical 3D calculations of magnetic fluxes on the surface of nonrotating F0–M0 main-sequence stars. The estimates of X-ray fluxes derived based on these calculations proved to be in good agreement with observations, as well as estimates of the expected chromospheric emission flux in the MgII lines of K dwarfs derived from X-ray observations.

Recently, Kochukhov et al. (2020) performed the magnetometry of solar-type stars, measuring the enhancement of spectral lines by the magnetic field. This method differs from the wide-spread magnetometry from spectropolarimetric observations by the fact that the total magnetic flux is measured from the modules of local field components, whereas the spectropolarimetric signals of different signs are mutually suppressed, which at the numerous small magnetic structures leads to the underestimation of $fB$. To overcome the basic difficulty of magnetometry for the enhancement of spectral lines — the separation of magnetic and nonmagnetic mechanisms for the line broadening in ordinary spectra, Kochukhov et al.

developed a methodology for the use of spectral lines with different Lande factors and performed magnetometry of 14 G dwarfs and one K dwarf of different age and with different activity levels based on three lines of the same neutral iron multiplet in the region of 5500 Å. As a result, the authors found that the values of *fB* decreased from 1.3–2.0 kG for stars younger than 120 million years up to 0.2–0.8 kG for older stars, detected anticorrelation of the mean field and rotation period or the Rossby number, and yielded calibration of *fB* (Ro). Kochukhov et al. concluded that all the considered fields are with the strength $B \approx 3.2$ kG, while the growth of *fB* is caused by the increase of the filling factor from 10% to 50% for a star with growing activity. The stars under investigation revealed a distinct correlation between the mean magnetic field and indices of coronal and chromospheric activity $L_X/L_{bol}$ and $\log R'_{HK}$, respectively. Having compared the derived data with the results of spectropolarimetric magnetometry, the authors estimated that the latter yielded about 1% of the total magnetic field energy for the most active stars and about 0.01% for the least active ones. Let us recall that earlier Johns-Krull and Valenti (1996) performed magnetometry based on the broadening of the neutral iron spectral line for the one of the most active red dwarfs EV Lac (see Fig. 12 and the text therein).

A concept of the decisive dependence of the solar-type activity level of a star on its spottedness degree has recently been developed by Katsova et al. (2022). The authors show that the occurrence rate of weak X-ray flares on the Sun does not practically depend on its spottedness degree, whereas the occurrence rate of powerful flares of classes M and X considerably depends on this value. Since a filling factor of spots of even the most active Sun does not exceed a small fraction of a percent of its surface, whereas on the most active dwarfs it reaches tens of percent, then it becomes clear both the absence of superflares on the Sun and their presence on active stars. The authors estimate the magnetic field strength in starspots to be 2 kG and the maximum energy of stellar superflares to be $(1–3) \cdot 10^{36}$ erg.

## 4.3. Stellar Dynamo Models

Since a solar magnetograph was invented by the father and son Babcocks in 1952, solar observations have clearly demonstrated a relation between different phenomena of solar activity and local magnetic fields. On the other hand, the measurements of large-scale solar magnetic fields — which in itself is a nontrivial challenge — have shown their relation with global properties of the solar cycle such as 11- or 22-year cyclicity of spot formation. To ascertain the nature of celestial body magnetism, to estimate its level, and to describe its structure, a theory of magnetic field generation by the motions of the conductive liquid was created, for the case of stars it was plasma. It is important to bear in mind that, speaking about magnetic field generation (self-excitement), one implies *a growth of the magnetic flux* rather than merely magnetic energy. It is this theory that is known as dynamo theory. The dynamo problem seemed to be first formulated in a brief pioneer letter of Larmor (1919), which was called "How can a rotating body, for instance the Sun, become a magnet?". Thus, from the very beginning the Sun was an object of the dynamo theory. Not all the problems associated with the dynamics of magnetic fields are attributed, strictly speaking, to the dynamo theory. For example, the problems of magnetic concentrated tubes that are buoyed up (Parker, 1955a) and/or their formation (Kleeorin et al., 1989, 1990; Kitchatinov and Mazur, 2000) do not refer to the dynamo theory, since these processes are not accompanied by a growth of the magnetic flux but only its redistribution. However, to apply the dynamo theory to concrete stars and the Sun, these processes are extremely important because they determine to a great extent the local magnetic fields on the solar surface. The complete solution of the dynamo problem includes the induction equation, the Navier Stokes equation, the continuity equation, the entropy transfer equation, the radiative transfer equation, and several equations of state: the Mendeleev–Clapeyron equation, the Saha ionization equation, equation for the Rosseland mean absorption coefficient of the stellar plasma electromagnetic radiation, and others. In such a statement, the problem seems to be absolutely overwhelming. Therefore, any solvable dynamo problem represents only some dynamo model. Historically, the first successful dynamo model was the Parker model (1955b) of the axisymmetric large-scale solar magnetic field. In this model, Parker combined the "winding" of the poloidal field into the toroidal one due to stellar differential rotation with helicity of convective-cyclonic flows arising under the influence of the Coriolis force and inhomogeneity of turbulent convection. Parker called this effect cyclonic; however, after the papers of Steenbek et al. (1966) it became widely known as the alpha effect, and the Parker model was accepted as the $\alpha$–$\omega$ dynamo. This model was significantly developed for both the Sun and stars; some of these models taking into account the appreciable difference of the stellar structure from the solar one were mentioned above. Unfortunately, our knowledge on the influence of the Coriolis force on the stellar interior is insufficient to consider these appreciable differences properly, and thus, decreases the reliability of these models. All these models: $\alpha$–$\omega$, $\alpha^2$, $\alpha^2\omega$, and even nonaxisymmetric models are somehow associated with stellar rotation and are accompanied by a growth of the mean field. A number of dynamo schemes formulated before the appearance of the mean field concept have lost their importance so far. However, there is one more dynamo type, the so-called small-scale dynamo of Kazantsev (1967). Contrary to the $\alpha$–$\omega$-, $\alpha^2$-, and $\alpha^2\omega$ models, this dynamo type is not associated with rotation, though the presence of the alpha effect can change the properties of this dynamo: without the alpha effect these fields have a random small-scale character, for the generation of this type it is sufficient only to overcome a small generation threshold in the magnetic Reynolds number, $\mathrm{Rm} > 412$–$730$ depending on a degree of compressibility of the stellar plasma (Rogachevskii and Kleeorin, 1997). Certainly, the magnetic Reynolds number in stars far exceeds this threshold. Nonetheless, the transverse size of magnetic tubes, being produced by this mechanism, may increase from a few tens of kilometers up to, at least, the basic turbulence scale, if a hydrodynamic alpha effect appears. In the conditions of the Sun, this mechanism seems to be responsible for the formation of magnetic pores. However, for other stars, which rotate very slowly or, on the contrary, very rapidly, the ordinary large-scale dynamo mechanisms may turn off, and in this case, the Kazantsev dynamo becomes most relevant. It seems to be referred to when observers discuss an alternative model of the distributed dynamo with a weak association with rotation.

The stellar dynamo theory is extensively discussed in the literature. This chapter briefly but strictly outlines the general physical concept of stellar dynamo and some recent results of the nonlinear $\alpha$–$\omega$ dynamo, applying to the main-sequence stars. Furthermore, we discuss the problems of formation and buoying of magnetic concentrated tubes, though this problem is not, strictly speaking, related to the dynamo theory.

### 4.3.1. Basic Notions on the Stellar Dynamo Theory, Antidynamo Theorems

Any main-sequence star, for instance the Sun, is a gaseous sphere whose equilibrium in the first approximation is determined by a balance of the gravity force and the pressure gradient, i.e., by the hydrostatic equilibrium. For many main-sequence stars of late spectral types F, G, K, and M, the turbulent convection is typical, and it develops in their outer layers. According to the present-day notions, the magnetic fields are excited in these convective envelopes. Rotation also plays a significant role in stellar magnetic field generation.

In normal stars, rotation and magnetic fields are weak, and they practically have no impact on their total equilibrium. However, magnetic fields play a crucial role in stellar activity. Thus, various manifestations of solar and stellar activity — spots, faculae, flocculi, chromospheric flares, prominences, etc. — are associated with solar and stellar magnetic fields. Although the detailed mechanisms of many manifestations of activity have not been sufficiently studied, it is widely accepted that activity is governed by the magnetic field (see, e.g., Bakulin et al. 1983; Gibson, 1977). Hence, it is clear that any serious prediction of activity should be based on studying properties of solar and stellar magnetic fields. This makes the problem of constructing the theory of stellar (solar) magnetism highly relevant from both the purely scientific and the practical point of view.

The magnetism of stars, as well as other celestial bodies, finds it explanation in the hydromagnetic dynamo theory. The fundamental equilibrium underlying the stellar dynamo theory is the induction equilibrium

$$\frac{\partial \mathbf{H}}{\partial t} = \mathrm{rot}\left(\mathbf{v}\times\mathbf{H} - \nu_m \mathrm{rot}\mathbf{H}\right) \cdot \qquad (75)$$

Here $\mathbf{H}$ is the magnetic field of stellar plasma, $\mathbf{v}$ — the velocity of hydrodynamic stellar plasma motions, $\nu_m = c^2/4\pi\sigma$ — the magnetic field diffusion coefficient in the CGS unit system, $c$ — the speed of light, $\sigma$ — the plasma conductivity. At $\nu_m = 0$, the first term in (75) $\mathrm{rot}(\mathbf{v} \times \mathbf{H})$ describes the freezing of the magnetic field $\mathbf{H}$ into the plasma of a celestial body moving with velocity $\mathbf{v}$. If the plasma is quiescent ($\mathbf{v} = 0$), then the remaining term $-\mathrm{rot}\left(\nu_m \mathrm{rot}\mathbf{H}\right) \approx \nu_m \Delta\mathbf{H}$ describes the magnetic field diffusion through the conducting plasma. Note that from the estimate of $\nu_m = c^2/4\pi\sigma$ it follows that the higher the plasma conductivity $\sigma$, the lower the diffusion coefficient. For our aims, it is worth noting the estimate of diffusion time: $t_d = L_H^2/\nu_m$, where $L_H$ is the characteristic scale of magnetic field variation. To estimate $\nu_m$ in the fully ionized hydrogen plasma at the temperature $\sim 10^5$–$10^6$ K, we can take

$$\nu_m = \frac{5.2\times10^{12}}{T^{3/2}} \ . \tag{75a}$$

Here $T$ is the temperature in kelvins (Braginsky, 1963). The estimate for the middle of the convective zone ($\sim 10^6$ K) yields $\nu_m \approx 5 \cdot 10^3$ cm$^2$/s. Near the outer boundary of the solar convective zone the temperature reaches $\sim 6 \cdot 10^3$ K; the ionization degree significantly decreases as compared to the middle of the convective zone and turns out to be equal to $3 \cdot 10^{-4}$. Nonetheless, the estimate of Braginsky for this temperature accounts for $\nu_m \approx 10^7$ cm$^2$/s and proves to be surprisingly close to the more strict estimate from the book of Vainshtein et al. (1980): $\nu_m \approx 7 \cdot 10^6$ cm$^2$/s. The other important characteristic time within Equation (75) is the time of magnetic field transport by plasma $t_v = L_H/v$. Obviously, if the transport time $t_v$ is much shorter than the diffusion time $t_d$, then over the time $t_v$ no diffusion can occur. Thus, the condition of freezing of the magnetic field into the plasma is $t_v \ll t_d$. The inverse condition means the predominance of diffusion over freezing and, in general, the weakness of turbulence effects in the magnetic field evolution. In this context, it is convenient to introduce the dimensionless magnetic Reynolds number

$$\mathrm{Rm} = \frac{t_d}{t_\mathrm{v}} = \frac{L_H^2 \mathrm{v}}{\nu_m L_H} = \frac{L_H \mathrm{v}}{\nu_m}$$

According to the dynamo theory, the magnetic field generation occurs due to hydrodynamic motions of the stellar plasma in the presence of low but final diffusion $\nu_m$, i.e., there is a growth of the magnetic flux. In fact, for the magnetic field generation the magnetic Reynolds number should be high, but ultimate! Under astrophysical conditions, the hydrodynamic motions usually have a turbulence character with a scale that is significantly lower than the characteristic sizes of a celestial body. The typical example of such motions is turbulent convection in stars. At first glance, it seems that turbulent-convective flows can generate fields of only a scale of convective cells or smaller. Such magnetic fields are commonly referred to as small-scale ones. And they are indeed generated and seem to be observed. Certainly, such fields can be observed only on the Sun, for instance, in the form of pores. However, if a celestial body has differential (or solidbody) rotation with the linear velocity $\mathbf{U} \propto \Omega$, then the large-scale magnetic field generation is possible. Apparently, the large-scale fields govern stellar activity across the entire star.

The equations that describe the large-scale magnetic field evolution are obtained from the magnetic hydrodynamic equations by averaging the latter based on turbulent pulsations of $\mathbf{u}$: $\mathbf{v} = \mathbf{u} + \mathbf{U}$, $\mathbf{U}$ — the mean velocity associated with differential rotation, $\langle \mathbf{v} \rangle = \mathbf{U}$. This averaging represented by angular brackets yields

$$\frac{\partial \mathbf{B}}{\partial t} = \mathrm{rot}\left(\mathbf{U}\times\mathbf{B} + \boldsymbol{\mathcal{E}} - \nu_m\, \mathrm{rot}\mathbf{B}\right) \ . \tag{76}$$

Equation (76) is called the mean-field dynamo equation or sometimes the Steenbek equation; it is fairly similar to Equation (75) with a replacement of **H** ⇒ **B** and **v** ⇒ **U**, excluding the new term $\boldsymbol{\mathcal{E}} \equiv \langle \mathbf{u} \times \mathbf{b} \rangle$ that is commonly referred to as the inaccurate but well-established term ”turbulent electromotive force”. Here the fields **B** and **b** are the mean and fluctuation components of the total magnetic field **H** = **B** + **b**, respectively. Obviously, $\langle \mathbf{H} \rangle = \mathbf{B}$, $\langle \mathbf{v} \rangle = \mathbf{U}$. In fact, the basic problem of the linear mean-field dynamo theory is to establish a link between $\boldsymbol{\mathcal{E}} \equiv \langle \mathbf{u} \times \mathbf{b} \rangle$ and the mean magnetic field **B**.

In the pioneer paper of Steenbek, Krause, and Radler (1966), such a link was established for the isotropic and helical turbulent medium in the linear mean-field approximation. In this case, the turbulent electromotive force is

$$\boldsymbol{\mathcal{E}} \approx \alpha \mathbf{B} - \frac{1}{2}\left[\nabla \eta_T \times \mathbf{B}\right] - \eta_T \operatorname{rot} \mathbf{B}, \tag{77}$$

where $\eta_T$ is the turbulent magnetic viscosity, $-\nabla\eta_T/2$ — the velocity of diamagnetic pumping (pushing out) of the mean field **B** from turbulence, $\alpha$ — the value that characterizes the so-called alpha effect (see the discussion below), which hereafter will be called the alpha effect value. Let us discuss (77) in detail. Note that in the helical turbulent medium, where $\langle \mathbf{u} \cdot \operatorname{rot}(\mathbf{u}) \rangle \neq 0$, a random turn of the turbulent element is coordinated with its random velocity: for instance, the rising turbulent elements turn in a counterclockwise direction, while the sinking ones do vice versa. Notice that the hydrodynamic helicity density $\chi_u = \rho_0 \langle \mathbf{u} \cdot \operatorname{rot} \mathbf{u} \rangle$ in the turbulent nonmagnetic medium along with the turbulence energy density $E_u = \rho_0 \langle \mathbf{u}^2 \rangle / 2$ is a quasi-integral of motion. This means that the spatial spectra coincide for these values, as well as the response (relaxation) times to changes in the state of large-scale quantities in a turbulent medium. By an order of magnitude this time is $\tau_0 = \ell_0 \big/ \sqrt{\langle \mathbf{u}^2 \rangle}$. Here $\ell_0$ is the so-called integral scale of turbulence or the size of an energy-containing eddy. It is usually compared to the mixing path length $L_0$ of the turbulent stellar convection theory, supposing that it is of the same order of the height of a typical convective cell. This mixing path length, in its turn, is usually compared to the pressure height scale $h_p$. For example, in the known solar convective zone model of Spruit (1974), it is accepted that $L_0 \approx 1.5h_p$. If the latter is likely close to the true statement, then the estimate of $\ell_0 \approx L_0$ seems to be overvalued, following the results of laboratory studies of turbulent convection (Bukai et al., 2009) on the assumption that the Rayleigh number Ra $\simeq 10^8$, $\ell_0 \approx (0.25 \div 0.3) L_0$. This fact will be taken into account when constructing stellar dynamo models.

Equation (77) is well-reasoned at both the magnetic Reynolds number Rm $\ll$ 1 (Steenbek et al., 1966) and Rm $\gg$ 1 (Vainshtein, 1970; Vainshtein and Zeldovich, 1972; Molchanov et al., 1983; Dittrich et al., 1984). According to these studies, at the parameter Rm $\gg$ 1, which is characteristic of stars, the expressions for $\eta_T$ and $\alpha$ in the simplest case have the following form

$$\alpha = -\frac{\tau_0}{3}\langle \mathbf{u}\cdot \mathrm{rot}\mathbf{u}\rangle$$
$$\eta_T = \frac{\tau_0}{3}\langle \mathbf{u}^2\rangle = \frac{\ell_0}{3}\sqrt{\langle \mathbf{u}^2\rangle} .$$

Substituting (77) into (76), we obtain

$$\frac{\partial \mathbf{B}}{\partial t} = \mathrm{rot}\left[\left(\mathbf{U} - 0.5\nabla\eta_T\right)\times\mathbf{B} + \alpha\mathbf{B} - \left(\nu_m + \eta_T\right)\mathrm{rot}\mathbf{B}\right] . \quad (78)$$

For reference, we also give here an alternative form of Equation (78) for the vector potential of the magnetic field **B** = rot**A**

$$\frac{\partial \mathbf{A}}{\partial t} = \left(\mathbf{U} - 0.5\nabla\eta_T\right)\times \mathrm{rot}\mathbf{A} + \alpha\mathbf{B} - \left(\nu_m + \eta_T\right)\mathrm{rot}\left(\mathrm{rot}\mathbf{A}\right) + \nabla\psi .$$

(78a)

Here $\psi$ is the scalar potential, which is found from the so-called calibration condition (see, e.g., Landau and Lifshits, 1976). In the nonrelativistic physics, the following calibration conditions are usually used: $\psi$ = 0 or div**A** = 0 (Coulomb calibration), which is equivalent to

$$\Delta\psi = -\nabla\alpha\cdot\mathbf{B} - \mathrm{div}\left[\left(\mathbf{U} - 0.5\nabla\eta_T\right)\times\mathbf{B}\right] + \left[\nabla\left(\nu_m + \eta_T\right)\right]\cdot \mathrm{rot}\mathbf{B} .$$

Thus, it follows from a comparison of (78) and (75) that the mean-field diffusion abruptly increases under the influence of turbulence. Indeed, the ratio of diffusion coefficients in Equations (77) and (74) is

$$\frac{\nu_m + \eta_T}{\nu_m} = \frac{\ell_0\sqrt{\langle \mathbf{u}^2\rangle}}{3\nu_m} \approx \frac{\eta_T}{\nu_m} = \frac{1}{3}\mathrm{Rm} .$$

This yields a very simple estimate for the magnetic Reynolds number

$$\mathrm{Rm} = \frac{3\eta_T}{\nu_m} . \quad (78b)$$

As can be seen in Fig. 98, practically for all the spectral types of stars, except for M5 and F5, $\eta_T$ is almost constant inside the outer convective envelope and accounts for $\eta_T^{(G2)}$ = $1.2 \cdot 10^{13}$ cm$^2$/s for G2 stars and $\eta_T^{(M0)} = 2 \cdot 10^{12}$ cm$^2$/s for M0 stars. For these spectral types, $\nu_m^{(G2)} \approx 10^7$ cm$^2$/s and $\nu_m^{(M0)} \approx 10^7 = 2 \cdot 10^7$ cm$^2$/s. Here, to estimate the temperature of near-surface layers of such stars, their effective temperatures are used (see Table 25). This yields

$$\mathrm{Rm}^{(G2)} = 3\eta_T^{(G2)}/\eta^{(G2)} \approx 3.6\times10^6 \text{ and } \mathrm{Rm}^{(M0)} \approx 3\times10^5 .$$

Thus, the magnetic Reynolds numbers for these stars differ approximately by a factor of 10 and in both cases are rather high. Therefore, in stellar dynamo models, the contribution of ohmic diffusion $\nu_m$ into diffusion of the mean field **B** is negligible.

As it follows from Equation (77), inhomogeneous turbulence produces the effective velocity $\mathbf{v}_{\mathrm{eff}} = -0.5\nabla\eta_T$ directed toward lowering the turbulence level. It follows from Fig. 98 that for all main-sequence stars (except for M5), the turbulent magnetic viscosity $\eta_T$ is practically constant except for the bottom and upper layers of the convective zone, where the velocity $\mathbf{v}_{\mathrm{eff}}$ is directed toward the radiative core and the photosphere, respectively. This property of the convective zones became known as stellar diamagnetism or the diamagnetic pushing out of the mean field from the convective zone. The velocities of this pushing out are usually low. For instance, for the Sun, following the estimates of Vainshtein et al. (1980), they accounts for 1.2 km/s at the top of the convective zone and 13 m/s at its base.

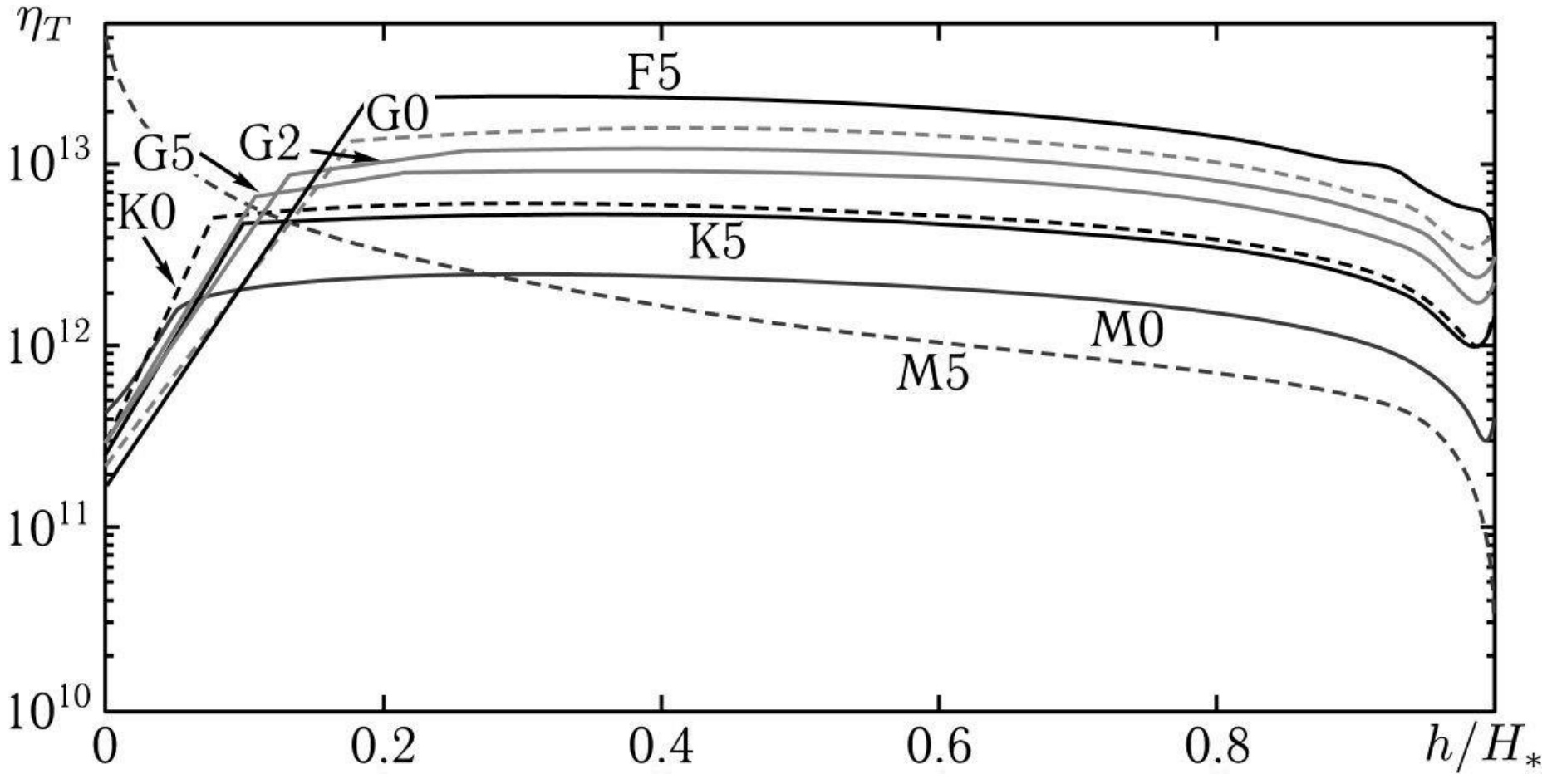

Fig. 98. Turbulent diffusion coefficient (in cm$^2$/s) in convective envelopes of stars of late spectral types from F5 to M5 calculated based on the theory of the mixing path length. The horizontal axis denotes a height of the point above the convective zone bottom attributed to its thickness

Table 25. Some characteristics of main-sequence stars significant for stellar magnetism

| Spectral type | $T_{eff}$, K | $R_*/R_\odot$ | $g_*/g_\odot$ | $H_c^{(*)}/R_*$ | $\mu_*$ | $\eta_T$,cm$^2$/s | $\Omega_\odot\tau_0$ | $\lvert D_{cr}\rvert$ |
|---|---|---|---|---|---|---|---|---|
| F0 | 7240 | 1.3 | 1.01 | ≈ 0 | ∞ | — | — | ∞ |
| F5 | 6540 | 1.2 | 0.903 | 0.1617 | 6.18 | $2.27 \cdot 10^{13}$ | 0.1–1.77 | 6026 |
| G0 | 5920 | 1.05 | 0.999 | 0.2499 | 4 | $1.57 \cdot 10^{13}$ | 0.23–2.3 | 1634 |
| G2(⊙) | 5780 | 1 | 1 | 0.2995 | 3.34 | $1.2 \cdot 10^{13}$ | 0.4–6.8 | 1142 |
| G5 | 5610 | 0.93 | 1.07 | 0.3162 | 3.16 | $8.9 \cdot 10^{12}$ | 0.78–10.02 | 1054 |
| K0 | 5240 | 0.85 | 1.08 | 0.3204 | 3.12 | $5.7 \cdot 10^{12}$ | 1.19–15.2 | 1054 |
| K5 | 4410 | 0.74 | 1.26 | 0.2888 | 3.47 | $5.4 \cdot 10^{12}$ | 0.78–10.2 | 1236 |
| M0 | 3920 | 0.63 | 1.18 | 0.4422 | 2.26 | $2.31 \cdot 10^{12}$ | 4.9–201 | 628 |
| M5 | 3120 | 0.32 | 2.05 | 1 | 1 | $(7.1–480) \cdot 10^{11}$ | $(12.5–1.5) \cdot 10^3$ | 21 |

Finally, let us focus on the most unusual term on the left-hand part of the mean-field equation (77) $\mathrm{rot}(\alpha B)$, which describes the alpha effect itself. Using the averaged Maxwell equations in the quasi-stationary ("electrotechnical", i.e., without shift current) approximation

$$\frac{1}{c}\frac{\partial \mathbf{B}}{\partial t} = -\mathrm{rot}\mathbf{E},$$

$$\mathrm{rot}\mathbf{B} \approx \frac{4\pi}{c}\mathbf{J}$$

let us write Equation (77) in the form of the effective Ohm law in plasma

$$\frac{c}{4\pi}\mathrm{rot}\mathbf{B} \approx \mathbf{J} = \sigma_T \left\{ \frac{1}{c}\left[ \left( \mathbf{U} - 0.5\nabla\eta_T \right) \times \mathbf{B} + \alpha\mathbf{B} \right] + \mathbf{E} \right\}. \qquad (79)$$

Here $\mathbf{E} = \langle \boldsymbol{\varepsilon} \rangle$ is the mean electric field, whereas $\mathbf{J} = \langle \mathbf{j} \rangle$ is the average current density, $\boldsymbol{\varepsilon}$ and $\mathbf{j}$ are the immediate fields and currents in plasma, turbulent conductivity $\sigma_T = 3\sigma/\mathrm{Rm}$ is far lower than ohmic; therefore, the mean field is frozen into the plasma only partially. This leads to the well-known paradox from the book of Moffatt (1978), which is as follows. The typical size of a spot is $L_{\odot Sp} \approx 3.5\times10^9$ cm, the typical coefficient of ohmic diffusion is $\nu_m^{(G2)} \approx 10^7$ cm$^2$/s. Estimating the time of ohmic diffusion, we get $t_d = L^2_{\odot Sp}/\eta^{(G2)} \approx 1.2\times10^{12}\,\mathrm{s} \approx 4\times10^4$ years. Thus, the time of ohmic evolution of a separate spot proves to be unreasonably high. This paradox is taken off by turbulent diffusion. The transition to turbulent conductivity decreases conductivity of the medium and diffusion time by a factor of $1.2 \cdot 10^6$. This yields the time of about twelve days. Such a time is in good agreement with the typical time of development of solar spots — of an order of a couple of weeks.

As for the alpha effect, $\mathbf{J} \sim \alpha\mathbf{B}$, then, as it follows from (79), it produces current that is parallel to the magnetic field! This phenomenon is rather rare and occurs, except for the turbulent dynamo theory, for instance, in the early nonturbulent Universe (Vilenkin, 1979) — the chiral effect generating current is parallel to the field solely due to quantum relativistic effects. On the other hand, Sokoloff (2000) supposes that the failure to include a similar term in the classical Ohm's law $\mathbf{j} = \sigma\left(\mathbf{v}\times\mathbf{H}/c + \boldsymbol{\varepsilon}\right)$ is a "serious miscalculation of classical electrodynamics". Because if there is a part of the current that is proportional to the electric field $\mathbf{j} \propto \boldsymbol{\varepsilon}$, then why is there no part of the current that is proportional to the magnetic field $\mathbf{j} \propto \mathbf{H}$?!

Let us show qualitatively that the alpha effect combined with stellar differential rotation — the omega effect — leads to the dynamo effect of the axisymmetric stellar field. Let us start from the omega effect.

**A.** Imagine a star having the poloidal field $\mathbf{B}_{\mathrm{pol}}$ (Fig. 99) and latitudinal differential rotation $\partial\Omega/\partial\vartheta \neq 0$. Suppose for certainty that the equator rotates faster than the poles, then, due to the partial freezing, the lines of force of the poloidal field start to wind around a star similar to a string wound around a spool (Fig. 99).

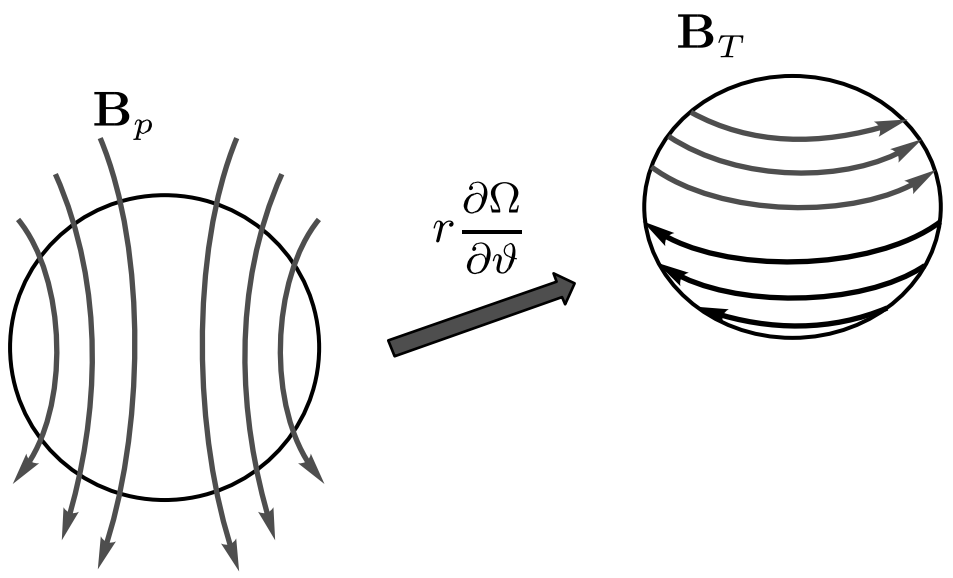

Fig. 99. The omega effect: lines of force of the poloidal field start to wind around a star similar to a string wound around a spool, generating the toroidal field

Since such a string can be wound many times until turbulent diffusion stops the process, the generated toroidal field $\mathbf{B}_{\text{tor}}$ may be far larger than the initial poloidal field $\mathbf{B}_{\text{pol}}$. Note that the fields in opposite hemispheres have an opposite sign. It is important to bear in mind that the omega effect itself cannot ensure an exponential steady growth of the magnetic flux. The reason is that, although the toroidal field $\mathbf{B}_{\text{tor}}$ increases, the poloidal field $\mathbf{B}_{\text{pol}}$ having no source in the presence of only the omega effect decays under the influence of turbulent diffusion. After this, the toroidal fields $\mathbf{B}_{\text{tor}}$ decays as well. Note that the discussed decay occurs in the star that is totally deprived of turbulence, since the equation proves to be the same as that in the turbulent case but with molecular diffusion instead of turbulent one.

For the Sun, the time of such diffusion $t_d$ is far longer than the cycle period: $t_d = R_\odot^2 / \nu_m^{(G2)} \approx 1.63$ million years. This statement is a particular case of the Cowling anti-dynamo theorem (1957) asserting that "in axisymmetric flows the axisymmetric magnetic field generation is impossible". It is obvious that turbulent flows are non-axisymmetric. This violates the condition of the Cowling anti-dynamo theorem. Thus, the omega effect is able to ensure only a temporary increase of the toroidal field $\mathbf{B}_{\text{tor}}$ rather than dynamo, since there is no source maintaining the poloidal field $B_{\text{pol}}$ in the omega mechanism.

**B.** In this context, a necessary source of the poloidal field is the alpha effect: it generates mean electric current $\mathbf{J} \sim \alpha\mathbf{B}$, in particular, mean electric toroidal current that is proportional to the toroidal field $\mathbf{B}_{\text{tor}}$: $\mathbf{J}_{\text{tor}} \sim \alpha\mathbf{B}_{\text{tor}}$ (Fig. 100). Such current in accordance with the approximated Maxwell equation $\text{rot}(\mathbf{B}_{\text{pol}}) \approx 4\pi\mathbf{J}_{\text{tor}}/c$ generates the poloidal field, which is added to initial one, enhancing the poloidal field:

$$\mathbf{B}_p(\mathbf{r}) = \frac{4\pi}{c}\text{rot}^{-1}(\mathbf{J}_T) = \frac{1}{c}\int \frac{(\mathbf{r}-\mathbf{r}')\times\mathbf{J}_T(\mathbf{r}')}{|\mathbf{r}-\mathbf{r}'|^3} d^3\mathbf{r}' \sim \frac{1}{4\pi\eta_T}\int \frac{(\mathbf{r}-\mathbf{r}')\times(\alpha\mathbf{B}_T)_{\mathbf{r}'}}{|\mathbf{r}-\mathbf{r}'|^3} d^3\mathbf{r}' \qquad (80)$$

Here $\mathbf{r}$ and $\mathbf{r}'$ are the points inside the stellar convective zone and magnetic field observations, respectively. The latter equality in the chain (80) is only approximate (on an order of magnitude), since not all toroidal current $\mathbf{J}_{\text{tor}}$ is generated by the alpha effect.

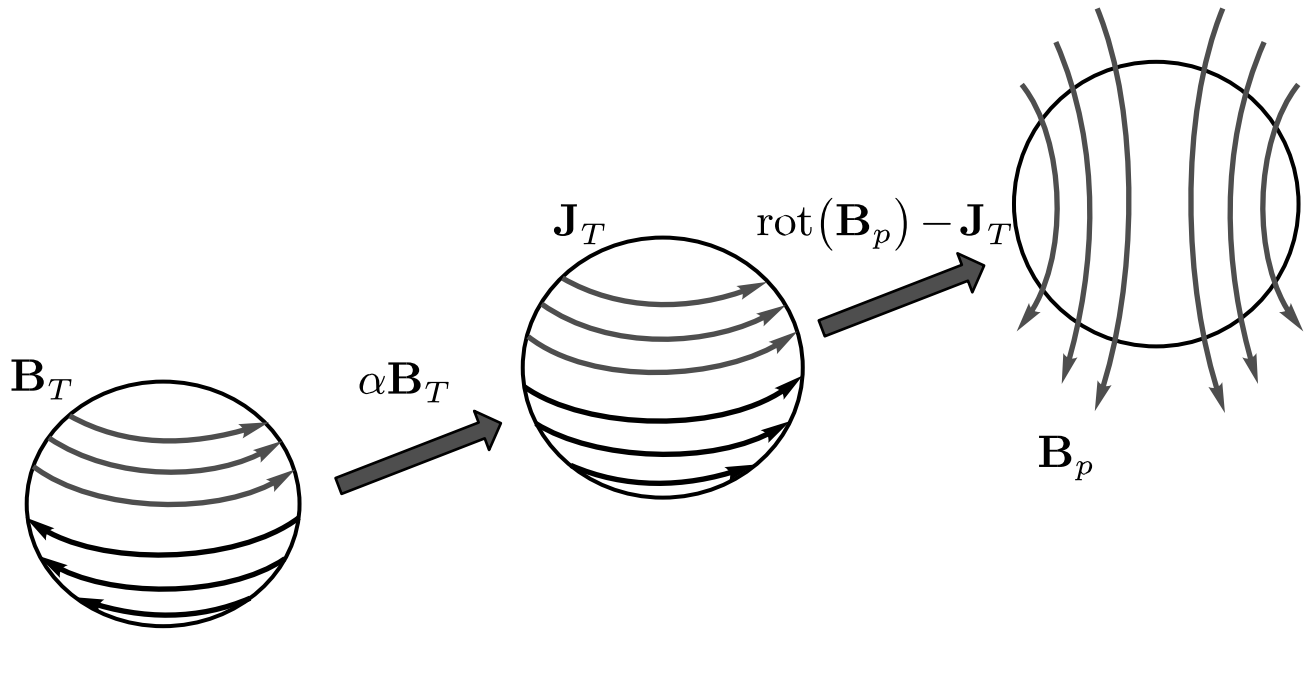

Fig. 100. A scheme of transforming the toroidal field $\mathbf{B}_{\text{tor}}$ into the poloidal field $\mathbf{B}_{\text{pol}}$ due to the alpha effect

The combined action of the omega and alpha effects encloses the instability loop and creates the $\alpha$–$\omega$ dynamo. This mechanism is graphically displayed in Fig. 101.

Notice that the alpha effect is able to maintain dynamo by its own, i.e., without the omega effect. The reason is that, according to the relation $\mathbf{J} \sim \alpha\mathbf{B}$, the toroidal field generates toroidal current $\mathbf{J}_{\mathrm{tor}} \sim \alpha\mathbf{B}_{\mathrm{tor}}$, whereas the poloidal field — poloidal current $J_{\mathrm{pol}} \sim \alpha\mathbf{B}_{\mathrm{pol}}$. Thus, according to Equation (80) and analogous equation for the toroidal field, there arises the following chain of generation:

$$\begin{array}{ccc} \mathbf{B}_T & \xrightarrow{\mathbf{J}_T \sim \alpha \mathbf{B}_T} & \mathbf{J}_T \\ \Uparrow & & \Downarrow \\ \mathbf{J}_p & \xleftarrow{\mathbf{J}_p \sim \alpha \mathbf{B}_p} & \mathbf{B}_p \end{array}$$

This chain, obviously, comprises the generation loop which generates the $\alpha^2$ dynamo.

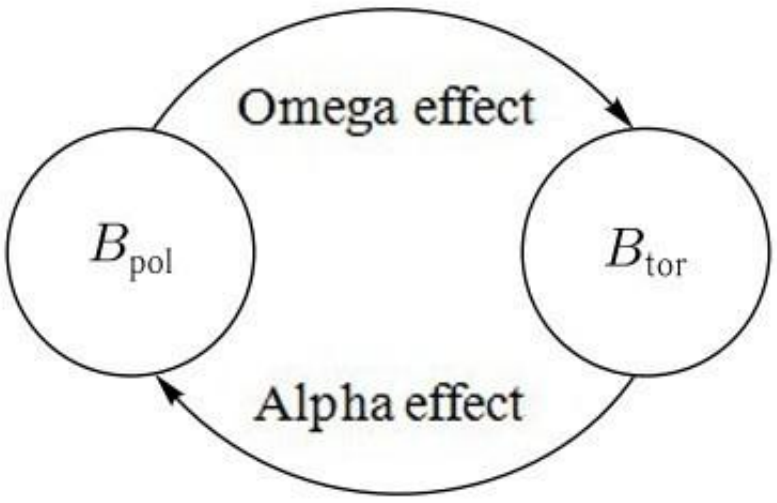

Fig. 101. A scheme of the $\alpha$–$\omega$ dynamo mechanism: the toroidal field $\mathbf{B}_{\mathrm{tor}}$ due to the alpha effect transforms into the poloidal field $\mathbf{B}_{\mathrm{pol}}$ (see Fig. 100), and the latter, in its turn, through the omega effect transforms into toroidal

The power of any dynamo mechanism is described through the dimensionless parameters $\mathrm{R}_\alpha$ for the alpha effect and $\mathrm{R}_\omega$ for the omega effect, which are analogous to the magnetic Reynolds number (see Sect. 4.3.2):

$$\mathrm{R}_\Omega = \frac{R_*^3 |\nabla\Omega|}{\eta_T}; \ \mathrm{R}_\alpha = \frac{R_*\alpha}{\eta_T},$$

where $R_*$ is the stellar radius. In most cases, $\mathrm{R}_\Omega >> \mathrm{R}_\alpha$; therefore, the $\alpha$–$\omega$ dynamo prevails. In the following section, we consider in detail the description of these mechanisms.

## 4.3.2. Dynamo Numbers and Dynamo Waves

**4.3.2.1. Dynamo number.** Equation (79) is still extremely complicated and cannot be used directly as an instrument for analysis of stellar dynamo. To move forward, it is useful to recall that div**B** = 0, i.e., the magnetic field is a solenoidal field: **B** = rot**A**, where **A** is the magnetic field vector potential. Note that we are interested in the behavior of the magnetic field on a spherical body — a star. Then, according to Vainshtein et al. (1980), the magnetic field vector potential **A** has the form

$$\mathbf{A} = R_* \operatorname{rot}\left(\hat{\mathbf{r}}\Psi(r,\vartheta,\varphi,t)\right) + \hat{\mathbf{r}}\,\Phi\,(r,\vartheta,\varphi,t) = R_* \nabla\,\Psi \times \hat{\mathbf{r}} + \hat{\mathbf{r}}\,\Phi\ ,$$

where $\hat{\mathbf{r}} = \mathbf{r}/r$ is the unit vector directed along the stellar radius, $\vartheta$ and $\varphi$ are the standard spherical angular coordinates associated with stellar centric coordinates with the latitude $\phi$ and the longitude $\lambda$ as follows: $\vartheta = \pi/2 - \phi$ and $\varphi = \lambda$. In the expression, for the vector potential **A** the first summand is the poloidal field vector potential, whereas the second summand is the toroidal field vector potential; the stellar radius $R_*$ is introduced into the first summand for the both functions $\Psi$ and $\Phi$ to have dimension of the magnetic field vector potential. Substituting this form of the vector potential into Equation (78), we can obtain for functions $\Psi$ and $\Phi$ an extremely cumbersome non-axisymmetric system of equations (see Vainshtein et al., 1980). Taking into account that for stars the "acquisition" of information on the non-axisymmetric magnetic field component from observations is rather complicated, let us consider an axisymmetric version of the mean-field dynamo problem: $\partial\Psi/\partial\varphi = \partial\Phi/\partial\varphi = 0$. Then for the magnetic field vector potential **A** we have

$$\mathbf{A} = R_* \nabla \Psi \times \hat{\mathbf{r}} + \hat{\mathbf{r}}\Phi = -\hat{\boldsymbol{\varphi}} R_*/r\left(\partial\Psi/\partial\theta\right) + \hat{\mathbf{r}}\Phi = \hat{\boldsymbol{\varphi}} A_\varphi + \hat{\mathbf{r}}\Phi .$$

Here $\hat{\boldsymbol{\varphi}}$ is the unit vector of the toroidal field that is parallel to the stellar equator. Calculating the magnetic field at such a vector potential, we find

$$\mathbf{B} = \mathrm{rot}(\mathbf{A}) = \mathrm{rot}\left(\hat{\boldsymbol{\varphi}}A_\varphi + \hat{\mathbf{r}}\Phi\right) = \mathrm{rot}\left(\hat{\boldsymbol{\varphi}}A_\varphi\right) - \hat{\mathbf{r}} \times \nabla\Phi = \mathrm{rot}\left(\hat{\boldsymbol{\varphi}}A_\varphi\right) - \hat{\boldsymbol{\varphi}}/r\left(\partial\Phi/\partial\theta\right)$$
$$= \mathrm{rot}\left(\hat{\boldsymbol{\varphi}}A_\varphi(r,\vartheta,t)\right) + \hat{\boldsymbol{\varphi}} B_\varphi(r,\vartheta,t) .$$

Thus, we look for a solution of the axisymmetric dynamo problem in the form of a sum of the poloidal and toroidal fields determined by the expressions

$$\mathbf{B}_p = \mathrm{rot}\left(\hat{\boldsymbol{\varphi}}A_\varphi\right) = \nabla A_\varphi \times \hat{\boldsymbol{\varphi}} - \hat{\mathbf{r}}A_\varphi/r = \hat{\mathbf{r}}/r\sin\vartheta\left[\partial A_\varphi \sin(\vartheta)/\partial\vartheta\right] - \hat{\boldsymbol{\theta}}/r\left[\partial(A_\varphi r)\ /\partial r\right],$$

$$\mathbf{B}_T = \hat{\boldsymbol{\varphi}}B_\varphi(r,\vartheta,t) .$$

Let us substitute these solutions into Equations (78) and (78a) and get their $\varphi$ components. This yields

$$\frac{\partial B}{\partial t} = r\sin(\vartheta)\left[\nabla\Omega \times \nabla A\right]_\varphi + \frac{1}{r}\left[\frac{\partial}{\partial r}\left(\eta_T \frac{\partial(rB)}{\partial r}\right) + \frac{\eta_T}{r}\frac{\partial}{\partial\vartheta}\left(\frac{1}{\sin\vartheta}\frac{\partial(\sin\vartheta B)}{\partial\vartheta}\right)\right]$$
$$-\frac{1}{r}\left[\frac{\partial}{\partial r}\left(\alpha\frac{\partial(rA)}{\partial r}\right) + \frac{1}{r}\frac{\partial}{\partial\vartheta}\left(\frac{\alpha}{\sin\vartheta}\frac{\partial(\sin\vartheta A)}{\partial\vartheta}\right)\right] , \qquad (81)$$
$$\frac{\partial A}{\partial t} = \alpha B + \eta_T \Delta_s A$$

where $\Delta_s = \Delta - \left(r\sin\vartheta\right)^{-2}$ is the Stokes operator, which for toroidal fields replaces the Laplace operator in spherical geometry, since $\mathrm{rot}\left[\mathrm{rot}\left(\hat{\boldsymbol{\varphi}}A\right)\right] = -\hat{\boldsymbol{\varphi}}\Delta_s A$. Note that

$$\frac{1}{r}\left[\frac{\partial^2(rB)}{\partial r^2} + \frac{1}{r}\frac{\partial}{\partial\vartheta}\left(\frac{1}{\sin\vartheta}\frac{\partial(\sin\vartheta B)}{\partial\vartheta}\right)\right] = \Delta_s B \ .$$

Notice that the first term in the equation for the toroidal field of the system (81) describes the omega effect, i.e., generation of the toroidal magnetic field from the poloidal one by differential rotation: $U_\varphi = r\Omega(r,\vartheta)\sin(\vartheta),\ \nabla\Omega \neq 0$. This term is as follows:

$$r\sin(\vartheta)\left[\nabla\Omega\times\nabla A\right]_{\varphi}=\frac{1}{r}\left(\frac{\partial\Omega}{\partial r}\frac{\partial}{\partial\vartheta}-\frac{\partial\Omega}{\partial\vartheta}\frac{\partial}{\partial r}\right)(rA\sin(\vartheta))$$

Here the turbulent diffusion coefficient is assumed to depend on depth only, i.e., $\eta_T=\eta_T(r)$, whereas the coefficient $\alpha$ having, by the way, the velocity dimension from the depth and latitude $\phi$: $\alpha=\alpha(r,\vartheta)$ ($\vartheta=\pi/2-\phi$). In the system of equations (81), there is no contribution of the scalar potential $\psi(r,\vartheta)$ since in the axisymmetric problem there is no longitudinal dependence and $(\nabla\psi)_\varphi\sim\partial\psi/\partial\varphi=0$ (see Equation (78a)). For simplicity of writing formulas, $\varphi$ indices are omitted for magnetic field and vector potential toroidal components. For the further analysis it is useful to make the system (81) dimensionless. To this end, let us represent $\Omega=\Omega_*+\delta\Omega_*\omega(r,\vartheta)$, $\eta_T=\eta_T^{(0)}\bar{\eta}(r)$, $\alpha=\alpha_0\bar{\alpha}(r,\vartheta)$. In these expressions, $\eta_T^{(0)},\alpha_0$ are the maximum values of the turbulent diffusion coefficient and the alpha effect in the stellar convective zone, respectively; $\Omega_*$ is the angular rotation velocity of the stellar equator; $\bar{\eta}(r)$, $\bar{\alpha}(r,\vartheta)$ are the dimensionless functions describing distributions of the turbulent diffusion coefficient and the alpha effect value on the stellar convective zone, respectively. The maximum values of these functions are equal to unity. The difference in the angular rotation velocity $\delta\Omega_*$ and the dimensionless function $\omega(r,\vartheta)$ are selected so that the maximum function value $\left|R_*\nabla\omega(r,\vartheta)\right|$ is equal to unity. Let us measure the toroidal magnetic field in units of $B_0: B=B_0\bar{B}(\bar{r},\vartheta)$, $\bar{r}=r/R_*$ is the dimensionless radial coordinate inside a star and on its surface $\bar{r}=1$, $\bar{B}(\bar{r},\vartheta)$ is the dimensionless toroidal field. In this case, it is useful to measure a vector potential of the poloidal magnetic field in units of $A_0=R_*\mathrm{R}_\alpha B_0$, i.e., $A=A_0\bar{A}(\bar{r},\vartheta)$, $\bar{A}(\bar{r},\vartheta)$ is the dimensionless vector potential of the poloidal magnetic field. We will also measure the time in units of the characteristic time of turbulent diffusion $t_d=R_*^2/\eta_T^{(0)}$, $t=\bar{t}\,t_d$.

After such dimensionlessness the system (81) is as follows:

$$\begin{aligned}\frac{\partial B}{\partial t}&=D\left\{r\sin\vartheta\left[\nabla\omega\times\nabla A\right]_\varphi\right\}+\frac{1}{r}\left[\frac{\partial}{\partial r}\left(\bar{\eta}_T\frac{\partial(rB)}{\partial r}\right)+\frac{\bar{\eta}_T}{r}\frac{\partial}{\partial\vartheta}\left(\frac{1}{\sin\vartheta}\frac{\partial(\sin\vartheta B)}{\partial\vartheta}\right)\right]\\&-\frac{(\mathrm{R}_\alpha)^2}{r}\left[\frac{\partial}{\partial r}\left(\bar{\alpha}\frac{\partial(rA)}{\partial r}\right)+\frac{1}{r}\frac{\partial}{\partial\vartheta}\left(\frac{\bar{\alpha}}{\sin\vartheta}\frac{\partial(\sin\vartheta A)}{\partial\vartheta}\right)\right]\\\frac{\partial A}{\partial t}&=\bar{\alpha}B+\bar{\eta}_T\Delta_s A\end{aligned}\qquad(82)$$

Here for simplicity of denotations the horizontal line that means dimensionless of values is omitted in expressions containing $\bar{B},\bar{A},\bar{r},\bar{t}$ and neglected by diamagnetic pushing out of the field. The first equation of the system (82) describes the behavior of the toroidal field and comprises two dimensionless parameters: $\mathrm{R}_\alpha$ and the *dynamo number* $D=\mathrm{R}_\alpha\mathrm{R}_\omega$, $\mathrm{R}_\omega=R_*^2\delta\Omega_*/\eta_T^{(0)}$, $\mathrm{R}_\alpha=R_*\alpha_0/\eta_T^{(0)}$. Thus, the *dynamo number* is a product of two effective Reynolds numbers that are based on the alpha effect value and the differential rotation velocity $\delta\Omega_* R_*$. The obtained system of equations (82) is still complicated for a direct analytical study. Significant simplifications are obtained if one takes into consideration the practical constancy of the turbulent diffusion coefficient inside almost all the stellar convective envelopes, i.e., $\bar{\eta}_T = 1$, and considers $\mathrm{R}_\alpha << \mathrm{R}_\omega$; then the $\alpha$–$\omega$ dynamo mechanism dominates on a star. As a result, we get

$$\begin{aligned} \frac{\partial B}{\partial t} &= D\left\{ r \sin\vartheta \left[ \nabla\omega \times \nabla A \right]_\varphi \right\} + \Delta_s B \\ \frac{\partial A}{\partial t} &= \bar{\alpha} B + \Delta_s A \end{aligned} \tag{83}$$

**4.3.2.2. Dynamo Numbers of Stars and Properties of the Stellar Interior.** If we measure stellar differential rotation in angular velocities of solar rotation $\delta\Omega_* = \Omega_\odot \delta\bar{\Omega}_*$ ($\Omega_\odot \simeq 2.78\times10^{-6}$ rad/s), its radius — in solar radii — $R_* = R_\odot \bar{R}_*$, whereas the turbulent diffusion coefficient — in units of $\eta_0 = 10^{12}$ cm$^2$/s and $\eta_T^{(0)} = \eta_0 \bar{\eta}_T^{(0)}$ (see Table 25), then it is easy to get

$$\mathrm{R}_\omega = \frac{\bar{R}_\odot^2 \Omega_\odot}{\eta_0} \frac{\bar{R}_*^2 \delta\bar{\Omega}_*}{\bar{\eta}_T^{(0)}} \simeq 1.36\times10^4 \frac{\bar{R}_*^2 \delta\bar{\Omega}_*}{\bar{\eta}_T^{(0)}} \,.$$

Let us estimate, as an example, this parameter for the Sun. As it follows from observations, $\delta\bar{\Omega}_\odot \simeq 0.2$ — this is a 20% difference between the equator and the near-polar zone. As for turbulent diffusion, then the simple theory of the mixing path length yields $\eta_T^\odot = 12\times10^{12}$ (see Table 25). Given the laboratoty measurements (Bukai et al., 2009), this value proves to be by a factor of 16 lower, which yeilds for $\bar{\eta}_T^{(\odot)} \approx 0.75$. Finally, we get $\mathrm{R}_\omega^{(\odot)} = 3.63\times10^3$. Notice that, contrary to $\mathrm{R}_\omega$, the estimate of $\mathrm{R}_\alpha$ is far less obvious. The reason is that hydrodynamics of the convective zones of *rotating* stars is still weakly developed. In other words, even the most modern calculations based on the mixing length theory, unfortunately, cannot be applied directly to the stellar dynamo theory, since they do not take into account the influence of the Coriolis forces on turbulence. The point is that as rotation increases, these forces produce two-dimensionalization of turbulence (Busse, 1970) and at the Coriolis parameter $2\Omega_* \tau_0^{(*)}(r) \geq 1$ (see Table 25) make it significantly two-dimensional, i.e., significantly anisotropic. It is true at least for stars of all spectral types later than F5. The attempts to overcome this problem were undertaken since the 60s of the XX$^{\text{th}}$ century, and the Krause formula was obtained (Steenbek et al., 1966)

$$\alpha = -\tau_0 \frac{\langle \mathbf{u} \cdot \mathrm{rot}\,(\mathbf{u}) \rangle}{3} \approx \frac{2}{3} \ell_0^2 \mathbf{\Omega}_* \cdot \frac{\nabla\left(\rho_0 \sqrt{\langle u^2 \rangle}\,\right)}{\rho_0 \langle u^2 \rangle} \,.$$

This relation is valid at $2\Omega_* \tau_0^{(*)}(r) << 1$. Since the mixing length $\ell_0$ increases with depth inside stellar convective zones, the coefficient $\alpha$ should also increase with depth. Further investigations (Dyrney and Robinson, 1982; Kichatinov and Rüdiger, 1993; Kleeorin and Rogachevskii, 2003) showed that the Krause formula is disturbed at $2\Omega_* \tau_0^{(*)}(r) \geq 1$. This is manifested in the fact that, firstly, at the depth where the Coriolis parameter $2\Omega_* \tau_0^{(*)}(r)$ achieves unity, the alpha effect value reaches its maximum; secondly, proportionality of the alpha effect value and helicity $\chi_u = \langle \mathbf{u} \cdot \mathrm{rot}(\mathbf{u}) \rangle$ is disturbed. Results of the calculation of the alpha effect value as the function $\Omega_* \tau_0^{(*)}(r)$ from Kleeorin and Rogachevskii (2003) are given in Fig. 102. Here, for comparison, the value of $-\tau_0 \chi_u / 3 = -\tau_0 \langle \mathbf{u} \cdot \mathrm{rot}(\mathbf{u}) \rangle / 3$ is presented as a function of the Coriolis parameter.

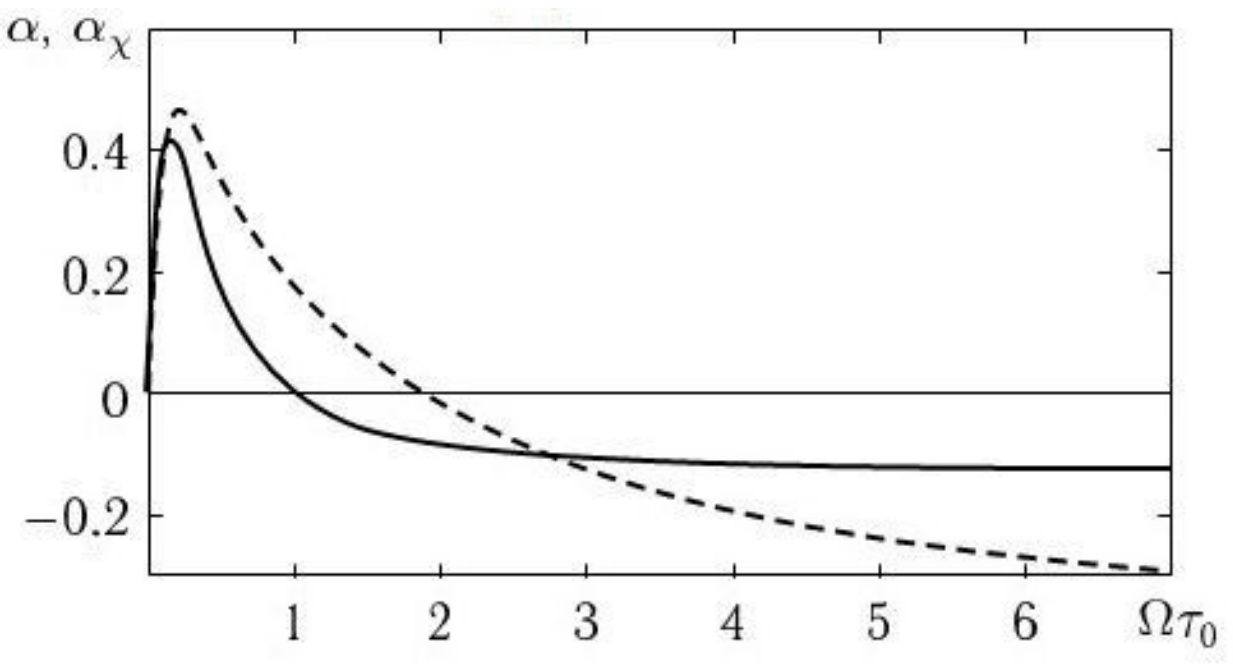

Fig. 102. Dependences of the alpha effect (*solid line*) $\alpha_\chi = -\tau_0 \chi_u / 3$ (*dotted line*) as a function of the Coriolis parameter $\Omega_* \tau_0^{(*)}(r)$ in dimensionless units near the stellar pole. Values are normalized on $\ell_0 \sqrt{\langle u^2 \rangle}\, |\nabla \rho(r)| / 6\rho(r)$ (Kleeorin and Rogachevskii)

Since this relates to the case of mainly fast rotation accompanied by two-dimensionalization of turbulence, the appropriate anisotropy is taken into account in calculations. In this case, it was assumed that energy of the two-dimensional part of turbulence accounts for 60% of the energy of its three-dimensional part. As for the form of thermal disturbances (thermics), their horizontal size is one and a half times larger than the vertical size.

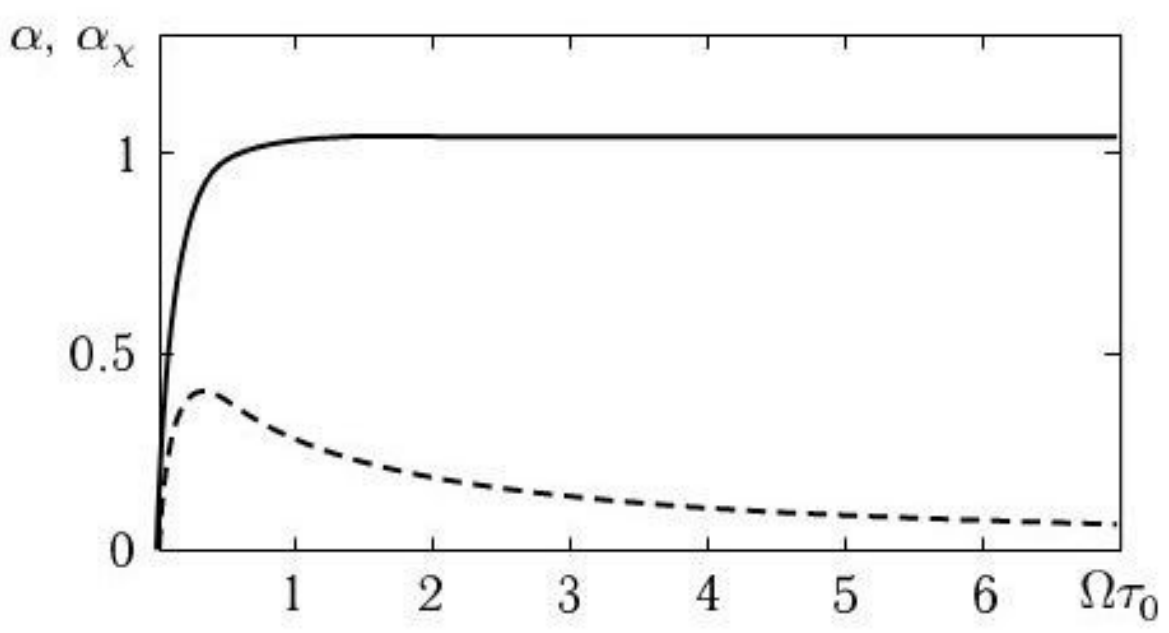

Fig. 103. The same as in Fig. 102 but in a fully isotropic case

Taking into account that, on the one hand, the alpha effect maximum falls on the radius in the convective zone at which $\Omega_* \tau_0^{(*)}(r) \approx 0.5$ (see Figs. 102 and 103) and, on the other hand, the constancy of turbulent viscosity inside the stellar convective zone $\eta_T(r) = \langle [u(r)]^2 \, \tau_0^{(*)}(r) \rangle / 3$ , for the maximum alpha effect we obtain

$$\alpha_*^{(\max)} = \bar{\alpha}_0 \frac{\ell_0 \sqrt{\Omega_* \eta_T^{(0)}}}{\sqrt{6}\rho(r)} |\nabla \rho(r)| \approx \frac{3\bar{\alpha}_0}{20} \sqrt{\Omega_* \, \eta_T^{(0)}}$$

Here $\bar{\alpha}_0$ is the dimensionless coefficient varying between 0.3 and unity (Kleeorin and Rogachevskii, 2003). Estimating

$$\mathrm{R}_\alpha = \frac{\alpha_*^{(\max)} R_*}{\eta_T^{(0)}} = \frac{3\bar{\alpha}_0 R_*}{20} \sqrt{\frac{\Omega_*}{\eta_T^{(0)}}} = \frac{3\bar{\alpha}_0}{20} \sqrt{\frac{R_\odot^2 \Omega_\odot}{\eta_0}} \sqrt{\frac{\bar{R}_*^2 \bar{\Omega}_*}{\bar{\eta}_T^{(0)}}} \simeq 18\bar{\alpha}_0 \sqrt{\frac{\bar{R}_*^2 \bar{\Omega}_*}{\bar{\eta}_T^{(0)}}} ,$$

we get, for instance for the Sun, $\mathrm{R}_\alpha = 5.2\bar{\alpha}_0$. Taking into account the obtained result, for the dynamo number we have

$$D = \mathrm{R}_\omega \mathrm{R}_\alpha \simeq 6.12 \times 10^4 \bar{\alpha}_0 \delta \bar{\bar{\Omega}}_* \bar{R}_*^3 \sqrt{\frac{\bar{\Omega}_*^3}{\left(\bar{\eta}_T^{(0)}\right)^3}} , \tag{83a}$$

where $\delta \bar{\bar{\Omega}}_* = \delta \Omega_* / \Omega_*$ (it should be recalled that $\delta \bar{\Omega}_* = \delta \Omega_* / \Omega_\odot$ ). There is a relation between them: $\delta \bar{\bar{\Omega}}_* = \delta \bar{\Omega}_* / \bar{\Omega}_*$ . For the Sun this estimate yields $D_\odot \simeq 0.94 \times 10^5 \bar{\alpha}_0$. Let us note one more important circumstance: as rotation increases, the position of the alpha effect maximum is shifted to the stellar surface. This can be clear on the example of the Sun (Fig. 104).

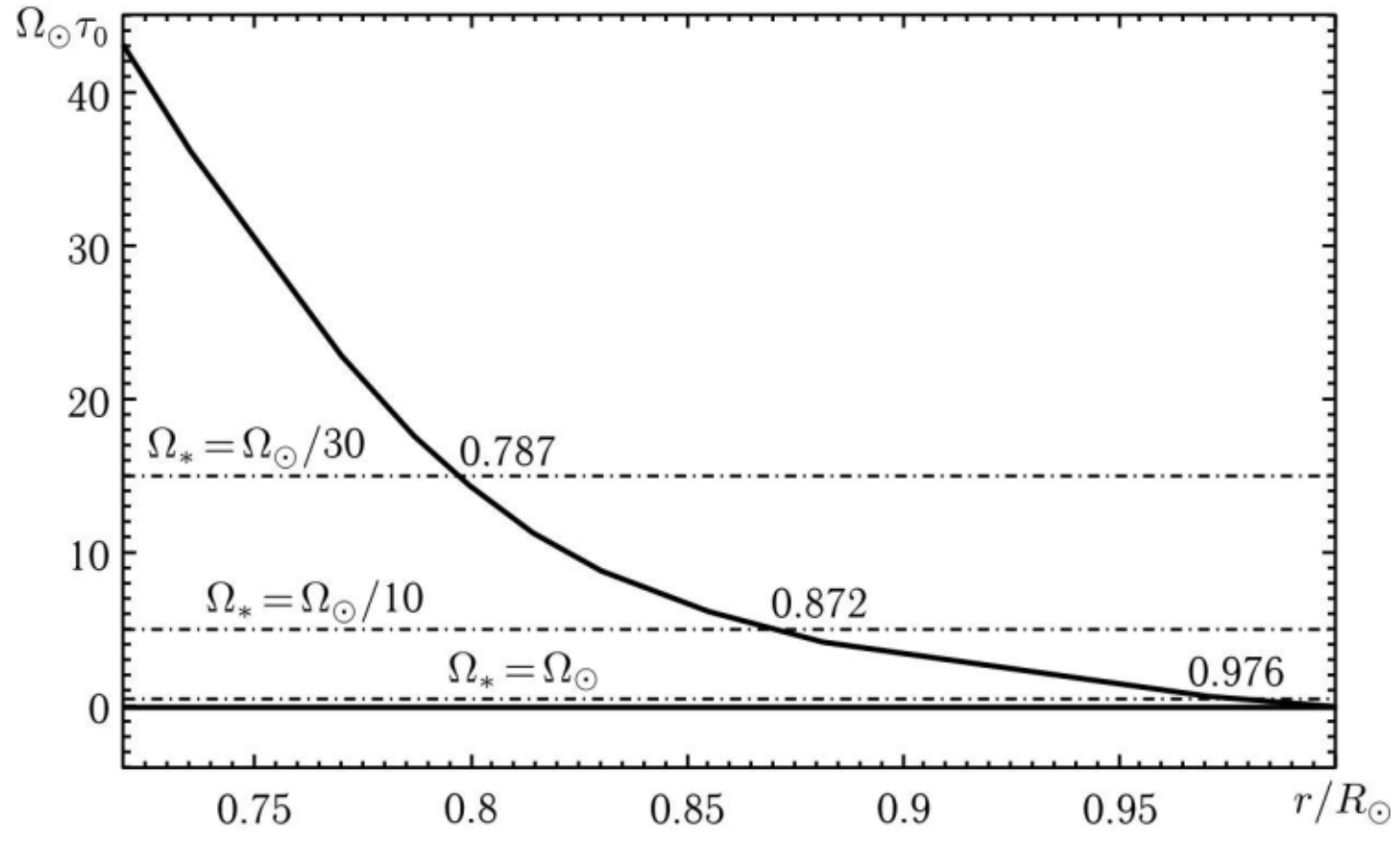

Fig. 104. The solid line denotes a dependence of the dimensionless parameter $\Omega_\odot \tau_0(r)$ in the convection zone on the Sun on the dimensionless radial component $r/R_\odot$ following Spruit (1974). Dash-dotted lines correspond to the solutions of the equation $2\Omega_* \tau_0^{(\odot)}(r) = 1$ at different stellar rotations. The angular rotation velocity of a star $\Omega_*$ in this graph varies from $\Omega_\odot/30$ to $\Omega_\odot$. Whereas the position of the alpha effect maximum from a depth of 140 Mm ( $\approx 0.8R_\odot$ ) up to a depth of 17 Mm (an approximate depth of supergranules)(Rogachevskii and Kleeorin, 2018)

There are no doubts that along with the radial position of the alpha effect maximum, it is also of importance the alpha effect distribution in longitude, which also depends on depth, since this distribution depends on the parameter $\Omega_* \tau_0$. The latter parameter, obviously, is a function of the radial component in the stellar interior. The paper of Kleeorin and Rogachevskii (2003) is devoted to the calculation of these dependences. Results of these calculations are presented in Fig. 105. The solid line corresponds to the case when the energy of the two-dimensional part of turbulence exceeds by a factor of 6.5 the energy of the three-dimensional part of turbulence, the vertical size of thermics is 1.54 times larger than the horizontal size; the dotted line denotes the case when the energy of the two-dimensional part of turbulence accounts for 60% of the energy of the three-dimensional part of turbulence, the horizontal size of thermics is 1.48 times larger than the vertical size; the dash-dotted line corresponds to the case when the energy of the two-dimensional part of turbulence exceeds by a factor of 6.5 the energy of the three-dimensional part of turbulence, the horizontal size of thermics is 1.6 times larger than the vertical size; dots correspond to the case when the energy of the two-dimensional part of turbulence exceeds by a factor of 6.5 the energy of the three-dimensional part of turbulence, the vertical size of thermics is 1.28 times larger than the horizontal size. The dash-dotted curve denotes the dependence $\alpha/5$. The values are normalized on $\ell_0 \sqrt{\langle u^2 \rangle} \, |\nabla \rho(r)| / 6\rho(r)$. Notice that in the case of slow rotation there is no such diversity of behavior.

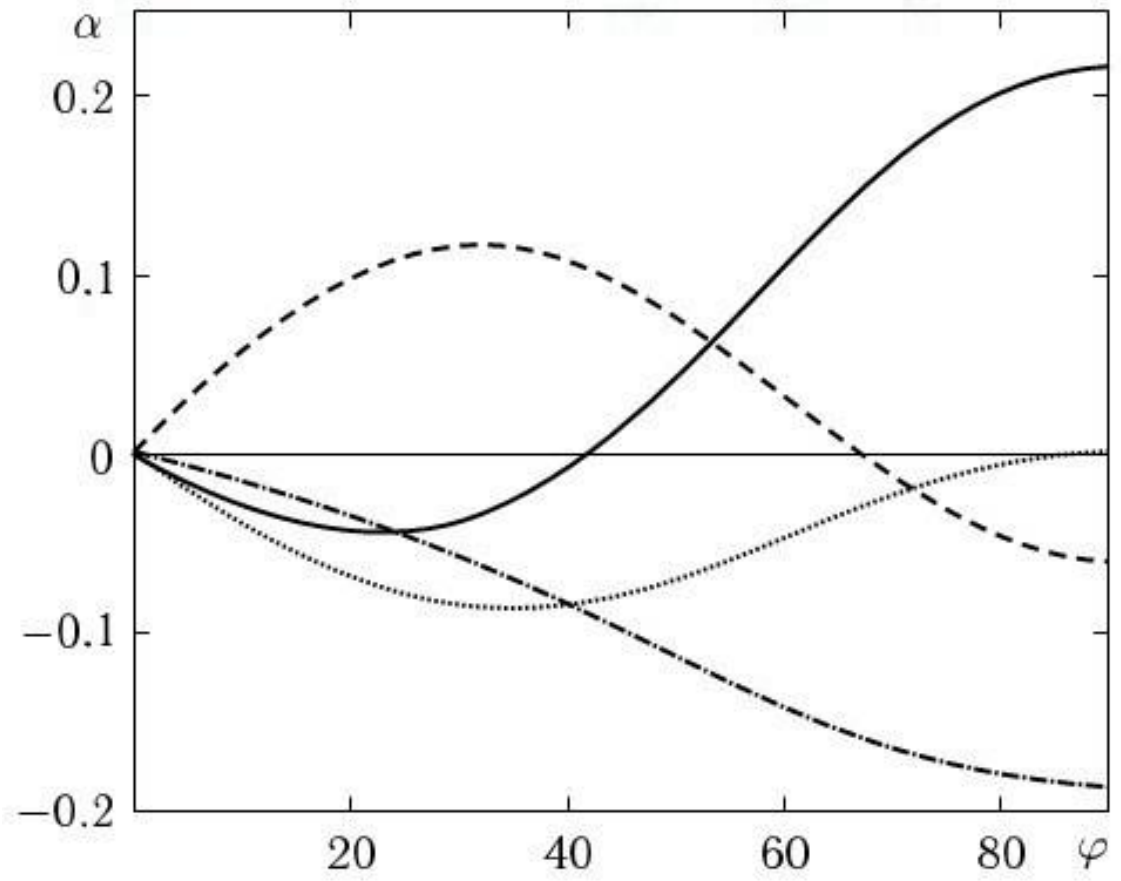

Fig. 105. Dependences of the alpha effect on latitude at fast rotation of a star $\Omega_* \tau_0^{(*)} = 5$ and at different values of mechanical and thermal anisotropy of turbulent convection (Kleeorin and Rogachevskii, 2003)

It follows from Fig. 105 that the alpha effect behavior at fast rotations, i.e., in the stellar interior, is quite sensitive to the details of both mechanical and thermal anisotropy of turbulent convection. It is this information that is practically completely unavailable in observations and absolutely poor-studied theoretically. Furthermore, this is true not only for stars but even for the Sun! Apparently, future studies of the stellar interior should be devoted to studying anisotropy of turbulent convection, including its two-dimensionalization. Only then it makes sense to construct detailed models of the stellar dynamo of certain stars. Therefore, in the next section, wherever possible, the general properties of the linear solutions to dynamo equations will be considered in detail.

**4.3.2.3. Stellar Dynamo Waves and Critical Dynamo Numbers.** The system (83) is still complicated for analytical studying, although within this system Parker in his pioneer work (1955) detected linear dynamo waves. Mathematical details of this study can be found in Parker's book (1979). A method of solution proposed by Parker (1955) is rather complicated and cumbersome. Therefore, for ease of presentation, we will act in the spirit of quasiclassical approximation of quantum mechanics, i.e., we will suppose that the field generation occurs near some fixed latitude $\phi_0$ (somewhat similar to the royal sunspot zone), and the dependence of coefficients of Equation (83) on latitude will not be taken into account. The second simplification is a neglect of the curvature. Specifically, this means that the Stokes operator is replaced into a simplified Laplacian:

$$\Delta_s B = \frac{1}{r}\left[\frac{\partial^2 (rB)}{\partial r^2} + \frac{1}{r}\frac{\partial}{\partial \vartheta}\left(\frac{1}{\sin\vartheta}\frac{\partial(\sin\vartheta B)}{\partial\vartheta}\right)\right] \Rightarrow$$
$$\Rightarrow \frac{\partial^2 B}{\partial r^2} + \frac{\partial^2 B}{\partial \phi^2}.$$

After this, the system is as follows

$$\frac{\partial B}{\partial t} - \left(\frac{\partial^2 B}{\partial r^2} + \frac{\partial^2 B}{\partial \phi^2}\right) = D\cos\phi_0\left(\left[\nabla\omega\times\nabla A\right]\cdot\hat{\boldsymbol{\varphi}}\right),$$
$$\frac{\partial A}{\partial t} - \left(\frac{\partial^2 A}{\partial r^2} + \frac{\partial^2 A}{\partial \phi^2}\right) = \sin\phi_0\, B \qquad (84)$$

where $\nabla = \hat{\boldsymbol{\phi}}\,\partial/\partial\phi + \hat{\mathbf{r}}\,\partial/\partial r$ and the unit vectors $\hat{\boldsymbol{\phi}}$, $\hat{\mathbf{r}}$, $\hat{\boldsymbol{\varphi}}$ directed along the latitude, radius, and longitude, respectively, form the right triple, i.e., $\hat{\boldsymbol{\varphi}} = \left[\hat{\boldsymbol{\phi}}\times\hat{\mathbf{r}}\right]$. In this approximation, $\mathbf{B}_p = -\hat{\mathbf{r}}\,\partial A/\partial\phi + \hat{\boldsymbol{\phi}}\,\partial A/\partial r$. From two equations (84), one can obtain one equation for the toroidal component of the vector potential $A$

$$\left[\frac{\partial}{\partial t} - \left(\frac{\partial^2}{\partial r^2} + \frac{\partial^2}{\partial \phi^2}\right)\right]^2 A = -\frac{D}{2}\sin 2\phi_0\left(\hat{\mathbf{e}}_\perp\cdot\nabla A\right), \qquad (85)$$

where $\hat{\mathbf{e}}_\perp$ is the unit vector perpendicular to $\nabla\omega$: $\hat{\mathbf{e}}_\perp = -\hat{\mathbf{r}}\,\partial\omega/\partial\phi + \hat{\boldsymbol{\phi}}\,\partial\omega/\partial r$, i.e., parallel to the surface of constant angular velocity. This includes that at $|\nabla\omega| \approx \text{const}$ (see above) in dimensionless Equation (84) one should suppose $|\nabla\omega| = 1$.

A. Dynamo waves in the infinite space. Let us look for a solution to Equation (85) in the form of a flat dynamo wave (Parker, 1955b):

$$A = \mathrm{Re}\left\{A_0 \exp\left[\lambda t + i\left(k_r r + k_\phi \phi\right)\right]\right\}, \qquad (85a)$$

where $\lambda = \gamma + 2\pi i / P_*$, $\gamma$ is the increment of the exponential growth of the magnetic field, $P_*$ is the period of stellar cyclic activity, which takes into account the field sign (without taking into account the sign of $\bar{P}_* = P_*/2$). Substituting this solution into Equation (85), we obtain

$$\lambda = \frac{1 - i\,\mathrm{sgn}\left(D \sin\left(2\phi_0\right)\left(\hat{\mathbf{e}}_\perp \cdot \mathbf{k}\right)\right)}{2} \sqrt{\left|D \sin\left(2\phi_0\right)\left(\hat{\mathbf{e}}_\perp \cdot \mathbf{k}\right)\right| - k^2} \,.$$

Here $\mathbf{k} = \hat{\mathbf{r}} k_r + \hat{\boldsymbol{\phi}} k_\phi$ and $i^2 = -1$. This yields the following expressions for the increment of the exponential growth of the magnetic field and the stellar cyclic activity period

$$\gamma = \frac{1}{2} \sqrt{\left|D \sin\left(2\phi_0\right)\left(\hat{\mathbf{e}}_\perp \cdot \mathbf{k}\right)\right| - k^2} \quad (86a)$$

$$\bar{P}_* = -\pi\,\mathrm{sgn}\left(D \sin\left(2\phi_0\right)\left(\hat{\mathbf{e}}_\perp \cdot \mathbf{k}\right)\right) \sqrt{\frac{1}{\left|D \sin\left(2\phi_0\right)\left(\hat{\mathbf{e}}_\perp \cdot \mathbf{k}\right)\right|}} \,. \quad (86b)$$

Several well-known results of the linear dynamo theory concerning properties of the most rapidly growing stellar dynamo mode follow from these expressions. The following statements are among them.

- The Yoshimura theorem (1975): the most rapidly growing dynamo wave propagates along the surface of constant angular velocity. This is well seen from the fact that $\gamma \sim \sqrt{\left|\left(\hat{\mathbf{e}}_\perp \cdot \mathbf{k}\right)\right|}$ is maximum when $\hat{\mathbf{e}}_\perp$ and $\mathbf{k}$ are parallel, i.e., the dynamo wave propagates along the surface $\omega \approx$ const.

- The period of cyclic activity of a star is proportional to the period of its rotation around the axis for the most rapidly growing dynamo wave. This result can be obtained finding from (86a) the $k$ value of the wave vector $\mathbf{k}$ at which the increment of growth (86a) is maximum. Substituting the found $k$ value into (86b), we get

$$\bar{P}_* = -4\pi\,\mathrm{sgn}\left(D \sin\left(2\phi_0\right)\left(\hat{\mathbf{e}}_\perp \cdot \mathbf{k}\right)\right)\left|D \sin\left(2\phi_0\right)\right|^{-2/3} = = 7 \times 10^{-3} \left[\frac{1}{\sin^2\left(2\phi_0\right)}\right]^{1/3} \left(\bar{\alpha}_0 \delta \bar{\bar{\Omega}}_*\right)^{-2/3} \frac{\bar{\eta}_T^{(0)}}{\bar{R}_*^2} \bar{T}_* \,,$$

where the dynamo number is already substituted above $D = \mathrm{R}_\omega \mathrm{R}_\alpha \simeq 6.12 \times 10^4\ \bar{\alpha}_0 \delta \bar{\bar{\Omega}}_* \bar{R}_*^3 \sqrt{\bar{\Omega}_*^3 / \left(\bar{\eta}_T^{(0)}\right)^3}$. Referring to dimensional variables, we obtain the activity period in years $P_*^{(dm)} = 7 \times 10^{-3}\, \bar{T}_*\, R_\odot^2 / \eta_0 \left(\bar{\alpha}_0 \delta \bar{\bar{\Omega}}_* \left|\sin\left(2\phi_0\right)\right|\right)^{-2/3} =$ $1.1 \bar{T}_* \left(\bar{\alpha}_0 \delta \bar{\bar{\Omega}}_* \left|\sin\left(2\phi_0\right)\right|\right)^{-2/3}$.

For the Sun this estimate yields $P_{\odot}^{(dm)} = 1.1\bar{T}_* \left(\bar{\alpha}_0 \delta\bar{\bar{\Omega}}_* \left|\sin(2\phi_0)\right|\right)^{-2/3} = 3.5(\bar{\alpha}_0)^{-2/3}$ years, where it is taken that $\delta\bar{\bar{\Omega}}_* \simeq 0.2,\ \phi_0 \simeq 30^\circ$. To obtain the observed period of the solar cycle, it is sufficient to suppose $\bar{\alpha}_0 = 0.2$, which yields a period value of 10.2 years.

B. Dynamo waves with boundary conditions in the radial direction. The critical dynamo number. It is obvious that in real stars the solution corresponding to (85a) cannot exist. The reason is the presence of inner and outer boundaries of the stellar convective zone. The real boundary conditions at both inner and especially outer boundaries of the convective zone are very complicated. The presence of starspots, flare activity at the boundary of the convective zone and the photosphere, the penetration of turbulent convection into the radial core of a star are just several complications of the dynamo problem from the point of view of boundary conditions. Nonetheless, the most general tendency that can be spoken about is the strongest decrease in effective resistivity of the stellar plasma due to the shutdown of turbulent convection in the mentioned regions. Specifically, we suppose $B = \partial A/\partial r\big|_{r=1} = \partial A/\partial r\big|_{r=1-1/\mu} = 0$ to be the boundary condition of Yoshimura (1975), where $\mu_* = R_*/H_*$, whereas $H_*$ is the convective zone depth of a star. Physically, such a boundary condition means that at boundaries of the convective zone the poloidal field has only a radial component. Let us look for a solution to Equation (85) in the form: $A = \mathrm{Re}\left\{\left[A_0 \exp(\lambda t + ik\phi)\right] \cos\left\{k_r\left[1-\mu_*(1-r)\right]\right\}\right\}$. It follows from the boundary conditions of Yoshimura that $k_r = n\pi$. Substituting this solution into Equation (85), we obtain

$$\gamma = \sqrt{\frac{\left|kD\sin(2\phi_0)\right|}{4}} - k^2 - \mu_*^2\pi^2 n^2 \quad (87a)$$

$$\bar{P}_* = -\frac{\pi\,\mathrm{sgn}\left(kD\sin(2\phi_0)\right)}{\sqrt{\left|D\sin(2\phi_0)k\right|}} \quad (87b)$$

Finding the most rapidly growing dynamo wave, we get

$$\gamma = \frac{3}{4}\left[\sqrt[3]{\left(\frac{D}{2}\sin(2\phi_0)\right)^2} - \frac{4}{3}\mu_*^2\pi^2 n^2\right]. \quad (87c)$$

It follows from this relation that there exists a minimal dynamo number just starting from which the field generation is possible, i.e., $\gamma > 0$. This dynamo number is called the critical dynamo number $D_{cr}$. It follows from (87c) that

$$D > D_{cr} = \left[2\mu_*\pi n/\sqrt{3}\right]^3 2/\sin(2\phi_0) . \quad (88)$$

For example, if we take $\phi_0 \simeq 30^\circ$; $\mu_\odot \approx 3$ (the case of the Sun), then $D_{cr} \simeq 3000$, which is in good agreement with more complicated models. Relation (88) shows that the critical dynamo number grows as $D_{cr} \propto \mu^3 = \left(R_*/H_*\right)^3$. This means that for M5 stars $D_{cr}^{(\mathrm{M5})} \approx 100$, whereas for F5 stars $D_{cr}^{(\mathrm{F5})} \approx 700000$. Two more important properties follow from the given solution.

- The distribution of dynamo waves has the following feature. For the most rapidly growing waves to propagate in the direction of decreasing latitude, i.e., toward the equator in the northern hemisphere, as it is mainly observed, one should require in Parker's solution (1955) that $\bar{P}_* > 0$ or $-\mathrm{sgn}\left(D\sin\left(2\phi_0\right)k\right) > 0$. Since $\sin\left(2\phi_0\right) > 0$ for the northern hemisphere, we get that $D < 0$. In the southern hemisphere $\sin\left(2\phi_0\right) < 0$. Therefore, to get a positive frequency in the southern hemisphere at the same negative dynamo number, it is sufficient to suppose $\left(\hat{\boldsymbol{\phi}}\cdot\mathbf{k}\right) < 0$. The latter means that the wave vector **k** changes its sign in the southern hemisphere. In other words, if in the northern hemisphere dynamo waves propagate toward the south pole, then in the southern one toward the north pole or, what is the same, to the equator in both hemispheres.
- According to the dynamo theory, the shift of phases between components of the field corresponds to a *lag* of the toroidal field from radial one by $\pi/4$. On the other hand, the latitudinal field directed on $-\hat{\boldsymbol{\phi}}$ lags behind the toroidal one by the same $\pi/4$. To confirm this, it is sufficient to calculate components of the toroidal $B = \left(\sin\phi_0\right)^{-1}\left[\partial/\partial t - \left(\partial^2 A/\partial r^2 + \partial^2/\partial\phi^2\right)\right]A$ and poloidal $\mathbf{B}_p = -\hat{\mathbf{r}}\,\partial A/\partial\phi + \hat{\boldsymbol{\phi}}\,\partial A/\partial r$ fields

$$B_r = -k_\phi \,\mathrm{Re}\left\{A_0 \exp\left[\gamma t + i(\pi t/\bar{P}_* + k_\phi\phi + \pi/2)\right]\right\}\cos\left\{\pi n\left[1-\mu(1-r)\right]\right\}$$

$$B = \left[2|D|\cos\left(\phi_0\right)\Big/\sqrt{\sin\left(\phi_0\right)}\right]^{2/3} \mathrm{Re}\left\{A_0 \exp\left[\gamma t + i(\pi t/\bar{P}_* + k_\phi\phi + \pi/4)\right]\right\}\cos\left\{\pi n\left[1-\mu(1-r)\right]\right\}$$

$$B_\phi = \mu\pi n \,\mathrm{Re}\left\{A_0 \exp\left[\gamma t + i(\pi t/\bar{P}_* + k_\phi\phi)\right]\right\}\sin\left\{\pi n\left[1-\mu(1-r)\right]\right\}$$

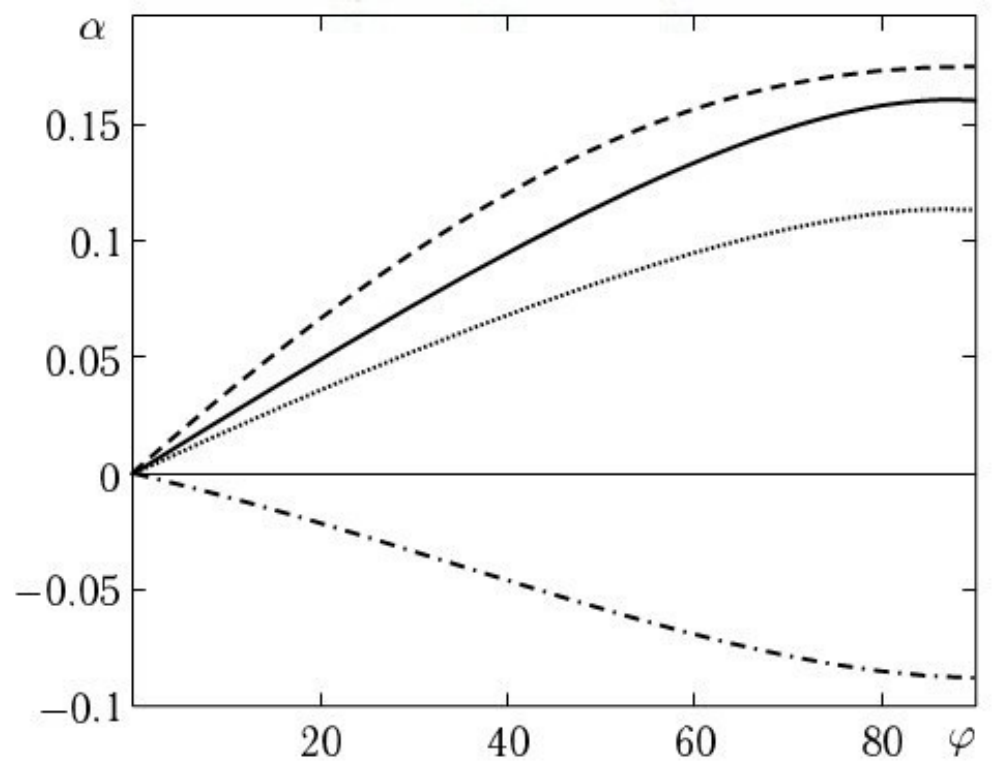

Fig. 106. The same as in Fig. 105 but at slow rotation of a star $\Omega_*\tau_0^{(*)} = 0.1$. At slow rotation, the alpha effect is seen to be proportional to sin$\phi$, which is in agreement with the Krause formula (see above)

It is interesting that approximately the same phase shift is observed in the solar case (Moffatt, 1978). At the same time, it is significant that the dynamo number should be negative, whereas $\alpha$ is positive in the northern hemisphere. It follows necessarily from the latter that $\delta\bar{\Omega}_* < 0$. According to Fig. 106, at weak rotations (i.e., at the stellar surface) the coefficient $\alpha$ is indeed positive, if only thermics of turbulent convection are not too flattened. As to the condition $\delta\bar{\Omega}_* < 0$, then on the upper boundary of the solar convective zone, following helioseismology, this condition is met at all the latitudes (Fig. 107).

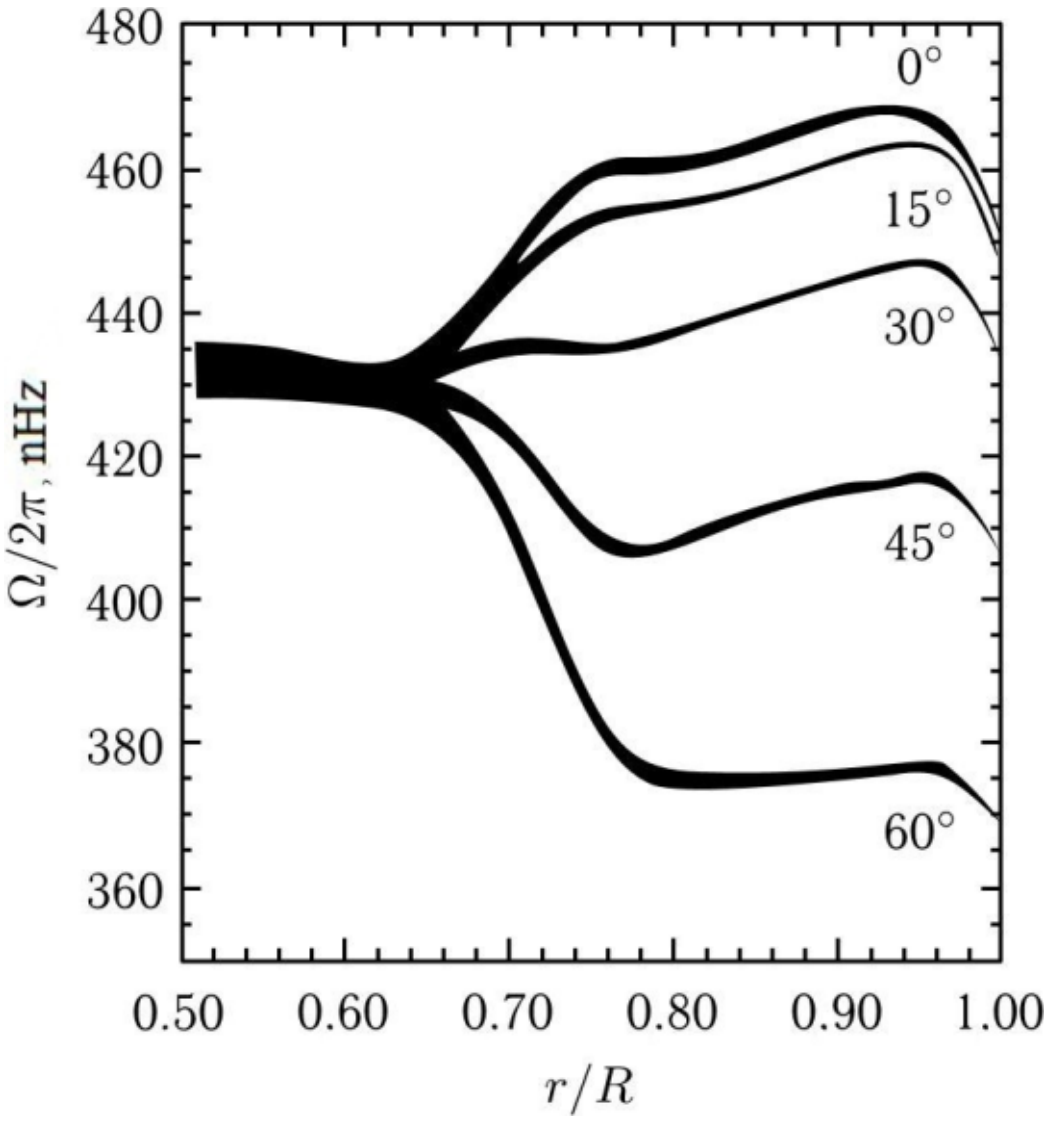

Fig. 107. Radial frequency profiles of the solar rotation $\Omega_\odot(r/R_\odot)/2\pi$ for different latitudes according to the data of helioseismology (Kosovichev et al., 1997)

Figure 107 shows that at the surface ($r/R_\odot > 0.94$) the differential rotation of $\partial\omega/\partial r$ at all the latitudes is *negative*, whereas in most part of the convective zone $0.7 < r/R_\odot < 0.94$ it is *positive*. Note that the zones of strong differential rotation (tachoclines) are observed at both the surface (upper tachocline) and the base of the solar convective zone (lower tachocline — $r/R_\odot \simeq 0.7$).

Furthermore, the lower tachocline is pronounced weakly at latitudes of the royal zone and stronger at higher latitudes. The modern theory of stellar differential rotation cannot answer the question regarding typicality of these profiles for other main-sequence stars. It is also not clear on the nature of the lower tachocline. However, the theory yields a confident rotation growth in most part of the convective zone, which is a result of the $\Lambda$ effect (Kichatinov and Rüdiger, 1993), and the upper tachocline as a result of the direct influence of the Coriolis force on the convective flow (Kleeorin and Rogachevskii, 2006). An example of such a profile for the latitude $\approx 30°$ is given in Rogachevskii and Kleeorin (2018) (see Fig. 108). The feature of this latitude (the latitude 28°, to be exact) is in the fact that the equations for differential rotation permit an accurate analytical solution, since at this latitude the contributions dependent on latitude turn into zero.

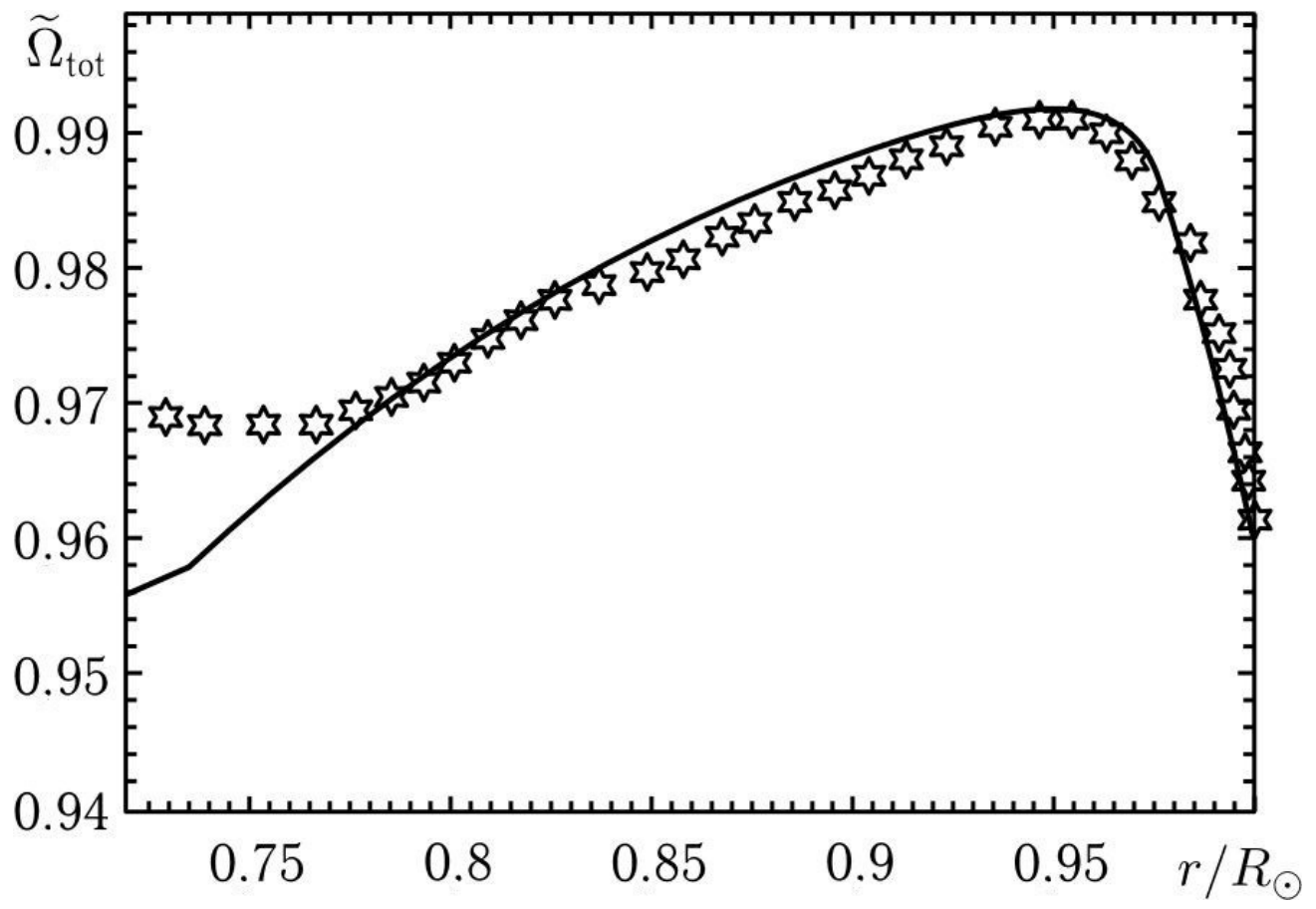

Fig. 108. Radial profile of the solar rotation for the latitude ≈ 30° $\tilde{\Omega}_{tot} = \Omega_\odot(r/R_\odot, 30^\circ)/\Omega_\odot(1, \phi = 0)$. This profile is normalized to the solar equatorial rotation. Asterisks denote the data of helioseismological observations from Fig. 107 (after Kosovichev et al., 1997). The solid line marks the theoretical calculation made by Rogachevskii and Kleeorin (2018)

The theoretical curve is seen to describe well the observations up to the lower tachocline $r/R_\odot < 0.752$. From the point of view of the dynamo theory, two results of the differential rotation theory seem to be important:

— there exists an upper tachocline in which *always* $\delta\bar{\Omega}_* < 0$;

— the profiles normalized to equatorial rotation do not depend on absolute rotation but only on the profile of $\Omega_* \tau_0^{(*)}(r)$.

Dependence on the absolute rotation value arises at the very fast rotations when the rotation period of a star is about a few days.

Only overlapping generation sources have been considered so far — the profiles of these sources were simply not taken into account. For slowly rotating stars, this seems to be always valid. But for the stars of spectral types earlier than G0, at the rotation with $\Omega_* > 4\Omega_\odot$, it turns out that in all the convective envelope of a star $\Omega_* \tau_0^{(*)}(r) \geq 1$ (see Table 25). This means that the upper tachocline disappears, the lower tachocline becomes the main one, and the maxima of alpha and omega effects turn out to be separated in space. As a result, we have come to the dynamo pattern with nonoverlapping sources of generation.

C. Dynamo waves with nonoverlapping sources of generation spaced in the radial direction. This model was reviewed in detail by Kleeorin and Ruzmaikin (1981). It includes the following: let the value of $\partial\omega/\partial r$ be different from zero near the bottom of the convective zone, whereas the region of localizations of $\alpha(r)$ is shifted to the surface. Specifically, suppose at $r_\omega^{(1)} = r_\omega - L_\omega/2 < r < r_\omega + L_\omega/2 = r_\omega^{(2)}$ $\partial\omega/\partial r \neq 0$, whereas at $r_\alpha^{(1)} = r_\alpha - L_\alpha/2 < r < r_\alpha + L_\alpha/2 = r_\alpha^{(2)}$ $\alpha(r) \neq 0$ ($H_{\alpha\omega} = r_\omega - r_\alpha$, $H_{\alpha\omega} > L_\omega + L_\alpha$ or even better $H_{\alpha\omega} \gg L_\omega + L_\alpha$). Thus, $L_\omega, L_\alpha$ are the widths of sources and $r_\omega, r_\alpha$ are the positions of their maxima. In other regions of the convective envelope the sources are absent. Let us assume a fast decrease of the magnetic field outside the generation region as boundary conditions. The solution outside the generation regions is as follows: $A,B\propto\exp(\lambda t+ik\phi+qr)$ at $r<r_\omega^{(1)}$ and $A,B\propto\exp(\lambda t+ik\phi-qr)$ at $r>r_\alpha^{(2)}$, the complex wave vector $q$ ($\mathrm{Re}(q)>0$) is defined by the integral equation

$$q^2=\frac{ikD}{4\mu'^2}\int_{r_\omega^{(1)}}^{r_\omega^{(2)}}\frac{d\omega}{dr}e^{2qr}dr\int_{r_\alpha^{(1)}}^{r_\alpha^{(2)}}\alpha(r)e^{-2qr}dr\ \cdot$$

Here $\mu'=R_*/H_{\alpha\omega}$. This equation is drastically simplified if $H_{\alpha\omega}\gg L_\omega+L_\alpha$. In this case, we obtain (Kleeorin and Ruzmaikin, 1981)

$$q'^2=\frac{ikD\ell_\alpha}{4}e^{-2q'}, \tag{89}$$

where $\ell_\alpha=\int_{r_\alpha^{(1)}}^{r_\alpha^{(2)}}\alpha(r)dr$, $q'=q\mu'$. It follows from the behavior of the solution outside the generation regions that $\lambda=q^2-k^2$. At $D\gg1$ from (89), such expressions follow for the increment of the exponential magnetic-field growth and the period of stellar cyclic activity:

$$\gamma\approx\mu'^2\left\{\ln^2\left[\frac{|kD|\ell_\alpha}{\ln\left(|kD|\ell_\alpha\right)}\right]-\frac{\pi^2}{4}-k^2\right\} \tag{89a}$$

$$\bar{P}_*\approx-\frac{4\,\mathrm{sgn}(kD)}{\mu'^2\left|\ln\left[|kD|\ell_\alpha/\ln\left(|kD|\ell_\alpha\right)\right]\right|} \tag{89b}$$

Finding in the usual way the most rapidly growing mode, we obtain $k_{\gamma_{\max}}\approx\ell_\alpha\ln D/\sqrt{D}$. Taking into account that $D>D_{cr}\gg1$, we see that the appropriate wavelength $\lambda_{\gamma_{\max}}=2\pi k_{\gamma_{\max}}^{-1}\approx2\pi\sqrt{D}/(\ell_\alpha\ln D)\gg\pi$, i.e., less than the angular distance from pole to pole of a star. Roughly speaking, the most rapidly growing mode of this model just "doesn't fit" the star. In such a case, the minimum possible $k$ should be chosen on the basis that between the pole and the equator n waves should be arranged. Thus, $k=2\pi/\pi(2n)^{-1}=4n$. From the condition $\gamma=0$, this yields the critical dynamo number

$$D_{cr}=\frac{\exp\left(\sqrt{16n^2+\pi^2/4}\right)}{4n\ell_\alpha}\sqrt{16n^2+\frac{\pi^2}{4}}\ \cdot$$

Let us give the critical dynamo number for the two first modes $\ell_\alpha\sim0.1$: $D_{cr}^{(1)}\approx800,\ D_{cr}^{(2)}\approx3.5\times10^4$. Thus, nonoverlapping generation sources easily produce generation of only epy first mode. Substituting the found $k=2\pi/\pi(2n)^{-1}=4n$, we obtain

$$\bar{P}_*\approx\frac{4}{\mu'^2\left|\ln\left[|4nD|\ell_\alpha/\ln\left(|4nD|\ell_\alpha\right)\right]\right|}\propto\left(\frac{H_{\alpha\omega}}{R_*}\right)^2. \tag{90}$$

Thus, for the stars with nonoverlapping generation sources, the following result is valid. Since, according to (90), the dependence of the cycle period on the dynamo number $D$ has a logarithmic character, the distance between the maxima of generation sources $H_{\alpha\omega}$ mainly defines the dependence of the cycle period on the stellar rotation. As shown in Item B, the distance between the lower tachocline and the alpha effect maximum $H_{\alpha\omega}$ increases with increasing angular velocity and, consequently, should cause a growth of the stellar cycle period $\bar{P}_*$ with increasing angular velocity. For instance, with decreasing angular rotation velocity of the Sun by a factor of 10, the cycle period decreases by a factor of 2 if the upper tachocline is insignificant. This rather fine effect seems to be really observed: it follows from Fig. 82*a* in Chapter 3.1 that when stellar rotation periods change from ~ 2 to ~ 22 days, the activity cycle period varies from 18 to 2.5 years, which qualitatively agrees with the estimate given above.

In previous sections, it was shown that dynamos depend significantly on the details of the construction of stellar convective zones. In other words, to construct such a dynamo, one should know in detail the distribution of generation sources over the stellar radius. Since, as stated above, such information is unavailable, then a tempting idea arises to average Equations (83) over the convective zone depth and proceed to the so-called no-r stellar dynamo model suggested by D.D. Sokoloff in Kleeorin et al. (2003).

D. No-r stellar dynamo model. Let us average the system of equations (83) according the convective zone depth of a star. This yields

$$\begin{aligned}\frac{\partial B(\phi,t)}{\partial t} &= -D\cos\phi\frac{\partial A(\phi,t)}{\partial\phi}+\left(\frac{\partial^2}{\partial\phi^2}-\mu_*^2\right)B(\phi,t)\\ \frac{\partial A(\phi,t)}{\partial t} &= \sin\phi\, B(\phi,t)+\left(\frac{\partial^2 A}{\partial\phi^2}-\mu_*^2\right)A(\phi,t)\end{aligned}. \tag{91}$$

Here $\mu_*^2 = -\int_{R_1}^{1}\partial^2 A(r,\phi)/\partial r^2\,dr \Big/ \int_{R_1}^{1} A(r,\phi)\,dr \approx -\int_{R_1}^{1}\partial^2 B(r,\phi)/\partial r^2\,dr \Big/ \int_{R_1}^{1} B(r,\phi)\,dr$,

$A(\phi,t) = R_1 = 1 - H_*/R_*$, $\int_{R_1}^{1} A(r,\phi,t)\,dr \Big/ (1-R_1)$.

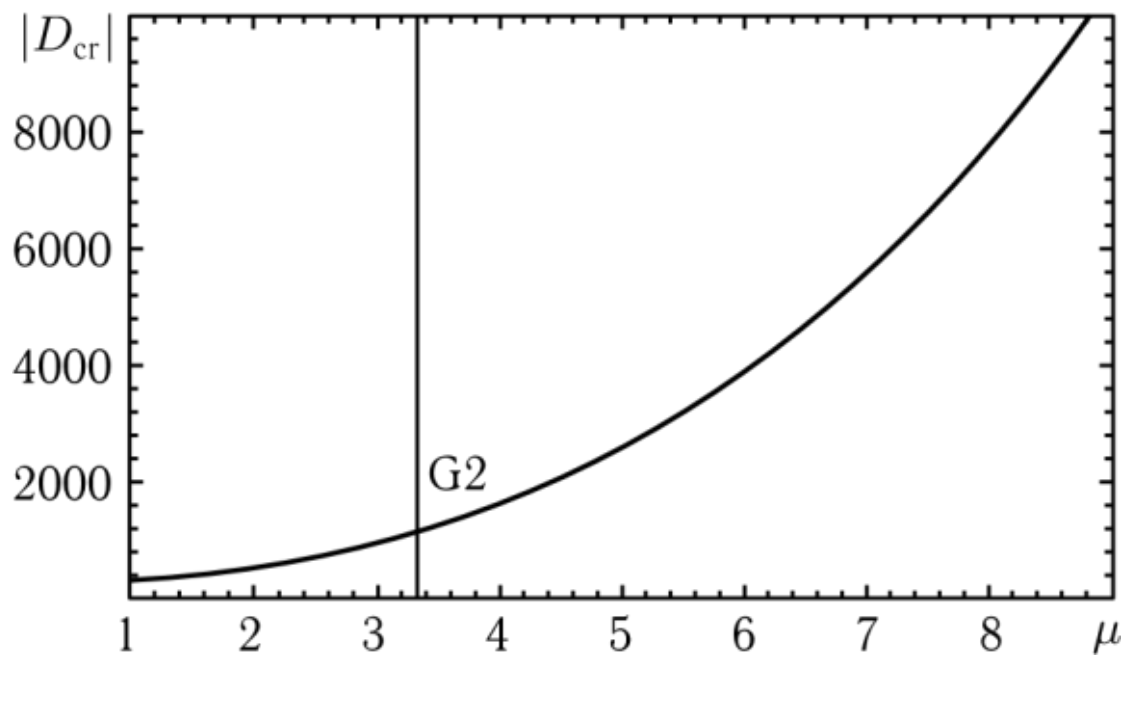

Fig. 109. Dependence of the critical dynamo number $|D_{cr}(\mu)|$ from the paper of Kleeorin et al. (2016). The values of these critical dynamo numbers from the graph are given for stars of some spectral types in Table 25

For the simplest estimate, one should take that $\mu^2 = \left(R_*/H_*\right)^2$. The obtained system, strictly speaking, can be solved only numerically. But its tremendous advantage is the presence of only one parameter $\mu$ dependent on the stellar spectral type. Meanwhile, the solution of the system (91) by its properties is close to the solutions discussed in Items A–C. It is useful to present a graph of the dependence of the critical dynamo number $\left|D_{cr}(\mu)\right|$ (Fig. 109) found in Kleeorin et al. (2016).

Concluding this item, note that the exceptional stability of the linear version to nonphysical divergences of the solution can be referred to advantages of this model. This allows one to study deeper the properties of secular behavior of the stellar cycle and chaotic increases (decreases) in stellar activity.

E. Other methods for studying the linear (kinematic) stellar dynamo. Immediately after the justification of Equation (83) Steenbek, Krause, and Radler (1966) undertook an attempt to integrate it numerically within the total axisymmetric linear (without an inverse effect) dynamo model. The result of this calculation for $D = D_{cr}$ is given in Fig. 110.

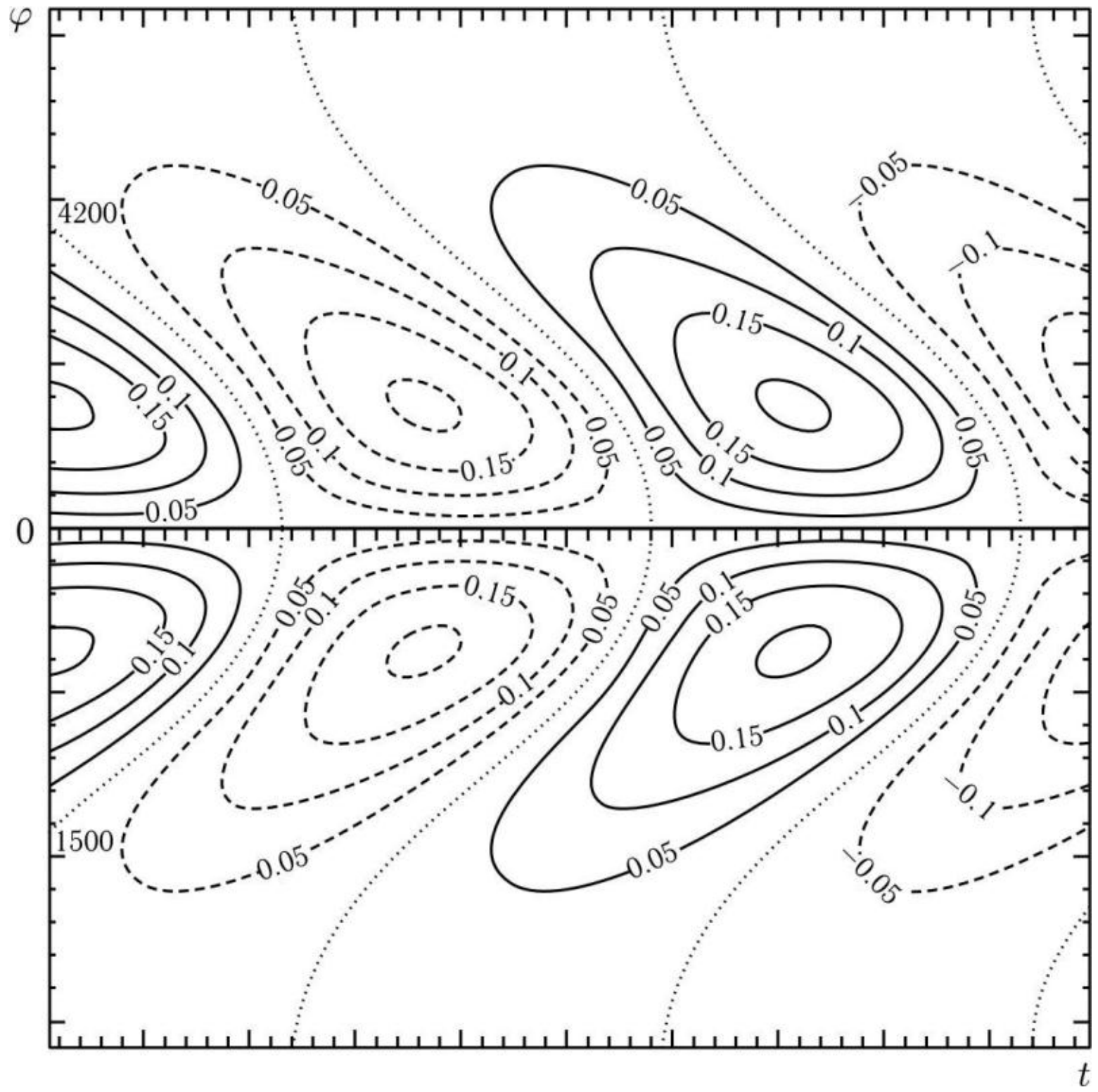

Fig. 110. Isolines of the toroidal magnetic field $B$ — solutions of Equation (83) in the latitude sinus–time diagram (Butterfly diagram). The calculation was performed for the critical dynamo number $D_{cr}$ (from the paper of Steenbek, Krause, and Radler (1966)). Numbers on isolines denote the field levels in these lines. Solid lines correspond to positive values of the toroidal field in the northern hemisphere, the dotted lines — to the negative ones. Dots denote the lines of a change in sign of the toroidal field

Similar calculations were subsequently performed by other authors (see, e.g., the paper of Ivanova and Ruzmaikin, 1976). They acquired the results analogous to those of Steenbek et al. (1966). The main features of these results are a strictly sinusoidal behavior of the magnetic field in time, which corresponds to a linear problem, and the absence of the polar activity branch.

However, it is possible that its absence is an artifact of insufficient accuracy of calculation methods, since the presence of the polar activity branch was demonstrated by asymptotic methods by Kuzanyan and Sokoloff (1995, 1997), who referred to the fact that the dynamo mode needs higher values of the dynamo number $D > D_{cr} \geq 1000$. This allows one to implement a decomposition of the solution over a small value of $D^{-1/3}$. In this case, sphericity of the problem inevitably generates the polar branch, though very weak — only several percent of the toroidal field amplitude near the stellar equator. Along with this work, we should note an asymptotic study of Makarov et al. (1987) in which the presence of the polar branch is associated with a variation in sign of the dynamo number $D$ due to a change in sign of the alpha effect with depth (see Fig. 111) or a sign of differential rotation (see Fig. 106) with depth. A change in sign of the alpha effect with latitude (see Fig. 105) will probably leads to the same thing. Note that numerical studies of the linear spherical dynamo were almost ceased in recent decades. On the one hand, this is due to the fact that for further progress one should construct the detailed models of generation sources of stellar magnetic fields, taking into account the effect of the Coriolis forces on turbulent convection, which is difficult. Essentially, only the dependence of the Coriolis parameter of a star $\Omega_{\odot}\tau(r)$ on the radial coordinate is currently available (Fig. 111).

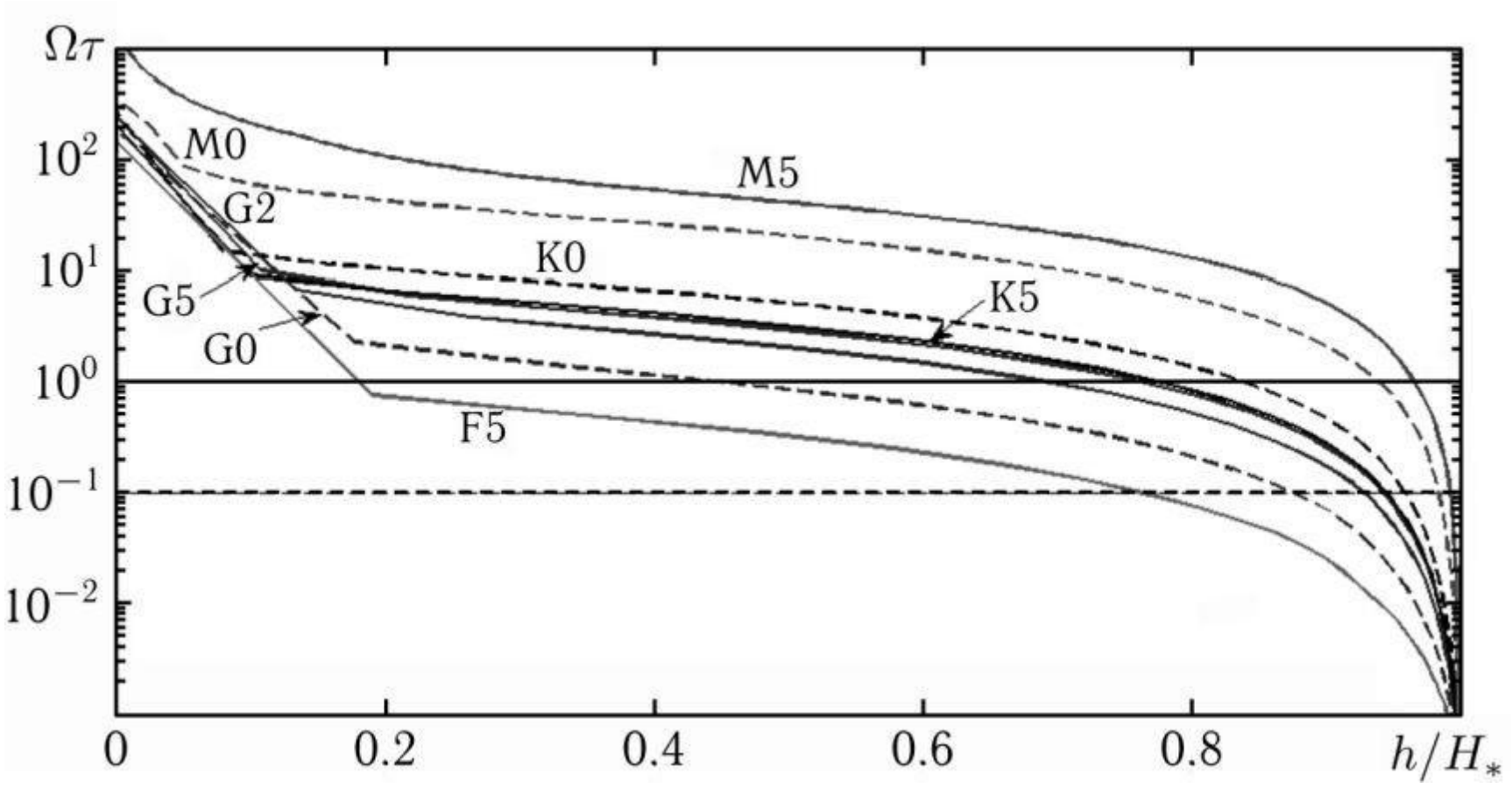

Fig. 111. Dependences of the Coriolis parameters $\Omega_{\odot}\tau(r)$ in convective envelopes of stars of late spectral types from F5 to M5 calculated using the mixing length theory. Stellar rotation velocity in this calculation is equal to solar. The horizontal axis is the height of the point above the bottom of the convective zone attributed to its thickness

On the other hand, from available observations only the dependence of the cycle period on stellar parameters may be weakly associated with the level of the magnetic field generated by the stellar dynamo. The other observed values are directly defined by the magnitude of the stellar magnetic field. Therefore, the real stellar dynamo is certainly the nonlinear dynamo.

## 4.3.3. Nonlinear Dynamo and Its Relation with Observations

**4.3.3.1. Physical Basics of Nonlinear Dynamo.** To construct a stellar nonlinear dynamo**,** the induction equation (75) is not enough. It is necessary to add the Navier-Stokes equations, the continuity equation, entropy transport equations, the radiation transfer equation, and several equations of state: the Mendeleev–Clapeyron equation, the Saha ionization equation, the equation for the Rosseland mean absorption coefficient of electromagnetic radiation of stellar plasma, and others. Since we are interested in considering the stellar dynamo on the background of stellar convection, let us give these equations in the stellar convection approximation (Vandakurov, 1976):

$$\frac{\partial \mathbf{H}}{\partial t} = \mathrm{rot}\left(\mathbf{v}\times\mathbf{H} - \nu_m \mathrm{rot}\mathbf{H}\right), \tag{92}$$

$$\frac{\partial \mathbf{v}}{\partial t} = -\vec{\nabla}\left(\frac{p_{tot}}{\rho_0}\right) - \mathbf{g}\,s + \frac{\mathbf{F}_M}{\rho_0} + \mathbf{v}\times\left(2\mathbf{\Omega}_* + \mathbf{w}\right) + \mathbf{f}_\nu(\mathbf{v}), \tag{93}$$

$$\frac{\partial s}{\partial t} + \left(\mathbf{v}\cdot\nabla s\right) = -\frac{\Omega_B^2}{g}\mathrm{v}_z + \frac{E}{c_p\rho_0 T} + \frac{1}{T\rho_0}\mathrm{div}\left(\kappa\rho_0\nabla T\right), \tag{94}$$

$$\mathrm{div}(\mathbf{v}) = \mathbf{\Lambda}\cdot\mathbf{v} \equiv -\frac{\nabla\rho_0}{\rho_0}\cdot\mathbf{v}\,. \tag{95}$$

Here $p_{tot} = \bar{P} + \rho_0 \mathrm{v}^2/2 + H^2/8\pi$ is the total pressure, $\mathbf{F}_M = \mathbf{\Lambda} H^2/8\pi + \left(\mathbf{H}\cdot\nabla\right)\mathbf{H}/4\pi$ — the magnetic force, $\mathbf{\Omega}_*$ — the angular rotation velocity of a star, $\mathbf{w} = \mathrm{rot}\left(\mathbf{v}\right)$, $\mathbf{f}_\nu(\mathbf{v}) = \nu\left(\Delta\mathbf{v} + \frac{1}{3}\vec{\nabla}\left(\mathbf{\Lambda}\cdot\mathbf{v}\right)\right)$ — the viscous force per unit mass, $s = \ln\left[T/T_0\left(p_0/p\right)^{(\Gamma-1)/\Gamma}\right]$ — the specific dimensionless entropy, $T_0$ and $p_0$ — the reference values of temperature and pressure, for instance, at the base of the stellar convective zone, $\kappa = 4a_0c_0T^3/3\chi\rho_0$ cm$^2$/s — the radiation diffusion coefficient of temperature, where $c_0$ — the speed of light in vacuum, $a_0$ = $7.55\cdot10^{-15}$ K$^{-3}$, $\chi\rho_0$ cm$^{-1}$ — the Rosseland mean absorption coefficient of electromagnetic radiation of stellar plasma,

$$\Omega_B^2 = -\frac{1}{\Gamma}\left(\frac{\mathbf{g}\cdot\nabla\bar{T}}{\bar{T}} - (\Gamma-1)\frac{\mathbf{g}\cdot\nabla\rho_0}{\rho_0}\right) \equiv -\mathbf{g}\cdot\mathbf{y}$$

is the square of the Brunt-Vaisala frequency, $\bar{T}$ and $\rho_0$ — the hydrostatic equilibrium profiles of temperature and density, $\Gamma = C_p/C_V$ — the effective adiabatic index. The square of frequency can change in sign: $\Omega^2{}_B < 0$ corresponds to the required condition of convective instability and the superadiabatic logarithmic temperature gradient (Schwarzschild criterion $-y_r > 0$), $\Omega^2{}_B > 0$ corresponds to internal gravitational waves. Taking viscosity and diffusion of entropy into account, the excitement of convection requires that

$$\mathrm{R}a > \mathrm{R}a_{cr},$$

$$\mathrm{R}a = -\frac{\Omega_B^2 L_p^4}{\nu\kappa} \equiv \frac{\mathbf{g}\cdot\mathbf{y}L_p^4}{\nu\kappa},$$

$$L_p^{-1} = \left|\frac{\nabla p}{p}\right|$$

(sufficient condition). The dimensionless parameter Ra is named the Rayleigh number. The analytical estimates and laboratory experiments show that the critical Rayleigh numbers $\mathrm{Ra}_{\mathrm{cr}}$ vary within ≈ 660 and ≈ 1710, depending on boundary conditions. The stellar Rayleigh numbers $\mathrm{R}a_* \simeq 10^{18} \div 10^{20} \ggg \mathrm{R}a_{cr}$. Therefore, for stars the Schwarzschild criterion practically coincides with the sufficient condition. The energy release value per unit mass $E = E_{\nu m} + E_\nu$ is the sum of ohmic $E_{\nu m}$ and viscous $E_\nu$ energy release per unit mass. According to Vandakurov (1976), $E_{\nu_m} = \nu_m \left(\mathrm{rot}\mathbf{H}\right)^2 / 4\pi$, $E_\nu = \nu\rho_0 \tilde{\tilde{\boldsymbol{\omega}}}^2$, where $\nu \simeq 2.2\times10^{-16} T^{5/2}/\rho_0$ cm$^2$/s (Spitzer, 1956),

$$\tilde{\tilde{\boldsymbol{\omega}}}^2 = \frac{1}{2}\left(\tilde{\tilde{\omega}}_{rr}^2 + \tilde{\tilde{\omega}}_{\vartheta\vartheta}^2 + \tilde{\tilde{\omega}}_{\varphi\varphi}^2\right) + \tilde{\tilde{\omega}}_{r\vartheta}^2 + \tilde{\tilde{\omega}}_{r\varphi}^2 + \tilde{\tilde{\omega}}_{\vartheta\varphi}^2 - \frac{2}{3}\left[\vec{\nabla}\left(\mathbf{\Lambda}\cdot\mathbf{v}\right)\right]^2,$$

where $\omega_{rr}, \omega_{\vartheta\vartheta}, \omega_{\phi\phi}, \omega_{r\vartheta}, \omega_{r\phi}, \omega_{\vartheta\phi}$ are the components of the velocity shift tensor in spherical geometry:

$$\tilde{\tilde{\omega}}_{rr} = 2\frac{\partial \mathrm{v}_r}{\partial r}, \quad \tilde{\tilde{\omega}}_{\vartheta\vartheta} = \frac{2}{r}\left(\frac{\partial \mathrm{v}_\phi}{\partial \phi} + \mathrm{v}_r\right), \quad \tilde{\tilde{\omega}}_{\varphi\varphi} = \frac{2}{r\cos\phi}\left(\frac{\partial \mathrm{v}_\varphi}{\partial \varphi} - \sin\phi \mathrm{v}_\phi + \cos\phi \mathrm{v}_r\right),$$

$$\tilde{\tilde{\omega}}_{r\vartheta} = \frac{1}{r}\left[-\frac{\partial \mathrm{v}_r}{\partial \phi} + r^2\frac{\partial}{\partial r}\left(\frac{\mathrm{v}_\phi}{r}\right)\right], \quad \tilde{\tilde{\omega}}_{r\varphi} = \frac{2}{r\cos\phi}\left(\frac{\partial \mathrm{v}_r}{\partial \varphi} + \cos\phi r^2 \frac{\partial}{\partial r}\left(\frac{\mathrm{v}_\varphi}{r}\right)\right),$$

$$\tilde{\tilde{\omega}}_{\vartheta\varphi} = -\frac{1}{r\cos\phi}\left[\frac{\partial \mathrm{v}_\phi}{\partial \varphi} + \cos^2\phi \frac{\partial}{\partial \phi}\left(\frac{\mathrm{v}_\varphi}{\cos\phi}\right)\right].$$

The condition $\Omega_B^2(r) < 0$ roughly outlines the boundaries of stellar convective zones. Since this relates to outer convective envelopes of stars, then here, contrary to Vandakurov (1976), the energy release is not included, occurring in the course of thermonuclear reactions in the stellar interior rather than in outer convective envelopes, since they require a temperature of the medium of at least $T \ge 1.5\times10^6\,K$. The system (92)–(95) is called the system of thermo-magneto-hydrodynamic equations (TMH) written in the anelastic Boussinesq approximation. (In the literature this approximation is called the anelastic ideal gas equation approximation.) It works as follows: the dynamo mode in the induction equation (92), producing a growth of the field, leads to the emergence of additional magnetic force $F_m$ in the Navier-Stokes equation which in combination with Equation (95), being an analog of the continuity equation in the anelastic Boussinesq approximation, describes a variation of the hydrodynamic plasma velocity. This variation of velocity, following (94), modifies the entropy transfer; this modifies the buoyancy force $-\mathbf{g}s$ and additionally changes the hydrodynamic (caused by convection) velocity of stellar plasma in such a way that the generation process is stabilized or there arises a complicated nonlinear behavior of the magnetic field and the appropriate behavior of stellar activity that is similar to the formation of starspots, the Maunder minimum, or the secular activity cycle. Within the framework of such an approach all the observed activity (even spot formation) is supposed to be described by mean fields. Therefore, the system of equations (92)–(95) is required to be averaged over turbulent pulsations of velocity and magnetic field **u** and **b**, respectively (see Sect. 4.3.1), and turbulent pulsations of entropy. This results in a system of equations, similar to (92)–(95), but for the mean values of **U**, **B**, and *S* — the mean velocity, the mean magnetic field, and the mean entropy, respectively. These equations are as follows:

$$\frac{\partial \mathbf{B}}{\partial t} = \mathrm{rot}\left(\mathbf{U}\times\mathbf{B} + \boldsymbol{E}_T - \nu_m\,\mathrm{rot}\mathbf{B}\right), \tag{96}$$

$$\frac{\partial \mathbf{U}}{\partial t} = -\vec{\nabla}\left(\frac{P_{tot}}{\rho_0}\right) - \mathbf{g}S + \frac{\mathbf{F}_M^{(\mathbf{B})} + \mathbf{F}^{(T)}}{\rho_0} + \mathbf{U}\times\left(2\boldsymbol{\Omega}_* + \mathbf{W}\right) + \mathbf{f}_\nu(\mathbf{U}), \tag{97}$$

$$\frac{\partial S}{\partial t} + \left(\mathbf{U}\cdot\nabla S\right) = -\frac{\Omega_B^2}{g}U_z + \frac{\bar{E}+E_T}{c_p\rho_0\bar{T}} + \frac{1}{\bar{T}\rho_0}\mathrm{div}\left(\kappa\rho_0\nabla\bar{T}\right) - \frac{1}{\rho_0}\mathrm{div}\left(\rho_0\boldsymbol{\Phi}_T\right), \tag{98}$$

$$\mathrm{div}(\mathbf{U}) = -\frac{\nabla\rho_0}{\rho_0}\cdot\mathbf{U} \equiv \boldsymbol{\Lambda}\cdot\mathbf{U}\cdot \tag{99}$$

The additional terms in the system (96)–(99) marked by the index *T* describe the influence of turbulence on the mean fields and flows. In the induction equation (96), the "turbulent electromotive force" is still valid $\boldsymbol{E}_T \equiv \langle \mathbf{u}\times\mathbf{b}\rangle$, the true turbulent force per unit volume appeared in the Navier-Stokes equation: $F_i^{(T)} = \partial\sigma_{ij}^T/\partial x_j$, where $\sigma_{ij}^T$ is the tensor of Maxwell-Reynolds turbulent stresses:

$$\sigma_{ij}^T = -\rho_0\left\langle u_i u_j\right\rangle + \frac{1}{4\pi}\left(\left\langle b_i b_j\right\rangle - \frac{\delta_{ij}}{2}\left\langle b^2\right\rangle\right), \tag{100}$$

whereas in the mean entropy equation *S* its turbulent flow $\boldsymbol{\Phi}_T = \langle\theta\mathbf{u}\rangle$ is valid. Indices *i*, *j*, *k*, ... take the values 1, 2, 3, or *x*, *y*, *z*, or $\vartheta$, $\varphi$, *r*, whereas the order of indices in these triples is agreed: they all are right triples. Recall that **u**, **b**, and *s* are turbulent velocity fluctuations of the stellar plasma, magnetic field, and entropy, respectively; angle brackets $\langle\ldots\rangle$ denote the statistical averaging, $\delta_{ij}$ — the unitary diagonal tensor (a matrix of 3 × 3). Let us note again that the basic objective of the mean field theory is the establishment of a link between $E_T$, $\sigma_{ij}{}^T$, $\boldsymbol{\Phi}_T$, and mean fields **U**, **B**, and *S*. The system (96)–(99) is absolutely useless without it. A similar problem for the linear induction equation (76) was solved (for the electromotive force $E_T$) under the assumption that the field of velocity fluctuations **u** is set by its own momenta. Then, using accurate formal solutions of Equation (75) and turbulence models, one can obtain Equation (77) for $E_T$ (Steenbek, Krause, and Radler, 1966; Vainshtein, 1970; Vainshtein and Zeldovich, 1972; Molchanov et al., 1983; Dittrich et al., 1984). For the nonlinear system (92)–(95) none of the formally accurate solutions was known. Therefore, it is impossible, generally speaking, to strictly substantiate, for example, the nonlinear generalization of Equation (77). In this case, it is naturally to refer to approximate methods. The simplest and the most effective among all these methods is the so-called $\tau$ approximation developed within the dynamo theory by Vainshtein (Vainshtein, 1970; Vainshtein, Zeldovich, and Ruzmaikin, 1980), Kitchatinov (1988), Kleeorin, Rogachevskii, and Ruzmaikin (1990). The $\tau$ approximation with accounting of rotation for the calculation of $\mathsf{E}_T$ was developed in detail by Radler et al. (2003), Kleeorin and Rogachevskii (2003). The main points of the $\tau$ approximation are as follows. Since the correlators defining $(\boldsymbol{E}_T)_i = \varepsilon_{ijk}\langle u_j b_k\rangle$, $\mathbf{\Phi}_T = \langle \theta \mathbf{u}\rangle$, and $\sigma_{ij}^T$ are calculated here (see (100)), i.e., $\langle u_i b_j\rangle$, $\langle u_i \theta\rangle$, $\rho_0 \langle u_i u_j\rangle$, and $\langle b_i b_j\rangle$, the equations for fluctuations **b**, **u**, and $\theta$ are required. For this aim, it is sufficient to subtract equations of the system (96)–(99) from equations of the system (92)–(95). This yields

$$\frac{D\mathbf{b}}{Dt} = \mathbf{I} + \mathbf{N}_{\mathbf{b},\mathbf{u}}, \tag{101}$$

$$\frac{D\mathbf{u}}{Dt} = \mathbf{I}_{\mathbf{u}} + \mathbf{N}_{\mathbf{u},\mathbf{b},\mathbf{u}}, \tag{102}$$

$$\frac{D\theta}{Dt} = I + N_{\mathbf{u},\theta}, \tag{103}$$

$$\operatorname{div}(\mathbf{u}) = \mathbf{\Lambda}\cdot\mathbf{u} \equiv -\frac{\nabla\rho_0}{\rho_0}\cdot\mathbf{u} \tag{104}$$

Here

$$\mathbf{N}_{\mathbf{b},\mathbf{u}} = \operatorname{rot}\left(\mathbf{u}\times\mathbf{b} - \nu_m \operatorname{rot}\mathbf{b}\right) - \operatorname{rot}\boldsymbol{E}_T$$

$$\mathbf{N}_{\mathbf{u},\mathbf{b},\mathbf{u}} = -\frac{\mathbf{F}^{(T)}}{\rho_0} - \vec{\nabla}\left(\frac{\tilde{p}_{tot}}{\rho_0}\right) - \mathbf{g}\theta + \frac{\mathbf{f}_M}{\rho_0} + \mathbf{u}\times\left(2\mathbf{\Omega}_* + \boldsymbol{\omega}\right) + \mathbf{f}_\nu(\mathbf{u})$$

$$N_{\mathbf{u},\theta} = \frac{1}{\rho_0}\operatorname{div}\left(\rho_0\mathbf{\Phi}_T\right) - \frac{\Omega_B^2}{g}u_z - \left(\mathbf{u}\cdot\nabla\theta\right) + \frac{e}{c_p\rho_0\bar{T}} + \frac{1}{\bar{T}\rho_0}\operatorname{div}\left(\kappa\rho_0\nabla\tilde{T}\right)$$

where $\tilde{p}_{tot} = \tilde{p} + \rho_0 u^2/2 + b^2/8\pi + \mathbf{B}\cdot\mathbf{b}/4\pi$, $\mathbf{f}_M = \mathbf{\Lambda} b^2/8\pi + \left(\mathbf{b}\cdot\nabla\right)\mathbf{b}/4\pi$, $e = E - \bar{E} - E_T$, $\tilde{T} = T - \langle T\rangle$.

All the equations of this system have a universal form:

— on the left side of the equation, instead of partial derivatives, as for instance in the system (92)–(95), there are *material* derivatives with respect to the large-scale velocity **U**, for example $D\theta/Dt = \partial\theta/\partial t + \left(\mathbf{U}\cdot\nabla\right)\theta$. This means, first, Galilean invariance of these equations with respect to the large-scale velocity *U*. Second, the transition to the accompanying coordinate system moving with the velocity **U** restores partial derivatives on the left sides of Equations (101)–(104) in the accompanying system, where certainly **U** = 0. In general, in accordance with these equations, all the “turbulent life” meaningfully occurs in the accompanying reference system, whereas its large-scale velocity somewhat “passively” carries this “turbulent life”;

— the right side of equations involves two contributions — terms in square brackets and sources which are bilinear differential forms of the first order (involve only the first derivatives). Specifically:

$$I(\mathbf{u},S) = -(\mathbf{u}\cdot\nabla S),$$

$$\mathbf{I}(\mathbf{B},\mathbf{U},\mathbf{b},\mathbf{u}) = (\mathbf{B}\cdot\nabla)\mathbf{u} - (\mathbf{u}\cdot\nabla)\mathbf{B} - \mathbf{B}\,\mathrm{div}\,\mathbf{u} + (\mathbf{b}\cdot\nabla)\mathbf{U} - \mathbf{b}\,\mathrm{div}\,\mathbf{U},$$

$$\mathbf{I_u}(\mathbf{B},\mathbf{U},\mathbf{b},\mathbf{u}) = -(\mathbf{u}\cdot\nabla)\mathbf{U} + (\mathbf{B}\cdot\nabla)b/4\pi\rho_0 + (\mathbf{b}\cdot\nabla)\mathbf{B}/4\pi\rho_0 + \mathbf{\Lambda}(\mathbf{B}\cdot\mathbf{b})/4\pi\rho_0 + 2\mathbf{u}\times\mathbf{\Omega}_*$$

are the sources of turbulent pulsations of entropy, magnetic field, and velocity, respectively. One can directly make sure that the average values of sources are equal to zero. These sources are identically set to zero at the homogeneous mean velocity, entropy, and in the absence of the mean field and rotation. As to the terms $N_{\mathbf{u},\theta}$, $\mathbf{N}_{\mathbf{b},\mathbf{u}}$, $\mathbf{N}_{\mathbf{u},\mathbf{b},\mathbf{u}}$, then without sources one can obtain that they yield the system which coincides with the system (92)–(95) with replacement of $\mathbf{H}\to\mathbf{b}$, $\mathbf{v}\to\mathbf{u}$, $s\to\theta$ in the accompanying reference system and with additional compensating terms $-\mathrm{rot}\boldsymbol{E}_T, -\mathbf{F}^{(T)}/\rho_0, \mathrm{div}(\rho_0\mathbf{\Phi}_T)/\rho_0$:

$$\frac{\partial\mathbf{b}}{\partial t} = \mathrm{rot}(\mathbf{u}\times\mathbf{b} - \nu_m\mathrm{rot}\mathbf{b}) - \mathrm{rot}\boldsymbol{E}_T, \tag{105}$$

$$\frac{\partial\mathbf{u}}{\partial t} = -\vec{\nabla}\left(\frac{\tilde{p}_{tot}}{\rho_0}\right) - \mathbf{g}\theta + \frac{\mathbf{f}_M}{\rho_0} + \mathbf{u}\times\boldsymbol{\omega} + \mathbf{f}_\nu(\mathbf{u}) - \frac{\mathbf{F}^{(T)}}{\rho_0}, \tag{106}$$

$$\frac{\partial\theta}{\partial t} = -\frac{\Omega_B^2}{g}u_z - (\mathbf{u}\cdot\nabla\theta) + \frac{e}{c_p\rho_0\bar{\bar{T}}} + \frac{1}{\bar{\bar{T}}\rho_0}\mathrm{div}(\kappa\rho_0\nabla\tilde{T}) + \frac{1}{\rho_0}\mathrm{div}(\rho_0\mathbf{\Phi}_T), \tag{107}$$

$$\mathrm{div}(\mathbf{u}) = \mathbf{\Lambda}\cdot\mathbf{u} \equiv -\frac{\nabla\rho_0}{\rho_0}\cdot\mathbf{u}. \tag{108}$$

Note that among solutions of (92)–(95) the solutions of (105)–(108) are only those for which $\langle\mathbf{b}\rangle = 0, \langle\mathbf{u}\rangle = 0, \langle\theta\rangle = 0$. Further, these solutions will be called background turbulence $\mathbf{b}_0$, $\mathbf{u}_0$, $\theta_0$.

Let us exemplarily compose the equation for a correlator $\langle u_i u_j\rangle \equiv f_{ij}(\mathbf{r},t), \langle u_i u_i\rangle \equiv f_{ii}(\mathbf{r},t) = \langle u^2\rangle = \langle u_x^2\rangle + \langle u_y^2\rangle + \langle u_z^2\rangle$. (Here from twice occurring indices one implies the summing up — the so-called Einstein rule.) For this aim, we compose equations for $i$ and $j$ components of Equation (102), multiply $i$ component by $u_j$, and $j$ component by $u_i$, add and average over turbulent fluctuations. As a result, we obtain

$$\frac{\partial f_{ij}}{\partial t} = \overleftrightarrow{\mathbf{I}}_{ij}^{\mathbf{u}} + \overleftrightarrow{\mathbf{N}}_{ij}^{\mathbf{u},\mathbf{b},\mathbf{u}} \tag{108a}$$

in the reference system accompanying the mean velocity **U**. Here $\overleftrightarrow{\mathbf{I}}_{ij}^{\mathbf{u}} = \langle u_j(\mathbf{I_u})_i\rangle + \langle u_i(\mathbf{I_u})_j\rangle, \overleftrightarrow{\mathbf{N}}_{ij}^{\mathbf{u},\mathbf{b},\mathbf{u}} = \left\langle(\mathbf{N}_{\mathbf{u},\mathbf{b},\mathbf{u}})_i u_j\right\rangle + \left\langle(\mathbf{N}_{\mathbf{u},\mathbf{b},\mathbf{u}})_j u_i\right\rangle$. When the mean field is absent, the mean velocity is homogeneous, as well as entropy — $\mathbf{I}^{\mathbf{u}}_{ij} = 0$; hence, Equation (108a) has the form $\partial f_{ij}^{(0)}/\partial t = \overleftrightarrow{\mathbf{N}}_{ij}^{\mathbf{u},\mathbf{b},\mathbf{u}}(\mathbf{u}_0,\mathbf{b}_0,s_0)$. Thus, in accordance with the formulated above definition of background turbulence (solution of the system (105)–(108)), one can suppose that the latter equation is an equation for background hydrodynamic turbulence. Notice that totally for the description of turbulence, in addition to (108a), five more equations are required: two equations for tensors $g_{ij} = \langle u_i b_j \rangle, h_{ij} = \langle b_i b_j \rangle$, two for vectors $(\mathbf{\Phi}_T)_i = \langle u_i \theta \rangle, (\boldsymbol{\beta}_T)_i = \langle b_i \theta \rangle$, and one for scalars $\Theta = \langle \theta^2 \rangle$. Importance of three tensors and the vector $\mathbf{\Phi}_T$ is obvious from expressions for turbulent contributions into the system (96)–(99), i.e., for electromotive and turbulent forces $(\boldsymbol{E}_T)_i = \varepsilon_{ijk} g_{jk}, F_i^{(T)} = \partial \sigma_{ij}^T / \partial x_j$, as well as for the entropy flow $(\mathbf{\Phi}_T)_i = \langle u_i \theta \rangle$. The other two correlators are also needed to calculate these main and obvious ones. For instance, on the right side of the equation for $g_{ij}$ there is the summand $\langle (\mathbf{N}_{\mathbf{u},\mathbf{b},\mathbf{u}})_i b_j \rangle$, with a contribution of the form $(-\mathbf{g})_i \langle \theta b_j \rangle \equiv (-\mathbf{g})_i (\boldsymbol{\beta}_T)_i$. Analogously, the scalar $\Theta = \langle \theta^2 \rangle$ is involved into the right side of the equation for the entropy flow through the summand $\langle \theta \mathbf{N}_{\mathbf{u},\mathbf{b},\mathbf{u}} \rangle$, with a contribution of $-\mathbf{g} \langle \theta^2 \rangle \equiv -\mathbf{g}\Theta$. The system of five equations, similar to (108a), is rather complicated and unclosed. For example, on the right side of Equation (108a) in the term $\overleftrightarrow{\mathbf{N}}_{ij}^{\mathbf{u},\mathbf{b},\mathbf{u}}$ there are contributions of the form $\langle (\mathbf{u} \times \boldsymbol{\omega})_i u_j \rangle \equiv \langle (\mathbf{u} \times \mathrm{rot}\, \mathbf{u})_i u_j \rangle$, i.e. the third momenta $\langle (\mathbf{u} \times \mathrm{rot}\mathbf{u})_i u_j \rangle = \langle u_j \nabla_i u^2 / 2 \rangle - \langle u_j (\mathbf{u} \nabla) u_i \rangle$ are involved into the equations for the second momenta $f_{ij}(\mathbf{r}, t) = \langle u_i u_j \rangle$. An attempt to construct equations for the third momenta leads to the equations comprising the forth momenta, and so on. Thus, the system of equations proves to be unclosed all the time, whereas the number of equations in it rapidly grows. Therefore, the only possibility to get a compact and foreseeable answer for turbulent contributions is to break the system substituting, for instance, the third momenta for the second ones. In order to make it, let us formulate two hypotheses of Vainshtein (1970):

1) background turbulence $f_{ij}^{(0)}$ is *stationary*, i.e., $\partial f_{ij}^{(0)} / \partial t = \overleftrightarrow{\mathbf{N}}_{ij}^{\mathbf{u},\mathbf{b},\mathbf{u}}(\mathbf{u}_0, \mathbf{b}_0, s_0) = 0$;
2) background turbulence $f_{ij}^{(0)}$ is *stable in statistical sense*. If one "turns off immediately" the sources of turbulence, for instance $\overleftrightarrow{\mathbf{I}}_{ij}^{\mathbf{u}}$, then over several lifetimes of energy-carrying vortices the turbulence should relax to the background turbulence $f_{ij}^{(0)}$. In other words, the simplest form of the solution ensuring such behavior can be as follows:

$$f_{ij}(t) = f_{ij}^{(0)} + \left[ f_{ij}(t = t_0) - f_{ij}^{(0)} \right] \exp\left( -\frac{t - t_0}{\tau_0} \right) \tag{109}$$

In this solution, $t_0$ is the time of turning off the source $\overleftrightarrow{\mathbf{I}}_{ij}^{\mathbf{u}}$, $\tau_0$ is the turnover time of an energy-carrying eddy. After turning off the source $\overleftrightarrow{\mathbf{I}}_{ij}^{\mathbf{u}}$ at the moment of time $t_0$, the solution, which at the moment $t_0$ was equal to $f_{ij}(t = t_0)$, exponentially relaxes to the background turbulence $f_{ij}^{(0)}$. To construct the $\tau$ approximation, let us subtract the equation for background turbulence from Equation (108), taking into account the hypothesis 1 — $\partial f_{ij}^{(0)} / \partial t = 0$. As a result, we get

$$\frac{\partial f_{ij}}{\partial t} = \overleftrightarrow{\mathbf{I}}_{ij}^{\mathbf{u}} + \overleftrightarrow{\mathbf{N}}_{ij}^{\mathbf{u},\mathbf{b},\mathbf{u}} - \overleftrightarrow{\mathbf{N}}_{ij}^{\mathbf{u},\mathbf{b},\mathbf{u}}(\mathbf{u}_0, \mathbf{b}_0, s_0)\,. \tag{110}$$

Turning off the source in this equation, we obtain the situation described by the solution of (109). Substituting this solution into the latter equation, we get

$$\overleftrightarrow{\mathbf{N}}_{ij}^{\mathbf{u},\mathbf{b},\mathbf{u}} - \overleftrightarrow{\mathbf{N}}_{ij}^{\mathbf{u},\mathbf{b},\mathbf{u}}(\mathbf{u}_0, \mathbf{b}_0, s_0) = -\frac{f_{ij} - f_{ij}^{(0)}}{\tau_0}\,.$$

Substituting this relation into (110), we obtain

$$\frac{\partial f_{ij}}{\partial t} = \overleftrightarrow{\mathbf{I}}_{ij}^{\mathbf{u},\mathbf{\Phi}} - \frac{f_{ij} - f_{ij}^{(0)}}{\tau_0}, \tag{111}$$

where $\overleftrightarrow{\mathbf{I}}_{ij}^{\mathbf{u},\mathbf{\Phi}} = \overleftrightarrow{\mathbf{I}}_{ij}^{\mathbf{u}} + \langle(2\mathbf{u}\times\mathbf{\Omega}_*)_i u_j\rangle + \langle(2\mathbf{u}\times\mathbf{\Omega}_*)_j u_i\rangle + g_i\left(\Phi_j^{(T)} - \Phi_j^{(T,0)}\right) + g_j\left(\Phi_i^{(T)} - \Phi_i^{(T,0)}\right)$, $\Phi_i^{(T,0)}$ is the entropy flow in the absence of the mean magnetic field, rotation, inhomogeneous mean flow, and so on. Usually, $\tau_0$ is much shorter than the characteristic time of the regular magnetic field variation. For example, following Fig. 104, at the bottom of the solar convective zone $\tau_0$ achieves seven months, which is approximately 38 times shorter than the solar 22-year cycle. Figure 111 shows that for other main-sequence stars, except for M5 types, this condition is also fulfilled. Then one can be restricted by a formal quasistationary solution for Equation (111):

$$f_{ij} = f_{ij}^{(0)} + \tau_0 \overleftrightarrow{\mathbf{I}}_{ij}^{\mathbf{u},\mathbf{\Phi}}\,. \tag{112}$$

Unfortunately, the solution of Equation (112) is not closed. As it follows from the source $\overleftrightarrow{\mathbf{I}}_{ij}^{\mathbf{u},\mathbf{\Phi}}$, it comprises, for example, the turbulent flow of entropy $\Phi_T$. Therefore, in is necessary to add equations for other correlators $\mathbf{\Phi}_T = \langle\theta\mathbf{u}\rangle, \boldsymbol{\beta}_T = \langle\theta\mathbf{b}_j\rangle$, $h_{ij} = \langle b_i b_j\rangle$, and $g_{ij} = \langle u_i b_j\rangle$ to Equation (112). Since these equations are interrelated, they form a lengthy system of linear equations which require a joint solution. The case is complicated by the fact that, for instance, Equation (112) includes contributions of $\langle(2\mathbf{u}\times\mathbf{\Omega}_*)_i u_j\rangle + \langle(2\mathbf{u}\times\mathbf{\Omega}_*)_j u_i\rangle$ in the source $\overleftrightarrow{\mathbf{I}}_{ij}^{\mathbf{u},\mathbf{\Phi}}$, i.e., containing again the correlation $f_{ij} = \langle u_i u_j\rangle$. An analogous phenomenon is observed in equations for $\mathbf{\Phi}_T$ and $\boldsymbol{\beta}_T$. Therefore, to solve equations for $\mathbf{\Phi}_T$, $\boldsymbol{\beta}_T$, and $f_{ij}$ with respect to these values, it proves to be necessary to turn matrices containing the angular rotation velocity of a star $\mathbf{\Omega}_*$ of the second and forth order, respectively. Note that in these calculations we have limited for simplicity to the only time of turbulence relaxation $\tau_0$. Such approximation is indeed widely used in the nonlinear dynamo theory (see, for example, Field et al., 1999; Pipin, 1999). Indeed, following Kolmogorov (1941), the time $\tau_0$ should depend on the eddy scale. This problem can be solved if turbulence is described in the physical rather than in the Fourier space. The additional advantage of such a description is a transformation of differentiation and integration operators into multiplication and division operations, as well as a significant simplification of matrices describing the influence of rotation on turbulence. It is unlikely to be relevant to provide details of all these calculations within the framework of this chapter. One can find them in the papers of Field et al. (1999), Vainshtein (1970, 1971, 1980), Vainshtein et al. (1972, 1980), Kitchatinov (1986, 1987), Kitchatinov and Rüdiger (1993, 1995, 1997, 2005), Kleeorin and Ruzmaikin (1982), Kleeorin et al. (1990), Pipin (1999), Radler et al. (2003), Rogachevskii and Kleeorin (1997, 1999, 2007). In this list it is important to note the exceptionally important direct numerical verifications of analytical results of the mentioned above authors by A. Brandenburg and his colleagues with both the direct numerical modeling of unaveraged equations (92)–(95) and the developed test-field method (see, for example, Rheinhardt and Brandenburg, 2010). This method explores the response of turbulence to infinitesimal mean-field perturbations and makes it possible to measure turbulent diffusion of the magnetic field, the alpha effect value, and so on practically in the same way as in the laboratory if it would be possible in this laboratory to reproduce the highly conducting medium at large magnetic and hydrodynamic Reynolds numbers. In the course of these experiments, the new MHD phenomena omitted by theoreticians were detected, as well as some undetected phenomena predicted by them. However, it is not excluded that to detect them, the higher Reynolds numbers are required; this can be achieved in the direct numerical modeling today and, probably, in the coming years.

**4.3.3.2. Basic Results of the Nonlinear Dynamo Theory Important for the Theory of Mean Magnetic Fields of Stars.** In this section, we outline the expressions acquired for correlators $\rho_0 \langle u_i u_j \rangle, \langle b_i b_j \rangle, (\boldsymbol{E}_T)_i = \varepsilon_{ijk} \langle u_j b_k \rangle$ and $\sigma_{ij}^T = -\rho_0 \langle u_i u_j \rangle + \left( \langle b_i b_j \rangle - \delta_{ij} \langle b^2 \rangle / 2 \right) / 4\pi$ that are important in the context of the nonlinear stellar dynamo. First of all, let us provide the expression for the "electromotive" force $\mathsf{E}_T$, its nonlinear version is as follows:

$$\boldsymbol{E}_T \equiv \langle \mathbf{u} \times \mathbf{b} \rangle = \ddot{\boldsymbol{\alpha}}(\mathbf{B})\mathbf{B} - \eta_T(\mathbf{B}) \nabla \times \mathbf{B} + \left[ \mathbf{V}_{eff}(\mathbf{B}) \times \mathbf{B} \right].$$

Here, such additional effects as kappa and delta effects $\kappa(\mathbf{B})$ и $\delta(\mathbf{B})$ are not considered (Steenbek et al., 1966). The nonlinear alpha effect $\ddot{\boldsymbol{\alpha}}(\mathbf{B})$ has a matrix character, and here the dyadic designations are appropriate for the tensor (matrix) of the alpha effect. Thus, in the nonlinear problem, $(\ddot{\boldsymbol{\alpha}})_{ij}$ acquires a tensor character, which certainly significantly complicates the mathematical form of stellar dynamo equations. Fortunately, within the axisymmetric spherical $\alpha$–$\omega$ dynamo, the only one component is important — tensor $\alpha_{\varphi\varphi}(\mathbf{B})$, since part of the longitudinal component of the electromotive force associated with the alpha effect has the form $\left( \boldsymbol{E}_T^{(\alpha)} \right)_\varphi = \alpha_{\varphi\varphi}(\mathbf{B}) B_\varphi$ . This component $\alpha_{\varphi\varphi}(\mathbf{B})$ is as follows:

$$\alpha_{\varphi\varphi}(\mathbf{B}) = \psi_h(\mathbf{B})\alpha_h + \psi_m(\mathbf{B})\alpha_m. \tag{113}$$

Here $\alpha_h$ is the value of the hydrodynamic part of the alpha effect — in the first approximation the appropriate coefficient of the linear theory (see Sect. 4.1.2 and Fig. 102, 103, 105, 106). In the relation (113) there appears a principally new contribution that is proportional to the magnetic current helicity of the turbulent magnetic field, the so-called magnetic part of the alpha effect revealed by Frish et al. (1975) (see also Pouquet et al., 1976):

$$\alpha_m = \frac{\tau_0}{12\pi\rho}\langle \mathbf{b}\cdot \mathrm{rot}\,\mathbf{b}\rangle \equiv \frac{\tau_0 \chi_c}{12\pi\rho}\,.$$

The studies have shown that if the mean magnetic field **B** is equal to zero then the magnetic part of the alpha effect turns to zero. Moreover, the association of $\alpha_m$ with **B** was revealed to have the form of not an algebraic equation but defined by the *dynamic* evolutionaty equation for the magnetic part of the alpha effect. Therefore, $\alpha_m$ was named dynamic nonlinearity. As to functions $\psi_h(\mathbf{B})$ and $\psi_m(\mathbf{B})$, they were named algebraic nonlinearities. Functions $\psi_m(\mathbf{B})$ were found in Field, Blackman, and Chou (1999), whereas $\psi_h(\mathbf{B})$ — in Rogachevskii and Kleeorin (2000, 2001, 2004), Fig. 112. These are as follows:

$$\psi_m = 3\left[1-\arctan(\sqrt{8}\bar{B})/\sqrt{8}\bar{B}\right]/8\bar{B}^2\,,$$

$$\psi_h = \left[4\psi_m(\mathbf{B}) + 3L(\mathbf{B})\right]/7\,,$$

where $L(\mathbf{B}) = 1-16\bar{B}^2 + 128\bar{B}^4 \ln\left(1+1/8\bar{B}^2\right)$, $\bar{B} = |\mathbf{B}|/B_{eq} = |\mathbf{B}|/u_0\sqrt{4\pi\rho_0}$ is the value of the field normalized over the field $B_{eq} = u_0\sqrt{4\pi\rho_0}$, whose energy is equal to the kinetic energy of turbulence.

#### 4.3.3.3. Basic Nonlinear Mechanisms Revealing Within the Nonlinear Alpha Effect

A. Alfvénization of turbulence. In Fig. 112, the thin lines denote the dependences of algebraic nonlinearities approximated with the Iroshnikov function (1971) (see also Rüdiger (1974), Roberts and Soward (1975):

$$\psi = \frac{1}{1+\zeta\bar{B}^2}\,.$$

Specifically, it turned out that for $\psi_m$ $\zeta \approx 3.738$, whereas for $L(B)$— $\zeta = 12$. The Iroshnikov functions describe, roughly speaking, the process of stellar plasma Alfvénization as the magnetic field increases. The nature of Alfvénization consists in the gradual replacement of ordinary Kolmogorov turbulence by an ensemble of nonlinear Alfvén waves (see, e.g., Moffett (1980)) as the mean magnetic field **B** increases.

Nonetheless, with the typical turbulent time there occurs the following: if **B** = 0, the turnover time of the turbulent vortex $\tau_h \simeq \ell_0/u_0$ is the turbulent time $\tau_0$. Conversely, in the very strong fields $\tau_0$ transforms into the Alfvén time $\tau_A \simeq \ell_0/V_A$, where $V_A = B/\sqrt{4\pi\rho}$. In the intermediate situation, certainly, the faster one of these two mechanisms dominates, i.e., $\tau_0^{-2} \simeq \zeta\tau_A^{-2} + \tau_h^{-2}$ or

$$\tau_0 \simeq \frac{\tau_h\tau_A}{\sqrt{\zeta\tau_h^2 + \tau_A^2}} \to \frac{\tau_0}{\sqrt{1+\zeta\bar{B}^2}}\,. \tag{114}$$

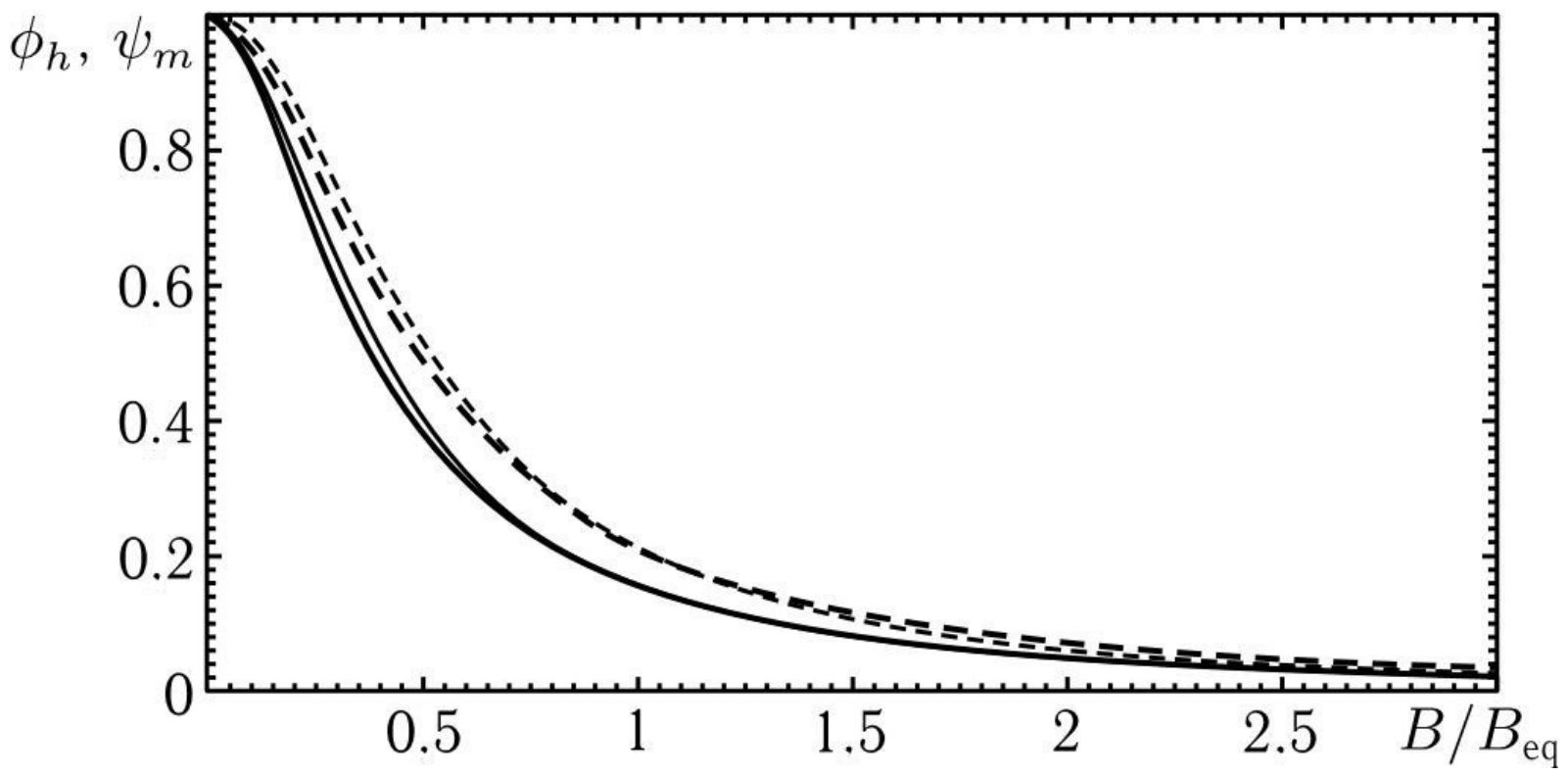

Fig. 112. Dependences of algebraic nonlinearities $\psi_h(\mathbf{B})$ (*solid thick line*) and $\psi_m(\mathbf{B})$ (*dashed thick line*) on the magnetic field value; *thin lines* — dependences of algebraic nonlinearities approximated with the Iroshnikov functions (1971)

If we assume that for the hydrodynamic helicity $\chi_h = \langle \mathbf{u} \cdot \mathrm{rot}\mathbf{u} \rangle$ the evolution equation of the type of (111),

$$\frac{\partial \langle \mathbf{u} \cdot \mathrm{rot}\mathbf{u} \rangle}{\partial t} = I_{\langle \mathbf{u} \cdot \mathrm{rot}\mathbf{u} \rangle} - \frac{\langle \mathbf{u} \cdot \mathrm{rot}\mathbf{u} \rangle}{\tau_0},$$

then we immediately find $\langle \mathbf{u} \cdot \mathrm{rot}\mathbf{u} \rangle = \tau_0 I_{\langle \mathbf{u} \cdot \mathrm{rot}\mathbf{u} \rangle}$. This indeed yields the Iroshnikov function $\alpha = -\tau_0^2 I_{\langle \mathbf{u} \cdot \mathrm{rot}\mathbf{u} \rangle} / 3 \to \alpha \simeq \alpha_0 / \left(1 + \zeta \bar{B}^2\right)$ for the alpha effect.

Let us move from the algebraic nonlinearity to dynamic one that is associated with current helicity $\chi_c = \langle \mathbf{b} \cdot \mathrm{rot}\,\mathbf{b} \rangle$ (see Equation (113)).

B. Dynamic nonlinearity and magnetic current helicity. Note that magnetic helicity was so named since this helicity is proportional to the current: $\mathrm{rot}\,\mathbf{b} \simeq 4\pi \mathbf{j}/c$. In fact, the condition $\chi_m = \langle \mathbf{b} \cdot \mathrm{rot}\,\mathbf{b} \rangle \neq 0$ means that the lines of force of magnetic field fluctuations have a form of helices, whose step is $\ell_{\chi_c} \approx \langle b^2 \rangle / \langle \mathbf{b} \cdot \mathrm{rot}\,\mathbf{b} \rangle$ and $\ell_{\chi_c} \geq \ell_0$, where $\ell_0$ is the typical size of turbulent vortices. If this helicity is small, $\ell_{\chi_c} \gg \ell_0$ and lines of force turn out to be not helical. We can roughly imagine that the force lines of magnetic field fluctuations represent a set of thin closed ($\mathrm{div}\mathbf{b} = 0$!) “weak magnetic mainsprings” of the size $\geq \ell_0$ along which the current flows with the density $\mathbf{j} \simeq c\ \mathrm{rot}\,\mathbf{b}/4\pi$. The studied mean magnetic field **B** is dipped into such a “magnetic medium”. Intuitively, it is almost obviously that in such a medium it is harder for the viscosity and Coriolis forces to turn around and rise (dip) the elements of force lines of the large-scale magnetic field, i.e., to implement motions constituting the essence of the kinematic alpha effect. Nonetheless, it is obvious that the “intuitive glance” outlined here is rather insufficient for understanding properties of the magnetic part of the alpha effect. To understand these properties, let us refer to the other helicity — magnetic $\chi_m = \langle \mathbf{a}\cdot\mathbf{b}\rangle$, where **a** is the vector potential of the magnetic field, i.e., **b** = rot**a**. First of all, one can show that the value $\chi_m$ is associated with the magnetic part of the alpha effect even tighter than $\chi_c$. Indeed, the precise expression for the magnetic part of the alpha effect is as follows: $\alpha_{ij}^{(m)} = \psi_m(\mathbf{B})\left(\varepsilon_{imn}\langle\tau b_n\nabla_j b_m\rangle + \varepsilon_{jmn}\langle\tau b_n\nabla_i b_m\rangle\right)/8\pi\rho$ (Rogachevskii and Kleeorin, 1999). Referring to the Fourier transforms for the fields in this equation, we get

$$\alpha_{ij}^{(m)} = \frac{1}{4\pi\rho}\int\tau(k)k_ik_j\langle\mathbf{a}(\mathbf{k})\cdot\mathbf{b}(-\mathbf{k})\rangle d\mathbf{k}. \quad (115)$$

Equation (115) was obtained within the Coulomb calibration div**a** = 0 (see Equation (78a)), whereas **a**(**k**) and **b**(**k**) are the Fourier transforms for appropriate fields. Notice that in the Fourier space (or in the **k** space) the condition div**a** = 0 has the form **k** · **a** = 0. This condition was directly used to obtain Equation (115). For a further study of dynamic nonlinearity it is crucial that the helicity $\mathrm{X}_m = \mathbf{H}\cdot\mathbf{A}_{\mathrm{tot}}$ is, as well as energy, the quasiintegral of motion. Here $\mathbf{A}_{\mathrm{tot}} = \mathbf{A} + \mathbf{a}$ is the vector potential of the total magnetic field **H** = **B** + **b** = rot $\mathbf{A}_{\mathrm{tot}}$. To prove this fact, one can apply the induction equation (75). Substituting **H** = rot $\mathbf{A}_{\mathrm{tot}}$ into it and eliminating the operator rot, we get the equation for the vector potential $\mathbf{A}_{\mathrm{tot}}$:

$$\frac{\partial\mathbf{A}_{tot}}{\partial t} = \mathbf{v}\times\mathbf{H} - \nu_m\,\mathrm{rot}\left(\mathrm{rot}\mathbf{A}_{tot}\right) - \nabla\Phi. \quad (116)$$

Not defining concretely the calibration, we multiply Equation (116) by **H**, Equation (75) by $\mathbf{A}_{\mathrm{tot}}$, and then sum them. As a result, we obtain the equation for the helicity $\mathrm{X}_m$:

$$\frac{\partial\mathrm{X}_m}{\partial t} = -\mathrm{div}\left(\mathbf{\Phi}_m\right) - 2\nu_m\mathrm{X}_c. \quad (117)$$

Here $\mathbf{\Phi}_m = \mathbf{A}_{tot}\times(\mathbf{v}\times\mathbf{H}) - \nu_m\mathbf{A}_{tot}\times\mathrm{rot}\mathbf{H} + \mathbf{H}\Phi$ is the magnetic helicity flux, the magnetic helicity losses $D_{\mathrm{X}_m}$ account for $D_{\mathrm{X}_m} = 2\nu_m\mathbf{H}\cdot\mathrm{rot}\,\mathbf{H} = 2\nu_m\mathrm{X}_c$ and are proportional to current helicity. If we take the Coulomb calibration div$\mathbf{A}_{\mathrm{tot}}$ = 0 for $\mathbf{A}_{\mathrm{tot}}$, then it can be shown that $\mathrm{X}_c$ and $\mathrm{X}_m$ have the same signs, i.e., $D_{\mathrm{X}_m}$ indeed leads to the dissipation of helicity. A helicity flux does not produce helicity and does not destroy it but just redistributes it in space. Thus, $\mathrm{X}_m$ behaves as a quasiintegral of motion with the zero source. At the zero magnetic diffusion $\nu_m$ the total value of magnetic helicity in all the space $X_m$ is maintained since $dX_m/dt = \int\partial\mathrm{X}_m/\partial t\,d\mathbf{r} = 0$ or $X_m = \int\mathrm{X}_m d\mathbf{r} = \mathrm{const}$. If magnetic diffusion is positive, then, due to the coincidence of signs of $\mathrm{X}_c$ and $\mathrm{X}_m$, the helicity loss occurs at every point in space. This means that in the stationary magnetic fields the only one asymptotic state is possible (at $t \to \infty$): $\mathrm{X}_m$ = 0. Averaging this condition, we obtain $\langle\mathrm{X}_m\rangle = \mathbf{A}\cdot\mathbf{B} + \langle\mathbf{a}\cdot\mathbf{b}\rangle = 0$ or $\langle \mathbf{a} \cdot \mathbf{b} \rangle = -\mathbf{A} \cdot \mathbf{B}$. From Equations (78a) and (78) we can obtain the following equation for the mean field helicity $\mathbf{A} \cdot \mathbf{B}$:

$$\frac{\partial \mathbf{A} \cdot \mathbf{B}}{\partial t} = -\operatorname{div}\left(\mathbf{\Phi}_{\mathbf{A} \cdot \mathbf{B}}\right) + 2\left(\alpha \mathbf{B}^2 - \left(\nu_m + \eta_T\right)\mathbf{B} \cdot \operatorname{rot}\mathbf{B}\right). \tag{118}$$

Here $\mathbf{\Phi}_{\mathbf{A} \cdot \mathbf{B}} = \mathbf{A} \times \left(\mathbf{U} \times \mathbf{B}\right) - \mathbf{B}\psi - \left(\nu_m + \eta_T\right)\mathbf{A} \times \operatorname{rot}\mathbf{B} + \alpha \mathbf{A} \times \mathbf{B}$ is the mean field helicity flux, and $\alpha$ should be understood to mean (113). If one can neglect the divergence of the mean field flux, then it follows from (118) that

$$\mathbf{B} \cdot \operatorname{rot}\mathbf{B} \approx \frac{1}{\nu_m + \eta_T}\left(\alpha \mathbf{B}^2 - \frac{\partial \mathbf{A} \cdot \mathbf{B}}{\partial t}\right). \tag{119}$$

In the case of the stationary dynamo (as in galaxies) the summand $\partial \mathbf{A} \cdot \mathbf{B} / \partial t$ turns into zero. In the case of the stellar (quasiperiodic) dynamo, this summand can be omitted approximately by averaging (119) over several binary cycles of stellar activity. Thus, taking into account that rot **B**, we obtain for stars:

$$\overline{\alpha_m} \propto \overline{\langle \mathbf{a} \cdot \mathbf{b} \rangle} \propto -\overline{\mathbf{B} \cdot \operatorname{rot}\mathbf{B}} \approx -\frac{\overline{\alpha \mathbf{B}^2}}{\nu_m + \eta_T}, \tag{120}$$

where it is accounted that, according to (115), $\alpha_m \propto \langle \mathbf{a} \cdot \mathbf{b} \rangle$, and the lines above the values mean the averaging over several binary cycles of stellar activity. Thus, from (120) it clearly follows that signs of the alpha effect and the magnetic part are obviously opposite and, consequently, the magnetic part of the alpha effect can on average stabilize, for example, the $\alpha$–$\omega$ dynamo through several cycles. For further progress in the development of the dynamic nonlinearity theory it is required: 1) to obtain the equation for mean magnetic helicity of fluctuation magnetic fields $\chi_m = \langle \mathbf{a} \cdot \mathbf{b} \rangle$; 2) using Equation (115), to calculate the magnetic part of the alpha effect taking into account the spectral properties of this helicity. For this aim, we should multiply Equation (92) by $\mathbf{b}$ and Equation (75) by $\mathbf{a}$, to sum them and average. This yields

$$\frac{\partial \chi_m}{\partial t} = -2\boldsymbol{E} \cdot \mathbf{B} - \operatorname{div}\left(\mathbf{\Phi}_{\chi_m}\right) - 2\nu_m \langle \mathbf{b} \cdot \operatorname{rot}\mathbf{b} \rangle, \tag{121}$$

where the small-scale helicity flux is $\mathbf{\Phi}_{\chi_m} = \mathbf{U}\chi_m - \left\langle \mathbf{b}\left(\mathbf{U} \cdot \mathbf{a}\right)\right\rangle + \langle \mathbf{u}\left(\mathbf{b} \cdot \mathbf{a}\right)\rangle - \langle \mathbf{b}\left(\mathbf{u} \cdot \mathbf{a}\right)\rangle - \mathbf{B}\langle \mathbf{u} \cdot \mathbf{a}\rangle + \left\langle \mathbf{u}\left(\mathbf{B} \cdot \mathbf{a}\right)\right\rangle$ $+ \langle \mathbf{b}\phi \rangle - \nu_m \langle \mathbf{a} \times \operatorname{rot}\mathbf{b} \rangle$. It should be reviewed the question concerning the decay or helicity relaxation $\chi_m$ defined by its spectral behavior. Since the spectral properties of this helicity are not ordinary, they should be considered in detail. First of all, according to (121), the mean magnetic helicity of fluctuation magnetic fields $\chi_m$ is a quasiintegral of motion: it has the source $I_{\chi_m} = -2\boldsymbol{E} \cdot \mathbf{B} = -2\left(\alpha \mathbf{B}^2 + \eta_T \mathbf{B} \cdot \operatorname{rot}\mathbf{B}\right)$ defined by the mean field, there is a term yielding the helicity dissipation $D_{\chi_m} = 2\nu_m \langle \mathbf{b} \cdot \operatorname{rot}\mathbf{b} \rangle \equiv 2\nu_m \langle \operatorname{rot}\mathbf{a} \cdot \operatorname{rot}\mathbf{b} \rangle$ and the spatial flux $\mathbf{\Phi}_{\chi_m}$ redistributing helicity in space. According to the general theory (Moffett, 1978;

Vainshtein et al., 1980) such a quasiintegral should have the spectrum $\chi_m(k) \propto (k/k_0)^{-q}$ in the inertial interval of turbulence, i.e., for the scales $k_0^{-1} \sim \ell_0 \gg \ell_k \sim k^{-1} \gg \ell_\nu \sim k_0^{-1}\,\mathrm{Rm}^{-1/(3-q)}$. According to Kolmogorov (1941), $q \approx 5/3$. Certainly, it is impossible to create helicity without generating a certain portion of energy. This simple fact is expressed by the condition of existence of helicity (see, e.g., Moffett (1978)) — the so-called condition of implementability:

$$\chi_m(k) \le 2E_m(k)/k\,. \tag{122}$$

Here $E_m(k)$ is the spectral energy density of magnetic fluctuations. The theoretical reasons (see, e.g., Moffett (1978)) and solar observations (see, e.g., Stenflo (1978)) show that this spectrum of magnetic energy is as follows: $E_m(k) \propto (k/k_0)^{-q_E}$, where $1 < q_E < 5/3$. Substituting this into (122) and taking into account that $q \approx 5/3$, we obtain

$$1 \ge C \le (k/k_0)^{2/3-q_E}\,, \tag{123}$$

where $C$ is some constant that is less than unity. Taking into account that $1 < q_E < 5/3$, we obtain that the condition (123) is practically immediately disturbed. The latter means that instead of the normal integral of motion with the spectrum $\chi_m(k) \propto (k/k_0)^{-5/3}$ there appears a value with the turned-off flux over the spectrum, when all helicity is accumulated in the integral scale of turbulence $\ell_0 \sim k_0^{-1}$. In particular, the recent direct numerical modeling of Brandenburg et al. (2018) showed that $\chi_m(k) \propto (k/k_0)^{-11/3}$. This allows us to estimate the dissipation of magnetic helicity as

$$D_{\chi_m} = 2\nu_m\langle \mathbf{b}\cdot \mathrm{rot}\mathbf{b}\rangle \equiv 2\nu_m \langle \mathrm{rot}\mathbf{a}\cdot\mathrm{rot}\mathbf{b}\rangle \sim \nu_m \frac{\chi_m}{\ell_0^2} \approx \frac{\chi_m}{\tau_0\,\mathrm{Rm}} = \frac{\chi_m}{T}\,. \tag{124}$$

In fact, the point is in the limitation of the magnetic helicity spectrum $\chi_m(k) \propto (k/k_0)^{-11/3}$. Due to this, only external scales take part in the dissipation of magnetic helicity. Using Equation (115) and taking the helicity spectrum into account, we obtain the relation between the magnetic helicity $\chi_m$ with the magnetic part of the alpha effect $\alpha_{ij}^{(m)}$ (Kleeorin and Ruzmaikin, 1982):

$$\begin{aligned} \alpha_{ij}^{(m)} &= \frac{1}{4\pi\rho}\int \tau(k) k_i k_j \langle \mathbf{a}(\mathbf{k})\cdot\mathbf{b}(-\mathbf{k})\rangle d\mathbf{k} \approx \delta_{ij}\alpha_m \\ \alpha_m &= \frac{\tau_0\chi_m}{12\pi\rho_0\ell_0^2} = \frac{\mu}{4\pi\rho_0\eta_T}\chi_m \end{aligned}\,, \tag{125}$$

where $\mu = 1/9$ and the equation $\tau_0/\ell_0^2 = 1/3\eta_T$ is used. Combining Equations (121), (124), and (125), we get the equation for the magnetic part of the alpha effect $\alpha_m$ (Kleeorin and Ruzmaikin, 1982):

$$\frac{\partial \alpha_m}{\partial t} = \frac{\mu}{2\pi\rho_0}\left(\mathbf{B}\cdot\mathrm{rot}\,\mathbf{B} - \frac{\alpha\mathbf{B}^2}{\eta_T}\right) - \frac{\mu}{4\pi\rho_0\eta_T}\mathrm{div}\left(\mathbf{\Phi}_{\chi_m}\right) - \frac{\alpha_m}{T}\,. \tag{126}$$

The main feature of this equation is the extremely long relaxation time of the magnetic part of the alpha effect $T = \tau_0\mathrm{Rm}$, whereas the hydrodynamic part of the alpha effect relaxes over the time $\tau_0$. For example, for the Sun, $\tau_0$ on the upper boundary of granulation accounts for 5–6 minutes, Rm ~ $10^5$, and for the relaxation time of the magnetic part of the alpha effect we obtain $T$ ~ 1 year, which is significantly lower than the solar activity period (~ 11 years). However, at a depth of ~ $1.5 \cdot 10^3$ km $\tau_0$ is 17 min, Rm ~ $3 \times 10^6$, and the relaxation time of the magnetic part of the alpha effect $T$ achieves 90 years, which is almost 10 times higher than the cyclic activity period. On the whole, with increasing depth inside the convective zone $T$ dramatically grows: at the depth of supergranulation (~ 10000 km) $T$ proves to be of an order of twenty thousand years. It should be expected that the relaxation time of the magnetic part of the alpha effect $T$ will behave analogously on the solar-type stars as well.

Let us estimate the magnetic field level following from Equation (126). For this aim, we average this equation by averaging (119) over several binary cycles of stellar activity, which yields the stationary version of (126). Combining this stationary version of Equation (126) with Expression (113) for the full alpha effect, we get

$$\alpha = \frac{3\psi_h\alpha_h + \psi_m\mathrm{Rm}\left[2\eta_T\bar{\mathrm{X}}_{\mathbf{B}\cdot\mathrm{rot}\mathbf{B}} - \mathrm{div}\left(\bar{\mathbf{\Phi}}_{\chi_m}\right)\right]}{\left(3 + 2\psi_m\,\mathrm{Rm}\,\bar{B}^2\right)}\,.$$

Here $\bar{\mathrm{X}}_{\mathbf{B}\cdot\mathrm{rot}\mathbf{B}} = \mathbf{B}\cdot\mathrm{rot}\,\mathbf{B}/B_{eq}^2$ is the normalized current helicity of the mean field, $\bar{\mathbf{\Phi}}_{\chi_m} = \mathbf{\Phi}_{\chi_m}/B_{eq}^2$ is the mean normalized helicity flux of turbulent magnetic field fluctuations. Nonlinear numerical solutions of the solar (stellar) dynamo problems show that at the nonlinear stage $\alpha \approx \alpha_h D_{\mathrm{cr}}/D$. This yields the following estimate of the magnetic field:

$$B \approx B_{eq}\sqrt{\left[\frac{3}{2\,\mathrm{Rm}} + \frac{\eta_T\bar{\mathrm{X}}_{\mathbf{B}\cdot\mathrm{rot}\mathbf{B}}}{\alpha_h} - \frac{\mathrm{div}\left(\bar{\mathbf{\Phi}}_{\chi_m}\right)}{2\alpha_h}\right]\frac{D}{D_{cr}}}\;, \tag{127}$$

where it is taken for the estimate that $\psi_h(\mathbf{B})/\psi_m(\mathbf{B}) \approx 1$ (see Fig. 112). Let us estimate the field in accordance with (127). Consider the simplest case of the homogeneous field $\mathbf{B}$ (with $\bar{\mathrm{X}}_{\mathbf{B}\cdot\mathrm{rot}\mathbf{B}} = 0$) and the helicity flux of turbulent magnetic field fluctuations $\mathbf{\Phi}_{\chi_m}$ (with $\mathrm{div}\left(\bar{\mathbf{\Phi}}_{\chi_m}\right) = 0$). This yields $B \approx B_{eq}\sqrt{3D/\left(2D_{cr}\,\mathrm{Rm}\right)}$, and for the middle part of the solar convective zone (the depth $d$ ~ $10^{10}$ cm) at $D/D_{\mathrm{cr}} \approx 4$ we obtain $B \approx 10$ G. Although $B_{\mathrm{eq}} \approx$ 5300 G! The physical reason of such strange behavior is the significant accumulation of magnetic helicity over the huge relaxation time $T = \tau_0\mathrm{Rm}$, if, for example, it is not taken aside by the helicity flux $\mathbf{\Phi}_{\chi m}$. This case of paradoxically weak magnetic field generation was named in the literature as the catastrophic (nonlinear) suppression of magnetic field generation and made a lot of noise at its time (see, e.g., Gruzinov and Diamond, 1994). As a result, during almost 10 years (1996–2005) the American *Astrophysical Journal* refused to publish papers on the dynamo theory since it (dynamo) is ineffective! Note that the notion about the catastrophic suppression of magnetic field generation appeared already in the late 80s and proceeded for about 15 years basically under the influence of S.I. Vainshtein (Vainshtein and Cattaneo, 1992). The matter was got off the ground when the role of helicity flux $\mathbf{\Phi}_{\chi m}$ was figured out (Kleeorin et al., 2000, Blackman and Field, 2000, Vishniac and Cho, 2001). Indeed, for the homogeneous field **B** and the inhomogeneous helicity flux of turbulent magnetic field fluctuations $\mathbf{\Phi}_{\chi m}$ (with $\mathrm{div}\left(\bar{\mathbf{\Phi}}_{\chi_m}\right) \neq 0$) within Rm → ∞ we get

$$B \approx \sqrt{\left|\frac{\mathrm{div}\left(\mathbf{\Phi}_{\chi_m}\right)}{2\alpha_h}\right| \frac{D}{D_{cr}}} .$$

It clearly follows from the latter expression that no hints of the catastrophic suppression left. However, there is no any meaningful estimate of the field as well. Even if we assume that $\mathbf{\Phi}_{\chi_m} = -\kappa_T \nabla \chi_m$, then for the meaningful estimate of the field we can do nothing without solving Equation (126) in combination with the system of equations (81).

C. Reynolds-Maxwell stresses and negative effective magnetic pressure. Consider the nonlinear effects arising within the averaged Navier-Stokes equation. For this, it is required to calculate the effective Reynolds-Maxwell stresses

$$\sigma_{ij}^{T} = -\rho_0 \left\langle u_i u_j \right\rangle + \frac{1}{4\pi}\left(\left\langle b_i b_j \right\rangle - \frac{\delta_{ij}}{2}\left\langle b^2 \right\rangle\right) . \tag{128}$$

Let us suppose first for the simplicity of estimating that the turbulent fluctuations both hydrodynamic and magnetic are isotropic and $-\rho_0 \left\langle u_i u_j \right\rangle = -2 E_K \delta_{ij}/3, \left\langle b_i b_j \right\rangle / 4\pi = 2 E_M \delta_{ij}/3 = \delta_{ij} \left\langle b^2 \right\rangle / 8\pi$. Here $E_k$ and $E_m$ are the kinetic and magnetic energies of turbulence, respectively. Substituting these correlators into (128), we obtain

$$\sigma_{ij}^{T} = -\frac{2E_K}{3}\delta_{ij} - \frac{E_M}{3}\delta_{ij} = -\frac{2}{3}\left(E_{(tot)} - \frac{E_M}{2}\right)\delta_{ij} = -\left(P_T - \frac{E_M}{3}\right)\delta_{ij} . \tag{129}$$

Here $P_T = 2E_{(tot)}/3$; $E_{(tot)} = \langle u^2 \rangle / 2 + \langle b^2 \rangle / 8\pi$. As expected, Reynolds-Maxwell stresses in isotropic turbulence turn out to be isotropic. These stresses represent effective turbulent pressure $P_T^{(\mathrm{eff})} = P_T - E_M/3$ — the coefficient with the negative sign in front of the "individual" tensor $\delta_{ij}$ in (129). It particularly follows from this relation that at the same ordinary turbulent pressure $P_T$, the effective turbulent pressure $P_T^{(\mathrm{eff})}$ decreases with increasing energy of magnetic turbulent pulsations $E_m$. Suppose that in the absence of the mean field **B** the energy of turbulence and appropriate turbulent pressure are equal to $E_0^{(tot)}$ and $P_0^{(T)} = 2E_0^{(tot)}/3$, respectively. Let us superimpose the *homogeneous* field **B** onto the system. Since such a field cannot do work, then the total energy of turbulence does not change, but the effective turbulent pressure $P_T^{(\mathrm{eff})}$ decreases! Indeed, in the presence of the mean field the energy of turbulent pulsations certainly increases by $\delta E_m$ that is proportional to the mean field energy $B^2/8\pi$ : $\delta E_M = aB^2/8\pi$ . According to (129), this leads to the variation in effective turbulent pressure $\delta P_T^{(\text{eff})} = -\delta E_M/3 = -aB^2/24\pi$ . Adding ordinary magnetic pressure of the mean field $B^2/8\pi$ to the variation in effective turbulent pressure $\delta P_T^{(\text{eff})}$, we obtain that the effective pressure of the mean field in turbulence is (Kleeorin et al., 1989, 1990)

$$P_B^{(\text{eff})} = \left(1 - \frac{a}{3}\right)\frac{B^2}{8\pi} = \left(1 - q_p\right)\frac{B^2}{8\pi} \equiv Q_p \frac{B^2}{8\pi} . \qquad (129a)$$

In the absence of turbulence $a$ and $q_p$ tend to zero, $Q_p = 1$, whereas the effective pressure of the mean field $P_B^{(\text{eff})}$ coincides with ordinary magnetic pressure. Conversely, in strongly developed turbulence $q_p$ proves to be higher or even much higher than unity, in fact, $q_p \sim \ln(\text{Rm})$ (see below). This means that, for instance, at $\text{Rm} \approx 10^6$ $Q_p \sim -10$. Thus, $P_B^{(\text{eff})}$ indeed turns out to be negative and ten times higher in absolute value than ordinary magnetic pressure $B^2/8\pi$! As the field **B** grows, $q_p$ decreases and $Q_p$ initially achieves zero and then unity.

The described picture of isotropic turbulence at any magnetic fields is very simplified. Indeed, even if at the beginning (at **B** = 0) the turbulence is isotropic, then under the action of the field it becomes significantly anisotropic (see Kleeorin et al., 1990, 1996; Kleeorin and Rogachevskii, 1994; Rogachevskii and Kleeorin, 2007):

$$\begin{aligned} \left\langle b_i b_j \right\rangle &= \frac{\delta_{ij}}{3}\left\langle b_{B=0}^2 \right\rangle + \left(q_p - \frac{q_s}{2}\right) B^2 \delta_{ij} - \frac{q_s}{2} B_i B_j \\ \left\langle u_i u_j \right\rangle &= \frac{\delta_{ij}}{3}\left\langle u_{B=0}^2 \right\rangle - \frac{1}{4\pi\rho_0}\left[\left(q_p - \frac{q_s}{2}\right) B^2 \delta_{ij} - \frac{q_s}{2} B_i B_j\right] \end{aligned} . \qquad (130)$$

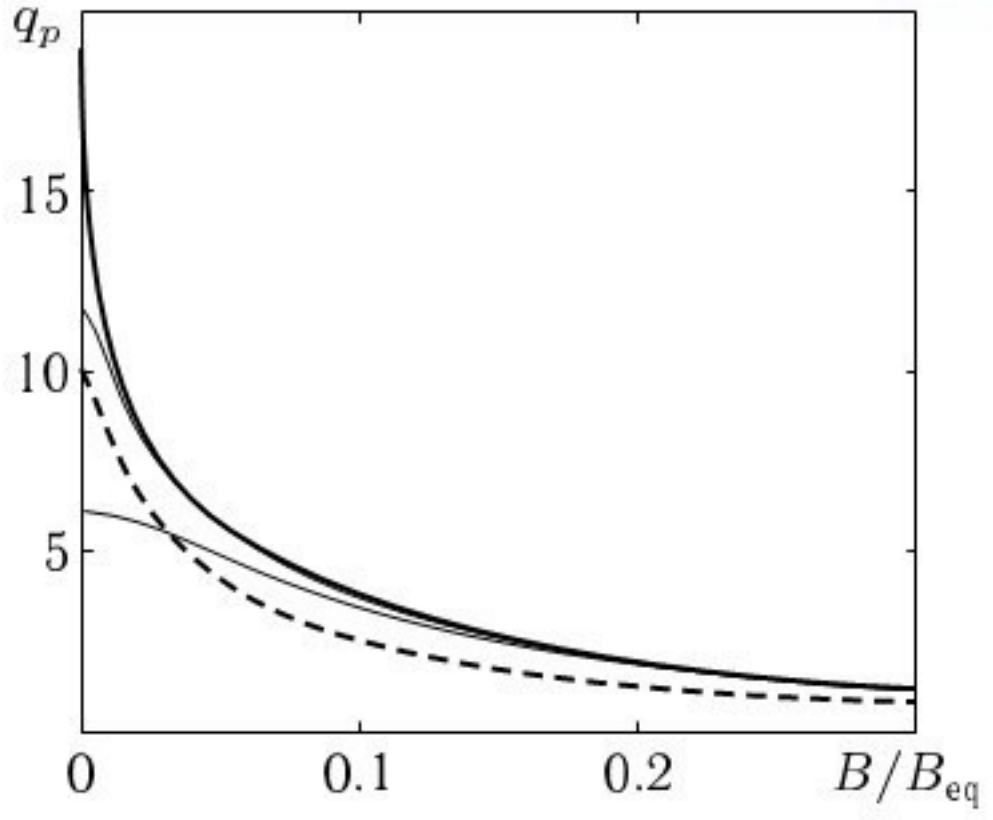

Fig. 113. Dependences of $q_p(\mathbf{B})$ at different values of the magnetic Reynolds number: $\text{Rm} \approx 10^{10}$ (*thick solid line*), $\text{Rm} \approx 10^6$ (*upper thin solid line*), $\text{Rm} \approx 10^3$ (*lower thin solid line*). The thick dashed line denotes the case when 70 % of turbulence energy is produced due to the forces of small-scale viscosity and $\text{Rm} \approx 10^6$. In all cases it was supposed that $\left\langle b_{B=0}^2 \right\rangle \ll 4\pi\rho_0 \left\langle u_{B=0}^2 \right\rangle$ , i.e., $\lambda = 1$ (Rogachevskii and Kleeorin, 2007)

A rather lengthy derivation of this equation within the $\tau$ approximation is not given here. We just note that coefficients $q_p$ and $q_s$ do not depend on the field **B** (Fig. 113), the parameter $\lambda = 1 - \left\langle b_{B=0}^2 \right\rangle / 4\pi\rho_0 \left\langle u_{B=0}^2 \right\rangle$, and the magnetic Reynolds number Rm. At $B \ll B_{eq} / 4\sqrt[4]{\mathrm{Rm}}$ $q_p \approx 4\lambda/5(\ln \mathrm{Rm} + 4/45)$. Note that the effect of negative magnetic pressure disappears if the equipartition is observed in turbulence without the magnetic field.

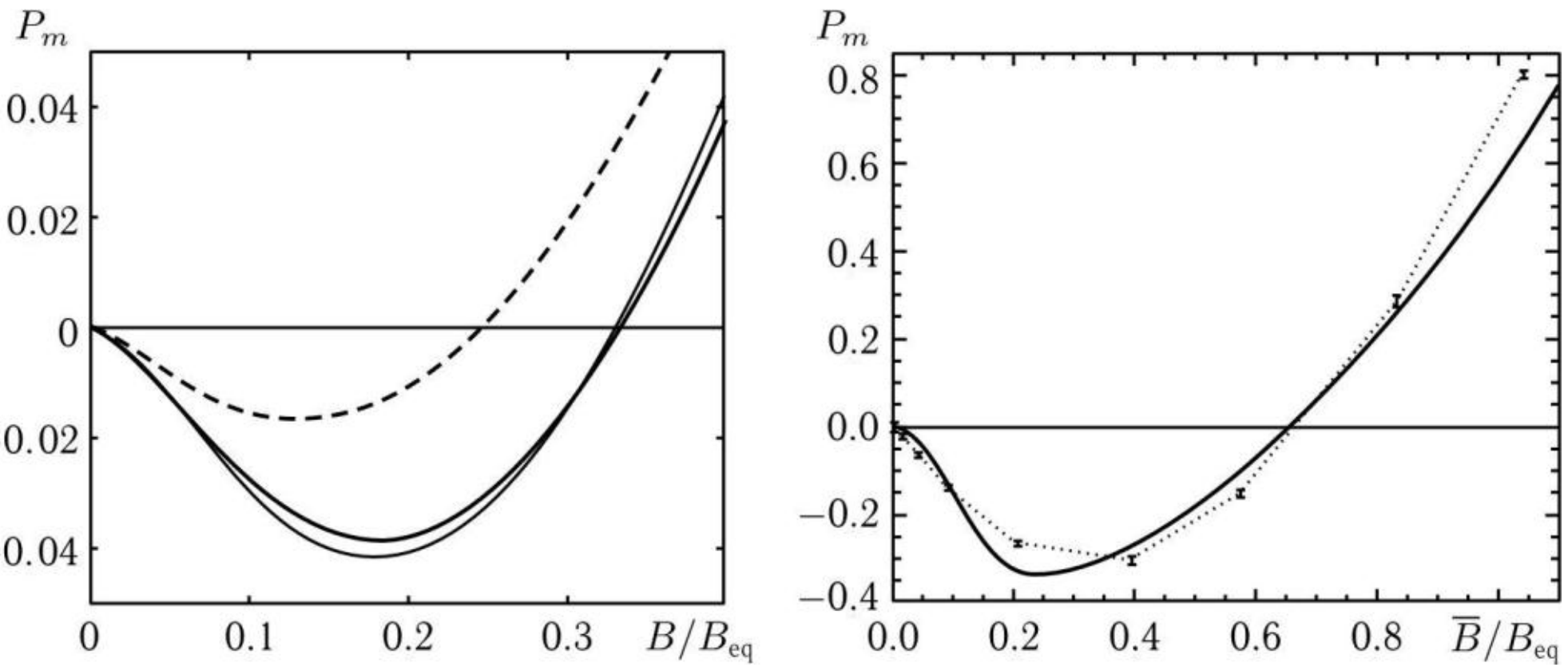

Fig. 114. Dependences of $P_B^{(\mathrm{eff})}(B/B_{eq})/B_{eq}^2$. The right panel is the result of direct numerical modeling from the paper of Brandenburg et al. (2010). The calculations were carried out at Re = 1130 and Rm = 283. The *dotted line* denotes the real behavior of $P_m$ in numerical experiments of Brandenburg et al. The *thin solid line* is the approximation of this dependence by the function (129a), where $q_p = q_{p0}\left[1 - 2\arctan\left(B^2/B_{eq}^2\right)/\pi\right]$ ($q_{p0} \approx 21$). The left panel shows the behavior of $P_m$ at different values of the magnetic Reynolds number: Rm ≈ $10^{10}$ (*thick solid line*), Rm ≈ $10^3$ (*thin solid line*). The *thick dashed line* denotes the case when 70 % of turbulence energy is produced due to the action of forces of small-scale viscosity and Rm ≈ $10^6$. In all cases it was supposed that $\left\langle b_{B=0}^2 \right\rangle \ll 4\pi\rho_0 \left\langle u_{B=0}^2 \right\rangle$, i.e., $\lambda$ = 1 (Rogachevskii and Kleeorin, 2007)

Figure 114 shows the dependence of the ratio of effective pressure of the mean field $P_B^{(\mathrm{eff})} = \delta P_T^{(\mathrm{eff})} + B^2/8\pi$ on $B_{eq}^2/8\pi$ as a function of equipartition of the mean field $B/B_{\rm eq}$ normalized to the field. The left panel shows the theoretical results from the paper of Rogachevskii and Kleeorin (2007), whereas the right panel exhibits the results of direct numerical modeling from the paper of Brandenburg et al. (2010).

A comparison of the left and right panels shows the following: in numerical experiments the effect of negative pressure proved to be significantly stronger than its theoretical estimates. First, the region of negative pressure in numerical experiments has the width ≈ $0.7B_{\rm eq}$ (right panel), which is twice wider than its theoretical estimates ≈ $0.35B_{\rm eq}$ (left panel). Second, the maximum of negative magnetic pressure in numerical experiments $\left|\left(P_B^{(\mathrm{eff})}(B/B_{eq})\right)_{\min}\right| \approx 0.4B_{eq}^2$ is by an order of magnitude higher (!) than that in the theory $\left|\left(P_B^{(\mathrm{eff})}(B/B_{eq})\right)_{\min}\right| \approx 0.04B_{eq}^2$. This circumstance can have quite far-reaching consequences for the spot-formation process.

D. Reynolds-Maxwell stresses and stellar differential rotation. Taking into account the derived results, the equation for the average rotation velocity of a star (97) is as follows:

$$\frac{\partial \mathbf{U}}{\partial t} = -\vec{\nabla}\left(\frac{P_{tot}}{\rho_0}\right) + \frac{1}{4\pi\rho_0}\left(\mathbf{B}\cdot\vec{\nabla}\right)Q_s\mathbf{B} + \mathbf{\Lambda}\frac{B^2}{8\pi\rho_0}Q_p + \mathbf{U}\times\left(2\mathbf{\Omega}_* + \mathbf{W}\right) + \frac{1}{\rho_0}\vec{\nabla}\cdot\hat{\tau}^T \cdot \qquad (131)$$

Here $\bar{P}_{tot} = \bar{P} + \left(\langle \mathbf{u}^2\rangle/3 + 2\nu_T\vec{\nabla}\cdot\bar{\mathbf{U}}/3\right)\rho_0 + Q_p B^2/8\pi$,

$\hat{\tau}^T_{ij} = \rho_0\nu_T\left(\nabla_i U_j + \nabla_j U_i\right) + \hat{\tau}^{\mathbf{\Lambda}}_{ij} + \hat{\tau}^{\mathbf{\Phi}}_{ij}$, where $\hat{\tau}^{\mathbf{\Lambda}}_{ij} \sim \rho_0\Omega_*\nu_T\left(\ell_0\Lambda\right)^2 V_{ij}\left(\Omega_*\tau_0(r)\right)$ and $\hat{\tau}^{\mathbf{\Phi}}_{ij} \sim \rho_0 g\Phi_T\Omega_*\tau_0^2(r)\, \mathit{\Phi}_{ij}\left(\Omega_*\tau_0(r)\right), \Lambda = \left|\nabla\rho_0/\rho_0\right|$) — the lambda effect (proportional to $\Lambda^2$) detected, apparently, by Kitchatinov (1987) and the flow effect (proportional to $\Phi_T \sim \langle u_z\theta\rangle$) revealed by Kleeorin and Rogachevskii (2006), respectively. In addition, in Equation (131) the convective zone was supposed to be strictly isentropic, i.e., the superadiabatic gradient is strictly equal to zero. Therefore, in Equation (131) the viscosity force $-\mathbf{g}S$ is absent. Rotation is the axisymmetric flow that is directed parallel to the stellar equator. Therefore, the velocity of such a flow can always be written as $U_\varphi = [\Omega_{*0} + \delta\Omega(r,\vartheta)]r\sin\vartheta$, where $\delta\Omega(r,\vartheta)$ describes differential rotation. Making a projection of Equation (131) onto the stellar rotation velocity by neglecting meridional circulations in its convective zone and adding Equations (81) and (126), we obtain

$$\frac{\partial B}{\partial t} = \frac{1}{r}\left(\frac{\partial\Omega_*}{\partial r}\frac{\partial}{\partial\vartheta} - \frac{\partial\Omega_*}{\partial\vartheta}\frac{\partial}{\partial r}\right)(rA\sin(\vartheta)) + \\ + \frac{1}{r}\left[\frac{\partial}{\partial r}\left(\eta_T\frac{\partial(rB)}{\partial r}\right) + \frac{\eta_T}{r}\frac{\partial}{\partial\vartheta}\left(\frac{1}{\sin\vartheta}\frac{\partial(\sin\vartheta B)}{\partial\vartheta}\right)\right], \qquad (131a)$$

$$\frac{\partial A}{\partial t} = \alpha B + \eta_T\Delta_s A \qquad (131b)$$

$$\frac{\partial\alpha_m}{\partial t} + \frac{\alpha_m}{T} = \frac{\mu}{2\pi\rho_0}\left\{-B\Delta_s A + \frac{1}{r^2}\left(\frac{1}{\sin^2\vartheta}\frac{\partial A\sin(\vartheta)}{\partial\vartheta}\frac{\partial B\sin(\vartheta)}{\partial\vartheta} + \frac{\partial rA}{\partial r}\frac{\partial rB}{\partial r}\right)\right. \\ \left. -\frac{\alpha}{\eta_T}\left[B^2 + \frac{1}{r^2}\left(\frac{\partial rA}{\partial r}\right)^2 + \frac{1}{r^2\sin^2\vartheta}\left(\frac{\partial A\sin(\vartheta)}{\partial\vartheta}\right)^2\right]\right\} - \frac{\mu}{4\pi\rho_0\eta_T}\mathrm{div}\left(\mathbf{\Phi}_{\chi_m}\right), \qquad (132)$$

$$\alpha(\mathbf{B}) = \psi_h(\mathbf{B})\alpha_h + \psi_m(\mathbf{B})\alpha_m$$

$$\frac{\partial\Omega_*}{\partial t} = \frac{1}{\rho_0 r^4}\frac{\partial}{\partial r}r^4\left[\rho_0\nu_T\frac{\partial\Omega_*}{\partial r} + \frac{1}{r^2}\left(r\hat{\tau}^{\mathbf{\Phi}+\mathbf{\Lambda}}_{r\varphi} + \frac{BQ_s}{4\pi\sin^2\vartheta}\frac{\partial A\sin(\vartheta)}{\partial\vartheta}\right)\right] \\ + \frac{1}{r^2\sin^3\vartheta}\frac{\partial}{\partial\vartheta}\left[\nu_T\sin^3\vartheta\frac{\partial\Omega_*}{\partial\vartheta} + r\sin^2\vartheta\left(\hat{\tau}^{\mathbf{\Phi}+\mathbf{\Lambda}}_{\vartheta\varphi} - \frac{BQ_s}{4\pi\rho_0}\frac{\partial(Ar)}{\partial r}\right)\right]. \qquad (133)$$

Here $\hat{\boldsymbol{\tau}}^{\Phi+\Lambda} = \hat{\boldsymbol{\tau}}^{\Lambda} + \hat{\boldsymbol{\tau}}^{\Phi}$ is the sum of stresses associated with the lambda and flow effects, respectively. The accurate expressions for $\hat{\tau}^{\Lambda}_{r\varphi}$, $\hat{\tau}^{\Phi}_{r\varphi}$, $\hat{\tau}^{\Lambda}_{\vartheta\varphi}$, and $\hat{\tau}^{\Phi}_{\vartheta\varphi}$ are given in Kitchatinov and Rüdiger (2005) for $\hat{\boldsymbol{\tau}}^{\Lambda}$, Kleeorin and Rogachevskii (2006), and Rogachevskii and Kleeorin (2018) for $\hat{\boldsymbol{\tau}}^{\Phi}$. The system (131)–(133) is closed if one specifies the magnetic helicity flux $\boldsymbol{\Phi}_{\chi_m}$. The simplest choice for the magnetic helicity flux is an assumption that this flux has a diffusion character (Covas et al., 1998):

$$\boldsymbol{\Phi}_{\chi_m} = -\kappa_T \nabla \chi_m . \tag{133a}$$

The coefficient of magnetic helicity diffusion $\kappa_T$ is usually chosen to be several times lower than the turbulent viscosity $\nu_T$. Such a choice is based on the experience of working with the equation for the laboratory and atmospheric turbulence energy (see, e.g., Elperin et al., 2002, 2006). The system (131)–(133) describes the stellar nonlinear dynamo of mean fields starting from the differential rotation generation and ending by the calculation of the toroidal $\mathbf{B}_{\text{tor}}$ and poloidal $\mathbf{B}_{\text{pol}}$ fields with the vector potential $A$:

$$\mathbf{B}_p = \frac{\hat{\mathbf{r}}}{r\sin\vartheta}\frac{\partial A\sin(\vartheta)}{\partial\vartheta} - \frac{\hat{\boldsymbol{\theta}}}{r}\frac{\partial Ar}{\partial r} .$$

At first glance, the solution of Equations (131)–(133) seems to be an absolutely monumental problem. However, fortunately, Equation (133) does not practically undergo the influence of the system (131)–(132). We represent the solution of Equation (133) as $\Omega_* = \Omega_{*0} + \delta\Omega(r,\vartheta) + U_B / r\sin\vartheta$. Here $\Omega_{*0} + \delta\Omega(r,\vartheta)$ is the solution of Equation (133) at $Q_s$ being equal to zero, i.e., without influence of the magnetic field. Substituting this solution into (133), we obtain the equation for $U_B$:

$$\begin{aligned} \frac{\partial U_B}{\partial t} &= \frac{1}{\rho_0 r^3}\frac{\partial}{\partial r} r^3 \left[\rho_0 \nu_T r \frac{\partial}{\partial r}\left(\frac{U_B}{r}\right) + \frac{BQ_s}{4\pi r\sin\vartheta}\frac{\partial A\sin(\vartheta)}{\partial\vartheta}\right] \\ &+ \frac{1}{r\sin^2\vartheta}\frac{\partial}{\partial\vartheta}\sin^2\vartheta\left[\nu_T \sin\vartheta \frac{\partial}{\partial\vartheta}\left(\frac{U_B}{\sin\vartheta}\right) - \frac{BQ_s}{4\pi r\rho_0}\frac{\partial(Ar)}{\partial r}\right] . \end{aligned} \tag{134}$$

The approximate solution of this equation was found by Kleeorin and Ruzmaikin (1984, 1991):

$$U_B \approx -\frac{Q_s B\ell_0}{4\pi r\nu_T\rho_0\sin\vartheta}\frac{\partial A\sin(\vartheta)}{\partial\vartheta} = -Q_s\frac{\ell_0}{4\pi\rho_0}\frac{BB_r}{\nu_T} .$$

For the Sun, this yields $|U_B| \sim 4$ m/s, whereas $\delta\Omega R_\odot \sim 4$ km/s, i.e., the magnetic contribution into differential rotation accounts for just 0.1%. For the magnetic effect to be at least 10% of the stellar differential rotation, it is required to lower the rotation of a star by a factor of 100, which, according to (83a) and Table 25, yields $D^{(\mathrm{M0})} < 25 \ll D_{cr}^{(\mathrm{M0})} = 628$, $D^{(\mathrm{M5})} < 3.1 \ll D_{cr}^{(\mathrm{M5})} = 26$ for M stars, i.e., such a star cannot have the magnetic field at all.

Table 26. Three most important magnetic parameters $B_{s2}$, $B_0$, and $\bar{\Omega}_{cr}$ and some other fundamental parameters of main-sequence stars

| Parameters | Spectral type | | | | | | | | |
|---|---|---|---|---|---|---|---|---|---|
| | F0 | F5 | G0 | G2 | G5 | K0 | K5 | M0 | M5 |
| $M_*/M_\odot$ | 1.7 | 1.3 | 1.1 | 1 | 0.91 | 0.78 | 0.7 | 0.47 | 0.21 |
| $R_*/R_\odot$ | 1.3 | 1.2 | 1.05 | 1 | 0.933 | 0.85 | 0.74 | 0.63 | 0.32 |
| $\eta_T^{(*)}/\eta_T^{(\odot)}$ | — | 1.89 | 1.31 | 1 | 0.74 | 0.48 | 0.45 | 0.2 | ≈ 0.5–4 |
| $\rho_*/\rho_\odot$ | 0.77 | 0.75 | 0.95 | 1 | 0.929 | 1.27 | 1.73 | 1.88 | 6.4 |
| $B_0$, G | — | 75 | 67 | 55 | 51 | 41 | 35 | 23 | 217 |
| $\bar{T}$ | 1.25 | 1.13 | 1.024 | 1 | 0.971 | 0.907 | 0.763 | 0.678 | 0.540 |
| $B_{s2}$, G | — | 23 | 56 | 74 | 118 | 180 | 345 | 3.560 | $188 \cdot 10^3$? |
| $\mu_*$ | ∞ | 6.18 | 4 | 3.34 | 3.16 | 3.12 | 3.47 | 2.26 | 1 |
| $\bar{H}_*$ | 0 | 0.649 | 0.877 | 1 | 0.986 | 0.909 | 0.712 | 0.695 | 1.069 |
| $\bar{\Omega}_{cr}$ | ∞ | 1.238 | 0.470 | 0.276 | 0.210 | 0.160 | 0.245 | 0.114 | 0.543 |

For stars of earlier spectral types such a weak rotation makes the magnetic field generation more hopeless. See also Table 26. Thus, Equation (133) is solved separately from the system (131)–(132), after which its solution $\Omega_*(r, \vartheta)$ is substituted into this system that is solved independently. We should note that it would be incorrect to suppose that magnetic fields do not have influence on stellar rotation at all. The point is that the circumstellar magnetic field that is basically generated by a nearby star in the course of the interaction with the interstellar medium triggers the electromagnetic rotation braking of this star. Along with the stellar wind, this effect causes the loss of angular momentum by the star and rotation deceleration.

We proceed to the analysis of Equations (131)–(132). For this aim, it is convenient to use dimensionless variables introduced in Subsect. 4.3.2.1: $\Omega = \Omega_* + \delta\Omega_*\omega(r,\vartheta)$, $\eta_T = \eta_T^{(0)}\bar{\eta}(r)$, $\alpha_h = \alpha_0\bar{\alpha}(r,\vartheta)$, $\alpha_m = \alpha_0\bar{\alpha}_m$, $B_0 = \eta_T/R_* \sqrt{2\pi\rho_*/\mu}$: $A_0 = R_* \mathrm{R}_\alpha B_0$, $B = B_0\bar{B}(\bar{r},\vartheta)$, $A = A_0\,\bar{A}(\bar{r},\vartheta)$ ($\bar{r} = r/R_*$ is the dimensionless radial coordinate inside a star, whereas $\bar{r} = 1$ is on its surface, $\bar{A}(\bar{r},\vartheta)$, $\bar{B}(\bar{r},\vartheta)$, $\bar{\alpha}_m(\bar{r},\vartheta)$ are the dimensionless toroidal vector potential of the poloidal magnetic field, the toroidal field, and the magnetic part of the alpha effect, respectively. The values of the characteristic field $B_0$ for late main-sequence stars are given in Table 26. After the nondimensionalization, the system (131)–(132) is as follows (Kleeorin and Ruzmaikin, 1982):

$$\frac{\partial B}{\partial t} = \frac{D}{r}\left(\frac{\partial \Omega_*}{\partial r}\frac{\partial}{\partial \vartheta} - \frac{\partial \Omega_*}{\partial \vartheta}\frac{\partial}{\partial r}\right)(rA\sin(\vartheta)) + \Delta_s B \qquad (134a)$$

,

$$\frac{\partial A}{\partial t} = \left(\bar{\alpha}\psi_h(B) + \frac{\bar{\alpha}_m}{\bar{\rho}}\psi_m(B)\right)B + \Delta_s A \qquad (134b)$$

$$\frac{\partial \bar{\alpha}_m}{\partial t} + \frac{\bar{\alpha}_m}{\bar{T}} = \left\{ -B\Delta_s A + \frac{1}{r^2}\left( \frac{1}{\sin^2\vartheta} \frac{\partial A \sin(\vartheta)}{\partial \vartheta} \frac{\partial B \sin(\vartheta)}{\partial \vartheta} + \frac{\partial rA}{\partial r}\frac{\partial rB}{\partial r} \right) \right.$$
$$\left. -\left( \bar{\alpha}\psi_h(B) + \frac{\bar{\alpha}_m}{\bar{\rho}}\psi_m(B) \right)\left[ B^2 + \frac{\mathrm{R}_\alpha^2}{r^2}\left(\frac{\partial rA}{\partial r}\right)^2 + \frac{\mathrm{R}_\alpha^2}{r^2\sin^2\vartheta}\left(\frac{\partial A\sin(\vartheta)}{\partial\vartheta}\right)^2 \right] \right\}. \tag{135}$$
$$+\bar{\kappa}\Delta\bar{\alpha}_m$$

Here $\bar{T} = \mathrm{Rm}\,\ell_0^2/3R_*^2$, $\bar{\kappa} = \kappa_T/\eta_T$, $\bar{\rho} = \rho_0/\bar{\rho}_*$ is the mean stellar density. The difference of this system from the system in Kleeorin and Ruzmaikin (1982) is accounting of algebraic nonlinearity and a distinct diffusion type of the helicity flux (133a). Note that here for simplicity of designations for dimensionless fields and values the line above is omitted, except for the hydrodynamic and magnetic parts of the alpha effect $\bar{\alpha}$, $\bar{\alpha}_m$, the dimensionless density $\bar{\rho}$, and the dimensionless diffusion coefficient $\bar{\kappa}$ for the magnetic part of the alpha effect. On the whole, this system is a nonlinear generalization of the system (83a) comprising Parker's dynamo waves (1955) in the case of algebra dynamic nonlinearity (Kleeorin and Ruzmaikin, 1982). At first glance, it seems that the best way to solve the stellar dynamo problem is to just investigate the system (134)–(135) for different spectral types and at different rotations (dynamo numbers associated numerically with stellar rotation according to (83a)). Although our notions on the profiles of stellar differential rotations and alpha effects are rather poor, but this is not the main problem of the system (134)–(135). The main problem is a variation in density $\bar{\rho}$ by six (for the Sun) and more (for other stars) orders (!) inside the stellar convective zone. *Any modern numerical methods cannot allow one to get a reliable solution to the system of nonlinear differential equations in partial derivatives at such variations of coefficients!* Therefore, at this stage, the analytical and asymptotic investigations are preferential.

Let us now proceed to the analytical estimate of the mean field level of the rotating main-sequence stars.

E. Estimate of the mean field level of the rotating main-sequence stars. Acting similar to what is in the case with the $\alpha$–$\omega$ dynamo, we discard in (135) terms that are proportional to $\mathrm{R}^2_\alpha$. Moreover, we make use of Equation (134b) to simplify drastically Equation (135):

$$\frac{\partial \alpha_m}{\partial t} + \frac{\alpha_m}{\bar{T}} = -\frac{B}{\bar{\rho}}\frac{\partial A}{\partial t} + \frac{1}{\bar{\rho} r^2}\left( \frac{1}{\sin^2\vartheta}\frac{\partial A\sin(\vartheta)}{\partial\vartheta}\frac{\partial B\sin(\vartheta)}{\partial\vartheta} + \frac{\partial rA}{\partial r}\frac{\partial rB}{\partial r} \right) + \bar{\kappa}\Delta\bar{\alpha}_m. \tag{136}$$

Afterwards, it is convenient to represent this system (134a), (134b) in the operator form:

$$\frac{\partial}{\partial t}\begin{pmatrix} A \\ B \end{pmatrix} = \left(\hat{L} + \hat{N}\right)\begin{pmatrix} A \\ B \end{pmatrix}. \tag{137}$$

Here the operators $\hat{L}$ and $\hat{N}$ are introduced:

$$\hat{L}=\begin{pmatrix}\Delta_s & \bar{\alpha}\\ D\hat{\Omega} & \Delta_s\end{pmatrix},\quad \hat{N}=\begin{pmatrix}0 & \alpha_m\\ 0 & 0\end{pmatrix},$$

whereas the differential operator of differential rotation $\hat{\Omega}$ is as follows:

$$\hat{\Omega}=\frac{1}{r}\left(\frac{\partial\Omega_*}{\partial r}\frac{\partial}{\partial\vartheta}-\frac{\partial\Omega_*}{\partial\vartheta}\frac{\partial}{\partial r}\right).$$

Note that for simplicity we neglected algebraic nonlinearities.

To solve (136)–(137), we consider the supporting linear system for eigenvalues (proper numbers) $p_n$ and for eigenvectors $\mathbf{e}_n=\begin{pmatrix}a_n\\ b_n\end{pmatrix}$, which are the kinematic part of the problem (137):

$$\hat{L}\mathbf{e}_n=p_n\mathbf{e}_n. \tag{137a}$$

Here $a_m(r,\vartheta)=|a_m(r,\vartheta)|\exp\left[-ig_m^{(A)}(r,\vartheta)\right], b_m(r,\vartheta)=|b_m(r,\vartheta)|\exp\left[-ig_m^{(B)}(r,\vartheta)\right]$ are the complex functions. As shown by Kleeorin and Ruzmaikin (1984), the operator $\hat{L}\mathbf{e}_n=p_n\mathbf{e}_n$ represents a non-self-conjugate operator with, generally speaking, complex eigenvalues $p_n=\gamma_n+i\omega_n$ and nonorthogonal eigenvectors $(\mathbf{e}_n\cdot\mathbf{e}_m)\equiv\iint\left(a_n^*a_m+b_n^*b_m\right)r^2\sin\vartheta d\vartheta dr\neq 0$ at $m\neq n$. The asterisk denotes the complex conjunction. The general mathematical properties of non-self-conjugate operators, their eigenvectors, and proper numbers can be found in the book of Gohberg and Krein (1965). Some of complex proper numbers can have a positive real part. This means that such modes describe the dynamo mode. Indeed, according to the general theory of Gohberg and Krein (1965), these vectors along with the attached vectors (the latter ones exist only if the operator $\hat{L}$ has a degeneration) form the total, though nonorthogonal basis from which it is possible to separate any two-component function as $\begin{pmatrix}A(r,\vartheta)\\ B(r,\vartheta)\end{pmatrix}$. In other words, any solution of the nondegenerate linear problem of the $\alpha$–$\omega$ dynamo as

$$\frac{\partial}{\partial t}\begin{pmatrix}A\\ B\end{pmatrix}=\hat{L}\begin{pmatrix}A\\ B\end{pmatrix} \tag{138}$$

at a given dynamo number $D$ can be represented as $\begin{pmatrix}A\\ B\end{pmatrix}=\sum_{m=1}^{\infty}C^m\mathbf{e}_m\exp(p_mt)\equiv\sum_{m=1}^{\infty}C^m\begin{pmatrix}a_m\\ b_m\end{pmatrix}\exp(p_mt)$, where $C^m=|C^m|\exp\left(i\varphi_{C^m}\right)$. Note that in some concrete (approximated) tasks for the simplified operator $\hat{L}$ vectors $\mathbf{e}_n$ have already been found (see Subsect. 4.3.2.3). Suppose for definiteness only for the first mode $\gamma_1>0$. If such a mode does not exist, then there is no dynamo. This means that $D>D_{cr}^{(1)}$ only for it, for others $D < D_{cr}^{(m)}$ ($m > 1$). Then, for example, for the real fields Re $A \to A$, Re $B \to B$, which are the solution of Equation (138), we have

$$\begin{pmatrix} A \\ B \end{pmatrix} = |C^1| \exp(\gamma_1 t) \begin{pmatrix} |a_1(r,\vartheta)| \cos\left[\omega_1 t - g_1^{(A)}(r,\vartheta) + \varphi_{C^1}\right] \\ |b_1(r,\vartheta)| \cos\left[\omega_1 t - g_1^{(B)}(r,\vartheta) + \varphi_{C^1}\right] \end{pmatrix} + \sum_{m=2}^{\infty} |C^m| \exp(-|\gamma_m| t) \begin{pmatrix} |a_m(r,\vartheta)| \cos\left[\omega_m t - g_m^{(A)}(r,\vartheta) + \varphi_{C^m}\right] \\ |b_m(r,\vartheta)| \cos\left[\omega_m t - g_m^{(B)}(r,\vartheta) + \varphi_{C^m}\right] \end{pmatrix}. \tag{139}$$

Hence, we can see that the general valid solution of dynamo Equation (138) represents a set of nonstationary harmonic waves, which have amplitudes growing and dropping exponentially in time. The waves with exponentially growing amplitude correspond to dynamo waves. If one arranges the proper numbers so that $\gamma_1 > \gamma_2 > \gamma_3 > ...\gamma_m ...$, then it can be claimed that after the time $\sim 6|\gamma_2^{-1}|$ all the contributions, except for the first one, decrease by more than a factor of 400 and, in fact, in the sum (133) only the first summand is left. The coefficients of the mentioned expansion are defined by the initial field $\begin{pmatrix} A \\ B \end{pmatrix}_{t=0}$ and can be found from the condition

$$\begin{pmatrix} A \\ B \end{pmatrix}_{t=0} = \sum_{m=1}^{\infty} C^m \begin{pmatrix} a_m \\ b_m \end{pmatrix}. \tag{130a}$$

To find these coefficients, one needs to construct a basis that is dual to $\mathbf{e}_n = \begin{pmatrix} a_n \\ b_n \end{pmatrix}$. Such a basis is formed by eigenvectors of the operator conjugated to $\hat{L}$

$$\hat{L}^\dagger = \begin{pmatrix} \Delta_s & -D\hat{\Omega} \\ \bar{\alpha} & \Delta_s \end{pmatrix},$$

Proper numbers $p_m^\dagger = \gamma_m - i\omega_m = p_m^*$ and eigenvectors $\mathbf{e}^m = \begin{pmatrix} a^m & b^m \end{pmatrix}$ of the operator $L^\dagger$ satisfy the equation

$$\mathbf{e}^m \hat{L}^\dagger = p_m^\dagger \mathbf{e}^m .$$

As Kleeorin and Ruzmaikin (1984) showed, the condition of mutual orthogonality and normalization of these bases has the form

$$\left(\mathbf{e}^n \cdot \mathbf{e}_m\right) \equiv \iint \left((a^n)^* a_m + (b^n)^* b_m\right) r^2 \sin\vartheta d\vartheta dr = \begin{cases} 0, \ n \neq m \\ \left(D\dfrac{dp_m}{dD}\right)^{-1} \int (a^m)^* \bar{\alpha} b_m d^3r, \ \ n = m \end{cases}.$$

Therefore, in particular, we acquire the expression for coefficients of the expansion $C^n$, multiplying (139a) scalarly by $\mathbf{e}^n$ to the left:

$$C^n = D\frac{dp_n}{dD}\frac{\iint\left[(a^n)^* A(t=0) + (b^n)^* B(t=0)\right] r^2 \sin\vartheta d\vartheta dr}{\iint (a^n)^* \bar{\alpha}\left(r,\vartheta\right) b_n r^2 \sin\vartheta d\vartheta dr}.$$

To solve the nonlinear problem, we do the following way. Let us expand the solution of the nonlinear problem into a series according to the basis $\mathbf{e}_m$, which corresponds to $D = D_{cr}$, i.e., $\gamma_1 = 0$:

$$\begin{pmatrix} A \\ B \end{pmatrix} = \sum_{m=1}^{\infty} F^n(t)\mathbf{e}_n. \tag{139b}$$

We substitute this expansion into (137) and multiply scalarly the both parts of (137) by $\mathbf{e}^n$ to the left. This yields the equation for $F^n(t)$ (Kleeorin et al., 1994, 1995):

$$\frac{dF^n}{dt} - F^n(t)p_n = D_{cr}\left(\frac{dp_n}{dD}\right)_{cr}\sum_{p=1}^{\infty} F^p(t)\left[\left(\frac{D}{D_{cr}} - 1\right)\frac{G_p^n}{G_n^n} + \alpha_p^n(t)\right]. \tag{140}$$

Here $G_p^n = \int (a^n)^* \bar{\alpha} b_p \, d^3r + \left(p_p - p_n\right)\int (b^n)^* b_p \, d^3r$, whereas $\alpha_p^n(t) = \int (a^n)^* \ \alpha_m(t,r,\vartheta) b_p \, d^3r / G_n^n$ are the time functions associated with the magnetic part of the alpha effect. Equations for $\alpha_p^n(t)$ can be found from the equation for the magnetic part of the alpha effect (136), if we substitute the solution (139b) into it and neglect by the diffusion term $\bar{\kappa}\Delta\bar{\alpha}_m$. This results in

$$\frac{d\alpha_p^n}{dt} + \frac{\alpha_p^n}{T} = \sum_{k,s=0}^{\infty} F^k(t)\left(F^s M_{ksp}^n - \frac{dF^s}{dt} S_{ksp}^n\right). \tag{141}$$

Here

$$M_{ksp}^n = \frac{1}{G_n^n}\iint \frac{(a^n)^*}{\bar{\rho}(r)}\left(\frac{1}{\sin\vartheta}\frac{\partial a_s \sin(\vartheta)}{\partial\vartheta}\frac{\partial b_k \sin(\vartheta)}{\partial\vartheta} + \sin\vartheta\frac{\partial r a_s}{\partial r}\frac{\partial r b_k}{\partial r}\right) b_p d\vartheta dr$$

$$S_{ksp}^n = \frac{1}{G_n^n}\int \frac{(a^n)^*}{\bar{\rho}(r)} b_k b_p a_s d^3r.$$

The systems (140), (141) describe an extremely abundant behavior of the stellar magnetic field — from the one-mode, ideally periodic stellar cycle (Kleeorin et al., 1995) up to chaotic behavior (Kitiashvili and Kosovichev, 2008, 2009) in the manner of the unbalanced solar cycle. Since based on the available observations it is difficult to judge on the chaotic behavior of stellar cycles, we restrict ourselves to estimating by the one-mode approximation at which the basic contribution is made by the first (oscillatory) mode with eigenvectors $\mathbf{e}_1 = \begin{pmatrix} a_1 \\ b_1 \end{pmatrix}$ and $\mathbf{e}_1^* = \begin{pmatrix} a_1^* \\ b_1^* \end{pmatrix}$ and proper numbers $p = \pm i\omega$ ($\gamma = \mathrm{Re}\, p = 0$). Following (140) and (141), we obtain then the following equations (Kleeorin et al., 1995):

$$\begin{aligned} &\frac{dF}{dt} - iF\omega_{cr} = \Delta p \left\{ F\left[ \left( \frac{D}{D_{cr}} - 1 \right) + \alpha_1 \right] + F^*\alpha_2 \right\} \exp\left(i\beta_p\right) \\ &\frac{d\alpha_1}{dt} + \frac{\alpha_1}{T} = (m + m^*)FF^* - sF\frac{dF^*}{dt} - s^*F^*\frac{dF}{dt} \\ &\frac{d\alpha_2}{dt} + \frac{\alpha_2}{T} = 3m^*F^2 - s^*F\frac{dF}{dt} \end{aligned} \quad . \qquad (142)$$

The following designations are used in (142):

$$\begin{aligned} &\Delta p \exp(i\beta_p) \equiv D_{cr}\left( \frac{dp}{dD} \right)_{cr} \\ &m = \frac{1}{G_1^1} \iint \frac{(a^1)^*}{\bar{\rho}(r)} \left( \frac{1}{\sin\vartheta} \frac{\partial a_1^* \sin(\vartheta)}{\partial\vartheta} \frac{\partial b_1 \sin(\vartheta)}{\partial\vartheta} + \sin\vartheta \frac{\partial r a_s}{\partial r} \frac{\partial r b_k}{\partial r} \right) b_1 d\vartheta dr \\ &s = \frac{1}{G_1^1} \int \frac{(a^1)^*}{\bar{\rho}(r)} a_1^* b_1^2 d^3 r \quad G_1^1 = \int (a^1)^* \bar{\alpha} b_1 \, d^3 r \end{aligned} .$$

Resolving Equation (142) with respect to $F$ and calculating the mean square $\left|\bar{F}\right|^2 \approx \left(t_2 - t_1\right)^{-1} \int_{t_1}^{t_2} \left|F\right|^2 dt$ for the value of the stellar surface field $B_s$, we obtain (Kleeorin et al., 1995):

$$B_s = B_0 \sqrt{\frac{\rho_s}{\rho_*} \left|\bar{F}\right|^2} \approx B_0 \sqrt{\frac{3R_*^2 \rho_s \left(D - D_{cr}\right)}{4D_{cr}\, \mathrm{Rm}\, \ell_s^2 \left|\mathrm{Re}(s)\right| \rho_* (1 + 3d)}} \, , \qquad (143)$$

where $\rho_s/\rho_*$ is the ratio of the surface and mean densities of a star. Note that the time of averaging the squared field $t_2 - t_1$ should be at least several times higher than the stellar cycle period. For further consideration we note that according to Kleeorin et al. (1995) $\left|\mathrm{Re}(s)\right| \simeq \left(b_s^4/b_m^2\right) H_\rho \Big/ \left(H_* \sqrt{D_{cr}}\right)$, $H_\rho$ is the near-surface density height scale,

$b_s = b_1(1, \vartheta)$ and $b_m = b_1(1-\mu_*^{-1}, \vartheta)$ are the normalized fields on the surface and at the base of the stellar convective zone. The parameter $d$ is a value of about unity. The accurate value of this parameter can be found in Kleeorin et al. (1995). We proceed in (143) to such fundamental quantities as the stellar mass, its effective temperature, surface density, relative thickness of the convective zone, etc.:

$$B_s \approx B_0 \left(\mu_*\right)^{\frac{1}{4}} \left(\frac{\bar{\rho}_s}{\bar{\eta}_T}\right)^{\frac{1}{2}} \left(\bar{M}_* \sqrt{\bar{T}^3}\right)^{-3} \left(\frac{D}{D_{cr}} - 1\right)^{\frac{1}{2}} \equiv B_{s2} \left(\frac{D}{D_{cr}} - 1\right)^{\frac{1}{2}}. \qquad (144)$$

The sense of the field $B_{s2}$ is the value of the stellar surface field at $D = 2D_{cr}$.

Let us discuss the dependence of the stellar surface field on rotation. As it follows from (144), to find this dependence, it is sufficient to find the ratio $D/D_{cr}$, i.e., to substitute (83a) and (88) into it and take into account the data from Table 25:

$$\frac{D}{D_{cr}} = 80.6 \sin\left(2\phi_0\right) \bar{\alpha}_0 \delta \bar{\bar{\Omega}}_* \bar{H}_*^3 \sqrt{\frac{\bar{\Omega}_*^3}{\left(\bar{\eta}_T\right)^3}}.$$

$$\bar{H}_* = \frac{\bar{R}_*}{\bar{\mu}_*} = \frac{H_*}{H_\odot}$$

As a result, introducing the critical stellar rotation $\bar{\Omega}_{cr} = 5.4\times10^{-2} \bar{\eta}_T / \bar{H}_*^2 \left(\bar{\alpha}_0 \delta \bar{\bar{\Omega}}_* \sin\left(2\phi_0\right)\right)^{2/3}$ — the minimum rotation starting with which the star has a regular magnetic field, we rewrite Equation (144) in the quite elegant form:

$$B_s \approx B_{s2} \left(\sqrt{\frac{\bar{\Omega}_*^3}{\bar{\Omega}_{cr}^3}} - 1\right)^{\frac{1}{2}}. \qquad (145)$$

Table 26 lists three important parameters that determine, along with rotation, the stellar magnetic field and the stars dependent on spectral types. Together with the mentioned parameters, Table 26 involves all the required values determining these basic parameters. This is done for the reader to repeat these presentations by oneself. Note that in the given estimates there is still some uncertainty associated with two circumstances. First, when calculating $B_{s2}$, we suppose that $\bar{\rho}_s = \rho_s / \rho_s^{(\odot)} = 1$, which can possibly be not the case. Second, when calculating $\bar{\Omega}_{cr}$, the accurate values of all three parameters $\sin\left(2\phi_0\right)$, $\bar{\alpha}_0$, and $\delta\bar{\bar{\Omega}}_*$ as the functions of the stellar spectral type are not known yet. For estimating it was accepted that for all stars $\bar{\alpha}_0 = 0.5$, whereas $\sin\left(2\phi_0\right)$ and $\delta\bar{\bar{\Omega}}_*$ have their solar values, i.e., 0.87 and 0.2, respectively. Nonetheless, we can draw the following general conclusion on stellar activity: at variations of the stellar spectral type from F to M $B_{s2}$ grows, whereas $\Omega_{cr}$ drops. This means that activity grows toward later spectral types of stars. In this context, it is useful to compare the data from Table 26 and consequences of Equation (145) with the data on concrete stars from Table 4 (see Chapter 1.2). The results of this comparison are compiled in Table 27. Note that Table 27 includes not all the stars from Table 4: for some types of stars, exemplarily for G8, there are no data at all in Table 26; on the other hand, for stars of spectral types M0 and later the divergence between the theory and observations with respect to $B_s$ is tens and even hundreds of times! This seems to be all the more surprising that for stars of spectral types G0–K5 rotating with the angular velocities of > $5.8\Omega_{\odot}$, the correspondence between observations is quite satisfying. The interpretation of this is apparently reduced to the fact that the "constants" $\bar{\alpha}_0$, $\sin(2\phi_0)$, and $\delta\bar{\bar{\Omega}}_*$ start to depend on rotation — they decrease. Evidently, $\sin(2\phi_0)$ and $\delta\bar{\bar{\Omega}}_*$ are the most sensitive to this. For instance, if one supposes that for the star G1 171.2A both $\sin(2\phi_0)$ and $\delta\bar{\bar{\Omega}}_*$ decrease by a factor of 1.44 due to fast rotation, then $\Omega_{cr}$ = 0.735 (instead of 0.245) and the field $B_s$ proves to be of an order of 3 kG. On the other hand, for the main-sequence stars of spectral types later than M3.5, the convective zone extends from the center of a star up to the surface; and in this case, rotation can easily become solidbody, whereas the dynamo mechanism changes its type from $\alpha$–$\omega$ to $\alpha^2$. The property of such a dynamo principally differs from the properties of the $\alpha$–$\omega$ dynamo. In particular, the poloidal and toroidal fields are of the same order. In this context, the appearance of the strong magnetic field is possible at both poles (the dipole mode) and poles and the equator (the quadrupole mode). However, if there are both modes of similar power, then the situation is possible when the strong field will be only at one of the poles and the equator. More extended notions on the properties of the $\alpha^2$ dynamo mechanism can be found in the book of Moffatt (1978) and in papers of Brandenburg (2017, 2019). As to the properties of slowly rotating ($\bar{\Omega}_* \sim 0.7 \div 1$) stars of spectral types G0–K5, they demonstrate in observations the fields which are significantly larger than those predicted by theory.

Table 27. Surface magnetic fields of some rotating main-sequence stars: observations and theory

| Star | Spectral type | $\bar{\Omega}_*$ | $B_s$, kG (observations) | $B_s$, kG (theory) |
|---|---|---|---|---|
| Sun | G2 V | 1 | 1.5 | 0.193 |
| HD 115383 | G0 V | 5.2 | 1.0 | 0.335 |
| HD 20630 | G5 V | 2.7 | 1.8 | 0.792 |
| HD 131511 | K1 V | 2.82 | 1.7 | 1.5 |
| HD 26965 | K1 V | 0.686 | 1.7 | 0.506 |
| HD 185114 | K1 V | 0.933 | 1.36 | 0.651 |
| G1 171.2 A | K5 Ve | 13.7 | 2.8 | 7.0 |
| EQ Vir | K5 Ve | 6.51 | 2.5 | 4.0 |

The simplest interpretation of this phenomenon is the concentration of weak surface fields into small, as compared to the stellar size, spots (like the solar ones). For example, if we suppose that the observed field on the Sun of 1.5 kG is a result of accumulation of the magnetic flux of the calculated field of 193 G on the small area from the entire area of the “royal zone” (between 10° and 40° heliolatitude), then we obtain that the concentrated fields should occupy about 1.5% of the solar surface in the total accordance with the data of Table 4 (see Chapter 1.2). Certainly, such a correct estimate of the area of stellar spottedness cannot be derived for other stars from Table 27 since the estimate of the calculated field $B_S$ is not sufficiently reliable. A possible mechanism of the magnetic field concentration in spots is considered in the next section.

### 4.3.4. Formation of Active Regions and Starspots

The dynamo theory of large-scale stellar magnetic fields as a basis of stellar activity “without any doubt is attributed to the most impressive achievements of cosmic electrodynamics and physics of cosmic plasma” (Priest, 1982). However, on the background of these achievements there are a number of questions which have not principally been answered by the dynamo theory of mean stellar fields: in the pure form the dynamo theory is not able to describe neither the formation of active regions nor starspots. This can be very clearly seen from the comparison of the right and left panels in Fig. 115.

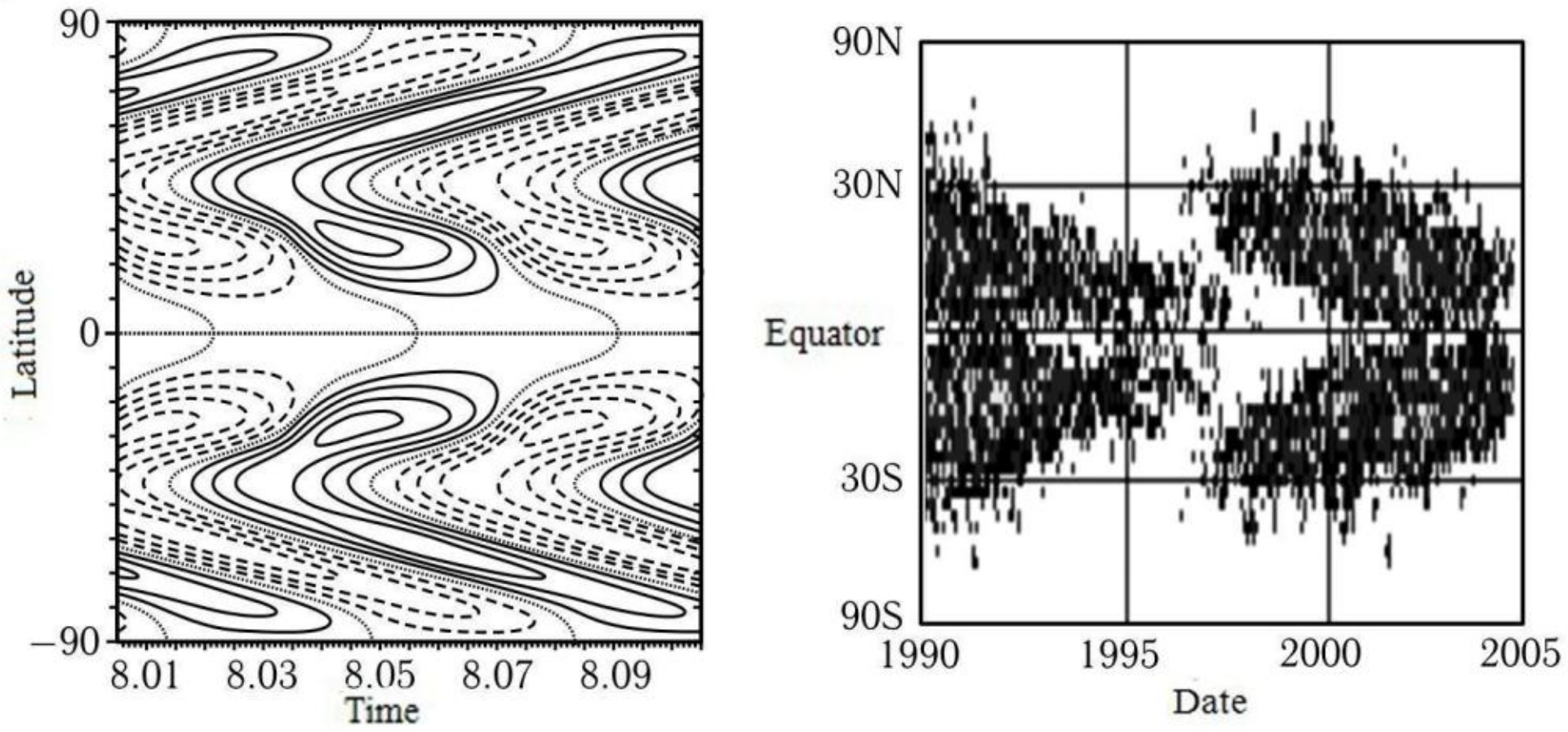

Fig. 115. *Left panel*: isolines of the toroidal magnetic field at a depth of 42000 km in the solar dynamo model in which the system (134), (135) was numerically integrated. *Solid lines* denote isolines of the toroidal magnetic field directed with the solar rotation, the dashed lines — against it. A density difference at the depth of the convective zone was taken of an order of 1000 in the model (Zhang et al., 2006). *Right panel*: a fragment of the observed butterfly diagram demonstrating at what latitude, at what moment of time, and with what intensity the exit of the new flux occurred into the solar photosphere

In some similarity of these two panels many supporters of the mean-field dynamo theory can see the triumph of this theory. Meanwhile, what is represented in panels to the right and left is absolutely not the same. The only thing these panels have in common is that both describe some kind of wave process. And if in 1966 the result that is similar to that represented in Fig. 115 could enthuse, then today one needs to clearly explain how the fields displayed in the left panel are associated with an exit of the new field flux into the photosphere shown in the panel to the right. The main difference between these two panels is that the left panel displays the smooth regular field, whereas the right one — the structured field bunched into the ropes buoying from under the photosphere. The general idea, which stands behind the buoying mechanism, belongs to Parker (1955a). He asserted that the tube (or the rope) of the horizontal magnetic field, which is in thermal equilibrium with the surrounding stellar plasma, will definitely buoy since its density $\rho_i = \rho_e - \gamma_{term} B^2 / 8\pi C_s^2$ turns out to be lower than that of the medium surrounding it $\rho_e$. Here $C_S$ is the adiabatic speed of sound in the medium, $\gamma_{\rm term} = C_P/C_V$ is the thermodynamic ratio of plasma heat capacity at constant density and at constant volume, respectively. It remains unclear how such a magnetic tube with the abrupt border can pass through the turbulent convective zone under the photosphere and cannot be destroyed by turbulent magnetic and hydrodynamic viscosities? And anyway, where do such magnetic ropes come from? Within the system (134), (135) it is very unlikely to get such fields: indeed, the sunspot scale (radius) is of about $0.036R_\odot \approx R_\odot/28$. For such a field to excite at the kinematic stage, it is necessary that according to (88) $D > D_{\mathrm{cr}(n=28)} = 8.8 \cdot 10^7$, which corresponds to the rotation of a solar-like star of an order of one (!) hour following (145). Thus, the fields of similar scale cannot be directly relevant to dynamo processes. So what mechanisms for the formation of similar ropes have been suggested in the literature? Let us start with the ideas of Gudzenko and Chertoprud (1980). According to their assumptions, there exists a "magnetic crystal" inside the solar convective zone — some hypothetic object composed of magnetic tori, each of them has the size similar to that of an active region. Every 10–11 years (possibly, under the influence of Jupiter's rotation around the Sun) these magnetic tori start falling off the "magnetic crystal" and, buoying into the photosphere due to their buoyancy effect, form the active regions which gradually undergo degradation being replaced by new ones from the same "magnetic crystal". Gradually, this "falling" of tori goes away, weakens, and the Sun appears in the activity minimum. About 11 years later, this "falling-off" starts again, and solar activity returns to its maximum. The hypothesis is not without some elegance, we should just answer the question as to where this "magnetic crystal" comes from. And what will be after all these tori run out in the "magnetic crystal"? A very rough estimate shows that there is enough number of tori for 40 cycles if we take into consideration that all the cycles in their maxima have the Wolf number $W \simeq 121$ and tori fill entirely all the solar convective zone. Ideologically, this model is essentially close to the model of Choudhuri and Gilman (1987). This also refers to the toroidal tube (only thin), whose long radius depends on heliolatitude and amounts to $R_\odot \sin\phi$. The tube is "put" on the Sun similar to a hat put on the head. The second radius of the tube is small. According to Choudhuri and Gilman (1987), this tube buoys from the base of the convective zone to the surface. However, its buoyancy under the influence of stellar rotation occurs not at the same heliolatitude where the tube lay at the bottom of the convective zone. If the field value does not exceed several kilogauss, then its buoyancy causes the appearance of a tube at the latitudes that are close to the pole. The latter obviously contradicts the observations since there are practically no spots at high latitudes. As shown by Choudhuri and Gilman (1987), for the buoyancy at the latitudes not higher than 40°, there should be a field of not less than 50 kG inside the tube! This field exceeds by a factor of 10 the maximum field of equipartition in the solar convective zone (at a depth of about $10^{10}$ cm, i.e., in the middle of the solar convective zone) and by a factor of 16 the equipartition field at the base of the zone at a depth of about $2 \cdot 10^{10}$ cm! Where such a field can be taken from? Probably, in the same place where the "magnetic crystal" is… To interpret such a field, the so-called flux tube dynamo is used (Schüssler, 1980). In fact, this is an ordinary mean-field dynamo in which mean magnetic fields are considered as being composed of individual force line tubes, inside which the field is high, but if one "smears out" it uniformly over the space, it will turn out to be not so high, and therefore can be described with ordinary dynamo equations. Certainly, such a qualitative understanding of nature of the field emergence has the right to exist. Nonetheless, the question "where do the thin tubes of the giant field come from?" remains open. And it cannot be solved through the philological discussions. The answer to this question can be just a presentation of the solution to the system of equations (92)–(95) describing the set of magnetic tubes on the background of turbulence. Such a solution is actually unknown. The only thing that is actually known is the solution corresponding to the small-scale dynamo in which one scale corresponds to the general scale of turbulence $\ell_0$ and the second one is far less: $\ell_f \approx \ell_0 \,\mathrm{Rm}^{-3/4} \exp(\pi/0.6)$ (Rogachevskii and Kleeorin, 1997). For the Sun, one can take the supergranule size ($2 \cdot 10^9$ cm) as $\ell_0$, Rm $\approx 10^7$, whereas $\ell_f \approx 2\times10^6$ cm. The obtained objects can be interpreted as a two-scale structure, whose one size is 1000 times larger than the other one. Despite the fact that the structure considered by Rogachevskii and Kleeorin represents an isotropic, in the statistical sense, object, one can easily imagine the following ensemble of magnetic fields — magnetic tori randomly oriented in space, whereas the large torus diameter is 20000 km and the small one is 20 km. However, this does not look like a sunspot. This, at the very least, is similar to a pore! The correct model of spot formation should certainly include two following mechanisms: 1) the mechanism of magnetic flux concentration produced by the mean-field dynamo into a relatively thin rope, whose size nevertheless exceeds significantly the integral turbulence scale $\ell_0$ and 2) the mechanism of carrying this rope out of the stellar convective zone into the photosphere with the formation of active regions and/or spots. There are two such models known in the literature: the Kitchatinov–Mazur model (2000) associated with magneto-thermal instability and the model of magnetic tube formation associated with instability of negative effective magnetic pressure (Kleeorin et al., 1989). Note that both models work at the stellar surface rather than in the stellar interior as in previous models. Let us describe qualitatively the way how these models work.

A. The Kitchatinov–Mazur model (2000). Suppose that the upper part of the convective zone is in equilibrium in the presence of the constant magnetic field *B*. Then the turbulent thermal flux at the surface $\Phi_r^{(T)}$ is equal to $\Phi_r^{(T)} = -c_p T \kappa_T (B^2)\, dS/dr$. Let us superimpose the sinusoidal magnetic field disturbance onto the system. Since the entropy turbulent diffusion coefficient $\kappa_T(B^2)$ decreases with growing field, the thermal flux proves to be modulated. For instance, according to Rogachevskii and Kleeorin (2007), at the fields $B \gg 0.5 B_{eq}\,\mathrm{Rm}^{-1/4}$ and at significantly lower than the equipartition level

$$\kappa_T(B^2) \approx \kappa_T(0)\left[1 - \frac{8B^2}{5B_{eq}^2}\left|\ln\left(\frac{B^2}{B_{eq}^2}\right) + \ln 8\right|\right].$$

In particular, the derivative of the logarithm of the entropy turbulent diffusion coefficient on dimensionless magnetic energy is

$$\kappa_B = B_{eq}^2 \kappa_T^{-1} \partial \kappa_T / \partial B^2 = -B_{eq}^{-2} \left[ 8/5 \left| \ln\left( B^2 / B_{eq}^2 \right) + \ln 8 \right| - .2\,\mathrm{sgn}\left( \ln\left( 8B^2 / B_{eq}^2 \right) \right) \right].$$

In Rogachevskii and Kleeorin (2007), one can find the full expression for $\kappa_T$ at arbitrary magnetic fields **B**. In the places where the flux is lower, the temperature decreases. As a result, it turns out that in the places where the field is stronger, the temperature is lower, and vice versa. This can lead to the development of viscosity forces and appropriate flows which can principally result in the formation of tubes. After the formation, they undergo, according to Parker (1955a), the buoyancy process and after the exit into the photosphere can form an active region.

B. The Kleeorin, Rogachevskii, and Ruzmaikin model (1989, 1990). Let us consider again the situation when the upper part of the stellar convective zone is in equilibrium in the presence of the constant horizontal magnetic field **B**. Imagine that the plasma volume (hereinafter the tube) extended along the horizontal field (perpendicular to the surface gravity) is shifted as a whole downward. Since the magnetic field both inside and outside the "tube" (as well as temperature) is the same, whereas the density downward is higher due to stratification, the pressure outside the tube turns out to be higher than inside it. As a result, there emerges a compression of the tube across the field, and, due to a partial freezing of the field into the plasma, the field enhancement at the expense of the magnetic flux concentration. However, due to the negative magnetic pressure ($Q_p < 0$), the pressure inside does not grow but drops! Certainly, the tube cannot be neither too thin — the process is stopped by turbulent diffusion — nor too big (due to the stratification in density). The process of accumulating the flux in the tube will proceed until the field in it achieves the value ~ $0.7B_{\mathrm{eq}}$ at which the effective magnetic pressure achieves its minimum (see Fig. 114, right panel). Afterwards, the arisen tubes can start to be buoyant according to the Parker mechanism (1955a) and form active regions. Certainly, if both mechanisms work cooperatively, then the accumulation effect is stronger, the tube has a more pronounced boundary, whereas the buoyancy effect is faster. Obviously, the described mechanisms have a threshold character for the mean magnetic field. In order to find this threshold, one needs to proceed to the quantitative description of tube formation instability. Let us take as an example the model of Kleeorin et al. (1989, 1990). For this aim, we consider the following problem: there is a horizontal toroidal field which is constant both in the horizontal and vertical planes. Let us choose Equations (96), (97), and (99) as the equations for describing the phenomenon we are interested in. Neglecting the Coriolis effects and turbulent viscosities, we obtain the following equations of motion and induction:

$$\frac{\partial \mathbf{U}}{\partial t} = -\vec{\nabla}\left( \frac{\bar{P}_{tot}}{\rho_0} \right) + \frac{1}{4\pi\rho_0}\left( \mathbf{B} \cdot \vec{\nabla} \right) Q_s \mathbf{B} + \mathbf{\Lambda} \frac{B^2}{8\pi\rho_0} Q_p \qquad (146a)$$

$$\frac{\partial \mathbf{B}}{\partial t} = \mathrm{rot}\left( \mathbf{U} \times \mathbf{B} \right) \equiv \left( \mathbf{B} \cdot \vec{\nabla} \right) \mathbf{U} - \left( \mathbf{U} \cdot \vec{\nabla} \right) \mathbf{B} - \mathbf{B}\left( \mathbf{\Lambda} \cdot \mathbf{U} \right) \qquad (146b)$$

$$\mathrm{div}\,\mathbf{U} = \mathbf{\Lambda} \cdot \mathbf{U} \qquad (146c)$$

Note that in the induction equation (146b) we neglect for simplicity the turbulent electromotive force and $\mathbf{\Lambda} = -\nabla\rho_0 / \rho_0 = \mathbf{e}_z / H_\rho$. The equilibrium solution to the system (146a)–(146c) has the form:

$$\mathbf{U}=0\,,\mathbf{B}=\mathbf{e}_y B_0\,,\ \bar{P}_{tot}=-\frac{B_0^2\rho_0}{8\pi}\int_0^z \frac{Q_p}{H_\rho \rho_0(z')}dz'.$$

We seek the disturbed linear solution to Equation (146a) in the following form:

$$\delta U_z=\partial\xi/\partial t=\lambda V\ \exp(0.5\Lambda z)\exp\left(\lambda t+i\left(k_\perp x+k_\| z\right)\right)\ \text{and}$$

$$\delta U_y=\lambda U\ \exp(0.5\Lambda z)\exp\left(\lambda t+i\left(k_\perp x+k_\| z\right)\right).$$

The relation between $V$ and $U$ is found from the condition (146c). Using the variable $\xi$, Equation (146b) is easily integrated and yields the magnetic field disturbance. To exclude pressure, we take twice the rotor from the left and right sides of Equation (146a) and project the derived equation onto the $z$ axis. This yields

$$\frac{\partial^2}{\partial t^2}\left(-\Delta+\frac{1}{4H_\rho^2}\right)\bar{V}(x,z)=2\frac{V_A^2}{H_\rho^2}F\left(\frac{B^2}{B_{eq}^2}\right)\Delta_\perp\bar{V}\,, \tag{147}$$

where $\bar{V}(x,z)=\lambda V\exp\left(\lambda t+i(k_\perp x+k_\| z)\right),\xi=\exp(0.5\Lambda z)\bar{V}(x,z)$ and $F=Q_p\left[1+\partial\ln\left|Q_p\right|/\partial\ln(\bar{B}^2)\right]$. Substituting these solutions into (146a)–(146c), we obtain the expression for the increment of instability growth of negative magnetic pressure:

$$\lambda=\frac{V_A k_\perp}{H_\rho}\sqrt{\frac{-2F\left(B^2/B_{eq}^2\right)}{k_\perp^2+k_\|^2+1/4H_\rho^2}}\ . \tag{148}$$

We can show that the account of turbulent viscosities, both magnetic and hydrodynamic, can lead to a decrease in the increment (148). If these viscosities are equal to each other, then the answer is simple:

$$\lambda=\frac{V_A k_\perp}{H_\rho}\sqrt{\frac{-2F\left(B^2/B_{eq}^2\right)}{k_\perp^2+k_\|^2+1/4H_\rho^2}}-\eta_T\left(k_\perp^2+k_\|^2+\frac{1}{4H_\rho^2}\right). \tag{148a}$$

It follows from (148a) that for exciting instability the condition $k_\| \to 0$ is optimal. Then the required conditions of instability are as follows:

$$F\left(B^2/B_{eq}^2\right)<0\ \text{and}\ a^2-\frac{\left(Z^2+1\right)^3}{Z^2}>0\ , \tag{148b}$$

where $Z=2k_\perp H_\rho$ and $a=8H_\rho V_A\sqrt{-2F\left(B^2/B_{eq}^2\right)}/\eta_T$. The beginning of instability corresponds to such a value of $a_{cr}=\sqrt{f_{\min}}$ (critical field value) and $Z_{cr}$ (the horizontal size of the tube) at which the function $f(Z)=(Z^2+1)^3/Z^2$ has a minimum in the point $Z_{cr}$. These values are determined analytically by approximately the same way as it was done in the Rayleigh theory for laminar convection — from the condition of function $f(Z)$ minimum, and account for (Fig. 116) $a_{cr}=\sqrt{27}/2\approx 2.6$, whereas $a_{cr}=\sqrt{27}/2\approx 2.6$ and $Z_{cr}=1/\sqrt{2}\approx 0.7$. The appropriate critical field is

$$B_{cr} \approx \frac{0.46\eta_T}{H_\rho}\sqrt{\frac{\pi\rho_0}{|F|}} = 0.136 B_{eq}\frac{\ell_0}{H_\rho\sqrt{|F|}}\ , \tag{149}$$

whereas the critical size (a half of the horizontal wavelength) $L_{cr} = 2\pi H_\rho\sqrt{2} \approx 8.9 H_\rho$.

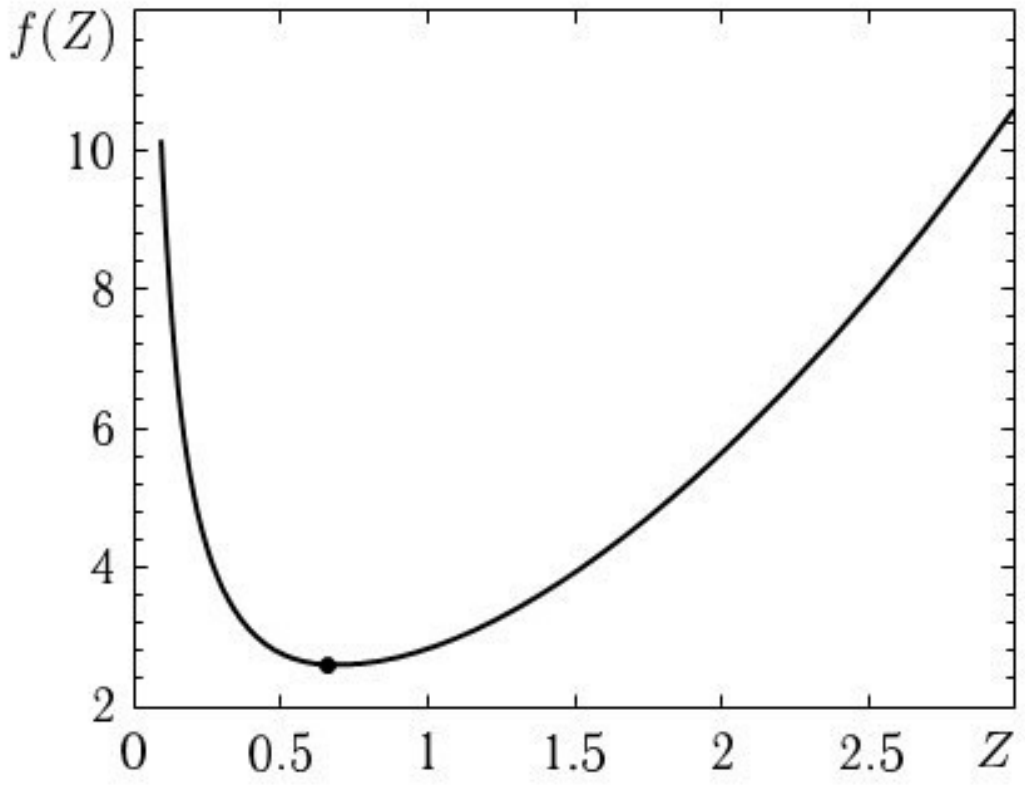

Fig. 116. A neutral curve of $f(Z)$. The black dot denotes the point corresponding to $a_{cr} = f(Z_{cr}) = \sqrt{27}/2 \approx 2.6$ and $Z_{cr} = 1/\sqrt{2} \approx 0.7$

If the magnetic field is somewhat higher than the critical value, then, according to the unequality (148b), there arises an interval of tube sizes from a few less to a few more than $L_{cr}$. If one takes for estimate the solar parameters at a depth of $10^9$ cm, then it turns out that $B_{eq} \approx$ 800 G and $\ell_0/H_\rho \approx 0.72$. Supposing for estimate $F \approx -1 \div -2$, we obtain $B_{cr} \approx$ 80–60 G which, according to Table 27, is approximately 3.5 times lower than the surface toroidal field of the Sun. Note that the condition $B_s > B_{cr}$ cannot be fulfilled for any heliolatitude, since at the poles and equator the dynamo wave amplitudes turn into zero due to boundary conditions. Therefore, a certain range of latitudes is distinguished in which the condition $B_s(r,\vartheta) > B_{cr}$ is indeed fulfilled. This property can possibly explain the existence on the Sun of the so-called royal spot formation zone. This analysis did not take into account a decrease in the Alfvén velocity with depth due to the pressure growing, which almost for sure complicates the formation of tubes far from the surface. A significant growth of equipartition with the field depth makes the fragmentation of the toroidal field into tubes impossible at great depths. In this context, it seems likely that the depth of formation of active regions is close to $10^9$ cm. At this depth, $H_\rho \approx 3.9 \cdot 10^8$ cm. This yields $L_{cr} \approx 2.83\pi H_\rho \approx 3.5 \cdot 10^9$ cm for the critical size of an active region. Such a size is in agreement with the observed sizes of sunspots, which do not exceed 35000 km. These estimates seem to be able to constrain the development of the analytical linear theory of spot formation within the model of Kleeorin, Rogachevskii, and Ruzmaikin (1989, 1990). Therefore, for further progress of the stellar spot formation theory one needs to refer to numerical simulations. Such numerical simulations within the mentioned model were performed in the series of papers by Brandenburg et al. (2010, 2011, 2012, 2013, 2014, 2016); Käpylä et al. (2012, 2016); Kemel et al. (2012a, 2012b, 2013a, 2013b); Jabbari et al. (2013, 2014, 2015, 2016, 2017); Losada et al. (2012, 2013, 2014, 2019); Warnecke et al. (2013, 2016) using indirect numerical simulations, large-eddy simulations, mean-field simulations. Some results of these calculations are presented in Fig. 117.

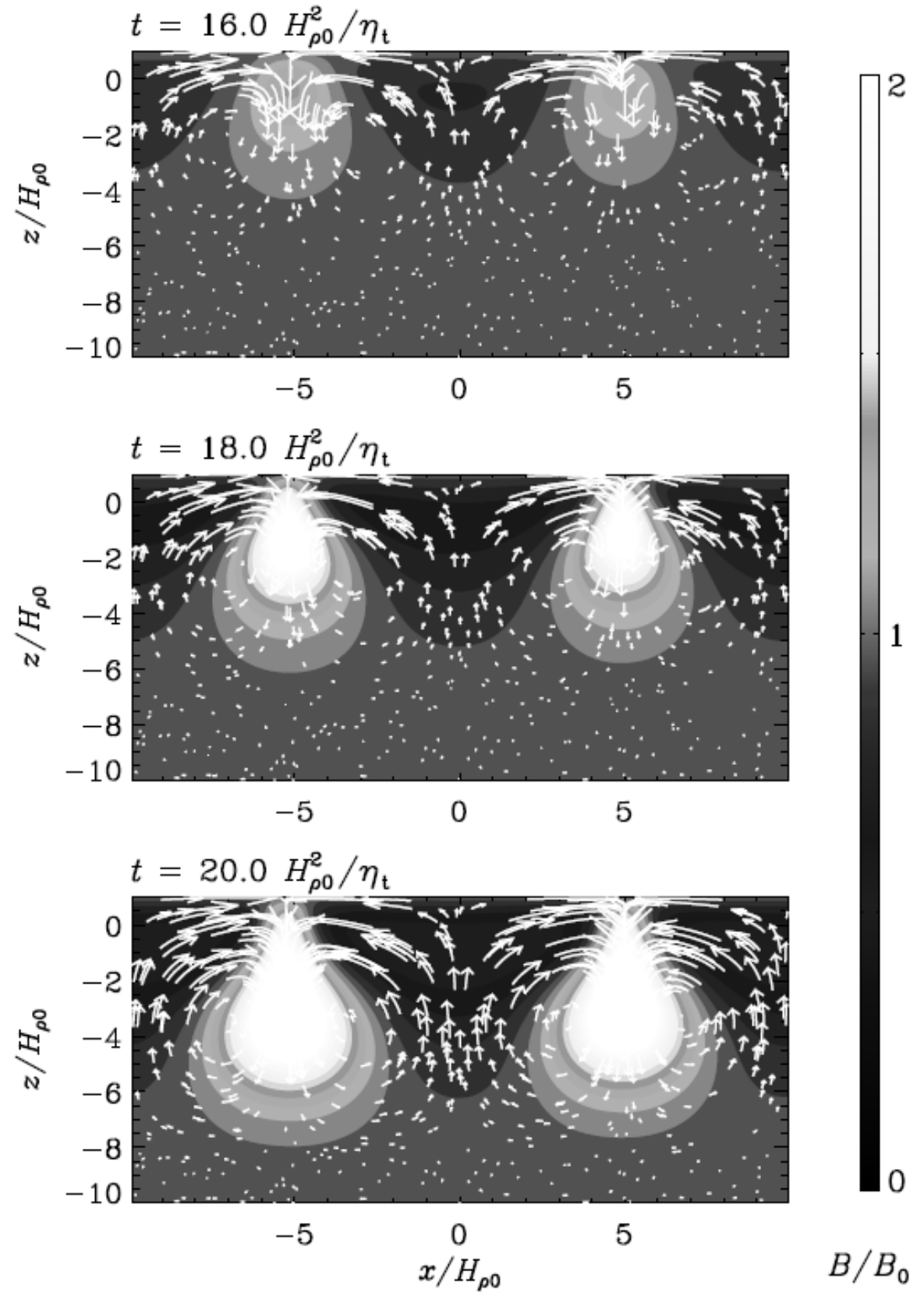

Fig. 117. Results of nonlinear evolution of the horizontal field under the influence of negative effective magnetic pressure instability (NEMPI). The process of magnetic flux concentration is shown in two horizontal tubes of the enhanced (*white color*) magnetic field surrounded from almost all the sides by the zero field regions (*black and dark grey colors*) (Brandenburg et al., 2010)

The region under calculation represents a vertical section of the turbulent region ("stellar convective zone") with a depth and width of 10 density scales of $H_{\rho 0}$. The account has a two-dimensional character — the velocity components $U_z$ and $U_x$ are not equal to zero, $U_y = 0$. In Fig. 117, the magnetic field is directed along the $Y$ axis into the plane of the figure. At the initial moment the system was near the generation threshold, i.e., $B_0 \geq B_{cr}$, therefore the tube size was close to $3.5H_{\rho 0}$. A quite slow evolution of the configuration at the beginning indicates that the system was near the instability excitement threshold NEMPI: the enhancement of the initial field less than by a factor of 1.5 occurs over 16 characteristic times of magnetic field turbulent diffusion $\eta_T/H_\rho{}^2{}_0$. It is noticeable that evolution accelerates as the magnetic field grows. Such calculations were carried out in the three-dimensional region. Their results are shown in Fig. 118.

Figure 118 shows the result of the magnetic flux concentration in the horizontal plane $XY$ not far from the upper boundary of the region under calculation at a depth of about $0.45H_{\rho 0}$. White and light-grey colors correspond to the field directed upward, dark and dark-grey colors — the field directed downward, grey color — the zero field. White arrows in both this and the previous figure denote hydrodynamic flows. These are quite similar to an active region on the Sun. For comparison, Fig. 119 displays the same solar active region.

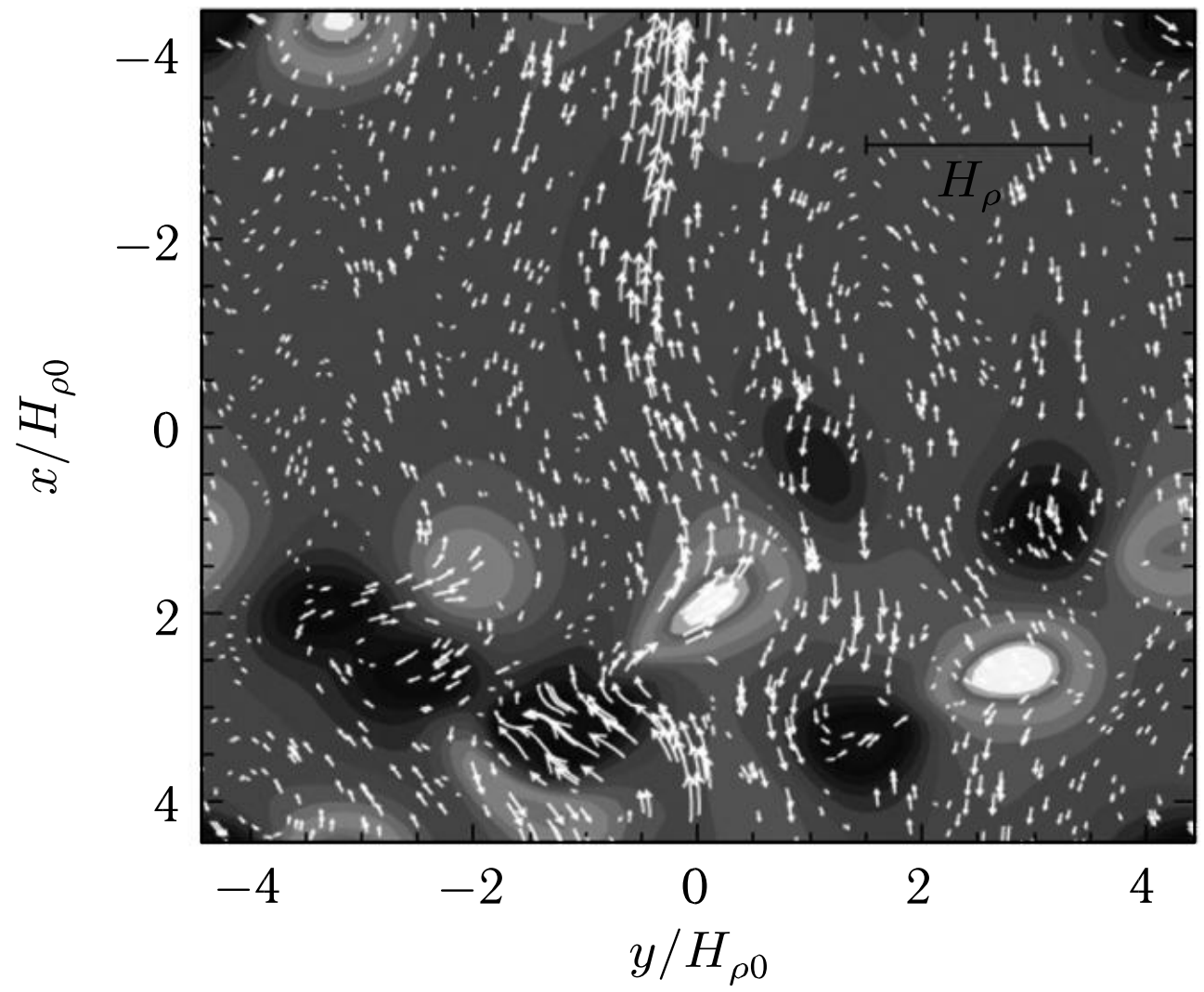

Fig. 118. Results of nonlinear evolution of the initially horizontal field under the influence of NEMPI in the three-dimensional region $9H_{\rho 0} \times 9H_{\rho 0} \times 9H_{\rho 0}$ with the boundary condition $U_z = B_z = 0$ at $z = 0$ (Brandenburg et al., 2010)

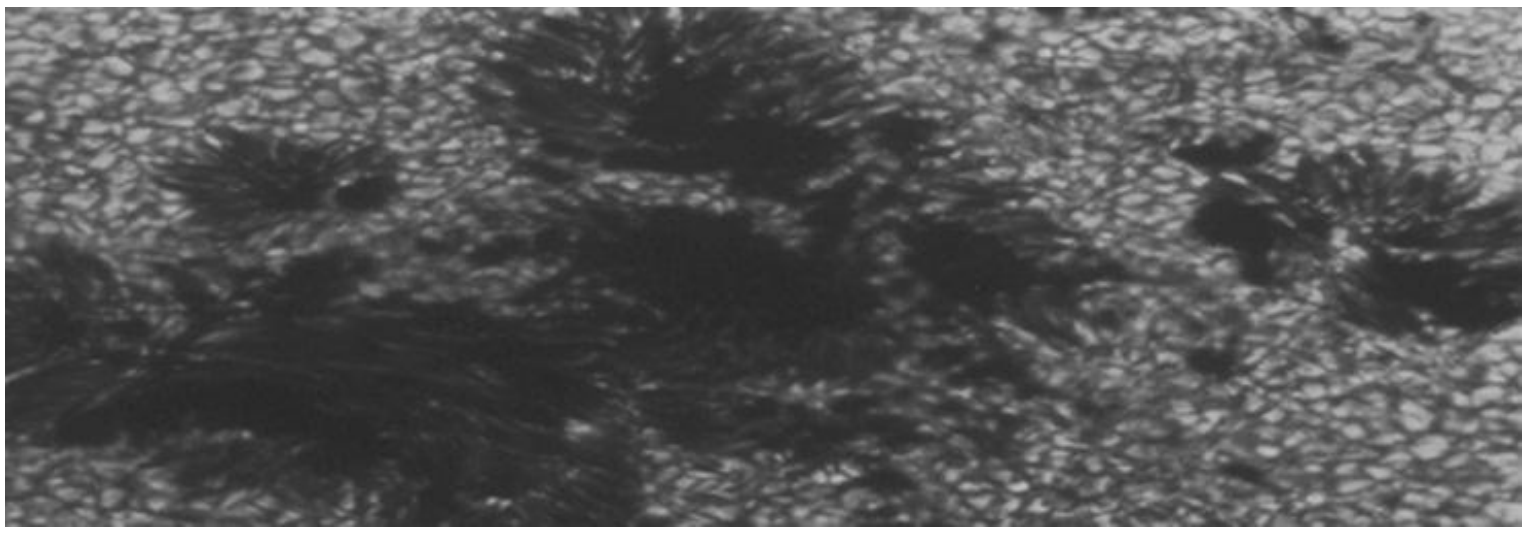

Fig. 119. Typical solar active region. Similarity with the calculated active region in Fig. 118 is obvious

However, despite external similarity the results of calculations in Fig. 117 drastically differ from the properties of real solar active regions: the magnetic field of an active region is larger — of about the equipartition field, whereas the calculated field does not exceed one percent of the equipartition field. The reason lies in the specific boundary condition $U_z = B_z = 0$. Refusing this condition drastically changes the results of calculations. Figure 120 displays the results of calculations within the mean-field model at the weak initial vertical field $B_z(t = 0) = 0.05B_{\rm eq}$ (Brandenburg et al., 2014). The calculated region has sizes of $6.5H_{\rho 0} \times 6.5H_{\rho 0} \times 20H_{\rho 0}$. In the direction $z$ from the depth $-10H_{\rho 0}$ to zero there is a region of high densities and pressures (model of the convective zone), whereas from zero to $10H_{\rho 0}$ — the zone of low densities and pressures (model of the photosphere, chromosphere, and/or even corona). The choice of the initial field precluded at once the condition $B_z = 0$. The results lost no time in being detected. The magnetic field at the end proved to be of an order of $B_{eq}$. The field structure is seen in Fig. 120 (left panel). It looks like a "torchere".

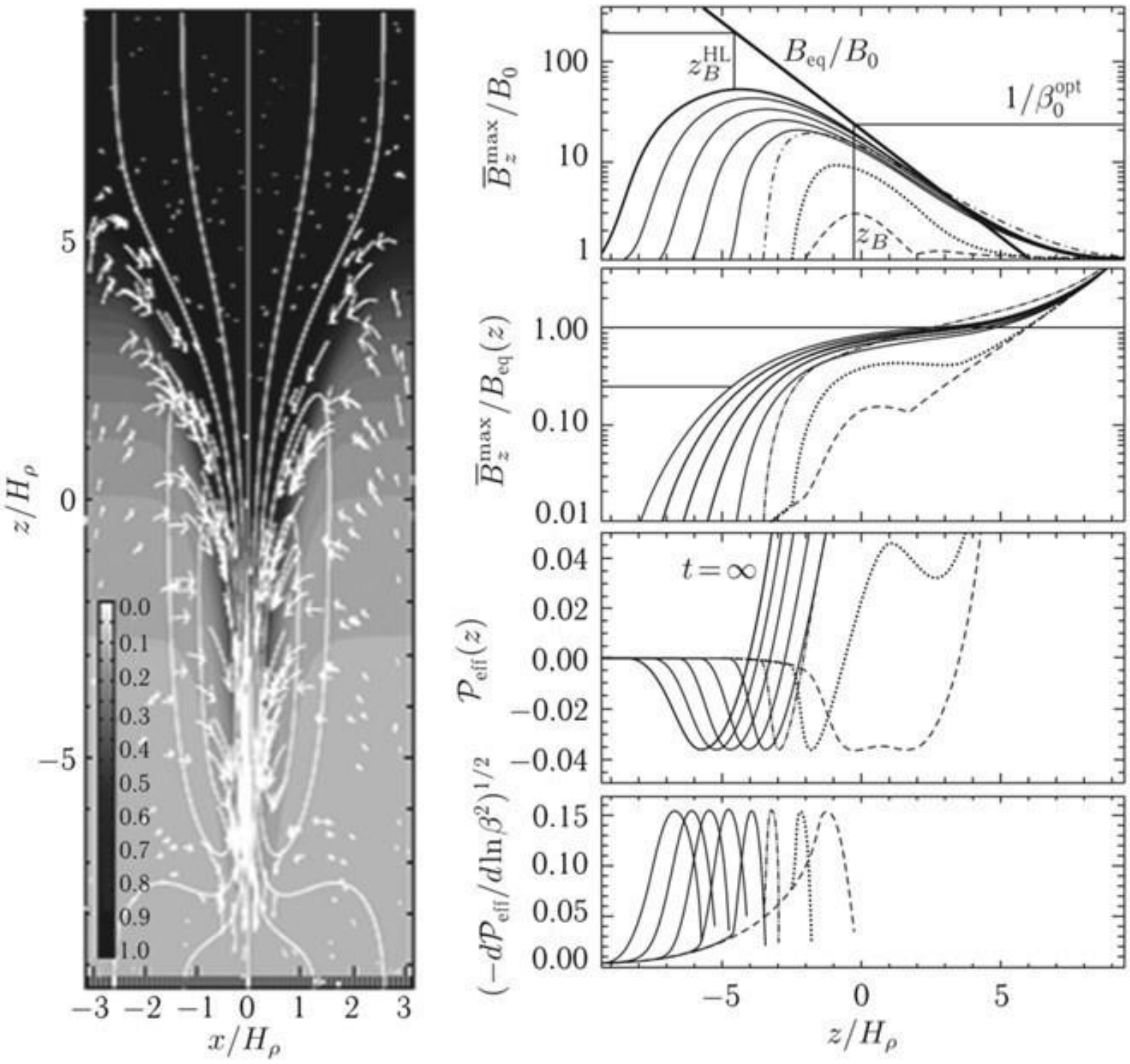

Fig. 120. The vertical field structure arisen at the nonlinear stage of instability development NEMPI (*left panel*) and graphs explaining the process of instability development and its saturation (*right panel*) (Brandenburg et al., 2014)

The nature of this structure is elucidated in the left panel of Fig. 120, which shows that the process of tube formation starts on the surface from the weak field achieving at the end of the process of field concentration its maximum value of about $\sim 5H_{\rho 0}$ of the field value of about $\sim 0.6B_{eq}$ (right panel). If one identifies $H_\rho \approx 3.9 \times 10^8$ cm and $H_\rho$ in calculations of Brandenburg et al. (2014), then it is obtained that the depth $\sim 5H_{\rho 0}$ accounts for about $2 \cdot 10^9$ cm, i.e., 20000 km on the Sun. At this depth the equipartition field accounts for 2000–3000 G, and, consequently, the maximum field should be 1200–1800 G. It is significant that, as in real spots, the downward flows are directed near the "convective zone boundary" inside a "spot". White solid lines denote force lines of the mean magnetic field. It is also noticeable that the beam of force lines (left panel) at the end of the process development is retained by negative magnetic pressure $-P_{eff}$ having the maximum at a depth of $\sim 6H_{\rho 0}$, whereas at the beginning of the process the magnetic pressure at this depth was absolutely scant (see the right panel). It is worthy to note that the direct numerical modeling of such a two-layer model of turbulence also demonstrated the formation of the similar magnetic field (Fig. 121).

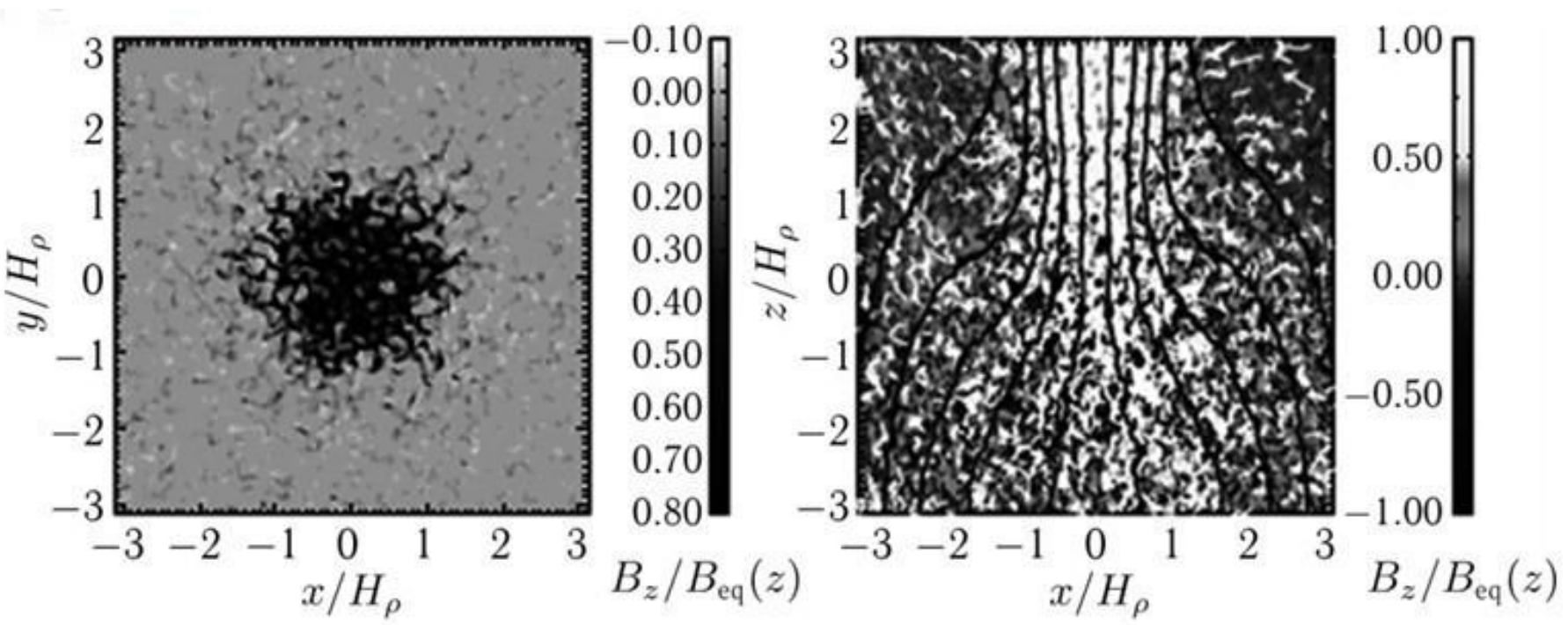

Fig. 121. Results of the direct numerical modeling of the magnetic field concentration ensured by NEMPI. *Shades of black and grey colors* display the magnetic field levels at a height of about $\sim \pi H_\rho$ above the upper "convective zone boundary" (*left panel*). The right panel shows the structure of the magnetic field and flows in the section. *Black solid lines* denote force lines of the averaged magnetic field, *white arrows* mark downward flows, *shades of black and grey colors* — the magnetic field levels (Brandenburg et al., 2013)

In the "spot" under calculation in Fig. 121, considerable attention is paid to inhomogeneity of the boundary, which is very similar to the boundary of a real spot. The field intermittency is also obvious: regions of the weak field are included as small contaminations inside the strong field region. Note also that using the mean-field theory (see Figs. 118 and 119), it is very difficult, if possible at all, to obtain the same realistic magnetic model of a "spot", as in Fig. 121. Further development of this model is the modeling of the horizontal magnetic field behavior within the direct numerical experiment. The results of these calculations are shown in Fig. 122.

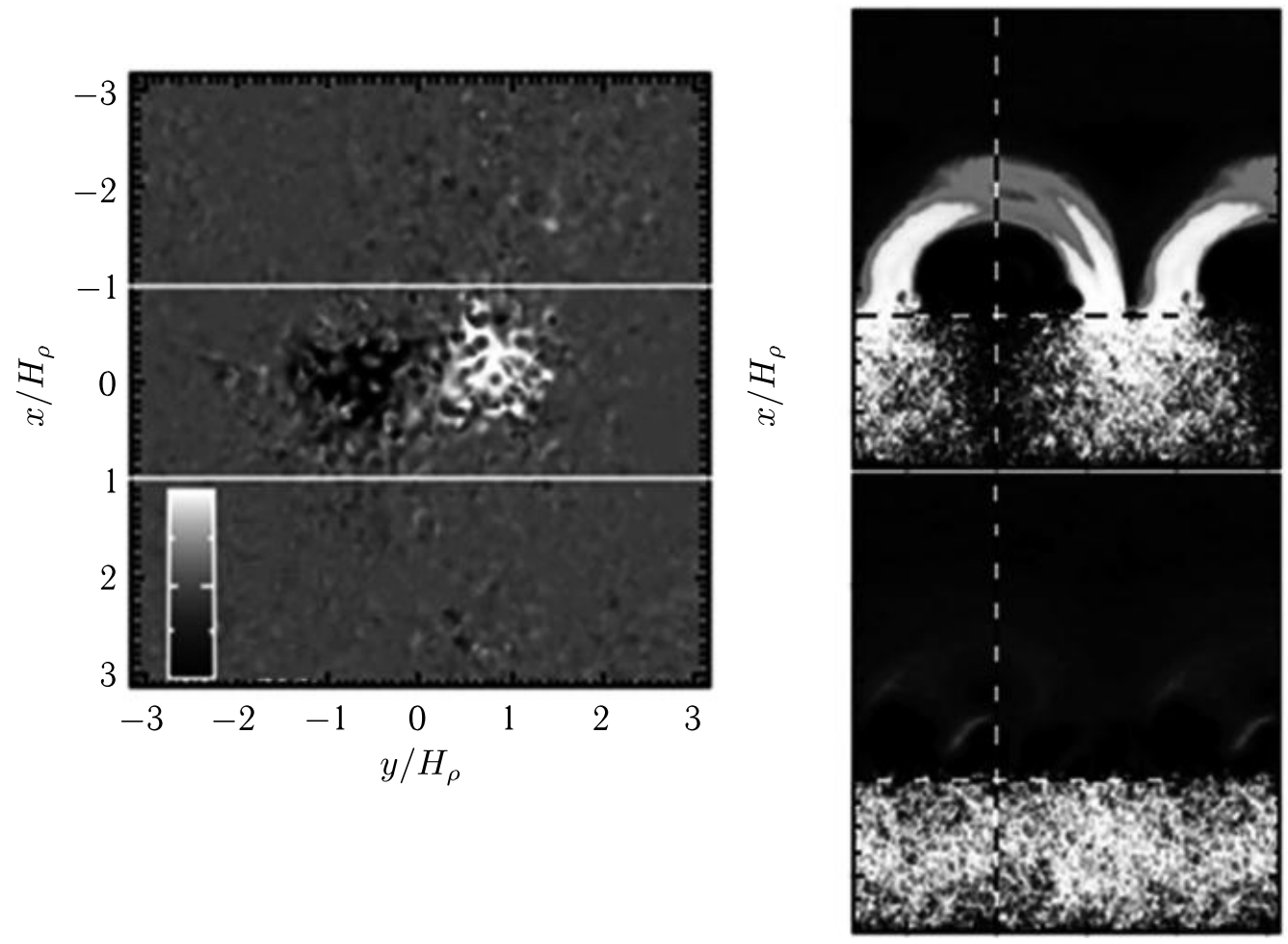

Fig. 122. *Left panel:* the bipolar group of the vertical magnetic field generated from the initially horizontal weak field at the moment $t = 2H_\rho^2/\eta_T$. *Right panel:* the arch structures above the bipolar region — the arch structures at the moment of time $t = 2H_\rho^2/\eta_T$ (*top*), the same but splitting structures at the moment of time $t = 3.1H_\rho^2/\eta_T$ (*bottom*) (Warnecke et al., 2016)

Particular attention is paid to a fairly abrupt boundary of polarity division in the left panel of Fig. 122. This suggests an idea of the existence of this current sheet boundary. And such a sheet has indeed been found!

Figure 123 represents a vertical section of the magnetic field. The powerful current sheet is seen to start at a height of about $\sim 1H_\rho$ and extends up to the end of the region under calculation, up to the height $\sim 3H_\rho$, whereas the field achieves a fairly high level of $\sim 2B_{eq}$, and a vastly abrupt transition from one polarity to another is observed.

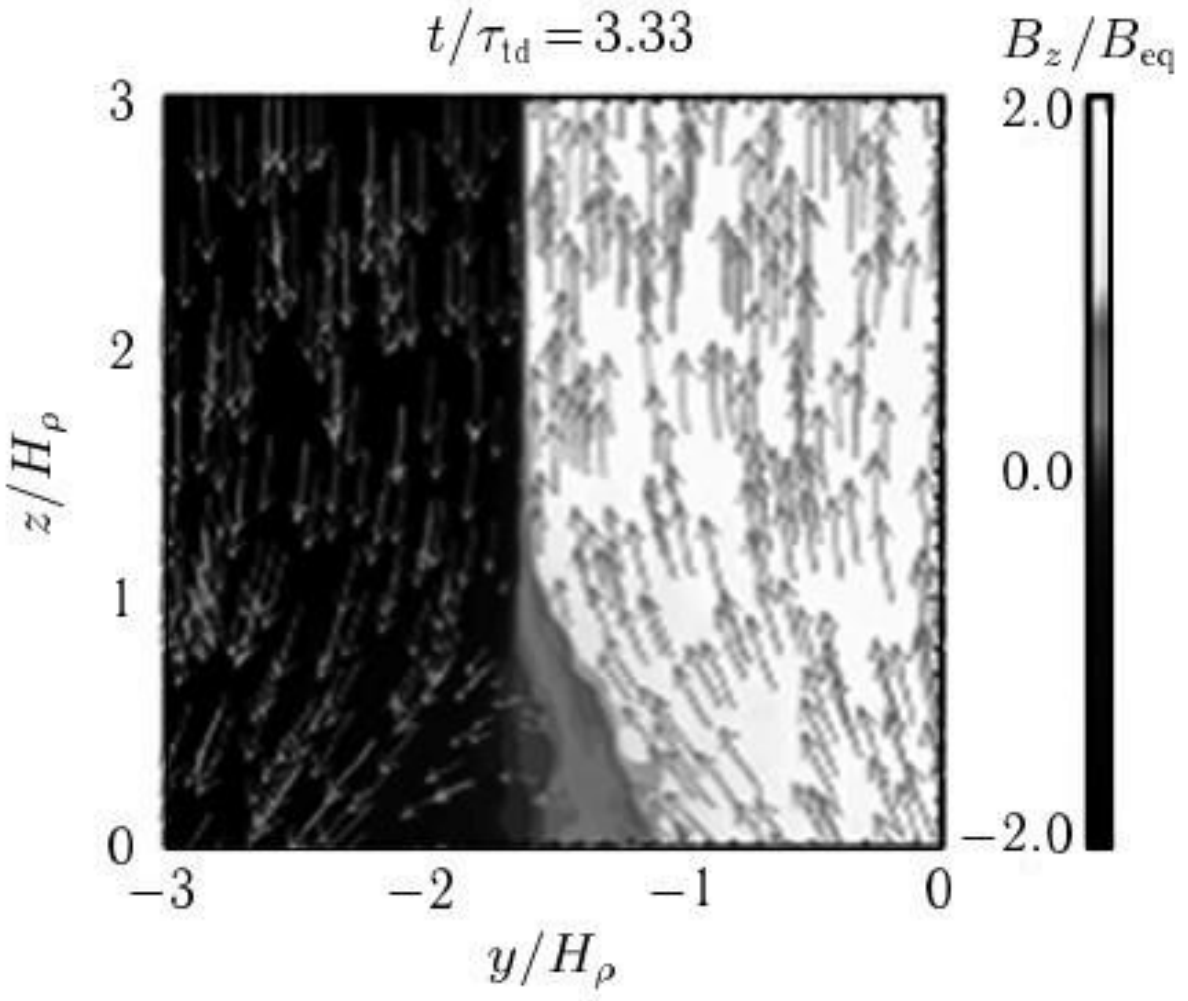

Fig. 123. The vertical section of the magnetic field. Arrows denote the lines of hydrodynamic motions (Jabbari et al., 2016)

Note that in these calculations not the external superimposed field was used but the field arisen as a result of the $\alpha^2$ dynamo. Certainly, any phenomena of the type of flares, radio bursts, flows of elementary particles, etc. cannot be obtained in such current sheets. For this aim, it is necessary to take plasma phenomena into account. However, even without this, the result is impressive.

* * *

In the next small part of the monograph, we consider the impact of stars with solar-type activity on the exoplanets orbiting them. Here, discussing the stellar magnetic field generation, we should note the paper of Kashyap et al. (2008) who found that the X-ray radiation of stars with nearby large planets is stronger by a factor of 4 than that in stars with distant planets. Hence, the stellar magnetism is not always a purely "internal affair" of a star. The physical basis for this phenomenon seems to be the multisphere dynamo of Herzenberg (1958). A brief review of this mechanism can be found in Moffatt (1978). The important peculiarity of this dynamo is a possibility of forming the common magnetosphere for several (minimum two) conducting rotators which even rotate as a solidbody. The required conditions of this dynamo effect are as follows:

a) the rotation axes of rotators are not parallel;

b) the rotation axes of rotators and lines that merge their centers are not in the same plane.

A sufficient condition for this effect is the required conditions plus a fairly fast rotation of rotators. The presence of turbulent convection on any of the conducting rotators can change generation through the Herzenberg mechanism, in particular, can enhance it. On the other hand, the global magnetic circulations in binary stars of early spectral types can without convection lead to enhancement of the field generation. However, these issues are not sufficiently covered in the literature in the context of stellar dynamo and need to be further studied.

* * *

Finally, we would like to formulate several general questions on the solar-type magnetism — both already solved by the available theory and those that wait for being addressed in its further development — and to give brief answers.

1. *Is the observed stellar magnetism a result of developing relic magnetism or a pure effect of the stellar dynamo?*

Both these: during the main-sequence stage of a star the relic magnetic field is likely to play a decisive role because it seems to make a general contribution into the seed field for dynamo. When a star being compressed to the main sequence, depending on the initial mass, there occurs a formation of either the convective core or the convective envelope out of which there is either a radial zone or a radial core. Turbulent convection forces the relic magnetic field out in these radial regions (Vainshtein et al., 1980). Thus, in the convective zones of stars later than, let's say, F5, the contribution of the relic field $B_{re\ell}^{(C.Z.)}$ is not significant: according to Vainshtein et al (1980) it is of an order of $B_{re\ell}^{(C.Z.)} \simeq B_{re\ell}^{(0)} \big/ \sqrt{\mathrm{Rm}}$, where $B_{re\ell}^{(0)}$ is the relic field in the radial zone. For stars similar to the Sun this field does not exceed a few gauss. In earlier quasilinear models (see, e.g., Pudovkin and Benevolenskaya, 1982), there were attempts to explain through $B_{\mathrm{rel}}{}^{(C.Z.)}$ such properties of solar activity as the Gnevyshev-Ohl rule. Further studies of more realistic nonlinear models, taking into account the evolution of magnetic helicity, showed that the presence of the residual field is not sufficient (Kleeorin et al., 2016) for the Gnevyshev-Ohl rule.

2. *Is there a unified widely accepted stellar dynamo theory?*

As stated above, a complete solution of the dynamo problem includes the induction equation, the Navier-Stokes equation, the continuity equation, entropy transport equations, the radiation transfer equation, and several equations of state: the Mendeleev–Clapeyron equation, the Saha ionization equation, the equation for the Rosseland mean absorption coefficient of electromagnetic radiation of stellar plasma, and others. This system of equations comprises a scientific basis for the widely accepted theory of stellar dynamo. In such a statement, the problem seems to be absolutely unmanageable. Nonetheless, there are several stellar dynamo models with quite different levels of elaboration, some of them are additive, the other ones are alternative. The quite developed models of the $\alpha$–$\omega$ dynamo are described above, partially the $\alpha^2$ dynamo, whose nonlinear versions allow one to understand many phenomena of stellar magnetism.

3. *Is there a theory of formation of poloidal and toroidal fields on stellar surfaces, their mutual switching, as some observers of M dwarfs suppose?*

No, there is no such a theory. However, the similar pattern can be provided by nonaxisymmetric modes of the $\alpha^2$ dynamo. Indeed, the modes of the pure axisymmetric $\alpha^2$ dynamo, as stated above, can generate toroidal and poloidal magnetic fields comparable in order of magnitude. The nonaxisymmetric $\alpha^2$ dynamo, as it follows from the paper of Brandenburg (2017), maintains this property, but there appears a wave running along the stellar equator. Since a phase shift between the toroidal and poloidal magnetic field components is close to $\pi/2$, then the observed picture represents an interchange of the toroidal and poloidal magnetic field components that are comparable in order of magnitude, in space and time. Note that for stars later than M0, the convective envelope occupies more than a half of a star, and after M4 the stars are fully convective. Therefore, as stated above, a decrease in differential rotation and a gradual transition into the $\alpha^2$ dynamo mode, i.e., into the mode of “switchings”, are likely typical of these stars.

4. *The maximum magnetic field is observed in spots. But on the Sun they occupy just fractions of a percent of the surface, on stars — tens of percent, up to a half of the surface. Is this a result of different inner structures or different initial conditions, or different age, or different dynamo?*

This is a result of different inner structures (spectral type), different initial conditions (initial angular momentum), different age (current angular momentum) and, as a consequence, generally speaking, different types of the mean-field dynamo. This can be seen from the following equation obtained when identifying a magnetic flux generated by the dynamo in the stellar “royal sunspot zone” and a flux accumulated in spots by a certain mechanism of field concentration:

$$\frac{S_{Sp}}{\pi R_*^2} \approx 2mN \frac{\left(B_s - B_{cr}\right) H_{form}}{\pi R_* B_{Sp} \cos\bar{\varphi}} \left[\varphi_{\max} - \varphi_{\min}\right].$$

Here $B_s$ is the stellar surface field determined by Equation (145), $B_{\rm cr}$ is the critical field for the spot-formation beginning defined by Equation (149). $B_{\rm Sp}$ — the typical field in a spot, $m$ — the typical number of spots in an active region, $N$ — the average number of active regions at a given latitude, $\varphi_{\max}$, $\varphi_{\min}$ — the upper boundary of the “royal sunspot zone”, respectively, $\bar{\varphi} = \left(\varphi_{\max} + \varphi_{\min}\right)/2$, $H_{form}$ — the depth of formation of magnetic ropes. Thus, the conditions for the formation of spots are as follows: $B_{\rm Sp} \sim B_{\rm eq} > B_s > B_{\rm cr}$. If at increasing rotation and/or at decreasing effective temperature of a star it turns out to be that $B_s > B_{\rm Sp} \sim B_{\rm eq}$, then the spot formation of small spots will cease, but in this case the extended stellar surfaces will be covered by the field $B_s \geq B_{eq}$.

5. *Is it clear why do spot formation bands of cooler stars systematically have lower latitudes?*

Not quite. We can just suggest a cautious hypothesis that the strong mean field at the equator is capable of enhancing the spot formation near the equator. i.e., at lower latitudes than, let’s say, on the Sun. Two independent mechanisms of existence of the strong field at the equator can be suggested: in the case of dynamo, the strong (toroidal) field at the equator is a result of significant contribution of the quadrupole symmetry mode into the poloidal field. This can be possible, on the one hand, at certain evolutionary stages of the magnetic field, for example, at its chaotic motion (as on the Sun). Interestingly, during the Maunder minimum rare spots appeared just at the equator (Baiada and Merighi, 1982). This suggests the predominance of the quadrupole component at that epoch. Currently, contrarily, the component of the dipole type prevails. The other possibility is the fast rotation of stars of late spectral types. In this case, it is possible the coexistence of the dipole and quardupole components of comparable value and the appearance of spots near the equator on a continuous basis. The dynamo of the $\alpha^2$ type can also yield a similar picture. The latter should be characteristic of spectral types M2–M5.

6. *There exist stars with a drift of the spot formation band during an activity cycle toward both the equator and the pole. What determines the value and direction of this drift? Why does motion toward the equator prevail, as on the Sun? Is it possible that the change of drift directions occur at some critical spectral type?*

A change of drift directions is evidently possible. As stated above, within the $\alpha$–$\omega$ dynamo the latitudinal drift direction of the main branch of stellar activity is determined by the sign of product of $\alpha d\omega_*/dr$. The drift to the equator requires the negative sign of this product in the northern hemisphere. Helioseismology confidently indicates that in the upper and lower tachoclines $d\omega_*/dr < 0$, whereas between them $d\omega_*/dr > 0$. It is difficult to say how typical it is for other stars, since observations of the stellar interior are practically impossible, whereas the theory of stellar differential rotation is insufficiently developed. As for the alpha effect, it is known that at certain anisotropy of turbulent convection and at rather fast rotation it can change its sign. Hence, a change in the drift direction of the spot formation band is generally possible. It is different if the pattern of differential rotation (while it is available) is indeed universal for all solar-type stars. Then this is about a sign of the alpha effect at the stellar surface. This sign is defined, on the one hand, by the anisotropy degree (both hydrodynamic and thermal) of turbulent convection and, on the other hand, by the Coriolis parameter $\Omega_*\tau(r)$ (see Figs. 102–106). Dependence $\Omega_*\tau(r)$ for stars of different spectral types is given in Fig. 111. It follows from these figures that at fairly fast rotations a sign of the alpha effect and, consequently, the drift direction of the spot formation band can change. However, there is a possibility that the drift direction of the spot formation band can literally change in front of the eyes of an observer over one-two activity cycles! The reason is that in the nonlinear regime the ongoing alpha effect is important, which comprises both hydrodynamic and magnetic contributions. The signs of these contributions are opposite. Therefore, in strong magnetic fields, i.e., at very high levels of stellar activity, a spontaneous switching of the drift direction of the spot formation band, for example, by a few cycles is possible. The reality of such a scenario is demonstrated by the recent calculation of possible behavior of solar activity throughout the coming 230 years (Fig. 124). It clearly shows that the epoch of low activity, which has currently started, should finish by 2060.

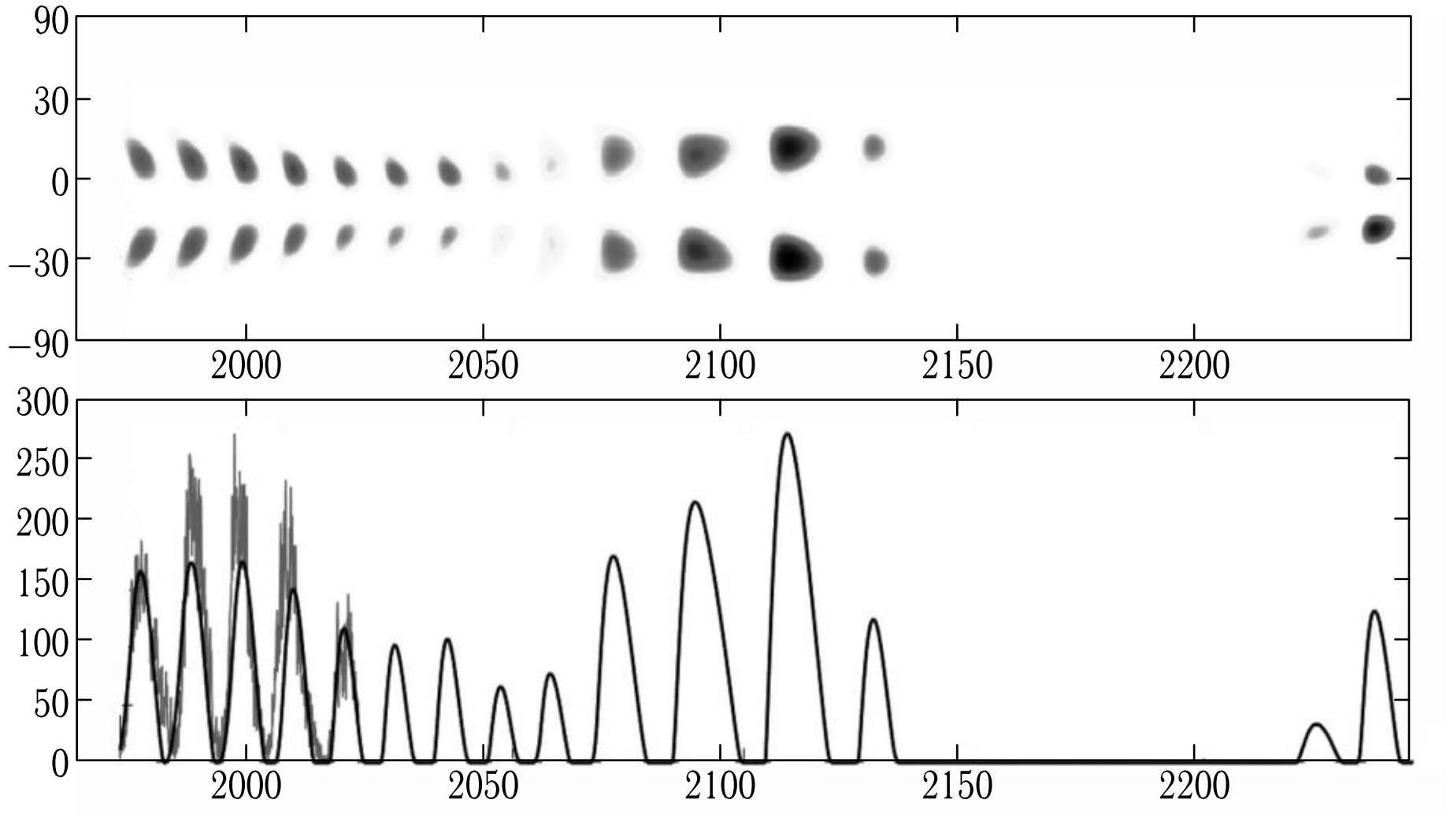

Fig. 124. Possible behavior of solar activity during the coming 230 years. *Top panel*: the butterfly diagram of solar activity approximately between 1950 and 2280. *Bottom panel*: Wolf numbers over the same period (Safiullin et al., 2018)

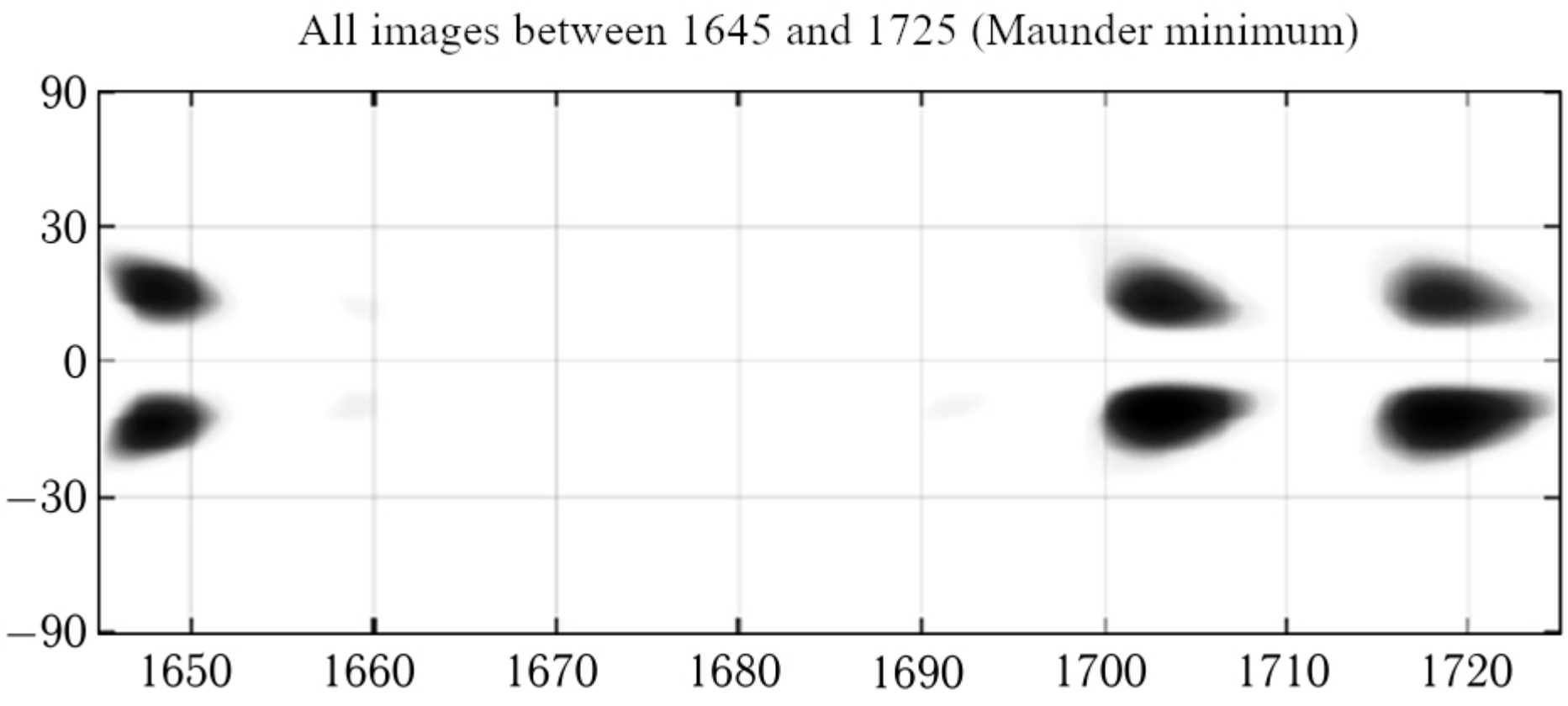

Fig. 125. Butterfly diagram of solar activity according to the model of Safiullin et al. (2018) calculated for the Maunder minimum period (1640–1725)

According to the given calculation, the epoch of low activity will be followed by an unprecedentedly powerful rise of activity when the average monthly Wolf numbers reach 290! Obviously, at that time the magnetic field should be especially high. Thus, a spontaneous change of the drift direction of the spot formation band can be possible. As it follows from Fig. 124, it will pass approximately between 2060 and 2075. After this, the drift to the pole will continue during 3–4 decades.

Within the model of Safiullin et al. (2018), there was also performed a calculation of the butterfly diagram of solar activity for direct counting the estimated time of the system of equations (91)–(136) “back” to the period between 1645 and 1725. Certainly, if we take into account the chaotic nature of a solar cycle, counting the estimated time “back” will not be a fully correct procedure. Therefore, strictly speaking, we cannot trust this result. Nonetheless, a practically precise fit of the “calculated” Maunder minimum with the real one is impressive (Fig. 125).

*7. Why does the difference in temperatures of the photosphere and spots is lower for cooler stars than that for hotter ones?*

This question is rather complicated, and a detailed quantitative (nonlinear) spot formation theory is required for its quantitative resolving. However, at the level of semiqualitative estimates the answer to this question seems to be quite obvious. Indeed, the condition for equilibrium of the tube of the vertical field which forms a spot means that the difference in gas pressures outside and inside the tube is compensated by the difference in dynamic pressures which involve the difference in both hydrodynamic pressure and magnetic one. This yields the following estimate for decreasing temperature in a spot $\delta T$:

$$\delta T \propto (k^2 + \xi^2 - 1) u_c^2 \frac{\mu_g}{R_g} \ .$$

Here $u_c$ is the convective velocity outside a spot, $\mu_g$ — the molecular mass, $R_g$ — the universal gas constant, $k = B_{\mathrm{Sp}}/B_{eq}$ — the ratio of the spot field to the equipartition field, $\xi = u_c^{(\mathrm{Sp})}/u_c$ — the ratio of the convective velocity inside a spot $u_c^{(\mathrm{Sp})}$ to the convective velocity outside it. It follows from the estimate that convective velocity is crucial for the decreasing temperature value in a spot $\delta T$. Its value is shown in Fig. 126.

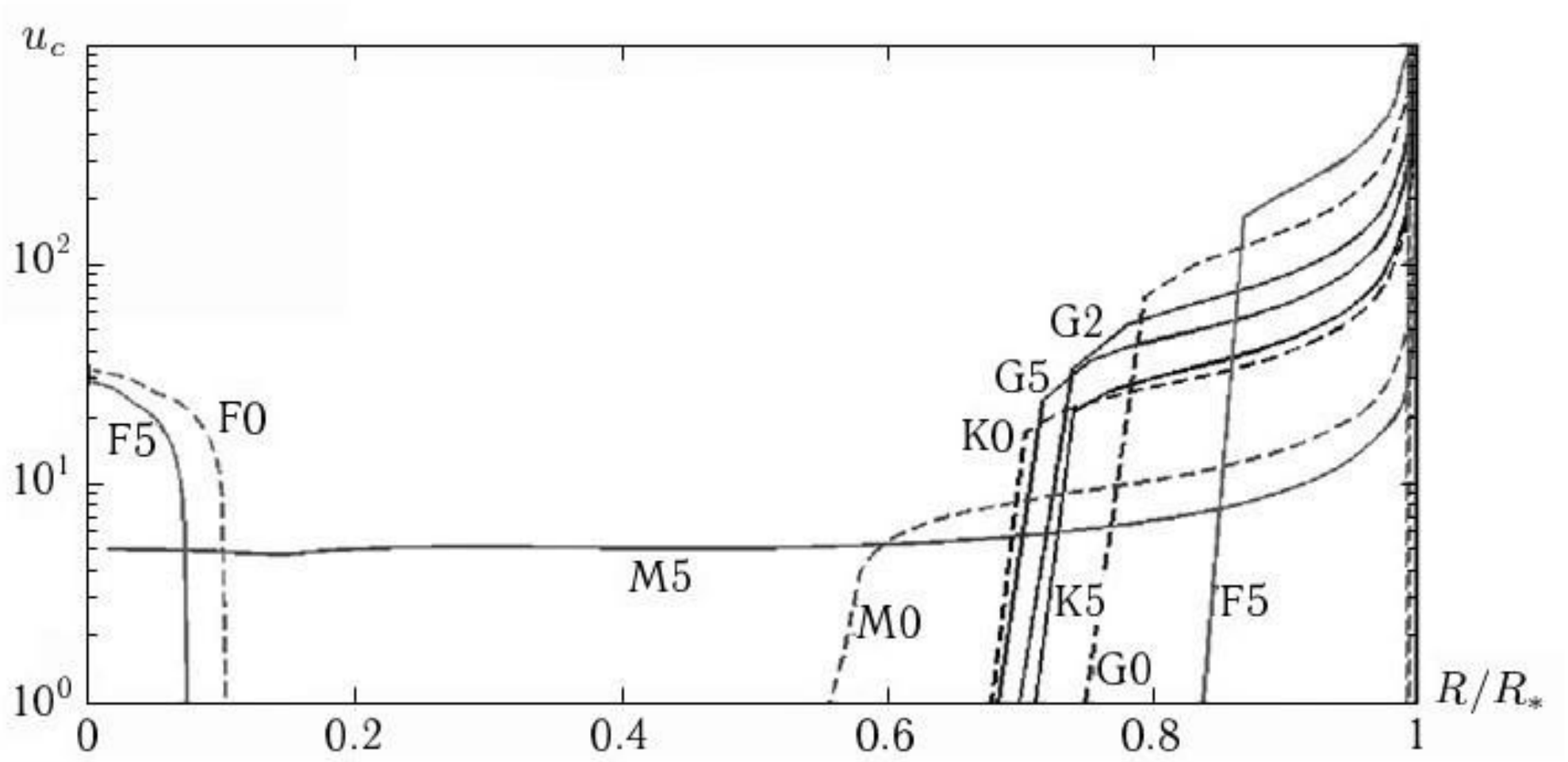

Fig. 126. Dependence of the convective velocity inside a star $u_c$ (cm/s) on the radial component normalized to the stellar radius for stars of spectral types from F0 to M5 (Kleeorin, 2019)

The figure shows that at varying spectral type of stars from G2 to M0 the convective velocity varies by an order of magnitude. This seems to explain the decrease $\delta T$ for stars of late spectral types (see Alekseev and Gershberg, 2021).

8. *Does the magnetic field have effect on mono- and multicyclicity of stellar activity*?

Yes, it does, since the multicyclicity of the mean magnetic field determines the multicyclicity of stellar activity (see Fig. 124). It follows from the figure that the nonlinear dynamo determines the basic cycle (~ 11 years — the whole figure), 22-year (the Gnevyshev-Ohl rule) in the interval between 1970 and 2065, two periods of the 80–100-year cycle between 1970 and 2165, which will be finished by a new Maunder minimum of 70–90 years. Thus, practically all the cycles observed on the Sun are reproduced by the nonlinear magnetic dynamo model of Safiullin et al. (2018).

*9. Does stellar magnetism impact on the ratio of axial rotation durations to the activity cycle?*

Yes, it does, since the presence of the large-scale (including interplanetary) magnetic field causes stellar braking and, hence, a variation in differential rotation.

*10. Are there any apparent manifestations of stellar magnetism?*

Certainly, if we associate all stellar activity with magnetism, then the photometrically detected spottedness and spectrally detected $H_{\alpha}$ emission should be considered as direct evidence for the existence of magnetic fields on a star.

*11. Is the modern theory of the inner structure of stars developed enough for constructing dynamo theories for concrete stars?*

No, not enough: for the dynamo theory one needs to know how anisotropy of turbulent convection varies under the influence of fast stellar rotation and what its differential rotation is. The modern theory of the inner structure of stars has no answers to these questions.

# Part 5

# Impact of Stellar Activity on Exoplanetary Environments

## 5.1. Introduction

The explosion of discoveries of extra solar planets (referred to as exoplanets) in our Galaxy by space and ground-based missions has created the field of exoplanetary science as a part of astrobiology, the interdisciplinary field of modern science. Currently, astronomers can detect at least one strange and exotic exoplanetary world per week, which provides a boost to scientifically addressing one of the major questions of modern science "Are we alone in the Universe?". Ninety-nine percent of known exoplanets were found around the low- and mid-mass F, G, K and M dwarf stars in our Galaxy, the major subject of Part 1 through 4 of this book.

A new era of our understanding of exoplanetary systems and their environments started 26 years ago with the observational detection of the first exoplanet orbiting a typical Sun-like (G2IV, 6 Gyr old) main sequence star, 51 Pegasi by Swiss astronomers Michel Mayor and Didier Queloz (Mayor and Queloz, 1995). Unlike Jupiter in the Solar System orbiting the Sun at 5 AU, 51 Peg b is half mass of Jupiter but 100 times closer from its Sun-like star. Because the conventional planet formation theories could not explain how such a massive planet could form so close to its parent star, such a giant planet was not unexpected to exist. The discovery was made by applying the ground based radial velocity (RV) method by measuring the periodic Doppler shifts of stellar spectra at about 70 m/s caused by a "wobbling" of planet hosting stars due to the gravitational tug of its planetary component. Since 1995, this technique has been commonly used to characterize over 5000 confirmed exoplanets with masses comparable or greater than the Earth's mass. Another observational methodology, the transit method, was used by Charbonneau et al. (2000) to detect the first exoplanet around a Sun-like star, HD 209458. Unlike the RV method, the transit method that measures the starlight dimming caused by the passage of the planet across the stellar disk and provides information about the planet's size. The RV method complemented with transit observations provide a reliable way to derive the planet's orbital inclination, and thus its true mass, radii (and thus average density), surface gravity along with the orbital distances and eccentricities. This information has shaped our understanding of the formation and architectures of other worlds with the launch of the Kepler Space Telescope in 2009 (Borucki et al., 2010) and discoveries of 1972 exoplanetary systems containing over 3700 exoplanets. The success of this mission was complemented by the observations of the Transiting Exoplanets Survey Satellite (TESS) that discovered over 1200 exoplanets along with the Hubble Space Telescope and ground-based observations with sizes smaller than Earth to larger than Jupiter, transiting stars of a wide variety of spectral classes, from F to M dwarfs, with about 50 of them orbiting subgiant and giant stars. The current exoplanetary demographics suggest that about 1/3 of discovered exoplanets have Earth size, while ~1400 rocky exoplanets larger than Earth and smaller than mini-Neptunes (known as super-Earths) with the sizes between 1.25 and 2 $R_E$ have no analogs in our Solar system (https://exoplanets.nasa.gov/; Fulton and Petigura, 2018).

The first statistical sample of small ($R \leq 4\ R_\oplus$), short period ($P < 100$ days), exoplanets with measured radii provided important data for population studies. Spectroscopic measurements of the planet hosting stars, along with precision distances from the European Space Agency *Gaia* mission, allowed the radii of ~1000 of these planets to be constrained to a precision of 5% (Fulton and Petigura, 2018). Statistical analysis of the resulting high-precision exoplanet sample revealed a bimodal radius distribution, with a population of rocky super-Earths and a population of gaseous sub-Neptunes separated by a gap in radius spanning approximately 1.5–2 $R_\oplus$. These data revealed the radius gap referred to as the Fulton gap (see Fig. 127) that is indicative of the formation and evolution history of planets. The shape of the distribution can be attributed to exoplanet atmosphere evolution under the impact of host star irradiation. Two important parameters that sculpt the distribution are host star mass and orbital separation. Both parameters are related to the X-ray (1–100 Å) and Extreme UV (EUV, 100–1240 Å) irradiation history of the planet and are supported by the observed trend of a shift in the bimodal distribution toward smaller planets as host star mass decreases (and stellar activity increases). This trend, the slope, and the width of the observed gap are consistent with photoevaporation of atmospheres driven by host star irradiation as the dominant factor determining the radius distribution of small exoplanets. The observed radius trends are also in line with theoretical predictions of photoevaporative atmosphere loss (Lopez and Fortney, 2013, Owen et al., 2020).

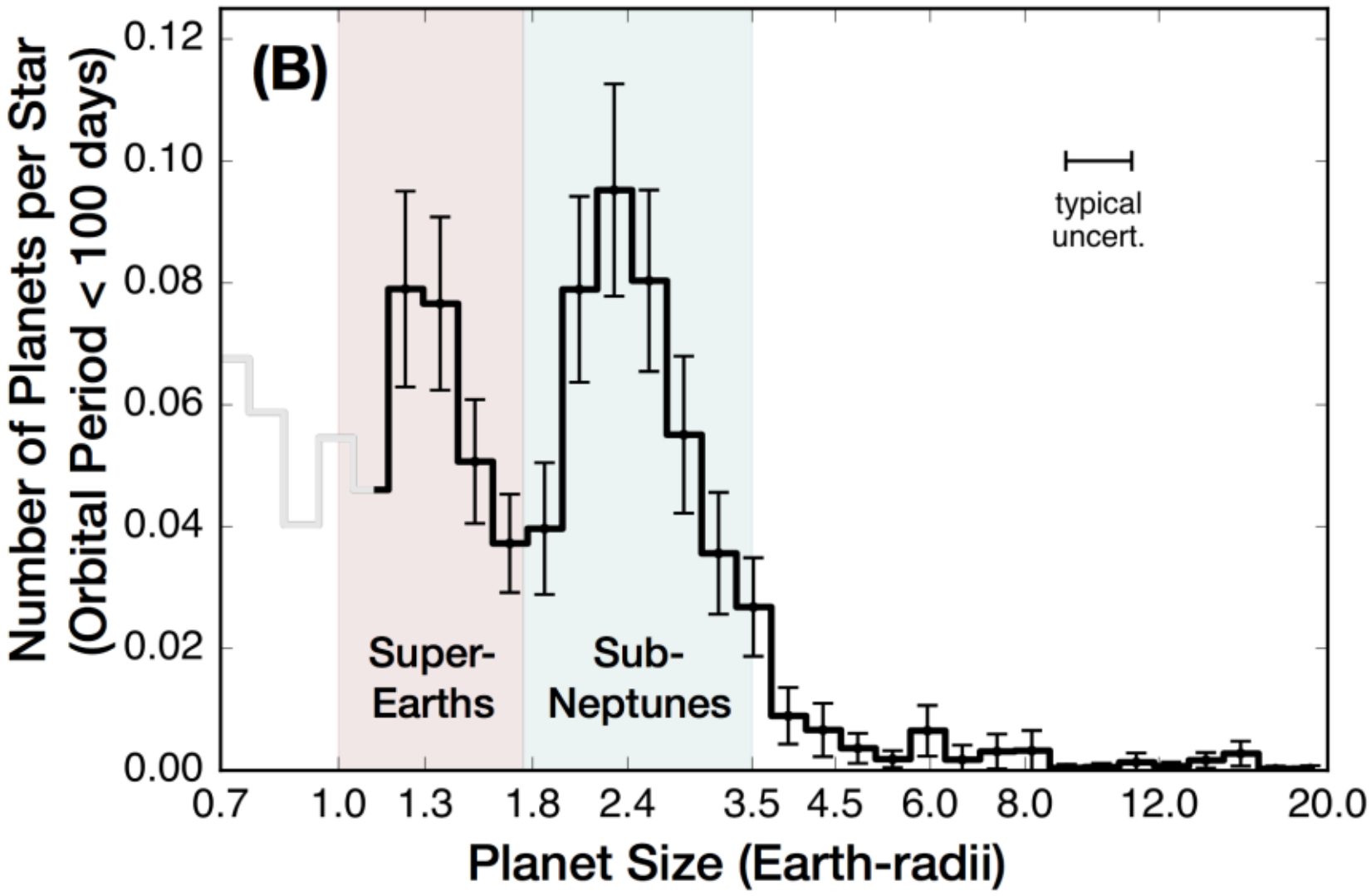

Fig. 127. Statistics of the planet size versus the number of planets per star with an orbital period of < 100 days reveals the planet radius gap of planets with radii between 1.5 and 2 of the Earth's radius. These observed trends are consistent with irradiation driven photoevaporative mass-loss being the dominant factor in determining the radius distribution of small, close-in planets. The figure is adapted from Fulton and Petigura (2018)

Kepler observations suggest that that low-mass exoplanets with the mass 0.3–30 $M_E$ are more common than more massive, Jupiter-like exoplanets. In average, between 37 and 60 percent of Sun-like stars host a terrestrial type (rocky) exoplanet in their habitable zones (Bryson et al., 2020). Because M-type dwarfs comprise over 75% of stellar population in the Galaxy, most of the exoplanets are found around these low-mass stars. Specifically, early M dwarf stars are more likely to have smaller, terrestrial-sized planets than Sun-like, G-dwarf stars (Mulders et al., 2015). Additionally, they harbor a greater number of planets in their exoplanetary systems (Ballard and Johnson, 2016).

These discoveries have ignited a search for potentially habitable exoplanets located within the habitable zone (HZ), the region around a planet hosting star within which a planet can sustain a standing body of liquid water on its surface (Kasting et al., 1993). Currently, astrophysicists have discovered over 200 rocky exoplanets within HZs mostly around M dwarfs, because the detections of small exoplanets are favored around the low-mass (0.1–0.5 solar mass), and thus low-luminosity stars (as low as 2000 times lower than the Sun's luminosity) with the radius of the inner boundary of the HZ around them varying between 0.04–0.16 times of the solar HZ. Specifically, the shorter orbital period and larger planet-to-star size ratio of a planet within the HZ of an M dwarf allow for easier transit detections.

For example, two closest M dwarfs to the Sun, Proxima Centauri has one Earth-like exoplanet (Proxima b) within a habitable zone, and the TRAPPIST-1 exoplanetary system contains seven low mass (rocky) exoplanets with three of them within the (conservative) habitable zone (Gillon et al., 2017).

As we learned from Part 1–3, M dwarfs are magnetically active stars that retain their activity for billions of years. Thus, the proximity of HZs to these planet hosts suggests that exoplanets should receive up to 2000 times greater fluxes of quiescent (steady state chromospheric, transition region and coronal) form of ionizing radiation than that received by Earth from the Sun (Wilson et al., 2021). This begs the important question: Can these fluxes be detrimental for the habitability of exoplanetary environments?

To assess the factors of habitability of rocky planets, we should thus gain the understanding of the detailed interaction between stars and exoplanets over geological timescales, the dynamical evolution of planetary systems, and atmospheric and internal dynamics. While this is a complex and multi-faceted problem, researchers made great strides in understanding the role of high-energy radiation, stellar winds, coronal mass ejections (CME) and stellar energetic particle (StPE) environments around planet hosting stars (of G, K and M dwarfs) on atmospheric chemistry, the rate of atmospheric escape from nitrogen rich secondary atmospheres and the impact on exoplanetary climates (Cohen et al., 2014; Airapetian et al., 2016, 2017a; Garcia-Sage et al., 2017; Dong et al., 2018). The outputs of magnetic activity specified as the perturbations traveling from stars to planets in the form of flares, winds, CMEs and StEP events are referred to as astrospheric space weather (see a comprehensive review in Airapetian et al., 2020).

The question of impact of stars on exoplanets requires an interdisciplinary approach, because to address it, we must start with the host star itself to determine its outputs and effects on the exoplanet environment, all the way from its magnetosphere to its surface. To understand whether an exoplanet is habitable at its surface, not only do we need to understand the changes in the chemistry of its atmosphere due to the penetration of energetic particles and their interaction with constituent molecules but also the loss of neutral and ionic species, and the addition of molecules due to outgassing from volcanic and tectonic activity. These effects will produce a net gain or loss to the surface pressure, and this will affect the surface temperature, as well as a net change in the molecular chemistry. Therefore, due to the complexity of the problem, we should study a set of interlinked research questions, all of which contribute pieces to the answer, with contributions from various disciplines involved in each topic.

With growing numbers of NASA (National Aeronautics and Space Administration) and ESA (European Space Agency) exoplanetary missions such as the Transiting Exoplanet Survey Satellite (TESS), the upcoming James Webb Space Telescope (JWST), the Characterizing ExOPlanet Satellite (CHEOPS), the PLAnetary Transits and Oscillations of stars (PLATO), the Atmospheric Remote-sensing Infrared Exoplanet Large-Survey (ARIEL) missions, in the relatively near term we will be better equipped with high quality observations to move from the phase of exoplanetary discovery to that of physical and chemical characterization of exoplanets suitable for life.

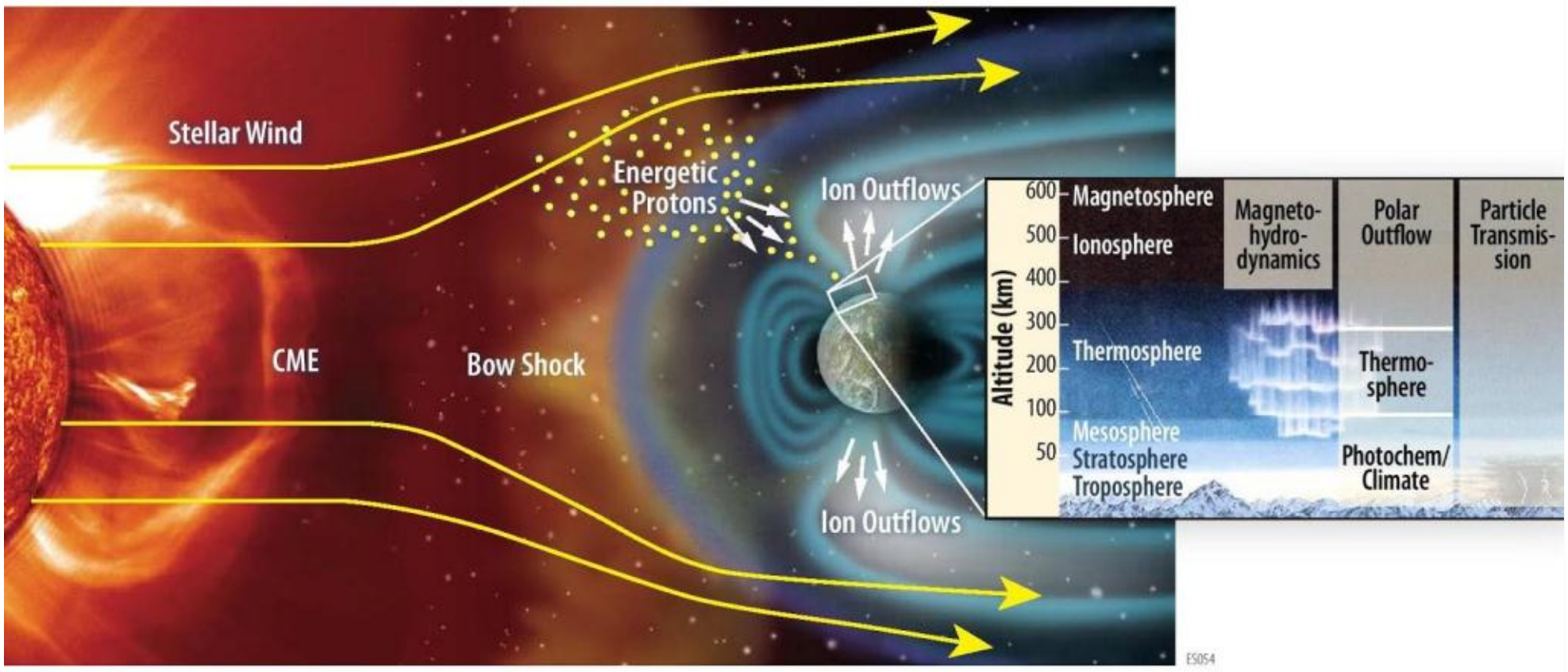

Fig. 128. Schematic view of the complex exoplanetary SW system that incorporates the physical processes driving stellar activity and associated SW including stellar flares, CMEs and their interactions with an exoplanetary atmosphere driven by its internal dynamics. While stellar winds and CMEs affect the shape of an exoplanetary magnetosphere, XUV and energetic particles accelerated on CME-driven shocks enter the atmosphere. The combined effects of XUV, stellar winds and CMEs drive outflows from the exoplanetary atmosphere. These processes are controlling factors of exoplanetary climate and habitability (Airapetian et al., 2020)

This Part describes various aspects of star-planet interactions in a global exoplanetary system environment with a systematic, integrated approach using theoretical, observational and laboratory methods combining tools and methodologies of four science disciplines: astrophysics, heliophysics, planetary and Earth science. The various forms of stellar magnetic activity and associated space weather processes in the form of stellar coronal X-ray and EUV emission, stellar winds, stellar flares, coronal mass ejections, stellar energetic particles collectively known as astrospheric space weather on a magnetized exoplanet are presented in Fig. 128.

## 5.2. Low- and Mid-Mass Stars as Planet Hosts

### 5.2.1. X-ray and EUV fluxes

With the advent of Kepler and TESS missions, researchers obtained unprecedented access to long-term data of active planet hosts that provided the first glance on variability of low and mid mass planet hosts. This variability includes short-term enhancements of stellar fluxes in the optical and X-ray and EUV bands on a scale of minutes to hours superimposed on rotational modulation caused by active regions lasting for tens of days to months (see Parts 1–2 of this book; Maehara et al., 2012; Notsu et al., 2019; Namekata et al., 2019; Herbst et al., 2021; Toriumi and Airapetian, 2022; Toriumi et al., 2022; Johnstone et al., 2021). Thus, to address the question of impacts of these factors on habitability, we need to characterize the contribution of these sources including the flaring and starspot activity to the overall flux of ionizing radiation (Yamashiki et al., 2019; Toriumi and Airapetian, 2022; Bogner et al., 2022).

The interaction of XUV flare emission, associated coronal mass ejections and stellar energetic particles causes heating, dissociation and ionization of atmospheric species of rocky exoplanets, and thus may affect their evolution (via the interplay between the outgassing and escape processes) and chemistry. This, in turn, can modify physico-chemical conditions required for the origin, development and support of life as we know it within the so-called "Biogenic Zones" (Airapetian et al., 2016) or "Abiogenesis Zones" (Rimmer et al., 2018), which represent a modification of the HZ defined earlier with the account of conditions for the thick atmosphere and formation of biologically relevant molecules required for the origin of life.

While the knowledge of stellar XUV fluxes is critically important for the exoplanetary field, it is nearly impossible to detect the emission from the full EUV band (10–92 nm), even for closest stars to the Sun, such as α Centauri. This is because the interstellar medium (ISM) absorption by hydrogen, helium and other resonance lines of many other important metal lines species contribute to the absorption of the longward EUV emission (>36 nm), while the short XUV bands (<36 nm) is relatively unaffected, and thus can be detected from close by stars by the Extreme Ultraviolet Explorer (EUVE), ROentgen SATellite (ROSAT) and Chandra X-ray Observatory (e.g., Ribas et al., 2005; Johnstone et al., 2021). Thus, estimation of these elusive stellar EUV fluxes from stellar observations is an important task.

The stellar XUV emission can be reconstructed using empirical and theoretical approaches. Empirical reconstructions have already provided valuable insights on the level of ionizing radiation from F, K, G and M dwarfs (Cuntz and Guinan, 2016; France et al., 2016, 2018; Loyd et al., 2016; Youngblood et al., 2017). The latter approach is based on the analysis of HST-STIS and COS based stellar far-ultraviolet (FUV) observations as proxies for reconstructing the EUV flux from cool stars, either through the use of solar scaling relations (Linsky et al., 2014; Youngblood et al., 2016) or more detailed differential emission measure techniques (e.g., Louden et al., 2017). Using these datasets, the authors found that the exoplanet host stars, on average, display factors of 5–10 lower UV activity levels compared with the non-planet-hosting sample. The data also suggest that the UV activity-rotation relation in the full F–M star sample is characterized by a power-law decline (with index $\alpha \approx -1.1$), starting at rotation periods 3.5 days. France et al. (2018) used N V or Si IV spectra and knowledge of the star's bolometric flux to estimate the intrinsic stellar EUV irradiance in the 90–360Å band with an accuracy of roughly a factor of $\approx 2$. The data suggest that many active

K, G and most of “quiet” M dwarfs generate high XUV fluxes from their magnetically driven chromospheres, transition regions and coronae. Another method developed by Lecavelier Des Etangs (2007) derives the emission from the whole EUV band from the ratio of observable solar and stellar short-EUV fluxes (<36 nm) and multiplying it by the solar EUV spectrum/scaling relation (also used in Johnstone et al., 2021). These early approaches could be valid for inactive Sun-like stars, but they would not hold for active Sun-like stars and M-dwarfs that have much intense and hotter coronae than the Sun. Another approach to reconstruct the XUV flux from TRAPPIST-1 is based on the semi-empirical non-LTE modeling of stellar spectra using radiative transfer codes constrained by HST Ly-α and GALEX FUV and NUV (Peacock et al., 2019).

The recently developed “Sun-as-a-Star” empirical model provides another viable way to characterize X-ray and EUV emission from coronae of solar-type stars (Toriumi and Airapetian 2022; Toriumi et al., 2022). By analyzing 10 years of multiwavelength synoptic observations of the Sun, they found that the solar irradiance and the total unsigned magnetic flux show power-law relations with an exponent decreasing from above unity to below as the temperature decreases from the corona to the chromosphere. Namekata et al. (2023a) applied the scaling relations to active young Sun-like stars (G-dwarfs), EK Dra (G1.5V), $\pi^1$ Uma (G1.5V) and $\kappa^1$ Cet (G5V) and found that the observed spectra (except for the unobservable longward EUV wavelength) are roughly consistent with the extension of the derived power-law relations. This suggests that this empirical model is a reliable new method to derive the XUV/FUV fluxes of Sun-like stars including at the unobservable EUV band longward of 36 nm. Is this method potentially applicable for reconstruction of the XUV fluxes from M dwarfs? The recent X-ray-magnetic flux scaling relations for 292 M dwarfs (Reiners et al., 2022) suggest that the power law index is much steeper than that of the solar-like stars (1.58 vs 1.15) possible suggesting the different heating conditions of highly stratified and magnetized coronal environments of these stars.

## 5.2.2. Stellar Winds

Stellar winds represent an extension of global stellar corona into the interplanetary space and are fundamental property of F–M dwarf stars. Stellar coronal winds are weak and no reliable detection of a wind from another star other than the Sun was reported. Stellar winds have a significant impact on the stellar evolution, on surrounding planetary systems, and on the evolution of gas and dust in galaxies (see Puls et al., 2008). For example, the mass loss from the young Sun (< 0.6 Gyr old) was an important factor in the early evolution of atmospheres of Venus, Earth and Mars (e.g., Airapetian et al., 2020). Understanding these impacts requires the physical mechanisms that drives winds and their evolution with time (Vidotto, 2021), which are based on observational constraints on the winds of solar-like stars of various ages to constrain the solar wind evolution (Suzuki et al., 2013; Airapetian and Usmanov, 2016; Réville et al., 2016; Shoda et al., 2020; Airapetian et al., 2021).

The empirical method of detection of stellar wind relies on the high-resolution observations of HI Ly-α absorption from the stellar astrosphere forming due to the wind interaction with the surrounding interstellar medium (Wood, 2018). The mass loss rates from young G and K dwarfs are well correlated with X-ray coronal flux as $\dot{M} \propto F_X^{1.29}$ reaching the maximum rates of 100 times of the current Sun’s rate. However, this relation fails for stars with the greater X-ray flux suggesting the saturation of surface magnetic flux. Another method is based on fitting rotational evolution models to observational constraints (Matt et al., 2015; Johnstone et al., 2015). This method suffers from unconstrained wind model parameters and lacks the ways to test them observationally. Another way to deduce mass loss rates from cool stars is via observations of radio free-free emissions due to the presence of stellar winds in solar-like stars (Villadsen et al., 2014; Fichtinger et al., 2017). So far, only upper limits on the wind mass loss rates of a handful of stars have been derived. Also, X-ray emission due to charge exchange between ionized stellar winds and the neutral interstellar hydrogen have been also used to provide upper limits on the mass loss rate due to stellar winds (Wargelin and Drake, 2002).

Data-constrained MHD simulations represent another viable way to reconstruct XUV and wind mass fluxes from solar-type stars. This approach suggests the availability of stellar photospheric magnetic maps inputs and the knowledge of energy fluxes in Alfvén waves. The constrains on these inputs can be derived from coordinated multi-wavelength multi-observatory efforts based on HST, XMM-Newton, Chandra, NICER, and ground-based photometric, spectropolarimetric observations provide crucial information for the characterization of input magnetic fields and energy fluxes from G–M dwarfs required for the theoretical (magnetohydrodynamic, MHD) modeling of their atmospheres. Stellar surface magnetic fields are derived from spectropolarimetric observations of G, K, and M dwarf stars. They provide information about large scale structures of the surface magnetic field and revealed field strengths much greater than that observed on the Sun (e.g., Rosén et al., 2016; See et al., 2019). The photospheric magnetic field strengths span the range between 10 G to several 100 G in young Sun-like stars, up to a few 1000 G for M dwarfs (Kochukhov et al., 2020). The reconstruction of surface magnetic fields of these stars also shows a great diversity of geometry of large-scale stellar magnetic fields varying from poloidal to toroidal configurations (Vidotto et al., 2014b). Recent observations revealed that the geometry of global magnetic fields and associated Poynting fluxes may vary on time scales of a few days to years such as BE Cet, HN Peg, $\epsilon$ Eri, $\kappa^1$ Cet, EK Dra, AU Mic, YZ CMi (Boro Saikia et al., 2015; 2018; Rosén et al., 2016; Waite et al., 2017). They also suggest that the unsigned magnetic field accounts for over 90% of the magnetic flux and is mostly concentrated in stellar starspots associated with stellar active regions (Kochukhov et al., 2020; Toriumi et al., 2020).

Recently, Alvarado-Gómez et al. (2016, 2018), do Nascimento et al. (2016), Boro Saikia et al. (2020) used Alfvén Wave Solar Model (AWSoM) constrained with ZDI magnetograms of active stars as boundary for realistic three-dimensional magnetohydrodynamic models of the coronae and winds of active solar-like stars. Airapetian et al. (2021) have extended these models by using the fully resolved profiles of chromospheric lines derived from the Hubble Space Telescope observations to constrain the energy flux of Alfvén waves at the upper chromosphere that propagate upward and dissipate energy in the upper layers forming the stellar corona. Figure 129 shows that the converged (steady) solution for the global magnetic coronal structure of $\kappa^1$ Cet at 2012.8 epoch, which is mostly represented by a dominant dipolar field tilted at 90 (dipole strength of 15.38 G vs 5.1 G and 7.99 G for quadrupole and octupole respectively at 2013.7) and resembles the current Sun's coronal state at minimum of the solar cycle. The right panel demonstrates the global magnetic field eleven months later suggesting that it has undergone a dramatic transition from a simple dipole to $45^0$ tilted dipolar magnetic field. The global wind shows the two-component stellar wind of $\kappa^1$ Cet. At 2012.8 epoch, the fast wind reaches its terminal velocity of 1152 km/s within the first 15 $R_{\rm star}$, while the slow and dense components of the wind originate from the regions associated with the equatorial streamer belt structures at 696 km/s. The simulated stellar wind density is ~ 50 times greater than that of the current Sun's wind and faster by a factor of 2, which produces a massive wind with the mass loss rates, which are 100 times greater than that observed from the current Sun.

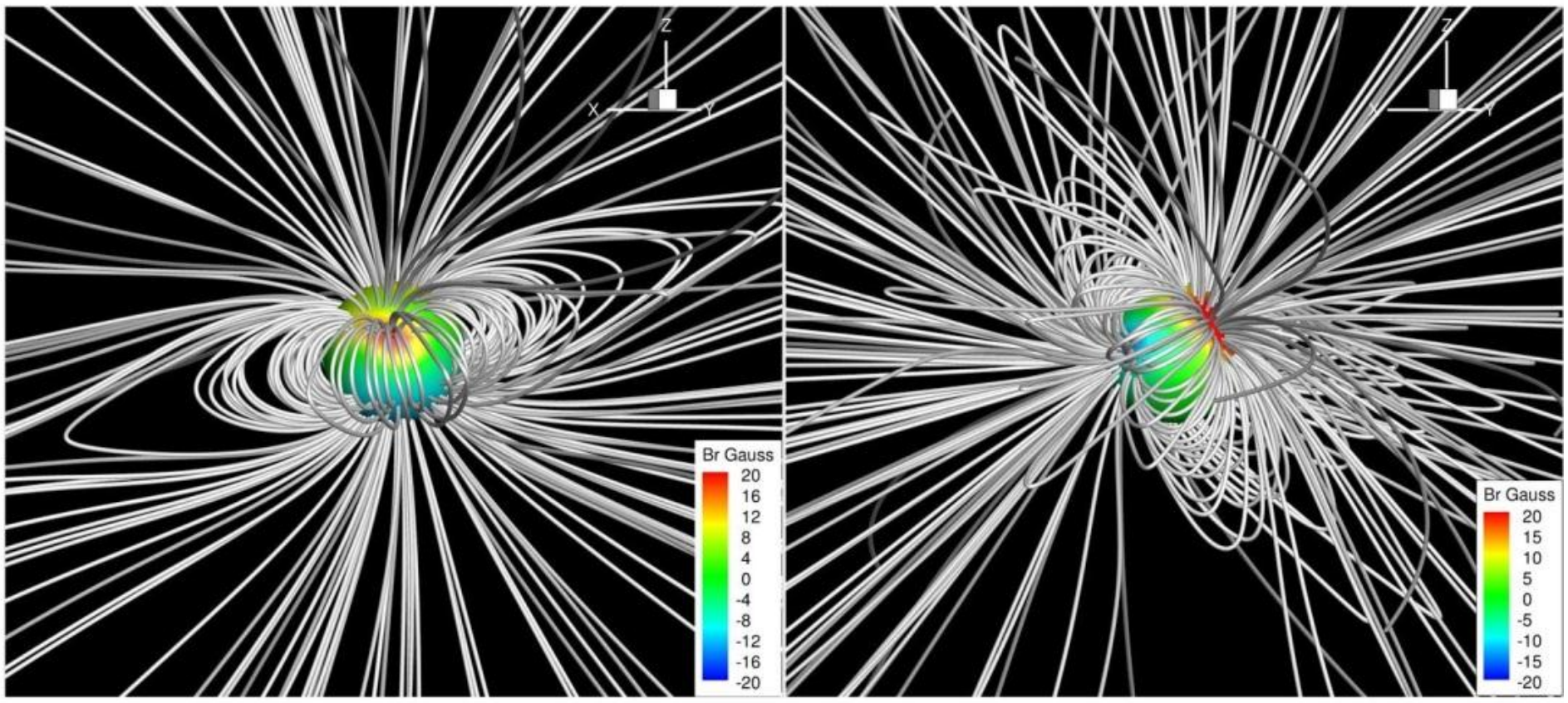

Fig. 129. The evolution of the 0.7 Gyr Sun proxy's global magnetic field over the course of 11 months, at 2012.9 (left) and 2013.8 (right) (from Airapetian et al., 2021)

The winds from M dwarfs are of particular interest due in part to the abundance of such stars in the galaxy. Also, the habitable zones of M dwarfs are much closer to the planet hosts than for earlier type stars like the Sun, and this makes their exoplanets in such locations to be potentially more vulnerable to much greater particle fluxes from stellar winds. Several numerical MHD models have been developed to study the properties of winds from M dwarfs to explore the interplanetary environment and impact on exoplanets. The estimated mass loss rates from these stars are comparable or smaller than that of the current Sun. For example, the mass loss rate from TRAPPIST1 is $4.1 \cdot 10^{-15}\, M_{\odot}$/yr (Dong et al., 2018), while the simulated mass loss rate from EV Lac is $3 \cdot 10^{-14}\, M_{\odot}$/yr (Cohen et al., 2014).

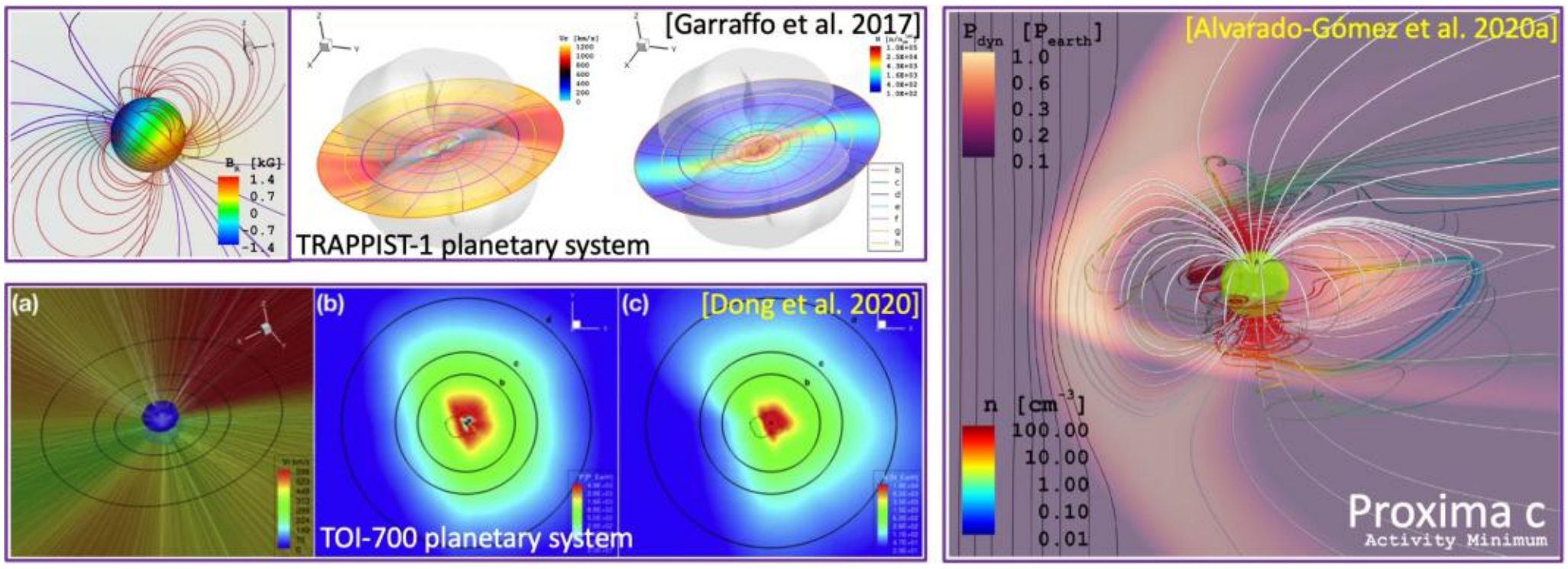

Fig. 130. Modeling stellar winds and star–planet interactions: TRAPPIST-1 system (upper left; Garraffo et al., 2017), TOI-700 system (lower left; Dong et al., 2020), and for Proxima c (right; Alvarado-Gómez et al., 2020a) (from Lynch et al., 2022)

The impact of these winds on planetary environments around M dwarfs have been presented in a number of studies (e.g., Vidotto et al., 2013; Garraffo et al., 2016; Dong et al., 2018; Alvarado-Gómez et al., 2019b, 2020a). Figure 130 shows three examples of star-to-planet MHD modeling to explore the impact of stellar winds on exoplanets in the TRAPPIST-19 (upper left; Garraffo et al., 2017; Dong et al., 2018), TOI-700 (lower left; Dong et al., 2020), and Proxima Centari (right; Alvarado-Gómez et al., 2020a) systems.

## 5.2.3. Coronal Mass Ejections: Observational Data

The multi-wavelength observations suggest that energetic (> X5.5 class flares with the bolometric radiated energy of ~ $10^{32}$ erg) solar flares are usually accompanied by the ejection of coronal magnetized clouds or CMEs (Gopalswamy et al., 2005). CMEs represent the most powerful solar eruptions in the Solar System and are produced in the corona due to magnetic reconnection process, expelling fast (up to 3000 km/s) and massive ($10^{16}$ g) cloud of charged particles and a magnetized cloud (Cane and Richardson, 2003). They represent one of the major hazards of space weather on Earth's environments and infrastructure (Schrijver et al., 2015).

As CMEs propagate out from the solar corona into interplanetary space, they drive shocks. These shocks produce energetic protons and heavier ions at their fronts with energy over 1 GeV via the diffusive shock acceleration also known as the first-order Fermi acceleration (Zank et al., 2000). This process produces gradual SEP events that can last for a few days affecting the chemistry of the middle atmosphere of Earth.

Daily occurrence rates of CMEs vary with the phase of the solar cycle (or sunspot numbers) with ~0.1 event per day during solar minimum and rising to ~6 events per day during solar maximum (Gopalswamy et al., 2022). Solar data suggest that CME-flare association frequency is nearly 100% for flares with energy greater than $7.5 \cdot 10^{31}$ erg (X5 class) and reduces to less than 20% for flare energy of $3 \cdot 10^{30}$ erg (C3 class) (Yashiro et al., 2004).

Also, filament and prominence ejections signatures are related to the formation and ejections of magnetic flux ropes in the solar and stellar corona; these phenomena were widely researched in solar studies. Solar active region filament eruptions have a high flare association rate of 95% and a comparable CME association rate (Jing et al., 2004). Moreover, Seki et al. (2021) reported another empirical condition suggesting that if the product of the maximum radial velocity and the filament length is greater then $8 \cdot 10^6$ km$^2$/s, then the probability of the filament to become a CME is 93%.

The first detection of stellar prominence eruption was reported by Houdebine et al. (1990). They observed a very broad blue wing enhancement in the $H_\gamma$ emission line and a weaker blueshifted component of $H_\delta$ with the projected velocity of 5800 km/s in the young dMe star AD Leo associated with the onset of a large flare. Other two studies detected weak line asymmetries on M dwarfs with the inferred velocities mostly below the escape threshold of the observed stars (Vida et al., 2019; Muheki et al., 2020). Recent observational studies concluded stellar superflares that are common on young G, K and M stars, but show less than 40 associated CME events. They have been detected using various methods including the signatures of Doppler shifts, X-ray absorption, or coronal dimming in extreme ultraviolet and X-ray (e.g., Argiroffi et al., 2019; Moschou et al., 2019; Veronig et al., 2021; Namekata, 2021).

Significant progress has been made in the characterization and interpretation of stellar CME signatures by utilizing the results of multi-wavelength Sun-as-a-star analyses of large

CME events (e.g., Leitzinger and Odert, 2022; Leitzinger et al., 2022; Namekata et al., 2022a,b; Xu et al., 2022).

Recent spectroscopic observations of a superflare with the energy of $10^{33}$ erg on young (~100 Myr old) solar analog, EK Dra, with TESS and SEIMEI provided the first signatures of a giant stellar filament eruption observed in the blueshifted Hα absorption line at 510 km/s CMEs (Namekata et al. 2022a). The latest observational campaign of EK Dra conducted in April of 2022 with NICER X-ray observatory, TESS and SEIMEI discovered two powerful prominence eruptions signified by the blueshifted $H_\alpha$ emission line components at 430 km/s and 690 km/s that were associated with powerful superflares with the white-light bolometric energies of $1.5 \cdot 10^{33}$ erg and $1.22 \cdot 10^{34}$ erg, respectively (Namekata et al., 2023b). Such events are often occurred on the Sun with filament and prominence eruptions associated with the cores of CME that propagate out of the solar corona forming interplanetary CMEs, the sources of space weather (Cliver et al., 2022).

Observational campaigns in radio-optical bands present one case of a type-IV radio burst (Zic et al., 2020) and no type-II radio bursts (Crosley and Osten, 2018a, b; Villadsen and Hallinan, 2019). Coronal dimming is another promising CME detection technique to search for stellar CMEs. Full-disk Solar Dynamics Observatory/EVE observations show dimming of EUV coronal lines like Fe IX 171A that are associated with CMEs (Mason et al., 2016; see also Harra et al., 2016; Jin et al., 2022). A few possible post-flare coronal dimmings that have been detected on K and M dwarfs were suspected as possible CME events (Veronig et al., 2021; Loyd et al., 2022). However, MHD simulations suggest that confined eruptions or failed CME events can also show dimming profiles at coronal temperatures (Alvarado-Gómez et al., 2019a; Jin, 2022).

Future campaign observations of Doppler shifts together with the coronal dimming and possibly low-frequency radio observations should be taken for reliable detections of elusive stellar CMEs.

## 5.2.4. Coronal Mass Ejections: Models

If we follow the statistical correlations between the solar flare-CME association rate as the function of the flare energy, then the frequency of CMEs should be comparable to the frequency of accompanying superflares on active stars. Extrapolation from solar to stellar regimes for active stars suggests that such stars should have winds hundreds or thousands of times stronger than the solar wind simply due to CMEs alone (e.g., Drake et al., 2013; Odert et al., 2017). Thus, this will have major implications on the stellar mass-loss and angular momentum loss rates (Drake et al., 2013; Odert et al., 2017; Wood et al., 2021) as well as on exoplanet habitability (Khodachenko et al., 2007; Airapetian et al., 2020). The relatively low frequency of CME detections from cool stars is hard to understand from this perspective.

One clue to resolve this problem comes from the fact that some powerful solar flares (X-class events) forming in large active regions are not accompanied by CMEs (Thalmann et al., 2015). The lack of CMEs probably linked to the existence of strong overlying magnetic fields within such active regions confining the ejections, and thus producing failed CMEs. Because magnetic flux associated with large starspots (over 10 times larger than the largest sunspot of 1947) on young solar analogs is 1-2 orders of magnitude greater than that observed on the Sun, this factor can play a dominant role in producing CMEs at lower rates than expected from the solar-CME association rate.

Using a global 3D MHD model, Alvarado-Gomez et al. (2018) concluded that a global stellar dipole field of 75 Gauss is capable of fully suppressing CMEs with energies up to $3 \cdot 10^{32}$ erg comparable to one of the largest solar X-class flares (Alvarado-Gomez et al., 2018).

Using a potential field source surface model, a new study (Sun et al., 2021) estimates the torus-stable zone (TSZ) above a bipolar stellar active region embedded in a global dipole field and finds that in the active cool star conditions, the TSZ can extend significantly higher compared to solar conditions. This effect is believed to be one of the reasons causing so far sparse detection of CMEs on active stars (Drake et al., 2016; Odert et al., 2017).

Also, it is known that large stellar starspots have relatively short lifetime compared to sunspots (Namekata et al., 2019). This is important to mention because active regions need time to accumulate non-potential magnetic energy to drive powerful superflares and CMEs. To clarify these questions, we need to develop observationally constrained models of energy storage and initiation of unstable flux rope driven eruptions in the lower stellar coronae that are hotter, denser and more magnetized and dynamically evolving at shorter time scales than the solar corona (Lynch et al., 2022). The first MHD models of these transient events in the corona of $k^1$ *Cet* suggest that stellar halo-type CMEs can produce low-frequency bursts rapidly drifting to frequencies < 10 MHz that cannot be detected on Earth (Lynch et al., 2019; 2022).

Lynch et al. (2016) have shown that even in bipolar streamer distributions, the overlying, restraining closed flux is removed in a breakout-like way through an opening into the solar wind, thus enabling the eruption of the low-lying energized flux. Therefore, the evolution and interaction between the low-lying energized and overlying restraining fields are extremely important aspects of modeling eruption processes in solar and stellar coronae to correctly estimate the CME ejecta properties and energetics. Figure 131 illustrates recent ARMS 3D simulation results by Lynch et al. (2019) of a massive halo-type (width of $360^0$) energetic CME eruption based on the observationally derived $k^1$ *Cet* magnetogram. The entire stellar streamer-belt visible in this figure is energized via radial field-preserving shearing flows and the eruption releases ~ $7 \cdot 10^{33}$ erg of magnetic free energy in ~10 hours. Magnetic reconnection during the stellar flare creates the twisted flux rope structure of the ejecta, and the ~2000 km/s eruption creates a CME-driven strongly magnetized shock.

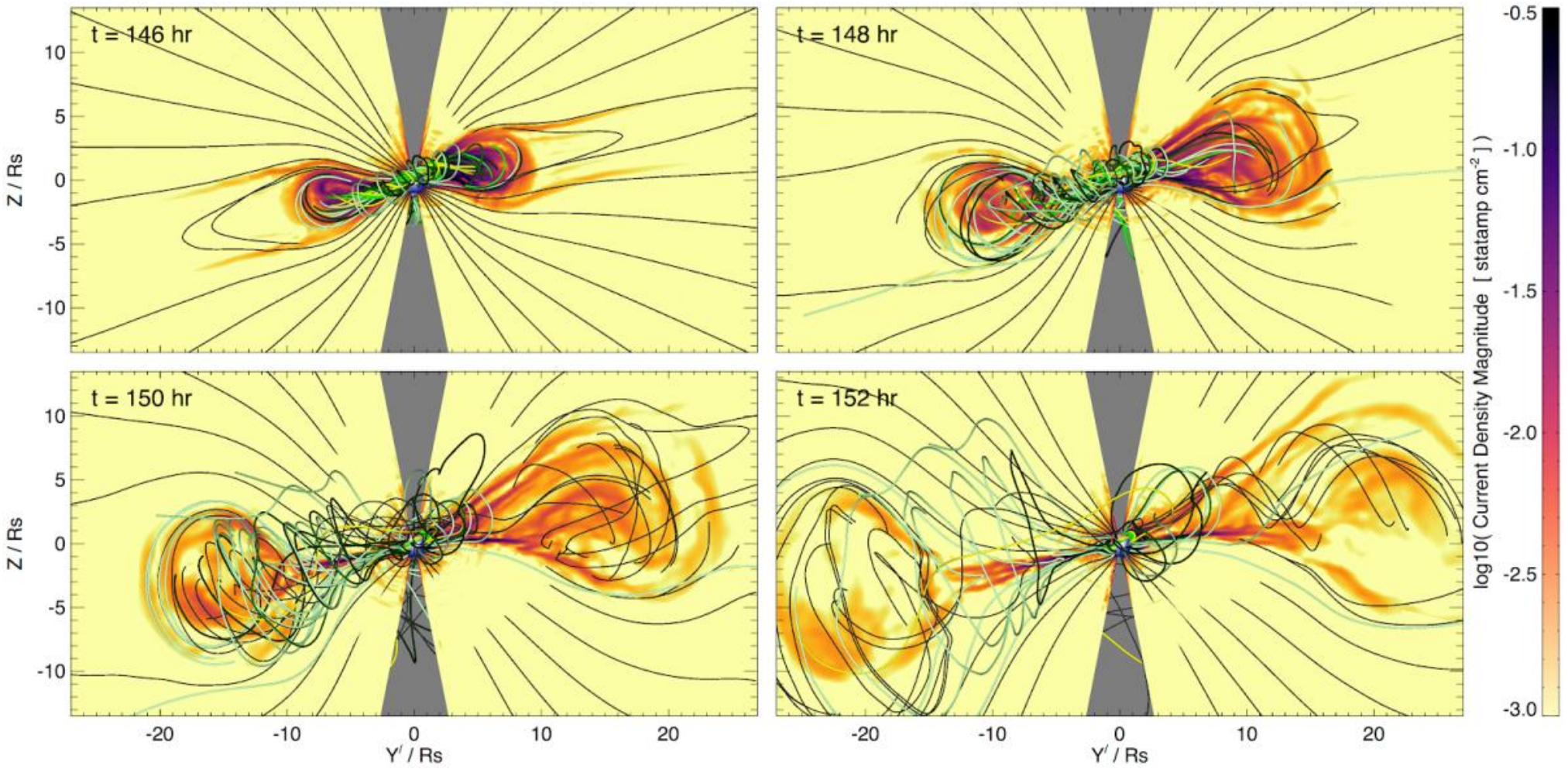

Fig. 131. ARMS 3D simulation of a massive (Carrington scale) stellar coronal eruption initiated from the $k^1$ *Cet* magnetogram. The contour planes show current density magnitude in the erupting flux rope (Lynch et al., 2019)

The maximum increase in total kinetic energy during the eruption was ~ $2.8 \cdot 10^{33}$ erg – on the order of the 1859 Carrington flare event. The meridional planes plot the logarithmic current density magnitude showing the circular cross-section the CME and representative magnetic field lines illustrate 3D flux rope structure.

When a stellar eruption has energy that is strong enough to escape the stellar magnetic field confinement, stellar CMEs will be driven by a similar physical process as solar CMEs but at much greater energy. For example, a recently modeled stellar CME eruption occurs under a stronger surface magnetic flux density (5 times the solar case) (Jin et al., 2022). The initial CME flux rope energy is about $10^{33}$ erg, which is an order of magnitude more energetic than the strongest eruptions observed on the Sun. Due to the strong magnetic confinement in the stellar corona, the resulting CME speed is ~ 3000 km/s, which is only slightly higher than a fast solar CME.

## 5.3. Impact of Stellar XUV Flux on Atmospheric Escape

Direct observations of atmospheric signatures of transiting exoplanets opened a new way to characterize the escaping atmospheres of giant exoplanets. Hydrogen-rich atmospheres of hot close-in giant planets (at distances from 0.02–0.05 AU) are exposed to the extremely high EUV fluxes which are two-three orders of magnitude greater than those experienced by the Earth's atmosphere during the solar maximum. EUV driven dissociation of molecular hydrogen produces the atomic hydrogen-rich upper atmospheric layers heated to temperatures up to 11000 K (e.g., Owen et al., 2020). At such high temperatures, the pressure gradient in the upper atmospheric regions can push the hydrogen gas to exceed the escape velocity of the planet, and thus to escape the planet's gravity into outer space.

The mass-loss processes can be identified by analyzing the stellar UV absorption lines as the planet transits in front of the star. To confirm the presence of atmospheric escape one needs to show that any escaping atmosphere extends beyond the Roche lobe radius of the planet. Thus, a transit signature that coincides in time with the passage of the planet in front of the star indicates an obscuring area that places it outside the planet's Roche lobe (so the atmosphere is no longer bound to the planet).

The Hubble Space Telescope Imaging Spectrograph (STIS) observations of a solar-like (G0V) star, HD 209458, discovered the first transiting hot Jupiter, HD 209458b, receives 10000 times more bolometric energy from the star than Jupiter, which justifies its name a hot Jupiter, and losses mass from its atmosphere. The absorption signatures of stellar Ly-$\alpha$ emission passed through the exoplanetary atmospheres allowed to measure the rates of atmospheric escape in several hot Jupiters exposed to extreme X-ray and EUV fluxes via photoionization and possibly Joule heating via stellar winds interaction with exoplanetary magnetospheres. The high spectral resolution of the Ly-$\alpha$ line in HD 209458 in and out of transit revealed the Doppler shift at the equivalent velocities of $\pm 100$ km/s (Vidal-Madjar et al., 2003). This line shift has been interpreted as the absorption by neutral atomic hydrogen-rich atmospheric region expanding into outer space at the velocity exceeding the planet's escape velocity. These observations suggested a mass loss rate of the cloud of atomic hydrogen at a rate of $10^{10}$g/s. Vidal-Madjar et al. (2004) and later one-dimensional hydrodynamic models of thermal winds heated by UV driven photoionization have been able to derive mass loss rates that are a few times $10^{10}$ g/s (Yelle, 2004) about 1% of its mass in 1 Gyr (Murray-Clay et al., 2009). Atmospheric escape of hydrogen and oxygen ions has been observed from Venus, Earth and Mars and is known to increase with the level of solar EUV fluxes during solar maximum (Airapetian et al., 2020).

The atmospheric escape from close-in exoplanets around active stars can be driven by thermal and/or nonthermal processes as shown in Fig. 132 (Gronoff et al., 2020). The thermal processes are represented by Jean escape and hydrodynamic escape, while nonthermal processes include ion escape, photochemical escape and ion pick up mechanism. Jean's escape occurs in thermal gas above the exobase, the region known as exosphere, where the mean free path is greater than the pressure scale height. The particles in the high energy tail of the thermal Maxwellian distribution with upward speeds greater than the planet's escape velocity, can escape the planet. This mechanism is efficient for lower mass particles such as hydrogen. However, it has been shown that if the UV energy deposition is high, the hotter exobase can expand to many planetary radii and thus reduce the effective escape velocity from the planet making this also important for higher mass particles (Johnstone et al., 2018, 2019; Airapetian et al., 2023). At much greater XUV radiative energy input from the star into the exoplanetary atmosphere, the heating rate of certain layers drives a thermal outflow in the form of transonic Parker winds (Parker 1958; Tian et al., 2008). This type of escape usually occurs in the hydrogen dominated atmospheres of close-in gas giants and such as HD 209458b and possibly super Earths (Schneiter et al., 2007).

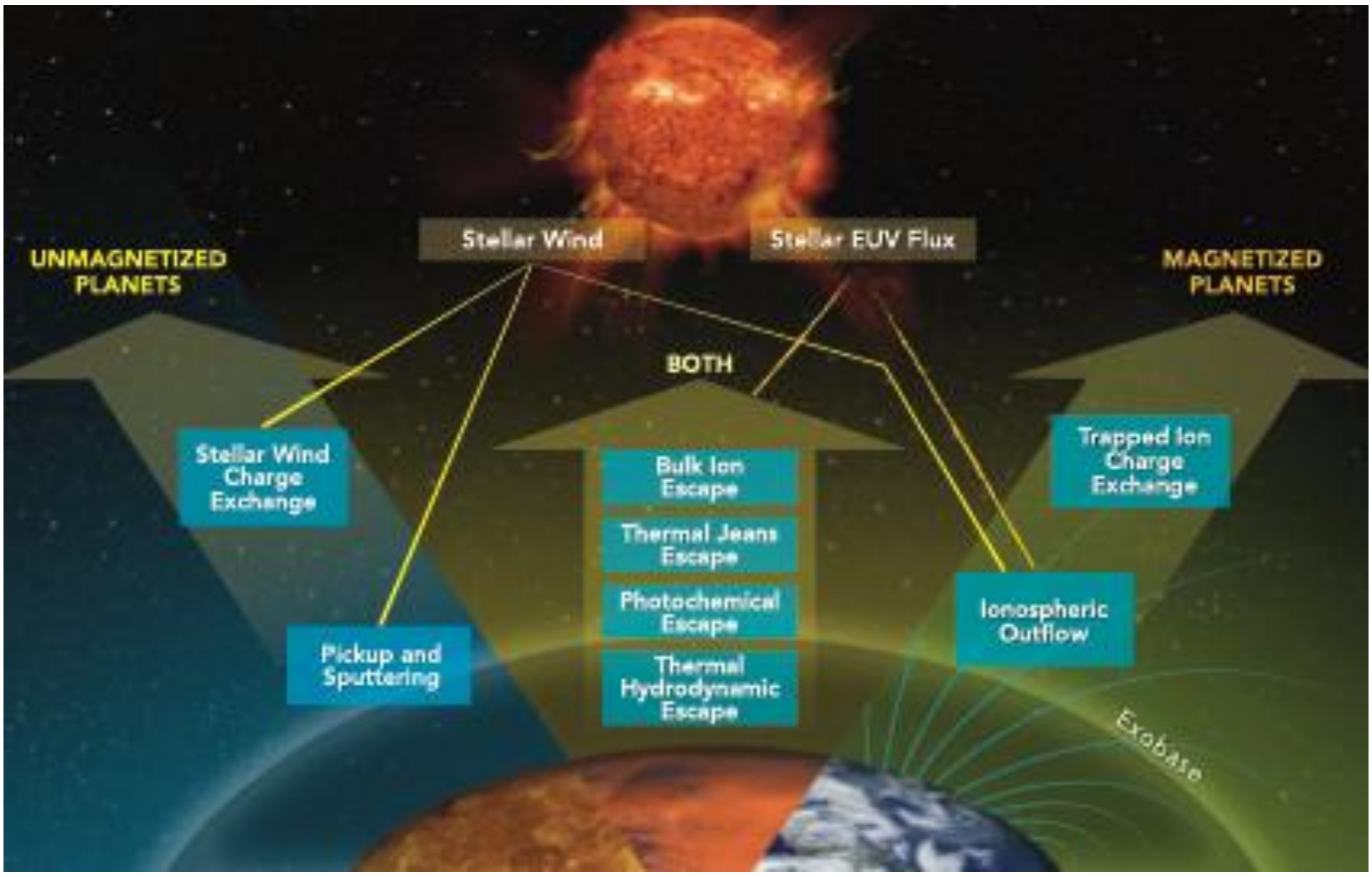

Fig. 132. Mechanisms of thermal and nonthermal atmospheric escape (from Gronoff et al., 2020)

Stellar wind can also affect the atmospheric escape rate. If the planet is unmagnetized, then the stellar wind pressure may suppress the exoplanetary wind, thus reduce the escape rate from an exoplanet (Shaikhislamov et al., 2016; Vidotto and Cleary, 2020). However, the situation can be opposite if the strong (fast and massive) stellar wind pushes on a planet's magnetosphere. If a giant exoplanet has a Jupiter-like magnetic moment, then the interaction of dense and fast stellar winds with close-in planetary magnetospheres can introduce electric currents that dissipate in their upper layers causing additional heating referred to as Joule heating that can exceed the UV photoionization driven heating (Cohen et al., 2014). In cases of relatively low atmospheric conductance, the stellar wind can then enhance the atmospheric escape rate.

Usually, the transition between Jean's escape from a static atmosphere to the dynamic atmosphere with an outflow or hydrodynamic escape is described by the Jeans escape flux $\propto (1+\lambda_j)e^{\lambda_j}$, where $\lambda_j$ is the Jeans escape parameter given defined as the ratio of the square of the ratio of the escape velocity to the thermal velocity of the gas particles or

$$\lambda_j = \frac{G M_p m}{k_B \, T_{ex} \, R_{ex}},$$

where *m* is the mean molecular mass of the atmosphere, $T_{ex}$ and $R_{ex}$ are the atmospheric temperature and the radius of the exobase, respectively. The atmosphere is typically hydrostatic at $\lambda_j > 30$, while it starts to escape for a molecular hydrogen dominated atmosphere at $\lambda_j \sim 2.5$ (Erkaev et al., 2015).

The enhanced XUV fluxes encountered by close-in exoplanets around active young G-K-M stars have significant consequences for atmospheric mass loss. The energy deposited by this flux through the absorption of XUV radiation can drive thermal, nonthermal and chemical escape processes forming ionosphere and thermosphere. The thermal escape process is driven by the temperature at the exobase, the atmospheric layer, where the particle mean free path is comparable to the pressure scale height. In this layer, high fast particles from the tail of Maxwellian distribution moving with outgoing velocity exceeding the plant's escape velocity will escape into space (Jean's escape). Its escape rate is controlled by the Jean's escape parameter, $\lambda_c$, represented by the ratio of gravitational energy to the mean particle's thermal energy. Tian et al., (2008) has applied an 1D hydrodynamic thermosphere model to study the effects on high XUV fluxes (up to 20 times the current Sun's flux, $F_0$) from the young Sun on the early Earth. They found that at the XUV fluxes below $5.3F_0$, photoionization driven heating increases the exobase temperature to ~ 8000 K moving it to 7700 km with the bulk velocity of > 10 m/s. However, above this critical XUV flux, very high Jeans escape at the exobase drives the upward flow producing adiabatic cooling, and thus reducing the exobase temperature. Lichtenegger et al. (2010) expanded Tian's model to calculate the atmospheric escape due to ionization and solar wind pick to be on the order of $10^7$g/s, thus removing the Earth's 1-bar atmosphere in 10 Myr. Recent hydrodynamic outflow model developed for an Earth-like planet ($N_2 – O_2$ dominated atmosphere) suggests that at XUV fluxes 60 times the current Sun's flux, the neutral atmosphere undergoes extreme hydrodynamic escape at the mass loss rate of $1.8 \cdot 10^9$ g/s (Johnstone et al., 2019). This suggests that the atmospheres of Earth-like planets around active stars should be depleted at the time scale of millions of years. More studies with multi-dimensional fully thermodynamic atmospheric models are required to study the initiation of hydrodynamic escape in atmospheres with various levels of $CO_2$ and NO that serve as the major atmospheric coolants (Airapetian et al., 2017b).

In nonthermal escape processes, escaping particles acquire energy through energy from nonthermal sources. For example, this may occur when energetic ion $A^+$ collides with a clod neutral particle, M, through the charge exchange process $A^+$ (*energetic*) + M(*cold*) –> A(*energetic*) + $M^+$(*cold*), for example, $H/H^+$ or $O^+/O$ charge exchange processes. Lighter species, such as hydrogen, tend to escape more easily through thermal escape, but strong XUV fluxes can also stimulate the loss of light and heavy ions through the ionospheric outflow and exospheric pick-up by stellar winds (Airapetian et al., 2017a; Kislyakova et al., 2014; Dong et., al. 2018; Lammer et al., 2009; 2018). In this process, the incident XUV photons ionize the atmospheric neutral particles yielding ions and electrons. The electrons generated by photoionization, known as photoelectrons, are much less gravitationally constrained then the much heavier ions, and in the absence of collisions would be largely free to escape to space. This is prevented, however, by a polarization electric field which serves to restrain the free escape of electrons while simultaneously enhancing the escape of ions .

The role of large XUV fluxes for generating enhanced ion escape for close-in planets was recently evaluated by Airapetian et al. (2017a, 2023). The authors extended their previous models of ions escape by the one-dimensional exoplanetary multi-fluid Global Ionosphere Thermosphere Model (exo-GITM) that comprehensively treats nonthermal and thermal escape processes from an Earth-like planetary atmosphere and explored the range of stellar quiescent EUV fluxes that initiate the transition from ions escape to the hydrodynamic escape regime using a multi-species treatment of the most important ions and neutrals and full thermodynamic treatment of the ionosphere and thermosphere. Airapetian et al. (2023) simulated three scenarios for an Earth mass and radius exoplanetary $N_2 – O_2$ atmosphere with a rotation period of 24 h and irradiated by an XUV emission from an active star and discussed these scenarios for specific exoplanetary systems including TRAPPIST-1 and TOI-700. They found that the atmospheric escape via ion outflow is the major escape mechanism for the incident XUV fluxes less than 60 XUV fluxes of the current Sun. The mass flow of the ion escape rate from an unmagnetized Earth-like planet is estimated to be $10^5$ kg/s at the exobase ion temperature of 10000 K, which is about two orders of magnitude greater than that estimated for 10 × XUV case in Airapetian et al. (2017a). This difference can be explained by the fixed exospheric temperature calculated in the single fluid atmospheric model with the exospheric base temperature of 2000 K. For XUV fluxes exceeding 60 times of the Sun's flux, the fully hydrodynamic regime takes over the nonthermal escape regime. The total atmospheric escape rate in the hydrodynamic regime is $8 \cdot 10^6$ kg/s, which is a factor of 2 smaller than that calculated in the single-fluid model of Johnstone et al. (2019).

TOI-700d rocky exoplanet. These results were applied for the TOI 700 exoplanetary system. TOI 700 is a slowly rotating M2 dwarf with the estimated age of > 1.5 Gyr. The star has relatively low magnetic activity with the XUV flux less than $2.4 \cdot 10^{27}$ erg (Gilbert et al., 2020). Thus, the TOI-700 system has three rocky exoplanets with the outer one with a size of 1.19 Earth's radius located in the habitable zone around the star. Dong et al. (2020) and Cohen et al. (2020) have studied the impact of the stellar wind from TOI 700 on the outer planet atmospheric loss and estimated the expected Joule heating rate due to induced ionospheric current dissipation should be lower than that expected from a regular CME from the current Sun, and thus it is possible that TOI 700 d can sustain an atmosphere. The XUV flux at the orbital location of this exoplanet is about 10 XUV fluxes of the current Sun with the associated rate of $10^5$ kg/s for an unmagnetized case. The authors used the scaling of the escape rates derived earlier by Airapetian et al. (2017a) and concluded that the ion escape rates are expected to be ~ 100 kg/s. This rate suggests the loss of 1 bar $N_2 – CO_2$ atmosphere expected for the late Hadean Earth within 1.5 Gyr, which would suggest that TOI-700 d has a great chance to retain a thick atmosphere given the outgassing rates comparable to the current Earth. Additionally, recent MHD simulations of the ion escape due to the interaction of the planetary magnetosphere with the tenuous stellar wind suggest that the escape rate is of the order of 30 kg/s (Dong et al., 2020; Cohen et al., 2020). Thus, the detection of $N_2 – CO_2$ atmosphere has good chances given the sensitivity of the JWST that needs to be calculated for a future search of its atmosphere and potential habitability.

TRAPPIST 1b-h exoplanets. Seven Earth-size exoplanets orbiting at distances ranging 0.011 to 0.066 AU from a magnetically active M8 ultracool dwarf, TRAPPIST 1, represent one of the most extreme XUV irradiated bodies. The closest planet, TRAPPIST b exoplanet is irradiated by ~ 2000 XUV fluxes, while the TRAPPIST d planet within the HZ experiences 500 XUV fluxes and the outer TRAPPIST h planet receives ~ 65 XUV fluxes (Wilson et al., 2021). The authors found that if all seven exoplanets had original $N_2 – CO_2$ rich atmosphere, at the observed XUV fluxes from the star, they should be subject to hydrodynamic escape. The lowest escape rate at ~ $8 \cdot 10^6$ kg/s is expected for the TRAPPIST-1h outer unmagnetized planet. This suggests over the lifetime of the exoplanetary system that is estimated to be around 7 Gyr, the planet is expected to lose ~ $1.6 \cdot 10^{24}$ kg or 80% of the planet's mass. This suggests that even the farthest planet should have lost all its volatiles gained from the protoplanetary disk since its formation, and thus all TRAPPIST 1b-h planets should be represented by bare rocky exoplanets.

If we assume that all seven exoplanets have water vapor atmospheres, we can apply the results of hydrodynamic escape simulations by Johnstone (2020). They suggest the minimum escape rate at $\sim 3.5 \cdot 10^6$ kg/s expected for the TRAPPIST 1h exoplanet, which is comparable to our results for $N_2 - CO_2$ rich unmagnetized atmosphere, and thus do not change our conclusions about the fate of TRAPPIST 1 exoplanets. We should also note that the XUV flux used in our atmospheric model represents the quiescent flux from the host star and does not account for frequent flare and associated coronal mass ejection activity from the star that can enhance the escape rate due to the quiescent flux.

## 5.4. Impact of Coronal Mass Ejections on a Magnetized Exoplanet

The effects of CMEs on atmospheres of close-in rocky planets have been studied previously by Khodachenko et al. (2007) and Lammer et al. (2007). They found that the interaction of a CME with an exoplanet result in stripping the 1-bar atmosphere of an Earth-like planet by a dense CME within a 1-Gyr. While these models were based on scaling an Earth-like interaction between planets and CMEs to a close-in orbit scenario, Cohen et al. (2011) studied the effects of CMEs on HD189733b's magnetosphere using 3D MHD models of interaction of CMEs with a magnetized hot Jupiter. They concluded that the CME hits the magnetosphere of a close-in exoplanet from the side and forms the fast orbital motion of the planet creates a long comet-like magnetospheric tail. They concluded that the magnetosphere of the hot Jupiter can be significantly impacted by the CME event and that the energization of the planetary magnetospheric–ionospheric system might be much higher than in the Earth. However, the modelling of the planetary atmosphere neglected a more self-consistent physical process to drive the flow from the exoplanetary atmosphere driven stellar radiation pressure. A more realistic modelling of the planetary atmosphere to study the influence of CMEs on the mass loss rates of hot Jupiter was carried out by Cherenkov et al. (2018).

If a giant exoplanet has a magnetic moment comparable to Jupiter's, then the stellar wind dynamic pressure compresses and convects the planetary magnetospheric field inducing the convective electric field and associated ionospheric current (Cohen et al., 2014; Airapetian et al., 2016, Airapetian, 2017). The dissipation of this current can significantly contribute to the atmospheric heating that is controlled by the atmospheric conductance (Cohen et al., 2014).

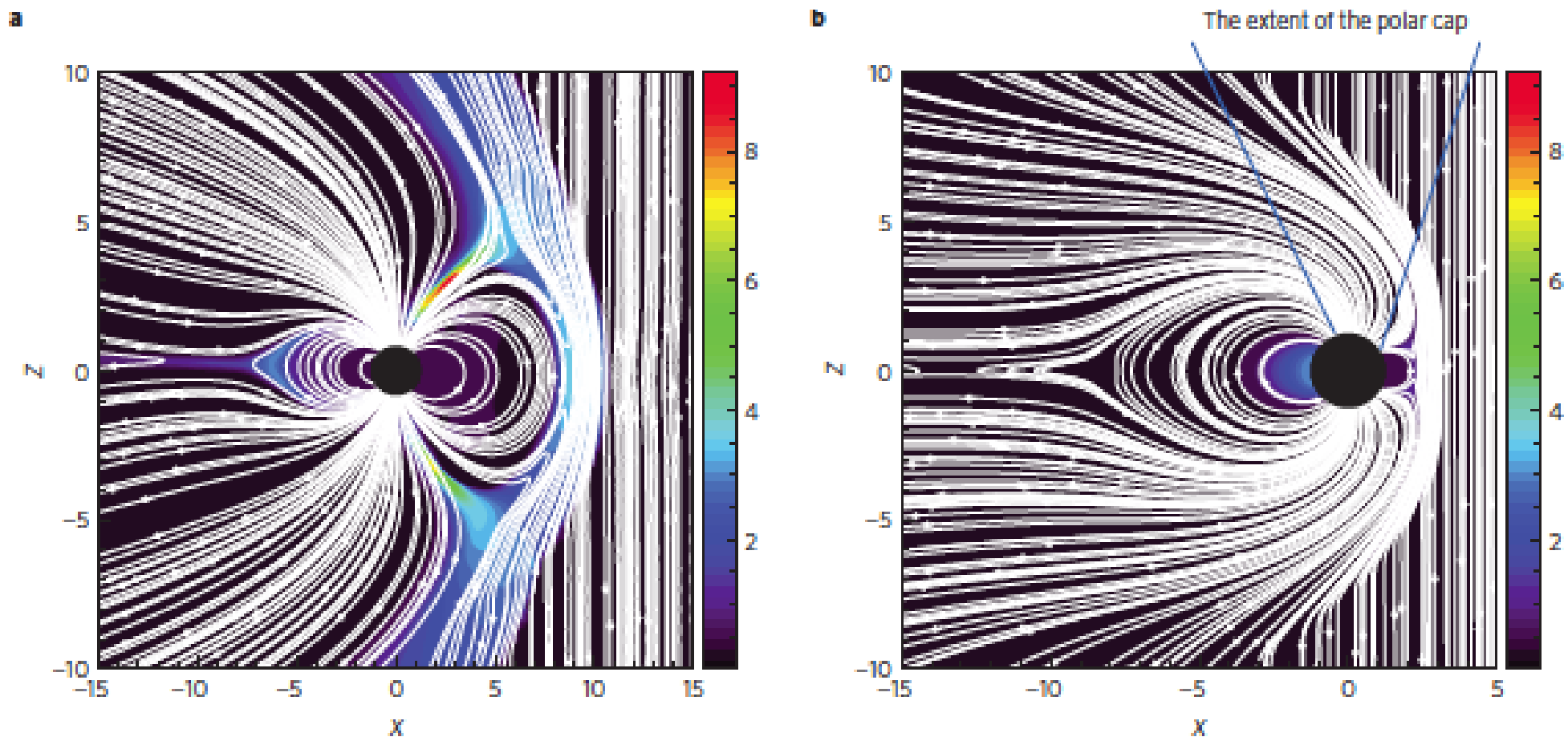

Fig. 133. Magnetic field lines and plasma pressure in the Earth's magnetosphere due to a CME event. (a) Initial state, (b) Final state. Magnetic field lines (white) and plasma pressure in nPa (color map). Axes represent distance from the Earth's center in units of the Earth's radius (Airapetian et al., 2016)

The simulation of the interaction of an energetic (Carrington-type) CME with the magnetosphere of a terrestrial-type planet is shown in Fig. 133 (Airapetian et al., 2016). The distance of the subsolar magnetosphere from the planetary surface varies in times in response to changing dynamic pressure from the solar wind and a CME. Its boundary known as the standoff distance is determined from the balance between the magnetic pressure of the planetary magnetosphere and the wind dynamic pressure. The larger the dynamic pressure, the smaller the standoff distance. The left panel of Fig. 133 shows that the magnetospheric standoff distance moves from ~ 10 RE (right panel) to ~ 1.5 RE as the Carrington scale CME from the young Sun with the southward Bz component of the magnetic field as a result of the combined effect of magnetic reconnection between two fields and the CME dynamic pressure (Airapetian et al., 2016; Airapetian, 2017). The boundary of the magnetospheric open-closed field shifts to 36 degrees in latitude opening 70% of the Earth's magnetic field. The CME dynamic pressure drives large field aligned electric currents that dissipate in the ionosphere-thermosphere region (at ~ 110 km) of the Earth-like planet via ion-neutral frictional resistivity producing Joule heating rate reaching ~ 4 W/m2. This is over 20 times larger than that observed during the largest geomagnetic storms and will thus increase the upper atmospheric temperature and ignite enhanced ionospheric outflow (Airapetian, 2017; Airapetian et al., 2017a). The models of associated atmospheric escape require the extension of sophisticated tools to consider multi-fluid effects for consistent treatment of neutral species in the ionosphere-thermosphere dynamics developed for the current and early Earth, Mars and Titan (Johnstone et al., 2018; Glocer et al., 2024).

The solar wind and a CME also compress the night-side magnetosphere and ignite magnetic reconnection at the night-side of the magnetosphere causing the magnetospheric storm as particles penetrate the polar regions of the planet. Also, the orientation of the magnetic field of the wind and CME as compared to that of the magnetospheric field controls the energy transfer from the wind to the planet, as it drives magnetic reconnection which in turn ignites particle acceleration and particle precipitation in the planetary ionosphere and thermosphere. This process is driven by electric fields that accelerate the ions against the neutrals resulting in current dissipation (Ohmic or Joule heating). In addition, the particle precipitation in the upper atmosphere impacts the local ionization and modifies Joule heating processes and atmospheric line excitation (i.e., auroral excitation, Schunk and Nagy, 1980).

Can the environments of such close-in terrestrial type planets within CHZ from their parent stars be hospitable to life? This answer requires the conditions of space weather around red dwarf stars including quiescent and flare driven XUV fluxes and their properties of stellar winds.

The impact of XUV emission from Proxima Centauri on its close-in rocky exoplanet was discussed by Garcia-Sage et al. (2017). They used the reconstructed XUV fluxes that appear to be over 2 orders of magnitude greater at the planet location than that received by Earth to evaluate the associated ion escape. The calculated escaping O+ mass flux from Proxima Cen b appears to be high as presented in Figure 134 and is consistent with the results of Airapetian et al. (2017a). They also discussed the impact of thermospheric temperature and polar cap area on the escape fluxes. A hotter thermosphere is obviously expected to result in stronger outflows, but a larger polar cap area (the area of the open magnetic flux connected to the stellar wind) also results in more net mass flux of ionized particles lost to space. The thermospheric temperature and the polar cap size have strong implications for the mass loss rate at Proxima Cen b and the ability for this exoplanet to retain its atmosphere on geological time scale.

In summary, due to its proximity to the star, Proxima Cen b resides in an extremely hostile and extreme space environment that is likely to cause high atmospheric loss rates. If it is not clear whether the planet can sustain an atmosphere at all, it is less likely that the planet is habitable, even though it resides in the CHZ around the star.

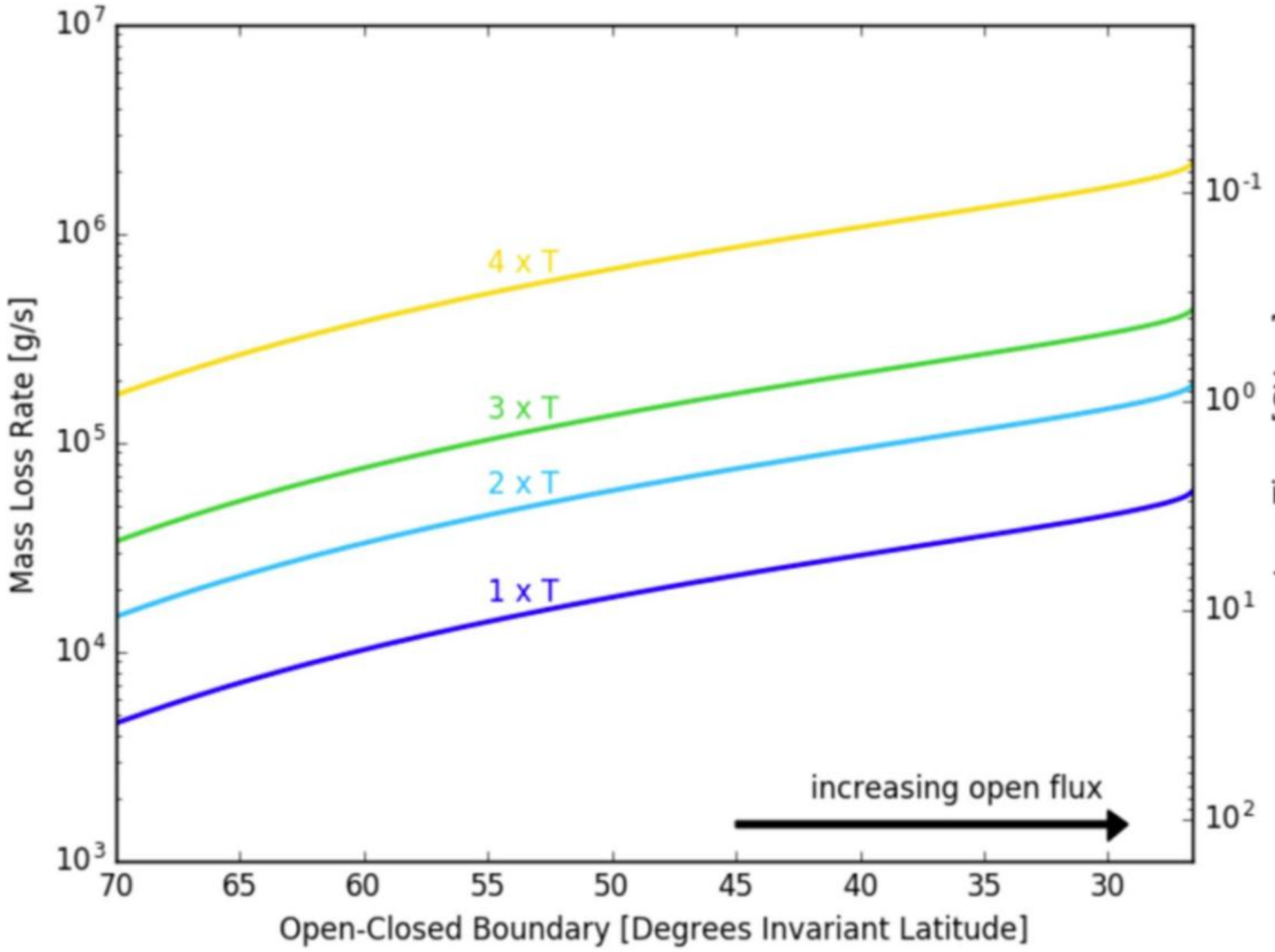

Fig. 134. The ion escape rate calculated for 4 different neutral thermospheric temperatures of Proxima b driven by XUV fluxes from Proxima Cen (Garcia-Sage et al., 2017).

The effects of the stellar wind from Proxima Centauri on its exoplanet were characterized by Garraffo et al. (2016). They have theoretically reconstructed the stellar wind from Proxima Centauri and modelled the space environment conditions around Proxima Cen b. The model of the stellar coronal wind was driven by magnetic field observations (ZDI map) of the star GJ 51, which was used as a proxy for Proxima Cen for which ZDI observations are not available. Following the limited information about the magnetic field of Proxima Cen (Reiners and Basri, 2008a), they used two cases of the average stellar magnetic field of 300 G and 600 G. They found that the stellar wind's dynamic and the magnetic pressure along the orbit of Proxima Cen b are extremely large, up to 20000 times that of the solar wind at 1AU. In addition, the planet experiences fast and very large variations in the ambient pressure along its orbit. As a result, the magnetosphere surrounding the planet undergoes huge variations within the relatively short period of the orbit, which potentially drive strong currents in the upper atmosphere and result in extreme atmospheric heating.

The recent simulation results by Varela et al. (2022) suggest that rocky exoplanets with an Earth-like magnetosphere around an M dwarf at 0.1 AU should be protected from the stellar wind during quiescent cent and CME-like space weather conditions if the rotation period of the star is 12 days or larger. However, for exoplanets orbiting closer than 0.1 AU from the star, stronger magnetospheric field would be required to avoid the direct impact of the stellar wind at low latitudes. The planet becomes particularly vulnerable during CME impacts that can exert at least 1000 times greater dynamic pressure on the planetary magnetosphere than that at quiescent conditions. This alone would move the stand-off distance by a factor of $10001/6 \approx 3.16$ times closer to the planet, which can create devastating conditions for exoplanetary habitability.

The interaction of stellar winds and CMEs can also provide observational signatures of exoplanetary magnetic fields due to the conversion of internal, bulk flow kinetic and magnetic energy of the CME flow into the kinetic energy required to accelerate the magnetospheric electrons along the magnetic field lines, and producing cyclotron-maser radiation emission at the low radio frequency range (Varela et al., 2018; Kavanagh, 2022; Sciola et al., 2021) . For example, if the exoplanetary magnetic field intensity is much lower than 4 G, then the frequency of produced radio emission is expected to be lower than 10 MHz, below the highest magnitude of frequency above which the waves penetrate the ionosphere or the ionospheric cut-off frequency. Thus, high sensitivity space-based radio telescopes would be required to detect radio emission from exoplanetary magnetic fields.

The combination of extreme XUV fluxes and the ambient wind pressure, along with the fast and strong variations, potentially make the atmosphere of Proxima Cen b highly vulnerable to strong atmospheric stripping by the stellar wind, and atmospheric escape because of the additional strong heating. The simulations suggest that Proxima Cen b may reside in the sub-Alfvénic stellar wind at least part of the orbit, which removes the magnetopause completely, and exposes the planetary atmosphere to direct impact by stellar wind particles.

## 5.5. Impact of Space Weather on Exoplanetary Atmospheric Chemistry

Exoplanetary atmospheric chemistry is affected by many internal and external factors including the combined effect of XUV-UV and stellar energetic particle driven precipitation into an exoplanetary atmosphere. The initial atmospheric chemical composition is defined by the interplay between the atmospheric escape and volcanic outgassing of species including $SO_2$, $H_2$, $CO_2$, $H_2S$, and $H_2O$, among other species, driven by the internal planetary dynamics (Airapetian et al., 2020). While XUV radiation mostly ionizes the upper atmosphere and contributes to the formation of the Earth's ionosphere and thermosphere (at up to 90 km from the ground), auroral electrons can reach the altitudes of ~ 80 km, while relativistic electrons penetrate to the mesosphere at 40 km. Solar energetic protons with energies up to a few GeV can precipitate through the troposphere and reach the ground causing Ground Level Enhancement events as measured by ground monitors of neutrons, the byproducts of interactions of high-energy protons with atmospheric species. Impact electrons formed because of photo and collisional ionization can then interact with molecular nitrogen, carbon dioxide, methane and water vapor igniting a reactive chemistry in the Earth's lower atmosphere.

What would be the effects of such ionizing sources on the atmospheres of early Earth, Mars and young rocky exoplanets? Could these sources have been beneficial in starting life on these (exo)planets? The first clues on the relation between the sources of ionizing radiation and the production of biological molecules were obtained in revolutionary experiments by Miller, 1953). In this experiment, a highly reducing gas mixture containing a water vapor, ammonia, $NO_3$, methane, $CH_4$ and $H_2$ was exposed to a spark discharge. The chemical reactions initiated by the discharge promoted the formation of simple organic molecules including hydrogen cyanide, HCN, formaldehyde, $CH_2O$. These molecules and methane react via Strecker synthesis forming amino acids, the building blocks of proteins (amino acids) and other macromolecules.

However, recent studies suggest that the early Earth's atmosphere was weakly reducing consisting of $N_2$, $CO_2$, CO and $H_2O$ with only minor abundance of $H_2$, $CH_4$ and $H_2S$ (Kasting et al., 1993; Lammer, 2018) or neutral (Schaefer and Fegley, 2017). Follow-up experiments showed that nonthermal energy input in the form of lightning (spark discharge between centers of positive and negative charge) in a weakly reducing atmosphere does not efficiently produce abundant amino acids as reported in experiments in highly reducing environments (Cleaves et al., 2008 and references therein). Later, Patel et al. (2015) developed photochemically driven chemical networks that produce abundant amino acids, nucleosides, the building blocks of RNA and DNA molecules and lipids. Lipids are complex biomolecules that are used to store energy and serve as structural units of cell membranes. Recent experiments suggest that near-UV (NUV at $\lambda$ = 2000–2800 Å) irradiation of the gas mixture is a beneficial factor for the formation of building blocks of life (Ranjan and Sasselov, 2016; Rimmer et al., 2018). However, while UV emission promotes biochemical reactions with participation of HCN, it cannot break triple bonds of molecular nitrogen, $N_2$, that require 10 eV to create odd nitrogen (N), which is required to produce abundant HCN, the requirement for the chemical network by Patel et al. (2015).

HCN is the most important feedstock molecule for prebiotic chemistry that is crucial for the formation of amino acids, nucleobases and complex sugars. In order to produce HCN and other biologically important molecules, molecular nitrogen needs to be converted into odd nitrogen and follow up $NO_x$ (NO, $NO_2$) molecules, the process known as the nitrogen fixation. In order to efficiently precipitate these molecules down to the ground for further polymerization, the nitrogen fixation and subsequent production of HCN should occur in the lower atmosphere (stratosphere–troposphere layers) of the early Earth (Airapetian et al., 2016). This is an important requirement for the efficient delivery of HCN to the ground for subsequent synthesis into complex molecules.

The alternative energy source, energetic particles from Galactic Cosmic Rays or stellar energetic particle events can provide a mechanism to deliver the particles directly to the lower atmosphere. The penetration depth depends on the frequency of collisions of protons with ambient molecules and for the Earth's atmosphere is scaled with the proton's energy (Jackman et al., 1980). This suggests that the protons with energy of 300 MeV can penetrate to the heights of 4.5 km above the ground. The collisions with molecules produce enhanced ionization of the atmosphere and form a broad energy distribution of secondary electrons at > 35 eV. These electrons then thermalize to lower energies in the atmosphere and as soon their energy reaches 10 eV, they become very efficient in breaking $N_2$ into odd nitrogen with subsequent formation of $NO_x$. To study the pathways to complex organic molecules driven by energetic protons, Kobayashi et al. (1990, 1995, 1998, 2017, 2018, 2022) and Miyakawa et al. (2002) have performed laboratory experiments exposing gas mixtures of $CO/CO_2$, CO, $N_2$ and $H_2O$ to 2.5 MeV protons. They reported production of amino acids precursors and nucleic acid bases as the results of secondary electron driven reactive chemistry. These experiments may have a direct relevance to the atmospheric chemistry of young rocky exoplanets irradiated by flare and CME associated high energy protons.

Young Sun-like stars are sources of frequent and energetic flares and associated CMEs. The CME driven shocks are the sites of efficient accelerations of SEPs that can accelerate particles to high energies (Fu et al., 2019; Hu et al., 2022). To study the role of SEPs from the young Sun-like stars as the source of high energy protons in gas phase prebiotic chemistry in the atmosphere of early Earth, Airapetian et al. (2016, 2020) have applied a photo-collisional atmospheric chemistry model driven by the XUV flux and frequent stellar energetic particle events. Protons with energies of > 0.3 GeV (at 0.5 bar atmosphere) precipitate into the middle and lower atmosphere (stratosphere and troposphere) and produce enhanced ionization, dissociation, dissociative ionization, and excitation of atmospheric species. The destruction of $N_2$ into ground state atomic nitrogen, $N(^4S)$, and the excited state of atomic nitrogen, $N(^2D)$, as the first key step toward production of bio-relevant molecules. Reactions of these species with the products of subsequent dissociation of $CO_2$, $CH_4$, and $H_2O$ produce nitrogen oxides, $NO_x$, CO and NH in the polar regions of the atmosphere. $NO_x$ then converts in the stratosphere into $HNO_2$, $HNO_3$ and its products including nitrates and ammonia.

The atmospheric model also predicts an efficient production of nitrous oxide, $N_2O$, driven primarily through $N(^4S) + NO_2 \rightarrow N_2O + O$; $NO + NH \rightarrow N_2O + H$. The recent model updated with the hard-energy spectrum of protons, vertical diffusion and Rayleigh scattering outputs the $N_2O$ production in the lower atmosphere by a factor > 300 greater than the earlier model output of Airapetian et al. (2016). The particle acceleration via diffusive shock acceleration mechanisms on quasi-parallel strong shocks produces mostly SEPs with harder spectra (Fu et al., 2019; Hu et al., 2022).

Laboratory experiments suggest that StEP driven chemistry predicts much more efficient production of HCN than that produced by lightning events for a weakly reducing ($N_2$ – $CO/CO_2$ – $CH_4$ – $H_2O$) atmosphere on early Earth (Kobayashi et al., 1998, 2018). HCN is formed primarily due to neutral reactions including $NO + CH \rightarrow HCN + O$, $CH_2 + N(^4S) \rightarrow HCN + H$, $CH_3 + N(^4S) \rightarrow HCN + H + H$, and $CH + CN \rightarrow HCN + H$. As it forms in the

stratosphere–troposphere region, HCN may subsequently rain out into surface reservoirs and initiate higher order chemistry producing more complex organics. For example, the hydrolysis of HCN through reactions with water cloud droplets produces formamide, $HCONH_2$, that can rain out to the surface. Formamide serves as an important precursor of complex biomolecules that can produce amino acids, the building blocks of proteins and nucleobases, sugars and nucleotides, the constituents of RNA and DNA molecules (Saladino et al., 2015; Hud, 2018).

To study the effects of SEPs on atmospheres of Earth-like exoplanets (with the current Earth's chemistry) around M dwarfs, Grenfell et al. (2012, 2013) applied an air shower approach in a coupled climate-photochemical column model considering formation of nitrogen oxides ($NO_x$). Results suggested strong removal of the atmospheric biosignature ozone due to catalytic removal involving $NO_x$ cycles. Tabataba-Vakili et al. (2016) applied an updated version of the same model including hydrogen oxides ($HO_x$) generated from stellar particle events. Results suggested that including the $HO_x$ effect led to some $NO_x$ removal into its unreactive reservoirs hence weaker loss of ozone. However, the assumptions about expected slopes and fluences of SEPs from M dwarfs as well as the atmospheric thickness of exoplanetary atmospheres need to be justified in the currently emerging and future physics-based models of StEP initiation from active M dwarfs (Hu et al., 2022).

Airapetian et al. (2017b) have applied the 2D GSFC atmospheric model to study the chemical response from strong stellar energetic particle events on nitrogen-rich Earth-like planets and found that $NO_x$ and $OH_x$ are efficiently produced in the thermosphere as a result of photodissociation (via X-ray and EUV emission) and collisional dissociation (via secondary electrons) of molecular nitrogen and water vapor. The mixing ratios of NO and OH molecules increase by a factor of 100 during strong magnetic storms as compared to the quiet time. The drastic enhancement of NO caused the depletion of stratospheric ozone. This suggests that storms initiate time-varying emissions from broad-band molecular bands of NO at 5.3-μm, OH at 1.6 and 2 μm and $O_2$ ($1\Delta$), the lowest electronically excited state of the $O_2$ molecule, at 1.27 μm, $N_2O$ at 3.7, 4.5, 7.8, 8.6 μm and $CO_2$ at 16 μm. These time-dependent chemical ingredients of abiotically enhanced concentrations of these molecules referred to as the "beacons of life" would provide information about the presence of thick nitrogen atmosphere and atmospheric water from terrestrial type exoplanets around active G, K and early M dwarfs. Their spectral signatures would be searched for in the upcoming space missions including James Webb Space Telescope.

## 5.6. Impact of Stellar Energetic Particles on Surface Dosages of Ionizing Radiation

Space weather from active stars can have a critical impact not only on exoplanetary atmospheres, but also on exoplanetary surfaces, and thus has the potential to create detrimental conditions on exoplanetary habitability. Recent studies found that SEPs on close-in terrestrial-type planets can significantly enhance the surface radiation dose and adversely impact their habitability (Atri, 2017, 2020; Yamashiki et al., 2019). This impact is controlled by the flare frequency, energy spectrum of particles accelerated with flare associated CME shocks, exoplanetary magnetic field, atmospheric pressure, and its chemistry. As argued by Hu et al. (2022), fast and energetic stellar CME events can produce SEPs with hard energy spectra of particles that can penetrate atmospheres of the Earth and rocky exoplanets forming ground-level enhancement events (Poluianov et al., 2017). Since the 1940s, over 70 GLE events with good magnetic connection to large solar flares have been detected and characterized (see https:/gle.oulu.fi). The energy of accelerated protons in such events have hard energy spectra with the power-law index between 1 and 2. Hu et al. (2022) has studied the formation of such hard spectra StEP events due to acceleration of protons in fast CME shocks. The energy spectra in extreme stellar events associated with superflares are harder (indicated by the > 430 MeV to > 200 MeV fluence ratio) than that observed in large SEPs from the current Sun.

The fast CME associated hard-spectra stellar energetic particles (StEP) with energies of > 430 MeV (the rigidity greater than 1 GV) would efficiently penetrate the lower layers of exoplanetary atmospheres (at the atmospheric pressure $\leq$ 1 bar), induce ionization of atmospheric species via collisions, and ignite chemical changes via collisional dissociation and excitation and reach the planetary surface.

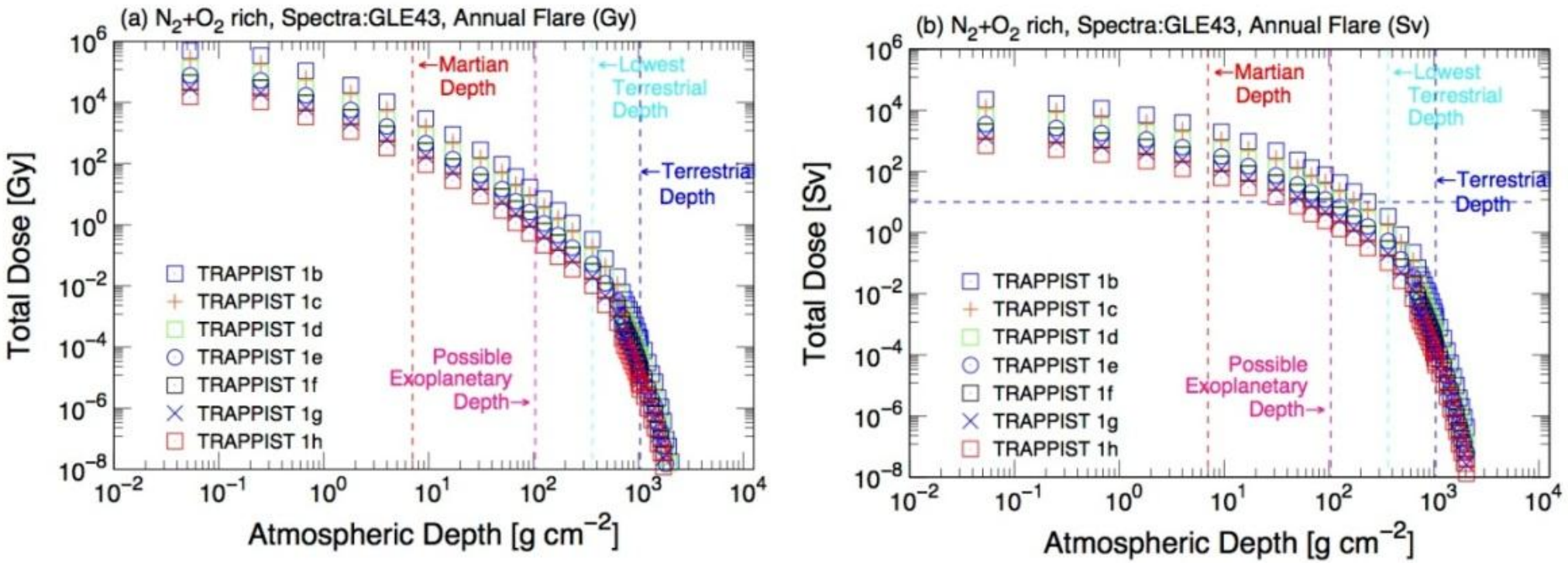

Fig. 135. Left: Vertical profile of radiation dose (in Gy) Right: (in Sv), caused by hard proton spectrum imitating GLE 43 penetrating $N_2$+$H_2$ rich terrestrial type atmosphere of TRAPPIST-1b (blue square), c(red cross),d(green square), e(blue circle), f(black square), g(blue cross) and h(red square) in logarithmic scale under Annual Maximum flare energy, calculating using PHITS and ExoKyoto tools. The vertical red dotted line represents Martian equivalent atmospheric depth, the pink dotted line represents the depth 0.1 bar atmosphere, the blue dotted line represents the lowest terrestrial atmospheric depth observed at the summit of Himalaya, and blue dotted line as terrestrial atmospheric depth. The horizontal blue dotted line represents 10 Sv, which may be considered as a critical dose for complex terrestrial-type lifeform. The figure is adapted from Yamashiki et al. (2019)

In order to estimate the energy in SEPs associated with flares, Yamashiki et al. (2019) used the maximum flare energy from stellar star spot sizes as input for the Particle and Heavy Ion

Transport code System (PHITS) suggest to calculate the ground-level dose for rocky exoplanets with atmospheres of different atmospheric pressures. Large fluxes of coronal X-ray and ultraviolet radiation from active stars induce high atmospheric escape rates from close-in exoplanets, suggesting that the atmospheric depth (column density) can be substantially smaller than that on Earth (Airapetian et al., 2017a). In a scenario with the atmospheric thickness of one-tenth of Earth's, the radiation dose from close-in planets including Proxima Centauri b and TRAPPIST-1e reaches near (few Sv) to over fatal levels (>10 Sv) for complex life forms (see Fig. 135). However, primitive bacteria including Deinococcus radiodurans can withstand much greater doses if they could effectively repair damage caused by ionizing radiation. In such harsh environments, organisms need an efficient repair mechanism leaving less energy for basic metabolic activities, growth, and reproduction (Makarova et al., 2001).

# Subject Index

# SUPPLEMENT

## Catalog of Stars with Solar-Type Activity — CSSTA[1]

## Introduction

The first lists of flare red dwarfs of UV Cet type appeared in the middle of the past century have comprised 2–3 dozen stars each. The first catalog of such stars was compiled at the Crimean Astrophysical Observatory and presented at the meeting of the International Astronomical Union in 1971. It comprised 53 objects and was placed in the VizieR database at number II/55 (Shakhovskaya, 1971).

At the end of the century, Hawley et al. (1996) carried out a spectral classification of about 2000 M dwarfs that are close to the Sun and found that 105 of them were emission M0–M3 dwarfs and 208 were M4–M8 stars with emission. Taking into account these and other results of those years in the Crimea, the GKL99 catalog was compiled (Gershberg et al., 1999) comprising 462 flare UV Cet stars and related objects in the vicinity of the Sun. In the VizieR database, the catalog is designated as J/A+AS/139/555, whereas in the search system SIMBAD it is represented as GKL99.

The GKL99 catalog was the basis for compiling a new list of stars with solar-type activity. A change of the name from “flare UV Cet stars” into “stars with solar-type activity” has marked certain progress in understanding the physics of activity. This was the GTSh10 catalog comprising 5535 objects. A detailed description of the GTSh10 catalog is given in Issue 1, V. 107 of the *Izvestiya Krymskoi Astrofizicheskoi Observatorii* (Gershberg et al., 2011).

Supplement to the monograph of R. E. Gershberg, N. I. Kleeorin, L. A. Pustilnik, and A. A. Shlyapnikov *Physics of Middle- and Low-Mass Stars with Solar-Type Activity* (M.: FIZMATLIT, 2020, 768 p., ISBN 978-5-9221-1881-1) provides a description of the second version of the catalog of stars with solar-type activity prepared at CrAO in 2019. This catalog has already included 29046 objects.

Below, we provide a description of the third version of the Catalog (CSSTA-3), present its structure and filling with data as at September 20, 2021.

## 1. Input Catalog

As in the previous version of the Catalog of Stars with Solar-Type Activity, the input list based on which the objects were selected is GAIA[2] Data Release 2 (GAIA DR2) (GAIA collaboration, 2018).

Interest to the project data in the context of compiling the current Catalog is caused by the following information contained in GAIA DR2. This is the two-color photometry of objects $G_{BP}$ (3300–6800 Å) and $G_{RP}$ (6300–10500 Å) of more than $1\times10^9$ objects, the detailed characteristic of light curves (~ 400 thousand objects), effective temperature (> 160 million objects), interstellar extinction (> 87 million), color indices (> 87 million), radii and luminosities (> 76 million).

The basic criterion for including stars into the CSSTA-3 catalog was a detection of at least one of characteristic phenomena of solar activity — sporadic flares, cool spots, chromospheric emission of hydrogen and ionized calcium, X-ray and radio emission, and their location in the lower main sequence (Fig. A1).

[1] The old version of the abbreviation is CSAST

[2] GAIA – https://www.esa.int/Science_Exploration/Space_Science/Gaia

The stars from GAIA, part of which was included into CSSTA-3, should thus meet the following criteria: $T_* < 7000°$ K, i.e., the stars should be of spectral type F5 and cooler (the left vertical border of the hatched region), $L_* < 1.1\,L_\odot$ (upper border), $L_* \geq 6.136\times10^{-6}\times T_* - 0.022$ (the equation that describes the cutting off the stars which fall into the region of white dwarfs — the left oblique border). There are 21321430 such stars in the GAIA DR2 catalog. Such a significant dataset can serve for searching for new objects with solar-type activity; however, in our case, it is used as an input catalog only and on its basis a selection of stars is implemented.

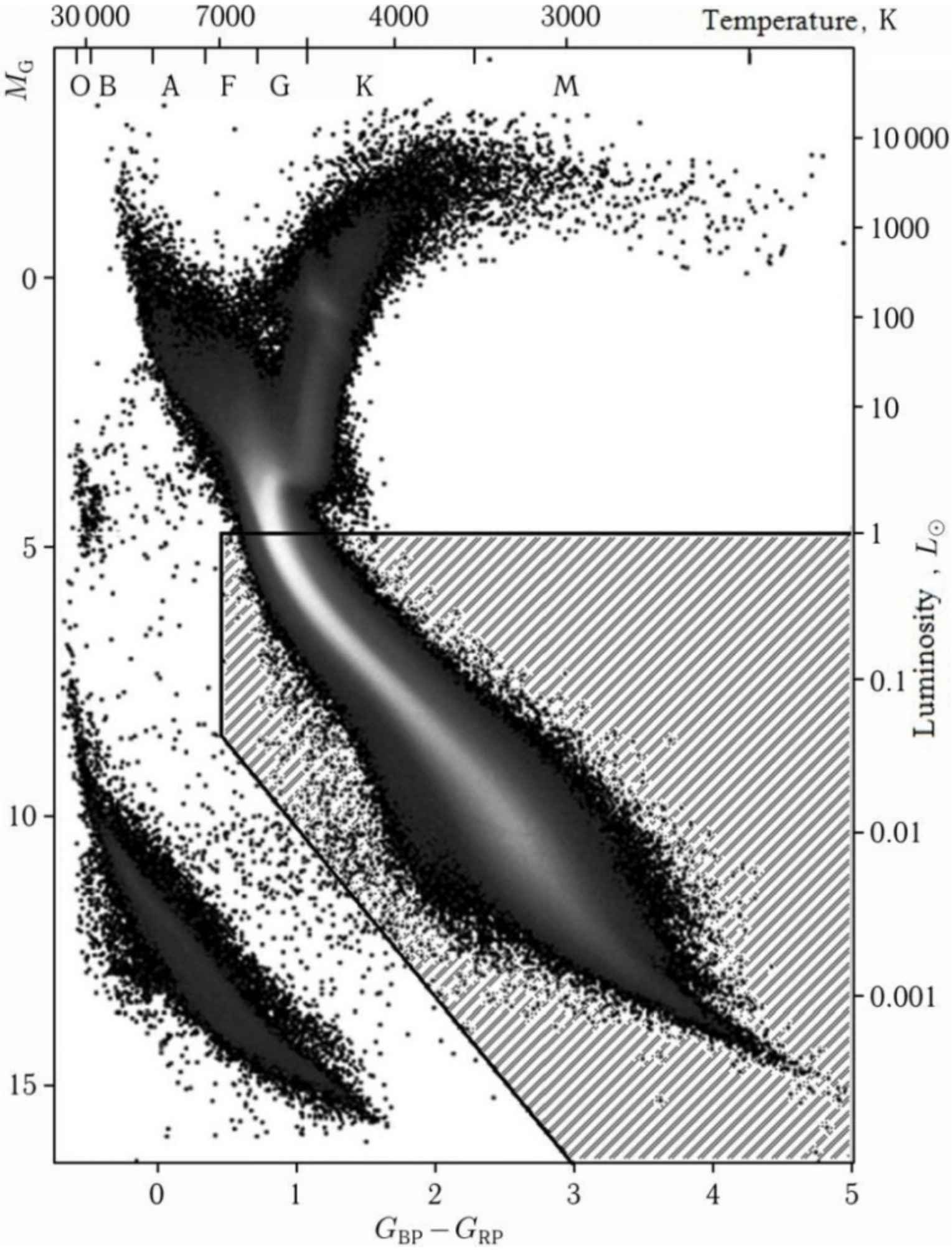

Fig. A1. A marked hatched fragment of the Hertzsprung-Russell diagram inside which the stars for CSSTA-3 were selected

With the aim of maximal using of data from the ground-based and space reviews, the input catalog was limited by $17^m$ and comprised a list of 13554156 objects.

## 2. Structure of CSSTA-3

CSSTA-3 contains 33 columns whose description is given in Table 1.

*Table 1*

| Column | Bytes | Format | Units | Label | Explanations |
|---|---|---|---|---|---|
| 1 | 1- 6 | I6 | --- | Record | Record from CSSTA-3 |
| 2 | 8- 16 | F9.5 | deg | RAJ(2000) | Right ascension (J2000) |
| 3 | 18- 26 | F9.5 | deg | DEJ(2000) | Declination (J2000) |
| 4 | 28- 64 | A37 | --- | SIM/Cat_ID | Main designation of the object according to SIMBAD or according to the source catalog |
| 5 | 66- 74 | A9 | --- | SIMBADTyp | Basic type from SIMBAD |
| 6 | 76- 79 | A4 | --- | VSXT | Variability type, as in the GCVS catalog |
| 7 | 81- 86 | F6.3 | mag | VmagSI | V magnitude from SIMBAD |
| 8 | 88- 93 | F6.3 | mag | VmagGA | V magnitude from GAIA (1) |
| 9 | 95-100 | F6.3 | mag | VmagSD | V magnitude from SDSS (2) |
| 10 | 102-107 | F6.3 | mag | maxVSX | ? Magnitude at maximum |
| 11 | 109-114 | F6.3 | mag | minVSX | ? Magnitude at minimum |
| 12 | 116-121 | F6.3 | mag | difVSX | Amplitude of variability from VSX (3) |
| 13 | 123-125 | A3 | --- | Pab | Passband from VSX |
| 14 | 127-129 | A3 | --- | Fl | Detected flares |
| 15 | 131-143 | A13 | --- | SIMBAD_Sp | Spectral type from SIMBAD |
| 16 | 145-149 | A5 | --- | GASpT | Spectral type from GAIA (4) |
| 17 | 151-153 | A3 | --- | SDS | Spectral type from SDSS |
| 18 | 155-160 | A6 | --- | W_emis | Lines emission (5) |
| 19 | 162-163 | A2 | --- | CA | Presence of the index of Chromospheric Activity (S) |
| 20 | 165-167 | A3 | --- | Spo | The presence of Spots (Spo) or their parameters (Spa) |
| 21 | 169-170 | A2 | --- | XF | X-ray emission (X) or registered flares (XF) |
| 22 | 171-172 | A2 | --- | UV | Ultraviolet radiation (6) |
| 23 | 174-175 | A2 | --- | IR | Infrared radiation |
| 24 | 177-178 | A2 | --- | RF | Radio emission (R) or registered flares (RF) |
| 25 | 181-187 | F7.2 | | Teffect | Stellar effective temperature |
| 26 | 189-192 | F4.2 | solRad | Rsol | Estimate of radius |
| 27 | 194-198 | F5.3 | solLum | Lumin | Estimate of luminosity |
| 28 | 200-211 | F16.10 | d | P_VSX | Period of a variable in days from VSX |
| 29 | 213-224 | F16.10 | d | P_KELT | Period of a variable in days from KELT |
| 30 | 226-237 | F16.10 | d | P_Kepler | Period of a variable in days from Kepler |
| 31 | 239-245 | F7.4 | y | CyclPer | Period of cyclic variability per year |
| 32 | 247-248 | A2 | --- | Pl | Presence of exoplanets (Y) and their number (Yn), or candidates (Y?) |
| 33 | 250-253 | I4 | --- | Note | Note |

Note (1): Calculated V magnitude from GAIA.
Note (2): Calculated V magnitude from SDSS.
Note (3): Difference in V magnitudes from maximum to minimum.
Note (4): Calculated spectral type from GAIA.
Note (5): Emission in the $H_\alpha$, $H_\beta$, $H_\gamma$, $H_\delta$ or CaII-K (C) lines. The + sign corresponds to the strongest line.
Note (6): Ultraviolet radiation (UV): far-UV (F), near-UV (N), far- and near-UV (FN).

## 3. Data from SIMBAD in CSSTA-3

To include the SIMBAD (Wenger et al., 2000) data into this Catalog, a series of selection criteria for objects was used according to their type, which emphasizes their physical nature. Thus, all the previous versions of the prepared at CrAO analogous catalogs included flare UV Cet stars and BY Dra variables. According to the SIMBAD data, 2259 objects were attributed to the Fl* (UV Cet) stars in 2019. In the latest version of the classification of objects in SIMBAD[1] in 2021, the stars of such a type were attributed to eruptive (!?). 1133 objects (1008 in 2019) are currently attributed to BY* (BY Dra), which are also included into the catalogs prepared earlier. Some of the indicated above stars have several determinations of variability types; among them there are T Tau, RS CVn, and other objects. All of them were deleted in the process of subsequent compilation.

After the cross identification based on coordinates in a radius of 1 arcsec of the input catalog with the SIMBAD database, 287462 objects were included into the preliminary list for further use. After excluding stars with main types that did not correspond to the required ones, 264611 objects left in the list. Taking into account that for further work with the current Catalog the single stars are preferred rather than binary or multiple stars, during the subsequent data filtration these objects were singled out into a separate list. As a result, we obtained a file containing 255191 objects.

As a final filtration of the given section of CSSTA was the singling out, into a separate list, of objects with unambiguous representation of luminosity classes, following the information from SIMBAD as compared to those determined from the GAIA data. After excluding these objects, the resultant file for the subsequent compilation into the Catalog comprises 252930 stars.

## 4. Index Catalog of Variable Stars, Types and Amplitudes

By the end of July 2021, the VSX[2] database (Watson et al., 2006) comprised information on 2114232 variable stars; among them there were 86379 BY Dra stars, 2003 UV Cet stars, and 59358 stars with detected rotational modulation.

To include into CSSTA-3, a series of selection criteria of objects was used according to their type, which emphasizes their physical nature. Thus, all the previous versions of the prepared at CrAO analogous catalogs comprised flare UV Cet stars and BY Dra variables, and all stars of these types from VSX were included into the current version of the Catalog, ruling out the objects that had several determinations of variability types and being involved into the binary or multiple systems. Particular attention was given to stars with detected rotational modulation.

After the cross identification based on coordinates in a radius of 1 arcsec of the input catalog with the VSX database, 14938 objects were involved into a preliminary list for further use. After excluding 767 stars of the main types that did not correspond to the required ones, a list prepared for CSSTA-3, with the VSX data included, comprised 252163 objects.

---

[1] SIMBAD – http://simbad.cds.unistra.fr/simbad/

[2] VSX – https://www.aavso.org/vsx/index.php

## 5. Spectral Data for CSSTA-3

At the end of March 2019, Mamajek[1] published an updated version of the database of color indices and effective temperatures for dwarf stars of the lower main sequence. The database described earlier in Pecaut et al. (2012), Pecaut and Mamajek (2013) was used during the compilation of CSSTA-3 to determine spectral types of some objects.

The database is constructed on the basis of the independent review of literature and information from different catalogs and reflects the modern system of the Morgan–Keenan (MK) spectral classification. It represents the spectral types of dwarfs from O3V to Y2V by the effective temperature, bolometric luminosity (normalized to the solar one), bolometric stellar magnitude, bolometric correction in the V band, absolute stellar magnitude in the V band, and corresponding color indices.

Color indices are based on the photometry in the Johnson *U, B, V* and Cousins $R_C$, $I_C$ systems, the Tycho catalogs of *Bt, Vt,* and *G, Bp, Rp* of GAIA DR2, data on *i, z, y* from Sloan (SDSS), *J, H, K* from 2MASS, and *W1, W2, W3, W4* from WISE projects.

Checking consistency of spectral types from the literature sources was implemented based on the optical observations of stars of spectral types from F3 to M4 carried out with the 1.5-meter telescope SMARTS mounted on the Sierra del Toro in Chile.

When refining and determining the spectral types of objects from CSSTA-3, the General Catalog of Stellar Spectral Classification (Skiff, 2014) was used containing 888588 bibliographic references to the spectral classification of 660614 stars. The number of stars later than F5V is 45547.

The previous version of the Catalog comprised a sample of information on the spectroscopy of M dwarfs from the SDSS Data Release 7 (West et al., 2011). To obtain information, the 2.5-meter wide-angle telescope with a multi-aperture spectrograph was used belonging to several universities of the USA, which is located at the Apache Point Observatory (New Mexico, USA).

In 2011, on the basis of data of the 7th implementation of the SDSS project, the catalog of spectral observations of 70841 M dwarfs was published. It involved five-color photometry in the SDSS system, determinations of spectral types of stars, data on the $H_\alpha$, $H_\beta$, $H_\gamma$, $H_\delta$, and CII-K lines and other information. After their detailed analysis, these data were partially integrated into CSSTA-3.

After the cross identification based on coordinates in a radius of 1 arcsec of the input catalog with the SDSS database, 63783 objects were involved into a preliminary list for further use. After excluding 721 objects of main types that did not correspond to the required ones and 600 binary objects from the SDSS database, 62462 objects were included into a list prepared for the current Catalog.

## 6. Observational Data from the Kepler Orbital Station

Kepler[2] is a space observatory of NASA launched in 2009 with the aim of searching for planets that orbit stars. Including the K2 project as a result of the mission, 3032 candidates for exoplanets were recorded; among them 3253 had been confirmed by mid-November 2022. Data of the Kepler project were of interest for the current version of the Catalog in two directions. First, the search for flare activity and, second, the presence of exoplanets in dwarf stars of the lower main sequence.

---

[1] Mamajek, 2019 – http://www.pas.rochester.edu/˜emamajek/EEM_dwarf_UBVIJHK_colors_Teff.txt
[2] Kepler – https://www.nasa.gov/mission_pages/kepler/main/index.html

A homogeneous search for stellar flares was carried out using all the available light curves obtained from Kepler. Within the first observing set of Kepler in 2011, 23000 K and M dwarfs were investigated. Thus, 373 flaring stars were identified; some of them exhibited several events throughout the whole observational period (Walkowicz et al., 2011). In 2012, Maehara et al. (2012) reported on the observation of 8300 stars from Kepler throughout 120 days and on the detection of 148 G stars for which flares with energies of $10^{29}$ to $10^{32}$ erg were recorded on time intervals of hours. A 2016 publication describes 851168 random outbursts detected from Kepler observations of 4041 stars (Davenport, 2016). However, a revision of Davenport's data was reported in 2019, which indicates that his list includes various pulsating stars, rapidly rotating stars, and stars with transit events that can cause either flares or rapidly changing brightness, leading to false positive observational processing signals (Yang and Liu, 2019). Comparison of Davenport's catalogs with those of Yang and Liu showed that only 396 stars from the first list are contained in the second one.

In 2017, Van Doorsselaere with colleagues published information on the detection of 16850 flares on 6662 stars observed by the Kepler observatory (Van Doorsselaere et al., 2017). A cross identification of the list with the data of Yang and Liu has already ensured the coincidence of 2223 objects. The rest objects, in the opinion of the Catalog’s authors, being compared to the data of Van Doorsselaere et al., were burdened with the same shortcomings as Davenport’s catalog.

## 7. Observational Data from the TESS Orbital Station

The orbital observatory for performing surveys of the sky with the aim of searching for exoplanets by the transit method (Transiting Exoplanet Survey Satellite — TESS[1]) analyses the brightness of bright and nearby stars (Ricker et al., 2014).

A number of papers are devoted to the analysis of observations from TESS, in particular, to the search for flare activity of red dwarfs. Let us consider a procedure for complementing CSSTA-3 with the data of these studies on the example of two publications “Stellar flares from the first TESS data release: exploring a new sample of M dwarfs” (Günther et al., 2020) and “A catalog of cool dwarf targets for the Transiting Exoplanet Survey Satellite” (Muirhead et al., 2018).

The first publication represents observations of 24809 stars with a 2-min exposure during the first two months of the TESS mission. The authors identified 1228 flare stars, 673 of them were M dwarfs. The final list involved 764 stars that exhibit flare activity; 632 of them were M dwarfs.

The results of these studies are placed in the VizieR database and represented by two tables: the catalog of individual flares found from the observational data of TESS in the first and second sectors (8695 flares) and the catalog of flare stars found by authors (1228 objects).

Among objects of the second table, 1131 are designated as dwarfs, 39 — as giants, and 58 objects have an undetermined luminosity class. To include into CSSTA-3, a cross identification of objects from the second table with the current version of the Catalog was carried out. CSSTA was supplemented with information on flares on 60 stars (Gorbachev, 2023).

---

[1] TESS – https://www.nasa.gov/tess-transiting-exoplanet-survey-satellite

## 8. The KELT Project and Stellar Rotation Periods for CSSTA-3

A very small “thousand-degree” telescope[1] consists of two instruments whose optics includes 80 mm objectives Mamiya 645 (f/1.9 with an effective aperture of 42 mm). The field of view of telescopes on the celestial sphere covers an area of 26°×26°. Instruments are located at the Winer Observatory in Sonoita, Arizona (USA) and at the South African Astronomical Observatory near Sutherland (Oelkers et al., 2018).

For almost 10 years of observations within the project, there was recorded information on 4000000 sources in the range of magnitudes $7^m < V < 13^m$ located in more than 70% of the sky. The basic scientific objective of the study is a detection of transit phenomena during the passage of planets with large radii on the background of bright host stars. In 2018, there was published a catalog of observations carried out within the KELT project containing information on 52741 objects which exhibit significant brightness variations of high amplitude probably caused by variability, including flaring one, as well as data for 62229 objects identified with probable stellar rotation periods. In CSSTA-3, the information from the KELT project is used to involve flare stars and data on rotation periods.

In the VizieR database, the results of observations within the KELT project are represented by three tables: 1 — contains astrometric information, information on stellar magnitudes and brightness variations for variables from KELT; 2 — upper limits of variability for non-varying sources in the TESS Input Catalog (3873790 objects), and 3 — information on the periodic brightness variations for the dwarf candidates from the TESS Input Catalog (TIC). To include into the current version of the Catalog, a cross identification of objects from the third table of the KELT catalog with the Input Catalog was performed.

The cross identification resulted in a preliminary list of 7022 stars from the KELT catalog to include into CSSTA-3. The list format corresponds to the structure of this Catalog according to the sections involved into the compilation.

The inclusion of information from the KELT database into CSSTA-3 made it possible to perform a comparative analysis of stellar rotation periods determined from the VSX and KELT data. It should be noted a significant number of periods from the KELT data that are close to one day, which is obviously a result of their automatic search and is not true. The periods presented in VSX were determined in the process of observational data reduction, which makes them more reliable. The problem of “similar periods” for a significant number of objects is well illustrated as the horizontal lines overlapping the whole range of effective temperatures, radii, and luminosities. The probability that tens and hundreds of stars have similar rotational periods seems to be very doubtful.

## 9. Supplementation of CSSTA-3 with the GTSh10 Data

After creating GTSh10[2], there appeared the data which allowed us to revise information contained in this catalog. The Catalog of Stars with Solar-Type Activity designated as GTSh10 was prepared in 2010. The data from publications of the preceding 10–15 years were used for its compilation. The Catalog mainly included dwarf stars with different manifestations of solar-type activity. It was composed of objects with dark spots, hydrogen and calcium chromospheric emission, fast flares in different wavelength ranges, and with radio and X-ray emission of stellar coronae. The compiled list comprises 5535 objects.

---

[1] KELT – https://keltsurvey.org/telescopes

[2] GTSh10 – http://www.crao.ru/~aas/CATALOGUEs/G+2010/eCat/G+2010.html

To include into CSSTA-3, a cross identification of the objects from GTSh10 with the compiled catalog was performed. As a result of the cross identification with a radius of 1 arcsec, 4733 objects from GTSh10 were found among the objects of the CSSTA-3 Input Catalog, whereas 802 objects were absent.

Since the cross identification was performed based on the coordinates, whereas most objects from GTSh10 have significant proper motions, then it is possible that for 802 stars the coordinates were specified with insufficient accuracy.

For 169 objects of the cross identification, according to the GAIA data, there were no values of luminosity and stellar radius, whereas for 171 — no values of temperature. Two objects have a temperature of more than 8000 K. 1444 stars have radii and luminosities in solar units exceeding those specified when compiling the Input Catalog. Taking into account that objects were included into GTSh10 based on the results of the analysis of publications with their detailed description, after a closer examination, all the stars that correspond to the specified criteria were involved into CSSTA-3.

## 10. X-Ray Sources for CSSTA-3

This section presents the identification of objects from CSSTA-3 in the X-ray wavelength range including that for objects of the optical range in the hard emission regions in proximity to red dwarf stars.

Various catalogs and databases were used to include information on the presence of X-ray emission in stars into the Catalog. In the previous version of CSSTA-3, 2909 objects with registered X-ray fluxes were presented. Independent processing of the “first light” image of the eROSITA telescope (Predehl et al., 2021) aboard the SRG observatory (Sunyaev et al., 2021) made it possible to identify 2485 X-ray objects.

Taking into account that the pixel size in the working image is ~ 15", the search for candidates for identifying isolated X-ray sources with optical objects was carried out in the area less than 40" to avoid edge processing effects. After the cross-identification of identified X-ray sources with the GAIA DR2 catalog, a list of matches amounted to 2868 objects. It is obvious that when the size of the search for GAIA DR2 objects in areas with a radius of < 20", there are non-single coincidences. In this case, up to 6 GAIA DR2 objects fell into the X-ray emission region. In order to obtain a one-to-one correspondence between X-ray sources and red dwarfs, regions with non-single coincidences were excluded from further consideration. After their exclusion, 899 objects remained in the list, 67 of which meet the criteria for temperature, radius, and luminosity for including into the Catalog (Shlyapnikov, 2021a).

The study of the equatorial region of the sky (eFEDS — eROSITA Final Equatorial Depth Survey) became the longest observation in the period of testing the capabilities of the eROSITA telescope instruments. In total, about 100 hours were spent. The eFEDS field has an area of approximately 140 deg$^2$ and consists of four separate rectangular scan areas of 35 deg$^2$. The site was chosen due to the presence of a significant number of multiwave observations of this region of the sky.

Based on the results of the eFEDS deep survey, two catalogs of X-ray sources were published: unambiguously detected sources in the 0.2–2.3 keV range and hard (2.3–5 keV) sources observed in the multiband mode (Brunner H. et al., 2022 ). To identify objects from CSSTA, 27910 X-ray sources with a high degree of detection probability were selected, which make up the main catalog of eFEDS.

Taking into account that the average radius of errors in determining the eFEDS coordinates is 4.8 arcsec, the search for identifying candidates was carried out in a threefold radius. As a result, among the objects of a deep survey of the equatorial region of the sky, the stars that fall into 110 fields limited by the radius of errors in determining X-ray coordinates were identified. Twelve stars were previously classified as X-ray sources, which confirms the correctness of independent identification. Galaxies were found in the identification regions of two stars, one of which is a known X-ray object. Several identities contain closely spaced objects. All areas were analyzed visually, and the presence of X-ray radiation was indicated in the Catalog for objects that met the selection criteria (Shlyapnikov, 2021b).

The previous version of CSSTA contained data on the registered X-ray emission from 2909 objects. To supplement the Catalog with information on stellar X-ray emission, the MORX compilation catalog (Flesch, 2023) was used. It included data related to the observation of objects in the X-ray range by the XMM-Newton[1], ROSAT[2], Chandra[3], and Swift[4] observatories.

In total, the catalog contains 3115575 optical objects. Each object has coordinates for epoch 2000, optical and radio/X-ray identifiers, photometry in the red and blue regions of the spectrum, and values of the calculated match probabilities.

The description of the MORX catalog contains information about the classification of objects made by the author. In total, 18 types of sources were classified. Among them there are extragalactic (actual galaxies, galaxies with active nuclei, objects of the BL Lac type, quasars and others) and galactic (star formation regions, cataclysmic variables, white dwarfs), including objects of an unknown type but with a predetermined redshift according to the SDSS data.

For analysis and subsequent inclusion into CSSTA, out of 3115575 MORX objects, 1357332 objects were selected that have classification as stars, X-ray or radio sources, as well as sources of unknown type. For the purpose of independent identification in the optical range of the spectrum, as well as extracting information about the proper motions of objects (to exclude possible extragalactic sources), the cross-identification of selected objects with stars from the GAIA DR2 catalog was performed.

After cross-identification with a radius of 5 arcsec (based on the fact that the PSF of the image of a point optical object should be at least 3 arcsec at a level of the half-width of approximation, and the search is performed in a region with a triple radius), out of 1357332 MORX sources in the GAIA DR2 catalog, 525337 objects were detected.

Considering that a significant number of the GAIA DR2 objects have the same coordinates but are not close pairs, they were excluded from further consideration. Note that most of these objects lack information about the effective temperature, radius, and luminosity. Objects that did not meet the selection criteria were also removed. As a result, the number of remaining objects amounted to 73031.

For identification by visual control, a special interactive interface was developed that allowed opening the area of the object under study in a new browser window with mapping information from the SIMBAD and NED[5] databases. The need for visual control was due to the presence of a large number of galaxies near some stars, one of which could be a source of X-rays.

---

[1] XMM-Newton – https://www.cosmos.esa.int/web/xmm-newton

[2] ROSAT – https://www.mpe.mpg.de/ROSAT

[3] Chandra – https://chandra.harvard.edu

[4] Swift – https://www.nasa.gov/mission_pages/swift/main

[5] NED – https://ned.ipac.caltech.edu/

Note that most of the objects under consideration are poorly studied, and there is no information about them in SIMBAD.

After monitoring all 73031 objects, only those objects that uniquely correspond to stars were added to CSSTA. At the beginning of November 2023, the cross-identification of CSSTA and selection from MORX gave a match for 4448 objects. Among them, there are 1371 stars with X-ray emission and 3077 X-ray sources identified with stars.

## 11. Ultraviolet Sources for CSSTA-3

The basic information on ultraviolet radiation for objects from the current Catalog was acquired through the cross identification with the observational data of the GALEX (Galaxy Evolution Explorer)[1] observatory. The NASA mission was started on April 28, 2003 and operated until June 28, 2013. The observatory was equipped with the 50-cm Ritchey-Chrétien Telescope (f/6.0) with a field of view of 1.2° operating in two ultraviolet ranges, far (FUV, $\lambda_{eff}$~1528 Å) and near ultraviolet (NUV, $\lambda_{eff}$~2310 Å). The GALEX database contains FUV and NUV images, about 500 million measurements, and more than 100000 low-resolution ultraviolet spectra.

For the cross identification of CSSTA-3 with the GALEX data, an updated version of the catalog of ultraviolet sources was chosen (Bianchi et al., 2017). The catalog includes all observations from the survey with the largest coverage of the sky regions recorded with both FUV and NUV detectors. This is more than 28700 fields and, in total, 57000 observations. The catalog is composed of two parts that contain 82992086 and 69772677 sources with the typical penetrating value in FUV = $19^m.9$ and NUV = $20^m.8$. The second part of the catalog is a more limited version that uses only a central part of each observed field of 1°.

As a result of the cross identification with a radius of 1 arcsec, 105542 stars were detected among the GALEX sources; the information on them was included into CSSTA-3.

## 12. Radio Sources for CSSTA-3

To supplement CSSTA-3 with information about the radio emission of stars, the MORX catalog described in Section 10 was used. Earlier, CSSTA indicated the registration of radio emission from 95 stars. The MORX catalog contains information obtained in the radio band in the NVSS[2], FIRST[3], SUMSS[4], and other projects. Cross-identification based on the coordinates of MORX objects and stars from CSSTA yielded a match for 361 objects.

## 13. Other Additions to CSSTA-3

### 13.1. Flares on Red Dwarf Stars

The detailed analysis of red dwarf stars with manifestations of flare activity included into CSSTA-3 is described in Sections 6 and 7. Since information on the studies was published before 2018, then by the moment of preparing the current version several new papers had been published and their data were added into CSSTA-3.

---

[1] GALEX – http://www.galex.caltech.edu/

[2] NVSS – https://www.cv.nrao.edu/nvss/

[3] FIRST – http://sundog.stsci.edu/

[4] SUMSS – http://www.astrop.physics.usyd.edu.au/sumss/

## 13.2. Cycles of Lower Main-Sequence Stars

### 13.2.1. Long Cycles

Monitoring of chromospheric activity yields valuable information on the stellar magnetic activity and its dependence on the fundamental parameters, such as effective temperature and rotation. Boro Saikia et al. (2018) represent a catalog of chromospheric activity of 4454 cool stars based on the combination of archival HARPS spectra and a number of other surveys including the Ca II H&K data from the Mount Wilson Observatory project. On the basis of this work, the current Catalog was supplemented with the mean S indices of chromospheric activity for 2559 objects with an identification radius of 1 arcsec and 53 data on the detected rotation periods and cycles.

### 13.2.2. Short Cycles

Along with studying long activity cycles over the past decades, owing to such projects as Kepler and TESS, there appeared a possibility of measuring cyclic variations of the stellar light curve amplitude and rotation period on timescales of a few years. Thus, using the Kepler data for four years (Reinhold et al., 2017), 23601 stars were studied. The periodic amplitude in the 0.5–6 year interval with a rotation period of 1–40 days was detected for 3203 stars.

The performed cross identification in a radius of 1 arcsec with the CSSTA-3 data made it possible to detect 874 objects involved into the new Catalog.

## 13.3. Exoplanets

Information about stars with confirmed or suspected exoplanets was added to the catalog. Analysis of the presence of X-ray and radio emission, if any, should contribute to understanding the possibility of the existence of “life” in the habitable zone. Flare activity also imposes certain restrictions on the development of life. Changes in brightness associated with the orbital periods of exoplanets around stars are superimposed on the overall light curve of a star. They must be taken into account when analyzing rotation and the period of possible cyclic activity.

The inclusion of information on exoplanets into the catalog was preceded by their studying at CrAO, which has been conducted since 2000. A detailed description of the studies is presented on the website[1].

By mid-November 2023, according to the NASA Exoplanet Archive[2], it had been confirmed the existence of 5550 planets discovered by 11 methods and 410 found by the TESS observatory. The discovery of 6977 possibly existing exoplanets according to TESS observations needs to be confirmed. Among the detection methods, the leader is the observation of the passage (transit) of the planet against the background of a star — 4176 discoveries. 1070 planets were discovered by changing the radial velocities of lines in the spectra of stars. 204 planets were detected by using the microlensing method. From 1 to 69 planets were found by using other methods.

---

[1] CrAVO exoplanet – https://sites.google.com/view/cravo-exop-psa

[2] NASA exoplanet archive – https://exoplanetarchive.ipac.caltech.edu

After cross-identification by coordinates between the Catalog of Confirmed Planets (CCP) and the Data Base of Candidates (DBC), it was found that 945 stars from the CCP and 771 from the DBC fall into CSSTA.

An analysis of the distribution of stars from the CCP by magnitude showed that the maximum distribution falls on the range from $12^m$ to $13^m$. This is due to the use of short-focus lenses for panoramic surveys of the sky in order to search for transits of exoplanets, as well as long-focus instruments when searching for changes in radial velocities. In the first case, restrictions are imposed on the penetrating power of astrographs, and, in the second case, on threshold restrictions on the signal-to-noise ratio in spectroscopy.

The maximum in the distribution of spectral types of stars falls on K1. Note that this sample of objects identified by CSSTA does not contradict the spectral classification of all stars with exoplanets described in Gorbachev et al. (2019), where the first maximum of the distribution is also in the region of spectral types K.

When analyzing the distributions of the periods of variability of the considered stars with exoplanets according to the VSX data and the orbital periods of planets around stars, it was recorded that the maxima in both distributions fall on 3 days. At the same time, it should be noted that 25 objects have the type of variability BY Dra according to the SIMBAD database, and 467 — RotV*, i.e., in the first and second cases, their variability is caused by the rotational modulation of the spotted surface rather than by the presence of exoplanets. Considering that the brightness variation amplitude for these stars lies in the range from $0^m.001$ to $0^m.1$, it is possible that most of the objects included in the Catalog are incorrectly classified according to the type of variability.

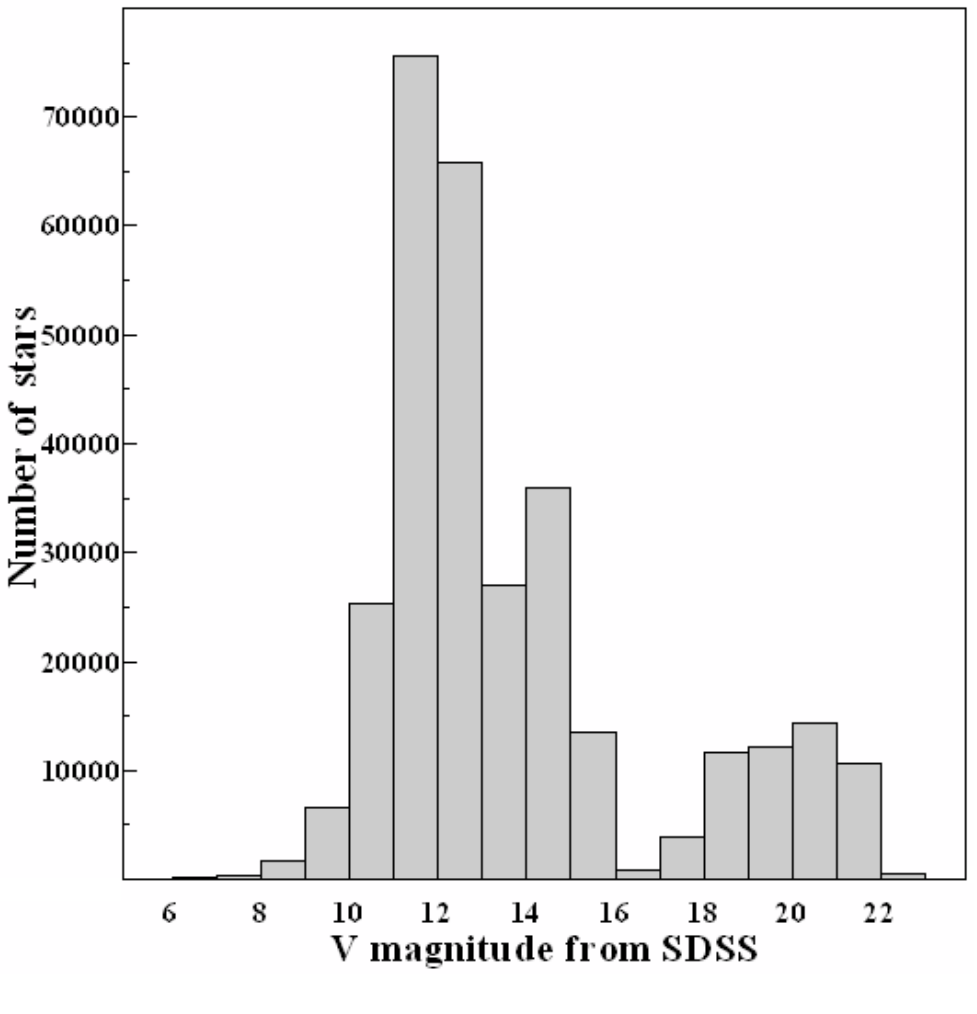

Fig. A2. Distribution of the number of stars from CSSTA-3 by the stellar magnitude V

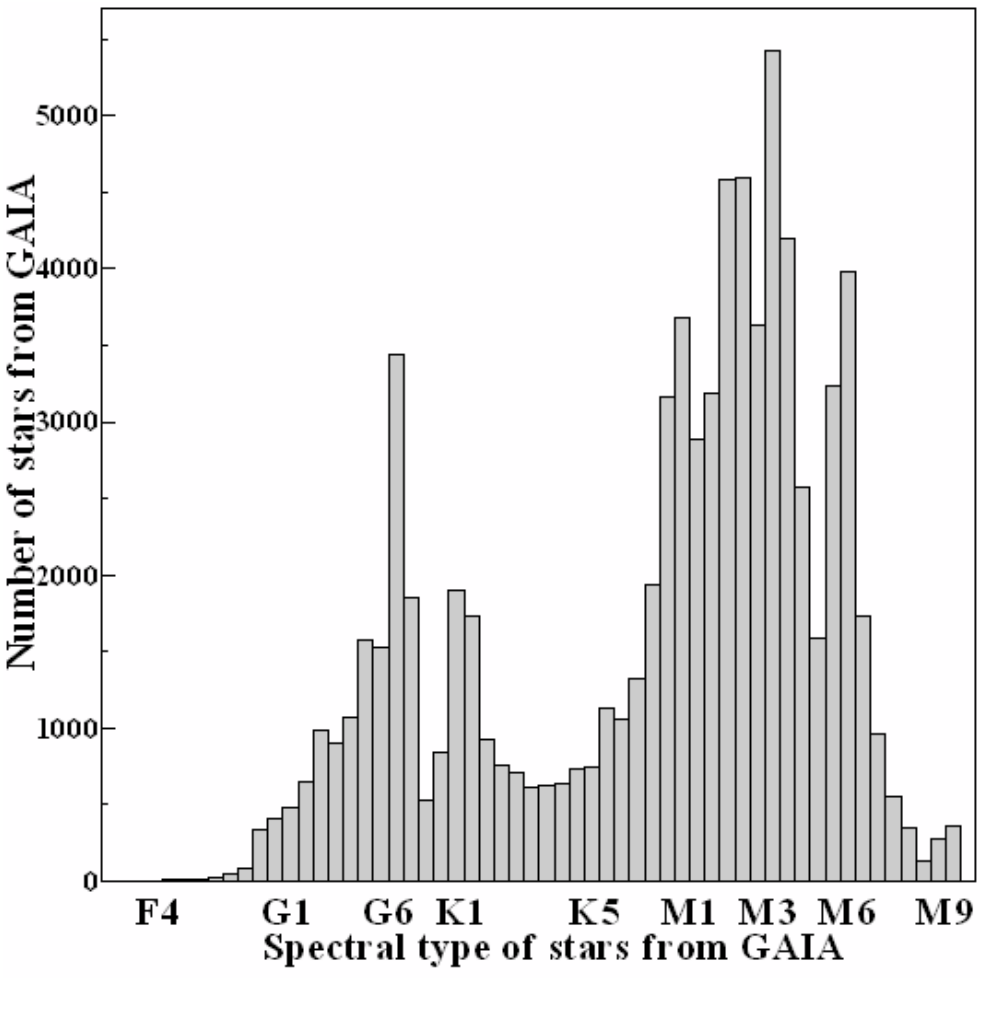

Fig. A3. Distribution of the number of stars from CSSTA-3 by spectral types

## 14. Version of CSSTA-3 as at 03.12.2023

The total number of stars in CSSTA-3 as at December 03, 2023 is 314618.

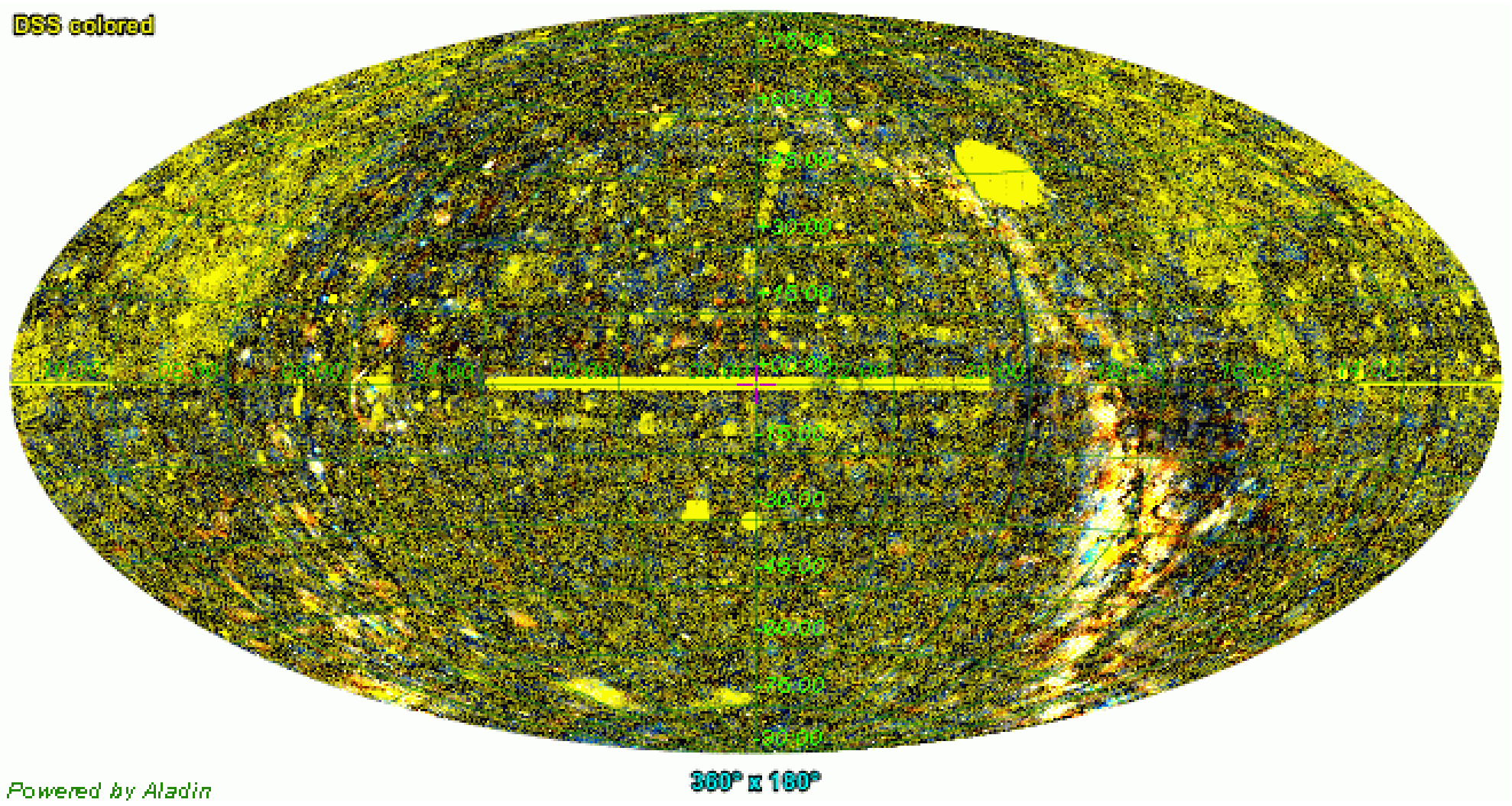

Fig. A4. Illustration of the distribution of 314618 objects from CSSTA-3 in the sky

## 15. Conclusions

After compiling all the described above sources, excluding coincidences and a number of other operations, the current CSSTA-3 catalog of the lower main-sequence stars with solar-type activity was obtained. It comprises 314618 objects, and the database that is realized on its basis is a developing project that contains hyperlinks to the original photometric and spectral observations. The Catalog is available on the website of the Crimean Astrophysical Observatory at http://craocrimea.ru/~aas/CATALOGUEs/S-2019/eCat/S-2019.html. More information about CSSTA-3 can be seen in ADS/NASA or VizieR.

To provide a wider access to the Catalog and the database of stars with solar-type activity, a copy of the information from the CrAO server is available via the Google resource at https://sites.google.com/view/csast.

## References

Note: all references include a bibcode of the SAO/NASA Astrophysical Data System (ADS).

*Научное издание*

**R.E. GERSHBERG**
**N.I. KLEEORIN**
**L.A. PUSTILNIK**
**V.S. AIRAPETIAN**
**A.A. SHLYAPNIKOV**

# PHYSICS OF MID- AND LOW-MASS STARS WITH SOLAR-TYPE ACTIVITY AND THEIR IMPACT ON EXOPLANETARY ENVIRONMENTS

*(на английском языке)*

Translated by Svetlana Knyazeva and Yana Poklad